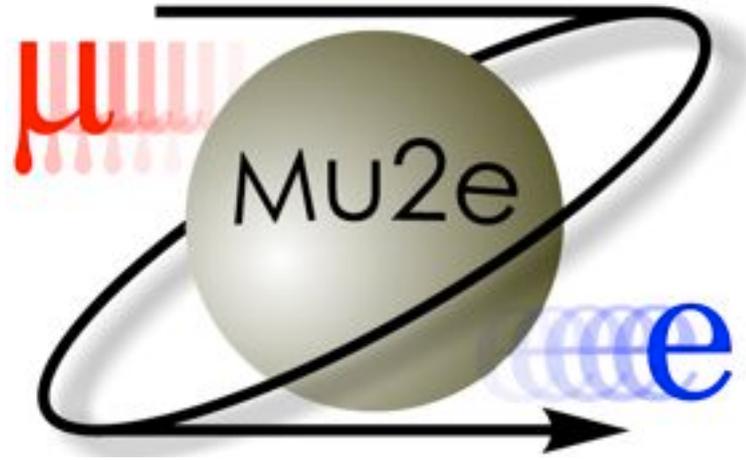

# Mu2e Technical Design Report

October 2014

Fermi National Accelerator Laboratory
Batavia, IL 60510
www.fnal.gov

**Managed by**
Fermi Research Alliance, FRA
For the United States Department of Energy under
Contract No. DE-AC02-07-CH-11359

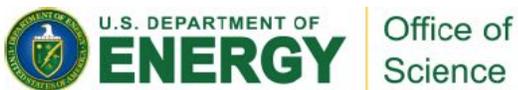
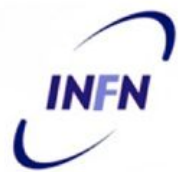

**Fermilab**
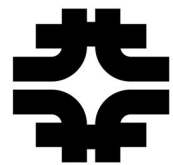

## DISCLAIMER

# Mu2e

L. Bartoszek
*Bartoszek Engineering, Aurora IL*

E. Barnes, J.P. Miller, J. Mott, A. Palladino, J. Quirk, B.L. Roberts
*Boston University, Boston Massachusetts*

J. Crnkovic, V. Polychronakos, V. Tishchenko, P. Yamin
*Brookhaven National Laboratory, Upton, New York*

C.-h. Cheng, B. Echenard, K. Flood, D.G. Hitlin, J.H. Kim, T.S. Miyashita, F.C. Porter,
M. Röhrken, J. Trevor, R.-Y. Zhu
*California Institute of Technology, Pasadena, California*

E. Heckmaier, T.I. Kang, G. Lim, W. Molzon, Z. You
*University of California, Irvine California*

A.M. Artikov, J.A. Budagov, Yu.I. Davydov, V.V. Glagolev,
A.V. Simonenko, Z.U. Usubov
*Joint Institute for Nuclear Research, Dubna, Russia*

S.H. Oh, C. Wang
*Duke University, Durham, North Carolina*

G. Ambrosio, N. Andreev, D. Arnold, M. Ball, R.H. Bernstein, A. Bianchi, K. Biery, R. Bossert,
M. Bowden, J. Brandt, G. Brown, H. Brown, M. Buehler, M. Campbell, S. Cheban, M. Chen,
J. Coghill, R. Coleman, C. Crowley, A. Deshpande, G. Deuerling, J. Dey, N. Dhanaraj,
M. Dinnon, S. Dixon, B. Drendel, N. Eddy, R. Evans, D. Evbota, J. Fagan, S. Feher, B. Fellenz,
H. Friedsam, G. Gallo, A. Gaponenko, M. Gardner, S. Gaugel, K. Genser, G. Ginther, H. Glass,
D. Glenzinski, D. Hahn, S. Hansen, B. Hartsell, S. Hays, J.A. Hocker, E. Huedem, D. Huffman,
A. Ibrahim, C. Johnstone, V. Kashikhin, V.V. Kashikhin, P. Kasper, T. Kiper, D. Knapp,
K. Knoepfel, L. Kokoska, M. Kozlovsky, G. Krafczyk, M. Kramp, S. Krave, K. Krempetz,
R.K. Kutschke, R. Kwarciany, T. Lackowski, M.J. Lamm, M. Larwill, F. Leavell, D. Leeb,
A. Leveling, D. Lincoln, V. Logashenko, V. Lombardo, M.L. Lopes, A. Makulski, A. Martinez,
D. McArthur, F. McConologue, L. Michelotti, N. Mokhov, J. Morgan, A. Mukherjee, P. Murat,
V. Nagaslaev, D.V. Neuffer, T. Nicol, J. Niehoff, J. Nogiec, M. Olson, D. Orris, R. Ostojic,
T. Page, C. Park, T. Peterson, R. Pilipenko, A. Pla-Dalmau, V. Poloubotko, M. Popovic,
E. Prebys, P. Prieto, V. Pronskikh, D. Pushka, R. Rabehl, R.E. Ray†, R. Rechenmacher, R.
Rivera, W. Robotham, P. Rubinov, V.L. Rusu, V. Scarpine, W. Schappert, D. Schoo,

A. Stefanik, D. Still, Z. Tang, N. Tanovic, M. Tartaglia, G. Tassotto, D. Tinsley, R.S. Tschirhart, G. Vogel, R. Wagner, R. Wands, M. Wang, S. Werkema, H.B. White Jr., J. Whitmore, R. Wielgos, R. Woods, C. Worel, R. Zifko
*Fermi National Accelerator Laboratory, Batavia, Illinois*

P. Ciambrone, F. Colao, M. Cordelli, G. Corradi, E. Dane`, S. Giovannella, F. Happacher, A. Luca`, S. Miscetti, B. Ponzio, G. Pileggi, A. Saputi, I. Sarra, R.S. Soleti, V. Stomaci
*Laboratori Nazionali di Frascati of INFN, Frascati, Italy*

M. Martini
*Laboratori Nazionali di Frascati of INFN, Frascati and Universita' Marconi, Roma, Italy*

P. Fabbricatore, S. Farinon, R. Musenich
*Istituto Nazionale di Fisica Nucleare, Genova, Italy*

D. Alexander, A. Daniel, A. Empl, E.V. Hungerford, K. Lau
*University of Houston, Texas*

G.D. Gollin, C. Huang, D. Roderick, B. Trundy
*University of Illinois, Urbana-Champaign, Illinois*

D.N. Brown, D. Ding, Yu.G. Kolomensky, M.J. Lee
*Lawrence Berkeley National Laboratory and University of California, Berkeley*

M. Cascella, F. Grancagnolo, F. Ignatov, A. Innocente, A. L'Erario, A. Miccoli, A. Maffezzoli, P. Mazzotta, G. Onorato, G.M. Piacentino, S. Rella, F. Rossetti, M. Spedicato, G. Tassielli, A. Taurino, G. Zavarise
*Istituto Nazionale di Fisica Nucleare and Università del Salento, Lecce, Italy*

R. Hooper
*Lewis University, Romeoville, Illinois*

D.N. Brown
*University of Louisville, Louisville KY*

R. Djilkibaev, V. Matushko
*Institute for Nuclear Research, Moscow, Russia*

C. Ankenbrandt
*Muons Inc., Batavia, Illinois*

S. Boi, A. Dychkant, D. Hedin, Z. Hodge, V. Khalatian, R. Majewski, L. Martin,
U. Okafor, N. Pohlman, R.S. Riddel, A. Shellito
*Northern Illinois University, DeKalb, Illinois*

A.L. de Gouvea
*Northwestern University, Evanston, Illinois*

F. Cervelli, R. Carosi, S. Di Falco, S. Donati, T. Lomtadze, G. Pezzullo, L. Ristori, F. Spinella
*Istituto Nazionale di Fisica Nucleare, Pisa, Italy*

M. Jones
*Purdue University, Lafayette, Indiana*

M.D. Corcoran, J. Orduna, D. Rivera
*Rice University, Houston, Texas*

R. Bennett, O. Caretta, T. Davenne, C. Densham, P. Loveridge, J. Odell
*Rutherford Appelton Laboratory, Oxfordshire, U.K.*

R. Bomgardner, E.C. Dukes, R. Ehrlich, M. Frank, S. Goadhouse, R. Group, E. Ho, H. Ma,
Y. Oksuzian, J. Purvis, Y. Wu
*University of Virginia, Charlottesville, Virginia*

D.W. Hertzog, P. Kammel
*University of Washington, Seattle*

K.R. Lynch, J.L. Popp
*York College of the City University of New York*

†Contact: rray@fnal.gov

This page intentionally left blank

This page intentionally left blank

# 1    Executive Summary

## 1.1    Introduction

Fermi National Accelerator Laboratory and the Mu2e Collaboration, composed of about 155 scientists and engineers from 28 universities and laboratories around the world, have collaborated to create this technical design for a new facility to study charged lepton flavor violation using the existing Department of Energy investment in the Fermilab accelerator complex.

Mu2e proposes to measure the ratio of the rate of the neutrinoless, coherent conversion of muons into electrons in the field of a nucleus, relative to the rate of ordinary muon capture on the nucleus:

$$R_{\mu e} = \frac{\mu^- + A(Z,N) \rightarrow e^- + A(Z,N)}{\mu^- + A(Z,N) \rightarrow \nu_\mu + A(Z-1,N)}.$$

The conversion process is an example of charged lepton flavor violation (CLFV), a process that has never been observed experimentally. The significant motivation behind the search for muon-to-electron conversion is discussed in Chapter 3. The current best experimental limit on muon-to-electron conversion, $R_{\mu e} < 7 \times 10^{-13}$ (90% CL), is from the SINDRUM II experiment [1]. With $3.6 \times 10^{20}$ delivered protons Mu2e will probe four orders of magnitude beyond the SINDRUM II sensitivity, measuring $R_{\mu e}$ with a single event sensitivity of $2.87 \times 10^{-17}$. Observation of this process would provide unambiguous evidence for physics beyond the Standard Model and can help to illuminate discoveries made at the LHC or point to new physics beyond the reach of the LHC.

The conversion of a muon to an electron in the field of a nucleus occurs coherently, resulting in a monoenergetic electron near the muon rest energy that recoils off of the nucleus in a two-body interaction. This distinctive signature has several experimental advantages including the near-absence of background from accidentals and the suppression of background electrons near the conversion energy from muon decays.

At the proposed Mu2e sensitivity there are a number of processes that can mimic a muon-to-electron conversion signal. Controlling these potential backgrounds drives the overall design of Mu2e. These backgrounds result principally from five sources:

1.  Intrinsic processes that scale with beam intensity and include muon decay in orbit (DIO) and radiative muon capture (RMC).
2.  Processes that are delayed because of particles that spiral slowly down the muon

  beamline, such as antiprotons.

3. Prompt processes where the detected electron is nearly coincident in time with the arrival of a beam particle at the muon stopping target.
4. Processes that mimic conversion electrons that are initiated by cosmic rays.
5. Events that result from reconstruction errors induced by additional activity in the detector from conventional processes.

A general description of these backgrounds can be found in Section 3.2 and a detailed description combined with estimates of background rates in Mu2e can be found in Section 3.6.

## 1.2 Scope

To achieve the sensitivity goal cited above a high intensity, low energy muon beam coupled with a detector capable of efficiently identifying 105 MeV electrons while minimizing background from conventional processes will be required. The muon beam is created by an 8 GeV proton beam striking a production target and a system of superconducting solenoids that efficiently collect pions and transport their daughter muons to a stopping target. The scope of work required to meet the scientific and technical objectives of Mu2e is listed below.

- Modify the accelerator complex to transfer 8 GeV protons from the Fermilab Booster to the Mu2e detector while the 120 GeV neutrino program is operating. To accomplish this the existing Recycler and Debuncher Rings will be modified to re-bunch batches of protons from the Booster and then slow extract beam to the Mu2e detector.
- Design and construct a new beamline from the Debuncher Ring to the Mu2e detector. The beamline includes an *extinction insert* that removes residual out-of-time protons.
- Design and construct the Mu2e superconducting solenoid system (Figure 1.1) consisting of a *Production Solenoid* that contains the target for the primary proton beam, an S-shaped *Transport Solenoid* that serves as a magnetic channel for pions and muons of the correct charge and momentum range and a *Detector Solenoid* that houses the muon stopping target and the detector elements.
- Design and construct the Mu2e detector (Figure 1.1) consisting of a tracker, a calorimeter, a stopping target monitor, a cosmic ray veto, an extinction monitor and the electronics, trigger and data acquisition required to read out, select and store the data. The tracker accurately measures the trajectory of charged particles, the calorimeter provides independent measurements of energy, position and time, the cosmic ray veto identifies cosmic ray muons traversing the detector region that can cause backgrounds and the extinction monitor detects scattered protons from the production target to monitor the fraction of out-of-time beam.

- Design and construct a facility to house the Mu2e detector and the associated infrastructure (see Figure 1.2). This includes an underground detector enclosure and a surface building to house necessary equipment and infrastructure that can be accessed while beam is being delivered to the detector.

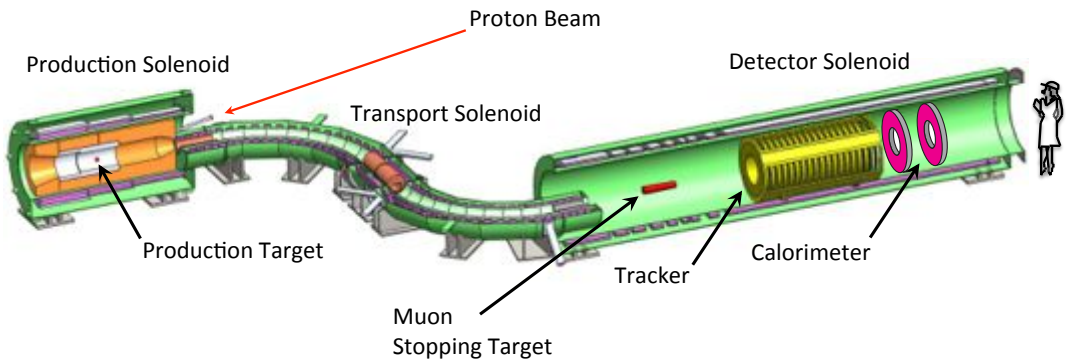

Figure 1.1. The Mu2e Detector.  The cosmic ray veto that surrounds the Detector Solenoid is not shown.

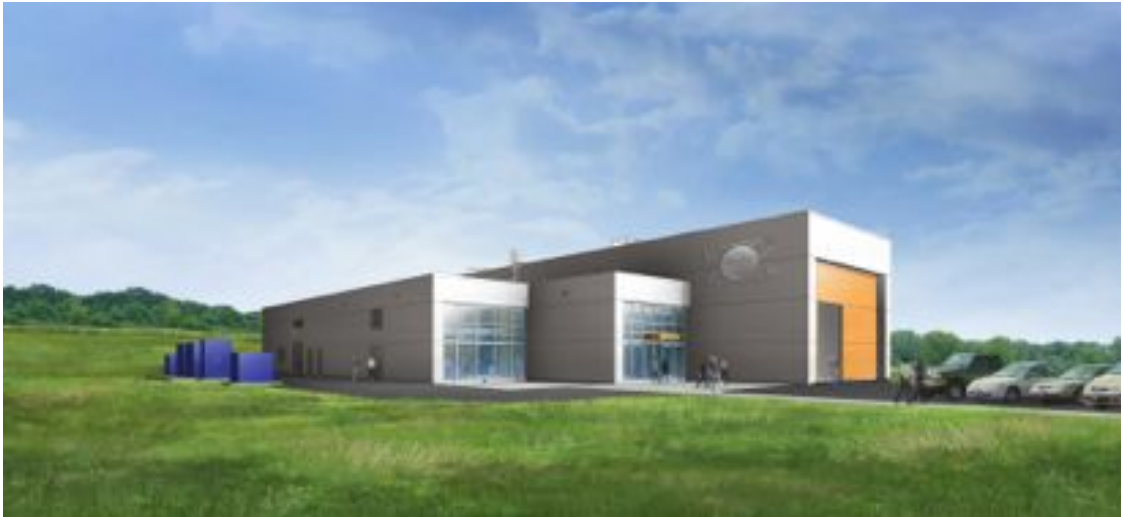

Figure 1.2. Depiction of the above-grade portion of the Mu2e facility.

Mu2e is integrated into Fermilab's overall science program that includes many experiments that use the same machines and facilities, though often in different ways. Because of the overlapping needs of several experimental programs, the scope of work described above will be accomplished through a variety of mechanisms.  The NOvA and g-2 experiments both require upgrades to the Recycler Ring that will be used by Mu2e. Infrastructure required by both Mu2e and g-2 will be funded as common Accelerator Improvement Projects (AIPs) and General Plant Projects (GPPs). These common projects will be managed by Fermilab to ensure completion on a time scale consistent with the Lab's overall program plan and to guarantee that the needs of the overall program are satisfied.

## 1.3  Cost and Schedule

The total project cost for Mu2e is $271M. This includes the base cost plus contingency and overhead, escalated to actual year dollars. A duration of 76 months, from CD-3a to CD-4, is required for the construction phase of the Project.

## 1.4  Acquisition Strategy

The acquisition strategy relies on Fermi Research Alliance (FRA), the Department of Energy Managing and Operating (M&O) contractor for Fermi National Accelerator Laboratory (Fermilab), to directly manage the Mu2e acquisition. The design, fabrication, assembly, installation, testing and commissioning for the Mu2e Project will be performed by the Mu2e Project scientific and technical staff provided by Fermilab and the various Mu2e collaborating institutions. Much of the subcontracted work to be performed for Mu2e consists of hardware fabrication and conventional facilities construction.

# 2    Project Overview

## 2.1    Project Mission

The primary mission of the Mu2e Project is to design and construct a facility that will enable the most sensitive search ever made for the coherent conversion of muons into electrons in the field of a nucleus, an example of Charged Lepton Flavor Violation (CLFV). Mu2e proposes to measure the ratio of the rate of the neutrinoless, coherent conversion of muons into electrons in the field of a nucleus, relative to the rate of ordinary muon capture on the nucleus:

$$R_{\mu e} = \frac{\mu^- + A(Z,N) \rightarrow e^- + A(Z,N)}{\mu^- + A(Z,N) \rightarrow \nu_\mu + A(Z-1,N)}.$$

The signature of this process is a monoenergetic electron with an energy nearly equal to the muon rest mass. The significant motivation behind the search for muon-to-electron conversion is discussed in Chapter 3. The best experimental limit on muon-to-electron conversion, $R_{\mu e} < 7 \times 10^{-13}$ (90% CL), is from the SINDRUM II experiment [1]. Mu2e intends to probe four orders of magnitude beyond the SINDRUM II sensitivity, measuring $R_{\mu e}$ with a single-event sensitivity of $2.87 \times 10^{-17}$.

To achieve this significant leap in sensitivity, Mu2e requires an intense low energy muon beam and a state-of-the-art detector capable of efficiently identifying, reconstructing and analyzing conversion electrons with momenta near 105 MeV/c.

## 2.2    Scope Required to Accomplish the Project Mission

A technical design has been developed for the Mu2e Project that meets the Mission Requirements described in Section 2.1. The scope includes the following:

- A proton beam that can produce an intense secondary muon beam with a structure that allows time for stopped muons to decay before the next pulse arrives.
- A pion capture and muon transport system that efficiently captures charged pions and transports negatively charged secondary muons to a stopping target. The momentum spectrum of the transported muon beam must be low enough to ensure that a significant fraction of the muons can be brought to rest in a thin target.
- A detector that is capable of efficiently and accurately identifying and analyzing conversion electrons with momenta near 105 MeV/c while rejecting backgrounds from conventional processes and cosmic rays.
- A detector hall facility to house the experimental apparatus.

### 2.2.1   Proton Beam

Mu2e requires a high intensity, pulsed proton beam to produce an intense beam of low energy muons with the time structure required by the experiment. Protons designated for Mu2e are acquired from the Booster during the available portions of the Main Injector timeline when slip-stacking operations are underway for NOvA. Two Booster proton batches, each containing $4.0 \times 10^{12}$ protons with a kinetic energy of 8 GeV, are extracted into the MI-8 beamline and injected into the Recycler Ring. After each injection, the beam circulates for 90 msec while a 2.5 MHz bunch formation RF sequence is performed. This RF manipulation coalesces each proton batch into four 2.5 MHz bunches occupying one seventh of the circumference of the Recycler Ring. Each of these bunches will be synchronously transferred, one at-a-time, through existing transfer lines to the Delivery Ring, where the beam is held in a 2.4 MHz RF bucket during resonant extraction to the experiment through a new external beamline. To help control the spill rate uniformity during resonant extraction a technique known as RF knockout will be used. RF knockout will allow for fast transverse heating of the beam. It will also serve as a feedback tool for fine control of the spill rate. The resonant extraction process will not completely remove the entire beam, so what remains must be disposed of in a controlled way. Therefore, a beam abort system will be required for the Delivery Ring to "clean up" beam that remains after resonant extraction is complete. The resonant extraction system will inject $\sim 3 \times 10^7$ protons into the external beamline every 1.7 μs (the revolution period of the Delivery Ring). An *extinction system*, in the form of a high frequency *AC dipole* (see Section 4.9), is required to suppress unwanted beam between successive pulses that can generate experimental backgrounds (see Section 3.6). After transiting the extinction system the proton pulses are delivered to the production target located in the evacuated warm bore of a high-field superconducting solenoid. The proton beam will have a transverse radius of about 1 mm (rms) and will be about 250 ns in duration. The proton beam deflects in the magnetic field of the solenoid before striking the production target, complicating the final focus beamline optics and steering. The production target is a radiatively cooled tungsten rod about the size and shape of a pencil. Not all of the proton beam interacts in the production target. The unspent beam is absorbed in an air-cooled beam absorber downstream of the production target. A monitor, located above the beam absorber, will measure scattered protons as a function of time to provide a statistical measure of the residual beam between pulses that traverses the extinction system. The proton delivery scheme is shown in Figure 2.1. The Mu2e proton beam requirements are described in [2].

Two Booster batches can be sequentially processed as described above during the part of the 1.33 second Main Injector cycle when the Recycler is not being used by NOvA. This corresponds to $8 \times 10^{12}$ protons per cycle for an average of $6 \times 10^{12}$ protons per second and a total of $1.2 \times 10^{20}$ protons per year ($2 \times 10^7$ sec.).

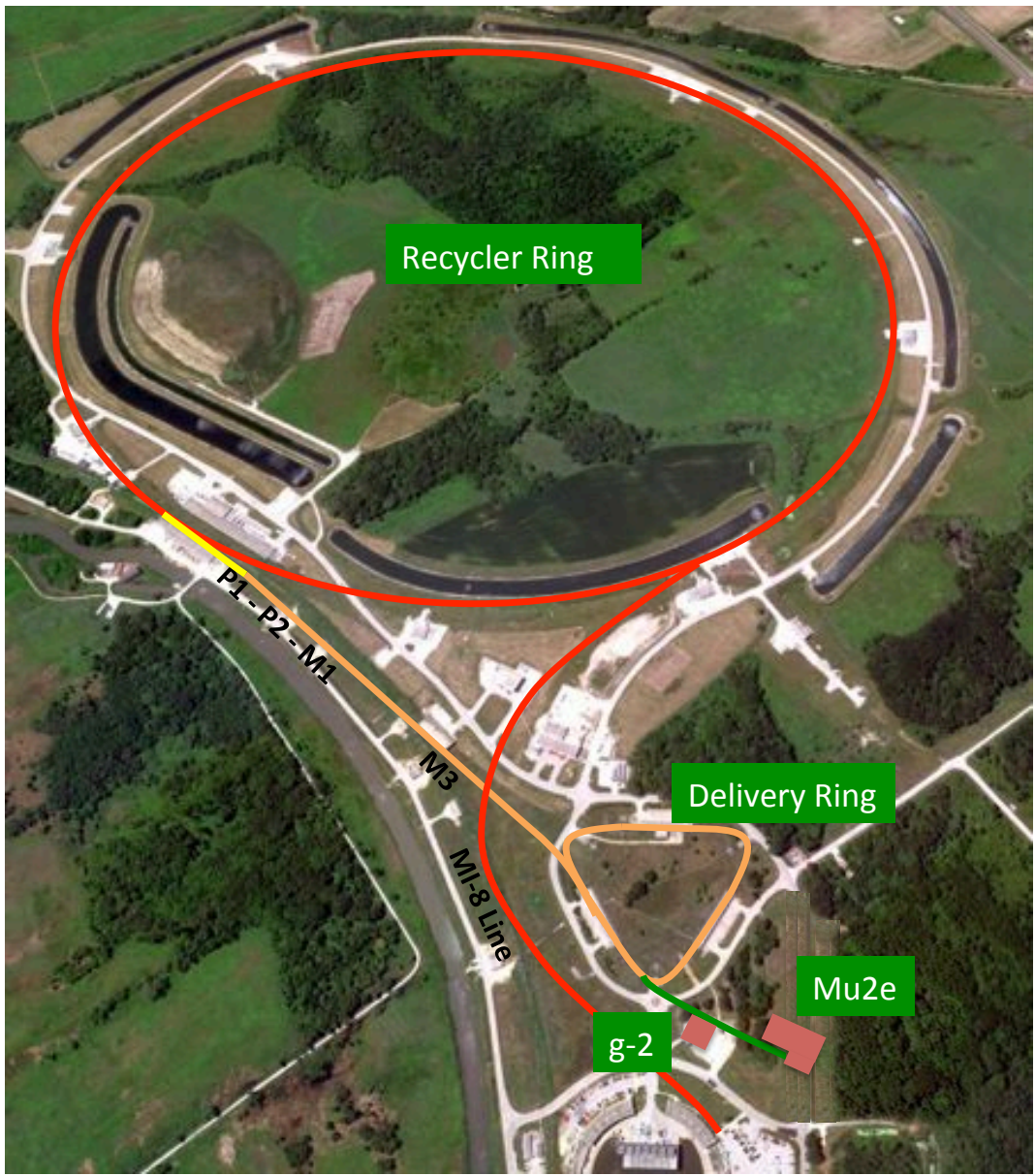

Figure 2.1. Layout of the Mu2e facility (lower right) relative to the accelerator complex that provides proton beam to the detector. Protons are transported from the Booster through the MI-8 beamline to the Recycler Ring where they will circulate while they are re-bunched by a 2.5 MHz RF system. The reformatted bunches are kicked into the P1 line and transported to the Delivery Ring where they are slow extracted to the Mu2e detector through a new external beamline.

Most of the infrastructure required to deliver proton beam to the Mu2e production target already exists or will exist before Mu2e needs it. The g-2 experiment, scheduled to take data before Mu2e, requires much of the same infrastructure. To satisfy the common needs of both projects a program to develop a Muon Campus through a series of Accelerator Improvement Plans (AIP) and General Plant Projects (GPP) has been initiated. The accelerator infrastructure required exclusively by Mu2e is part of the Mu2e Project and includes:

- Resonant Extraction System
- MHz Delivery Ring RF system
- Mu2e external beamline
- Extinction System
- Extinction Monitor System
- Production Target
- Radiation Safety and Shielding
- Beamline instrumentation and controls
- Diagnostic Beam Absorber
- Proton Target Beam Absorber.

These elements are described in detail in this Technical Design Report.

### 2.2.2    Superconducting Solenoids

The Mu2e superconducting solenoid system performs a number of essential functions that enable execution of the experiment, including

- Capture of pions from the production target
- Formation of the secondary muon beam
- Background rejection by shifting the pitch of high energy particles in the muon beamline before they reach the Tracker
- Provision of a relatively uniform field for momentum analysis of conversion electrons.

The solenoid system is divided into 3 functional units that have to operate as a single, integrated magnetic system, shown with their associated infrastructure in Figure 2.2. The magnetic field specifications for the solenoids (see Section 6.2) are derived from the Mu2e physics requirements and define a very specific configuration. The fringe field from one solenoid impacts the magnetic field in adjacent solenoids and there are significant forces between magnets, so the solenoids have to be designed as a system even though they will be constructed independently.

#### Production Solenoid

The *Production Solenoid*, shown in Figure 2.3, is a high field magnet with a graded solenoidal field varying smoothly from 4.6 Tesla to 2.5 Tesla. The gradient will be formed by 3 axial coils with a decreasing number of windings, made of aluminum stabilized NbTi. The solenoid is approximately 4 m long with an inner bore diameter of approximately 1.5 m that is evacuated to $10^{-5}$ Torr. The Production Solenoid is designed to capture pions and the muons into which they decay and guide them downstream to the Transport Solenoid. This process is initiated by 8 GeV protons striking a production target near the center of the Production Solenoid. A heat and radiation shield, constructed from bronze, will line the inside of the Production Solenoid to limit the heat load in the cold mass from secondaries produced in the production target and to limit radiation damage to the superconducting cable.

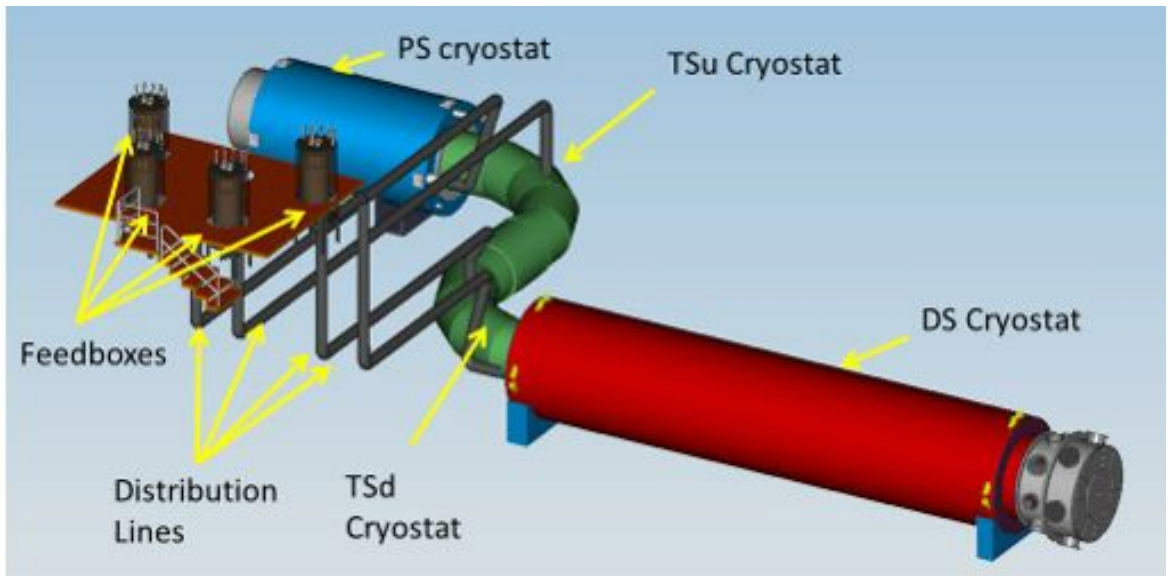

Figure 2.2. The Mu2e superconducting solenoid system, including the Production, Transport and Detector Solenoids and the cryogenic distribution system. Not shown are the power supply and quench protection systems.

Protons enter the Production Solenoid through a small port on the low field side of the solenoid before intercepting the production target. Remnant protons that are not absorbed by the target and very forward-produced secondary particles exit at the high field end of the solenoid. Pions in the forward direction with angles greater than ~30°, relative to the solenoid axis, are reflected back by the higher field and move along with the backward produced particles in helical trajectories towards the Transport Solenoid.

The Production Solenoid must generate an axially graded field varying smoothly from 4.6 Tesla to 2.5 Tesla. This axial field change is accomplished using three solenoid coils with 3, 2 and 2 layers of high-current, low-inductance aluminum-stabilized NbTi cable that allows for efficient energy extraction during a quench, requires fewer layers to achieve the required field strength and minimizes thermal barriers between the conductor and cooling channels. Aluminum stabilizer is used for several reasons. Nuclear heating from the large flux of secondaries produced in the production target is reduced in aluminum compared to copper stabilizer, the other alternative. Aluminum is less dense than copper, reducing the weight of the Production Solenoid. Aluminum can also be annealed at room temperature to reverse the impact of atomic displacements, primarily from neutrons, that degrade performance over time. This is further described in the section that follows.

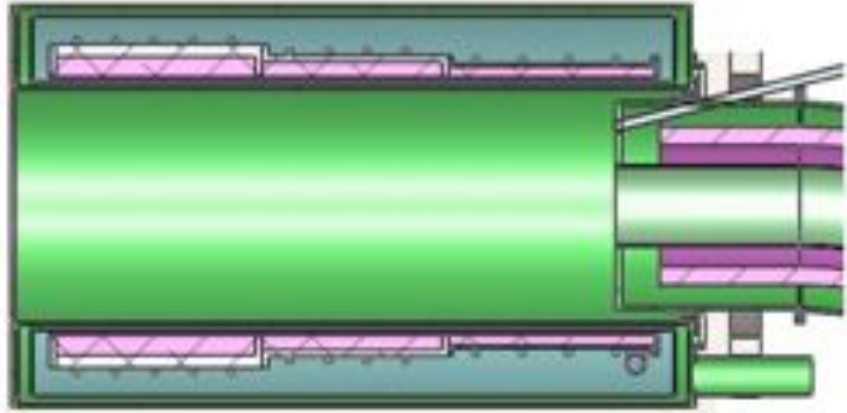

Figure 2.3. Plan view of the Mu2e Production Solenoid. The beam tube for the incoming proton beam is shown in the upper right.

### Heat and Radiation Shield

Lining the inside of the Production Solenoid warm bore is a heat and radiation shield designed to protect the solenoid's superconducting coils. The Heat and Radiation Shield is designed to limit the heat load in the cold mass to prevent quenching, limit radiation damage to superconductor insulation and epoxy and limit the damage to the superconductor's aluminum stabilizer. The shield is constructed primarily from bronze. Because the proton beam is incident from one side of the Production Solenoid, the pattern of energy deposition in the heat shield is asymmetric with the largest depositions being near the target and collinear with the incoming proton beam direction. Even with the protection of the Heat and Radiation Shield, a significant number of atomic displacements will occur over time in the aluminum stabilizer surrounding the superconductor. The Residual Resistivity Ratio (RRR) of the aluminum, the ratio of the electrical resistance at room temperature of a conductor to that at 4.5 K, will decrease to the point where the stabilizer cannot adequately protect the superconductor in the event of a quench. The RRR can be completely recovered by warming the aluminum stabilizer to room temperature. Based on models of neutron production and energy deposition, it is anticipated that it will only be necessary to warm up once per year, coincident with annual accelerator shutdowns. The Heat and Radiation Shield is shown in Figure 2.4 and described in detail in Section 4.11.3.

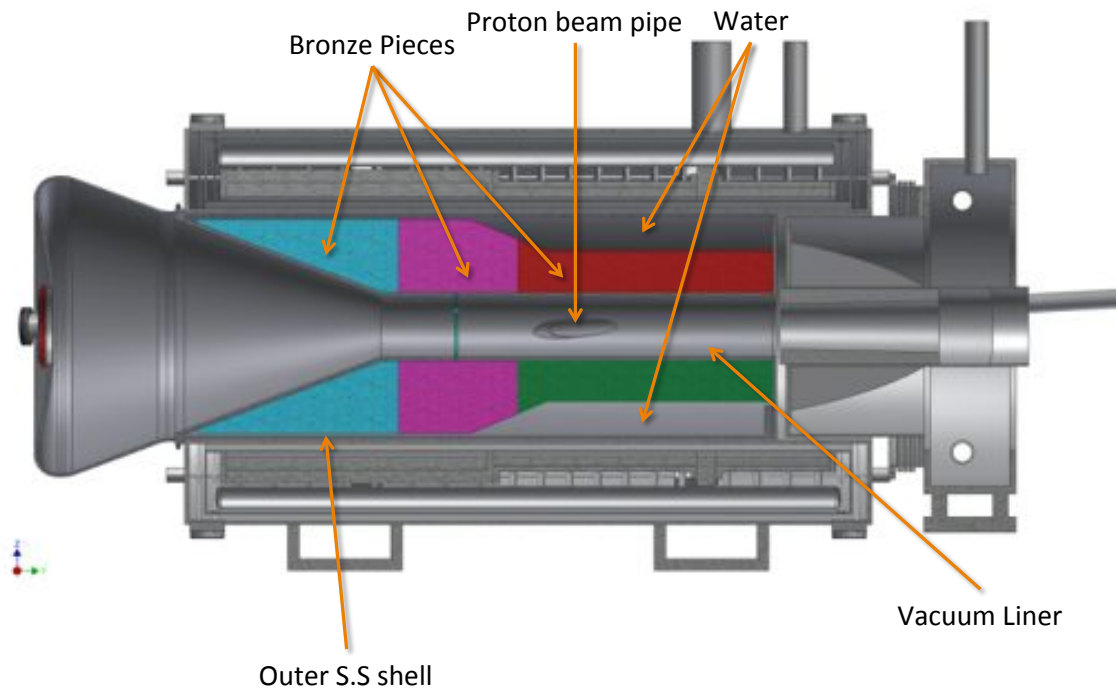

Figure 2.4. Elevation view of the Heat and Radiation Shield (HRS), designed to protect the Production Solenoid from secondaries produced in the production target. The HRS is constructed from three large pieces of forged bronze and cooled by water flowing around its periphery. Beam enters from the right.

***Transport Solenoid***

The S-shaped *Transport Solenoid* consists of a set of superconducting solenoids and toroids that form a magnetic channel that efficiently transmits low energy negatively charged muons from the Production Solenoid to the Detector Solenoid. Negatively charged particles with high energy, positively charged particles and line-of-sight neutral particles are nearly all eliminated by absorbers and collimators before reaching the Detector Solenoid. Selection of negatively charged muons is accomplished by taking advantage of the fact that a charged particle beam traversing a toroid will drift perpendicular to the toroid axis, with positives and negatives drifting in opposite directions. Most of the positively charged particles are absorbed in the central collimator. The Transport Solenoid consists of five distinct regions: a 1 m long straight section, a 90° curved section, a second straight section about 2 m long, a second 90° curved section that brings the beam back to its original direction, and a third straight section of 1 m length. The major radius of the two curved sections is about 3 m and the resulting total magnetic length of the Transport Solenoid along its axis is about 13 m. The inner warm bore of the Transport Solenoid cryostat has a diameter of about 0.5 m. The Transport Solenoid is shown in Figure 2.5.

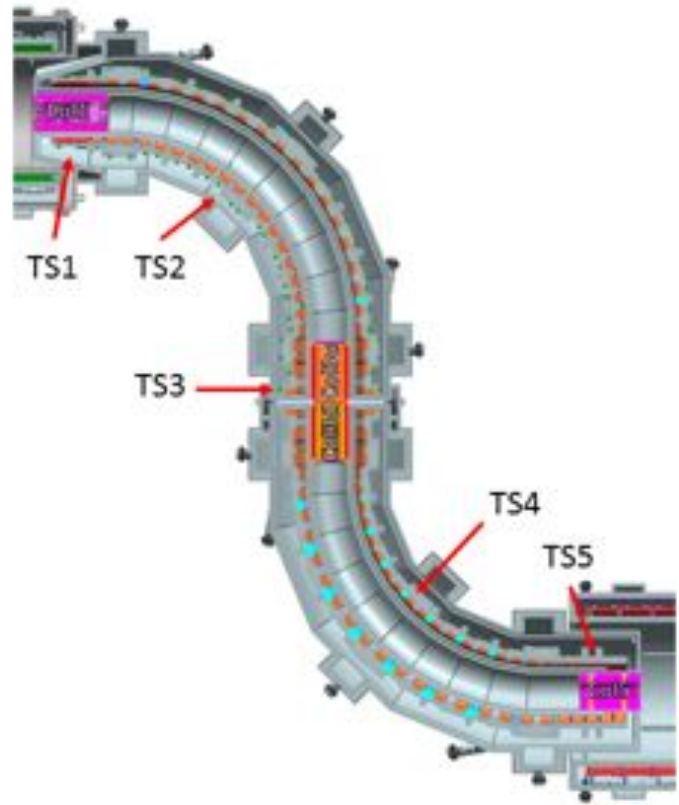

Figure 2.5. The Transport Solenoid consists of a set of superconducting solenoids and toroids that form a magnetic channel that efficiently transmits low energy negatively charged muons from the Production Solenoid.

Late arriving particles are a serious potential background for Mu2e (Section 3.6). To minimize the transport of particles that spend a long time in the magnet system the magnetic field in the straight sections is designed to always have a negative gradient that accelerates particles from the Production Solenoid through the Detector Solenoid. This eliminates traps, where particles bounce between local maxima in the field until they eventually scatter out and travel to the Detector Solenoid where they arrive late compared to the beam pulse. The requirement on a negative gradient is relaxed in the curved sections of the TS because trapped particles will eventually drift vertically out of the clear bore and be absorbed by surrounding material.

### Detector Solenoid

The Detector Solenoid is a large, low field magnet that houses the muon stopping target and the components required to identify and analyze conversion electrons from the stopping target. It is nearly 11 m long with a clear bore diameter of about 2 m. The muon stopping target resides in a graded field that varies from 2 Tesla to 1 Tesla. The graded field captures conversion electrons that are emitted in the direction opposite the detector components causing them to reflect back towards the detector. The graded field also plays an important role in reducing background from high energy electrons that are transported to the Detector Solenoid by steadily increasing their pitch as they are accelerated towards the downstream detectors. The resulting pitch angle of these beam electrons is inconsistent with the pitch of a conversion electron from the stopping target. The actual detector components reside in a field region that is relatively uniform. The inner bore of the Detector Solenoid is evacuated to $10^{-4}$ Torr to limit backgrounds from muons that might stop on gas atoms. The graded and uniform field sections of the Detector Solenoid are wound on separate mandrels but housed in a common cryostat. The conductor is aluminum stabilized NbTi. The gradient is achieved by introducing spacers to effectively change the winding density of the superconducting cable. The Detector Solenoid is shown in Figure 2.6.

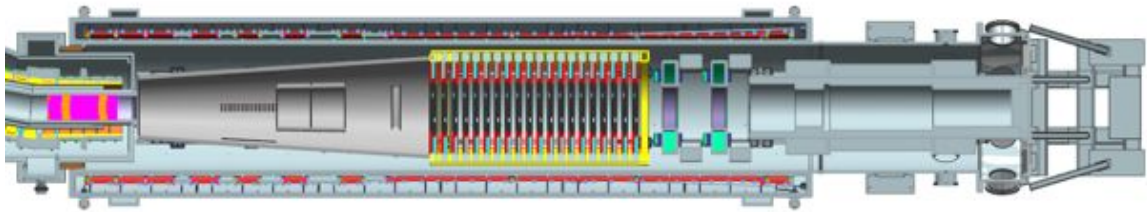

Figure 2.6. The Detector Solenoid is a large, low field magnet that houses the muon stopping target and the components required to identify and analyze conversion electrons from the stopping target.

The solenoids are the cost and schedule driver for the Project. The Production and Detector Solenoids will be constructed in industry. The relatively unique Transport Solenoid will be designed and fabricated at Fermilab, though many of the components (superconducting cable, cryostats, etc.) will be procured from industry. The make-buy decisions are based on the similarity of the Production and Detector Solenoids to other solenoids fabricated in industry and to the limited availability of resources at Fermilab. The superconducting cable required for the solenoids are long-lead items that must be procured early.

Significant infrastructure is required to support the operation of the solenoids. This includes power, quench protection, cryogens (liquid nitrogen and liquid helium), control and safety systems as well as mechanical supports to resist the significant magnetic forces on the magnets.

### *2.2.3*   **Secondary Muon Beam**

To reach the required experimental sensitivity Mu2e requires a significant number of negatively charged muons to be stopped in a thin target. To efficiently transport muons, minimize scattering off of residual gas molecules, minimize multiple scattering of conversion electrons and prevent electrical discharge from detector high voltage the Muon Beamline must be evacuated to the level of at least $10^{-4}$ Torr. In the Production Solenoid the vacuum must be maintained to better than $10^{-5}$ Torr to minimize oxidation of the tungsten production target. Tungsten is prone to oxidation at elevated temperatures.

The muon stopping target must be massive enough to stop a significant fraction of the incident muon beam but not so massive that it corrupts the momentum measurement of conversion electrons that emerge. Lower energy muons allow for a thinner target to help alleviate these concerns. The momentum distribution of muons at the Mu2e stopping target is shown in Figure 2.7. The number of muons that reach and stop in the stopping target depends on a number of factors. These include the proton beam energy, the magnetic field in the Production and Transport Solenoids, the clear bore of the solenoids the design of the collimators, the stopping target material and geometry.

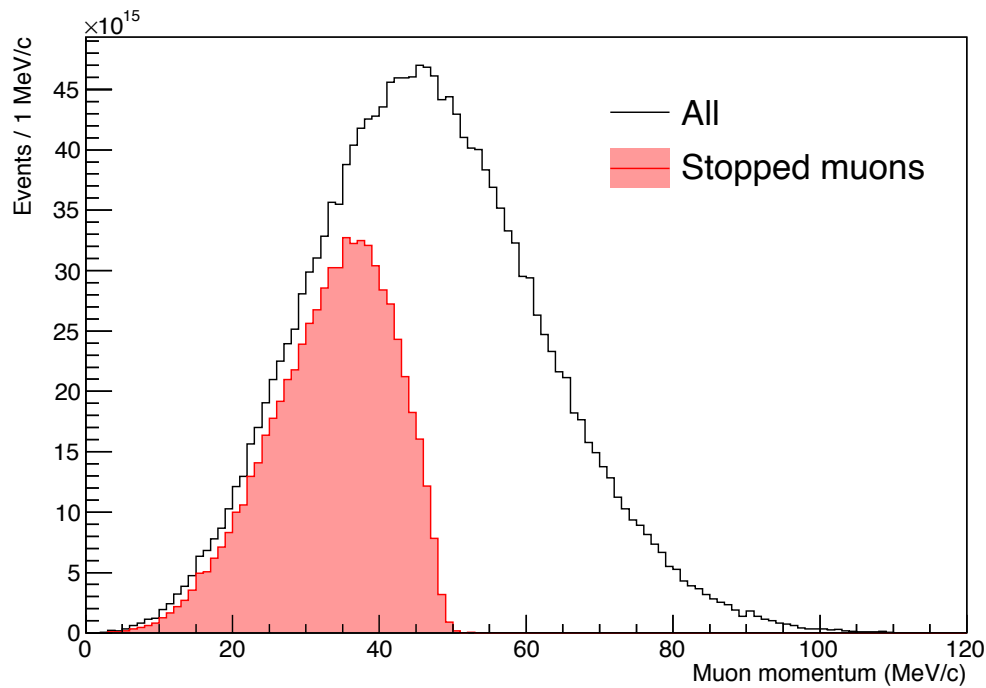

Figure 2.7. Momentum distribution of muons delivered to the stopping target as well as the distribution of muons that stop in the target.

Protons enter the Production Solenoid, a relatively high field solenoid with an axial field that varies from 4.6 Tesla to 2.5 Tesla, through a small port on the low field side of the solenoid before intercepting the production target, mounted in the evacuated warm bore of the Production Solenoid. Remnant protons that are not absorbed by the target and very forward-produced secondary particles exit at the high field end of the solenoid. The remainder of the charged particles, which are primarily pions, are reflected back by the higher field and move in helical trajectories towards the Transport Solenoid. The size and shape of the production target, the target supports and the clearance inside the Production

Solenoid warm bore have all been designed to maximize the yield of stopped muons. The production target is made from a high Z material (tungsten) to maximize pion production while the geometry is designed to minimize pion reabsorption. The target supports are designed with a small physical profile to minimize scattering and absorption of pions and muons and the diameter of the warm bore of the Production Solenoid is large enough to allow pions and muons within the acceptance of the Transport Solenoid to pass through unobstructed.

To optimize the number of stopped muons a detailed simulation package with an accurate particle production model is required. The calculated values of particle fluxes in the secondary muon beam are based on GEANT4 simulations of proton interactions in a tungsten target. GEANT4 has a variety of hadron interaction codes and the cross sections and kinematic distributions can vary significantly between them. In order to reduce exposure to the uncertainty in the hadronic models of low energy hadron production, the results from GEANT4 have been normalized to data from the HARP experiment [3]. HARP measured the double differential cross-section for production of charged pions emitted at large production angles in proton-tantalum collisions at 8 GeV/c. The data from HARP does not cover the full kinematic range required for Mu2e. To cover the full range required for Mu2e the QGSP-BERT hadronic model [4] is used. QGSP-BERT and HARP are consistent in the region where they overlap. As a crosscheck, the production model is compared to the results from a Novosibirsk experiment [5] where measurements of pion production are reported in 10 GeV/c proton-tantalum interactions with more coverage in the backward direction than provided by HARP. This results in 0.0019 stopped $\mu^-$ per proton on target when all of the material in the muon beamline is included. Errors on the double differential cross-section measurements by HARP are in the 10% range. The QGSP-BERT model and the difference between tungsten and tantalum introduce additional uncertainty. The overall uncertainty on the stopped muon rate is conservatively estimated to be 30%.

The Transport Solenoid is designed to maximize the stopped muon yield by efficiently focusing the charged particles created in the Production Solenoid towards the stopping target located in the Detector Solenoid. High energy negatively charged particles, positively charged particles and line-of-sight neutral particles will nearly all be eliminated by the two 90° bends combined with a series of absorbers and collimators. To minimize the transport of particles that spend a long time in the solenoid system the magnetic field in the straight sections must have a continuous negative gradient. This eliminates traps, where particles bounce between local maxima in the field until they eventually scatter out and travel to the Detector Solenoid where they arrive late compared to the beam pulse. The requirement of a negative gradient is relaxed in the curved sections of the Transport Solenoid because bouncing particles will eventually drift vertically out of the clear bore and be absorbed by surrounding material.

As the charged particle beam traverses the first curved toroid section of the Transport Solenoid it will disperse vertically, fanning out by charge and momentum as shown in Figure 2.8. A collimator with a vertically displaced aperture resides in the central straight section and performs a sign and momentum selection, resulting in a low energy, negatively charged beam. The second toroid section in the Transport Solenoid nearly undoes the vertical dispersion, returning the beam close to the solenoid axis. The beam does not return exactly to the solenoid axis because of the smaller magnetic field in the second bend resulting from the negative field gradient.

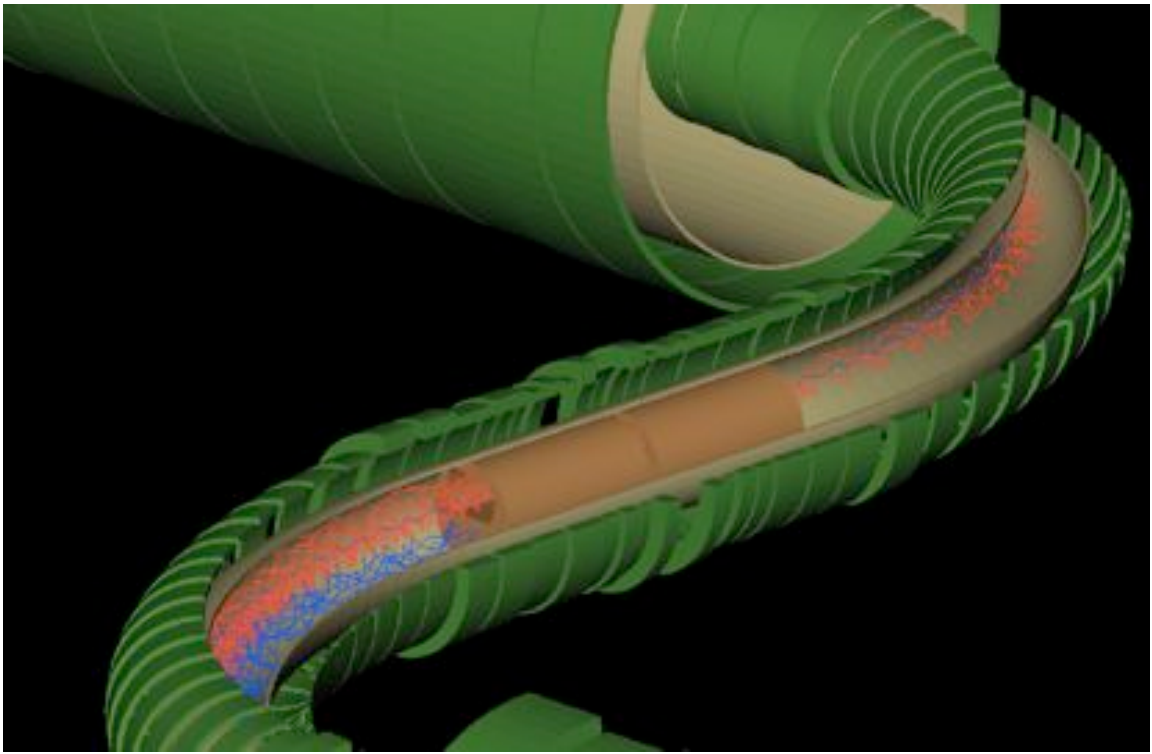

Figure 2.8. As the charged particle beam traverses the first curved toroid section of the Transport Solenoid it will disperse vertically, fanning out by charge and momentum. The central collimator absorbs the positively charged particles (blue) while allowing the negatively charged particles (red) within a particular momentum window to pass through.

The central collimator that performs momentum selection can be rotated, allowing positively charged beam to be delivered to the Detector Solenoid for purposes of calibration. Embedded in the middle of the central collimator is a thin window made of low-Z material to absorb slow moving antiprotons created in the production target. Antiprotons that reach the detector solenoid and annihilate can be a source of background. The vast majority of antiprotons have momenta below 200 MeV/c, but in rare instances antiprotons with momenta exceeding 300 MeV/c are produced, requiring a more massive window to ensure annihilation. However, this window also reduces the yield of stopped muons, so the thickness must be carefully optimized. A wedge shaped antiproton window simultaneously optimizes attenuation of antiprotons and transmission of muons by taking advantage of the correlation between momentum and vertical displacement in the Transport Solenoid (see Figure 2.9). High momentum antiprotons must penetrate more material while the low momentum muons that are most likely to stop in the stopping target pass through the thinnest part of the wedge. The low Z window also separates the upstream and downstream vacuum volumes to prevent radioactive ions or atoms from the production target from contaminating the detector solenoid volume. Studies of antiproton production and propagation are still underway and an additional window upstream of the central collimator may also be necessary.

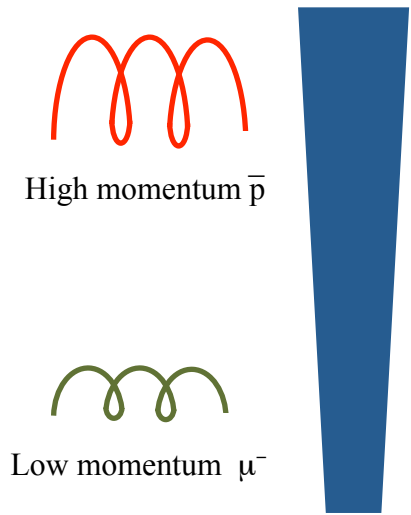

Figure 2.9. A wedge shaped antiproton window simultaneously optimizes attenuation of antiprotons and transmission of muons by taking advantage of the correlation between momentum and vertical displacement in the Transport Solenoid.

***Muon Stopping Target***
The muon stopping target consists of a series of thin aluminum discs arranged coaxially along the Detector Solenoid axis in a graded field that varies from 2 Tesla to 1 Tesla (Figure 2.10). Energy loss and straggling in the stopping target are significant contributors to the momentum resolution function. The distributed, tapered target is designed to stop as many muons as possible while minimizing the amount of material traversed by conversion electrons that are within the acceptance of the downstream tracker. The graded field captures conversion electrons that are emitted in the direction opposite the detector components causing them to reflect back towards the detector. Not all of these reflected electrons will be used in the final data sample as many of them will pass through nearby material, lose energy or scatter and fail the analysis cuts. More importantly, the graded field also shifts the pitch of beam particles that enter the Detector Solenoid and travel to the tracker, playing an important role in background suppression. Because of the diffuse nature of the muon beam a significant number of muons can strike the structure supporting the stopping target, producing electrons from muon Decays In Orbit (DIO) at large radius where the acceptance for reconstruction in the detector is high. Prompt, low-energy DIO electrons cannot contribute background to Mu2e. The DIO endpoint energy and the muon lifetime both depend on the Z of the target nucleus. The muon lifetime decreases with increasing Z as does the DIO endpoint, so high Z materials are preferred for the support materials to reduce backgrounds. For this reason, tungsten wires have been chosen for the target support.

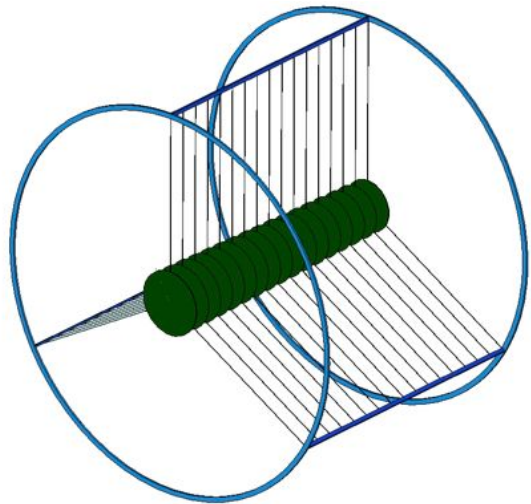

Figure 2.10. The muon stopping target and its mechanical support.

### Muon Beam Stop

Muons that do not stop in the stopping target pass through apertures in the detectors and are transported to the muon beam-stop at the downstream end of the Detector Solenoid. The Muon Beam Stop is designed to absorb the energy of muons that reach the end of the Detector Solenoid. This is required to reduce activity in the detectors from muon decays and captures in the beam stop. This is especially important during the signal measurement period that begins about 700 ns after the proton microbunch hits the production target. Near the downstream end of the Detector Solenoid the uniform magnetic field transitions to a graded field that drops off along the beam direction. The field gradient reflects most low energy charged particles produced in the beam stop away from the detectors. The muon beam stop is constructed from a combination of high-Z materials in which muons have a relatively short lifetime, so decays and captures take place well before the signal measurement period begins, and polyethylene, intended to reduce neutron rates.

### 2.2.4   The Detector

The Mu2e detector is located inside the evacuated warm bore of the Detector Solenoid in a nearly uniform 1 Tesla magnetic field and is designed to efficiently and accurately identify and analyze the helical trajectories of ~105 MeV electrons in the high-rate time-varying environment of Mu2e. The detector consists of a tracker and a calorimeter that provide redundant energy/momentum, timing, and trajectory measurements. A cosmic ray veto, consisting of both active and passive elements, surrounds the Detector Solenoid and nearly half of the Transport Solenoid.

#### Tracker

The Mu2e tracker is designed to accurately measure the trajectory of electrons in a uniform 1 Tesla magnetic field in order to determine their momenta. The limiting factor in accurately determining the trajectory of electrons is multiple scattering in the tracker. High rates in the detector may lead to errors in pattern recognition that can reduce the acceptance for signal events and possibly generate backgrounds if hits from lower energy particles combine to create accidental trajectories that are consistent with conversion electrons.  A low mass, highly segmented detector is required to minimize multiple scattering and handle the high rates.

The Mu2e tracker is a low mass array of straw drift tubes aligned transverse to the axis of the Detector Solenoid. The basic detector element is a 25 µm sense wire inside a 5 mm diameter tube made of 15 µm thick metalized Mylar®. The tracker will have ~23,000 straws distributed into 20 measurement stations across a ~3 m length. Planes are constructed from two layers of straws, as shown in Figure 2.11, to improve efficiency and help determine on which side of the sense wire a track passes (the classic "left-right" ambiguity). A 1.25 mm gap is maintained between straws to allow for manufacturing tolerance and expansion due to gas pressure. The straws are designed to withstand changes in differential pressure ranging from 0 to 1 atmosphere for operation in vacuum. The straws are supported at their ends by a ring at large radius, outside of the active detector region.  The tracker is shown in Figure 2.12.

Each straw will be instrumented on both ends with preamps and TDCs that will be used to measure the drift time to determine the distance of approach of charged tracks relative to the drift wire. The arrival time of the signal at each end of the straw will be compared in order to determine the location of the track intercept along the length of the straw. Each straw will also be instrumented with an ADC for dE/dx capability to separate electrons from highly ionizing protons. To minimize penetrations into the vacuum, digitization will be done at the detector, with readout via optical fibers. A liquid cooling system will be required for the electronics to maintain an appropriate operating temperature in vacuum.

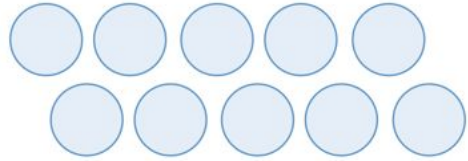

Figure 2.11 A section of a two-layer tracker straw plane. The two layers are required for full efficiency and help resolve the left-right ambiguity.

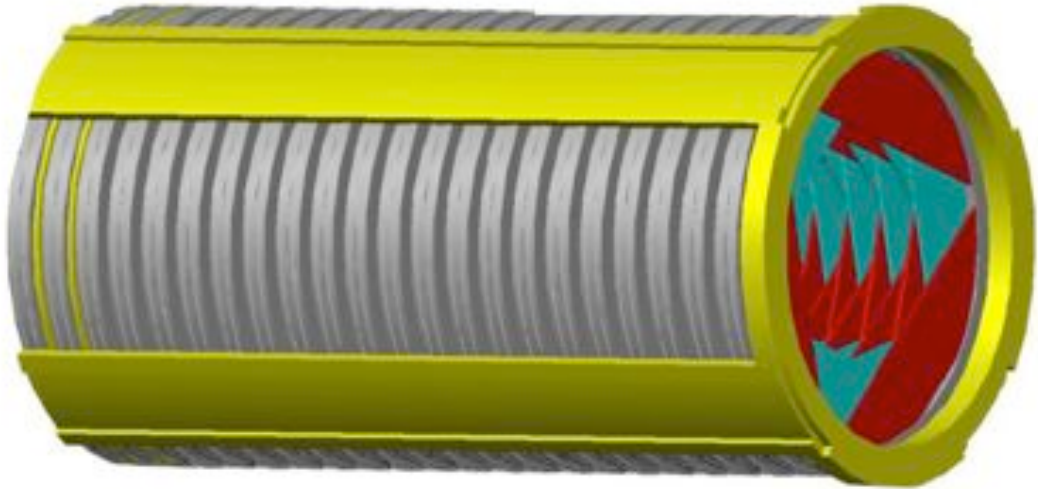

Figure 2.12 The Mu2e straw tube tracker. The straws are oriented transverse to the solenoid axis.

The tracker is designed to intercept only a small fraction of the significant flux of electrons from muon decays-in-orbit. The vast majority of electrons from muon decay in orbit are below 60 MeV in energy (Figure 3.7). Only electrons with energies greater than about 53 MeV, representing a small fraction of the rate (about 3%) will be observed in the tracker. Lower energy electrons will curl in the field of the Detector Solenoid and pass unobstructed through the hole in the center of the tracker. This is illustrated in Figure 2.13.

Tracker resolution is an important component in determining the level of several critical backgrounds. The tracker is required to have a high-side resolution of $\sigma < 180$ keV [6]. The requirement on the low side tail is less stringent since it smears background away from the signal region while a high-side tail smears background into the signal region. Current simulations indicate that the high side resolution of the Mu2e tracker can be well represented by the sum of two Gaussians. The high-side resolution, which is the most important for distinguishing conversion electrons from backgrounds, has a core component sigma of 115 KeV/c, and a significant tail sigma of 176 KeV/c. The net resolution is significantly less than the estimated resolution due to energy loss in the upstream material. The Tracker is described in detail in Chapter 8.

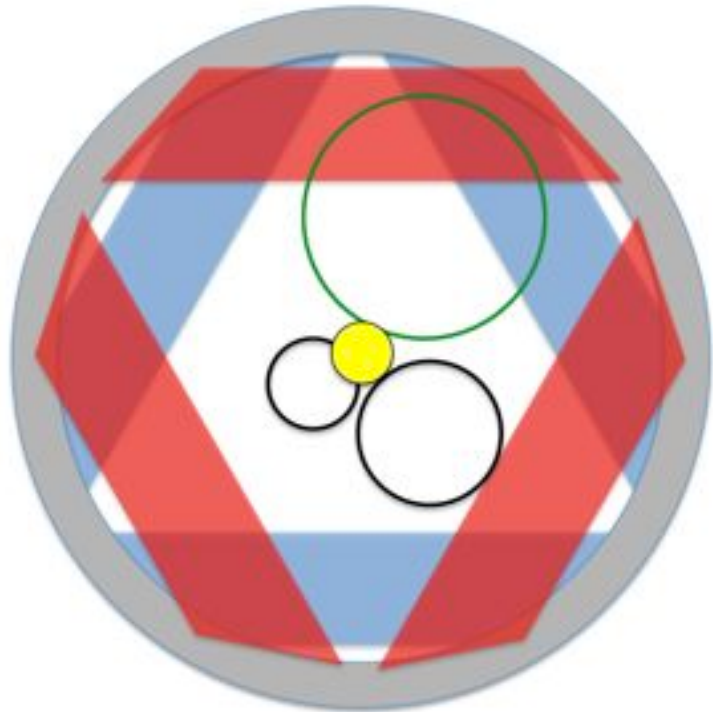

Figure 2.13. Cross sectional view of the Mu2e tracker with the trajectories of a 105 MeV conversion electron (top) and a 53 MeV Michel electron (lower right) superimposed. The disk in the center is the stopping target. Electrons with energies smaller than 53 MeV (lower left), representing most of the rate from muon decays-in-orbit, miss the tracker entirely.

*Calorimeter*

High rates of hits in the tracker may cause pattern recognition errors that add tails to the resolution function and result in background. Accidental hits can combine with or obscure hits from lower energy particles to leave behind a set of hits that might reconstruct to a trajectory consistent with a higher energy conversion electron. Extrapolating the fitted trajectory to the downstream calorimeter and comparing the calculated intercept with the measured position in the calorimeter may help to identify backgrounds that result from reconstruction errors. Another source of background is cosmic ray muons, not vetoed by the CRV system. Cosmic rays generate two distinct categories of background events: muons trapped in the magnetic field of the Detector Solenoid and electrons produced in a cosmic muon interaction with detector material. The energy and timing measurements from the Mu2e calorimeter provide information critical for efficient separation of electrons and muons in the detector (Section 9.4.1).

The calorimeter may also be used in a software or firmware trigger to reduce the volume of data-to-storage. The calorimeter consists of 1860 $BaF_2$ crystals located downstream of the tracker and arranged in two disks (Figure 2.14). The crystals are of hexagonal shape, 33 mm across flats and are 200 mm long. Each crystal is read out by two large-area APDs; solid-state photo-detectors are required because the calorimeter resides in a 1 Tesla magnetic field. Front-end electronics is mounted on the rear of each disk, while voltage distribution, slow controls and digitizer electronics are mounted behind each disk. A laser flasher system provides light to each crystal for relative calibration and monitoring purposes. A circulating liquid radioactive source system provides absolute calibration and an energy scale. The crystals are supported by a lightweight carbon fiber support structure.

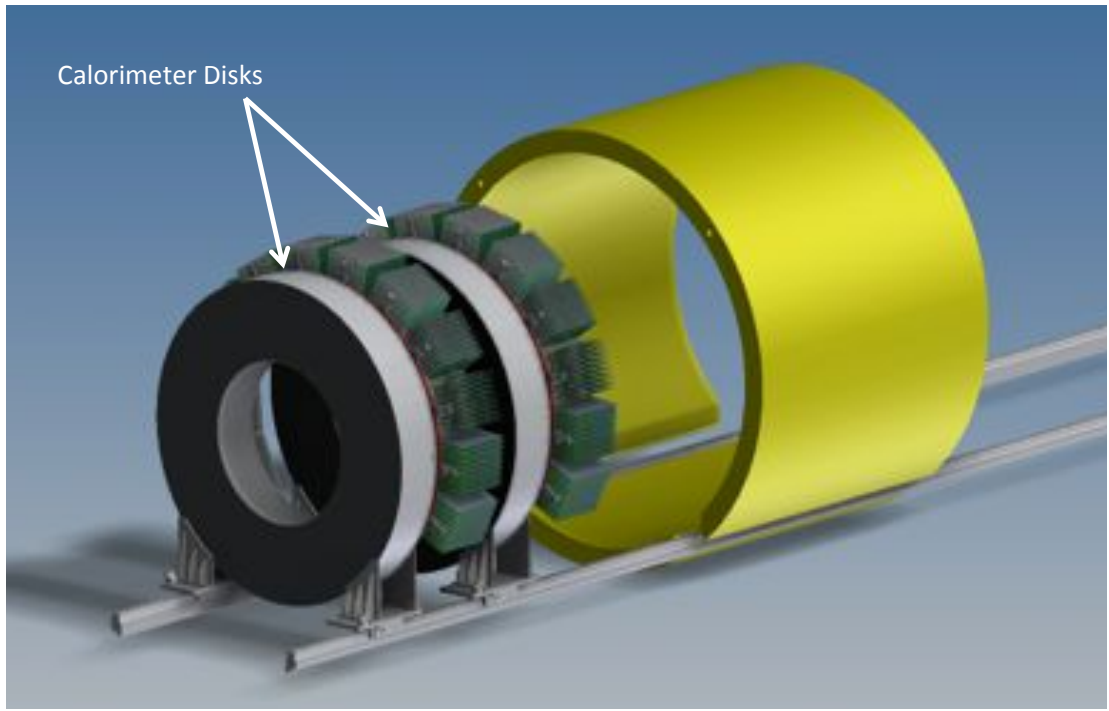

Figure 2.14. The Mu2e calorimeter consisting of an array of $BaF_2$ crystals arranged in two annular disks. Electrons spiral into the upstream faces.

### Cosmic Ray Veto

Cosmic-ray muons are a known source of potential background for muon-to-electron conversion experiments like Mu2e. A number of processes initiated by cosmic-ray muons can produce 105 MeV particles that appear to emanate from the stopping target. These muons can produce 105 MeV electrons and positrons through secondary and delta-ray production in the material within the solenoids, as well as from muon decay-in-flight. The muons themselves can, in certain cases, be misidentified as electrons. Such background events, which will occur at a rate of about one per day, must be suppressed in order to achieve the sensitivity required by Mu2e. Backgrounds induced by cosmic rays are defeated by both passive shielding, including the overburden above and to the sides of the detector hall, as well as the shielding concrete surrounding the Detector Solenoid, by particle identification criteria using the tracker and calorimeter, and, most importantly, by an active veto detector whose purpose is to detect penetrating cosmic-ray muons.

The cosmic ray veto consists of four layers of long extruded scintillator strips, with aluminum absorbers between each layer. The scintillator surrounds the top and sides of the Detector Solenoid (DS) and the downstream end of the Transport Solenoid (TSd), as shown in Figure 2.15. The strips are 2.0 cm thick, providing ample light to allow a high enough light threshold to be set to suppress most of the backgrounds. Aluminum absorbers between the layers are designed to suppress punch through from electrons. The scintillation light is captured by embedded wavelength-shifting fibers, whose light is detected by silicon photomultipliers (SiPMs) at each end (except those counters closest to the TSd).

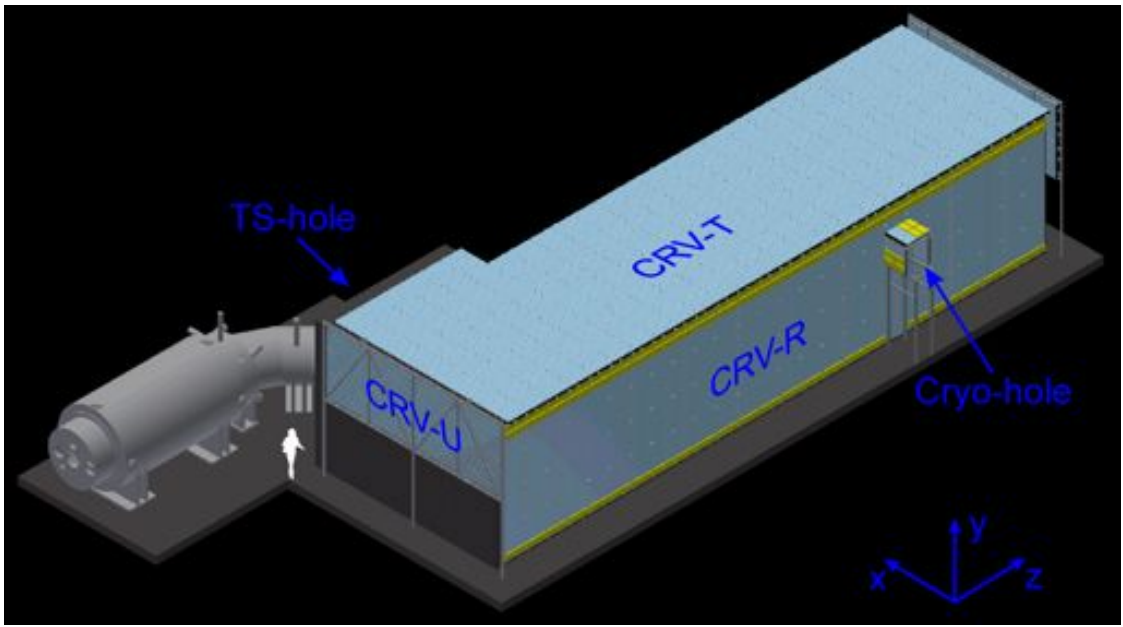

Figure 2.15. The cosmic ray veto covering the Detector Solenoid and half of the Transport Solenoid.

### *2.2.5* **Conventional Facilities**

The conventional facilities for the Mu2e Project include the site preparation, Mu2e surface building and the underground enclosure to house the Mu2e detector. Routing of utilities from nearby locations and installation of new transformers to power the facility are included in the scope of the conventional facilities work. Together the conventional facilities comprise approximately 23,000 ft$^2$ of new construction space. An entry view of the Mu2e facility is shown in Figure 2.16.

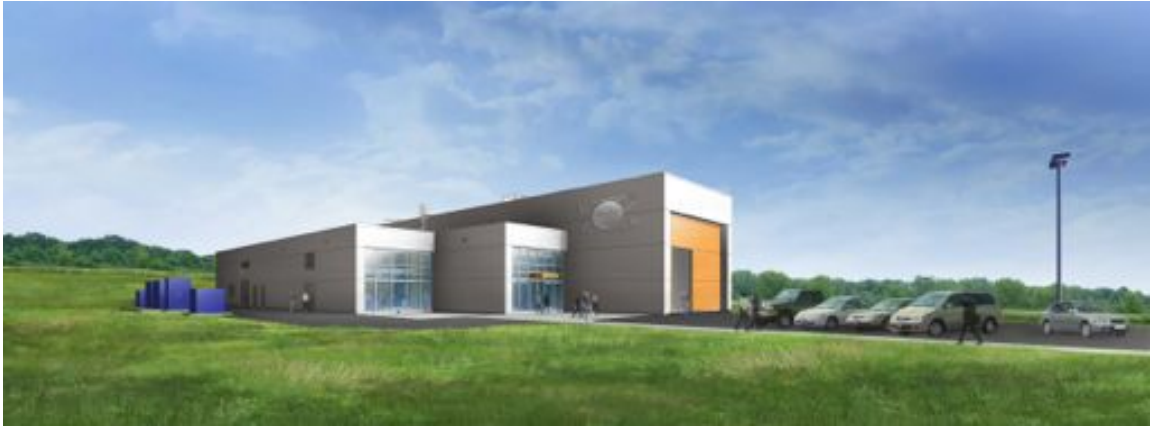

Figure 2.16. Entry View of the Mu2e facility, looking northwest.

***Sustainable Design and LEED***

The Mu2e Project is not required to meet the Leadership in Energy and Environmental Design (LEED)-Gold certification due to the function and operation of the facility. Specifically, the facility will not be occupied on a regular basis. In lieu of LEED-Gold certification, the Project plans to utilize guiding principles and ASHRAE recommendations to meet sustainability goals [7].

### *2.2.6* **Off Project Scope**

There is significant scope that is required by both Mu2e and the g-2 experiment. This common scope has been collected into a series of Accelerator Improvement Projects (AIP) and General Plant Projects (GPP) that are funded independent of the Mu2e and g-2 Projects. These AIPs and GPPs are described below.

Much of the off-project scope was the result of value engineering on the part of Fermilab, the Mu2e Project and the g-2 Project. For example, the Muon Campus provides a common cryo facility for both Mu2e and g-2, replacing individual cryo plants for each Project. At one time Mu2e planned to use both the Antiproton Debuncher and Accumulator Rings to re-bunch and slow extract protons from the Booster to the Mu2e apparatus. It was realized that both Mu2e and g-2 could re-bunch protons using a new RF system in the Recycler. This allowed Mu2e to eliminate its need for the Accumulator Ring, freeing up the Accumulator Ring magnets for reuse in the M4 beamline and eliminating the need for additional RF systems, kickers and injection systems.

***Recycler RF AIP***

The scope of the Recycler RF AIP is to design, assemble and install 7 full 2.5 MHz RF cavities in the existing Recycler Ring. These cavities will be used to re-bunch batches of protons from the Booster, necessary for both Mu2e and g-2. As of August 1, 2014 the Recycler RF AIP was 12% percent complete.

### Muon Campus Infrastructure GPP

The scope of the Muon Campus Infrastructure GPP includes an upgraded cooling system for cryo compressors and extension of the existing MI-52 service building to provide room for power supplies for the new Recycler extraction kickers.

### Beam Transport AIP

The scope of the Beam Transport AIP is to provide upgrades to support extraction of primary protons from the Recycler Ring and transport to either the g-2 target station or the Delivery Ring, the later being necessary for Mu2e. The extraction system includes a Lambertson magnet and kickers. The transport system includes upgrades and aperture improvements to existing beamlines, powers supplies and instrumentation. As of August 1, 2014 the Beam Transport AIP was 23% percent complete.

### Cryo AIP

The Cryo AIP provides the common cryogenic system required to support both Mu2e and g-2. The cryo plant will be installed in the MC-1 building that will house the g-2 experiment. Four refurbished Tevatron satellite refrigerators will be re-purposed for this task. Much of the cryo piping will also be re-purposed from the Tevatron. Figure 2.17 shows two of the four Tevatron refrigerators that have been installed. As of August 1, 2014, the Cryo AIP was about 50% complete.

### Delivery Ring AIP

The Delivery Ring AIP provides upgrades to the Delivery Ring (formerly the antiproton debuncher) to support both Mu2e and g-2. The upgrades include an injection and abort system, electrical infrastructure and rerouting of the controls system. As of August 1, 2014, the Delivery Ring AIP was 10% complete.

### Beamline Enclosure GPP

The Beamline Enclosure GPP provides the underground beamline enclosure and above ground berm between the Delivery Ring and both the g-2 and Mu2e buildings. The design of the Beamline Enclosure is complete and construction is expected to commence in the Fall of 2014. The Beamline Enclosure and the Mu2e building will be a single construction package executed by the same subcontractor. This will allow a common, integrated approach to site prep, grading, storm water runoff, ESH&Q and allow for a coordinated set of construction activities. The purchase order will be split between Mu2e Project funds for the Mu2e building and operating funds for the GPP.

Table 2.1 lists the scope required by Mu2e and, in each case, identifies the funding source.

Table 2.1. The scope of work required to produce an operational Mu2e experiment. This includes the scope of the Mu2e Project as well as the scope included in the Muon Campus AIPs and GPPs.

| Item | Description | Funding Source |
|------|-------------|----------------|
| Recycler upgrades | • MI-8 to Recycler Connection<br>• Recycler injection kicker | NOvA |
| Recycler upgrades | • 2.5 MHz RF system | Recycler RF AIP |
| Transfer Line Modifications | • Recycler extraction system<br>• Beam transport from Recycler Ring to Delivery Ring | Beam Transport AIP |
| Recycler upgrades | • MI-52 extension for Recycler extraction kicker power supplies | Muon Campus Infrastructure GPP |
| Delivery Ring Modifications | • Debuncher injection kicker<br>• Proton abort system<br>• Removal of Collider equipment | Delivery Ring AIP |
| Delivery Ring modifications | • 2.4 MHz RF system<br>• Resonant extraction system | Mu2e Project |
| Muon Campus Beamline Enclosure | • Beamline tunnel to house M4 beamline elements for g-2 and Mu2e | Beamline Enclosure GPP |
| M4 Beamline | • Beamline elements<br>• Extinction system<br>• Diagnostic Absorber<br>• Proton Beam Absorber | Mu2e Project |
| Conventional Construction | • Surface Building<br>• Underground enclosure to house detector<br>• Utilities | Mu2e Project |
| Solenoids | • Production Solenoid<br>• Transport Solenoid<br>• Detector Solenoid<br>• Power<br>• Quench protection<br>• Cryo distribution | Mu2e Project |
| Muon Campus Cryo Plant | • Cryo refrigerators<br>• Warm lines for compressed Helium<br>• Cold lines to Mu2e and g-2 detector halls | Cryo AIP |
| Detector | • Tracker<br>• Calorimeter<br>• Cosmic Ray Veto<br>• Stopping Target Monitor<br>• Trigger and Data Acquisition system | Mu2e Project |

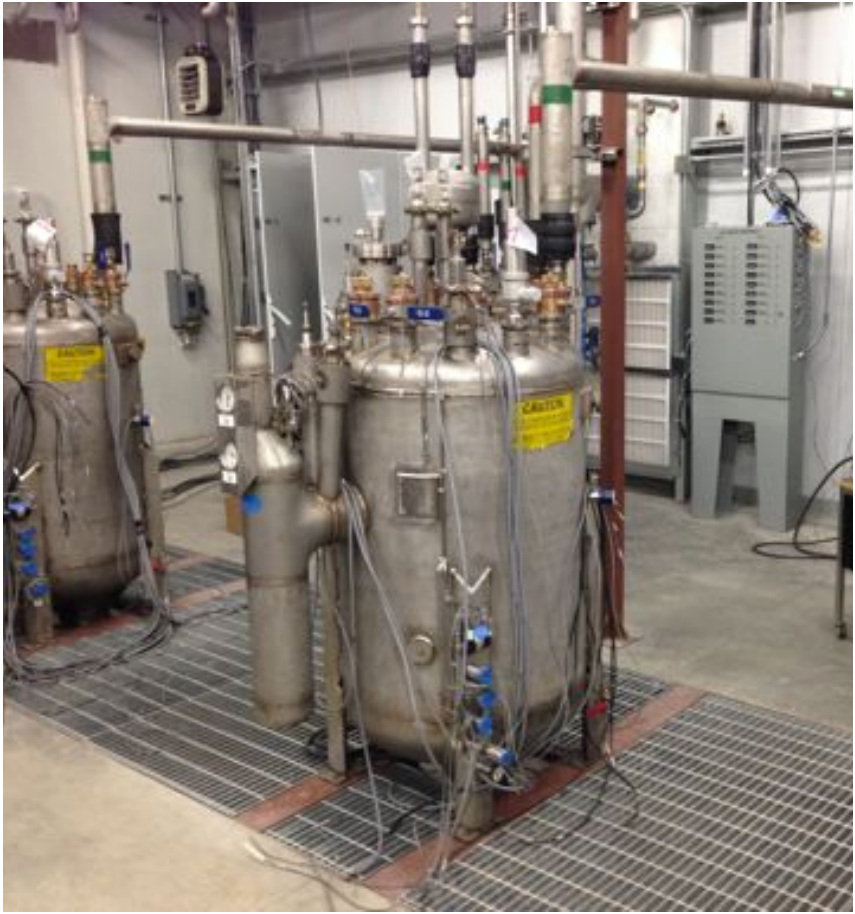

Figure 2.17. Two of the four Tevatron satellite refrigerators that have been installed in the MC-1 building to provide cryogenic helium for the Mu2e and g-2 experiments.

## 2.3    Project Organization

The Mu2e Project consists of nine subprojects coordinated by a central Project Office located at Fermilab.  The subprojects, or Level 2 systems, are:

1. Project Management
2. Accelerator Systems
3. Conventional Construction
4. Solenoids
5. Muon Beamline
6. Tracker
7. Calorimeter
8. Cosmic Ray Veto
9. Trigger and DAQ.

The Fermilab Project Office is headed by the Project Manager and assisted by a Deputy Project Manager and two Project Engineers. Project office support staff includes a Financial Manager, Project Controls Specialists, an ES&H Coordinator, a Risk Manager, a Quality Assurance Manager, a Configuration Control Manager and administrative support. Fermilab provides additional ES&H support and oversight. The Mu2e Project Office has developed detailed plans for project management, risk management, configuration control and quality assurance [8][9][10][11].

### 2.3.1   Work Breakdown Structure

The Mu2e Project has been organized into a Work Breakdown Structure (WBS). The WBS contains a complete definition of the Project's scope and forms the basis for planning, executing and controlling project activities. The Project WBS is shown in Figure 2.18 and Figure 2.19 down to level 3. Items are defined as specific deliverables (WBS 2 – 9) or Project Management (WBS 1).

1. *Project Management* – Project Office administrative and management activities that integrate across the entire project (management, regulatory compliance, quality assurance, safety, project controls, budget, risk management, etc.)
2. *Accelerator* – All phases of R&D, design, procurement, installation, integration and testing of the accelerator systems that are part of the Mu2e Project.
3. *Conventional Construction* - All phases of design, procurement, construction and integration of the conventional construction facilities including site preparation and access to utility systems.
4. *Solenoids* – All phases of R&D, design, procurement, installation, integration, testing and commissioning of the superconducting solenoid system and associated infrastructure including quench protection and systems to distribute cryogens and power.
5. *Muon Beamline* – All phases of R&D, design, procurement, integration, testing and commissioning of the series of deliverables associated with the Muon Beamline system.
6. *Tracker* – All phases of R&D, design, procurement, assembly, installation, integration, testing and commissioning of the tracker, tracker electronics and associated support infrastructure.
7. *Calorimeter* – The project scope for the calorimeter includes procurement, testing and processing of 2/3 of the crystals, 1/2 of the photosensors, 1/2 of the digitizers as well as the R&D, design, construction and installation of the calibration system and the front end electronics. The rest of the calorimeter will be provided in-kind by INFN.
8. *Cosmic Ray Veto* - All phases of R&D, design, procurement, assembly, integration and testing of the cosmic ray veto, the veto electronics and associated support infrastructure.

9. *Trigger and DAQ* - All phases of R&D, design, procurement, assembly, installation, integration, testing and commissioning of the data acquisition system.

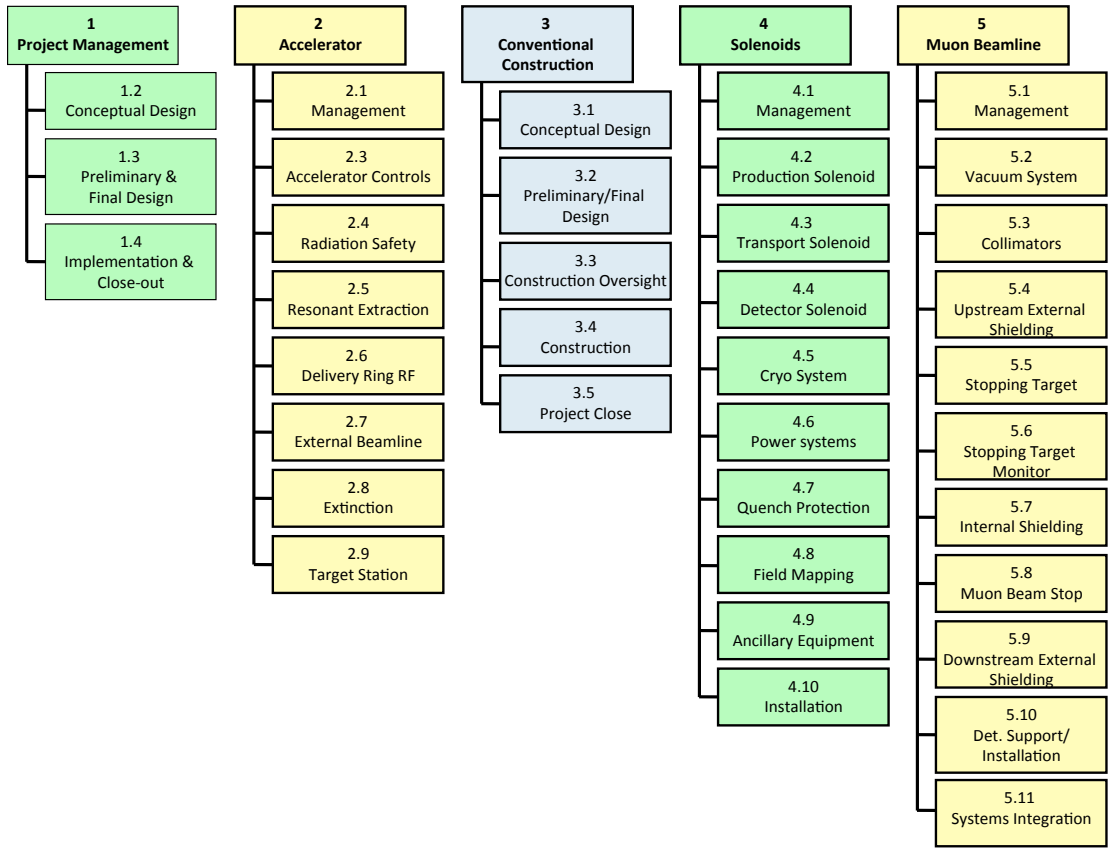

Figure 2.18. Mu2e Project WBS to Level 3 for Project Management, Accelerator, Conventional Construction, Solenoids and Muon Beamline.

## 2.4    Project Management

### 2.4.1    Project Controls

The Mu2e Project is in full compliance with the DOE certified FRA Earned Value Management System (EVMS). The Earned Value Management System is used to monitor, analyze, and report project performance. Mu2e's EVMS implementation uses Primavera P6 scheduling software for the resource loaded cost and schedule, Cobra for escalation, burdening, and earned value reporting and analysis and Fermilab's Oracle Project Accounting system for tracking obligations and actual costs. The Fermilab EVMS description can be accessed through the Fermilab Office of Project Management website [12].

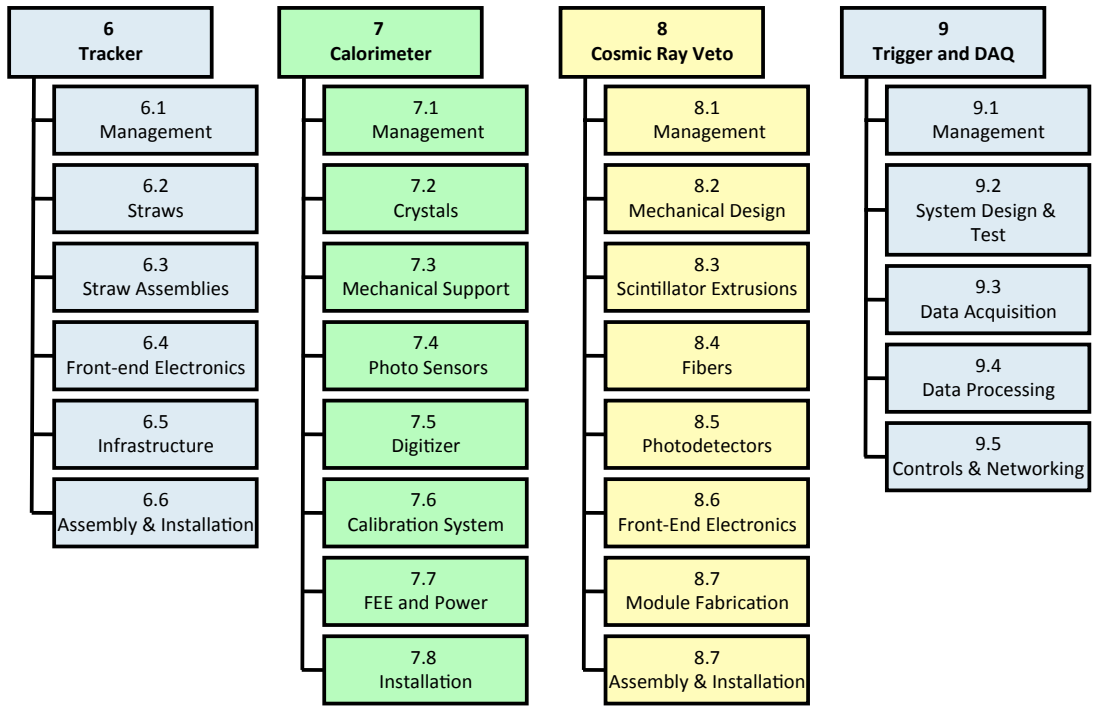

Figure 2.19. Mu2e Project WBS to Level 3 for the Tracker, Calorimeter, Cosmic Ray Veto and the Trigger and DAQ.

### 2.4.2 ES&H Management

The Laboratory Director has overall responsibility for establishing and maintaining Fermilab's ES&H policy. It is Fermilab's policy to integrate environment, safety and health protection into all aspects of work, utilizing the principles and core functions of the Integrated ES&H Management System and implemented through the appropriate lines of management. The Mu2e Project Manager reports to the Fermilab Director or his designee and is responsible for implementing Fermilab's ES&H policies into all aspects of the Mu2e Project. The Mu2e Project Management Plan [8] includes a section on Integrated Safety Management that describes how the Mu2e Project ES&H policies fit within the DOE approved Fermilab ES&H program.

The philosophy of Integrated Safety Management (ISM) will be incorporated into all work on Mu2e, including any work done on the Fermilab site by subcontractors and sub-tier contractors. Integrated Safety Management is a system for performing work safely and in an environmentally responsible manner. The term "integrated" is used to indicate that the ES&H management systems are normal and natural elements of doing work. The intent is to integrate the management of ES&H with the management of the other primary elements of work: quality, cost and schedule.

### 2.4.3   Quality Assurance

Quality Assurance and Quality Control (QA/QC) systems are designed, as part of the Quality Management Program, to ensure that the components of the Mu2e Project meet the design specifications and operate within the parameters mandated by the requirements of the Mu2e physics program.  The Mu2e Project Manager is responsible for achieving performance goals. The Mu2e Quality Assurance Manager is responsible for ensuring that a quality system is established, implemented, and maintained in accordance with requirements.  The Quality Assurance manager will provide oversight and support to the project participants to ensure a consistent quality program.

The QA/QC elements in place for the Mu2e Project draw heavily on the experience gained from similar projects in the past. Senior management recognizes that prompt identification and documentation of deficiencies, coupled with the identification and correction of the root causes, are key aspects of any effective QA/QC Program.  The Project Manager endorses and promotes an environment in which all personnel are expected to identify nonconforming items or activities and potential areas for improvement.

### 2.4.4   Configuration Management

Configuration Management is a formalized process to manage proposed system changes and provide an audit trail to manage and maintain the evolution of system configurations. A Configuration Management Plan establishes a baseline, defines the rules for changing that baseline and records changes as they occur.  The origin of changes and their status at any subsequent point should be readily identifiable.

The Mu2e Project uses several tools to achieve this objective, including a document control system that supports versioning and document signoff to "approve" a version, drawing management systems, and software control with a versioning and release system based on a software repository.

### 2.4.5   Risk Management

Project risk in Mu2e is mitigated through a structured and integrated process for identifying, evaluating, tracking, abating and managing risks in terms of three risk categories: cost, schedule and technical performance. A Risk Management Board, chaired by the Project Manager, meets regularly to identify risks and develop mitigation plans.

Any project faces both threats and opportunities and must strive to exploit the opportunities while ensuring that the threats do not derail the project.  Numerous informal and formal approaches are used to identify threats and opportunities, assessing their likelihood and prioritizing them for possible mitigation or exploitation. The key to successful risk management is to implement a deliberate approach to accepting, preventing, mitigating or avoiding them. The Mu2e Project becomes aware of potential risks in many ways, notably during work planning, meetings and reviews as well as via lessons learned from others. Risk is managed during the planning and design phase by implementing appropriate actions, such as ensuring adequate contingency and schedule float, pursuing multiple parallel approaches and/or developing backup options. Every effort must be made to specify these actions in a manner that reduces the risk to an acceptably low level.

Risks that are identified will be managed as early as possible to assure that they do not delay the timely completion of the project or stress its budget in unexpected ways. The Mu2e Risk Management Plan [9] is under configuration management.

## 2.5 Key Performance Parameters Required to Obtain the Expected Outcome

Project completion (CD-4) will be accomplished when the scope defined in the WBS dictionary has been completed and the apparatus has been demonstrated to be functioning to the required level by achieving Key Performance Parameters (KPP) described in the Project Execution Plan [13]. The WBS dictionary is under change control. After achieving the Key Performance Parameters, the Project Manager will request acceptance and approval of CD-4. The Key Performance Parameters are those that demonstrate functionality of the system while achievement of beam parameters required for the experimental program will be obtained after routine tuning and operation of the accelerator complex.

Two sets of KPPs have been defined. A set of *Threshold* parameters define the minimum acceptable performance for CD-4 while a set of *Objective* parameters define the desired outcome. The Objective KPPs are costed in the performance baseline.

## 2.6 Cost and Schedule

The Mu2e Project cost was developed from a bottoms-up, resource-loaded cost and schedule developed in Primavera P6. The total Project cost for Mu2e is $271M. The cost, broken down by Level 2 subsystem, is shown in Table 2.2.

The overall estimate uncertainty for the technical component of the Project (not counting the Level-of-Effort Project Management tasks) is 37%. The contingency is a combination of design maturity, applied activity-by-activity, and a statistical evaluation of risks and opportunities [14]. A Tier 0 milestone schedule to construct Mu2e is shown in Table 2.3.

24 months of programmatic float has been added to the estimated Project completion date to arrive at a CD-4 milestone.

Table 2.2. The Mu2e Total Project Cost, by level 2 subsystem. Costs are fully burdened and escalated into actual year $k. Estimate uncertainty percentage is for the work remaining. Completed work has an estimate uncertainty of 0%. Work performed is through September 2014.

|  | Performed | ETC | Estimate Uncertainty | % EU on ETC | Total |
|---|---|---|---|---|---|
| Project Management | 9,447 | 11,221 | 860 | 8% | 21,528 |
| Accelerator | 11,533 | 29,272 | 8,619 | 29% | 49,424 |
| Conventional Construction | 2,470 | 18,775 | 3,462 | 18% | 24,707 |
| Solenoids | 21,307 | 66,661 | 20,867 | 31% | 108,835 |
| Muon Beamline | 4,353 | 15,214 | 5,454 | 36% | 25,021 |
| Tracker | 3,020 | 8,503 | 3,204 | 38% | 14,727 |
| Calorimeter | 520 | 4,407 | 1,113 | 25% | 6,040 |
| Cosmic Ray Veto | 1,926 | 4,847 | 1,645 | 34% | 8,418 |
| Trigger & DAQ | 1,789 | 3,011 | 963 | 32% | 5,763 |
| Risk-based Contingency |  |  | 6537 |  | 6537 |
| **Total** | **56,366** | **161,910** | **52,724** | **33%** | **271,000** |

Table 2.3 Preliminary CD Milestone Schedule. Note that satisfying the Key Performance Parameters is not a CD milestone but is added to provide context for the CD-4 milestone.

| Major Milestone Events | Preliminary Schedule |
|---|---|
| CD-0 (Approve Mission Need) | 1st Qtr, FY10 (A) |
| CD-1 (Approve Alternative Selection and Cost Range) | 3d Qtr, FY12 (A) |
| CD-3a (Approve start of Long-Lead Procurement) | 3d Qtr, FY14 (A) |
| CD-2 (Approve Performance Baseline) | 1st Qtr, FY15 |
| CD-3b (Approve start of Phased Construction/Fabrication) | 1st Qtr, FY15 |
| CD-3c (Approve start of Construction) | 2d Qtr, FY16 |
| Key Performance Parameters Satisfied | 1st Qtr, FY21 |
| CD-4 (Includes 24 months of programmatic float) | 1st Qtr, FY23 |

# 3    Muon to Electron Conversion

## 3.1    Physics Motivation

Before the discovery of neutrino oscillations, it was generally understood that lepton flavor changing processes were forbidden in the Standard Model (SM) and that the lepton flavor numbers $L_e$, $L_\mu$ and $L_\tau$ were conserved. Since neutrinos were taken to be massless, the mass matrices for the charged leptons and the charged-current weak interactions could be simultaneously diagonalized. Neutrino oscillations are, however, *prima facie* evidence of mixing between lepton families, and are manifestly lepton flavor-violating (LFV) processes. This will be the case as well for any model having a mechanism for generating neutrino masses. The rate at which LFV processes occur in the neutrino sector is constrained by the measured neutrino mixing parameters, but the rate at which charged lepton flavor violation (CLFV) occurs is model-dependent and can vary over many orders of magnitude. For example, in the minimal extension to the Standard Model, in which neutrino masses are generated by introducing three right-handed SU(2) singlet fields and three new Yukawa couplings [1], the CLFV process $\mu^- N \rightarrow e^- N$ can occur only through loop diagrams whose amplitudes are proportional to $(\Delta m^2_{ij} / M_w^2)^2$ where $\Delta m^2_{ij}$ is the mass-squared difference between the $i^{th}$ and $j^{th}$ neutrino mass eigenstates. Because the neutrino mass differences are so small relative to $M_w$, the rates of CLFV decays in this modified version of the SM are effectively zero (e.g., $< 10^{-50}$ for both $\mu^+ \rightarrow e^+\gamma$ and $\mu^- N \rightarrow e^- N$). Many New Physics (NP) models, however, predict significant enhancements to CLFV rates, and in particular to that of the muon to electron conversion process. Many well-motivated physics models predict rates for CLFV processes that are within a few orders of magnitude of the current experimental bounds. These include the MSSM with right-handed neutrinos, SUSY with *R*-parity violation as well as models with leptoquarks, new gauge bosons, large extra-dimensions, and a non-minimal Higgs sector [2]. The Mu2e experiment, with a single-event sensitivity of a few $10^{-17}$ for the ratio of $\mu^- N \rightarrow e^- N$ conversions to conventional muon capture, has real discovery potential over a wide range of New Physics models and may prove to be a powerful discriminant among models.

### 3.1.1    Charged Lepton Flavor Violation – Model Independent Searches

There is an active global program searching for CLFV processes using rare decays of muons, taus, kaons, and *B* mesons. The ratio of rates among various CLFV processes is model-dependent and varies widely, depending on the underlying physics. It is therefore important to pursue experiments sensitive to different processes in order to elucidate the mechanism responsible for flavor-violating effects. Because of the existence of intense muon sources, the most stringent limits currently come, and will continue to come, from the muon sector. There are three rare muon processes that stand out: $\mu^+ \rightarrow e^+\gamma$, $\mu^+ \to e^+ e^+ e^-$ and $\mu^- N \to e^- N$, with $\mu^- N \to e^- N$ offering the greatest potential sensitivity. Searches for these processes have thus far yielded only upper limits on the corresponding rates. The current experimental limits (all at 90% CL) on the branching ratios are: BR($\mu^+ \to e^+ \gamma$)< $5.7 \times 10^{-13}$ [3], BR($\mu^+ \to e^+ e^+ e^-$) < $1.0 \times 10^{-12}$ [4], and $R_{\mu e}$(Au) ($\mu \to e$ conversion on gold) < $7 \times 10^{-13}$ [5], where $R$ is the conversion rate normalized to the capture rate. In the coming decade significant improvement is expected in the sensitivity of searches for all three processes.

The MEG experiment [3] at PSI has already reached a limit of $5.7 \times 10^{-13}$ for the branching ratio of $\mu^+ \to e^+ \gamma$ and is currently being upgraded to an expected sensitivity of $6 \times 10^{-14}$ [6]. Mu2e at Fermilab, as well as the COMET [7] experiment at JPARC, aims at sensitivities of $10^{-16} - 10^{-17}$ on $R_{\mu e}$(Al). These two processes have complementary sensitivity to New Physics effects; results from both are important in untangling the underlying physics. To illustrate this, one can estimate the sensitivity of a given CLFV process in a model-independent manner by adding two qualitatively different lepton-flavor-violating effective operators to the Standard Model Lagrangian parameterized by $\Lambda$, the effective mass scale of New Physics, and $\kappa$, a dimensionless parameter that controls the relative contribution of the two terms [8]:

$$\mathcal{L}_{\text{CLFV}} = \frac{m_\mu}{(1+\kappa)\Lambda^2} \bar{\mu}_R \sigma_{\mu\nu} e_L F^{\mu\nu} + \frac{\kappa}{(1+\kappa)\Lambda^2} \bar{\mu}_L \gamma_\mu e_L \left( \sum_{q=u,d} \bar{q}_L \gamma^\mu q_L \right).$$

Most New Physics contributions to muon to electron conversion and $\mu^+ \to e^+ \gamma$ are accounted for in these two classes of effective operators. If $\kappa \ll 1$, the first term, a flavor-changing magnetic moment operator, is dominant. If $\kappa \gg 1$, the second term, a four-fermion interaction operator, is dominant. Simply put, the first term arises from loops with an emitted photon and can mediate all three rare muon processes. The photon is real in $\mu^+ \to e\gamma$ and virtual in $\mu^- N \to e^- N$ and $\mu^+ \to e^+ e^+ e^-$. The second term includes contact terms and a variety of other processes not resulting in an on-mass-shell photon. Therefore, the $\mu^- N \to e^- N$ and $\mu^+ \to e^+ e^+ e^-$ processes are sensitive to New Physics regardless of the relative contributions of the first and second terms. The New Physics scale, $\Lambda$, to which these two processes are sensitive is shown as a function of $\kappa$ in Figure 3.1 and [8]. The projected sensitivity of the MEG experiment will probe $\Lambda$ values up to 1000 - 2000 TeV for $\kappa \ll 1$, but has little sensitivity for $\kappa \gg 1$. The projected sensitivity of the Mu2e experiment will probe $\Lambda$ values from 2000 to nearly 10,000 TeV over *all* values of $\kappa$. It is important to emphasize that these effective operators provide a general description of most New Physics scenarios in which large CLFV effects might appear in $\mu^+ \to e^+ \gamma$ decay and $\mu^- N \to e^- N$ conversion; the conclusions on relative sensitivity are generically true. Thus the Mu2e experiment's sensitivity in the range of $10^{-16} - 10^{-17}$ remains relevant and important in all outcomes of MEG. If MEG observes a signal, then Mu2e should as well; the ratio of measured rates can then be used to simultaneously constrain $\Lambda$ and $\kappa$ (determining which types of new physics models are favored). A null result from MEG does not preclude a Mu2e discovery, since the New Physics may lead to effective interactions to which the $\mu^+ \to e^+ \gamma$ process is largely insensitive.

It is important to emphasize that $\Lambda$ is an effective mass scale and not immediately comparable to the mass scale reach of direct searches, such as those at the LHC. In the case of the magnetic moment interaction, $\Lambda$ is related to the mass $M$ of the new particles via a loop factor and the new couplings $g$ of the new interactions, e.g. $1/\Lambda^2 \sim g^2 e/(16\pi^2 M^2)$, where $e$ is the electromagnetic coupling. In the case of the four-fermion operator, $\Lambda$ could be more directly related to the masses of the new particles, e.g., $1/\Lambda^2 \sim g^2/M^2$.

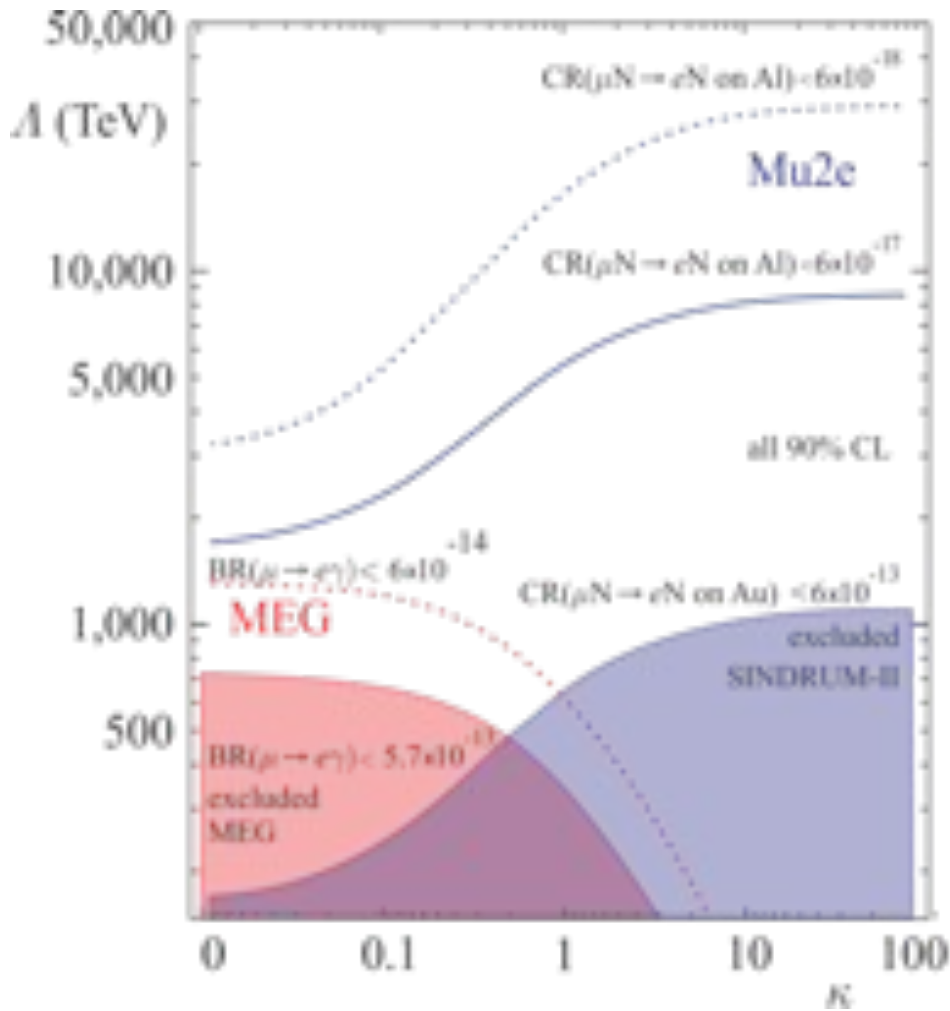

Figure 3.1. The sensitivity to the scale of new physics, $\Lambda$, as a function of $\kappa$, for a muon to electron conversion experiment with a sensitivity in the range of $10^{-16} - 10^{-17}$ is compared to that for a muon-to-electron-gamma experiment with a sensitivity in the range of $10^{-13} - 10^{-14}$. See the text for a definition of $\kappa$. The current and projected exclusion regions of parameter space for $\mu \to e\gamma$ are indicated by red contours, those for $\mu \to e$ conversion by blue contours.

### 3.1.2   Specific Models of CLFV

All models of new physics at the electroweak scale constructed to address outstanding issues in particle physics – the origin of the dark matter, the gauge hierarchy problem, etc. – contain CLFV effects at different levels. Some predict that CLFV processes are rare, while others are already severely constrained by existing CLFV bounds and would be ruled out if CLFV effects are not observed in the near future. We discuss some specific examples below.

***An SO(10) Type I See-Saw Grand Unified Model***

Supersymmetric versions of the Standard Model with weak-scale supersymmetry (SUSY)-breaking parameters are very popular extensions of the SM that address the gauge hierarchy problem and naturally accommodate a dark matter candidate. Very often, they also lead to large rates for CLFV processes. Concrete predictions depend on the mechanism behind SUSY breaking and other phenomenologically well-motivated assumptions. As an example, the model discussed in detail in [9] and [10] allows one to compute the $\mu^- N \to e^- N$ rate in titanium as a function of SUSY breaking parameters in the context of an SO(10) SUSY GUT model with very massive right-handed neutrinos, after one considers different hypothesis for the neutrino Yukawa couplings ("PMNS-like" or "CKM-like"). This recent study takes into account the light Higgs mass, the recently observed value of $\theta_{13}$, and recent limits on LHC searches for superpartners [10].

Figure 3.2 shows that Mu2e and potential upgrades will be able to test most of the PMNS-like parameter space, particularly for large tan$\beta$, and will be able to explore a portion of the CKM-like parameter space.

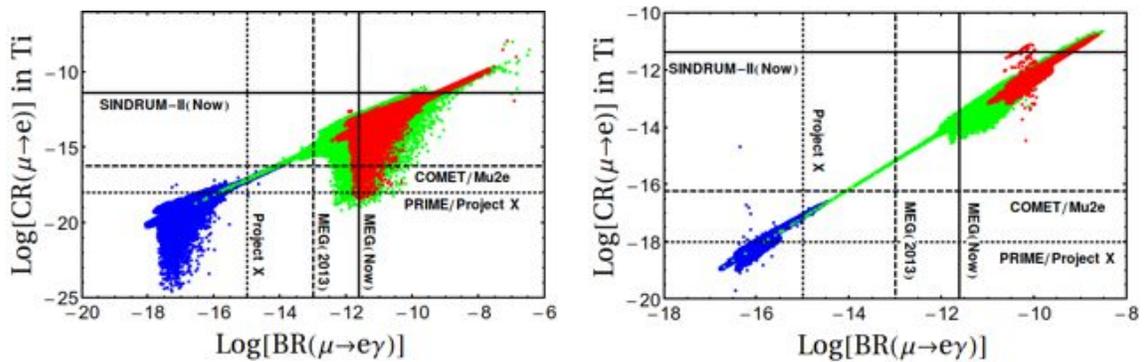

Figure 3.2. Muon Conversion Rate CR($\mu \to e$ in Ti) versus BR($\mu \to e\gamma$) for the PMNS case in mSUGRA (red), the Non-Universal Higgs Mass (green) and the CKM case (blue). The left plot is for tan$\beta$=10 and the right for tan$\beta$=40. The different horizontal and vertical lines correspond to A Scalar Leptoquark Model

### Scalar leptoquarks

Models with scalar leptoquarks at the TeV scale can, through top mass enhancement, modify the $\mu \to e$ conversion rate and BR($\mu \to e\gamma$) while satisfying all known experimental constraints from collider and quark flavor physics [11]. Figure 3.3 compares the reach in the new coupling $\lambda$ for a range of scalar leptoquark masses for the $\mu \to e$ conversion rate with the sensitivity of Mu2e and BR($\mu \to e\gamma$) at the sensivity of the MEG upgrade.

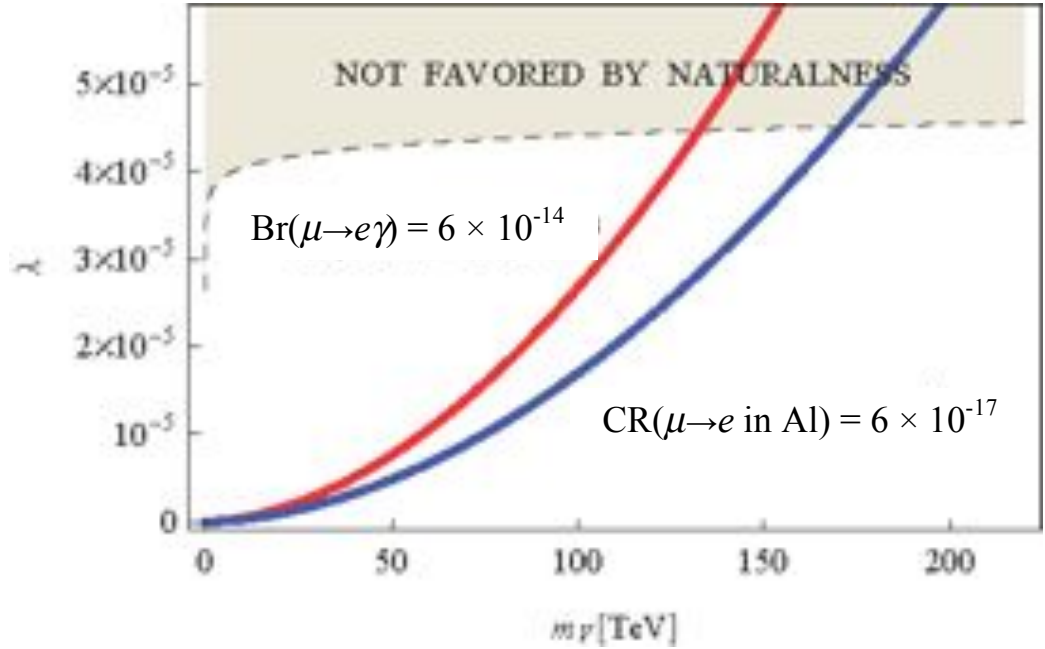

Figure 3.3. The combination of couplings $\lambda$ from Eq. (14) of [11] as a function of the scalar leptoquark mass for the $\mu \to e$ conversion rate (CR) in Al at the sensitivity of Mu2e and the branching fraction BR($\mu \to e\gamma$) at the sensitivity of the MEG upgrade (courtesy B. Fornal). The shaded region consists of points that do not satisfy a naturalness criterion defined in Eq. 7 of [11].

### Flavor-violating Higgs decays

One of the highest priorities in particle physics is to study the newly discovered Higgs boson, and measure all of its properties. Non-standard flavor-violating decays of the 125 GeV Higgs to quarks and leptons are a very interesting probe of New Physics [12]. Constraints from CLFV on new interactions that lead to $h \to e\mu$, $e\tau$, $\mu\tau$ severely outweigh the sensitivity of collider experiments. Current $\mu \to e$ conversion (see Figure 3.4) implies $\sqrt{|Y_{\mu e}|^2 + |Y_{e\mu}|^2} < 4.6 \times 10^{-5}$; Mu2e is expected to be sensitive to $\sqrt{|Y_{\mu e}|^2 + |Y_{e\mu}|^2} >$ few $\times 10^{-7}$. In these types of scenarios, constraints involving muon couplings are substantially stronger than those involving $\tau$ couplings.

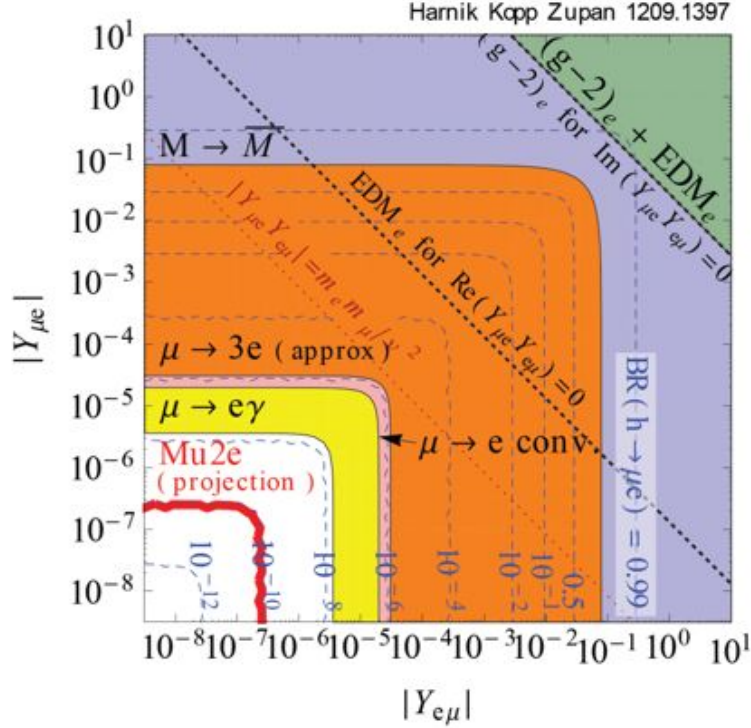

Figure 3.4. Constraints on the flavor-violating Yukawa couplings $|Y_{e\mu}|$, $|Y_{\mu e}|$ for a 125 GeV Higgs boson [12]. The diagonal Yukawa couplings are approximated by their SM values. Thin blue dashed lines are contours of constant BR for $h \to \mu e$, while the thick red line is the projected Mu2e limit.

### Left-Right Symmetric Models

Left-right symmetric models are attractive extensions of the Standard Model that restore parity at short-distances, are potential remnants of grand unification at very short-distances, and allow one to build potentially testable neutrino mass models. A recent study [13] discusses predictions of left-right models assuming the new breaking scale is around 5 TeV. The expected rates for muon-to-electron conversion and $\mu \to e\gamma$ are correlated and expected to lie within the sensitivity reach of both the MEG upgrade and Mu2e, as depicted in Figure 3.5. If this model is correct, the observation of $\mu \to e\gamma$ with a branching ratio of $10^{-13}$ would imply a muon-to-electron conversion rate around of $10^{-14}$ and many hundreds of events in the Mu2e experiment.

### 3.1.3   CLFV and Neutrino Masses

In spite of the fact that neutrino oscillations imply CLFV, measurements of neutrino oscillation processes do not allow us to reliably estimate the rate for the various CLFV processes. The reason is that while neutrino oscillation phenomena depend only on neutrino masses and lepton mixing angles, the rates for the various CLFV processes depend dramatically on the mechanism behind neutrino masses and lepton mixing, currently unknown. Different neutrino mass-generating Lagrangians lead to very different rates for CLFV. The observation of CLFV, therefore, will potentially provide vital information when it comes to revealing the physics responsible for neutrino masses.

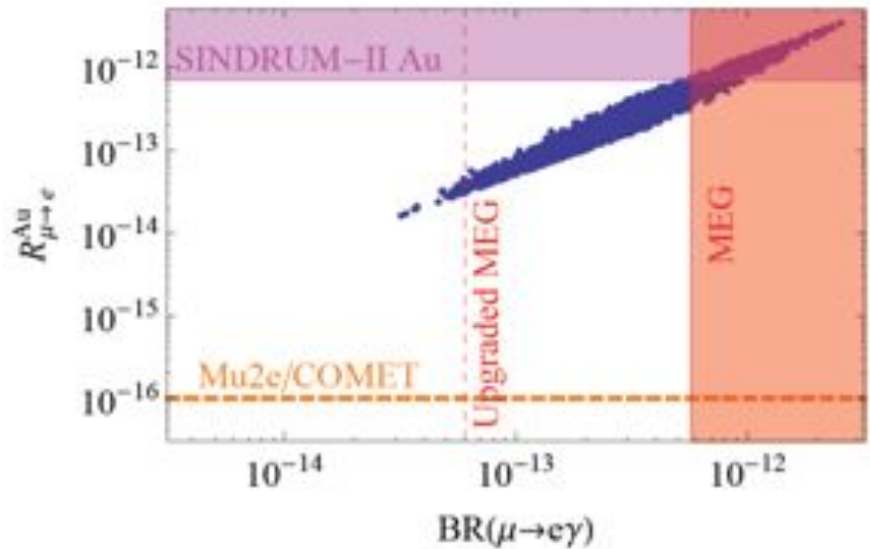

Figure 3.5. Correlation between the branching ratio for $\mu \rightarrow e\gamma$ and the muon-to-electron conversion rate in gold in a left-right symmetric model. See [13] for details. The red and magenta shaded regions are constrained by MEG [3] and SINDRUM-II [5], respectively and the vertical and horizontal dashed lines indicate projected limits from the MEG upgrade and Mu2e/COMET, respectively.

The same physics that determines the pattern of the leptonic mixing matrix may also influence the relative rates of different CLFV process. In the case of purely Dirac neutrinos, for example, the ratio of $\mu \rightarrow e\gamma$ to $\tau \rightarrow \mu\gamma$ is completely determined in terms of known neutrino mixing parameters and is (approximately) proportional to $tan^2\theta_{13}/cos^2\theta_{23} \sim 1/20$. Two concrete examples are discussed below of the interplay between neutrino observables, the physics responsible for neutrino masses, and CLFV.

In the MSSM with MSUGRA boundary conditions, as discussed earlier, large CLFV rates are a consequence of the seesaw mechanism. In more detail, neutrino Yukawa couplings lead to off-diagonal mass-squared parameters for the scalar leptons (for a detailed discussion see, for example, [15]). Hence, CLFV processes probe some combination of the neutrino Yukawa couplings and the right-handed neutrino masses, providing non-trivial information regarding the neutrino mass sector (this statement is very dependent on the physics of SUSY breaking, which must be well understood). A different linear combination of Yukawa couplings and heavy masses determines the observed active neutrino masses. Combined, CLFV, SUSY searches, and neutrino experiments allow one to begin to reconstruct the physics responsible for nonzero neutrino masses. Knowledge of this physics is fundamental when it comes to determining whether leptogenesis [16] is the mechanism responsible for the matter–antimatter asymmetry of the universe. For a concrete example, see [17]. If thermal leptogenesis is ever to be tested experimentally, CLFV will certainly play a fundamental role [18].

In the case discussed above, the relation between neutrino mixing parameters and CLFV is indirect. There are scenarios where the neutrino masses and the lepton mixing angles can be directly related to the rates of several CLFV processes, including models of large extra-dimensional Dirac neutrinos [19], and models where neutrino Majorana masses are a consequence of the existence of SU(2) triplet Higgs fields. Figure 3.6 from [20] depicts the rate for different muon CLFV processes as a function $|U_{e3}| \cos \delta$, for different hypothesis regarding the neutrino mass ordering (normal or inverted) [20]. Here $|U_{e3}| = \sin\theta_{13} \sim 0.15$ is the "reactor" mixing angle and $\delta$ is the "Dirac" $CP$-odd phase that can be measured in next-generation neutrino oscillation experiments. The overall expectation for the transition rates depends on parameters external to the neutrino mass matrix, like the triplet mass and vacuum expectation value. The combination of data from neutrino oscillation experiments, high energy collider experiments (like the LHC) and CLFV should ultimately allow one to thoroughly test particular Higgs triplet models and, if these turn out to be correct, unambiguously reveal the physics behind neutrino masses.

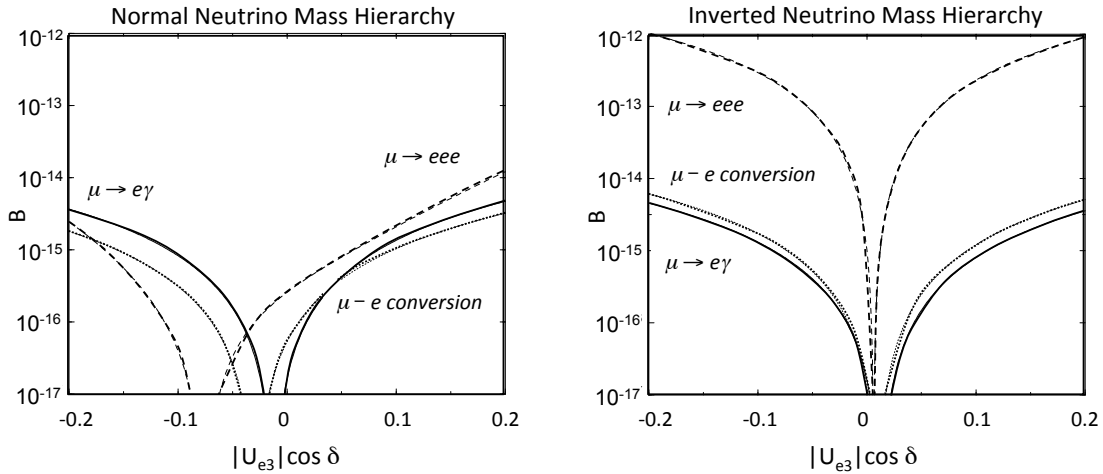

Figure 3.6. The branching ratios $B$ for $\mu \to e$ (solid line) and $\mu \to eee$ (dashed line), and the normalized capture rate $B$ for $\mu \to e$ conversion in Ti (dotted line) as a function of $|U_{e3}| \cos \delta$ in a scenario where neutrino masses arise as a consequence of the presence of a triplet Higgs field with a small vacuum expectation value. The lightest neutrino mass is assumed to be negligible while the neutrino mass hierarchy is assumed to be normal (left-hand side) and inverted (right-hand side). See [20] for details.

### 3.1.4   Negative muon to positron conversion

The Mu2e set-up allows for the concurrent search of muon to positron transitions: $\mu^- + (Z, A) \to e^+ + (Z\text{-}2, A)$. These are lepton-flavor and lepton-number violating processes that are guaranteed to occur if neutrinos are Majorana fermions, not unlike searches for neutrinoless double-beta decay. Like CLFV processes, the expected rates for muon-to-positron transition are very dependent on the mechanism behind neutrino masses. Unlike CLFV, muon-to-positron conversion directly tests the physics responsible for lepton number violation, and any observation of muon-to-positron conversion implies that neutrinos are Majorana fermions.

In the standard high-energy seesaw picture, the expected rate of muon-to-positron conversion is tiny, proportional to the square of the $e$-$\mu$ element of the neutrino mass matrix, $m_{e\mu} = \sum_{i=1,2,3} U_{ei} U_{\mu i} m_i \leq 1$ eV, where $m_i$ are the neutrino masses and $U$ are the elements of the leptonic mixing matrix. While bounds from neutrinoless double-beta decay are, naively, much more constraining, the $e$-$\mu$ character of the muon-to-positron transition provides independent information. For more details see, for example, [14]. Current bounds on muon-to-positron transitions are competitive with those from rare kaon decays. Mu2e should provide, absent a discovery, bounds that are several orders of magnitude more stringent.

### 3.1.5   Charged Lepton Flavor Violation in the LHC Era

By the time the next generation of CLFV experiments reach their target sensitivities, the LHC experiments are expected to have analyzed many fb$^{-1}$ of data collected at center-of-mass energies of 7 to 14 TeV, thus exploring New Physics at the TeV scale. The importance and impact of pursuing next generation CLFV experiments is independent of what the LHC data might reveal over the next several years.  As discussed in the previous sections, these and other ultra-rare processes probe New Physics scales that are significantly beyond the direct reach of the LHC, and thus may offer the only evidence of New Physics phenomena should it lie at mass scales significantly above the TeV range. A more optimistic scenario would assume New Physics discoveries at the LHC do occur and ask to what degree do measurements of CLFV processes complement the LHC experiments? The LHC experiments do not, aside from specially-tuned cases, have sensitivities to CLFV processes that approach that of next-generation $\mu^+ \rightarrow e^+\gamma$ and $\mu^- N \rightarrow e^- N$ experiments. Thus, Mu2e probes the underlying physics in a unique manner with a sensitivity that is significantly better than any other CFLV process can hope to accomplish on a similar timescale. Moreover, many of the new physics scenarios for which the LHC has discovery potential predict rates for $\mu^- N \rightarrow e^- N$ in the discovery range for Mu2e (*i.e.,* larger than $10^{-16}$). The points that make up Figure 3.2, for example, correspond to scenarios in which the LHC would discover new phenomena. We have only examined a few specific models here, but the results are representative of the power of muon-to-electron conversion. It is generally understood that an experiment sensitive to $\mu^- N \rightarrow e^- N$ rates at the level of $10^{-16}$ to $10^{-17}$ would have discovery potential that overlaps the parameter space to which the LHC is sensitive and would help constrain that parameter space in a manner complementary to what the LHC experiments can accomplish on their own [1].

## 3.2   Signal   and   Backgrounds   for   Muon   Conversion Experiments

The conversion of a muon to an electron in the field of a nucleus is coherent: the muon recoils off the entire nucleus and the kinematics are those of two-body decay. The mass of a nucleus is large compared to the electron mass so the recoil terms are small. A conversion electron is therefore monoenergetic with energy slightly less than the muon rest mass (more detail is given below). The muon energy of 105.6 MeV is well above the maximum energy of the electron from muon decay (given by the Michel spectrum) at 52.8 MeV; hence, the vast majority of muon decays do not contribute background, subject to an important qualification for negative muons bound in atomic orbit, as discussed below. This distinctive signature has several experimental advantages including the near-absence of background from accidentals and the suppression of background electrons near the conversion energy from muon decays.

When a negatively charged muon stops in a target it rapidly cascades down to the 1S state [21]. Capture, decay or conversion of the muon takes place with a mean lifetime that has been measured in various materials and ranges from less than ~100 ns (high-Z nuclei) to over 2 μs (low-Z nuclei) [22]. Neutrinoless conversion of a muon will produce an electron with an energy that is slightly less than the rest mass of the muon and depends on the target nucleus:

$$E_e = m_\mu c^2 - B_\mu(Z) - C(A),$$

where Z and A are the number of protons and nucleons in the nucleus, $B_\mu$ is the atomic binding energy of the muon and C(A) is the nuclear recoil energy. In the case of muonic aluminum, the energy of the conversion electron is 104.97 MeV and the muon lifetime is 864 ns [22]. An electron of this energy signals the conversion.

At the proposed Mu2e sensitivity there are a number of processes that can mimic a muon-to-electron conversion signal. Controlling these potential backgrounds drives the overall design of Mu2e. These backgrounds result principally from five sources:

1. Intrinsic processes that scale with beam intensity; these include muon decay-in-orbit (DIO) and radiative muon capture (RMC).
2. Processes that are delayed because of particles that spiral slowly down the muon beam line, such as antiprotons.
3. Prompt processes where the detected electron is nearly coincident in time with the arrival of a beam particle at the muon stopping target (e.g. radiative pion capture, RPC).

4.  Electrons or muons that are initiated by cosmic rays.
5.  Events that result from reconstruction errors induced by additional activity in the detector from conventional processes.

A free muon decays according to the Michel spectrum with a peak probability at the maximum energy at about half the muon rest energy (52.8 MeV) and far from the 105 MeV conversion electron energy. If the muon is bound in atomic orbit, the outgoing electron can exchange momentum with the nucleus, resulting in an electron with a maximum possible energy (ignoring the neutrino mass) equal to that of a conversion electron, however with very small probability. At the kinematic limit of the bound decay, the two neutrinos carry away no momentum and the electron recoils against the nucleus, simulating the two-body final state of muon to electron conversion. The differential energy spectrum of electrons from muon decay-in-orbit falls rapidly as the energy approaches the endpoint, approximately as $(E_{endpoint} - E_e)^5$. The spectrum of electron energies that results from muon decays in orbit in aluminum, our target of choice, is illustrated in Figure 3.7 where the most prominent feature is the Michel peak. As described above, the nuclear recoil slightly distorts the Michel peak and gives rise to a small tail that extends out to the conversion energy. Because of the rapid decrease in the DIO rate as the electron energy approaches the endpoint, the background can be suppressed through adequate resolution on the electron momentum. To reduce the DIO background, the central part of the energy resolution function must be narrow and high energy tails must be suppressed. This depends on the intrinsic resolution of the tracker detector as well as the amount of material traversed by conversion electrons.

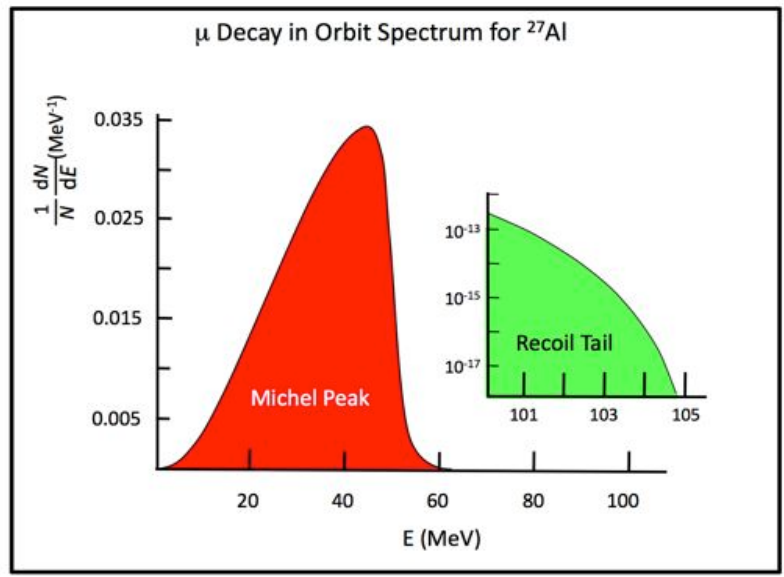

Figure 3.7 The electron energy spectrum from muon decay-in-orbit in aluminum. The recoiling nucleus results in a small tail (inset) that extends out to the conversion energy.

To date, there have been no experimental measurements of the DIO spectrum with sufficient sensitivity near the endpoint energy. The rate is very low and therefore far more stopped muons than in previous experiments are required to see it. The shape of the spectrum near the endpoint is dominated by phase space considerations that are generally understood but important corrections to account for nuclear effects must also be included. The veracity of these corrections is untested by experiment. However, a number of theoretical calculations of the DIO spectra of various nuclei have been done over the years, in particular a recent one by Czarnecki, *et al* [23]. The uncertainty in the rate versus energy near the endpoint is estimated at less than 20%.

Radiative muon capture (RMC) on the nucleus ($\mu^- \, \mathrm{Al} \rightarrow \gamma \nu \, \mathrm{Mg}$) is an intrinsic source of high energy photons that can convert to an electron-positron pair in the stopping target or other surrounding material, producing an electron near the conversion electron energy. Photons can also convert internally. These internal and external rates, by numerical accident, are approximately equal for the Mu2e stopping target configuration. Radiative muon capture can produce photons with an endpoint energy close to the conversion electron energy but shifted because of the difference in mass of the initial and final nuclear states. Ideally, the stopping target is chosen so that the minimum masses of daughter nuclei are all at least a couple of MeV/c$^2$ above the rest mass of the stopping target nucleus, in order to push the RMC photon energy below the conversion electron energy; for aluminum the RMC endpoint energy is 101.9 MeV, about 3.1 MeV below the conversion electron energy. The shape of the photon spectrum and the rate of radiative muon capture are not well known for medium mass nuclei and experiments have not had enough data to observe events near the kinematic endpoint. The electrons that result from photon conversions cannot exceed the RMC kinematic endpoint for the energy of the radiated photon, so the planned energy resolution of the conversion peak (on the order of 1 MeV FWHM including energy straggling and tracking uncertainties) can render this background negligible.

Most low-energy muon beams have significant pion contamination. Pions can produce background when they are captured in the stopping target or surrounding material and produce a high energy photon through radiative pion capture (RPC):

$$\pi^- N \rightarrow \gamma N^*$$

RPC occurs in 2.1% of pion captures for an aluminum target [24]. The kinematic endpoint is near the pion rest mass energy with a broad distribution that peaks at about 110 MeV. If the photon then converts in the stopping material, one sees an electron-positron pair and in the case of an asymmetric conversion, the outgoing electron can be near the conversion energy, thus appearing to be a conversion electron. In addition, the photon can internally convert:

$$\pi^- N \to e^+ e^- N^*$$

and by numerical accident, the internal and external conversion rates are about equal. Thus electrons resulting from photon conversions, both internal and external, can produce background. RPC background can be suppressed with a pulsed proton beam: the search for conversion electrons is delayed until virtually all pions have decayed or annihilated in material. Beam electrons near the conversion energy that scatter in the target, along with the in-flight decay of a muon or pion in the region of the stopping target are other examples of prompt backgrounds.

Cosmic rays (electrons, muons, photons) are a potential source of electrons near the conversion electron energy. If such electrons have trajectories that appear to originate in the stopping target they can fake a muon conversion electron. Identifying an incoming cosmic ray particle can reject these events. Passive shielding and veto counters around the spectrometer and particle identification help to suppress this background. Note that this background scales with the experiment's live time rather than with beam intensity.

Track reconstruction can be affected by other activity in the detector, causing tails in the energy resolution response function that can move low-energy DIO electrons into the signal momentum window. Additional activity in the detector primarily originates from the muon beam, from multiple DIO electrons within a narrow time window, and from muon capture on a target nucleus that results in the emission of photons, neutrons and protons. The protons ejected from the nucleus following muon capture have a very small kinetic energy and are highly ionizing, so the large pulses they leave behind in tracking detectors can shadow hits from low energy electrons, potentially adding to the likelihood of reconstruction errors. Ejected neutrons can be captured on hydrogen or other atoms and produce low-energy photons. Low-momentum electrons can be created in the tracker by photons that undergo Compton scattering, photo-production, or pair production, and by delta-ray emission from electrons and protons. Because of the low mass of the tracker, these electrons can spiral a considerable distance through the detector before they range out, generating a substantial number of in-time hits. Electron-generated hits caused by neutron-generated photons are the most common and difficult to remove form of background activity. Our simulations include this additional activity and its effect on the momentum resolution tails is included in the background estimates described below. The rate of background activity scales linearly with beam intensity. Systematic uncertainties are assigned to account for the uncertainties in the rate of this background activity. The momentum resolution tails are controlled through careful design of the detector and reconstruction software, and by using estimates of track reconstruction quality when selecting physics samples.

## 3.3    Previous Muon Conversion Experiments

Since the time of the discovery of the muon there has been a rich history of searches for charged lepton flavor violation [25].   Experiments using muons to search for charged-lepton-flavor violation are some of the most promising.  They have been constructed to search for muons decaying into an electron and a gamma ($\mu \to e\gamma$), muons decaying into three electrons ($\mu^+ \to e^+e^-e^+$), and the coherent muon to electron conversion process in nuclei ($\mu^-N \to e^-N$).   The present constraints (at 90% CL) from these CLFV searches using muon decay are $Br(\mu \to e\gamma) < 5.7 \times 10^{-13}$ [3], $Br(\mu^+ \to e^+e^-e^+) < 1 \times 10^{-12}$ [4] and from muon to electron conversion $R_{\mu e} < 7 \times 10^{-13}$ [5]. Searches in $\mu \to e\gamma$, $\mu^+ \to e^+e^-e^+$, and $\mu^-N \to e^-N$ are complimentary in that their sensitivity to CLFV is different depending on the underlying new physics model [8]. In fact, if a signal of charged lepton flavor violation is observed, then the relative rates of $\mu \to e\gamma$, $\mu^+ \to e^+e^-e^+$, and $\mu^-N \to e^-N$ can constrain the underlying physics responsible for the observed CLFV interactions.

Steinberger and Wolfe first searched for muon to electron conversion in 1955 [26]. Many other searches have been performed since ([27] - [33]). The techniques employed in the most recent experiments are particularly noteworthy and provide important input for the design of more sensitive experiments such as Mu2e.

In 1988 a search for muon to electron conversion was performed at TRIUMF [33].  A 73 MeV/c muon beam was stopped in a titanium target at a rate of $10^6$ $\mu^-$/sec. A hexagonal time projection chamber located in a uniform 0.9 Tesla axial field was used to measure the energy of electrons. Scintillation counters were used to tag those electron candidates coincident with the arrival of a particle at the stopping target as prompt background. No events were observed with energies consistent with the muon-to-electron conversion hypothesis. However, nine events with momenta exceeding 106 MeV/c were observed. The source of these events was thought to be cosmic rays, a hypothesis that was later confirmed in a separate experiment that measured the cosmic ray induced background with the beam turned off.  The limit from the TRIUMF search was $4.6 \times 10^{-12}$ (90% CL).

The 1993 SINDRUM II experiment, performed at PSI, focused negative muons with a momentum of 88 MeV/c and an intensity of $1.2 \times 10^7$ $\mu^-$/sec on a Titanium target [34]. During a 25 day run a total of $4.9 \times 10^{12}$ muons were stopped.  The electron energy was measured with a spectrometer inside a superconducting solenoid with a 1 Tesla field. The spectrometer consisted of several cylindrical detectors surrounding the target on the beam axis. Two drift chambers provided the tracking while scintillation and Cerenkov hodoscopes were used for the timing of the track elements and electron identification. A scintillation beam counter in front of the target helped to recognize prompt background electrons produced by radiative capture of beam pions or beam electrons scattering off the target. The pion contamination was reduced by a factor of $10^6$ by passing the beam through a thin moderator that reduced the muon flux by 30%. The few surviving pions had very low momenta and a simulation showed that ~ 99.9% of them decayed before reaching the target. Electrons from radiative pion capture in the moderator could reach the target and scatter into the detector solid angle. This background was easily recognized since it was strongly peaked in the forward direction and had a characteristic time correlation with the cyclotron RF. The electron spectrum agreed well with the predictions for muon decay-in-orbit. No events were observed with energies consistent with the muon-to-electron conversion hypothesis resulting in a limit of $4.3 \times 10^{-12}$ (90% CL).

In 2000 SINDRUM II performed a new search for muon to electron conversion using a 53 MeV/c muon beam and a gold target [5]. The conversion energy for gold is 95.6 MeV. During a 75-day live time $4.4 \times 10^{13}$ muons were stopped. After removing forward prompt events, the electron spectrum was well described by muon decays in orbit and no events were observed in the signal region. One electron event, thought to be pion induced, was identified at higher energy. A final limit on muon to electron conversion in gold was set at $7 \times 10^{-13}$ (90% CL).

## 3.4   Overview of Mu2e

Previous muon to electron conversion experiments have not observed events in the signal region, though events at higher energies have been observed and attributed to pion background and cosmic rays. Based on these results there would appear to be considerable room for improvement for an experiment with sufficient muon intensity, momentum resolution and rate capability so long as prompt backgrounds and cosmic rays are controlled. Mu2e proposes to improve on previous measurements by a factor of approximately 10,000 by deployment of a highly efficient solenoidal muon beam channel and a state-of-the-art detector combined with the power and flexibility of Fermilab's accelerator complex. The major improvements implemented for Mu2e that make this significant leap in sensitivity possible are discussed below. The Mu2e apparatus is shown in Figure 3.8.

An integrated array of superconducting solenoids forms a graded magnetic system that includes the Production Solenoid, the Transport Solenoid and the Detector Solenoid. The Production Solenoid contains the Production Target that intercepts an 8 GeV kinetic energy, high intensity, pulsed proton beam. The S-shaped Transport Solenoid transports low energy $\mu^-$ from the Production Solenoid to the Detector Solenoid and allows sufficient path length for a large fraction of the pions to decay to muons. Additionally, the Transport Solenoid attenuates nearly all high energy negatively charged particles, positively charged particles and line-of-sight neutral particles. The upstream section of the Detector Solenoid houses the muon stopping target and has a graded magnetic field. The graded field increases the acceptance for conversion electrons and plays a key role in rejecting beam-related backgrounds. The downstream section of the Detector Solenoid has a nearly uniform field in the region occupied by the tracker and the calorimeter.

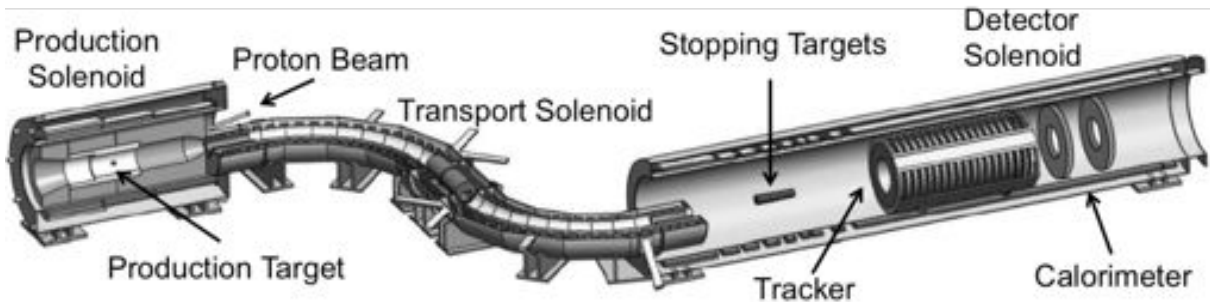

Figure 3.8. The proposed Mu2e apparatus. Shielding and the Cosmic Ray Veto that surround the Detector Solenoid, absorbers inside the Detector Solenoid, and the extinction and stopping target monitors are not shown.

The tracking detector is made from low mass straw tubes oriented transverse to the solenoid axis.  The momentum resolution is dominated by fluctuations in the energy lost in the stopping target and proton absorber, by multiple scattering, and by bremsstrahlung of the electron in the tracker. The calorimeter consists of about 1900 crystals arranged in two disks oriented transverse to the solenoid axis. The calorimeter provides timing and energy information important for providing a fast trigger and efficient particle identification.  The tracker and calorimeter are discussed in detail later in this report.

To increase the sensitivity to muon-to-electron conversion by a factor of 10,000 the intensity of stopped muons will be increased to about $1.5 \times 10^{10}$ Hz. This significant increase in stopped muons is achieved by placing the production target in a graded solenoidal field that varies from $2.5 - 4.6$ Tesla. A proton beam enters the Production Solenoid moving in the direction of increasing field strength, opposite the outgoing muon beam direction and away from the detectors. A large fraction of the confined pions decay, producing muons. The graded field steadily increases the pitch of the muons, effectively accelerating them into the lower field of the Transport Solenoid that transports negatively charged muons within the desired momentum range to the stopping target.[1] The MuSIC R&D effort at Osaka University has validated this approach, demonstrating the principle of high muon yields from a target in a superconducting solenoid for the first time [38].

---

[1] This overall scheme was first suggested by Djilkibaev, Lobashev and collaborators in an earlier proposal called MELC [35] and subsequently adopted in the BNL MECO proposal. Proponents of the muon collider have subsequently adopted their ideas for muon collection in graded solenoids [36][37].

For the Mu2e system, using the QGSP-BERT model of particle production in a GEANT4 simulation, the resulting efficiency is ~0.0019 stopped muons per incident proton.

The SINDRUM method of using beam counters to tag and veto prompt backgrounds can no longer be used at the rates required for Mu2e. Those prompt backgrounds are dominated by pion-capture processes in the stopping target. The relevant timescale was the pion lifetime of 26 nsec. The PSI beam of SINDRUM and SINDRUM-II was a continuous stream of short beam bursts every 20 nsec. Therefore the timescales are comparable and this process limited the experiment. Instead of relying on beam veto counters, we plan to use a pulsed beam with a large separation between pulses compared to the pion lifetime. With such a beam one can wait for pions to decay or interact in matter and thereby largely eliminate the pion-capture background. Since the muon lifetime in a stopping target like Al is long (864 nsec) the loss of muons is acceptable if the time between pulses is about twice the muon lifetime and one simply waits for the pions to decay. Mu2e will therefore search for conversion electrons between proton pulses during times when the flux of particles in the secondary muon beam is relatively low and after the pion-capture process has been reduced by about $10^9$. Fermilab provides a nearly perfect ring for such an experiment. The Fermilab Debuncher, unused in the post-collider era, will be re-purposed (and renamed the Delivery Ring). It will supply a single circulating bunch that will be slow-extracted, providing a pulsed beam to Mu2e every cyclotron period of 1695 nsec for 8 GeV protons. This circumference of 1695 nsec is about twice the muon lifetime in aluminum, and the storage and extraction process can be made to have little or no beam between pulses. Figure 3.9 shows the beam structure and the delayed search window.

The muon stopping target will be located in a graded solenoidal field that varies smoothly from 2.0 to 1.0 Tesla. The active detector will be displaced downstream of the stopping target in a uniform field region. This configuration increases the acceptance for conversion electrons, suppresses backgrounds, and allows for a reduction of rates in the active detector.

The 105 MeV conversion electrons (along with decay-in-orbit electrons from normal Michel Decay) are produced isotropically in the stopping target. The tracker surrounds a central region with no instrumentation: the vast majority of electrons from Michel Decay have radii in the 1 Tesla field that are too small to intercept the tracker and are thus essentially invisible; the few remaining are a source of background we will discuss at length. Here, a final gradient field in the region of the stopping target and before the tracker plays a critical role. Conversion electrons, at 105 MeV, emitted at $90° \pm 30°$ with respect to the solenoid axis ($p_T > 90$ MeV/c) are projected forward and pitched by the gradient into helical trajectories with large radii that intercept the tracking detector.

Electrons in this range that emerge from the target in the direction opposite the tracking detector (upstream) see an increasing magnetic field that reflects them back towards the detector. In addition to nearly doubling the geometric acceptance for conversion electrons, the graded field helps to reject background by shifting the transverse momentum of electrons passing through it. Conversion electrons within the acceptance of the tracker originate from the stopping target with transverse momenta > 90 MeV/c. The graded magnetic field shifts the transverse momentum of the conversion electrons into the range between 75 - 86 MeV/c by the time they reach the tracker. Electrons with a total momentum of 105 MeV/c that are generated upstream of the stopping target, at the entrance to the Detector Solenoid, cannot reach the tracking detector with more than 75 MeV/c of transverse momentum because of the effect of the graded field, eliminating many potential sources of beam-related background.

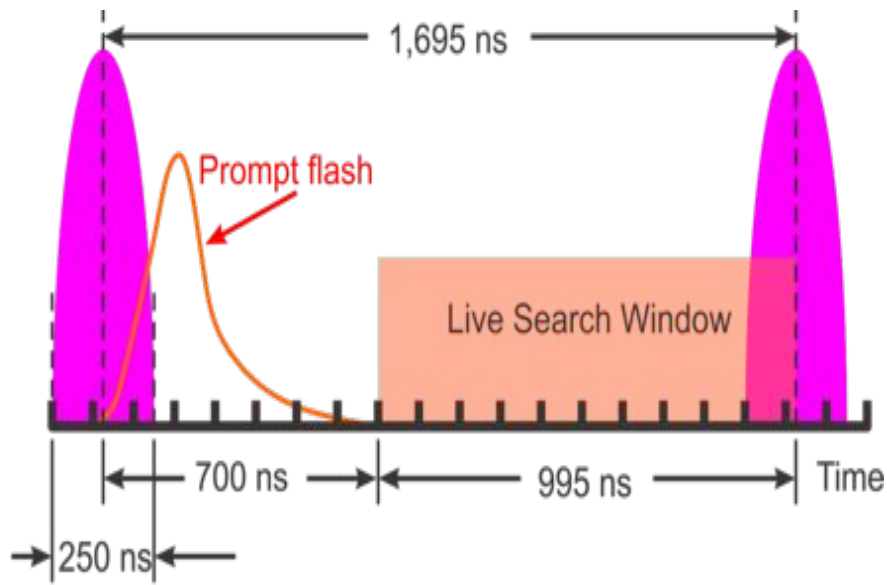

Figure 3.9. The Mu2e spill cycle for the proton beam and the delayed search window that allows for the effective elimination of prompt backgrounds when the number of protons between pulses is suppressed to the required level. The start time and duration of the search window shown in the figure is illustrative. The exact delay and duration are optimized to keep pion-capture backgrounds at a low level.

The detector is displaced downstream of the stopping target in order to:

- reduce the acceptance for neutrons and photons emitted from the stopping target and to allow space for absorbers to attenuate protons ejected by nuclei as part of the muon capture process in the stopping target.
- provide a region for the aforementioned gradient field to pitch conversion electrons into a region of good acceptance. This helps to reduce accidental activity in the detector from beam particles entering the Detector Solenoid and, as stated, from the vast majority of electrons from muon decay-in-orbit, which go undisturbed through the evacuated center of the tracker and calorimeter. The extent and gradient of the field, the size of the central evacuated region, and their geometric locations, are therefore jointly designed to maximize acceptance for conversion electrons while also greatly reducing decay-in-orbit and beam-related backgrounds and occupancy.

• reduce activity in the detector from the remnant muon beam, about half the intensity entering the Detector Solenoid (stopping more of it would require more stopping material, yielding more accidental activity and smearing out the conversion peak because of energy loss straggling). The remnant muon beam enters an absorber (the muon beam stop) designed to minimize albedo that could increase accidental activity in the detectors.

Shielding and a plastic scintillator based cosmic ray veto (CRV) system surrounds the Detector Solenoid. Cosmic rays have been a limiting factor in past experiments and the CRV is designed to ensure that cosmic-ray-induced backgrounds are a subdominant background for the Mu2e experiment.

The next section describes the simulation and reconstruction software used and the selection criteria applied to reduce backgrounds to acceptable levels. We then describe how the methodology of the Mu2e experiment summarized above will limit the total background to less than one event and improve the sensitivity to CLFV by several orders of magnitude relative to past experiments.

## 3.5   Mu2e Simulation, Reconstruction, and Selection

The known processes that may create backgrounds for muon conversion experiments were discussed in general in Section 3.2. In this section we describe the simulation and reconstruction software used to emulate the known background processes and simulate the detector response. The algorithms used to reconstruct the simulated data are also described. The selection criteria used to estimate the background yields and conversion electron acceptance are detailed.

### 3.5.1   Simulation

A common suite of simulation software is used to compute the background yields and signal efficiency described below. The simulation is based on the GEANT4 package [39][40]. Common set-up files are provided that describe the default Mu2e geometry and materials. Detailed magnetic field maps are produced using the OPERA 3D [41] and SOLCALC [42] software packages and assuming the nominal conductor geometries and coil positions. Additional field maps that include variations from the nominal field, based on the fabrication tolerances for the conductor and coils, are generated. The additional field maps are used to evaluate systematic uncertainties.

The geometric description of the setup, illustrated in Figure 3.10, includes details on multiple scales in a single model. They range from the dirt surrounding the detector hall, the building walls and shielding, to individual coils of the solenoids, and to the multiple material layers in the wall of each individual tracker straw. All elements of the muon beam line are included, the HRS, the production target and its supports, the anti-proton absorbers, the collimators, the stopping target foils and their supports, and the muon-beam stop located at the end of the DS. In addition to a detailed description of the straws, the manifolds and support shells of the tracker are modeled. The front-end readout boards are represented but the cables and services are not. Since they are located outside the active region of the tracker, at large radius, we do not expect them to significantly affect the accidental hit occupancy in the tracker. The calorimeter crystals are represented separately and the support structure and readout electronics are also modeled. The CRV model includes the scintillator counters, the aluminum spacers, and the aluminum shell that forms the outside of the modules.

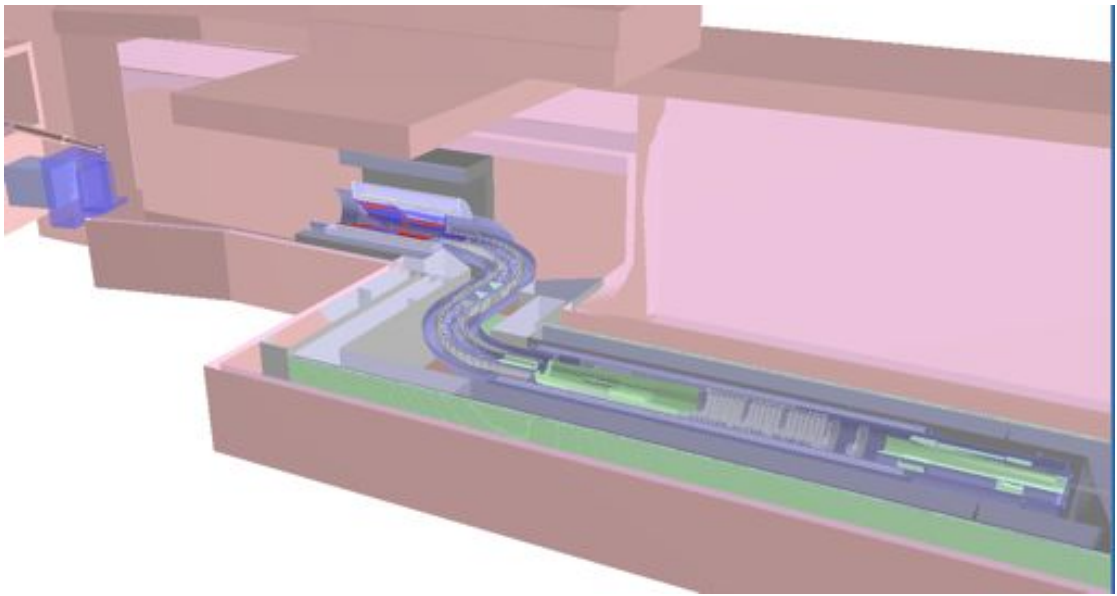

Figure 3.10 A cut-out view of the GEANT4 model of Mu2e used for the simulation results reported.

***Simulating physics processes***

The standard "Shielding" physics list provided with GEANT4 is intended for use by low background experiments and shielding applications. It uses the Bertini Cascade (BERT) model for low energy hadron-nucleus interactions and the Fritiof (FTF) model for high energy hadron-nucleus interactions. The de-excitation of nuclei (P) and the radioactive decay of long-lived isotopes are included. The high precision modeling of neutron interactions (HP) is used for neutrons below 20 MeV. The chiral invariant phase space model (CHIPS) is used for hyperon and anti-baryon production. For most of the studies described below we use a custom physics list "Shielding_MU2E01" for which we change the transition from the Bertini model to the Fritiof model so that it occurs in the 9.5-9.9 GeV range instead of the default 4-5 GeV region. This change improves the agreement between the simulated pion production cross sections and the HARP data.

When muons stop in our aluminum stopping target, they are captured in an atomic excited state. They promptly fall to the ground state. For aluminum, about 39% of the muons will decay in orbit, while the remaining 61% will be captured on the nucleus. The electrons and photons from the atomic cascade to the ground state are modeled in GEANT4, but the products of muon decay in orbit and muon nuclear capture are modeled using custom code. The decay in orbit is modeled using the latest calculations as described in [23]. The spectra and particle multiplicity for the photons, neutrons, and protons emitted in the muon nuclear capture process are based on published data. The custom code is used for muon stopped in our aluminum target. For muon stops in other locations, the GEANT4 models are used for all aspects. Variations of these models are investigated as part of the systematic uncertainties.

To make it practical to perform optimizations of the detector and shielding geometry and materials, the simulations have been subdivided into several stages. The first stage begins the simulation with 8 GeV protons interacting in the production target and tracks all resulting particles to the mid-point of the transport solenoid. We save the location, arrival time, and four-momenta of any particle making it that far. The second stage begins with the output of the first stage and tracks particles to the entrance of the detector solenoid. The third stage propagates the surviving particles from stage two through the upstream portion of the detector solenoid vacuum and records muon and pion stops in the aluminum stopping target. Particles that intersect the tracker or calorimeter are recorded. The fourth and final stage describes the complete Mu2e setup and tracks all surviving particles through the tracker and calorimeter. In this manner we significantly reduce the amount of CPU necessary to study variations in the tracker, for example, since only the fourth stage would have to be re-run. This is much faster than re-doing simulations from primary protons. In addition, the stage approach allows us to employ resampling techniques where appropriate, significantly increasing the effective statistics of the sample in a very CPU efficient manner. The cosmic-ray and antiproton background studies use the same staged technique but with definitions of the specific stages that are better suited to their particular study. The main physics simulation uses range cuts of 1 mm in stage 1, 100 μm in stages 2 and 3, and 10 μm in stage 4, which is the only stage where particles are propagated through the tracker straws. Stages 1 and 2 do not track electrons below 1 MeV. In the subsequent stages all particles are simulated regardless of their energy. $2.1 \times 10^9$ primary proton events have been simulated for the main sample, which corresponds to about 70 micro-pulses. This is about a factor of 100 more statistics than available for the Mu2e Conceptual Design Report (CDR) studies.

The protons are modeled as a delta-function with an arrival time at the production target of t=0. The resulting arrival time of a particle at any given point is later smeared using a realistic timing distribution for the original proton pulse. This allows us to incorporate new estimates of the proton pulse shape (e.g. as part of assigning systematic uncertainties or due to changes in the accelerator beamline design) without having to re-simulate the samples.

To study detector performance with realistic hit occupancies we produce "mixed" events in which we model the accidental occupancy in the tracker and calorimeter over 300-1695 ns of a micro-pulse. All known contributions to the accidental occupancy are included in the mixed events: the "beam flash" particles that come through the transport solenoid or penetrate the detector shielding from the outside, electrons from muon decay in orbit, the neutrons, photons and protons produced by muon nuclear capture in the stopping target, and the muon decay and capture products from muons stopped elsewhere inside the detector solenoid. We start at 300 ns because the occupancies at early times are prohibitively large. The appropriate timing distribution is used for each source of the mixed sample and contributions from neighboring micro-pulses are included. This is primarily important for the neutron contributions since the time it takes for some of the neutrons to thermalize is long compared to the 1695 ns micro-pulse spacing. Standard samples of mixed events contain only the accidental occupancy, so that the same mixed event can be re-used to overlay a signal electron from a conversion event or a track from any of the background processes in different studies. The arrival time of each particle in the event overlay is randomized using the appropriate timing distribution for the process under study. The reconstruction algorithms described below take as input the full mixed event, and not just the subset that contains the signal event overlay. In this manner the efficiencies and yields presented include the effects of the accidental occupancy as well as the effects of an a priori unknown $t_0$ from which to seed the reconstruction algorithms.

To increase the effective MC statistics a re-sampling technique is used at various stages whenever possible. For example, the stop positions and times of muons stopped in the aluminum target are re-used multiple times to generate accidental hits for the mixed events described above. Since the muons are at rest, their decay and capture products are isotropically distributed. By randomly sampling the direction and energy of the decay-in-orbit and nuclear-capture products for each use of a given stop position, large re-sampling factors can be safely used. Muons stopped outside the aluminum target are handled similarly. The beam flash component is also re-used, by randomly sampling the arrival time from the proton pulse time distribution.

***Simulating detector response***

Energy depositions in active detector elements (e.g. tracker straw volume, calorimeter crystal, or a CRV scintillator counter) from the GEANT4 samples described above are used as input to a detector simulation. Since the tracker is the single most precise detector element, and thus drives the experimental sensitivity, a complete simulation – from ionization in the straws to digitization output of the ADCs and TDCs on the tracker frontend – was developed. The simulations for the calorimeter and cosmic veto are less sophisticated, but capture the significant effects.

The tracker signal simulation is described briefly here and documented in detail elsewhere [43]. The tracker signal simulation starts with the energy deposits in the straw gas predicted by GEANT4, for all particles that pass through the tracker including the accidental hit occupancy from the overlay of the mixed events described above. These energy deposits are subdivided into individual ionizations, and processed through a parameterized simulation of the straw drift, the gas amplification, the signal transit, and the electronic amplification and shaping. Electronic signals arriving at the straw ends from all processes are coherently added, and the resultant waveform is digitized at both ends. The resultant simulated digitized signals ('digis') have a discrimination threshold applied and are used as input for the reconstruction.

For the calorimeter and CRV the GEANT4 energy depositions from all particles traversing the active regions of the sub-detectors (including the accidental occupancy from the overlay of mixed events) are summed separately for each crystal and each scintillator counter. The calorimeter energy is smeared to mimic resolutions measured in test beam or estimated from calculations. Readout thresholds are applied to these smeared quantities, which are then used as input for the reconstruction. The energy depositions in the CRV scintillators have readout thresholds applied after the inclusion of the Burke suppression factor, if applicable.

### 3.5.2   Reconstruction

A common suite of reconstruction software is used to compute the background yields and signal efficiency described below. The output of the detector simulation is used as input to the reconstruction software. The track reconstruction software takes as input digitized information from the tracker frontend, just as it will for real data. Since the tracker is the single most precise detector element, and thus drives the experimental sensitivity, sophisticated and robust algorithms for pattern recognition and track fitting have been developed.

***Track reconstruction***

The Mu2e track reconstruction algorithm is described briefly here, and in detail elsewhere [44]. First, 'digis' are converted to hits, by re-interpreting the simulated digital data as physical time and energy. The time distribution of tracker hits in simulated events that include a single signal conversion electron (CE) overlayed with the mixed events is shown in Figure 3.11 from 300-1695 ns, broken down by the origin of the particle that ultimately produced the hits. Tracker hits produced before 300 ns are suppressed at the front end, to keep from saturating the DAQ transmission. A typical CE within the geometric acceptance of the tracker produces 40 hits on average. The initial signal/noise of hits produced by the CE in the live window of a microbunch is roughly 1/85.

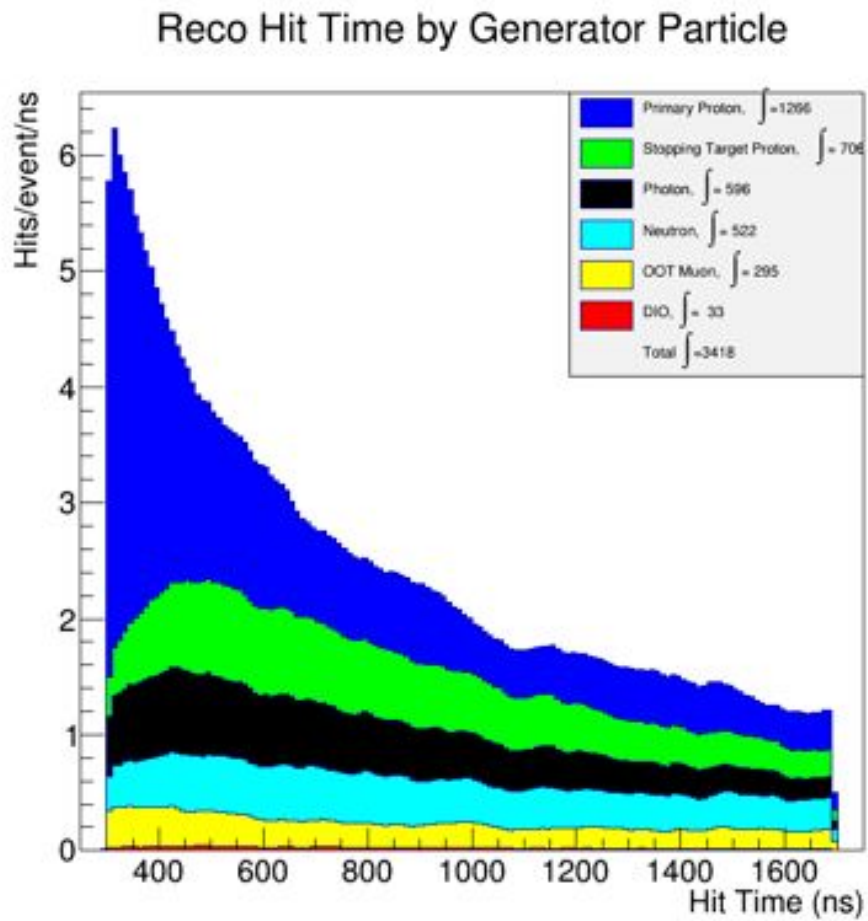

Figure 3.11 Distribution of reconstructed tracker hits from simulated events that include a CE overlayed with the mixed events. Primary proton hits originate from particles not coming from a stopped muon. Stopping target proton, photon, and neutron hits come from particles ejected in muon nuclear capture at the stopping target. Muons that stop "out of target" (OOT) also produce hits via nuclear capture and decay in orbit processes.

Hits are filtered based on their time, energy and position. These simple cuts remove most of the proton and DIO background hits. The principle remaining hit background comes from Compton electrons produced by photons hitting the straw walls. The initial photons are produced both directly from nuclear de-excitation following muon capture, and indirectly from capture of neutrons on material in the DS produced following muon capture. These low-energy electrons move in tight spirals through the tracker, producing tight clusters of hits when projected in the plane transverse to the magnetic field axis. Roughly 90% of these hits are removed by a dedicated algorithm which searches for such clusters of hits. About 90% of the CE hits survive this hit selection and the CE hit signal/noise improves to 1/10.

The remaining tracker hits are grouped together into time clusters, within a window of ~50 nsec, corresponding to the drift + transit time spread of hits produced by the CE. A representative sample of individual CE event hit time distributions is shown in Figure 3.12, with the hits produced by CE tracks and the identified clusters shown. Over 95% of CE events producing at least 20 hits in the tracker are identified as a time peak, and the final signal/noise of the selected hits in a time peak containing CE hits is roughly 6/1.

Time peaks containing at least 15 hits are passed to a geometric pattern recognition algorithm. Approximate 3-d hit positions are assigned to each hit using either the time difference between the signal arrival at the straw ends, or by stereo matching of hit pairs in adjacent panels of a station. The typical position resolution achieved is ~1 mm perpendicular to the straw (the straw radius/$\sqrt{12}$), and ~1 cm along it. A robust helix fit using the 3-d hit positions is used to define an initial track. The helix fit result is used to initialize a least-squares fit, which uses only the perpendicular position information to constrain the fit. The least squares fit uses the wire as hit position, and the straw radius/$\sqrt{12}$ as hit error. The least squares fit produces initial helix parameters and covariance matrix, and an initial track $t_0$. An iterative Kalman filter track fit is seeded with the least-squares parameters, covariance, and $t_0$, using the $t_0$ to define straw drift circles. The Kalman fit accounts for scattering and energy loss in the straw material, as well as the inhomogeneity in the DS field. Left-right hit ambiguity resolution, track $t_0$ refinement, and an outlier hit search are performed using a simulated annealing algorithm while iterating the Kalman filter fit. The final reconstructed momentum is extracted from the Kalman filter fit result, evaluated at the upstream entrance to the tracker.

We note that there is room for improvement in the track reconstruction, in ways that affect both the resolution and the efficiency. For instance, the reconstruction currently uses a linear model of the relationship between the measured drift time and the most probable distance of closest approach to the wire, while simulation includes realistic non-linear effects. The hit error is also assumed to be independent of drift radius. These and other deficiencies will be addressed in planned upgrades of the Mu2e software before commissioning with data begins.

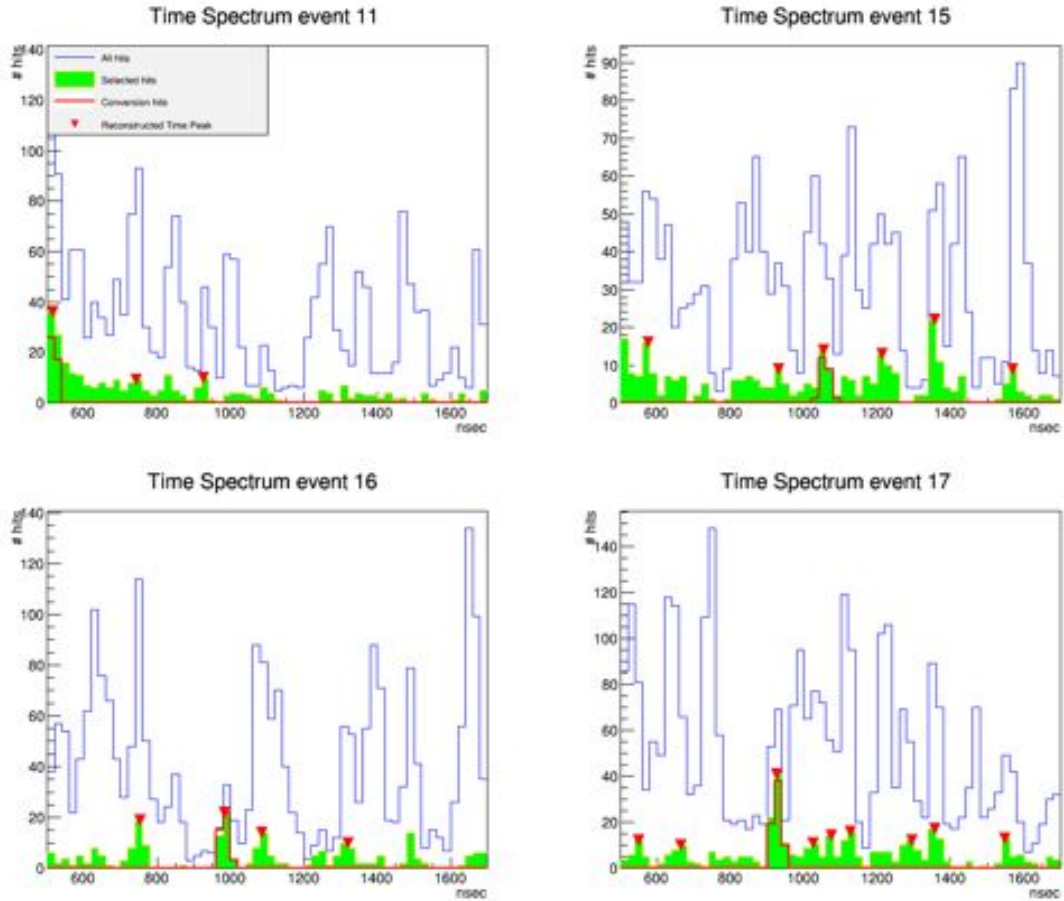

Figure 3.12 Time spectra for tracker hits from 4 separate events that include a CE overlayed with a mixed event. The x-axis range, 500-1695 ns, corresponds to the time window used in the reconstruction software. The blue histogram corresponds to the spectra for all hits, the green histogram are those hits surviving the hit-selection criteria, and the red triangles are the time peaks identified by the reconstruction algorithm. The red histogram corresponds to the hits produced by the CE in the event.

The tracker presents approximately 1% of a radiation length of material to the average CE, with most of the material in the straw walls. This material, while less than the stopping target or internal proton absorbers, also impacts the resolution due to energy loss straggling. The bigger effect of the tracker internal material however is to cause multiple Coulomb scattering. This limits the intrinsic resolution of the spectrometer. Reconstruction effects also impact the resolution. In particular, non-Gaussian tails in the multiple scattering and individual hit resolution functions, errors in assigning the left-right ambiguity to hits, and pattern recognition errors, all contribute to a non-Gaussian high-side tail.

***Calorimeter cluster and cosmic veto reconstruction***

The energy depositions in the calorimeter crystals are clustered using a seed-and-shoulder algorithm adopted from BaBar. Crystals with an energy deposition larger than 10 MeV become "seeds" to which additional crystals may be added to form a cluster. Crystals that are physically contiguous with a seed cluster and that have an energy deposition coincident (<10 ns) with the seed and larger than 1 MeV are added to the cluster. Cluster merging algorithms are applied. The total cluster energy is determined as the sum over all included crystals. The cluster time and position are determined using energy-weighted averages.

A cosmic veto is formed using energy depositions in the CRV scintillators that are coincident with one another. A localized coincidence in three of the four CRV layers is required. Because cosmic rays are incident at a variety of angles and because one layer is allowed to miss, a large number of scintillator hit patterns can qualify as a veto. The allowed hit patterns were determined using dedicated simulations of the four-layer CRV modules and muons incident at angles from 0 to 90 degrees relative to the normal.

### 3.5.3   Selection Criteria

Candidate conversion electron events are required to satisfy requirements using the reconstructed information from the tracker, calorimeter, and CRV. While a rigorous multidimensional optimization has not been performed, the selection requirements are chosen to reduce backgrounds to a low level while maintaining a high signal efficiency.

The intrinsic quality of reconstructed tracks in Mu2e varies widely, due to the variation in production angle of the particles, the random effects of energy loss and scattering in material, the geometric acceptance of the tracker, and the effect of background hits on the pattern recognition and track fit. To select a minimum-quality sample of tracks for analysis, a set of requirements are applied to quantities measured in the reconstruction. No selections based on MC truth are applied. The quality selections include requirements on the number of hits that are active in the fit (outlier hits are de-activated as part of the simulated annealing), the estimated uncertainty on the momentum and track $t_0$, and most importantly, the chi-squared consistency of the fit.

To reduce physics backgrounds, additional requirements are made. We require the measured track pitch lie in a range that excludes electrons from muon and pion decay-in-flight, and high-energy beam electrons entering the DS from the TS. To reduce the cosmic ray background, we require the track origin be consistent with coming from the target and that its maximum radius not intersect the Outer Proton Absorber. To reduce backgrounds from pion-capture processes, we require a minimum time for the track $t_0$ relative to the proton pulse. Backgrounds from muons traversing the tracker are eliminated by employing a particle identification algorithm that combines calorimeter and tracker information.

The full set of selection criteria are listed in Table 3.1. The acceptance × efficiency for the track selection criteria alone is 9.3% for conversion electrons originating in the stopping target. For tracks satisfying the track requirements, the calorimeter and particle identification criteria are 96% efficient. Requiring there be no cosmic veto from the CRV reduces the acceptance by 4.5%. The total signal acceptance is then 0.093*0.96*(1-0.045) = 8.5%.

Table 3.1 Selection criteria used to determine the background yields and signal acceptance. The criteria are successively applied.

| Parameter | Requirement |
|---|---|
| *Track quality and background rejection criteria* | |
| Kalman Fit Status | Successful Fit |
| Number of active hits | $N_{active} \geq 25$ |
| Fit consistency | $\chi^2$ consistency $> 2 \times 10^{-3}$ |
| Estimated reconstructed momentum uncertainty | $\sigma_p < 250$ keV/c |
| Estimated track $t_0$ uncertainty | $\sigma_t < 0.9$ nsec |
| Track $t_0$ (livegate) | 700 ns $< t_0 <$ 1695 ns |
| Polar angle range (pitch) | $45° < \theta < 60°$ |
| Minimum track transverse radius | -80 mm $< d_0 <$ 105 mm |
| Maximum track transverse radius | 450 mm $< d_0 + 2/\omega <$ 680 mm |
| Track momentum | 103.75 $< p <$ 105.0 MeV/c |
| *Calorimeter matching and particle identification criteria* | |
| Track match to a calorimeter cluster | $E_{cluster} > 10$ MeV |
| | $\chi^2$ (track-calo match) $< 100$ |
| Ratio of cluster energy to track momentum | E/P $< 1.15$ |
| Difference in track $t_0$ to calorimeter $t_0$ | $\Delta t = |t_{track} - t_{calo}| < 3$ ns from peak |
| Particle identification | $\log(L(e)/L(\mu)) < 1.5$ |

Figure 3.13 shows the effect of the track selection requirements on the acceptance × efficiency. The first column is normalization. The acceptance selections in the next three columns are loose cuts made on Monte Carlo truth information, just to demonstrate what intrinsic limits the structure of the experiment imposes, independent of any algorithm. The next two columns are the quality selections, followed by three physics selections, described in Table 3.1. The final selection is for a nominal momentum window to separate conversion electrons from DIO.

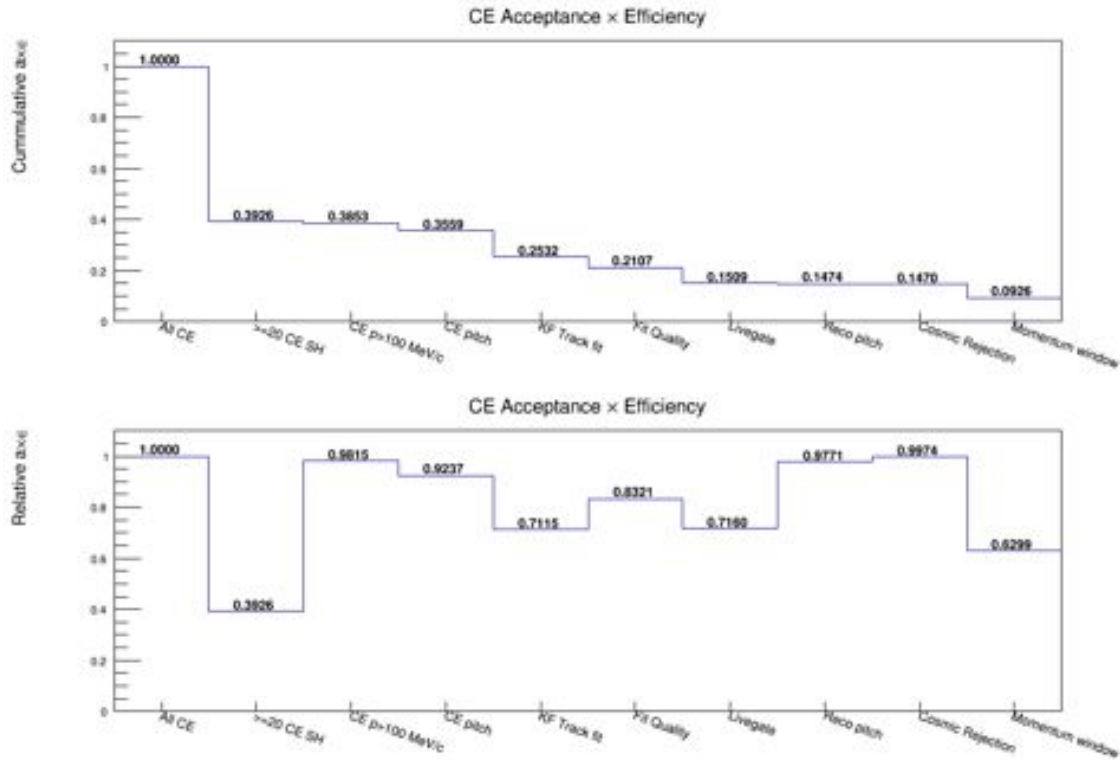

Figure 3.13 The cumulative (top) and relative (bottom) acceptance × efficiency for the track selection criteria as a function of sequential selections.

We note that the quality of the tracks, and hence momentum resolution, varies continuously with the quality parameters. For instance, tracks with at least 30 active hits and a chisquared consistency greater than 0.01 have a 10% narrower core resolution and 20% smaller high-side resolution tail than the standard selection in Table 3.1. However, for a cut-based analysis, the better momentum resolution does not compensate for the reduced CE efficiency and these more restrictive requirements result in a degraded experimental sensitivity. Similarly, tracks outside the selection range of the physics backgrounds cuts will be useful sideband samples for studying and measuring those backgrounds. To maximally exploit the information content of the Mu2e tracker data, we plan to use more sophisticated techniques than simple cuts when performing our final analysis. For instance, we might weight events according to quality, or separate the data into quality-selection bands. We can also reduce the physics background uncertainties using likelihood-based analysis of the physics backgrounds as a function of parameters such as track $t_0$ and polar angle. The criteria listed in Table 3.1, and the results presented below after applying them, should therefore be considered as temporary placeholders for what our actual analysis will be once the experiment has data, and we have had time to develop optimal algorithms.

### 3.5.4    Track momentum resolution

The intrinsic momentum resolution of the tracker, including material and reconstruction effects, is shown in Figure 3.14 for signal conversion electrons that satisfy the track selection criteria. The difference between the momentum predicted by the Kalman track-fit and the MC true momentum, evaluated at the entrance to the tracker, is displayed. The resolution is fit to a Crystal Ball function, which models the (Gaussian) core resolution and the negative bremsstrahlung tail, together with an exponential positive resolution tail. As the core resolution of 118 keV/c is much less than the RMS equivalent spread due to upstream passive materials, the tracker core intrinsic resolution makes a small contribution to the overall experimental resolution. However, the roughly 2% high-side exponential tail has a disproportionate influence on the DIO background yield, as it shifts the fast-falling DIO spectrum to larger momentum.

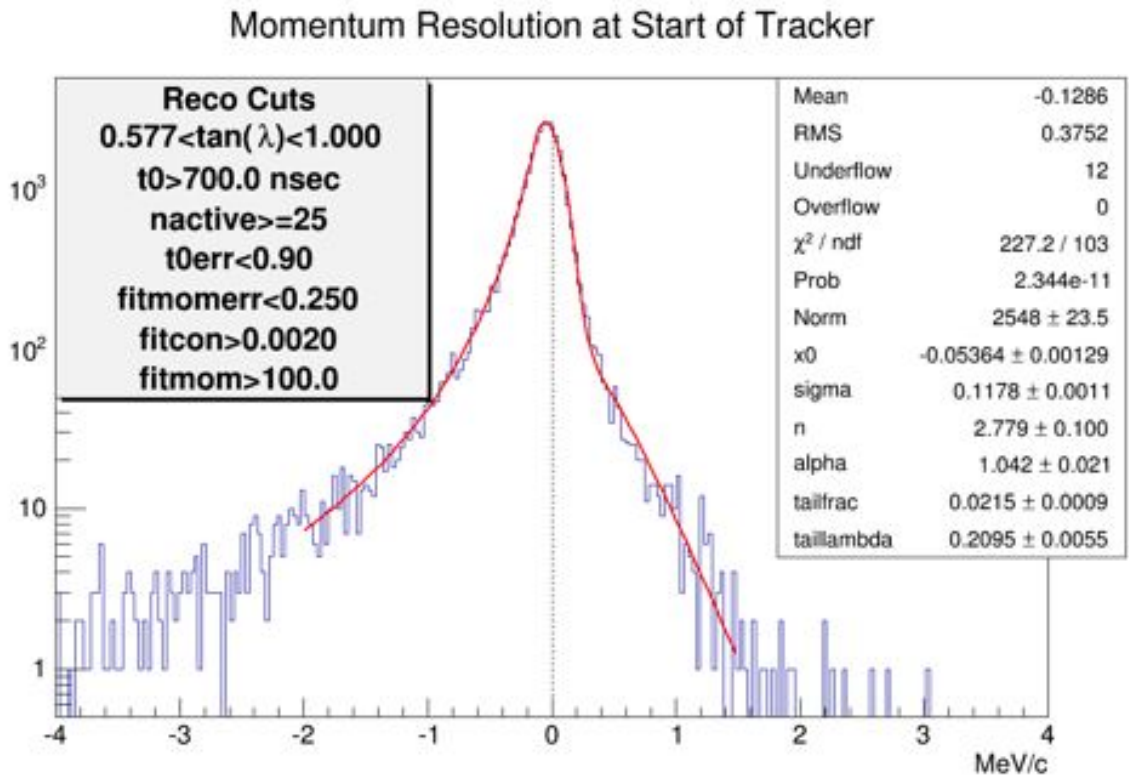

Figure 3.14 The intrinsic resolution of the tracker for conversion electrons surviving the track selection criteria of Table 3.1 and shown in the upper left inset. The red line is a fit to a crystal ball function. The resulting mean (x0), core resolution (sigma), high-side tail fraction (tailfrac), and high-side tail resolution (taillambda) are given in the upper right inset.

# 3.6    Mu2e Background and Sensitivity

The known processes that may create backgrounds for muon conversion experiments were discussed in general in Section 3.2. In this section we explicitly estimate the backgrounds expected by Mu2e. Eliminating potential backgrounds drives many of the design features of the Mu2e detector. We begin with a brief description of the methodology used to estimate each background source. We conclude with our current estimate of the experimental sensitivity.

### *3.6.1*    **Decay-in-orbit Background Yield and Conversion Electron Acceptance**

Muons that capture on Aluminum atoms have a 39.1% chance of decaying while orbiting the Al nucleus [22].  Most of these decay in orbit (DIO) muons produce a high-energy electron, originating from the stopping target.  Since DIO electrons come from stopped muons, their timing distribution is identical to that of a potential conversion electron (CE). The only measureable difference between a DIO electron and a CE is the energy, which is reduced by the energy carried off by the two neutrinos.  The overlap between the DIO energy spectrum and the CE selection window creates an irreducible physics background for Mu2e.  To keep the DIO background within tolerable limits, Mu2e must have very good energy resolution.  To quantitatively estimate the DIO background, Mu2e must have a good understanding of the experimental energy spectrum of the DIO.  These issues are discussed in detail below.  When discussing the DIO background we also discuss the CE sensitivity, as the yield of DIO and CE are strongly coupled.

### *Electron energy spectrum from muon decay-in-orbit*

The energy spectrum of electrons produced by muons decaying in free space has been carefully studied, and was found to agree with the Standard Model predictions [45][46]. To balance energy and momentum, the free muon decaying at rest can give at most half its available energy to the electron. When bound to an atom, interactions with the nucleus distort the muon decay (DIO) electron energy spectrum. In particular, recoil against the nucleus allows the electron energy to exceed the kinematic limit of the free muon decay.

A detailed theoretical prediction of the DIO electron energy spectrum for muons captured on aluminum is shown in Figure 3.15.  The calculation includes relativistic effects and nuclear size and recoil effects [23].  As can be seen, the Michel edge at half the muon mass is softened by the interaction with the nucleus, producing a long tail that extends up to the kinematic endpoint (= predicted CE energy).  The tail of the spectrum falls rapidly due to the limited phase space available to produce neutrinos with energy close to zero. In [23] the authors state that the principal uncertainty in their prediction comes from higher-order radiative corrections, which they estimate to be 'small'.  Detailed numerical estimates of the DIO radiative corrections and estimates of their uncertainty are ongoing.

In [23] the authors show that a 5th order polynomial in the energy difference $\Delta E = E_{DIO} - E_{CE}$ provides a very good approximation to the detailed theoretical prediction, in the range 85 MeV < $\Delta E$ < 105 MeV. As described below, this parameterization is used in the simulations from which the Mu2e DIO background yield is estimated.

Note that the calculation in [23] assumes the DIO occurs after the muon has reached the 1S atomic orbit. If the muon decays while in a higher orbit, the decay electron would have a higher energy, due to a lower Coulomb barrier. Given the estimated time of ~$10^{-13}$ seconds for the muon to reach the 1S state, and a muon decay lifetime of ~$10^{-6}$ seconds [21], the probability that the muon decays before reaching the 1S state is ~$10^{-7}$. The binding energy of the Al 1S state is 464 keV. If we assume all of that energy is available to the decay, we can approximate the decay spectrum of non-1S muon decay by shifting the predicted 1S DIO spectrum near the endpoint by this amount. This shift results in approximately a 10-fold increase in rate 1 MeV from the endpoint energy. Thus the background from DIO from muons not in the 1S state will be negligible (~$10^{-6}$) compared to the background from 1S DIO muons.

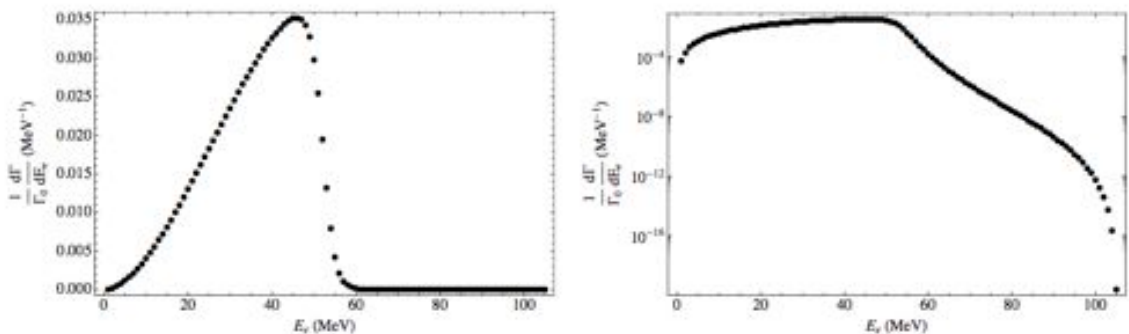

Figure 3.15 Predicted energy spectrum for electrons from muon DIO on Aluminum on a linear (left) and log (right) scale.

***Experimental effects on the DIO spectrum***

The spectrum of Figure 3.15 is affected by experimental effects such as energy loss, Coulumb scattering, and reconstruction efficiencies and resolutions. This is illustrated in Figure 3.16, which shows the results of a simple parametric simulation of the momentum spectrum of DIO background and signal CE. While this simple simulation is not used to estimate the event yields, it is useful for illustrating the individual impact of the various experimental effects on the final momentum spectra. The upper left plot shows theoretical predictions of the momentum spectrum with no experimental effects applied. The upper right plot shows the momentum spectrum after a realistic acceptance, determined from a detailed simulation of the Mu2e tracker, is applied. The acceptance is close to flat over the momentum range relevant to the Mu2e measurement, so the main effect is to reduce the rate of both DIO and CE by the same factor, without affecting their separation. Similarly, the (average) energy loss in material simply shifts both the DIO and CE spectrum by the same amount, maintaining their separation, as shown in the lower left plot. Detector resolution however smears the spectra, creating a broad region of overlap, as shown in the lower right plot. This overlap in reconstructed momentum defines both the DIO background yield, and impacts the final acceptance for CE.

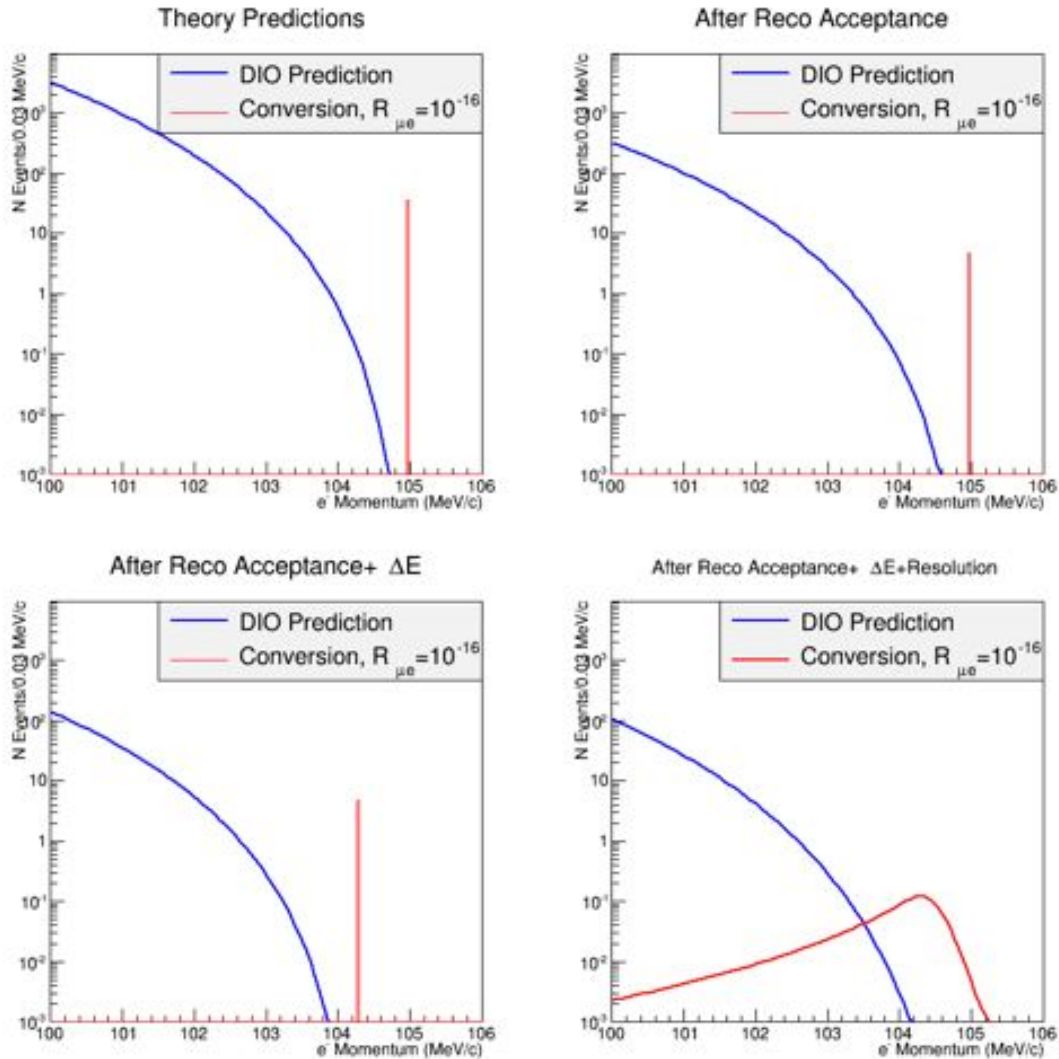

Figure 3.16 Parametric simulations to illustrate how various experimental effects affect the momentum spectrum of electrons from DIO and muon conversion. These plots are normalized to the nominal $6.7 \times 10^{17}$ stopped muons expected for the full Mu2e program. Upper left are theoretical predictions. Upper right are after applying realistic detector acceptances. Lower left is after acceptance and average energy loss effects. Lower right is after all reconstruction effects. Note the logarithmic scales.

Interactions in material upstream of the tracker are the dominant contribution to the energy loss and resolution smearing depicted in Figure 3.16. The Mu2e spectrometer volume contains passive materials that are necessary for the operation of the experiment.

Electrons produced in the stopping target can interact with those materials, which reduces and degrades their energy. Figure 3.17 shows the momentum spectrum of electrons from CE originating in the stopping target, at a point just upstream of the tracker, as predicted by our detailed GEANT4 simulation. The long negative tail comes from radiative energy loss (Bremsstrahlung), while the core width comes mostly from straggling in the ionization energy loss. The FWHM of the momentum of 700 keV/c corresponds to a (Gaussian) 300 keV/c RMS. The largest material effect comes from the stopping target itself. A careful optimization of the stopping target mass and geometry has been performed to balance the effects of stopping power and energy degradation. The other passive material upstream of the tracker is the Inner Proton Absorber (IPA). This material slows or stops some of the protons produced in nuclear decay following muon capture, thereby protecting the tracker from high hit rates and large charge deposition. Compared to the CDR, the IPA has been reduced in mass by a factor of 2, which was found to improve the overall sensitivity to $R\mu e$. The effect of residual gas in the DS vacuum ($10^{-4}$ Torr) is negligible.

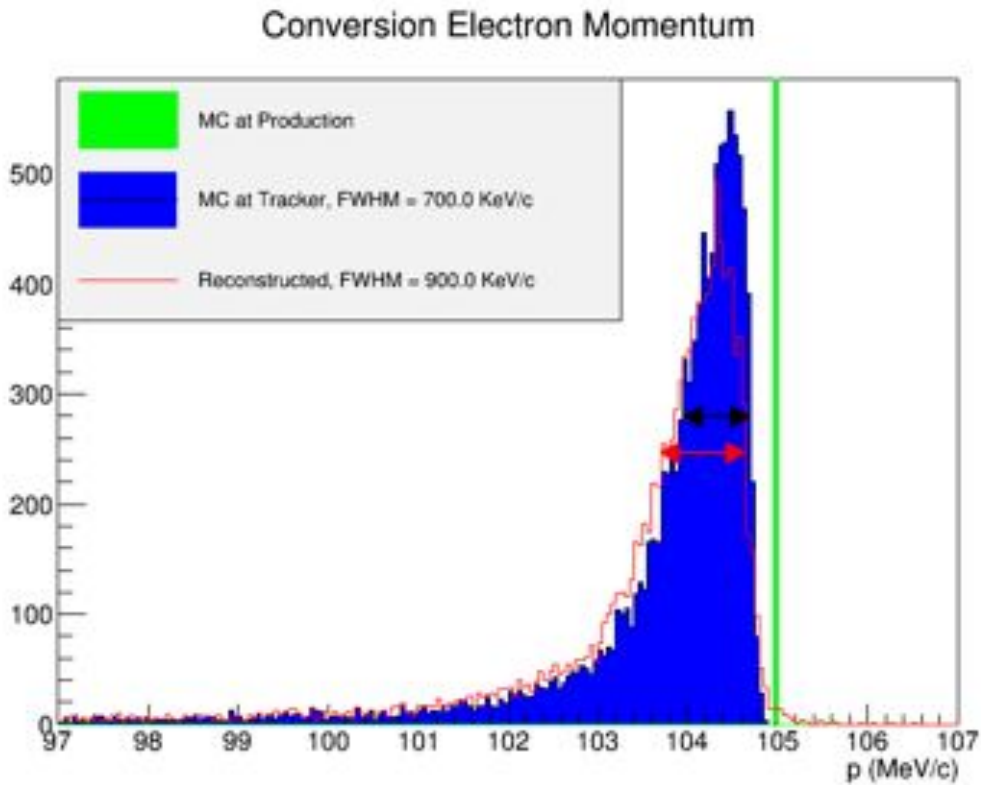

Figure 3.17 The momentum spectrum of conversion electrons at the entrance to the tracker as estimated by a detailed GEANT4 simulation of the Mu2e apparatus. The spectra at production (green), after interaction in upstream material (blue), and after reconstruction (red) are shown.

Figure 3.17 also shows the momentum spectrum from the same CE reconstructed in the tracker. Material in the tracker itself further shifts and broadens the spectrum. In addition, reconstruction effects introduce a high-side tail.

***Estimate of the DIO background yield and CE acceptance***

We predict the DIO background and CE yield using the same detailed simulation and reconstruction software for both the DIO and CE events as described in Section 3.5. To improve the statistical resolution, the DIO momentum is generated flat between 95 and 105 MeV, and events are weighted according to the cross section predicted by the formula in [23]. Flat generation plus weighting provides better statistical precision in the high-momentum part of the spectrum, where the background tracks are most likely to originate. To emulate realistic tracker occupancies, the DIO and CE events are overlaid with the mixed events (cf. Section 3.5.1) and the track reconstruction algorithm is run, exactly the same for DIO and CE events. The selection criteria of Section 3.5.3 are applied.

Figure 3.18 shows the reconstructed momentum spectrum of selected tracks, measured at the entrance of the tracker, from the DIO background. Overlaid is the expected signal from conversion electrons assuming $R\mu e = 1 \times 10^{-16}$, predicted by the full Mu2e simulation. Both plots contain many hundreds of times more data than are expected for Mu2e, but are normalized to the $6.7 \times 10^{17}$ muon stops expected in the nominal Mu2e run. Selecting tracks with momentum between 103.75 and 105 MeV/c results in a DIO background of $0.22 \pm 0.03$ events, and a CE Single Event Sensitivity (SES) of $2.6 \pm 0.07 \times 10^{-17}$, where the quoted uncertainties are due to limited Monte Carlo statistics and corrections for particle-ID and cosmic veto requirements have not yet been included.

The DIO histogram in Figure 3.18 shows significant single-bin fluctuations, in spite of the large number of simulated events. These fluctuations come from very rare single events in the far high-side tail of the momentum resolution. This is demonstrated in Figure 3.19, which plots the difference between the reconstructed and true momentum of DIO electrons in the signal momentum window defined above. The tail portion of the resolution (defined as $\Delta p > 500$ keV/c) constitutes over 2/3 of the DIO entries in the signal window. Improving the track reconstruction resolution tails will be a priority for future optimization studies – particularly through better handling of non-linear effects that degrade the single-hit resolution.

***Systematic Uncertainties***

Several effects introduce systematic uncertainty affecting the estimated DIO background yield or the conversion electron acceptance. The largest direct effects come from the theoretical prediction of the DIO spectrum, and the determination of the absolute momentum scale of reconstructed tracks. Indirect effects from uncertainties in the rates of accidental hits in the tracker, and uncertainties in the tracker simulation, also contribute.

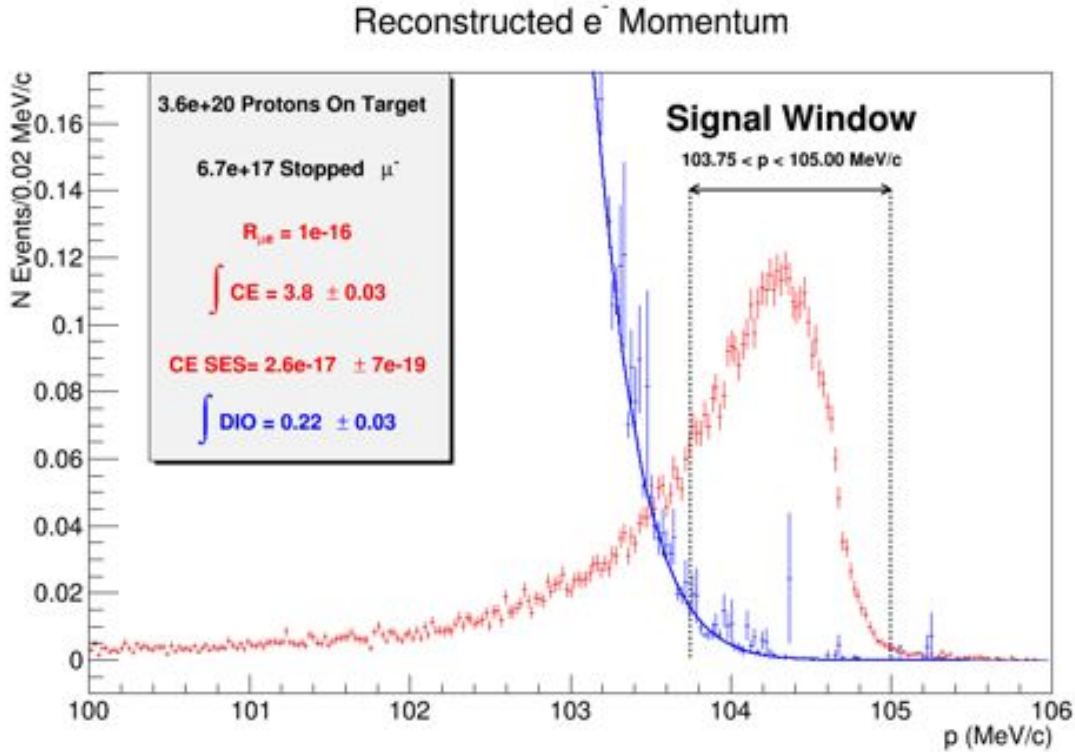

Figure 3.18 The simulated reconstructed momentum spectrum for DIO events (blue) and conversion electron (CE) events surviving the track selection criteria and assuming $R_{\mu e}=10^{-16}$. The distributions are each normalized to the total number of muon stops expected for $3.6 \times 10^{20}$ protons on target.

### *Uncertainties in the predicted DIO spectrum*

The predicted DIO spectrum in [23] does not include higher-order radiative effects, which may affect the shape of the spectrum near the endpoint, thereby changing the DIO background estimate. Naively, we expect the radiative effects to be of order α, and to move the DIO events out of the signal momentum region, not into it, as the radiated photon will take energy away from the DIO electron. A detailed higher-order calculation of the DIO spectrum including radiative effects is in progress but there are no results yet. In the meantime, we set the systematic uncertainty associated with the predicted DIO spectrum by drawing on comparisons with radiative effects in Kaon physics [47] that indicate the DIO background will increase by no more than 20%, or 0.04 events after all selection criteria have been applied.

### *Uncertainty from the absolute momentum scale*

An uncertainty in the momentum scale of the tracker affects both the CE acceptance and the DIO background yield. This is shown in Figure 3.20, which plots the change in the integral yields for DIO and CE (for $R\mu e = 1 \times 10^{-16}$) in the nominal momentum signal window from Table 3.1, as a function of a putative shift in the momentum. Mu2e requires an independent verification of the tracker momentum scale with an accuracy of 1/1000, or 100 keV/c at the CE momentum [48]. Propagating this momentum scale uncertainty results in a change of $^{+0.09}_{-0.06}$ in the DIO integral rate, and $^{+0.20}_{-0.23}$ in the CE integral rate, as shown in Figure 3.20. To reduce this uncertainty in the DIO background, we can intentionally set the momentum selection window 100 keV/c higher than the optimal value. As shown in Figure 3.21 this reduces the uncertainty on the DIO background by a factor of 1.5, and keeps the expected DIO background below 0.2 expected events (at the 1 $\sigma$ level). This protection comes at the cost of losing on average 7% (relative) of the expected CE yield, or up to 15% (relative) of the expected CE yield, in the case that the momentum scale is under-estimated by 100 keV/c (1 $\sigma$).

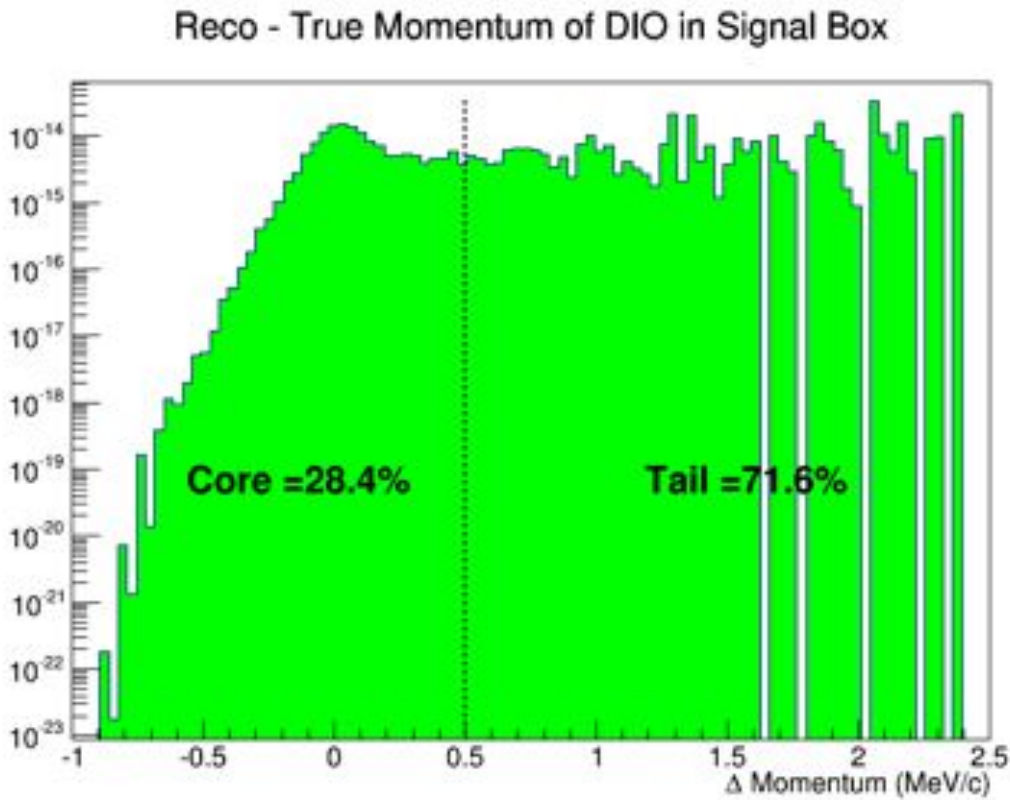

Figure 3.19 The difference ($\Delta$) between the reconstructed momentum and the true momentum for DIO electrons that survive the selection criteria and fall into the signal momentum window. The contribution from the core resolution is defined to be those electrons for which $\Delta p < 500$ keV/c, while the tail is defined to be those for which $\Delta p > 500$ keV/c.

An alternate strategy for dealing with momentum scale uncertainty is to use the DIO spectrum itself as an internal reference, by fitting the measured spectrum outside the signal window. In this case, the momentum scale becomes a nuisance parameter of the experiment, and the DIO background in the signal window is directly estimated by extrapolating the fit into the signal window. A toy Monte Carlo study of this strategy is shown in Figure 3.22, which displays the results of several independent toy experiments, each made with the full statistics expected by Mu2e. Each experiment is conducted by making a random sampling of the DIO and CE spectra from detailed GEANT4 simulations after the application of all selection criteria and normalizing to the number of stopped muons expected for the full Mu2e run. The DIO spectrum in the range 100 MeV/c < p < 102.5 MeV/c is fit to a polynomial function inspired by the approximation to the theoretical DIO endpoint spectrum given in [23], but including parameters to model experimental acceptance, efficiency, and resolution effects. The DIO fit is then extrapolated into the signal window to provide an estimate of the DIO background in each experiment. The average estimated DIO background value over an ensemble of 1000 toy experiments is consistent with the (input) central value from the GEANT4 simulation, and has an average statistical uncertainty of 0.06 events.

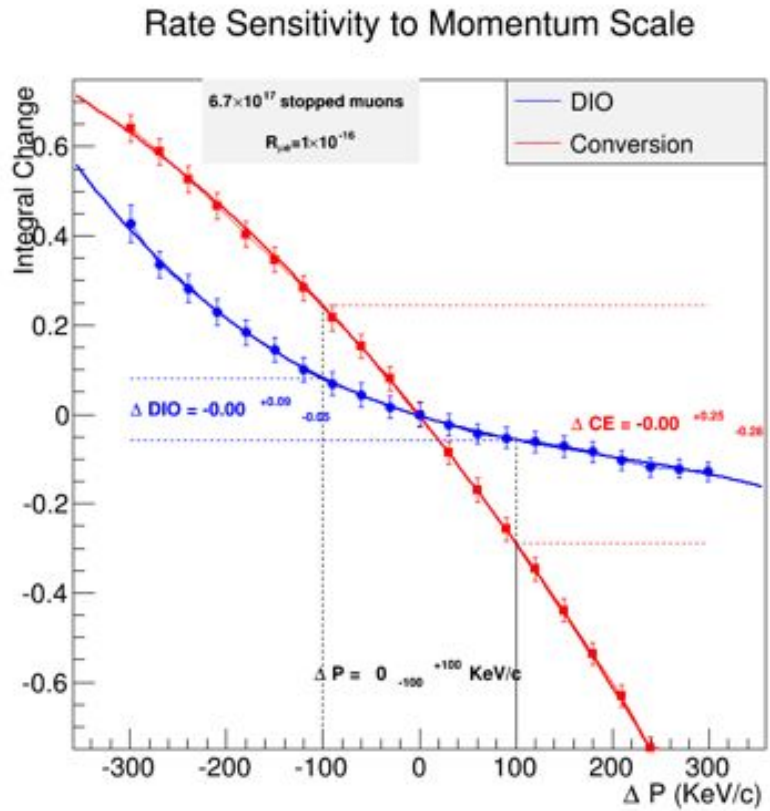

Figure 3.20 Expected change in the DIO and CE yields for the nominal momentum signal window. The changes corresponding to an uncertainty of 100 keV/c on the momentum scale are illustrated.

The DIO background self-calibration strategy relies on accurate estimates of the DIO spectrum shape, the experimental distortions to that spectrum coming from momentum resolution and acceptance, and the backgrounds to that spectrum. Preliminary studies show that in-situ measurements of the momentum resolution function performed with electrons produced from cosmic rays, combined with Monte Carlo estimates of the acceptance dependence on momentum, are adequate for estimating the DIO background with a precision comparable to the theoretical error (0.04 events). This self-calibration method can also provide an estimate of the absolute acceptance × efficiency. Further studies are needed to understand the full systematic uncertainties implied by this method.

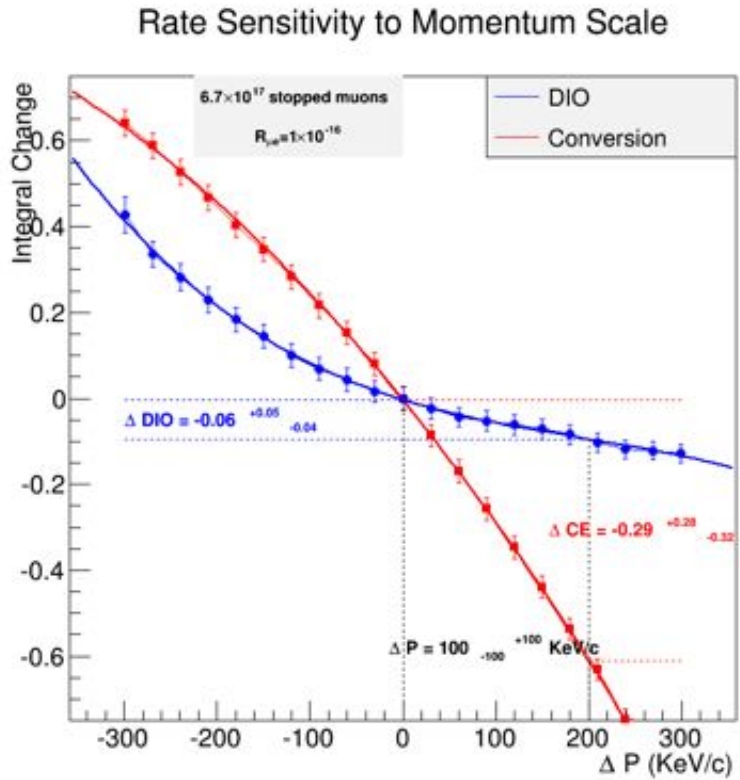

Figure 3.21 Expected change in the DIO and CE yields for a momentum signal window that has been shifted up by 100 keV/c. The changes corresponding to an uncertainty of 100 keV/c on the momentum scale are illustrated.

### *Uncertainties in the accidental hit rate*

Many of the processes that contribute accidental hits in the tracker have large uncertainties in either the rate or spectrum. To obtain robust estimates of the DIO background, we must evaluate the tracker performance for a reasonable range of those uncertainties.

Figure 3.23 shows the response of the track reconstruction to coherent increases (and decreases) in the rate of all the processes that contribute accidental tracker hits. The scale factor '0' corresponds to no accidental hits (ie. just the CE without the overlay of a mixed event), and '1' corresponds to the nominal accidental hit rates in the tracker, as shown in Figure 3.11. The upper plot of Figure 3.23 shows the decrease in the CE track reconstruction efficiency as the rate of accidental hits is scaled. The efficiency response is roughly linear, with a loss of 0.7% absolute efficiency for each unit factor in the accidental hit rate. The lower plot shows how the momentum resolution parameters vary with the rate of accidental hits in the tracker. The parameters are extracted from a fit to the momentum resolution, as depicted in Figure 3.14. Both the core resolution and the resolution tails are seen to be constant within uncertainties over the range of accidental hit rates explored.

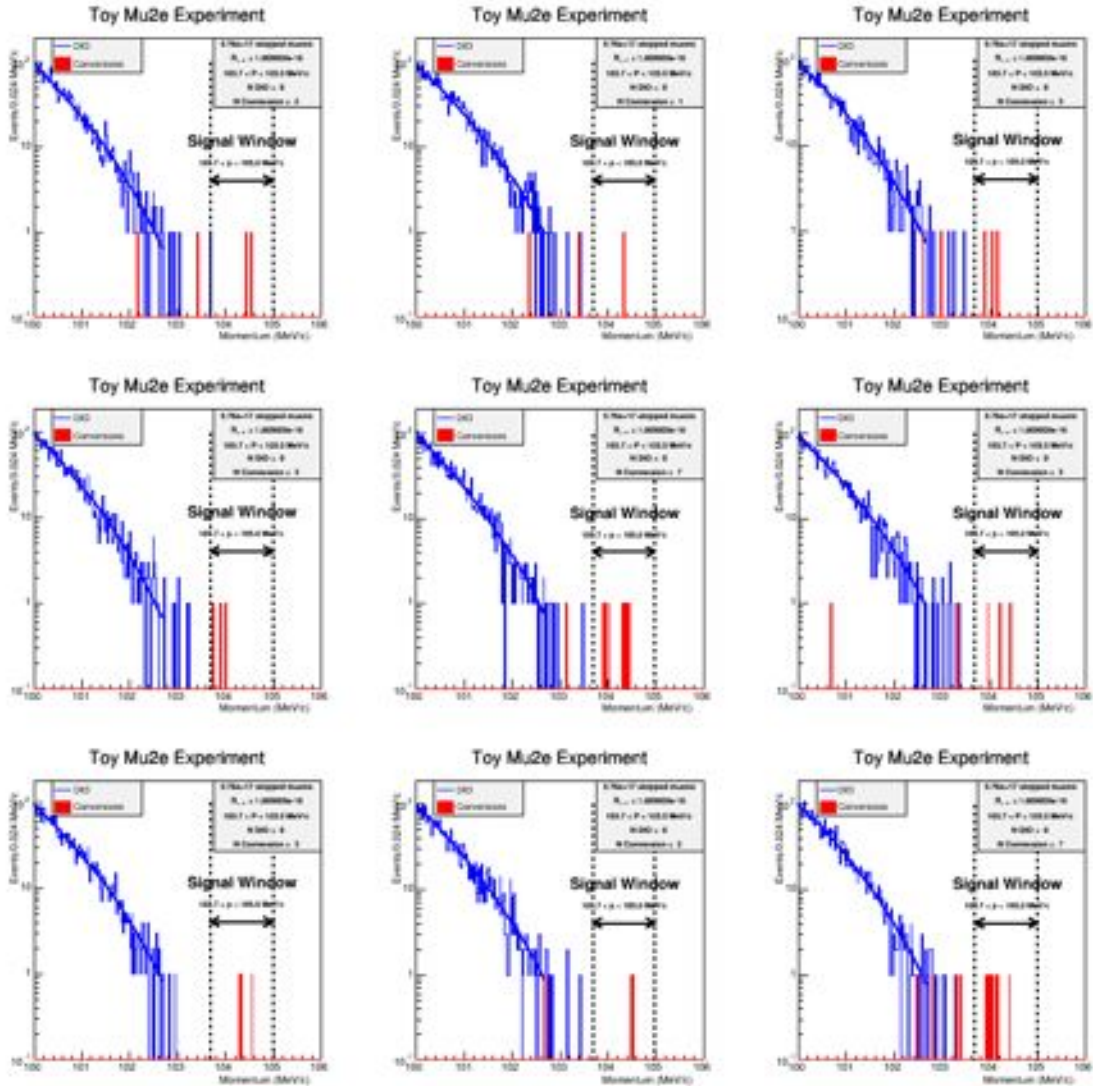

Figure 3.22 Nine toy Mu2e experiments, based on GEANT4 simulations assuming for $R\mu e = 10^{-16}$, each with the full expected Mu2e statistics. The blue histogram is from DIO, the red from CE events. The DIO spectrum in the range 100 MeV/c < p < 102.5 MeV/c is fit to a polynomial, and extrapolated into the signal window to estimate the DIO background.

The coherent scaling of the tracker hit rates depicted in Figure 3.23 is an overly conservative thing to do since the various processes contributing the accidental hits have independent sources of uncertainty that affect the track reconstruction performance in differing ways. To quantify the effect of these uncertainties on the DIO background yield and CE acceptance, we perform dedicated simulation studies, where the rate of each individual physical source accidental hits is varied within its uncertainties. For instance, the rate of neutrons produced in muon capture in aluminum was measured to be 1.26 ± 0.06 [49]. The spectrum of those neutrons however is uncertain, and we have evaluated several possible spectra to understand the range of accidental hit rates induced by those neutrons, finding a variation of < 20%. In addition, the process by which neutrons produced in muon capture result in a tracker hits relies on the GEANT4 modeling of neutron interactions. Different models of neutron interactions are known to predict neutron capture rates that differ by less than a factor of two [50]. Consequently, we estimate the uncertainty associated with the neutron-induced accidental hits by simulating CE reconstruction with the neutron-induced hit rate increased by a factor of two.

Similarly, we have propagated the effect of uncertainties in the energy spectra and rates of photons and protons emitted following muon capture on Al. The photon spectrum and rate is well measured [21], and the proton rate and spectrum doesn't have a large impact on track reconstruction, as the hits from protons are mostly eliminated early in the tracker hit selection. We assume a 100% uncertainty in the rate of accidental hits induced from muons that stop outside the aluminum target. The results of the individual scaling tests are shown in Figure 3.24. The background process being scaled and the scaling factor used are listed on the x axis. The top plot shows the impact on the CE acceptance × efficiency, the bottom the impact on the momentum resolution parameters. No change is seen on the momentum resolution due to these changes in accidental hit rates. The change in CE efficiency is assessed as a systematic uncertainty as recorded in Table 3.2.

We note that the AlCap experiment [51] at PSI, a collaboration which includes Mu2e members, is studying the rate and energy spectrum of particles produced in the nuclear capture of muons on aluminum. This experiment was designed to greatly improve our understanding of the processes relevant to Mu2e, in time to be useful for Mu2e analysis. The proton, neutron, and photon spectra and rates measured by AlCap will constrain the uncertainties affecting the estimated rates of accidental hits, which will reduce the systematic uncertainties shown in Table 3.2.

The nominal Mu2e simulation assumes 100% of the straws are functioning. The tracker requirements [52] allow up to one station's worth of dead straws during normal operation, due to the sum of all potential causes of straw inefficiency. We model the degraded tracker by disabling the hit reconstruction for all the straws in a randomly-selected single plane (6 sectors), which simulates the effect of a major gas leak. We also disable 576 randomly-selected straws throughout the rest of the tracker, to simulate electronics failure or inefficiency. Figure 3.24 also shows the effect of simulating the 'degraded' tracker on the CE efficiency and resolution. The change in CE efficiency is assessed as a systematic uncertainty as recorded in Table 3.2.

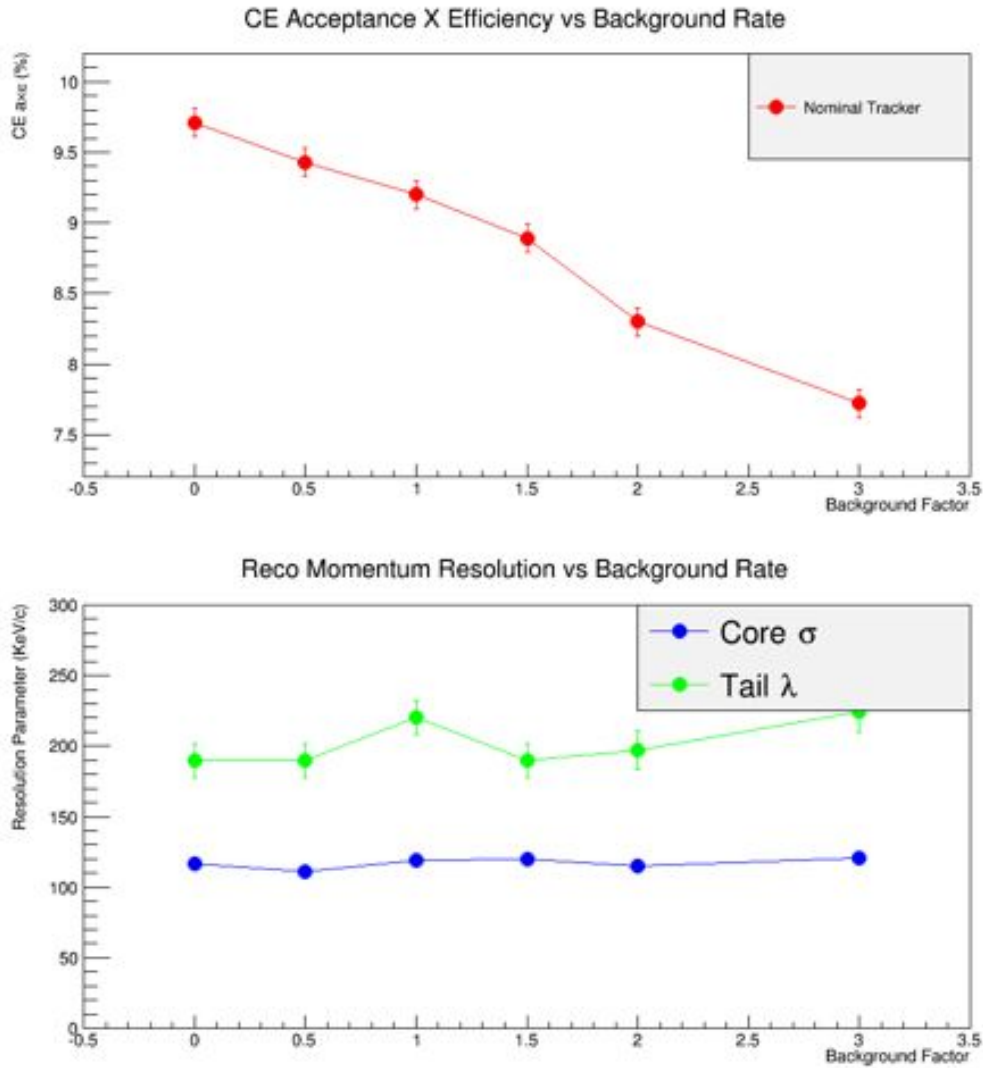

Figure 3.23 Changes to the track reconstruction efficiency (top) and resolution (bottom) as a function of the rate of accidental (background) hits in the tracker. A factor of '1' corresponds to the nominal accidental rates expected for Mu2e and used to construct the mixed events (cf. 3.5.1), while '0' corresponds to no additional activity and '2' corresponds to twice the nominal rate of accidental hits. The resolutions in the bottom plot are determined from a fit to a crystal ball function as shown in Figure 3.14.

### Uncertainties due to variations of the magnetic field

The specifications of the Mu2e solenoids allow for up to 5% variations in the absolute field, as well as variations in the gradient regions, due to fabrication tolerances associated with the conductor geometry and with the placement of the coils. To evaluate the impact of these variations, we calculated 100 alternate field maps by randomly sampling each of the relevant fabrication tolerances. Additional field maps were calculated assuming particular systematic effects (e.g. assuming the conductor was systematically wider or thicker than nominal, but still within specifications). Among these many alternate field maps, those that gave the largest excursions from the nominal field were identified. The standard CE simulation and reconstruction was performed using these alternate field maps, and the CE acceptance × efficiency was evaluated under those conditions. No significant change in the CE acceptance or resolution was observed, within the statistical limits of the test (1% relative). We conclude that possible variations in the magnetic field will not impact the reconstruction, provided the magnetic field is accurately mapped.

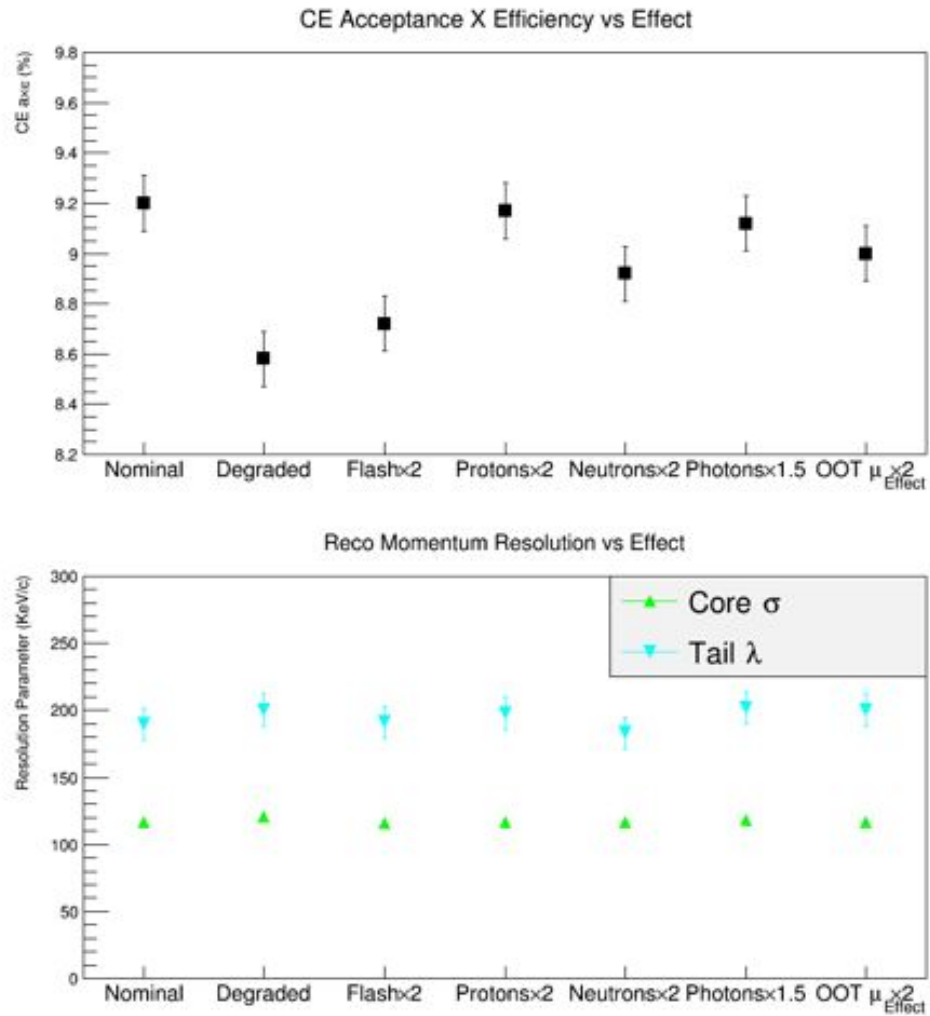

Figure 3.24 Changes to the track reconstruction efficiency (top) and resolution (bottom) for variations in the accidental hit rate induced by various underlying processes. The effect of simulated a 'degraded' tracker is also shown.

Table 3.2 Summary of systematic uncertainties on the DIO background yield and the conversion electron (CE) single-event-sensitivity. These were quantified using the methodologies described in the text.

| Effect | Uncertainty in DIO background yield | Uncertainty in CE single-event-sensitivity ($\times 10^{-17}$) |
|---|---|---|
| MC Statistics | ±0.02 | ±0.07 |
| Theoretical Uncertainty | ±0.04 | - |
| Tracker Acceptance | ±0.002 | ±0.03 |
| Reconstruction Efficiency | ±0.01 | ±0.15 |
| Momentum Scale | +0.09, -0.06 | ±0.07 |
| μ-bunch Intensity Variation | ±0.007 | ±0.1 |
| Beam Flash Uncertainty | ±0.011 | ±0.17 |
| μ-capture Proton Uncertainty | ±0.01 | ±0.016 |
| μ-capture Neutron Uncertainty | ±0.006 | ±0.093 |
| μ-capture Photon Uncertainty | ±0.002 | ±0.028 |
| Out-Of-Target μ Stops | ±0.004 | ±0.055 |
| Degraded Tracker | -0.013 | +0.191 |
| Total (in quadrature) | +0.10, -0.08 | +0.35, -0.29 |

### *Results*

Using the detailed simulation, reconstruction, and selection described in Section 3.5 and assuming the nominal $3.6 \times 10^{20}$ protons on target, the DIO background yield and CE acceptance have been estimated. For nominal conditions the predicted DIO background yield is $0.22 \pm 0.03$, and the CE SES is $(2.6 \pm 0.07) \times 10^{-17}$, where the uncertainties are statistical only. Correcting for the predicted electron particle ID (PID) efficiency (96%), the dead time due to the Cosmic Ray Veto rejection (4.5%), and including the systematic uncertainties described above and summarized in Table 3.2, we arrive at a final prediction of $0.2 \pm 0.03^{+0.09}_{-0.07}$ DIO events, and a CE single-event-sensitivity of $(2.8 \pm 0.07^{+0.32}_{-0.27}) \times 10^{-17}$, where the first error is statistical and the second is systematic. Several strategies for improving the reconstruction algorithms and analysis techniques that should reduce the DIO background, improve the sensitivity, and reduce the uncertainties in future, have been discussed and are being pursued.

### 3.6.2 Pion-Capture Background Yields

Pions that survive to arrive at the aluminum stopping target during the delayed live gate can potentially give rise to a large background from the $\pi^- + \mathrm{Al} \rightarrow \gamma^{(*)} + X$ process. This radiative pion capture (RPC) process occurs promptly as the pion stops in the aluminum. Since the pion lifetime (26 ns) is short relative to the lifetime of muons captured on aluminum (864 ns), these backgrounds can be suppressed by delivering a pulsed proton beam, minimizing the tails of the proton bunch, eliminating out-of-time protons, and by employing a delayed live gate. In combination these mitigations achieve a suppression of about 17 orders of magnitude.

Backgrounds from the $\pi^- + \mathrm{Al} \rightarrow \gamma^{(*)} + X$ process come from two sources, which we will refer to collectively as "RPC" background:

- radiative processes where the stopped pion is captured by the Al nucleus and radiates an on-shell photon $\gamma$, which then undergoes external electron-positron pair production, and
- internal conversion processes, where a virtual photon $\gamma^*$ is emitted upon pion capture, internally converting to an electron-positron pair.

The resulting electron will contribute as background if it is reconstructed within the delayed live gate and with a momentum that lies within the chosen signal region of the analysis. In contrast to the internal-conversion process, the background contribution from the radiative process is driven by the amount of material that can induce the external electron-positron pair conversion. Thus, the theoretical descriptions for each of the pion-capture sources are largely independent of one another and we treat them separately.

Because the final state $X$ is not unique but can assume any number of nuclidic states, the radiated photon energy is not monochromatic, but rather follows a spectrum that must be predicted or measured. The radiated photon energy spectrum for pion capture on aluminum has never been measured. Instead we use the measured photon energy spectrum for pion capture on magnesium [24], which is not expected to significantly differ from the aluminum spectrum. The rate of radiative captures is measured on a wide range of nuclei and is $(2.15 \pm 0.20)\%$ for pion captures on magnesium, which is the assumed RPC rate for aluminum.

The theoretical framework for internal conversions has been presented in References [53][54]. The relevant quantity is the conversion coefficient, which is expressed as a double-differential quantity with respect to the virtual photon mass and the energy asymmetry of the produced electron-positron pair. The coefficient depends on two parameters, the mass of the nuclear final state $X$ and the virtual photon energy, the latter of which is assumed to follow the same distribution as for the radiative process described in the previous paragraph. The internal-conversion background is normalized to the radiative process by the integrated internal conversion coefficient, which is on the order of the fine structure constant $\alpha$ (approximately 0.007).

A full description of the methodology used to estimate the RPC background is presented in [55]. To summarize, we begin by simulating 8 GeV protons interacting in the production target to produce many particles, including negatively charged pions, which are propagated down the transport solenoid to the aluminum stopping target. We record the stop time and stop position for all pions stopped in the aluminum target. To make more efficient use of computing resources, the pion lifetime is set to infinity. We record the proper time for each pion and use it to accurately take into account the finite pion lifetime by applying appropriate event weights. The stopped-pion stop times and stop positions are used to seed the next stage of the simulation, which tracks the final state photons from radiative processes and the final state electron-positron pairs from the internal conversion processes through the Mu2e detector volume. The radiative and internal-conversion processes are simulated separately.

For the radiative process, the photons are isotropically produced with an energy spectrum randomly sampled from the measured spectrum of Reference [24] and with the creation time and position randomly sampled from the stopped-pion stop time and position distributions. The random sampling properly accounts for the correlations between stop time and stop position. The internal-conversion $e^+e^-$ pairs are simulated similarly except that their initial momenta are randomly sampled using the double-differential formula of [54]. In each case the final state particles are then propagated through the detector solenoid region using GEANT4, including the full detector simulation and the beam-related occupancies. The reconstruction software performs pattern recognition and track fitting using this hit-level information as input. The RPC background yield is estimated using the tracks surviving the selection requirements described in Section 3.5.3 and accounting for the finite pion lifetime as described above. The contributions from the radiative and internal-conversion processes are summed.

The stopped-pion stop-time distribution depends on the arrival-time distribution of the protons at the production target in the initial simulation stage. Two contributions are considered: an "in-time" contribution corresponding to protons that arrive at the production target as part of the 1695 ns micropulse structure, and an "out-of-time" contribution corresponding to protons that arrive at the production target in between the micropulses. For the in-time component, the proton arrival-time distribution is taken from the average expected proton pulse shape derived from detailed beamline simulations, while the normalization is taken to be 3.6 x $10^{20}$ protons on target. For the out-of-time component, the proton arrival time distribution is taken to be uniformly distributed between 0 and 1695 ns, while the normalization is taken to be (3.6 x $10^{20}$)$e$, where $e$ is the extinction, defined to be the ratio of out-of-time protons to the total number of protons. Variations of the proton arrival-time distributions are included in the systematic uncertainties as discussed below.

The in-time contribution to the RPC background as a function of the start of the live gate is shown in Figure 3.25. We choose the start of the live gate so that this in-time contribution is well below 0.1 events. The out-of-time contribution is then calculated as a function of the extinction for that same live gate window. We specify an extinction requirement so that the out-of-time contribution to the RPC background is no larger than the in-time contribution. For a livegate of $700 < t_0 < 1695$ ns the RPC background is $0.0084 + 0.0148 \times e/10^{-10}$ for $3.6 \times 10^{20}$ protons on target. The relative contributions from radiative and internal-conversion processes are 51% and 49%, respectively. The statistical uncertainty on the background estimate is 13% while the total systematic uncertainty is 21% as described below. Assuming an extinction of $10^{-10}$, the RPC background is $0.023 \pm 0.006$ where the statistical and systematic uncertainties have been combined in quadrature. Pion-capture contributions from antiproton interactions are estimated as described in Section 3.6.4 and have not been included here so as to avoid double-counting.

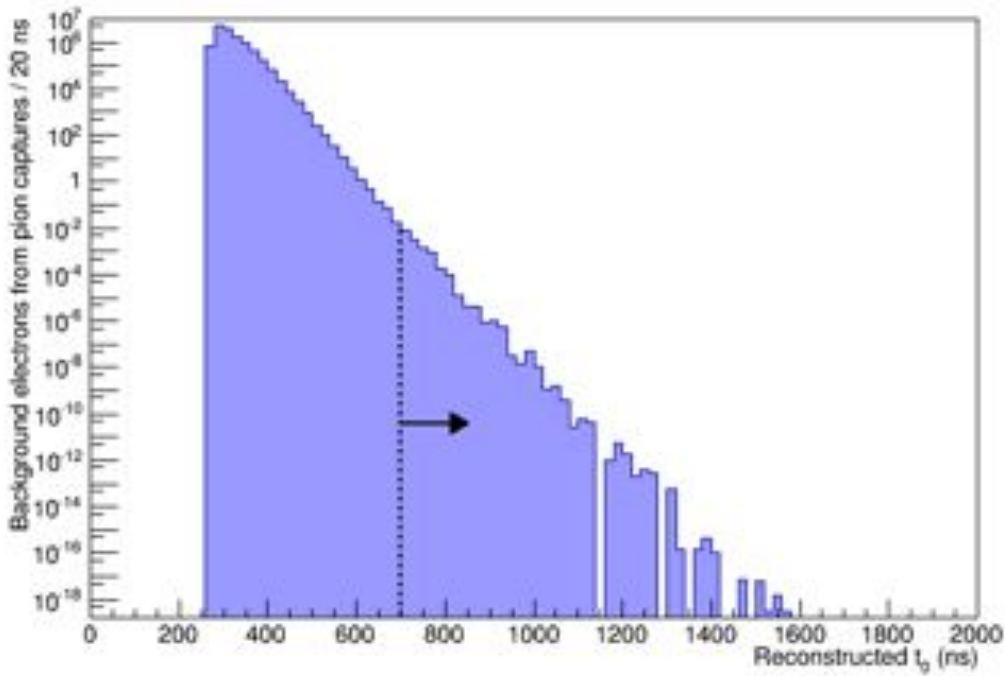

Figure 3.25 Reconstructed $t_0$ distribution of electrons from the pion-capture processes. The track selection criteria described in Section 3.5.3 have been required (except for the livegate cut). The figure is normalized to $3.6 \times 10^{20}$ protons on target.

### Systematic uncertainties

The systematic uncertainty was assessed by quantifying the change in the RPC background estimate for a given variation in the input parameters or assumptions. The simulation was rerun using a different proton pulse time distribution, corresponding to a shape with maximum pulse width; the change in yields was 10%, which is assigned as a systematic uncertainty. For the internal-conversion subsample, a systematic uncertainty of 5.5% is assigned for variations in the assumed internal conversion coefficient, based on a range of measurements and theoretical predictions. In addition, even though the deviation of the virtual photon energy spectrum from the radiative spectrum is expected to be slight, we replaced the spectrum with a uniform distribution, normalized to the same area as the sampled on-shell distribution. The change in internal-conversion backgrounds is on the order of 30%, which we assign as a conservative systematic uncertainty. An overall RPC rate uncertainty of 9.3% is also assumed based on the measurement uncertainty from Reference [24]. The overall systematic uncertainty is thus 21%.

### 3.6.3 Muon-capture Background yields

Whereas the ordinary muon capture process $\mu^- + \text{Al} \rightarrow \nu_\mu + \text{Mg}$ enters as the denominator in the conversion rate $R_{\mu e}$, the radiative muon capture (RMC) process $\mu^- + \text{Al} \rightarrow \gamma^{(*)} + \nu_\mu + \text{Mg}$ can produce background contributions in a way similar to the RPC backgrounds discussed in Section 3.6.2. Backgrounds from RMC processes result from intrinsic physics interactions and cannot necessarily be effectively mitigated by components external to the stopping target. The choice of stopping target material is thus important. In particular, if the daughter nucleus is sufficiently high in mass, the kinematic endpoint of RMC photons can be significantly separated from the DIO endpoint, thus removing a potentially large source of background. The kinematic endpoint for on-shell photons is determined by the expression

$$k_{\max} = m_\mu c^2 - |E_{\text{b}}| - E_{\text{recoil}} - \Delta M$$

where $E_{\text{b}}$ is the binding energy of a muon on Al (0.47 MeV), $E_{\text{recoil}}$ is the recoil of the Mg daughter nucleus (0.21 MeV), and $\Delta M$ is the nuclear mass difference of Mg and Al (3.11 MeV). For the muon mass of 105.66 MeV, the maximum kinetic energy the photon can acquire is 101.9 MeV, which is roughly 3 MeV lower than the DIO endpoint. Electrons from radiative muon capture are thus unlikely to be a major source of background, given the estimated momentum resolution (cf. Section 3.5.4).

Any potential RMC contributions originate from both external and internal (virtual) photon conversion to an $e^+e^-$ pair. Although theoretical descriptions of the process exist for electron-positron pair production from on-shell photons [56][57][58], none exist for the internal-conversion process. We thus do not explicitly simulate the internal RMC conversion process but assume that its contribution equals that of external $e^+e^-$ pair production (as is the case for the RPC background discussed in Section 3.6.2). References [59] and [60] provide the measured RMC photon energy spectrum for aluminum. The spectrum is fitted to a functional form motivated by the closure approximation [56][57][58]:

$$\frac{d\Lambda_\gamma\left(E_\gamma\right)}{dE_\gamma} = N\left(1 - 2x + x^2\right)x\left(1 - x\right)^2$$

where $x$ is the $E_\gamma / k_{max}$, and $N$ is a normalization factor that ensures the integral from 57 MeV to the kinematic endpoint equals the overall RMC rate measurement of $(1.43 \pm 0.15) \times 10^{-5}$, normalized relative to the ordinary muon capture rate [59]. The value of $k_{max}$ is treated as a parameter to be varied in the fit. Even though the formal kinematic endpoint for photons is 101.9 MeV for Al, the measured endpoint is $90 \pm 2$ MeV; results statistically consistent with 90 MeV are obtained for a variety of elements [59] [60].

Assuming the 90 MeV endpoint, the background yields from RMC are inconsequential as the probability to mis-reconstruct a corresponding track with a momentum that lies within the signal region of 103.75 to 105 MeV/$c$ is negligible. The assumed statistical uncertainty is 0 on this background yield. To be conservative, however, we estimate the RMC background assuming the closure approximation expression can be used with an endpoint value of 101.9 MeV, although we acknowledge such an assumption is inconsistent with the measured data of Reference [59].

The simulation procedure is very similar to that used for the RPCs, described in Section 3.6.2. A collection of stopped muons is sampled to obtain creation positions and times for RMC photons. As in the RPC case, the sampling accounts for correlations in muon stop time and stop position. The conversion of photons into electron-positron pairs in the detector solenoid region is performed by GEANT4, including the full detector simulation and beam-related occupancies, the entire track reconstruction chain, and the track selection requirements of Section 3.5.3 are applied. In contrast to the RPC background, only in-time proton contributions are relevant for RMC background contributions.

For a livegate of $700 < t_0 < 1695$ ns the on-shell photon RMC background is estimated to be roughly $2 \times 10^{-3}$, normalized to $3.6 \times 10^{20}$ protons on target. Assuming internal-conversion contributions are equivalent to the on-shell background, the total RMC background is estimated to be $4 \times 10^{-3}$. We use this value as an extremely conservative estimate of the systematic uncertainty associated with the RMC background yield. For the central RMC background yield, we use the 90 MeV kinematic endpoint, as supported by experimental measurement, which gives a null background result. The 101.9 MeV kinematic endpoint background estimate of $4 \times 10^{-3}$ is used to assign a systematic uncertainty. Due to the small central value, we do not consider any further systematic uncertainties. The RMC background estimate is thus $0.000^{+0.004}_{-0.000}$.

### 3.6.4   Antiproton-induced Background Yields

The 8 GeV kinetic energy protons are above the production energy threshold for antiprotons that are a serious background in Mu2e. Antiproton-induced backgrounds arise because:

- Antiprotons do not decay and carry a negative electric charge. Those with momenta less than 100 MeV/c can propagate through the Transport Solenoid and reach the stopping target.
- Antiprotons with momenta less than 100 MeV/c travel slowly, with speeds less than 0.1c; they spiral slowly through the Transport Solenoid and can take up to several micro-seconds to reach the Detector Solenoid. Consequently the expected flux of antiprotons at the stopping target is nearly constant in time so that the delayed live gate and the extinction systems do not effectively mitigate the resulting backgrounds.
- Antiprotons will annihilate on nuclei, releasing significant energy and producing a significant number of secondary particles. These secondaries can include electrons themselves, or they can produce electrons in tertiary interactions such as capture or decay.

The most effective mitigation against antiproton-induced backgrounds is to limit the number of antiprotons reaching the stopping target region. This is accomplished by placing thin absorbers upstream in the Transport Solenoid. The absorbers are kept thin, so as to minimize the number of muons lost.  With the absorbers in place, there are two major sources of background from antiprotons:

- Antiprotons enter the Detector Solenoid and annihilate in the stopping target to produce secondary particles, including $\pi^0$, $\pi^-$ that can produce background electrons with energy around 105 MeV.
- Antiprotons annihilate in the thin absorbers or somewhere else in the Transport Solenoid and produce secondary particles, some of which can propagate through the Transport Solenoid and reach the stopping target. Of particular note are $\pi^-$ since they will produce background electrons via pion-capture as discussed in Section 3.6.2.  However, in this instance the delayed live gate is not as effective in reducing the pion-capture background events because the pions typically arrive late at the stopping target due to the slowly traveling parent antiproton.

A large-scale simulation based on the GEANT4 framework is used to study the background from antiprotons. Several improvements in the methodology, relative to that employed for the Conceptual Design Report (CDR), have been implemented. In particular, by artificially increasing the antiproton production cross section in the simulation and by employing the staged approach and re-sampling techniques described in Section 3.5.1 we were able to significantly increase the statistics of the sample. In addition the modeling of antiproton production and annihilation were significantly improved in GEANT [61][62].

The simulation begins by modeling 8 GeV protons interacting in the production target to produce antiprotons. The FTFP physics list was used with modifications that artificially increased the antiproton production cross-section without changing the momentum or angular distributions of the resulting antiprotons [63]. The GEANT4 predicted differential cross section is compared to published results for a variety of targets and incident proton energies in Figure 3.26. These comparisons were used to normalize the GEANT4 predicted total cross section to the experimental data [61][64]. The final sample of antiproton-induced particles at the entrance to the DS, from which the background yield is estimated, corresponds to $3.6 \times 10^{20}$ protons on target. These particles are tracked through the volume of the DS and reconstructed (cf. Section 3.5) and the selection criteria of Section 3.5.3 are required.

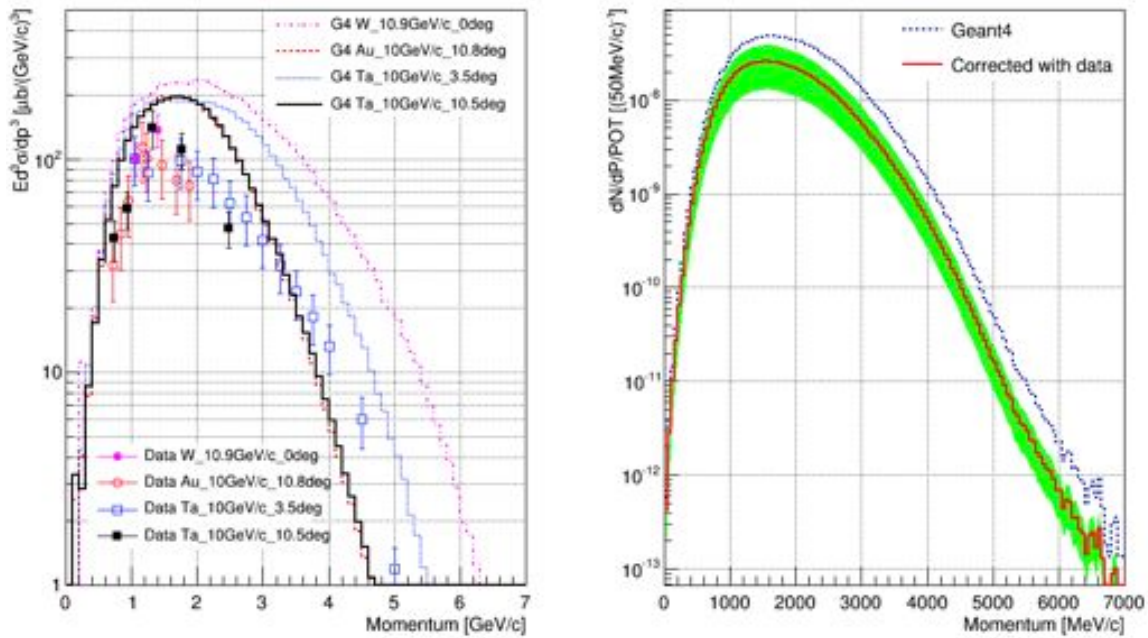

Figure 3.26. The GEANT4 predicted differential production cross-section for antiprotons is compared with experimental data points (left). These comparisons are used to form bin-by-bin corrections. On the right, the GEANT4 predicted distribution of antiproton momentum at the production vertex is shown before (dashed line) and after (solid line) the corrections have been applied. The shaded area represents a 50% uncertainty, which is the quadrature sum of the experimental uncertainties and uncertainties associated with the correction procedure.

Using a single antiproton absorber in the middle of the TS, as was initially proposed in the CDR, yields an estimated background of about 0.5 events, which is about five times larger than the previous study. The difference is understood to come primarily from two effects. Firstly, the new antiproton production model predicts a larger number of high momentum antiprotons (p > 1 GeV/c).  Secondly, there is a significant contribution from antiprotons produced in the forward direction at the production target that interact and backscatter into the acceptance of the TS.  The probability for this to occur is quite small (~$10^{-12}$) and the previous simulations, which used weighted events, did not have enough statistics to observe this effect.  The small transport probability is offset by the much larger forward-production cross-section.  This is demonstrated in Figure 3.27. From this Figure the contribution from backward produced antiprotons is also clear, where the smaller backward-production cross-section is compensated by the larger transport probability (~$10^{-6}$).

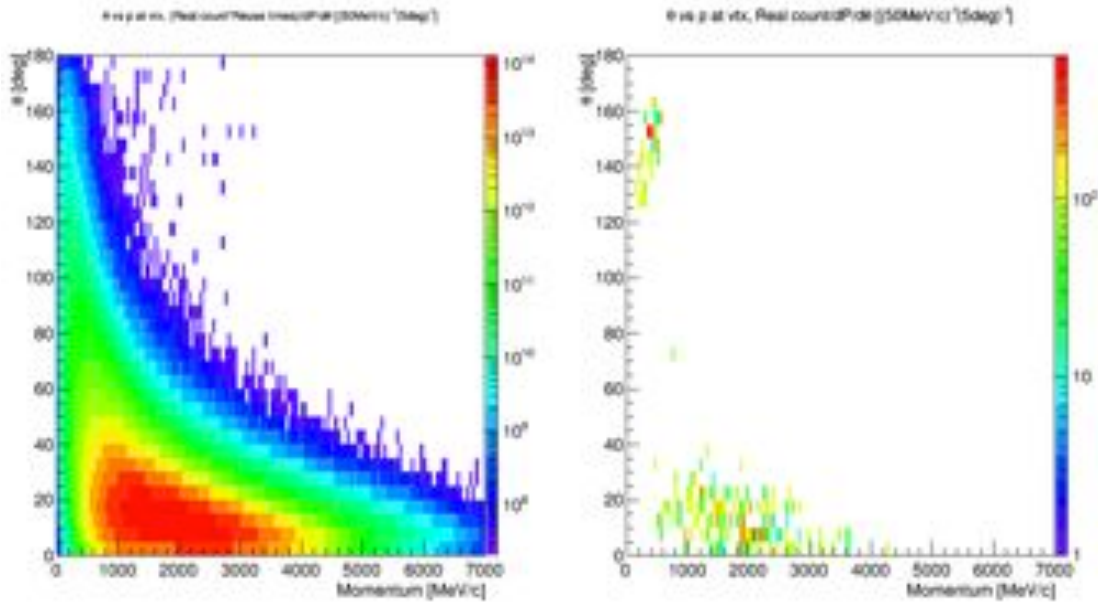

Figure 3.27 Scatter plots of the antiproton direction (θ) relative to the incoming proton direction and momentum at production (left) and, for those that survive, at the entrance of the DS (right). Here, only a single absorber window at the center of the TS was employed.

To reduce the background caused by antiprotons to an acceptable level, several modifications are made to the TS beamline:

- A thin absorber (350 μm kapton) is placed at the entrance to the TS.
- An arc shaped absorber (200 mm long, 30 mm thick carbon covering 140° in azimuth) is placed at the bottom of the first TS collimator along the inner radius.
- The thickness of the original absorber at the center of the TS is increased (+80 μm kapton).

By placing an absorber further upstream in the TS, most of the antiprotons are annihilated earlier than in the original design, so that the resulting π- either decay en-route to the DS or reach the stopping target more quickly so that the delayed live gate is more effective at discriminating against them. The arc shaped absorber is designed to absorb most of the high momentum antiprotons while minimally affecting the yield of muons stopping in the aluminum stopping target. The increased thickness of the original absorber removes antiprotons that scatter in the first absorber but are not annihilated.

The shape, thickness and position of the new absorbers were optimized to eliminate antiprotons while minimally affecting muons. These modifications reduce the antiproton-induced background by about a factor of 10 while reducing the yield of stopped-muons by only 7%. The effect of these additional absorbers has been included in all the background yield and CE single-event-sensitivity estimates presented in this report.
The final parameters and material choice for these absorbers is still being investigated as we continue to consider fabrication, installation, and operational maintenance issues.

Including the modifications to the TS beamline, and using a full simulation and reconstruction, the total antiproton-induced background for the selection criteria of Section 3.5.3 is $0.047 \pm 0.024$ events, where the uncertainty is the quadrature sum of the statistical (3%) and the systematic uncertainty (50%, described below). This total includes contributions from antiprotons that survive and annihilate in the stopping target (0.022), pion-captures in the stopping target originating from pions produced in antiproton annihilations upstream in the TS (0.021), and high energy electrons produced in upstream antiproton annihilations that scatter in the stopping target and are reconstructed (0.005).

The systematic uncertainty is dominated by uncertainties in the antiproton production cross section. The experimental data used to normalize the GEANT4 prediction has an uncertainty of about 35%. In addition, the normalization was applied as a single scale factor. Comparisons of differential cross sections between data and the normalized simulation show residual discrepancies at the level of about 35%, which is assigned as an additional systematic uncertainty. Adding these in quadrature gives a total systematic uncertainty of 50%.

### *3.6.5*   **Muon Decay-in-flight Background Yield**

While the kinematic endpoint of the electron spectrum from free muon decay is well below the 105 MeV expected for conversion electrons, in-flight muon decays can boost the energy of the resulting electron into the signal region. For example, a muon moving at $0.6c$ (approximately 79 MeV/c) can produce electrons at 105.6 MeV.

We define the muon decay-in-flight ($\mu$-DIF) background to correspond to those electrons that originate from muons decaying inside the volume of the DS. Muon decays upstream of the DS are included in the beam electron background discussed in Section 3.6.7. The estimate is made starting with the standard simulation sample discussed in Section 3.5.1 and corresponding to 2.08 x $10^9$ protons on target. Muons that survive into the DS have their arrival times, arrival positions, and momenta saved. We re-sample this saved sample by randomizing the decay to make a large sample of $\mu \rightarrow e\nu\nu$ decays in the DS volume. In order to significantly reduce the amount of cpu-time required to make the estimate, approximations are made that systematically overestimate the background. We separately calculate the contribution to this background from in-time protons and out-of-time protons.

Figure 3.28 - Figure 3.30 show the momentum, arrival time, and average momentum versus arrival time for muons surviving to the DS. One can see that while muons continue to arrive at the DS during the observation period, their typical momentum drops with time, so that a smaller fraction of them will be energetic enough to produce a background electron. The sample contains no muons with an arrival time $t_0 > 700$ ns and with a momentum large enough to create a background electron (p > 75 MeV/c). Therefore we set an upper limit on the number of background electrons from $\mu$-DIF from in-time protons using $t_0 > 200$ ns and extrapolate to later times.

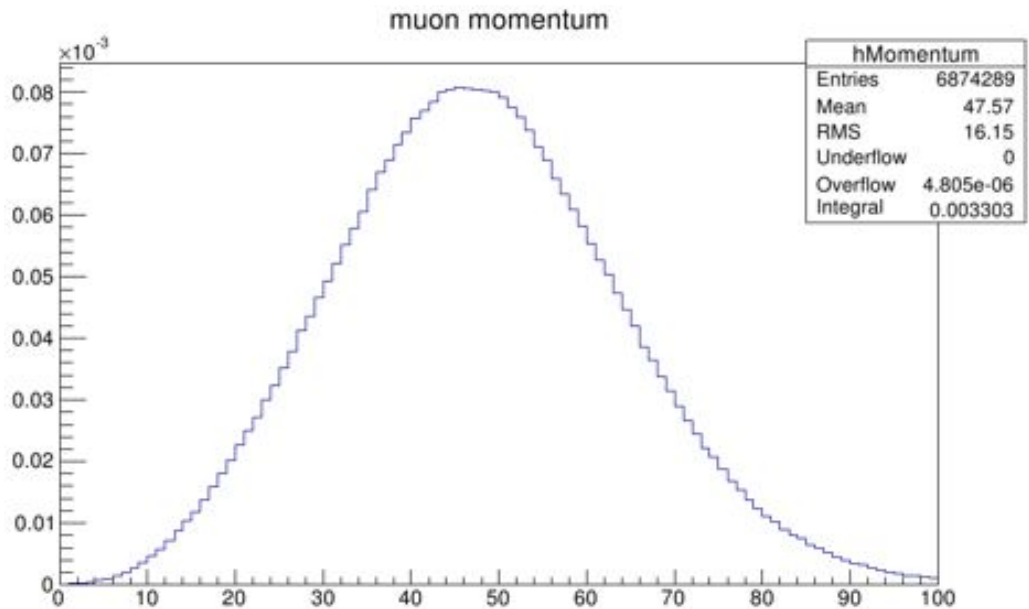

Figure 3.28 The momentum distribution of all muons surviving to the DS volume.

To extrapolate from the 200 ns start time to later times we use a dedicated simulation sample, where muon decay is turned off, and only energetic (p >68 MeV/c) muons are tracked. A total of 2.4 x $10^{10}$ protons on target have been simulated with these settings. The proton pulse time profile was not applied to the dedicated sample: all protons hit the target at t=0. The arrival time of muons in this sample, weighted by their survival probability, is shown in Figure 3.31. The distribution drops by approximately 2 orders of magnitude every 50 ns in the region where we have events. Assuming that this trend continues we expect the yield of background electrons from μ-DIF from in-time protons to be $< 10^{-3}$ for $t_0 > 450$ ns, and to be negligible during the signal search window.

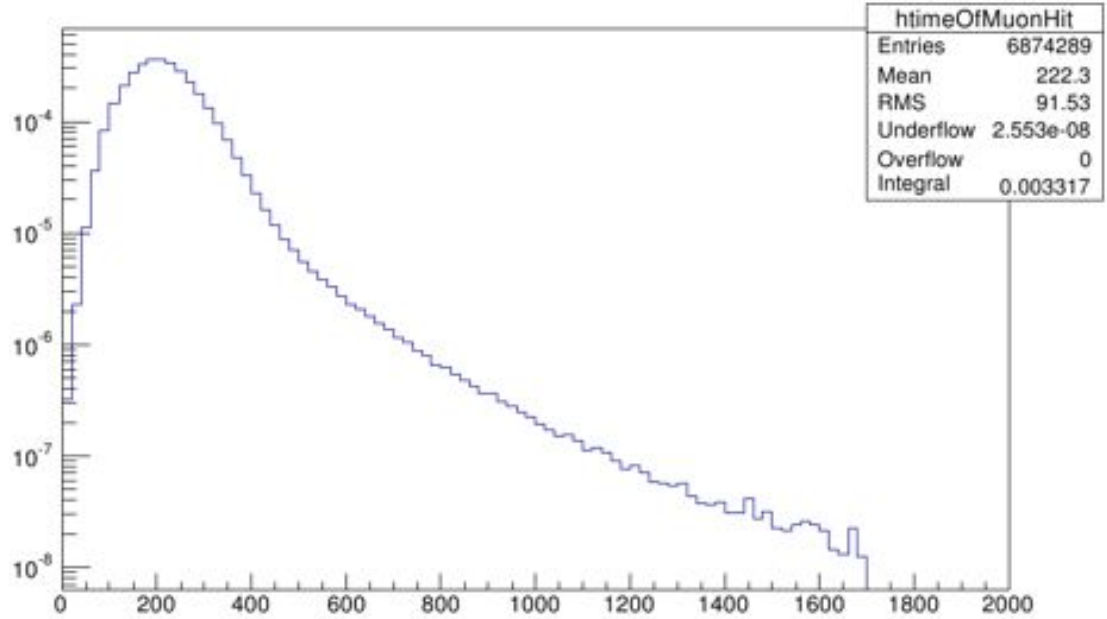

Figure 3.29 The arrival time (ns) distribution for all muons surviving to the DS volume.

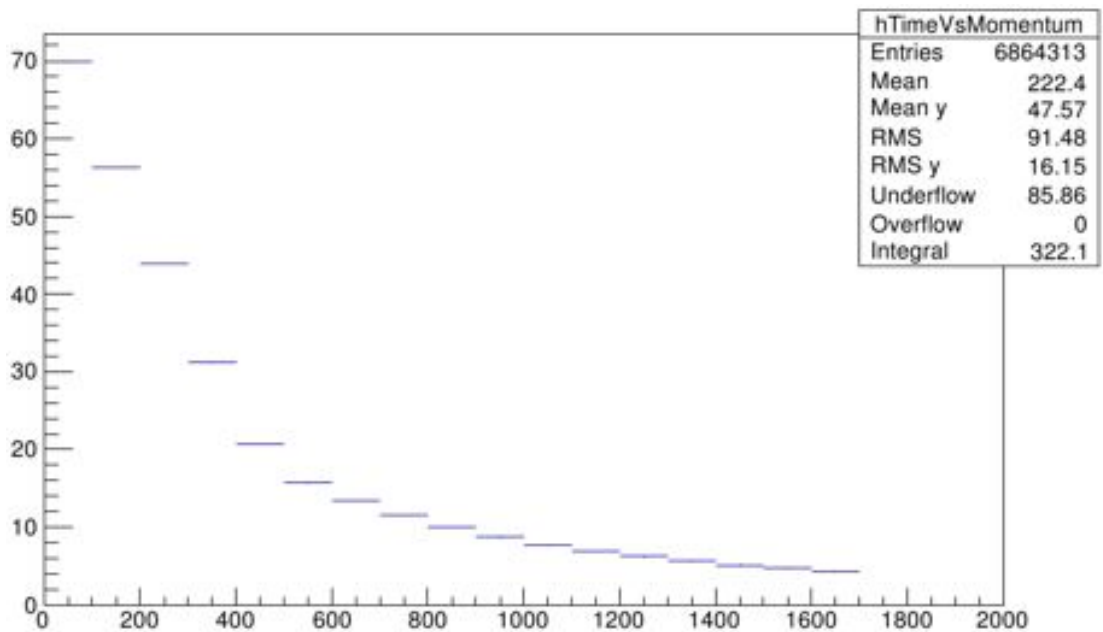

Figure 3.30 A profile plot of the average momentum (MeV/c) as a function of the arrival time (ns) for all muons surviving to the DS volume.

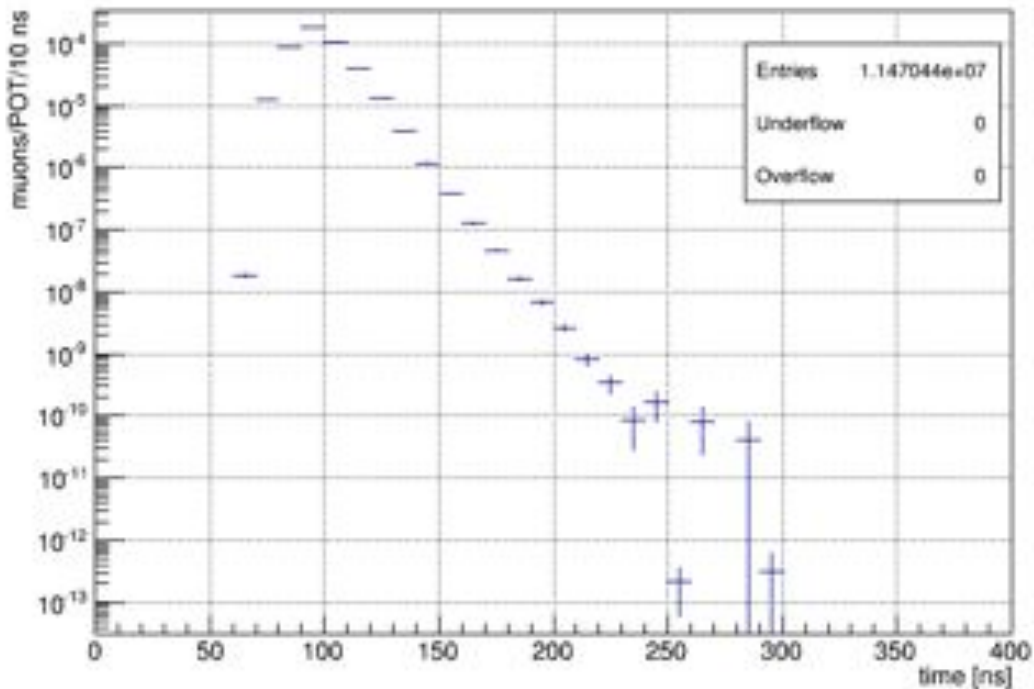

Figure 3.31. Arrival time of muons with p > 68 MeV/c at the upstream end of the DS. The protons at the production target were all generated at t=0. This plot is used to extrapolate the yield of energetic muons to larger times.

We separately estimate the background yield from in-flight decays of muons produced by out-of-time protons surviving the extinction channel. Using the procedure described above, but with the $t_0$ requirement removed, we estimate $9 \times 10^{-14}$ background-like electrons per out-of-time proton-on-target. Assuming an extinction of $10^{-10}$ this yields an upper limit of 0.003 background electrons from μ-DIF from out-of-time protons.

Summing the two contributions gives a total μ-DIF background yield of < 0.003 events.

### 3.6.6 Pion Decay-in-flight Background Yield

The two-body $\pi \rightarrow ev$ decay produces electrons with approximately 70 MeV/c momentum. Electrons from in-flight pion decay can be boosted to higher momentum and create background in Mu2e. For example a pion with momentum of about 58 MeV/c can decay to produce a 105 MeV/c electron.

We define the pion decay-in-flight (π-DIF) background to correspond to those electrons that originate from pions decaying inside the volume of the DS. Pion decays upstream of the DS are included in the beam electron background discussed in Section 3.6.7. The estimate is made starting with a simulation sample corresponding to $5 \times 10^9$ protons on target with pion decays disabled. The proper time of the pions is recorded and used to weight candidate background electrons by the survival probability of their parent pion.

Due to the short pion lifetime, the yield of background electrons from decay in flight of pions that originate from in-time protons is significantly suppressed by the delayed live gate. To estimate the contribution from the in-time protons, the momentum vector of pions surviving to the DS is used to calculate the kinematics of daughter electrons assuming a two-body $e\nu$ decay. Electrons surviving relaxed requirements on MC truth quantities are summed after accounting for the parent pion survival probability and the $\pi \rightarrow e\nu$ branching fraction. The expected number of candidate background electrons from $\pi$-DIF originating from in-time protons is estimated to be less than $10^{-5}$ for $t > 400$ ns and is negligible for $t > 700$ ns.

The yield of background electrons from decay in flight pions originating from out-of-time protons is estimated using a full simulation. The sample uses pions that survive to the DS volume and are forced to decay via the $\pi \rightarrow e\nu$ channel. Using the standard track reconstruction and selection criteria (cf. Section 3.5) and correcting for the survival probability of the parent pion and the $\pi \rightarrow e\nu$ branching fraction, the expected yield of electrons from $\pi$-DIF originating from out-of-time protons is $0.0011 \pm 0.0001$ assuming a beam extinction of $10^{-10}$, where the uncertainty arises from the limited statistics of the simulation sample.

Summing the two contributions gives a total $\pi$-DIF background yield of 0.001 events.

### *3.6.7*   **Beam Electron Background Yield**

Electrons produced in the Production and Transport Solenoids are a potential source of background. These beam electrons can be produced in the production target, primarily through $\pi^0$ production followed by conversion of the decay photons. They can also be produced by decays or interactions of beam particles anywhere upstream of the muon stopping target. Electrons produced from upstream antiproton annihilations are included in the estimate of Section 3.6.4 and are excluded here to avoid double-counting them.

The principal means of mitigating backgrounds from beam electrons are twofold. The collimators in the Transport Solenoid are designed to suppress the transport of particles with momenta above 100 MeV/c and the magnetic field in the upstream section of the Detector Solenoid is graded, which trades $p_T$ for $p_z$ and pitches forward particles entering the DS from the beamline (i.e. towards smaller $\theta$). The graded field in the upstream portion of the DS is designed to pitch forward beam particles so that they fall outside the acceptable pitch range specified in Section 3.5.3 provided the particle does not scatter in the stopping target.

To estimate the background from beam electrons we begin with the output of the stage 2 simulation sample discussed in Section 3.5.1. The secondary particles were transported through the muon beamline to the entrance of the DS, with all physics processes enabled. High momentum electrons arrive within 200 ns of protons striking the production target, which is much earlier than the start of the delayed live gate. For $2.2 \times 10^{9}$ simulated protons on target, there are only 165 high energy electrons (p > 95 MeV/c) entering the DS. The angular distribution of beam electrons is concentrated at small angles with respect to the detector axis, as is illustrated in Figure 3.32. Consequently, the electrons must undergo a large-angle scatter in materials upstream of the tracker in order to satisfy the pitch angle criteria of Section 3.5.3.

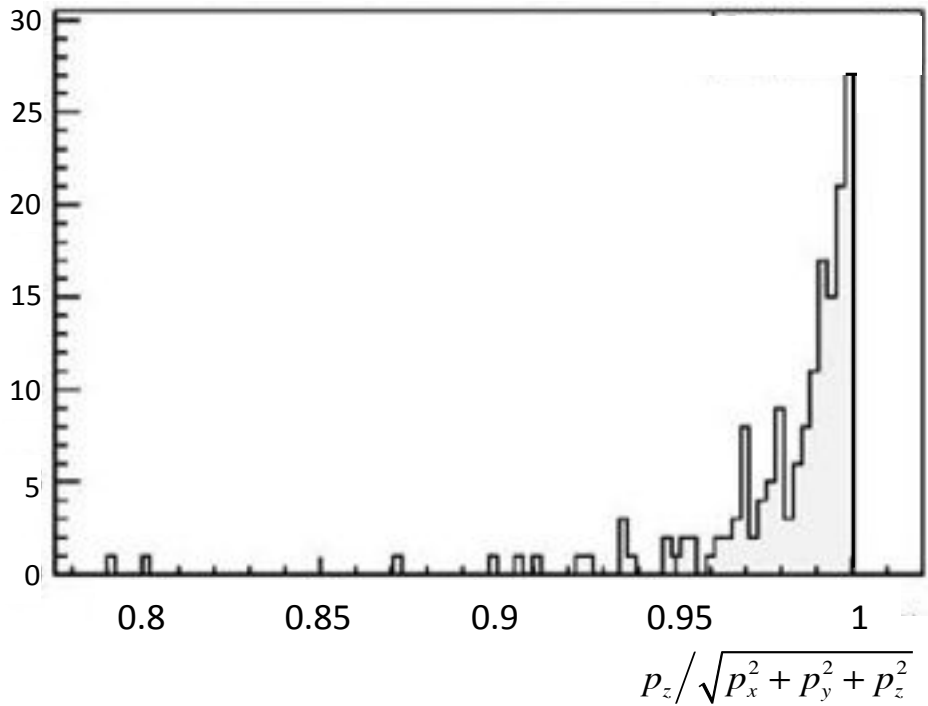

$$p_z / \sqrt{p_x^2 + p_y^2 + p_z^2}$$

Figure 3.32 Distribution of cos(θ) for high-momentum electrons (p > 95 MeV/c) entering the DS. Note that our selection criteria require $45^{o} < \theta < 60^{o}$, corresponding to 0.500 < cos(θ) < 0.707.

To estimate how often beam electrons will experience a large-angle scatter, a dedicated simulation was performed. The high momentum electrons arriving at the DS from the stage 2 simulation were resampled using a kernel density function approach. Each high momentum electron is replaced by a "cloud" of 1000 electrons whose position, energy, and direction were randomized according to probability distribution functions (PDF) determined from fits to the original sample. The resulting dataset is passed through the GEANT4 simulation 100 times, varying the random number seed, to accumulate the rare large-angle scattering events. Electrons that intersect the geometric volume of the tracker were recorded. This dedicated simulation sample corresponded to 2.2 x $10^{14}$ protons on target and yielded 16 electrons that intersected the tracker with momentum $100 < p < 110$ MeV/c and pitch $0.4 < p_z/p < 0.7$. All high momentum electrons arrived at the DS within 150 ns after their parent proton hit the production target. Thus, the delayed live gate completely eliminates this background for beam electrons originating from in-time protons. Out-of-time protons can, however, produce beam electrons in the delayed live gate, but these are suppressed by the extinction channel. For $3.6 \times 10^{20}$ protons on target and an extinction of $10^{-10}$, the beam-electron background yield is estimated to be $(2.6 \pm 1.4) \times 10^{-3}$, where the uncertainty includes statistical (25%) and systematic (50%) contributions added in quadrature. The systematic uncertainty was determined by varying the fit used to determine the PDF for the kernel density re-sampling, repeating the simulation, and recalculating the background. The full scale of the observed variations relative to the nominal background estimate ($\pm 50\%$) is assigned as the systematic uncertainty.

### 3.6.8 Cosmic Ray Induced Background Yield

Backgrounds from cosmic ray interaction or decay are a potentially limiting background and must therefore be vetoed using detectors that cover a large portion of the solid angle around the Detector Solenoid and a portion of the Transport Solenoid (the Cosmic Ray Veto system, CRV, is discussed in Section 10). These detectors must perform with high efficiency despite a hostile environment that includes a large flux of neutrons emanating from the muon stopping target, muon beam stop, the production target, and the TS collimators. Note that the cosmic ray background scales with live running time rather than with the number of protons on target.

Cosmic rays can induce background through a number of mechanisms, including

- Muon decay in the Detector Solenoid.
- Muon interactions in the stopping target, proton absorber, tracker, calorimeter, or other nearby material that produces electrons.
- Muons that enter the Detector Solenoid, scatter in the stopping target and are misidentified as electrons.
- Muons at shallow angles that enter the Transport Solenoid, scatter in a collimator, and produce electrons or traverse the DS and are misidentified as electrons.

Simulation studies were done to determine the various types of backgrounds induced by cosmic-ray muons, the required coverage and efficiency of the CRV, and the required calorimeter particle-identification efficiency in rejecting events that mimic conversion electrons.

Two types of simulations have been done: a "general" simulation, in which cosmic rays are generated over the entire DS and TS regions, and "targeted" simulations, in which the cosmic rays are generated within a limited phase space in order to target specific regions of limited CRV coverage with high statistics. These regions include the start of the CRV in the middle of the TS, and the DS-upstream and DS-downstream sections of the CRV.

Only the muon component of the cosmic-ray flux is simulated. The energy spectrum and angular distribution is based on the Daya Bay code [66]. The Daya Bay predictions are found to agree to within 20% when compared to the predictions of the CRY [67] generator and to data [68]. The 20% is assigned as a systematic uncertainty.

### General Simulation

In the general simulation the cosmic-ray muons are generated uniformly in a horizontal production plane centered at the middle of the tracker. They are propagated from the surface plane above the detector hall to the tracker. The GEANT4 detector simulation and reconstruction algorithms described in Section 3.5 are employed. We record tracks surviving the track selection criteria of Section 3.5.3 with an extended momentum window $100 < p < 110$ MeV/c in order to increase statistics. Note that the acceptance of the tracker is relatively uniform over this range of momenta, so a simple scaling, based on the ratio of the widths of the relevant momentum windows, can be applied to estimate the final background yield.

By default the track reconstruction algorithm performs pattern recognition assuming an electron traveling in the downstream direction (ie. away from the stopping target towards the calorimeter). The simulation reveals a special class of background events that originate with muons that enter the tracker from the downstream end of the DS, – for example, after scattering in the calorimeter - move upstream through the tracker toward the stopping target, get reflected by the graded magnetic field in the region of the stopping target, and then re-enter the tracker traveling in the downstream direction. An example of such an event is depicted in Figure 3.33. A large fraction of these events can be identified and removed by re-running the track reconstruction algorithm assuming an electron traveling in the upstream direction (ie. away from the calorimeter towards the stopping target). Events for which an upstream track is reconstructed are rejected. This selection is >99% efficient for conversion electrons satisfying the track selection criteria of Section 3.5.3.

About two-thirds of the surviving events are electrons with $\mu^+$, $\mu^-$, and $e^+$ accounting for the other one-third. The application of the calorimeter and particle-identification criteria of Section 3.5.3 removes the non-electron tracks. The $e^+$ and $\mu^+$ fail to satisfy the $\Delta t$ requirement since they originate in the calorimeter and travel upstream through the tracker, the μ⁻ fail the particle-identification likelihood-ratio requirement.

A total of 27.909 billion cosmic-ray events were generated. This corresponds to a veto live time of $2.98 \times 10^5$ seconds, which is about 2% of the total veto live time [69]. Out of the generated events, 61,199 events could be reconstructed with a downstream electron hypothesis. Table 3.3 below lists the number of events surviving the various requirements. It also lists the types of particles responsible for the reconstructed tracks. The production processes and the production volumes of the events surviving the track selection criteria are shown in Figure 3.34.

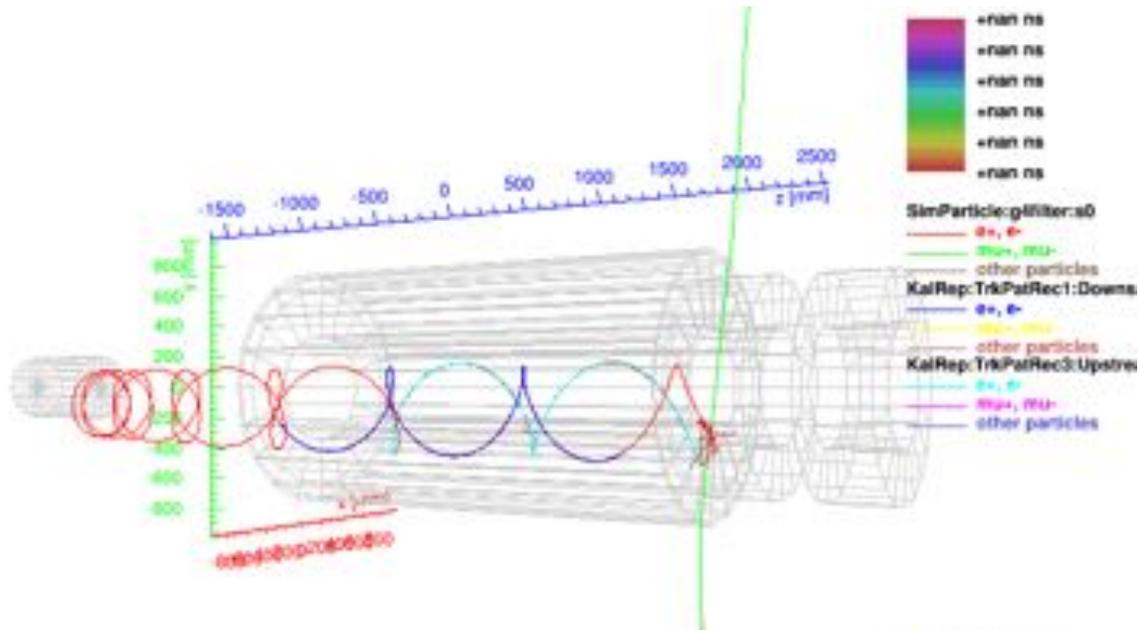

Figure 3.33 An event display from simulation showing a background candidate induced from a through-going cosmic ray that interacts in the calorimeter to create an electron. The electron, shown in red, first travels upstream, then gets reflected and travels downstream through the tracker. Both the upstream and downstream segments are reconstructed (light blue and dark blue).

Table 3.3: Number of events surviving different requirements from the "general" simulation described in the text. The sample represents about 2% of the total veto live time.

|  | Events | e⁻ | e⁺ | μ⁻ | μ⁺ | π⁻ | π⁺ |
|---|---|---|---|---|---|---|---|
| Reconstructed as downstream-going e⁻ | 61,199 | 18,429 | 3,766 | 18,686 | 20,316 | 1 | 1 |
| Survive track-selection criteria (100<p<110 MeV/c) | 441 | 243 | 23 | 85 | 90 | 0 | 0 |
| Veto events reconstructed as upstream-going e- | 201 | 140 | 4 | 21 | 33 | 0 | 0 |
| Survive calorimeter and particle-id criteria | 131 | 131 | 0 | 0 | 0 | 0 | 0 |

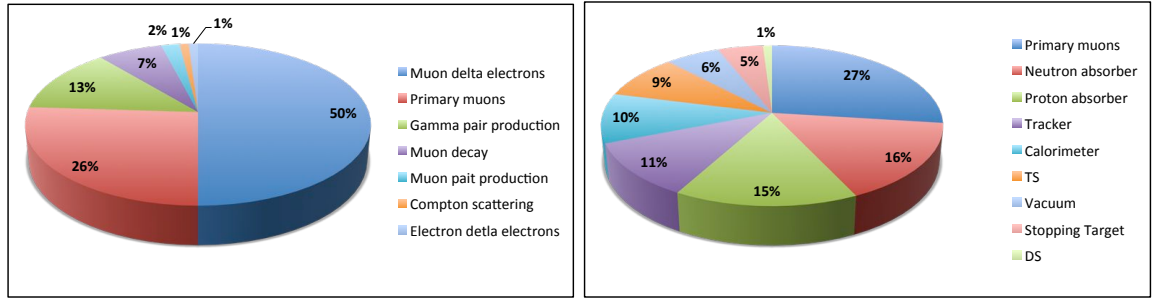

Figure 3.34: Production processes (left) and production volumes (right) of the cosmic-ray muon induced particles that are reconstructed as downstream going electrons and surviving the track selection criteria of Section 3.5.3.

All of the surviving events (even excluding the calorimeter requirements) were induced by cosmic rays that traverse at least one of the CRV sections and can thus be vetoed. About 15% traverse more than one CRV section.

Assuming a CRV inefficiency of $10^{-4}$, scaling to the momentum window in Section 3.5.3, and normalizing to the full expected veto live time of Mu2e gives a total cosmic-ray induced background of $0.078 \pm 0.017$, where the uncertainty is the quadrature sum from the statistical (8%) and systematic (20%) uncertainties.

***Targeted simulations***

The targeted simulations employ the same detector simulation and reconstruction algorithms as the general simulations, but begin with cosmic-rays generated over a limited phase-space so as to target regions of Mu2e with limited CRV coverage, where the $10^{-4}$ suppression from the veto does not apply. The purpose of these simulations is to identify potential sources of cosmic-ray induced background that may be under-represented in the general simulation sample described above because they occur with very small probability. Three separate targeted simulations are performed: cosmic rays incident on the *y-z* plane located at *x* = -104 mm (which sits at the opening of the CRV where it surrounds the middle of the TS), cosmic rays incident on the portion of the CRV located at the downstream end of the DS (CRV-D), and cosmic rays incident on the portion of the CRV located at the upstream end of the DS (CRV-U). Each of these targeted simulations corresponded to a veto live time of about 100% of the Mu2e expected veto live time.

The TS and CRV-U targeted samples yield 6 $\mu^-$ that survive the track reconstruction selection criteria and are candidate background events. These are induced by very shallow angle cosmic rays that enter the TS and scatter in the collimator in the middle of the TS. None of them can be vetoed by the CRV. Scaling to the momentum window of Section 3.5.3 and to the expected Mu2e veto live time gives a background estimate of 0.77 events surviving the track selection criteria and upstream-veto. In the simulation none of the 6 events survive the calorimeter and particle-identification criteria. The expected muon rejection of these criteria is 200.

After the application of all selection criteria these targeted simulations predict an additional cosmic-ray induced background of $0.004 \pm 0.002$ events, where the total uncertainty includes the statistical (41%) and systematic (20%) uncertainties added in quadrature.

The CRV-D targeted sample yields 1 primary $\mu^+$ that survives the track reconstruction selection criteria and is a candidate background event. This cosmic ray enters in the lower half of the downstream DS, where, currently, there are no CRV modules. This event is rejected by the calorimeter and particle-identification requirements. As a consequence of this targeted simulation, we plan to extend the CRV coverage in this region so that events of this type will be additionally vetoed by the CRV. Assuming this additional CRV coverage, events like this contribute negligibly to the cosmic-ray induced background.

***Results***

Summing over the results of the general and targeted simulations yields a total cosmic-ray induced background of $0.082 \pm 0.018$ events, assuming a CRV inefficiency of $10^{-4}$ (as required) and a particle identification muon-rejection factor of 200 (as required). The total uncertainty includes statistical (8%) and systematic (20%) uncertainties added in quadrature.

It is important to note that Mu2e will be able to directly measure the cosmic-ray induced background using data collected when the beam is not being delivered.

## 3.7    Summary of Background Yields and Signal Sensitivity

The background estimates from this Chapter are summarized in Table 3.4. Using the selection criteria of Section 3.5.3 yields a total background estimate of $0.36 \pm 0.10$ events. The uncertainty includes contributions from the limited statistics available in the simulation samples used to make the estimates and contributions from systematic uncertainties that quantify the effect of various modeling uncertainties as discussed in the text. Some of these uncertainties will likely be reduced once measurements with data can be made.

The expected single-event sensitivity for a three-year run is $\left(2.87 \pm^{0.32}_{0.27}\right) \times 10^{-17}$ as set out in Table 3.5. We assume three-years worth of physics running at $2 \times 10^7$ seconds of running per year at an average beam power of 8 kW, corresponding to two batches of $4 \times 10^{12}$ protons from the booster every 1.33 seconds. For planning purposes, the actual run duration is assumed to last an additional year – 4 years total – to accommodate calibration runs, cosmic-ray veto studies, and dedicated background runs.

Table 3.4 A summary of the estimated background yields using the selection criteria of Section 3.5.3. The total run time and corresponding number of protons on target are provided in Table 3.5. An extinction of $10^{-10}$, a cosmic ray veto inefficiency of $10^{-4}$, and particle-identification with a muon-rejection of 200 are used. 'Intrinsic' backgrounds are those that scale with the number of stopped muons, 'Late Arriving' backgrounds are those with a strong dependence on the achieved extinction, and 'Miscellaneous' backgrounds are those that don't fall into the previous two categories.

| Category | Background process | Estimated yield (events) |
|---|---|---|
| Intrinsic | Muon decay-in-orbit (DIO) | $0.199 \pm 0.092$ |
| | Muon capture (RMC) | $0.000^{+0.004}_{-0.000}$ |
| Late Arriving | Pion capture (RPC) | $0.023 \pm 0.006$ |
| | Muon decay-in-flight ($\mu$-DIF) | $<0.003$ |
| | Pion decay-in-flight ($\pi$-DIF) | $0.001 \pm {<}0.001$ |
| | Beam electrons | $0.003 \pm 0.001$ |
| Miscellaneous | Antiproton induced | $0.047 \pm 0.024$ |
| | Cosmic ray induced | $0.082 \pm 0.018$ |
| | Total | $0.36 \pm 0.10$ |

Relative to the sensitivity reported in the Mu2e Conceptual Design Report, the CE single-event sensitivity has improved by about a factor of two while keeping the background unchanged (within uncertainties). Most of the improvement is due to improvements in the reconstruction algorithms and to optimizations of detector material. It's worth noting that this improvement was achieved in the face of adding realism to the detector simulations, which often degraded the sensitivity. Additional improvements can be expected, including using the calorimeter to seed the pattern recognition algorithm. Initial studies show that such an algorithm can add 5-10% (relative) efficiency to the total acceptance x efficiency after all selection criteria and that the combination of track reconstruction algorithms has an efficiency that is less dependent on the rate of accidental hits in the tracker. In addition, the two algorithms can be played-off of one another and together provide for a more efficient and more robust reconstruction.

Table 3.5 The expected sensitivity for three years worth of physics running. The single-event sensitivity shown here is about a factor of two better than what was achieved for the CDR. This improvement is mostly due to improvements in the reconstruction algorithms and other minor optimizations. These improvements and optimizations will continue and the sensitivity is expected to reach the indicated goal.

| Parameter | Value |
|---|---|
| Physics run time @ $2 \times 10^7$ s/yr. | 3 years |
| Protons on target per year | $1.2 \times 10^{20}$ |
| $\mu^-$ stops in stopping target per proton on target | 0.0019 |
| $\mu^-$ capture probability | 0.609 |
| Total acceptance x efficiency for the selection criteria of Section 3.5.3 | $\left(8.5 \pm^{1.1}_{0.9}\right)\%$ |
| Single-event sensitivity with Current Algorithms | $\left(2.87 \pm^{0.32}_{0.27}\right) \times 10^{-17}$ |
| Goal | $2.4 \times 10^{-17}$ |

## 3.8   Summary of Physics Requirements

The physics requirements necessary to achieve the sensitivity estimated in Table 3.5 are detailed in [70] and are summarized here.

- To suppress prompt backgrounds from beam electrons, muon decay-in-flight, pion decay-in-flight and radiative pion capture requires a pulsed beam where the ratio of beam between pulses to the beam contained in a pulse is less than $10^{-10}$. This ratio is defined as the beam *extinction*. The spacing between beam pulses should be about twice the lifetime of muonic aluminum (>864 ns) and the beam pulse should not be wider than 250 ns.

- In order to suppress backgrounds from decays of muons in atomic orbit in the stopping target, the reconstructed width of the conversion electron energy peak, including energy loss and resolution effects, should be on the order of 1 MeV FWHM or better with no significant high energy tails.

- In order to suppress backgrounds from beam electrons, the field in the upstream section of the Detector Solenoid must be graded so that the field decreases toward the downstream end. This graded field also serves to increase the acceptance for conversion electrons.

- Suppression of backgrounds from cosmic rays requires a veto surrounding the detector. The cosmic ray veto should be nearly hermetic on the top and sides in the region of the collimator at the entrance to the Detector Solenoid, the muon stopping target, tracker, and calorimeter. The overall efficiency of the cosmic ray veto should be 0.9999 or better.

- Suppression of long transit time backgrounds places requirements on the magnetic field in the Production Solenoid, the Detector Solenoid, and the straight sections of the Transport Solenoid. The Production Solenoid must have a field that, as one moves downstream, rises to a maximum upstream of the target then decreases uniformly, which results in the target being located in a region of negative gradient. From the production target to the downstream end of the Production Solenoid, there must be no local positive gradients in the particle transport portion of the magnetic field volume (as opposed to regions where there is shielding material). The negative gradient causes the pitch of helices to increase with time (pitch is proportional to p(longitudinal)/p(total)), generally reducing transit times of particles, preventing local trapping of particles, and increasing the number of low energy charged particles traveling downstream. The field gradients in the three straight sections of the Transport Solenoid must be continuously slightly negative and relatively uniform in order to avoid trapped particles and reduce the downstream transit times of particles.
- The ability to separate muons and pions from electrons with high reliability and high efficiency is required to eliminate backgrounds from ~105 MeV/c muons and pions.
- To mitigate backgrounds induced from antiproton annihilation, thin windows in appropriate places along the muon beam line are required to absorb antiprotons.
- The capacity to identify and record events of interest with high efficiency must exist.
- The capacity to take data outside of the search window time interval must exist.
- The capacity to collect calibration electrons from $\pi^+ \rightarrow e^+ \nu$ is required.
- The capacity to measure the beam extinction to a level of $10^{-10}$ with a precision of about 10% over about a one hour time span must exist.
- The capacity to determine the number of ordinary muon captures with a precision of order 10% must exist.
- The muon beam line is required to have high efficiency for the transport of low energy muons (~0.002 stopped negative muons per 8 GeV proton on target). To mitigate backgrounds from muon and pion decay-in-flight, it must suppress transport of high-energy muons and pions. It must also greatly suppress the transport of high energy electrons.
- The muon beam line should avoid a direct line-of-sight path of neutral particles (mainly photons and neutrons) from the production target to the muon stopping target.
- The detectors must be able to perform in a high-rate, high-radiation environment.
- The muon beam line should be evacuated.

# 3.9 Experimental Reach with Evolution of Fermilab Accelerator Complex

It is worth considering whether there exist plausible upgrade paths for Mu2e. One possible scenario would utilize an upgraded Fermilab accelerator complex to further probe charged-lepton flavor violation with an improved Mu2e.

Fermilab is pursuing a "Proton Improvement Plan" (PIP) campaign to improve the reliability, beam-power and flexibility of the accelerator complex. The first stage of this campaign is now underway (PIP-I) and will support the proton beam requirements of the Mu2e experiment defined in this Technical Design Report. The next phase of this campaign, PIP-II [71], is a mature concept that envisions a number of additional improvements and modifications to the Fermilab accelerator complex, providing the U.S. with the opportunity to establish long-term world leadership in particle physics research based on intense beams. The primary goals of PIP-II are to provide unique capabilities in delivering proton beam power of greater than 1 MW to the neutrino production target at the initiation of LBNF (Long Baseline Neutrino Facility) operations, and to establish a flexible platform for future development of the Fermilab complex to multi-MW capabilities in support of a broader research program.

PIP-II's capabilities to improve Mu2e sensitivity have been studied by the collaboration [72]. PIP-II would replace the 8kW, 8-GeV proton beam of PIP-I with an 80kW 1-GeV beam with flexible proton pulse timing and a proton pulse width of 100 nsec, about half the proton width of PIP-I. The lower beam energy of PIP-II eliminates backgrounds from antiproton production and will reduce radiation damage to the Production Solenoid per stopped muon. The narrower proton pulse width will reduce backgrounds from Radiative Pion Capture (RPC) and facilitate pulse frequency optimization for other stopping targets (such as titanium) with shorter capture lifetimes. A modest upgrade of the cryogenic cooling capacity for the PIP-II superconducting linac will permit the proton beam to operate with nearly a 100% duty factor, reducing instantaneous detector rates per stopped muon by a factor of three with respect to PIP-I.

In the instance of no conversion electron signal observed by Mu2e in the PIP-I era, operations of a suitably upgraded Mu2e detector in the PIP-II era can further increase the search sensitivity with an aluminum stopping target by nearly an order of magnitude. If Mu2e observes a conversion electron signal, then operations in the PIP-II era can dramatically increase the statistical significance of the observation and provide an opportunity to search for another conversion signal with a different stopping target such as titanium. Establishing a conversion signal on two different nuclei will be an important test of detector response systematics and can begin to discriminate the character of the new physics driving the electron conversion as depicted in Figure 3.35 [73].

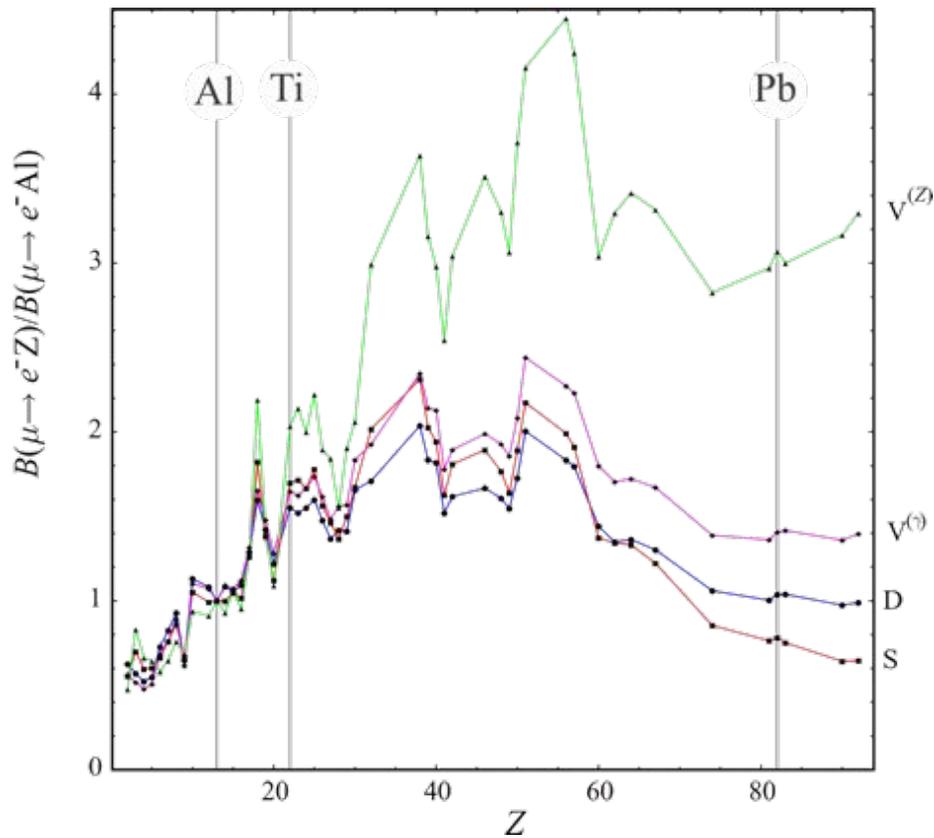

Figure 3.35. Target dependence of the $\mu \rightarrow e$ conversion rate in different single-operator dominance models considered in [73]. The conversion rates are normalized to the rate in aluminum ($Z = 13$) versus the atomic number $Z$ for the four theoretical models described therein: D (blue), S (red), V($\gamma$) (magenta), V($Z$) (green). The vertical lines correspond to $Z = 13$ (Al), $Z = 22$ (Ti), and $Z = 82$ (Pb). See [73] for details.

This page intentionally left blank

# 4  Accelerator Systems

## 4.1  Introduction

The secondary muon beam on the Mu2e stopping target is derived from the decay of pions produced by the interaction of an intense 8 GeV kinetic energy proton beam with a tungsten target. This chapter describes the upgrades to the existing Fermilab Accelerator facilities required for the delivery and targeting of the primary proton beam.

### 4.1.1 Accelerator Systems project scope

The Mu2e Accelerator Systems upgrades (WBS 475.02) are divided into eight Level 3 sub-projects that are shown in Table 4.1. These sub-projects are briefly described in this introduction and will be elaborated in detail in the remainder of this chapter.

Table 4.1 Mu2e Accelerator Systems Level 3 sub-projects

| WBS | Name |
| --- | --- |
| 475.02.01 | Project Management |
| 475.02.03[1] | Instrumentation and Controls |
| 475.02.04 | Radiation Safety Improvements |
| 475.02.05 | Resonant Extraction System |
| 475.02.06 | Delivery Ring RF System |
| 475.02.07 | External Beamline |
| 475.02.08 | Extinction Systems |
| 475.02.09 | Target Station |

#### 4.1.1.1  Project Management

The Project Management WBS item contains the project management tasks for the Mu2e Accelerator Upgrades. These tasks include reviews, reports, supervision of the Accelerator Upgrades management team, Technical Board meetings, Accelerator Level 3 management meetings, standards preparation, EVMS tracking and analysis, cost estimates, schedule preparation, and change control.

#### 4.1.1.2  Instrumentation and Controls

The Instrumentation and Controls WBS item contains the tasks necessary for the design and fabrication of the accelerator controls and instrumentation upgrades required for operation of the Fermilab Accelerator Complex for the Mu2e experiment. These tasks include:

---

[1] WBS 475.02.02 contains conceptual design work for Recycler Ring extraction upgrades that are no longer part of the scope of the Mu2e project.

- Design and implementation of required controls systems upgrades for beam transport to the Recycler Ring and the Delivery Ring

- Design and implementation of Delivery Ring Abort controls

- Design and implementation of the M4 beamline control system

- Design and implementation of Delivery Ring instrumentation upgrades for Mu2e

- Design and implementation of instrumentation for the M4 beamline

The Instrumentation design and implementation is described in detail in section 4.3. The design and implementation of control system upgrades are described in detail in section 4.4.

### 4.1.1.3   Radiation Safety Improvements

The Radiation Safety Improvements WBS item contains the tasks required for the design and implementation of the Radiation Safety upgrades that are required to maintain the level of radiation protection required by the Fermilab Radiological Control Manual. These tasks include the following:

- Design, fabrication, and implementation of an M1 beamline to Delivery Ring Total Loss Monitor (TLM) radiation safety system

- Design, fabrication, and implementation of the Delivery Ring radiation safety system upgrades, which includes a Delivery Ring TLM radiation safety system and in-tunnel shielding of known beam loss points

- Design, fabrication, and implementation of the external (M4) beamline radiation safety system, which includes M4 beamline safety system interlocks, M4 beamline TLM system, and M4 beamline in-tunnel shielding

- Design, fabrication, and implementation of the Mu2e proton service building radiation safety interlock system

Radiation Safety Improvements are described in detail in section 4.5.

### 4.1.1.4   Resonant Extraction System

The Resonant Extraction System WBS item contains the tasks required for the design, fabrication, and installation of the systems necessary for the resonant extraction of Mu2e beam from the Delivery Ring synchrotron. These tasks include the following:

- General engineering design of the Delivery Ring resonant extraction system

- Design, manufacture, and installation of the resonant extraction electrostatic septum (ESS) modules (two modules) and power supply

- Design, procurement, and installation of the resonant extraction tune quadrupole magnets and power supplies

- Design, manufacture, and installation of the resonant extraction harmonic sextupole magnets and power supplies

- Design, procurement/manufacture, and installation of the resonant extraction dynamic bump magnets and power supplies

- Design, manufacture, and installation of the resonant RF knock out (RFKO) kicker and power supply

- Design, manufacture, and installation of the resonant extraction fast feedback devices and electronics.

The Resonant Extraction System design is described in detail in section 4.6.

### 4.1.1.5   Delivery Ring RF System

The Delivery Ring RF System WBS item contains the tasks required for the design and construction of a 2.4 MHz RF system that synchronously captures beam from the Recycler Ring and holds it in a stationary RF bucket during resonant extraction. These tasks include the following:

- Design, construction, and installation of the Delivery Ring 2.4 MHz low level RF system

- Design, manufacture, and installation of an LCW cooling system for the 2.4 MHz RF cavity

- Procurement and installation of an 8 kW driver amplifier for the 2.4 MHz RF cavity

- Installation of the 2.4 MHz RF Cavity[2]

- Longitudinal tracking simulation model of the Delivery Ring 2.4 MHz RF system that predicts the longitudinal phase space distribution of protons during the Mu2e spill.

The Delivery Ring RF System design is described in detail in section 4.7.

### 4.1.1.6   External Beamline

The External Beamline WBS item contains the tasks required for the design, fabrication, and installation of the systems necessary for the delivery of Mu2e beam from the Delivery Ring synchrotron to the Mu2e proton target. These tasks include:

---

[2] The 2.4 MHz RF cavity is provided by the Recycler RF AIP

- Design of the external (M4) beamline optics that includes: vertical bend out of the Delivery Ring, the left bend section, extinction dipole insert, extinction collimation section, diagnostic absorber line, and final focus section

- Magnet and power supply selection, procurement, and installation for the M4 beamline

- Design, fabrication, and installation of the M4 beamline vacuum and mechanical systems

- Design, fabrication, and installation of the M4 diagnostic absorber and associated beamline.

The External Beamline design is described in detail in section 4.8.

### 4.1.1.7   Extinction Systems

The Extinction Systems WBS item contains the tasks required for the design, fabrication, and installation of the systems necessary for the extinction of out-of-time particles en route to the proton target as well as the systems for monitoring the level of extinction achieved. These tasks include:

- Design, fabrication, and installation of the extinction 300 kHz and 5 MHz AC dipole magnets and power supplies

- Design, fabrication, and installation of the extinction collimation system in the M4 beamline.

- Design, fabrication, and installation of the target extinction monitoring system.

Extinction design is described in detail in section 4.9. Extinction monitoring is described in section 4.10

### 4.1.1.8   Target Station

The Target Station WBS item contains the tasks required for the design, fabrication, and installation of the systems necessary for the proton target station. These tasks include the following:

- Design and fabrication of the Mu2e proton target and target support system

- Design, fabrication, and installation of the Production Solenoid (PS) Heat and Radiation Shield (HRS)

- Design, fabrication, and installation of the proton target beam absorber

- Design and fabrication of the PS and HRS protection collimator

- Design, fabrication, and implementation of the proton target handling system.

The design of Target Station components is described in detail in section 4.11.

## 4.1.2 Muon Campus Projects

The upgrades to the Fermilab Accelerator complex necessary to run the Mu2e experiment are distributed over several projects. These projects will transform the Fermilab Antiproton Source into what is now called the Muon Campus [1]. In the near term, the Muon Campus will support the operation of the Muon g-2 and the Mu2e experiments.

Many of the accelerator upgrades required for the Mu2e experiment are also necessary for Muon g-2. Since the g-2 experiment will run before Mu2e, these upgrades will be installed and commissioned well before they are needed for Mu2e beam operations. Table 4.2 gives a summary of the various accelerator upgrades that are required for the successful operation of the Mu2e experiment, but are not within the scope of the Mu2e Project.

## 4.1.3 Accelerator Requirements

The Mu2e Accelerator upgrades are governed by seven requirements documents. These documents are:

- Mu2e Proton Beam Requirements [2]
- Beam Extinction Requirement for Mu2e [3]
- Production Target Requirements [4]
- Extinction Monitor Requirements [5]
- Requirements for the Mu2e Production Solenoid Heat and Radiation Shield [6]
- Mu2e Proton Beam Absorber Requirements [7]
- Protection Collimator Requirements [8].

The Beam Extinction and Extinction Monitoring requirements are described in sections 4.9 and 4.10 of this chapter. The Production Target, Heat and Radiation Shield, Proton Beam Absorber, and Protection Collimator requirements are handled in the Target Station section (section 4.11). The Mu2e Proton Beam Requirements will be described here.

Table 4.2 Accelerator Upgrades required for the Mu2e Experiment

| Accelerator Upgrade | Project |
|---|---|
| MI-8 beamline to Recycler Ring Injection | NOvA Project |
| Recycler Ring 2.5 MHz RF system | Recycler RF AIP[3] |
| Delivery Ring 2.4 MHz RF Cavities | Recycler RF AIP |
| Single bunch extraction from Recycler Ring | Beam Transport AIP |
| Beamline aperture upgrades | Beam Transport AIP |
| AP1, AP2, AP3 to M1, M2, M3 conversion | Beam Transport AIP |
| Beam transport instrumentation & infrastructure | Beam Transport AIP |
| Beam transport controls | Delivery Ring AIP |
| Delivery Ring Injection | Delivery Ring AIP |
| Delivery Ring Abort | Delivery Ring AIP |
| Delivery Ring infrastructure | Delivery Ring AIP |
| Delivery Ring Controls and Instrumentation | Delivery Ring AIP |
| D30 straight section reconfiguration | g-2 Project |
| Delivery Ring Extraction (except ESS) | g-2 Project |
| Extraction line (M4) to M5 split | g-2 Project |
| M4 beamline enclosure | MC Beamline Enclosure GPP[4] |

The Mu2e experiment requires a total of approximately $3.6 \times 10^{20}$ protons delivered to the production target over the course of a 3 to 4 year run. The proton beam consists of a train of narrow pulses that must be separated by an interval that is longer than the lifetime of a $\mu^-$ captured in the Aluminum stopping target (864 nsec). Furthermore, the proton beam must be extinguished between these pulses such that the ratio of out-of-time beam to in-time beam is less than $10^{-10}$. The narrow pulse and extinction level requirements are necessary for the reduction of prompt background events to an acceptable level. The Mu2e Proton Beam Requirements are summarized in Table 4.3. These requirements are illustrated pictorially in Figure 4.1.

## 4.1.4 Mu2e Operating Scenario

The proton beam will require considerable manipulation to produce the longitudinal structure required by the Mu2e experiment. These manipulations are performed in the Recycler and the Delivery storage rings and in the beamline that connects the Delivery Ring to the target. Figure 4.2 shows the layout of the Fermilab accelerator systems used to accomplish the Mu2e beam manipulations (see also Figure 4.6).

---

[3] An AIP is an Accelerator Improvement Project
[4] A GPP is a General Plant Project

Table 4.3. Summary of the Mu2e Proton Beam Requirements

| Parameter | Design Value | Requirement | Unit |
|-----------|-------------|-------------|------|
| Total protons on target | $3.6 \times 10^{20}$ | $3.6 \times 10^{20}$ | protons |
| Time between beam pulses[5] | 1695 | >864 | nsec |
| Maximum variation in pulse separation | < 1 | 10 | nsec |
| Spill duration | 54 | >20 | msec |
| Beamline Transmission Window | 230 | 250 | nsec |
| Transmission Window Jitter (rms) | 5 | <10 | nsec |
| Out-of-time extinction factor | $10^{-10}$ | $\leq 10^{-10}$ | |
| Average proton intensity per pulse | $3.1 \times 10^7$ | $< 5.0 \times 10^7$ | protons/pulse |
| Maximum Pulse to Pulse intensity variation | 50 | 50 | % |
| Minimum Target rms spot size[6] | 1 | 0.5 | mm |
| Maximum Target rms spot size[6] | 1 | 1.5 | mm |
| Target rms beam divergence | 0.5 | < 4.0 | mrad |

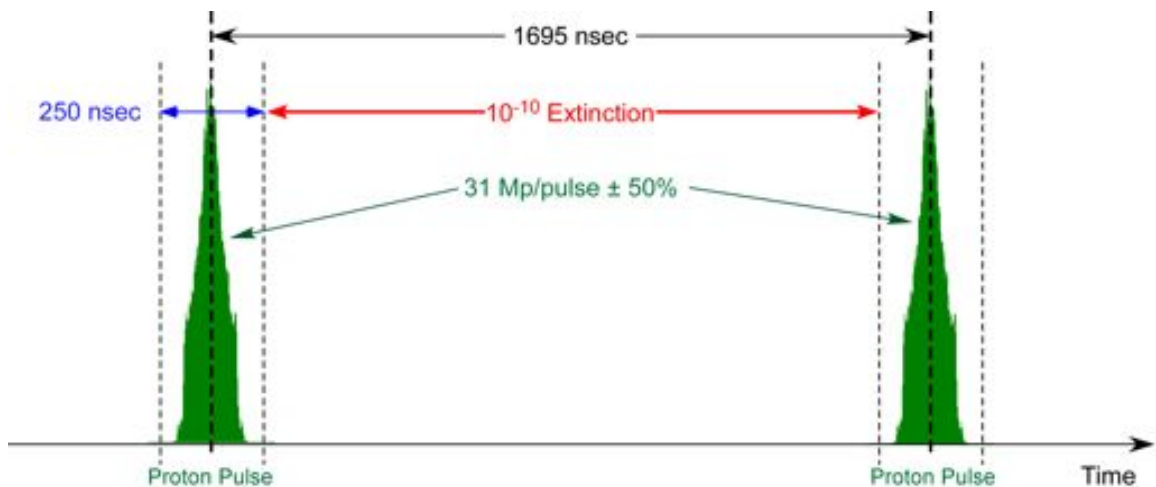

Figure 4.1. Longitudinal structure of the proton beam delivered to the Mu2e production target. The green shapes show the time profile of a beam pulse. The extinction and beam pulse intensity requirements are illustrated.

Protons designated for Mu2e are acquired from the Booster synchrotron by utilizing the unused portions of the Main Injector timeline during slip-stacking operations for NOvA (see Figure 4.3). Booster protons containing 81 batches of 53 MHz bunches, are extracted into the MI-8 beamline and injected into the Recycler Ring. As each batch circulates in the Recycler Ring it is re-bunched with a 2.5 MHz RF system [9] to form four bunches with the bunch characteristics required by the Mu2e experiment (see Section 4.7). After the 2.5 MHz bunch formation, the beam is extracted from the Recycler, one bunch at a time, and transported to the Delivery Ring. The beam is then resonantly extracted into the M4 beamline where it is transported to the Mu2e production target (see Sections 4.6 and 4.8). After the resonant extraction sequence is complete, a cleanup abort kicker is fired to remove any remaining beam.

---

[5] This is the Delivery Ring revolution period.
[6] Assumes a round beam

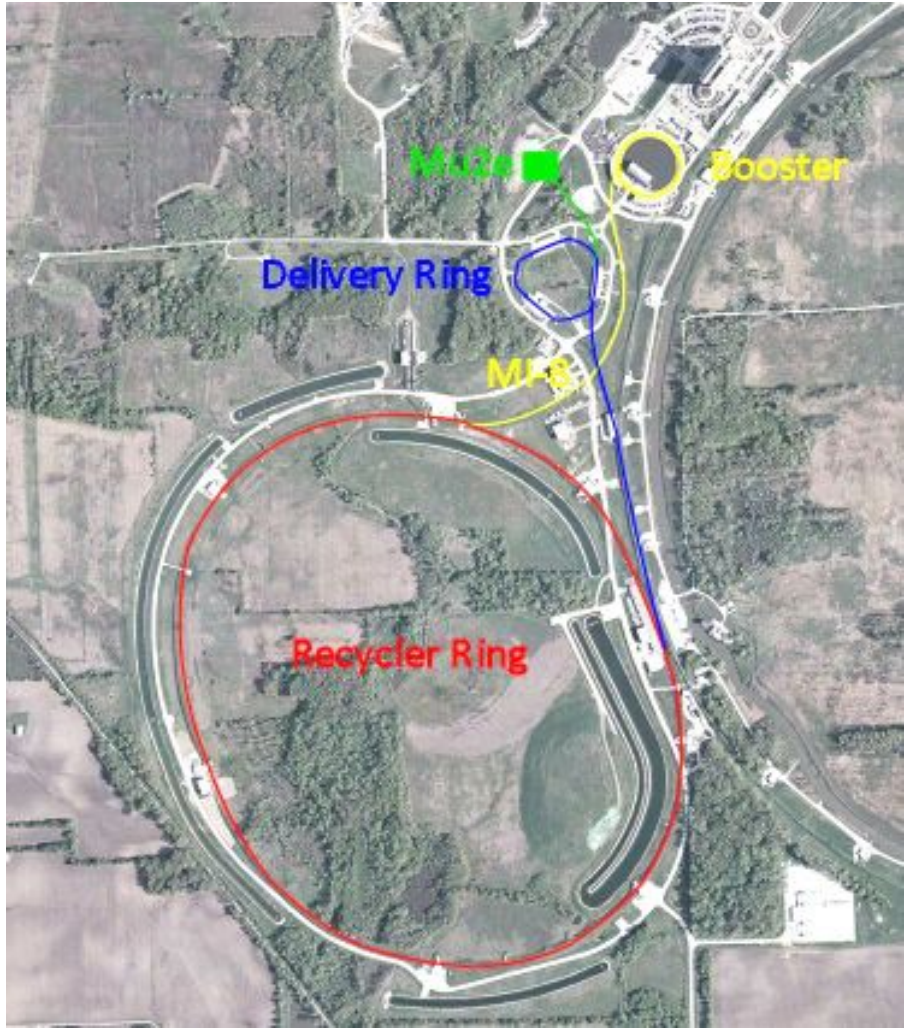

Figure 4.2. The components of the Fermilab accelerator complex used to acquire protons for the Mu2e experiment. The proton beam path from Booster to Recycler is shown in yellow. The beam path in the Recycler is in red. The beam path from Recycler to Delivery Ring is in blue, and the beam path from Delivery Ring to Mu2e target is in green.

The Delivery Ring to Mu2e target external beamline (called the M4 beamline) is a new facility that transports the proton beam to the Mu2e production target (Section 4.8). The M4 beamline contains a beam extinction insert that removes out-of-time beam to the required level (Section 4.9). Upon arrival at the production target, the beam interacts with a tungsten target inside the shielded super-conducting Production Solenoid (Section 4.11).

The resulting pions decay, producing the muons that will ultimately constitute the muon beam for the experiment.

### 4.1.5 Macro Time Structure of the Proton Beam

The Mu2e experiment must share the Recycler Ring with the NOvA experiment, which uses the Recycler for proton slip-stacking. This sharing is accomplished by performing the required Mu2e beam manipulations in the Recycler prior to the injection of the first proton batch designated for NOvA. There are a total of twenty possible proton batch injections into the Recycler Ring from the Booster within each Main Injector cycle. These proton injections will occur at a maximum rate of 15 Hz (one batch every 67 msec)[7]. Of the twenty available proton batches, NOvA requires twelve batches for slip-stacking. That leaves eight injection ticks (533 msec) for Mu2e to acquire its beam and complete the 2.5 MHz bunch formation process (see Figure 4.3).

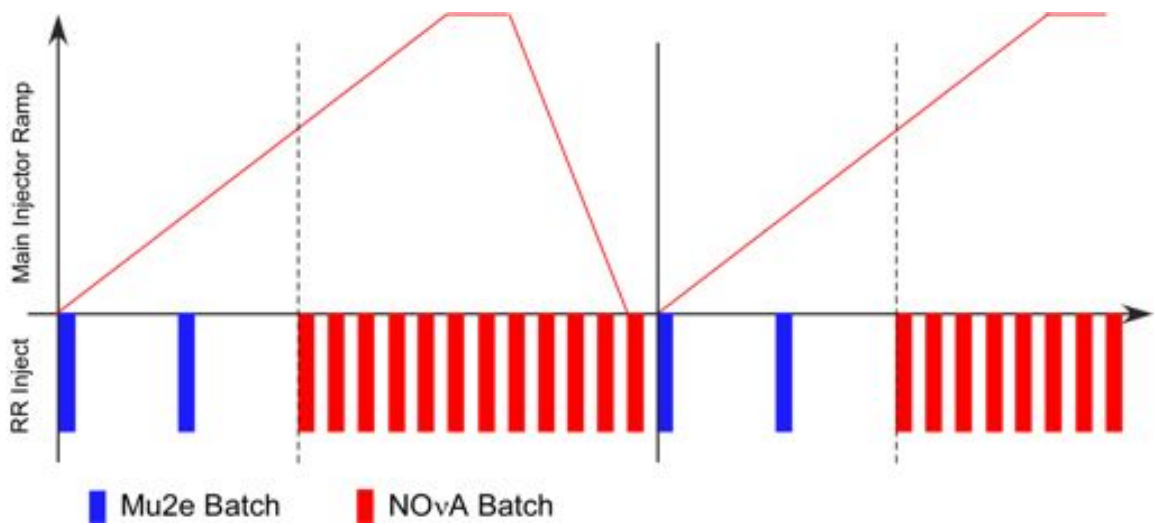

Figure 4.3. The accelerator timeline is shared between Mu2e and NOvA. The blue and red bars represent Mu2e and NOvA proton batch injections respectively. Mu2e beam manipulations in the Recycler Ring occur in the first eight 15 Hz ticks[8]. NOvA proton batches are slip-stacked during the remaining twelve 15 Hz ticks. The total length of a cycle is 20 ticks = 1.333 sec.

Figure 4.4 shows the utilization of the Mu2e portion of the Main Injector cycle. Two proton batches are injected into the Recycler, one at the beginning of the cycle and one four Booster cycles (ticks) later. Each batch occupies one seventh of the circumference of the Recycler Ring. After each injection, the beam circulates for 90 msec while the 2.5 MHz bunch formation RF sequence is performed. This RF manipulation coalesces the proton batch into four 2.5 MHz bunches. These bunches are transferred, one bunch at a time, to the Delivery Ring where the beam is slow-spilled to the experiment. Table 4.4 gives the parameters of the spill.

---

[7] This statement assumes the successful implementation of the Proton Improvement Plan (PIP).
[8] One Booster tick = 1/15 sec = 66.7 msec.

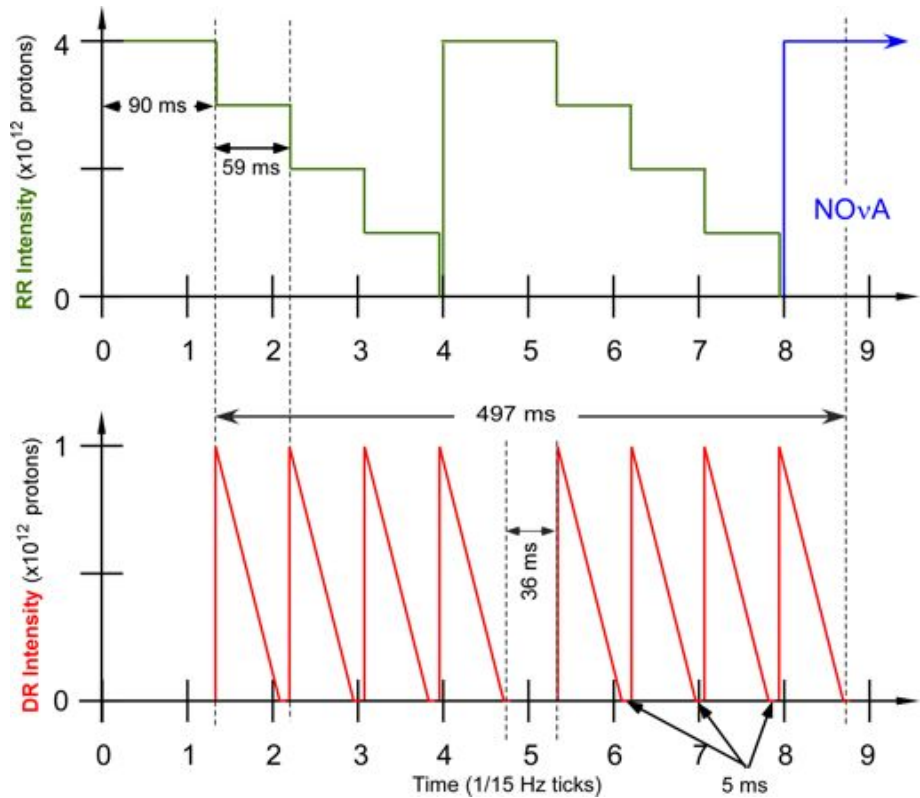

Figure 4.4. This figure shows the first eight Booster ticks of a Main Injector cycle. The top graph shows the Recycler Ring beam intensity as a function of time. The bottom plot shows the Delivery Ring beam intensity as a function of time. Proton batches are injected into the Recycler at the beginning of the cycle and again at the fourth tick.  After each injection, the beam is bunched with 2.5 MHz RF and is synchronously transferred to the Delivery Ring one bunch at a time. The beam is then resonantly extracted from the Delivery Ring over a period of 54 msec.

### 4.1.6 Accelerator Parameters

Table 4.5 gives a list of accelerator parameters pertinent to the Mu2e accelerator configuration. Unless otherwise stated, these values are used in the calculations and simulations described in this chapter.

## 4.2    Beam Physics Issues

The beam intensities anticipated for Mu2e operation far exceed the intensities seen in the Antiproton Source during Collider running. Thus, intensity dependent effects must be given careful consideration. We discuss separately the impact of high intensity on the transverse and longitudinal degrees of freedom. Transverse effects predominantly manifest themselves in beam self-defocusing, which causes incoherent shifts in the betatron tunes of the circulating particles. In the longitudinal degree of freedom, we consider the synchrotron tune shift and space charge induced beam self-impedance. Longitudinal beam dynamics may cause collective beam instabilities in both longitudinal and transverse directions when certain intensity thresholds are exceeded.

Table 4.4. Delivery Ring Spill Parameters

| Parameter | Value | Units |
|---|---|---|
| MI Cycle time | 1.333 | sec |
| Number of spills per MI cycle | 8 | |
| Number of protons per micro-pulse | $3.1\times10^{7}$ | protons |
| Maximum Delivery Ring Beam Intensity | $1.0\times10^{12}$ | protons |
| Instantaneous spill rate | $18.5\times10^{12}$ | protons/sec |
| Average spill rate | $6.0\times10^{12}$ | protons/sec |
| Duty Factor (Total Spill Time ÷ MI Cycle Length) | 32 | % |
| Duration of each spill | 54 | msec |
| Spill On Time per MI cycle | 497 | msec |
| Spill Off Time per MI cycle | 836 | msec |
| Time Gap between $1^{st}$ set of 4 and $2^{nd}$ set of 4 spills | 36 | msec |
| Time Gap between spills | 5 | msec |
| Pulse-to-pulse intensity variation[9] | ±50 | % |

## 4.2.1 Space Charge

At Mu2e beam intensities the self-defocusing space charge field of the circulating beam is not small in comparison to the external focusing field of the lattice quadrupole magnets. Space charge defocusing shifts the betatron tune downward relative to the bare lattice tune. Furthermore, the amount of tune shift depends on the betatron amplitude of a circulating particle. Small amplitude particles near the core of the beam charge distribution are subject to the largest tune shifts while large amplitude particles in the tails of the beam distribution undergo the smallest tune shifts. Thus, an intense beam containing a strong core of particles with small betatron amplitudes will present a wide distribution of tune shifts.

---

[9] The pulse intensity is expected to be approximately uniform on short time scales (< 1 msec). The time scale of the variation in pulse intensity is expected to be of order a few msec.

Table 4.5. Accelerator Parameters for Mu2e operations.

| Parameter | Value | Units |
|-----------|-------|-------|
| **Booster** | | |
| Beam Energy | 8.9 | GeV |
| Intensity per batch | $4\times10^{12}$ | protons |
| 53 MHz Bunches per batch | 81 | |
| Repetition rate | 15 | Hz |
| Average Repetition rate for Mu2e Beam | 1.5 | Hz |
| Transverse emittance | $15\pi$ | mm-mrad |
| Longitudinal emittance per 53 MHz bunch | 0.12 | eV-sec |
| **Recycler Ring** | | |
| Maximum Beam Intensity (for Mu2e) | $4\times10^{12}$ | protons |
| Revolution Frequency | 89.824 | kHz |
| $\eta$ | -0.00876 | |
| 2.5 MHz Re-bunch time | 90 | msec |
| 2.5 MHz bunches/batch | 4 | |
| Average Mu2e beam power | 7.69 | kW |
| Transverse emittance | $16\pi$ | mm-mrad |
| **Delivery Ring** | | |
| Maximum Intensity | $1\times10^{12}$ | protons |
| Revolution Frequency (central orbit) | 590018 | Hz |
| $\eta$ | 0.00607 | |
| Orbit Length (central orbit) | 505.294 | m |
| Average Injection Frequency | 6.0 | Hz |
| Peak Injection Frequency | 17.0 | Hz |
| $\nu_x$ | 9.650 | |
| $\nu_y$ | 9.735 | |
| Average $\beta_x$ | 9.5 | m |
| Average $\beta_y$ | 9.5 | m |
| Horizontal Admittance | $35\pi$ | mm-mrad |
| Vertical Admittance | $35\pi$ | mm-mrad |
| Peak Laslett space charge tune shift | 0.0097 | |
| Peak Space charge tune shift from tracking simulations | 0.0070 | |
| Bunch Length (rms) | 35 | nsec |
| Synchrotron tune | $5.9\times10^{-5}$ | |
| Transverse emittance | $16\pi$ | mm-mrad |
| Maximum Extracted Beam Power | 7.69 | kW |

The overall effect of space charge is the creation of a tune spread that extends from the bare lattice tune downward by an amount that depends on beam intensity and particle amplitude. The overall extent of the tune distribution is equal to the tune spread of the low amplitude particles at the core of the beam. This maximum tune shift can be estimated using the Laslett formula [10]:

$$\Delta v_x = \frac{3 r_p N_{tot} L_R}{2 \pi \gamma^2 \varepsilon_N L_b} = 0.0097 \tag{4-1}$$

where $r_p$ is the classical proton radius, $N_{tot}$ is the total number of particles, $L_R$ is the orbit length, $L_b$ is the effective bunch length, and $\varepsilon_N$ is the normalized horizontal emittance.

Equation (4-1) assumes that the beam is round and that particle amplitudes are dominated by their betatron oscillations. The precise nature of the space charge tune shift must be obtained from the more accurate integration around the ring inherent in tracking simulations. Tracking simulations will also properly account for the effects of beam broadening in the high dispersion regions of the lattice.

Substituting the Delivery Ring parameter values from Table 4.5 into Equation (4-1) yields a Laslett tune shift for the Delivery Ring of $\Delta v_x = 0.0097$. The Delivery Ring tune footprint from an ORBIT tracking simulation [11] is shown in Figure 4.5. The smaller tune shift shown in the tracking simulation (~0.007) is a consequence of the large energy spread of the beam after bunch formation. The beam spreads transversely in the arcs reducing the defocusing field felt by each particle. Since the arcs constitute a relatively large part of the Delivery Ring circumference, the effect is significant.

The increased tune footprint of the beam due to space charge constrains the choice of the operating point such that the entire tune footprint must lie to the right of the $2 v_x + v_y = 29$ resonance line. The tune footprint also must be in the vicinity, but to the left of, the $3 v_x = 29$ line, which is the line used for resonant extraction. The greatest impact of space charge induced effects is on resonant extraction. This is discussed further in Section 4.6.

## 4.2.2 Coherent instabilities

Transverse stability in the Delivery Ring for Mu2e operating conditions is studied in References [12][10] and [13]. Reference [13] also treats longitudinal stability and accounts for space charge effects. The conclusion of these studies is that the Delivery Ring is longitudinally and transversely stable for Mu2e beam conditions.

---

[10] The analysis of reference [12] does not include the effects of space charge.

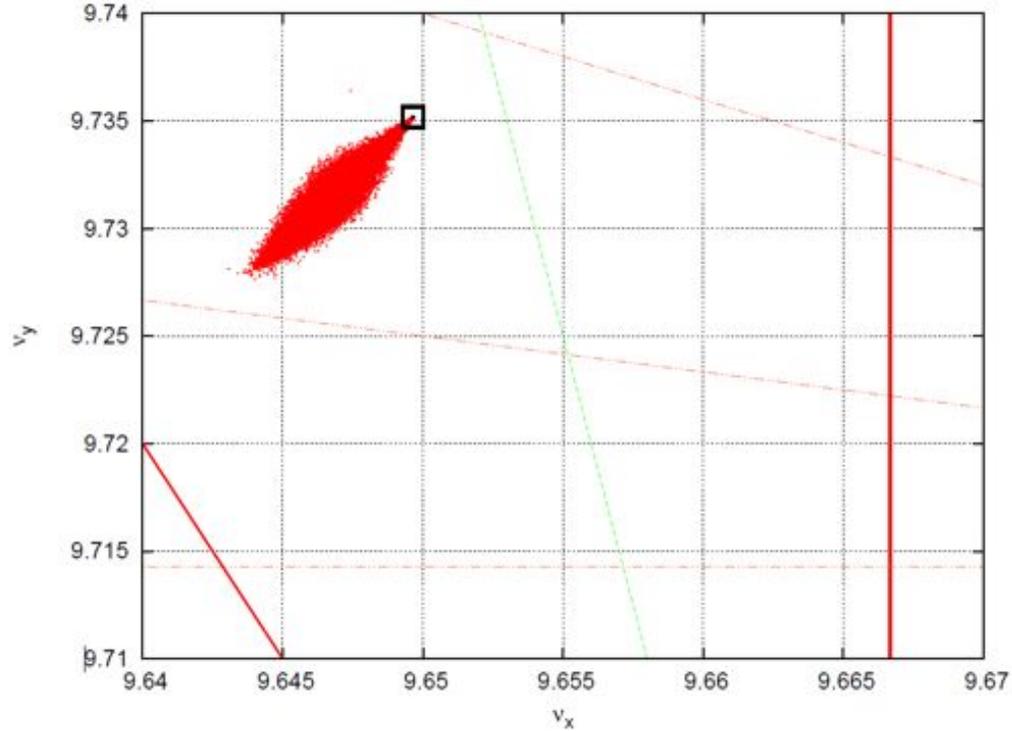

Figure 4.5. Delivery Ring tune footprint from an ORBIT [11] simulation for a beam intensity of $1 \times 10^{12}$ protons. The black box in the upper right of the plot indicates the bare lattice tunes. The thick red lines are 3[rd] order resonance lines; the dashed green lines are 6[th] order resonance lines; and the orange dot-dashed lines are 7[th] order resonance lines. The resonance line used for third integer extraction is the $3\nu_x = 29$ line at the right of the plot.

Betatron tune shifts due to space charge play a significant role in the treatment of the transverse stability. The space charge tune shift in the Delivery Ring significantly exceeds the synchrotron tune. In this case, for zero chromaticity, a bunch is stable up to the transverse mode coupling instability (TMCI) threshold [13], [14]. In Reference [14] it is shown that the TMCI threshold for a Gaussian bunch of rms length $\tau_b$ in a round chamber with conductivity $\sigma$ and radius $b$ occurs at[11]

$$ K \equiv \frac{N_b\, r_p \beta_x\, R_0\, Q_{sc}\, \eta_{occ}}{4\pi^2 \gamma Q_s^2\, b^3 \sqrt{\sigma\, \tau_b}} \approx 100 \qquad (4\text{-}2) $$

where $N_b$ is the number of protons in the bunch, $\beta_x$ is the average Delivery Ring beta-function, $Q_{sc}$ is the space charge tune spread, $\eta_{occ}$ is the dipole occupancy (~25%), $Q_s$ is the synchrotron tune, and $\sigma = 1.3 \times 10^{16}$ sec$^{-1}$. The main contribution to the impedance comes from the fraction of the circumference occupied by the dipoles where the vertical aperture is $b = 2.6$ cm (the remaining 75% of the ring is a round chamber with a 6.4 cm radius). These Delivery Ring parameters yield $K \approx 4$. This value is significantly smaller than the threshold value of Equation (4-2). Thus, under Mu2e operating conditions, the beam should be well below the TMCI threshold.

---

[11] Figure 4 of Reference [14] shows that for $K \lesssim 100$ the modes are uncoupled and therefore below the onset of TMCI.

If the chromaticity is not zero, weak head-tail instability may be possible. The maximum growth rate of the weak head-tail instability is given in Reference [14]:

$$\Gamma \, T_0 = 0.1 \frac{N_b \, r_p \beta_x \, R_0 \, \eta_{occ}}{\pi \gamma \, b^3 \sqrt{\sigma \, \tau_b}} \left[ \text{turn}^{-1} \right]. \tag{4-3}$$

This yields $\Gamma T_0 N_t \sim 0.1$ for a spill duration of 54 msec ($N_t = 3.2 \times 10^4$ turns). Thus the weak head-tail instability should not be an issue. In the unlikely event that head-tail is an issue, the insertion of a small amount of chromaticity should be sufficient to damp this instability.

### 4.2.3 Other Intensity Dependent Effects

The bunched beam should not be affected by space charge or resistive wall longitudinal impedances. The space charge synchrotron tune shift is estimated to be as small as ~1% of the synchrotron tune, while the space charge resistive wall tune shift is even smaller.

The electron cloud instability should not be a big concern, since each bunch is short compared to the zero-current time for any visible cloud to be built [13].

An analysis of intra-beam scattering shows that it is too slow to be seen, given the relatively short time the beam circulates in the Delivery Ring [15].

## 4.3   Accelerator Instrumentation

### 4.3.1 Instrumentation Requirements

Mu2e Instrumentation can be divided into four categories.

- **Beam Line**:  This includes all instrumentation used to measure single-pass primary proton beam in the beam lines between the Recycler and Delivery Ring.

- **Delivery Ring**:  This includes all instrumentation used to measure circulating beam in the Delivery Ring.

- **Abort Line**:  This includes all instrumentation used to measure the beam in the Delivery Ring abort line.

- **External Beam Line**:  This includes all instrumentation used to measure the slow spill beam in the M4 line.

The requirements of the accelerator instrumentation are to measure the beam intensities, positions, profiles and losses in all four of the above categories. Table 4.6 summarizes the beam condition requirements for each category. The layout of the Muon Campus beam lines is shown in Figure 4.6.

Table 4.6. Beam requirements for Accelerator Instrumentation.

| | Beam Lines | Delivery Ring | Abort Line | External beam Line |
|---|---|---|---|---|
| Beam Line Names | P1 Stub, P1, P2, M1, and M3 | Delivery Ring | Abort Line | M4 |
| Particles | Protons | Protons | Protons | Protons |
| Momentum (GeV/c) | 8.88626 | 8.88626 | 8.88626 | 8.88626 |
| # of Particles | $1 \times 10^{12}$ | $1 \times 10^{12}$ (spill start) to $2 \times 10^{10}$ (spill end) over 54 msec | $2 \times 10^{10}$ at the end of every spill or up to $1 \times 10^{12}$ when beam permit is pulled. | Slices of $3 \times 10^{7}$ every 1.695 µsec totaling $1 \times 10^{12}$ over the 54 msec slow spill cycle. |
| Bunch Length (FW) | 250 nsec | 250 nsec | 250 nsec | 250 nsec |
| Transverse Emittance (mm-mrad) | 15π | 16π | 30π | 30π |
| Beam Line Length | ~975 m | 505 m | 72 m | 244 m |

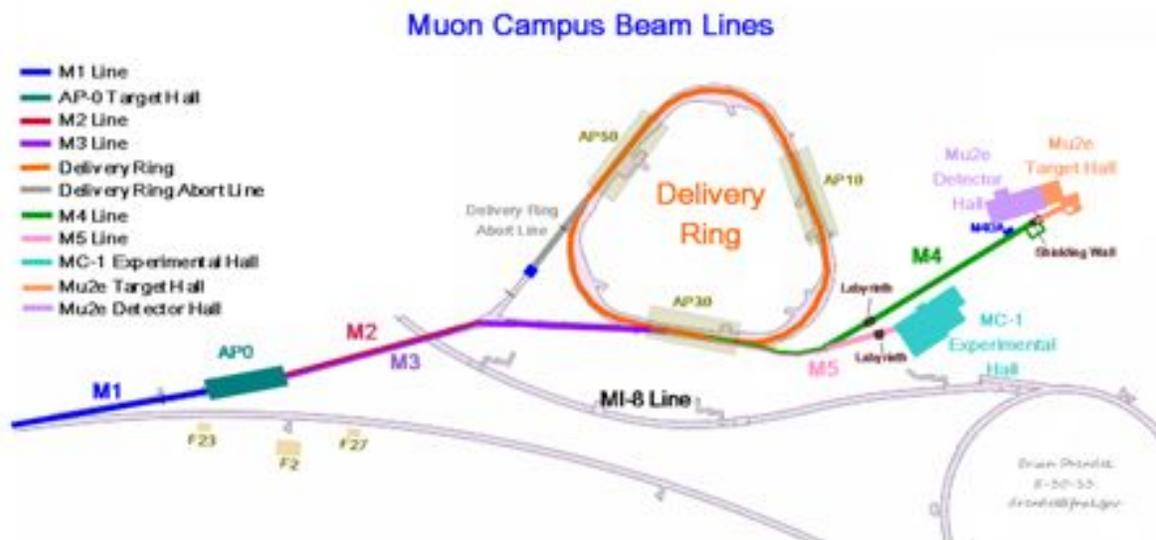

Figure 4.6. Muon Campus beam lines.

The accelerator instrumentation required for the operation of Mu2e is funded by the Beam Line AIP, the Delivery Ring AIP and the Mu2e project. Table 4.7 outlines the funding source for each type of instrumentation.

Table 4.7. The funding for the accelerator instrumentation required for Mu2e operations comes from multiple sources. This table identifies the funding source for each type of instrumentation.

| Category | Instrumentation Type | Funding Source |
| --- | --- | --- |
| Beam Line | Toroid | Beam Line AIP |
| | Beam Position Monitor | Beam Line AIP |
| | Beam Loss Monitors | Beam Line AIP |
| | Profile Monitors | Beam Line AIP |
| Delivery Ring | DCCT[12] | Mu2e Project |
| | Beam Position Monitor | Delivery Ring AIP |
| | Beam Loss Monitor | Delivery Ring AIP |
| | Tune Measurement System | Mu2e Project |
| Abort Line | Toroid/Ion Chamber | Mu2e Project |
| | Profile Monitor | Mu2e Project |
| | Beam Loss Monitors | Delivery Ring AIP |
| External Beam Line | Ion Chamber | Mu2e Project |
| | Profile Monitor | Mu2e Project |
| | Beam Loss Monitor | Mu2e Project |

## 4.3.2 Instrumentation Technical Design

### 4.3.2.1 Beam Line Instrumentation

Single-pass primary proton beam will traverse the P1 stub[13], and the P1, P2, M1, and M3 beamlines. Much of the instrumentation needed to measure the primary proton beam during Mu2e operation already exists but must be modified for use with the faster cycle times and 2.5 MHz RF beam structure [32] [33]. The overall beam intensity is similar to that seen in Pbar stacking operations, and in many cases requires that only small calibration changes be made to the instrumentation. Toroids will be used to monitor beam intensity and will be used in conjunction with Beam Loss Monitors (BLMs) to maintain good transmission efficiency in the beam lines. Multiwires and Secondary Emission Monitors (SEMs) will provide beam profiles in both transverse planes. Beam Position

---

[12] DCCT = DC Current Transformer (see Section 4.3.2.1.1)
[13] The P1 stub is the short line that connects the Recycler Ring to the P1 beamline.

Monitors (BPMs) will provide real-time orbit information and will be used by auto-steering software to maintain desired beam positions in the beam lines.

#### 4.3.2.1.1   Beamline Toroids

Toroids are beam transformers that produce a signal that is proportional to the beam intensity. There are two toroids in the P1 line, one in the P2 line, two in the M1 line and one in the M3 line. With the exception of 2 toroids in the M2 line, all the transformers are 3100 models from Pearson Electronics. These toroids have a 3.5" inner diameter, 1 V/A sensitivity in high impedance, and a 0.04%/μsec droop rate. The exceptions in the M2 line are a custom 7737 model from Pearson Electronics. Having 1 V/A sensitivity and a 10.75" inner diameter, these toroids are a special version of the 1010 model from Pearson Electronics. They will continue to be used in Mu2e operation to measure the primary proton beam.

A block diagram of a toroid system is shown in Figure 4.7. The electronics for these toroids are comprised of legacy analog processing inside of NIM crates. Filters, chokes, and preamps will be added for analog conditioning. Electronics will be modified, where necessary, to calibrate the toroids for Mu2e operations [18].

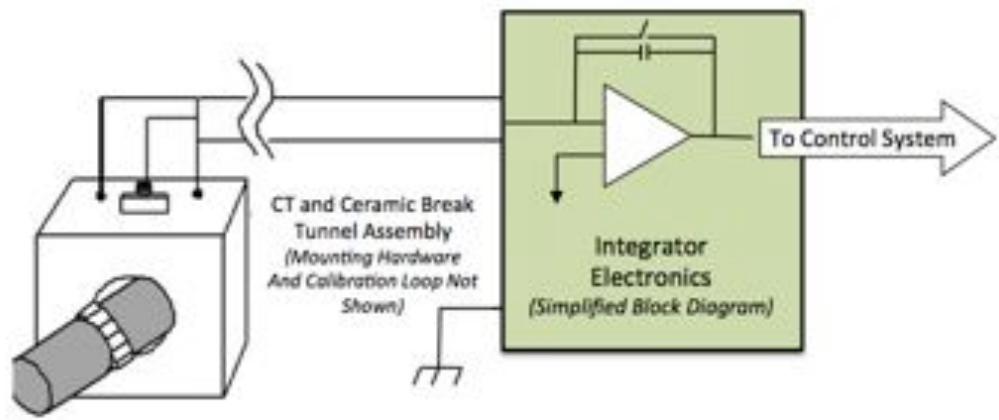

Figure 4.7. Simplified Toroid system block diagram (not to scale) [18].

#### 4.3.2.1.2   Beamline BPMs

Beam line BPMs provide single pass orbit position information with sub-millimeter resolution, and will continue to be the primary beam position devices in the P1, P2, M1 and M3 lines. All BPMs share the Echotek style of electronics that were built as part of the Rapid Transfers Run II upgrade [17], and is the current standard for beam line BPMs. A functional diagram of the BPM hardware is shown in Figure 4.8. These BPMs were designed to detect 7 to 84 consecutive 53 MHz proton bunches and four 2.5 MHz antiproton bunches for Collider Run II operations. Minimal electronics modifications will be required to measure the single 2.5 MHz bunches of $1 \times 10^{12}$ particles expected during Mu2e operations [19]. Two additional BPMs will be installed in the P1 stub.

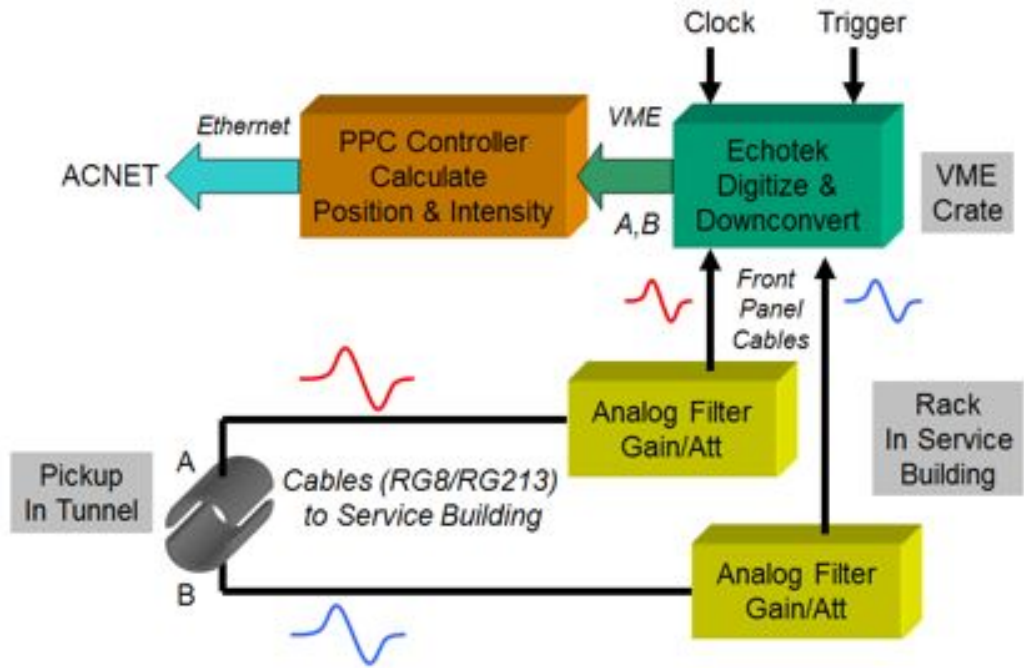

Figure 4.8. BPMs with Echotek processing electronics will be used to measure the transverse beam position of the 2.5 MHz primary proton beam in the P1, P2, M1 and M3 lines for Mu2e operations [17].

### 4.3.2.1.3   Beamline BLMs

BLMs are already in place in the P1, P2, M1 and M3 beam lines. Existing ion chamber detectors will be utilized for Mu2e operation. BLMs will be upgraded to modern BLM log monitor electronics, repurposing unused components from the Tevatron to minimize cost. Two additional BLMs will be installed in the P1 stub [20].

### 4.3.2.1.4   Beamline Profile Monitors

There are two types of beam profile monitors in the beam lines, multiwires in the P1 and P2 lines, and SEMs in the other beam lines. The profile monitors will primarily be used for commissioning, studies, and documentation of the beam lines. General maintenance will be performed on the hardware and electronics to ensure proper functionality. The current location and wire spacing of the monitors will be reviewed and modified accordingly. Two additional multiwires will be installed in the P1 stub [21].

### 4.3.2.2   Delivery Ring Instrumentation

Primary proton beam will circulate in the Delivery Ring as it is slow spilled to the M4 line over a period of 54 msec. Like with the beam transport lines, most of the instrumentation needed for operation of the Mu2e Delivery Ring already exists but must be modified or upgraded to accommodate the faster cycle times. The existing DC current transformer (DCCT) will be used to monitor beam intensity through the slow spill cycle. Beam position monitors (BPMs) and beam loss monitors (BLMs) will be used to monitor the positions and losses in the line. Both systems will need significant hardware and electronics modifications to work under Mu2e operational conditions; however, much of the needed equipment can be repurposed from the collider. To regulate and optimize the Delivery Ring resonant extraction process a Delivery Ring tune measurement scheme will be required. Schottky Detector hardware and electronics will be recycled from the Tevatron to construct this system.

### 4.3.2.2.1 Delivery Ring Beam Intensity Monitors (DCCT)

A DCCT is a device used to measure the quantity of circulating beam with high precision. D:BEAM[14] will become the beam intensity read back for the Delivery Ring. The Accumulator DCCT becomes a spare [18] [22] [35].

Both systems have a full-scale range of 400 mA ($400 \times 10^{10}$ protons). Measurements have shown nonlinear errors approaching ~2% for DC currents greater than 200 mA. To improve linearity, the calibration procedure will rely on a least squares fit between 0-200 mA. The system has an accuracy of one part in $10^5$ over the range of $1 \times 10^{10}$ to $2 \times 10^{12}$ particles with a noise floor of $2 \times 10^9$ (see Figure 4.9).

The Delivery Ring DCCT will not require any specific changes to the physical detector, but will need its analog conditioning and VME electronics modified for Mu2e operation.

Figure 4.10 is a block diagram of the DCCT system. The pickups consist of two sets of supermalloy tape-wound toroidal cores with laminations. The laminations act to reduce eddy currents. The beam passes through the center of the toroidal core and acts as a single turn on both toroid sets. The beam sensing electronics are attached to wire windings on each of the toroid sets. One set of cores (T1 & T2) is driven into saturation with an 800 Hz sinusoidal drive signal. The passing beam induces a net magnetic flux on this set of cores, and a second harmonic of the drive signal is detected in the sense winding. The electronics processes this signal and produces an equal and opposite current that minimizes the harmonic and thus keeps the net toroid flux at zero. The second set of cores (T3) is not driven into saturation and follows classical transformer theory. This signal is detected and processed by the electronics to extend the bandwidth of the entire system. The combination of the resulting signal from processing each sense winding produces the complete feedback signal.

---

[14] D:BEAM is the accelerator control system designation of a particular readout of the Delivery Ring DCCT. The "D" refers to the Delivery Ring. There can be several control system parameters associated with a single instrument. For example, D:BEAM and D:BEAMB are different readouts of the Delivery Ring DCCT.

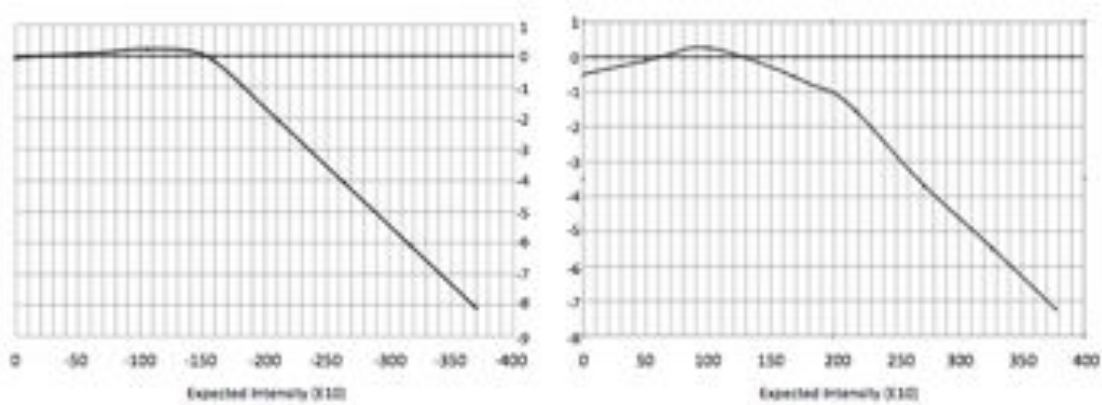

Figure 4.9. Resulting errors in Debuncher (left) and Accumulator (right) DCCT signals when fitting calibration data to provide minimal errors for intensities below 200 mA. In both plots the x-axis is the beam intensity in units of $1\times10^{10}$ particles, and the y-axis is the expected error between measured and actual beam intensity in units of $1\times10^{10}$ particles. The Debuncher DCCT has an error value less than $2\times10^{9}$ for all beam intensity values less than $150\times10^{10}$ particles. The Accumulator DCCT has an error value less than $5\times10^{9}$ over the same range.

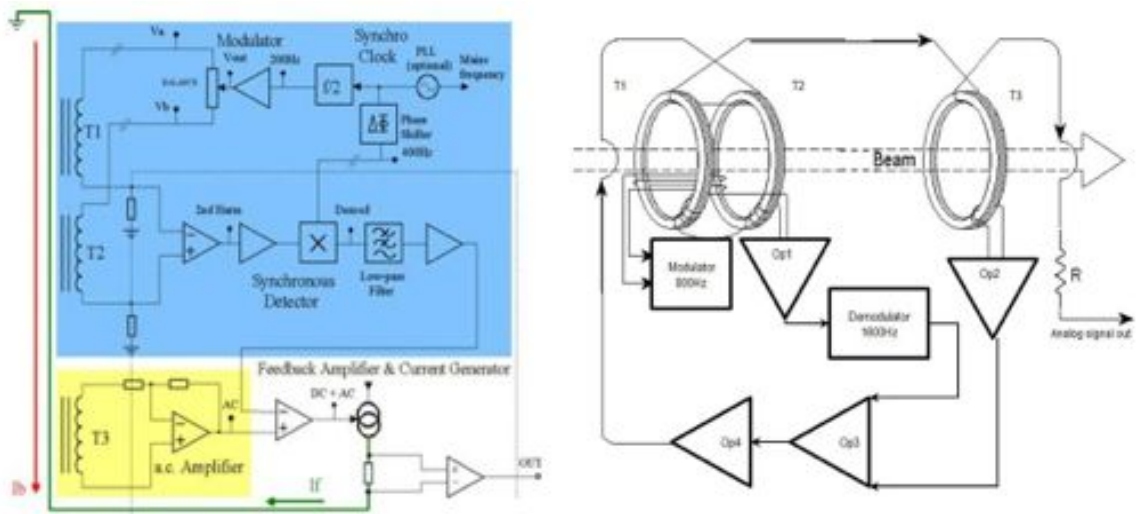

Figure 4.10. Functional block diagrams of DCCT operation.

As shown in Figure 4.11, the receiver chassis provides three different sets of analog outputs. Each output has a different voltage range and bandwidth, based on modifications in the receiver electronics. To best measure both the injected and leftover Delivery Ring beam, the following configurations were chosen:

- 1 Hz bandwidth with 40 mA/V scale. This output is routed to a Keithley digital voltmeter (DVM) located in a rack in the AP10 control room. The Keithley DVM is a GPIB device that communicates with the control system through the AP1001 front end, resulting in the D:IBEAM read back that updates once per second with a scale in the mA particle range. Due to the slow 1 Hz sample rate, D:IBEAM is not useful to measure Mu2e slow extracted beam, but will be used for circulating beam studies, diagnostic studies, or calibrations. Maintaining this device also will provide backward compatibility to other intensity algorithms running on AP1001.

- 400 Hz bandwidth with 40 mA/V scale. This output will simultaneously provide two read back devices in the control system. One device, D:IBEAMB, is processed through an MADC, using the standard CAMAC 190 card communicating through the Pbar CAMAC front end. At a 720 Hz update rate, this 12 bit read back effectively provides 40 samples of beam intensity measurements in the mA particle range over the 54 msec spill. Although this device would allow us to measure both the injected beam and the beam as it evolves through the slow spill cycle, D:IBEAMB is primarily intended for diagnostic purposes. In addition, this output will provide a second device, D:BEAM. Its output is sampled by a VME front end after it is conditioned thru a buffer amplifier (see the later description of PBEAM). D:BEAM is intended to be the primary measurement of the injected beam and the beam through the slow spill cycle.

- 400 Hz bandwidth with 5 mA/V scale. This output, D:IBEAMV will drive a buffer amplifier, whose output is sampled by a VME front end (see the later description of PBEAM). This intensity read back, in the µA range, will provide the best measurement of the leftover beam in the Delivery Ring after the slow spill has finished.

The DCCT noise levels are essentially the same in these outputs. The noise of the Debuncher DCCT is about 1.4 µA rms for both the 5 and 40 mA/V outputs (square root of the sum of the squares from 1 to 58 Hz). The Accumulator DCCT measures 2.2 µA rms. A plot of typical Debuncher DCCT noise levels is shown in Figure 4.12.

PBEAM is a VME front end located in AP10. The front end includes a five-slot VME crate with a Motorola MVME-2401 controller. An ICS-110BL-8B provides 4 differential 24-bit ADC channels delivering 18 bits of accuracy. The complete specifications of the PBEAM DCCT front end are given in Table 4.8. A PMC-UCD provides TCLK[15] and allows ACNET channels that track beam intensity. (MDAT can also be supported if desired.) Originally, it was instrumented such that a single input to the digitizer originated from each of the pbar DCCT systems (i.e. one for Debuncher and another for Accumulator).

---

[15] TCLK refers to the 10 MHz Tevatron Clock system

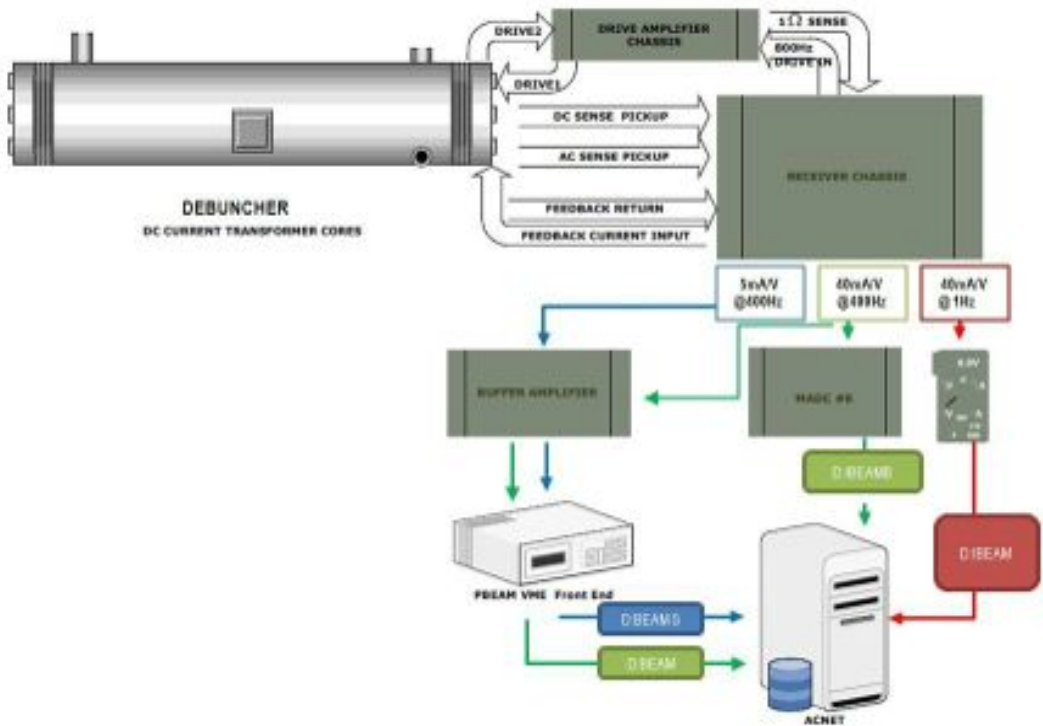

Figure 4.11. Delivery Ring DCCT system block diagram. The former Accumulator DCCT pickup, drive amplifier chassis, and receiver chassis will serve as a spare. It will be modified to be identical.

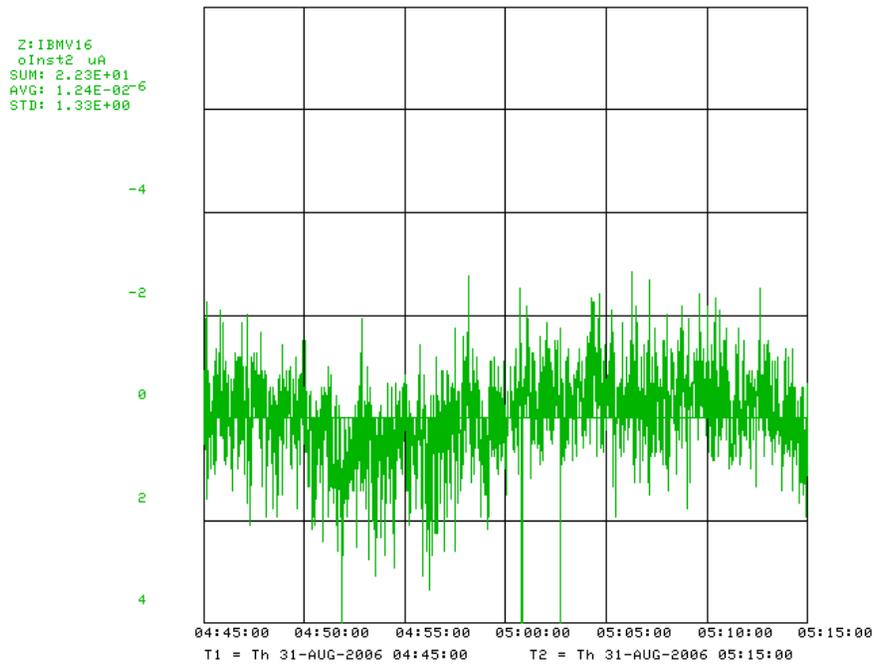

Figure 4.12. DCCT noise levels. Plot shows DCCT output in μA versus time for 30 min.

Since the Accumulator now becomes a spare system, the VME front end will be modified so that both inputs to the digitizer will be from one DCCT system (i.e. the Delivery Ring DCCT system). One channel will be fed from the 40 mA/V receiver chassis output, while the other will be fed from the 5 mA/V output. A conditioning amplifier will be used for filtering and to provide differential inputs to the ADC. The Delivery Ring signals will be filtered, amplified and driven differentially to the ADC about 50 feet away, located above the Accumulator DCCT. The digitizer oversamples the 400 Hz DCCT outputs at 720 Hz and can provide read backs with a resolution on the order of a sliding average of twelve 720 Hz samples. Digital signal processing will allow the measurement of the injected beam and beam through the slow spill cycle as well as the measurement of the leftover beam after slow spill has finished. The proposed digitizer has the dynamic range to measure the full 400 mA range while providing ½ µA rms resolution. Processing the signals digitally will also improve accuracy. The DCCT's provide between 1 and 2 µA rms provided the 60 Hz harmonics are removed.

PBEAM was designed to provide stable read back that is fast enough to sample Delivery Ring beam at various times during the slow spill cycle. D:BEAM is the Delivery Ring Beam Current or main intensity device and is used to derive other arrayed data. PBEAM also provides two other sets of devices that record beam measurements. Fixed-event devices will be labeled as D: BEAMXX, where XX is the TCLK reset event. Variable-triggered-event devices will be labeled as D:BEAMx. Each element will have a settable TCLK event and delay, measured in seconds. The TCLK event is set via the "Timer Reference" selection on an accelerator console parameter page. The delay is the D/A setting and the intensity read back is the A/D reading.

#### 4.3.2.2.2   Delivery Ring Beam Position Monitors (BPM)

The primary system used to measure the beam orbit in the storage rings will be a set of beam position monitors (BPM) distributed along the rings. The existing split-plate BPM pick-ups are suitable for Mu2e operation and will not require modifications.

The BPM read-out hardware is based on an analog differential receiver-filter module for analog signal conditioning, and a digital signal processing system, reusing the Echotek 8-channel 80MSPS digital down-converter and other VME hardware from the Recycler BPMs. This system provides beam position and intensity measurements with a dynamic range of 55 dB and an orbit measurement resolution of ±10 µm. The position measurements can be performed on 2.5 MHz bunched beam, as well as on a 53 MHz bunched Booster batch. Data buffers are maintained for each of the acquisition events and support flash, closed orbit and turn-by-turn measurements. A calibration system provides automatic gain correction of the BPM signal path. The software will need to be modified to handle specific events and data acquisition for Mu2e operation [23].

Table 4.8. ICS-110B Motherboard Specifications for PBEAM DCCT front end.

| DCCICS-110 Mother Board Specifications for DCCT | |
| --- | --- |
| No. of Diff. Analog Inputs | 4,8,16 or 32 |
| Input Impedance | 10 kΩ |
| Full Scale Input | 2 V pp differential |
| Max. Input Signal BW | 40 kHz |
| Input Sample Rate | 128 X Output Rate for BW < 22 kHz<br>64 X Output Rate for BW > 22 kHz |
| Output Rate (Effective Sample Rate) | Max. 100 kHz/channel<br>Min.     2 kHz/channel |
| Internal Sample Clock | Programmable in steps of 20 Hz |
| Dynamic Range | >110 dB in 128 X oversampling mode<br>>105 dB in 64 X oversampling mode |
| Total Harmonic Distortion | <-105 dB |
| Crosstalk | <-105 dB |
| On Board Storage | 64 K Words |
| Output Word Length | 32 bits packed for 2 channels or 24 bits for 1 channel on both VME and VSB<br>24 bits only on FPDP |
| VMEBus Interface | A32/24/16 D32 BLT Slave<br>Vectored Interrupts |
| VSBBus Interface | A32 D32 BLT Slave<br>Polled Interrupts |
| FPDP Interface | Refer to ICS Input Technical Note #15<br>Programmable Word Rate up to 20Mwords/s |
| Power | 6.0 Amps @ + 5V<br>0.42 Amps @ + 12V<br>0.25 Amps @ - 12V |
| Operating Temp | 0 to +50°C |
| Storage Temp | -40 to +85°C |
| Humidity | 95% Rel. Humidity, non-condensing |
| Board Size | 6U VMEbus Standard |

#### 4.3.2.2.3    Delivery Ring Beam Loss Monitors

The beam loss monitors (BLMs) are used to locate and measure beam losses in the storage rings. Although there is already a BLM system in place in the Delivery Ring, it will require significant upgrades for Mu2e operation. The existing photomultiplier tubes will be too sensitive for the expected Mu2e radiation environment, so they will be replaced by ion chambers repurposed from the Tevatron. The electronics have to be re-designed to accommodate the fast cycle time planned for Mu2e. The system will provide a sample-and-hold acquisition technology on individual beam pulses [20].

#### 4.3.2.2.4    Delivery Ring Tune Measurement

The Delivery Ring is a resonant extraction machine that will require a tune measurement system [36]. The delivery ring tune measurement system will need to measure the average tune and the tune spectrum through the entire 54 msec resonant extraction cycle. These tune measurements will have a resolution of 0.001 at 600 Hz and 0.0001 using averaging. In addition to measuring the tune, the system will measure transverse emittance using information from the tune spectrum.

The delivery ring tune measurements will consist of two parts, (1) a Schottky detector system and (2) a direct diode detection base-band Q (3D BBQ) measurement. Both systems have the advantage of being non-destructive to the beam. While the Schottky system has the additional advantage that it can measure transverse emittance, it has the challenge of achieving the desired accuracy at an adequately fast update rate. Averaging of the 3D BBQ system will alleviate this problem.

The Schottky detector system consists of 21.4 MHz resonant pickups taken from the decommissioned Tevatron as well as its receiver electronics [28]. The 21.4 MHz resonant pickups are two separate horizontal and vertical units with one meter long copper electrodes with a stepper motor driven adjustable aperture. Figure 4.13 shows the vertical pickup unit with attached 21.4 MHz resonator. The Schottky system will measure the tune spectrum and transverse emittance.

The second tune measurement system is based on the technique of direct diode detection base-band Q [29]. Previous 3D BBQ systems have reached a sensitivity allowing observation of beam betatron oscillations with amplitudes below one micron [30]. Since the 3D BBQ electronics are relatively cheap, the delivery ring tune system will utilize one or more BPMs as signal pickups, thus allowing for tune averaging to improve the tune resolution with an adequately fast update rate. Figure 4.16 shows a functional block diagram of a 3D BBQ system.

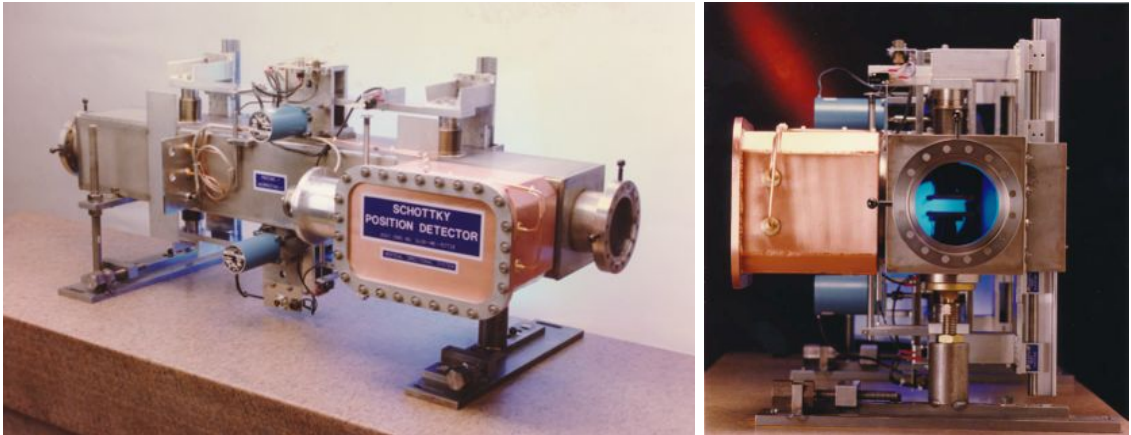

Figure 4.13. Tevatron 21.4 MHz Schottky pickup with resonator.  Side view (left) and Beam view (right).

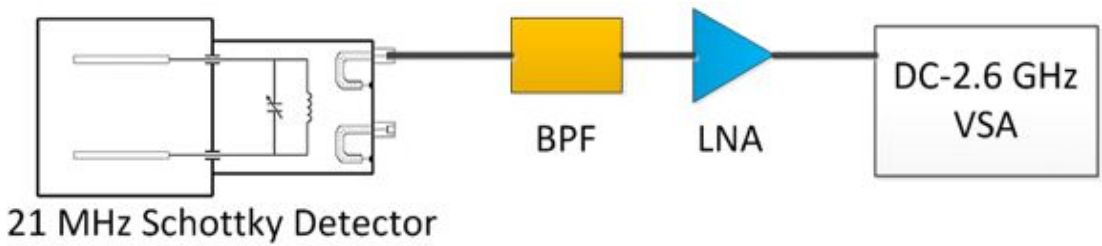

Figure 4.14. Setup for Schottky beam commissioning studies.

The RF Schottky diodes act as sample and hold peak detectors for each electrode signal with the subsequent filter setting the decay rate.  After applying DC suppression, only the amplitude of the turn-by-turn change in the signal remains. The resulting difference signal, which is the betatron modulation, can be further filtered and amplified to provide very high sensitivity to the betatron motion. This signal is then sampled by low rate high precision ADCs.

### 4.3.2.3   Abort Line Instrumentation

Leftover primary proton beam from each spill from the Delivery Ring, as well as Delivery Ring beam that remains when the beam permit goes away, will be sent to the Delivery Ring abort located in the former AP2 line. As with the beam transport lines, most of the instrumentation needed for operation of the Abort Line already exists, but needs to be modified or upgraded to accommodate the faster cycle times.  An existing Toroid and Ion Chamber will be used to monitor beam intensity in the abort line.  Beam position monitors (BPMs) and beam loss monitors (BLMs) will be used to monitor the positions and losses in the line. Much of the needed equipment can be repurposed from unused collider equipment; however both systems will require significant hardware and electronics modifications to work under Mu2e operational conditions. Beam profiles will be measured by using two existing SEMs.

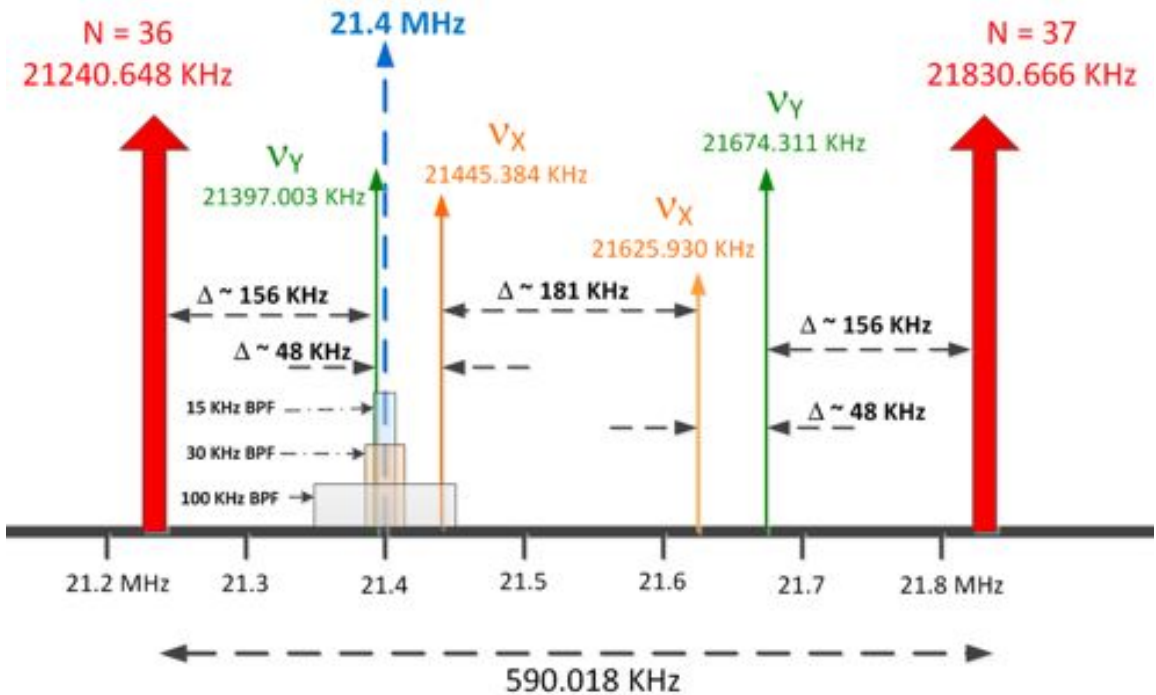

Figure 4.15. Schottky Spectrum for the Delivery Ring. Vertical tune sidebands are shown in green and horizontal tune sidebands are shown in orange. The grey boxes indicate the ranges of available band pass filters that will be used during Schottky commissioning studies. Additional filters will need to be acquired in order to measure the vertical and horizontal upper sidebands.

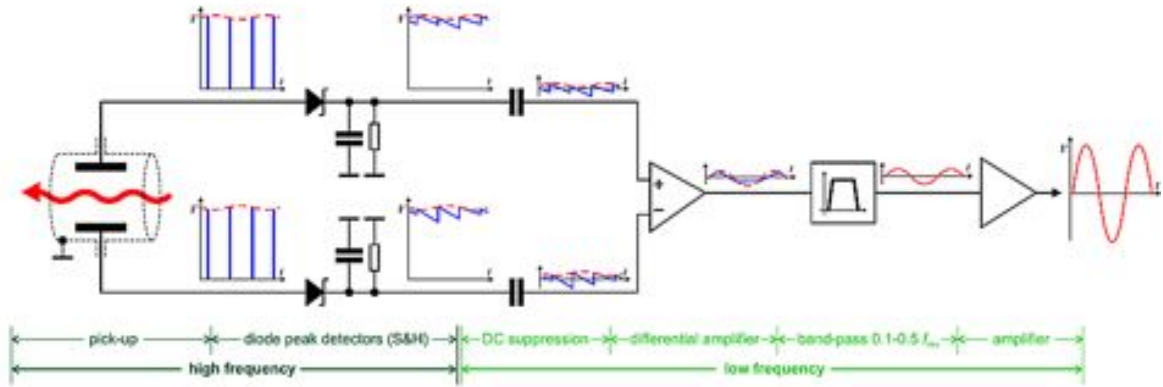

Figure 4.16. Functional block diagram of a direct diode detection base-band Q detector system [30].

### 4.3.2.4   M4 Line Instrumentation

Slow extracted proton beam from the Delivery Ring will traverse the newly constructed M4 line before arriving at the Mu2e target. Instrumentation hardware and electronics will be recycled from other areas. The instantaneous intensity of the slow spill beam is too small to be measured with Toroids, so retractable ion chambers will be constructed to measure the beam intensity. A Beam Loss Monitor (BLM) system, based on the hardware and electronics that exist in the P1 and P2 lines, will be implemented to help maintain good transmission efficiency in the M4 line. Secondary Emission Monitors (SEMs) and Segmented Wire Ion Chambers (SWICs) will provide beam profiles in both transverse planes.

### 4.3.2.4.1   Ion Chambers

Ion chambers will be the primary intensity measurement devices in the M4 line. A photo and engineering drawing of a Fermilab ion chamber is shown in Figure 4.17. Each ion chamber consists of three signal foils interleaved between four bias foils, each spaced 1/4" apart. The foils are sealed in an aluminum chamber 10 inches in diameter by 4.5 inches long, continuously purged with an 80% argon - 20% carbon dioxide gas mix. Protons passing through $ArCO_2$ gas generate 96 $e$/ion pairs or about $1.6 \times 10^{-17}$ charges/cm, which equals about 1.6 pC for $1.0 \times 10^5$ protons [21].

Three ion chambers will be installed in the M4 line and one in the Diagnostic Absorber line.  Since the ion chamber vessel contains $ArCO_2$ gas, it must be separated from beam tube vacuum.  The ion chamber in the diagnostic absorber line will be installed in a gap in the beam line with one 0.003" titanium vacuum window on each side of the ion chamber.

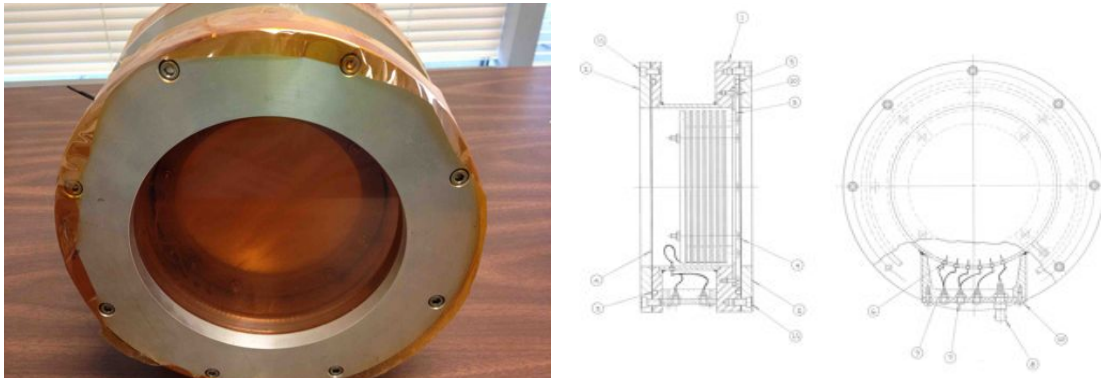

Figure 4.17. Fermilab Ion Chamber

The ion chambers in the M4 line will not be installed in the above described manner because the beam going through both M4 line ion chambers and vacuum windows would result in excessive coulomb scattering during high intensity operations [31].  The solution is to make the ion chamber retractable.

The gas filled ion chamber will be isolated from beam tube vacuum by packaging in an anti-vacuum box. An anti-vacuum box is a sturdy machined aluminum shell with a .003 inch thick titanium foil window mounted on each side for the beam to pass through. The anti-vacuum box allows the detector to be mounted in a beam line vacuum chamber while the ion chamber inside the box remains at atmospheric pressure. There is a vacuum tight duct attached to the box in which the gas tubing, signal and high voltage cables are routed in order to get them to atmosphere outside the vacuum chamber.

To save engineering and assembly costs, the anti-vacuum boxes will be installed inside of bayonet vacuum vessels that are being repurposed from Switchyard. The bayonet type drive slides the ion chamber linearly into and out of the beam with a screw drive system. Bayonet drives use a 72 RPM Superior Electric Slo-Syn AC synchronous stepping motor coupled directly to the screw shaft. The detector linear drive shaft is housed in a collapsible bellows that seals it from atmosphere. Figure 4.18 shows an ion chamber assembly, the anti-vacuum box and the bayonet vacuum can.

### 4.3.2.4.2    Multiwires

Eighteen beam profile monitors will be used to measure beam profiles and positions at key locations in the M4 line. They will be used for both orbit diagnostics as well as automated orbit correction.   As a result, these devices will need to be in the beam path when measuring the beam, as well as have the ability to move in and out of the beam as needed by the automated orbit correction system.

A Multiwire system was chosen the possible alternatives (SWICs[16] or SEMs[17]) because they can be left in the beam without introducing unacceptably high multiple scattering of the beam particles in the materials of the instrument. Unlike SWICs, the multiwire wire planes are installed inside of a vacuum can that is common to the beam tube vacuum. The wires do not provide excessive mass and do not require isolating vacuum windows. The existing NuMI extraction multiwire design (shown in Figure 4.19) will be used in the M4 beamline.

The wire planes are under vacuum and mounted on ceramic boards that can be moved in and out of the beam with a motor drive assembly.   In addition, the ceramic boards are slotted to allow them to be moved into and out of the beam while beam is present.  To accommodate the slots, the ceramic boards must be installed in vacuum cans at 45° as shown in the left picture of Figure 4.20.

Multiwires use two planar arrays of fine diameter wires to produce a profile signal. A set of 48 signal and one or more ground wires are arrayed in the vertical and horizontal axis. The wires will be 0.002 inches in diameter tungsten with wire pitches of 0.5 mm and 1 mm, depending on the location and lattice requirements.

---

[16] SWIC = Segmented Wire Ion Chamber
[17] SEM = Secondary Emission Monitor

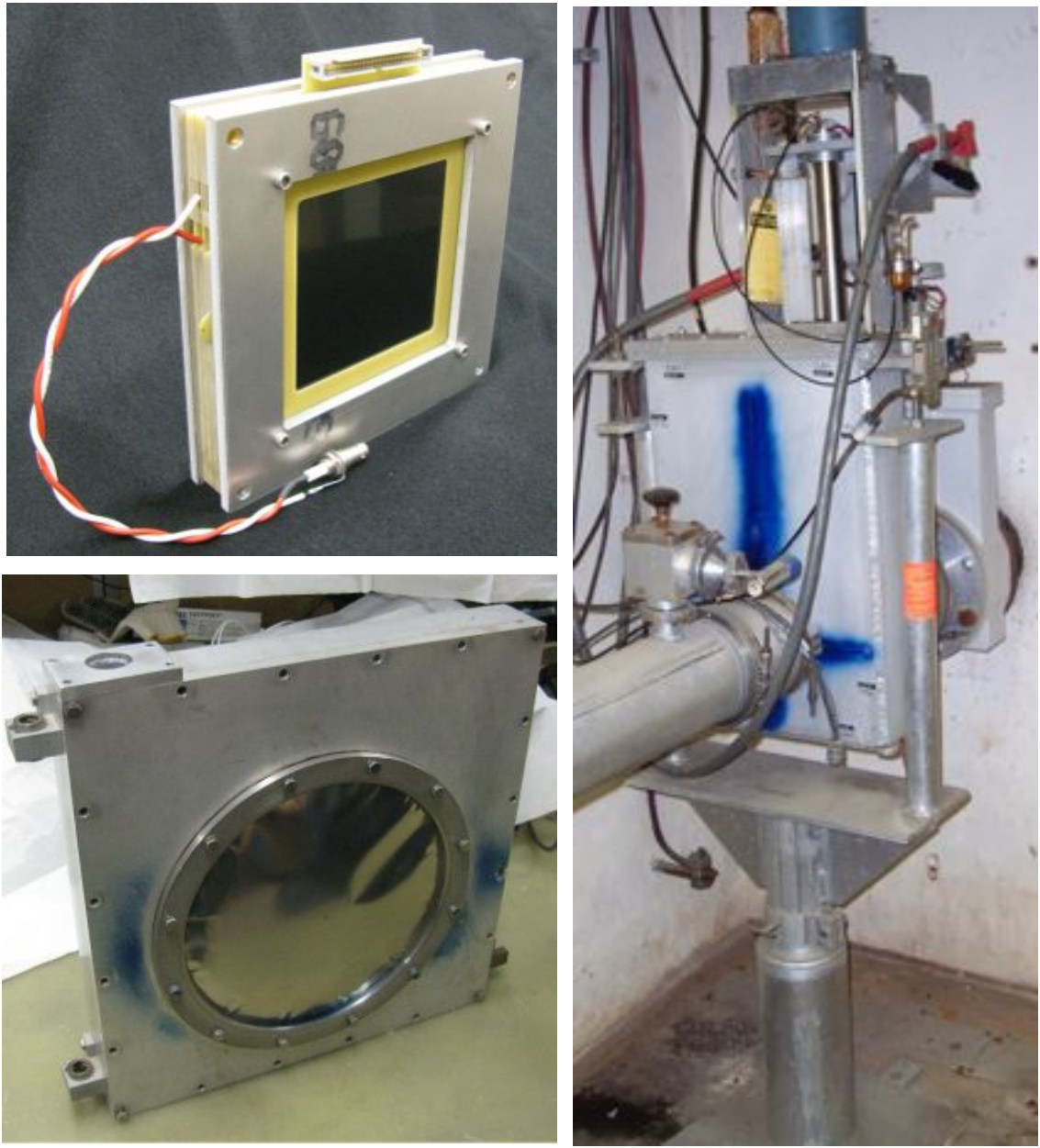

Figure 4.18. Retractable Ion Chamber: The ion chamber design has been modified to fit inside of a Proportional Wire Chamber (PWC) assembly (top left). The signal connection is at the top and the high voltage connection comes out the left side. The ion chamber is installed in an anti-vacuum box (lower left). ArCO2 gas is pumped into this chamber, and there is a vacuum window on both front and back of this module. The anti-vacuum chamber is installed inside of the bayonet can (right) which is pumped down to beam tube vacuum. The ion chamber foil can be lowered into the beam or raised out of the beam via a motor drive.

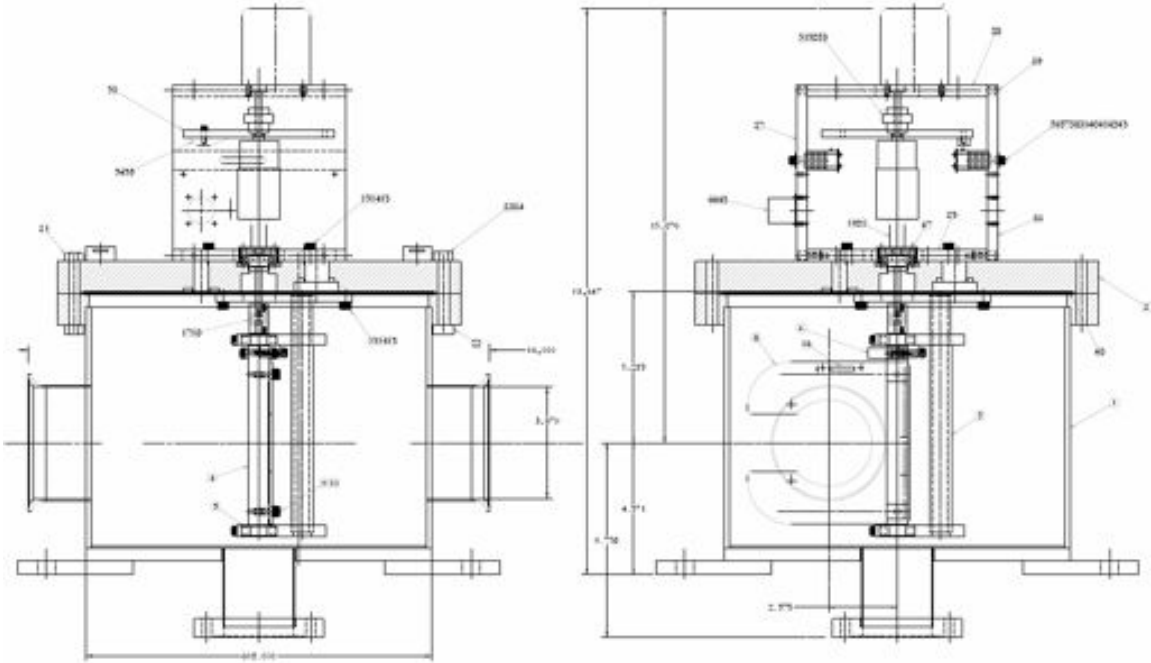

Figure 4.19. Multiwire vacuum can design used for the NuMI extraction multiwires will be used in the M4 line.

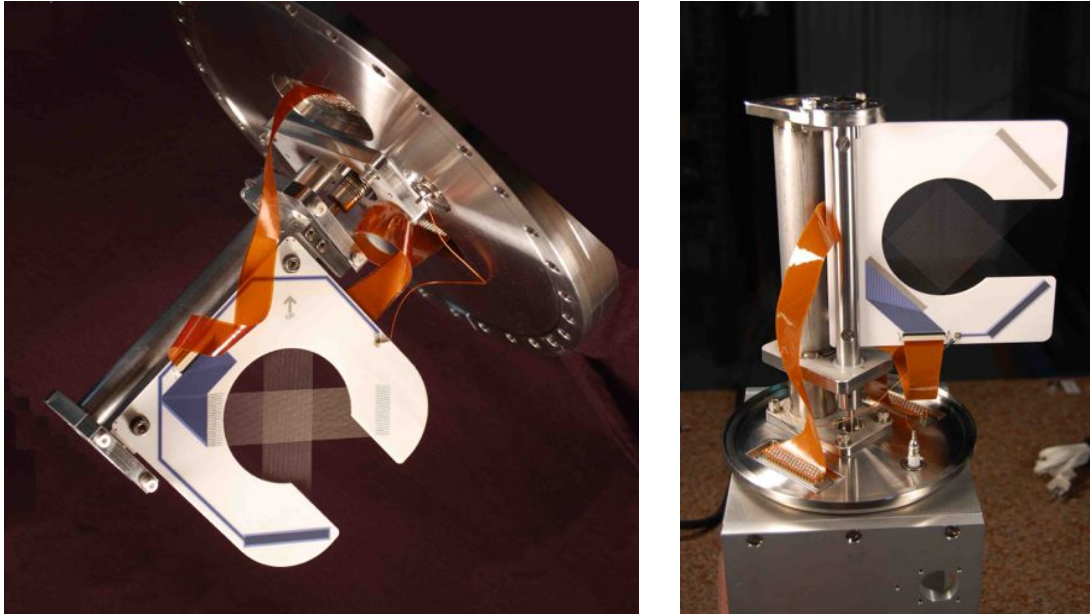

Figure 4.20. Slotted ceramic board holds both horizontal and vertical wire planes.  The ceramic is slotted to allow it to be moved into and out of the beam path with beam present.

The ceramic frame holding the wires in place is rotated into the beam and comes to rest on a hard stop in a position perpendicular to the beam axis. Particle beams passing through the wire planes produce secondary electron emission in each of the 48 signal wires in proportion to the beam intensity. The collected charge on each wire is integrated in a Fermilab generation 3 profile monitor scanner. The scanner communicates to the accelerator controls system via Ethernet where a local application generates an X-Y plot of charge versus wire position in each transverse plane. The resulting plot indicates the size, shape, intensity and position of the beam as shown in Figure 4.21.

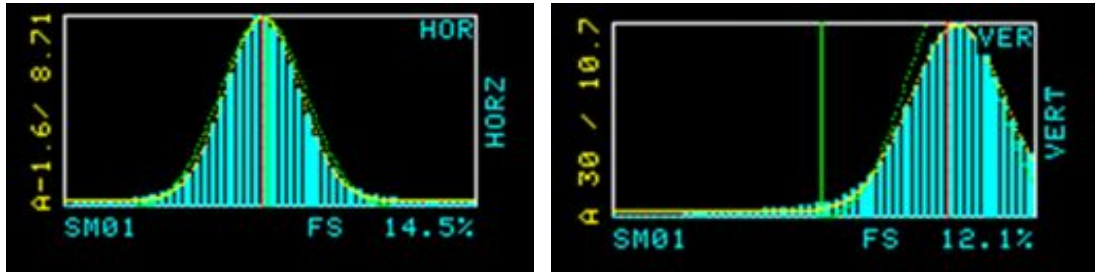

Figure 4.21. Sample horizontal and vertical beam produced by Fermilab standard profile monitor software.

The SWIC scanner is at the center of all beam profile monitor data acquisition electronics. The SWIC scanner collects the charge from each of the detector wires and converts the values of the charges to a set of digital numbers. The data are transferred to the Accelerator Control System and used to graph the beam profile or for other purposes.

The SWIC scanner consists of five printed circuit boards, one controller board and four analog integrator boards. It has a set of 96 integrator circuits, 48 for horizontal and 48 for vertical. The integrators collect the charge from each of the detector wires and convert it to a voltage value proportional to the total charge collected. The basic integration capacitor value for most SWIC scanners is 100 pF. This value provides the most sensitivity. Other values commonly in use are 1000 pF and 10,000 pF for integration time constants of ten or one hundred times the basic 100 pF value, for use in higher intensity beams. Longer time constants require more charge from the detector to reach the same voltage output from the integrators. Larger capacitors are used in higher intensity beams to minimize the possibility of overloading the integrators.

The integrators collect charge until they reach the end of the integration time that is set up in the plot application program by the user, or until at least one wire reaches the preset threshold voltage. At the end of the integration period the integrators are switched from sample mode to hold mode. The integrated voltages on each channel are measured one-by-one. The voltages are converted to digital values. After conversion from analog to digital the integrators are all reset to zero and the scanner is ready to take another sample.

The "next generation" SWIC scanner is an evolution of the previous design. A block diagram is shown in Figure 4.22. The SWIC interfaces to the scanner through integrator boards (96 channels total), which are a direct carry-over. The control board is centered on an Altera Cyclone III FPGA, which handles sequence control, ADC conversion, TCLK decoding, and timing. Communications and data handling are performed by a Rabbit Semiconductor RCM3209 module. The Rabbit module includes the microprocessor and Ethernet interface. New features include Ethernet communications, advanced triggering options, and background subtraction.

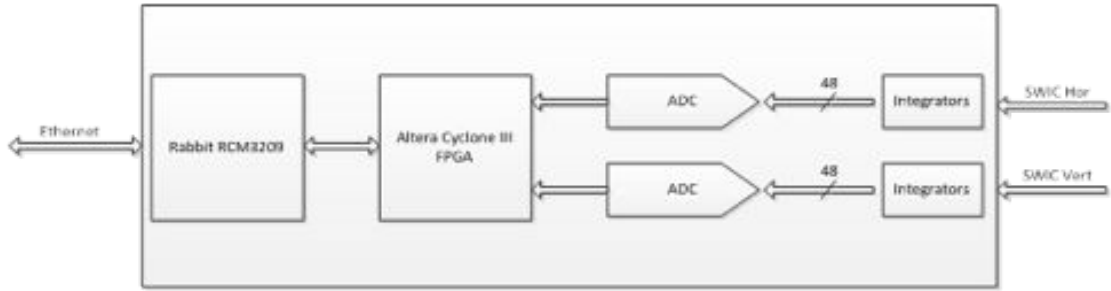

Figure 4.22. SWIC scanner block diagram.

**Preamps:**

For a high intensity spill, preamps will not be required for SWICs to measure slow extracted beam with a sampling window opened up to 1/3 of the spill length. For the less sensitive SEMs and for SWICs where a shorter fraction of the spill is desired, a preamp will be added to the profile monitor.

The new preamp is, at its core, a simple non-inverting op-amp circuit (see Figure 4.23). The output of the SEM is modeled as a current source. This current is converted to a voltage by running it through a resistor (R_I2V in Figure 4.23). The op-amp amplifies this voltage (gain = 1 + RA/RB) and converts it back to a current by running it out through another resistor (R_V2I in Figure 4.23) [27].

There is a DC blocking capacitor at the output of the amplifier. This is to prevent any amplifier offset voltage from washing out our signal. This offset voltage could be reduced with a trim pot. However, the offset will never be zero, and the trim pot adds about $3 to the cost per channel. The downside of the DC blocking capacitor is that it requires the use of a very short integration window. In this configuration, it may not be possible to just turn on the integrator and accumulate for tens or hundreds of turns. Experimentation may show that the DC blocking capacitor is either unnecessary or undesirable.

The SWIC scanner uses a TI/Burr-Brown ACF2101 integrator. The solid-state switch on its input has a resistance of 1.5 kΩ ("typical", per the datasheet). This extra 1.5 kΩ must be accounted for in the gain equation (R_INT in Figure 4.23).

The SEM signal is based on a bunch intensity of $2 \times 10^7$, with a 3% total capture rate, which is spread evenly over 60 foils, arriving in a 120 nsec timeframe. The integrator ends up with about 100 mV of signal. This can be amplified in the SWIC scanner by $10\times$ or $100\times$ if necessary.

If necessary, the gain of the amplifier can be easily increased or decreased by choosing different resistor values. For early studies, the gain was left at a fairly moderate level, hoping to avoid noise and stability issues. Early attempts at bench testing gave outputs on the order of volts, not millivolts. Beam studies will be required to fine-tune the gain levels for optimal functionality [27].

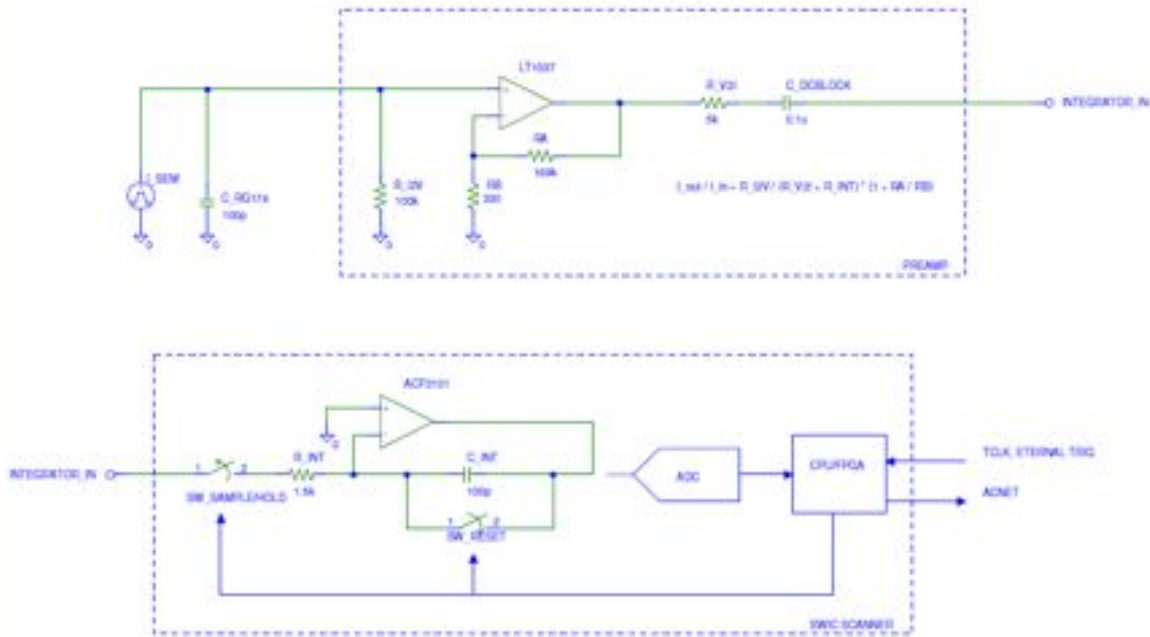

Figure 4.23. Profile Monitor preamp design [27].

### 4.3.2.4.3   Beam Loss Monitors (BLMs)

The M4 Line Beam Loss Monitor (BLM) system has been designed to measure a 0.2% localized loss with a microsecond integration time. This will allow observation of losses developing during a single slow spill. 20 BLMs will be placed at key locations along the 245 m beam line. This system design is identical to the existing Main Injector, P1, P2, M1 and M3 BLM systems. There is not a sufficient pool of spare hardware and electronics to instrument the entire M4 line, so new BLMs will have to be constructed [34].

BLM chassis will be installed at AP30 and the Mu2e Experimental Hall. Each chassis supports 12 BLMs units. The system uses ion chamber loss monitors originally designed for the Tevatron. The ion chamber's high voltage supply is contained within the BLM chassis. It is capable of supplying 2500 V at 500 μA. The High Voltage is set to 2000 V during chassis testing and no further adjustment is required as the Ion Chamber has a flat high voltage to gain response. An Abort Demand signal is produced by feeding the analog "or" of the 12 daughter card outputs to a voltage comparator circuit. The comparator reference level is set by adjusting a rear panel pot. The daughter card can be set up to operate as a Set-Reset integrator, a fast amplifier or as a Log Amplifier (Figure 4.24). The Integrator has a full-scale range of either 0.14 rad or 0.014 rad dependent upon which branch is used. The fast amp has a rise and fall time of 1 μsec and a full-scale range of 0.014 rad. The Log Amp has a 6 Decade dynamic range allowing loss readings from 0.001 rad/second to 1000 rad/second (Figure 4.25). One and only one of these signals can be jumper configured to provide the output to the MADC. The Daughter Card also has a track and hold circuit that can be inserted in the signal path before outputting to the MADC. A CAMAC 377 card provides the Start Track and Start Hold triggers. A timing card within the BLM chassis fans out these triggers to the 12 Daughter cards. The Start Track signal also resets the integrator circuit when the daughter card is configured as an integrator.

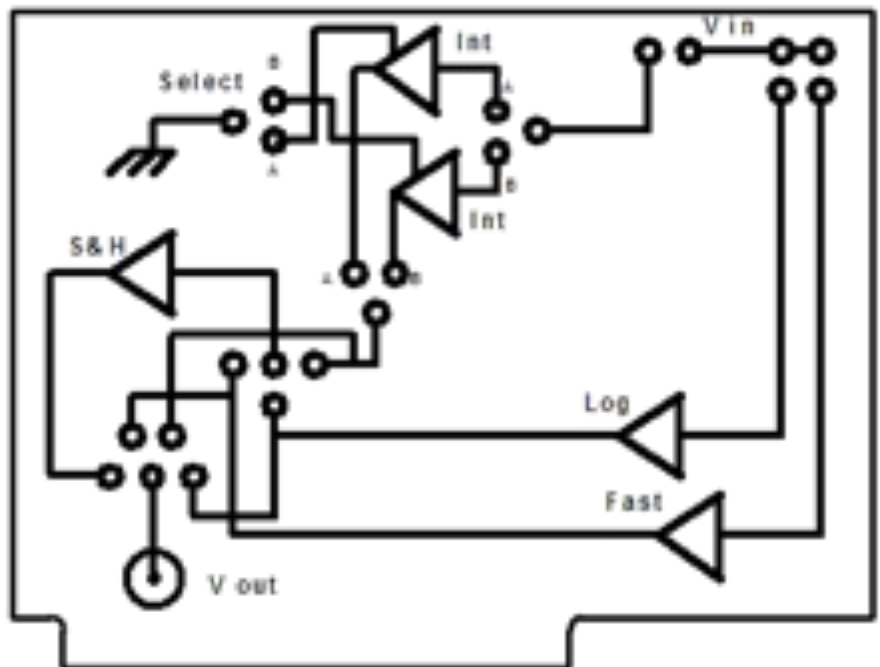

Figure 4.24. Beam Line Daughter card jumpers. This drawing is taken from Main Injector Note-208 [16].

The daughter cards are set up to function as a Log amp with a track and hold output. Track and hold triggers are provided via a CAMAC 377 card. The daughter card outputs are feed into an MADC. From there they can be read from an accelerator console parameter page, Fast Time plotted, or displayed graphically via a control system console application program (see, for example, Figure 4.26).

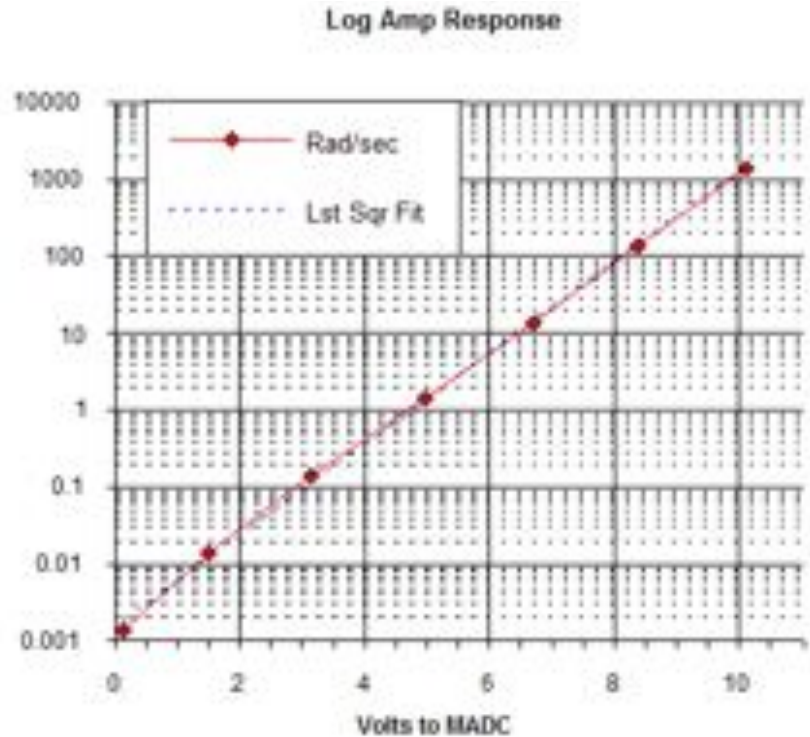

Figure 4.25. Log amp response on profile monitor preamp.

### 4.3.3 Instrumentation Risks

#### 4.3.3.1 Delivery Ring Injection Damper Required

Initial calculations show that an injection damper will not be required for the Delivery Ring. However, beam commissioning will determine if orbit control is sufficient to control the beam trajectory. If not, excessive emittance dilution is possible. Mitigation of this contingency would consist of building an injection damper system [37] [38].

#### 4.3.3.2 Inadequate low intensity Delivery Ring Tune Measurement

The Delivery Ring tune measurement will require fast measurement of a low intensity signal. Initial calculations show that the proposed systems will be able to measure the Delivery Tune with the desired accuracy and rate; however, if that is not the case then mitigation of this risk would consist of designing and building a new Delivery Ring tune measurement system [24] [25] [26].

### 4.3.4 Instrumentation Quality Assurance

Installation and commissioning of the M4 beamline instrumentation will be performed by qualified Accelerator Division staff. All necessary parts will be procured by Fermilab personnel and inspected by qualified Instrumentation engineers or technicians prior to installation. Final testing and commissioning of instrumentation devices will be performed by Fermilab technical staff.

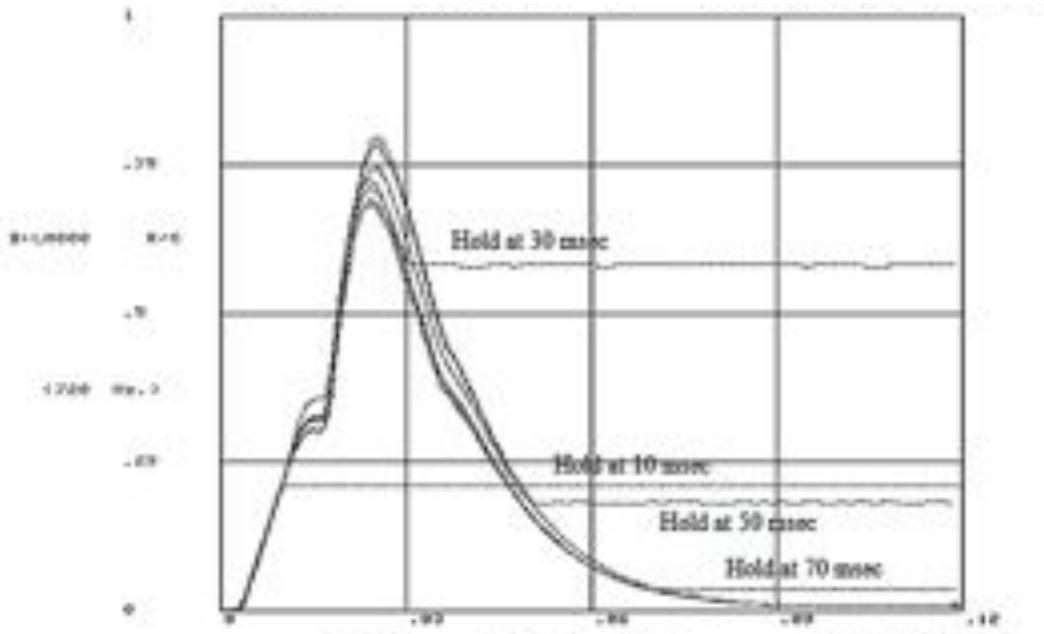

Figure 4.26. Beam Loss Monitor response over time. This figure shows the output of a Booster loss monitor plotted using the accelerator control system Fast Time Plot facility. The vertical axis is beam loss measured in Rad/sec. The horizontal axis is time relative to the BLM trigger event.

### 4.3.5 Instrumentation Installation and Commissioning

The Delivery Ring tune system is being repurposed from the Tevatron and will require minor modification to become operational in the Delivery Ring. The first stage of commissioning will be to install the existing Schottky and BBQ hardware in the Delivery Ring and existing electronics in the upstairs service buildings. The devices will be commissioned during beam studies in 2014. This initial set of measurements will be used to determine if any modifications are necessary to achieve the final requirements.

Cable pulls to the beamline instrumentation will be made once beneficial occupancy is obtained for the M4 beamline enclosure and cable trays are in place. All M4 beamline instrumentation will be installed at the time of magnet installation. Electronics will be installed in the AP30 service building, MC-1 Experimental Hall, and Mu2e Experimental Hall once beneficial occupancy is obtained for each building.

## 4.4    Accelerator Controls

### 4.4.1 Controls Requirements

The central controls system is located in the Accelerator Division cross gallery where a number of controls and communications signals arrive from various sources from across the accelerator complex. Table 4.9 summarizes the signals required at each Muon Campus service building and the media requirements for each system. Each system in the table will be further explained in the technical design portion of this report (4.4.2).

Table 4.9. Controls and Communications requirements for each Muon Campus service building

| Required System | MI60 ,F0, F1, F2 | F23,AP0, F27 | AP10, AP30, AP50 | Mu2e |
|---|---|---|---|---|
| **Ethernet** | Single-mode Fiber 100-1000 Mbps | Wireless/Thicknet 10 Mbps | Single-mode Fiber 100-1000 Mbps | Single-mode Fiber 100-1000 Mbps |
| **CAMAC or HRM** | CAMAC | CAMAC | CAMAC | HRM |
| **Serial Timing Links** | Multi-mode Fiber | Multi-mode Fiber | Multi-mode Fiber | Multi-mode Fiber |
| **Beam Synch** | Multi-mode Fiber (RRBS & MIBS) | Multi-mode Fiber (RRBS) | Multi-mode Fiber (RRBS) | Multi-mode Fiber (RRBS) |
| **Permit Loop** | Multi-mode Fiber (Muon) | Multi-mode Fiber (Muon) | Multi-mode Fiber (Muon) | Multi-mode Fiber (Muon) |
| **FIRUS** | Single-mode Fiber | Single-mode Fiber | Single-mode Fiber | Single-mode Fiber |
| **Safety System** | 4/6/20 Conductor | 4/6/20 Conductor | 4/6/20 Conductor | 4/6/20 Conductor |
| **SEWs** | Single-mode Fiber | Single-mode Fiber | Single-mode Fiber | Paging System |
| **Radmux** | 18 AWG 16×30 | 18 AWG 16×30 | 18 AWG 16×30 | 18 AWG 16×30 |
| **Phones** | 100/400 Conductor | 100/400 Conductor | 100/400 Conductor | 100/400 Conductor |

For existing Muon Campus service buildings, the communications ducts will be removed to accommodate construction of the M4 and M5 beamlines. Restoration of the controls and communications signals will be performed as part of the Delivery Ring AIP [41]. Completion of this work is required prior to Mu2e commissioning and operations.

### 4.4.2 Controls Technical Design

Control and communications lines to the existing Muon Campus service buildings will use existing infrastructure. New control and communications lines will be required for the Mu2e experimental hall.

#### 4.4.2.1 CAMAC and Links

The existing accelerator service buildings will continue to use the legacy controls infrastructure that is currently in place. These service buildings include all of the Main Injector service buildings, as well as F0, F1, F2, F23, F27, AP0, AP10, AP30 and AP50. Future Muon Campus service buildings, including MC-1 and the Mu2e building, will be upgraded to a more modern controls infrastructure that will be discussed later in this section.

CAMAC crates exist in each service building and communicate with the control system through a VME style front-end computer over a 10 MHz serial link, as shown in Figure 4.27. Both digital and analog status and control of many accelerator devices occur through the CAMAC front ends. There should be no need to install additional CAMAC crates, as there is excess capacity in most of the existing crates. An inventory of existing CAMAC crates in the Muon Department service buildings shows that about 25% of the slots are unoccupied and could be used for additional CAMAC cards [39]. In addition, further slots have become available that were used to interface devices that became obsolete with the conclusion of Collider Run II operations. It is anticipated that there will be ample CAMAC crate coverage for Mu2e operation in the existing Muon Department service buildings, and very few crates will need to be added or moved.

There are serial links that are distributed between the service buildings via the accelerator enclosures that provide the necessary communications paths for CAMAC signals as well as other necessary signals such as clock signals, the beam permit loop and the Fire and Utilities System (FIRUS). Controls serial links can be run over multimode fiber optic cable or copper Heliax cable. Most Muon Department links that run through accelerator enclosures are run over Heliax cable that should function normally in the radiation environment expected during Mu2e operations.

Accelerator device timing that does not require synchronization to the RF buckets will remain on the existing 10 MHz Tevatron Clock (TCLK) system. The existing TCLK infrastructure will remain in existing service buildings and new TCLK link feeds will be run via multimode fiber optic cable to the Mu2e Experimental Hall.

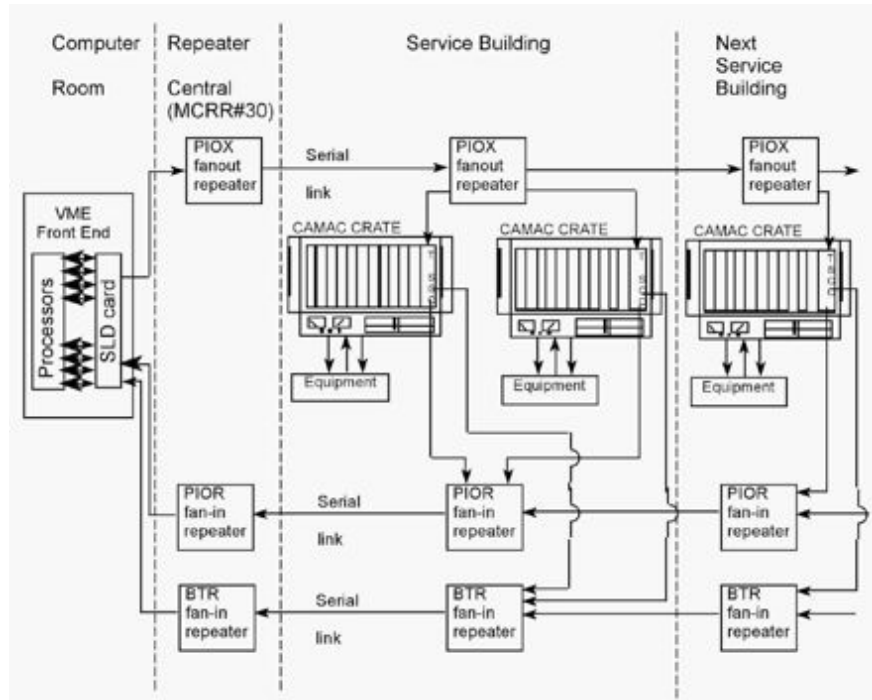

Figure 4.27. Legacy CAMAC crates interfacing VME front ends via serial links provide both analog and digital status and control of accelerator devices and will continue to be used in existing Muon Department Service Buildings [40]. This drawing is taken from the AD Operations Controls Rookie Book [45].

Accelerator device timing for devices that require synchronization to the RF buckets will continue to be handled through the Beam Synch Clocks; however, a few changes will be required to maintain functionality. The F0, F1 and F2 service buildings will need both 53 MHz Main Injector beam synch ) for SY120 operations and 2.5 MHz Recycler beam synch (RRBS) for g-2 and Mu2e operations. These buildings already support multiple beam synch clocks, so the addition of RRBS will require minimal effort.  An obsolete 53 MHz Tevatron beam synch (TVBS) feed in the MI60 control room will be replaced with a 2.5 MHz RRBS feed to provide the necessary functionality. The remaining Muon Department service buildings currently use 53 MHz MIBS, but will require 2.5 MHz RRBS for g-2 and Mu2e operations. This functionality can be obtained by replacing the MIBS feed at F0 with RRBS and using the existing infrastructure.  Further upgrades and cable pulls will only be required if it is later determined that both MIBS and RRBS are required in these service buildings.  New beam synch feeds to the g-2 and Mu2e service buildings will be run via multimode fiber optic cable.

The Delivery Ring permit loop provides a means of inhibiting incoming beam when there is a problem with the beam delivery system. The existing Pbar beam permit infrastructure will be used in the existing buildings.  The CAMAC 201 and 479 cards that provide the 50 MHz abort loop signal and monitor timing will need to be moved from the MAC Room to AP50 to accommodate the addition of the abort kicker at AP50. Existing

CAMAC 200 modules in each CAMAC crate can accommodate up to eight abort inputs each. If additional abort inputs are required, spare CAMAC 200 modules will be repurposed from the Tevatron and will only require an EPROM or PAL change to bring them into operation. The permit loop will be extended to the MC-1 and Mu2e service buildings via multimode fiber optic cable from the Mac Room. CAMAC will not be available in the MC-1 and Mu2e Experimental Halls, so the abort permit signal for these buildings will be extended via an additional cable pull to the AP30 service building.

Permit scenarios are being developed to support necessary operational scenarios that include running beam to the Delivery Ring abort dump when Mu2e and g-2 are down, and running beam to either experiment while the other is down.

### 4.4.2.2   Hot-Link Rack Monitor

New controls installations in the Mu2e Experimental Hall will use Hot-Link Rack Monitors (HRM's) in place of CAMAC. An HRM runs on a VME platform that communicates with the control system over Ethernet, as shown in Figure 4.28. Unlike CAMAC, no external serial link is required, minimizing the need for cable pulls between buildings. Each HRM installation provides 64 analog input channels, eight analog output channels, eight TCLK timer channels and eight bytes of digital I/O. This incorporates the features of multiple CAMAC cards into a single-compact chassis. Like CAMAC, when additional functionality or controls channels are needed, additional units can be added. Two HRMs will be installed in both MC-1 and Mu2e Experimental Halls and should provide ample controls coverage for both accelerator and experimental devices [43] [44].

### 4.4.2.3   Ethernet

Many modern devices have some form of Ethernet user interface. In addition, many devices and remote front ends use Ethernet to interface the control system instead of using the traditional CAMAC. The results are an increasing demand on the Controls Ethernet. Figure 4.29 is a map of the Muon Controls network. All of the current Muon Ring service buildings have Gigabit fiber optic connections from the cross-gallery computer room to Cisco network switches that are centrally located in each service building. These will provide ample network bandwidth and connections after the reconfiguration for g-2 and Mu2e operations. A central Ethernet switch that fans out to the other Muon Campus buildings is currently located in AP10, but will need to be moved to AP30 as will be discussed later in this document [48].

Ethernet connectivity between the Delivery Ring service buildings is provided by multimode fiber optic cable traversing the Ring enclosure. This will be upgraded to single mode fiber optic cable in order to support rerouting of the site emergency warning system as well as to provide a more robust fiber infrastructure for the higher radiation levels anticipated for Mu2e operations.

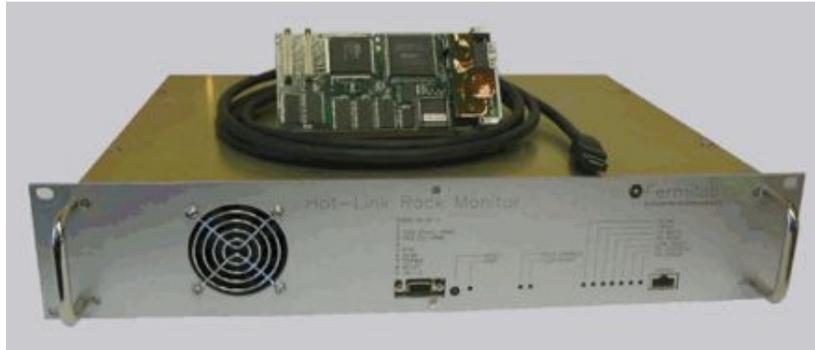

Figure 4.28. A Hot-Link Rack Monitor is a flexible data acquisition system composed of a remote unit and a PCI Mezzanine card that resides in a VME crate. Each HRM provides sixty four 16 bit analog input channels, 8 analog output channels, 8 TCLK timer channels and 8 bytes of digital I/O. HRM's will eventually replace all of the functionality of CAMAC [43].

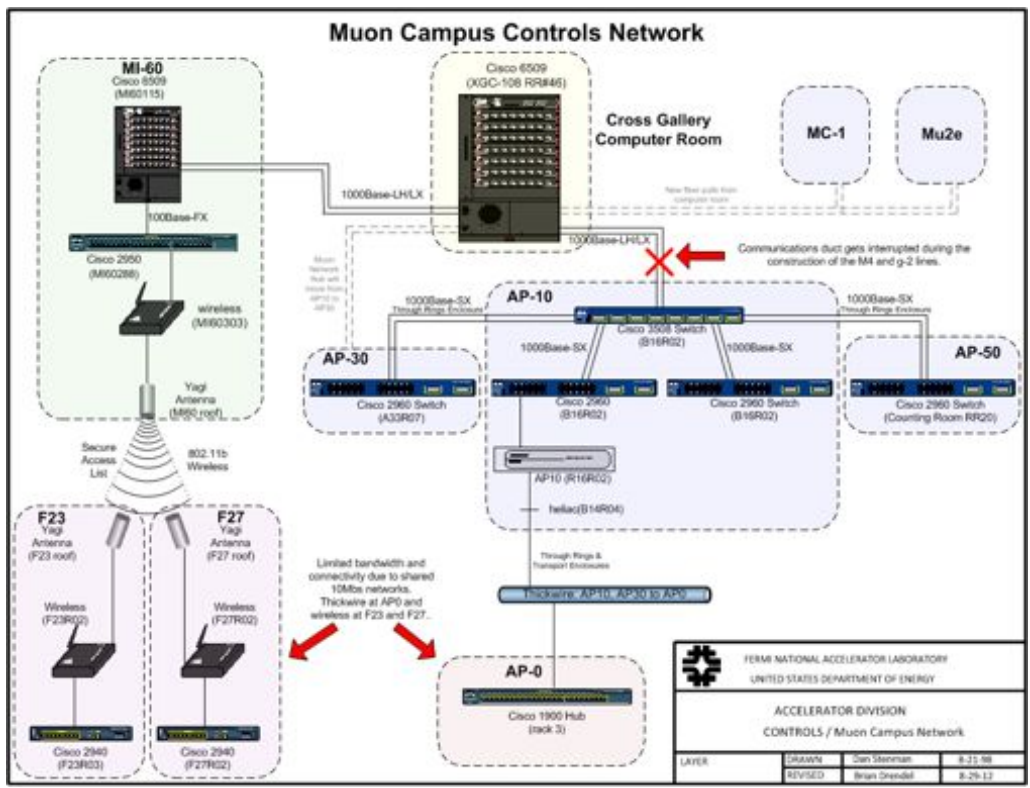

Figure 4.29. Controls Ethernet to the Muon Department Service Buildings should be adequate for Mu2e operations. The central switch at AP10 will be moved to AP30. Legacy networks at AP0, F23 and F27 have limited bandwidth and connectivity, but should be sufficient for Mu2e operations.

Most beamline service buildings have gigabit fiber connected to centrally located network switches that provide ample network bandwidth and connections. AP0, F23, and F27 are the only three buildings that do not have this functionality. AP0 runs off a 10 Mbps hub that connects to 10Base5 "Thicknet" that runs through the Transport and Rings enclosures back to AP10, while F23 and F27 run off 802.11b wireless from MI60. Both are 10 Mbps shared networks with limited bandwidth and connectivity. It is anticipated that the network in these three buildings will be sufficient for Mu2e operations.

### 4.4.2.4   *Restoring Controls Connectivity:*

Construction of the M4 and M5 beamline enclosures will require removal of the underground controls communication duct that provides connectivity between the Accelerator Controls NETwork (ACNET) and the Muon Campus [47].  Included in this communication duct is the fiber optic cable that provides Ethernet connectivity as well as 18 Heliax cables that provide the controls serial links and other signals including the Fire and Utility System (FIRUS) [49]. These cables currently traverse this communications duct to the center of the D20 location in the Rings enclosure, and travel through cable trays on the Delivery Ring side to the AP10 service building. After removal of the communications duct, FESS will construct new communications ducts from the existing manholes. These communications ducts will go directly to AP30, MC-1 and Mu2e service buildings without going through accelerator enclosures. See Figure 4.30 for drawings of the current and future controls connectivity paths.

#### 4.4.2.4.1   Restoring Connectivity

When the Heliax and fiber optic cables are cut during the above mentioned communications duct removal, controls connectivity will be lost. The base plan for restoring both Ethernet and controls link connectivity is to pull new fiber optic cable from the cross gallery to the manhole outside of Booster Tower West and on to AP30 via the new communications duct.  As a result of the new fiber pull, the Ethernet and controls links will fan-out from AP30 instead of AP10.  This will require some additional controls hardware configuration and labor.  Efforts will be made to minimize the disruption by pulling the fiber and staging the new hardware at AP30 before the communication duct is cut. This is especially important for FIRUS, which is a necessary safety system [49].

Single-mode fiber will be needed for the Ethernet and FIRUS connectivity and multimode fiber will be needed for the controls serial links. Bundles of 96 pair single mode and 36 pair multi-mode fiber optic cable will be run to AP30.  This provides the necessary connectivity in a minimal amount of space. Similar fiber bundles will also be pulled to MC-1 and Mu2e.

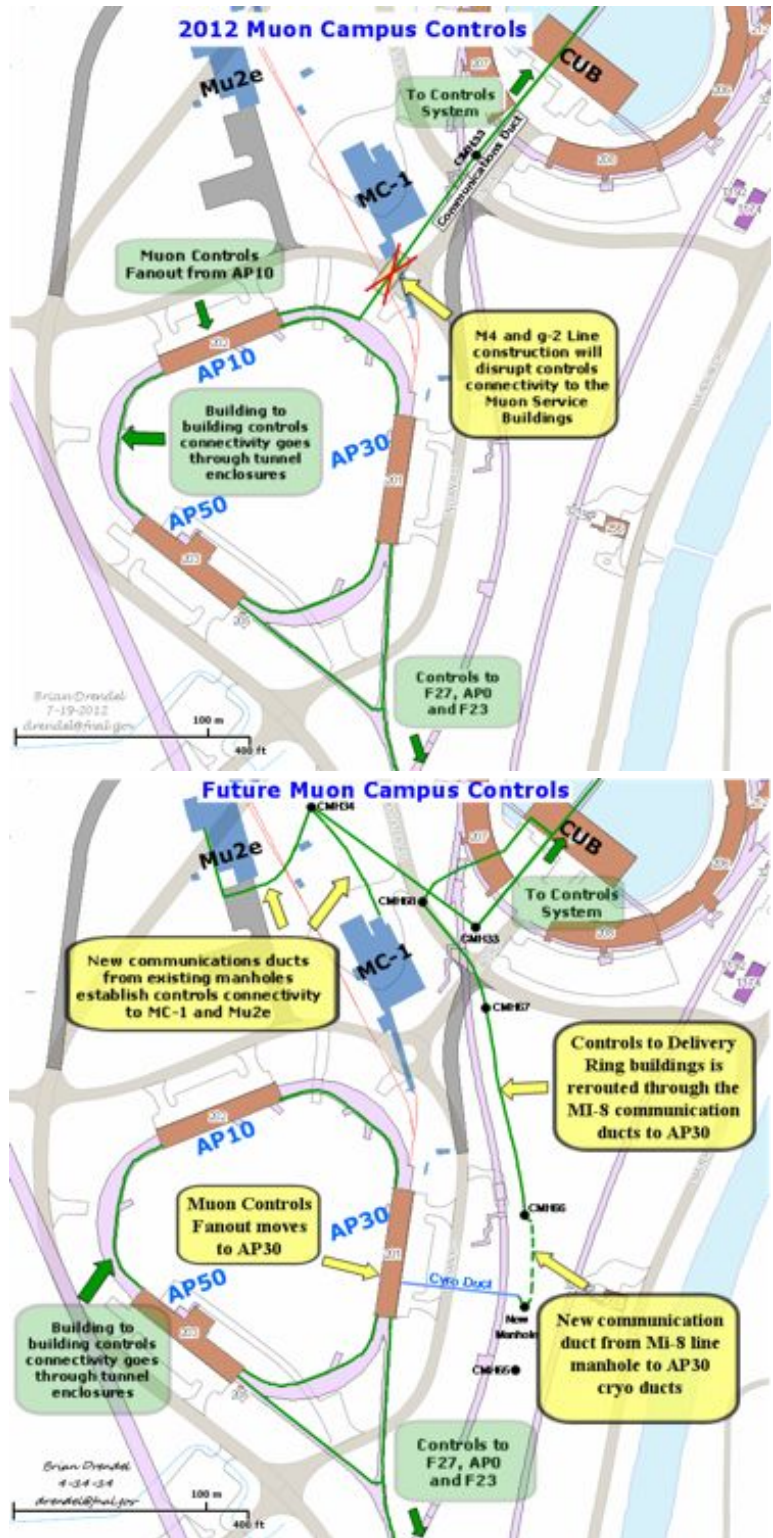

Figure 4.30. Muon campus controls paths. During construction of the M4 and M5 beamlines, the communications duct that provides controls connectivity to the Muon Campus will be interrupted (top). A new communications duct will be built to restore controls connectivity to the Muon Service Buildings (bottom). New controls will be established to the MC-1 and Mu2e Experimental Halls.

#### 4.4.2.4.2   Establish Connectivity to Mu2e

New fiber optic cable will be pulled from the MAC Room to the Mu2e Experimental Hall. Single-mode fiber is needed for Ethernet and FIRUS and multimode fiber is needed for the timing links and the abort permit loop. Bundles of 96 pair single mode and 36 pair multi-mode fiber optic cable will be run to Mu2e. The fiber pulls will provide ample connectivity for all Ethernet and controls signals for both the accelerator and experiment. The Mu2e experiment anticipates requiring network rates approaching 100 MB/sec during production data taking that can be handled easily with the proposed infrastructure.

### 4.4.2.5   Safety System Interlocks

Safety system interlocks will need copper cable pulled to AP30, MC-1 and Mu2e [41]. The existing Safety System signal trunk lines, which consist of seven 20 conductor #18 AWG cables that run from the safety system vault room XGC-005 through the Central Utility Building (CUB) to AP10, will be interrupted due to the Muon Campus construction. These trunk lines will need to be spliced at CUB and replaced with new cables from CUB to the AP30 Building. These cables will be pulled at the same time as the Control System fiber to minimize contract electrician costs. Figure 4.31 gives a pictorial representation of each of the required cable pulls. It should be noted that the costs outlined here deal only with the Safety System trunk line cables for the above mentioned areas and does not include the necessary Safety System assemblies, cable and hardware needed for the individual enclosure interlocks.

### 4.4.2.6   Muon Rings Interlocks

The safety system interlocks will have to be reestablished to the existing Muon Campus areas when the seven 20-conductor cables are interrupted. The existing safety system splice junction box below the CUB outside stairwell will be removed, and the seven 20-conductor cables will be pulled back to inside the double doors that separate the CUB outside stairwell from the utility tunnel, where the cables will be terminated in a new junction box. The seven 20-conductor cables running from the removed junction box to the communication manhole CMH33 heading to AP10 will be pulled out and scraped.

At the new utility tunnel junction box six 20-conductor cables will be pulled to the AP30 Service Building via the MI-8 line communications ducts to AP30 Cryo Room. In the AP30 Cryo Room a junction box will be installed to terminate these cables and for use as a break-out point. A six pair twisted shielded cable will be pulled into AP30 for the safety system audio system. This six pair cable along with a new 20 conductor cable will be pulled from the safety system vault room XGC-005 to the utility tunnel junction box and terminated there. From this utility tunnel junction box a six pair twisted shielded and 20-conductor cable will be pulled into the new junction box at AP30.

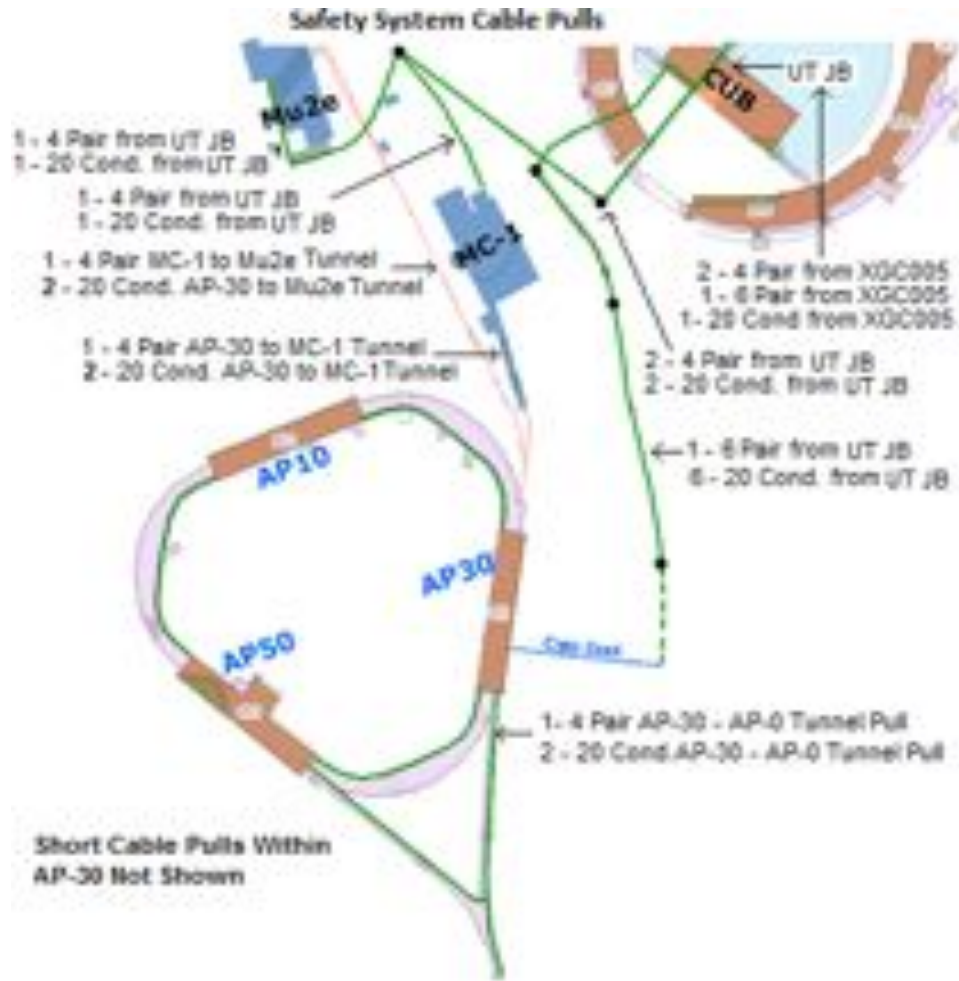

Figure 4.31. Safety System Interlock Cable pulls [50].

From the new junction box in AP30 two 20-conductor and one 4-pair twisted shielded cable will be pulled to the AP0 safety system junction box. These cables will be pulled through the Muon Rings and Transport tunnel enclosures. Pulled within the AP30 service building will be two 20-conductor and one 4-pair twisted shielded cables between the new AP30 junction box and the existing AP30 junction box.

Two 20-conductor cables will be pulled from the new AP30 junction box and a new safety system end rack that will be installed to accommodate a critical device controller installation for g-2 and Mu2e operations [41].

Making the transition to the new system will be made as efficiently as possible, but will take some time due to the required cable terminations and rerouting of the safety system signals.

### 4.4.2.7 Mu2e Interlocks

For the Mu2e interlock system one 4-pair twisted shielded cable will be pulled from the safety system vault room, XGC-005, to the new utility tunnel junction box. One 20-conductor and one 4-pair twisted shielded cable will be pulled from the new utility tunnel junction box to the Mu2e building. Two 20-conductor cables will be pulled from the Mu2e building to the new safety system end rack located at AP30. One 4-pair twisted shielded cable will be pulled from the MC-1 building to the Mu2e building. The cables pulled between MC-1, Mu2e and AP30 will be pulled through the enclosure tunnel [41].

### 4.4.2.8 Radmux

The Multiplexed Radiation Monitoring Data Collection System (MUX) is operated by the ES&H / Radiation Protection / Instrumentation Team. The MUX system is used to collect data from radiation monitors throughout the accelerator and beamline areas. The system provides an interface between the radiation monitors and the hardware network, collects real-time data for its various users, logs the raw data, processes and archives the data. Additionally the archived data serves as the legal record of radiation levels throughout the laboratory [40]. Radmux connectivity will be restored to the existing muon buildings and established to the Mu2e Experimental Hall via new cable pulls [51].

### 4.4.2.9 Phone

Phone connections to the existing Muon service buildings will be reestablished by splicing into the 400-pair cable in the MI-8 communications duct. A new section of 100-pair cable will be run from the splice via a new communications duct path established by the Delivery Ring AIP to the AP30 service building.

Phone connections to the Mu2e Experimental Hall will be established by splicing into 400-conductor pair phone line in the CMH33 Manhole and running new 100-conductor pair phone line to the MC-1 and Mu2e Experimental Halls [52].

### 4.4.2.10 Site Emergency Warning System

The Site Emergency Warning System (SEWS) currently runs to the Muon Campus buildings over the CATV system. When the communications duct is cut, the CATV system will not be reestablished to the Muon Campus buildings. Instead, the SEWS will be run over single mode optical cable to AP30 and then through the Delivery Ring enclosure to AP10, where a connection will be made to the existing system.

No cabling infrastructure will be needed for the SEWS in the Mu2e and MC-1 service buildings. A paging system internal to each building and will be tied to a radio receiver (called a TAR) that receives the SEWS radio broadcast. The messages will be broadcast over the paging system [53].

### 4.4.3 Controls Risks

#### 4.4.3.1 Legacy Networks

If the legacy Ethernet networks at AP0, F23, and F27 provide insufficient connectivity or bandwidth for Mu2e operations, they can be most cost effectively upgraded by replacing the current 10Base5 "thicknet" with single-mode fiber optic cable. The cable path would be from the AP30 service building to the Delivery Ring enclosure, along the cable trays toward the M3 beam line, and down the transport enclosure. From the transport enclosure, the fiber optic cable can be run to F27 and AP0. An additional fiber optic cable pull from AP0 through the PreVault enclosure provides a path to F23.

#### 4.4.3.2 Radiation Damage

The largest risk associated with legacy network upgrade proposed in the previous section is the susceptibility of single-mode optical cable to radiation damage. If the radiation environment in the accelerator enclosures does not allow for single-mode optical cable, then higher cost rad-hardened fiber optic cable will have to be pulled. Standard 96 count single-mode fiber costs approximately $1.50/foot, whereas 24 count rad hardened fiber costs approximately $22/foot. Upgrading to the radiation-hardened cable would add approximately $50k to the cost of the cable pull. Other fiber optic cable path options have been considered, but prove to be more costly to implement.

### 4.4.4 Controls Quality Assurance

Contract electricians under the direction of Accelerator Division management will complete all cable pulls and complete fiber optic terminations. Safety system cable terminations will be managed by Fermilab ES&H personnel, and phone cable termination will be managed by the Telecommunications Department. All controls links, FIRUS configuration and network connections work will be managed by Accelerator Division Controls Department personnel. All parts will be procured by Fermilab personnel and inspected before being installed. Final testing and calibration of controls devices will be performed by Fermilab technical staff before locating equipment in the service buildings.

### 4.4.5 Controls Installation and Commissioning

Installation of control systems will occur once construction of the communications ducts has been completed and beneficial occupancy of the Mu2e Experimental hall is established. Installation will occur in the following order. Further details can be found in the Muon Campus Controls Cost Estimates documentation [40].

- All Fiber optic and copper cable will be ordered as per the specifications determined by the engineers for each system [41].

- Unused Heliax cable between the Central Utility Building and Manhole CMH-33 will be removed to make room for the necessary pulls to the Mu2e Experimental Hall.

- Contract electricians will pull innerduct from the cross gallery to the Mu2e Experimental Hall. The innerduct will reserve the space for the fiber optic cable pulls.

- Contract electricians will pull all fiber optic and copper cables from the cross gallery to the Mu2e Experimental Hall.

- Contract Electricians will complete all cable terminations in both the Cross Gallery and Mu2e Experiment Hall.

## 4.5    Radiation Safety Plan

### 4.5.1 Radiation Safety Requirements

Radiation Safety Plan requirements come from two sources: the Mu2e Project mission need and the Fermilab Radiological Control Manual (FRCM) [54]. The Mu2e Project requires delivery of an 8 kW proton beam by $3^{rd}$ integer, slow resonant extraction to the production target located inside of the Production Solenoid. This requirement is discussed in conjunction with the FRCM in the Technical Design section below. Some of the principle FRCM requirements are briefly introduced in this section while their applications are discussed in conjunction with the stated physics goal below in the Technical Design section.

Fermilab has a mature radiological controls program that will be applied to the operation of Mu2e accelerator, beam line, and experimental facilities. The details of the program related to entry controls, posting of radiological areas and control of radiological work are not discussed here, except where unusual circumstances related to Mu2e facilities warrant additional discussion.

#### 4.5.1.1    *Prompt Effective Dose Control*

The FRCM requirement to control the prompt effective dose rate outside of accelerator and beamline tunnels fall into two broad categories: the normal condition and the accident condition. The permitted effective dose rates cover a wide range of values depending upon the controls that can be implemented on a location-by-location basis. Table 4.10 and Table 4.11 contain the range of dose rate limits relevant to these conditions, and are reproduced from the FRCM for convenience.

Table 4.10. Control of Accelerator/Beamline Areas for Prompt Radiation under <u>Normal Operating Conditions</u> (from Table 2-6 of FRCM).

| Dose Rate (DR) Under Normal Operating Conditions | Controls |
|---|---|
| DR < 0.05 mrem/hr | No precautions needed. |
| 0.05 ≤ DR < 0.25 mrem/hr | Signs (CAUTION -- Controlled Area). No occupancy limits imposed. |
| 0.25 ≤ DR < 5 mrem/hr | Signs (CAUTION -- Controlled Area) and minimal occupancy (occupancy duration of less than 1 hr). |
| 5 ≤ DR < 100 mrem/hr | Signs (CAUTION -- Radiation Area) and rigid barriers (at least 4' high) with locked gates. For beam-on radiation, access restricted to authorized personnel. Radiological Worker Training required. |
| 100 ≤ DR < 500 mrem/hr | Signs (DANGER -- High Radiation Area) and 8 ft. high rigid barriers with interlocked gates or doors and visible flashing lights warning of the hazard. Rigid barriers with no gates or doors are a permitted alternate. No beam-on access permitted. Radiological Worker Training required. |
| DR ≥ 500 mrem/hr | Prior approval of SRSO[18] required with control measures specified on a case-by-case basis. |

### 4.5.1.2 *Effective Dose Control from Sources of Residual Activity*

The control of effective dose due to residual radioactivity is a function that the FRCM assigns to the relevant Division or Section ES&H Department. Residual activation is variable depending upon factors such as beam power, irradiation time, and cool-down time and is addressed by ES&H on an ad hoc basis. A calculation showing expected residual dose rates for components in the slow resonant extraction region at AP30 has been completed [55].

### 4.5.1.3 *Air Activation*

The production, control, and release of airborne radioactivity are permitted processes that are regulated by the State of Illinois and are monitored by the ESH&Q Section. The release of activated air must be anticipated and controlled such that the emissions from all sources at the laboratory are below limits imposed by permit.

### 4.5.1.4 *Ground Water Activation*

Ground water activation is regulated by the Code of Federal Regulations and is monitored by the ESH&Q Section.

---

[18] SRSO = Fermilab's Senior Radiation Safety Officer

Table 4.11. Control of Accelerator/Beamline Areas for Prompt Radiation under <u>Accident Conditions</u> – When It is Likely that the Maximum Dose Can Be Delivered (From Table 2-7 of FRCM)

| Maximum Dose (D) Expected in 1 hour | Controls |
|---|---|
| D < 1 mrem | No precautions needed. |
| 1 < D ≤ 10 mrem | Minimal occupancy only (duration of credible occupancy < 1 hr) no posting |
| 1 ≤ D < 5 mrem | Signs (CAUTION -- Controlled Area). No occupancy limits imposed. Radiological Worker Training required. |
| 5 ≤ D < 100 mrem | Signs (CAUTION -- Radiation Area) and minimal occupancy (duration of occupancy of less than1 hr). The Division/Section/Center RSO[19] has the option of imposing additional controls in accordance with Article 231 to ensure personnel entry control is maintained. Radiological Worker Training required. |
| 100 ≤ D < 500 mrem | Signs (DANGER -- High Radiation Area) and rigid barriers (at least 4' high) with locked gates. For beam-on radiation, access restricted to authorized personnel. Radiological Worker Training required. |
| 500 ≤ D < 1000 mrem | Signs (DANGER -- High Radiation Area) and 8 ft. high rigid barriers with interlocked gates or doors and visible flashing lights warning of the hazard. Rigid barriers with no gates or doors are a permitted alternate. No beam-on access permitted. Radiological Worker Training required. |
| D ≥ 1000 mrem | Prior approval of SRSO required with control measures specified on a case-by-case basis. |

### 4.5.1.5   Radiation and Electrical Safety System Interlocks

The Radiation and Electrical Safety Systems currently employed at Fermilab will also be used for the Mu2e facilities. The Critical Devices to be employed are described below in the Technical Design section.

### 4.5.1.6   Interlocked radiation detectors

A partial set of laboratory standard shielding requirements [56] for an 8 kW, 8 GeV proton beam is given in Table 4.12. The Categories 1A through 5A provide an upper limit of effective dose rate for an 8 kW continuous beam loss if the given shield thickness is present. For example, a location where a 20.6 foot shield is present while an 8 kW, 8 GeV proton beam is continuously lost at a single point, would result in an effective dose rate of up to 5 mrem/hr. Application of Categories 1A through 5A are for situations in which sufficient passive shielding exists to provide adequate protection.

---

[19] RSO = Radiation Safety Officer

Categories 6A through 10A relate the effective dose that could be received for a single pulse of $6.25 \times 10^{12}$ protons[20] for a particular shield thickness. Categories 6A through 10A are applied in conjunction with interlocked radiation detectors. For example, a single beam pulse of $6.25 \times 10^{12}$ protons lost at a location where the shield is 11.4 feet thick would result in a delivered effective dose of up to 1 mrem. Interlocked radiation detectors are used for Categories 6A through 10A to limit the duration and severity of beam loss conditions when insufficient passive protection (shielding) exists.

Table 4.12. Partial list of shield criteria for 8 kW, 8 GeV proton beam

| Magnet in Enclosure | | |
|---|---|---|
| Dose (D) | Category | Shield Thickness (ft.) |
| D < 1mrem | 1A | 23.0 |
| $1 \leq D \leq 5$ mrem | 2A | 20.6 |
| $5 \leq D \leq 100$ mrem | 3A | 16.2 |
| $100 \leq D \leq 500$ mrem | 4A | 13.8 |
| $500 \leq D \leq 1000$ mrem | 5A | 12.8 |
| Interlocked Detectors (1 pulse – $6.25 \times 10^{12}$ protons) | | |
| D < 1mrem | 6A | 11.4 |
| $1 \leq D \leq 5$ mrem | 7A | 9.0 |
| $5 \leq D \leq 100$ mrem | 8A | 4.6 |
| $100 \leq D \leq 500$ mrem | 9A | 3.0 |
| $500 \leq D \leq 1000$ mrem | 10A | 3.0 |

A summary of the radiation shielding thicknesses for the Mu2e facilities is given in Table 4.13. The basis for the Radiation Safety Plan derives from a comparison of Table 4.12 and Table 4.13. It would be possible to operate most of the Mu2e facilities with passive shielding if, as required by Table 4.10 and Table 4.11, four or eight foot fences where to be installed around the entire facility. Interlocked radiation detectors would be required for the Transport Enclosure at the AP0 Service Building and at all Delivery Ring Service Buildings because those locations have less than 12.8 feet of shielding. However, Indian Road, the major thoroughfare between the Main Injector and the remainder of Fermilab, would have to be closed off with fences. Since both closing Indian Road and adding shielding to the entire complex are impractical options, the use of interlocked radiation detectors is imperative. Consequently, the Radiation Safety shielding plan for Mu2e facilities is based solely upon the Categories 6A through 10A in Table 4.12.

---

[20] An 8 kW, 8 GeV beam delivers $6.25 \times 10^{12}$ protons/sec to any given location

Table 4.13. List of Mu2e accelerator and beam line facilities along with the nominal shield

| Location | Nominal shield (ft.) |
|---|---|
| Pre-Vault Enclosure at AP0 Service Building – M1 line | 13 |
| Transport Enclosure at AP0 Service Building – M3 line | 12.5 |
| Transport Enclosure Shielded Tunnel – M3 line | 14 |
| Transport Enclosure under Indian Road – M3 line | 13 |
| Transport Enclosure to Delivery Ring – M3 line | 13 |
| Delivery Ring at Arcs | 13 |
| Delivery Ring at Service Buildings | 10 |
| Beam Transport from Delivery Ring to Target Hall – M4 line | 16 |

The interlocked radiation detector currently approved for use as a credited safety system at Fermilab is the Chipmunk ion chamber. A Chipmunk has a nominal detector length of less than one foot. A total of about 235 Chipmunk ion chambers would be required to adequately cover the shielded locations listed in Table 4.13, assuming that a spacing of 15 feet is sufficient. About 45 Chipmunks are presently installed at these locations, primarily at the Service Buildings. Consequently, about 190 additional chipmunks would be required.

The costs to develop, build, install, and maintain such a number of Chipmunks were considered extraordinary. Consequently, the Mu2e Project received a suggestion to consider the development of an alternative long detector from the ESH&Q Section in May of 2011.

The development of the long detector ion chamber, referred to as a Total Loss Monitor (TLM), shown in Figure 4.32, has been ongoing since May 2011. The TLM system consists of two main parts: the detector [65] and the electrometer [66] [67]. The detector response has been characterized utilizing the TLM electrometer for proton beam loss under a variety of conditions [58] [59] [60]. The TLM electrometer has been developed outside the scope of the Mu2e project.

The TLM system design was submitted to the ESH&Q Section for preliminary approval as a credited safety system in October 2013 [69].

The trip level for an integrating style interlocked radiation detector becomes the nominal upper limit of the normal operating condition. Since the time-weighted average effective dose is limited by the Radiation Safety System (RSS), a trip level must be chosen that is high enough to permit normal operating losses with some reasonable margin without exceeding the normal effective dose rate limits established in Table 4.10.

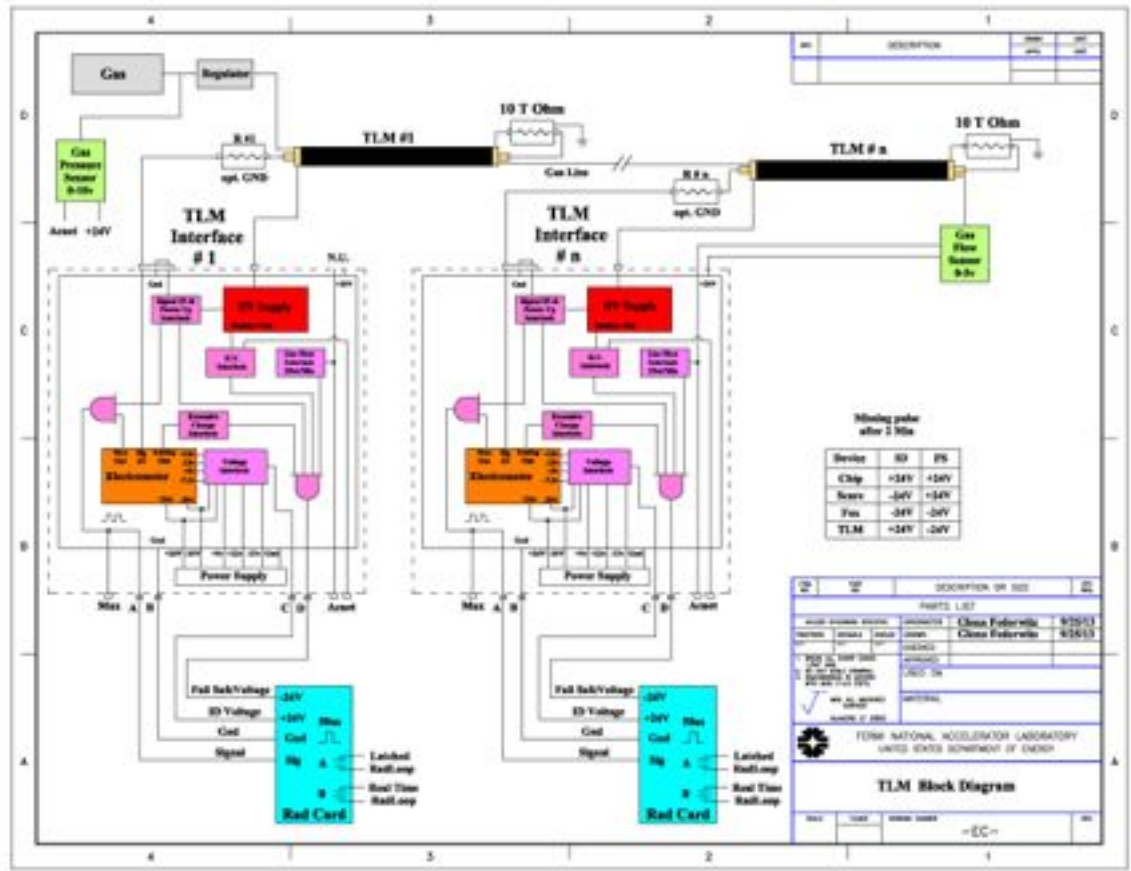

Figure 4.32. TLM System Schematic. The TLM detector gas volumes may be connected in series to share detector gas systems. However, each detector is connected to an individual TLM electrometer.

## 4.5.2 Radiation Safety Technical Design

### 4.5.2.1 Delivery Ring Extraction Losses

As can be determined from Table 4.13, the most challenging shield design for the Mu2e project is at the AP service buildings. In particular, the slow resonant extraction process occurs at the AP30 service building. Previous controlled beam loss measurements and shield calculations have been made to characterize the situation without a realistic model of operational beam loss mechanisms [61] [62] [63]. In more recent work, a model of the slow resonant extraction system including the Electrostatic septa, Extraction Lambertson, C-magnet, quadrupole magnets and a subset of extraction line magnets has been developed to more accurately assess the nature of beam losses in the AP30 straight section[64]. As shown in Figure 4.33, the model includes the AP30 service building and the nearby Type 2[21], exit stairway. An in-tunnel shielding system [68], as shown in Figure 4.34 has been developed to supplement the existing 10-foot service building shield. The complete in-tunnel shielding system, incorporated into the MARS simulation model, is shown in Figure 4.35. This figure also shows a sample of 8 GeV extracted proton beam tracks in the extraction and circulating beam channels.

A comparison of the result of the MARS simulations both without and with in-tunnel shielding is shown in Figure 4.36. The peak normal effective dose rate in the AP30 service building with in-tunnel shielding is just under 40 mrem per hour. From Table 4.10 it can be seen that this is within the allowable operating range for normal beam loss. The building would be posted as a Radiation Area and access would be restricted through the entry control program to authorized personnel during beam operations.

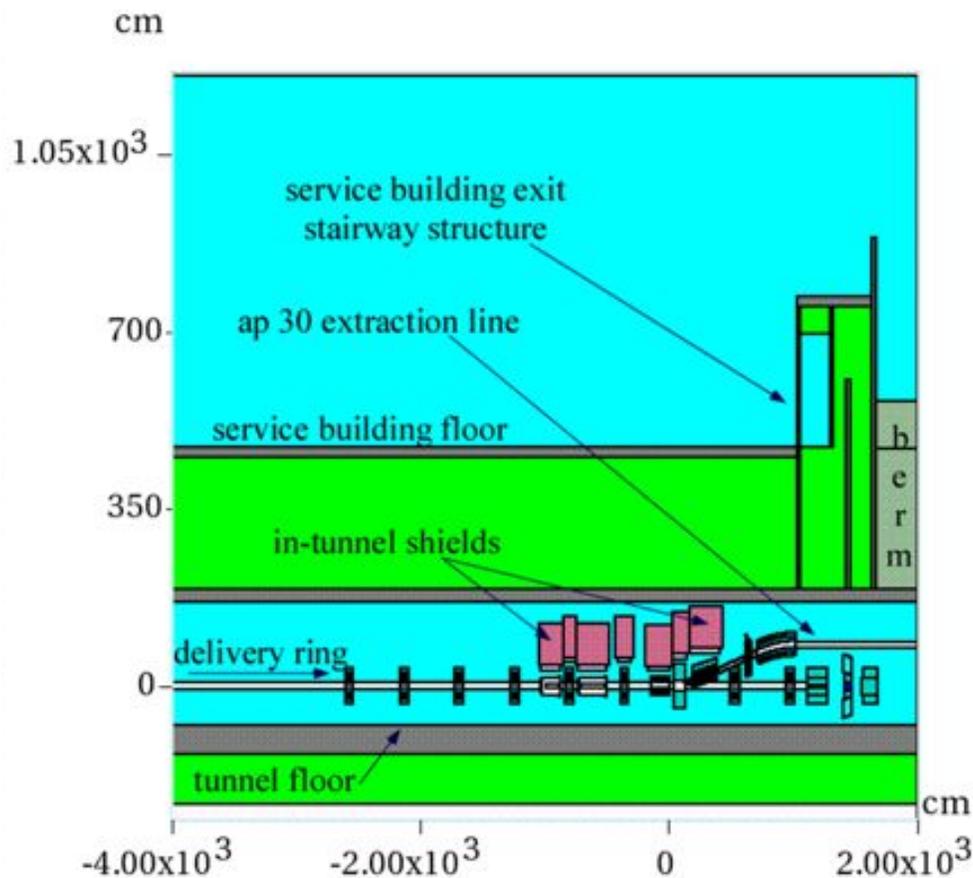

Figure 4.33. Elevation view depicting MARS model of AP30 service building, including the exit stairway and the slow resonant extraction system.

---

[21] A Type 1 stairway has an associated elevator; a Type 2 stairway does not have an associated elevator. The presence of an elevator shaft is relevant for shielding considerations.

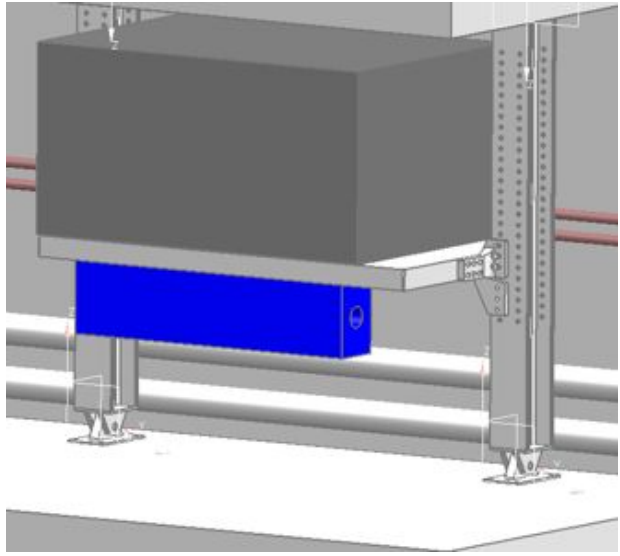

Figure 4.34. Example of in-tunnel shield design at the Extraction Lambertson location.

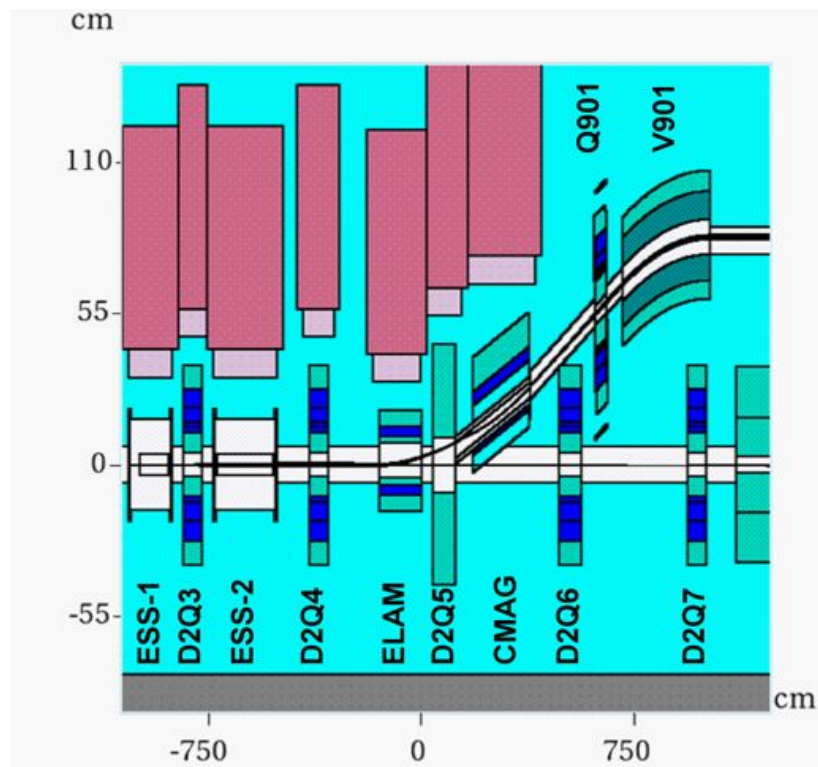

Figure 4.35. MARS model of the Delivery Ring extraction region. Several extraction region devices are shown with 3 foot thick in-tunnel steel shielding (red blocks). The Electrostatic Septa (ESS-1 and ESS-2), various quadrupoles, the Extraction Lambertson (ELAM), and the C-magnet (CMAG) are the devices to be shielded. The first extraction line quadrupole (Q901) and the downward vertical bend (V901) do not require in-tunnel shielding. The tracks of an 8 GeV proton beam sample directed at the upstream electrostatic septum wires are shown in black. Part of the beam is diverted by the ESS field into the extraction channel and part of the beam remains near

the circulating orbit of the Delivery Ring. Tracks of shower particles resulting from proton beam loss are suppressed.

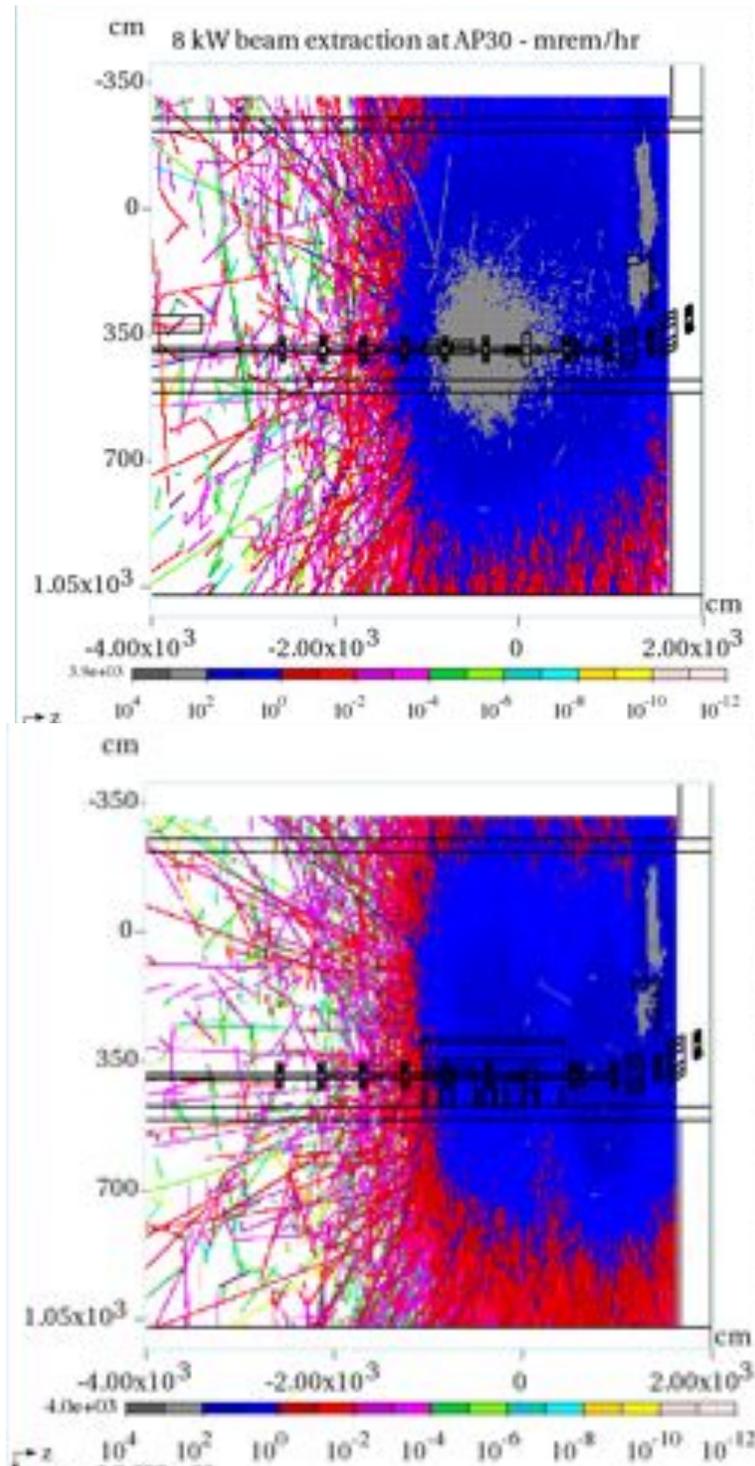

Figure 4.36. A comparison of MARS simulation histograms in plan views at the elevation of the AP30 service building floor. The top histogram shows the effective dose rate without in-tunnel shielding. The bottom histogram shows the dose rate with in-tunnel shielding. The in-tunnel shielding reduces the effective dose rate to acceptable levels. The elevated level shown at the upper right side of the figures is within the exit stairway that is inaccessible during beam on operation. The lower black lines at Y = 508 cm indicates the tunnel concrete shield wall. The parking lot adjacent to the AP30 service building begins at Y = 750 cm.

An additional MARS simulation was performed to determine effective dose rates in the parking lot adjacent to the AP30 service building, along Indian Road that passes by the AP30 service building, and at greater distances due to radiation skyshine. The result of the calculation for the parking lot and Indian Road is shown in Figure 4.37.

The peak effective dose rate in the parking lot is generally less than 1 mrem/hr. The effective dose rate at Indian Road is typically less than 0.05 mrem/hr. Therefore, the occupancy of the roadway will not be restricted.

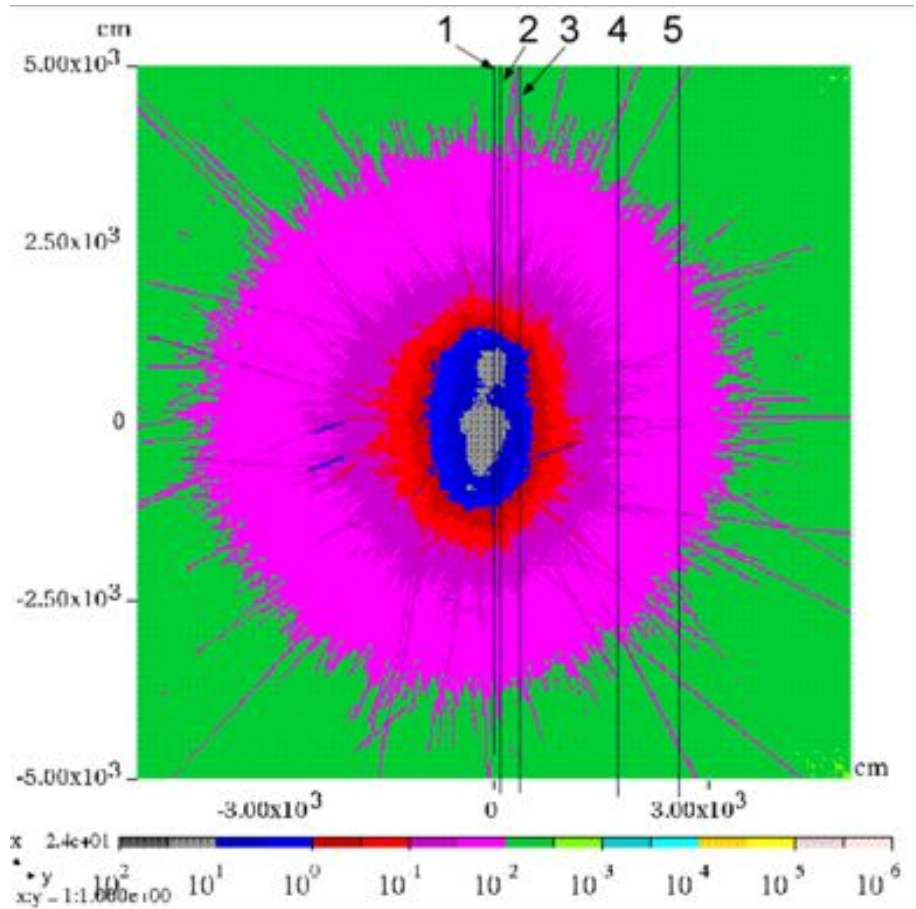

Figure 4.37. This image shows the result of a MARS skyshine simulation for the in-tunnel shielding case. The effective dose rate (mrem/hr) includes contributions from all particle fluences, both direct and reflected from the atmosphere. Line 1: Delivery Ring centerline; Line 2: Outer surface of tunnel concrete vertical wall; Line 3: AP30 service building outer edge; Region between lines 3 and 4: AP30 Parking Lot; Line 5: edge of Indian Road.

The result of the skyshine calculation is shown in Figure 4.38. In this calculation, the effective dose rate is due solely to skyshine radiation. The average effective dose rate at 500 meters, the nominal distance to Wilson Hall, is less than 0.2 mrem/year.

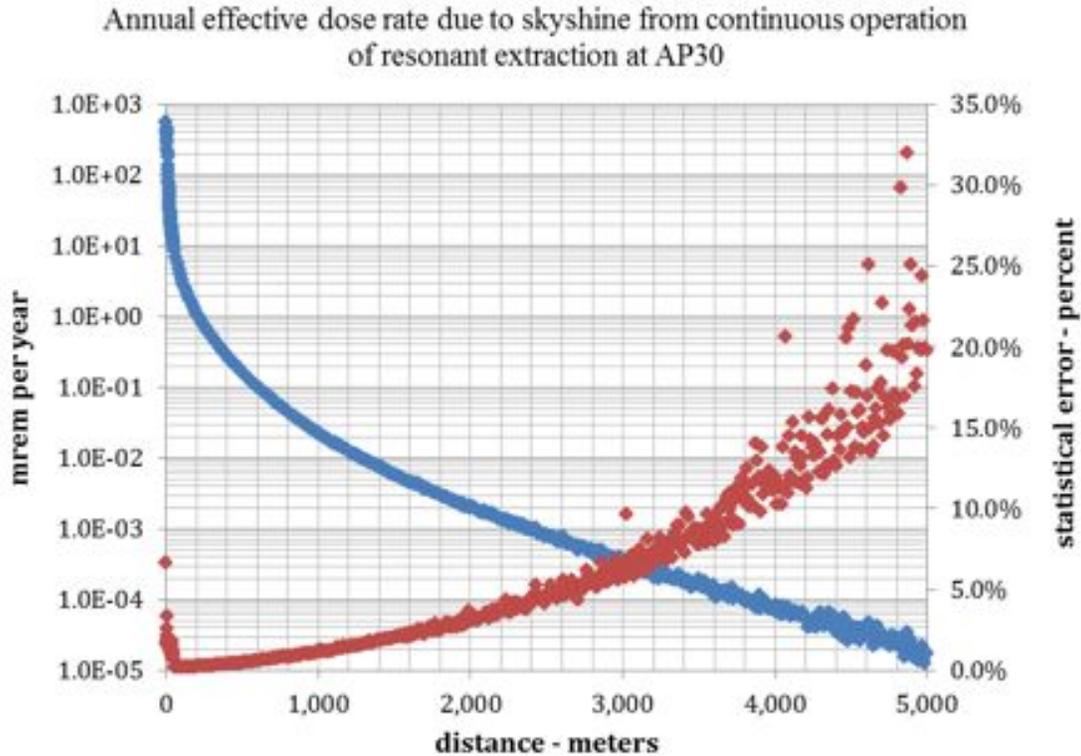

Figure 4.38. Radiation effective dose due to skyshine as a function of distance from the AP30 service building. The statistical errors (shown in red) as a function of distance are influenced by the volume of tissue equivalent detector used in the simulation[22]. The effective annual dose rate at 500 meters, the nominal distance to Wilson Hall, for continuous occupancy, is less than 0.2 mrem/year.

The final concern associated with normal extraction losses at AP30 is the direct radiation exposure to occupants of Wilson Hall. Since the source of radiation is underground, an observer on the ground floor of Wilson Hall will receive a lower direct radiation dose than an observer on a floor that is sufficiently high that there is a direct line of sight to the source that does not pass through the ground. To calculate the direct dose, a MARS simulation was made with a cylindrical tissue equivalent detector centered at AP30. This cylindrical detector was constructed to be 0.3 meters thick, 70 meters high, with a radius of 500 meters (the distance from AP30 to Wilson Hall). This simulation was used to predict the annual effective direct radiation dose rate from Delivery Ring extraction losses as a function of floor elevation. The azimuthal angle of Wilson Hall in this detector geometry is shown in Figure 4.39. The result of the simulation for the entire cylinder is shown in Figure 4.40.

---

[22] A small volume at a large distance intercepts a relatively small sample of particle tracks increasing the statistical error in the mean dose rate measured in that volume.

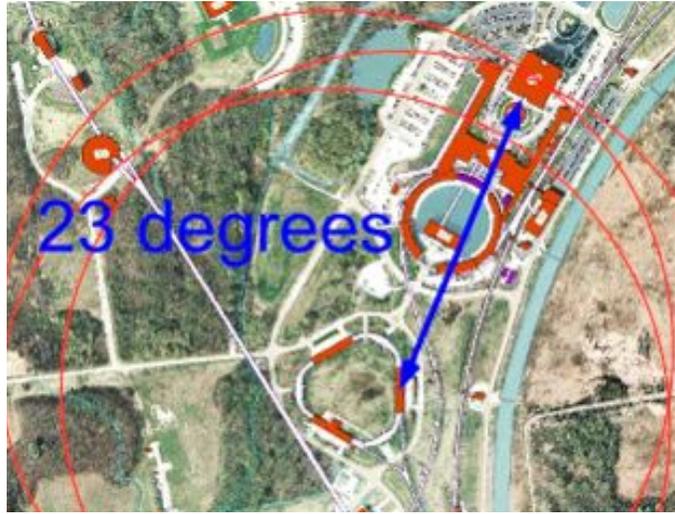

Figure 4.39. The azimuthal angle between the z axis of the MARS simulation at AP30 service building and the center of Wilson Hall

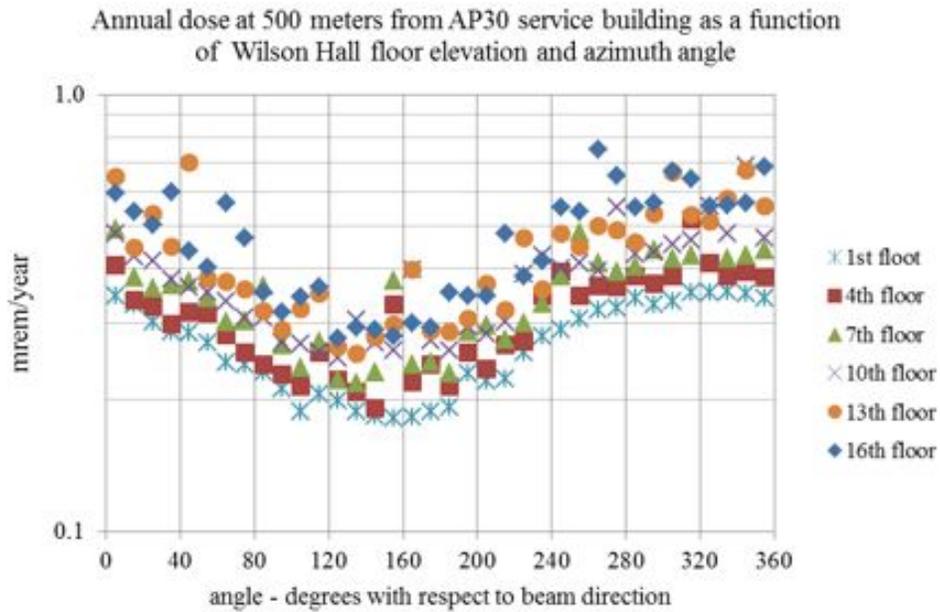

Figure 4.40. Total effective annual dose rate as a function of floor elevation in Wilson Hall from direct radiation exposure originating from the AP30 service building plus skyshine. The azimuth angle of Wilson Hall is 23°.

A TLM installed in the AP30 straight section will be employed to limit beam losses to those expected at nominal levels. A trip level margin commensurate with limitations for the control level chosen from Table 4.10 will apply. As a result, it would not be possible for effective dose rate delivered under any conditions, including accidents, to exceed permit limits under normal conditions.

The AP10 and AP50 service buildings are not expected to have significant beam losses, though the abort kickers under the AP50 straight section will require monitoring. The control of beam losses for the AP10 and AP50 service buildings is discussed in the next section.

### 4.5.2.2  *General Protection Scheme for Limiting Prompt Effective Dose Rate*

The preceding section covered the consequences and control features for the AP30 extraction region. In this section, the remainder of the Mu2e facilities up to the target hall are addressed.

In general, normal operating beam losses are expected to be minimal and will not require additional control measures, such as in-tunnel shielding. As discussed in section 4.5.1.6, interlocked radiation detectors will be required to ensure the effective dose rate limits are observed for all tunnel enclosures. TLMs are to be used for this purpose.

Since interlocked radiation detectors are to be used and the detector trip level setting defines the limiting condition, a control level from Table 4.10 must be chosen. There are two categories in Table 4.10 that are relevant to the Mu2e facility. These categories are identified with a control level ID number and duplicated in Table 4.14. TLM locations and trip levels are described in Table 4.15.

The TLM response for a controlled beam loss has been determined at the Booster for 8 GeV proton beam loss using an $Ar/CO_2$ detector system [59]. The experimentally determined value of 2.6 nC/$10^{10}$ protons is used in Table 4.15 to calculate a TLM charge collection level.

Table 4.14. Control level indices for use in Table 4.15

| Control level | Dose range | Required controls |
|---|---|---|
| 1 | $0.25 < DR < 5$ mrem/hr | Signs (CAUTION -- Controlled Area) and minimal occupancy  (occupancy duration of less than 1 hr) |
| 2 | $5 < DR < 100$  mrem/hr | Signs (CAUTION -- Radiation Area) and rigid barriers (at least 4' high) with locked gates.  For beam-on radiation, access restricted to authorized personnel. Radiological Worker Training required. |

Table 4.15. TLM locations and trip levels. The numbers in the Control Level column are drawn from Table 4.14.

| TLM | Location | Control Level | shield (ft.) | beam loss scaling factor (protons per mrem) | lower hourly dose rate for control level (mrem/hr) | upper hourly dose rate for control level (mrem/hr) | Project Suggested TLM limiting rate (mrem/hr) | extended limit - protons per hour | TLM trip level (nC/min) |
|---|---|---|---|---|---|---|---|---|---|
| 1 | Pre-Vault Enclosure at AP0 Service Building – M1 line | 2 | 13 | 1.89E+13 | 5 | 100 | 50 | 9.43E+14 | 604 |
| 2 | Transport Enclosure at AP0 Service Building – M3 line | 2 | 12.5 | 1.34E+13 | 5.00 | 100 | 50 | 6.71E+14 | 430 |
| 2 | Transport Enclosure Shielded Tunnel – M3 line | 1 | 14 | 3.73E+13 | 0.25 | 5 | 5 | 1.86E+14 | 119 |
| 3 | Transport Enclosure under Indian Road – M3 line | 1 | 13 | 1.89E+13 | 0.25 | 5 | 0.25 | 4.72E+12 | 3 |
| 4 | Transport Enclosure to Delivery Ring – M3 line | 1 | 13 | 1.89E+13 | 0.25 | 5 | 5 | 9.43E+13 | 60 |
| 5 | Delivery Ring 20 Arc | 1 | 13 | 1.89E+13 | 0.25 | 5 | 5 | 9.43E+13 | 60 |
| 6 | Delivery Ring 40 Arc | 1 | 13 | 1.89E+13 | 0.25 | 5 | 5 | 9.43E+13 | 60 |
| 7 | Delivery Ring 60 Arc | 1 | 13 | 1.89E+13 | 0.25 | 5 | 5 | 9.43E+13 | 60 |
| 8 | Delivery Ring at AP10 Service Buildings | 2 | 10 | 2.44E+12 | 5.00 | 100 | 50 | 1.22E+14 | 78 |
| 9 | Delivery Ring at AP30 Service Buildings | 2 | 10 | 2.44E+12 | 5.00 | 100 | 50 | 1.22E+14 | 78 |
| 10 | Delivery Ring at AP50 Service Buildings | 2 | 10 | 2.44E+12 | 5.00 | 100 | 50 | 1.22E+14 | 78 |
| 11 | Upstream Beam Transport from Delivery Ring to Target Hall – M4 line | 1 | 16 | 1.46E+14 | 0.25 | 5 | 5 | 7.28E+14 | 467 |
| 12 | Downstream Beam Transport from Delivery Ring to Target Hall – M4 line | 1 | 16 | 1.46E+14 | 0.25 | 5 | 5 | 7.28E+14 | 467 |

A description of the entries in Table 4.15 includes the following:

- Location – The section of tunnel covered by a common TLM detector. The section is uniformly shielded so that the limiting effective dose per lost proton is nominally constant throughout the region.

- Shield (ft.) – the minimum number of feet of shielding at the TLM installation.

- Beam loss scaling factor – the number of protons lost per mrem for the given shielding thickness. The factor is determined from the standard shield scaling criteria [56].

- Lower hourly dose rate for control level – this is the lower range value for the given control level.

- Upper hourly dose rate for control level - this is the upper range value for the given control level.

- Project suggested TLM limiting rate – this is the trip level determined by the Mu2e Project that should be achievable while meeting the physics goals for the Mu2e experiment.

- Extended limit – the product of the beam loss scaling factor and the Project suggested TLM limiting rate.

- TLM trip level – the average charge collected in nanocoulombs per minute that would result in an interlocked radiation detector trip.

The suggested TLM trip levels are determined assuming a beam loss that occurs at a single point. Since the TLM system cannot distinguish how the collected charge was distributed, the actual effective dose rate at any location along the shielded location will generally be lower than the Project suggested limit. This is because beam losses are more likely to be distributed over macroscopic distances rather than a single point. Consequently, the TLM system is a conservative protection system.

Residual activation should also be considered in setting TLM trip levels. One watt per meter of prompt beam loss is the accepted level that allows worker access without the need for extraordinary controls. The trip levels given in Table 4.15 are modified in Table 4.16 to limit beam losses to 1 watt/ meter. The maximum effective dose rate outside the shield is reduced accordingly where modified trip levels apply.

### 4.5.2.3   M4 Beam Line Shield Wall and Diagnostic Absorber

A 170-watt beam absorber is required in the M4 beamline for commissioning the beamline and Delivery Ring resonant extraction. The initial beam commissioning period will take place while the Mu2e apparatus is being installed in the detector enclosure. Consequently, a shield wall is required to limit the radiation dose rate in the detector enclosure during beam commissioning. MARS simulations were used to develop the design of the Diagnostic Absorber and shield wall [71]. Figure 4.41 shows a plan view of the arrangement. The effective dose rate at the location of the Detector Solenoid is less than 0.05 mrem/hr during normal beam operation to the diagnostic absorber.

An accident condition in which the 170-watt proton beam is lost on the MDC switching magnet was also considered. The resulting dose rate at the Production Solenoid was calculated to be about 250 mrem/hr. A TLM will be located in the M4 line upstream of shield wall. The TLM trip level for this operating mode will have to be reduced from the nominal 248 nC/minute to about 6.5 nC/minute in order to permit non-radiation workers unrestricted access to the area around the Production Solenoid.

Table 4.16. Modified TLM trip levels to limit beam loss to 1 Watt/meter. The maximum effective dose rate by location is reduced commensurately with the reduced trip level. The trip level for TLM 9 (AP30) will be determined with consideration of the in-tunnel shielding.

| TLM | Location | Extended limit – protons per hour | TLM length | Watts | watts per meter | TLM trip level (nC/min) | Modified trip level (nC/min) | Maximum effective dose rate |
|---|---|---|---|---|---|---|---|---|
| 1 | Pre-Vault Enclosure at AP0 Service Building – M1 line | 9.43E+14 | 11.6 | 336 | 29.0 | 604 | 21 | 1.7 |
| 2 | Transport Enclosure Shielded Tunnel – M3 line | 1.86E+14 | 138 | 66 | 0.5 | 119 | 119 | 5.0 |
| 3 | Transport Enclosure under Indian Road – M3 line | 4.72E+12 | 10 | 2 | 0.2 | 3 | 3 | 0.3 |
| 4 | Transport Enclosure to Delivery Ring – M3 line | 9.43E+13 | 138 | 34 | 0.2 | 60 | 60 | 5.0 |
| 5 | Delivery Ring 20 Arc | 9.43E+13 | 118 | 34 | 0.3 | 60 | 60 | 5.0 |
| 6 | Delivery Ring 40 Arc | 9.43E+13 | 118 | 34 | 0.3 | 60 | 60 | 5.0 |
| 7 | Delivery Ring 60 Arc | 9.43E+13 | 118 | 34 | 0.3 | 60 | 60 | 5.0 |
| 8 | Delivery Ring at AP10 Service Buildings | 1.22E+14 | 51 | 44 | 0.9 | 78 | 78 | 50.0 |
| 9 | Delivery Ring at AP30 Service Buildings | | | | | | | |
| 10 | Delivery Ring at AP50 Service Buildings | 1.22E+14 | 51 | 44 | 0.9 | 78 | 78 | 50.0 |
| 11 | Upstream Beam Transport from Delivery Ring to Target Hall – M4 line | 7.28E+14 | 138 | 259 | 1.9 | 467 | 248 | 2.7 |
| 12 | Downstream Beam Transport from Delivery Ring to Target Hall – M4 line | 7.28E+14 | 138 | 259 | 1.9 | 467 | 248 | 2.7 |

### 4.5.2.4   M5 Beam Line Shield Wall

A shield wall, provided by the muon g-2 experiment is required in the M5 beam enclosure to permit personnel access to the MC-1 service building during Mu2e beam operation. MARS simulations were used to develop the design of the M5 beamline shield wall [74]. The simulations show that a TLM trip level of 520 nC/min will limit the effective dose rate downstream of the M5 shield wall to 0.25 mrem/hr. The limit in Table 4.16 for the upstream TLM is 248 nC/min, well below the specified trip level in the M5 analysis. The combination of the shield wall and this TLM trip level will be sufficient to permit trained radiation workers unlimited access to the MC-1 service building.

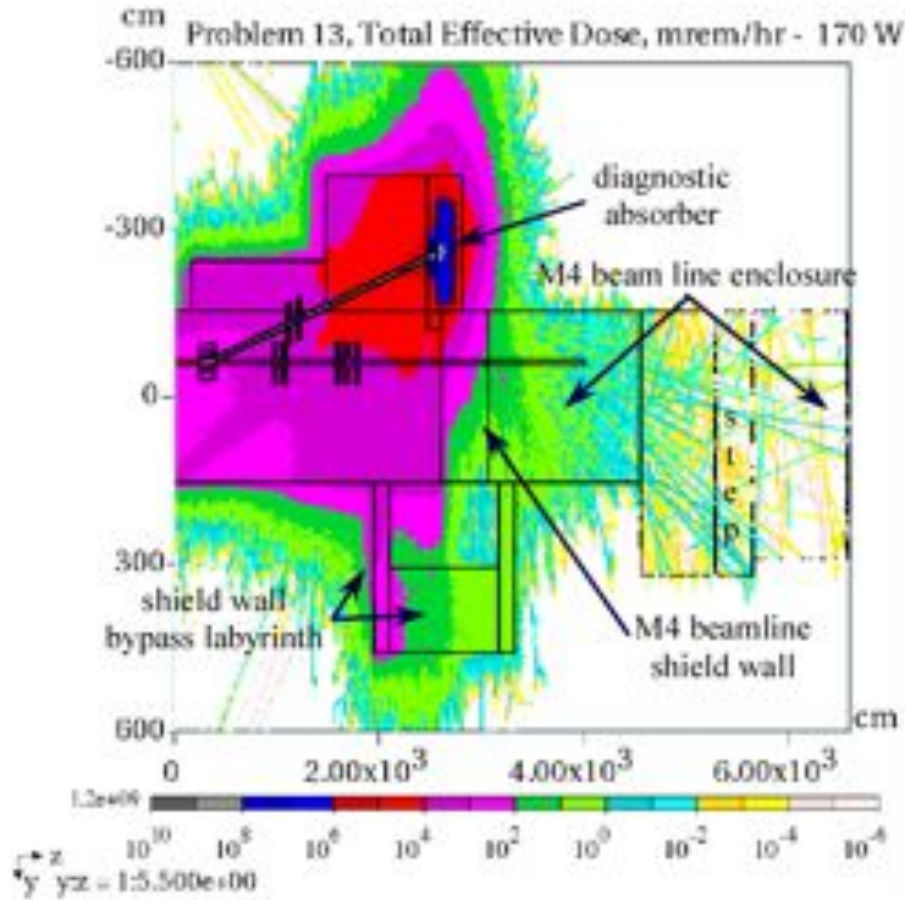

Figure 4.41. Plan view of MARS simulation showing the diagnostic absorber, shield wall, shield wall bypass labyrinth under the normal operating condition.

### 4.5.2.5 Exit Stairways and Penetrations

Penetrations through the passive radiation shielding including stairways, ducts and cable penetrations are considered in this section. TLMs described above also play a role in limiting the radiation dose rate for these penetrations through the radiation shield.

An Excel spreadsheet developed by the ES&H Section [72] was used to calculate the radiation dose rates at the exit of labyrinths and penetrations based upon user input parameters including the source term, aspect ratio, and length of each of the legs of the labyrinth or penetration. The evaluated penetrations are listed in Table 4.17 along with the resulting dose rate calculated for the 2000 Pbar shielding assessment [73]. The third column of the table shows the resulting dose rate by scaling to the 8 kW beam power required for Mu2e. The fourth column shows the maximum number of protons lost per hour as limited by the TLM trip levels established in Table 4.16. The fifth column shows the maximum possible dose rate (single point beam loss) at the exit of facility penetrations based upon the TLM trip levels established in Table 4.16. As indicated in Table 4.17, the resulting radiation effective dose rates at the exits of these penetrations are within limits prescribed by the FRCM. No additional remediation is required for the existing facility including the three elevator shafts at the type 1 stairways[23].

Table 4.17. 2000 Pbar shielding assessment penetration dose calculations scaled to proposed TLM trip levels. Radiation dose rates at penetration exits would require no addition mediation if TLMs are used as described above.

| Penetration Name | Calculated exit dose rate from 2000 pbar shielding assessment | Scaled to 8 kW, 8 GeV proton beam loss | TLM # | Max protons lost/hour limited by TLMs | Penetration dose rate limited by TLMs |
|---|---|---|---|---|---|
| Determined for $3.6 \times 10^{13}$, 8 GeV primary protons per hour | | | | | |
| ACC/DEB airshaft | 7.54E-02 | 47 | 5,6,7 | 9.43E+13 | 0 |
| ACC/DEB stairway type 2 | 1.85E-03 | 1 | 8,9,10 | 1.22E+14 | 0.0171 |
| Transport to AP0 penetrations | 9.62E-02 | 60 | 2 | 1.86E+14 | 0 |
| Stub Room Penetrations | 2.00E-01 | 125 | 8,9,10 | 1.22E+14 | 1 |
| AP0 water pipe penetrations | 8.21E-01 | 513 | 2 | 1.86E+14 | 4 |
| Transport air duct vent to AP0 | 4.01E-03 | 3 | 2 | 1.86E+14 | 0 |
| Transport to F27 Penetrations | 6.32E-14 | 0 | 2 | 1.86E+14 | 0 |
| ACC/DEB elevator shafts | 5.09E-01 | 318 | 8,9,10 | 1.22E+14 | 2 |
| Transport stairway | 4.47E-02 | 28 | 2 | 1.86E+14 | 0 |
| ACC/DEB stairway type 1 | 1.41E-05 | 0 | 8,9,10 | 1.22E+14 | 0 |
| AP50 Pit Vent | 7.63E-07 | 0 | 10 | 1.22E+14 | 0 |
| AP50 Pit Labyrinth | 1.78E-02 | 11 | 10 | 1.22E+14 | 0 |
| Determined for $1.8 \times 10^{16}$, 120 GeV primary protons per hour | | | | | |
| PreVault stairway | 1.58E-02 | 0 | 1 | 3.26E+13 | 0 |
| Sweeping Magnet Penetrations | 1.23E+00 | 0 | 1 | 3.26E+13 | 0 |
| PreVault to F23 Penetrations | 5.12E-04 | 0 | 1 | 3.26E+13 | 0 |

---

[23] A Type 1 stairway has an associated elevator; a Type 2 stairway does not have an associated elevator.

**4.5.2.6  Production Solenoid, Transport Solenoid, and Detector Solenoid Radiation Shielding**

The Production Solenoid, Transport Solenoid, and Detector Solenoid rooms are an integral part of the Mu2e experiment hall depicted in Figure 4.141. The 8 kW, 8 GeV delivered by the M4 line is directed to tungsten target located inside the Production Solenoid. The un-interacted primary beam plus the resulting shower exit the Production Solenoid, travel across an air gap, and are stopped in the main beam absorber. Since the beam is completely stopped in the experimental facility, the normal condition is, with one exception, the worst case condition. It should be possible to steer the primary 8 GeV beam off of the tungsten target. This will be an expected condition during target scans. The beam dump is designed to safely accept the full 8 GeV primary beam; the peak power density for this condition is just 9.7 mW/g [133].

The shielding design for the normal condition beam operation has been studied in some detail [136]. The resulting effective dose rate is shown in the 3D histogram shown in Figure 4.42. The peak effective dose rate, 50 mrem/hr, occurs at the Production Solenoid drop hatch. Consequently, the area will be required to be enclosed by a 4' high fence and the area will be required to posted as a "Radiation Area" in accordance with FRCM requirements.

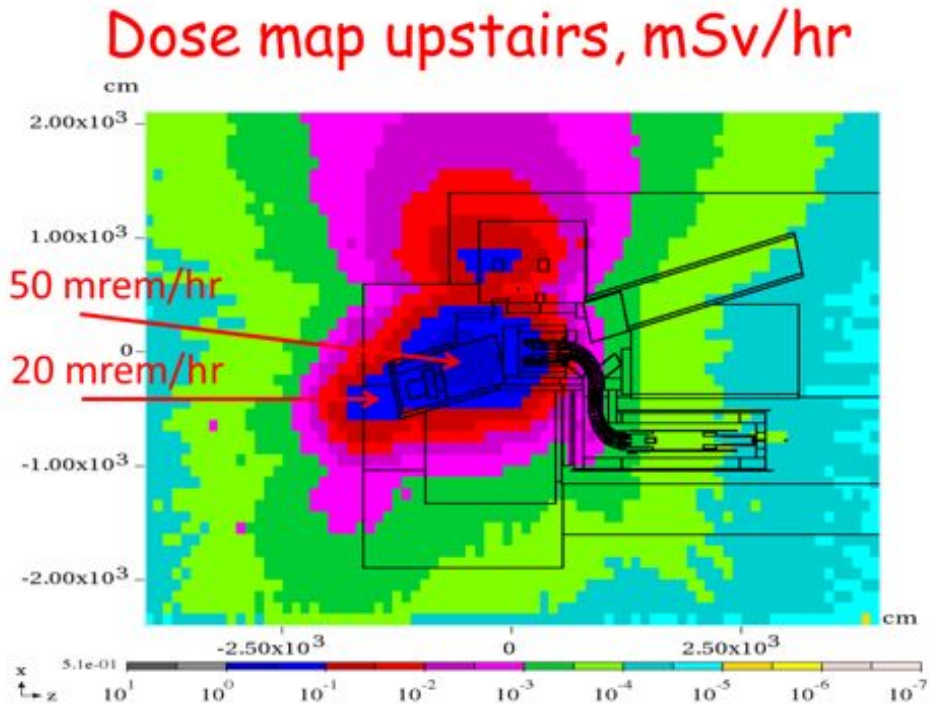

Figure 4.42. 3D histogram of radiation effective dose rate at ground level for the Mu2e experimental facility. The histogram scale is in mSv/hr while the peak dose rates above the Production Solenoid drop hatch and the extinction room drop hatch are given in units of mrem per hour.

The Mu2e experiment facility includes 5 drop hatches, 3 exit stairways, an elevator shaft, and 14 major penetrations. These design of these features has been evaluated [136] and found to meet all requirements of the Fermilab Radiological Control Manual [54].

### 4.5.2.7 Ground water activation

The major sources of ground water activation due to beam operations for the Mu2e experiment include losses at the following locations:

- Delivery Ring beam absorber

- M4 beamline Diagnostic absorber

- Proton target absorber (downstream of the Production Solenoid)

- Delivery Ring Injection

- Delivery Ring extraction

Detailed calculations for ground water activation for Mu2e operation of the Delivery Ring have been completed [57]. No ground water issues have been identified.

### 4.5.2.8 Surface water activation

The major sources of surface water activation due to beam operations for the Mu2e experiment are the same sources as those listed for ground water activation. Detailed calculations for surface water activation for Mu2e operation of the Muon Campus have been completed [57]. No surface water issues have been identified.

### 4.5.2.9 Airborne radioactivity

The major source of airborne radioactivity due to beam operations for the Mu2e experiment is from primary/secondary beam passing through the air volume between the Production Solenoid and the Proton Target Beam Absorber. The Production Solenoid Room will be under negative pressure relative to the outdoor environment and to the Detector Solenoid Room to prevent activated air infiltration to those areas. About 900 cfm of dry conditioned air will be injected into the Production Solenoid Room to minimize production of nitric acid. To minimize the release of airborne radioactivity, Production Solenoid Room air will be transported through the M4 beamline enclosure to the exit stairway at the upstream end of the M4 beamline enclosure. There, air will be released to the atmosphere through an exhaust fan. The exhaust fan speed will be set to ensure the Production Solenoid Room and M4 beamline enclosure are at negative pressure with respect to atmosphere. Air flow barriers in the M5 beamline and between the Delivery Ring and M4 beamline will be required in order to maintain negative pressure. In addition, various shield walls and access points will be sealed.

Detailed calculations for other, less significant, sources of airborne radioactivity for Mu2e operation have been completed [57]. The contribution of airborne radioactivity due to Mu2e operations will be a reasonably small fraction of the permitted annual emissions [57].

### 4.5.2.10 Radiation Safety System Critical Devices

The Muon Campus beam operations will serve multiple purposes, primarily to deliver beam to the MC-1 service building for the muon g-2 experiment and to the Mu2e proton target for the Mu2e experiment. Critical devices have been chosen for both purposes and are described in Reference [77].

## 4.5.3 Radiation Safety Risks

Two risks have been identified and are described below.

### 4.5.3.1   TLMs cannot be used to limit the intensity and duration of beam loss

The TLM system must be approved as an accredited safety system by the ESH&Q Section in order to be used to limit beam losses. The ESH&Q section gave preliminary approval to use the TLM system as a credited system in April, 2014 [137]. If a technical reason is found during the continuing review process that prohibits accreditation of the TLM system, an alternative approach using Chipmunk ion chambers, previously discussed in section 4.5.1.6 may be used. Assuming a 15-foot spacing is adequate, 190 chipmunks would be required to be produced and installed in place of the proposed TLM systems. The full consequences are discussed in Reference [75].

### 4.5.3.2   Radiation levels outside of the Mu2e facility are too high

The resonant extraction system at AP30 is a known beam loss point. If the effective radiation dose rate cannot be sufficiently attenuated by in-tunnel shielding, additional shielding or a reduction in beam intensity may be required to produce further reductions. The full consequences of this risk are discussed in Reference [76].

## 4.5.4 Radiation Safety Quality Assurance

### 4.5.4.1   In-tunnel shielding design

An engineering design has been completed for the in-tunnel shield support stands. An engineering note has been prepared and an AD Mechanical Support Department design review will be conducted. In addition, the design will be reviewed by the Facility Engineering Services Section to ensure compatibility of the design with tunnel enclosure structures.

### 4.5.4.2   TLM system design

The TLM system includes the TLM chassis, detector, detector gas systems and heartbeat resistor. The system has been designed by AD Electrical Engineers and Engineering Physicists. The design is subject to an intense review process by the AD ES&H Department and the Fermilab ESH&Q section.

If the TLM system is approved as a safety system, it will come under the purview of the AD ES&H Department Interlocks Group. The Interlocks Group will perform periodic calibration and system tests on the TLM system, similar to what is currently done to maintain other Radiation and Electrical Safety Systems. In addition to periodic testing, the TLM system has been designed to be fail-safe. In the event a system parameter goes out of tolerance, the Radiation Safety System will interrupt beam delivery to the affected area until the cause of the out-of-specification condition is found, repaired, tested, and returned to service.

The TLM system will be subject to review by the Shielding Review Committee (SRC). The SRC, a team of experts specializing in radiation protection, are called upon to review safety systems for conformance with FRCM requirements. The TLM system as described in this design report must receive the approval of the SRC before it may be deployed.

### 4.5.4.3   Radiation and Electrical Safety Systems (RSS and ESS)

The Radiation and Electrical Safety Systems that will be used for Mu2e are the same systems that have been in service for many decades at Fermilab. The RSS and ESS designs are subject to review and approval by the ESH&Q Section.

## 4.5.5 Radiation Safety Installation and Commissioning

### 4.5.5.1   In-tunnel shielding installation

The installation of the in-tunnel shielding will be performed by riggers, a specialty group adept at installing massive equipment in tight areas. The installation will be conducted under the supervision of AD Mechanical Support Department Engineers and in accordance with Engineering Design Notes.

### 4.5.5.2   TLM installation

The TLM signal and high voltage cables will be installed by contract electricians under the supervision of the AD Electrician Contract Coordinator. The TLM detector gas system will be installed by AD Mechanical Support Department Technicians. The TLM detectors will be terminated by the AD ES&H Department Interlocks Group. Testing and commissioning of the TLM system will involve a joint effort of the AD ES&H Department and the Muon Department.

### 4.5.5.3   Radiation Safety and Electrical Safety System interlock installation

Installation of the RSS and ESS systems will be accomplished by contractor electricians under the supervision of the AD Electrician Contract Coordinator. Final system terminations, subsystem interconnections and testing will be performed by the Interlocks Group of the AD ES&H Department [70].

## 4.6   Resonant Extraction System

### 4.6.1 Introduction

The Resonant Extraction System is used to provide the experiment with proton pulses with the specific time structure described in section 4.1.5. This structure is determined by the time structure of the circulating beam as, for each turn, a small fraction of the circulating beam is peeled off and redirected to the experiment through the extraction beamline. This is known as "slow extraction," or "slow spill extraction," because the single bunch circulating in the Delivery Ring is fully extracted over a relatively long duration (spill): 54 msec, or about 32,000 turns. After a spill ends, either the next bunch is injected into the Delivery Ring and a new spill begins, or there is a pause until the next Main Injector supercycle. As indicated in Figure 4.4, there are eight spills in each supercycle. The Delivery Ring is reset to its initial state during short intervals after each spill. There is also a substantial interval of no spill in the supercycle when the Recycler is busy with beam manipulations for the neutrino program.

To maintain uniform response of detector subsystems, the spill rate should be maintained as uniform as possible. Two subsystems control the spill process. The first is a circuit of quadrupole magnets that drives the horizontal tune to the resonance value of 29/3. This is a relatively slow system and provides a coarse regulation. The second system employs horizontal "heating" of the beam to assist extraction. This subsystem offers faster spill regulation and is used in a feedback loop to compensate small, fast spill rate variations. Known as the RF Knock-Out (RFKO) method, it will be discussed in detail later in Section 4.6.3.4. The use of both these subsystems by the spill regulation system will be discussed in Section 4.6.3.5.

### 4.6.2 Resonant Extraction Requirements

The requirements for resonant extraction are outlined in the Proton Beam Requirements document [2]. Table 4.18 contains the main specifications for the extracted beam.

### 4.6.3 Resonant Extraction Technical Design

Continuous beam extraction is realized by creating a resonance condition that destabilizes part of the beam in a controlled way. To accomplish this, the machine horizontal tune must be close to a third integer (m/3), and a strong sextupole field must be present to excite the resonance. Unstable particles stream away from the beam center and are intercepted in the Electrostatic Septum (ESS), where a thin plane of wires or foils separates the region of circulating beam from a region of high electrostatic field that deflects particles horizontally. The deflected beam then enters the field region of a Lambertson magnet[24], about 8 m downstream of the ESS, which gives it a vertical kick that sends it into the External Beam Line. The extraction process is driven by ramping the field in a dedicated circuit of quadrupole magnets to move the horizontal tune towards the resonance value.

Table 4.18. Main specifications for the Mu2e Resonant Extraction

| Parameter | Value |
|---|---|
| Spill duration | 54 msec |
| Number of spills in a supercycle | 8 |
| Full spill intensity | Up to $10^{12}$ protons |
| Number of protons extracted per pulse (turn) | $3 \times 10^7$ |
| Time between pulses (turns) | 1.695 μsec |
| Reset time between spills | 5 msec |
| Spill rate variations | < 50% |
| Beam leftover in the end of spill | < 5% |
| Beam losses in the extraction region | < 2% |
| Extraction time duty factor | 32% |

By the end of a spill, the beam must be completely removed from the machine. Any beam present in the ring during the reset time would result in high losses and contamination of the RF intra-bucket (extinction) space. Therefore the entire leftover beam is removed by the abort system at the end of the spill. Locations of all corresponding elements are shown in Figure 4.43.

The following sections will be organized as much as possible according to the Resonant Extraction WBS structure. This will be followed by a general discussion of Risks and Quality Assurance.

### 4.6.3.1   Resonant Extraction General

#### 4.6.3.1.1   Theory

The theoretical background for resonant extraction is described in publication [78], which was the basis of the Conceptual Design of resonant extraction for the Mu2e project. This conceptual work was further extended in reference [79] into an analytical model that can be used for calculations of the extraction efficiency as a function of time and various machine parameters, such as acceptance, initial beam emittance, and Twiss functions.

---

[24] A Lambertson magnet consists of two chambers: a field-free chamber for the beam circulating in the ring and a region containing a dipole magnetic field that vertically kicks the beam in the extraction channel into the extraction beamline.

This model is useful for a parameter space analysis, but is not intended to replace more rigorous tracking simulations that include effects of space charge, RF manipulations, feedback regulation and other complex details. The equations that follow in this section are expressed in the phase space of (complex) canonical, normalized, horizontal coordinates $a$ and $a^*$, which are related to the usual unnormalized, real coordinates, $(x, x')$, or angle-action coordinates, $(\phi, I)$, as follows:

$$a = \sqrt{I} \cdot e^{i(\pi/2 - \varphi)} = \frac{x + i(\alpha x + \beta x')}{\sqrt{2\beta}} \qquad (4\text{-}4)$$

where $\alpha$ and $\beta$ are the usual Twiss lattice functions. The leading order Hamiltonian of the third integer resonance is then presented as

$$H = \Delta \nu \cdot I - (g \cdot e^{-i3\phi} + g^* \cdot e^{i3\phi}) I^{\frac{3}{2}} + \ldots \qquad (4\text{-}5)$$

Here, $\Delta \nu \equiv \nu_x - 29/3 \approx 0$ is the difference between the machine horizontal tune and the resonant tune and is presumed to be small; the "resonance coupling constant," $g$, is a linear functional of the sextupole field strength distribution.

$$g = \frac{i}{6\sqrt{2}} \frac{1}{4\pi} \sum \frac{B'' l}{B\rho} \beta_x^{3/2}(\theta) e^{-i3(\psi_x(\theta) - \Delta \nu \theta)} \qquad (4\text{-}6)$$

where the sum is carried out over the locations of all the sextupole magnet components in the machine. This interaction divides the phase space into two zones of stable and unstable motion. The boundary between these zones is called the separatrix, which is an equilateral triangle in canonical phase space, as shown in Figure 4.44.

The size of this separatrix is determined by the separation between the origin and a vertex, $a_0$, which is related to the control parameters, $\Delta \nu$ and $g$, as follows:

$$|a_0| = \sqrt{I_0} = \left| \frac{\Delta \nu}{3g} \right| \qquad (4\text{-}7)$$

Therefore, to create an unstable motion within the beam, one needs to be close to the resonance tune and have a substantial sextupole field strength. Particles outside the separatrix move away from the center in the directions shown by the red lines in Figure 4.44.

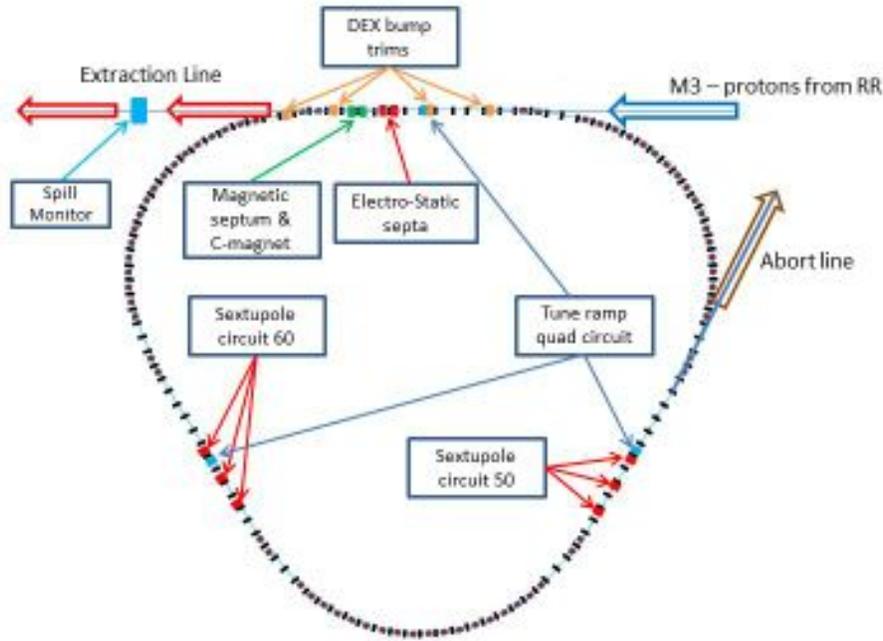

Figure 4.43. Location of the Resonant Extraction elements in the Delivery Ring.

The orientation of the separatrix is determined by the phase of the parameter *g*, which is determined by the positions of the sextupoles relative to the point of observation. To minimize losses at the septum, the angle $\phi_0$ should be made close to $\pi/3$ plus a multiple of $2\pi/3$, corresponding to motion of unstable particles parallel to the *x*-axis.

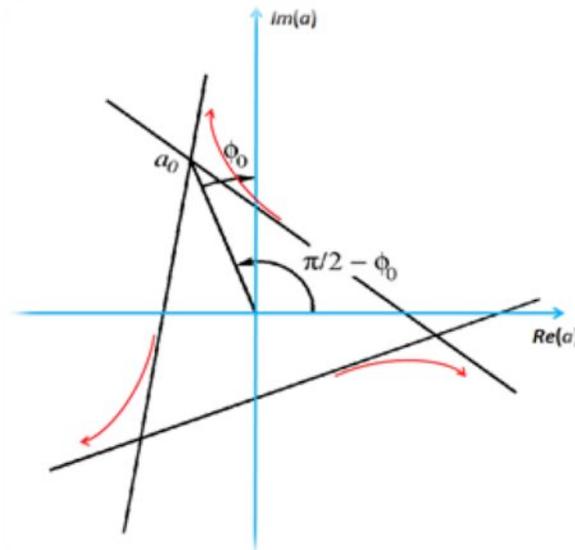

Figure 4.44. Third integer resonance separatrix shape and orientation.

During a spill the machine tune is changed by varying the excitation of ramped quadrupoles from their initial excitations toward excitation currents corresponding to a horizontal tune at the resonance value of 29/3. As $\Delta\nu$ decreases, the separatrix shrinks (see equation (4-7) ), pushing more particles into the unstable region. Thus, the natural sequence of extraction is to remove the particles with higher betatron amplitudes first.

When the beam possess a substantial tune spread, this natural sequence is complicated by the possibility that a beam particle's relative proximity to the separatrix is also a function of its position in the tune distribution in addition to its betatron amplitude. This gives rise to the undesireable prospect of extraction from multiple separatrices. The chromatic tune spread can be controlled by a careful choice of the machine chromaticity. However, the space charge tune spread is irreducible and becomes a serious issue at high beam intensities.

At 8 kW beam power in the Delivery Ring, the space charge tune spread is noticeable, as discussed earlier in section 4.2.1. To mitigate this effect, we approach the resonance from below ( $\Delta \nu < 0$ ) during the extraction sequence (see Figure 4.5). In this case, particles with higher betatron amplitudes (and smaller space charge tune shift) correspond to the upper side of the tune spectrum, which is closer to the resonance and therefore the natural order of extracting high amplitude particles first is preserved.

### 4.6.3.1.2    Septum Beam Loss: Analytical Model

At a fixed location in the ring an unstable particle appears at each consecutive turn in different outgoing branches of the separatrix. Only one direction is of an interest for us, the direction towards the septum. The change of coordinate in this direction is therefore calculated after three consecutive turns and is called the "step size." The step size is an important parameter because the fraction of beam lost on the septum foils ($R_L$) is approximately calculated as the ratio of the septum plane thickness to the step size[25]. In a more accurrate formulation [79], under optimal conditions, $R_L$ can be better approximated by:

$$R_L = \frac{d_w}{X_S^2 - X_0^2} \frac{2X_0}{\ln\left(\dfrac{X_{\max} - X_0}{X_{\max} + X_0} \dfrac{X_S + X_0}{X_S - X_0}\right)} \tag{4-8}$$

Here $d_w$ is the effective septum foil plane thickness, $X_0$ is the projection of the separatrix size, $a_0$, on the $X$-coordinate, $X_S$ is the optimum septum plane position and $X_{max}$ is the maximum "reach position" allowed by the machine acceptance. The $X_s$ and $X_{max}$ are calculated in the model [79], and $a_0$ is a known function of time. Equation (4-8) shows that as the separatrix shrinks during the spill, losses on the foil plane will monotonically decrease due to the reduction of $X_0$.

---

[25] The stepsize for Mu2e resonant extraction from the Delivery Ring is approximately 5 mm. With an effective foil plane thickness of 50 μm, $R_L \approx 1\%$.

Figure 4.45 shows results of septum foil plane loss calculation using this model. The time dependence of $a_0$ (separatrix size) was taken from Synergia2 tracking simulations [82], where the tune ramp, $\Delta\nu(t)$, had been optimized. The effective septum foil plane width was assumed to be 50 $\mu$m. The red line shows the model expectation for losses with the present Delivery Ring optics and $\beta_x = 9$ m at the ESS. The horizontal beta function at the ESS can be increased to improve extraction performance. For example, $\beta_x = 15$ m can be achieved by a local beta-bump that doubles the beta-function at the Q203 focusing quadrupole. Although this is possible, optics modifications are not currently included in the scope of the project. The plot on the bottom of Figure 4.45 shows efficiency curves for three different values of the machine acceptance: 25, 35, and 50 $\pi$ mm-mrad. The nominal acceptance of the Delivery Ring is 35 $\pi$ mm-mrad. Substantial opening of the machine aperture would be required to achieve the same effect as a local doubling of the beta-function.

### 4.6.3.1.3   Septum Beam Loss: Tracking Simulations

The details of beam transport through the extraction channel, and in particular its interaction with the septum wire/foil plane, were studied using the MARS simulation code [81]. The MARS code includes various types of particle interactions with material and the tracking of secondary particles. The model includes a geometrical description of lattice elements that can be hit by the showering particles: septa, quads, beam pipe, etc. This allows calculation of the beam losses and activation of lattice elements. Only the local losses caused by the scattering in the septum foil plane are taken into consideration. A particle is considered lost if it does not make it into the spatial and momentum aperture of the extracted beam. Those particles that are scattered back into the circulating beam area are not considered lost. However, about 0.3% of these particles do not fall within the ring acceptance. These particles will be lost eventually somewhere else in the ring.

The exact accounting of beam interaction with foils makes this simulation the most realistic method of beam loss calculation. The resulting beam losses obtained using the MARS simulation are slightly less than those from the analytical model, as expected. The detailed plots of losses will be presented in the ESS section (4.6.3.1.4) below. A full description of the beam loss simulation studies can be found in [84]. This model was also a part of the broader set of radiation studies that addressed in-tunnel shielding needs (see section 4.5.2.1).

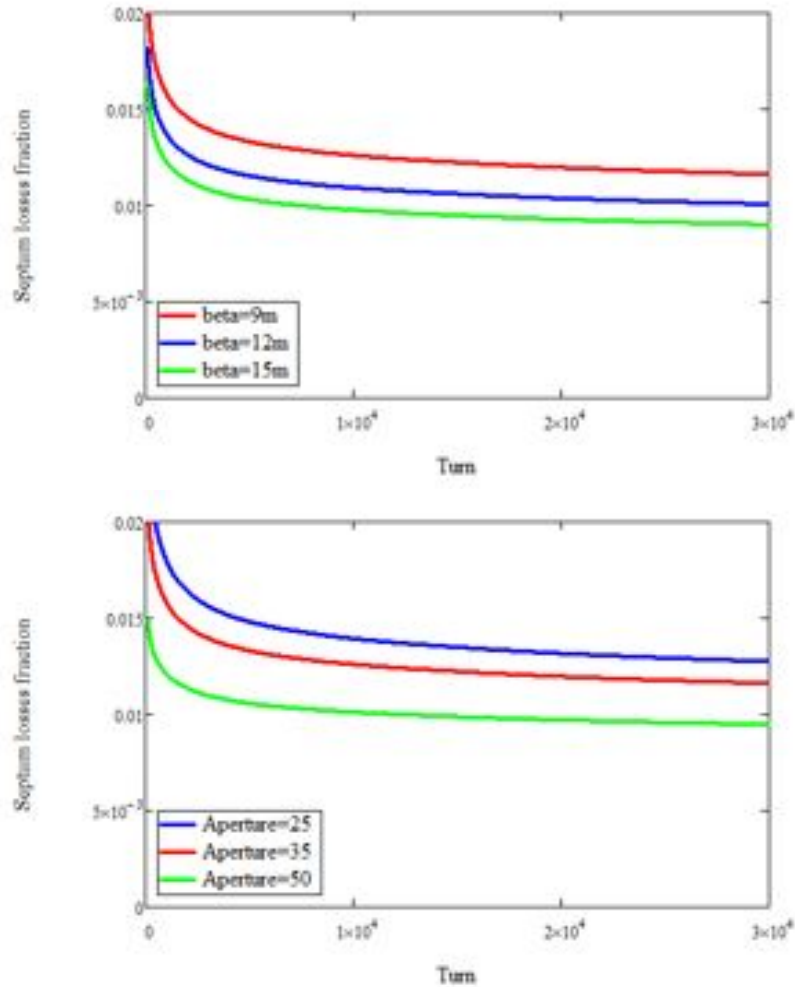

Figure 4.45. Model calculations of the loss fraction ($R_L$) as function of turn number: Top - for 3 different β-functions at the septum: 9, 12 and 15 m. Bottom - for 3 different values of the machine acceptance: 25, 35 and 50 π mm-mrad.

### 4.6.3.1.4    Extraction Tracking Simulations

A detailed study of the extraction process has been carried out with tracking simulations using the Synergia2 code [80]. The model for the Delivery Ring has been considerably improved since the Conceptual Design Report [83]. Calculations have been enhanced with realistic magnet ramping, aperture loss filters and spill feedback regulation using the RF knock-out technique (see Sec. 4.6.3.4). Operational procedures for the electrostatic septum alignment and dynamic compensation of the extracted beam angle were also included in the model. Figure 4.46 shows the footprint of the extracted beam in the phase space coordinates ($X$, $X'$) without using the dynamic bump (left) and with it (right) (the dynamic bump is discussed in section 4.6.3.3.2). The blue ellipses show the horizontal phase space ellipses that match the extracted beam with 99.9% containment. The extracted beam without the dynamic bump correction is matched to an (unnormalized) emittance of 1.1 π mm-mrad. The matching horizontal beta-function is 10 m. The dynamic bump corrected beam is matched to a 0.6 $\pi$ mm-mrad emittance and a beta-function of 18 m.

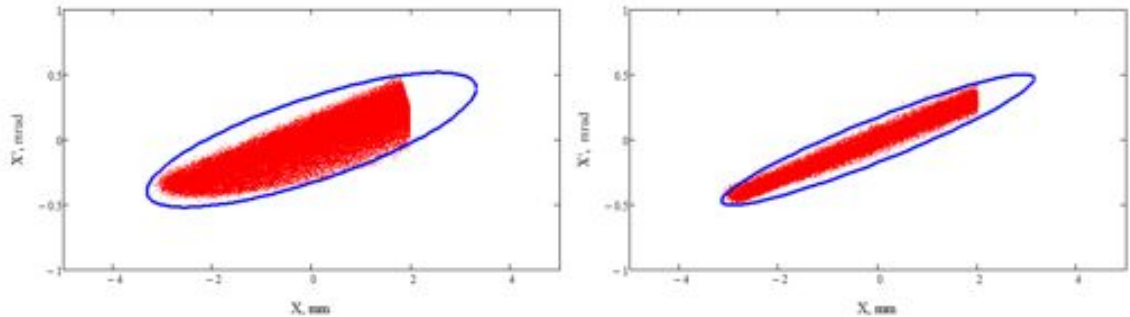

Figure 4.46. Extracted beam shapes and matching ellipses without the dynamic orbit angle correction (left) and with it (right). Matching total emittance and matching lattice functions are different.

The ultimate angle spread of the corrected extracted beam is determined by the tune spread in the circulating beam as that causes extraction from different separatrices. In our case the tune spread is dominated by the horizontal chromaticity, which is set to 1.0 to facilitate an adequate mixing of the coherent RFKO excitation (see Section 4.6.3.4). For the full description of the tracking simulation studies see Reference [82].

### 4.6.3.2   Electrostatic Septum (ESS)

Electrostatic septum design is the most complex and resource consuming element of the Mu2e Resonant Extraction subproject. The design is based on the previous experience of building and operating Electrostatic Septa (ESS) at Fermilab [85]. We also capitalize on a more recent experience with the Tevatron high voltage Electrostatic Separators.

There have been significant advances in ESS science and technology since the last Main Injector septum was manufactured at Fermilab. Moreover, rapidly growing beam intensities in a new generation of experiments and applications require more stringent limitations on beam losses and field stability. Therefore, we have undertaken a series of R&D studies and prototype assemblies aimed at achieving the best septum design in light of the recent progress in this field. As much as possible, we also use the recent experience of building and operating electrostatic septa at other labs.

Figure 4.47 shows the generic design of the ESS. The high electrostatic deflecting field region is created in the gap between the thin wire or foil plane (1) and the titanium cathode (2). Foils are kept at zero voltage and mounted on the grounded C-shaped aluminum frame (3) that overlaps with the circulating beam area. The cathode is connected to negative high voltage via feed-throughs (4) and is isolated from the rest of the structure with high voltage standoffs (not shown in Figure 4.47).

The stability of the deflecting field is very important as high voltage discharges can potentially destroy the foils. Foil breakage limits the lifetime of the septum. Broken foils must be immediately removed from the high field area to avoid short-circuiting the foil plane to the cathode. This is done with cantilevered retraction springs (5).

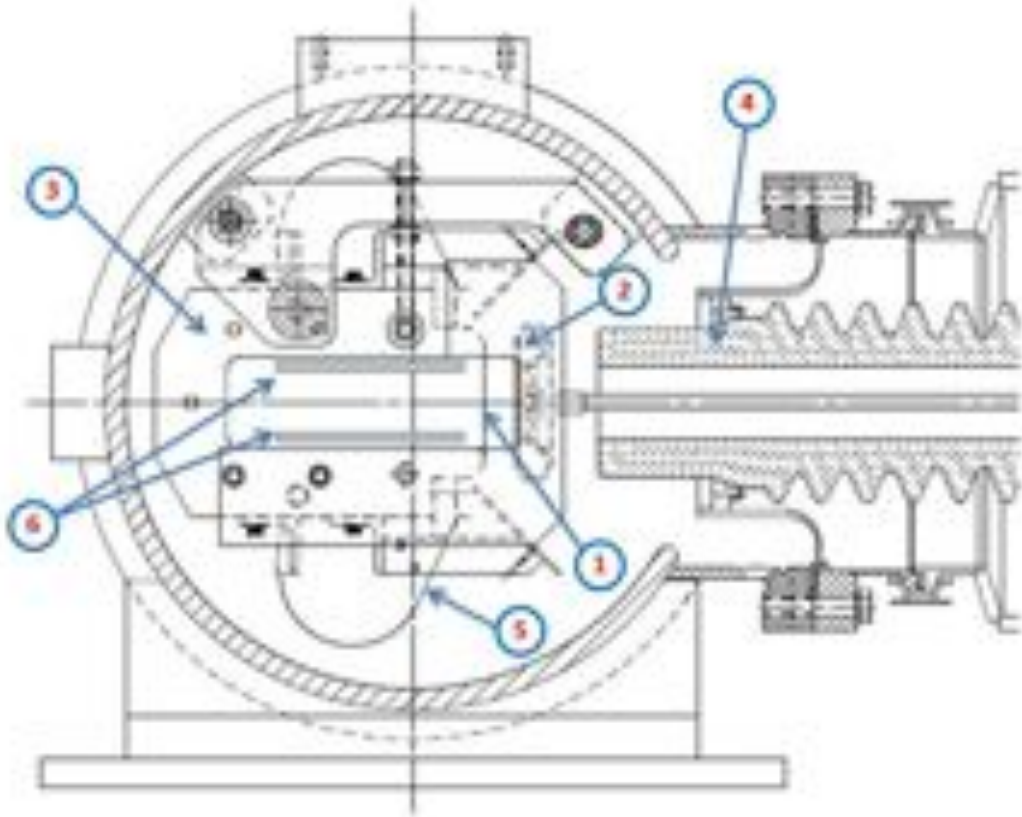

Figure 4.47. A cross section of the Fermilab Electrostatic Septum showing: ①- wire/foil plane, ②-cathode, ③-C-shaped frame, ④- high voltage feed-through, ⑤-retracting spring, ⑥-clearing electrodes.

The circulating beam is surrounded by two clearing electrodes (6) from top and bottom that create a low electric field to remove ionization in the residual gas. The flatness of the foil plane is critical to the extraction efficiency. When extraction is not necessary, the foil plane must be moved away from path of the beam. Fine alignment of the foil plane is also very important, so that both upstream and downstream sides can be moved independently with an accuracy of better than 100 μm. Movement of the foil plane can be accomplished by either placing motion devices inside the vacuum or by moving the septum vessel as a whole.

### 4.6.3.2.1   Septum geometry

The ESS consists of two modules that straddle a focusing quadrupole (Q203). This maximizes the kick efficiency in the septum. Another advantage of this geometry is that the deflection of the extracted beam in the first module creates a substantial gap between the circulating beam and the foil plane in the second module.

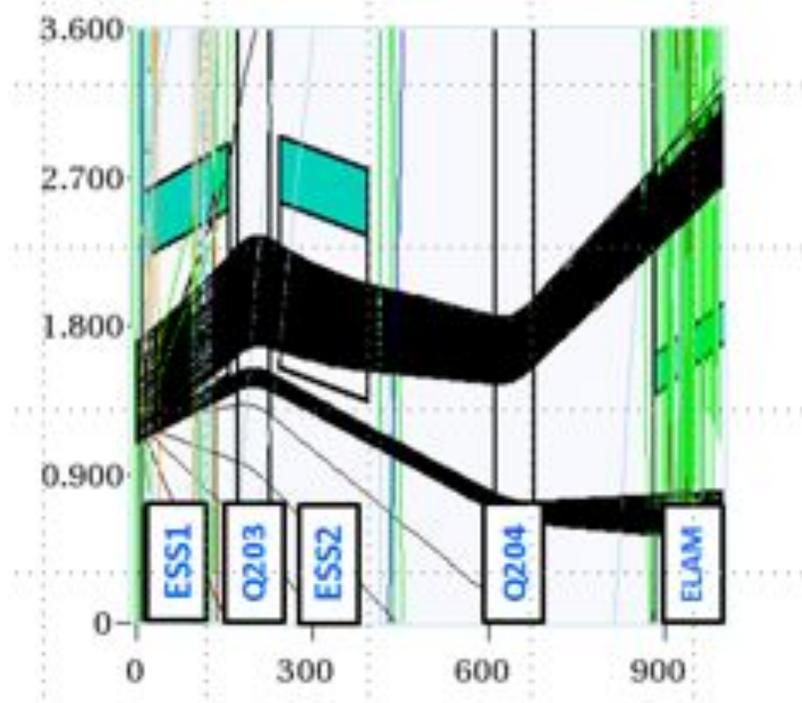

Figure 4.48. Plan view for the MARS tracking in the extraction channel. Main lattice elements are marked with labels. Only extracted beam and a small fraction of the circulating beam are shown. Black traces show protons, secondary particles are shown in other colors.

Figure 4.48 shows beam tracking with MARS code through the two electrostatic septa ESS1 and ESS2, quadrupoles Q203 and Q204, and the extraction Lambertson, ELAM. Only a small fraction of the circulating beam – beam originating close to the septum plane – is shown. The black traces correspond to protons; the colored traces represent secondary particles. Circulating beam and extracted beam start to separate in ESS1 due to deflection in the electrostatic field. This separation becomes substantial in ESS2, and consequently there are practically no losses there. The circulating beam in ESS1 remains close to the foil plane, continuing to contribute to losses due to its angular spread. This component of the losses can be minimized by reducing the length of ESS1 and increasing the ESS2 length so that the total kick remains unchanged. Figure 4.49 shows losses as a function of beam angle misalignment for different ESS1/ESS2 length combinations: the red points correspond to equal lengths, $L1/L2 = 1.5\text{m} / 1.5 \text{ m}$; the green points, to 1.0 m/2.0 m; the blue points to 0.7 m/2.3 m and the purple to 0.5 m/2.5 m. There is little difference in beam loss in the vicinity of the optimal beam angle for the different septa lengths. The losses increase for longer ESS1 lengths at larger angles. Therefore, reducing $L1$ gives more tolerance for the beam angle spread. There is little difference across the whole range between the 1 m and 0.7 m sets. From both performance and fabrication standpoints, it becomes impractical to increase the length of the ESS2 any further. Indeed, as the beam gap becomes smaller in ESS2 it becomes more difficult to maintain its alignment in this gap. At larger angles, the beam starts crossing the ESS2 foil plane and losses grow as manifested in the 4[th] set of points in Figure 4.49.

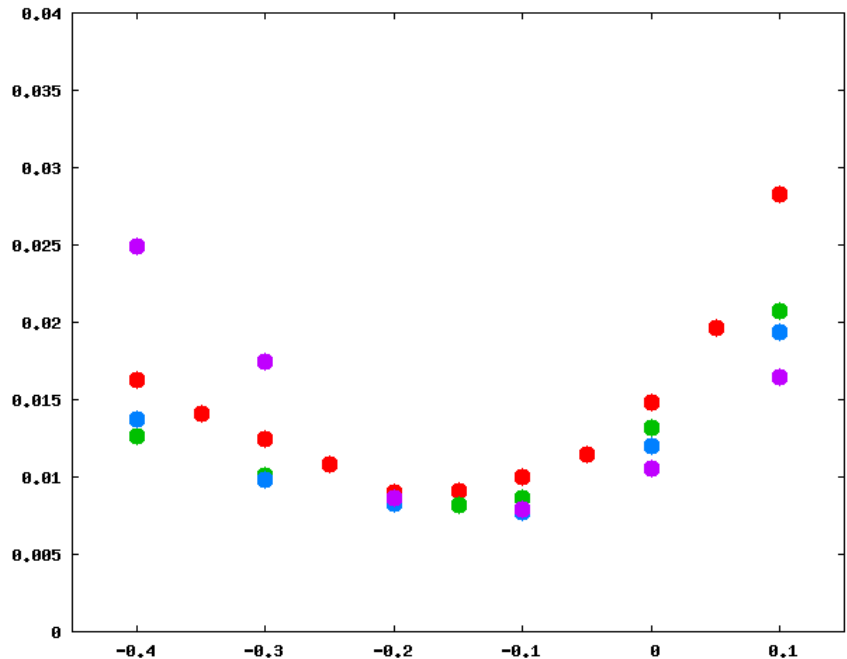

Figure 4.49. Total septum loss fraction versus beam incident angle misalignment for various ESS1 lengths: Red points: $L1 = 1.5$ m; Green – $L1 = 1.0$ m; Blue- $L1 = 0.7$ m; Purple- $L1 = 0.5$ m. The combined length of both septa is preserved: $L1 + L2 = 3$ m.

As there is no advantage of $L1 = 0.7$ m over $L1 = 1.0$ m, the latter is the preferred choice due to the physical and technical challenges of building and operating a longer ESS2.

Note that the optimal angle for all cases is different from zero. The reason for this is the asymmetry of beam conditions on different sides of the septum plane. For beam that is parallel to the foil plane in the field free region, particles have a chance to cross the septum due to the angular spread in the beam. This is not the case on the other side of the foil plane: particles are quickly deflected away from the plane by the strong electrostatic field. Therefore the fraction of particles crossing the plane can be reduced by aligning the beam slightly away from the septum plane in the field free region.

### 4.6.3.2.2 Septum foil choice

The most significant difference between the Mu2e ESS design and previous designs at Fermilab, is switching from a multi-wire septum plane to one made of thin foil strips. A wire plane minimizes the amount of material exposed to the beam, but wires are extremely delicate and are more difficult to handle than foil. Most importantly, wires are prone to breakdowns due to high voltage discharges. Foils are more stable mechanically and, therefore, should be preferred whenever possible. Moreover, it is not practical to use wires of less than 100 μm diameter due to their short lifetime. Tungsten and Molybdenum foils are readily available from industry in 50 μm and 25 μm thicknesses.

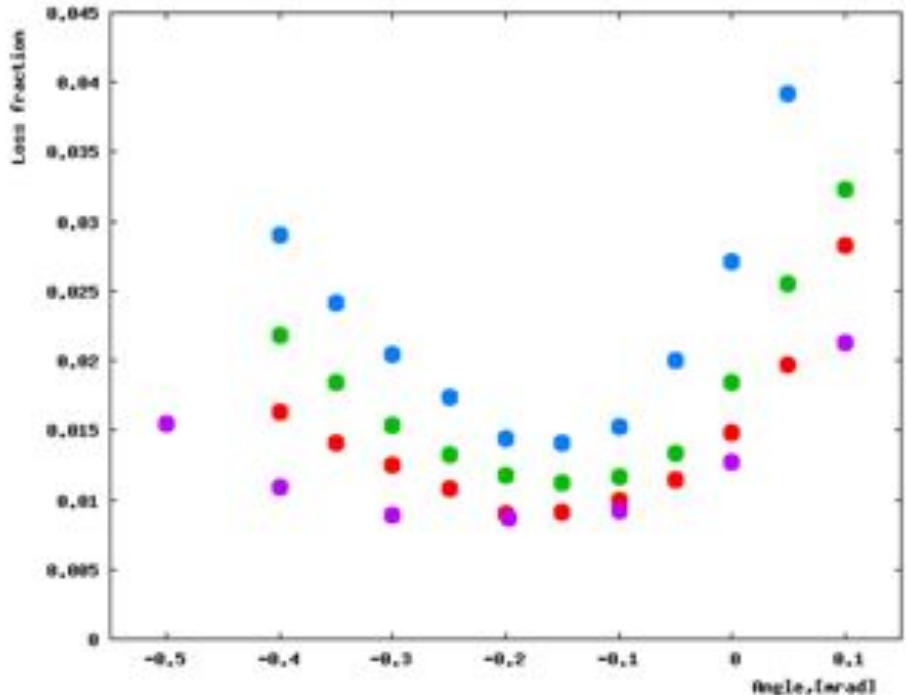

Figure 4.50. Fraction of beam lost in the septum plane versus the incident angle of the beam with respect to the septum plane. Red points: 50 μm W foil strips plane (2mm/4mm); Green points: 50 μm W solid foil plane; Blue points: 50 μm "black hole" plane; Purple points: 100 μm W wire plane with 2 mm spacing.

The implications of the septum plane material composition for beam loss have been studied with the extraction channel MARS model [84]. Figure 4.50 shows a conservative comparison of losses in various septum materials. The blue markers show fractional losses in a 50 μm thick layer of "black hole" – imaginary material that absorbs everything incident on it. Green points represent a solid tungsten plane of 50 μm thickness. Tungsten foil ribbons, 50 μm thick and 2 mm wide, spaced by 4 mm center-to-center are shown in red. The beam loss performance of 50 μm tungsten foils is very similar to that of 100 μm diameter tungsten wire (purple) at small angles. Therefore wires do not have an apparent advantage if the beam angle is well controlled.

### 4.6.3.2.3    Diffuser plane

According to Figure 4.49 and Figure 4.50, the best condition for the lowest septum losses is a narrow beam angular spread and correct alignment of the septum to the beam angle. In this case, losses are minimal and are dominated by those particles that travel a substantial distance through the foils. These losses can be reduced further with pre-scattering in light foils or wires in front of ESS1. If these pre-scattering foils are aligned with the foils of the septum, they will affect selectively only the small fraction of beam particles that are the most susceptible to becoming lost. The length of the diffuser must be sufficient to allow weakly scattered particles to gain sufficient deviation from their original path and miss the foils of the main septum. The particles that are scattered in the diffuser will then pass through the main septum area without scattering; either with the circulating beam or in the extraction channel where they will receive the full kick of the septum and become cleanly extracted.

Figure 4.51 shows an angle scan made in the MARS tracking simulation for 50 μm thick tungsten foils (2 mm width/4 mm pitch) with a 30 cm long diffuser (green points) and with no pre-scattering (red). Pre-scattering is the most effective for head-on beam that is perfectly aligned with the septum plane. Therefore the effect of the diffuser is reduced because of the non-zero optimal beam angle discussed above. Nevertheless, there is an approximate 25% reduction of losses for a narrow range of small angles. As seen in Figure 4.51, this method does not give any advantage at large angles; good alignment and narrow angular spread in the beam are required to take full advantage of pre-scattering.

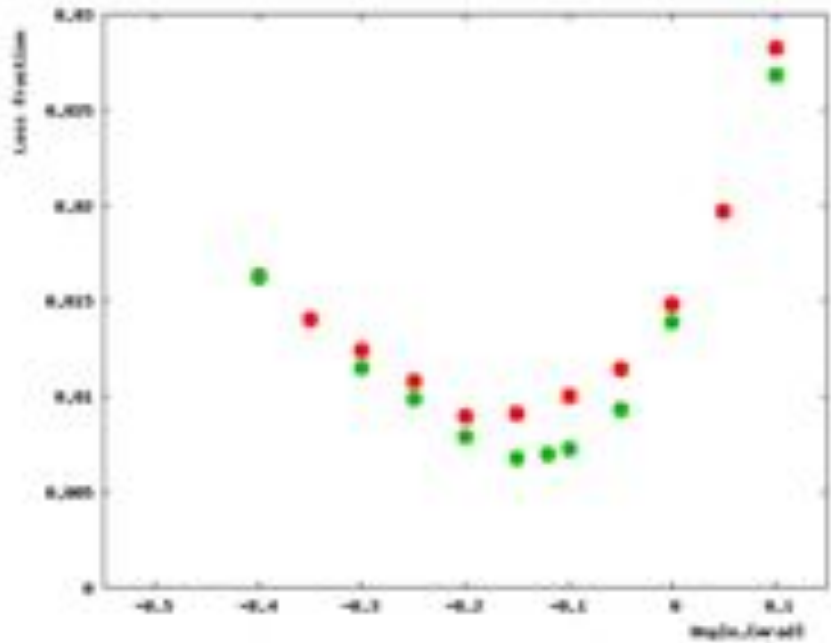

Figure 4.51. Fraction of beam losses in the 50 μm tungsten thick foil strip septum (2mm wide, 4mm pitch) with (green points) and without (red points) a 30cm long carbon foil strip diffuser in front of the septum plane.

### 4.6.3.2.4    Foil width and spacing choice

The dimensions of the foil strips are determined primarily to minimize the material in the path of the beam while keeping them strong enough to maintain stability and durability in an environment of strong mechanical and electrical forces. The possible foil shapes and dimensions are also constrained by the variety of products currently offered by industry. A width of 1 mm appears to be a good starting point as foils of different materials with this width are commonly available on the market in a wide range of thicknesses.

The amount of beam loss due to the beam interaction with the septum plane depends on the effective septum plane thickness, which is determined not only by the foil thickness itself, but also by its flatness, foil sag in the electrostatic field, curling, and tilting of the strips. To study such effects and the mechanical properties of foils, we are building a mechanical testing bench that we call the Prototype-I (see section 4.6.3.2.5 below).

Foil spacing is constrained by the depth of the electrostatic field penetration into the circulating beam region.  If the field penetration extends deep into the circulating beam area and reaches a region of high beam density, the positive ions that reside there by virtue of beam interactions with the residual gas will be directed into the field region by the penetrating field. The presence of ions in large quantities in the vicinity of the foils causes electric discharges between the cathode and the foil plane, which is the principal determiner of the maximum achievable electric field strength.

ANSYS v14 3D electrostatic field calculations were used to determine the field map and optimize the foil spacing. The ion clearing field formed by the electrodes above and below the circulating beam, held at constant voltages of -8 kV and -5 kV, were also included in these calculations.

Table 4.19 shows the depth of the high field penetration with the regular foil structure. Penetration is fairly shallow for center-to-center spacing below 4 mm. But we would like to take one further step and model the system with one missing foil.  This can help avoid the situation where a foil failure would result in creating discharge conditions that could lead to a cascade of failures at adjacent foils. Table 4.20 shows the distance of penetration in the case of one missing foil.

Table 4.19. Field penetration for a fully intact foil plane.

| Foil Spacing (mm, center-center) | Gap Size (mm) | Extension of cathode field into circulating beam region (mm) |
|---|---|---|
| 2 | 1 | 1.1 |
| 2.5 | 1.5 | 1.4 |
| 3 | 2 | 1.8 |
| 4 | 3 | 2.9 |

Table 4.20. Field penetration for a foil plane with a single missing foil.

| Foil Spacing (mm, center-center) | Gap Size (mm) | Extension of cathode field into circulating beam region with missing foil (mm) |
|---|---|---|
| 2 | 3 | 4.3 |
| 2.5 | 4 | 6.3 |
| 3 | 5 | 11.8 |
| 4 | 7 | 16.9 |

Based on these calculations, a center-to-center foil spacing of 2 - 2.5 mm should be chosen. Results shown in the tables are illustrated in Figure 4.52 and Figure 4.53, which show the field intensity at different distances away from the foil plane into the circulating beam region using two different clearing electrode voltages. Positive field strength pulls ions towards the cathode. When the field is negative, ions would stay in the circulating beam region and be eventually trapped in the clearing electrodes.

As the voltage on the clearing electrodes is increased, the high field penetration depth is decreased. When voltages on the clearing electrodes are substantially reduced, as seen in Figure 4.53, the high field region will penetrate a significant distance into the circulating beam region between the foils in the fully intact region when a foil is missing. This effect is best seen at 0.025 m and 0.075 m.

#### 4.6.3.2.5   ESS Prototypes
A first stage prototype of the foil plane frame has been constructed to study foil mechanical properties, plane flatness, and assembly methods. These studies are in progress. Foil alignment is measured with an optical profilometer that provides surface position measurements with submicron accuracy both across and along the ribbon directions. A view of the prototype at the optical stand is shown in Figure 4.54.

A second stage prototype (Prototype-II) is also in preparation. To save on the prototype cost, an existing Tevatron horizontal electrostatic separator will be reconfigured into an electrostatic septum. A cross section of the separator at the high voltage feed-throughs is shown in Figure 4.55. There are two ion pumps mounted on the vacuum chamber that will be reused.

The prototype will be connected to a high voltage power supply and tested under vacuum. Cathode conditioning techniques can be evaluated as well as the vacuum performance of the system. The test aims to achieve stable operation at -150 kV and higher with a 15 mm gap between the septum plane and the cathode. This setup can also be used to test the design of the septum moving mechanism in vacuum.

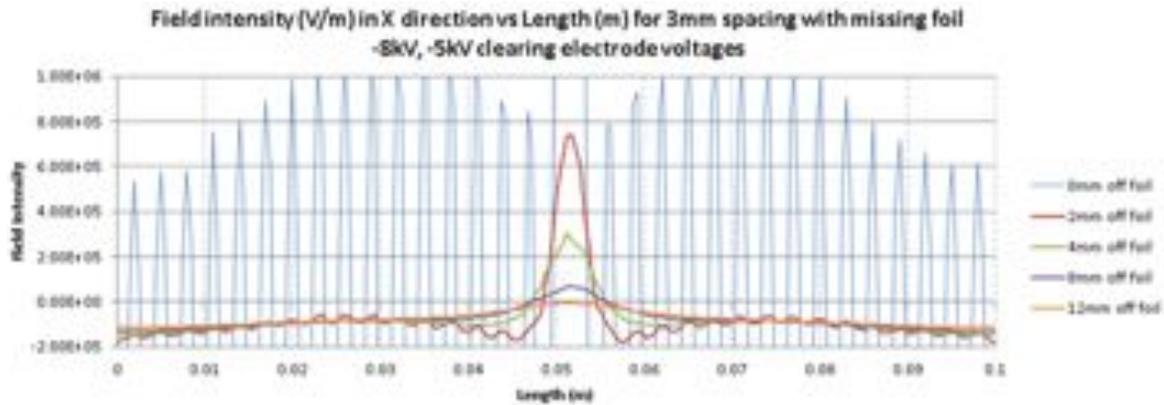

Figure 4.52. Field intensity for various distances off the foil plane with -8 kV/-5 kV clearing electrode voltages. The horizontal axis shows the position along the beam direction in a small region centered on a missing foil.

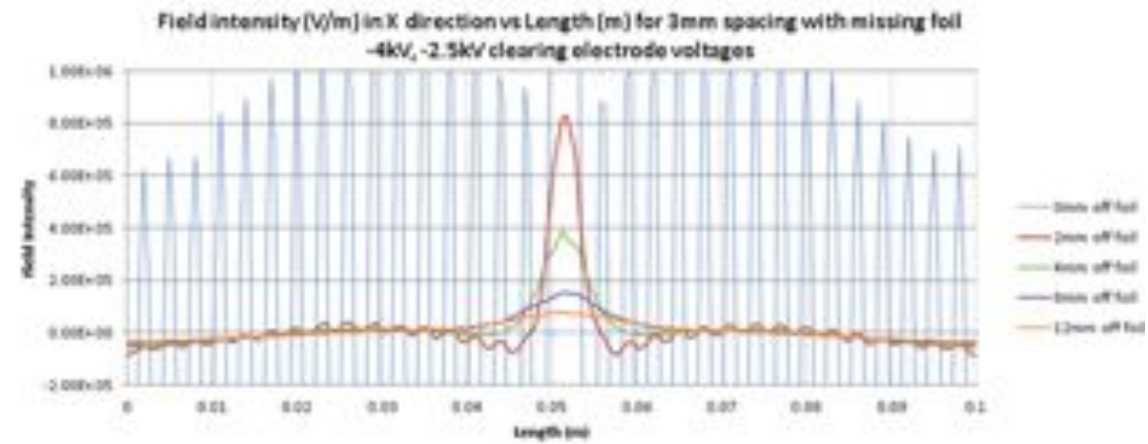

Figure 4.53. Field intensity for various distances off the foil plane with -4 kV/-2.5 kV clearing electrode voltages. The horizontal axis shows the position along the beam direction in a small region centered on a missing foil.

### 4.6.3.3  Ramped magnets and power supplies

Figure 4.43 shows the placement of the magnets that support slow extraction. There are three types: sextupoles to excite the resonance, quadrupoles to ramp the tune, and trim dipoles to create a dynamic bump. We shall discuss each of them separately. There is one common characteristic of all three types. All magnets are capable of supporting fast field changes through the spill and, in particular, returning to their initial field setting within a short reset time before start of the next spill. Table 4.21 summarizes the main specifications for all magnet types. Numerical analysis has been performed to study possible effects of the AC field on mechanical stresses, field quality degradation and residual field in the end of the spill. Results of these studies, reported in [86], confirm that the magnet choices are suitable for the designed mode of operation.

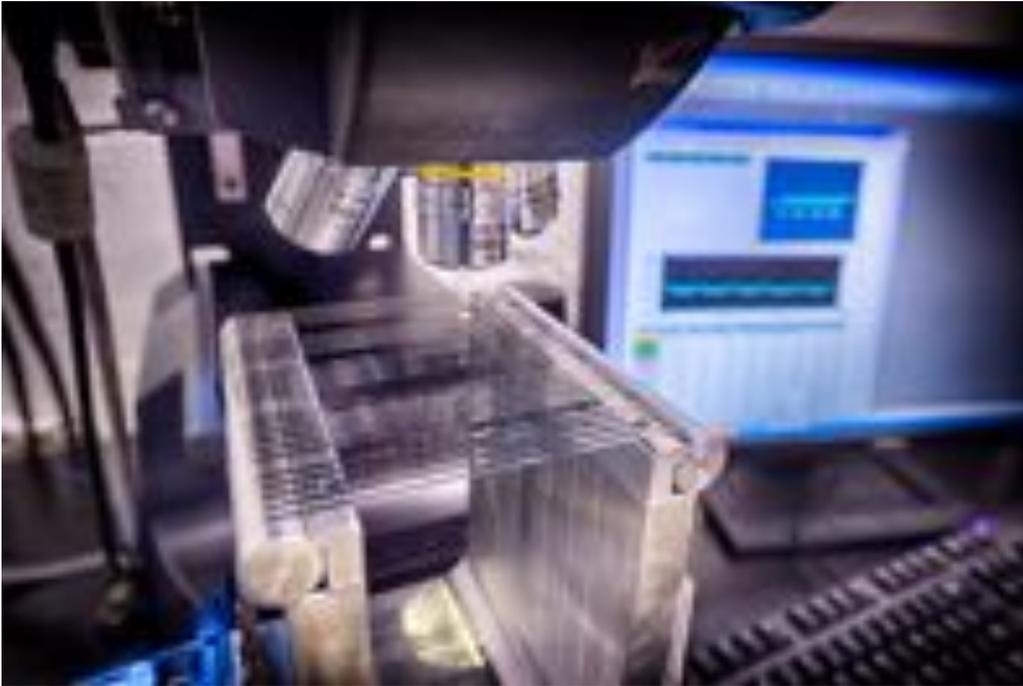

Figure 4.54. First stage prototype at the optical test stand. The stand provides surface elevation measurements along and across ribbons at sub-micron accuracy.

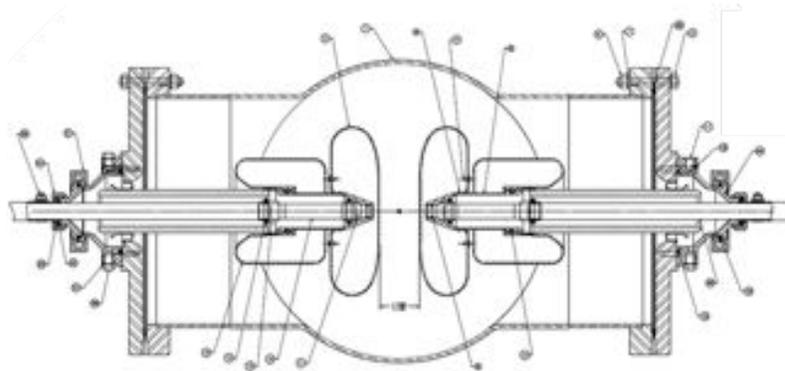

Figure 4.55. Tevatron horizontal separator cross-section at high voltage feed-throughs.

### 4.6.3.3.1   Tune squeeze magnets

Moving the machine tune to the resonance point during the spill and then resetting it to its $t = 0$ value is done by ramping a circuit of three fast quadrupole magnets. These quadrupoles are located near the middle of each straight section of the Delivery Ring. Due to their large separation, each magnet has its own power supply. The magnet specifications in Table 4.21 are satisfied by using the former Cooling Ring's CQA type quadrupoles. These magnets are available in storage and minimal refurbishment is needed to prepare them for operation. The tune ramp profile is selected to provide a uniform spill rate. The integrated quadrupole gradient dependence on time can be calculated from that curve according to

$$GL(t) = \frac{4\pi \cdot B\rho}{3\beta} \cdot \Delta\nu(t) \qquad (4\text{-}9)$$

where $B\rho$ is the beam rigidity, and $\beta$ is the horizontal beta-function at the magnet locations. The gradient is calculated using the $\Delta\nu(t)$ obtained from the Synergia2 tracking simulations. The resulting gradient curve is shown in Figure 4.56. The main specifications for the magnet power supplies are shown in Table 4.22. Full details of the power supply studies can be found in Ref. [87].

Table 4.21. Specifications for the Resonant Extraction magnets.

| Magnet | Quantity | Base* Integrated gradient | Integrated gradient excursion | Base* Current [A] | Max Current Excursion [A] | Max supply Current [A] | Max $dI/dt$ [A/sec] |
|---|---|---|---|---|---|---|---|
| Sextupole (ISA) | 6 | 80 T/m | 32 T/m | 200 | ±80 | 300 | 16,000 |
| Tune Quad (CQA) | 3 | 0 | 0.2 T | 0 | 80 | 100 | 16,000 |
| DEX Trim (NDB) | 4 | ND | 0.014 Tm | ND | 14 | ±40 | 1,400 |

*Base values are assumed at $t = 0$

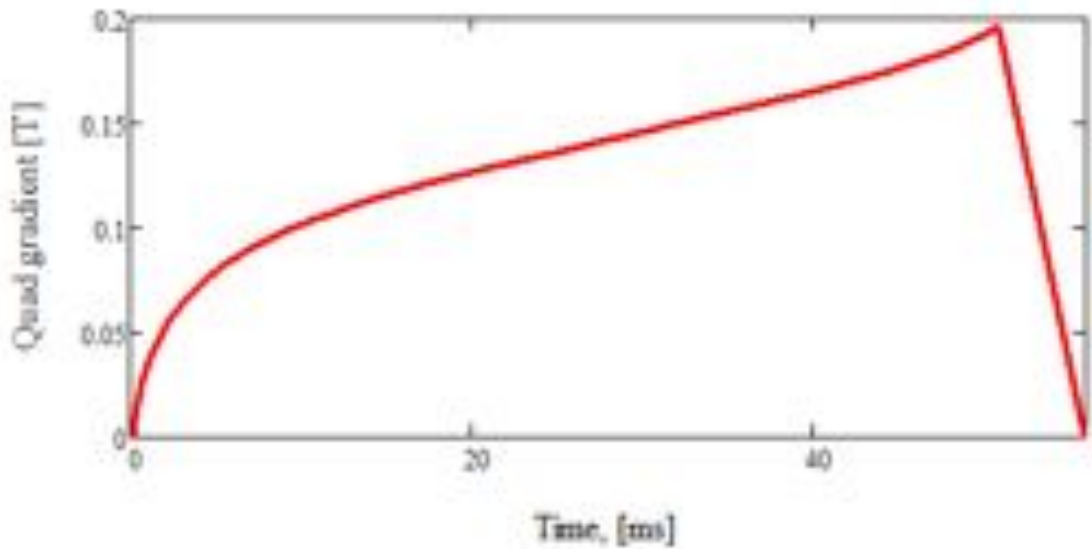

Figure 4.56. Excitation curve for the tune ramping quads calculated in the Synergia2 simulations.

Water cooling is not required for continuous operation in this mode. Due to the very sharp slopes at the beginning and end of the ramp, substantial driving voltage is required from the power supplies. The power supply choice is based on the booster switchers. Two 65 A units will be connected in parallel to make a $160-180$ V switcher with a 130 A maximum.

Each loop will need of order 1.7 kW. At 100 A the *IR* voltage will be ~17 V, leaving ~160 V for the necessary $L \cdot dI/dt$, allowing for a maximum $dI/dt$ of 74,000 A/sec at a peak current of 100 A.

Table 4.22. Specifications for the Resonant Extraction magnet power supplies.

| Magnet | N | $I_{max}$, [A] | I_base, [A] | $\Delta I$, [A] | Max $dI/dt$ [A/sec] | Regulation accuracy, % | Regulation stability | Curve accuracy | Ripple, % |
|---|---|---|---|---|---|---|---|---|---|
| Sextupole (ISA) | 6 | 300 | 200 | $\pm 80$ | 16,000 | <0.5% | <0.5% | <0.5% | <1% |
| Tune Quad (CQA) | 3 | 100 | 0 | 80 | 16,000 | <0.5% | <0.5% | <0.5% | <0.05% |
| DEX Trim (NDB) | 4 | $\pm 40$ | ND | $\pm 14$ | 2,800 | <1% | <1% | <1% | <1% |

### 4.6.3.3.2  Dynamic bump

Figure 4.57 illustrates the simulation's beam phase space outside the separatrix in normalized coordinates $(x, \alpha x' + \beta x')$ at different times in the spill cycle. As the separatrix is squeezed, the beam angle ($x'$) at the septum (foil) plane position is changing.

This results in the growth of losses at the septum. To keep beam at the optimal angle with the septum plane throughout the spill, an orbit correction can be dynamically applied using a local orbit angle bump, henceforth referred to as the Dynamic Bump or the Dynamic Extraction Bump (DEX bump). The required 4-bump can be built with four correctors positioned as shown in Figure 4.43. The orbit angle correction is zero at the beginning of extraction and reaches its maximum value of 0.4 mrad at the end of the spill. The DEX bum ramp is shown in Figure 4.58.

Debuncher style corrector magnets (NDB) will be used for the DEX bump. The plot shows the magnet current needed for the standard configuration. Due to a very fast ramp-down at the end of the cycle and high magnet inductance (0.4 H), the voltage needed to drive this curve is substantial. The Booster switch mode power supplies will be used for these magnets. The two magnet coils will be powered in parallel. In this case the total current needed will increase but the driving voltage will decrease by a factor of two. A standard 65 A unit will be upgraded by installing higher voltage FET's and filter components to produce a unit that is capable of 350 V output. Each magnet will need of order 1.0 kW. Thus, the four magnets can be supported from a single 350 V bulk power supply. At 40 A the *IR* voltage will be ~22 V, leaving ~328 V for the required *L·dI/dt,* allowing for a maximum *dI/dt* of 3,120 A/sec at a peak current of 40 A.

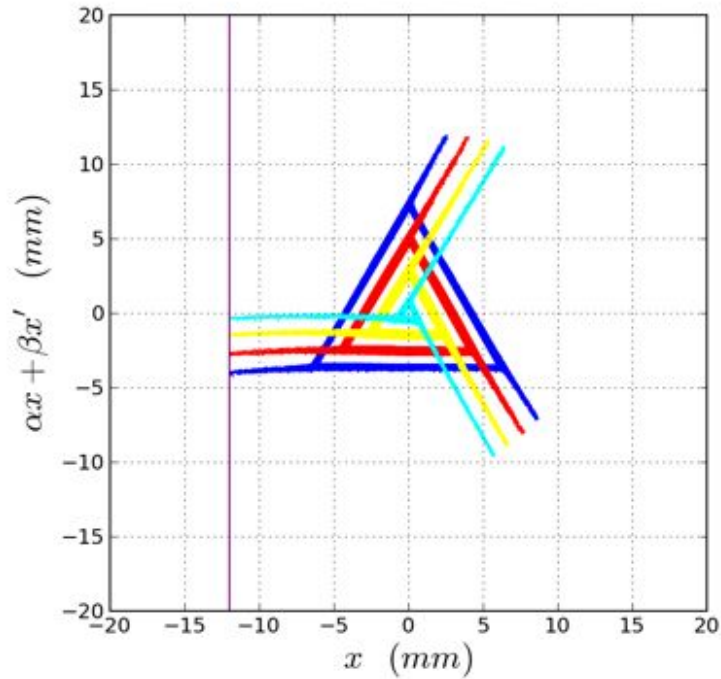

Figure 4.57. Path of unstable particles in the phase space of normalized coordinates at four different excitations of the tune squeeze quads during an extraction ramp as seen at the upstream end of ESS1. The change in ordinate value at which a path crosses the septum foil plane (vertical line) indicates a change of *x′* for particles entering ESS1 as the spill progresses.

### 4.6.3.3.3 Sextupole magnets

The strength of the resonance coupling constant *g*, defined in Equation (4-6), is controlled by two families of harmonic sextupole magnets. These sextupoles will be combined into two groups of three magnets in two of the Delivery Ring straight sections, as shown in Figure 4.43. These magnets are located at the main focusing quads, so the 60º phase FODO cell advance makes a convenient summation for the third harmonic strength over these locations. The phase contribution of each circuit to the coupling constant differs by about 90º, which is also very convenient for the phase adjustments.

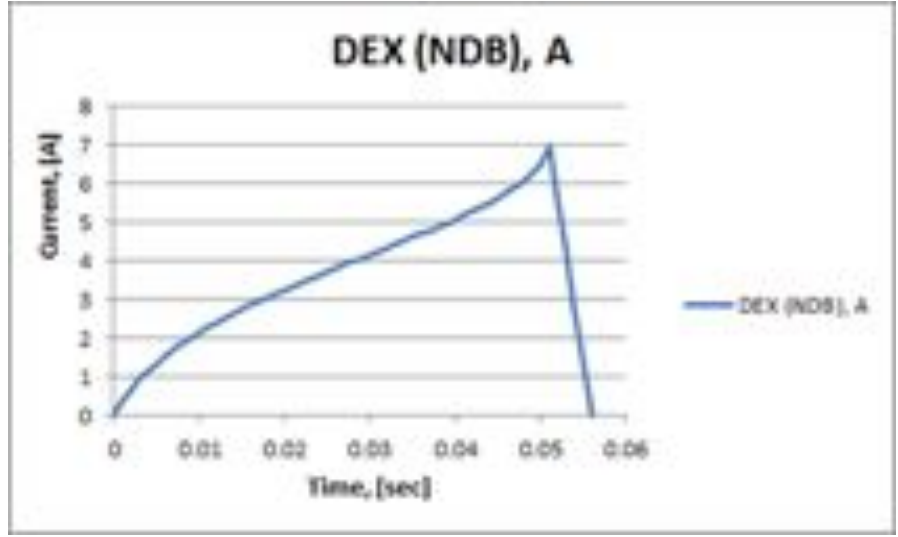

Figure 4.58. Current in the DEX bump corrector with the largest strength as function of spill time.

The required parameters for these sextupoles are satisfied with an ISA-type magnet, used in the Fermilab Main Injector. No spare magnets of this type are readily available, so these magnets will be built at Fermilab using existing designs and tooling. Each of the two sextupole groups is powered by separate supplies. In the minimal design scenario, sextupole magnets will be run in DC mode, providing a constant field. However more complex modes of operation may become desirable to improve slow extraction performance. For example, if the dynamic bump turns out to be destructive due to aperture limitations, an alternative solution would be to compensate the beam angle at the septum by rotating the separatrix: i.e. adjusting the phase of the sextupole harmonic strength, making the angle $\phi_0$ (in Figure 4.44) variable during the spill. Figure 4.59 shows the current curves calculated for each of the two sextupole circuits to provide this separatrix rotation and constant beam angle at the septum plane during the spill.

We would like to retain the ramping capability in the design of the magnets and power supplies. To accomplish this, magnet cores must be fabricated with laminated steel. Ramp curves shown in Figure 4.59 are representative of a wide range of applications and they have been used in both magnet and power supply studies in references [86] and [87]. The power supply design is based on the existing Booster trim 65 A switch mode supply. Six 65 A units will be connected in parallel to make a $160 - 180$ V switcher that provides a 300 A maximum current. Each circuit will need $2.8 - 3.0$ kW of power. At 257 A the $IR$ voltage will be 18 V, leaving ~150 V for the required $L \cdot dI/dt$, allowing for a maximum $dI/dt$ of 19,200 A/sec at a peak current of 300 A.

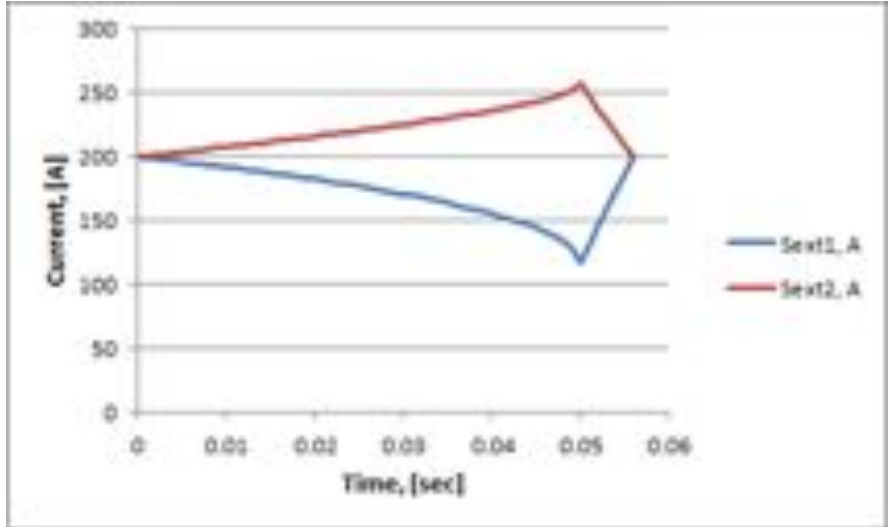

Figure 4.59. Sextupole ramp for the dynamic correction of the separatrix angle.

### *4.6.3.4  RF Knock-Out*

It is desirable that event rates in the Mu2e detector systems be uniform during the spill. Therefore, it is important to maintain the spill rate as constant as possible. The bulk of this requirement is satisfied by implementing the feed forward curve of the quadrupole circuit driving the machine tune to the resonance point. The ramp curve depends on many factors that are difficult to calculate, therefore it will be determined empirically through learning algorithms built into the regulation logic. This will provide a coarse regulation that can flatten the spill rate on average. However, there could be beam conditions and fast variations of the machine, such as machine tune ripples that cannot be regulated by these magnets due to their relatively slow response. A faster means of regulation, up to 1 kHz, is offered by the RFKO technique. This technique was recently proposed for use in medical applications in Ref. [88], [89] and since then has become an important tool for slow spill intensity control. This method uses the transverse heating of the beam as a way to control the beam phase space density at the separatrix boundary and, therefore, the extraction rate. Horizontal excitation of the beam in the Delivery Ring will be implemented with an old Tevatron style damper kicker, the main parameters of which are summarized in Table 4.23.

Table 4.23. Parameters of the RFKO kicker

| Parameter | Value |
|---|---|
| Strip line length | 1.4 m |
| Gap | 6.35 cm |
| Max power | 1 kW |
| Max kick angle | 0.8 μrad |

The kicker signal is FM modulated within a frequency range covering the beam tune spread. Different kinds of modulation strategies were studied in simulations [90]. Simulations show that there must be a non-zero chromatic tune spread (in the beam in addition to the space charge tune spread) to effectively transform (mix) beam coherent excitations into horizontal beam heating. However, large chromaticity leads to extraction coupling with the momentum distribution and requires a larger FM bandwidth that slows down the heating process. Another negative consequence of a large tune spread is that particles of differing energy in a non-monochromatic beam are extracted from different separatrices, which increases the effective angular spread of the extracted beam resulting in greater septum losses. Chromaticity will have to be carefully optimized to provide adequate mixing while keeping the negative effects small.

The RFKO high level electronics will be implemented with Amplifier Research solid state power amplifiers at up to 1 kW power. AM and FM modulation for these amplifiers will be done by the RFKO control logic, which is an essential part of the spill regulation system that will be discussed in Section 4.6.3.6.

### 4.6.3.5   Spill monitoring

Spill monitoring will be accomplished by a Wall Current Monitor (WCM) located in the external beamline.  A WCM was chosen because it is a passive device that detects the beam structure through its image current rather than direct interaction.  This design choice was made as part of the overall effort to minimize beam loss in the resonant extraction process.

A prototype of the WCM has been built and tested, and can be installed as a working unit with no further fabrication cost. The WCM utilizes a widely used technique for measuring pulsed beam intensities. Due to the very low intensities of the extracted beam ($3 \times 10^7$ protons per pulse) special consideration is required to increase its impedance and improve its sensitivity at low frequencies.

The WCM consists of a beam pipe gapped by a ceramic break. Resistors are placed across the gap to generate a voltage when the beam image current reaches the gap. The beam pipe-gap combination is surrounded by magnetic material primarily to dampen the current that is sourced from the ceramic gap. In addition, a stainless steel shell encloses the ceramic gap and magnetic cores to contain the stray fields. Figure 4.60 shows the WCM prototype as built.

The electrical model of the wall current monitor is a parallel RLC circuit that can be divided into two frequency regimes as follows. The high frequency response is dominated by the gap capacitance and the total equivalent resistance of the resistors across the ceramic gap. The low frequency response is dominated by the inductance of the cores in parallel with the resistors across the gap. It is important in our case that the flat part of the frequency response extends to the low frequency region down to the beam revolution frequency of approximately 590 kHz. The microwave model of the WCM was developed and studied with Microwave Studio software before designing the prototype.

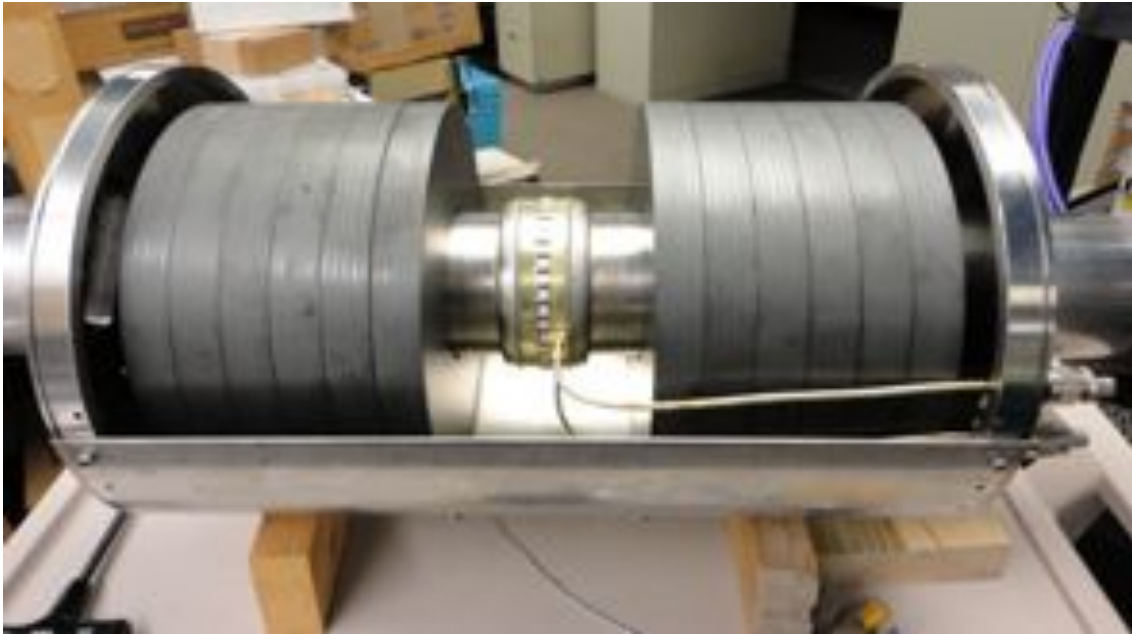

Figure 4.60. The WCM prototype on the test bench. The gap is surrounded by a stainless steel shell and 12 ferrite cores.

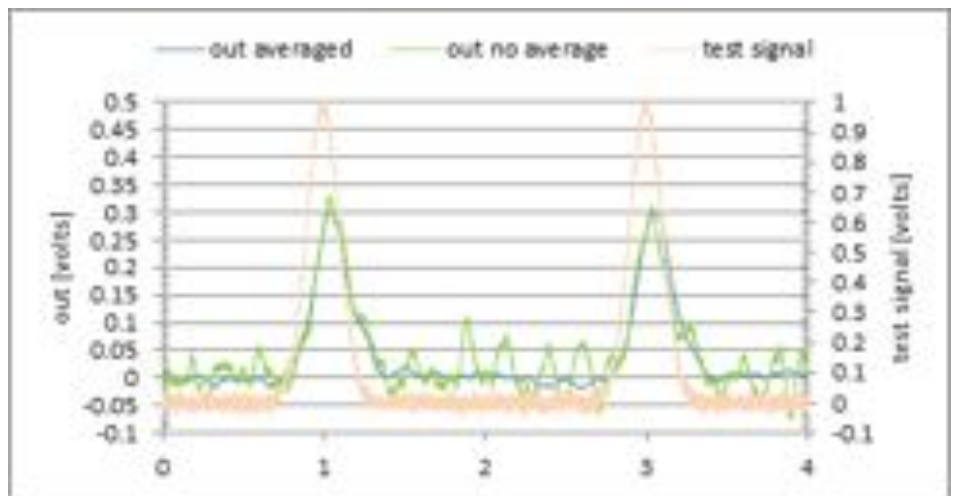

Figure 4.61. Signal response to 1 μA peak current pulses. Horizontal scale is time in μsec. The green trace is a single shot and blue trace represents averaging over 16 samples.

Based in the positive results of the simulations, a prototype was built and tested. The detailed results of the simulations and prototype testing are described in Reference [91].

Figure 4.61 shows the signal response of the prototype with an 80 dB amplifier to short pulses of 1 μA peak current separated by 2 μsec time periods. The green trace shows a single shot response, the blue trace is the response averaged over 16 samples and the yellow trace shows the input signal in Volts.

Although these lab tests represent ideal conditions and the noise floor in the tunnel environment is much higher, the test signal used was an order of magnitude smaller than the real bunch signal, so there is a good confidence that this design will satisfy the requirements of the Spill Monitor.

### *4.6.3.6  Spill regulation*

Spill rate error is defined as the difference between the actual spill rate measured by the Spill Monitor (SM) and the desired (ideal) spill rate. Spill regulation is achieved by feeding the spill error signal through the adaptive learning filter back to the RFKO kicker and the ramping quad circuit to correct the spill rate. The SM signal is amplified with a 60-80 dB amplifier in the tunnel. This output is sent to an analog signal conditioner that scales the signal to use the full dynamic range of a 14-bit ADC. The ADC board is a VME64X digitizer that has 4 channels of 125 MHz 14-bit ADC and 4 DAC channels. It also has an FPGA programmed to perform signal processing on each of the ADC channels. The signal processing components are an NCO, Cascaded Integral Comb (CIC) filter and Finite Impulse Response (FIR) filter. The output is an In Phase (I) and a Quadrature (Q) pair used for amplitude and phase computation of the processed signal. Before the spill signal is down-converted, a few cycles must be averaged to improve the signal-to-noise ratio. This operation is also performed in the FPGA.

The rest of the logic resides in the crate controller, implemented in the form of an MVME5500 processor running VXWORKS real-time operating system. The principal elements of the regulation system include:

- FM modulation.

- Amplitude Regulation of the FM signal using feedback.

- The difference between ideal (constant) spill and the measured spill rate.

- Integration of the spill rate after down-conversion to compare it with the ideal circulating beam intensity in the Delivery Ring. This portion could be used to regulate the quad circuit.

- Cycle-to-cycle learning algorithm based on adaptive learning whose output is summed with the feedback loop. Its inputs are the ideal beam intensity and the integration of the spill rate measured by the wall current monitor.

- The Amplitude regulation index value is set by the feedback and learning portions of the regulation loop.

- The learning loops are based on adaptive Least Mean Square (LMS) and Finite Impulse Response (FIR) filters whose outputs are summed with feedback loops.

The block diagram of the regulation logic is shown in Figure 4.62.

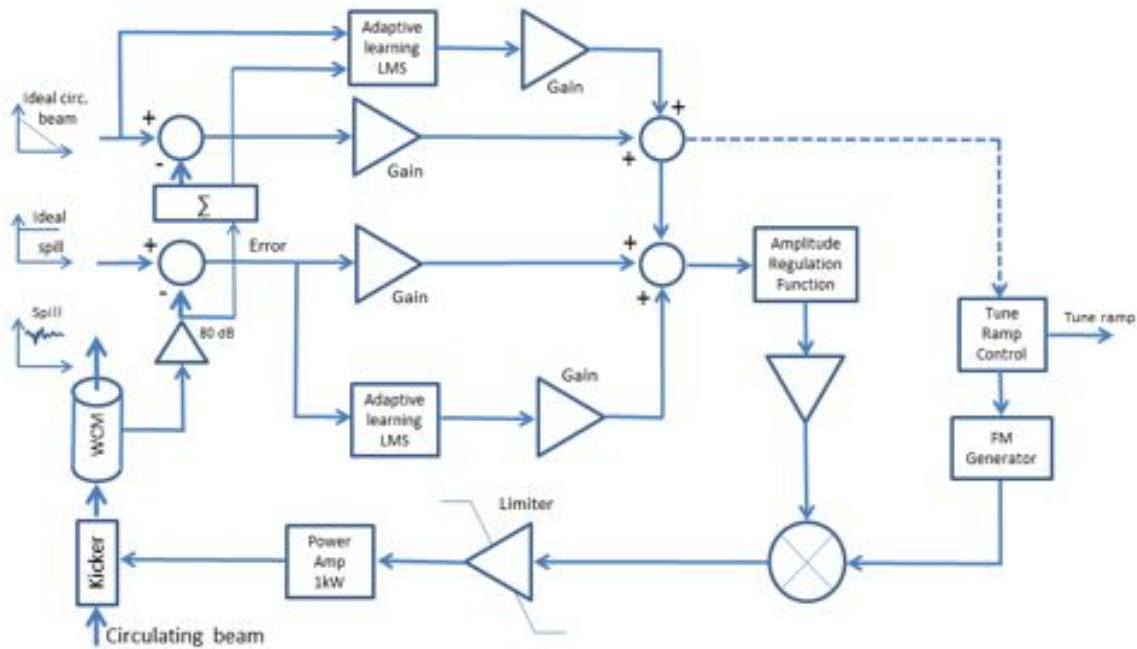

Figure 4.62. Block diagram of the spill regulation system.

The central frequency of the FM modulation has to track the betatron sideband frequency to heat the beam most effectively. Therefore, the central frequency depends on the quadrupole circuit ramp that drives the machine tune. The quadrupole ramp is shown as the right leg in the diagram of Figure 4.62. The tune ramp can be run open loop or it can be included into the regulation logic as shown by the dashed line. More detailed description of the regulation system can be found in Ref. [92].

### 4.6.4 Resonant Extraction Risks

#### 4.6.4.1  High Beam Loss

The principle risk for the Resonant Extraction subproject is that of poor extraction efficiency and the resulting higher than expected beam losses. The design value for extraction beam loss is 2% or less. Detailed extraction simulation studies show that for losses higher than 2%, the installation of costly additional shielding may be necessary to protect personnel at ground level in the vicinity of the extraction region [64] (see also section 4.5.2.1). Moreover, high residual radiation in the tunnel imposes a substantial radiation hazard to personnel conducting accelerator maintenance work. This hazard can be mitigated by a longer cool down time before the tunnel work is performed. Ultimately, higher than expected extraction beam losses would limit the beam intensity delivered to the Mu2e production target to a level that maintains prompt radiation dose rates within acceptable limits.

### 4.6.4.2   Reduction of Aperture due to Dynamic Extraction Bump

Another risk is also related to the beam losses caused by the potential conflict between machine acceptance and fast orbit deflections due to the dynamic extraction bump. This risk is largely mitigated by including a fast ramping capability into the design of the harmonic sextupole magnets, so that the function of the dynamic bump could be replaced by a phase rotation from the variation of the sextupole strength, as described earlier.

### 4.6.4.3   A Possible Opportunity to Reuse Spare Sextupole Magnets

Presently there are a number of ISA sextupole magnets in storage at Fermilab that are designated as special process spares for the Main Injector. These magnets are likely to be consumed by planned Recycler upgrades. However, considerable cost savings would be realized if one or more of these spare magnets became available for use by the Mu2e project.

## 4.6.5 Resonant Extraction Quality Assurance

The design of the Resonant Extraction system has been validated through numerous analytical and numerical simulation studies. Simulation tools used in these studies have also been benchmarked between each other and with other codes on similar tasks. This design has also been discussed and reviewed by peer experts on many occasions. The technical requirements determined in the design studies have been formulated in the set of specification documents for the hardware elements. Those hardware elements that contain any uncertainties have been or will be prototyped and tested. All magnets fabricated for the Resonant Extraction system will be measured in the Technical Division Magnet Test Facility using final versions or prototypes of the ramped power supplies. All hardware design, fabrication and installation will be compliant with standard Fermilab practices and the Fermilab Engineering Manual.

## 4.6.6 Resonant Extraction Installation and Commissioning

Installation of the Resonant Extraction tunnel hardware will proceed during and after beam operations for the g-2 experiment. During the g-2 run all magnets and the RFKO kicker can be installed because their apertures are compatible with the g-2 aperture requirements. The Spill Monitor will be installed in the AP30 portion of the beam line and can be used for the g-2 diagnostics as well. The Electrostatic septa present certain conflicts with the g-2 extraction kicker. Consequently the ESS installation schedule will be coordinated within the framework of the Muon Campus program. Commissioning of Mu2e Resonant Extraction will be a lengthy process and therefore must be started as soon as it can be accommodated in the Muon Campus schedule.

## 4.7    Delivery Ring RF System

### 4.7.1 RF Systems Overview

Figure 4.1 at the beginning of this chapter shows the required longitudinal structure of the beam to be delivered to the Mu2e production target. The required beam consists of a train of narrow (<250 nsec FW) pulses separated by the revolution period of the Delivery Ring (1.695 μsec). The interval between pulses will ultimately be cleared of beam such that the ratio of in-time beam to out-of-time beam is less than $10^{-10}$. This structure is necessary for the reduction of prompt backgrounds in the Mu2e experiment. Wider pulses would require a delay in the live gate for the Mu2e detectors thereby reducing the live-time of the experiment.

The narrow pulses of the required longitudinal structure are created by a 2.5 MHz RF system in the Recycler Ring and preserved in a 2.4 MHz RF system in the Delivery Ring. The extinction between pulses is accomplished by the extinction system, which will be covered in section 4.9. A summary of RF system parameters relevant to Mu2e operation is given in Table 4.24.

#### *4.7.1.1   Recycler Ring RF Manipulations*

The required longitudinal structure of the beam will be largely accomplished by a 2.5 MHz RF re-bunching sequence in the Recycler Ring. This RF sequence will re-bunch the train of 53 MHz bunches that constitute a proton batch into four 2.5 MHz bunches that occupy one seventh the circumference of the Recycler Ring. The Recycler Ring RF upgrades required to accomplish this are contained within the scope of the Recycler RF AIP [9].

Each of the four 2.5 MHz bunches will be synchronously transferred, one bunch at a time, to the Delivery Ring, where the beam is held in a 2.4 MHz RF stationary bucket during resonant extraction. The final longitudinal phase space beam distributions in the Recycler prior to transfer to the Delivery Ring are shown in Figure 4.63. The proton bunches that are transferred to the Delivery Ring show artifacts of the original 53 MHz bunch structure. As can be seen in Figure 4.63, the entire bunch length, including most of the tails, is contained within the 250 nsec full beam width requirement of Table 4.3[26].

---

[26] 99% of the beam is contained within ±80 nsec. The beam is extinguished to the $10^{-4}$ level outside of ±110 nsec.

Table 4.24. Recycler and Delivery Ring RF Parameters

| Parameter | Value | Units |
|---|---|---|
| **Recycler Ring 2.5 MHz Bunch Formation RF System** | | |
| Harmonic Number | 28 | |
| Frequency | 2.515 | MHz |
| Peak Total Voltage | 80 | kV |
| Number of Cavities | 7 | |
| Duty Factor | 40 | % |
| Bunch Formation time | 90 | msec |
| | | |
| **Delivery Ring 2.4 MHz RF System** | | |
| Harmonic Number | 4 | |
| Frequency | 2.360 | MHz |
| Peak Total Voltage | 10 | kV |
| Number of Cavities | 1 | |
| Duty Factor | 40 | % |

### 4.7.1.2  *Synchronous Transfer to the Delivery Ring*

To minimize beam loss during extinction, the Delivery Ring RF system must synchronously capture the proton bunches from the Recycler and maintain a matched RF bucket throughout resonant extraction. The Recycler Ring is roughly seven times the circumference of the Delivery Ring. Consequently the Recycler RF harmonic number is seven times greater than that of the Delivery Ring. Moreover, the Recycler Ring is operated farther from transition than is the Delivery Ring[27]. These two circumstances indicate that the Recycler RF bucket can be matched with a relatively small cavity voltage in the Delivery Ring. The flat-top voltage of the Recycler RF sequence is 80 kV. To match this, only 8 kV is required from the Delivery Ring RF cavity.

Beam transfer to the Delivery Ring will involve a synchronous bucket-to-bucket transfer with a frequency hop due to the fact that the ratio of Recycler to Delivery Ring circumference is not an integer. The digitally synthesized Low Level RF (LLRF) system will provide exact phase crossing alignment to facilitate these transfers. A similar system was used during Tevatron Collider Run II for Main Injector to Debuncher beam alignment during antiproton stacking.

---

[27] The Recycler Ring η = -0.00876, the Delivery Ring η = 0.00607.

### 4.7.1.3   Delivery Ring Longitudinal Dynamics

Since the proton bunch arrives in the Delivery Ring with a good deal of filamenting (see Figure 4.63), a precise match of the RF bucket to the incoming beam is not possible. Consequently, there will be significant variation of the bunch length and energy spread with time as the bunch tumbles in the Delivery Ring RF bucket (see Figure 4.64).

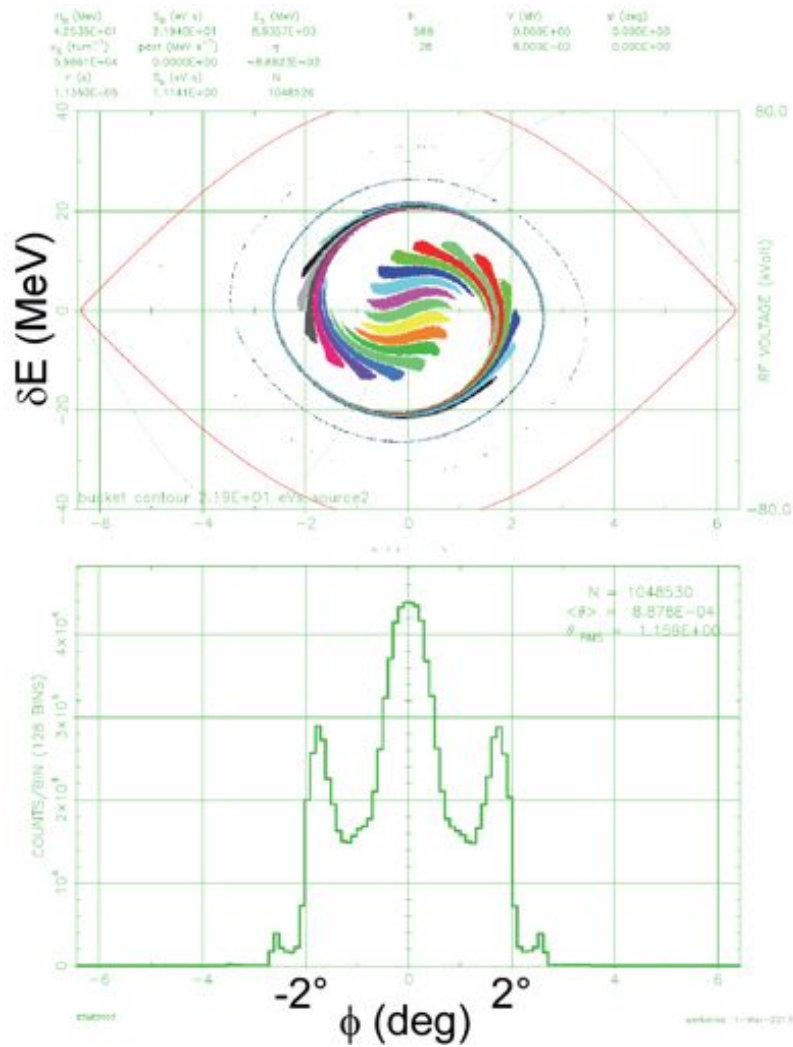

Figure 4.63. Longitudinal phase space distribution of protons in the Recycler Ring after the completion of the bunch formation sequence as calculated by an ESME [93] simulation. The top plot is a scatter-plot of the energy-phase coordinates of each proton. The individual 53 MHz bunches are separately colored. The bottom plot gives the projection of this distribution on the phase axis. Each degree of phase is approximately 31 nsec of pulse length. Artifacts of the original 53 MHz modulation of the beam are clearly visible.

This motion in the RF bucket throughout the spill causes substantial changes in the shape of the pulses delivered to the Mu2e target. Figure 4.65 shows the beam longitudinal phase space at two positions in the synchrotron motion a quarter of a synchrotron period apart.

The pulse time distributions are very different at these two extremes of the bunch orientation in longitudinal phase space.

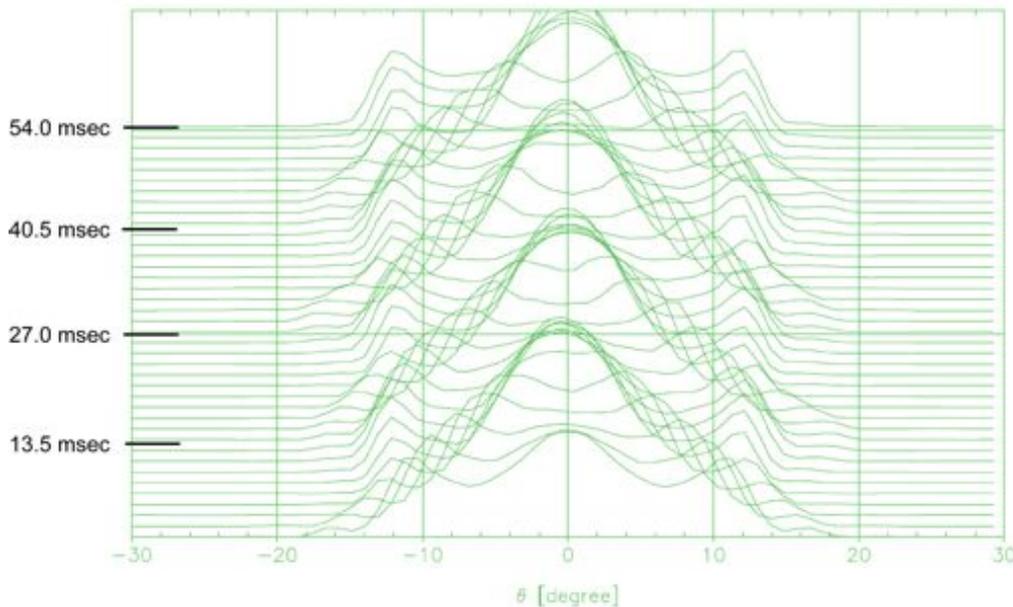

Figure 4.64. Water fall display of the variation of the proton bunch time profile as the bunch rotates in the 2.4 MHz RF bucket. A trace is plotted every 1.35 msec over the course of the spill. The vertical axis is time relative to the start of the spill. The horizontal axis is Delivery Ring phase (1° = 4.708 nsec). The period of the modulation is approximately one half of the small amplitude synchrotron period ($T_{sync}$ = 25.6 msec).

Interactions of the beam with itself and with the wakefield it leaves in the RF cavity have the effect of pushing some high amplitude protons into the tails of the longitudinal distribution. An ESME tracking simulation that includes space charge effects and the impedance of the RF cavity has been developed to study this effect [94]. Figure 4.66 shows the simulated proton time distribution of a single bunch 38 ms into the spill. Beam extracted from the Delivery Ring must have an out-of-time extinction level at or below $10^{-4}$ for the external extinction system to meet the requirements of the experiment (see section 4.9). Figure 4.66 demonstrates that this requirement is easily met.

### 4.7.1.4   Overview Delivery Ring 2.4 MHz RF Hardware

A peak RF voltage of approximately 10 kV with a 40% duty factor will be needed. The Delivery Ring 2.4 MHz cavity will be a single ferrite-load cavity of the type being manufactured for the Recycler Ring 2.5 MHz RF system. Transient beam loading will be pronounced as single bunches are circulating in the Delivery Ring. Power amplifier requirements are somewhat relaxed since the average beam current of a single bunch is only a quarter of a proton batch ($1 \times 10^{12}$ protons). The power amplifier for the Delivery Ring RF system will consist of a solid-state amplifier similar to what is presently used for bunch coalescing in the Main Injector. Table 4.25 shows the physical parameters of the Delivery Ring 2.4 MHz RF system.

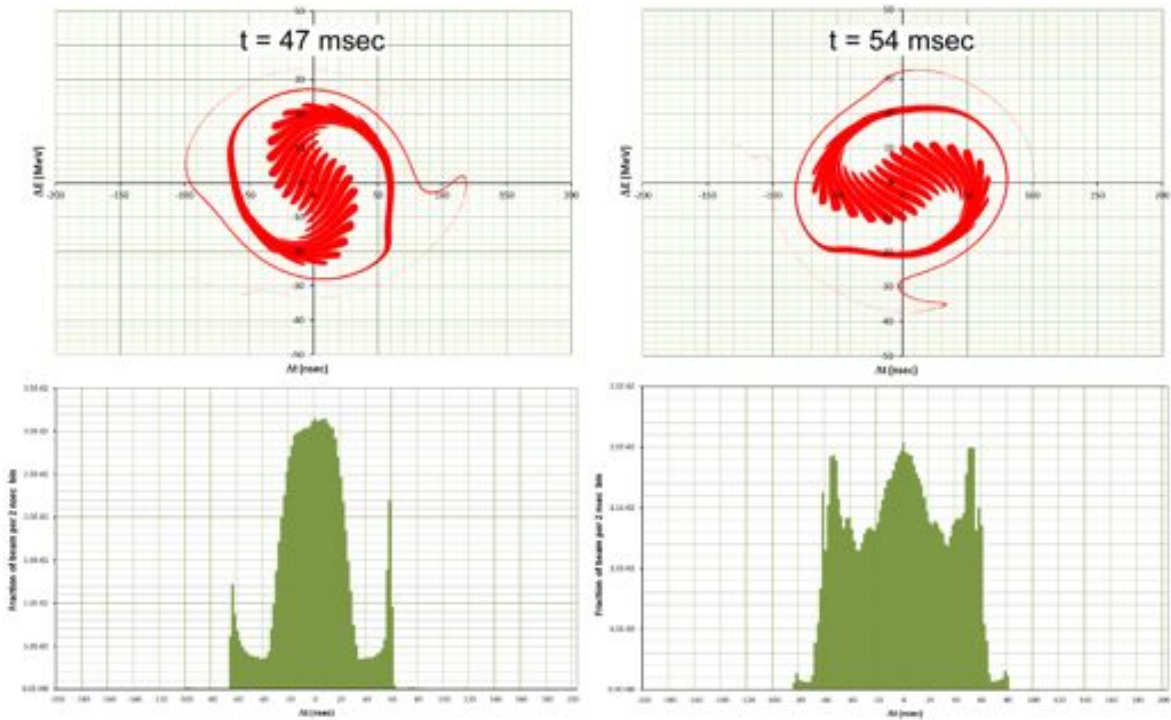

Figure 4.65. Results of an ESME longitudinal tracking simulation of the Delivery Ring. The extremes of the proton pulse time distributions are illustrated by two locations in the synchrotron motion of the beam a quarter of a synchrotron period apart. The plots on the left occur at a minimum ($t$ =47 msec) in the time profile width in Figure 4.64. The plots on the right correspond to a pulse width maximum ($t$ = 54 msec). The top plots show energy versus time for each proton tracked; the bottom plots show the projection of the phase space distribution on the time axis. The time axis on all plots goes from -200 nsec to +200 nsec.

### 4.7.2 Low Level RF System

The Mu2e low-level RF (LLRF) system will draw on experience from the Recycler LLRF system, as shown in Figure 4.67. The LLRF system will be used to operate one 2.4 MHz Ferrite loaded cavity and it will have precise control over the cavities amplitude, phase and frequency. The hardware used to do this will reside in a VXI crate. The RF clock signals will come from a stable 50 MHz master oscillator and will be used to provide the RF clock to the Delivery Ring System. The Radial Position of the beam will come from a Stripline detector.

Table 4.25. Parameters of the Delivery Ring 2.4 MHz RF system

| Quantity | Value | Units |
|---|---|---|
| Beam Current ($I_p$) | $178.564 \times 10^{-3}$ | A |
| Number of cavities | 1 | |
| R/Q | 400 | $\Omega$ |
| Q | 125 | |
| Cavity Voltage | 10 | kV |
| Power Loss per Cavity | $1.0 \times 10^3$ | W |
| Total Apparent Power | $1.04858 \times 10^3 \angle 17.5089°$ | VA |
| Total Current | $209.716 \times 10^{-3} \angle 17.5089°$ | A |
| Induced Mode Compensated | 3.786 dB = 35.3% | |
| Robinson Stable | 4 | |

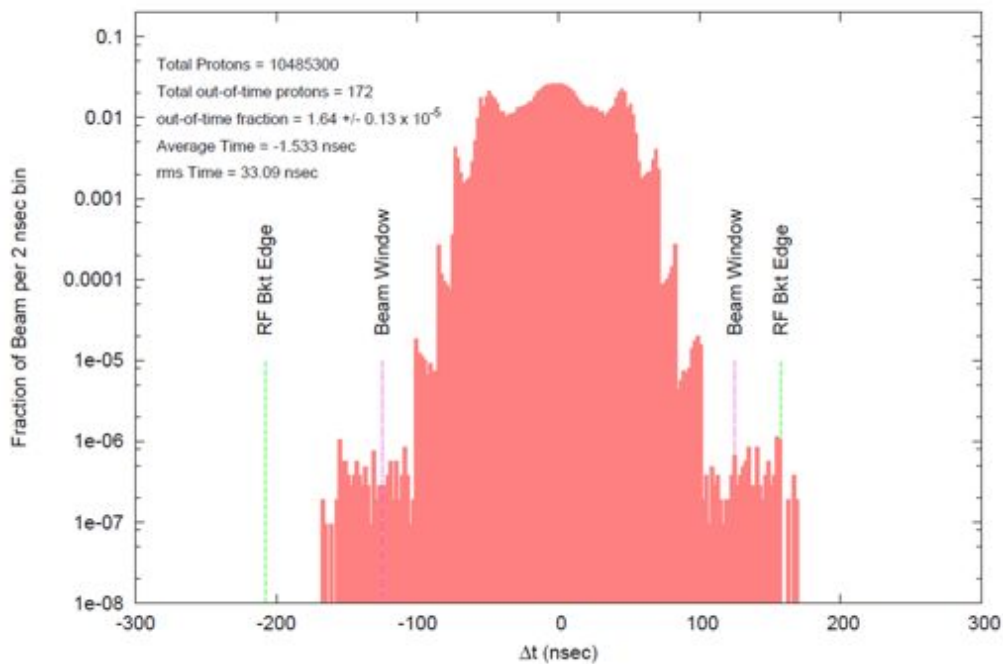

Figure 4.66. Time distribution of a single proton pulse as it appears in the M4 beamline upstream of the extinction insert. This pulse was extracted from the Delivery Ring 38 msec after the start of the spill. Time ($\Delta t$) is measured relative to the center of the RF bucket. The Beam Window, indicated by the magenta lines, is the $\pm 125$ nsec window that bounds the in-time beam. The fraction of beam outside of the Beam Window is $1.64 \pm 0.13 \times 10^{-5}$.

### 4.7.2.1   Low Level RF Requirements

The Delivery Ring low level RF system includes: a VXI based LLRF system, a Stripline detector radial beam position measurement, a Low Level accelerator console application page, and cabling for the LLRF system. The current Recycler Low Level console application page is shown in Figure 4.68. This application will be a template for the Delivery Ring Low Level Console application.

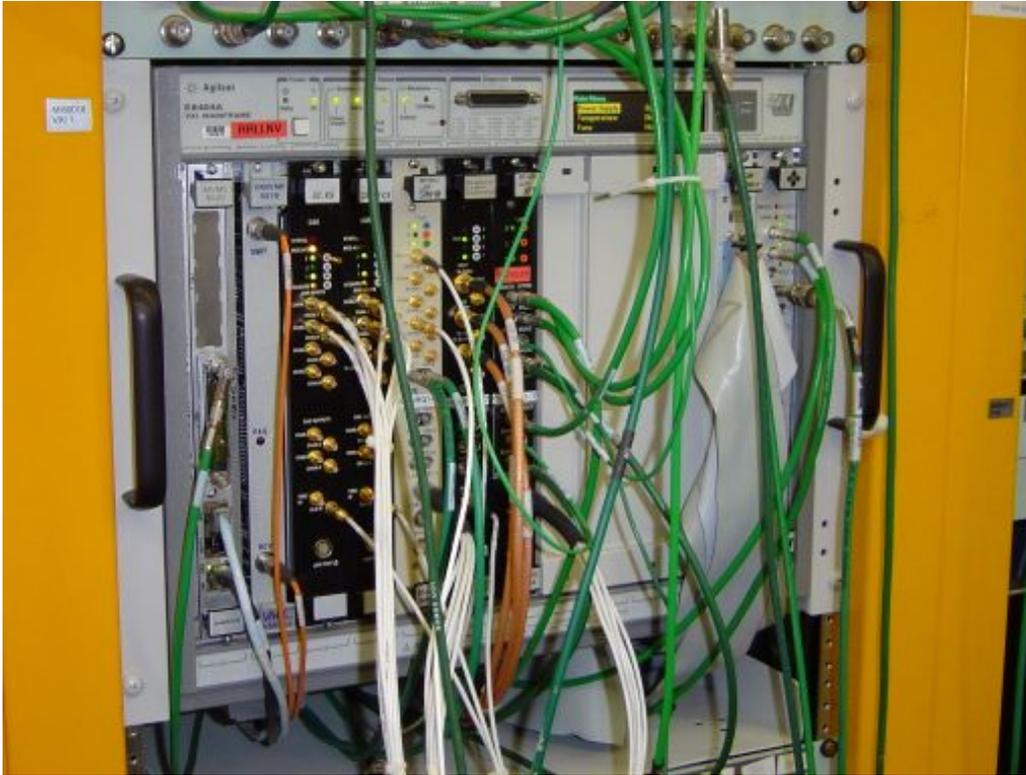

Figure 4.67. The Recycler LLRF system. Slot 0 is the controller. Slot 1 is the Reflective Memory. Slot 2 is the RF Switch (SWH). Slot 3 is the Direct Digital Synthesizer (DDS). Slot 4 and 5 is the Digital Signal Receiver (DSR). Slot 6 is the Transfer Synchronization (XFR). Slot 7 is the Direct Input/Output Module (IO 100). Slot 8 is the Multi-channel field controller (MFC).

The VXI LLRF system will require the same cards shown above in Figure 4.67 to keep maximum compatibility and flexibility between the existing Recycler and Main Injector VXI LLRF Systems.

- Controller
- Reflective Memory.
- RF Switch
- Direct Digital Synthesizer.
- Digital Signal Receiver.
- Transfer Synchronization.
- Direct Input/Output Module.
- Multi-channel Field Controller.

In addition, the Recycler Low Level Console Application page is to be ported over to the Delivery Ring. Only minor modifications should have to be implemented to the console application to make it compatible for Delivery Ring Operation.

The VXI LLRF System will also require a 50 MHz Master Oscillator that will need to be phase stable. The phase stability of the 2.4 MHz RF signal is required to be down to the 200 psec level of jitter.

Figure 4.68. Low Level Console Application Page

### 4.7.2.2 *Low Level RF Technical Design*

The VXI based LLRF system is to include the following (see Figure 4.69):

- Controller – PowerPC based microcontroller or better. The module provides a VXI interface, central processing unit, and Ethernet interface. The controller also loads software over the network to manage the control system interface. The inputs to this card are Ethernet and Tevatron Clock (TCLK) for timing.

- Reflective Memory – Reflective Memory/Optical Link between LLRF front-ends. Data written to the internal memory of the card will be visible to all cards on the link. A common piece of data on this card will be the Delivery Ring operational frequency. This card's input and output will communicate with the Recycler VXI crate via a fiber optic link.

- Digital Signal Receiver (DSR) – Digital Signal Receiver detector for Cavity RF signals. This card can calculate the radial position and phase. The input to this card will receive the fan-out to the cavity, the fan-back from the cavity, the wall current monitor (WCM), and the radial position (RPOS) of the beam.

- RF Switch (SWH in Figure 4.69) – 2-channel 4-input RF switch. Delivery Ring reference from one switch and Delivery Ring offset AA markers[28] from another switch. The outputs will be routed to the Recycler Ring LLRF system. The input to this card will receive the fan-out to the cavity, the fan-back from the cavity, the AA marker from the Recycler and reference RF from the Recycler. The outputs will be an AA marker and RF to the Recycler LLRF system for transfer synchronization.

- Direct Digital Synthesizer (DDS) – Supplies RF reference and RF output to cavity. This module will provide the primary RF outputs for the system. The module will receive its clock input from a 50 MHz phase stable oscillator. RF1 is the reference output and RF2 will go directly to the High-level RF (HLRF) section. The frequency and phase will be controlled to a very high resolution. The input to this card will be a 50 MHz master oscillator and the fan-out to the cavity and fan-back from the cavity. The outputs from this card will be the RF output to the cavity (RF1) and the RF reference signal (RF2) of RF1.

- Transfer Synchronization (XFR) – Used for transfer synchronization between Recycler and Delivery Ring. This module will serve to generate markers for transfer synchronization between Recycler RF and the Delivery Ring. The input to this card will be the AA marker from the Recycler, RF from the Recycler and the RF1 signal from the DDS card as input. The output will be the Delivery Ring AA marker.

- Multi-channel Field Control (MFC) – 32 input channels that can be digitized at up to 65 MHz. This module will have 32 input channels and have the capability to digitize at 65 MHz. With its floating point DSP, one can do calculations on the HLRF. The 14 bit 260 MHz DAC outputs will be able to give the responses of those calculations. Examples of the input to this card are the wall current monitor and cavity probe. Output from the card is left as optional at this point in time.

---

[28] An AA marker is a signal that gives the time of a 53 MHz bunch in the Booster batch currently circulating in the Recycler. It is reset at the time of injection of every new batch into the Recycler.

- Digital Input/Output Module (IO 100)– Can be used to flip output bits or write output bytes under software control. This is a general purpose I/O module that can be used to flip output bits or write output bytes under the software control. The input to this card will be machine data (MDAT). The output of this card will be used to write injection phase and transfer phase to the Recycler.

- VXI UCD – Universal Clock Detector for timing synchronization. The Universal Clock Detector inputs will come from TCLCK and MDAT.

A block diagram of the VXI based LLRF system is shown in Figure 4.69. The pertinent inputs and outputs for each card slot are also given in this figure.

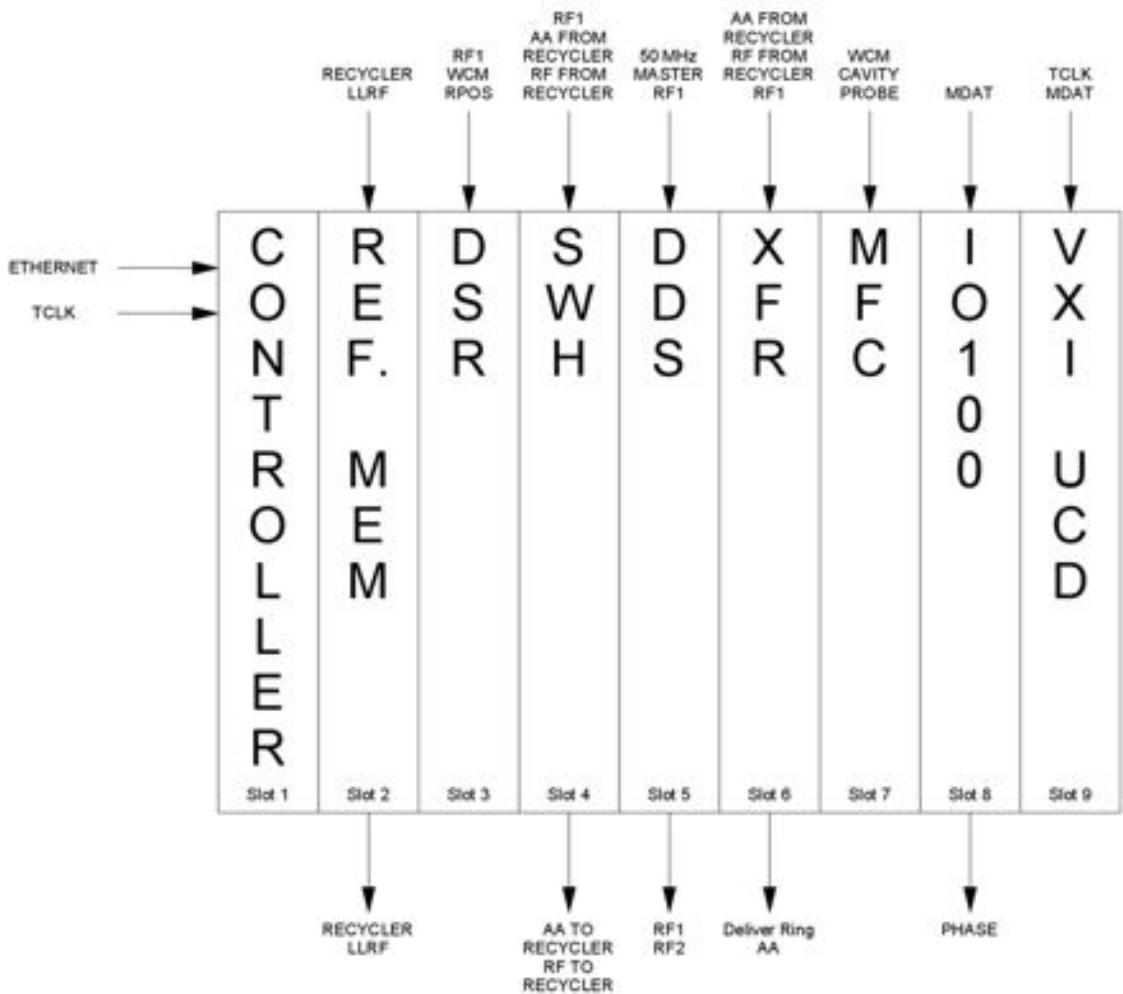

Figure 4.69. Block Diagram of the VXI LLRF System.

### 4.7.2.3  *Low Level RF Risks*

This system is to be based on the existing Recycler LLRF system, but will use the latest in hardware technology available at the time of construction. The present design assumes a straightforward upgrade of the existing software will accommodate the newer technology. The risk here will be in how much of the existing LLRF software code can be repurposed for the Delivery Ring's new hardware.

### 4.7.3 Delivery Ring RF Beam Studies & Tuning RF

The Delivery Ring Beam Studies and Tuning RF system will facilitate the capture of 2.5 MHz Recycler bunches for the purpose of beam studies. This RF system will allow Delivery Ring orbit measurements using the Delivery Ring BPM system. The Beam Studies RF system will also provide an interface by which the RF frequency can be changed for purposes of accelerating and decelerating the beam within the momentum aperture of the Delivery Ring.

#### 4.7.3.1  Beam Studies RF Requirements

Once the Delivery Ring RF system is installed, different states will be programed on the Low-level Console application.  A new state will be written for all possible study modes – for example: beam capture studies, BPM studies, momentum aperture studies, and beam bunch quality studies.  It should be noted that these studies will only require the programming of a state and will not require hardware changes on the Delivery Ring RF System.

#### 4.7.3.2  Beam Studies RF Technical Design

Figure 4.68 shows an example of a state for a study on the Low-level Console application. The studiers will program the state by populating it with LLRF MESSAGES. Each MESSAGE will have a unique time of implementation and a DATUM will be filled out when required for a particular MESSAGE. Meetings between the studiers will occur with the LLRF group and pertinent MESSAGES from Recycler and Main Injector will be ported over.  Any new MESSAGES that need to occur will be identified at this time and developed by the LLRF group.

#### 4.7.3.3  Beam Studies RF Risks

The beam studies RF system adds the capabilities described above to the Delivery Ring RF system in a way that is common and well established in RF systems throughout the Fermilab accelerator complex. No risks are foreseen in the implementation of these capabilities in the Delivery Ring RF system.

### 4.7.4 Delivery Rings RF Cooling System

The Delivery Ring RF Cooling System will provide deionized water to both the 2.4 MHz cavity and the 2.4 solid-state driver.

#### 4.7.4.1  RF Cooling Requirements

The Mu2e 2.4 MHz cavity will require 4 gpm of deionized water and the 2.4 MHz solid-state driver will require 20 gpm of deionized water. The cooling system must have the capability to support an additional 2.4 MHz cavity and 2.4 MHz solid-state driver if needed. An HLRF station will require 25 kW of heat dissipation. Knowing this, the cooling system should be able to support two stations and then have a factor of two in overhead.  The final design for the new cooling system should be able to support 100 kW of dissipation and be able to regulate the temperature at 90°F with only $\pm\frac{1}{2}°$ of swing.

### 4.7.4.2   RF Cooling Technical Design

The primary location of the piping for the RF cooling system is the AP50 Service Building. A cooling skid assembly, along with associated connecting stainless steel piping, will be located in the AP50 Service Building. Approximately 100 gpm/100 kW low conductivity water (LCW) will be required to meet the cooling needs of the RF cavity portion of the Mu2e beam line project.  A 2-1/2" Stainless steel piping header will deliver temperature controlled LCW to the Mu2e RF Cavity planned for installation in what is the existing portion of the Delivery Ring enclosure underneath the AP50 service building.  Also, a 2-1/2" Stainless steel piping header will deliver temperature controlled LCW to the Mu2e solid-state driver located in the ground level portion of the AP50 service building.  Deionizing bottles will be part of the RF LCW cooling skid assembly and used only for Mu2e RF LCW System polishing.  LCW makeup water will come from the existing nearby PBar LCW System. The expansion tank will serve as an emergency reservoir and provide positive system pressure and gas purging. The LCW will dissipate the heat through a plate and frame heat exchanger to the existing building chilled water system (CHW).  2" copper tubing will be needed for the CHW piping.

### 4.7.4.3   RF Cooling Risks

A nearly identical RF cooling system will be built for the Recycler as part of the Recycler RF AIP. The Recycler RF cooling system will be implemented and de-bugged well in advance of the time for implementation of the Delivery Ring RF cooling system. Thus, there are no risks anticipated for this system.

## 4.7.5 Delivery Ring 2.4 MHz RF

The High-level RF cavity system for $h = 4$ Delivery Ring will require a solid-state amplifier for operations.  This system will utilize the same solid-state driver planned for the Recycler Ring 2.5 MHz RF system. The existing Recycler Ring solid-state amplifier is shown in Figure 4.70.

### 4.7.5.1   2.4 MHz RF Requirements

A 2.4 MHz, 8 kW Solid State Amplifier is required to drive the 2.4 MHz ferrite loaded RF Cavity. The Cavity is required to operate a continuous waveform at 10 kV. For a perfectly matched system, 1 kW of drive power will be required.

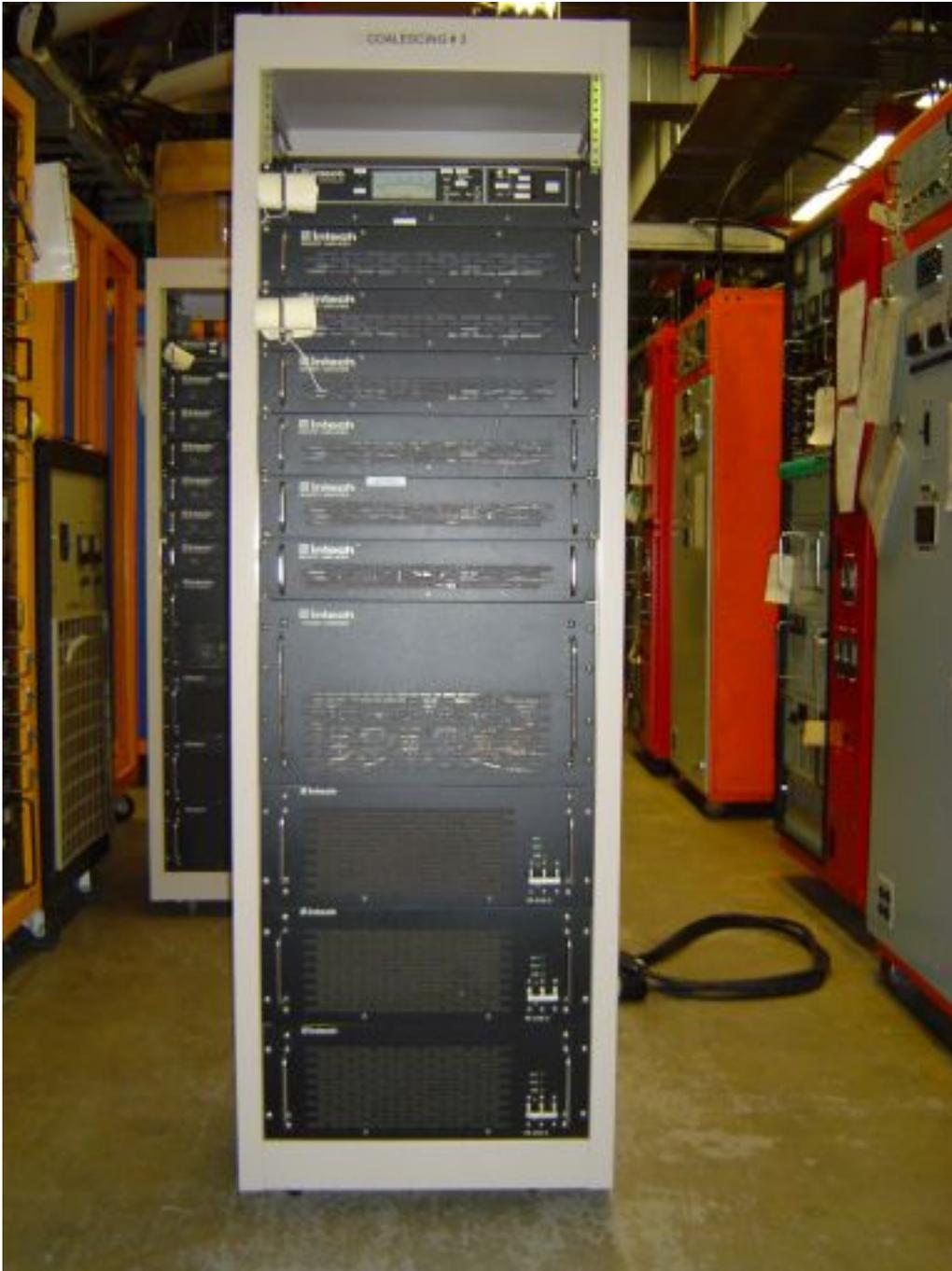

Figure 4.70. Recycler Ring 2.5 MHz Solid State Amplifier. The top rack mount is the controller. The controller also fans out the input from the LLRF to each of the solid state modules. Rack mounts two through seven are 1 kW 2.5 MHz solid-state modules. The fourth from the bottom rack mount is the output combiner for each of the 1 kW solid-state modules. The bottom three rack mounts are the 48 V DC power supplies to the 1 kW solid state modules.

8 kW Solid State Amplifier Design Specifications

- Amplifier must be all solid state, no tubes allowed.

- Frequency range: 2.4 – 5.0 MHz.

- Maximum continuous waveform output power greater than 7 kW into a 50 Ω load over the entire 2.4 – 5.0 MHz frequency range.

- Harmonics:  All harmonics must be greater than 25 dB below the fundamental at the 7 kW output power level.

- Amplifier must be cable of operating into any output load impedance without damage.  Automatic shutdown of the amplifier is allowed when the reflected power from the load exceeds 2 kW.

- Gain: greater than 60 dB

- Amplifier must withstand 1 $V_{rms}$ input signal levels for all load conditions.

- Protection:  All amplifiers must be thermally protected such that automatic shutdown will occur if the temperature of the RF power transistors exceed their safe operating temperature.

- Amplifier components must fit into a standard 19-inch relay rack.

- Input connector:  Type N.

- Output connector:  7/8.

- Remote monitoring capability:

  - On/Off status

  - Over temperature shutdown

  - Forward output power

  - Reflected power from load

### 4.7.5.2  2.4 MHz RF Technical Design

The 53 MHz, 1 kW modules used in the recent Booster Solid-State update will be modified to operate from 2.4 to 5 MHz. Preliminary and Final design have been completed and the proper magnetics have been identified to proceed with production. The 1 kW solid-state modules will be made to be directly interchangeable with the Recycler Ring 2.5 MHz 1 kW modules.  To control the new 8 kW Solid-State Amplifier the present controller used for the Main Injector Solid State drivers will be replaced. The LLRF signal will be amplified with a Mini-Circuits preamplifier and distributed to each of the 1 kW Solid State modules using a Mini-Circuits splitter.  A 10 kW combiner will be ordered from Werlatone.  The combiner will have eight Type N inputs and one 7/8 output and be able to work in the 2.4 to 5 MHz frequency range. Attached to the combiner will be an output directional coupler. It will have 40 dB of attenuation on the Forward and Reverse power ports. Werlatone will also supply the directional coupler. 48 V DC will be supplied to each of the 1 kW Solid State modules with one 10 kW TDK Lambda ESS power supply.  This is the same supply used in the recent Booster Solid-State upgrade. All of these components will reside in a 19-inch rack and the output directional coupler will slightly protrude from the back of the rack. Copper tubing in back of the 19-inch rack will be used to distribute cooling water to each of the 1 kW solid state modules and the Werlatone combiner.

### 4.7.5.3   2.4 MHz RF Risks
The Delivery Ring 2.4 MHz solid state amplifiers are similar to the 53 MHz, 4 kW Solid-State Drivers used for the upgrades recently completed for the Booster and the Recycler Rings. There are no significant changes between the Delivery Ring 2.4 MHz solid state amplifiers and those used in the Booster and Recycler. With the success of the amplifiers in the Booster and Recycler upgrades, we believe there is no significant risk associated with the Delivery Ring subsystem.

## 4.7.6 Delivery Ring RF Quality Assurance
The Low Level RF System will be debugged and tested in the LLRF lab before it is installed in the Delivery Ring.

Delivery Ring Beam Studies will be used for quality assurance and commissioning purposes as well as to study the interactions between HLRF and LLRF and the synchronization between the Recycler Ring and Delivery Ring.

The Delivery Ring RF Cooling System will be checked to verify that it can deliver the specified flow rate of the water, temperature control of the water and desired conductivity of the water.

The Delivery Ring 2.4 MHz RF solid-state driver will be tested with a 50 Ω dummy load to check that it meets full criteria.  The controls for the solid-state driver will be fully tested on the bench and during the high power tests into the dummy load.

## 4.7.7 Delivery Ring RF Installation and Commissioning
The Delivery Ring RF System will be installed at AP50.  Electricians will be used to install all of the new cables and power utilities.  Electrical Technicians will phase match the cables using a Network Analyzer.  Mechanical Technicians will supply the water hookups.  Once installation is complete, commissioning will be done during Delivery Ring Beam Studies.

# 4.8   External Beamline

## 4.8.1 Overview
The Mu2e external beamline, referred to as the M4 line, must cleanly separate and transport resonantly extracted beam from the Delivery Ring to the Mu2e production target, generate the required beam characteristics for the experiment and, additionally, perform inter-bunch extinction of out-of-time particles. This is accomplished through a number of specialized beam-optics insertions, achromatic modules, and an AC sweep dipole linked to a series of extinction collimators, respectively. These custom insertions were specifically designed to optimally interface with the local geography. Further, as part of the Muon Campus complex, the M4 beamline must support beam operations for the g-2 experiment.  An overview of the Muon Campus is shown in Figure 4.71.

A consequence of combining Mu2e and g-2 operation[29] is that the D30 straight section (underneath the AP30 service building) in the Delivery Ring must be shared for injection and extraction and both external beamlines must also share a common upstream section (M4/M5) following extraction. Specifically, injection systems for both experiments occupy the upstream end of the D30 straight section, and extraction systems occupy the downstream end. The g-2 beam is extracted from the Delivery Ring with magnetic kickers while Mu2e beam is resonantly extracted using two electrostatic septa (the g-2 extraction kicker is located upstream of the Mu2e septa). The 3 GeV muon beam required by g-2 and the Mu2e 8 GeV proton beam utilize common magnetic components in the D30 straight section to complete separation from the Delivery Ring. Extraction must occur vertically to accommodate the Delivery Ring enclosure, taking advantage of existing civil construction from a prior beamline.  After extraction from the Delivery Ring is complete, vertical steering magnets transport beam down either the Mu2e M4 line or the g-2 M5 line. The large differences in beam size and energy place difficult, sometimes conflicting, demands on the common extraction optics, especially the extraction Lambertson and vertical-bending dipoles.  The civil constraints of the local geography further complicate and restrict the layout of the two external beamlines.

## 4.8.2 Beamline Layout and Optics

The local geography for what is termed the Muon Campus complex is shown in Figure 4.72.  Civil and geographical constraints (avoidance of wetlands, for example) dictate a 30° - 40° bend after extraction from AP30 for the Mu2e M4 external beamline in order to maximize its length and optimize the location of the experimental hall. In this region, about 200 m is available for the external Mu2e beam line. Another civil restriction that significantly impacts the Mu2e line is the location of the g-2 ring and experiment. The g-2 experiment must be positioned to avoid even low-level stray magnetic fields from Mu2e components on the one side (maximal distance from the strong Mu2e experimental solenoids) and utility corridors on the other side, the latter setting the minimum amount of left bend required for the g-2 external beamline.

---

[29] NOTE: The combination of functionality in the D30 straight section for both Mu2e and g-2 does not imply the ability to simultaneously operate both experiments.  Simultaneous operation of g-2 and Mu2e will not be possible.

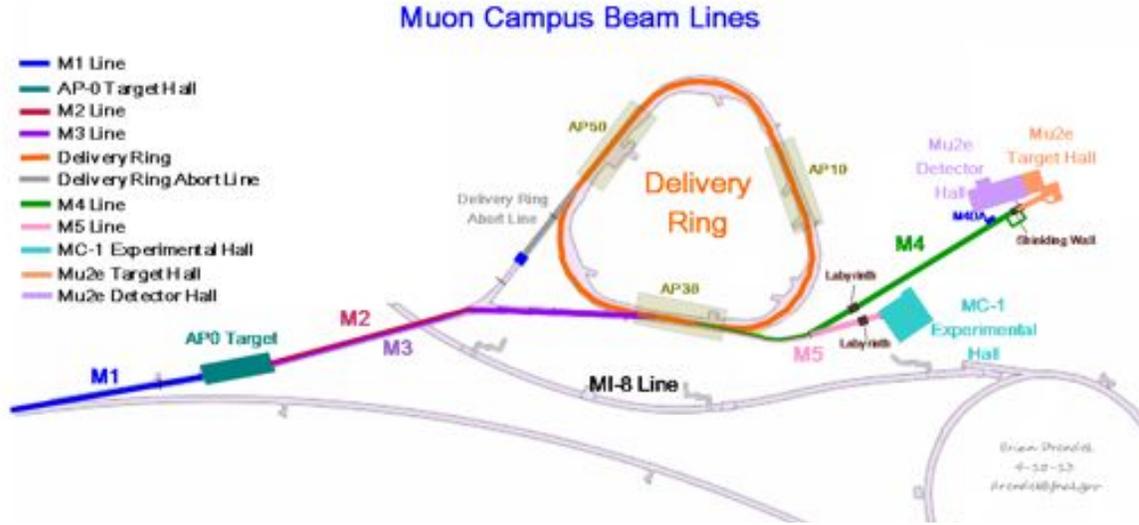

Figure 4.71. An overview of the Muon Campus.

As stated above, the length of the Mu2e beamline is limited by wetland avoidance and the much reduced g-2 beamline length by the ring enclosure location. The very short distance (~120 m) from the common extraction Lambertson to the g-2 ring mandates efficient, space-conserving separation of the two external lines. Since physical separation from the Delivery Ring must occur vertically, the most efficient separation of the two lines is also vertical. This is accomplished by reversing a vertical-bend dipole in this section. Strong, independent horizontal left-bend dipole strings then direct beam to either Mu2e or g-2 experiments. Final separation into independent civil enclosures is achieved by utilizing a large difference in the strengths of the left bends between the Mu2e and g-2 lines. These strong horizontal left-bend strings must immediately follow the vertical separation stage in both lines. Rapid separation is particularly important for the g-2 external beamline given the short distance to the experiment and the need for matching and tuning sections.

Even with 200 m to work with, accommodating two experiments along with the insertions required for extinction, collimation, and beam manipulations is difficult and requires combining functions when designing the insertions. Therefore, multi-function custom insertions and beamline sections have been designed when possible to meet the requirements outlined in the following section.

### 4.8.2.1   Beamline Optics Requirements

To appreciate the complexity of this beamline, the requirements that must be met are listed under the following major categories: civil, beam properties and capability, and specialty optics. The first and third sections are relevant to both Mu2e and g-2 and the second is germane only to Mu2e.

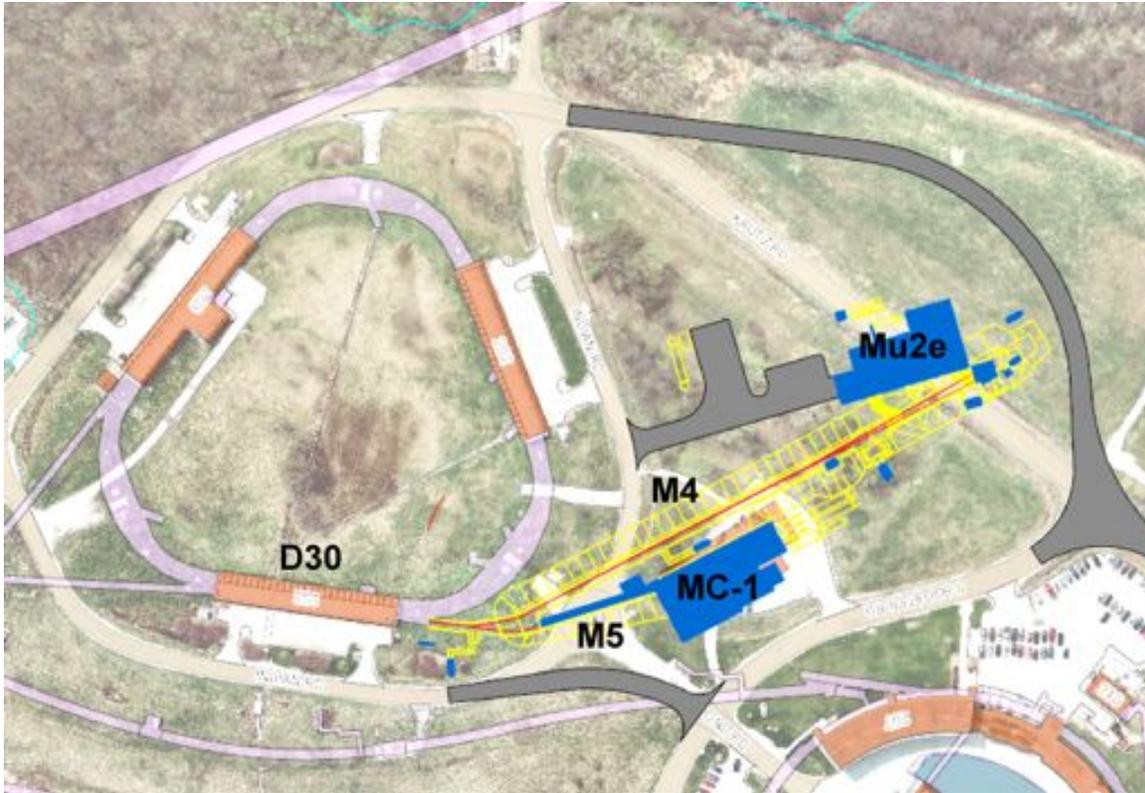

Figure 4.72. Site layout of the Muon Campus showing the Delivery Ring and the Mu2e and g-2 experimental halls and connecting beamlines. Note: the plan to divert Kautz Road around the right (west) end of the Mu2e building has been dropped.

#### 4.8.2.1.1   Civil Requirements

The following requirements are derived from the constraints imposed by existing facilities, conventional construction, and geography on the layout of the external beamline.

- **No civil modifications to Delivery Ring enclosure**
  This implies vertical separation of external beamline from the Delivery Ring.

- **A total of 4′ vertical elevation change (up) inside the Delivery Ring enclosure**
  This elevation change takes advantage of existing civil construction (from a previous beamline) and positions the M4 line for the final elevation and downward angle needed to traverse the production solenoid to the target. The elevation change must occur in two-stages in order to clear Delivery Ring components and simultaneously generate the necessary phase advance for vertical dispersion suppression. The first stage levels off at 32" to avoid conflicts with Delivery Ring components. The second stage elevates the line the remaining 16", leveling off at the M4 final beamline elevation.

- **Horizontal left bend**
  A 26.5° increase in bearing from the D30 straight section is required for g-2 beam and an additional 13.7° (total 40.2° left bend) is required for the M4 line. The first increase in bearing is required to match the direction of extracted beam from the D30 straight to the optimal geographic location established for the g-2 experimental building. The additional bearing for the M4 line is required to establish a separate beamline enclosure for Mu2e, allowing sufficient shielding to be installed between the Mu2e and g-2 experiments for beam operations, and also optimizes the geographic location of the experimental enclosure with respect to the surrounding site and existing utility and civil infrastructure.

- **Shield Wall**
  A shield wall is required to allow beam operations to the Mu2e diagnostic absorber during installation of the Mu2e solenoids and detectors.

- **A diagnostic beam line and absorber**
  A short diagnostic line after the extinction section is required for commissioning, tuning, and beam studies. Use of the diagnostic absorber will prevent radiation exposure to personnel in the Mu2e building and unnecessary activation of the experimental apparatus.

- **A 2.75° downward slope at the entrance to the solenoid**
  This angle is required to direct and level beam onto the target through the field of the production solenoid. To cancel dispersion at the target, the entire final focus section must be installed on a 1.375° slope; with equal vertical bends at the beginning and end of the final focus. Due to the length required for the final focus and the large slope involved, a step down is planned near the middle of the final focus section rather than a ramped floor. The total elevation change from the upstream end of the final focus section to the experimental target is about 3.9'. The length and value of the slope (~1.375°) requires a step down rather than a ramp for installation of components. This downward slope results in a total vertical elevation change of 3.9' from the upstream end of the final focus section to the experimental target.

- **A total length of not more than ~200 m from the Delivery Ring to the production target**
  This is the maximum length allowable for wetland avoidance, and to minimize environmental impacts and site modifications.

### 4.8.2.1.2    Beam Properties and Capability

The following is a set of beam requirements for the M4 beamline.

- Horizontal resonant extraction from the D30 straight section of the Delivery Ring.

- Beamline must transport a $30\pi$ mm-mrad emittance (95%) and a momentum spread up to $\pm1\%$ (95%) with small changes in beta functions (5% or less).

- Inter-pulse extinction of at least $10^{-10}$. The extinction AC dipole generates a kick between beam pulses, displacing out-of-time particles onto collimators (see section 4.9.2.1).

- Round beam on target, $\sigma = 1$ mm, no dispersion ( this implies ~2 m beta functions at target).

- Target beam position and angle scan capability of $\pm$ 1 cm, with no angle change, and $\pm0.8°$, with no position offset, both vertical and horizontal (this translates into large motion of final focus scan dipoles of up to $\pm14$ cm in the direction opposite to the bend plane). Stands must support automated motion on three of the target scan CDA dipole magnets.

- 6.5 m reserved from last beamline element to final focus for an HRS/PS protection collimator and the required distance through the production solenoid to the target.

### 4.8.2.1.3   Specialty Optics Insertions

The following is a list of required special insertions into the M4/M5 and M4 beamlines. The design details of these insertions will be given in the Beamline Optics Technical design section (4.8.2.2).

***M4/M5 special inserts:***

- **Dispersion suppression**
  Achromatic optics are required in the M4/M5 section of the Mu2e line to suppress vertical dispersion from the D30 vertical extraction system.  Dispersion must be suppressed upstream of the horizontal left-bend string to avoid coupling between the two planes.

- **Mu2e vertical achromat**
  The Mu2e vertical achromat is a complex 4-bend achromat. The vertical bends include the extraction Lambertson, the off-center quadrupole (D2Q5), the C-magnet, the first leveling bend (EDWA), and a final bend/reverse bend pair of MDC dipoles (see Figure 4.75).

- **Separation  of Mu2e and g-2**
  To separate the two lines physically and optically, the separation must occur vertically due to space constraints.  Although combined, the beamline to Mu2e and g-2 must be independently tunable in the vertical bend section downstream of Delivery Ring extraction to meet experimental beam specifications (emittance and energy change). Both tunes must satisfy conditions for a vertical achromat.

***M4 special inserts:***

- **Achromatic 40.2° horizontal left-bend insert**

  An achromatic module embedding the 40.2° horizontal left-bend dipole string is required.

- **Extinction section**

  The specifications of the AC dipole require a high beta in the horizontal (to maximize the horizontal kick) and a low beta in the vertical (to accommodate the small gap) that minimizes technical demands (see section 4.9.2.1). Two collimators are required upstream of the AC dipole to remove high amplitude halo and a single collimator is required downstream to absorb out of time beam which has been deflected by the dipole.

- **Final Focus section**

  Another achromatic module is required for the final focus design incorporating a pair of 1.375° vertical down bends to set the correct slope into the production solenoid and to provide simultaneous suppression of vertical dispersion at the target.

- **Matching**

  The beamline also requires the necessary matching sections between the custom insertions and achromatic modules.

### 4.8.2.2   Beamline Optics Technical Design

An overview layout of the final M4/M5, M4, and M5 beamlines that meet the requirements is depicted in Figure 4.73. As indicated in the optics requirements section above, the beamline is best described in terms of its modular functionality. Correspondingly, the following descriptions will detail the important sections, and discuss the rationale and justify the design approach for each section. The location of each section in the overall external beamline layout is shown in Figure 4.74.

#### 4.8.2.2.1   Extraction from the Delivery Ring

The incorporation of g-2 and Mu2e extraction systems into the Delivery Ring D30 straight section has been carefully designed. The extraction part of the straight section is considered to start at the center of D30Q (the center of the D30 straight). Figure 4.75 shows the layout of the extraction devices in the Delivery Ring.

A number of septa locations were studied (Mu2e requires two electrostatic septa). The optimal location for the two electrostatic Mu2e septa is ~0.3 m upstream and 0.45 m downstream of D2Q3 (this is considered the minimal spacing requirement for components). The first Mu2e electro-static septum (ESS) module is approximately a half FODO cell downstream from the g-2 extraction kicker.

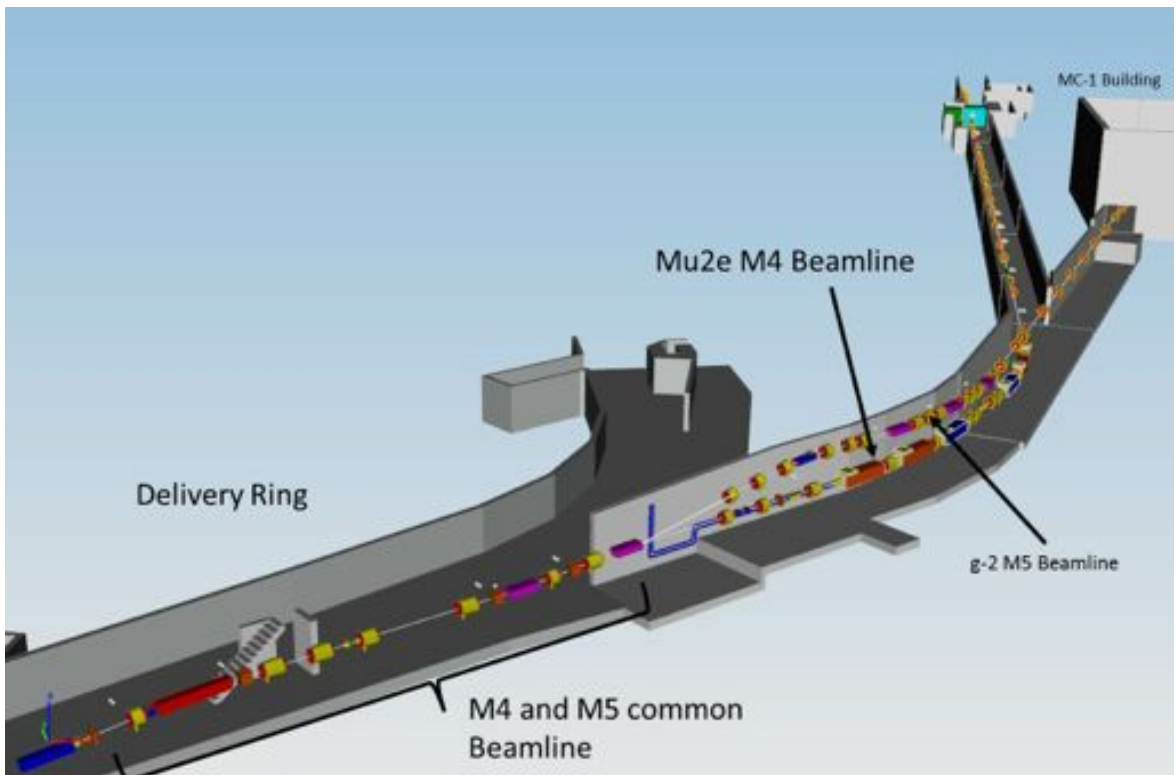

Figure 4.73. An overview of the Delivery Ring extraction region, the M4/M5, M4, and M5 lines.

The resonant extraction ESS modules will provide at least 1 mrad of outward horizontal kick (to beam right looking downstream). The downstream defocusing quadrupole, D2Q4, serves to enhance the effect of the kicker and maximize the beam separation at the entrance to a Lambertson magnet. This kicked beam is then tracked in coordinate space through the Lambertson and D2Q5 with an offset relative to the Delivery Ring's central reference orbit. At the entrance to the Lambertson, the horizontal offset generated from circulating Delivery beam is ~23 mm. The Lambertson is specified to be 1.5 m in length with a 0.8 T maximum field for Mu2e beam, and is located just upstream of D2Q5 (by 0.4 m). It is adjusted to deliver a 38 mrad upward bend for both Mu2e and g-2.

To clear downstream Delivery Ring components, however, additional bend is required. With the Lambertson positioned just upstream of the focusing quadrupole, D2Q5, this quadrupole then acts like a combined-function magnet for the off-center extracted beam and augments the upstream Lambertson kick by 17 mrad. (Since D2Q5 is a horizontally focusing quadrupole, a positive vertical offset generates an upward kick – a kick that is critical to efficient, low-loss separation of extracted beam from the Delivery Ring.) Just downstream of D2Q5 a 2 m-long C-magnet with a 57 mrad bend angle will be required because there is still insufficient separation to insert a full dipole. The combined kick from all three vertical bends (111.5 mrad) allows the extracted beam to clear the next

magnet – the last horizontally defocusing quadrupole in the D30 straight section (D2Q6). The beam pipe in the M4/M5 extraction line clears D2Q6 by a few inches and is 0.524 meter, extracted beam center to circulating beam center at the upstream end.  After beam exits the C-magnet, a "small" 4Q24 type quadrupole can be centered on the extracted beamline just upstream of D2Q6 and represents the first independent quadrupole in the Mu2e/g-2 line.

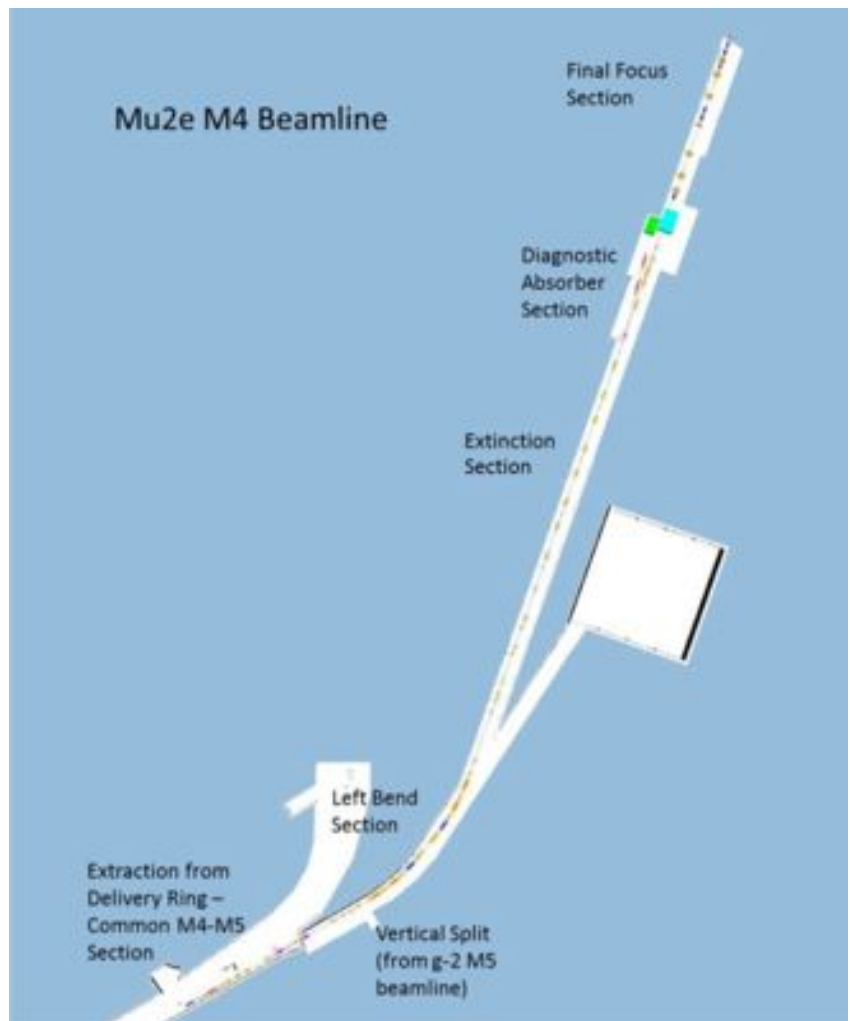

Figure 4.74. Location of the beamline specialty sections.

Once the beam clears the Delivery Ring components, it can be steered onto a centered mid-plane trajectory starting with the C-magnet that begins the combined Mu2e/g-2 or M4/M5 section of the external beamline.  Steering trim magnets have been strategically placed to correct for any differences between the g-2/Mu2e and kicker/septa forms of extraction. The exact extraction orbit depends sensitively on the D30 quadrupole strengths and these depend on the Delivery Ring tunes established for resonant extraction or muon beam delivery for Mu2e and g-2, respectively. It is unlikely these will be identical; however for the optics design described here, the 2004 Run II Debuncher operational strengths were used.

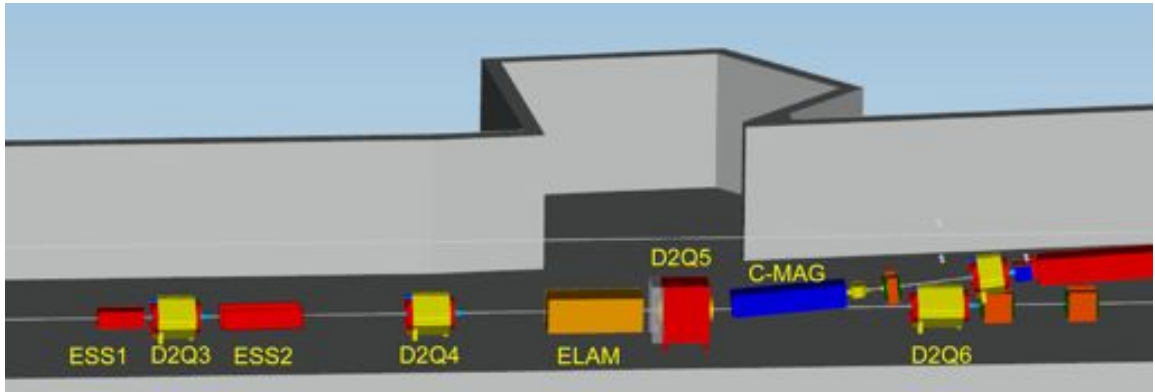

Figure 4.75. Layout of Delivery Ring extraction components.

The initial bend upwards is so strong (to clear the Delivery Ring components) that the beamline must be leveled before the final beamline elevation. This is necessary to allow sufficient space to implement a vertical achromat, which requires significant phase advance generated by quadrupoles. Leveling the beamline reference trajectory at an intermediate elevation allows a straight to be inserted with sufficient space for a sequence of quadrupoles that generate the needed phase advance to cancel vertical dispersion after the final set of vertical bends. Given the still limited vertical clearance, an EDWA type dipole, which has small core dimensions, can be installed after D2Q6 with a bend equal and opposite to the combined bends of the Lambertson, C-magnet, and D2Q5 focusing quadrupole. Leveling the line at 0.8128 m, or ~32", above the Delivery Ring centerline provides for a long elevated "straight" that allows SQ series quadrupoles to be installed without conflicts with the Delivery Ring below.

Downstream of the vertical leveling bend an achromat is implemented using four quadrupoles. This straight section is followed by two MDC dipoles for Mu2e with reverse bends (up/down) that elevate the Mu2e extracted beam to a final elevation of 1.22 m (4') above the Delivery Ring. The final elevation of the Mu2e line is 223.22430 m (732.36' in site coordinates); 3.9' above the M4 beamline enclosure floor and 8.1' from the enclosure ceiling (at 740.5').

For g-2 operation, the last vertical dipole in the M4/M5 section reverses polarity and increases in strength to switch beam delivery from the M4 to the M5 line. The common M4/M5 part of the external beamline thus extends from the C-magnet to the vertical dipole, V907, (the last vertical dipole in the Mu2e configuration) after which the two beamlines are completely separate as shown in Figure 4.76.

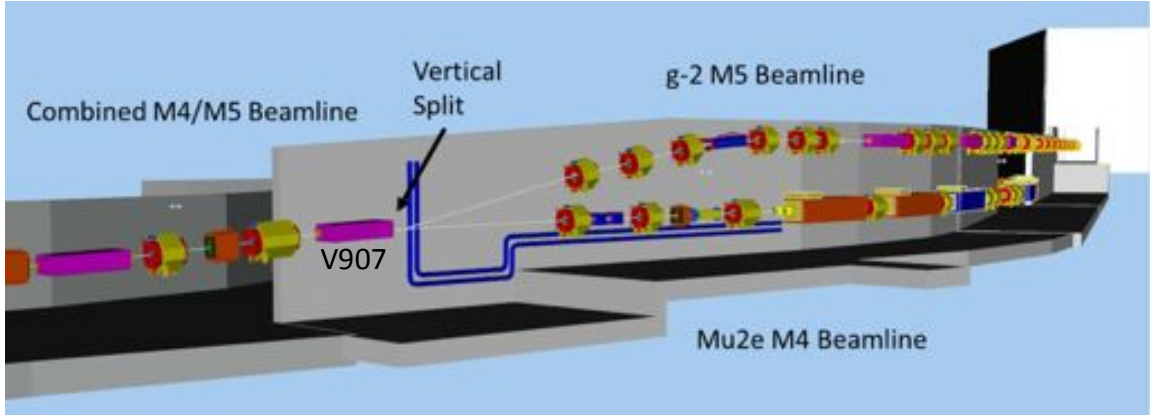

Figure 4.76. Schematic of the separation of the M4/M5 line from the Delivery Ring to the M4 and M5 line vertical separation.

Figure 4.77 displays the achromatic optics of Delivery Ring extraction from the upstream end of the first electro-static septum module (ESS1) through the extraction Lambertson (ELAM) and then to the end of the achromat at V907. Delivery Ring optical functions are assumed to be those that correspond to the Collider Run II optics of the Debuncher. Since extraction is horizontal, the vertical optical functions are likely to be close to their Run II values. However, due to the dynamics of resonant extraction, the horizontal phase space will not be elliptical (see Figure 4.46). Downstream matching sections must shape the beam as required for downstream specialized insertions; i.e. high beta, extinction collimation, and final focus. The horizontal kick of the extraction septa does not generate significant horizontal dispersion so it is neglected at this time. Extracted beam properties will differ significantly between g-2 and Mu2e. Therefore it is important that the two vertical extraction achromats for g-2 and Mu2e are different and can be independently tuned. This is realized in the optics design by employing a different and reversed setting for V907 and by adding an additional vertical bend in the M5 line.

#### 4.8.2.2.2   Left-Bends

Immediately downstream of the vertical section, a strong left bend is required to meet the constraints on the directionality of the beamline and to provide sufficient separation for the shielding needed between the two external beamlines and the g-2 ring enclosure to permit beam operation of one experiment and enclosure access for work on the other. The horizontal separation needed between the Mu2e M4 enclosure and both the g-2 M5 and MC1 enclosures requires that the horizontal bend module be as compact and located as close to the vertical section as possible. The net bend (or bearing) increases significantly with any additional downstream translation of this section (and the g-2 storage ring must also rotate in response) to compensate for the smaller separation distance.

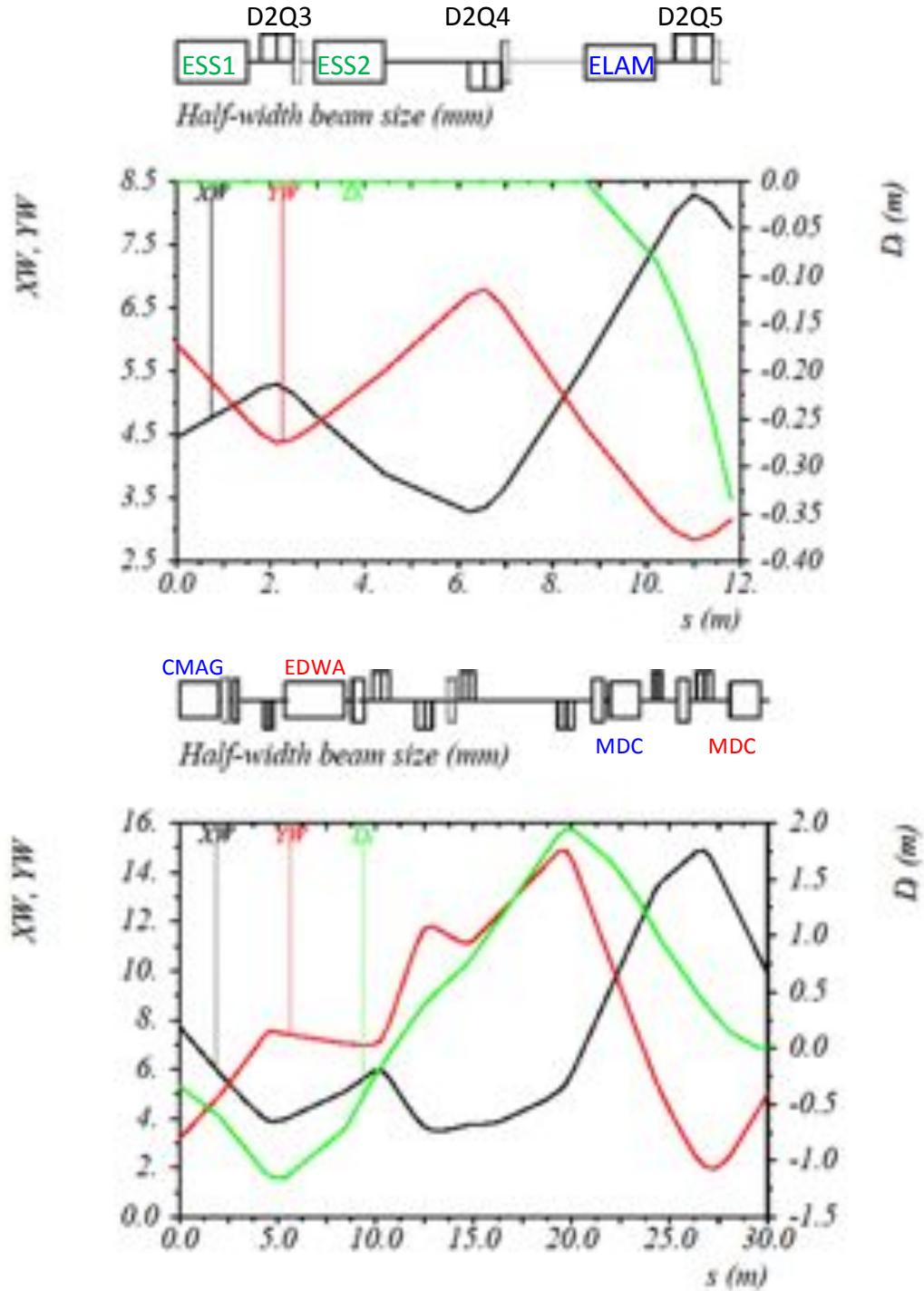

Figure 4.77. Half-width beam size (black: horizontal, red: vertical) through the vertical section and the vertical dispersion function (green). The Top plot is from the septum through the extraction Lambertson and the bottom plot is from the C-magnet through the last vertical dipole, V907. Horizontal right bends are labeled in green, vertical up bends in blue, and vertical down bends in red. Focusing quadrupoles are labeled in black and defocusing (horizontally) are in red.

The present approach employs six bends as shown in Figure 4.78. The first two bends are comprised of two 6-4-120 type dipole magnets powered in series followed by a stronger 6-3-120 dipole to deliver 2 × 6.2° and 7.7° of horizontal bend, respectively. The module has reflective symmetry so that the next three bends are in reverse order, giving a total bend of 40.2°. Outside of this left-bend module there must be no residual dispersion. Therefore, this section must fulfill conditions for a linear achromat. Phase advance and compact dipole placement for dispersion cancellation constrain the optics of this section. Optics matching must occur in matching sections upstream and downstream of the left-bend module.

### 4.8.2.2.3   High Beta AC Dipole Insertion

The optics required to make the Extinction AC dipole (see Section 4.9.2.1) technically feasible pose the most challenging optical design problem for the line. The layout of this section of the beamline is shown in Figure 4.79. This insert resembles to some degree the high beta upstream of a collider interaction region. In this case, however, the horizontal plane must have a very large beta but a small beam size in the vertical (low beta – the vertical beam size is about ±0.5 cm). The high horizontal beta enhances the extinction kick of the AC dipole and the small vertical beam size allows for a smaller dipole gap, thus reducing the required dipole excitation to acceptable power levels.

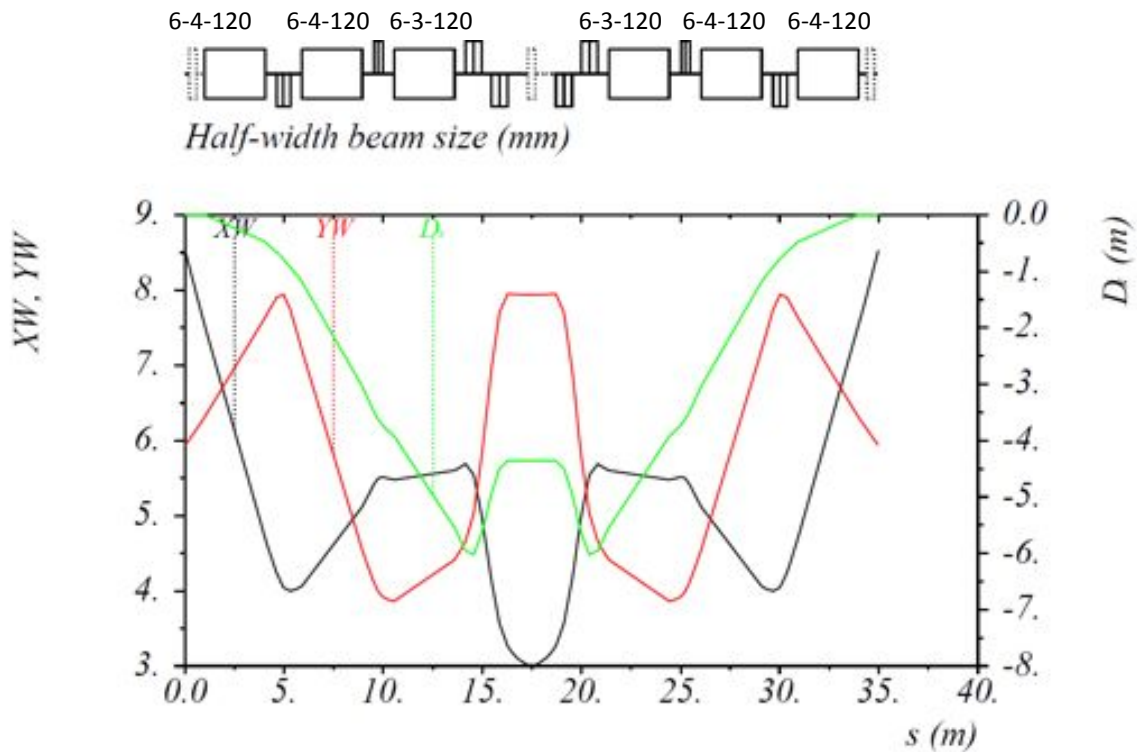

Figure 4.78. Optics of the left-bend section; black: horizontal beam size, red: vertical beam size, green: horizontal dispersion.

The high-beta insertion dominates, and largely determines, the physical length of the beamline. This can be seen in the unavoidable transition from the beta functions generally characteristic of the line to the high beta value at the extinction dipole (see Figure 4.79). Strong focusing is already employed to cause a rapid change in the horizontal beta function in the space of ~20 m. Increasing the focusing makes the high beta increasingly achromatic (chromatic effects are $\propto \beta kl$, with $kl$ the normalized quadrupole strength times its length). That is, increasing the focusing to shorten this section causes an unacceptable change in the beamline optics for off-momentum particles; >5% change in the beta functions for a 1% $dp/p$ change, which is outside of the requirements.

### 4.8.2.2.4   Extinction Collimation Section

The optics of the extinction collimation region is shown in Figure 4.79. Two horizontal collimators are required upstream of the AC dipole to remove high amplitude halo, which would otherwise be deflected *into* the transmission channel by the AC dipole. The first collimator is located 90° in phase advance upstream of the AC dipole, which the second is located just upstream of the AC dipole. The extinction collimator is located 90° downstream of the AC dipole. The design of the collimators and extinction efficiency are discussed in a later section (section 4.9.2.4).

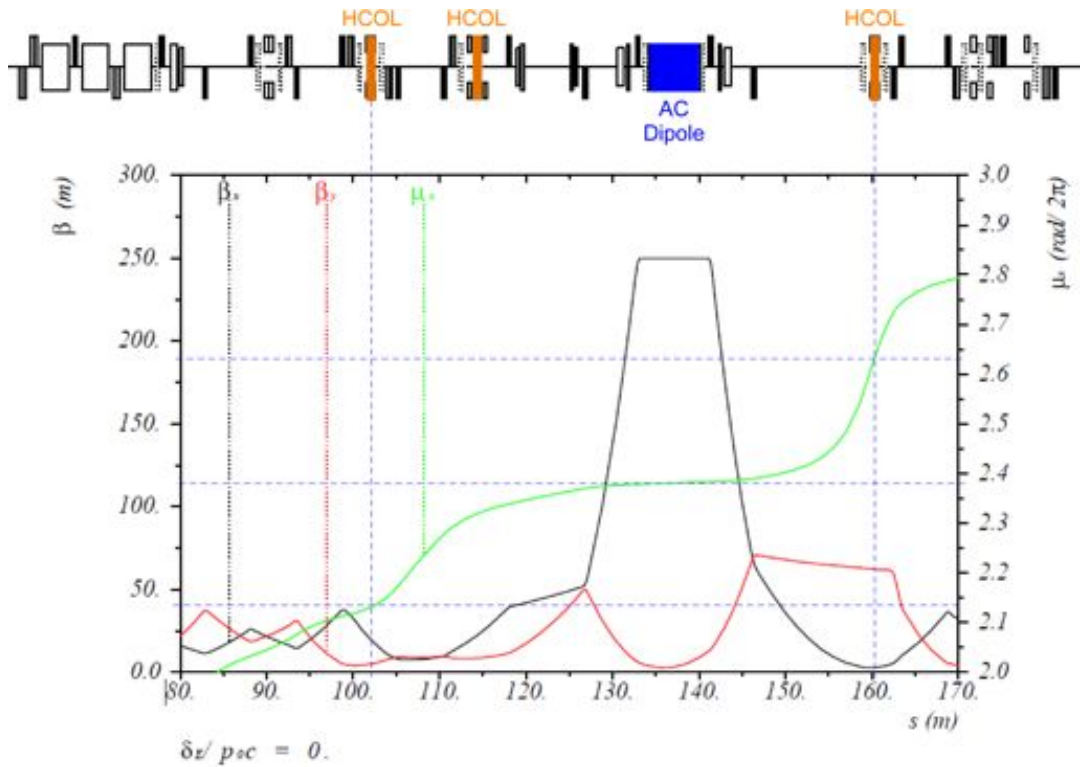

Figure 4.79. M4 beamline Extinction Section optics. Horizontal and vertical beta-functions are in black and red respectively. The horizontal phase advance is in green. The collimators are highlighted in orange above the graph; the AC dipole is highlighted in blue. The blue dotted lines indicate the horizontal collimator positions and $\pm\pi/2$ in horizontal phase advance relative to the center of the AC dipole.

Positioning the beam accurately in the collimator apertures is important; therefore, profile monitors are critical in this section. Centering the in-time beam in the collimators and profiling the beam after the action of the collimators is necessary to establish proper extinction and coordination with AC dipole operation. A multiwire is required downstream of each collimator to verify the centering of the beam as the collimator is closed. Since the optics through the collimator is also critical, an upstream multiwire is also planned.  (Since there is a waist at the center of each collimator, beam properties can be established with only two profile monitors.)  A beam loss monitor will be used in conjunction with the downstream multiwire to more finely monitor and tune the beam interaction with the collimators.

### 4.8.2.2.5   Shield Wall and Diagnostic Absorber

Figure 4.80 shows the layout of the M4 beamline diagnostic absorber and associated shield wall. A CDC dipole located downstream of the collimation section is used to divert beam to a standalone diagnostic absorber. Two spare 3Q120 quadrupole magnets control the optics to the diagnostic absorber (the optics in this line are not critical but diagnostic capability depends on the ability to change phase and observe phase space evolution). The diagnostic absorber design is discussed in section 4.8.5.

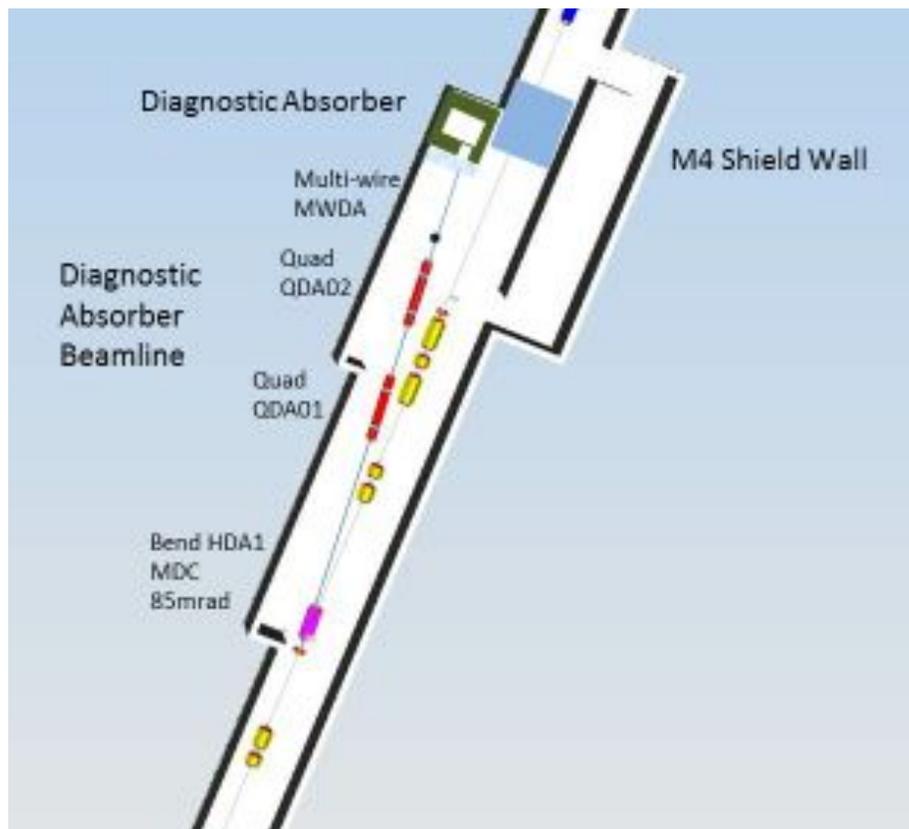

Figure 4.80. Layout of the shield wall and diagnostic absorber.

#### 4.8.2.2.6   Final Focus

The final focus (Figure 4.81) is a typical "collision" style optics region using a quadruplet. A complete module has been designed with reflection symmetry about the outer quadrupole or point-to-point focusing with no demagnification. A profile monitor is installed at the upstream image point to determine the beam profile on target (as no monitor can be located near the target). The telescope focuses the beam to a round, achromatic waist with a 2 m low-beta function at the location of the production target center, which is consistent with the 1 mm rms beam size required by the size of the production target for a normalized emittance of $30\pi$ mm-mrad (95%). Although point-to-point has almost double the chromaticity of parallel-to-point in a final-focus telescope, the chromaticity is still so low that chromatic and geometrical aberrations do not affect the quality of the beam spot size at the target. The target design radius is 3 mm, which corresponds to 2.5 - 5$\sigma$ depending on the extracted emittance. The optics and corresponding beam size are plotted in Figure 4.82.

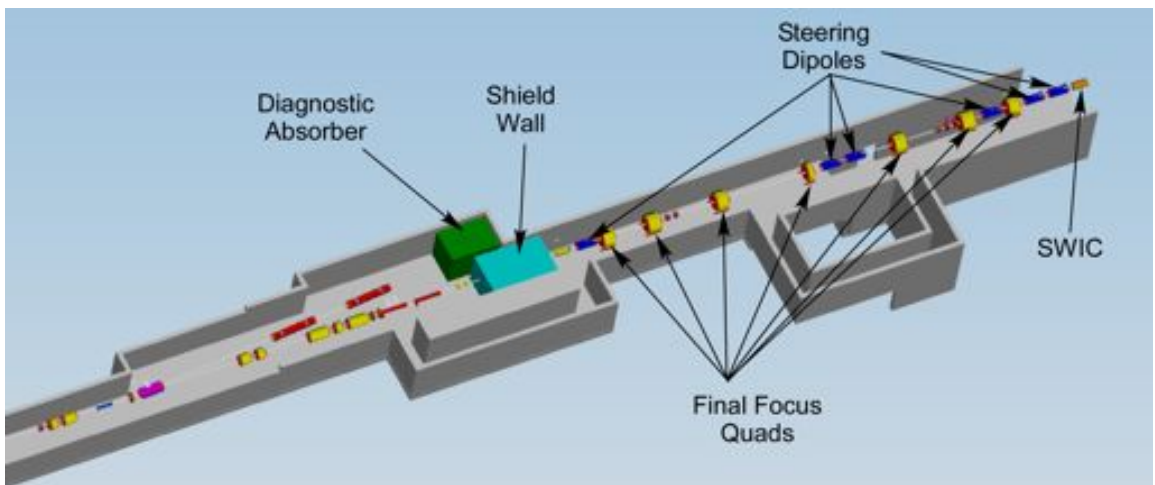

Figure 4.81. M4 enclosure layout showing the diagnostic absorber, the shield wall, and downstream final focus section.

The elevation change due to the vertical down angle required of beam entering the Production Solenoid as well as the concomitant dispersion suppression is accommodated in the final focus section.  A compact final focus and reasonable quadrupole apertures requires the 2.75° final decline to be half initiated further upstream in order to cancel vertical dispersion during nominal operating conditions and not impose additional achromatic constraints on the final focus. The entire final focus section is therefore mounted on a 1.375° slope generated by the first of a pair of 1.375° bend vertical dipoles just upstream of the first quadrupole in the final focus telescope. The module contains a – I transformation as required to cancel vertical dispersion.

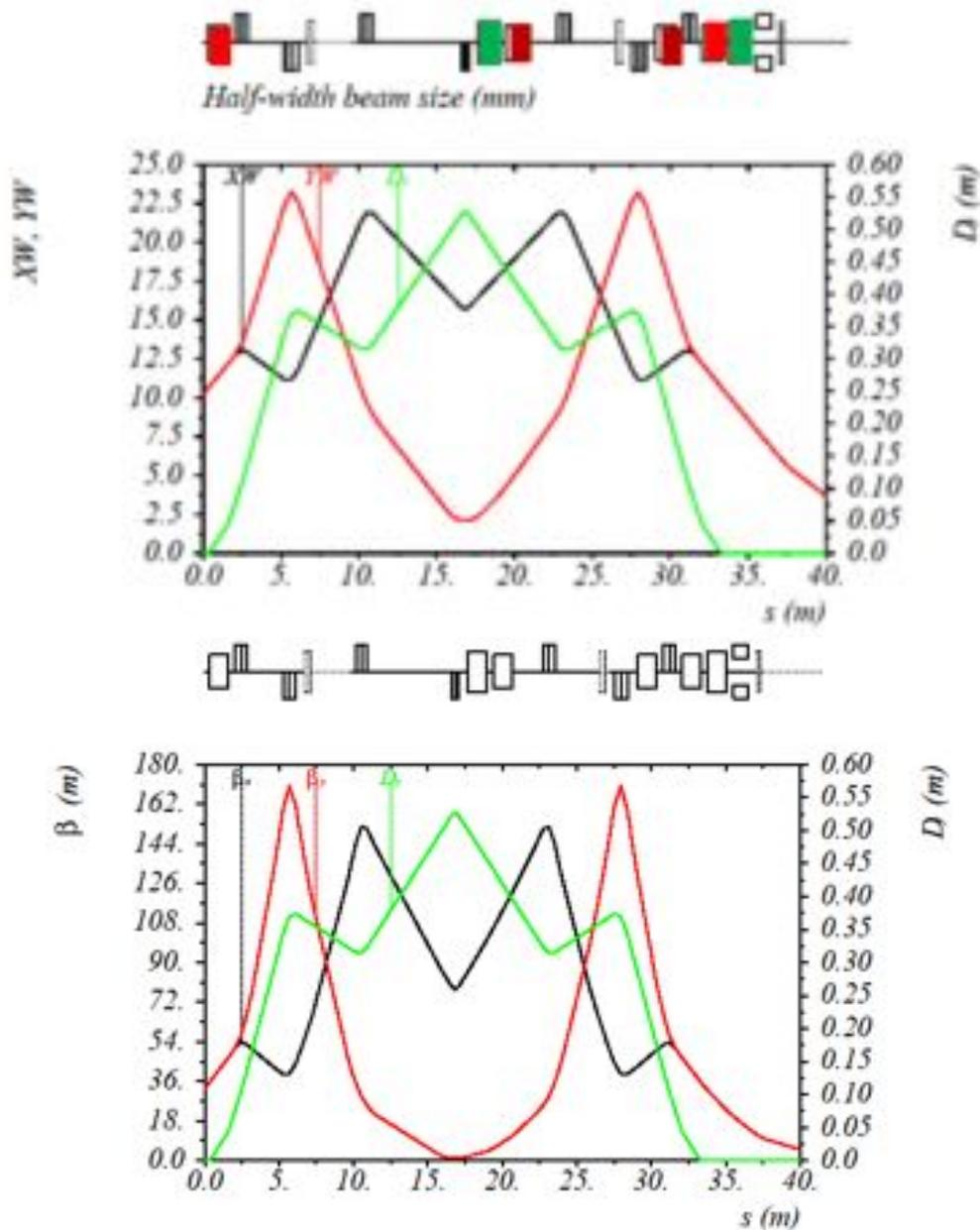

Figure 4.82. Achromatic optics of the final focus. The top plot shows the beam sizes (black: horizontal, red: vertical, green dispersion) and the bottom plot the lattice functions. Light red dipoles are the pair of 1.38° CDA down bends and the dark red and dark green dipoles are the target scan CDA dipole pairs vertical and horizontal, respectively.

The following summarizes the present demands on the final focus:

- 1 mm transverse σ on target (2 m beta function).

- Beam divergence at the target less than 10 mrad.

- A 2.75° vertical decline at the entrance to solenoid – must be achromatic, no vertical dispersion at target.

- Independent position/angle controls in BOTH vertical and horizontal.

- ±1 cm in horizontal/vertical for target scan (see section 4.8.2.2.7).

- ±0.8° in horizontal/vertical for target scan (see section 4.8.2.2.7).

#### 4.8.2.2.7   Target Scans

The steering magnets in the final focus must be capable of wide adjustment in order to scan the target over the required ± 0.8° angular range with no position offset. The required position scanning range of ±1 cm with no angle change is not the technical driver of this capability. This large variation in angle required for the target scan in turn implies large position offsets in the final focus quadrupoles due to the 6.5 m distance between the last steering magnet and the target (space for protection collimator, profile monitor, and trajectory through the solenoid). All of the final focus quadrupoles are therefore chosen from the LQ series (6.25" pole tip aperture). These quadrupoles will be outfitted with star-chambers to extend the transverse displacement acceptance by 1-2" (target scans cannot be executed diagonally).

The layout of the steering magnets is designed to minimize the orbit excursion and to satisfy all aperture requirements in all final focus quadrupoles. The steering magnets must be interleaved not only between each plane but also with final focus quadrupoles for compactness. Since independent position/angle scans require two magnets per plane at different phase advances they cannot be proximate. The steering magnets chosen are cooling ring dipoles (CDA) because of their extremely large aperture in the bend plane (8" good field region). However exceeding the aperture of opposite-bend plane dipoles during a target scan is unavoidable and beam trajectories cannot be contained within their small 3.25" gap.  Movement of intervening opposite-bend plane dipoles during an angle target scan is required to keep scanned beam within their limited gap aperture.  Further, to minimize the apertures required for target scans requires a polarity flip of all of the final focus quadrupoles for a vertical versus a horizontal scan.  With reflection-symmetric optics and round beams, no upstream re-matching is required so this approach does not present an operational problem.  Figure 4.83 and Figure 4.84 show the transverse scan offsets through the final focus.  These offsets exceed the available aperture in the out-of-plane CDA dipoles. Therefore, motion stands are required on the downstream CDA dipoles. The required ranges of motion are ±5 cm on VT951, ±9 cm V952, and ±14 cm on HT952.

It should be noted that although a strong vertical steering capability of ±0.8° is in place, the vertical dispersion cannot be canceled away from the nominal operating point. The steering magnets cannot be positioned in an achromatic configuration without extreme increases in aperture over most of the final focus. Given these constraints, the present final focus design accomplishes its purpose and is the result of optimizing the position of the steering dipoles from the standpoint of minimal quadrupole apertures and offsets in the gap direction.

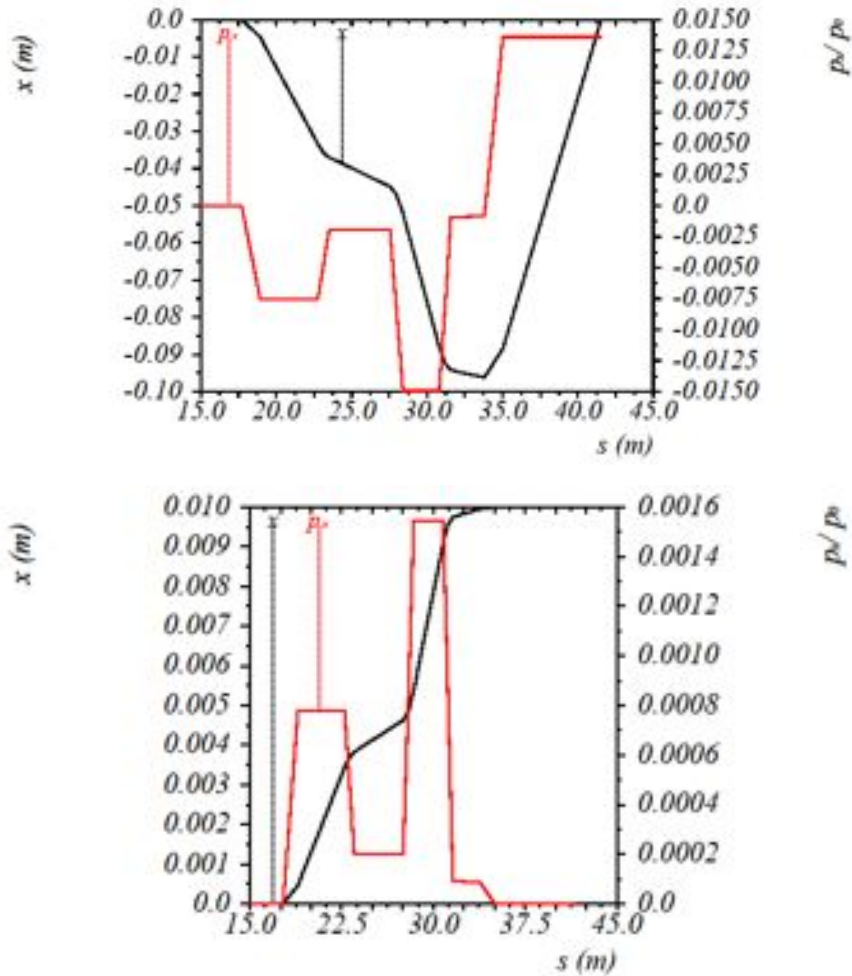

Figure 4.83. Horizontal angle (top) and position (bottom) target scans for +0.8° and +1 cm (the opposite direction just inverts the plot) with position: black and angle: red. Plots begin at the first steering magnet and end at the target position (~41.5 m).

### 4.8.2.2.8   Beamline Optics Performance

The optics of the full line from the C magnet through to the target is plotted in Figure 4.85. The ±1% $dp/p$ momentum performance is documented in Figure 4.86. The line optics is exceptionally stable as a function of momentum. Given the large number of matching sections, phase space changes of at least 20% and probably more can be accommodated and critical sections recovered such as the high beta, extinction collimation section and final focus beam properties on target.

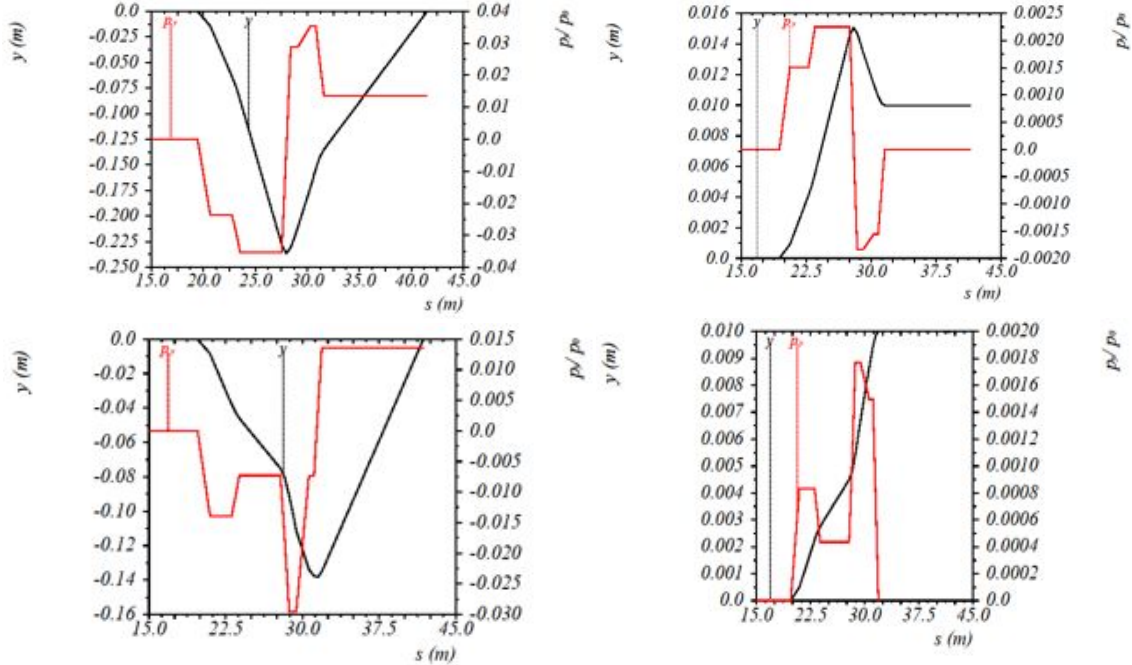

Figure 4.84. Vertical angle (left) and position (right) target scans for +0.8° and +1 cm (the opposite direction just inverts the plot) with position: black and angle: red. Plots begin at the first steering magnet and end at the target position (~41.5 m). Top plots have the nominal final focus quadrupole polarities. The bottom plots are for reversed final focus quadrupole polarities. Note the significant decrease in vertical offset requirement for the reversed polarity scans.

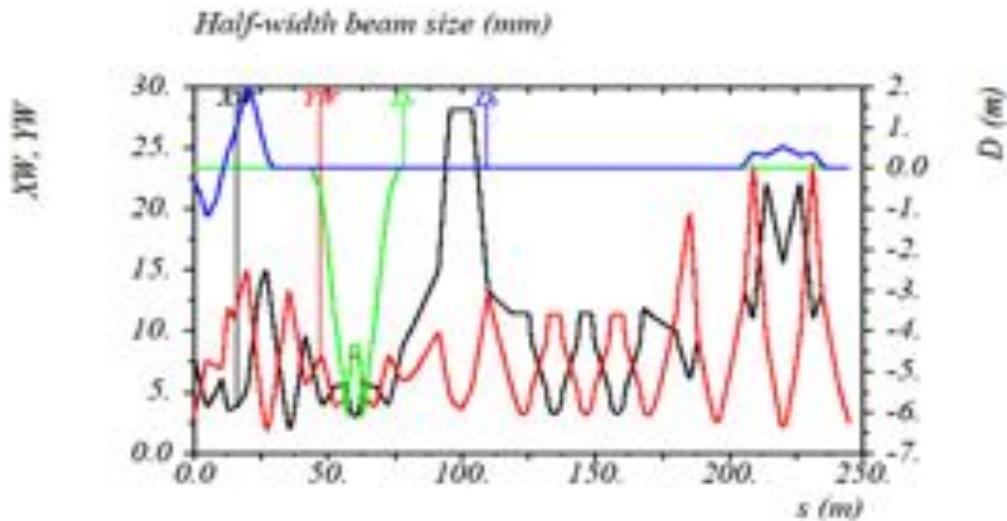

Figure 4.85. Beam size from C magnet to target. Black: horizontal beam size, red: vertical, green: horizontal dispersion, and blue: vertical dispersion.

### 4.8.2.2.9   Steering Errors and Correction

Steering through the collimation section and fixing beam on target is critical. Steering magnets with appropriate strengths have been inserted at critical locations throughout the line with the exception of the extinction collimation section. Due to the tight space constraints a viable steering magnet could not be implemented and the weak NDA type magnets do not provide sufficient bend to impact the beam locally. The extinction section consists of pairs of doublet quadrupoles that are very resistant to beam motion. If aligned properly, and beam is directed correctly in position and angle, then beam will traverse the five extinction collimators correctly. Additionally, the transverse positioning of the collimators can be adjusted. Thus, strong steering is provided upstream and downstream of the extinction collimator string and also upstream and downstream of the AC dipole. Beam stability during the spill is not known but will be measured using approximately 10 time slices on the multiwires and information provided to the extraction system.

Figure 4.87 shows an example of an error correction from position and then angle errors at extraction from the Delivery Ring.

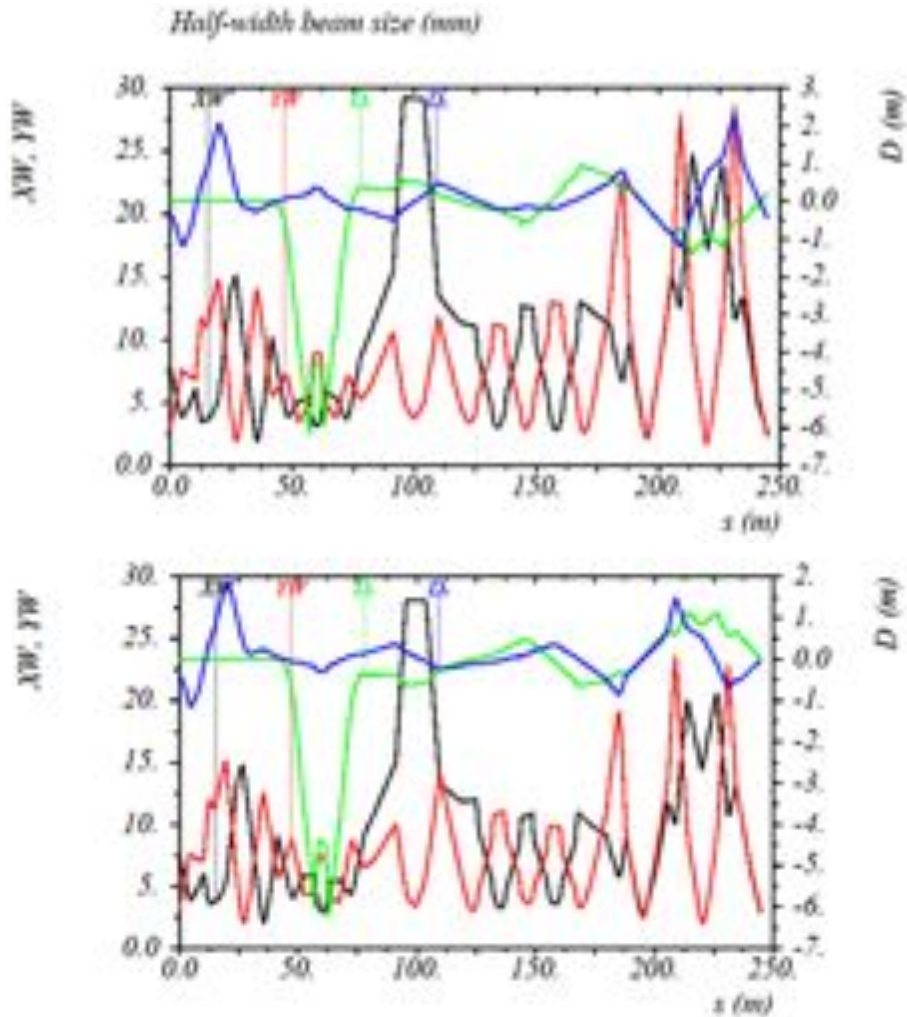

Figure 4.86. Beam size from C magnet to target for a *dp/p* of -1% (left) and +1% (right).

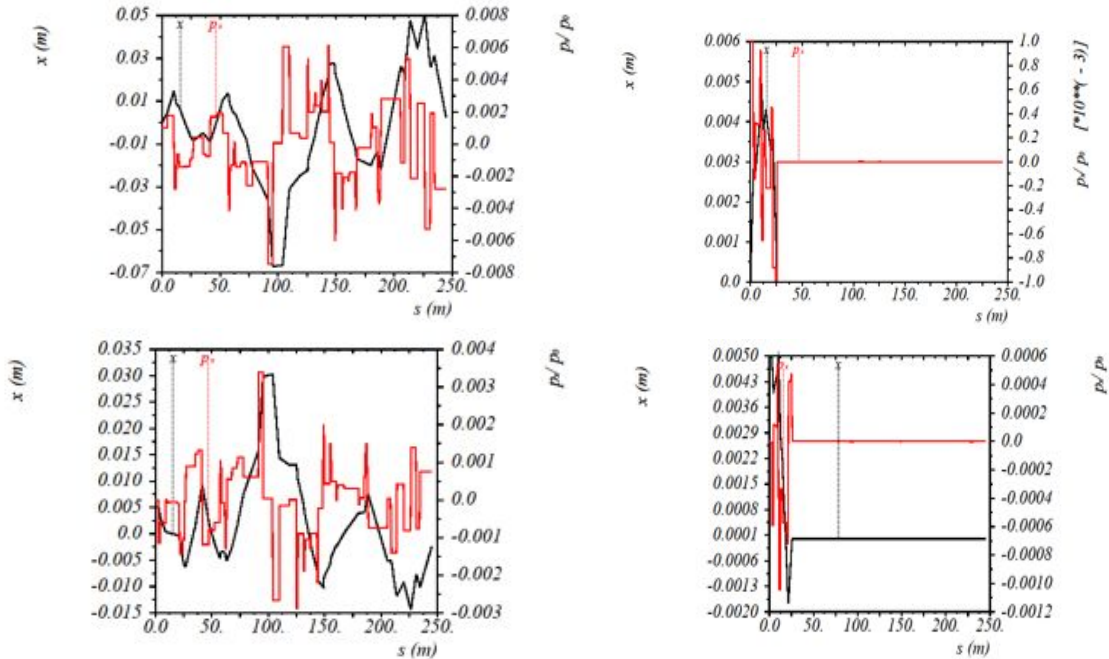

Figure 4.87. Impact of position and angle errors at extraction and correction capability (top: angle offset of 1 mrad and bottom position offset of 5 mm; black: position, red: angle).

### 4.8.2.3  Beamline Optics Risks

The large transverse beam excursions in the quadrupoles of the final focus section during target angle scans preclude any final beam placement requiring simultaneous large horizontal and vertical angle offsets from the nominal trajectory. If such angles are necessary to properly target the beam, magnet moves in the final focus section will be required to center the beam in the final focus aperture at the new nominal positioning.

### 4.8.2.4  Beamline Optics Quality Assurance

A beam-stay-clear region for low-loss operation of the M4/M5 beam line and M4 lines is implemented corresponding to at least one σ. Magnet parameters derived from the optics are translated into operational parameters (mainly currents) and checked against technical data for each component to insure compliance with required operation.  A minimum 10% safety margin in nominal versus maximum routine operation is maintained for all magnetic components. A final beamsheet detailing all component locations is derived directly from the optics design-code (survey) output. This master beamsheet specifies all component locations (as required by the optics design) translated into alignment coordinates consistent with existing alignment data (Delivery Ring components and enclosure) and new civil and experiment alignment data. This beamsheet is also the reference for mechanical design and all of the beam line drafting layouts.

### *4.8.2.5   Beamline Optics Installation and Commissioning*

Beamline optics and performance will be verified utilizing strategically positioned profile monitors throughout the beamline, which can be placed in the beam singly or together to verify optical parameters. The phase advance between monitors is such that non-elliptical phase space distributions generated by resonant extraction can be accurately determined during commissioning. To ensure compliance with the beam size requirement at the target, a specific monitor has been located at the upstream image point (no demagnification, a -I transformation). No profile monitor is feasible near the target.

A beam commissioning plan is under development and commissioning will occur at reduced intensity utilizing the diagnostic beam absorber line. All profile monitors can operate with a good signal to noise/background ratio up to two orders of magnitude less than the experimental spill intensity of $3 \times 10^7$ protons/bunch.

## 4.8.3 Beamline Components

Almost all of the components are magnets repurposed from the Accumulator ring (SQ and LQ series quadrupoles, MDC dipoles, and NDA and NDB steering trims) and other available magnets (cooling ring dipoles and VDPAs where stronger steering than the ND series trims is required). Many of the stands can also be reused with their accompanying quadrupole with a simple height compensator. No longitudinal adjustment is presently incorporated into the existing Accumulator stands and is not planned at this time for the M4 beam line. A few stands, particularly those that require motion control, will be new designs.

### *4.8.3.1   Beamline Components Requirements*

- **Integrated quadrupole strengths up to 11.2 T (13.5 T/m maximum gradient)**. Strong focusing quadrupoles are required in a number of sections and particularly in the final focus region to generate the low beta at the target.

- **Large, 8" aperture quadrupole to replace current D2Q5 in the Delivery Ring**. A quadrupole with an 8" aperture in the current D2Q5 (LQE) location is required to accommodate the vertical and horizontal offsets of extracted beam.

- **Large aperture quadrupoles in final focus for target scan**. Target scan specifications require up to 14 cm (5.5") offsets in one plane.

- **Dipole strengths up to 4.14 T-m (net bend of 8° at 8 GeV).** This bend is required to make the left-bend string and associated module compact.

- **Main Steering Dipole Correctors 0.01 T-m.** A bend strength of ~10% of a main left-bend dipole is required for steering corrections in this section. Other areas requiring strong steering correction to offset strong quadrupole steering or center on an aperture restrictions is the high beta region vertically and the long bore through the shield wall.

- **Steering Trim Dipoles ~1 mrad at 8 GeV.** Other sections require less steering strength to compensate alignment or field errors.

- **Stand height adjustment of ±1".** This tolerance allows for normal variations in the enclosure floor.

### 4.8.3.2    Beamline Component Risks

#### 4.8.3.2.1    Insufficient availability of SQ series quadrupoles
There may be an insufficient number of SQ series quadrupoles from the Accumulator to populate all the required Muon Campus beamlines.

#### 4.8.3.2.2    Required range of motion in target scan dipoles is too large
Another potential risk is the large range in motion of target scan magnets in both stand and inter-component connections (such as bellows).

#### 4.8.3.2.3    Beamline component performance in the fringe field of the Production Solenoid
A final risk includes the ferromagnetic properties of the protection collimator and bayonet SWIC in the strong fringe fields of the solenoid and whether any monitor will work in such fields and also the high radiation environment.

### 4.8.3.3    Quality Assurance
All components will be installed per the master beamsheet, which will be reviewed and checked by the alignment group, civil engineers, and mechanical engineers independently. This beamsheet is also confirmed in the master (drafting) layout of the line which includes not only all components, beampipe, and connections, but also all enclosures and potential conflicts (in 3D). Alignment data is provided after installation for final confirmation and approval.

Below is a checklist of QA plans to promote trouble-free execution of the project.

- All of the SQ and LQ type quadrupoles, steering trims (NDAs, NDBs), and left-bend magnets (6-3-120 and 6-4-120 type dipoles) repurposed from Accumulator will be preserved in operating condition within the Delivery Ring tunnel in a controlled environment.

- LQ quadrupoles outfitted with star chambers will allow the large horizontal or vertical offsets required for the target scans, increasing the offset achievable from 3.3" (half pole-tip aperture) to the required 5.5".

- External components, the VDPAs from the recycler will be taken to Technical Division for testing and minor refurbishment.

- CDAs (cooling ring dipoles) presently in storage will require coil refurbishment and complete testing.

- Components will be moved without the beam pipe inserted to avoid damage during rigging.

- Staging of major components in order of installation to minimize rigging and interference in preparation for eventual installation.

- Standard spool pieces are being implemented for each type of large magnets to insure adequate inter-magnet spacing for installation and also magnetic field isolation between components. Vacuum and leak testing of spool pieces will be performed in shop optimizing throughput and minimizing tunnel work. Flanges and bellows will be reused from Accumulator.

- Any components installed in the fringe field of the solenoid cannot be ferromagnetic. This includes the protection collimator, beampipe, SWIC and vacuum window. Ferromagnetic content will be predetermined before acceptance.

- Maintenance work required on Accumulator magnets directly transported to M4 tunnel will be identified and addressed after installation with routine water flow, high-pot and electrical testing. Outside of a catastrophic failure such as a coil, maintenance and repairs can be made in place in the tunnel.

- Upon arrival, stand parts will be quality control inspected. After assembly all welds will be inspected and verified.

- All diagnostics and motion stands will be tested prior to installation in the tunnel.

- Load testing of all the ceiling drop-in Hilti's and ceiling-hung stands will be done and documented. Civil engineering (FESS) approval will be requested in all situations that involve connections to concrete, floor loading, and concrete strength determinations.

### 4.8.3.4   Beamline Components Installation and Commissioning

Installation is carefully planned to minimize conflicts and unnecessary rigging and handling by staging the magnets in the correct installation order in the beamline tunnel. Cable tray installation, cable pulls and pipefitters for tunnel LCW lines can be performed in scheduled maintenance shutdowns. Stakeout of stands and stand installation can also be performed during shutdowns. Rigging of major components will follow after utility installation to avoid interferences. Spools will be installed in major components prior to installation on stands for ease of installation and to expedite interconnections. Rough alignment of major components will be scheduled and once rough aligned, minor components and diagnostics and their respective stands can be installed. Installation of minor components and beampipe can follow the rough alignment sequentially. Final alignment and as-founds for all minor components (steering trims are not normally installed on precision stands; alignment tolerances are much less restrictive.)

Specifications and installation procedures will be written to ensure that the parts and installation meet all technical requirements. Additionally, quality control checks will be done on the existing equipment to ensure that the parts are handled and checked during installation. Additionally, punch lists will be used during the installation so that all necessary steps are completed

Once installation of all components is complete roughing pumps must bring vacuum to below $5\times10^{-8}$ Torr to start ion pumps; leak testing will occur if necessary. Once ion pumps stay on the tunnel can be secured and all components exercised via the control system. A power-on access (usually at reduced current) will be performed last to check polarity and fields prior to beam delivery.

## 4.8.4 Magnet Power Supplies

### 4.8.4.1 Magnet Power Supply Requirements

Magnets, operating currents, regulation requirements and service building locations have all been described and specified in detail in reference [100], [101] and [102]. These documents also describe calculations of the power and cable requirements necessary to supply the necessary current to the beamline magnets. In addition to the power requirements, the regulation requirements are also specified. This defines the quality of the current feedback device (DCCT) and DAC needed for each system. These supporting requirement documents are used to group power supplies by size to reduce the number of unique units needed and to minimize the number of spare supplies that needed to be kept on hand.

### 4.8.4.2 Magnet Power Supply Technical Design

#### 4.8.4.2.1 System Layout

AD E/E Support has designed and developed a current regulation and controller system that is used in DC applications. The control portion of the system uses a PLC to manage input power, status read back and control to the accelerator controls system (ACNET). A variety of commercial power supplies are used in voltage mode to provide current to the magnets. This requires an interface circuit to convert the status and control to signal levels that can be provided to the controls system. Two versions of the interface chassis exist and can support 8-16 power supplies.

Each current regulator uses a PC-104 embedded micro controller to provide current regulation for up to four power supplies. This controller also collects the status and control information from the power supplies via the PLC and converts this information into a format that the ACNET controls systems can present to operations. Figure 4.88 shows a block diagram of the current regulator and control system.

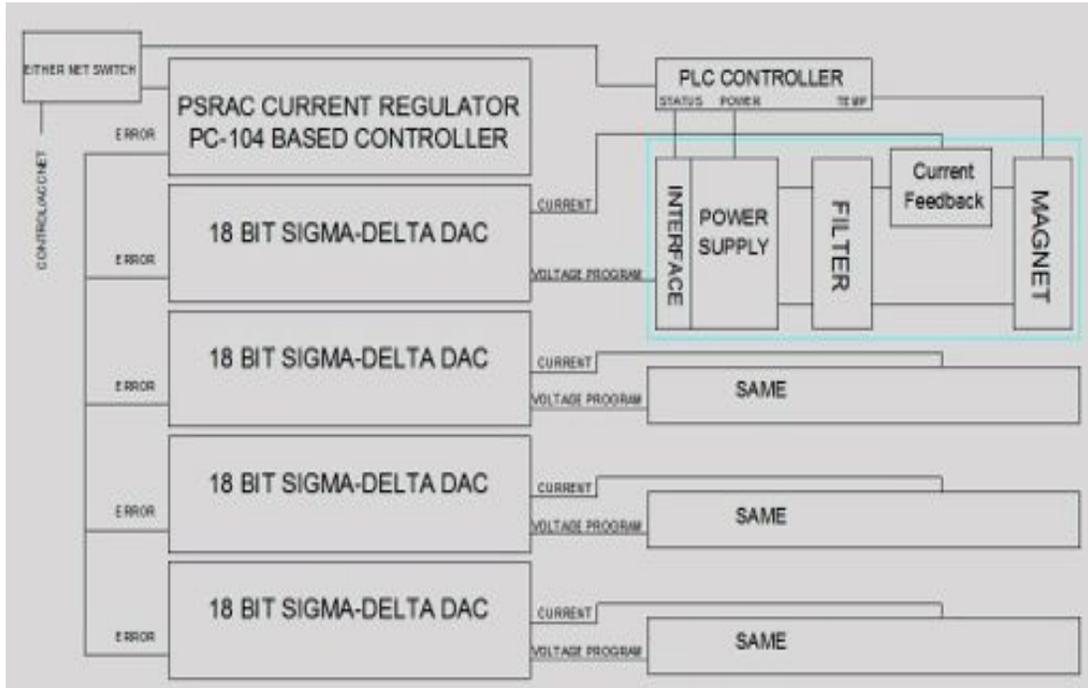

Figure 4.88. Block Diagram of the Current Regulator used for beamline power supplies.

### 4.8.4.2.2    Current Regulation

There are two items needed for a reliable current regulation system. One is the current measuring device and the other is a stable current reference. Each system will have a total current monitoring DCCT (DC Current Transformer) used to provide accurate and stable current feedback to the regulator.  The current regulation system will be the same design that was constructed for the last Main Injector (MI) and Ion source upgrades. This system is a Digital/Analog combined regulator built using a PC-104 embedded processor system that regulates current by providing a total voltage reference to the power supply. This regulator supports four power supplies in a single chassis and provides all of the voltage drive through the power supply for the main current, including any correction needed. The DCCT's used for the feedback system will be commercial devices. This regulation system can support operation at the $\pm 4$ ppm level but can be improved to the $\pm 0.25$ ppm level by procuring a high performance DCCT.  For power supplies in the Muon Campus upgrade it is expected that a system that provides regulation at the level of $\pm 300$ ppm will be used.  This will allow for a much lower cost current feedback device.  This system is not intended for fast ramping power supplies and has a *dI/dt* limiter to be used during startup.

### 4.8.4.2.3   Operation

The current regulator has one PC-104 processor that sends an 18-bit digital reference to each of four Sigma Delta DACs. Each DAC module receives the analog current from a DCCT/HAL probe and converts the current to a voltage using burden resistors in the temperature regulated module. The reference and current signals are subtracted to generate an error signal, which is then amplified 100×. This amplified error is then sent to the PC-104 module that adjusts the drive to the power supply to minimize the error signal. The amplification is employed to increase the sensitivity of the PC-104's AD converter; looking at small signals with only 16-bits is limiting – more bits are used for larger signals. The magnet parameters are loaded into the PC-104 along with the current loop bandwidth and maximum gain limit. This information is then used to provide the correction to each power supply.

An additional feature of this system is that the PC-104 has a built in Transient Recorder that will record trip events and provide data for analysis. This will be useful for the identification and diagnosis of infrequently occurring single event trips.

A second built in option for this regulator is a window detector that can be set up to monitor current, DAC settings, and the current error signals to ensure a proper set range. These limits are set up using an independent path into the processor. These type of limits are used in the MI to prevent beam transfers if the signals are not in the correct operational range but can be used as an indicator that one of the circuits is out of the operational range for the source. The PC-104 monitors and uses four analog signals, Current Reference, Current, Voltage, and Current error. These signals are stored in the Transient Recorder during a trip and can be plotted at 1,440 Hz using the standard Fermilab accelerator console Fast Time Plot facility.

### 4.8.4.2.4   PLC Controller

All of the control and status read backs are provided through the PC-104 current regulator using a single E-net connection to the ACNET system. The power supply system uses a Programmable Logic Controller to collect data from four power supplies over the E-net connection and passes it through the PC-104 to ACNET. The information is collected in the PC-104 that converts it to an ACNET format. The PC-104 also provides the ON, OFF and Reset functions to the power supplies through the PLC that manages common elements to all four supplies. The PLC allows for level shifting of signals from the many different types of power supplies that will be used. All of the signal and control information is provided to the individual power supplies with an E-net cable back to the controls system. This has the benefit of reducing the amount of controls cards for collecting data and the cabling that is needed to connect to the cards.

All of the power supplies will need monitors and control of the 480 VAC input power. The PLC is used to manage the control of this power and monitor signals common to all including safety systems, door interlocks, smoke detectors and magnet temperature. The PLC interfaces the 480 VAC input power to the supplies using a custom-built starter panel with a limit of 40 kW. Two of these 40 kW starter panels can be installed in a standard relay rack and power two supplies or groups of supplies. The 40 kW power limit allows up to four 10 kW power supplies in one half of a rack to reduce the amount of floor space needed for power supply racks. The Electrical Safety System (ESS) prefers to have a direct connection to line power to trip off supplies without passing through logic or programmable devices. The starter panel provides an interface for the ESS connection.

### 4.8.4.2.5   Interface Chassis

The PC-104 will be used to regulate current for many different sizes and types of power supplies. Some regulation systems use TTL for status and control while some use +24 VDC. Therefore, the interface chassis and cards provide a place to convert signals that can be sent back to the controls system. Using the PLC and interface chassis allows the PC-104 code to be identical for all systems, though some of the PLC code will possibly need to be unique.

### 4.8.4.2.6   Switch Mode (SM) Power Supply

The plan is to group SM power supplies by size to reduce the number of different sizes and the number of spares required. The specification for the SM style power supplies will define the voltage, current and power level for each size as well as a voltage regulation and ripple. Sharing a common line voltage for all supplies is desired in order to mix different power levels in a single rack. There are three or four manufactures in the US that can meet all of the specified needs. Figure 4.89 shows an example of a switch mode power supply that would be similar to that planned for the M4 beamline.

### 4.8.4.2.7   SCR Style Power Supply

The specification for the SCR style power supplies will be based on the present design of the 75 kW power supplies used in the MI. This specification requires the use of the FNAL AD E/E voltage regulator. Accelerator Division E/E Support Department will have these voltage regulators constructed and two copies are provided to the manufacturers to use for testing. SCR supplies in general support two quadrant operation. This capability is not needed in this installation but to improve the inventory of supplies this requirement will be retained in the specifications. Manufacturers of modern SCR power supplies use PLCs internal to the equipment rather than constructing custom circuit boards to provide control connections. The specification for the power supply will include the detailed information needed to ensure that any PLCs that might be used are compatible with maintenance tools on hand. Figure 4.90 shows an example of an SCR style power supply that is similar to that planned for the M4 beamline.

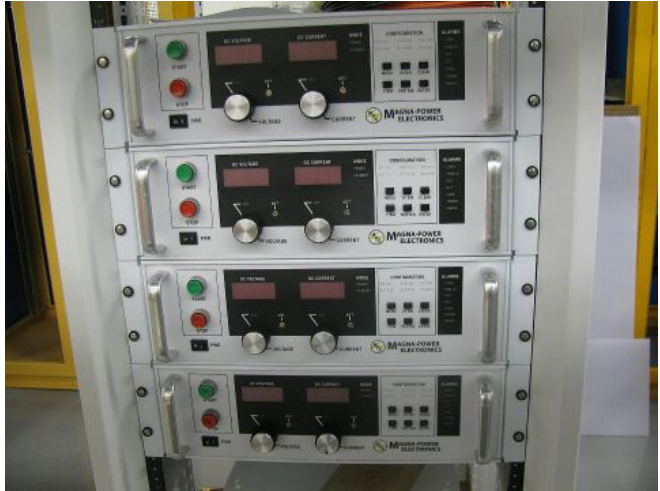

Figure 4.89. Example of a Switch Mode Power Supply.

### 4.8.4.2.8   LCW Cooling System

SCR supplies will need LCW cooling water for at least the SCR's and possibly the magnetics. The LCW cooling water will be sourced from the Central Utility Building via the M4 and M5 beamline tunnels up to the MC-1 and Mu2e buildings. 50 kW of LCW cooling are required with a minimum pressure range of 60-100 psi and a flow of 18 gpm in each service building solely for beamline power supplies. The limited number of manufacturers for equipment of this type will limit the number of offers and may have a direct influence on the pricing. When procuring the power supplies, a spare power supply that can work in multiple locations should be chosen to eliminate the need to procure a spare of every type.

### 4.8.4.2.9   Power Supply Electrical Infrastructure and Layout

The M4 beamline magnets and tunnel components will be powered from three of the Muon Campus service buildings. The service buildings will include the AP30 service building, MC-1 and Mu2e experimental building. The MC-1 and Mu2e buildings will be newly constructed and the electrical power distribution infrastructure will be sized according to power load demands. The MC-1 building will house magnet power supplies for the both the Mu2e and g-2 experiments and for both the M4 and M5 beamline power supplies. The Mu2e building will house additional power supplies for the M4 beamline Modification of the electrical distribution infrastructure will be needed, including new fan packs installed on the distribution transformer, upgraded cables on the transformer secondary and new upgraded switch gear to meet the increase in power demands. AP30 and its magnet power supplies will be sourced from the 1500 kW building transformer on feeder 42.

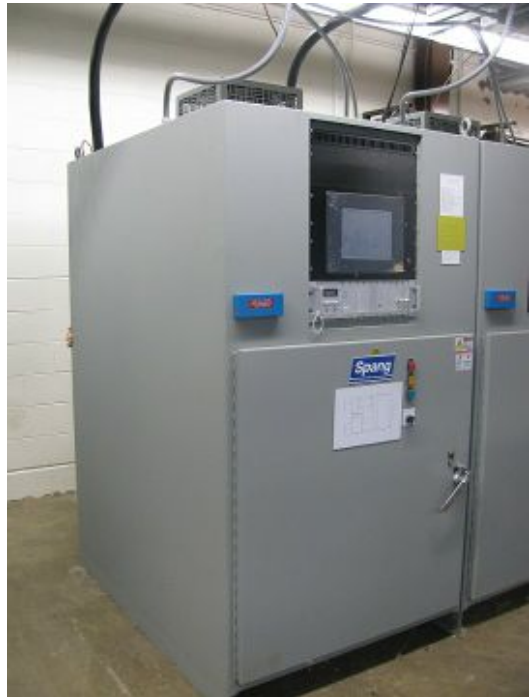

Figure 4.90. Example of an SCR type Power Supply.

### 4.8.4.2.10  AP30 Service Building Power Supply Layout

Figure 4.91 depicts the layout of the beamline power supplies and the power distribution in the AP30 service building. The Mu2e M4 beamline power supplies will power the first magnet in the M4 beamline proper. The Q908 quadrupole is the first magnet circuit downstream of the Mu2e/g-2 vertical dipole switch at location V907. Quadrupoles Q908 though Q918 are powered using seven switching mode (SM) type power supplies (Section 4.8.4.2.6).  Dipole magnets H910, H911, H912, H916, H917 and H918 are powered from a phase controlled (SCR) (Section 4.8.4.2.7) 500 kW power supply relocated to AP30 from the F27 Main Ring service building.  The eight power supply circuits sourced from AP30 will require a total of 24 500 MCM load cables and operate at an average power of approximately 271 kW.

On advisement of the Accelerator Division's Electrical Coordinator, and the need to route correctly sized triplex tray cable (type TC) AC power feeders to the Mu2e M4 beamline power supplies, all out-of-service feeders and load cabling will be removed from the uppermost power cable trays.  This task essentially clears both trays for unencumbered multiple-cable installation pulls.

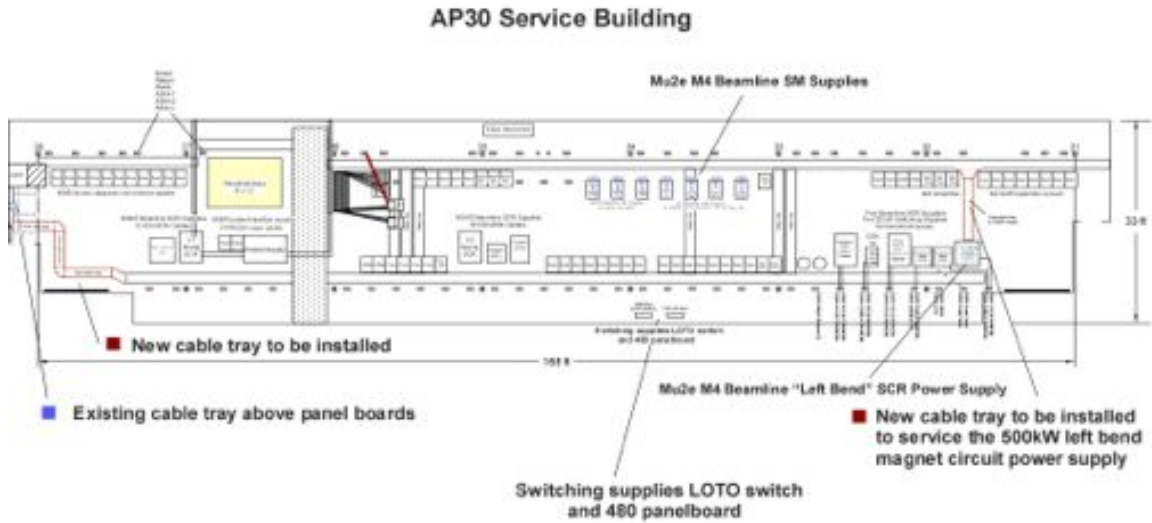

Figure 4.91. Layout of the AP30 Service Building

Located along the south wall of AP30 are the 2000 Amp DHP-AP-30-1 (Distribution High-Power) and 600 Amp PHP-AP-30-2 (Panel High-Power) panel boards that will be used to source the Mu2e M4 power supplies. To cleanly make the transition from the panel boards to the existing cable tray on the east side of the structure an additional cable tray system will be installed. This new section of cable tray will be quite full near the panel board area, but will become less dense as the cables travel northward through the tray and drop out to their various load locations. In an initial design, all AC feeders for the M4 beamline SCR power supplies, sited at the northeast end of the building, passed through this section of tray. However, the 500 kW left bend PEI supply will now have its three 350 MCM triplex AC power feeders routed along the inside wall power cable tray due to fill constraints and improved ventilation. A dedicated cross-over cable tray will be installed for access to the left bend 500 kW power supply.

In addition to the AC feeders mentioned above, a new 400 Amp panel board (PHP-AP-30-4) and LOTO disconnect switch sourced from the DHP will be sited midway along the east wall of the building to power the M4 beamline quadrupole switching-mode power supplies. The specification calls for a 14-position Square-D I-Line panel board. A typical Anti-Proton (AP) service building type Unistrut assembly will secure the equipment into position. A length of cable tray will be attached to the wall for access to the crossover tray for AC feeders to the power supply racks sited on the Accumulator ring-side of the building. Standard tray cable will be used in this installation.

### 4.8.4.2.11  MC-1 Building Power Supply Layout

As mentioned earlier, the MC-1 building's electrical infrastructure is new equipment included as part of the building construction. The Mu2e M4 beamline shares a transformer with the g-2 experiment's M5 beamline.  The power distribution is designed to power both the M4 and M5 beamlines if needed. Figure 4.92 shows the MC-1 power supply room layout as well as the electrical distribution.

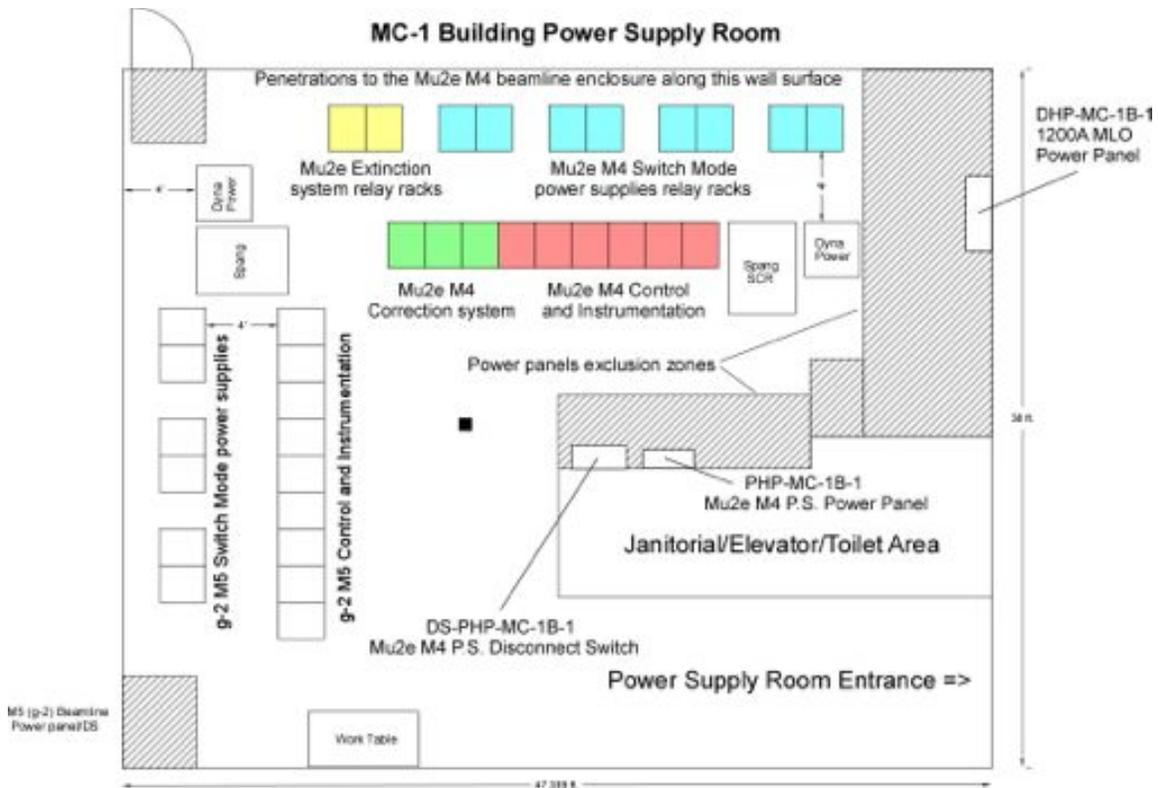

Figure 4.92. Layout of the MC-1 power supply room.

The transformer for the external power supplies sources a 1200 Amp MLO (Main Lug Only) panel board DHP-MC-1B-1 (Distribution High Power) located on the north wall of the power supply room. The panel has positions for four 400 Amp breakers. The Mu2e M4 beamline will use one of these breakers to source a 400 Amp LOTO disconnect switch (DS-PHP-MC-1B-1) and 400 Amp panel board PHP-MC-1B-1. This arrangement is consistent with the Accelerator Division's one-point LOTO system to de-energize a section of beamline or ring and is found in all new construction of this type.

The section of the Mu2e M4 beamline to be powered from the MC-1 building begins at quadrupole Q919 (the first magnet beyond the left bend dipoles) through Q945. It also includes the diagnostic absorber section dipole switch circuit D:HDA1 and absorber beamline quadrupoles D:QDA1&2 for a total of 21 power supply circuits using 40 500 MCM and six 1/0 load cables operating at an average power of 155 kW.

All but one of the aforementioned power supply circuits will be of the switch-mode type. The diagnostic absorber dipole switch circuit will use a phase controlled (SCR) power supply. Standard sized relay racks will be used for the switch mode power supplies, controls, communications and instrumentation. Each rack will have at least one single phase 120 V, 20 Amp power plug strip available.

A standard 4 inch square wire way, sourced from power panel PHP-MC-1B-1, will be used to route all AC feeders to the various switching mode power supply starter panels. This technique greatly simplifies the installation and reduces overall cost. The wire way attaches to the cable tray Unistrut support system and has flexible conduit drops to each relay rack as needed.

All SCR power supplies for the Mu2e M4 beamline will use appropriately sized electrical metallic tubing (EMT) for their AC feeders sourced from the same 480 VAC panel board as described above.

Having all the beamline supplies powered from a common panel board and safety disconnect switch provides a safe and convenient single-point LOTO isolation for this section of the M4 beamline.

### 4.8.4.2.12 Mu2e Building Power Supply Layout

The Mu2e building will have new electrical distribution equipment provided as part of the building construction. A dedicated transformer will drive the power supplies for the final focusing section of the Mu2e M4 beamline that will be located along the northeast wall of the high bay area. Figure 4.93 shows the Mu2e building high bay layout.

The seven quadrupoles, Q946 though Q952, are powered using four switching mode power supplies and the six dipoles, V945 through HT952, are powered using six switching mode power supplies for a total of ten magnet circuits sourced from the Mu2e service building. The ten power supply circuits will require a total of 62 500 MCM load cables and operate at an average power of approximately 194 kW.

Eight of the ten switching mode power supply circuits will have polarity reversing switches for bi-polar control. The polarity reversing switches will be recycled from the P150 beamline used during Collider running. In this arrangement the load cables exit the relay racks from the bottom through penetrations to the enclosure.

Two six-inch PVC ducts per rack can handle a total of eight load cables per penetration. A typical rack layout includes a Spang 1000 Amp, 500 V reversing switch placed near the bottom to facilitate cable termination and a 9U sized 45 kW switching power supply above the switch assembly.

Figure 4.93. Layout for the Mu2e building high bay showing the location of M4 beamline power supplies. The devices in the red oval are the M4 line supplies.

Two of the final focus quadrupole magnet circuits operate at an average of 2 and 4 kW and will be paired together in the same relay rack. This provides one dedicated rack from the set of five paired racks for power supply control and instrumentation in a central location with space available from the supplies rack mentioned above. The power supplies and associated relay racks are sourced from electrical panel boards located on the wall behind the relay rack locations. Figure 4.94 illustrates the power panel arrangement in the high bay area of the Mu2e building. The Mu2e M4 beamline 600 Amp LOTO disconnect switch is shown to the right of the figure and is sourced from switchboard SWBD-Mu2e-B1. Beamline power supplies panel boards PHP-Mu2e-B1-1 and PHP-Mu2e-B1-2 are both controlled from this isolation safety switch.

Figure 4.94. Layout of the Mu2e building electrical distribution panels

As in the MC-1 service building, the Mu2e building will utilize a standard 4 inch square wire way sourced from PHP-Mu2e-B1-1 that will be used to route all AC feeders to the various switching mode power supply starter panels using flexible conduit drops where needed.

Relay rack plug power will be sourced from the clean house power transformer through the 120/208 power panel PP-Mu2e-A1-4-A1. A separate dedicated wire way will be installed along the instrumentation cable tray with flexible 20 Amp single-phase conduit drops to each relay rack location.

### 4.8.4.2.13  Procurement Strategy

Recent experience with procurement of power supplies will be used to manage the procurements required for Mu2e.  The RFQ, Ion source, and NOvA beam lines have been upgraded to supply beam to the existing accelerator complex using new supplies and controls systems procured using many different contracts with many different vendors. Mu2e will take advantage of lessons learned from procurements of similar equipment to ensure project success. There will be a large number of procurements for the many small parts needed to construct the control systems and these parts can be managed independently of the physical power supplies. Because of the number of elements involved, Mu2e will contract will an assembly house to construct sub-assemblies. As the sub-assemblies are returned from the assembly house, Fermilab staff will test, calibrate and locate the equipment in the service buildings.

Specifications will be written for each power supply type based on recent design procurements in 2013 for ANU[30]. Mu2e will use the upgraded electronics systems developed by ANU and will procure identical equipment requiring little or no rework. The switch mode power supplies will be standard production units from vendors that will have a shorter lead time than the control electronics, therefore the control electronics components will be procured early with the physical supplies being procured just before they are required for installation. The large custom SCR supplies will have lead times similar to the control electronics and will also have to be procured early.

### 4.8.4.3   Magnet Power Supply Risks

The switch mode power supplies have multiple vendors that will be evaluated based on the specification for the many different sizes of supplies that are needed. For some manufacturers, most of the required power supply sizes are already in their product inventory; from others extended versions of their product inventory will be needed. Manufactures will not disclose the operating junction temperatures of devices internal to their power supplies. This has a direct impact on operation life, so a plan will be developed for de-rating the supplies being requested.

---

[30] ANU = Accelerator and NUMI Upgrades.  ANU is the part of the NOvA project that provides upgrades to the accelerator and neutrino beam lines.

The SCR style power supplies have a more limited number of available suppliers. The specification will be very detailed to ensure the quality of equipment being supplied. All of these supplies will be custom or semi-custom designs, posing the standard risks associated with unique equipment. These supplies will have to be tested with Fermilab staff present before shipment to Fermilab. The details of the required testing are specified in the specification for each type of supply.

### 4.8.4.4  Magnet Power Supply Quality Assurance

The assembly of the regulators will be contracted to a local company with final testing and calibration done by Fermilab electrical technicians. To maintain quality control of the components used in the electronics, Fermilab technicians will procure all components and inspect them before shipping them to the assembler. The switch mode power supplies will be inspected and tested by Fermilab technical staff as soon as they arrive. Inspections and testing will be witnessed at the vendor/s for the SCR type power supplies.

### 4.8.4.5  Magnet Power Supply Installation and Commissioning

The output of the power supplies will be connected to the magnets using THHN/TWNN power cable in a dedicated cable tray for DC power. All of the cable tray and power cables will be installed by local trade contractors led by a Fermilab cable installation expert.

#### 4.8.4.5.1  Installation coordination

There are many activities associated with installation and commissioning of the magnet power supplies that must be coordinated.

- As the enclosure and service buildings are completed the water system should be installed.

- After the water system is installed the cable tray and all power cable should be installed. Cable termination to the magnets will be performed after magnet installation.

- Magnets will be located in the enclosure.

- Beam Pipe and instrumentation will be installed.

- As the beam pipe is being completed the magnet and power supply water and cables can be connected.

- The controls electronics will be assembled and tested in-house.

- Location of electronics can be started as soon as the controls racks are installed in the service buildings and the control power is on.

- Most of the interconnect cable will be assembled by a local wire assembly house. Small quantities and special cable will be constructed by Fermilab technical staff.

### 4.8.4.5.2   Commissioning of equipment

- After final installation of equipment, all the controls and monitoring circuits will be tested by Fermilab engineering staff.

- Internal control information will be loaded into the Regulation electronics by Fermilab engineering staff that will verify the correct regulation for each magnet loop.

- Final data base and controls pages will be set up using Fermilab operations and Muon Campus experts.

### *4.8.4.6   Operations Lock Out Tag Out*

A 600 Amp disconnect will be installed for each of the new beamlines, allowing for lockout and tag-out for each individual enclosure. The disconnects have viewing windows to allow for verification. This system is identical to the MI LOTO system that is used now for enclosure access.

## 4.8.5 Diagnostic Absorber Beam Line and Absorber Core

### *4.8.5.1   Diagnostic Absorber Requirements*

A diagnostic beam absorber is required in the M4 beam line in order to commission extraction from the Delivery Ring and a portion of M4 beamline with proton beam while allowing for installation work to continue in the downstream M4 production target area. The beam commissioning and installation period could overlap for up to 2 years. Therefore, the requirements for determining the capacity of the beam absorber are based on two different operational running modes needed for beam commissioning.

**Mode 1:** Low intensity protons with intensity of $5 \times 10^{10}$ protons/pulse every 10 seconds at a kinetic energy of 8 GeV, which are single turn extracted from the Delivery Ring for commissioning the M4 beamline.

**Mode 2:** Resonantly extracted proton beam at a kinetic energy of 8 GeV with four standard Mu2e pulses separated by 10 msec injected into the Delivery Ring (DR) with an intensity of $1 \times 10^{12}$ protons/pulse. The four pulses are repeated every 30 seconds for the commissioning of the resonant extraction system.

Table 4.26 summarizes the beam parameters for the two modes of operation for beam commissioning. Under these conditions the requirement for the capacity of the M4 diagnostic absorber is 170 W. The absorber will be a passive core with no water cooling.

The position of the absorber is required to be to the left of the M4 beamline in order to avoid inference with the open aisle used for personnel travel and transport of magnets.

The Mu2e requirement document [97] describes the full set of requirements for the M4 diagnostic beam absorber.  Requirements for radiation safety are also provided and are discussed in section 4.5.2.3 in the Radiation Safety section of this chapter.

Table 4.26. Beam Parameters for 2 modes of operation for M4 beamline diagnostic absorber.

| Mode | protons/pulse | Pulse Rep Rate (sec) | Energy (GeV) | Beam Power (W) | Protons/Hr |
|------|------------|---------|-------|---------|------------|
| 1 | $5.0{\times}10^{10}$ | 10 | 8 | 6.4 | $1.80{\times}10^{13}$ |
| 2 | $4.0{\times}10^{12}$ | 30 | 8 | 170.7 | $4.80{\times}10^{14}$ |

### 4.8.5.2   Diagnostic Absorber Technical Design

#### 4.8.5.2.1   Diagnostic Beam Absorber Line

The main purpose of the diagnostic absorber is to allow low power beam commissioning with an 8 GeV proton beam while the installation of the downstream Mu2e solenoids and detectors takes place. The M4 diagnostic absorber will also allow for periods of quick tune up or a place to send low power beam during experimenter access periods during Mu2e operation. Therefore, a diagnostic absorber beamline has been designed upstream of the final focus section of the M4 beamline.

Beam will be extracted into the nominal M4 beamline and transported down the diagnostic beamline in order to hit the absorber core. Figure 4.95 shows the location of the diagnostic absorber block and the diagnostic absorber beamline relative to the M4 beamline.  The diagnostic absorber beamline begins 174.325 m upstream of the Delivery Ring Extraction Lambertson, and is downstream of the extinction collimation section.  At this location a horizontal dipole, HDA1, will be used to bend the beam into a diagnostic absorber beamline.  When this dipole is powered, the beam is bent 5° (87mrad) horizontally into the diagnostic absorber beam line.  HDA1 will operate at a current of 1070 A to produce this bend. The line is roughly 24 m long and uses two 3Q120 quadrupole magnets to focus the beam into the absorber block.  The two quadrupoles, QDA1 and QDA2, operate at currents of 310 A. The beamline is instrumented with a single multiwire to measure and ensure a proper beam profile up to the absorber.  The intensity of the beam impinging on the diagnostic absorber will be measured with an ion chamber just upstream of HDA1, at quadrupole Q938 in the primary line.  Figure 4.96 shows the diagnostic absorber beamline with the important elements.

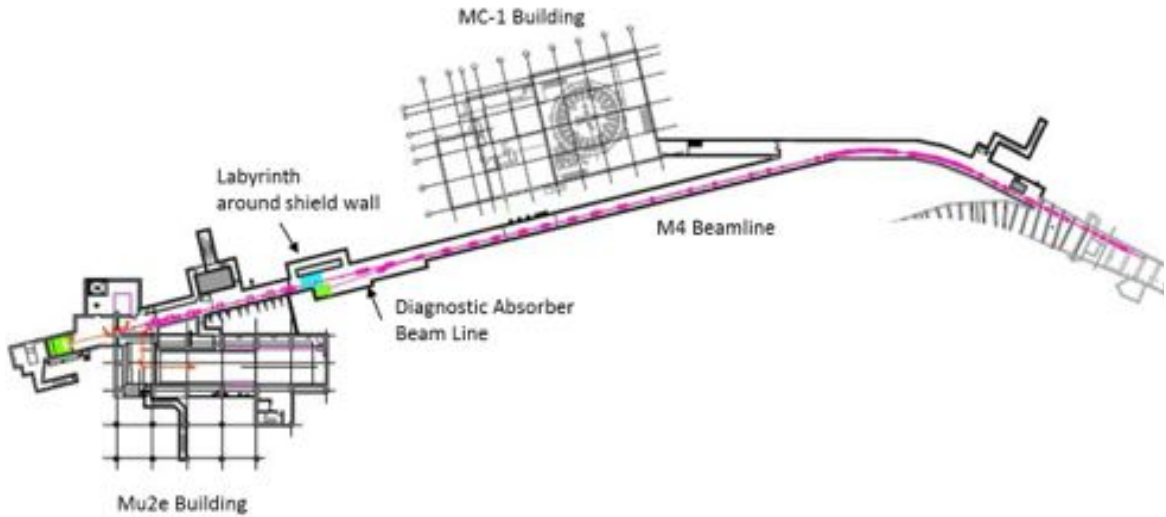

Figure 4.95. M4 Beamline showing location of the diagnostic beamline and absorber.

### 4.8.5.2.2   Diagnostic Absorber Core

The diagnostic absorber core is a series of 6 ft. × 6 ft. × 5 ft. long stacked steel plates surrounded by 1 ft. of concrete on the top and bottom.  There is 3 ft. of concrete on the back of the absorber and 2 ft. on the front face of the absorber.  A 1 ft. × 1 ft. × 3 ft. long albedo trap is located at the front entrance of the absorber.  A 4-inch beam pipe for the end of the absorber beamline will extend 2 ft. into the trap and be terminated by a vacuum window. The absorber has a capacity of 170 W and is located completely inside the M4 tunnel enclosure. Since there is a very small absorbed heat load value for the absorber, forced convective cooling is not necessary.  Figure 4.97 depicts the overall dimensions and layout of the absorber core with respect to the tunnel.

The absorber core is too large to be fabricated as a single piece, so it will be constructed out of stacked plates.  Fortunately, the 108 series steel plates at the Fermilab Railhead (a storage location) can be used, saving considerable material cost.

The design of the absorber's steel core requires construction of two stacks of 8 in. thick steel plates lined up one behind the other in the direction of the beam, corresponding to nine layers of steel. The steel plates in the first stack will have a length, measured in the direction of the beam, of 36 in.  The width of the steel will be 72 in. Similarly, the plates in the stack behind the first stack will have the same width, but their length will be 24 in. Figure 4.98 illustrates the stacked steel plate core of the beam absorber.

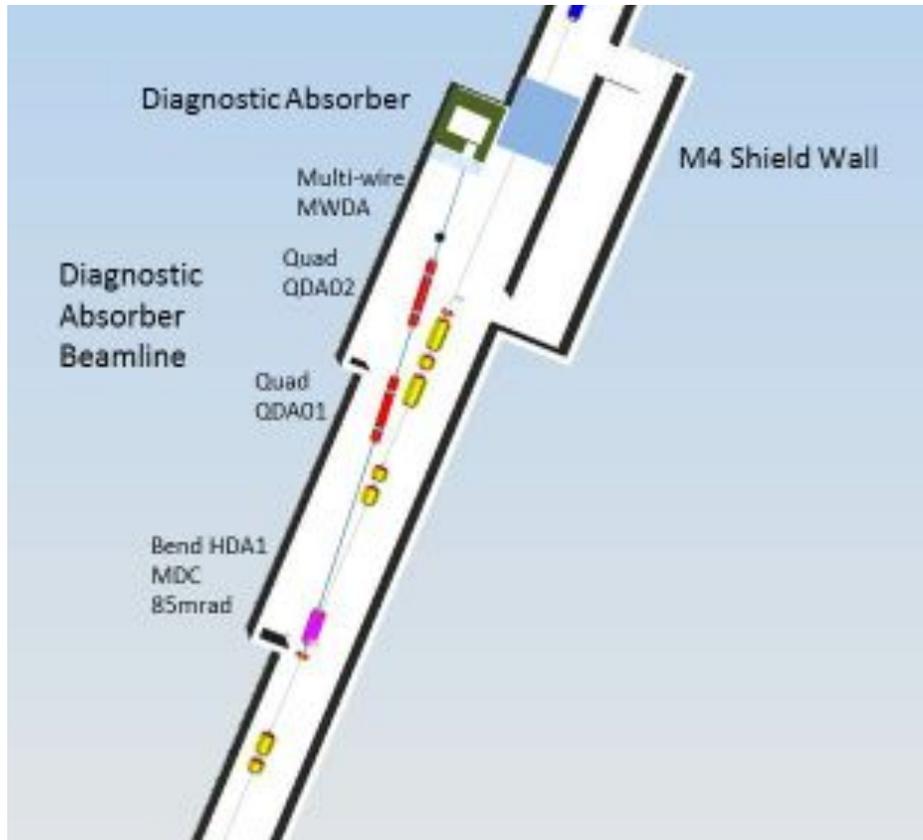

Figure 4.96. M4 diagnostic absorber beam line with components

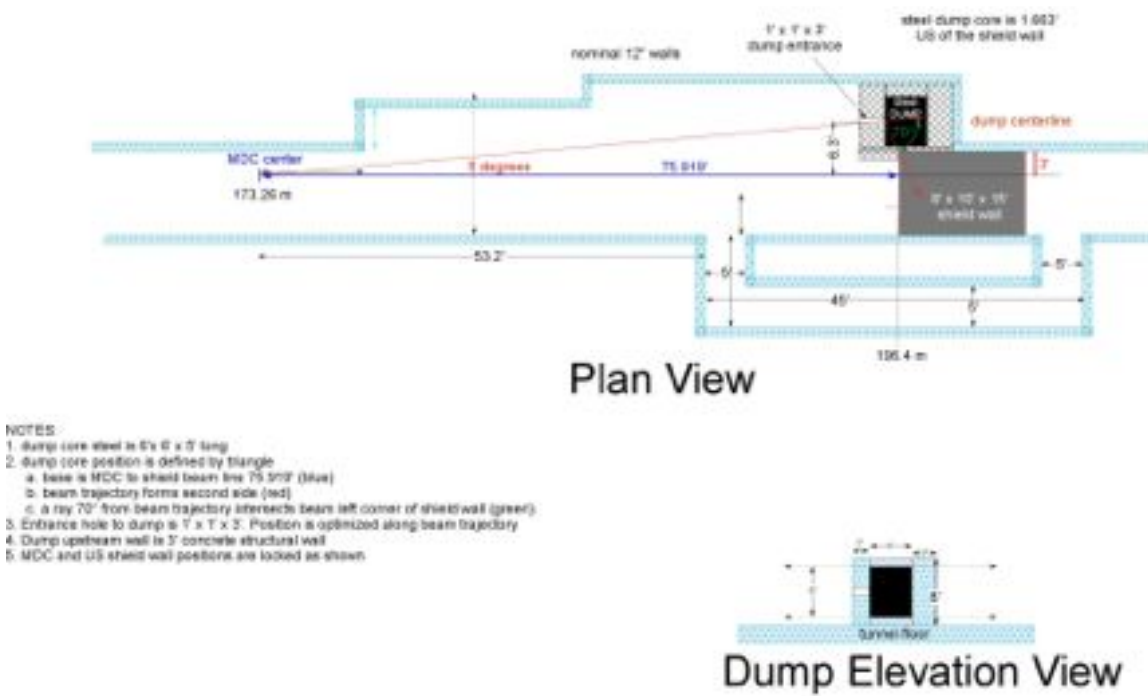

Figure 4.97. Layout and dimensions of the M4 diagnostic absorber.

The absorber will be placed on a concrete pad that will be 12 in. from the enclosure floor. Since, the beam height from the enclosure floor is 48 in. the beam center will be at the center of the fifth steel layer from the bottom. Since, the steel plates are 8 in. thick, the design allows considerable room for thickness variations due to wide variation in the thickness of the available 108 series steel plates. It is highly unlikely that the beam center would hit a seam between two steel plates. All the steel plates will have tapped holes at four locations on the top plane. These holes will facilitate the transportation and stacking of the plates.

A structural steel form will be designed and bolted to the front face of the absorber's core. This form would accommodate pouring of the shielding concrete. It will also act as an Albedo trap through which the beam would traverse and eventually hit the core.

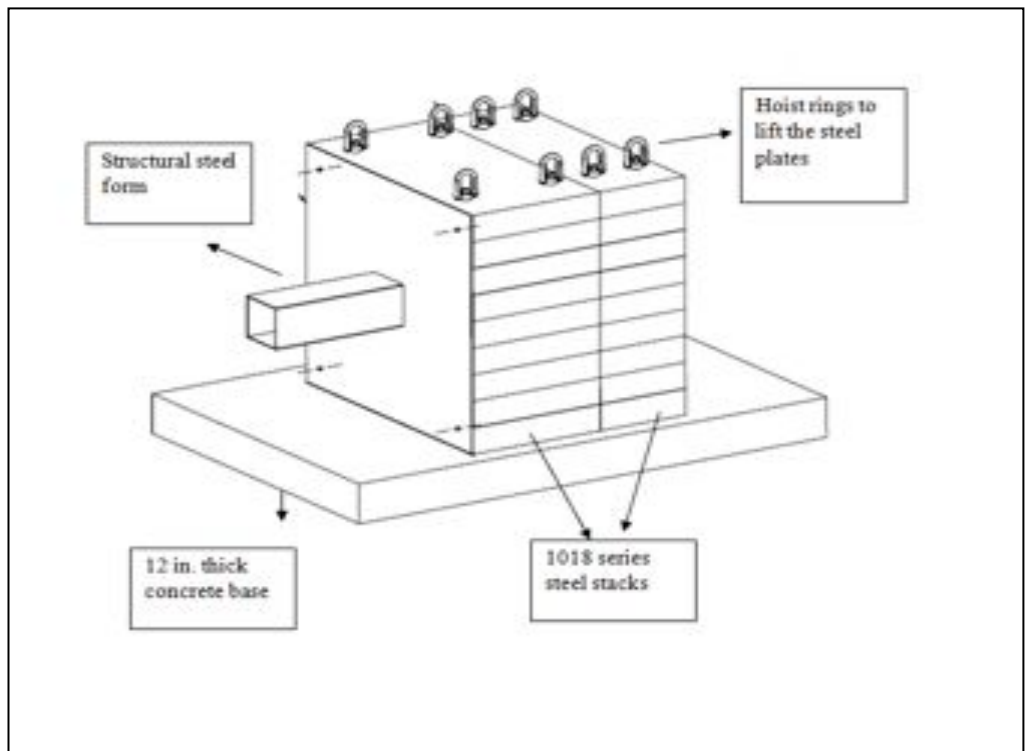

Figure 4.98. Stacked steel plates make up the core of the M4 diagnostic beam absorber.

### 4.8.5.2.3   MARS Simulations

MARS simulations have been performed to calculate the rates of particles produced from the back of the absorber that propagate to the Mu2e building areas where personnel could be present (see section 4.5.2.3).  The size and dimensions of the absorber stacked steel and surrounding concrete were determined from these simulations.  Radiological effects for ground and surface water activation were also estimated.  The results were all found to be under limits and standard Fermilab requirements. Details of these results can be found in the Radiation Safety Plan (section 4.5.2.3).

### 4.8.5.3  Diagnostic Absorber Risks

The quantity of steel required to build the M4 absorber core is approximately 90,000 lb. Quotations from steel vendors on the above amount of steel were found to be cost prohibitive. Thus, excess steel available on the Fermilab site was sought. Fermilab's Railhead has suitable steel slabs that can be used to build the absorber core, however, there are a few peculiarities associated with this steel.

Preliminary XRF (X-Ray Fluorescence) and surface analysis of these steel slabs showed that their outer surface is coated with lead based paint. When subjected to machining processes, owing to the intense heat generated due to friction, the paint could transform to particulate form. This is hazardous to the personnel working in close proximity to the steel. Thus, it is mandatory that each steel slab, before being subjected to any machining process, be sandblasted to eliminate the outer lead paint coating.

The lead paint alone is not the only cause of concern. Anecdotal evidence suggests that the steel taken from the Railhead is of inferior quality and difficult to machine.

To understand these issues, one steel slab will be subjected to the sandblasting procedure followed by machining. If the test slab can be cut according to specifications, then it is almost certain that the all the steel needed for constructing the absorber core can be taken from the Railhead, resulting in appreciable material cost savings.

Thus, a calculated risk of using the Railhead steel for constructing the absorber core is well justified. However, if it becomes evident that the steel is of inferior quality, one would be forced to seek outside vendors to build and transport the steel slabs as per engineering specifications. This avenue may prove to be quite expensive.

### 4.8.5.4  Diagnostic Absorber Quality Assurance

The Fermilab village machine shop is equipped to handle very large slabs of steel. Also, engineering drawings of the steel core were discussed with the machining specialists and they were very optimistic about meeting the engineering specifications in the lab's machine shop, provided the quality of steel is satisfactory.

It has to be noted that the M4 absorber is part steel and part concrete. Pouring the concrete is as important as constructing the steel core. Thus, the engineering drawings of the absorber were discussed with FESS. Meeting dimensional tolerances, especially with respect to elevations, are critical, for the beam has to hit the flat face of the steel slab and not the seam between two slabs. Thus, the pouring of the concrete base has to be monitored carefully, ideally in the presence of an alignment team.

#### 4.8.5.5  Diagnostic Absorber Installation and Commissioning

Installation and commissioning of the M4 diagnostic absorber will involve recourses from the Accelerator Department, FESS, and an alignment team. This is especially true while pouring the concrete base, for the position of the absorber is of paramount importance. Figure 4.97 illustrates the location and orientation of the diagnostic absorber in the M4 beam line.

Stacking the individual steel slabs would require a rehearsed installation and hazard analysis procedure. Also, the installation would occur during the enclosure construction phase of the Beamline Enclosure General Plant Project (GPP). The stacked steel will need to be processed and delivered prior to the time the enclosure walls and surrounding absorber concrete are poured. This would provide the rigging crew much needed flexibility while maneuvering the unwieldy steel slabs.

The mechanical design of the absorber core includes what are known as dovel pins. These pins would connect two vertically adjacent steel slabs. They would also prevent the slabs from shifting during the concrete pouring phase of the absorber construction. It is essential that the alignment team survey the positioning of all the slabs that make up the absorber core in order to ensure proper alignment to the absorber beamline.

Once the structural steel form is bolted to the front face of the core and the concrete is poured around and atop the core, the installation phase is complete. A final survey of the critical locations on the diagnostic absorber would conclude the commissioning phase of the project. Finally, if deemed necessary, the front face and the side face, the face closest to the enclosure aisle would be covered with a 4 in. thick slab of marble.

### 4.8.6 M4 Beamline Vacuum

#### 4.8.6.1  Beamline Vacuum Requirements

The M4 line is an 810 foot long single pass beamline that will require a beam tube to be under vacuum to limit beam-gas scattering, allow a common vacuum connection to the Delivery Ring and M5 beamline, and allow for the reuse of ion pumps from the Accumulator.  A detailed listing of the M4 beamline vacuum requirements can be found in reference [98].  A summary of the requirements are listed below.

- The beam line needs to maintain a pressure of $1.0 \times 10^{-8}$ Torr or better.

- All vacuum components being reused from the Antiproton Ring need to inspected and tested prior to installation into the beamline. Some refurbishing will be required.

- All components and devices need to be leak checked to a sensitivity of $2 \times 10^{-10}$ atm·cc/s with helium prior to installation into the beamline.

- All components should be ultrasonically cleaned prior to welding and again prior to installation.

- Assembly shall be performed using ultra-high vacuum handling practices.

- All components must be oil-free.

- Beam tubes shall be steam cleaned, blown dry, then the ends capped to maintain cleanliness until installation.

- Bake-out of the beamline is not necessary.

### 4.8.6.2  Beamline Vacuum Technical Design

The M4 line is a single pass beamline that transports 8 GeV protons from the Delivery Ring to the Production Solenoid and target. The upstream section of the M4 beamline is shared with the g-2 M5 beamline upon leaving the Delivery Ring proper. The common section of the beamline is referred to as the M4/M5 combined section and shares common beam tube, magnets and instrumentation. Since the M4 beamline is required to have an average vacuum pressure of $1\times10^{-8}$ Torr, the beamline will be divided into 3 distinct vacuum sections in order to provide isolation from the Delivery Ring, the M5 line and the downstream final focus section of the M4 beamline. Isolation from the final focus is put in place so that if the final focus section downstream of the diagnostic absorber is not complete with installation, continuous vacuum to the diagnostic absorber will be available for beam commissioning of the upstream part of the beamline. Also, there will be a need to isolate the beam tube vacuum with vacuum windows in two locations. The first location is at the end of the diagnostic absorber beamline. The beam tube in front of the absorber will extend into the albedo trap and will be isolated by a vacuum window.

The second location where a vacuum window will be needed is at the end of the M4 beamline just upstream of the Production Solenoid. The window will be located just downstream of the multiwire, the last M4 beamline element. Figure 4.99 shows the location of the M4 beam valves that will separate the beamline into the three vacuum sections and the vacuum isolation windows.

#### 4.8.6.2.1  Vacuum Components

Many of the vacuum components that will be installed in the M4 beamline will be reused from the Accumulator Ring, AP2, AP3 and D to A beamlines. This includes beam tube, pumps, beam valves, etc. Reuse of the equipment is part of the Project's value engineering program to reduce the overall cost of the Project. However, reused components will require testing and evaluation, and some will require refurbishment, to allow their extended use for the lifetime of the Mu2e experiment. Table 4.27 is a summary of the vacuum hardware that will be needed for the M4 beamline. The table includes columns for hardware that will be reused and hardware that will be purchased new.

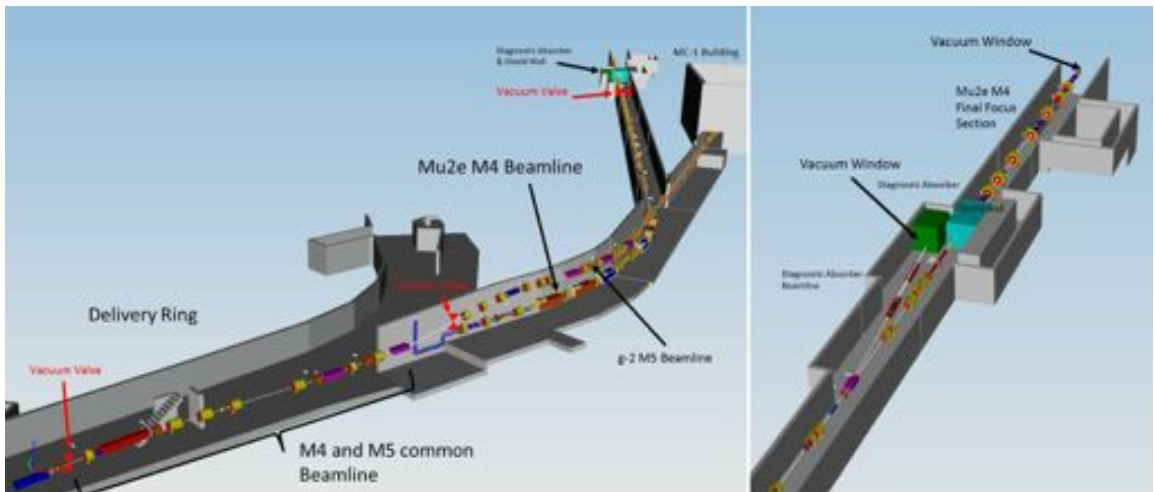

Figure 4.99. M4/M5 (left) and M4 (right) beamlines showing location of beam valves and vacuum windows.

### 4.8.6.2.2   Pumps

The required pumping capacity will be achieved through the reuse of 270 L/s and 30 L/s ion pumps from the Accumulator Ring. These will be installed directly into the M4 beamline roughly every 30 ft. to handle the gas load. The 30 L/s pumps will be used primarily in tight spaces like the horizontal bend section. Roughing pumps (oil free) and turbo molecular pumps will also be reused from the Accumulator Ring. They are free standing and movable and will be used for initial pump-down of the beamline from $1.0 \times 10^{-3}$ Torr and $1.0 \times 10^{-6}$ Torr respectively. Under normal vacuum operation ion pumps should be adequate to hold the required vacuum level of $1.0 \times 10^{-8}$ Torr.

The number of ion pumps required for the M4 line can be estimated by analogy with the existing AP-2 beamline. The AP-2 line beam tube diameter and length are similar to the new M4 line. The AP-2 line is 910 ft. long and has twelve 270 L/s ion pumps. It consists primarily of 5-1/2 inch OD SS beam tube with very few diagnostics or high gas load devices. Its vacuum pressure is $1 \times 10^{-8}$ Torr. This suggests that eleven 270 L/s ion pumps should provide similar vacuum pressure in the M4 line if all things were equal. However, the new M4 line will have additional diagnostic components with outgassing rates that are higher than that of plain Stainless Steel beam tube. To accommodate the higher gas loads the number of 270 L/s ion pumps has been increased to 20. Once the outgassing rates of these components have been more accurately determined, the pumping capacity will be adjusted accordingly.

Table 4.27. Components for M4 Vacuum System.

| Components | Existing | New |
|---|---|---|
| Ion Pump, 270 L/s | 20 | |
| Ion Pump, 30 L/s | 6 | |
| Roughing Pump, 20 CFH Scroll Pump | 2 | |
| Turbo Pump, 80 L/s Turbo Pump | 2 | |
| Valve, 1.5" Vent-up | 8 | |
| Valve, 4" Beam Valve, Pneumatic, Remote Controlled | 5 | |
| Valve, 4" Beam Valve, Hand Operated | 2 | |
| Window, Titanium | 3 | |
| Valve, Turbo Cart Pump-out | 8 | |
| Pirani Gauge | 8 | |
| Cold Cathode Gauge | 8 | |
| Ion Gauge | 8 | |
| Beam Tube, 4" Type 316L SS  (ft.) | 0 | 500 (est.) |
| Beam Tube 5-1/2" Type 316L SS (ft.) | 200 (est.) | |
| Beam Tube, 6" Type 316L SS (ft.) | 0 | 100 (est.) |

#### 4.8.6.2.3   Beam Tube

The M4/M5 common section of the beamline is 28.9 m (95 ft.) and the M4 beamline proper is 215.1 m (705 ft.).  Most of the beam tube that will be installed in the M4 proper section of the beamline will be 4" diameter.  There are some exceptions, such as the extinction section, where the beam size can potentially be bigger and 6" diameter beam tube will be used.  The end of the final focus will also require larger beam tube to allow for the required angle bumps on the target. The beam tube material will be stainless steel 316L welded.  Some beam tube will also be reused from the Antiproton source, which is 5.5" diameter. The M4/M5 combined section of the beamline will reuse the 5.5" diameter beam tube to meet the beam size requirements for both g-2 and Mu2e.

#### 4.8.6.2.4   Other Vacuum Considerations

There are a couple of elements installed in the M4 beamline that may require additional consideration. One is the AC Dipole located downstream of the major horizontal bend section. This device may need additional vacuum isolation, which will be accomplished by using hand operated beam valves to isolate that high gas load region during maintenance and pump-down of adjacent sectors. Another M4 beamline location that will require additional consideration is the protection collimator at the end of the beamline. One additional hand operated beam valve will be installed to isolate that high gas load region during maintenance and pump down of adjacent sectors. The end of the beamline vacuum window will act as a second valve.

### 4.8.6.2.5  Vacuum System Controls

The vacuum station located at the AP30 service building will be reused to house all of the vacuum controls, interlocking hardware, ion pump power supplies, etc. that will be needed for the M4 beamline vacuum system.  Figure 4.100 shows the vacuum station as is was used for the Accumulator Ring. Almost all of this equipment will be reused for the M4 beamline vacuum system.

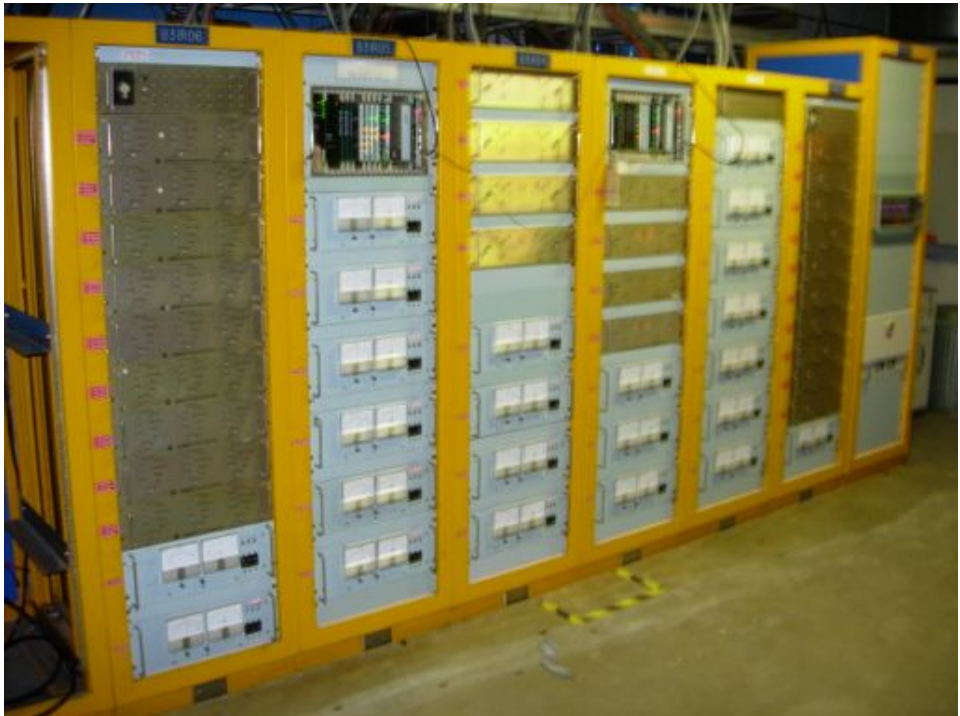

Figure 4.100. M4 beamline vacuum station located at AP30 service building will be reused.

### 4.8.6.3  *Beamline Vacuum Risks*

Currently the biggest risk for the vacuum system is the possible determination that it is necessary to change the beam tube material from stainless 316L welded to stainless 316L seamless.  There is presently a lack of available stainless 316L seamless, the cost of acquiring it is high and the quality is poor. There is a very small probability that there will be such a change in the requirement for the beam pipe. There are also potential risks of vendor delays that could impact the installation schedule. However, this risk has a low probability and a small impact to the overall M4 installation schedule.

### 4.8.6.4  *Beamline Vacuum Quality Assurance*

Specifications and installation procedures will be written to ensure that the acquisition and installation of parts will follow the technical requirements. Additionally, quality control checks will be done on the existing equipment to ensure that the parts are handled and checked along the way. Additionally, punch lists will be used during the installation so that all necessary steps are completed.  A Muon Campus installation coordinator has also been identified to aid in the installation schedule and flow of work.

### 4.8.6.5   Beamline Vacuum Installation and Commissioning

The beamline vacuum system will be installed toward the end of the beamline installation process after beneficial occupancy is granted for the tunnel enclosures. All magnet stands and magnets must be in place before any vacuum system can be installed. Only the cables for the vacuum system controls will be installed at the front end of the installation schedule.  Most beam tube components will be welded into place but bolted flanges and bellows will also be used. Upon completion of installation, the vacuum system will be leak checked and certified to the vacuum pressure required.  Commissioning will include fixing punch list items for the vacuum system and vacuum controls.

## 4.8.7 Beam Line Low Conductivity Water

### 4.8.7.1   Beamline Low Conductivity Water Requirements

The Low Conductivity Water (LCW) system is used as a standard water cooling system across the Fermilab accelerator complex. The LCW system is primarily responsible for cooling magnet and magnet power supply systems. Since the Mu2e experiment will reuse the Debuncher Antiproton production storage ring, the LCW system for this machine can also be reused, though with some modifications.  The newly built M4 external beamline and the new MC-1 and Mu2e buildings will all require standard LCW to cool magnets and power supplies. LCW is also required for the Production Solenoid Heat and Radiation Shield. The LCW system is sourced from the Central Utility Building (CUB) and must be piped to the Muon Campus.  Connections to the main LCW headers must be modified to provide service. Standard LCW cooling parameters for temperature, flow and conductivity can be seen in Table 4.28. Detailed requirements for sourcing, implementing and installing the LCW systems for the M4 beamline are provided in reference [95].

A tunnel compressed air system will also be required.  Tunnel compressed air is used to drive pneumatic control systems such as vacuum beam valves.  A system currently exists in the Delivery Ring and will need to be expanded for the new M4 beamline and service buildings.

Table 4.29 provides a summary of the LCW cooling demands for the external M4 beam lines and MC-1 and Mu2e buildings.

Table 4.28: Standard LCW Cooling Parameters

| Supply Temp (°F) | Flow (gpm) | Conductivity (MΩ-cm) | Supply Pressure (psi) | Return Pressure (psi) |
|---|---|---|---|---|
| 90 | 1610 | 15 | 215 | 45 |

Table 4.29. LCW Cooling Demand Requirements

| Location | Cooling Demand (kW) |
|---|---|
| M4 beamline | 710 |
| M5 beamline | 190 |
| MC-1 Building | 370 |
| Mu2e Building | 50 |
| M2 beamline | 115 |
| M3 beamline | 220 |
| Delivery Ring | 1160 |

### 4.8.7.2   Beamline LCW Technical Design

The Fermilab Muon Campus and associated Muon g-2 and Mu2e experiments require fluid system cooling. Because of the location of the Muon Campus, reconfiguring the existing 95° F low conductivity water (LCW) and central utility building (CUB) chilled water (CHW) piping systems to satisfy the Muon Campus's fluid cooling needs is the logical and least costly plan.

The new LCW cooling system will provide general process fluid cooling to the new Muon Campus accelerator infrastructure that includes the M4 beamline, the MC-1 and Mu2e buildings. The CHW system will provide comfort and limited process cooling to the Muon Campus service buildings (AP0, AP10, AP30, and AP50) as well as the new Muon Campus experimental halls.

The new Muon Campus LCW piping system layout is very similar to the existing 95°F LCW layout. However, in order to accommodate the cooling needs of the new Delivery Ring and M4 beam lattice configuration, some modification to the existing piping layout is needed. With the decommissioning of the Antiproton Accumulator Ring, all water-cooled Accumulator Ring magnets will be disconnected from the existing LCW system and relocated. The LCW piping network for the Delivery Ring and the AP-2 and AP-3 beamlines will remain largely unchanged, aside from the addition of a few new valve taps and minor piping header modifications necessary to provide physical clearance for the new Muon campus beam lattice.

The Muon Campus LCW system will include new piping lines in the M4 and M5 beamline enclosures. These general-purpose LCW piping lines will extend to the respective MC-1 and Mu2e buildings. In addition to the general LCW cooling needs, the MC-1 building will house electrical components that will require special LCW temperature setting and stability. A special heat exchanger configuration will be designed and installed to cool a portion of the incoming Muon campus LCW to the desired temperature setting using CHW provided directly to the MC-1 service building from CUB. A comprehensive thermal/flow simulation of the new Muon campus LCW system has been generated using AFT Fathom Pipe Flow Analysis & System Modeling Software. The heat load and flow summary is tabulated in Table 4.30.

The construction of the new M4 and M5 beam line enclosures present both design and construction challenges for Muon campus LCW and CHW piping. These underground beamline enclosures, once constructed, will interfere directly with the Muon campus 8" LCW and 6" CHW main supply and return piping header alignment, previously installed in this area.  As a result, these piping headers will have to be rerouted. The current design plan will require cutting the interfering piping sections and relocating them over the accelerator beamline enclosure. A sophisticated venting system where the cooling headers route over the beamline enclosure has to be designed to eliminate air pockets that may accumulate at this location. Additionally, in order to accommodate the water-cooling needs in the M4 and M5 accelerator beamline enclosures, pipe tees with isolation valves will be installed at suitable locations on the main supply and return piping headers between the CUB and the M4/M5 beamline enclosure. From the branches of the tees, the LCW piping headers will enter the M4/M5 beamline enclosure to provide cooling water to the various components. Figure 4.101 below illustrates the routing of the piping main headers from CUB and shows how the piping is distributed in the new beam line.

The new Muon campus will require approximately 2000 feet of new stainless steel pipe to accommodate the new LCW piping extending along the M4 and M5 beamlines and continuing to the MC-1 (Muon g-2) and Mu2e buildings. LCW supply piping will likely utilize 6" diameter NPS piping based on the current conservatively estimated cooling demands in this area shown in Table 4.30.  However, the size could be reduced to 4" NPS if the final estimated flow requirements are less than anticipated.

Table 4.30. Simulated flow and heat load summary for the Muon Campus LCW cooling system

| Heat Load Component | Inlet Temp. (°F) | Outlet Temp. (°F) | Temp. Difference (°F) | Avg. Temp. (°F) | Flow Rate (gpm) | Heat Load (kW) |
|---|---|---|---|---|---|---|
| **Delivery Ring Loads** | | | | | | |
| MAN 1[3] | 86.6 | 103.54 | 16.94 | 95.07 | 56.68 | 139.80 |
| MAN 2 | 86.6 | 101.80 | 15.20 | 94.20 | 41.90 | 92.73 |
| MAN 3 | 86.63 | 104.38 | 17.75 | 95.505 | 56.78 | 146.74 |
| MAN 4 | 86.64 | 103.72 | 17.08 | 95.18 | 50.19 | 124.81 |
| MAN 5 | 86.62 | 101.73 | 15.11 | 94.175 | 41.34 | 90.95 |
| MAN 6 | 86.63 | 103.61 | 16.98 | 95.12 | 76.11 | 188.17 |
| MAN 7 | 86.62 | 105.19 | 18.57 | 95.905 | 55.7 | 150.60 |
| MAN 8 | 86.6 | 97.79 | 11.19 | 92.195 | 35.26 | 57.45 |
| MAN 9 | 86.6 | 104.25 | 17.65 | 95.425 | 64.2 | 164.98 |
| **Delivery Ring Summary** | | | | | **478.16** | **1156.23** |
| **Remaining Loads** | | | | | | |
| M2 Line[1] | 86.64 | 92.85 | 6.21 | 89.74 | 127.54 | 115.32 |
| M3 Line | 86.62 | 97.15 | 10.53 | 91.88 | 142.48 | 218.45 |
| M4 Line | 86.59 | 103.85 | 17.26 | 95.22 | 279.69 | 702.88 |
| M5 Line | 86.58 | 93.91 | 7.33 | 90.24 | 178.72 | 190.74 |
| AP10[2] | 86.63 | 98.82 | 12.19 | 92.72 | 28.23 | 50.10 |
| AP30[2] | 86.65 | 96.58 | 9.93 | 91.61 | 69.17 | 100.01 |
| AP50[2] | 86.63 | 100.22 | 13.59 | 93.42 | 101.31 | 200.46 |
| MC-1 Bldg | 86.70 | 104.21 | 17.51 | 95.45 | 144.98 | 369.62 |
| Mu2e Bldg | 86.79 | 92.72 | 5.93 | 89.75 | 57.84 | 49.94 |
| **Total Summations** | | | | | **1608.12** | **3153.75** |
| P-Bar HX | 99.08 | 86.57 | 12.51 | 92.82 | 1609.35 | -2931.36[31] |

**Notes:**

1. Includes heat load for the abort lines

2. A conservative assumption of heat load is made

3. MAN stands for manifold

---

[31] Negative number reflects dissipated heat.

**General Notes:**

The production solenoid radiation heat shielding has a current estimated flow requirement of 3 gpm. This flow and associated heat load data is not included in the table above.

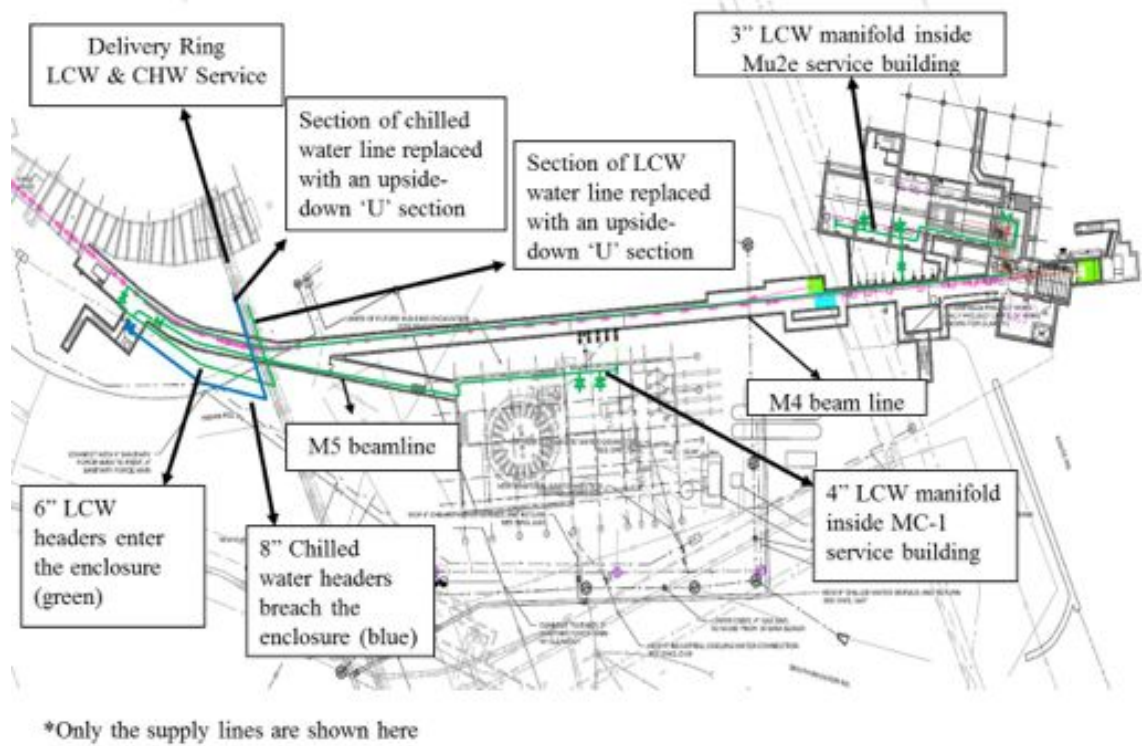

*Only the supply lines are shown here

Figure 4.101. LCW header routing for the M4 and M5 beamlines and the MC-1 and Mu2e service buildings.

### 4.8.7.2.1   Production Solenoid Heat and Radiation Shielding

The LCW system will also provide water for the Production Solenoid Heat and Radiation Shield (HRS). There is a vessel around the HRS that requires LCW at 3 gpm. Figure 4.102 shows how the LCW system will interface to the HRS.

### 4.8.7.2.2   Compressed Air System

In addition to the fluid cooling needs, the new Muon Campus will also have a need for a compressed air system. The compressed air system is drives pneumatic controls for tunnel systems such as beamline vacuum valves. The system also supports the use of pneumatic tools. The compressed air system for the existing Antiproton Source will remain essentially unchanged for the new Muon Delivery Ring footprint. However, a new and independent compressed air system will be specified and designed to accommodate the compressed air needs of the M4 and M5 beam line enclosures and the new Muon Campus experimental halls. The compressor will be located inside the MC-1 building. Figure 4.103 illustrates the Muon Campus compressed air system.

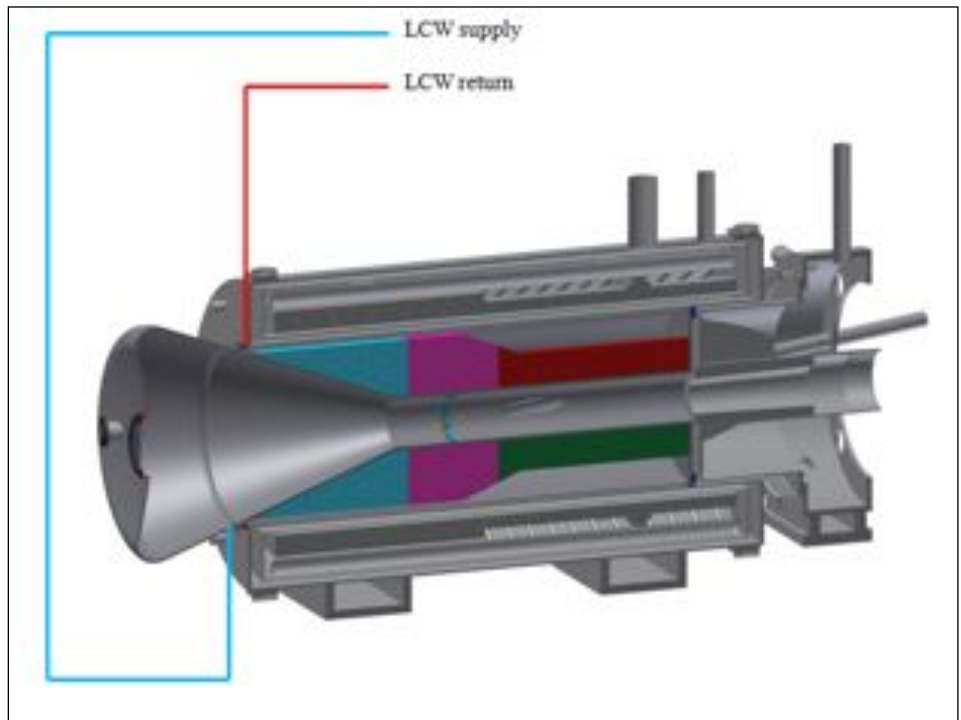

Figure 4.102. Model of the production solenoid showing LCW path.

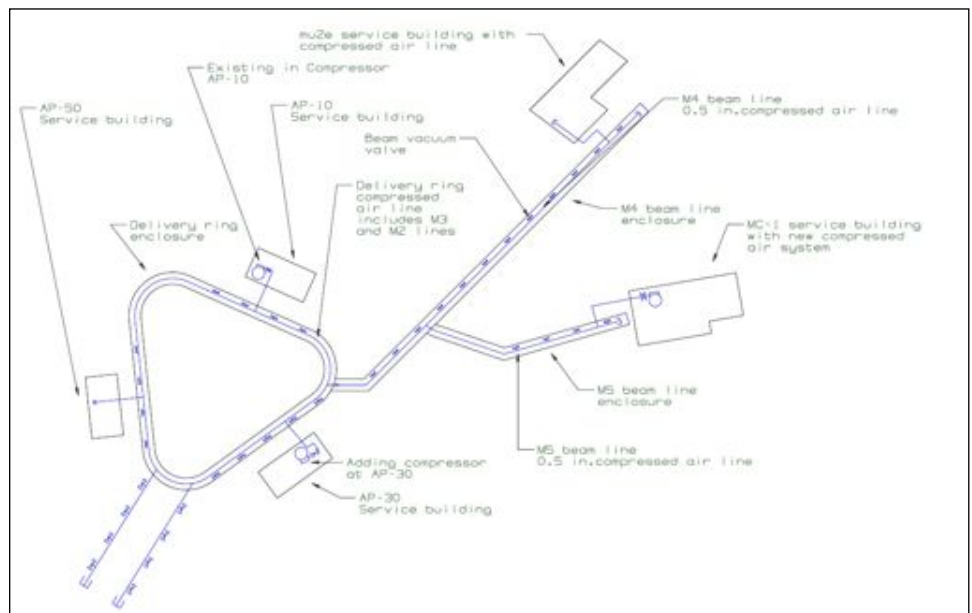

Figure 4.103. Illustration of Muon campus compressed air system.

### *4.8.7.3   Beamline Quality Assurance*

The Mu2e Project will adhere to strict construction and quality assurance codes, FESHM 5031 and ASME B31.3. The welders hired to construct the piping system will be ASME

Section IX and ASME B31.3 qualified. Further, all welders will be tested at the Fermilab weld shop on a welding procedure specification approved by the responsible engineer. In-process QA, which is required by the governing piping code, will be performed by the Task Manager to make sure that welds are of acceptable quality.

### 4.8.7.4 Beamline LCW Installation and Commissioning

The beamline LCW system will be one of the first systems installed in the M4 and M5 tunnels once beneficial occupancy is granted. FESS will reroute the main LCW and CHW headers from CUB and extend them into the tunnel enclosure at the M4/M5 alcove. Pipe fitters and welders will install the LCW header in the tunnel on the magnet side of the wall for the entire length of the M4 tunnel. From that point, the tunnel piping will be installed through dedicated penetrations into the MC-1 and Mu2e buildings. Commissioning the system will involve checking for leaks and venting all air out of the system. The LCW will be then tested to confirm required flow, temperate and pressure.

## 4.9 Extinction

The Mu2e experiment proposes to use a proton beam pulsed at approximately 0.6 MHz. The use of a pulsed beam is motivated by the fact that a significant background is produced by secondary beam particles (primarily pions) that reach the detection region during a time interval starting shortly after pulses hit the production target and extending for about 700 nsec. By detecting conversion electrons only for times later than 700 nsec, this background is significantly reduced.

Beam extinction is defined as the ratio of the number of protons striking the production target between beam pulses to the number striking it during the beam pulses. It has been established that an extinction of approximately $10^{-10}$ is required to reduce these backgrounds to an acceptable level [3]. The extinction requirement varies with the exact time that the proton strikes the target between the pulses, and $10^{-10}$ is a representative number assuming that the out-of-time particles are distributed uniformly between pulses.

Mu2e aims to achieve this extinction in two steps:

- The technique for generating the required bunch structure in the Recycler Ring will naturally lead to a high level of extinction. The fast kicker that transfers beam from the Recycler to the Delivery Ring should preserve this level of extinction. Some beam will leak out of the RF bucket in the Delivery Ring during slow extraction. Taking this into account, extinction of $10^{-4}$ or better is expected as beam is extracted to the M4 beamline.

- The M4 beamline line will incorporate a set of oscillating dipoles ("AC dipoles") to sweep out-of-time beam into a system of collimators. This will achieve an additional extinction factor of $10^{-7}$ or better.

## 4.9.1 Extinction Requirements

The extinction requirements are described elsewhere [3]. Protons hitting the production target during or slightly before the detection window can generate pions that can be captured in the stopping target. The subsequent radiative decay can produce a photon, which in turn pair produces to create a background electron, mimicking a conversion signal. An analysis of this background shows that an extinction level of $10^{-10}$ is required, where this as defined as the amount of beam between pulses divided by the beam in pulses, as illustrated in Figure 4.104.

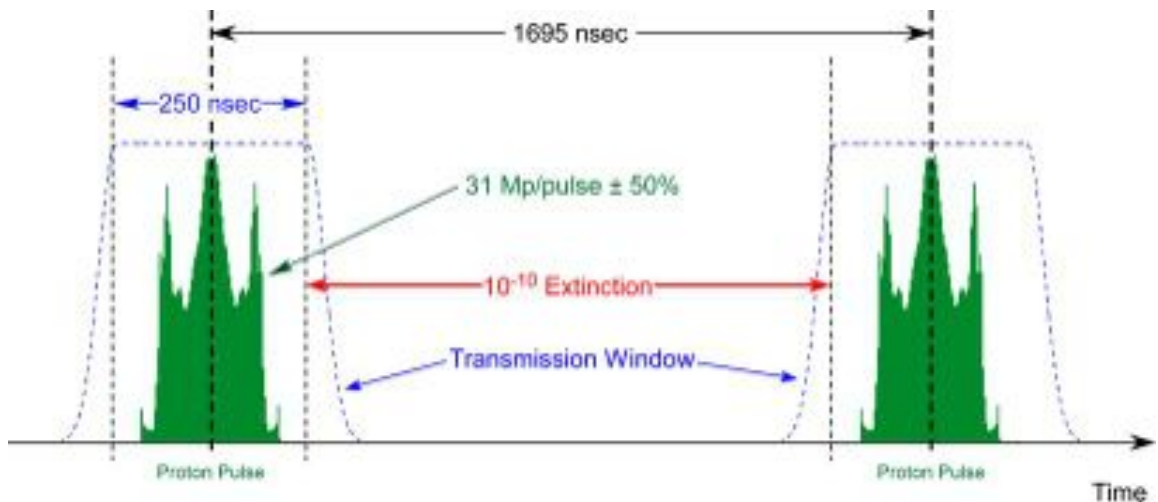

Figure 4.104. Time structure of the beam, indicating the required extinction.

## 4.9.2 Extinction Technical Design

### 4.9.2.1 Principles of Operation

The technique for generating the required bunch structure in the Recycler Ring will naturally lead to a high level of extinction. To preserve this level of extinction, extraction/injection kickers with short flat tops will be used to transfer beam from the Recycler, such that essentially no out-of-time beam is transferred to the Delivery Ring [105]. Out-of-time beam may develop in the Delivery Ring because of space charge, beam gas interaction, or beam loading effects on the RF, but this has been estimated to be at level below $1 \times 10^{-4}$ [106].

Extinction downstream of the Delivery ring is accomplished using deflecting magnets in the M4 beamline and a collimation system. In-time protons are transmitted to the production target with high efficiency and out-of-time protons are kicked into the collimators. The principle is illustrated in Figure 4.105. Ideally, one would like a kicker that would cleanly kick the out-of-time beam into an absorbing collimator, or equivalently kick in-time beam into the transport channel; however, such a kicker operating at the 600 kHz bunch rate is beyond the state of the art. Therefore, the system will utilize a pair of resonant dipoles to achieve the desired deflection.

These magnets still represent a technical challenge, so the design effort has focused on the optimization of the magnet specifications, which are tightly coupled to the optical parameters of the beam line [103]. To summarize, the minimum stored energy of the bending magnet scales roughly as

$$U \alpha \frac{1}{\sqrt{\beta_x L}}$$

where $\beta_x$ is the betatron function in the bend plane, and $L$ is the length of the magnet, assuming a waist in the non-bend plane. This leads to long, fairly weak magnets and a large betatron function in the bend plane. It was determined that a length of 6 m and a betatron function of 250 m were the largest practical values that could be achieved without unacceptable complications and increased length of the beam[104].

This design implicitly assumes there is no beam outside the specified admittance ellipse. Such beam risks being deflected into the transmission by the AC Dipole. It's therefore important to use collimators to precisely define the bounding beam admittance upstream of the AC dipole.

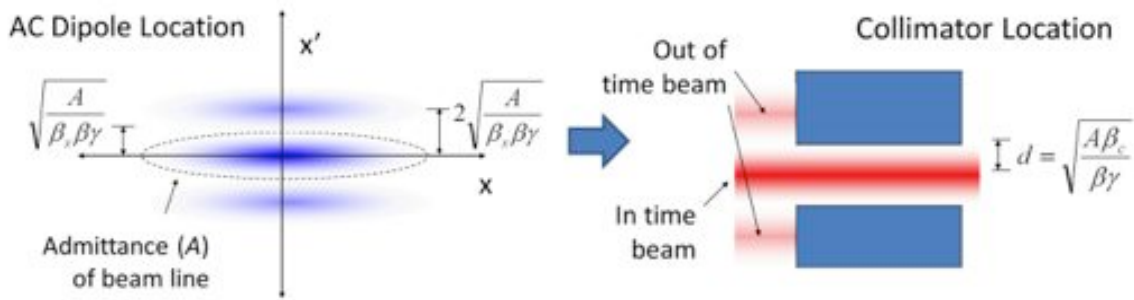

Figure 4.105. Principal of operation of the AC dipole/collimation system. The dipole system introduces an angular deflection, which causes a lateral transverse deflection 90° betatron phase advance downstream.

### 4.9.2.2   AC Dipole Design

The magnetic design of the AC dipole system is based on three harmonics, with the specifications shown in Table 4.31. The 300 kHz harmonic is phased such that in-time beam will pass through the collimator at the nodes. The 4.5 MHz harmonic reduces the slewing during transmission and has been optimized to minimize loss of in-time beam.

The 900 kHz harmonic is designed to reduce the maximum amplitude to prevent beam scraping upstream of the collimator. These harmonics generate the waveform shown in Figure 4.106. 50% of the beam is extinguished at roughly ±87 nsec, with 100% extinction at ±114 nsec. We are requiring the high frequency element to operate up to 5.1 MHz (17th harmonic), to accommodate shorter bunches, if the Delivery Ring is able to produce them.

Table 4.31. Specifications of the three harmonics of the Extinction AC dipole.

| Magnet | Frequency (kHz) | Length (cm) | Aperture | | Peak B Field (Gauss) |
|--------|-----------------|-------------|----------|----------|----------------------|
|        |                 |             | bend plane (cm) | non-bend (cm) |            |
| A      | 300             | 300         | 7.8      | 1.2      | 120                  |
| B      | 3800            | 300         | 7.3      | 1.2      | 15                   |

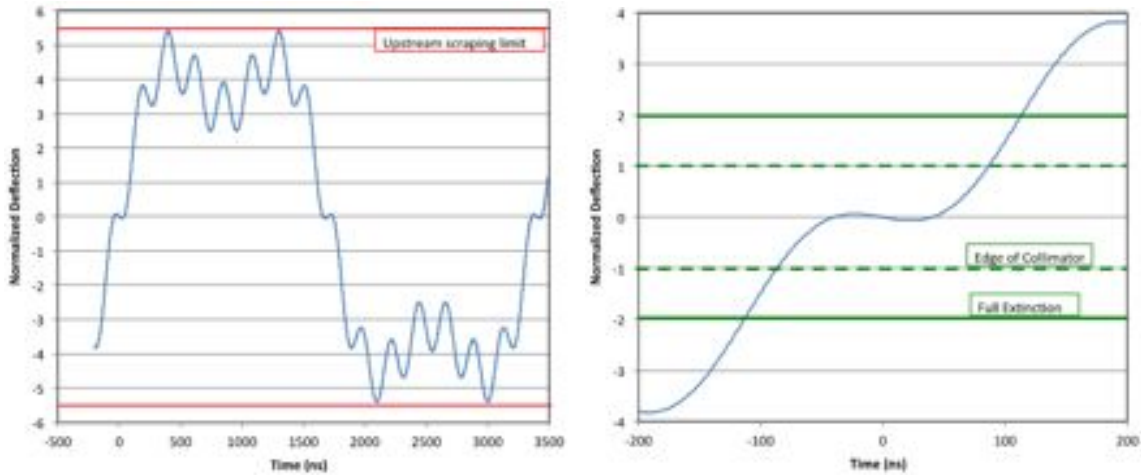

Figure 4.106. Final AC dipole waveform. On the left is the waveform over a complete cycle and on the right is the waveform over the transmission window.

The two harmonics use identical magnetic elements, each consisting of three 1 m long segments, as shown in Figure 4.107. The individual conductor leads are fed through for connection to the power and water supplies.

Tests were conducted using a half-meter prototype, shown in Figure 4.108. This prototype was tested at the requisite field at both 300 kHz and 5.1 MHz, as shown in Figure 4.109. Based on tests of the prototype, we have chosen CMD10, by Ceramic Magnetics, Inc., as the ferrite for the system. The properties of this ferrite are shown in Table 4.32. The low and high frequency components will dissipate 8 kW and 800 W of power respectively, and the total LCW requirement will be approximately 10 gpm.

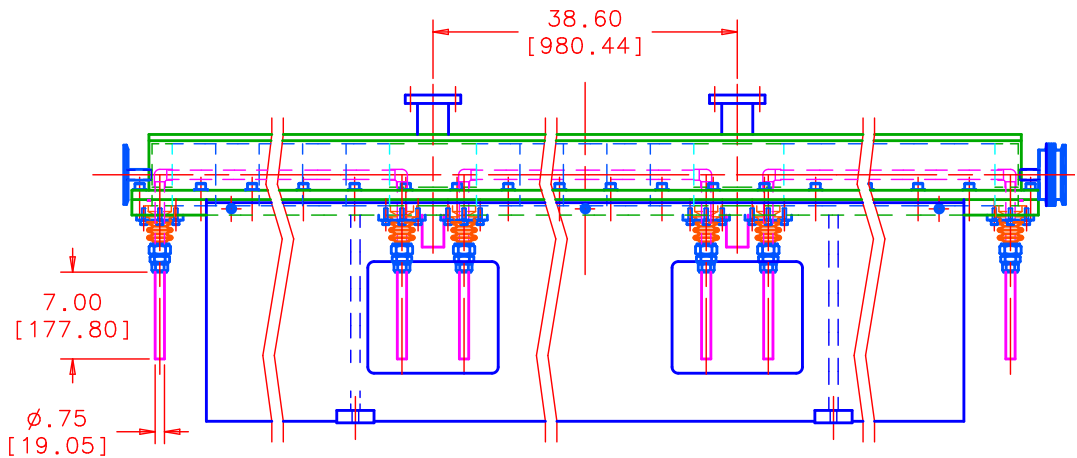

Figure 4.107. Design of three-element AC dipole module. The flange-to-flange length is 3 m. Two pump-out ports are shown on the top. Each 1 m long element has two separate water-cooled conductors, one on each side of the transmission channel. The entire ferrite element is under vacuum, and these conductors are connected to the outside via high-vacuum feed-throughs. Each 1 m segment can be independently excited.

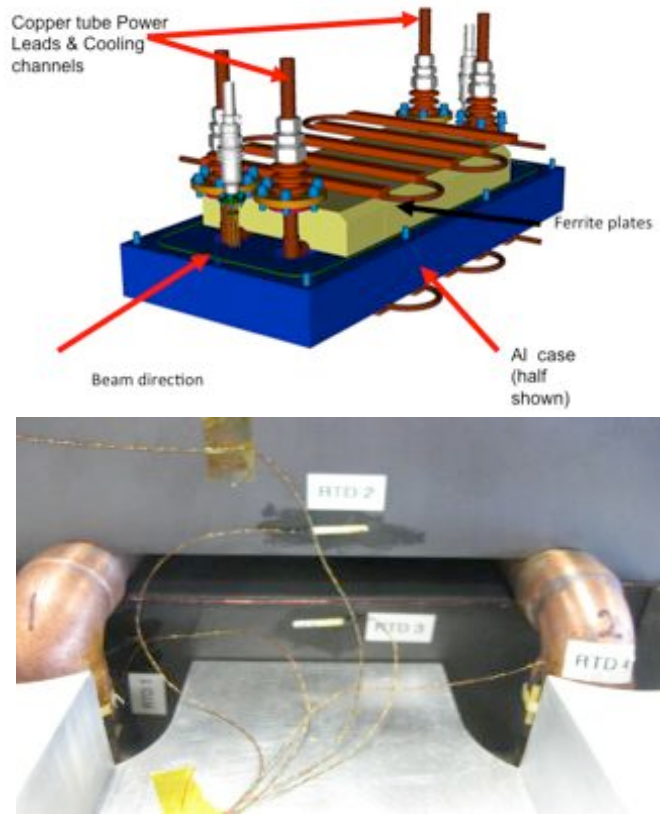

Figure 4.108. Drawing (top) and picture (bottom) of the half-meter prototype, shown to illustrate the details of the design. Single-pass water-cooled conductors pass on either side of the beam gap through the ferrite.

Table 4.32. Properties of CMD10 ferrite, measured in AC dipole prototype. Power dissipation values are for the half-meter prototype. Test showed that the required fields could be achieved with temperatures remaining below the Curie temperature of this ferrite.

| Coercive force | 0.12 Oersted | |
|---|---|---|
| Maximum permeability | 5500 | |
| Curie temperature | 130 °C | |
| | 300 kHz | 5.1 MHz |
| Current | 114 A | 8.6 A |
| Power | 1013 W | 190 W |
| P/I | 8.9 | 22.1 |
| $\mu'$ | 2100 | 500 |
| $\mu''$ | 220 | 580 |
| $\tan^{-1}(\mu''/\mu')$ | 6.0° | 49.2° |
| Q | 9 | 0.8 |

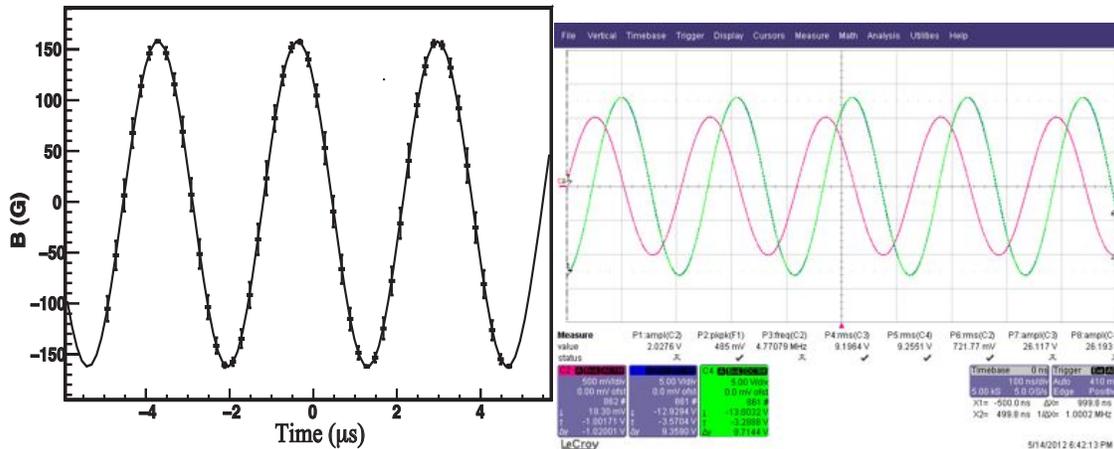

Figure 4.109. Response of the prototype at 300 kHz and 5 MHz. The plot on the left shows the field measured in the 300 kHz prototype, using an inductive pickup coil. The plot on the right shows the voltage and current at 5 MHz, indicating that the requisite field has been reached.

### 4.9.2.3   AC Dipole Power Supplies

The 300 kHz module will be powered by a slightly modified version of a switching power supply that has been frequently used at Fermilab [108]. The schematic of the circuit is shown in Figure 4.110. The high frequency module will be powered by a commercial RF power supply.

Figure 4.111 shows the conceptual layout of the control system. Electromechanical tuners will keep the magnets on resonance. These will be controlled by a feedback loop that monitors the relative phase between the voltage and the current. The high frequency component will be phase locked to the low frequency, and the entire system will be phase locked to the revolution frequency of the Delivery Ring.

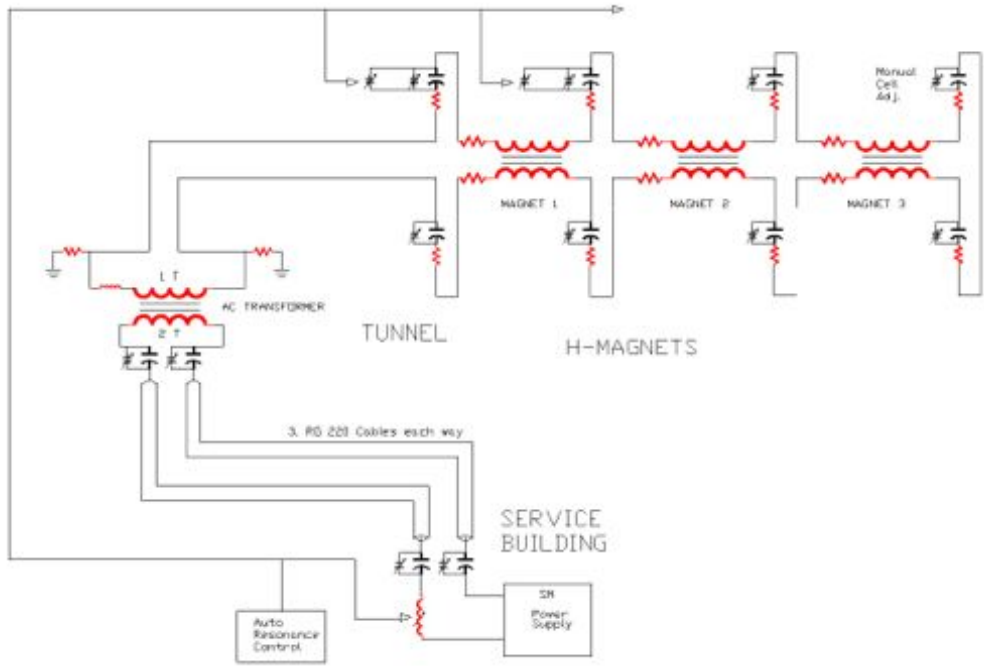

Figure 4.110. Modified (Krafczyk design) switching power supply and control loop for 300 kHz magnets.

### AC Dipole Power Supply and Control Conceptual Diagram

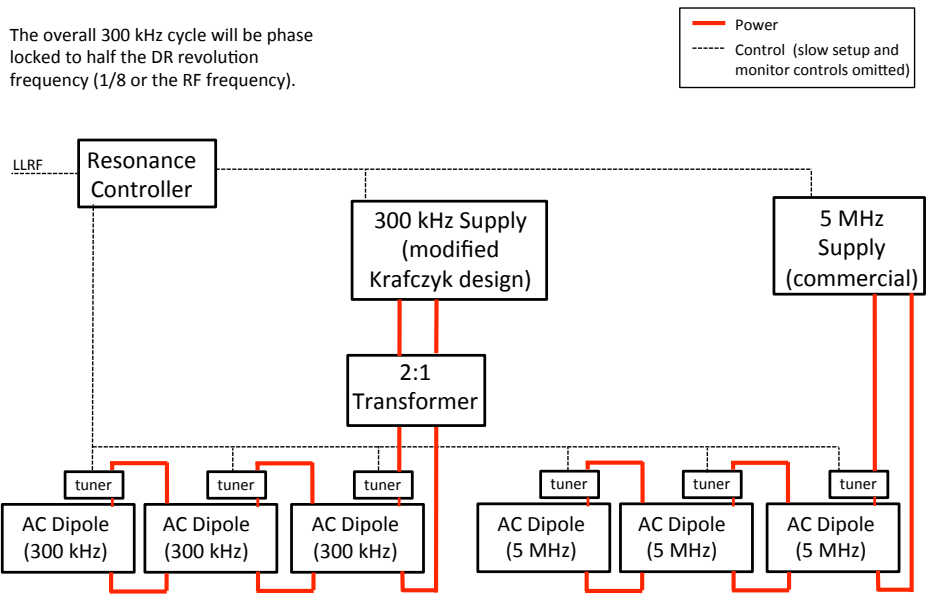

Figure 4.111. Conceptual layout of the power supply control system. The two harmonics shown are phase locked together and the entire system is phase locked to one of the four RF buckets in the Delivery Ring.

### 4.9.2.4   Collimator Placement and Design

The AC dipole works in conjunction with a system of five collimators, all operating in the horizontal plane. The first two collimators are upstream of the AC dipole to clean up any beam tails that are outside of the nominal 35 $\pi$ mm-mrad emittance of the beamline. The third collimator is approximately 90° in phase advance downstream of the AC dipole. This is designed to optimally absorb the out-of-time beam.

All collimators are of the same design, which is shown in Figure 4.112. Steel jaws open to a maximum of 3.3" and close to a minimum of 0.4". The ends of each jaw are independently moveable so the angle may be adjusted. The design allows for a 1 mm layer of Tungsten cladding, if it proves to be beneficial for the upstream collimators.

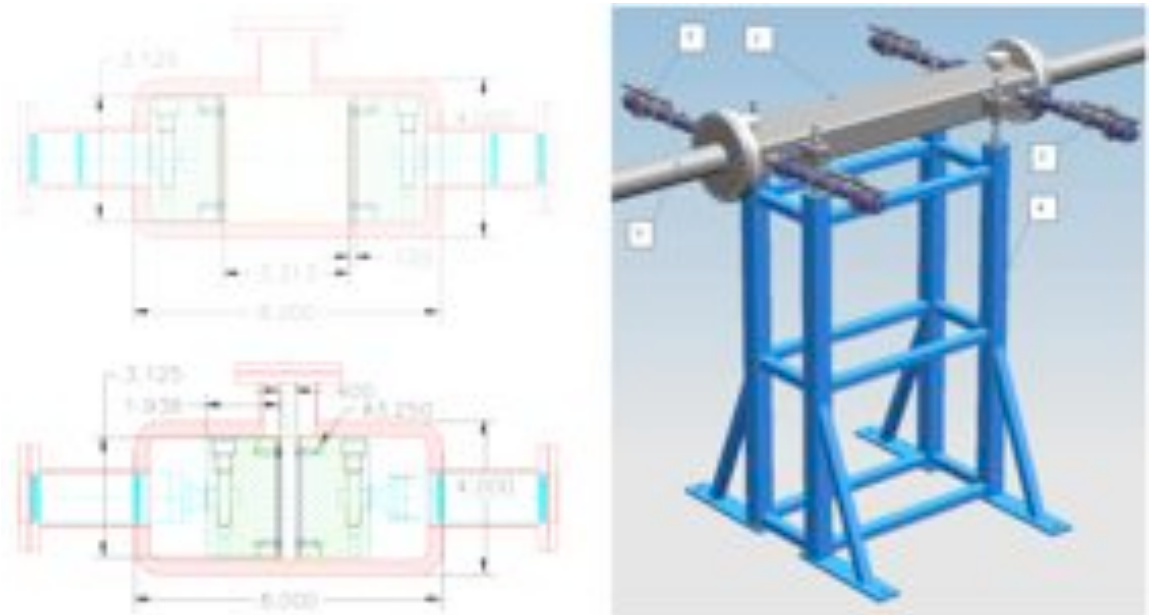

Figure 4.112. Collimator design.  Details shown at right are: 1-vacuum chamber; 2-motor drive mechanism; 3-LVTD; 4-frame; 5-beam tube.

Collimator movement and position read back are implemented through a standard Fermilab system of actuators, stepper motors, and LVDT position transducers. An example of the control system is illustrated in Figure 4.113.

### 4.9.2.5   System Performance

The total extinction of the system will be the product of the extinction level of the beam extracted from the Delivery Ring and the transmission window of the AC dipole.

A simulation of the delivery ring has been performed, including space charge and beam loading effects [106]. Figure 4.66 shows the time distribution of a proton bunch at the time of extraction from the Delivery Ring. The simulated extinction outside of the 125 nsec half-width, is $(1.64\pm0.13)\times10^{-5}$. The tails of the beam extend into a region a few tens of nsec after the transmission window; however, because this is convoluted with the decay time of the pions before the detection window, we are not sensitive to this beam [107].

The performance of the external extinction system was evaluated using G4Beamline [109]. The external extinction system achieves an extinction of $5\times10^{-8}$ or better for all beam outside of the nominal ± 125 nsec window. The performance of the entire system (Delivery Ring + collimators + AC dipole magnets) is shown in Figure 4.114. We see that the total extinction is better than $10^{-11}$ at all times for beam more than 10 nsec outside the nominal transmission window. The transmission of the in-time beam was found to be 99.7%.

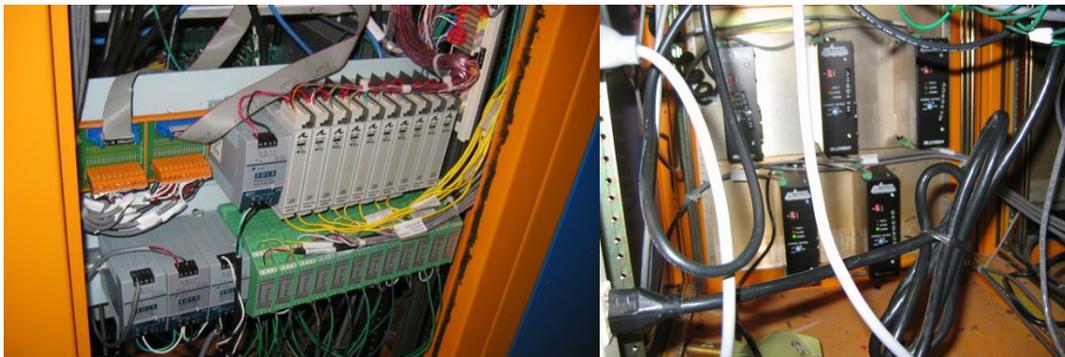

Figure 4.113. Example of control and readout system for collimators. The system is comprised of stepping motor controllers, shown at left, with interface cards to the ACNET control system, shown at right.

A detailed simulation of the system has been performed using both MARS and STRUCT. The performance of the system is shown in Figure 4.114 [109]. We see that the total extinction is better than $10^{-11}$ at all times for beam more than 10 nsec outside the nominal 230 nsec transmission window.

### 4.9.3 Extinction Risks

The primary risk is that the system will not achieve the required level of extinction for the beam that is extracted from the Delivery Ring, due to beam drifting out of the bucket during slow extraction. Because the simulations are believed to be accurate, this is considered to be an operational risk due to a malfunction or non-optimization of the RF system.

Problems with the AC dipole system could arise from incorrect phases or amplitudes in the magnetic elements. Again, this is considered an operational risk, as both will be monitored in the control system. Phase problems would be readily apparent as beam losses on the collimators increase, which would be observed in the beam loss monitors. More subtle problems would be detected by the Target Monitor System, which is described in section 4.10.2.1.

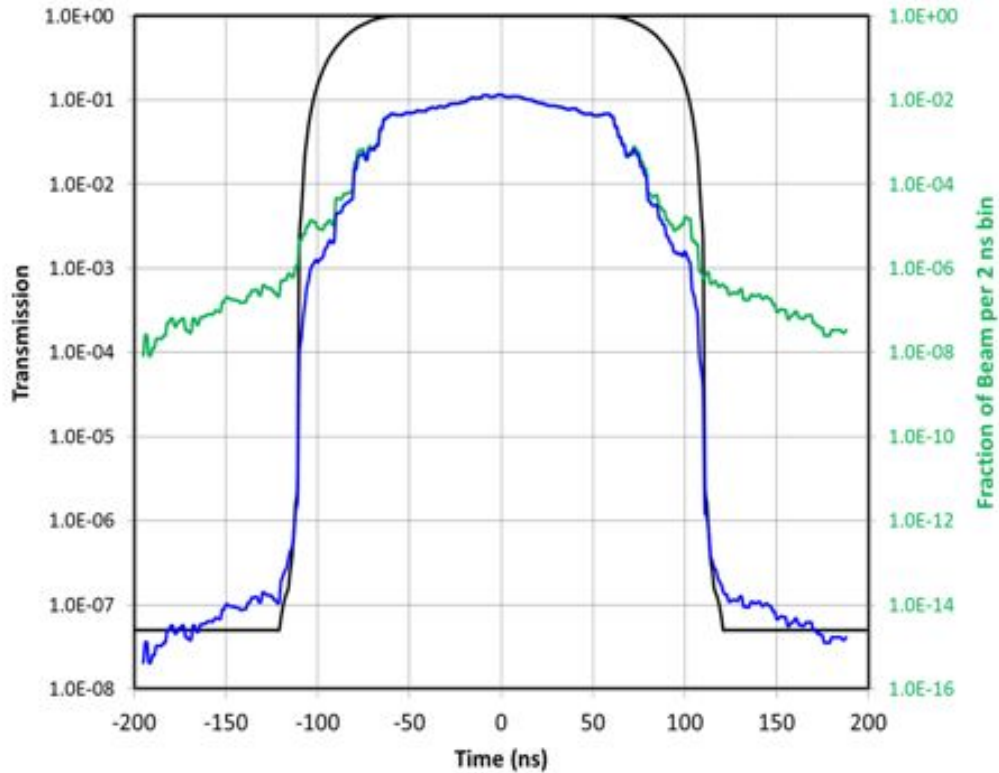

Figure 4.114. Performance of the combined system. The black line (scale at left) shows the transmission curve of the external dipole/collimator system, based on the G4Beamline simulation. The green curve (scale at right) shows the ESME simulation of the beam extracted from the Delivery ring. The blue curve shows the convolution of the two.

### 4.9.4 Extinction Quality Assurance

The magnets and power supplies will be tested together at full power, well before installation. Magnetic probes will be used to verify that the field meets specifications.

The collimators, actuators, stepping motors, and LVDTs will be assembled and tested prior to installation, to insure proper operation.

### 4.9.5 Extinction Installation and Commissioning

The AC dipole system and collimation system will be installed along with the other beam line elements. The system will be commissioned during the beam line commissioning to the dump.

The performance of the AC dipole system will be evaluated by varying the phase relative to the nominal bunch time. Thus the extinction factor can be measured at the beam dump using an ordinary hodoscope.

# 4.10  Extinction Monitoring

Measuring extinction at the $10^{-10}$ level will be very challenging. The Extinction Monitor described here will measure the extinction of beam that hits the production target, with the goal of getting a $10^{-10}$ measurement within roughly an hour, depending on the intensity and duty factor of the beam.

## 4.10.1 Extinction Monitoring Requirements

The requirements for extinction monitoring are summarized in Table 4.33. An explanation of these requirements is given elsewhere [5].

Table 4.33. Extinction monitoring requirements

| Specification | Target Monitor |
|---|---|
| Extinction sensitivity (90% CL) | $10^{-10}$ |
| Extinction accuracy (systematic) | 10% |
| Integration time | $6 \times 10^{16}$ POT (~1 hr at 100% duty factor) |
| Timing resolution (RMS) | <10 nsec |
| Rate-dependent error over dynamic range | <10% |
| Increase in beam emittance | N/A |
| Initial readiness | When the production target is ready |
| Access time (assuming monthly access is needed) | 4 hrs. |
| Radiation hardness (minimum protons delivered before replacement is required) | $4 \times 10^{20}$ POT |

The extinction monitor, also known as the "Target Monitor," will measure the overall performance of the extinction system by monitoring the extinction of the beam that is hitting the production target. The time scale of interest is the total time over which data is taken, but a shorter time has been chosen so that unforeseen problems may be detected before a significant amount of data is lost.

## 4.10.2 Extinction Monitoring Technical Design

### *4.10.2.1 Target Extinction Monitor*

The target extinction monitor design consists of a momentum-selecting filter consisting of collimators and a permanent dipole magnet, a magnetic spectrometer consisting of planes of silicon strips, scintillating trigger counters and a dipole magnet, and a range stack that will help establish the muon content of both in-time and out-of-time beams. It also includes a collimator located in the primary beam line upstream of the final focus section. This collimator limits the transverse size of the beam so that, in conjunction with the protection collimator, it eliminates beam halo that would otherwise interact with the Production Solenoid cryostat and the Heat and Radiation Shield (HRS) to produce a source of background events for the extinction monitor.

#### 4.10.2.1.1 Filter

Figure 4.115 shows the location of the filter with respect to the target and proton beam dump. The filter consists of three major components (see Figure 4.116). The first component is an entry collimator that selects secondary particles produced within a small solid angle from the production target and defines the location of the detector. In order to maximize signal rate, it is oriented close to the beam axis but at a sufficiently large angle to avoid the beam dump. The collimator is angled upwards in order to locate the detector above and behind the proton dump. This location minimizes civil construction costs. The length of the collimator is chosen to be just long enough to allow room for a meter of shielding above the dump core.

The second component is a permanent dipole magnet that is oriented to transport charged particles with an average momentum of 4.2 GeV/c. The charge and momentum were chosen to maximize signal rate (based on simulations). The magnet also serves the function of blocking line-of-sight down the channel, removing unwanted background from low energy neutral particles such as neutrons and photons. The magnet itself is a pre-existing permanent dipole that is being recycled for this purpose. The decision to use an existing permanent dipole not only reduces costs but minimizes maintenance and ensures stability of the signal rate. However, this means that the magnet aperture sets the useful maximum size of the collimator channels.

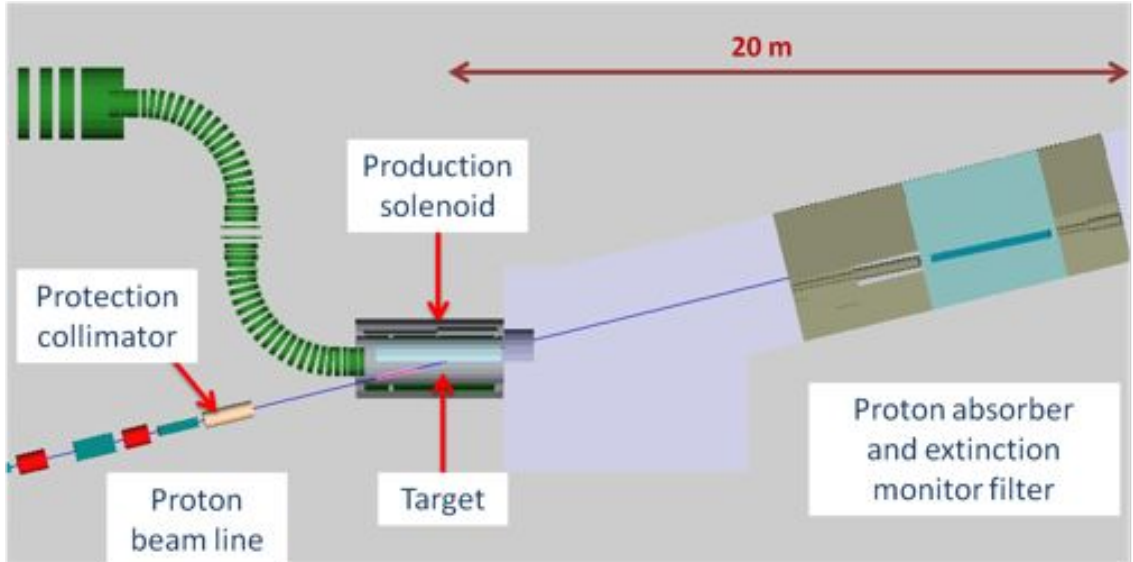

Figure 4.115. Location of the filter for the target extinction monitor.

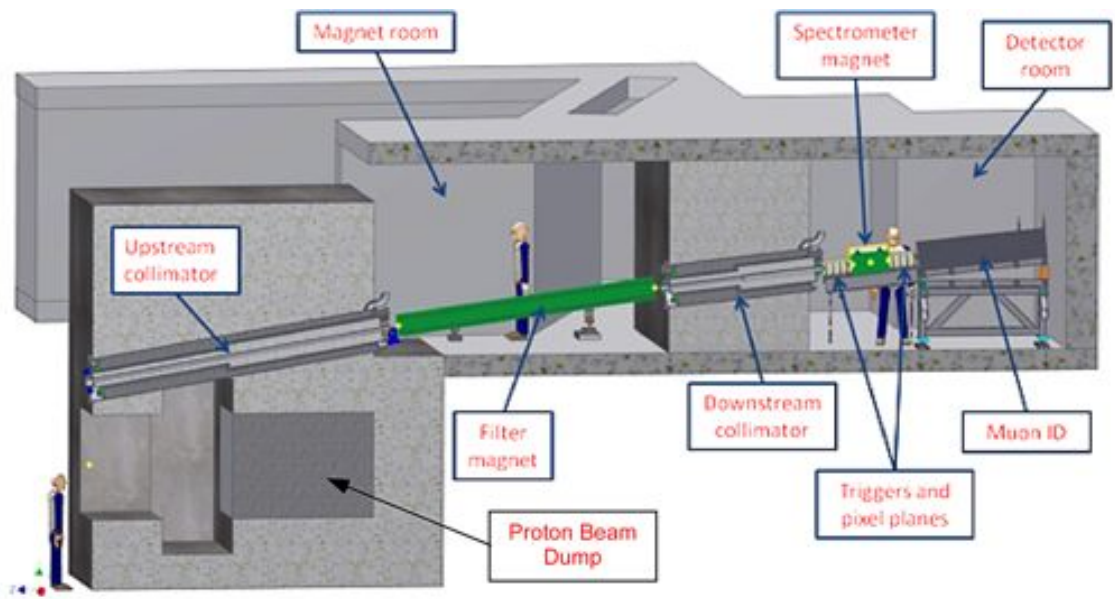

Figure 4.116. The components for the target extinction monitor

The third component is the exit collimator, which is embedded in a two-meter thick concrete wall that shields the detector from radiation produced by the particles that did not make it through the collimation system. The length is chosen to be just long enough to provide adequate shielding.

### 4.10.2.1.2  Filter Engineering Details

Figure 4.116 gives an overview of the target extinction monitor components.  At the far left is the proton absorber (proton beam dump) with the entrance collimator above it.  The green permanent magnet in the middle (magnet) room is the filter magnet. To the right is the exit collimator. In the detector room is an array of trigger scintillator paddles, pixel planes, a spectrometer permanent magnet, and a muon range stack. Visible at the top of the picture is the long narrow hatch through which all equipment in the middle and detector rooms will be installed.

The Extinction Monitor filter is made up of two alignable shielding plugs/collimators traversed by beam channels and separated by a permanent magnet. The plugs are steel pipe weldments filled with concrete. The plug collimators are surrounded by a larger pipe, called a fixed liner, and are adjustable by alignment mechanisms that will create the required pointing vectors in the final survey. There is a 0.50 inch gap between the plug collimator and the fixed liner. The alignment mechanisms can place either end of the plug anywhere inside the fixed liner pipe, changing the pitch and yaw angles of the beam channel.

Figure 4.117 shows a section through the entry collimator. The beam channel does not change in diameter, but the shielding diameter does change, creating a radiation labyrinth. The lighter gray section around the beam channel is concrete. Surrounding the plug is a half-inch air gap, followed by the larger pipes of the fixed liner. The darker gray material around the fixed liner is steel shot that will be poured in through the curved pipe at the upper right of the picture. The steel shot is contained by a larger pipe embedded in the concrete absorber.

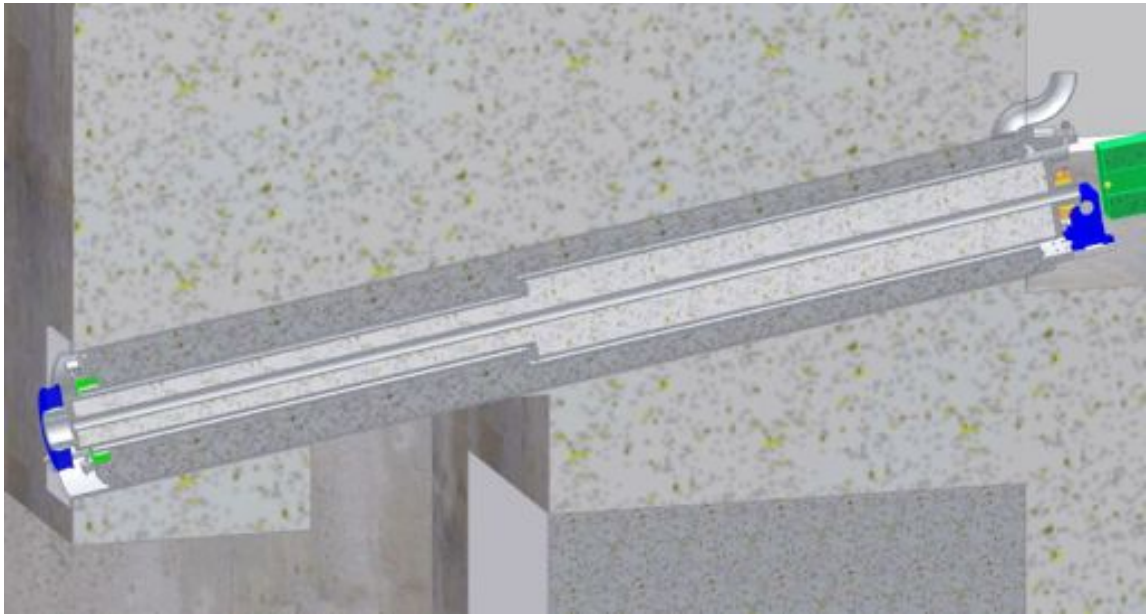

Figure 4.117. Cross section through the Entrance Collimator.

The radiation levels in the PS area (the left of the picture) will be relatively high. To avoid unnecessary exposure to personnel, adjustment cams for the upstream end of the entry collimator will be accessible from the magnet room using the blue gear boxes shown in the upper right. Both ends of the plug collimators are supported by custom spherical bearings to allow for angular alignment. The beam aperture through the entrance collimator is 50 mm. The entrance plug collimator weighs 2,022 lbs. The collimator beam channel is 4,286.25 mm long.

The collimator plugs may be adjusted at any time because of the air gap between the plugs and the fixed liner. The fixed liner may also be adjusted before the steel shot is installed. Figure 4.118 shows a close-up view of the upstream end of the entry collimator. The green rings are the inner and outer races of a spherical bearing that allows angular adjustment of the fixed liner. The blue rings are the inner and outer races of the spherical bearing that supports the upstream end of the plug collimator. The set-screws and blocks on the fixed liner allow it to be positioned within the pipe embedded in the concrete. Once it is positioned, there are rotating clamps that fix the position of the fixed liner flange relative to the flange welded to the embedded pipe. This seals the upstream end, allowing the steel shot to be poured in place.

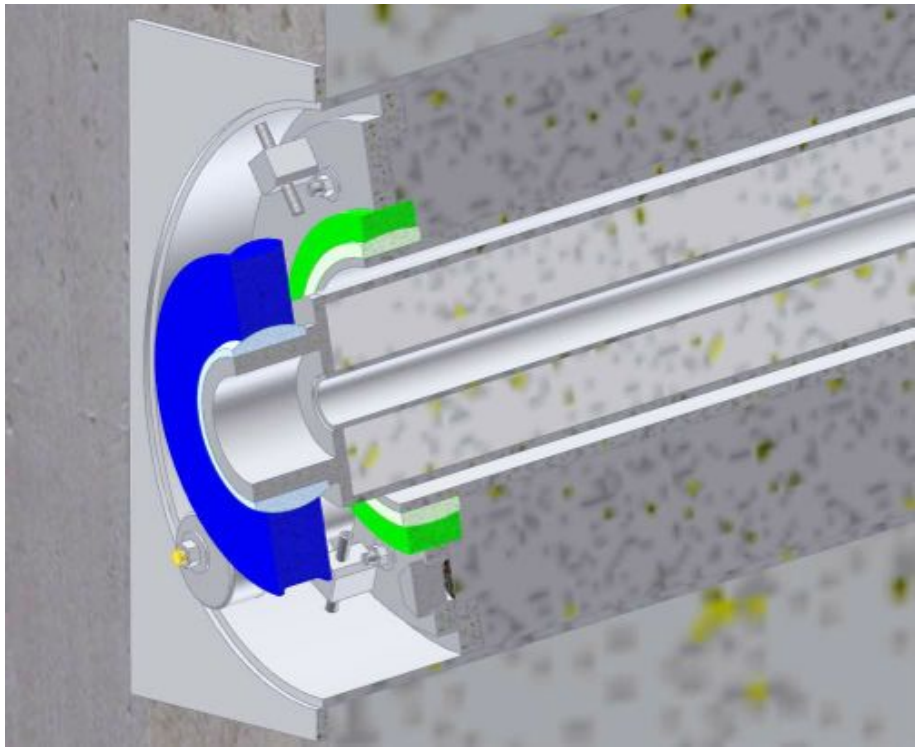

Figure 4.118. Close-up section view of the upstream end of the entrance collimator.

The entrance collimator plug is designed for alignment from within the magnet room, as the radiation level in the PS room will be too high to allow human access to the upstream end of the entrance collimator. The mechanism that provides this capability is shown in Figure 4.119. The dark blue ring is the outer race of the plain spherical bearing that supports the upstream end of the plug collimator. This race sits on two circular eccentric cams that are driven by shafts running through the space occupied by the steel shot up to the magnet room. The pitch and yaw angles of the entrance collimator can be changed by individually adjustable cams. Threaded rods on the flange allow for adjustment of the fixed liner to the required angle. The green ring is the spherical plain bearing on the fixed liner that allows for angular adjustment of the fixed liner.

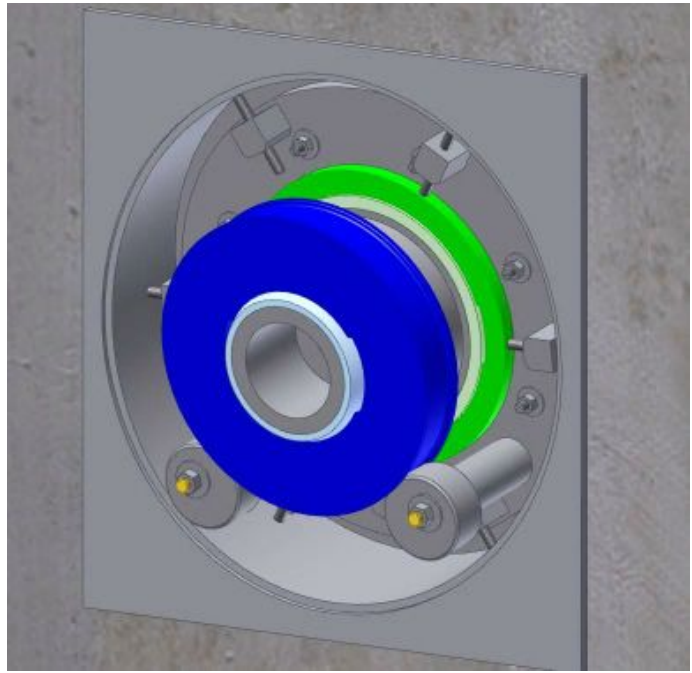

Figure 4.119. Front view of the upstream end of the entry collimator.

Figure 4.120 shows the downstream end of the collimator. The blue gear boxes allow the cam drive shafts to be adjusted and their worm and wheel design is self-locking. The mechanism on the shelf above the collimator allows the downstream end of the plug to be adjusted both horizontally and vertically. The orange and yellow rings are the inner and outer races of the spherical plain bearing that supports the downstream end of the plug collimator.

The filter magnet sits in between the entrance and exit collimator plugs on its own kinematic mount (see Figure 4.121). The magnet weighs 4,286 lbs. and is 3,683 mm long. The upstream end of the magnet support controls the position of the magnet along the beam direction because the ball and socket joint on which it sits (see Figure 4.122) has no freedom in that direction. The ball and socket is completely free to adjust in roll, pitch and yaw. The vertical threaded rods below the ball and socket joint control the height of the ball and socket. The socket is mounted on a horizontal slide that adjusts the transverse position. The fixture on which the downstream end of the magnet is mounted is shown in Figure 4.123. The blue plate adjusts this end of the magnet transversely and in yaw. The bronze bullet nose pins adjust for vertical, pitch and roll.

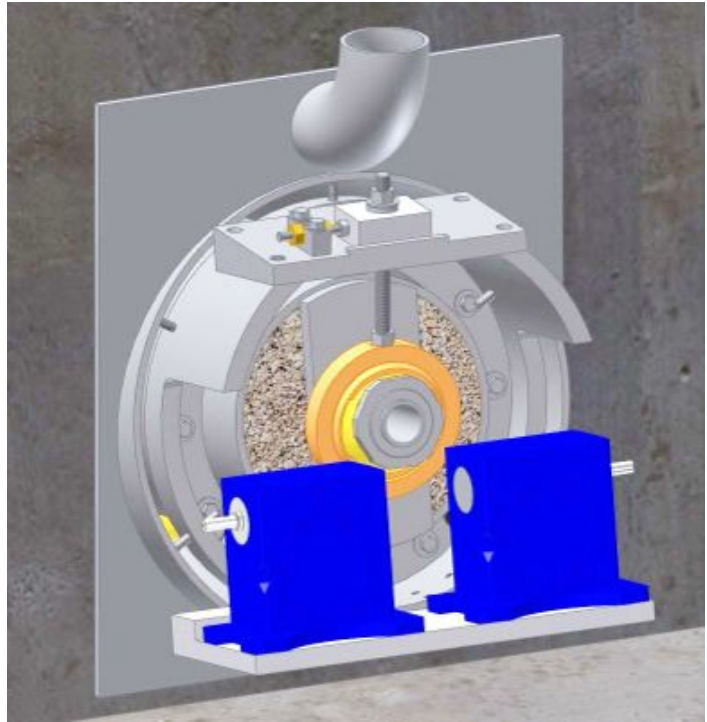

Figure 4.120. View of the downstream end of the entrance collimator.

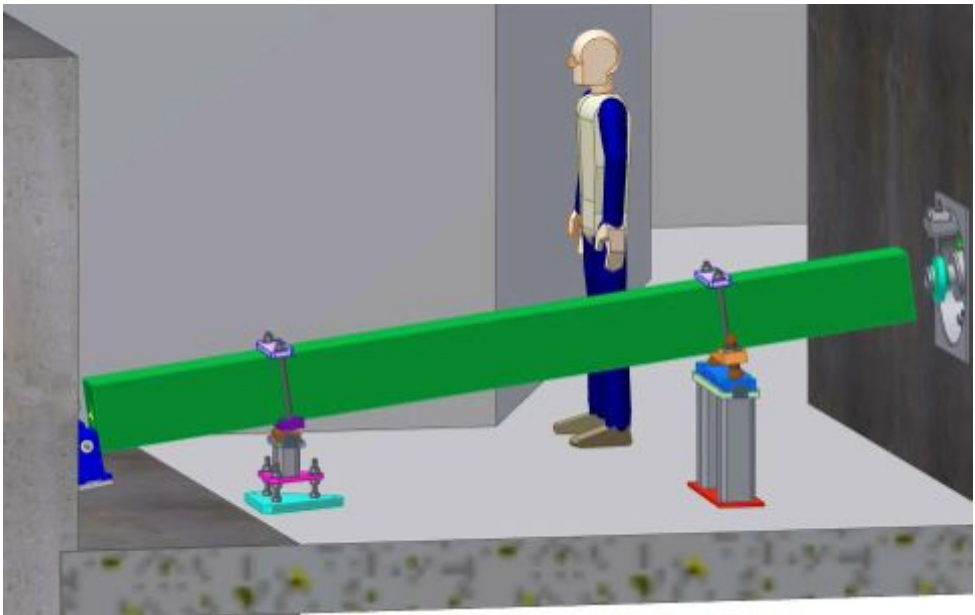

Figure 4.121. The filter magnet on its kinematic mount.

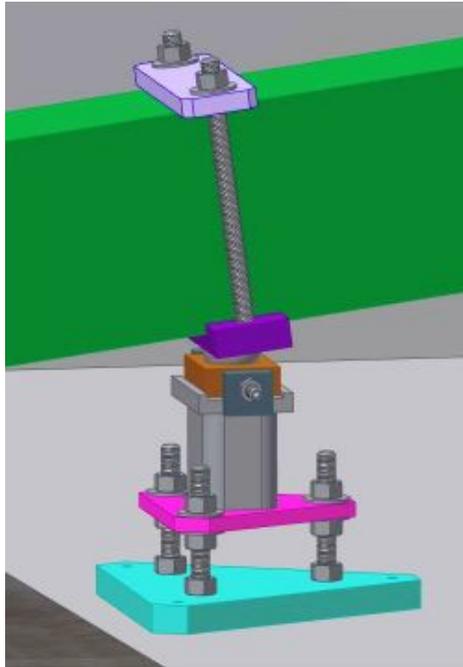

Figure 4.122. Close-up of the ball and socket joint of the filter magnet kinematic mount.

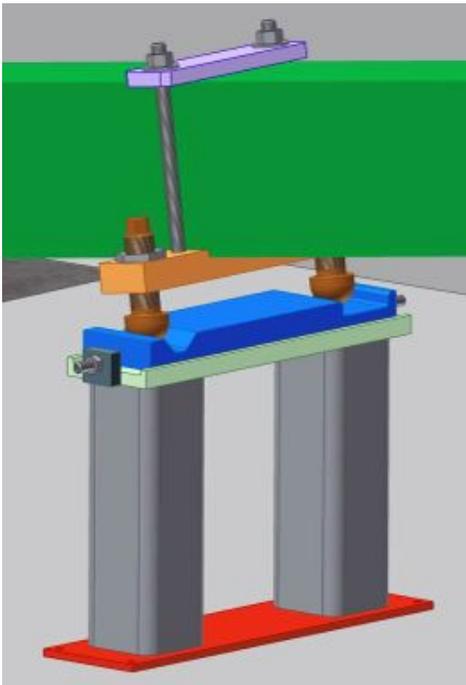

Figure 4.123: Close-up of the V-groove and flat joints of the kinematic mount.

The concept of the kinematic mount is that any adjustment can be made without interfering with other adjustments or causing strain to build up in the structure. If the horizontal position of the ball and socket joint at the other end is changed (changing the yaw of the magnet), the bullet nose pins simply slide on their mounting surfaces. Since the magnet rests on its stand with three ball joints, it can be picked up and replaced on the stand without changing its position. This concept has been employed on E760 and most recently on the LBNE target and horn modules. This allows the magnet to be lifted off its stand during surveying of the plugs and set aside to allow room for the surveyors to access the ends of the collimators. It can then be replaced to its original location.

The exit collimator shown in Figure 4.124 repeats most of the features of the entrance collimator in a shorter package, except for the cam drive. Since both ends of the exit collimator are accessible, both ends have simple horizontal and vertical screw adjustments. The beam channel opens up halfway through this collimator from the 50 mm aperture to 75 mm. This reduces the number of interactions occurring just upstream of detectors. The exit collimator weighs 1,015 lbs. and is 2,203 mm long.

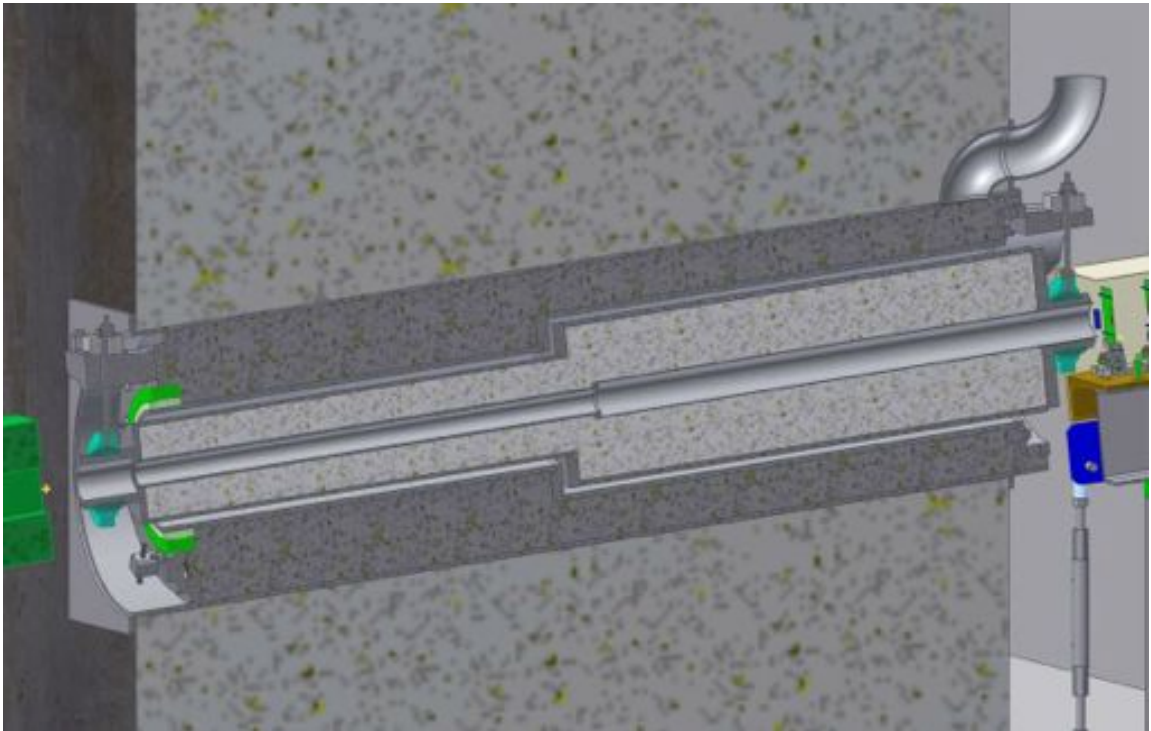

Figure 4.124. Section view through the exit collimator.

### 4.10.2.1.3  Pixels

The proposed detector is based on the FE-I4 silicon pixel chips developed for the ATLAS insertable B-layer upgrade using 130 nm CMOS technology. Each chip reads out 26,880 pixels arranged into 80 columns on a 250 µm pitch by 336 rows on a 50 µm pitch. Hits are digitized using a 24.9 nsec cycle synchronized to the accelerator, so that the number of clock ticks per Delivery Ring revolution period is an integer: (1694 ns)/(68 ticks) = 24.9 nsec. The distance between the peaks of two consecutive micro-bunches is 68 clock ticks: 10 "in-time" and 58 "out-of-time." Readout information identifies each pixel that was hit, the time of the hit and the time-over-threshold (ToT), which is the length of time the pixel remained in the high state (a proxy for the amount of deposited charge). Hit time and ToT are read out in units of clock ticks. More information about the chip can be found in [112] [113] [114] [115] [116].

The radiation performance and requirements of the extinction monitor depend on the expected flux of $\leq 2\times10^{12}$ neutrons/cm$^2$/year and $4\times10^{13}$ protons/cm$^2$/year (all without any re-weighting by known damage curves). The neutron flux was obtained from a MARS simulation of the extinction monitor shielding [117] whereas the proton flux was deduced from the signal rate simulations described earlier. The damage curve may significantly reduce the neutron equivalent fluence, but for protons with a momentum of 4 GeV/c, the damage factor is only about 0.6. Thus the effective flux is about $3\times10^{13}$/cm$^2$/year. Thus the radiation damage to a sensor will lead to a leakage current of order $I_0 + 24\ \mu A$ (per year) or an increase of 72 $\mu A$ after 3 years. This is well below the 800 $\mu A$/cm$^2$ FE-I4 maximum acceptable current. The chip itself is radiation hard to 250 Mrad and $5\times10^{15}$ neq/cm$^2$; also well above what we need.

A cooling system is needed to prevent the readout chips from overheating. The leakage current in a silicon sensor drops quickly with temperature, improving the signal to noise ratio. Therefore the sensors will be cooled to an operating temperature of a few degrees Centigrade, safely above the dew point of dry air. Heat from the sensors will be conducted by the TPG planes to the cooling pipes (shown in Figure 4.126 and Figure 4.128). An ethylene glycol based coolant will be pumped through the system and cooled using a recirculating chiller.

The Mu2e extinction monitor detector consists of two sensor stacks with four pixel planes each. A permanent magnet in-between the stacks allows reconstruction of the track momentum. The long pixel dimension is horizontal for all chips, providing a more precise measurement in the bend direction.

These pixel chips require an external trigger signal. However, for the extinction measurement, it is necessary to trigger on any out-of-time track passing through the pixels. This will be accomplished using scintillator trigger counters that are discussed below. In-time tracks will be pre-scaled to stay within the readout capabilities of the pixel chip. This is illustrated in Figure 4.125. A signal from each of the trigger scintillator counters is discriminated and fed into the TDAQ electronics, which computes a coincidence signal. The in-time signal is based on a clock counter. It defines the in-time and out-of-time regions. The bottom line in Figure 4.125 identifies a set of time bins that will be read out. The in-time hits are read out for only a fraction of proton micro-bunches.

A simulation has demonstrated that the chip can handle the data rate with a pre-scale factor of 10. This is sufficient to meet the measurement requirements, however further tests or simulations may indicate that an even lower pre-scale factor could be used. For example, the micro-bunch at $t = 0$ on Figure 4.125 is read out, but micro-bunches at $t = 68$ and $t = 136$ are not. During the out-of-time period, a trigger coincidence always generates a trigger signal (no pre-scaling). In addition to the prescaling of in-time trigger candidates, the trigger software widens the range of the clock to be read out by adding time margins before and after a trigger coincidence. This compensates for imperfect timing and asynchronous arrival of the particles.

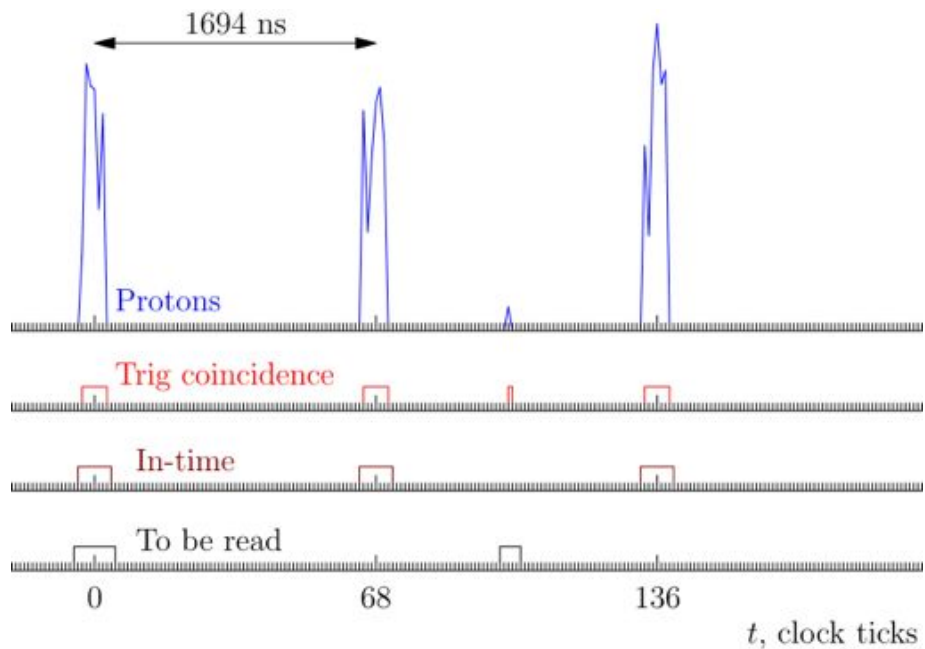

Figure 4.125. Trigger timing. Top line: trigger photo detector outputs; 2nd line: coincidence signal; 3rd line: in-time signal; 4th line: time bins for read-out.

Trigger commands will be sent over an LVDS input link. The minimal distance between commands is 5 clock ticks, but a static chip configuration allows the user to set a readout time interval for each command. The plan is to set the interval to 6 clock ticks. The chip requires 1 clock tick after the end of a readout cycle to get ready for the next trigger command. That amounts to a dead time. While each chip will have that 1-tick dead time after each readout, trigger commands to different pixel planes will be staggered, so that no more than one plane in each sensor stack in is in the "dead" state at any given moment. This arrangement guarantees that each track will be seen by at least 3 planes in each stack.

The mounting fixtures for the pixels are shown in Figure 4.126 and Figure 4.127. The three green externally threaded bushings give vertical, pitch and roll alignment. By rotating the green bushing, the aluminum triangular plate is adjusted in height and angle. The dark green mechanism under the pixel plane is for motion in the beam direction. The purple slide is for transverse motion. The orange assembly is for yaw alignment. The stack of adjustment mechanisms are commercially available items commonly used on optical benches.

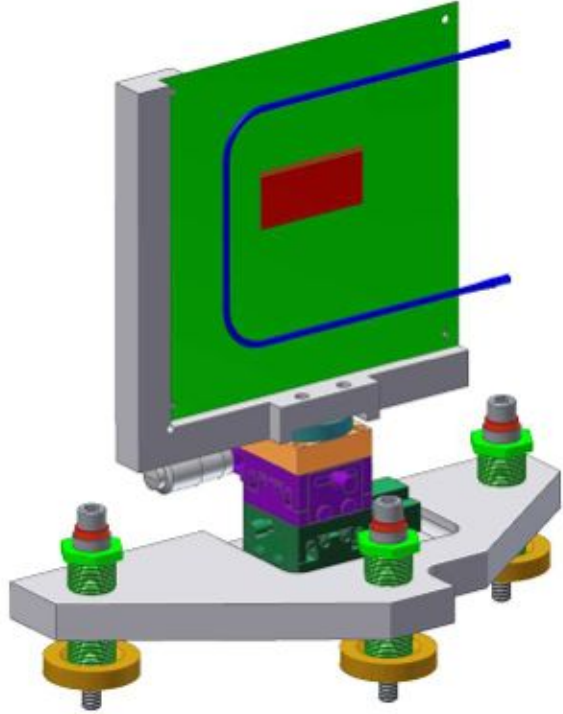

Figure 4.126. A single pixel plane support/alignment mechanism.

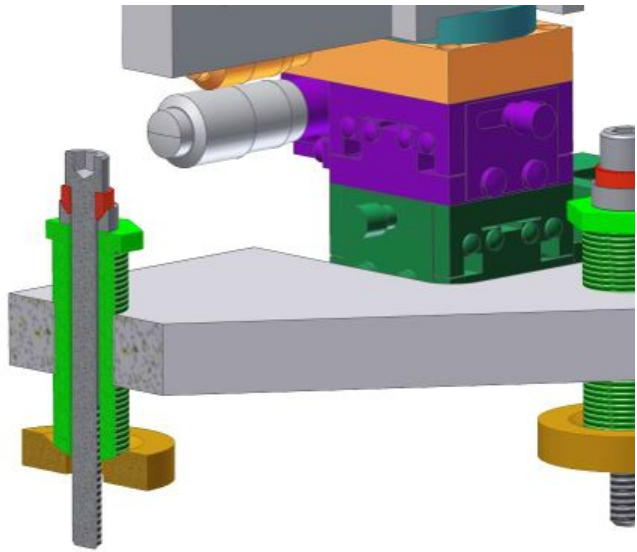

Figure 4.127. Section view showing the details of the vertical adjusters.

#### 4.10.2.1.4 Pixel Readout

The detection of individual charged particles arriving from the target is accomplished using a telescope comprised of eight planes of pixel sensors, four upstream and four downstream of a bending magnet. Each upstream plane makes use of two ATLAS sensors read out by four FE-I4B readout chips, while the downstream planes use three sensors read out by six ROC's. The sensor and readout chips are constructed as bump-bonded modules and are glued to a sheet of thermal pyrolytic graphite (TPG) to provide the mechanical support and cooling, while minimizing the mass of material in the path of the beam. The readout chips are wire bonded to a printed circuit board (the Pixel Hybrid) that will provide regulated power and interface the digital clock, control, and readout signals to a connector. Figure 4.128 shows the arrangement of the sensors and the pixel hybrid board on the TPG sheet.

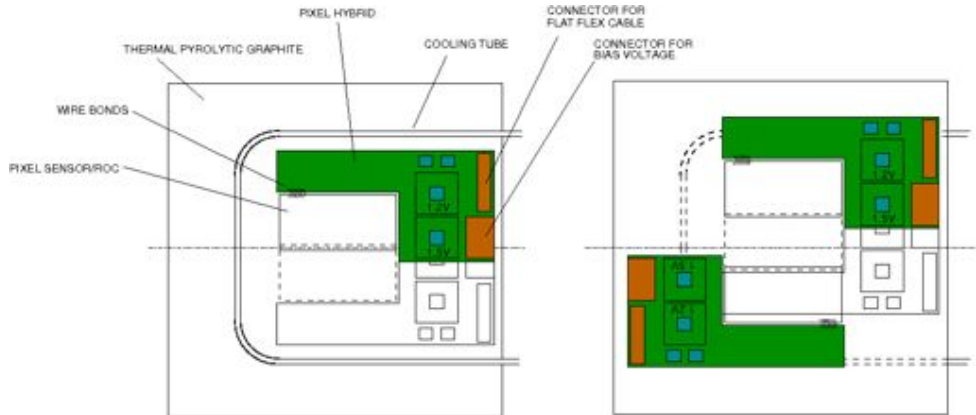

Figure 4.128. Pixel sensors and readout hybrid board arrangement on the TPG sheet. The two-sensor upstream plane is shown on the left and the three-sensor downstream plane is on the right.

This low-mass design will be constructed by first mounting the bump-bonded readout chip on an assembly fixture, where it will be tested using a probe card. The pixel hybrid PCB will then be clamped in place adjacent to the bump bonded module (BBM) and then wire bonded, followed by a repeat of the electrical tests. Epoxy will be deposited on the TPG sheet and a vacuum fixture will then hold the relative positions of the BBM and the hybrid module fixed while they are lifted from the fixture and precisely placed on the TPG sheet.

Not only does this approach reduce the amount of material in the sensor plane, but it also provides a way to thoroughly test each sensor assembly prior to incorporation into the completed telescope plane. Finally, the wire bonds will be encapsulated to protect them from mechanical stress and chemical attack from any moisture they may encounter over the life of the experiment.

The signals from the upstream and downstream sides of the telescope each interface to a printed circuit board referred to as the pixel-interface board. The role of the pixel interface board is to provide power distribution to the sensor planes, and to perform the logic level translation between the CML signals required by the FE-I4B chip and the LVDS signals used in the FPGA's in the DAQ system. Flat, flexible cable would be used to connect the pixel hybrid boards to the pixel-interface board. This low-cost option also simplifies the construction of the dry box used to contain the sensor planes that will be needed when operating them at temperatures below the dew point. The pixel-interface board does not contain any programmable logic and will be mounted close to the telescope planes to minimize the length of the FFC cables. Two 68-conductor VHDCI cables carry the signals from each pixel-interface board to the DAQ system.

### 4.10.2.1.5  Trigger Counters

The trigger counters serve three purposes. One is to provide a trigger for the pixel detectors during the out-of-time ticks. Between beam pulses, only a few particles per hour will be observed. The trigger counters will provide an external trigger for the pixel detectors that will measure the momentum of the particles. In addition, the trigger counters will determine the arrival time of the out-of-time particles. We expect very few particles during the out-of-time ticks, so a trigger efficiency of 99.9% is required. The fake trigger rate must be less 1 kHz.

The second function is to provide an accurate time stamp for any out-of-time particles that are observed. If we see more particles than expected, an important piece of diagnostic information will be their arrival time. The trigger counters will provide a time stamp with a resolution of better than a nanosecond for any particles present in the out-of-time clock ticks.

During the 250 nsec beam pulse, the counting rate will be too high for the trigger counters to observe each individual particle. However, they will have to be able to trace out the beginning and end of each beam pulse.

***Technical details:***

There will be six trigger counters, three upstream of the spectrometer magnet and three downstream. A trigger will require a coincidence of any two out of three counters upstream of the magnet, and any two out of three downstream of the magnet. The final trigger is then a coincidence of upstream and downstream requirements. If individual counters have an efficiency of 99%, the overall inefficiency of the trigger will be less than $1 \times 10^{-4}$, well within the requirements.

The counters will be 5 mm thick and slightly larger than the pixel detectors (45 mm $\times$ 40 mm upstream; 45 mm $\times$ 55 mm downstream). The scintillator used will be BC-404, a standard scintillator designed for fast timing. Pre-cut and polished pieces of scintillator can be purchased from commercial vendors. They will be read out through a standard light guide to a phototube. A Hamamatsu H6520, a ¾" tube with a built-in base, or similar, would be an acceptable phototube. The PMT signal will be input to a discriminator board present in the MicroTCA crate that will also produce the time stamp. The discriminated signals will then be passed to the FPGA in the MicroTCA crate for the necessary logic.

#### 4.10.2.1.6  Spectrometer Magnet

The spectrometer magnet is a repurposed prototype permanent dipole from the Fermilab Recycler Ring.  Its Mu2e purpose is to bend out low energy electrons generated from muons stopping in the upstream silicon. It was chosen because it was the shortest (0.5 m) available permanent dipole. Its 0.14 T-m integrated field provides enough bend to permit a ~10% momentum measurement, though that is not its primary purpose.

The spectrometer magnet mounting is shown in Figure 4.129. Its alignment is done crudely and then it is locked down in position by the purple corner brackets.  The magnet may be elevated above the channel if an offset is needed with respect to the pixel planes.

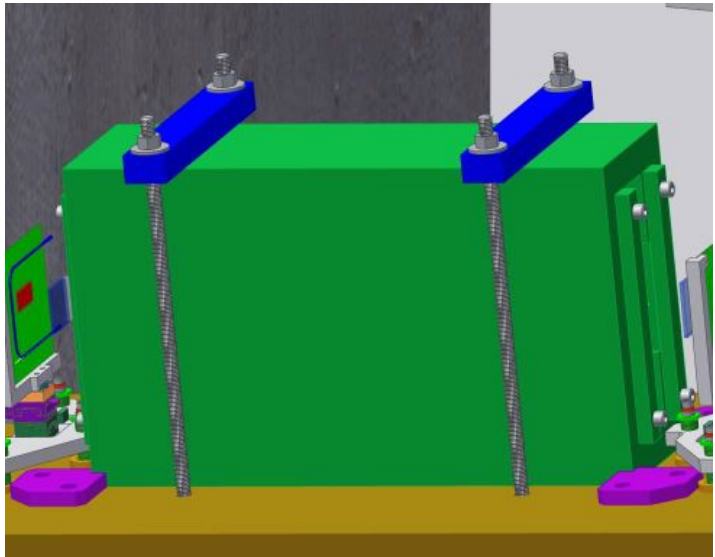

Figure 4.129. Close-up of the spectrometer magnet.

#### *4.10.2.2 Integrated Trigger/Pixel/Spectrometer support*

The extinction monitor filter is designed to be adjustable in the event of uneven settling of the proton dump with respect to the target hall. This means that the detector components must also be movable. However, in order to avoid the need to recalibrate reconstruction parameters and to facilitate the realignment effort, it is desirable that the pixel spectrometer and trigger counters be moved as a unit, maintaining their relative alignment. Thus, they will be mounted on a channel table as shown in Figure 4.130. The relative motions of the objects above the channel are small because the angles the pixel planes make with the spectrometer magnet are small and not likely to change much.

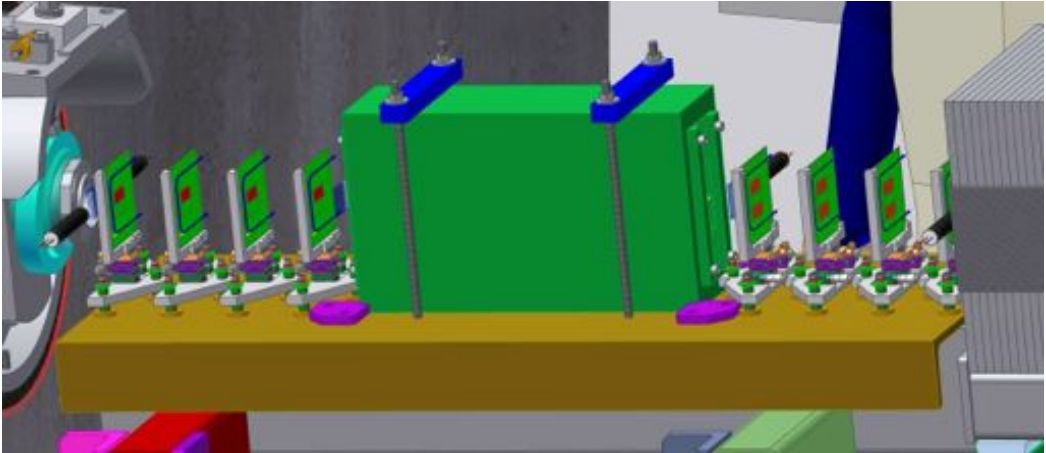

Figure 4.130. Triggers, pixels and spectrometer magnet on the channel table.

Figure 4.131 and Figure 4.132 show the six-strut support under the channel that does the overall alignment of everything above the table. The three vertical struts control the height, roll and pitch of the table. The two struts perpendicular to the beam axis control transverse motion and yaw. The single strut almost parallel to the beam controls the longitudinal position of the table. This design is a very cost effective way to align structures because the struts can be made from many identical, cheap parts. The struts act as turnbuckles that can extend or retract with rotation.

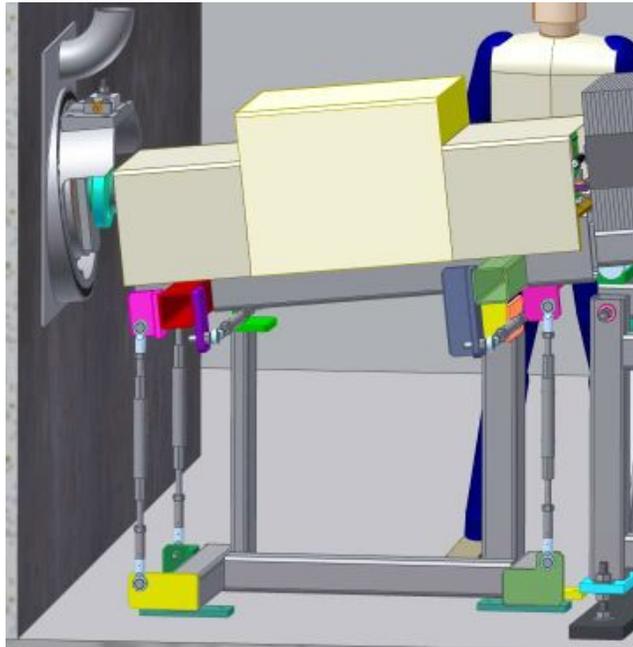

Figure 4.131. Six strut support structure to align the table on which the pixels, trigger, and spectrometer magnet reside.

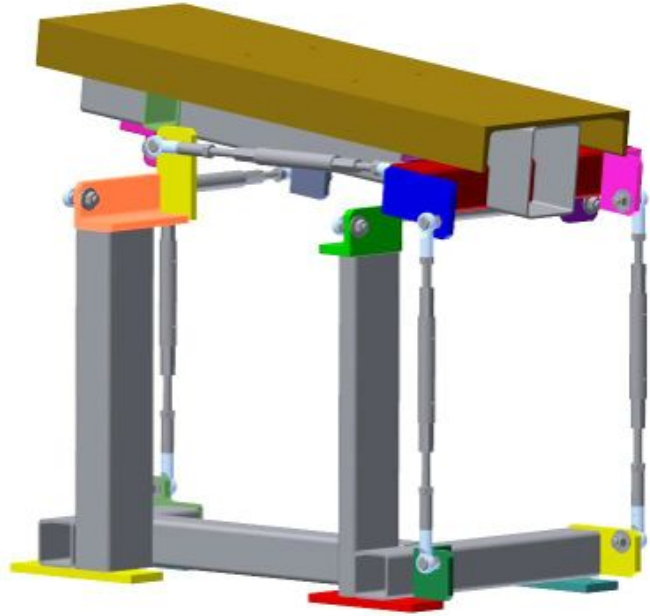

Figure 4.132. View of the 6-strut support without the components attached to the C-channel table at the top.

This same structure was built on a larger scale to support the MiniBooNE target and was remarkably rigid given all of the ball and socket joints. The struts do not show significant play when manufactured per the tolerances in the drawings. Any strut can be extended or retracted and all the other struts adapt as the table is reoriented. This does cause coupling between degrees of freedom, but the longer the strut the smaller the effect of coupling. The alignment is iterative, but as the displacements become smaller (as the object approaches the required position), the coupling becomes less significant.

The beige cover in Figure 4.131 is an enclosure that allows dry air to be circulated around the pixel planes. The pixel planes are cooled with glycol and the dry air is to prevent sweating of the cooling tubes under the cover.

#### 4.10.2.2.1  Muon ID Detectors
The muon range stack consists of a steel block, 40 cm square and 180 cm in depth. To facilitate construction, the steel will consist of a series of plates, 40 cm square and 1.2 cm in thick. These plates weigh approximately 35 pounds and are manageable without special lifting fixtures. These plates will be hand-stacked and then welded into place.

The range stack will include four scintillating planes (see Figure 4.133). Three scintillator counters will be placed as close to 145 cm, 162 cm and 180 cm as practical, subject to engineering considerations. The fourth scintillating plane will be located at a depth of about 41 cm. The scintillator will consist of 40 cm transverse sheets, a quarter inch thick. The scintillator will be BC-404 and read out by Y11 WLS optical fiber. The optical fibers will be affixed to the scintillator either by milling grooves into the scintillator or by taping it to the scintillator's surface. The optical fibers will then be bundled in an acrylic cookie, polished and attached to a phototube (see Figure 4.134).

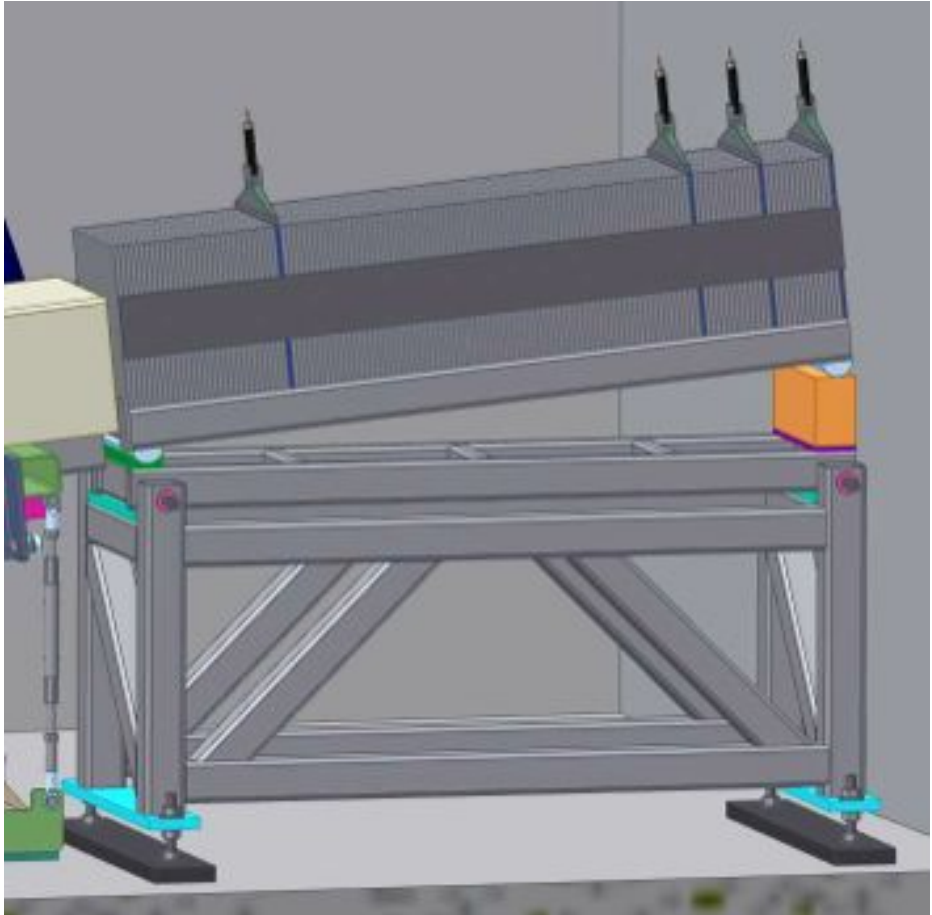

Figure 4.133. View of the Muon ID range stack.

Before the scintillator detectors will be inserted in the range stack, their efficiency to detect muons will be established either in beam tests or in a muon cosmic ray telescope. The dependence of efficiency will also be mapped out.

The phototubes will be read out by an ADC system specified in section 4.10.2.2.2, which will digitize the pulse height in 1 or 2 nsec intervals, resulting in the waveform of the signal from each detector.

The design of the support structure (shown in Figure 4.133) is based on a machinist's sine bar. During assembly the angled table frame is level. (The orange shim block is not in place.) ½" thick steel plates are stacked up along the table and welded to the side plates to make a monolithic block of steel. Four locations have slots for the insertion of scintillator tile/fiber assemblies. Once the stack is complete, hydraulic cylinders can be used to raise it to the required angle for the insertion of the orange shim. The support frame has horizontal adjustments at both ends to achieve horizontal and yaw alignment.

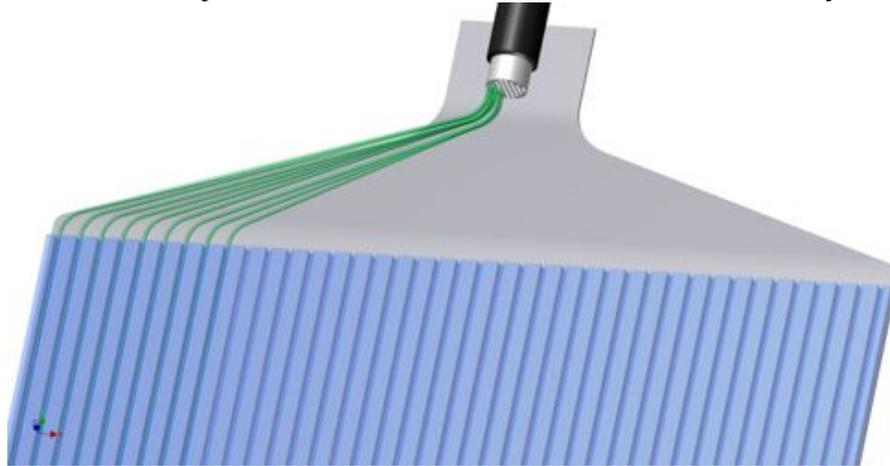

Figure 4.134. Close-up of the scintillator plate with grooves for the Y11 WLS fibers routed to a "cookie". The configuration shown is one of several possible arrangements of scintillator and fibers.

The vertical threaded rods near the floor provide vertical, additional pitch and roll degrees of freedom. The steel stack weighs 5,113 lbs. The support frame weighs 1,449 lbs. The width of the frame is 27 inches except for the cyan plates, which are 34 inches wide. This allows most of the frame to be lowered through the 28 inch wide hatch. However, the cyan plates at the base of the support frame may have to be welded on inside the detector room.

### 4.10.2.2.2 DAQ

A common architecture is used to control and read out the front-end electronics that instruments the beam extinction monitor and the target extinction monitor. On-line signal processing, trigger generation, and data formatting for readout are performed using FPGA's that receive data from the front-end electronics by means of FPGA Mezzanine Cards (FMC's). The FMC [121] provides a well-defined interface between the front-end electronics and the DAQ system that largely decouples the design of the front-end electronics from specific choices of hardware platforms used to implement the FPGA logic. However, the MTCA.4 (MicroTCA for High Energy Physics) extensions to the MicroTCA standard provide a natural framework in which to implement the DAQ system, since it leverages the availability of electronics modules previously developed for other high energy physics and accelerator applications.

As the beam and target extinction monitors are separated by hundreds of meters, their DAQ systems for the beam and target extinction monitors would be implemented in separate MicroTCA crates with identical underlying architecture but which would operate independently. The requirements of the system are satisfied by the use of a single type of AMC card that provides an FPGA with connectors and front-panel space that allow two FMC cards to be mounted. Three types of FMC cards are required to instrument the pixel telescope and the scintillators in the system with both fast timing and analog current readout.

The digital signals used to control and read out FE-I4B ROC's on the pixel planes are passed directly from the FPGA on the carrier AMC to the Pixel Interface Board via FMC cards that simply provide a connector for the 68-conductor VHDCI cables. Except for an inventory control PROM, this FMC would be completely passive, relying on the LVDS drivers and receivers in the FPGA for signal transmission and termination. As the nominal data rate from the FI-I4B readout chip is only 160 Mb/s, and these signals are buffered on the Pixel Interface Board, active drivers and receivers should not be required on the Pixel FMC board. The assignment of signals on the 68-pin VHDCI connectors allows two of the four pixel planes to be instrumented with one cable. Thus, two Pixel FMC cards are used to read out and control each side of the telescope.

The trigger scintillators are used to provide fast timing signals with which to initiate readout of the pixel planes between bunches. As the occupancy between bunches is low, it is sufficient to instrument these with leading-edge discriminators. It will be desirable to DC couple the PMT signals in order to mitigate effects of baseline shifts due to pulses received in-time with the previous bunch. The discriminator design is based on the use of high-speed comparators with differential LVDS outputs with thresholds controlled independently on each channel by means of by a low-speed serial DAC, programmed via the FPGA on the carrier AMC. The intrinsic timing resolution for this type of circuit is typically much better than one nanosecond and would be sampled on both edges of a 250 MHz clock signal in the FPGA to achieve a sampled timing resolution of 2 nsec. The FPGA firmware would provide the capability to dynamically adjust the relative timing of different channels, reducing the physical requirements on cable lengths used to instrument the PMT's. The available front-panel space allows for six input channels, an accelerator clock signal and a fast digital output that would be used to initiate readout of the pixel planes between bunches. This channel density allows the signals from three upstream and three downstream trigger scintillators to be processed by the same FPGA, which performs coincidence logic as well as continuous monitoring of efficiencies and rates of noise hits.

Within a bunch, the occupancy will be high enough that pulses from individual particles will overlap and it will not generally be possible to measure times of individual pulses. In order to measure the particle flux within each bunch, it will be necessary to digitize the waveforms from a subset of PMT's in order to measure the time-dependent current. By calibrating the response to single particles, the average current will provide a measure of the particle flux as a function of time within each bunch. The intrinsic shaping time of the PMT's will limit the effective bandwidth of the input signal and it will not be necessary to digitize the waveform with a sample rate beyond a few hundred MSPS. There are a wide range of commercially available waveform digitizer FMC modules that satisfy these requirements.

### 4.10.2.2.3  Simulations

This section describes the simulations that have been performed to assess how well the system meets the requirements.

***Simulations: Filter and collimators***

The expected signal rate in the extinction monitor produced by this filter design has been simulated using G4beamline. The results are summarized in Figure 4.135. The expected rate is $8.3 \times 10^{-7}$ per POT, which comfortably meets the requirements. Not unexpectedly, the signal is comprised mainly of protons with a 13% contribution from positively-charged pions.

The filter is also responsible for reducing beam related backgrounds, both accidental and off-target. The accidental background depends not only on the flux of low energy particles that survive the filter, but also the detector's sensitivity to that flux. Consequently, we defer discussion of this background to later and concentrate here on the off-target background.

The off-target background depends on how the orientation of the filter aligns with the distribution of out-of-time beam interactions inside the heat and radiation shield (HRS) that surrounds the target.  The latter, in turn, depends on the transverse structure of the out-of-time beam as it enters the HRS. This is not well understood, but a conservative assumption is that the beam is uniformly distributed across the available aperture as it enters the final focus section of the primary beam line. The final focus beam optics is designed assuming that the transverse beam size is at a minimum at this point and images this point source onto the target. This means that the transverse distribution of the out-of-time beam at the target will match that at the upstream waist point. A uniform distribution at this point will be imaged at the target location and result in a larger ratio of off-target to on-target interactions than a more Gaussian distribution.  By placing collimators at the upstream waist and at the point where the beam enters the HRS, one can restrict the transverse size of the beam such that it fits well within the beam entry opening. Thus, these collimators prevent the beam from interacting in the HRS and largely eliminate the off-target background.

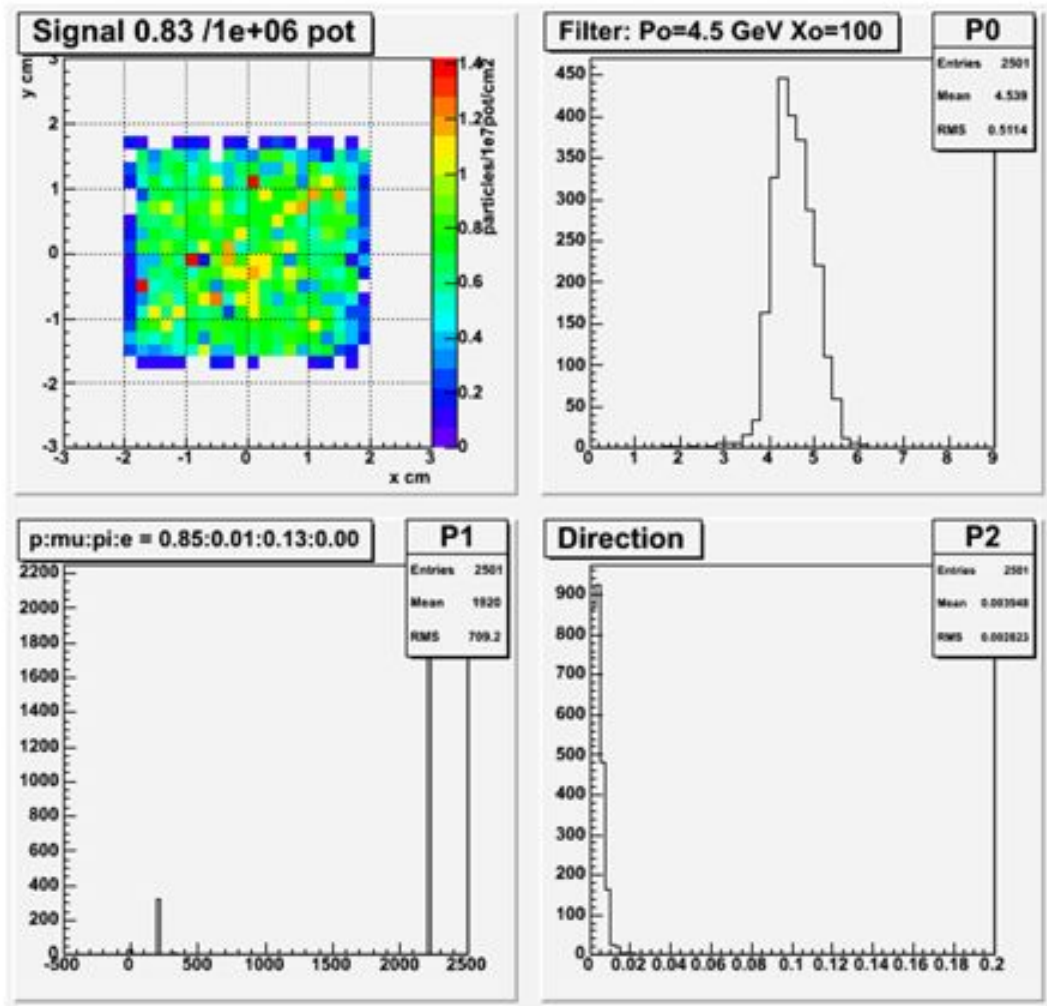

Figure 4.135. Results of a signal rate simulation for the target extinction monitor. The top left panel shows the expected distribution of signal tracks at the first pixel plane. The top right shows their momentum distribution. The bottom left shows the distribution of Geant4 particle ID's and the bottom right shows the angle in radians between their direction of travel and the filter axis.

To estimate the size of any residual off-target background, G4beamline simulations of the configuration described above have been performed. In these simulations, the apertures of the upstream and downstream collimators were varied. This allowed a prediction of the rate of signal events observed in the extinction monitor that originated from off-target interactions normalized to the number of beam protons that hit the target (about 0.5% of the beam that enters the HRS). It was found that, without the upstream collimator, this rate was about 10% of the true signal rate. With a 20 mm radius upstream collimator, the rate was too small to determine from the simulation. However, by comparing the distribution of interaction points with and without the upstream collimator (see Figure 4.136), it was possible to estimate a reduction factor. This was done by counting the number of interactions that occurred inside the HRS that generated a high momentum particle that reached the proton dump at an elevation corresponding to the filter entry point. A reduction factor of approximately 15 was obtained. Thus, it is concluded that the off-target background from interactions in the HRS will be less than 1% of the true signal rate, well within the requirements.

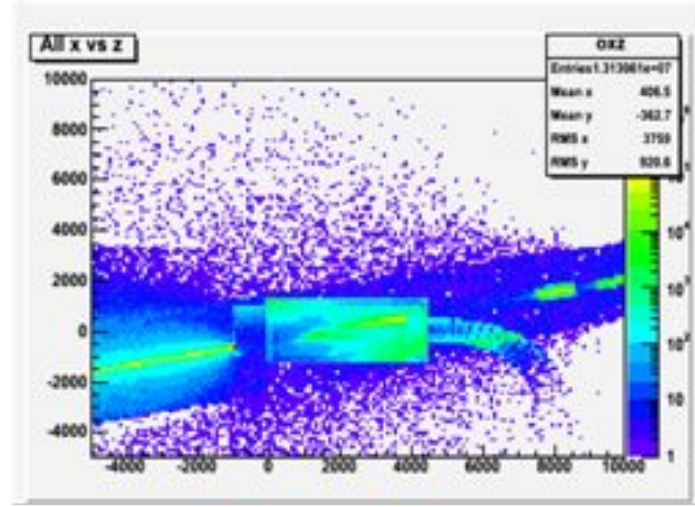

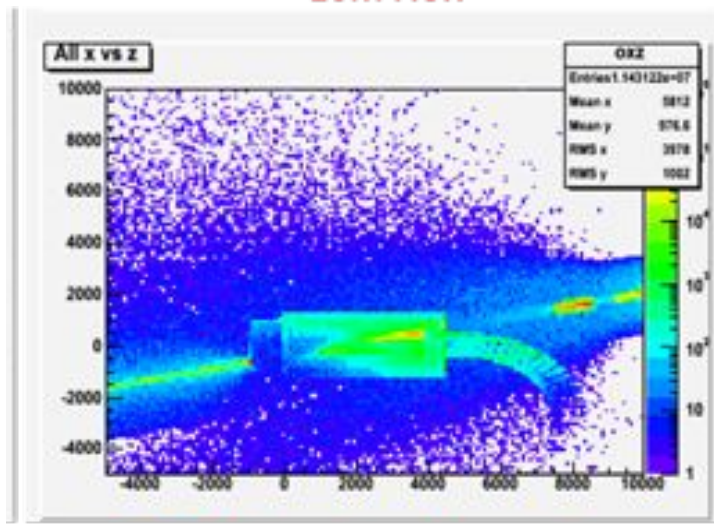

Figure 4.136. X vs. Z distributions of interactions that generate potential extinction background candidates with and without an upstream collimator. The outline of the production solenoid is clearly visible in both plots as are hot spots corresponding to the protection collimator and plumes where the beam passes through the air downstream of the solenoid. A collimator located at the beginning of the final focus optics greatly reduces the relative contribution from the upstream interactions.

***Simulations: Silicon spectrometer***

*Track reconstruction performance*
As explained above, each triggered track will have hits in at least 3 out of 4 planes read out in each of the pixel plane stacks. A detailed simulation of a 6-plane silicon pixel spectrometer (3 planes upstream and 3 downstream of the spectrometer magnet) was performed in GEANT4. Using information from only 3 + 3 planes represents the worst case, as most tracks will have hits recorded in all 4 + 4 planes. Charge drift and fluctuations in the sensor, and a realistic response model of the detector chip were implemented in the software [118]. Simulated and digitized hits that contained only detector-like data (no use of MC truth) were analyzed, and track candidates identified and fit [118].

Figure 4.137 shows the track reconstruction efficiency as a function of signal candidate multiplicity. It is calculated by simulating the passage of a number of 3 GeV/c protons through the detector. The efficiency is computed for MC particles that made hits in all 6 detector planes. There is no requirement that the particle is the generated proton; it may be a secondary. A particle is counted as reconstructed only if there is a track that shares all 6 clusters with the particle. At the nominal micro-bunch intensity less than 40 signals per micro-bunch are expected. The average efficiency is 97%, and it varies with multiplicity by less than 1% for an intensity range between 0 and 100 tracks/micro-bunch.

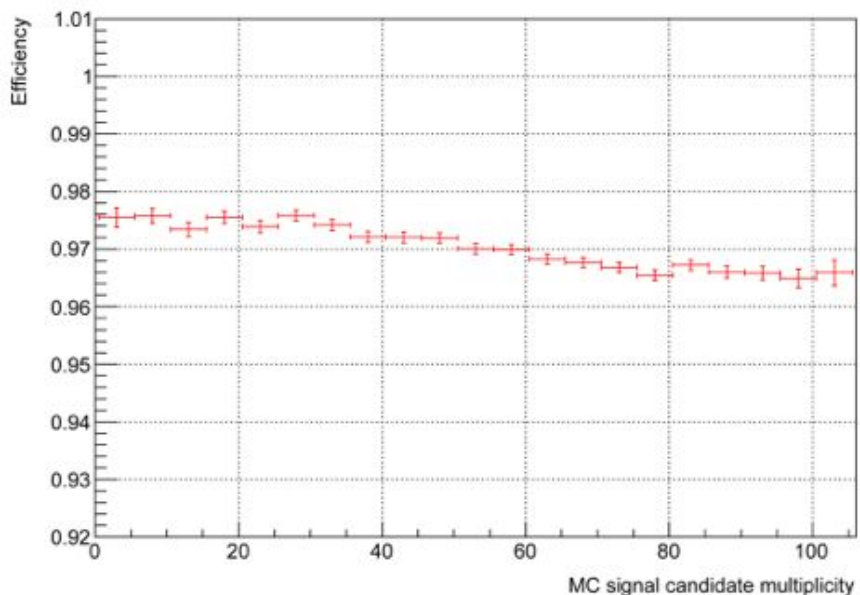

Figure 4.137. Track reconstruction efficiency as a function of signal candidate multiplicity

Figure 4.138 shows the ratio of the number of reconstructed tracks to the number of generated primaries for the same dataset. Its absolute value is dominated by the detector acceptance for the specific input proton distributions and is therefore somewhat arbitrary. However this ratio is sensitive to reconstruction efficiency and fake track rate. It does not depend on the truth matching procedure and its dependence on multiplicity is directly related to the systematic uncertainty on the extinction measurement coming from the intensity dependence of the reconstruction. The *x*-axis range corresponds to 0 to about $0.31 \times 360 \sim 113$ reconstructed tracks per micro-bunch. In this range, the ratio is constrained between 0.31 and 0.32. Therefore the total systematic uncertainty, including efficiency and fake rate, is 0.005/0.315 = 1.6% for this range.

*Extinction monitor backgrounds*

Extinction monitor signal tracks are a few GeV/c and arrive promptly after primary protons interact with the Mu2e production target. Any effects that create pixel hits after the end of the nominal proton micro-bunch are sources of background that may mimic signal from out-of-time protons. These are:

- Cosmic rays,

- Interactions of late arriving particles created by the proton beam,

- Radioactive decays in pixel sensors,

- Electronic noise.

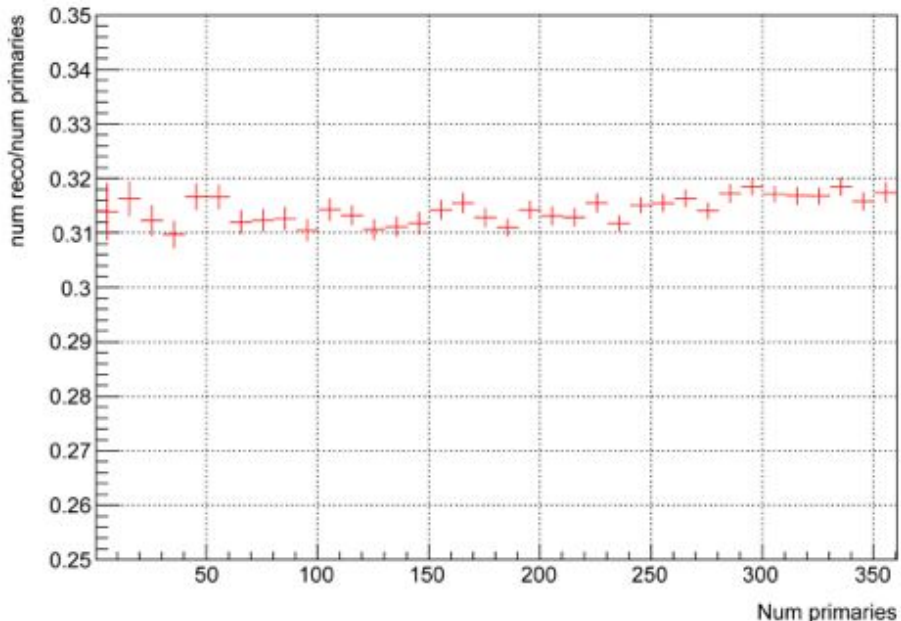

Figure 4.138. Reconstructed track ratio:

Among the backgrounds sources, only cosmic rays can produce out-of-time tracks with a few GeV/c momenta. Other classes of backgrounds generate lower momentum tracks that only reconstruct in one of the two sensor stacks, and individual pixel hits. The cosmic ray background track rate for a pixel based extinction monitor has been estimated using GEANT4 [119] as $0.030 \pm 0.007$ tracks/hour. This estimate is an upper bound as it does not take into account the momentum analyzing capability of the current detector design.

The background induced by late arriving particles from proton beam interactions was estimated using a high statistics, multi-stage simulation. The simulation used MARS to simulate proton interactions in the production target and the beam dump, and propagation of shower particles through the shielding. Particles entering the detector room volume were passed to the first stage of GEANT4 simulation. The particles were re-used multiple times, with their room entrance positions randomized using an adaptive randomization algorithm. The first GEANT4 stage recorded particles on the boundary of a box surrounding the pixel spectrometer and the second GEANT4 stage re-used them with another randomization, to achieve an equivalent statistics of $0.7 \times 10^9$ proton micro-bunches ($2 \times 10^{16}$ primary protons). Figure 4.139 (top) shows all simulated tracks in the vicinity of the pixel detector from a micro-bunch of $3\times10^7$ protons on target. Figure 4.139 (bottom) shows just out-of-time tracks for a micro-bunch.

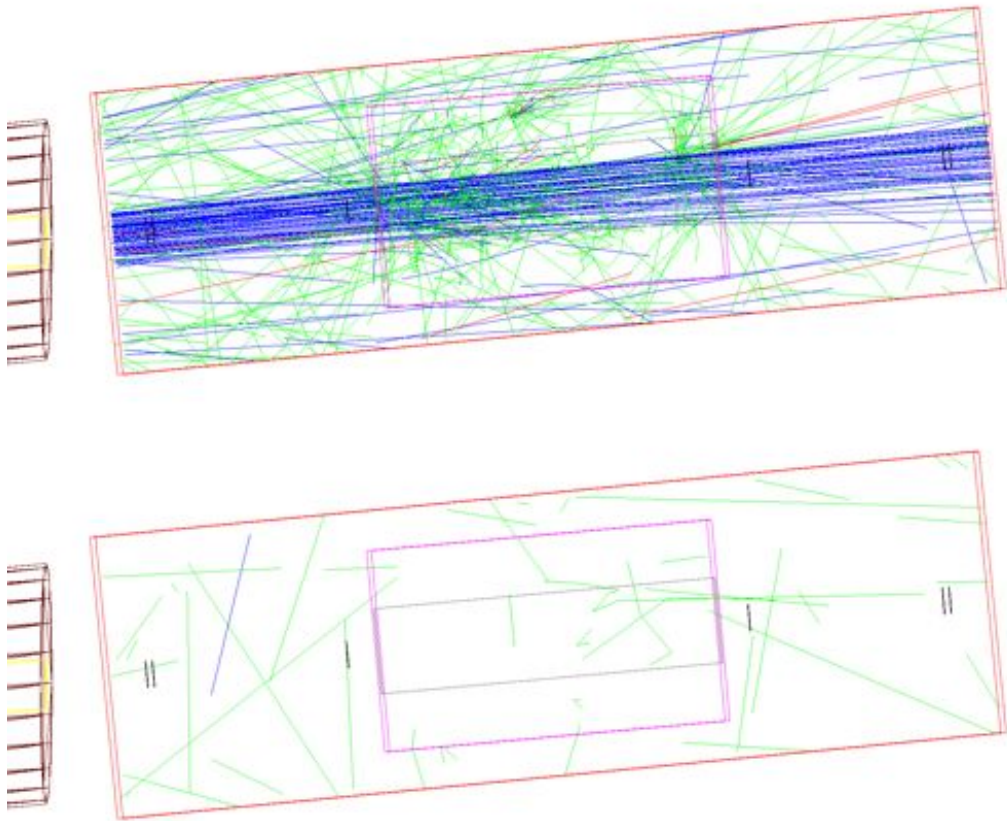

Figure 4.139. Simulated tracks in the extinction monitor detector from a single micro-bunch. The top plot shows all tracks while the bottom shows just the out-of-time plots.

The simulation determined that the out-of-time hit rate due to late particles is $2.9 \times 10^{-3}$ pixel hit/clock tick. The combined rate of random pixel hits from cosmic tracks, induced radioactivity, and electronic noise, does not exceed $3.6 \times 10^{-4}$ hits per pixel per clock tick [120]. This small contribution was added to the simulated hits from late particles, and all out-of-time hits in the simulated dataset were passed through the track reconstruction algorithm. No complete tracks were found. 1750 straight track segments were observed in just the upstream detector segment and 3107 segments in just the downstream half of the detector. The probability that an upstream and a downstream segment from independent sources occur in a given clock tick and combine into a fake track does not exceed $6 \times 10^{-15}$, which corresponds to a fake track rate of $7.4 \times 10^{-4}$ per hour.

In conclusion, the least restrictive bound on the pixel track background rate is $0.030 \pm 0.007$ tracks/hour, which is small compared to the signal rate at the nominal $10^{-10}$ extinction level.

### Simulations: Muon range stack

A possible background to the extinction measurement that cannot be reliably simulated is due to interactions upstream of the HRS that produce muons that penetrate the HRS and make their way into the filter's acceptance. Since the muon content of the real extinction signal is expected to be small, this background can be detected and corrected for by establishing the muon to hadron ratio of particles arriving both in-time and out-of-time with the standard extraction of protons to the production target. Of the possible technologies that might establish these ratios, a high absorption filter consisting of steel as a passive absorber, combined with scintillator sampling detectors was selected. This section describes the expected performance of the proposed detector using a standalone GEANT simulation. This detector is called the range stack.

The range stack sits behind a magnetic spectrometer that passes charged particles with a beam energy of $4 \pm 0.4$ GeV. During the 250 nsec proton pulse, $50 \pm 25$ particles will arrive, spread approximately uniformly across the width of the pulse. Muon contamination is predicted to be 1%. Thus, in order to fully explore the operational performance necessary for the range stack, particles were simulated at 3, 4 and 5 GeV. Further, the performance of 1, 50 and 100 protons arriving simultaneously was simulated.

To simply simulate detector performance, the three baseline beam configurations that were studied most thoroughly were: 4 GeV muons [1,000 events], 4 GeV protons [9,000 events], and 4 GeV protons with 100 arriving simultaneously [9,000 events]. Additional studies with 5 GeV beam particles were performed to establish that the additional energy did not materially affect the design needs.

To understand the best arrangement of steel and scintillator, a generic calorimeter configuration was used. The simulated detector consisted of 70 layers of 3.5 cm steel, interspersed with 6 mm scintillator. The dimensions of the detector were 40 cm in both transverse directions. The transverse dimensions were selected to be large enough to have more than 95% containment of proton-induced hadron showers at normal incidence in the center of the detector. The ionization energy deposited in all scintillator and steel layers was recorded. In order to characterize the detector for the arrival of 100 simultaneous protons, 100 single proton events were sampled from the single proton sample and the energy deposited in each layer was summed together. This choice means that the statistical uncertainties for the multi-proton sample cannot be easily interpreted; however these uncertainties have no substantial impact on the design.

Figure 4.140 shows the energy deposition in the scintillator layers for muons. The ionization energy exceeds 0.7 MeV for all layers. Accordingly, a 0.5 MeV threshold was chosen to give 100% efficiency for muons crossing scintillator.

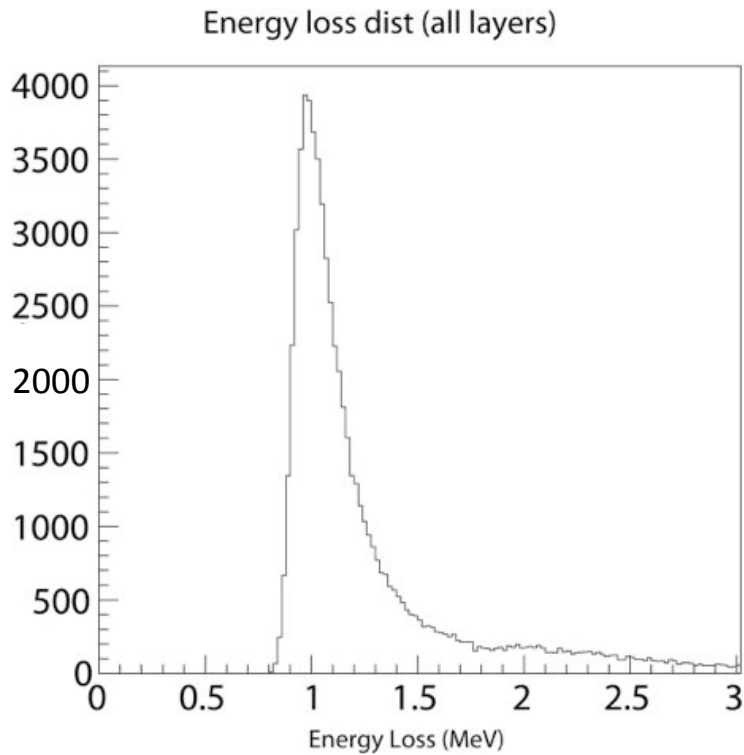

Figure 4.140. The energy deposition by single muons in scintillator in the Range Stack. All scintillators in the simulation are individually included in this figure. A cut at 0.5 MeV gives 100% efficiency for observing muons.

Figure 4.141 (left) shows the efficiency of a 0.5 MeV cut on the muon energy deposition for all scintillators in simulation. Below layer 40, the efficiency is 100%. At layer 50, the efficiency is 99.5**%,** dropping to 90% at the end of the simulated calorimeter. Figure 4.141 (right) shows the source of the efficiency loss. It is due to multiple scattering and muons leaving the transverse volume. While the simulated detector was 70-layers deep, the proposed range stack for Mu2e is equivalent to only 50 layers in the simulation.

Single protons do not penetrate the steel as easily. Figure 4.142 shows the percentage of scintillator layers that observe more than 0.5 MeV of energy for 4 GeV protons. Very early in the shower, all layers see this amount of energy; however by layer 40 the probability is 0.07%. At layer 50, the efficiency deviates from the simple fit, but is about 0.02%. Essentially, the chance of protons leaving more than 0.5 MeV of energy in layers 40 & 50 is very low.

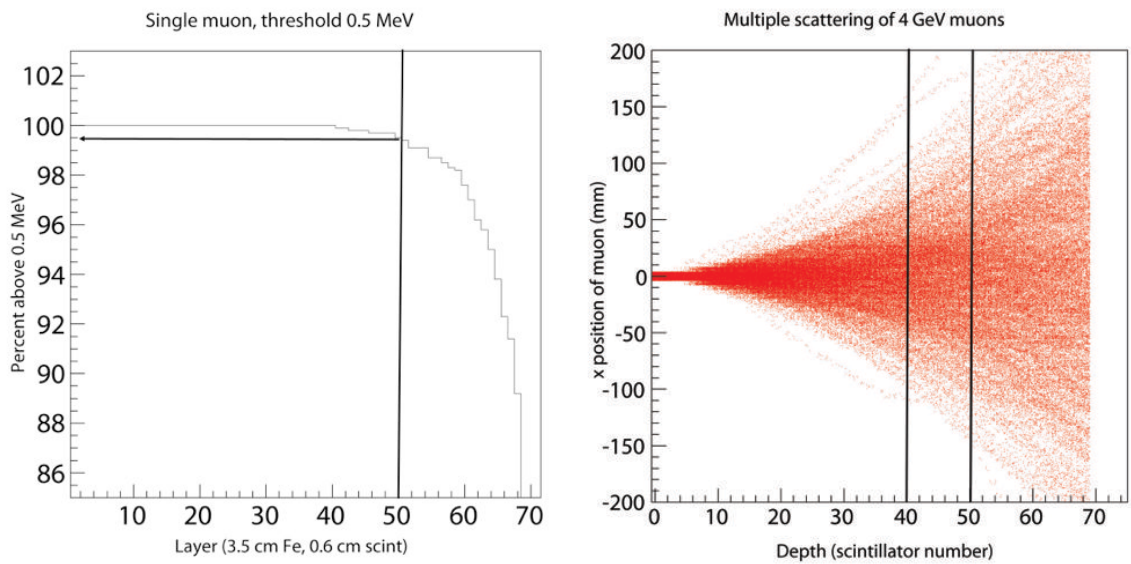

Figure 4.141. (left) Efficiency to see a single muon in scintillator with a 0.5 MeV cut (left). Multiple scattering of 4 GeV muons at normal incidence in the center of the detector (right). The vertical black lines identify the proposed location of scintillator detectors.

The final measure is the energy observed from 100 simultaneously-arriving protons. This is an upper limit on the performance that must be explored to ensure the required proton/muon discrimination. Figure 4.143 shows the percentage of scintillators that observe in excess of 0.5 MeV. Until approximately 30 layers into the detector, all scintillators see at least this much energy. By layer 40, the percentage of times that 100 simultaneously-arriving protons leaves 0.5 MeV is 10.4%, while by layer 50 the percentage is 0.9%.

The layer number can be converted to the amount of steel using the conversion of 3.5 cm/layer. To make the comparison slightly more accurate, the scintillator should be included. Since the density of scintillator and iron are 1 g/cm$^3$ and 8 g/cm$^3$ and the thicknesses are 3.5 cm and 0.6 cm, respectively, each layer of scintillator adds approximately $(0.6 \text{ cm}) \times (1 \text{ g/cm}^3 \div 8 \text{ g/cm}^3) = 0.075 \text{ cm}$ steel equivalent. These performance numbers suggest that the Range stack could consist of a large steel absorber with scintillator at 40 & 50 layers-equivalent, (i.e. 143 and 178.75 cm).

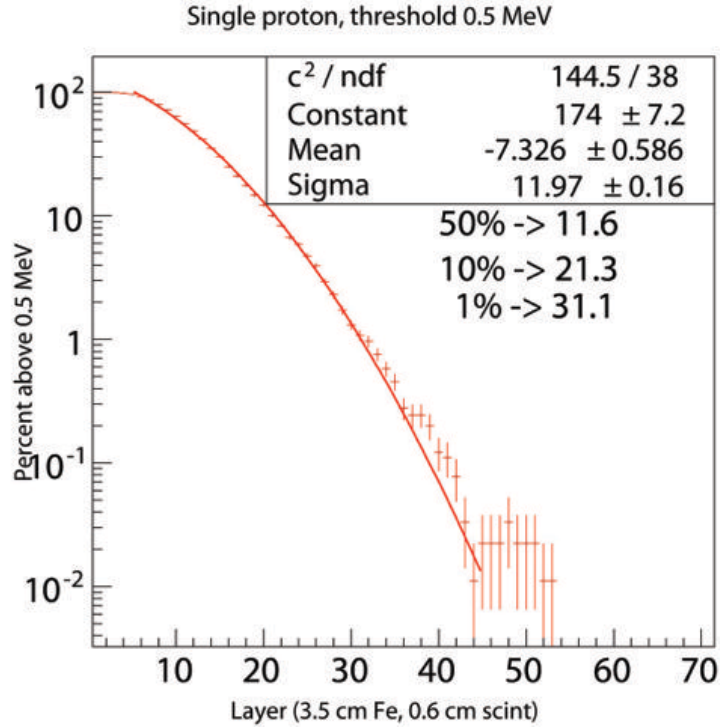

Figure 4.142. The percentage of the time that a specific scintillator has an energy deposition exceeding 0.5 MeV for single protons carrying 4 GeV at normal incidence at the center of the detector. The overlaid fit has no physical significance and is simply a Gaussian with reasonable agreement to the data curve. The numbers tell the interpolated layer number for 50%, 10% and 1% efficiency respectively.

The predicted performance would be that single protons essentially never fire these two scintillators, while single muons fire both of these scintillators 100% of the time (subject to the efficiencies inherent in light collection). Further, in the first layer, there is a 10% chance that the first layer will see more than 0.5 MeV from 100 simultaneously arriving protons, while in the final layer, the probability is just 1%.

Studies of the correlation in the energy deposition of layers 40 and 50 show that in the case of 100 simultaneously-arriving protons there is essentially no correlation in the energy deposited in these layers. This demonstrates that the energy deposited is not due to muons created in the hadron showers but instead reflects the variation in hadron energy deposition.

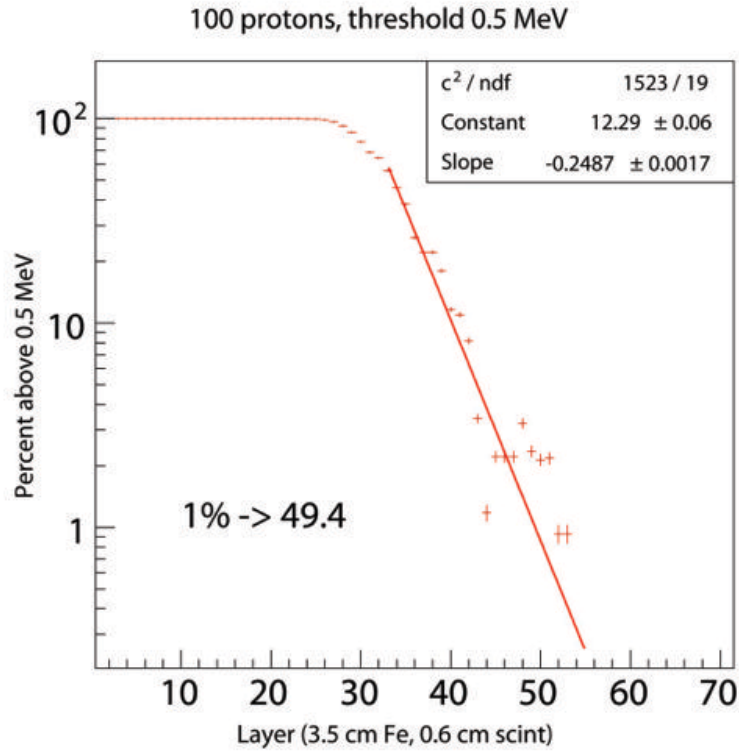

Figure 4.143. The percentage of time each layer of scintillator detects at least 0.5 MeV of energy when 100 protons with 4 GeV simultaneously hit the center of the detector. The unphysical uncertainties are an artifact of the methodology whereby these multiple proton events are generated. The fit is a non-physical exponential that allows for a sensible interpolation. The inset number reports the scintillator number for which 1% of the time these 100 protons deposit 0.5 MeV of energy.

In order for the detector to work as designed, the light collection efficiency should be high. For this scenario, coincidences between the two counters will identify muons. To do this, the charge from the photo detectors will be differentially digitized in 1 or 2 nsec increments throughout the 175 nsec spill duration. This design will count the muons incident on the calorimeter in-spill. Out-of-spill, isolated protons will not deposit energy in these counters, but muons will.

In order to reduce the risk associated potential detector inefficiencies, two additional planes will be inserted into the range stack. One will be between layers 40 & 50 as described above. The second will be placed at a more shallow depth such that it will be fired by incident protons about 50% of the time. This layer will be used to verify that Monte Carlo detector simulation correctly predicts the observed detector performance.

### 4.10.3 Extinction Monitoring Risks

Both the Upstream Extinction Monitor and the Target Monitor are based on very standard technology, so there are no risks beyond ordinary operational concerns.

### 4.10.4 Extinction Monitoring Quality Assurance

#### *4.10.4.1 Upstream Extinction Monitor*
The key quartz radiators will be connected to the photomultiplier tubes prior to installation in the detector assembly and the light yield will be tested using cosmic rays to verify the coupling to the phototube, the operation of the phototube, and light tightness.

Once the telescope is assembled, it can be connected to the DAQ and tested vertically, using a cosmic ray trigger, prior to being installed in the tunnel.

#### *4.10.4.2 Target Monitor*
The two critical aspects of the Target Monitor are the alignment of the filter channel and the operation of the pixels. The former is particularly important, since it would be very difficult to correct a mistake after the fact. An alignment network will be established in the detector hall and a survey target placed at the location where the production target will be. This will be used to align the filter channel prior to pouring concrete.

The pixel system, while complex, is based on the ATLAS FE-I4 pixel chips, for which there is a well-established quality control process [122], which we intend to follow.

### 4.10.5 Extinction Monitoring Installation and Commissioning

#### *4.10.5.1 Upstream Extinction Monitor*
The scattering foil will be installed when the beam line is assembled. It will be cabled to the readout system with the rest of the beam line instrumentation. The charged particle telescope will be installed after the beam line is complete. The system can be commissioned as part of the commissioning of the delivery beam.

#### *4.10.5.2 Target Monitor*
Assembly of the filter begins with insertion of pipes in the large poured shielding walls around the absorber and in between the magnet and detector rooms. Once the pipes are installed in both concrete shields, fixed liners are inserted into the pipes. The exit collimator fixed liner is shorter than the entrance collimator liner. The fixed liners can be adjusted by ±1 inch to accommodate the architectural tolerances of the building. They have a spherical plain bearing at their upstream ends and a cup weldment at their downstream ends that align that end with set-screws. The plug collimators are then inserted into the fixed liners. The plug collimator adjustment mechanisms are installed last, allowing the plugs to be surveyed into place.

The entrance collimator fixed liner is installed from the PS room as shown in Figure 4.144 and has to fit through the flange welded to the embedded pipe. Rigging will have to be set up in the PS area or from the PS hatch to accomplish this installation. The rotating clamps are turned 90° during the installation to fit through the flange. They are then rotated to hold the fixed liner to the embedded pipe flange. At the bottom left of Figure 4.145, the pipe plug that allows the steel shot to be removed can be seen. The fixed liner can only be adjusted when the steel shot is not present. If there is differential settling of the building between the PS and extinction monitor, the fixed liners can be repositioned.

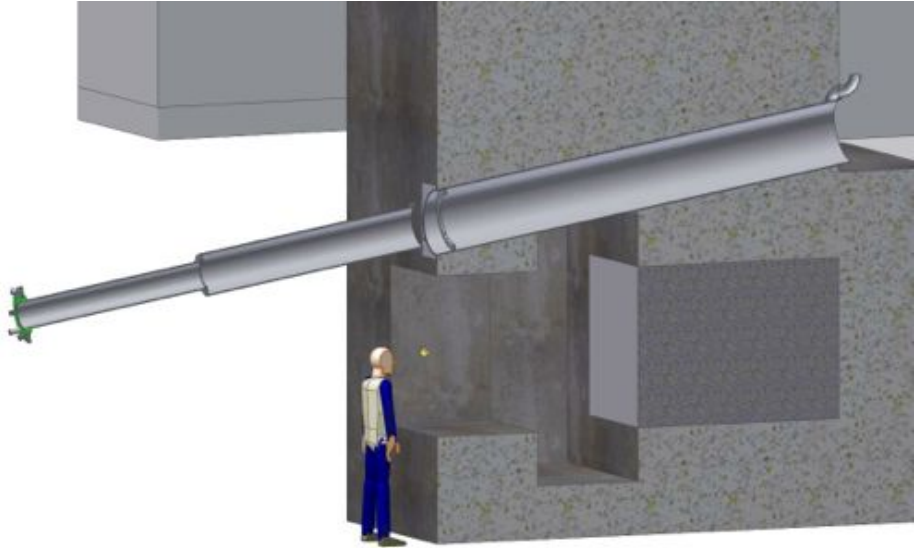

Figure 4.144. Installation of the entrance collimator fixed liner in the embedded pipe.

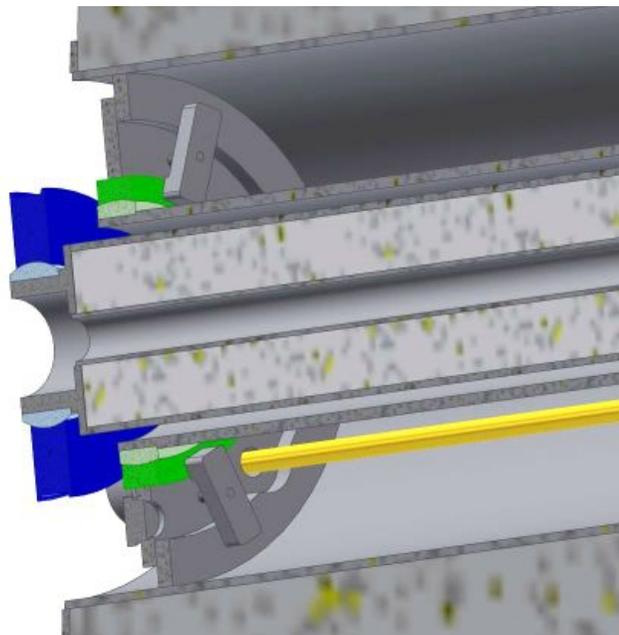

Figure 4.145. Section view of the entry collimator before the steel shot is installed, showing the rotating clamps and the cam drive shaft.

The procedure for installing the entry collimator into its fixed liner is depicted in Figure 4.146. These movements of the entrance collimator plug will require some agile rigging equipment to be designed. The hatch through which all the equipment in this area comes through is too narrow for any of the existing lifting hardware at Fermilab. Unistrut channels will be embedded in the ceiling of the magnet room and detector room to allow a temporary custom bridge crane to be assembled. The filter magnet will be the last part of the filter to be installed.

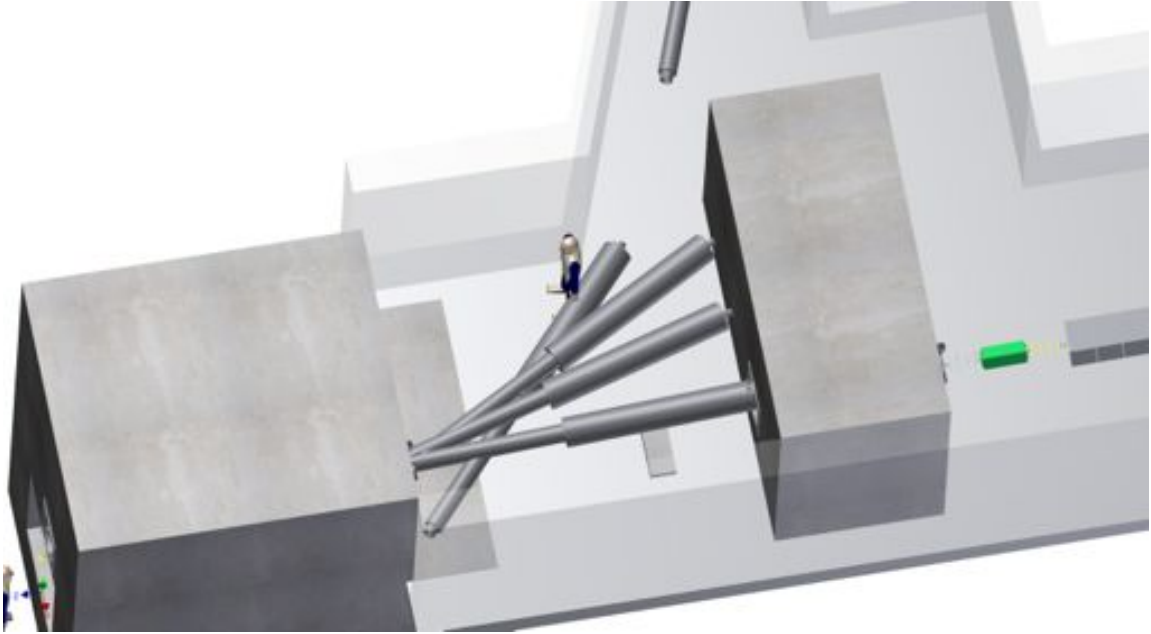

Figure 4.146. Motion of the entrance collimator during insertion into fixed liner.

## 4.11 Target Station

The Mu2e Production Target Station (Figure 4.147) consists of five components: the pion production target, the remote handling system to change targets, the production solenoid heat and radiation shield (HRS), the proton beam absorber, and the protection collimator.

Figure 4.148 shows the Target Station civil layout. The remote handling area is used to store the target change robot. The Production Solenoid (PS) hatch consists of a stack of shielding blocks that can be removed to allow access with an external crane. The proton beam absorber is unmovable and part of the concrete structure of the building.

### 4.11.1 Production Target

#### 4.11.1.1 Production Target Requirements

The requirements for the production target are given in Mu2e document 887 [4]. The Production Target generates pions that decay to muons as they are transported to the stopping target in the Detector Solenoid. The target is installed inside the bore of the Production Solenoid within a graded magnetic field, a configuration designed to maximize the production and capture of low-energy negative pions generated by interaction with the 8 GeV primary proton beam. The pion production cross section of the target material must be large enough to allow Mu2e to produce the required number of stopped muons. Pion production is maximized with a dense, high atomic number material to ensure a high rate of beam-target interactions. A compact target geometry minimizes pion reabsorption. The refractory metal[32] tungsten is ideal, since at the design beam power of 8 kW it is able to directly radiate the generated heat load to the solenoid shield without the need for a coolant.

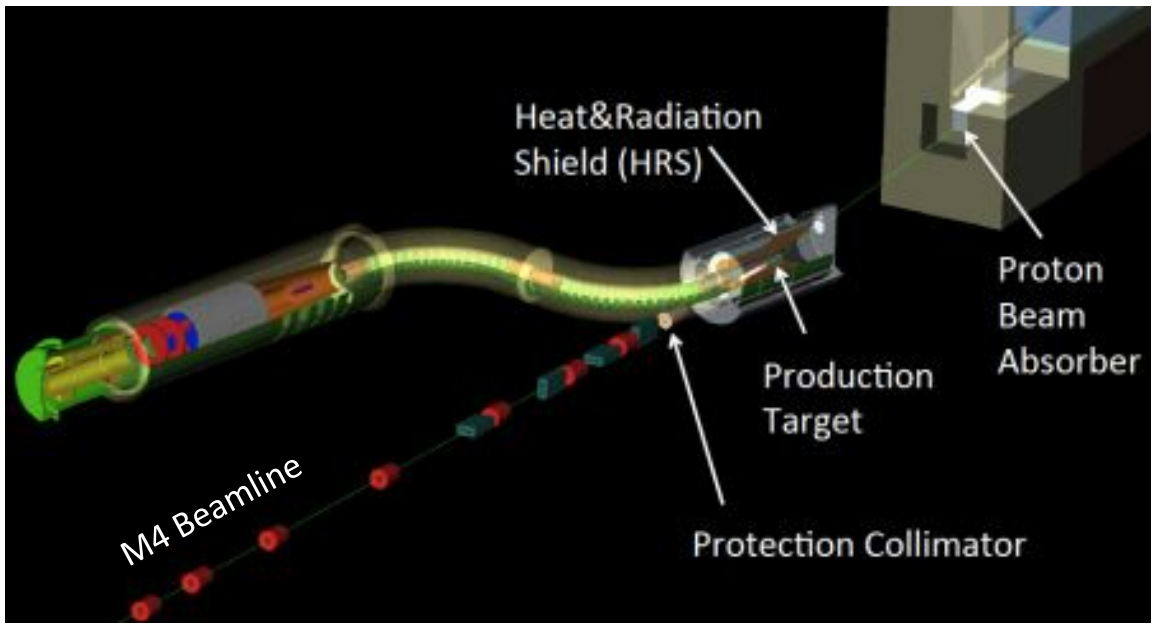

Figure 4.147. Layout of the Mu2e experiment and the Target Station components.

The production target must function in a harsh environment presenting a number of technical challenges including resistance to intense proton irradiation, continuous thermal cycling at ultra-high temperature[33] and resistance to chemical effects due to interaction with residual gasses in the vacuum. The minimum acceptable target lifetime has been specified as one year, during which the target must maintain its mechanical integrity and positional alignment to the beam. The specification for the vacuum level is $1 \times 10^{-5}$ Torr. The risk section below reviews the impact of the vacuum level on the target lifetime (section 4.11.1.3.1).

---

[32] The refractory metals are those elements of the periodic table that have exceptionally high melting points, are resistant to wear, corrosion, and deformation.
[33] >1500°C

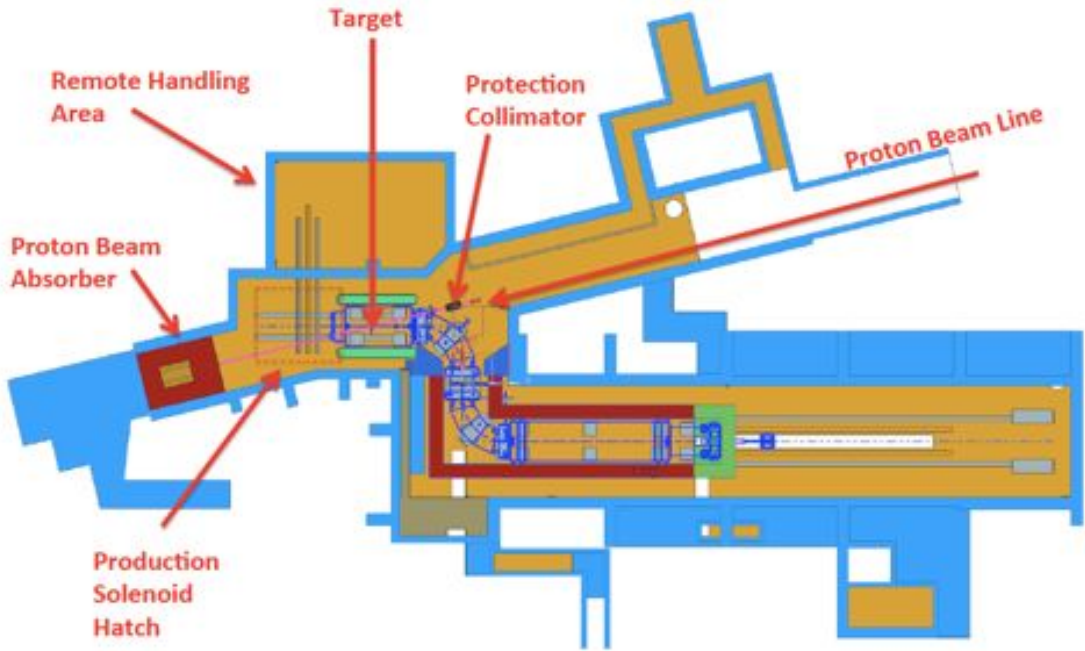

Figure 4.148. Civil layout of the Mu2e experiment and the proton beam line.

The target must be firmly supported by a structure that is stiff and minimizes vibrations. The impact of pion reabsorption on stopped muons suggests a structure that is low mass, and presents a small geometric profile to the spiral trajectories of pions within the Production Solenoid. The support system also enables remote installation, removal and exchange of the target and offline adjustment of the target rod position within its support structure.

Given the choice of beam and target size to optimize the stopped muon yield, overall alignment of the target rod with respect to the beam needs to be less than about 0.5 mm to avoid losing more than a few percent of the muon yield. When the first target is installed it must be placed close enough to the theoretical beam position that the beam can be steered onto the target. The target support structure and remote handling system must ensure replacement targets are placed within about ±0.25 mm of the first. In addition, the target position must be stable to about ±0.25 mm during operation, taking into account distortions due to thermal cycling from ambient conditions when the beam is turned on.

The incoming primary beam can be adjusted to give ±1 cm in both the vertical and horizontal directions at the target. The angle of the beam can also be varied in both planes to ±0.8°. Changes beyond this range will require an adjustment of the relative alignment of the beamline and the Production Solenoid.

### 4.11.1.2 Production Target Technical Design

The proposed Mu2e pion production target is a radiation-cooled tungsten rod, the size and shape of a pencil, and is described in detail in Mu2e document 2406 [123]. The 160 mm long, 6.3 mm diameter rod is mounted from a structure that resembles a bicycle wheel, consisting of conical tungsten target end 'hubs', spring-loaded refractory metal spokes and a titanium outer support ring as shown in Figure 4.149. Also shown are the three spring-loaded clamps and four handling lugs that form part of the remote handling/mounting system. The target assembly fits inside the 400 mm diameter clear bore of the production solenoid vacuum vessel.

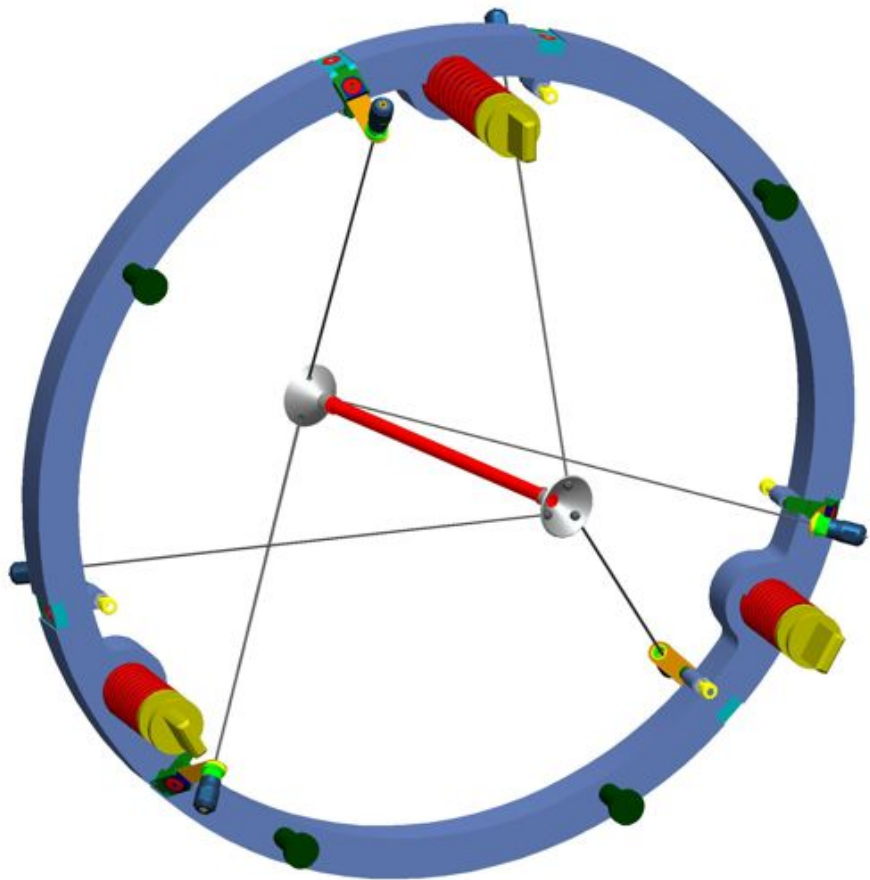

Figure 4.149. 3D view of the Mu2e target and 'bicycle wheel' support structure.

The target will be made from stock tungsten rod that is produced via the powder metallurgy process. Impurities will be limited such that the material will be at least 99.95% pure tungsten (as is the case in the present neutron spallation target at the ISIS facility in the UK) and will have a bulk solid fraction of at least 97%. During production the material is pressed, sintered then swaged and then ground to size. Tungsten is chosen as the target material primarily for its excellent high-temperature mechanical properties, which allow the target to run hot enough that it may be cooled radiatively. Of the pure refractory metals, tungsten has the highest melting point, highest tensile strength and lowest thermal expansion coefficient, making it extremely well suited to applications that involve thermal cycling to ultra-high temperatures. Its high thermal conductivity reduces temperature gradients and consequent thermal stresses and distortions. The purpose of the end hubs is to facilitate joining of the support spokes to the target rod. They are designed in such a way that this connection is placed outside the beam footprint to separate the end of the spoke from the hottest part of the target. The target end hubs will also be made from tungsten to avoid the potential for differential thermal expansion at the interface. The end hubs will either be separate parts that are mechanically fitted to the ends of the target rod, or possibly machined along with the target from a single piece of material to prevent any chemical corrosion/erosion from occurring at that interface.

Six refractory metal support spokes, that are approximately 1 mm in diameter and 250 mm long, form a low-mass target support structure that is designed to minimize the reabsorption of useful pions generated by the target. A pre-tension in the spokes ensures good rigidity in the structure. At their 'hot' end, each spoke is threaded through a hole in one of the hubs, and incorporates a 'ball' feature that mechanically stops the spoke from pulling through while allowing a range of angular positions. A collet chuck arrangement grips the plain 'cold' end of the spoke where it interfaces with a tensioner mechanism. The spoke materials under consideration are tungsten, tantalum, and alloys of those materials. While tungsten appears to have a superior resistance to high temperature creep, it also suffers from a ductile-to-brittle transition at around 300°C. Tantalum, on the other hand, remains ductile throughout the full range of temperatures of interest. A number of tungsten alloys that have been specially formulated to give high temperature creep resistance are also under consideration. The final choice will be influenced by practical manufacturing constraints as well as a need for good high-temperature creep resistance. A number of spoke manufacturing methods are under consideration, including a rotary wire EDM process whereby the spoke, complete with its spherical end feature, is produced from a single bar. Results from early prototyping of this technique are shown in Figure 4.150.

The spoke tensioners (Figure 4.151) are used to set the spoke lengths and to apply and maintain a specified preload in the spokes. A cantilever spring that is preloaded on assembly allows thermal expansions in the structural elements to take place without over-stressing the spokes, and provides some resilience to spoke creep. Coarse adjustment of the spoke length is provided through a collet chuck arrangement while fine adjustment is achieved using a tensioning screw. A spherical interface between the collet chuck and the cantilever spring provides a rotational degree of freedom that prevents bending of the spoke. The spoke tensioning mechanism is designed as far as possible to fit within the support ring footprint in order to minimize pion reabsorption.

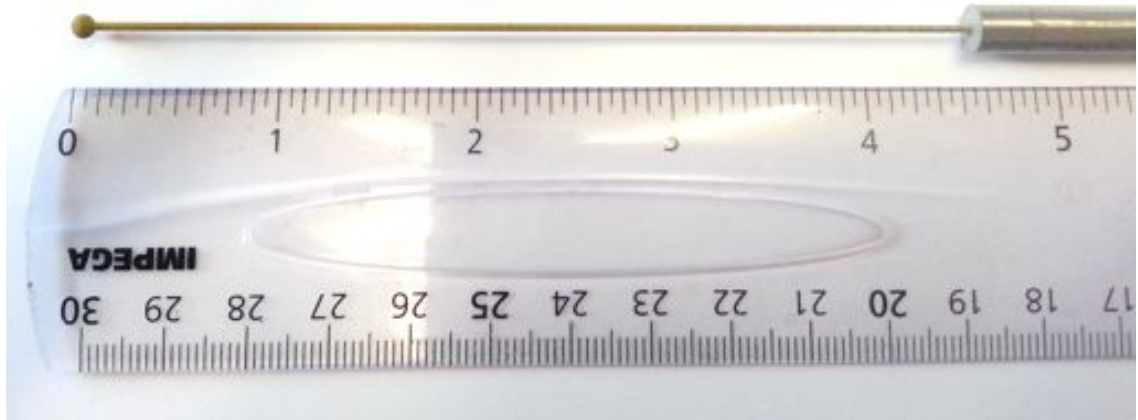

Figure 4.150. Prototype Spoke.

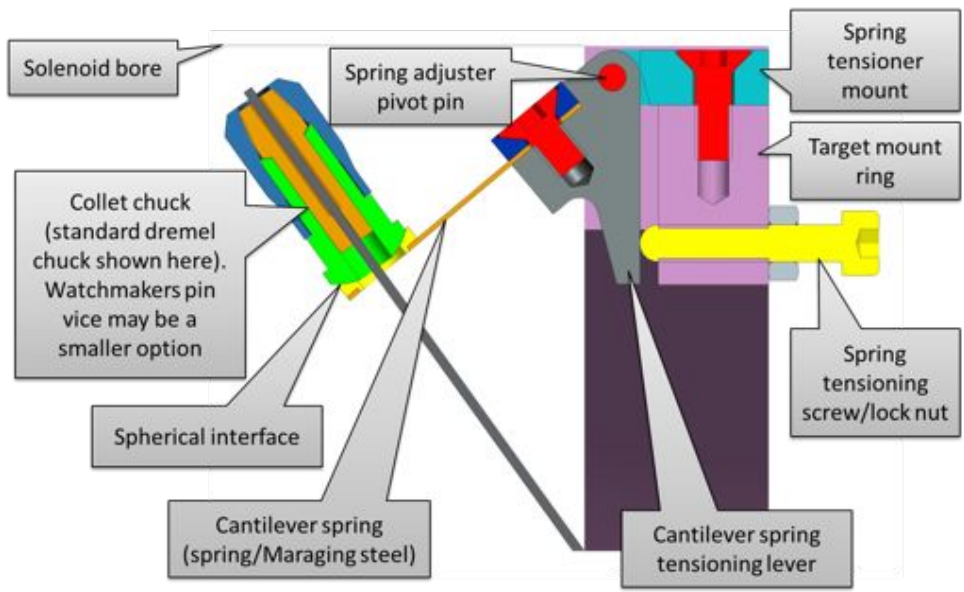

Figure 4.151. Spoke tensioning mechanics.

The target support ring is mounted directly from three lugs that protrude inside the HRS as shown in Figure 4.152. The HRS lugs have radial slots to facilitate mounting the target in such a way that it stays centered on the HRS bore regardless of thermal expansions in the support ring or HRS vessel.  The position of the HRS lugs with respect to the nominal beam position will be determined using a suitable survey fixture. Alignment of the target rod within its support ring is then carried out offline by an iterative process of survey and adjustment until the correct position is achieved. This method permits replacement targets to be pre-aligned and then installed in position using a remote target handling system described in section 4.11.2.

The target points horizontally off-axis by 14° relative to the support ring in order to be properly aligned with the incoming proton beam (see Figure 4.153).  This is achieved through the use of three different spoke lengths, although the six tensioning mechanisms are all identical. The structure is designed so that despite this asymmetry the target position remains stable to within the stated tolerance when the beam is on.

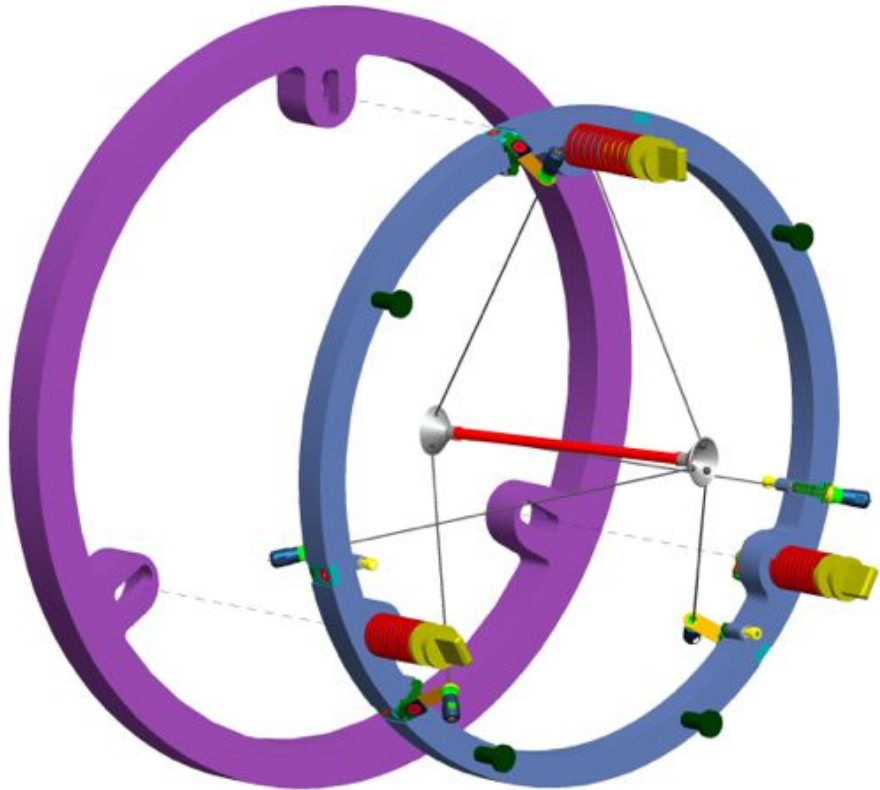

Figure 4.152. Target mount to HRS bore.

The mechanical stability of the target and its 'bicycle-wheel' support structure has been modeled in ANSYS. The parameters found to most significantly influence the rigidity of the structure were the spoke diameter and the spring constant. A large-diameter spoke makes the structure more rigid; however a small diameter is preferred to minimize unwanted particle interactions outside of the target. Therefore, a relatively small spoke diameter of 1 mm is suggested as the appropriate compromise. A large spring constant makes the structure more rigid; however it also exaggerates the increase in spoke tension that occurs when the target heats up. Again, a compromise of 10 N/mm is chosen in order to limit the tensile stress in the spokes to a reasonable level. The spring pre-extension is set such that, at all times, sufficient tension is present in the spokes to elevate their natural vibration frequency (violin mode) above that of the structure as a whole. Any additional tension adds little to the rigidity of the overall structure and has the disadvantage of inducing unnecessary additional tensile stress in the spokes. The vibration modes and frequencies of the structure were evaluated in both the 'hot' and 'cold' conditions and it was observed that there is little change in structural rigidity when the target is heated. The first natural frequency of the structure is around 50 Hz.

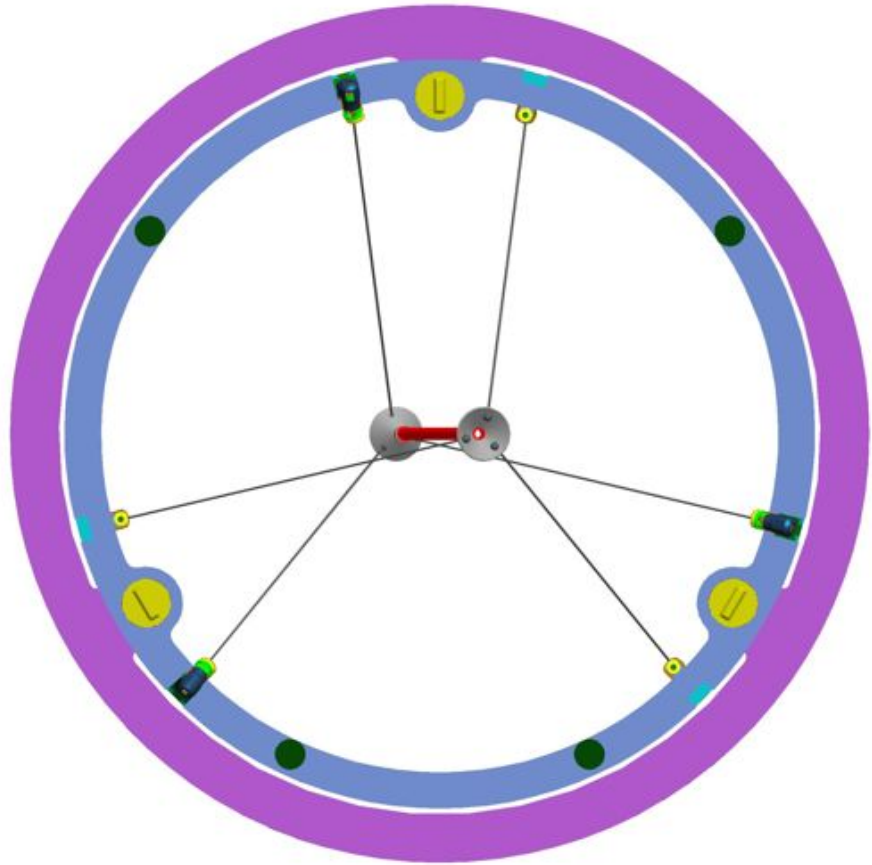

Figure 4.153. Target and support structure and HRS inner wall viewed along the solenoid axis.

A full description of the Mu2e beam parameters is given in Section 4.1 (see also Table 4.5). The heat load generated in the target by interaction with the beam has been calculated for the nominal beam parameters using both FLUKA and MARS. FLUKA calculates 560 W and MARS calculates 630 W for the design beam power of 7.7 kW. To account for the heat load uncertainty we have used the FLUKA value as the minimum bound and FLUKA x1.3 as a nominal 'worst case'. The power distribution within the target rises to a shower maximum two cm along the length and then decreases almost linearly to the end of the rod, as shown in Figure 4.154.

Although tungsten is a refractory metal, the material properties of interest (i.e. creep resistance, strength) and the resistance to chemical processes (i.e. corrosion/oxidation) degrade at elevated temperature. Factors that strongly influence the operating temperature of the radiation-cooled target include the magnitude of the beam heat load, the target surface area, and its emissivity. Figure 4.155 shows how the peak target temperature varies as a function of heat load and emissivity over the likely range of $0.2 < \varepsilon < 0.5$. It can be seen that the peak target temperature lies in the range $1400^0\mathrm{C} < T_{max} < 2000^0\mathrm{C}$.

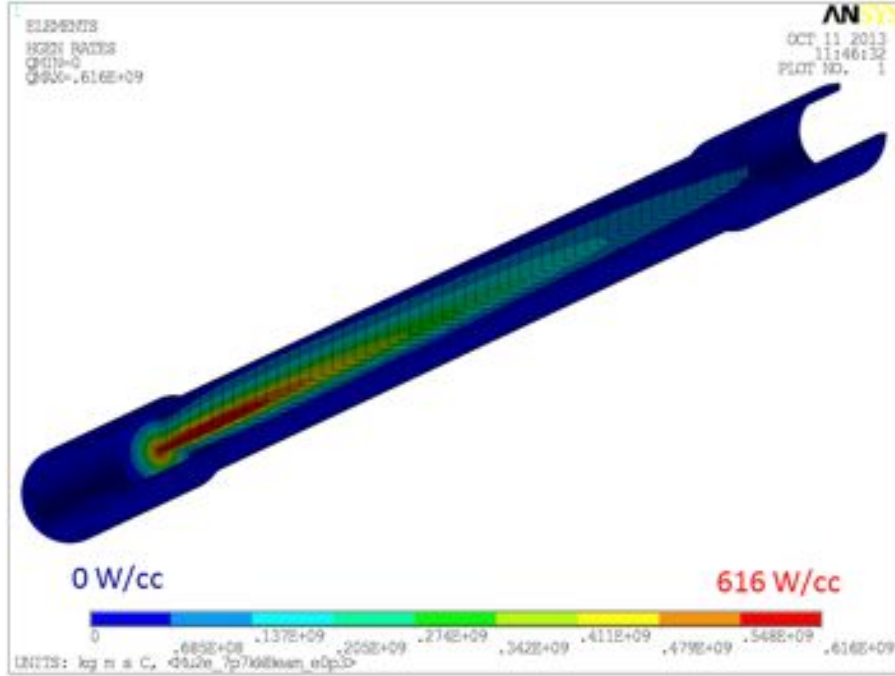

Figure 4.154. Cut-away view showing the 560 W beam heat load in the target.

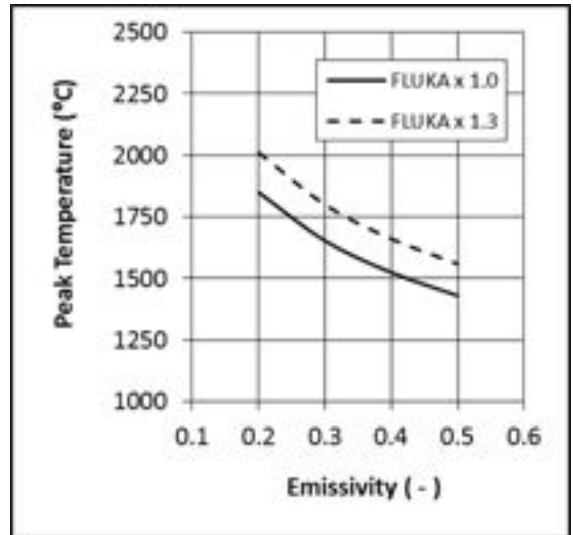

Figure 4.155. Peak target temperature versus emissivity for the energy deposition given by the Fluka model, and for 1.3× the energy deposition of the Fluka model.

The target temperature distribution for the nominal case of Q = 560 W and ε = 0.3 is shown in Figure 4.156. The target is designed so that it will operate successfully throughout the range of potential operating temperatures. Measurements of tungsten emissivity as a function of temperature are currently underway as part of the present RAL/Fermilab R&D. These measurements will help to narrow down the expected operating temperature range. It may be possible to artificially raise the emissivity of the target by applying a high emissivity coating or surface treatment of some kind. A number of potential high-temperature coatings are under investigation, including oxidation resistant silicon carbide (SiC) that we expect will produce a black surface that closely matches the thermal expansion characteristics of tungsten.

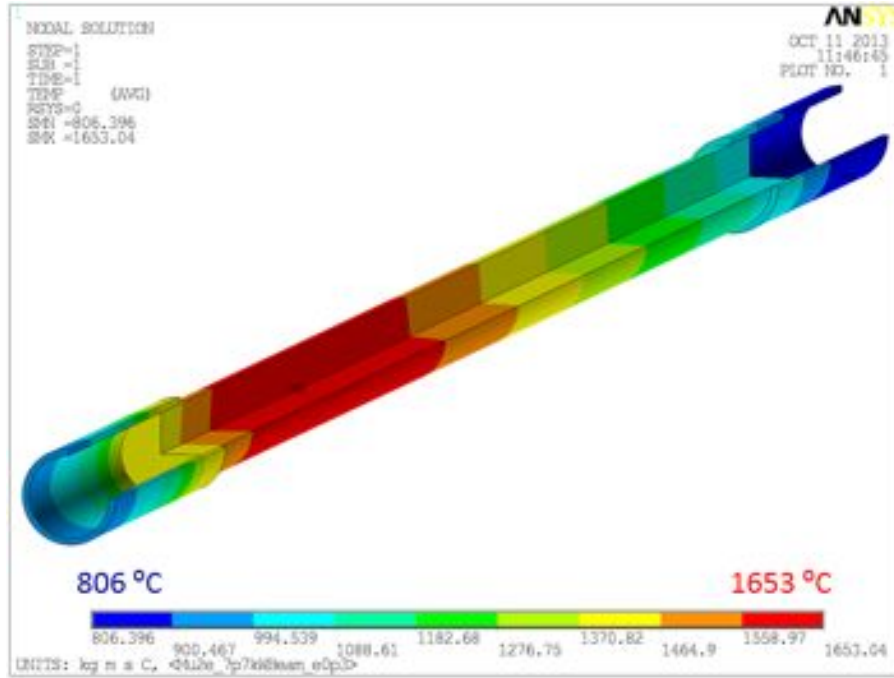

Figure 4.156. Target steady-state operating temperature assuming the FLUKA heat load of 560 W and an emissivity of 0.3.

The heat loads on the titanium target support ring come from direct and reflected thermal radiation (10 - 100 W depending on the emissivities of the various surfaces involved), from secondary particle interactions (less than 10 W), and from conduction along the spokes (less than 5 W). The temperature of the support ring, assuming it is cooled passively by radiating its heat to the surrounding water-cooled HRS vessel, is shown in Figure 4.157. That operating temperature depends largely on the emissivities of the vessel and support ring. For the range of emissivities considered, the support ring operating temperature is in the range $145 - 280^0$C.

For the case where all emissivities are 0.3 the titanium support ring heats up to around $180^0$C and expands radially by around 250 µm, tending to slightly increase the tension in the spokes. Titanium is a material ideally suited to the support ring application due to its favorable mechanical properties at the elevated temperatures predicted, coupled with a very low thermal expansion coefficient. Its high specific stiffness means that the overall mass of the target structure can be kept within reasonable limits for the remote handling system.

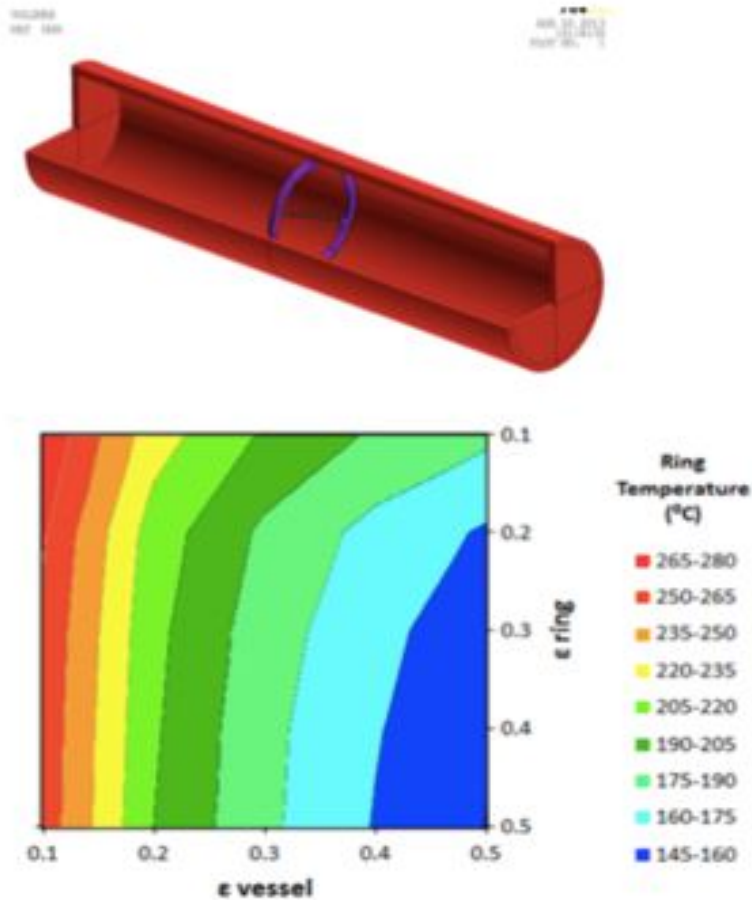

Figure 4.157. ANSYS heat transfer model (left), and support ring temperature as a function of emissivity (right).

The beam has macro cycle of 1.333 seconds that consists of 0.497 seconds of 'beam on' followed by 0.836 seconds 'beam off' (see section 4.1.5). The 'beam on' period is broken down into eight spills containing $1 \times 10^{12}$ protons and lasting for 54 msec. This time structure results in thermal stress cycling in the target as shown in Figure 4.158.

Although the stress range is relatively low, this stress cycling has the potential to cause premature failure due to fatigue as described in Mu2e document 2406 [123]. Literature studies and consultations with metallurgists have revealed a scarcity of data with which to predict the number of cycles to failure under the Mu2e operating conditions. A novel off-line thermal fatigue test has been devised as described in Mu2e document 2762 [124] to carry out accelerated lifetime tests of the target bar material under operating conditions representative of those in the Mu2e target. The test, which is now underway, utilizes a 1.6 kA, 1 msec pulsed electric current to set up time varying thermal gradients and stresses in a specially shaped sample. Multiphysics ANSYS simulations have been used to determine the test conditions and sample geometry. A new pulsed power supply comprising capacitor crates, charging power supply and switching unit has been specially designed and constructed for this purpose. Test samples have been manufactured from 6.3 mm diameter stock tungsten bar and a vacuum test rig with bespoke sample mounting fixture has been constructed to carry out these tests that are now underway. The equipment is shown in Figure 4.159.

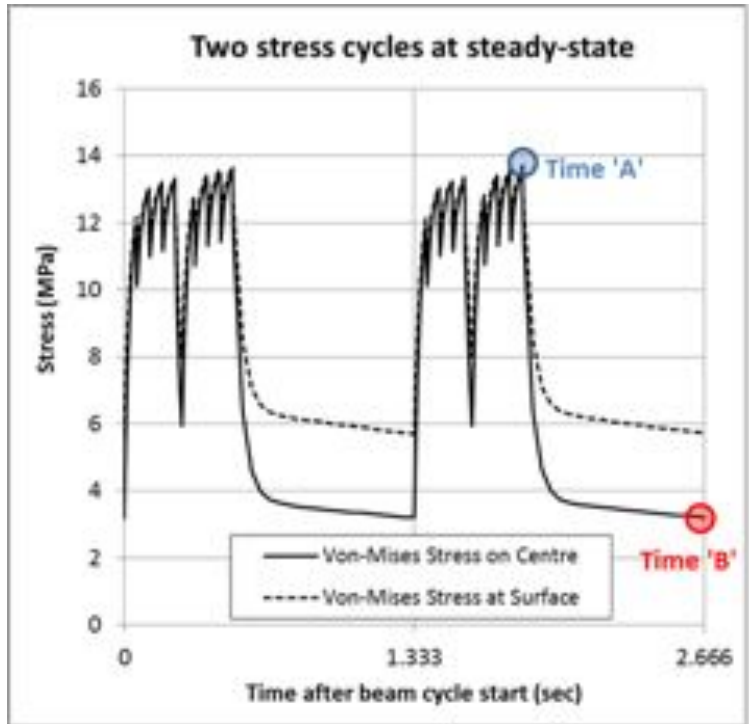

Figure 4.158. Stress-cycling in the target.

### 4.11.1.3 Production Target Risks

#### 4.11.1.3.1 Poor quality vacuum reduces target lifetime due to chemical processes.

Erosion of the tungsten rod, supports and spoke material is expected due to oxidation and water vapor-induced corrosion from impurities in the vacuum. It is difficult to predict the erosion rate by these processes at the expected achievable vacuum level of $1 \times 10^{-5}$ Torr due to uncertainties in the vacuum impurity constituents, uncertainties in the operating temperature as described above, and discrepancies and extrapolations in the data from different sources in the literature.

For example, assuming an oxygen partial pressure of $2 \times 10^{-6}$ mbar, the expected lifetime for a surface recession (i.e. erosion) of 0.1 mm of a tungsten target operating at 1750°C is between 70 and 1600 days.  However, there are a number of reasons to believe that this may be an under-estimate of the target life, including:

a)      It is likely that the partial pressure of oxygen in the vacuum will be much lower than $2 \times 10^{-6}$ mbar. Any residual oxygen leftover after pump-down will cause little damage as it will be burned up by the target. Impurities will then be dominated by outgassing from the vessel surfaces and components placed inside the vacuum. Measurements of the erosion rate of a hot tungsten wire in a vacuum poisoned by outgassing of presumably water-vapor and hydrocarbons from a large surface area of polythene showed much lower rates of surface recession than would have been the case in an equivalent partial pressure of oxygen.

b)   It may be that a surface recession of more than 0.1 mm could be tolerated prior to failure, depending on the precise location of the damage. The hottest part of the target, where we would expect to see the greatest rates of chemical erosion, is also the thickest part of the target and therefore inherently tolerant to erosion damage. The thinner, more delicate target end hubs and support wires will run at a lower temperature.

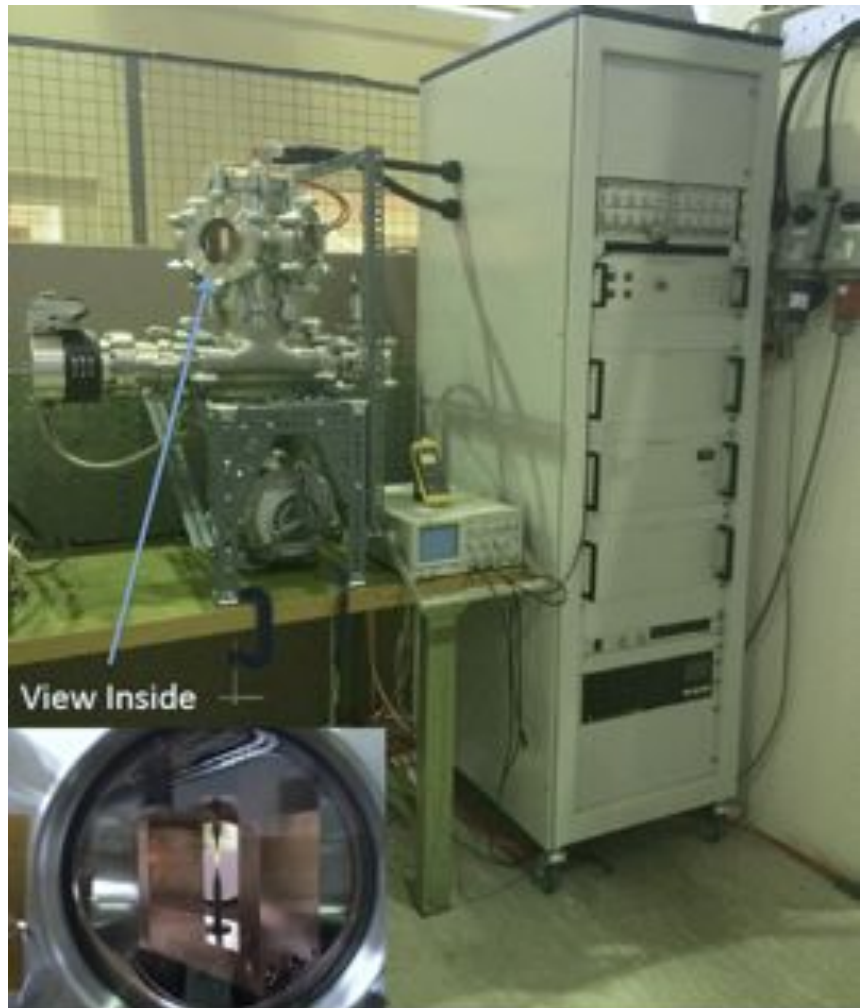

Figure 4.159. Vacuum test rig, with shaped tungsten test piece installed for pulsed current tests.

Methods of achieving a sufficient vacuum quality in the HRS area are being actively investigated. Use of polymeric substances will be minimized and the inner surface of the vacuum vessel will be cleaned. The effect of a contaminated vacuum on hot tungsten filaments is being investigated. A number of potential tungsten coatings are being actively investigated to test their efficacy in protecting tungsten at high temperatures, including iridium and SiC. An alternative may be to investigate iridium as a radiatively cooled target material. The target will be made slightly oversize to provide some corrosion allowance without significantly reducing pion yield.

### 4.11.1.3.2 Lower emissivity or higher heat load than expected resulting in excessive temperature of target at design beam power.

Experiments are underway to measure the emissivity of tungsten that has been manufactured in a similar way to the target material. If necessary, the emissivity may be enhanced either by a coating or by cutting grooves in the surface. A method of laser etching micron-sized grooves in tungsten has been successfully demonstrated and a number of coating systems are under investigation.

### 4.11.1.3.3 Beam induced thermal cycling leads to premature target failure.

A test program is underway to generate similar fatigue conditions in material samples as would be generated in a radiatively cooled Mu2e target.

### 4.11.1.3.4 Creep of spokes leads to target misalignment/failure.

The supports are designed to reduce the maximum spoke temperature. Creep resistant tungsten alloys will be investigated.

### 4.11.1.3.5 Target fails and jams inside bore of HRS making replacement impossible.

Care has been taken with design, material selection and connection to remote handling device to minimize the chance of occurrence. An early prototype of the target mount ring will be supplied to Fermilab for remote handling tests.

### 4.11.1.3.6 If any of the above risks materialize or there is an increase in the design beam power, then a radiation-cooled target is unfeasible.

A complete prototype of the target will be manufactured so that it can be thoroughly tested off-line. This will include heating tests using an induction heater that has already been purchased for this purpose. If the required target performance cannot be successfully demonstrated then it will be necessary to revert to a water-cooled target design or to investigate a helium-cooled design.

#### 4.11.1.4 Production Target Quality Assurance

Every part of the target will be thoroughly prototyped and tested off-line. A complete prototype will then be manufactured. All manufacturing and test methods will be thoroughly documented. The actual target supplied for experimental operation will use the same procedures as were developed for the prototype.

## 4.11.2 Target Remote Handling

The Mu2e target remote handling system provides a means to remove the downstream Production Solenoid window and the target, dispose of them both, and replace them with new components. A remote means of accomplishing these tasks is required because of the high radiation environment in and around the Production Solenoid after beam operations.

#### 4.11.2.1 Remote Handling Requirements

The target remote handling system shall perform the following tasks:

- Enter target hall from a separate side room

- Remove the target access window from the end of the PS

- Place target access window into waste storage cask (due to single-use vacuum seal)

- Detach and remove old target assembly from mounted position in HRS bore

- Place old target assembly into waste storage cask

- Obtain new target assembly

- Place new target assembly into HRS bore and latch into mounted position

- Obtain new target access window (with new vacuum seal)

- Place new target access window to the end of the PS and tighten all bolts

- Exit target hall and return to side room

The target remote handling system shall perform the above tasks either autonomously or via operator remote control with no human entry into the target hall.

The target remote handling system shall perform the above tasks even if building settlements (of typical magnitude) occur that adversely affect the as-installed alignment.

The target remote handling system shall include a stainless steel shielding door between the side room and target hall.

The target remote handling system shall include a waste storage cask with capacity to cover the expected life of the experiment.

The target remote handling system shall be built / developed into a fully operable system in a separate facility, then dismantled and re-installed into the Mu2e building.

The target remote handling system shall have its main robotic chassis, control system, and cask all installed in the Mu2e building by crane, and also will be removable by crane.

The target remote handling system shall incorporate safety factors, redundancy, and methods for manual intervention / override to minimize risk that the system becomes non-functional or cannot perform its tasks for various (reasonable) operational scenarios.

The target remote handling system shall be designed based on best practices with regard to radiation-tolerant components, corrosion resistance, durability, and reliability.

### 4.11.2.2 Remote Handling Technical Design

The target remote handling system is shown in Figure 4.160 below. The system utilizes a robotic chassis mounted to floor rails. On board the robotic chassis are servo-positioning X, Y, and dual Z-axes. After the shielding door is opened, the robot travels from the Remote Handling Room out into the target hall on a floor rail / linear drive system. Once the robot is in position behind the Production Solenoid, a pneumatic brake is engaged to hold it in place, and the Remote Handling Room door can be closed while the robot completes its tasks. Also included in the remote handling system are a staging frame (in the Remote Handling Room, which holds the new target and window), and the waste storage cask (in the target hall, which holds the old targets and PS access windows).

#### 4.11.2.2.1 Target Replacement Time

The total time to remove the old target and install a new one will be a superposition of the following (not necessarily sequential) tasks and durations:

A. Experiment shut-down / cool-down time (~ 1 week)

B. Crane removal of shielding blocks from access hatch into side room (~2 days in parallel with A)

C. Lower robot into room via crane, install robot & control system (~ 8 days)

D. Perform practice runs using staging frame (~2 days)

E. Robotic target exchange process (~1 day)

F. Disconnect and remove robot & control system from side room w/ crane (~4 days)

G. Re-install shielding blocks into access hatch (~2 days)

H. Experiment re-start, including vacuum pump-down time (~ 4 days in parallel with G)

Total target replacement time is approximately 4 weeks.

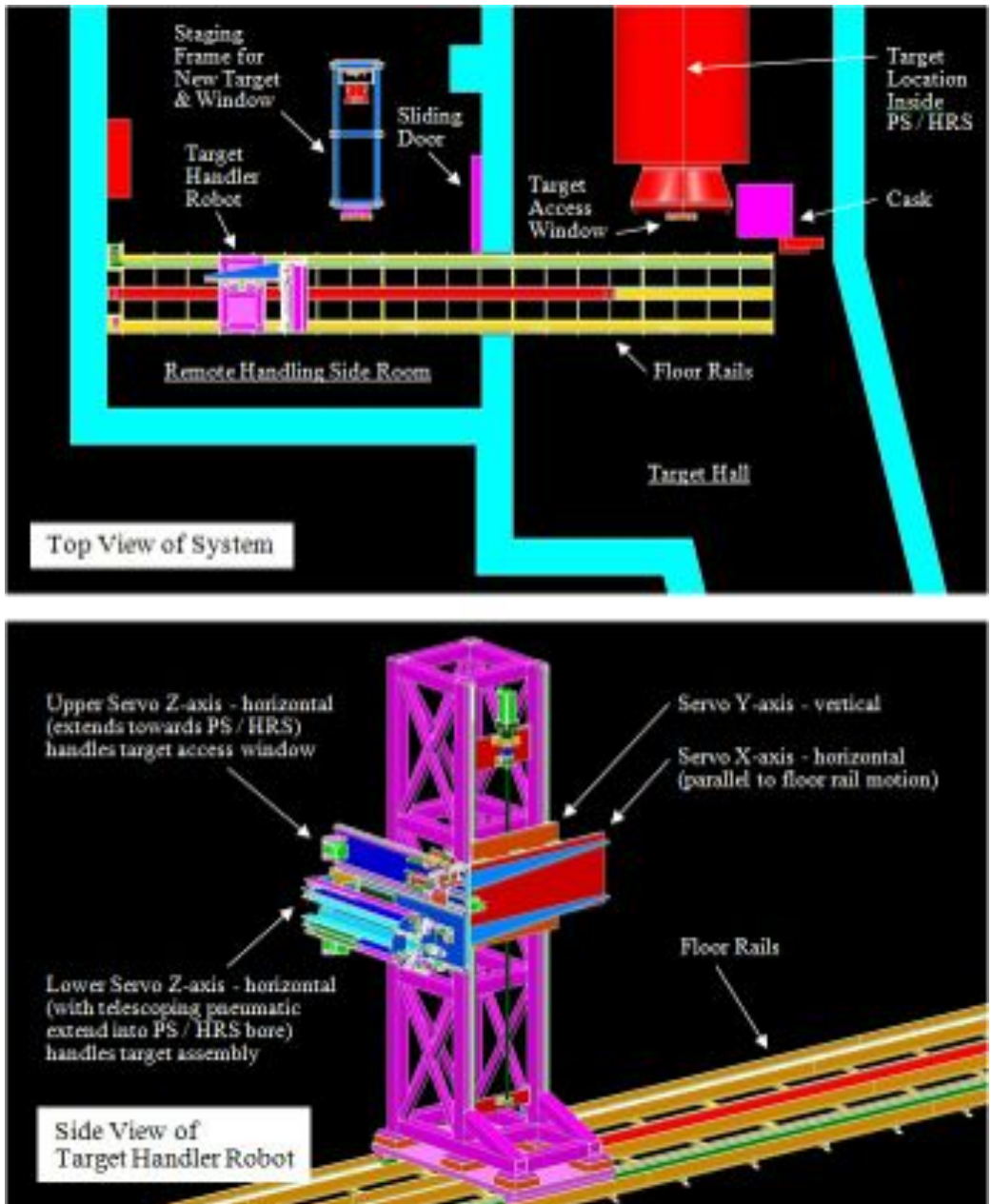

Figure 4.160. Top and side views of target remote handling system.

### 4.11.2.2.2  Target Handling System Design

The target access window and the robotic bolt driver / window grip mechanism are shown in Figure 4.161 below. The bolt circle that mounts the vacuum window to the PS end cap uses captured bolts, all spring-loaded to the retract position.  When all bolts are in the retract position, the window is held in place by two hanger pins extending from the window mounting flange on the end cap.

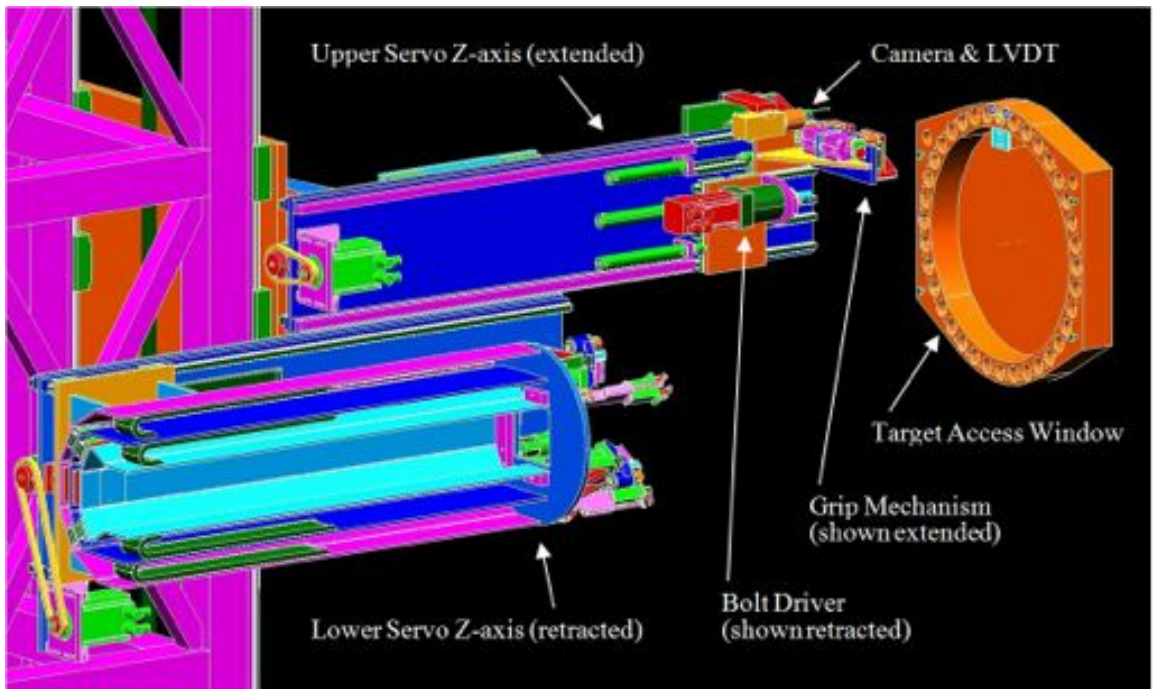

Figure 4.161. Robot upper Z-axis and target access window.

**Procedure for removing the old target access window:**

- Use video camera to locate all bolts in the bolt circle, also measure angular position of all hex-heads.

- Extend LVDT to measure Z-location of window.

- Extend bolt-driver on robot upper Z-axis, retract all 36x bolts in bolt circle.

- Move to 4 push-off bolt locations and drive each forward, forcing separation of window from mounting flange, retract bolt driver.

- Extend the window grip mechanism on robot upper Z-axis, grip and remove window.

- Retract upper Z-axis (holding window).

- Move robot forward on floor rails to position in front of cask.

- Open cask upper door, use robot's camera / vision system to locate cask opening.

- Extend upper Z-axis, placing window into cask, release grip, retract upper Z-axis, close cask door.

The target assembly mount inside the HRS bore and the target interface mechanism are both shown in Figure 4.162.

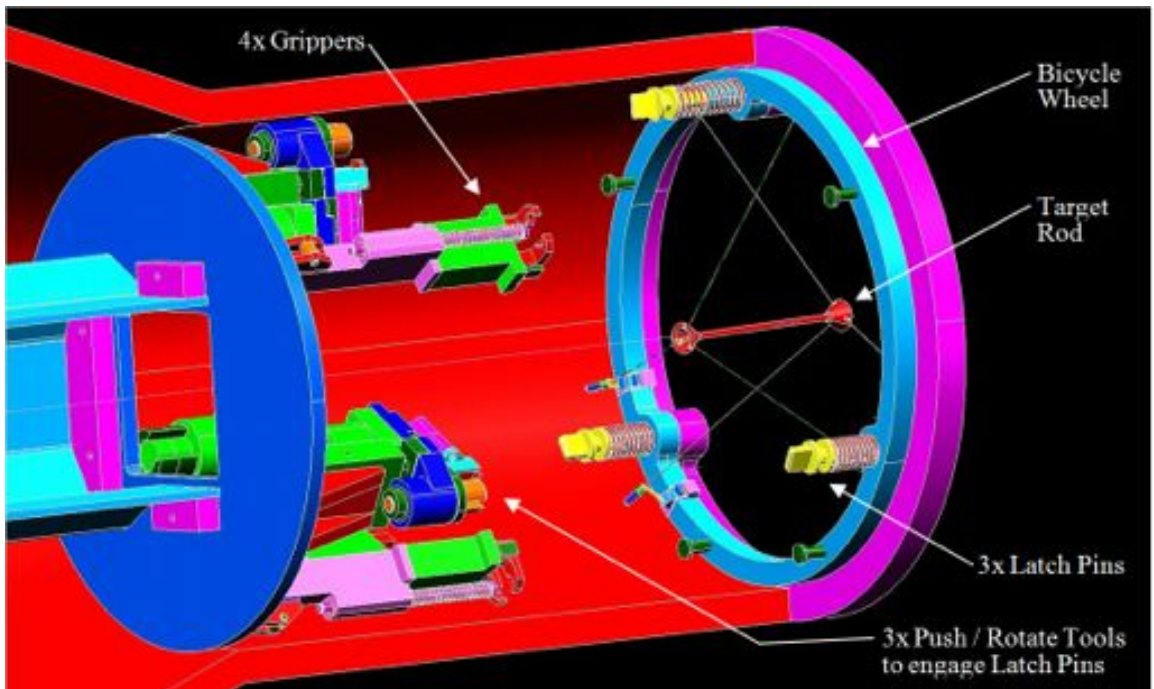

Figure 4.162. Target interface mechanism approaching the target assembly mounted in the HRS.

The outer ring of the target assembly (the "bicycle wheel") is located and held in place by three spring-loaded latch pins that interface to mating features on the datum / mounting surface inside the HRS bore. Three latch pins locate the target assembly by engaging in close-fitting slots. The latch pins hold the target assembly in place by pushing, then turning each one 90°, allowing small rollers to rotate-in behind the datum surface, thus clamping the bicycle wheel in place under spring pressure. The bicycle wheel also incorporates a helical cam in the pocket where each latch pin is mounted. Rotating the latch pin while in the retract position will cause the rollers to push forward slightly, forcing separation of the bicycle wheel from the datum surface with a large mechanical advantage.

The target interface mechanism is mounted to the end of an extending arm on the robot's lower Z-axis. The mechanism includes 3 pneumatic actuators to push and turn the 3 latch pins, and 4 pneumatic grippers to grasp the bicycle wheel. As the target interface mechanism is extended into the HRS bore, small rollers will pilot into the bore and help align the interface mechanism to the target assembly.

**Procedure for removing the old target assembly:**

- Now that the old window has been removed and disposed of, return robot to location behind PS, re-engage brake to hold in position.

- Use video camera to locate the exposed target access hole in the PS end cap.

- Extend LVDT to measure Z-location of window mounting flange on PS end cap.

- Position lower Z-axis on PS/HRS centerline, pneumatically extend telescoping robot arm into HRS bore (approx. 12' reach, with target interface mechanism mounted to its end).

- Final motion to engage the target assembly is via the servo-positioning Z-axis.

- Close 4 pneumatic grippers to grasp bicycle wheel.

- Push, turn 90°, retract upper latch pin. Turn 90° again, engaging cam feature to force separation (locally) of the bicycle wheel from the HRS mounting surface.

- Repeat above step for 2nd and 3rd latch pins.

- Return robot arm to full retract position (holding target assembly).

- Move robot forward on floor rails to position in front of cask.

- Open cask lower door, use robot's camera / vision system to locate cask opening.

- Extend lower Z-axis, placing target assembly into cask, release grip, retract lower Z-axis, close cask door.

**Procedure for placing the new target assembly and target access window:**

- Now that the old target assembly has been removed and disposed of, return robot to location behind the staging frame in the Target Handling Room, re-engage brake to hold in position.

- Use vision system and LVDT to locate new target assembly on frame.

- Extend robot arm to retrieve target assembly using same procedure as before.

- Return robot arm to full retract position (holding target assembly).

- Return robot to location behind PS; re-engage brake to hold in position.

- Use vision system and LVDT to locate target access hole in the PS end cap.

- Extend robot arm into HRS bore as before. Once target assembly is positioned against the HRS mounting surface, push and turn each of the 3 latch pins to hold in place.

- Release grippers and return robot arm to full retract position.

- Return robot to location behind the staging frame in Target Handling Room, re-engage brake to hold in position.

- Use vision system and LVDT to locate new target access window on frame.

- Extend upper Z-axis to retrieve window using same procedure as before.

- Return upper Z-axis to fully retracted position (holding new target access window).

- Return robot to location behind PS; re-engage brake to hold in position.

- Use vision system and LVDT to locate all bolt holes in the window mounting flange on the PS/HRS end cap.

- Extend upper Z-axis to place window onto hanger pins.

- Retract window grip mechanism and extend bolt driver.

- Install all 36 bolts in bolt circle using torqueing procedure (star pattern at half-torque, then star again at full-torque, process to be optimized per vacuum seal requirements).

- Once the new target assembly and access window have been installed, return robot to Target Handling Room and close the shielding door.

### 4.11.2.3 Remote Handling Risks

The following risks are being addressed for the target remote handling system:

#### 4.11.2.3.1 Target remote handling system needs to accommodate water-cooled target assembly.

If it becomes necessary to change from a radiation-cooled target to a water-cooled target, the increased complexity required by the remote handling system would increase the cost by approximately $3.3M [125]).

#### 4.11.2.3.2 Target access window or target assembly gets stuck in the installed position, and the target remote handling system is unable to remove it.

The target access window has been designed with this risk in mind. Once the bolt driver has backed-out all bolts in the main bolt-circle, the bolt driver will go to each of the 4 push-off bolts and drive them forward, forcing the window to separate from the mounting flange. The target assembly has been designed to incorporate 3 latch pins, each with an additional cam-forward feature. Once each latch pin has been rotated and retracted, the pin is rotated again, engaging a helical-cam that forces the bicycle wheel to separate from its mounting surface inside the HRS. This procedure will be tested on prototypes to determine if there are any potential failure mechanisms.

#### 4.11.2.3.3 Target rod and/or spokes break, causing debris in the HRS bore.

In this scenario, all possible shapes, sizes and locations of target and spoke fragments cannot be predicted. However, the target mount and target remote handling system should tolerate some debris being present. Any fragments behind the bicycle wheel will fall to the bottom of the HRS bore and remain there for the life of the experiment. Small fragments (approx. 1/8" in size, or less) in front of the bicycle wheel should also fall to the bottom of the HRS bore and remain there as they are simply passed over by the robot arm, target grippers, and bicycle wheel during a target change procedure. Larger fragments in front of the bicycle wheel will be swept forward by the bicycle wheel as it is pulled out. Once these fragments reach the tapered portion of the HRS bore, they will be free to slide downhill, coming to rest somewhere in the tapered section, or possibly sliding further - to the bottom of the PS end cap, where they will remain for the life of the experiment.

**4.11.2.3.4** Target remote handling system stops working while in the target hall with actuator arm extended and cannot fit back through doorway into the Remote Handling Room.

The extending arms of the target remote handling system will have manual modes to initiate a powered retract motion that is separate from the autonomous portion of the control program. Additionally, the pneumatic extend motions will be designed to allow the system to be physically pulled back with no air pressure present. The floor rail chain drive can be disconnected from its motor and manually pulled back.

**4.11.2.3.5** Target remote handling system stops working in target hall while holding target.

The target interface mechanism incorporates pneumatic grippers that are spring-loaded to the closed position, thus allowing the target assembly to remain held, even with loss of air pressure. Thus e-stop, reboot, control program modifications are all possible without dropping the target.

### 4.11.2.4 Remote Handling Quality Assurance
The target remote handling system will be designed and built in accordance with the requirements of the Fermilab Engineering Manual. The design of the target remote handling system is monitored by in-progress design review during weekly meetings of the Target Station Group. The in-progress review is followed by a final project review. The design is documented in the requirements and specification documents, CAD model, and drawings. As-built dimensions will be checked against the fabrication drawings. Operation is verified as part of the fabrication process and again after installation.

## 4.11.3 Heat and Radiation Shield

This section describes the Heat and Radiation Shield (HRS). It serves to protect the superconducting coils of the Production Solenoid (PS) from the intense heat and radiation generated by the primary (8 kW) 8 GeV kinetic energy proton beam striking the production target within the evacuated warm bore of the PS. The shield also protects the coils in the far upstream end of the Transport Solenoid (TS), a straight section of coils, called TS1, at the exit to the PS. The HRS shields downstream components from neutron and gamma background.

### 4.11.3.1 Heat and Radiation Shield Requirements

The heat shield is intended to prevent radiation damage to the magnet coil materials and ensure that quench protection is not significantly affected for the lifetime of the experiment. A detailed explanation of the HRS radiation damage requirements is given in Reference [6]. There are four primary performance parameters for the HRS:

1. The total allowed dynamic, i.e., instantaneous, heat load in the magnet coils.

2. The local peak power density in the superconducting coils.

3. The maximum local radiation dose to the superconductor insulation and epoxy over the lifetime of the experiment.

4. The radiation damage to the PS superconductor's aluminum stabilizer and copper matrix.

#### 4.11.3.1.1 Heat Load

An acceptable shield design should limit the heat load in the PS coils to less than 100 W for nominal operating conditions with the proton beam striking the target.

#### 4.11.3.1.2 Radiation Dose and Peak Power Density

The most radiation-sensitive material in the PS determines the lower limit of radiation tolerance. In particular, the epoxy used to bond the insulation to the superconducting cable can tolerate a maximum of 7 MGy before it experiences a 10% change in its shear modulus, so 350 kGy/yr is established as the limit that allows a conservative 20 years of operation. This limit of 350 kGy/yr is the equivalent of about 15 µW/gm. The requirement on power density is 30 µW/gm.

#### 4.11.3.1.3 Radiation Damage

The final parameter and most demanding requirement is related to transport in metals, in general, and the electrical conductivity of the component metals of the superconductor cable, in particular. At liquid helium temperature, damage to the atomic lattice of a superconducting cable, and its quench-stabilizing matrix made from normal conductor takes the form accumulating atomic displacements; i.e., tiny lattice defects. After exposing a metal sample to a given neutron flux spectrum, the sustained radiation damage can be characterized by the average number of displacements per atom (DPA). The DPA is directly related to electron transport in metals.

The Residual Resistivity Ratio (RRR) is defined as the ratio of the electrical resistance of a conductor at room temperature to that at 4.5°K. For a sample exposed to radiation, the RRR will decrease. However, warming such a sample to room temperature leads to recovery of the RRR [126] [127] but the degree of recovery depends on the metal. The PS utilizes superconducting cable embedded in an aluminum matrix for quench protection. Aluminum is one example of a material that shows complete recovery at 300 K. The annealing time is on the time scale of minutes at 300 K.

The HRS is designed to limit radiation to the aluminum stabilizer such that its RRR remains above 100 for at least a year of nominal beam operations. Reference [6] details how this translates to a DPA requirement of less than $4 - 6 \times 10^{-5}$ per year. Warming-up to anneal once per year is consistent with Fermilab Accelerator shutdowns that are typically scheduled once per year and last for several weeks.

A comparison of the power density and DPA requirements with the expected power deposition and DPA with the present HRS design is shown in Figure 4.163.

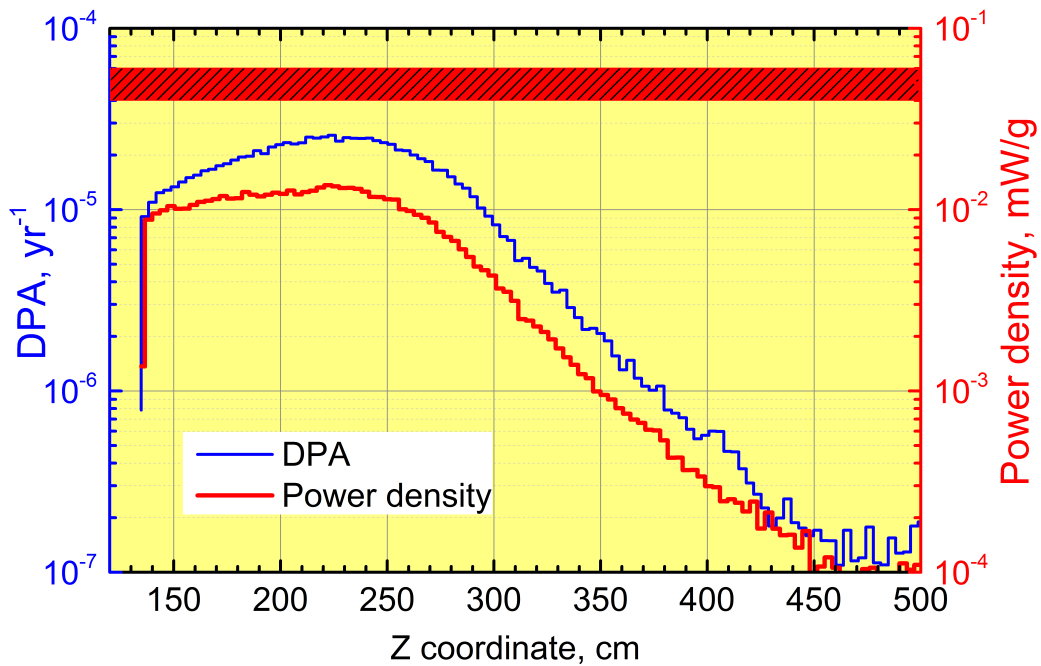

Figure 4.163. The MARS results for the peak DPA and peak power density at the PS coils.

### 4.11.3.1.4 Additional Requirements

The HRS must have sufficient inner aperture to allow good capture of pions and muons to maximize the stopping rate of negative muons in the Detector Solenoid stopping target. In addition, any acceptable shield design must avoid any line-of-sight cracks between components that point from the target to the PS inner cryostat wall and thus the magnet coils.

The materials used to construct the shield must not cause the magnetic field of the Production Solenoid to fall outside of specifications. This implies the use of non-magnetic materials with a magnetic permeability of less than 1.05.

The HRS will include stable mounting features that provide a mechanical connection between the target support structure and inner bore of the HRS vessel. In addition, to avoid overheating of the target support structure, the inner wall of the HRS vessel will be cooled such that its operating temperature stays below 100℃. Figure 4.164 shows the temperature distribution on the inner wall of an HRS modeled according to the present design.

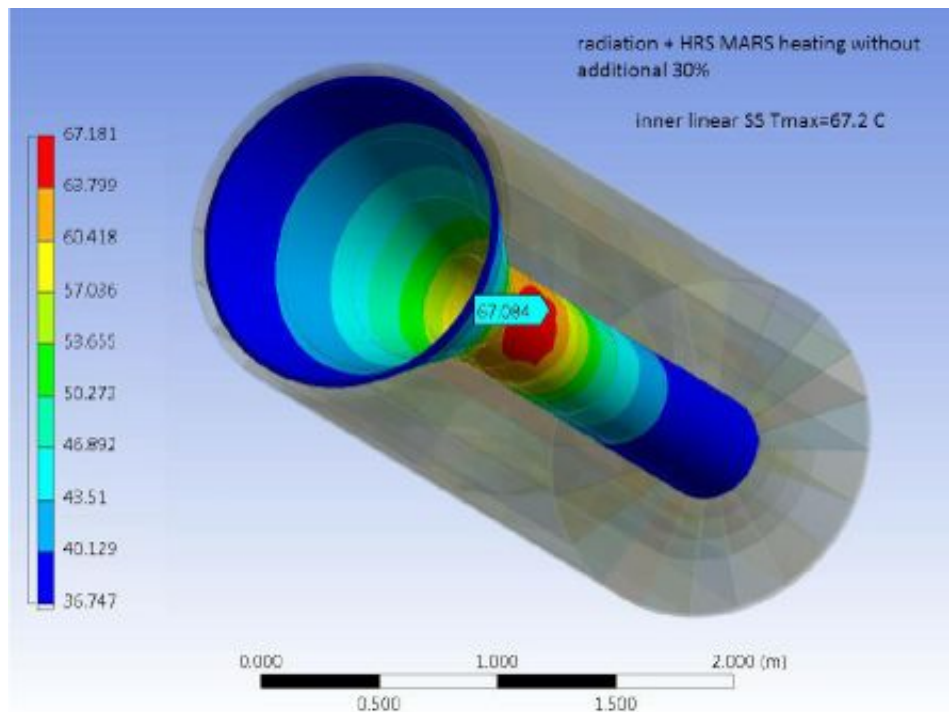

Figure 4.164. Temperature of the HRS inner liner taking into account the MARS energy deposition and the thermal radiation of the target. The maximum temperature of the inner liner is 67.2C.

### 4.11.3.2 Heat and Radiation Shield Technical Design

The HRS design has evolved over time based on numerous studies using MARS [128] and because interface requirements. The implementation of the PS+HRS geometry in MARS is shown in Figure 4.165. Extensive variations have been explored in both geometry and materials. Most recently the effect on the HRS design on the neutron and gamma backgrounds downstream of the production solenoid have also been considered.

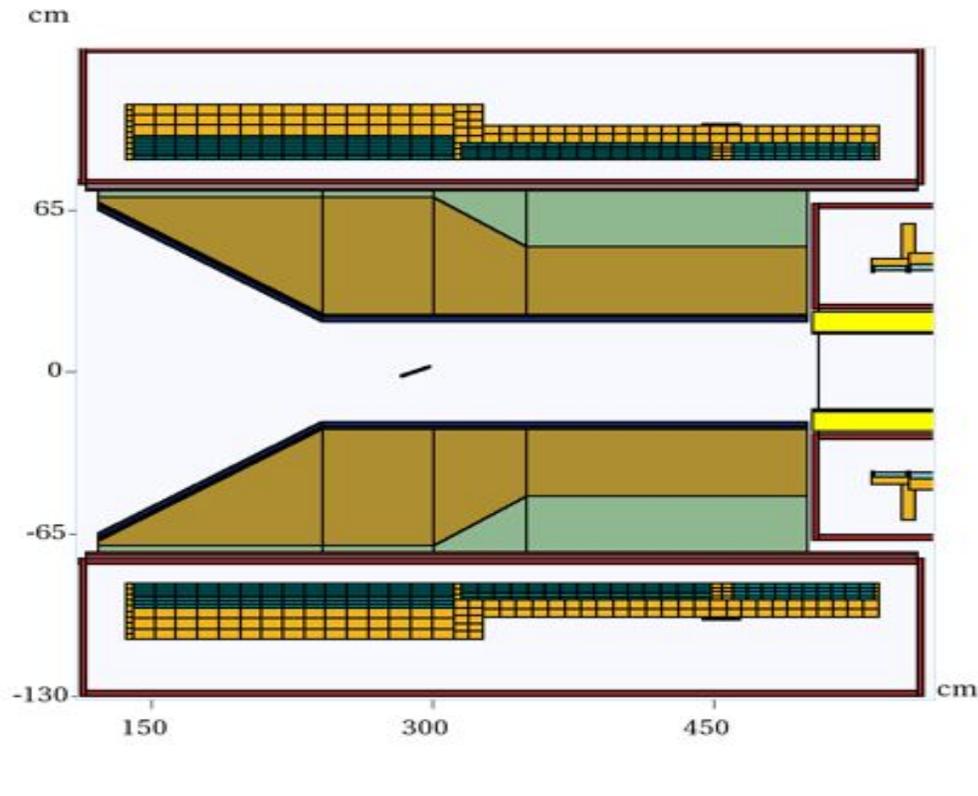

Figure 4.165. Heat and Radiation Shield geometry simulated in MARS15. This figure shows a horizontal slice at the elevation of the target. Note the target lies on the magnet axis in the plane shown. The target is tilted 14° away from the axis to align with the beam. Bronze is shown in brown and water is light green

Figure 4.166 shows the power density in a 20 cm thick slice of the HRS as a function of distance along the solenoid axis. Figure 4.163 shows the peak DPA and power density at the inner surface of PS coils.

The performance of the shield meets the requirements for the amount of heat and radiation permitted to reach the superconducting coils and is summarized in Table 4.34.

Table 4.34. Performance of HRS compared to specifications on the magnet coils.

|  | Peak DPA/yr* [$10^{-5}$] | Peak Power Density [$\mu$W/g] | Absorbed Dose [MGy/yr] | **Years Before 7 MGy | Dynamic Heat Load [Watts] |
|---|---|---|---|---|---|
| Specification | 4 to 6 | 30 | 0.35 | 20 | 100 |
| Performance | 2.4 | 13 | 0.26 | 27 | 24 |

* This is the DPA damage per year for which RRR degrades to 100. After this RRR reduction the PS must warm-up and anneal.
** 7 MGy is a conservative limit on coil epoxy exposure, 10% of shear modulus lost due to radiation damage.

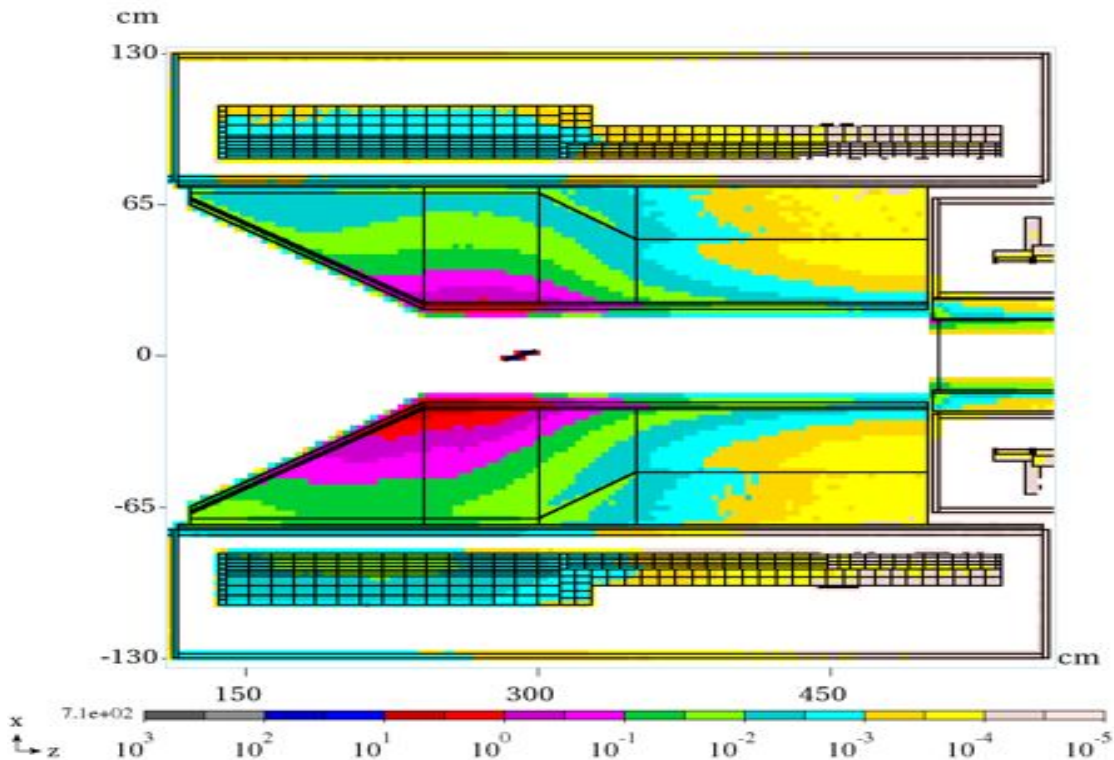

Figure 4.166. The MARS results for power density (μW/g). The proton beam enters from the right side at a 14° horizontal angle in this overhead view. The resulting shower in the HRS from the beam-target interaction peaks just downstream of the target and towards the proton beam direction.

Simulations of the HRS and Production Solenoid give the amount of shielding required, but there are interfaces with the beamline vacuum system and the solenoids that must also be considered in the design. The pion production target lifetime improves with better vacuum, so it was decided to exclude the HRS from the beam vacuum chamber. This results in less outgassing and a smaller vacuum chamber volume.

The HRS requires active cooling. The simplest and most reliable solution is to flow water around the entire HRS in tank. The HRS design includes a stainless steel liner that isolates the HRS volume, including the cooling water, from both the solenoid and the beamline vacuum. The water also provides good shielding for low-energy neutrons and gammas that are a source of noise in the Cosmic Ray Veto (CRV) detector.

The materials used to construct the HRS must not cause the magnetic field of the Production Solenoid to fall outside of specification, therefore, non-magnetic materials must be used. All conducting materials must be designed to reduce eddy current forces that might develop during a quench. Bronze C63200 satisfies these requirements and is the material of choice. It can also be manufactured in large forged pieces, simplifying assembly. The resulting design is shown in Figure 4.167, Figure 4.168, and Figure 4.169.

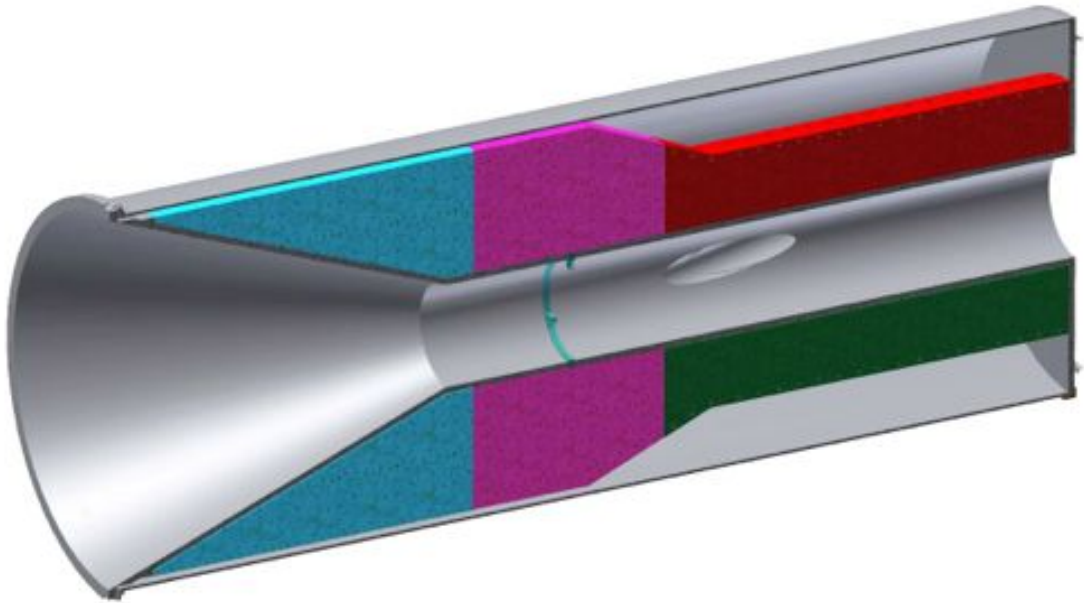

Figure 4.167. Section view of the HRS. Bronze pieces are shown in light blue, purple and red. A stainless steel liner surrounds the bronze on all sides and the vessel holds 600 gallons of water.

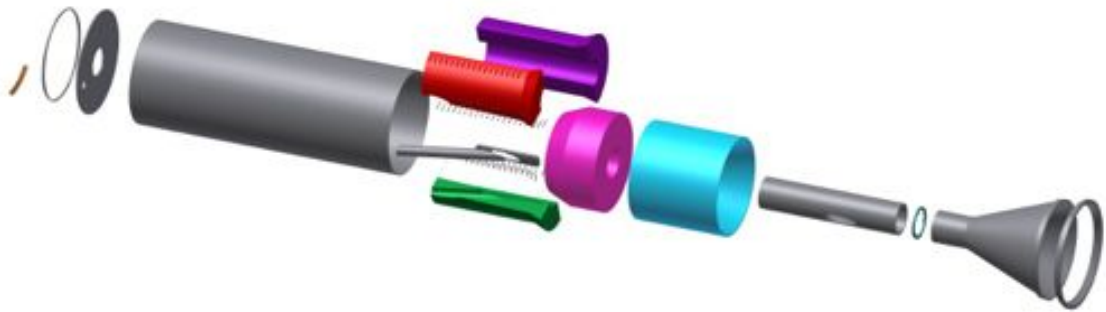

Figure 4.168. Exploded view of the HRS showing the stainless steel liner.

Figure 4.170 shows the HRS installed in the PS with the vacuum end cap welded in place. The outer tank wall has an outer radius of 73.66 cm. The PS inner cryostat wall, at a radius of 75 cm, will support the HRS. The length of the HRS is about 4 m and it has a minimum inner radius of 20 cm. The total assembly weight, including both the bronze and stainless steel liner, is 36 tons. The pion production target is located in this cavity, mounted on a support ring shown in Figure 4.171.

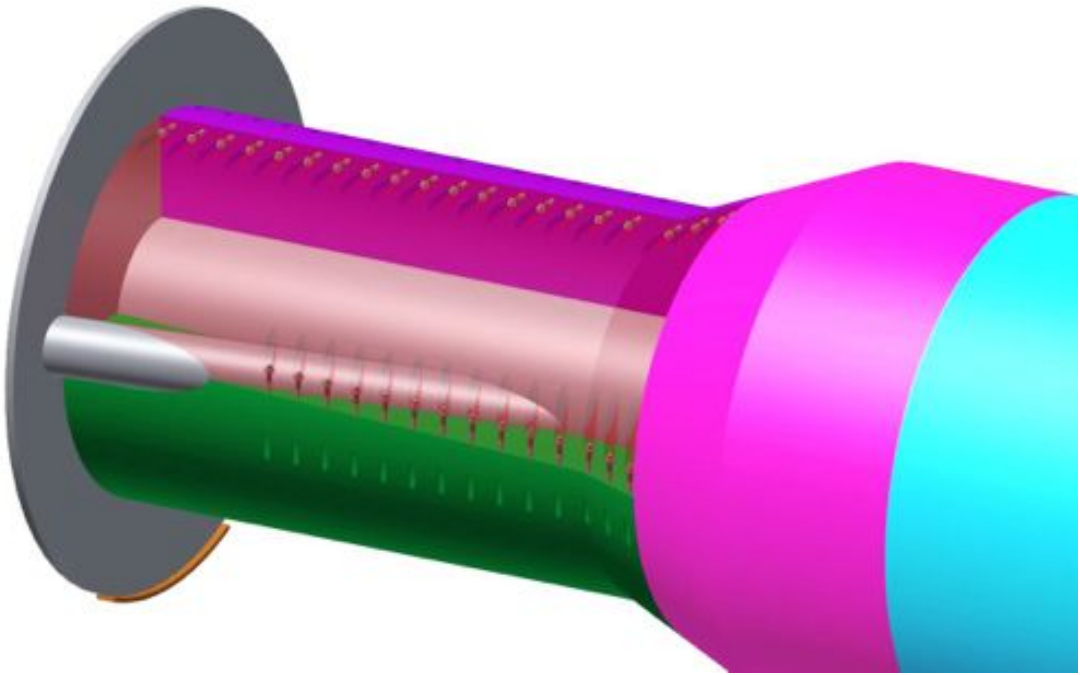

Figure 4.169. Cutaway of the upstream section of the HRS showing how the proton beam pipe passes through the HRS. The outer water tank shell is not shown and the red bronze block of Figure 4.168 is transparent. The three bronze blocks in this section are all bolted to each other to create a continuous cylinder that hangs from the stainless vacuum liner.

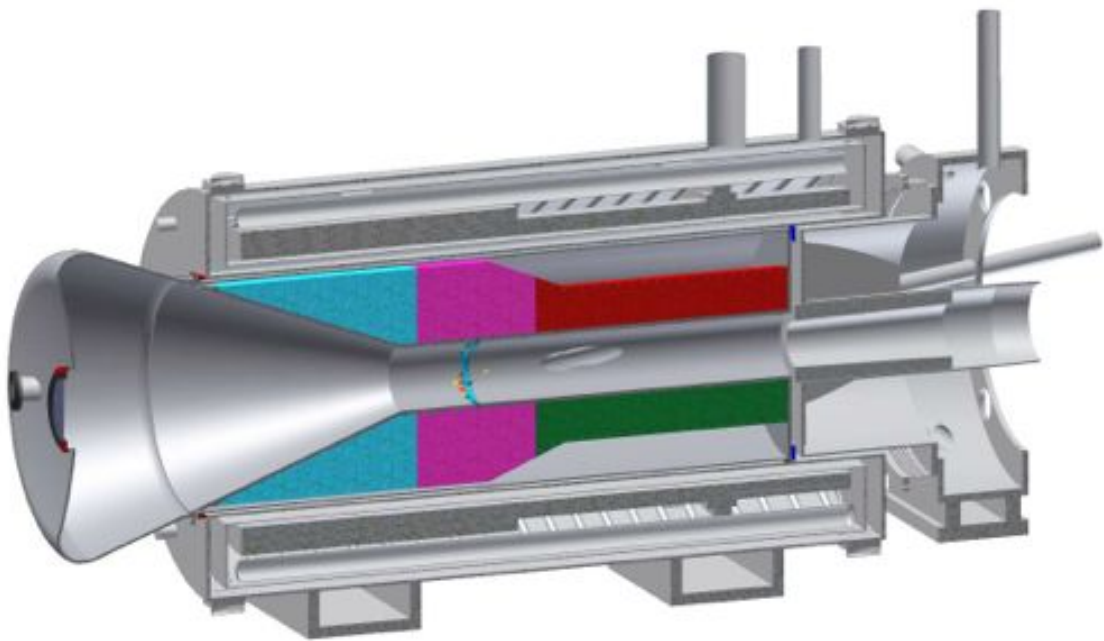

Figure 4.170. Section view of the HRS inside the PS. A portion of the TS is shown inserted into the upstream end of the PS and the vacuum end cap is welded to the downstream end of the HRS. The cyan, magenta, red, and green pieces are the bronze C63200 parts. The liner is stainless steel 316L.

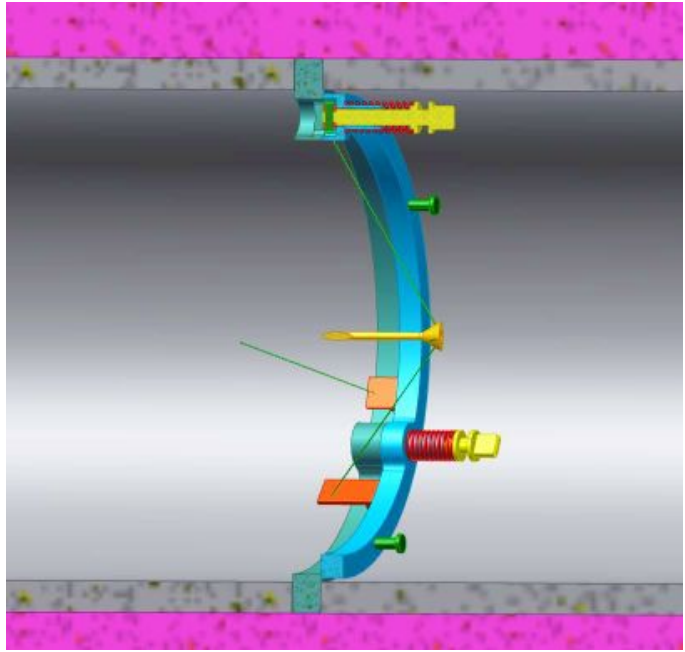

Figure 4.171. Elevation section view showing the target support ring and the twist lock fasteners.

The HRS must be mechanically stable. Current simulations suggest the shield will experience an average heat load of 3.3 kW. Since the HRS is inside the solenoid, radiation cooling (passive) leads to unacceptably high operating temperatures in the HRS; therefore, the shield must be actively cooled. The cooling is accomplished by immersing the bronze shielding in water. The water flows into and out of the HRS tank through ports at the top and bottom of the downstream flange. The source of cooling water will be the LCW system, which also supplies cooling water to the beamline magnets. The pressure of the LCW system, approximately 140 psig, forces the HRS to be designed as an ASME Boiler and Pressure coded vessel. Figure 4.172 – Figure 4.175 show in detail how the HRS is welded in place on both ends inside the PS.

Not shown in Figure 4.173 or Figure 4.174 are any ports in the red tube (weld adapter) to evacuate the annular space around the HRS. This vacuum space does not need a very good vacuum, but heat transfer to the PS from the HRS will be reduced if the area is evacuated.

A 3D thermal steady state analysis has been completed and the very acceptable results are shown in Figure 4.82 and described in [129]. The effect of the beam energy deposition as well as the thermal radiation from the target heating the inner wall of the liner has been accounted for. Figure 4.164 shows the temperature distribution of the inner liner from the energy deposition of the beam and the thermal radiation coming from the target. The calculated inner wall temperatures are well below the 100℃ required to preclude overheating the target support components mounted on the inner bore of the HRS. Figure 4.176 shows the temperature profile on the outer tank wall.

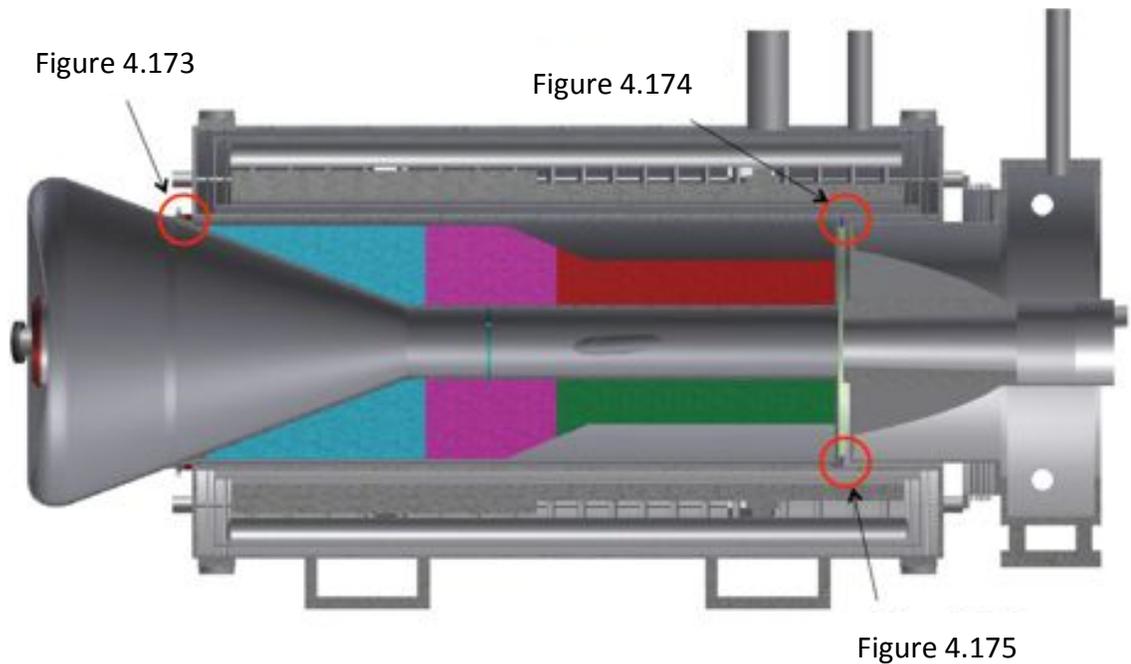

Figure 4.172. Section View through the PS, HRS and TS. The circled areas are shown in blow-ups on the next three figures.

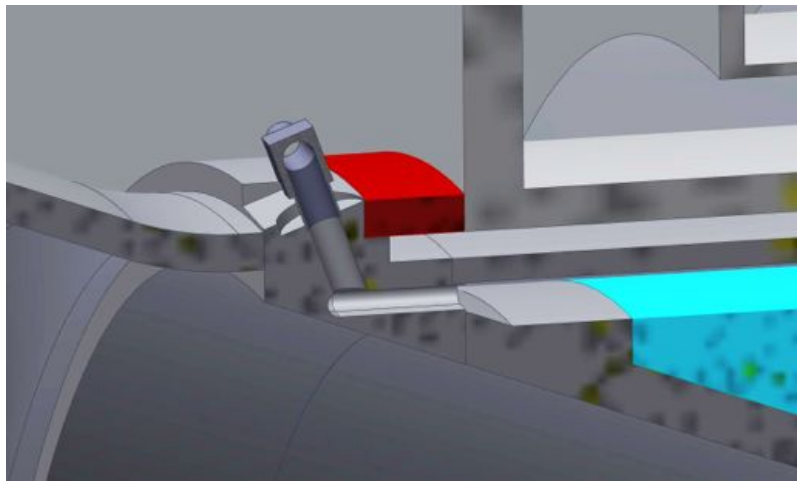

Figure 4.173. Close-up section view of the connection between the downstream end of the PS, the HRS and a water port. The water port has to be drilled in at an angle to the front face of the DS flange so that it does not penetrate the inside surface of the conical wall.  The port size is based on ¾" OD tube size. The annular space inside the HRS is smaller than ¾" in., so to compensate and create an adequate entrance area for the water, there is a slot milled into the downstream flange instead of a hole. This configuration should prevent any plugging from whatever debris might get into the LCW system. Multiple inlets and outlets will be included for redundancy. Water flow is expected to be around 3 gpm.

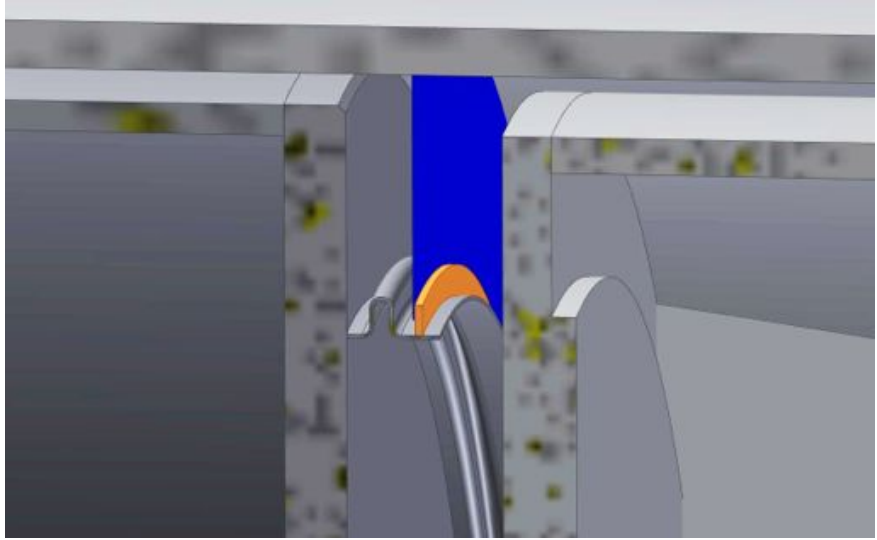

Figure 4.174. Close-up section view of the top of the upstream end of the HRS showing the welded connection to the PS. The blue ring is welded to the PS. The orange ring is welded in place after the HRS is inserted into the PS. This joint separates the annular space around the HRS water tank from the central vacuum space around the target. Since pumping speed to the annular space is very limited, separating this region should improve the vacuum around the target. This joint must be flexible to allow the HRS water tank to "breathe" with the application of internal water pressure.

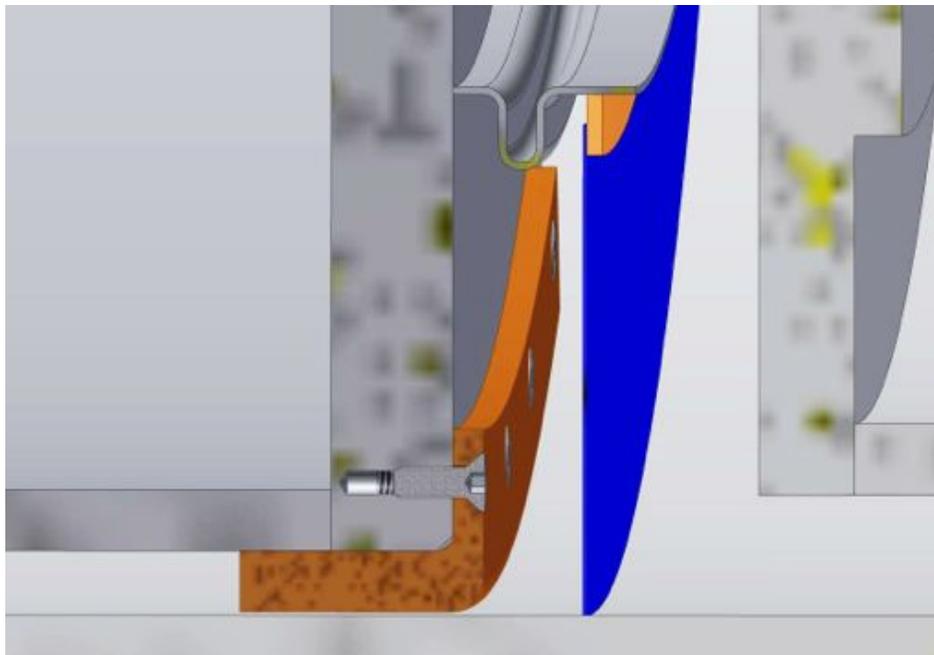

Figure 4.175. Close-up section view of the bottom of the upstream end of the HRS. The brown piece is a bronze arc segment sliding bearing. The HRS is supported by the weld to the downstream end, shown in the figure, and by this bearing at the upstream end. The HRS cannot be cantilevered off the downstream flange as FEA studies show that the deflection of the upstream end would be too large for assembly.

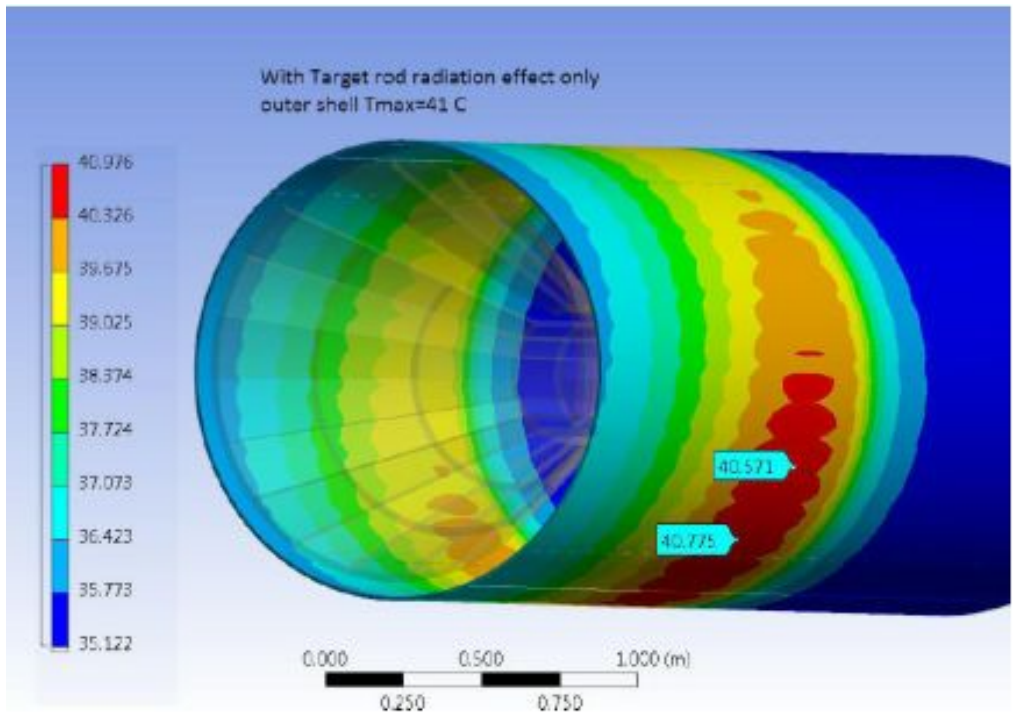

Figure 4.176. Temperature distribution on the outer tank shell of the HRS taking into account energy deposition from the beam and thermal radiation coming from the target.

The HRS will be assembled vertically on a rotating fixture with the downstream flange down near the floor. Successive pieces of bronze will be lowered around the inner liner. After the first two pieces of bronze are assembled, the beam pipe has to be welded to the inner liner. The next three pieces of bronze are assembled around the beam pipe and bolted to each other as shown in Figure 4.168 and Figure 4.169. The next step is to lower the outer tank wall and weld it to the downstream flange. Finally, the upstream end flange is lowered onto the assembly. The beam pipe, inner liner and outer tank wall are all welded to it.

### 4.11.3.3 Heat and Radiation Shield Risks

The HRS is designed conservatively, so the risks are considered low. There is a very large margin for magnet quenches due to beam heating during normal operations. During a quench there are large forces on the PS and HRS that have been accounted for in the design. The largest risk is that the calculation of the RRR degradation of the aluminum stabilizer in the coils due to neutron radiation has been underestimated and it would be necessary to warm-up to anneal the aluminum more than once per year. This would result in inefficiency in the operating time of the experiment.

### 4.11.3.4 Heat and Radiation Shield Quality Assurance

The HRS will be designed and built in accordance with the requirements of the Fermilab Engineering Manual. Design of the HRS is monitored by in-progress design review during weekly meetings of the Target Station Group. The in-progress review is followed by a final project review. The design is documented in the requirements and specification documents, FEA reports, CAD model, and drawings. As-built dimensions are checked against the fabrication drawings.

### 4.11.4 Proton Beam Absorber

#### *4.11.4.1 Proton Beam Absorber Requirements*

The Mu2e beam absorber [7] stops the unspent proton beam and secondary particles that make their way through and beyond the target in the forward direction. The beam power from the accelerator complex is 8 kW, and while 0.7 kW will be deposited into the target itself, and 3.3 kW will be absorbed by the Production Solenoid Heat and Radiation Shield, a significant amount of power is deposited in the beam absorber. The beam absorber must be shielded so that its prompt and residual radioactivity does not significantly contribute to the radiation dose rate at the downstream end of the production solenoid enclosure.

The absorber must be able to accept the entire beam power in the event that the target is missed, or during pre-targeting beam tests. The beam absorber must be placed outside of and well beyond the Production Solenoid to allow access to the crane hatch and room for remote target exchange equipment. The beam absorber must be compatible with the extinction monitor located above and behind the beam absorber [130].

#### *4.11.4.2 Proton Beam Absorber Technical Design*

The calculations in the following paragraph were performed with the 8 kW baseline beam power. Using the Revised Fermilab Concentration Model [131] at the Mu2e design beam intensity, the average concentration of radionuclides in the sump pump discharge will be 24 pCi/ml due to tritium and 2 pCi/ml due to sodium-22. This corresponds to 2% of the total surface water limit if the pumping is performed once a month (conservative scenario). Build-up of tritium and sodium-22 in ground water at $1.2 \times 10^{20}$ protons per year will be as low as $6.2 \times 10^{-8}$ % of the total limit over 3 years of operation. Air activation is acceptable ($< \sim$30 Ci/yr) with the planned 800 cfm air flow [132].

The beam absorber [133] must be able to accept the total number of protons required by the experiment, $3.6 \times 10^{20}$ over 3-4 years, plus an acceptable overhead to account for commissioning and tuning (100%), without replacement over the life of the experiment. The transverse dimensions of the absorber must be consistent with the beam properties after accounting for distance from the target and divergence of the beam, including scattering in the target. The transverse proton beam size at the absorber face has a sigma of 1.3 cm in both planes.

The absorber, shown in Figure 4.177, consists of a steel core with the dimensions $1.5 \times 1.5 \times 2$ m and concrete shielding with the dimensions $3.5 \times 3.5 \times 5$ m, so the core is surrounded by 1 m of concrete on the sides, the top, and the bottom. It has a $1.5 \times 1.5$ m opening toward the beam and also a $2.5 \times 2.5 \times 1$ m albedo trap to protect the downstream end of the Production Solenoid from the secondary particles generated by the spent proton beam in the core.

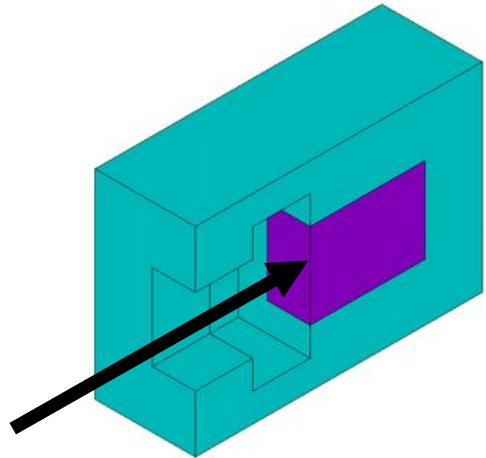

Figure 4.177. A cut-away view of the proton beam absorber. The purple rectangular volume is the steel absorber and the surrounding material and entrance are concrete.

Simulations were run for two absorber materials, steel and aluminum. Both materials are acceptable. The proton beam absorber will be built using steel that is being stored at the Fermilab railhead to eliminate the cost of purchasing new material. The core consists of ten 8" thick, 1.5 m by 1.5 m plates welded together for a total length of 2 m. Power depositions in the absorber are shown in Table 4.35. The simulations [133] show that power deposition in *accidental* mode (primary beam misses the target) will be 6.7 kW for the steel core. In the *normal operation* mode (primary beam hits the target) the power deposition will be 1.7 kW for the steel core. Residual dose on contact with the concrete shielding of the beam absorber after 30 days of irradiation and 1 day of cooling will be at the level of few mSv/hr.

The results of MARS simulations and ANSYS finite element analysis (FEA) show that for forced convection air cooling during normal operation mode, the peak power densities lead to a maximum absorber temperature of 70°C. The shielding concrete temperature is well below 100°C. The ANSYS temperature distribution plot for normal operation is shown in Figure 4.178. ANSYS FEA studies for the accident condition with an aluminum core show that the maximum temperature increases about 10°C if the accident condition persists for 15 minutes. Thus, based on the FEA result for the aluminum core, the steel core temperature is acceptable in the accident condition scenario.

Table 4.35. Power deposition in the proton beam absorber: The "accident" condition refers to the proton beam missing the target and the "operation" condition refers to the proton beam striking the target.

| Absorber material/mode | Q(kW) |
|---|---|
| Al, accident | 5.8 |
| Fe, accident | 6.7 |
| Al, operation | 1.4 |
| Fe, operation | 1.7 |

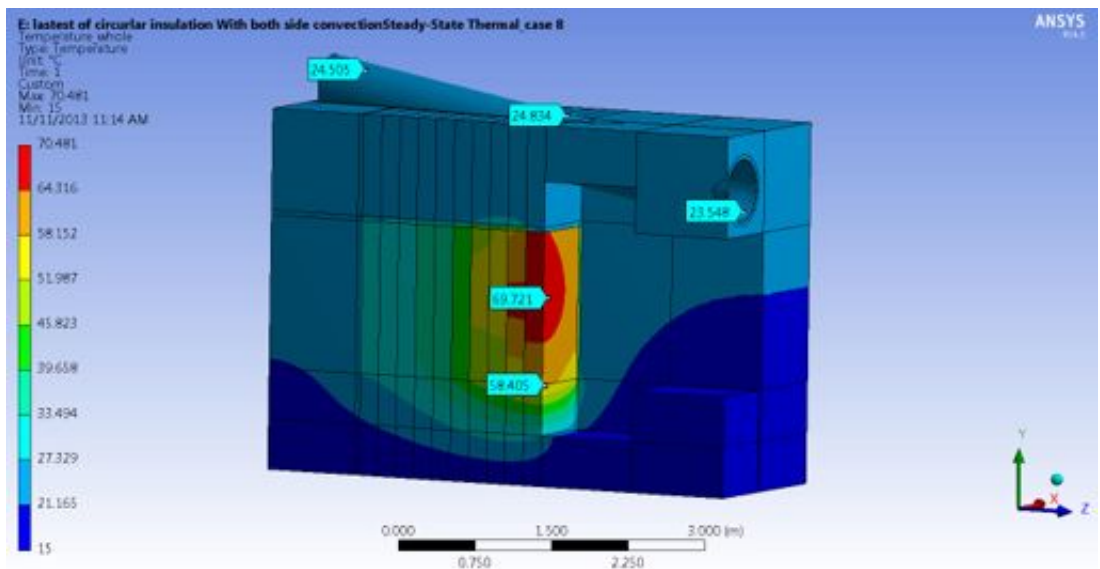

Figure 4.178. Temperature distribution in the absorber steel and shielding concrete for normal operation.

Extinction monitor components are installed in a steel pipe with a 24" outer diameter that runs through the shielding concrete above the absorber steel. The steel pipe is laid in a trough that is cast into the shielding concrete. The pipe is thermally insulated from the shielding concrete with ceramic insulation. This pipe is included in the ANSYS FEA of the proton beam absorber for normal operation. The ANSYS temperature distribution plot for the pipe is shown in Figure 4.179.

The proton beam absorber layout drawing with dimensions is in Mu2e Document 3589 [134]. Energy deposited in the absorber is removed by forced convection air-cooling. The airflow path through the absorber is shown in Figure 4.180 and Figure 4.181. 800 scfm of cooling air at 15.6°C is supplied to the absorber by a building air system through a 14" diameter pipe to cool the absorber. The cooling air exits the cooling passages at the top of the core at ~25°C after removing the 1.3 kW deposited by the beam in the core during steady-state operation.

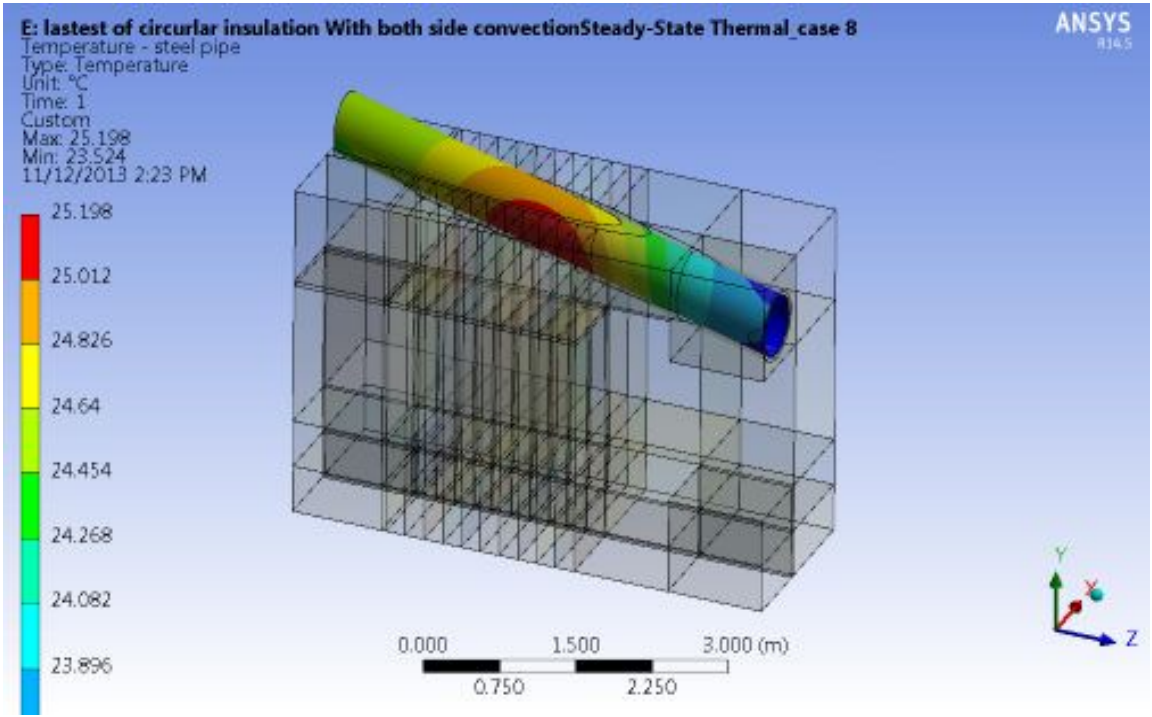

Figure 4.179. Temperature distribution for the steel extinction monitor pipe.

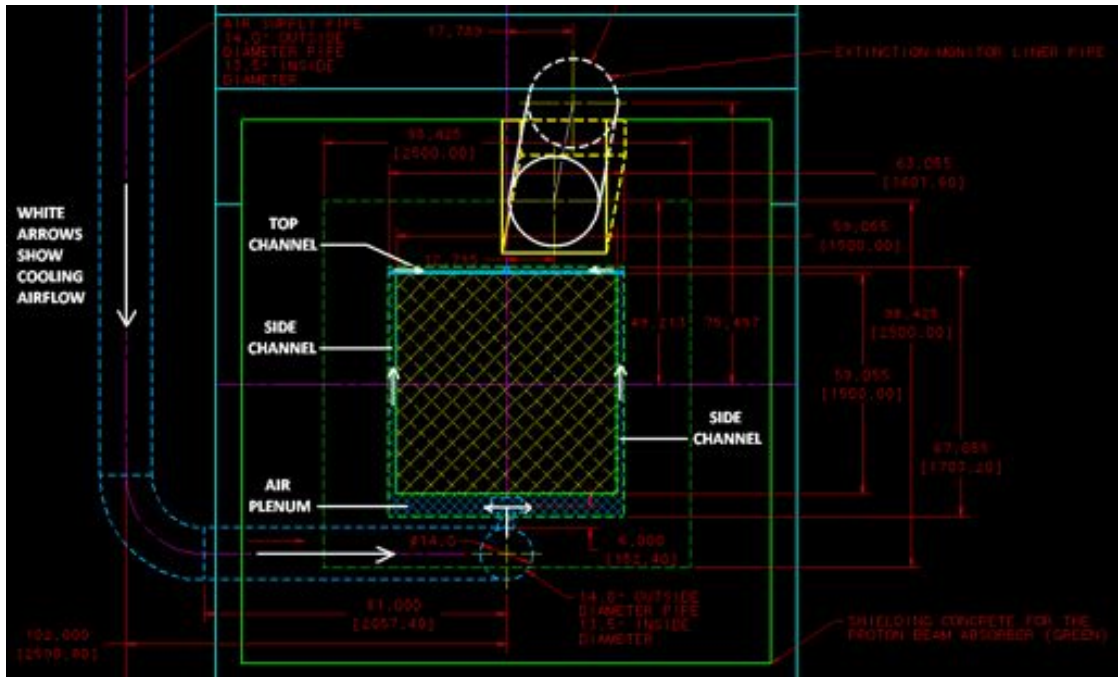

Figure 4.180. Airflow path through absorber – upstream view.

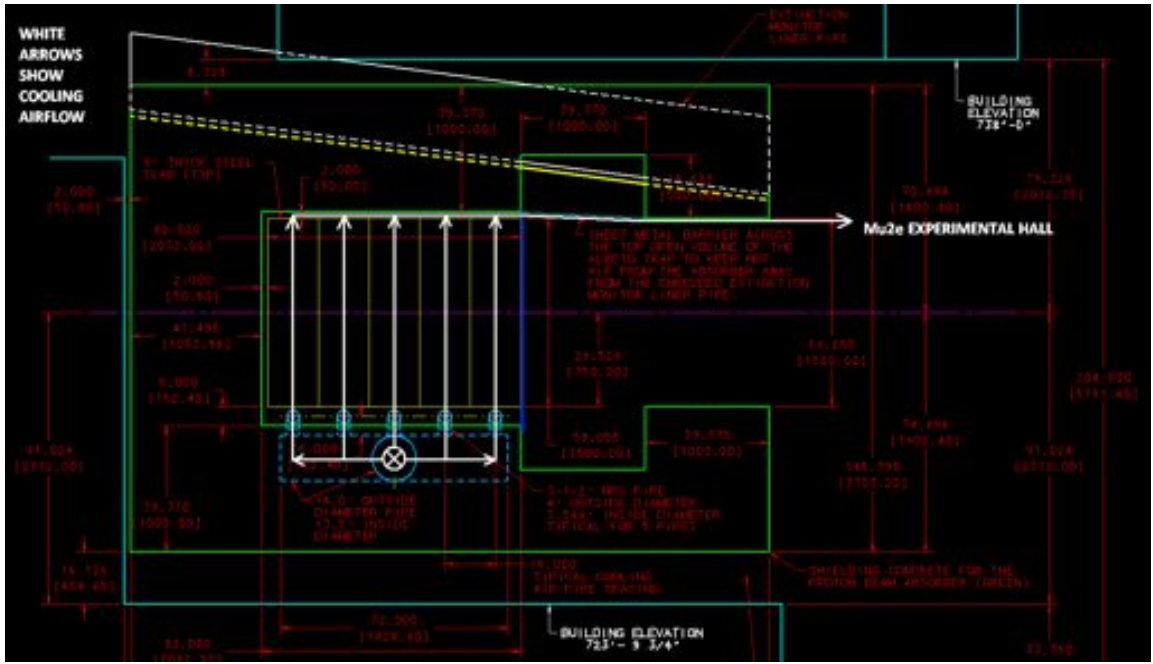

Figure 4.181. Airflow path through absorber – side view.

The cooling passages on the top and sides of the steel core are 2" wide. The passages are formed by the core steel on one side, i.e., the surface being cooled, and by ¼" thick plate on the other side. The ¼" plates are welded to the steel core using standoff rods. The ends of the side cooling passages at the upstream and downstream faces of the steel core are closed by welding ¼" thick plate across the opening at each end. The downstream end of the top cooling passage is also closed with ¼" thick steel plate. The upstream end of the top cooling passage is left open for the cooling air to exit. The top cooling passage is divided into two separate passages along the length of the core on the longitudinal centerline. An adjustable weir plate is installed at the exit of each of the two top channels to balance the cooling airflows during installation. There is a 6" high plenum under the steel core that is sealed on all four sides. The plenum floor is the horizontal face of the shielding concrete that the core sits on. The 6" high plenum is formed by installing the core on 6" thick steel pads placed on the shielding concrete. There is no cooling passage on the downstream face of the last 1.5 m × 1.5 m steel plate of the core. The 14" diameter air supply pipe runs under the 6" high plenum and terminates in a 14" air manifold pipe. The air manifold is slightly shorter than the steel core and it is located on the centerline of the core. Five 3-1/2" pipes on top of the manifold distribute the cooling air into the 6" high plenum uniformly along the length of the steel core. The 6" high plenum directs the cooling air into the side channels. Cooling air pressure drop is calculated at 1" WC (water column) and specified as 2" WC for the building air supply system.

In the ANSYS heat transfer FEAs, energy is distributed in the steel core according to the results of a MARS simulation. The shielding concrete is insulated on all external surfaces except the bottom; the bottom surface is held at a constant 15°C. Heat transfer coefficients in the cooling channels are calculated by hand and applied to the channel surfaces. The air heat sink temperature for each channel is the exit (not the average) temperature for that channel. The extinction monitor pipe is insulated from the shielding concrete with ceramic insulation. The extinction monitor pipe is perfectly insulated on its inside surface so no heat transfers to or from this surface. The heated exit airflow does not flow vertically upwards and into the extinction monitor pipe upon exiting the shielding concrete.

A steel door is mounted on the upstream face of the shielding concrete. The door is opened and closed manually with a long rod to protect personnel from the activated steel core during a beam-off access to the experimental enclosure.

### 4.11.4.3 Proton Beam Absorber Risks

The design of the proton beam absorber for Mu2e is similar to other beam absorbers that have long been in use at Fermilab. Thus, no design, fabrication, or operating risks have been identified. The only operating component, the cooling air fan, is located in a general-purpose area that can be accessed when the beam is operating.

### 4.11.4.4 Proton Beam Absorber Quality Assurance

The proton beam absorber will be designed and built in accordance with the requirements of the Fermilab Engineering Manual. Design of the proton beam absorber is monitored by in-progress design review during weekly meetings of the Target Station Group. The in-progress review is followed by a final project review. The design is documented in FEA reports, a CAD model and engineering drawings. As-built dimensions will be checked against the fabrication drawings. Balanced airflows are verified as part of the installation process.

## 4.11.5 Production Solenoid Protection Collimator

### 4.11.5.1 Protection Collimator Requirements

The Mu2e primary 8 GeV (KE) proton beam provides an average of 8 kW of beam power. Normally, this beam will interact with the production target located within the bore of the Production Solenoid (PS), with the remnant beam delivered to the proton beam absorber located well downstream of the production region. There are stringent limits for heat loads on components of the PS, the Transport Solenoid (TS), and the Heat and Radiation Shield (HRS). In addition, there are stringent limits on displacements per atom (DPA) for the aluminum-stabilized superconductor in the Production Solenoid. In particular, beam accident conditions that result in direct interaction of primary beam protons with the HRS, cryostat walls, or coils, could result in significant heat deposition or radiation damage to these components [8].

To ensure that the limits are not exceeded, even during beam accident conditions, Mu2e requires a Protection Collimator (PC) to be installed between the downstream end of the final beam element, and the entrance of the proton beam through the PS cryostat wall. To fit between the last trim dipole and beam profile monitor and the entrance to the PS cryostat, the PC must be no longer than 1.5 m.

During beam accident conditions, the PC must be able to absorb and dissipate the heat load from the full primary 8 kW (average) beam for at least 50 msec before the beam protection systems trip off the beam. To intercept the beam in any accident condition, the PC must have a minimum transverse dimension of 50 cm.

Vacuum consistent with the proton beamline vacuum requirements shall be maintained within the aperture of the PC.

During normal operations, the beam will have a transverse size within the PC aperture of ~1 cm full-width. To maximize targeting flexibility, the beam aperture through the PC should be as wide as possible, but the requirement to limit fake signals in the extinction monitor limits the aperture diameter to 80 mm [130].

The pion production target is 0.6 cm in diameter, thus the beam must move transversely by 1cm in each direction to ensure the beam can be swept across the target for targeting studies and optimization. Since the 80 mm aperture size violates the requirement to be able to swing the beam $\pm 0.8°$ about the geometric center of the target, the protection collimator must be able to move out of the proton beam within its vacuum volume.

Since it will be installed in close proximity to the PS, TS1, and TS2 coils, the PC must be constructed from non-magnetic materials so that it does not affect the field uniformity requirements of those systems. The PC and its supporting systems should be serviceable without removal from within the beam transport enclosure. As far as is practical, materials should be chosen to minimize residual activation for the protection of maintenance personnel and to minimize the radiation cool-down time to avoid significant downtimes that might impact the experimental sensitivity.

The Protection Collimator engineering requirements and specifications are defined in Mu2e document 2902 [135]. The primary requirements and specifications are:

- Core length of 1 meter.

- Core outer diameter of 50 cm.

- Beam aperture diameter through the center of the core of 8 cm.

- Beam aperture diameter for low-energy target scan of 20.8 cm.

- No science requirement for in-beam positioning accuracy as long as the repeatability requirement is met.

- In-beam positional repeatability is $\pm$0.25 mm.

- The core has two operating positions: (a) 8 kW beam operation and (b) Low-energy target scan. The time it takes to travel from one position to the next is not important but must be less than 10 minutes.

- Operating vacuum of $1\times10^{-8}$ Torr.

- Acceptable materials for construction are 300 series stainless steel and aluminum.

- Provide mounting points for beam loss monitor tubes on the top, bottom and sides of the Protection Collimator.

- Dissipate the heat load deposited during beam accident conditions.

### 4.11.5.2 Protection Collimator Technical Design

Figure 4.182 shows the layout of the Protection Collimator. The core material is stainless steel plate that weighs 3,517 pounds. The plates are welded together along the outer circumference. Virtual leaks can occur between plates that are pressed tightly together. To prevent potential virtual leaks in the core, a gap of roughly 0.1 to 0.25 mm is maintained between the plates after welding. The core has two positions depending on the two operating conditions: (1) 8 kW beam operation and (2) Low-energy target scan operation. In the 8 kW beam-on position the core is centered on the beamline. The core moves 358.14 mm laterally in the low-energy target scan position to create the beam aperture area required for the scan.

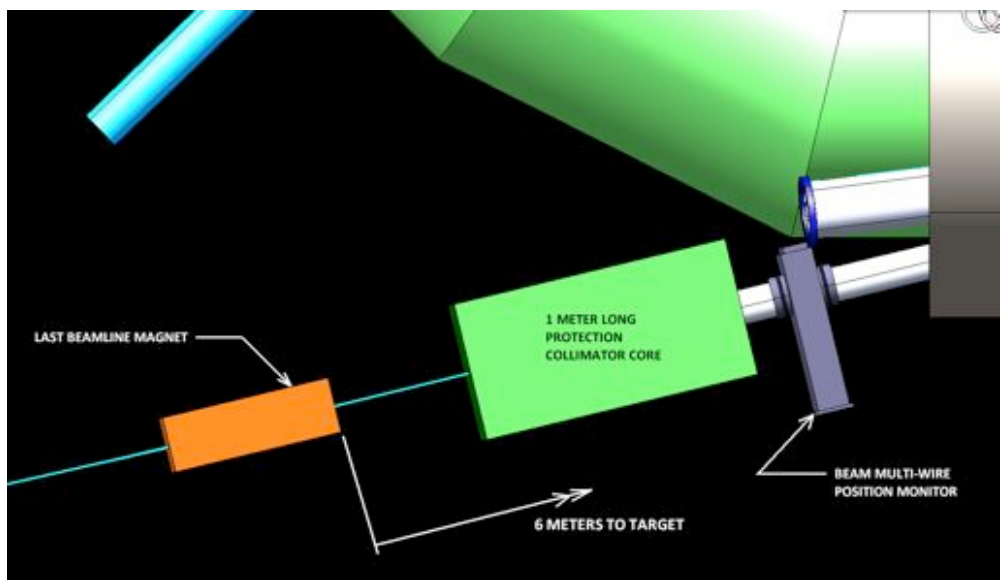

Figure 4.182. The Protection Collimator consists of the core, vacuum vessel, drive, and instrumentation.

The core moves on a welded, structural frame made with stainless steel tube. The bottom and lower sides of the vacuum vessel and the frame are a single welded assembly. The frame has forklift slots and also tapped holes for lifting fixtures to be used with an overhead crane.

The vacuum vessel is a rectangular box with removable access covers. As discussed above, the lower section of the vacuum vessel and the frame is an integral welded assembly. The bolted covers are vacuum-sealed using conflat seals (copper gasket and knife-edge flange). The beam pipe connections are sealed with conflat flanges. With the covers off, the entire internal volume of the vacuum vessel is accessible for assembly, upgrade, and maintenance.

The drive that moves the core is mounted on the vacuum vessel. The linear drive consists of a single worm gear ACME screw jack driven by an electric motor. One limit switch indicates the 8 kW beam operation core position and a second limit switch indicates the low-energy target scan position. The jack, motor, and limit switches are mounted outside the vacuum vessel for maintenance. The ACME screw travels in a bellows that seals the vacuum at the linear drive connection between the core and the ACME screw. The core is locked in position for transport to prevent core motion that could damage the drive system or create an unbalanced load.

The protection collimator is located between the last beamline element and the SWIC that is located just upstream of the 5" OD beam entry port into the Production Solenoid. It is mounted on a support stand, shimmed to the specified elevation, and bolted down at the specified location.

### 4.11.5.3 Protection Collimator Risks

The mechanical and electrical design of the protection collimator is straightforward and uncomplicated. Thus, no design, fabrication, or operating risks have been identified. There is always the risk that components can fail. The failure risk is mitigated by designing the protection collimator so all components are accessible for repair or replacement.

### 4.11.5.4 Protection Collimator Quality Assurance

The protection collimator will be designed and built in accordance with Fermilab Accelerator Division standards and guidelines (cleaning and mechanical details for high vacuum service, etc.) for beamline components. Design of the protection collimator is monitored by in-progress design review during weekly meetings of the Target Station Group. The in-progress review is followed by a final project review. The design is documented in the requirements and specification documents, CAD model, and drawings.

As-built dimensions are checked against the fabrication drawings. Operation is verified as part of the fabrication process and again after installation.

## 4.11.6 Target Station Installation and Commissioning

The proton beam absorber is the first target station component to be installed. Installation will commence shortly after beneficial occupancy of the Mu2e building and will be performed by an outside contractor. The proton beam absorber is installed in four steps. In the first step, the air distribution piping is installed and embedded in the lower section of the cast-in-place shielding concrete. The top surface of this lower concrete section is the floor of the 6" high air plenum under the absorber, as discussed above. In the second step, the core assembly is placed in position on the lower section of the shielding concrete that was poured in step one. In step three, the rest of the shielding concrete is poured. The ¼" plates that are welded to the steel core to form the air-cooling passages are also the innermost forms for the concrete poured in this step. In the fourth step, the extinction monitor pipe is installed and insulated.

The HRS is installed after the Production Solenoid installation. The HRS is assembled with its axis vertical (i.e. stacked); a rotating fixture is used to rotate the axis of the HRS to the horizontal position. The HRS and its rotating fixture are then lifted and attached to a transport/installation fixture. The transport/installation fixture supports the HRS as it is transported to the Mu2e site and allows it to be lowered by crane into the PS hall. There, the installation fixture must be aligned and locked to the PS, allowing the HRS to slide into the PS bore. Figure 4.183 shows a sketch of the HRS and the transport/installation fixture. Figure 4.184 shows the HRS in position in the PS, ready to be welded in place.

Field mapping of the PS will occur after the installation of the HRS. After the solenoid field mapping is complete, the target remote handling system will be used to install the target. The remote handling system will be commissioned outside of the Mu2e area with a mock setup to reduce the debugging time needed in the Mu2e area. It is desirable to practice the target change operation with the final setup.

Commissioning will begin with a low-intensity proton beam with a low repetition rate to minimize radiation levels. To aid in establishing the proton beam trajectory on the target, a multi-wire chamber will be located between the target and the beam dump to provide additional information. The adjustment available in the primary beam angle and position at the target will be used to verify the beam-target alignment. Section 4.8.2.2.7, in the External Beamline section, describes the adjustments that must be made to the M4 primary beam to provide the beam motion of the primary beam scans at the production target. The beam will be monitored by profile monitors and, initially, with an additional chamber between the target and proton beam absorber. The most direct rate information on the number of target interactions will come from the extinction monitor; however, additional information will be available from rates measured in the Mu2e detector.

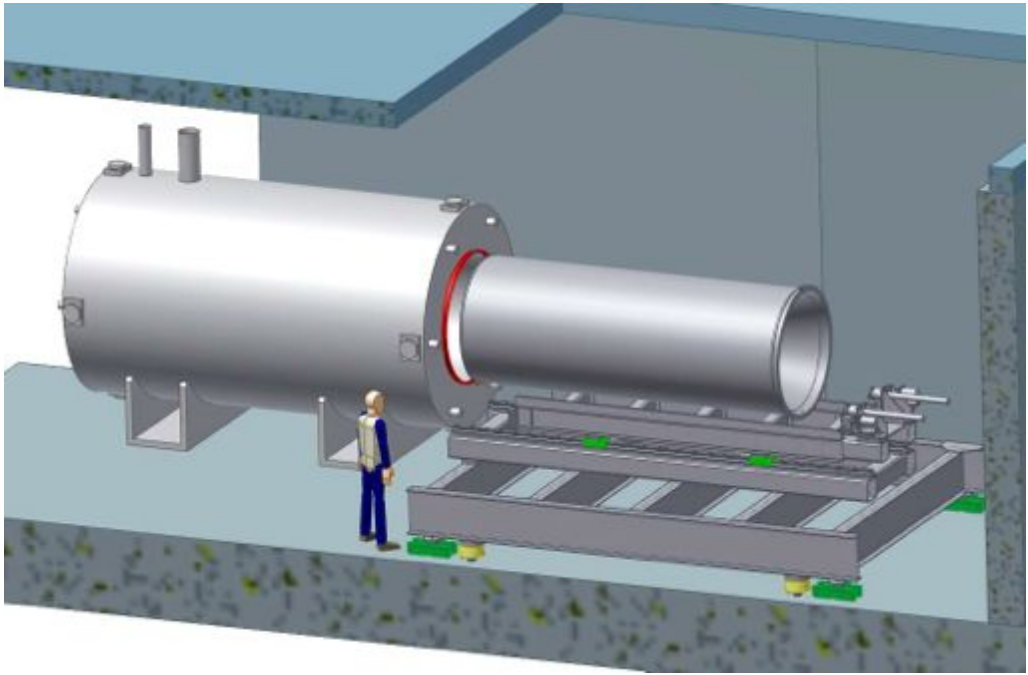

Figure 4.183. View of the installation of the HRS into the PS. The basic concept is that the installation fixture must be aligned and solidly connected to the PS. The solid connection is required because of the large force required to slide the HRS into the PS bore. Once the HRS is inserted it can be welded to the PS. The TS cannot be in place when the HRS is inserted because of the need for access to the upstream end of the HRS for welding.

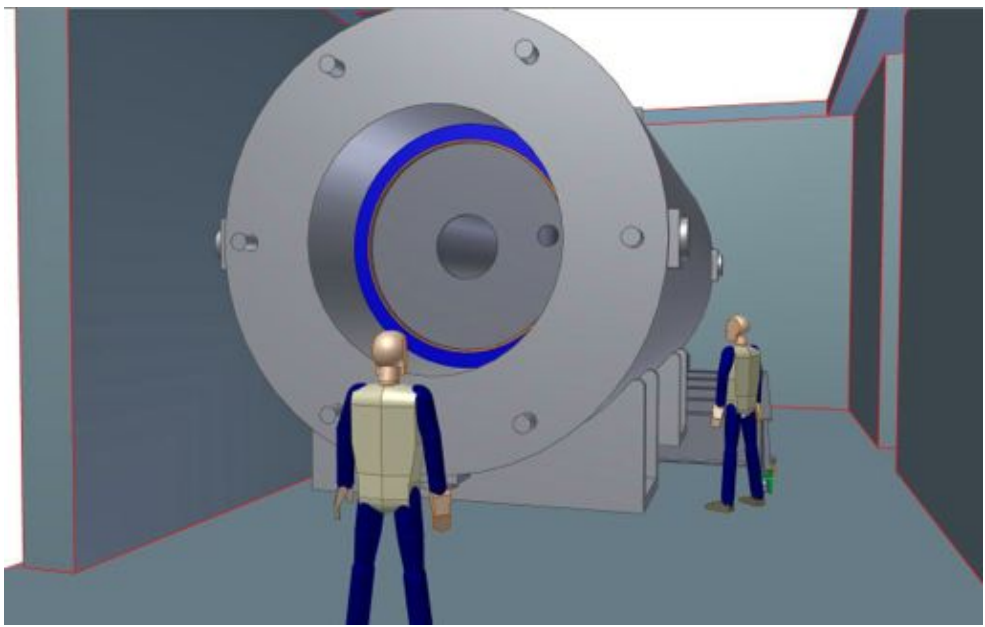

Figure 4.184. Perspective view showing the upstream end of the HRS after insertion into the PS. A welder will have to get inside the upstream end of the PS to weld the orange ring in place.

# 5  Conventional Construction

## 5.1  Introduction

The existing Delivery Ring (DR), part of the reprogrammed Fermilab Antiproton Source, and elements of the Muon Campus Program play an essential role in the beam delivery strategy for the Mu2e Experiment. No enclosure currently exists in the vicinity of the DR that would be suitable to house the Mu2e Experiment; therefore a new Mu2e conventional facility is required.

The Mu2e Conventional Construction subproject includes the management, planning, design, and construction of a new detector hall, modifications to the existing Delivery Ring electrical and HVAC systems and interfaces to the infrastructure provided by the Muon Campus projects. The Conventional Construction scope includes the elements of work normally included in civil construction projects such as earthwork, utilities, structural concrete, structural steel, architectural cladding, finishes, roofing, plumbing, process piping, HVAC, fire protection, lighting and electrical systems.

Design and construction activities have been packaged into two (2) fixed price sub-contract packages, T&M activities and several directly procured items, as listed below.

- <u>Mu2e Detector Hall</u>
  The Mu2e Detector Hall is comprised of a shielded below grade structure that will house the Mu2e experimental apparatus and a grade level service building that will provide space for assembly and support functions. This Detector Hall is the largest component in Conventional Facilities subproject. A view of the grade level service building is shown in Figure 5.1.

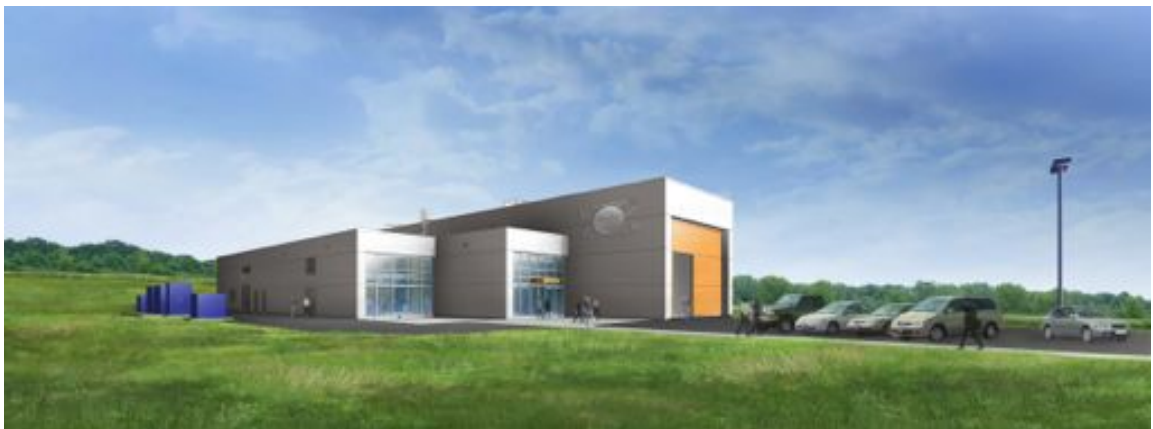

Figure 5.1. Entry View of the Mu2e facility, looking northwest.

- Delivery Ring Upgrades

  To accommodate the increased beam power required for the Mu2e Experiment, the electrical power systems at the AP-30 enclosure must be upgraded. Additionally, there will be a reconfiguration of the Delivery Ring ventilation system to control the air discharge during the time required for activated particles to decay to acceptable discharge levels.

- Time and Material (T&M) Activities

  The Extinction Monitor carrier pipes and part of the primary beam absorber will be installed by T&M after the majority of building settlement has occurred. T&M will be used for other items such as the FIRUS system and door access security systems.

- Direct Procured items

  Several long-lead items will be procured separately and provided to the subcontractor or, in the case of the concrete shield blocks, to one of the other sub-projects for installation. The items include the 30-ton overhead bridge cranes, the precast concrete shield blocks and the 15 kV electrical items such as the transformers, cable, switches, terminations and a stand by generator.

- Engineering, Design and Inspection

  The final design, technical specifications and drawings for the Mu2e Detector Hall has been procured from the Architectural & Engineering (A&E) firm of Middough, Inc., a firm currently under master contract with Fermilab. The Delivery Ring modifications will be designed in-house. The subproject manager for Conventional Construction will also serve as the Construction Manager during the construction phase. The Construction Management Office (CMO) consists of a Construction Manager (CM), Construction Coordinator(s) (CC), and Procurement Administrator (PA). While the CM is ultimately responsible for all coordination and correspondence with the Subcontractor, the CM may delegate certain daily responsibilities to the CC and PA. Line management runs from the CM to the Mu2e Project Manager.

- Muon Campus

  The Muon Campus is a collection of General Plant Projects (GPP) and Accelerator Improvement Projects (AIP) that are required by both Mu2e and g-2. General Plant Projects include the M-4 and M-5 beamline enclosures connecting the Delivery Ring to the MC-1 (g-2) and Mu2e buildings, an addition to the MI-52 kicker building and the Industrial Cooling Water (ICW) extension to the A-0 Compressor Service Building for cryogenic compressor cooling.

## 5.2    Requirements

### 5.2.1   Basis of Requirements

The requirements for the Mu2e Conventional Construction scope of work were developed from stakeholder input, organizational requirements and standards, and various requirements for safety and egress. The main source of stakeholder input was the other Mu2e subproject managers. Regularly scheduled meetings were held with the Mu2e Collaboration, Mu2e Technical Board, Mu2e Accelerator subproject leaders, Mu2e Solenoid subproject leaders, detector groups and the Mu2e Integration Group. The requirements employed in the design and construction of the Mu2e Conventional Construction is documented and controlled [1].

Based on the developed requirements and the Conceptual Design drawings, a two-day value engineering study was held with the Conventional Construction project team, representatives from several of the other Mu2e subprojects and discipline leads from the A&E. There were 62 proposals generated for the speculation list, 5 were found to be obvious cost savings, 22 had significant cost savings potential, 10 items were identified that increased performance, 7 items required additional evaluation by others on the project and 18 items were judged to provide no increase in value. After proposals were evaluated, those selected for implementation were incorporated into the requirements document and issued to the A&E [2].

The A&E firm provided documents for review at the 30, 60, 90 and 100 percent completion [3] levels. The 90% documents were issued for a Lab Wide Comment and Compliance Review. Space allocations for the technical components of the Mu2e apparatus were validated using 3D modeling tools. The A&E provided a cost estimate with each submittal, allowing for the cost impact of proposed changes to be understood and evaluated.

## 5.3    Technical Design

### 5.3.1   Overview

The Mu2e Conventional Construction subproject is responsible for the design and construction of the Detector Hall and service spaces required to assemble, house and operate the technical components required for the Mu2e Experiment. The Mu2e Detector Hall is an industrial-type structure, similar to others on the Fermilab site, with a built up roof and metal siding on a braced structural steel frame, with a cast-in-place reinforced concrete structure under the majority of the building. Contract documents fully defining the scope and details of the project can be found in reference [3].

The scope of work for the Delivery Ring includes the addition of exhaust fans to the existing 1500 kva transformer circuit, routing of new power lines into the AP-30 enclosure and replacement of the existing 2000 Amp switchboard with a new 2500 Amp switchboard.

## 5.3.2 Site Work

The placement of the detector building on the Fermilab site is determined by the lattice of the beamline that transports protons to the Mu2e production target, located inside the Mu2e apparatus. The well-defined size and shape of the Mu2e apparatus drives much of the building design. The Mu2e facility, the MC-1 building housing the g-2 experiment, the external beamline enclosure and the Delivery Ring form the nucleus of the Muon Campus, shown in Figure 5.2.

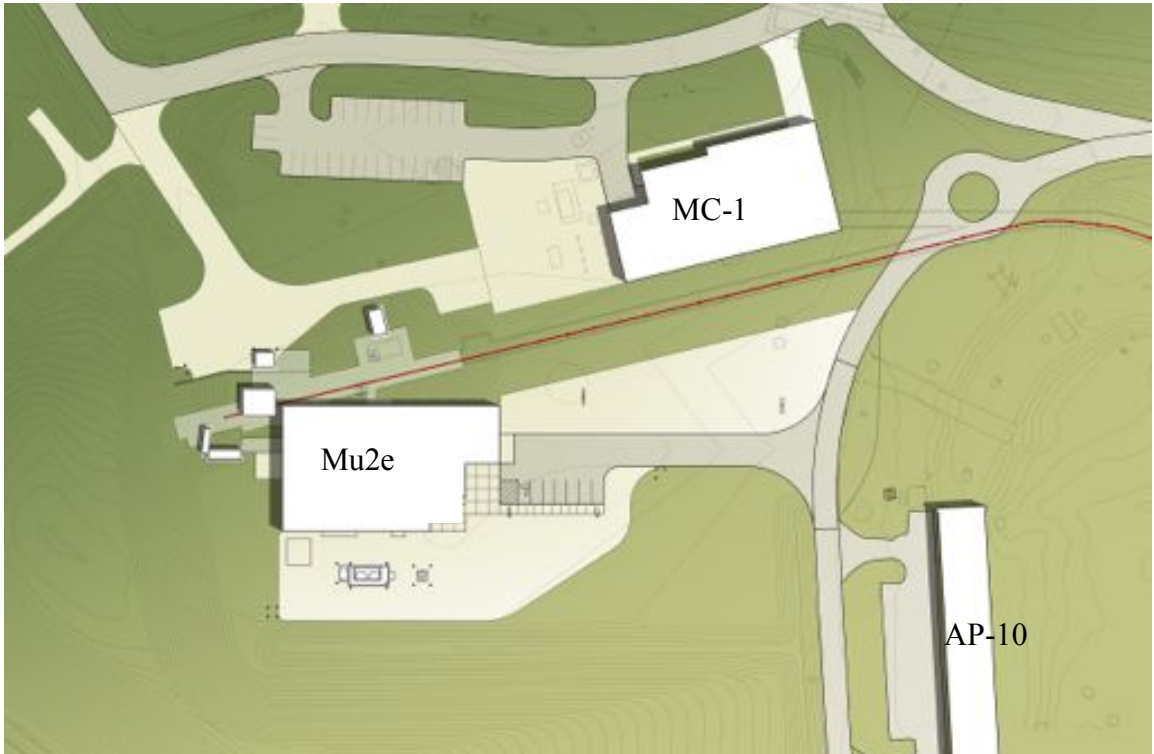

Figure 5.2. Muon Campus site plan, including the Mu2e building and the MC-1 building for the g-2 experiment. Part of the Delivery Ring is shown to the right.

The Mu2e building site is bounded on one side by a spoils pile, located west of the existing Kautz Road. Portions of Kautz Road will be eliminated between the South Booster Road and Giese Road and substantial portions of the existing spoils pile will be reshaped to allow for construction of the facility. Surface drainage ditches, swales and other structures have been designed to maintain the existing storm water flows through the site. Industrial Cooling Water (ICW), Domestic Water (DWS) and Natural Gas (G) are routed around the excavation to maintain service to the NuMi and Main Injector facilities with minimal interruptions. After completion of the below grade structure, underground installations of the various utilities, including Chilled Water (CW), Sanitary Sewer (SS), Domestic Water, ICW, Natural Gas, 13.8 kV electrical power and communications duct banks will be completed and backfilled. Final grading and seeding will stabilize any areas disturbed during construction.

The site design provides paved access to the building for personnel and equipment and crushed stone hardstand access to exterior equipment hatches and exterior equipment. Crushed stone hardstands are provided for both exterior staging and access for emergency vehicles around the outside of the structure.

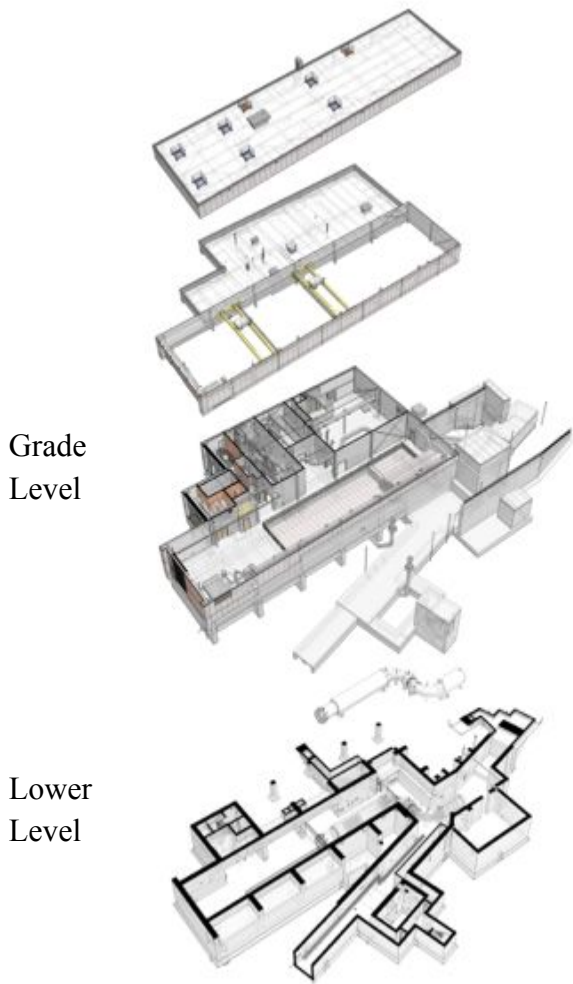

Grade
Level

Lower
Level

Figure 5.3. Exploded view of the Mu2e detector hall.

### 5.3.3  Architectural / Structural Design

The details of the Detector Hall configuration are derived from the requirements for the experiment's technical components. The facility is functionally divided between the shielded below-grade enclosure that houses the experimental apparatus and the grade-level service building that primarily provides support for the technical equipment below. An exploded view of the building, showing the two levels, appears in Figure 5.3. The primary beamline enclosure, provided as part of the Muon Campus scope, interfaces to the detector hall upstream of the Production Solenoid, as shown in Figure 5.3 and Figure 5.5. Sixteen feet of earth equivalent shielding is required between the primary beam and any areas that may be accessed by personnel when the beam is operating. This requires an above grade earth berm over the beamline and shielding between the beamline and the detector hall. The proton beam intersects the production target inside the Mu2e apparatus but a significant fraction of the beam is not absorbed by the target and continues on to a downstream absorber. Located above the absorber is a space to house the Extinction Monitor, part of the Mu2e detector that resides downstream of the production target and monitors the time structure of scattered protons.

The at-grade Detector Service Building will provide space for the various support services required for the Mu2e Experiment. The Mu2e Detector Service Building will be a braced frame, steel construction with prefinished metal siding and a built up roof system. The construction type and style will match that of other facilities at Fermilab.

Figure 5.4 provides a plan view of the Mu2e Detector Service Building that includes a high bay and an adjacent low bay. The high bay will be equipped with two 30-ton capacity overhead bridge cranes, similar in style and construction to cranes installed with the Main Injector project.

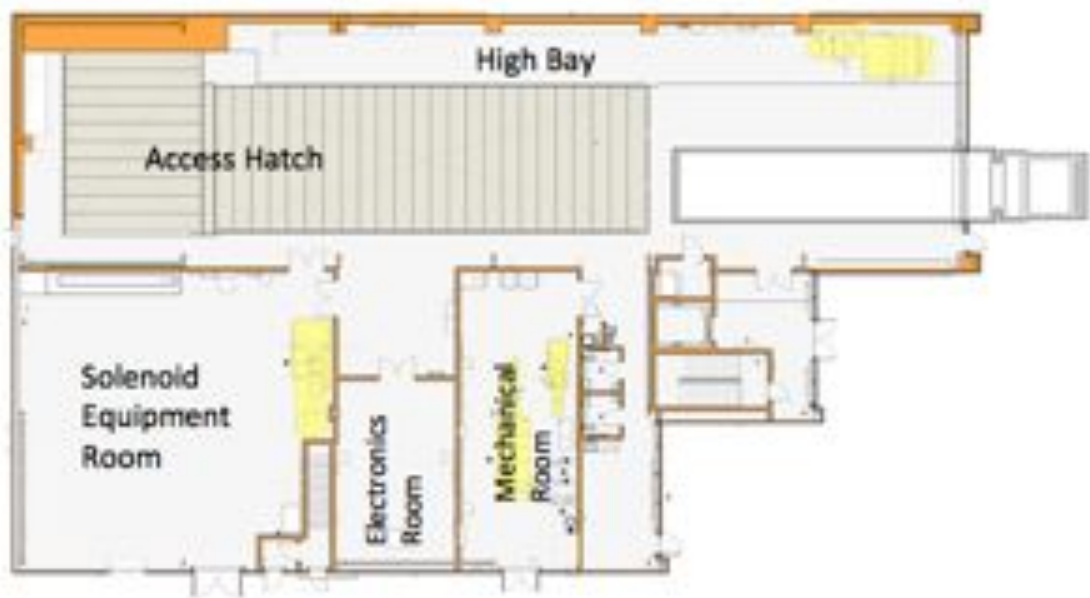

Figure 5.4 Plan view of the Mu2e above-grade Detector Service Building containing a high bay and an adjacent low bay. The low bay includes a room for solenoid power supplies and cryo distribution, an electronics room and a room for building mechanical services.

A portion of the high bay, shown in the upper right hand corner of Figure 5.4, will be used as a loading dock as well as a space for housing beamline power supplies. During beam operations the access hatch will be covered with 6 feet of modular, precast concrete shielding blocks to limit radiation exposure to acceptable levels. During installation the hatch will be open. Temporary railings will provide fall protection. The high bay contains space for temporary staging of the shield blocks should removal be required for maintenance and/or repair of the detector components.

The low bay area will contain the spaces required to support the assembly, installation and operation of the Mu2e detector including a solenoid equipment room, a mechanical room for mechanical and electrical equipment, an electronics room, toilet rooms, janitor's closet and exit stairs from the below grade enclosure. An elevator is provided for movement of personnel and equipment between levels.

The solenoid equipment room size is based on criteria provided by the Solenoid subproject. It will house the electrical equipment to power the solenoids as well as parts of the cryo system.

The Electronics Room will house the electronic racks for the solenoid controls and data acquisition system including the online processing farm, data controllers and networking equipment. This equipment, primarily computers, requires strict environmental controls for proper operation. This space will be conditioned with dedicated equipment to provide suitable environmental control. Clean, conditioned power will also be provided to this space. This space is intended to house equipment only and will not be occupied during normal operations. During normal operations the Mu2e Experiment will be run from a remote control room located within Wilson Hall.

The Mechanical Area will house the equipment that supports the assembly and operation of the detector and operation of the facility. This includes the cooling equipment for the detectors, ICW strainer, fire protection riser, heat exchangers, pumps and the huge number of meters and monitoring equipment required as part of the "smart lab" initiative. The Mechanical Area will also contain the electrical switchgear, transfer switch and panel boards to distribute the incoming electrical power. The electrical switchgear will serve conventional facilities equipment, the experimental apparatus and the 480 V HVAC systems. Electrical panels serving the lights, outlets and general house power will be included in the electrical power distribution system.

The below-grade Detector Enclosure will consist of a cast-in-place concrete structure located below the Mu2e Detector Service Building. The Detector Enclosure, shown in Figure 5.5, consists of the spaces required to house, support, assemble and install portions of the proton beamline, the Mu2e Solenoids and detector components as well as the detector support equipment.

The Detector Enclosure is functionally divided into a Production Solenoid Vault, a Transport Solenoid Hall and a Detector Solenoid Hall. The proton beam enters the Production Solenoid Vault and strikes the Production Target inside the Production Solenoid. The area around the Production Solenoid is filled with concrete shielding. The Production Solenoid vault has the highest radiation levels in the Mu2e facility.

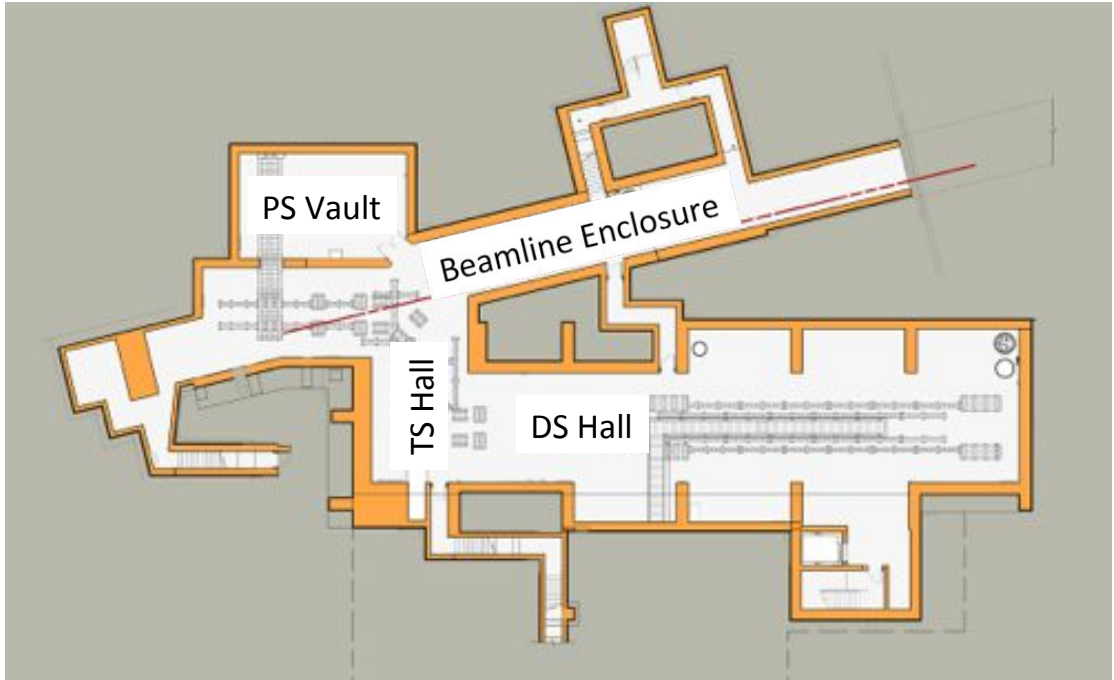

Figure 5.5 Plan view of the below grade Detector Enclosure.

A shielded hatch that opens to the outdoors, just West of the Detector Service Building, allows access to the underground enclosure in the vicinity of the Production Solenoid. The Production Solenoid, Heat and Radiation Shield and Proton Beam Absorber components could be lowered into the enclosure through this hatch if necessary. Steel track plates embedded in the concrete slab of the detector enclosure will facilitate movement of these heavy loads and anchor components to the slab to resist the large magnetic forces generated by the solenoids. Adjacent to the Production Solenoid is a remote handling area that will house equipment necessary to remotely remove and replace the radioactive production target. Radiation levels in this room during beam operations exceed tolerable levels for the remote handling equipment, so a small shielding hatch extending to grade level is provided to lower remote handling equipment into place only when needed.

The Transport Solenoid Hall contains the "S" shaped Transport Solenoid. The downstream half of the Transport Solenoid is surrounded on 3 sides by the Cosmic Ray Veto counters and it's associated shielding. Figure 5.6 shows the location of the Transport Solenoid in the detector hall.

The Detector Solenoid Hall contains the Detector Solenoid that is surrounded on 3 sides and one end by the Cosmic Ray Veto and it's associated shielding. Downstream of the Detector Solenoid is a detector staging area necessary initially for detector staging and installation and later for detector servicing. Figure 5.7 shows a longitudinal section through the Detector Solenoid Hall.

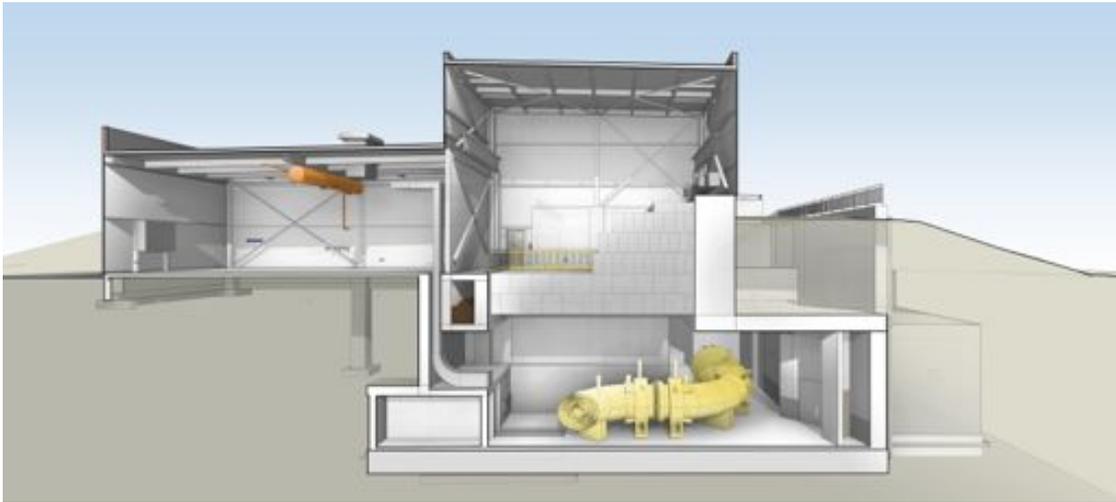

Figure 5.6 A transverse section of the detector hall showing the layout of the Transport Solenoid.

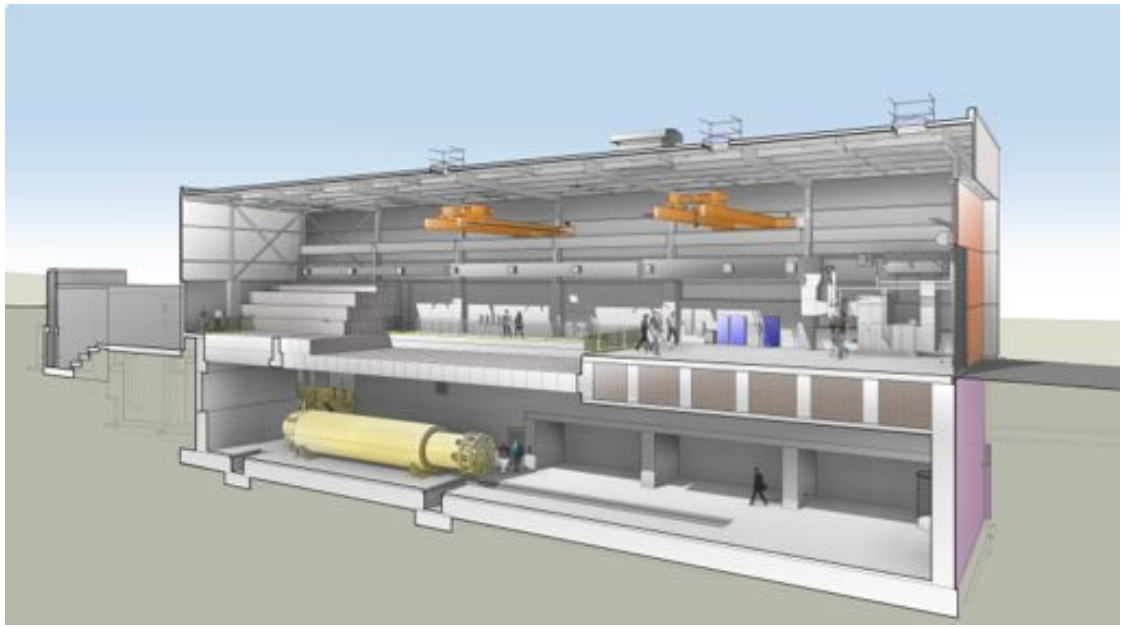

Figure 5.7 Longitudinal section through the Detector Solenoid Hall.

The Transport Solenoid and the Detector Solenoid are sited under large access hatches that will normally be filled with precast concrete shielding. The two 30-ton bridge cranes in the high bay will be used for installation of material and equipment in these areas, including the Detector and Transport Solenoids, neutron shielding and the various detector elements.

The below grade structure is constructed of cast-in-place reinforced concrete. The concrete structure around the beamline enclosure and Production Solenoid vault form a box type structure. In the area under the high bay the walls cantilever from the base mat and counterforts are added to resist the earth loading on the walls. The force on the steel rebar in the concrete walls from the Solenoid magnetic field has been calculated and determined to add only a few pounds per square foot to the concrete structure. Track plates are embedded into the base mat to facilitate the installation of the solenoids and to anchor them in place. The support plates are designed to resist the massive magnetic forces between the solenoids.

In addition to resisting earth loads, the counterforts support the high bay building columns and a support beam that carries the shield blocks. A ledge is designed into the top slab to carry six feet of concrete shielding. During maintenance periods it is expected that large portions of the shield block area will be removed to allow access to the equipment below. The ledge is designed to support a total of 15 feet of shield blocks, allowing blocks to be stacked inside the service building rather than being rigged outside.

All exposed interior concrete surfaces will be painted to reduce dust and increase lighting efficiency. The exterior concrete surfaces will be moisture-proofed. In the areas around the Production Solenoid, Proton Beam Absorber and Extinction Monitor where activation levels are expected to be high a waterproofing system will be employed to insure a dry interior and minimize the development of nitric acid.

The size of each space in the Mu2e facility is listed in Table 5.1.

### 5.3.4   Mechanical Systems

The conventional facility mechanical systems are comprised of:

- HVAC (heating, ventilating and air conditioning) equipment for each space to maintain acceptable indoor environmental conditions and to meet the stakeholder requirements.
- Building plumbing (Domestic Water, Natural Gas, Sump system, Fire Sprinkler, Drainage and Sewer)
- Process Utilities for system heat rejection (Chilled Water).

Detailed information describing indoor conditions, airflows, and temperatures are listed in the Basis of Design Documents [4], the civil drawing set [3], as well as the Conventional Facilities requirements document [1].

Each of the indoor HVAC units utilizes individual single-zone variable air volume (VAV) systems. Each system will incorporate an air-handling unit with a chilled water-cooling coil and a natural gas heating furnace. An airside economizer mode will be employed with each HVAC system to allow free cooling without the need for chilled water when outside air temperatures are sufficiently low. The HVAC systems for the upper level spaces will provide slight pressurization relative to the lower level spaces. The pressure differential is intended to restrict the migration of activated air from the lower level detector and beamline enclosures during beam operations.

Table 5.1 Area of the various spaces in the Mu2e facility.

|  |  | Square Feet | Square Meters |
|---|---|---|---|
| **Grade Level** | **Total** | **12600** | **1171** |
|  | Entry | 281 | 26 |
|  | Planning Room and Toilets | 485 | 45 |
|  | Mechanical Room | 975 | 91 |
|  | Machine Shop | 360 | 34 |
|  | Electronics Room | 630 | 59 |
|  | Solenoid Power Supply Room | 2400 | 223 |
|  | High Bay | 6390 | 594 |
| **Detector Enclosure** | **Total** | **9640** | **896** |
|  | DS Hall | 4485 | 417 |
|  | TS Hall | 890 | 83 |
|  | PS Hall | 1010 | 94 |
|  | Proton Absorber | 225 | 21 |
|  | Extinction Monitor Hall | 540 | 50 |
|  | Remote Handling Room | 980 | 91 |
|  | Beamline | 1200 | 112 |

The HVAC for the Electronics Room will utilize a constant-volume computer room air conditioning (CRAC) unit with chilled water cooling, electric heat / reheat, and an infrared humidifier.

An outdoor HVAC system will provide dehumidified 100% outside air to the Production Solenoid Vault through the absorber steel core. This air will eventually travel to the Beamline Enclosure and will be exhausted at a location between the MC-1 building and the Delivery Ring. The supply airflow is limited to about 900 cfm, matching the transit time of radionuclides prior to release to the atmosphere to their decay time. The required static pressure at the end of the supply duct is 2 inches of water column. To maintain the appropriate humidity levels, the spaces around the Production Solenoid Vault must be adequately sealed to minimize outdoor air infiltration, especially thru the hatch in the remote handling area. The walls and door between the Production Solenoid vault and the remote handling room, as well as the boundary between the Production Solenoid and Transport Solenoid must be sealed properly.

The mechanical room is provided with ventilation, utilizing a sidewall ventilation fan and a relief opening to the outdoors, and is heated using a natural gas fired heater. Miscellaneous passageways and stairs associated with the lower level will not be air-conditioned but will include localized electric heat. Chilled water valve taps will be provided in the mechanical room for detector cooling and dehumidification systems. Insulated chilled water piping will be routed to the remote handling room for vacuum pump cooling.

Ventilation equipment for the Oxygen Deficiency Hazard (ODH) safety systems shall be provided for the below-grade Detector Enclosure and the at-grade Solenoid Equipment Room. The control system will be designed and installed by the Fermilab Accelerator Division. The Solenoid Equipment Room will utilize a wall mounted exhaust fan to pull unconditioned ventilation air in through a wall louver in the case of an ODH alarm. The ODH ventilation systems serving the below-grade Detector Enclosure will include supply air systems, including indirect fired gas furnaces to temper the supply air during cold weather. No air conditioning will be included. The exhaust fans for all of these spaces will be direct drive and include non-motorized, gravity back draft dampers. The supply system's louvers will include motorized automatic dampers. The system will engage upon receipt of a 24 V DC signal from an ODH sensor.

A Building Automation System (BAS) will control the building HVAC units. Building Automation Systems are centralized, interlinked, networks of hardware and software that monitor and control the building environment. While managing various building systems, the automation system ensures the operational performance of the facility as well as the comfort and safety of building occupants. The system can be scheduled, controlled and monitored by the FESS Operations Group. Chilled water, domestic water and natural gas usage will be monitored through the BAS to comply with the DOE guiding principles for Leadership in High Performance and Sustainable Buildings.

The HVAC system serving the Solenoid Equipment Room, High Bay and Detector Solenoid Vault are designed for the following modes of operation:

- Experimental Mode: System will run whenever the experiment is running. Priority is pressurization then space conditioning.

- Installation Mode (Occupied): System will run based on schedule. Pressurization is not a priority. Unit to maintain occupied setpoints and can be shut down if setpoints are met. This is the assumed mode of operation during the four or five years of installation, immediately following beneficial occupancy.
- Installation Mode (Unoccupied): System will run based on schedule. Pressurization is not a priority. Unit to maintain unoccupied set points and can be shut down if setpoints are met.

A summary of the mechanical systems in the Mu2e facility is presented in Table 5.2.

### 5.3.5   Plumbing

The building will be outfitted with two toilet rooms, (each with a water closet, lavatory and floor drain) and one janitor's closet with mop basin and floor drain. An electric water cooler will be located just outside these rooms. Floor drains will also be provided in the upper mechanical room and at every HVAC location, except for the HVAC unit located in the high bay.  The plumbing fixtures at grade level will discharge to sanitary.  Floor drains will be positioned around the accessible areas of the lower level for incidental spills.  The water from the lower level floor drains will be directed to an isolated basin that can be tested and inspected.  If no contaminants are found then the water will be pumped to the underdrain sump basin for discharge, if contaminates are found, the water will be pumped into barrels and directed to the proper waste stream.  To collect the ground water perforated underdrain piping will be installed around the perimeter of the lower walls.  The underdrains will connect to two separated sump basins, one in the primary beam enclosure and on in the northeast corner of the DS Hall.  Each sump systems will be outfitted with duplex pumps, connected to the backup generator and alarms monitored by FIRUS.

### 5.3.6   Electrical

Two separate electrical systems will power the Mu2e facility. One will power the large loads for the experimental apparatus and the other will power the building and smaller experimental loads. This is shown in Figure 5.8. FESS Electrical Design Criteria will be used as a basis for the facility design.  Primary 13.8 kV electrical power from Feeder 24 will be extended from MC-1 and AP-10 facilities to the Mu2e facility. This will maintain a loop feed to allow for electrical system maintenance with minimal disruption throughout the Muon Campus. The power extension will include the construction of concrete encased duct banks, installation of 750 kcmil/15 kV primary conductor and 15 kV air switches.  The 15 kV four bay air switch located at the Mu2e building provides the connection between Feeder 24 and the disconnect switches for the building power transformers. An additional 15 kV disconnect switch is provided for the power transformer that supplies power to the solenoid and beamline power supplies.  This switch can be used as a lockout device for the technical equipment allowing for safe access during maintenance.

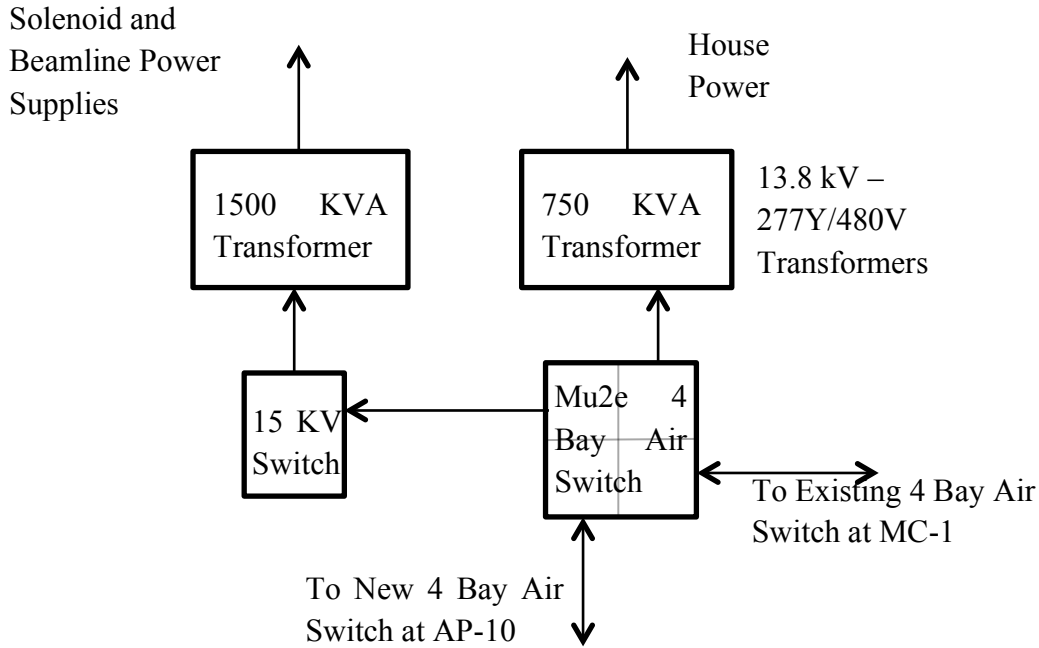

Figure 5.8 Primary Power distribution scheme for the Mu2e facility.

The building power transformers are located away from the building on a concrete pad with oil containment. Two 13.8 kV–277Y/480 Volt power transformers are to be installed, one 1500 KVA and one 750 KVA. The 1500 KVA power transformer will power the solenoids power supplies, beamline power supplies and other experimental apparatus. It has been sized based on stakeholder requirements with an adequate safety factor and will be recycled from the decommissioned TeVatron program. The 750 KVA power transformer will be purchased new and is sized based on the estimated loads of all other building equipment.

A 175 KW standby diesel generator will be installed on a concrete pad adjacent to the transformers. The generator fuel tank will have a double wall construction with leak detection. An automatic transfer switch is provided to switch loads to the generator upon normal power failure. The loads connected to the standby system include ODH protection equipment, elevator, sump pumps, cranes, and uninterruptable power supply (UPS) for emergency lighting and exiting signs.

Table 5.2 Summary of mechanical systems in the various spaces of the Mu2e facility.

| Space | Mechanical | Process Utilities | ODH | Dependencies |
|---|---|---|---|---|
| High Bay | HVAC | Chilled water for RAW skid | | Sealing of blocks in the center space |
| Planning Room and entry way | HVAC | | | |
| Upper Mechanical Room | Ventilation & Heating | Chilled water for Tracker Cooling | | Sealing of utility chase |
| Solenoid Equipment Room | HVAC | | ODH ventilation equipment | Sealing of utility chase |
| DS Hall | HVAC | | ODH ventilation equipment | Sealing of blocks in the center space |
| Lower Mechanical Room | HVAC | Chilled Water for dehumidifier | | Sealing of utility chase |
| Electronics Room | HVAC & humidification | | | |
| PS Vault | Dry Outdoor Air Supply thru Absorber | | ODH ventilation equipment | Supply duct interface to absorber. Wall separation between TS & PS / Wall separation between PS & remote handling |
| Beamline Tunnel | Ventilation Exhaust from MC tunnel | | | |
| Extinction Monitor | 100cfm dry OA | | | Duct continuation |
| Remote Handling | Heating only | Chilled Water for Vacuum Pumps | | Sealing of door between PS and ext. monitor |
| Stairways | Heating only | | | |

The power for the experimental apparatus is distributed by a 277Y/480V – 2,000 amp switchboard located in the Solenoid Equipment Room. This switchboard directly feeds the power supplies in that room as well as two additional panel boards for beamline power supplies in the high bay. The switchboard will be reclaimed from the decommissioned TeVatron and refurbished for use by Mu2e.

The power for the remaining building loads is provided by a 277Y/480V - 1,200 amp panel board. From this panel board, power is distributed to sub-panel boards and dry type transformers that serve loads throughout the facility. Convenience receptacles will be provided throughout the building. In areas with technical equipment, quad 120 V and dual 3 phase 208 V receptacles are provided. Welding receptacles providing 60 amps at 480 V are positioned throughout the building.

A central lighting control panel is provided to manage the building's lighting system. All switching of the lights will be done through this panel via local low voltage switches. The lighting system in the building's upper level is primarily fluorescent light fixtures. Emergency lighting on the grade level is provided by fixtures with individual 90 minute battery packs. Each emergency ballast is self-testing to comply with NFPA standards. These fixtures will connect to the generator and also configured as night lighting. In order to tolerate radiation exposure, the lower level lighting system is primarily incandescent light fixtures. Emergency lighting in the lower level is connected to an Uninterruptable Power Supply (UPS). The UPS will provide power to the designated emergency lights for a minimum of 90 minutes. The UPS is also connected to the standby generator for additional back up time. Table 5.3 lists the lighting levels provided in the various areas of the building.

Data and telecom services will also be routed to the facility. Receptacles for the data/telecom services will be located throughout the building. Each data/telecom receptacle will have two 1" conduits that extend above the ceiling for wiring. Alarm system conduits are installed for wiring from the various alarm stations to a junction box located above the Security Control Panel and FIRUS alarm panel.

Cable trays will be provided along the beamline. The trays will be 18" x 4" and mounted to the ceiling embedded channel inserts.

### 5.3.7  Life Safety

As with all Fermilab projects, environmental, safety and health will be integrated into all aspects of the Conventional Construction subproject. The primary set of building construction codes used to review the design of Mu2e project were the International Building Code (IBC) – 2009 Edition, including all referenced Codes within the IBC, and the 2009 version of National Fire Protection Association "Life Safety Code" (NFPA 101). All other applicable NFPA documents will be used to evaluate specific design features of Mu2e for compliance. The Mu2e facility will be classified by IBC as a Business Occupancy and by NFPA 101 as General Industrial Occupancy. The 2009 International Building Code (IBC) classifies this facility as a Use Group F-2 (Low Hazard Factory Industrial occupancies.  NFPA 101, Life Safety Code, 2009 Edition classifies this facility as an Industrial Occupancy – Special.   The single story building height of approximately 35 feet and area of approximately 12,500 sq. ft. are below the limits set-forth by IBC for Business Occupancy.  The above ground building will be constructed of Type IIB materials and the underground portion will be constructed of Type I materials.

Table 5.3 Lighting levels in the various spaces in the Mu2e facility.

| Area Description | Service luminance (foot-candles) |
|---|---|
| Receiving/Staging, high bay | 75 |
| Electronic/Data acquisition room | 50 |
| Solenoid and Power supply room | 50 |
| Mechanical and Electrical room | 30 |
| Temporary machine shop | 75 |
| Planning room | 50 |
| Detector solenoid area | 40 |
| Transport solenoid area | 30 |
| Production Solenoid area/Primary Absorber | 40 |
| M-5 beamline Enclosure area/Tunnel/Exit passage | 20 |
| Remote handling area | 20 |
| Elevator equipment room | 30 |
| Elevator lobby | 20 |
| Corridor / Entry / Vestibule | 20 |
| Stairs | 10 |
| Toilet room | 20 |
| Means of egress during emergency | 1 |

***Mu2e Fire Protection***

The maximum distance to an exit will not exceed 300 feet, in accordance with IBC and NFPA 101 regulations. Dead corridors will not exceed 50 feet. All corridors and aisles will be a minimum of 36 inches wide. The facility does not require a smoke control system, based on the depth of the enclosure and the time required to exit the building. Means of egress illumination and exit signage will be provided with at least 90 minutes of operation on emergency power. The elevator car lighting will also be operational for 90 minutes on emergency power.

All wall and ceiling finishes, along with movable partitions, will be Class A with a flame spread rating not to exceed 25 and a smoke development index not to exceed 50.

***Fire Protection Assessment***

The purpose of the Fire Protection Assessment (FPA) is to comprehensively and qualitatively assess the risk from fire within the Mu2e facility and to ensure that all DOE and Fermilab fire safety regulations are met. The FPA documents the overall design features of the facility and evaluates them against the applicable codes, standards, orders and directives. DOE fire protection criteria are outlined in DOE Order 420.1B and DOE Standard 1066. The FPA includes identification of the risks from fire and related hazards (direct flame, hot gases, smoke migration, fire-fighting water damage, etc.) in relation to the designed fire safety features to assure that the facility can be safely controlled and stabilized during and after a fire. The FPA has be updated as the design of the Mu2e facility has evolved.

The Mu2e site will be provided with two fire hydrants, spaced in accordance with NFPA 1 Fire Code regulations. The transformers will be located 25-feet away from the building and any combustable materials. There is dedicated access for fire department equipment and direct access to the fire sprinkler connection. Fire detection will be distributed throughout the facility. The above-grade building will be protected with a wet-type automatic sprinkler system. Manual pull stations will be provided at all exits. Portable fire extinguishers will be located throughout the facility in accordance with NFPA 10. The underground Detector Enclosure will be protected throughout with a pre-action (dry-type) automatic sprinkler system and monitored with an air sampling type smoke detection system. The pre-action sprinkler system requires dual action for water release. First, the detection system must identify a developing fire and open the pre-action valve. This allows water to flow into the system piping, which effectively creates a wet pipe sprinkler system. Second, individual sprinkler heads must release to permit water flow onto the fire. In addition, 2.5-inch fire hose connection valves will be available throughout the facility for use by the Fire Department. The building fire alarm system will be monitored by Fermilab's site-wide Facility Incident Reporting Utility System.

*Hazard Analysis Approach*

A principal component of an effective ES&H program is to ensure that all hazards have been properly identified and controlled through design and procedure. Conventional construction hazards pose the potential for serious injury, death, and damage to equipment and property. Due to the preponderance of operational controls rather than design controls, the post mitigation risk will be relatively high even though the probability of occurrence will be significantly reduced. Fermilab has a mature construction safety program with many recent experiences that provide input for future projects.  Lessons learned from these experiences combined with experience from other construction projects in the DOE complex will help manage the risk at the Mu2e facility. Typical construction hazards anticipated at the Mu2e facility site include:

- Site Clearing
- Excavation
- Work at elevations (steel, roofing)
- Utility interfaces, (electrical, ICW, DWS, Storm Sewer, Sewer, chilled water)
- Material Handling
- Finishing work
- Weather related conditions
- Transition to Operations.

One of the keys to controlling construction risks is to ensure that safety requirements effectively flow down to subcontractors and that sufficient supervision of subcontractors is in place.  The subcontractors performing construction work at the Mu2e facility will be pre-qualified and each will develop a hazard analysis and a safety plan for the work to be done. These plans will comply with Fermilab and applicable OSHA requirements. Fermilab will appoint Construction Coordinator(s), who will oversee the subcontractor's ESH&Q programs in accordance with FESHM 7010. They will fulfill an auditory function to ensure that all work is carried out in accordance with the subcontractor's ES&H and Quality Assurance plans.  Per FESHM Chapter 7010, the ESH&Q Section will appoint a construction safety expert, who is available to provide ESH&Q support for the Construction Coordinator. This individual will provide ESH&Q oversight of construction activities as well. The subcontractor will be required to have someone competent in appropriate safety procedures, with appropriate authority on site at all times while work is ongoing. The qualifications of this competent person will be commensurate with the hazards of the work activity in progress.  All subcontractors will perform daily work planning. All Construction Progress Meetings will have construction safety as a standing agenda item. Subcontractors will perform and document toolbox safety meetings. Daily safety inspections will be performed by the subcontractor and Fermilab

Construction Coordinators.  Safety performance will be assessed regularly and corrective action or incentive reduction assessed.

Any Fermilab employees or users seeking access to the construction site must have the appropriate safety training. The subcontractor will control access to the work areas and will set the minimum requirements for entry. The subcontractor will establish a training program to meet these requirements.

The Mu2e design team has considered relevant natural phenomena hazards. Those hazards of interest to this project include seismic events, flooding, high winds and snow loading.  All natural phenomena are normally considered to be low probability hazards.

### *Environmental*

During the construction phase of the Mu2e facility the disposition of spoils from excavation, dust, noise, impacts on ground and surface water, chemical use, frequent transport of components, spills and disposal of waste are issues that are addressed in the contract documents.

A full National Environmental Policy Act (NEPA) review of the Mu2e facility was performed in 2012 to ensure that there are no significant impacts to the site, the surrounding waterways or wildlife. A Categorical exclusion was granted and signed into effect on June 12, 2012. Fermilab has developed a Storm Water Pollution Prevention Plan (SWPPP) and submitted a Notice of Intent (NOI) to the IEPA to obtain a project specific permit issued under the Laboratory's General National Pollutant Discharge Elimination System (NPDES) Permit No. ILR10, issued from the IEPA to Fermilab for construction site activities. All sub-tier sub-contractors are required to sign and adhere to the provisions of the SWPPP. Domestic water and wastewater permits to construct have been submitted for approval.

### *Sustainability*

The project scope incorporates sustainable design principals into all phases of planning, design and construction. Sustainability is broadly defined as the design and implementation of projects to simultaneously minimize their adverse environmental impacts, maximize occupants' health and well being, and improve bottom line, life cycle and economic performance. Projects that exceed the $5M threshold that are not appropriate for Leadership in Energy & Environmental Design (LEED) certification, such as experimental and/or industrial buildings, must apply the principles of sustainability wherever appropriate and cost-effective and must describe the applied measures in project documents. The Mu2e Project utilizes the guiding principles and American Society of Heating, Refrigerating and Air Conditioning Engineers (ASHRAE) recommendations to meet sustainability goals.

## 5.4    Risks

Most of the risks associated with the Mu2e Conventional Construction sub-project are those typically seen with construction projects of this size and complexity, including the risk of cost overruns, unforeseen subsurface conditions, stakeholder requirement changes and delays in the Critical Decision process. Based on the ranking criteria defined in the Mu2e Project's Risk Management Plan [5], none of the identified risks have an overall ranking of "high".  A complete listing of the risks for Conventional Construction can be found in reference [6]. The top four ranked risks/opportunities at CD-1 included significant injury or death associated with conventional construction, cost overrun or underrun (opportunity) due to market conditions and inadequate capacity in the 15 kV electrical feeder systems.  The risk associated with the overall cost will be retired when bids for the civil construction contract are received in the Fall of 2014. The risk of exceeding the capacity of the 15 kV feeder has been transferred and accepted by the Muon Campus.

## 5.5    Quality Assurance

The construction subcontractor is responsible for all activities necessary to manage, plan and safely execute the activities associated with construction of the Mu2e facility, formally described in the scope-of-work section of contract documents. The Subcontractor's responsibilities include Quality Control of all work performed by the Subcontractor, suppliers, sub-tier contractors and consultants. This includes safety, submittal management, testing and inspection, and all other functions necessary to provide for quality construction that satisfies the Mu2e requirements and specifications.

Quality Assurance principles [7] will be applied to all aspects of Conventional Construction. Periodic Quality Assurance reviews of the entire project will be conducted from inception through closeout. Design and construction documents will be reviewed for appropriateness of the proposed systems, impact on existing systems, impact on operations, specific technical requirements and code compliance and compliance with best practices. This review process will be completed in accordance with the applicable portions of the Fermilab Director's Policy Manual, Section 10 [8]. The following elements will be included in the design and construction effort:

- Clear definition of responsibility levels as well as delineated lines of communication for exchange of information.
- Configuration management of design criteria and criteria changes and a record of all standards and codes used in the development of the criteria.
- Periodic review of the design process, including drawings and specifications, to ensure compliance with accepted design criteria.

- Identification of underground utilities and facility interface points prior to initiation of construction activities in affected areas.
- Conformance to procedures regarding updates to the project plan and compliance with the approved construction schedule;
- Conformance to procedures regarding the review and approval of shop drawings, sampling results and other required submittals.
- Conformance to procedures for site inspection by Fermilab personnel to record construction progress and adherence to the approved contract documents.
- Verification of project completion, satisfactory system start-up and final project acceptance.

Comments that result from a Comment and Compliance Review will be entered into an electronic database. This will clearly document the names, organizations and dates associated with comments and findings for specific projects and allow for formal tracking of comments.  All comments entered into the electronic database will elicit a response.

# 6   Solenoids

## 6.1   Introduction

The solenoids perform several critical functions for the Mu2e experiment.  Magnetic fields generated from these magnets are used to efficiently collect and transport muons from the production target to the muon stopping target while minimizing the transmission of other particles.  Electrons are transported from the stopping target to detector elements where a uniform and precisely measured magnetic field is used to measure the momentum of electrons.  The magnetic field values range from a peak of 4.6 T at the upstream end to 1 T at the downstream end.  In between is a complex field configuration consisting of graded fields, toroids and a uniform field region, each designed to satisfy a very specific set of criteria.

Mu2e creates this complex field configuration through the use of three magnetically coupled solenoid systems: the Production Solenoid (PS), the Transport Solenoid (TS) and the Detector Solenoid (DS). Requirements for the magnets, as well as a design concept that meets these requirements are described below. The Mu2e Solenoid system also includes all ancillary systems such as magnet power converters, magnetic field mapping, cryogenic distribution and quench protection instrumentation and electronics. The solenoid system is shown in Figure 6.1.

## 6.2   Requirements

### 6.2.1   General Requirements

The Mu2e solenoids and their supporting subsystems are designed to meet a complex set of requirements. The requirements are defined so that the deliverables will meet the physics goals of the experiment.  The requirements are explained in detail in several reference documents and summarized below [1] - [8].

Because of the high magnetic field and large amount of stored energy, the solenoids will be made from superconducting NbTi coils, indirectly cooled with liquid helium and stabilized with high conductivity aluminum. It must be possible to cool down and energize each solenoid independent of the state of the adjacent magnets. Individual magnets will have different schedules for installation and commissioning, requiring independent operation.  It will also be necessary to warm-up individual magnets to repair detector components housed inside or to anneal the conductor.  Furthermore, it may be required that the magnets be operated in special field configurations for detector calibration.

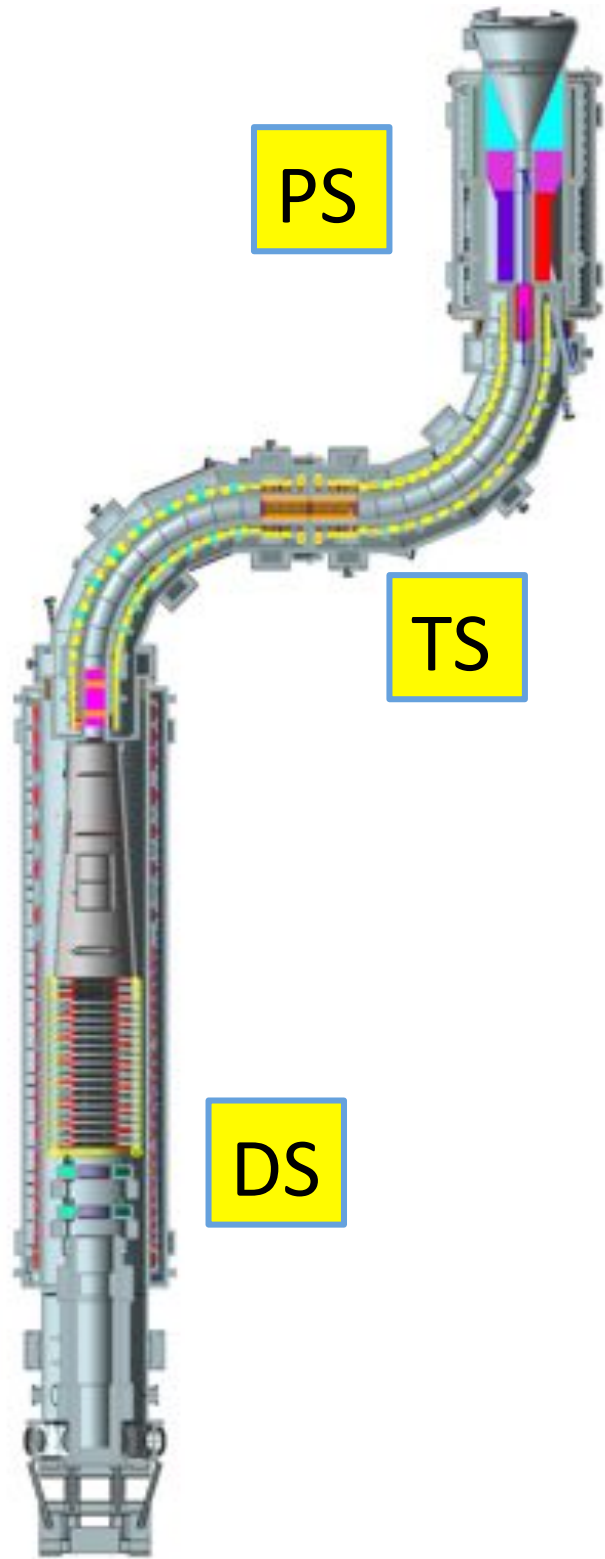

Figure 6.1. The Mu2e solenoid system.

Significant axial forces will be present between these magnetically coupled systems and these forces will change if the fields are changed. The magnets must be designed to withstand these forces when they are being operated in their standard configuration as well as in the various configurations described above. The mechanical support for each of the magnets will be independent and will not depend on adjacent magnets. This simplifies integration issues but complicates the mechanical support system. The bore of the magnets share a common beam vacuum but the magnet vacuums will be bridged with bellowed connections.

The solenoid coils must be designed for repeated full field quenches and thermal cycles, without degradation in performance, over the lifetime of the experiment. The expected duration of the experiment is 3 years at full luminosity; however, the magnets should be designed for the possibility of an extended physics run at the maximum design luminosity. The primary consequence of extended running will be the need for repeated thermal cycles to anneal the damage from irradiation. Quenches may occur during the initial campaign to full field as well as during normal operation conditions. The lifetime requirements are summarized in Table 6.1.

Table 6.1. Lifetime requirements for Solenoids.

| | |
|---|---|
| Design lifetime (min-max) | 5 – 20 years |
| Number of full thermal cycles after commissioning | Up to 100 |
| Number of full quenches after commissioning | > 100 |

## 6.2.2  Magnetic Field Requirements

Each of the solenoids performs a different set of functions and each has a unique set of field requirements. The requirements for each solenoid are described in the sections that follow and are summarized in Table 6.2.

Note that these requirements are necessary but not sufficient to fully specify the field. The magnetic field generated from the coil assemblies must keep experimental backgrounds at an acceptable limit as defined in [1]. The field is verified to meet the Mu2e requirements only after extensive computer background simulations supplied by the Mu2e experiment. These studies must cover the full range of possible field values that can result from coil manufacturing tolerances.

Table 6.2. Summary of Mu2e Solenoid field specifications. The "s" coordinate is the path length along the central axis, referenced from the geometric center of TS3; "r" refers to the perpendicular distance from the solenoid axis.  With the exception of the PS1-PS2 interface, the required magnetic field values at the beginning and end of each straight section are shown. *Curved sections TS2 and TS4 are defined by |dBs/dr| > 0.275 T/m for r = 0.

| Region | Length (m) | $s_{min}$ (m) | $s_{max}$ (m) | $B_{initial}$ (T) | $B_{final}$ (T) | $R_{max}$ (m) | Uniformity Requirement |
|---|---|---|---|---|---|---|---|
| PS1 | 1.5 | -10.58 | -9.08 | > 4.5 T at s=9.4m | n/a | ~0.5 | No local minimum anywhere |
| PS2 | 2.5 | -9.08 | -6.58 | n/a | 2.50 | 0.25 | On axis, dB/B < 0.05 about a uniform negative axial gradient from the peak field in PS1 to the TS $B_{initial}$ field. No local minima off axis. |
| TS1 | 1.0 | -6.58 | -5.58 | 2.50 | 2.40 | 0.15 | |dB/B| < 0.05 about a uniform negative axial gradient. |
| | | | | | | | dB/ds < -0.02T/m everywhere. |
| TS2* | 4.6 | -5.58 | -0.98 | n/a | n/a | 0.15 | Ripple |dB| < 0.02 T |
| | | | | | | | |dB/B| < 0.05 about a uniform negative axial gradient. |
| TS3 | 1.95 | -0.98 | 0.98 | 2.40 | 2.10 | 0.15 | dB/ds < -0.02T/m everywhere. |
| TS4* | 4.6 | 0.98 | 5.58 | n/a | n/a | 0.15 | Ripple |dB| < 0.02 T |
| | | | | | | | |dB/B| < 0.05 about a uniform negative axial gradient. |
| TS5 | 1.0 | 5.58 | 6.58 | 2.10 | 2.00 | 0.15 | dB/ds < -0.02 T/m everywhere. |
| DS1 Gradient | 3.0 | 6.58 | 9.58 | 2.00 | 1.18 | 0.3-0.7 cone | dBs/ds = 0.25+/- 0.05 T/m, |dB/B| < 0.05 |
| DS2 Transition | 1.2 | 9.58 | 10.78 | 1.18 | 1.00 | 0.7 | Magnitude of gradient decreasing |
| DS3 Uniform | 3.6 | 10.78 | 14.39 | 1.00 | 1.00 | 0.7 | |dB/B| < 0.01, dBs/ds negative |
| DS4 Uniform | 1.5 | 14.39 | 15.85 | 1.00 | 1.00 | 0.7 | |dB/B| < 0.05, dBs/ds negative |

### 6.2.2.1  PS Uniform axial gradient (PS2)

The Production Solenoid is a relatively high field solenoid with an axial grading that varies from 4.6 Tesla to 2.5 Tesla.  The purpose of the Production Solenoid is to trap charged pions from the production target and direct them towards the Transport Solenoid as they decay to muons.  The nominal peak field of 4.6 Tesla provides the high end of the field gradient while still allowing for sufficient operating margins for temperature and current density with NbTi superconductor. There is a ± 5% requirement on the deviation from a uniform gradient along the axis ($dB/B$). At radii less than 0.3 m there can be no local field minimums where particles might get trapped.

### 6.2.2.2  Solenoid straight sections in the transport solenoid (TS1, TS3, TS5)

Particles produced with a small pitch in a uniform field region can take a very long time to progress down the beamline toward the muon stopping target.  To suppress background from these late arriving particles, the three straight sections in the Transport Solenoid employ negative axial gradients for radii smaller than 0.15 m.  The radius is set by the geometry of the beam collimators.  This requirement is intended to eliminate traps, where particles bounce between local maxima in the field until they eventually scatter out and travel to the Detector Solenoid where they arrive late and may cause background.

### 6.2.2.3  Toroid sections (TS2 and TS4)

In the toroidal sections of the Transport Solenoid, the field varies as ~1/r, where r is the distance from the toroid center of curvature. In a toroid region, spiraling particles drift up or down depending on the sign of their charge, with a displacement that is proportional to their momentum and inversely proportional to their pitch. Particles with small pitch progress slowly through the toroid and drift to the wall where they are absorbed.  This allows for a relaxed gradient specification in the toroid sections, defined by dBs/dr > 0.275 T/m) where s is the coordinate along the beam path.  There is an additional requirement on the field ripple, dB, within a 0.15 m radius transverse to the central axis of the magnet system.  Large field ripples can trap particles.

### 6.2.2.4  Detector Solenoid Gradient region (DS1, DS2)

The muon stopping target resides in a graded field provided by the Detector Solenoid that varies from 2 Tesla to 1 Tesla [11].  On the Transport Solenoid side of the muon stopping target, the graded field captures conversion electrons that are emitted in the direction opposite the detector components causing them to reflect back towards the detector. On the other side of the stopping target, the graded field focuses electrons towards the tracker and calorimeter. The graded field also plays an important role in background suppression by shifting the pitch of beam particles that enter the Detector Solenoid out of the allowed range for conversion electrons before they reach the tracker. The muon stopping target is located approximately in the middle DS1.  In a conical volume defined by the proton absorber the uniformity requirement for the graded field is *dB/ds* = 0.25 T/m, where s is the direction along the solenoid axis. DS2 is the transition region between the graded and uniform fields. It should be as short as possible without introducing a local minimum in the axial field.

### 6.2.2.5   *Nearly Uniform field section (DS3 and DS4 Uniform)*

The field in this region has two competing requirements. First, the magnetic field is important for the measurement of the electron momentum and energy. The tracker-determined trajectory, along with the magnetic field map, determines the electron momentum. This requires the magnetic field to be as uniform as possible. The tracker-determined trajectory is then extrapolated into the downstream calorimeter and matched with calorimeter energy deposition. The energy/momentum match further validates the electron identification. The second design consideration is that local field minima are a potential source of backgrounds. This background is reduced by superimposing a small negative axial gradient in the tracker and calorimeter region.

## 6.2.3   Alignment Requirements

Magnetic elements must be properly aligned with respect to one another as well as external interfaces such as beam collimators, the proton beam line and internal detector elements in order to assure optimal muon transmission, suppression of backgrounds, minimization of forces amongst magnetically coupled systems, and minimization of radiation damage due to improperly located collimators. Generally speaking, alignment tolerances between cryostated magnetic elements (PS, TSu, TSd, DS) in their cold and electrically powered nominal positions are ~10 mm. Alignment tolerances between coils within a cryostated magnet are ~5 mm. Alignment between magnets and tracker elements is ~0.1 mm.

## 6.2.4   Radiation Requirements

Radiation damage to the solenoids must be carefully considered during the design process. The coil insulation, superconductor and superconductor stabilizer are the biggest concerns. The largest radiation dose is estimated to be 0.3 MGy/year at full beam intensity to the Production Solenoid [10]. Materials with poor radiation properties must be avoided. Irradiation of the conductor and insulation is not expected to be a major concern over a 20-year life cycle. There is a significant concern about damage to the superconductor stabilizer in the Production Solenoid, causing a significant reduction in electrical and thermal conductivity at low temperature [9].

For aluminum, the stabilizer room temperature resistivity ratio (RRR) must not fall below ~100 over the lifetime of the experiment. For copper the requirement is for the RRR to stay above 30.

The experimental hall in the vicinity of the PS will be highly activated and accessible only to highly trained personnel under strictly controlled conditions; therefore, every effort must be made to reduce the need for access.  Cryogenic valves, power connections and instrumentation connections should be located outside of this area.

## 6.2.5  Electrical Requirements

The following electrical requirements have been developed for the solenoid system.

- The magnets will be designed with sufficient superconductor margin to allow operation without quenching at full field during the delivery of peak beam intensities. The target operating $J_c$ margin is 30% and the required $T_c$ margin is 1.5 K.
- The superconductor will be standard copper stabilized NbTi strand.  In order to achieve the $J_c$ and $T_c$ margins for the Production Solenoid, the design $J_c$ (4.2 K, 5T) value must be greater than 2800 A/mm$^2$ for conductor in the peak field region of the PS. The magnet will be operated DC.
- The NbTi strand will be woven into a Rutherford cable. The cable will be further stabilized with low resistivity aluminum.
- The solenoids will be divided into several independent power circuits. Each circuit will have an external energy extraction resistor.  The value of the resistor will be chosen so that the peak voltages will be limited during a quench to less than 300 V to ground and 600 V across the magnet terminals.  For DS and PS depending on the intra-coil connection scheme, there is the possibility of ~100 V between adjacent coil layers.  Coil-to-ground and layer-to-layer insulations must be sized to meet these requirements. Turn-to-turn voltages are not expected to exceed 10 V; however, cable insulations must be designed conservatively as these potentially damaging turn-to turn electrical breakdowns will be difficult to detect during fabrication.
- The peak coil temperature must not exceed 130 K as the result of a quench. Stabilizer will be sized to meet this requirement.
- To insure that there is no net transfer of magnetic stored energy between magnets during a quench, the stand-alone time constant for energy extraction for each power circuit must be set to approximately 30 seconds.

## 6.2.6  Cryogenic Requirements

The solenoids will be divided into 4 cryogenic units.  All solenoid coils will be potted with epoxy and indirectly (conduction) cooled by liquid helium. The coils can be cooled using either a "force flow" or a "thermal siphoning" system.  80 K thermal intercepts in the cryostat will be cooled by liquid nitrogen.

Refrigerators (not included in the solenoid project scope) located in a separate cryo building will supply liquid helium for the entire solenoid system. Based on the estimated heat loads, the required liquid helium can be supplied by one Tevatron "satellite" refrigerator [12]. This means that during normal operations, the 4.5 K heat load cannot exceed 600 Watts. A second refrigerator will double the available capacity for use during initial cool-down and quench recovery. Cryogens will flow to and from each cryostat via a single cryo-link chimney. This chimney must be routed from the magnet cryostat through the concrete shielding and cosmic ray veto counters up to the cryogenic feed box located at ground level. Care must be taken in locating these gaps in the shielding to avoid "line of sight" paths for neutron and cosmic ray backgrounds. The chimney should be routed to minimize interference with utilities and crane coverage.

## 6.3 Technical Design

### 6.3.1 Production Solenoid

The first in the chain of the Mu2e magnets is the Production Solenoid (PS), shown in Figure 6.2. The role of PS is to collect and focus pions and muons generated in interactions of the 8-GeV proton beam with a tilted high-Z target by supplying a peak axial field between 4.6 T and 5.0 T and an axial gradient of ~1 T/m within the 1.5 m diameter warm bore.

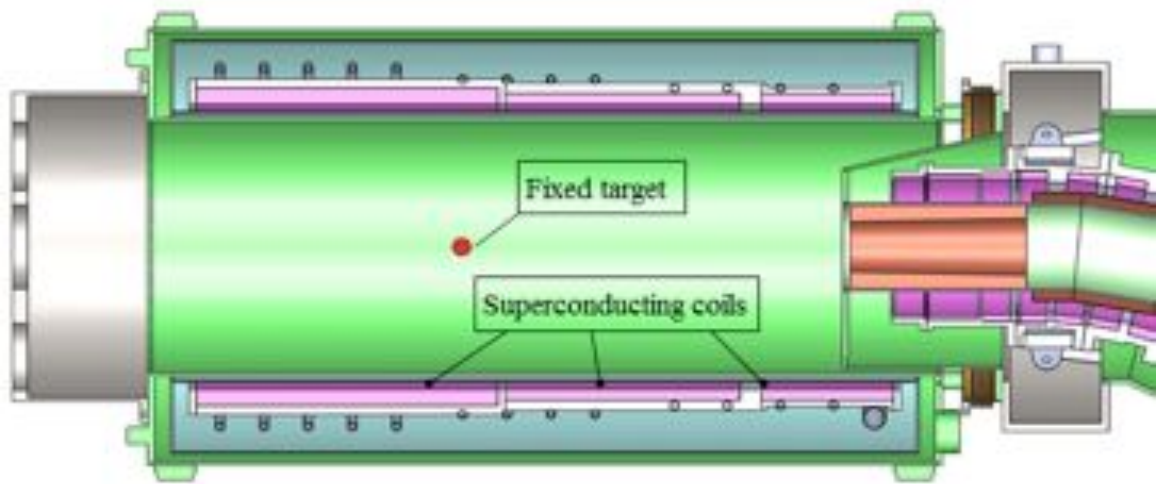

Figure 6.2. Cross-section through the PS cryostat with a part of the Transport Solenoid (TS) cryostat shown. HRS is not shown.

The PS is a challenging magnet because of the relatively high magnetic field and a harsh radiation environment that requires state-of-the-art conductor both in terms of the current-carrying capacity and structural strength. The PS coils are protected from radiation by a massive Heat and Radiation Shield (HRS) made of bronze, placed within the warm magnet bore. The HRS volume and cost were minimized while keeping the absorbed dose, peak power density, total power dissipation and number of displacements per atom (DPA) within the acceptable limits.

The magnet consists of three superconducting coils made of Al-stabilized NbTi cable surrounded by support shells made of structural Al and bolted together to form a single cold-mass assembly. The cold mass is supported inside the cryostat using a system of axial and radial support rods. During normal operation of the Mu2e magnet system, the axial force on the cold mass is about 1.3 MN acting toward the TS. The axial force reverses the direction in the case of a quench during stand-alone PS operation, which requires a 2-way axial support system.

In the case of a quench, the quench protection system extracts the stored energy to a dump resistor located outside of the magnet cryostat. Because of the relatively low resistivity of the Al support shells, a considerable fraction of the stored energy dissipates there due to eddy currents, accelerating the quench propagation in the coil volume.

The cold mass is instrumented with voltage taps, temperature gauges, strain gauges, and displacement sensors that monitor the magnet parameters during operation. Witness samples made of Al and Cu are placed at strategic locations on the inner cold mass surface to monitor the degradation of the cable stabilizer under irradiation. Once the critical level of degradation is detected, the magnet will be thermo-cycled to room temperature to restore the stabilizer resistivity.

### 6.3.1.1  General Requirements

The Mu2e Production Solenoid (PS) magnet must meet the specific design and operational requirements as defined in [2]. Because of the relatively high cost, complexity and difficulty to replace once installed into the system, the PS should have adequate operating margins and be as fail-resistant as practically achievable.

In particular, the magnet must be able to operate with the other adjacent magnets or standalone, survive the worst-case quench, tolerate reversal of the axial force direction, and return to the nominal operating parameters without retraining after each thermal cycle. The main magnet requirements to be used as the design key point are discussed below.

*Magnetic Field*

The magnetic field requirements for the Mu2e experiment are described in [1]. The PS field varies with the axial position. The maximum axial field on the axis is required to be at least 4.5 T. The axial field shall monotonically decrease with no more than ±5% non-linearity from the peak value to 2.5 T over the length of 2.8 m. In order to guarantee meeting the peak field requirement, the magnet shall be designed and fully operational at the maximum axial field of 5.0 T while meeting all other requirements under the static thermal load, excluding the particle radiation load. Note that energizing the magnet to 5.0 T requires using a trim power supply in TS to maintain the 2.5 T field at the interface with the TS magnet.

*Particle Radiation Load*
The magnet will experience particle irradiation due to interaction of the proton beam with the fixed target in the magnet aperture. The HRS is designed to limit the main radiation quantities to the following values: peak coil power density - 30 µW/g, total heat load on the cold mass - 100 W, peak lifetime absorbed dose - 7 MGy, and displacements per atom - (DPA) – $(4\text{-}6) \cdot 10^{-5}$ year$^{-1}$ [9]. The magnet shall be designed to operate under this maximum radiation load while meeting all other requirements.

*Degradation of RRR*
The residual resistivity ratio (RRR) of the cable stabilizer degrades under irradiation at cryogenic temperatures. The minimum allowable values for the RRR of Al and Cu stabilizers are 100 and 50, which will be reached in about a year of continuous operation for the given DPA values with a sufficient safety margin [9]. Once the critical degradation of RRR is detected, the magnet will be thermo-cycled to recover the RRR. The magnet shall be designed to operate at the minimum allowable values of RRR while meeting all other requirements.

*Cooling*
The magnet shall be cooled by heat conduction to the thermo-siphon system connected to the cryogenic plant, which provides the nominal liquid helium temperature of 4.70 K. In order to maintain the required 1.5 K temperature margin, the maximum allowable temperature of the superconducting coils is 5.10 K when operating at 4.6 T peak axial field and 4.85 K when operating at 5.0 T peak axial field.

*Quench Protection*
The magnet will be protected from quenches by an active quench protection system, which continuously monitors the coil voltages and extracts the energy to an external dump resistor once a quench is detected. The maximum coil temperatures and voltages during a protected quench must be limited to 130 K and 600 V.

*Structural Criteria*
The superconducting coils shall be reinforced against Lorentz forces by external support shells made of structural Al, and the cryostat shells will be made of austenitic stainless steel. The thicknesses of the coil support shells and cryostat shells must be chosen according to the 2013 ASME Boiler and Pressure Vessel Code for operation at 5.0 T peak axial field. Plastic deformation of the cable stabilizer shall be considered during calculation of the shell stresses.

The maximum allowable stress in all other structural elements, including the cold mass assembly bolts and the cryostat suspension rods, is the lesser of 2/3 of the minimum specified Yield Strength (Sy) or 1/3 of the minimum specified Ultimate Strength (Su). An additional safety factor of 2 will be included in calculation of the gravity loads on the cold mass suspension components.

A separate shipping restraint system shall be devised for supporting the cold mass during transportation. The cold mass suspension components shall not be used to support the shipping and handling loads.

### 6.3.1.2   Cold Mass

The PS cold mass consists of three epoxy-impregnated coils wound from an insulated Al-stabilized NbTi conductor and installed within the support shells made of structural Al. The shells are bolted together to form a single cold mass assembly as shown in Figure 6.3. The assembly procedure is outlined in [10]. The three separate coil modules have 3, 2 and 2 layers of Al-stabilized NbTi cable, wound in the "hard-way" around the aperture. The coil modules are reinforced by the shells made of structural Al.

The experiment relies on the magnet providing a precise magnetic field distribution in the aperture; thus, it is important to put the exact number of turns in each coil and to maintain the proper coil dimensions at room temperature and after cooldown. Figure 6.4 shows the necessary coil envelope dimensions at room temperature.

The cold mass cross-section is shown in Figure 6.5 and the details related to the component location and the warm to cold correlation are given in [10].

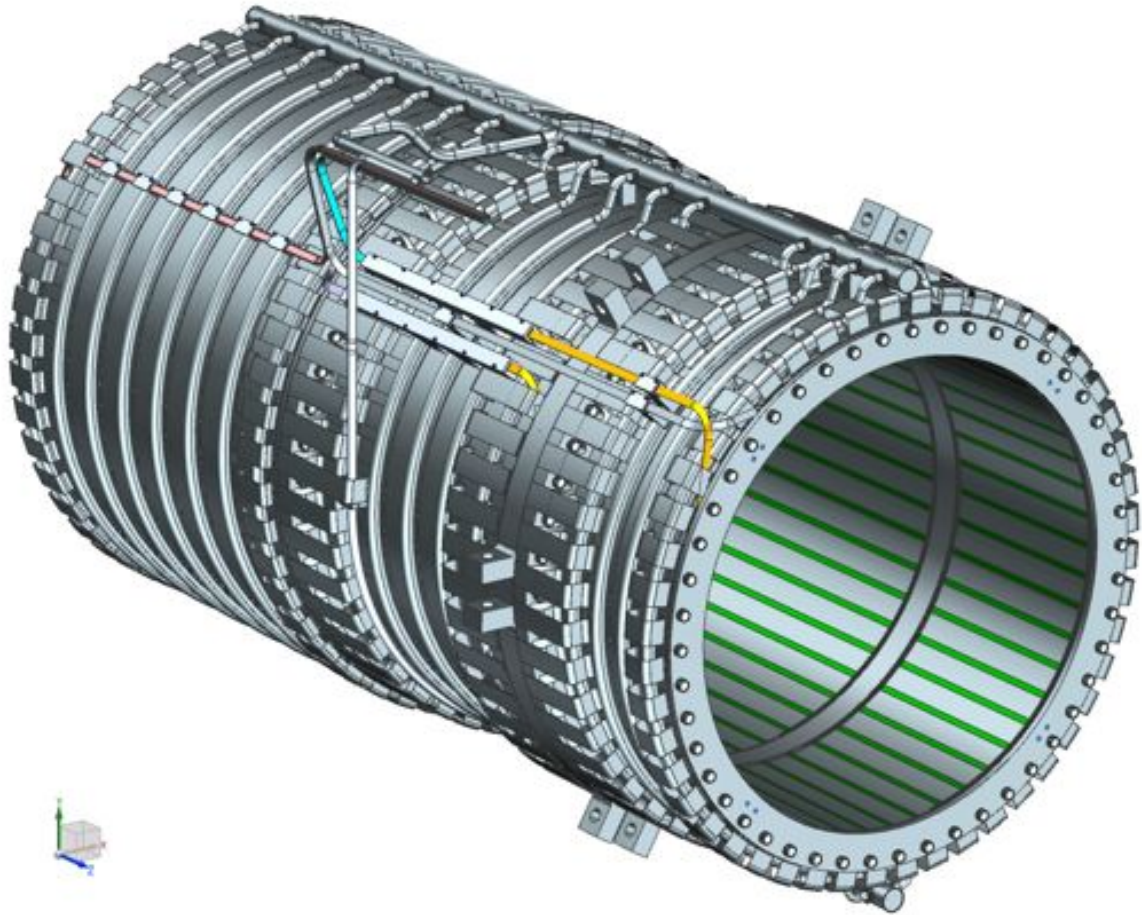

Figure 6.3. General cold mass view.

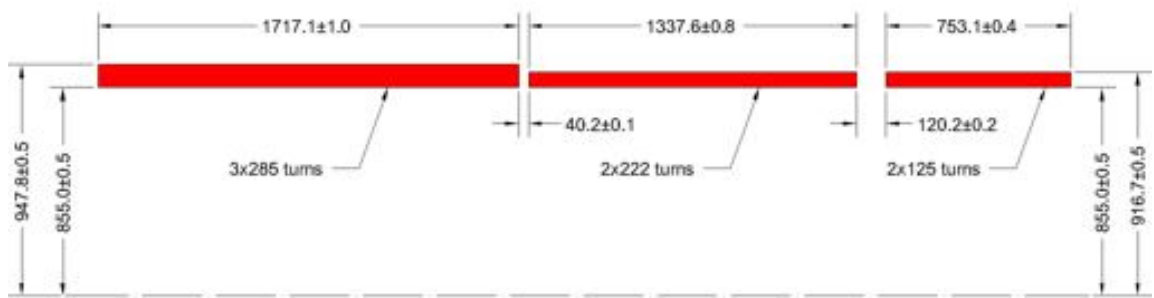

Figure 6.4. Room temperature dimensions in [mm] of the coil envelopes, which include the cables with the cable insulation and the inter-layer insulation. Other cold mass components are excluded from the coil envelopes.

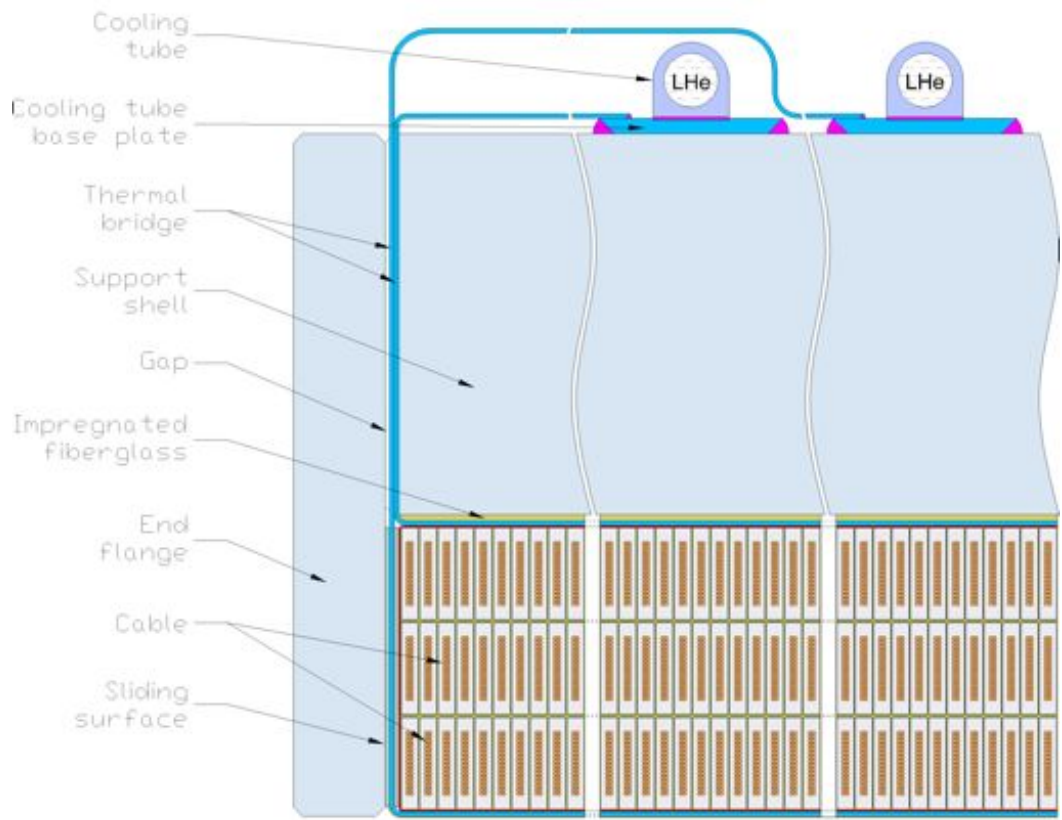

Figure 6.5. Cold mass cross-section though the thermal bridge strips at the non-TS end.

**Cold Mass Components**

*Cable*

The PS cable conforms to the specifications described in [13] - [15]. The cable employs aluminum stabilizer that is placed around the NbTi Rutherford type cable by either conforming or hot extrusion. Such cable technology has been used in nearly every large detector solenoid built in the past decades with the most recent examples being ATLAS Central Solenoid, ATLAS Toroids and CMS [16] - [19].

The PS cable cross-section is shown in Figure 6.6 and the cable parameters are summarized in Figure 6.7. A precipitation-hardened Al-0.1wt%Ni alloy is used for the stabilizer that in conjunction with cold working achieves the target 0.2% yield strength of > 80 MPa at 4.2 K and the RRR of > 600 [19].

The cable is structurally sound and bendable in the hard way, meaning the minor edge in, and in the easy way around the radii specified in Table 6.3 without degradation of the critical current, de-lamination of strands from Al stabilizer or buckling.

Table 6.3. PS Cable Parameters.

| Parameter | Unit | Value | Tolerance |
|---|---|---|---|
| Cable critical current at 5.0T, 4.22K | kA | >50.7 | |
| Cable critical current at 5.6T, 6.25K | kA | >10.0 | |
| NbTi filament diameter | μm | <40 | |
| RRR of Cu matrix | - | ≥80 | |
| RRR of Al stabilizer | - | ≥600 | |
| Strand Cu/non-Cu ratio | - | 0.95 | ±0.05 |
| 0.2% yield strength of Al stabilizer at 4.2K/293K | MPa | ≥80/60 | |
| Shear strength of Al-Cu bond at 4.2K | MPa | ≥50 | |
| Overall cable width at 4.2K/293K | mm | 30.0/30.1 | ±0.1 |
| Overall cable minor edge thickness at 4.2K/293K | mm | 5.40/5.42 | ±0.03 |
| Overall cable major edge thickness at 4.2K/293K | mm | 5.60/5.62 | ±0.03 |
| Strand diameter at 4.2K/293K | mm | 1.300/1.303 | ±0.005 |
| Number of strands | - | 32 | |
| No degradation/de-lamination/buckling after the hard-way bending with the radius of | m | 0.85 | -0.02 |
| No degradation/de-lamination/buckling after the easy-way bending with the radius of | m | 0.10 | -0.01 |
| Total delivered cable length | km | ≥10.7 | |

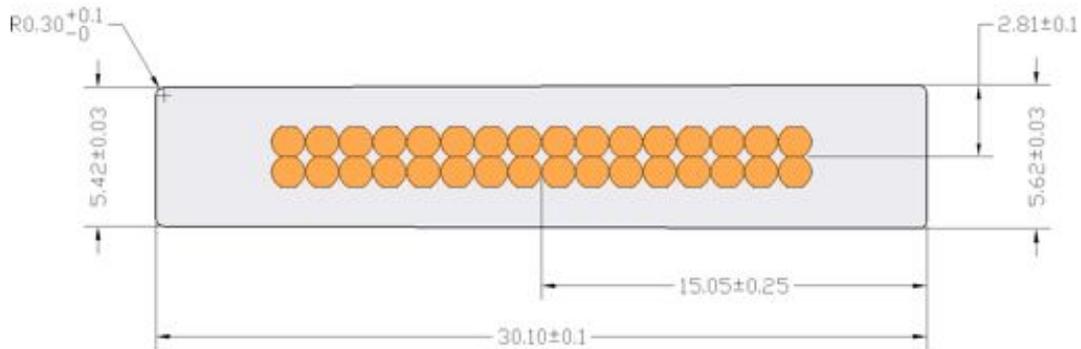

Figure 6.6. Cable cross-section with dimensions in (mm) at room temperature.

The specified minimum critical current is guaranteed by the cable vendor after the final cable heat treatment at (150 ± 2) °C for 24 hours that replicates the epoxy and insulation curing process. It is the responsibility of the magnet vendor to verify that the cable meets the minimum critical current requirement if the vendor selects a curing cycle with a different temperature or duration.

The minimum cable unit lengths to be delivered are specified in Table 6.4 after cutting all the samples required for tests. Each cable unit will be delivered on a separate spool and has a sufficient length to wind one of the coil layers without inner splices. Note that there is one extra piece length for the longest layer (4 pieces instead of 3 required) that is to be held in reserve, allowing replacement of any coil layer in the case of unforeseen fabrication difficulties.

Table 6.4. PS cable unit lengths.

| Type | Number of units | Minimum unit length, m |
|------|-----------------|------------------------|
| 1    | 4               | 1660                   |
| 2    | 2               | 1280                   |
| 3    | 2               | 750                    |

*Insulation*

Besides providing adequate electrical strength and radiation resistance, the coil insulation should have a relatively high thermal conductivity. This additional requirement, usually not as important for detector magnets of particle colliders, is due to the internal coil heating under the strong neutron radiation coming from the production target. The insulation design therefore plays a vital role in facilitating the heat extraction from the coil and reducing the peak coil temperature, especially in the case of conduction cooling, which requires careful analysis. A summary of the PS cable dimensions including insulation is shown in in Table 6.5.

The PS employs a composite cable insulation made of polyimide and pre-preg glass tapes. This type of insulation, originally developed for the TRISTAN/TOPAZ solenoid, was also used in the ATLAS Central Solenoid [16]. The nominal cable insulation thickness is 250 μm per side. It is made of two layers of 125 μm thick composite tape consisting of 25 μm of a semi-dry (BT) epoxy on one side of 25 μm polyimide (Kapton) tape and 75 μm of pre-preg E-glass on the other side as shown in Figure 6.7. The epoxy side should face the cable surface. The thickness of the pre-preg E-glass is a parameter that depends on the coil packing factor in the azimuthal direction, and shall be determined experimentally for the given winding setup. The E-glass thickness can be adjusted in 50-75 μm range in order to fit the required number of turns in the coil envelopes shown in Figure 6.4.

The ground insulation has a total thickness of 500 μm all around the coil envelopes. It is made of two layers of 250 μm thick composite sheet or wide tape each consisting of 25 μm polyimide (Kapton) in between the two layers of 112.5 μm pre-preg E-glass as shown in Figure 6.7.

Table 6.5. Cold and warm dimensions of the cable as wound in the coil.

| Number | Component name | Material | CTE (293K-4K), mm/m | Thickness, mm | |
|--------|----------------|----------|---------------------|---------------|---|
| | | | | Cold | Warm |
| *Radial direction* | | | | | |
| 1 | Cable insulation | G-10+Kapton | 8.650 | 0.250 | 0.252 |
| 2 | Cable | Al+Cu+NbTi | 3.356 | 30.000 | 30.101 |
| 3 | Cable insulation | G-10+Kapton | 8.650 | 0.250 | 0.252 |
| | Total | | 3.443 | 30.500 | 30.605 |
| *Axial direction* | | | | | |
| 1 | Cable insulation | G-10+Kapton | 8.650 | 0.250 | 0.252 |
| 2 | Cable | Al+Cu+NbTi | 3.759 | 5.500 | 5.521 |
| 3 | Cable insulation | G-10+Kapton | 8.650 | 0.250 | 0.252 |
| | Total | | 4.167 | 6.000 | 6.025 |

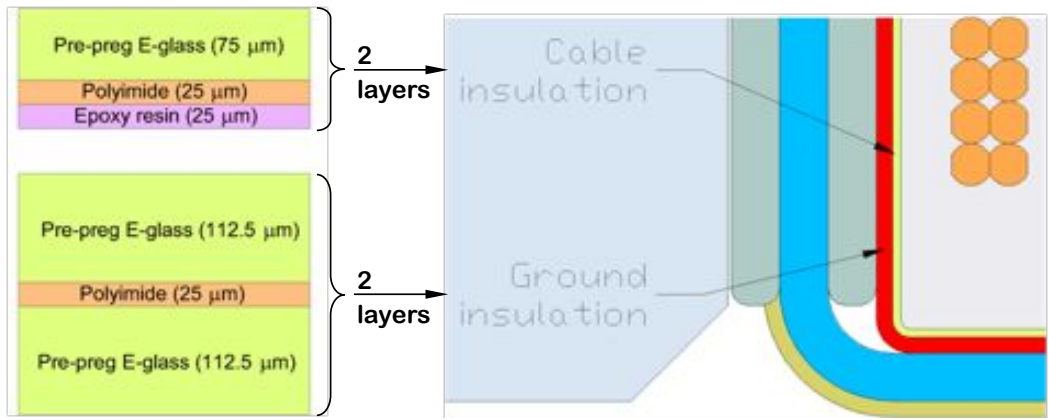

Figure 6.7. Coil insulation scheme.

To facilitate heat extraction from the coil and to increase structural integrity, all gaps between turns and layers are to be filled with epoxy resin. Depending on the vendor's experience, either vacuum impregnation or wet winding techniques can be used. Because the composite cable insulation is impermeable for epoxy, 500 µm thick layers of dry E glass are introduced between the coil layers to provide paths for epoxy penetration during the vacuum impregnation. If wet winding is used, the inter-layer insulation is to be saturated with liquid epoxy prior to the cable placement. To reduce the risk of having unfilled gaps between the layers, pre-preg interlayer insulation shall not be used.

*Thermal Bridges*

The cold mass is to be cooled via heat conduction to the thermosiphon system. To facilitate static and dynamic heat extraction and reduce the temperature increment in the coil, the cold mass is equipped with thermal bridges shown in Figure 6.8, bonded to the inner and outer coil surfaces using electrically non-conductive adhesive suitable for cryogenic applications.

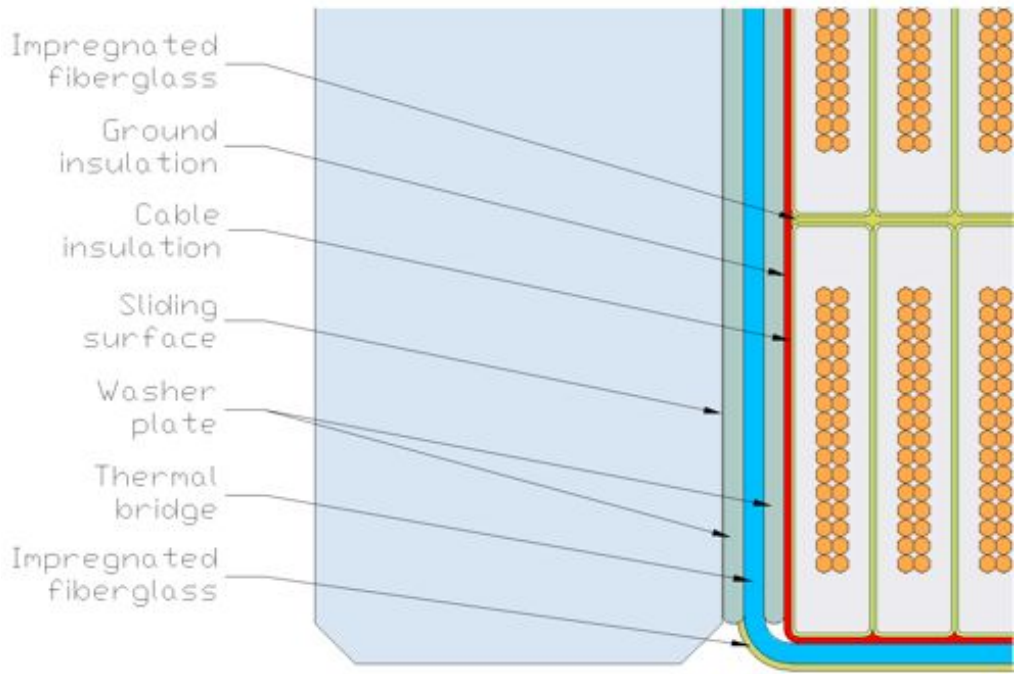

Figure 6.8. Coil to flange interface components.

The thermal bridges are strips of high purity 5N Al with a minimum residual resistivity ratio (RRR) of 1500 and 1.5±0.1 mm thickness. Each strip must be produced as a single continuous piece of sufficient length to make a connection with the cooling tubes. Joining several pieces by welding, soldering, gluing, etc. is not allowed. The strips are connected to the cooling tubes base plates by welding. The strips are routed to the outer cold mass surface between the bolts connecting the coil modules, hence the number of strips per inner or outer surface of the coil must be equal to the number of bolts per interface and the strip width must be selected to fit into the given space with sufficient clearance.

For the currently selected number and size of bolts, the strip width shall be in the 90-95 mm range. All strip edges must be rounded with ≥0.2 mm radius. In order to fit into the allocated space between the bolts, the thermal bridges must be separated from each other by equal distances. Since the inner and outer bridges share the same space between the bolts, it is crucial to put them at the same azimuthal positions. The gaps between the thermal bridges are to be filled with G-10CR.

To improve the strip surface adhesion to the coil insulation, the strips must be prepared with a uniform satin-matte finish by abrasive blasting to the roughness Ra of 5-10 μm, which represents an average of all the peaks and valleys in a standardized measurement area. Immediately prior to installation, the strips must be cleaned by Abzol VG or equivalent cleaner.

*Inner Coil Surface*

The inner coil surface is comprised of a 500 μm layer of dry fiberglass (E-glass), shown in Figure 6.5 to be impregnated with epoxy. The purpose of this material is to improve the bond strength between the thermal bridges and the coil surface by reducing the possibility of crack development and propagation at the edges of the strips.

*Buffer between the Coils and the Support Shells*

There is an extra layer of epoxy-impregnated fiberglass around the ground insulation that serves as a buffer material separating the coils from the support shells. The coils are to be wrapped with a sufficient amount of dry E-glass (5-10 mm thickness) prior to impregnation. After impregnation, the buffer material is to be machined down to an average of 2 mm to create a smooth, cylindrical mating surface.

*Washer Plates between the Coils and the Flanges*

The ground insulation at the coil ends shall be protected from possible damage by the edges of the thermal bridge strips during magnet fabrication and operation. This is accomplished by placing a 1.5 mm thick washer plate made of G-10CR between the ground insulation and the strips as shown in Figure 6.8. A second washer plate of the same thickness and material is placed on the other side of the strips to create a uniform sliding surface. The gaps between the thermal bridges are to be filled with G-10CR spacers. The washer plates will be permanently bonded to the strips, spacers between them and the ground insulation using non-electrically conductive adhesive resin suitable for cryogenic applications.

*Sliding Layers*

Due to presence of the NbTi/Cu strands in the Al-stabilized cable, the overall thermal contraction coefficient of the coil in the radial and azimuthal directions is less than that of the aluminum. Also, because of the structure bending under the Lorentz forces, small gaps tend to develop at some of the coil-flange interfaces.

To avoid accumulation of shear and tensile stresses at these interfaces, which can lead to cracks in insulation triggering magnet quenches, the coils will be allowed to slide with low friction and, if needed, separate from the end flanges. This is accomplished by placing a 100 μm layer of mica paper between the washer plates and the flanges as shown in Figure 6.8.

*Support Shells*

The support shells are cylinders made of Al 5083-O alloy that are placed around the coils in order to limit coil deformation and stresses under the Lorentz forces. Since they are the main structural elements of the cold mass, care must be taken to maintain the structural integrity and uniformity of material properties. Each shell is to be produced as a single solid piece using forging. Making the shell by welding several pieces is not allowed.

The shells are connected together using bolts passing through the flanges. The thermal bridges are made of soft aluminum that should not be stressed during assembly or operation. Hence, the faces of the shells have slots for routing the thermal bridges between the coils and the cooling tubes as shown in Figure 6.9. The number of slots is the same as the number of thermal bridges and the number of bolts per interface. The slot depth is 5±0.5 mm and the minimum slot width is 95 mm.

The forged shells shall have oversized dimensions to allow for machining of the inner and outer surfaces and the faces. The cooling tubes are welded to the outer surface prior to machining of the inner surface. The inner surface is to be machined after the coil outer surface is machined and measured. The required radial interference between the coils and the support shells is 0±0.5 mm at room temperature. The coil must be permanently bonded to the shell using adhesive resin suitable for cryogenic applications that must fill all the gaps between the coil and the shell.

To improve the coil-shell adhesion, the inner shell surface must be prepared with a uniform satin-matte finish by abrasive blasting to a roughness *Ra* of 5-10 μm, which represents an average of all peaks and valleys in a standardized measurement area. The mating coil and shell surfaces must be air dusted and cleaned by Abzol VG or equivalent cleaner.

*Flanges*

The flanges are disks made of Al 5083-O that cover the coil ends (outer flanges) or separate the coils from each other (inner flanges). The flanges shall be produced as single pieces by forging and machining. Alternatively, several flanges can be cut from a larger forged cylinder and machined to the proper dimensions. Just as is the case with the shell, joining several flange pieces by welding is not allowed.

The required axial interference between the coil faces and the flanges is 0±0.25 mm at room temperature, which may be finely tuned by adding/removing layers of mica paper between the coils and the flanges.

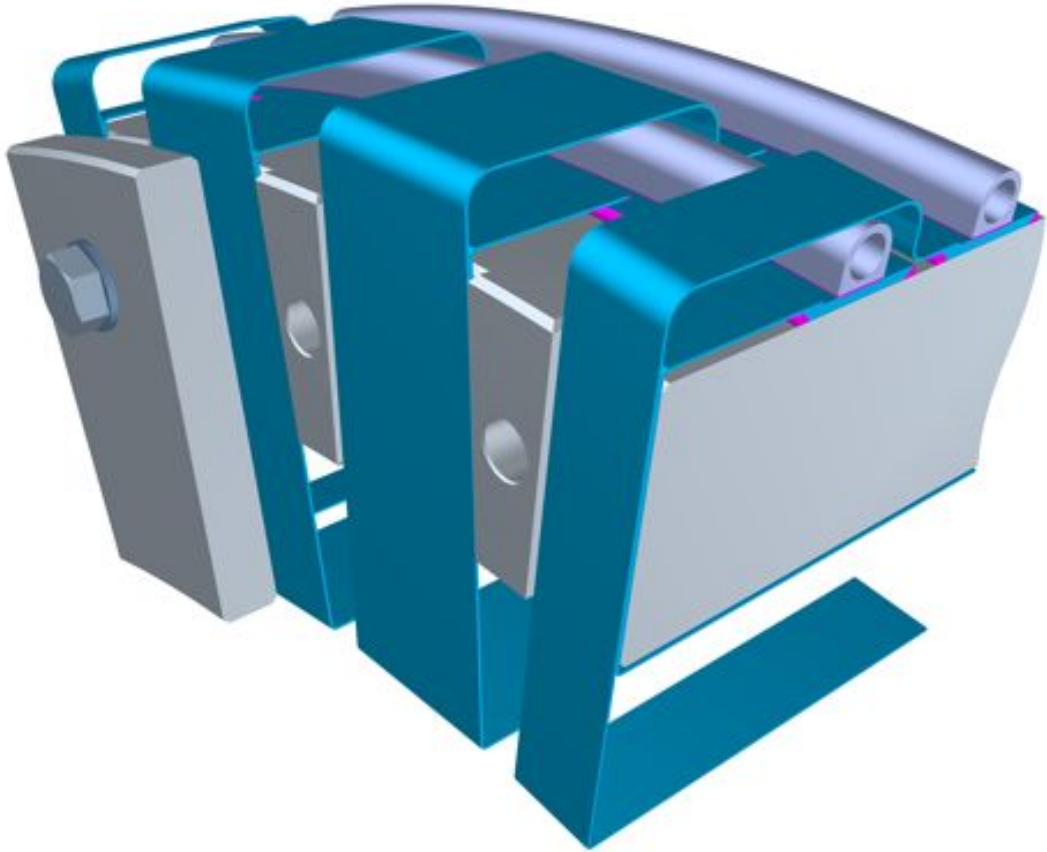

Figure 6.9. Shell to flange interface with the thermal bridges shown. Coil and insulation are not shown.

*Bolts*
The coil modules are connected using bolts passing through the flanges. The bolt material shall have the same or higher thermal contraction than the flange material in the 4 – 293 K temperature range to avoid losing the preload after cooling down. The material of choice for the bolts is Al 7075-T73, which has a slightly higher thermal contraction than Al 5083-O and a factor of ~1.7 higher tensile strength at the room temperature.

In order to meet the load requirements during assembly, transportation and operation, 48 equally spaced M24 bolts are used for each shell-flange interface. The bolt holes penetrate through the islands between the slots in the faces of the shells as shown in Figure 6.9. The joint strength benefits from the friction between the mating surfaces. In order to keep the shear load between the bolts and the mating surfaces uniform, the friction force under each bolt will be higher than the shear force; hence, the bolts shall be sufficiently preloaded during the assembly.

It is reasonable to assume some preload relaxation will occur days after the cold mass assembly due to local material yielding and creep. Each bolt preload shall be measured again at least 7 days after the cold mass assembly and adjusted if necessary prior to the cold mass installation in the cryostat.

*Venting Holes*

A general requirement for cold mass construction is that there may not be any partially closed volumes, for example, those closed by a threaded fastener that could considerably increase the cryostat evacuation time. Consequently, any closed volume larger than 0.5 cm$^3$ must be either permanently sealed by a vacuum-tight weld or vented to the cryostat vacuum space through an additional hole. The minimum venting hole diameter is 1 mm.

**Electrical Connections and Conductor Joints**

The electrical connection scheme of the PS coils is shown in Figure 6.10. The cable will be provided in sufficient piece-lengths to allow winding of complete layers without splices. Two types of joints are used within the cold mass as shown in Figure 6.11. The layer-to-layer joints are located at the ends of each coil layer next to the end flanges. The coil-to-coil and coil joints are located on the outer cold mass surface, similar to CMS [22]–[23].

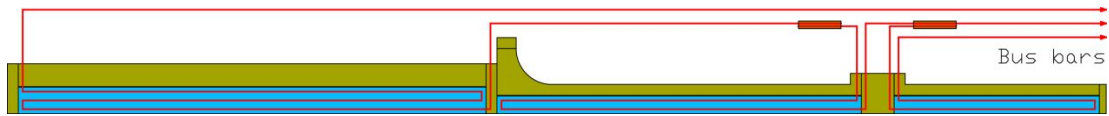

Figure 6.10. Electrical connections of the PS coil sections. Layer to layer joints are not shown.

All cable joints inside the cryostat are to be made without removing the Al stabilizer. The Al-to-Al joints shall be made by TIG welding of the adjoining cable edges. The acceptable weld filler materials are: 1199-O aluminum, 5N aluminum or 5N aluminum doped with 0.1wt% Ni (this is the material of the cable stabilizer).

The electrical resistance of any joint shall not exceed 1 nΩ at 4.2 K temperature and 5 T field. The minimum joint length will be determined experimentally by welding and testing cable samples. The critical current degradation in the superconducting cables due to joint welding shall not exceed 10%. Note above requirements were met in the CMS [22] and ATLAS CS [24] aluminum cable joints. A proposed starting point is the procedure developed for welding the CMS cables [22].

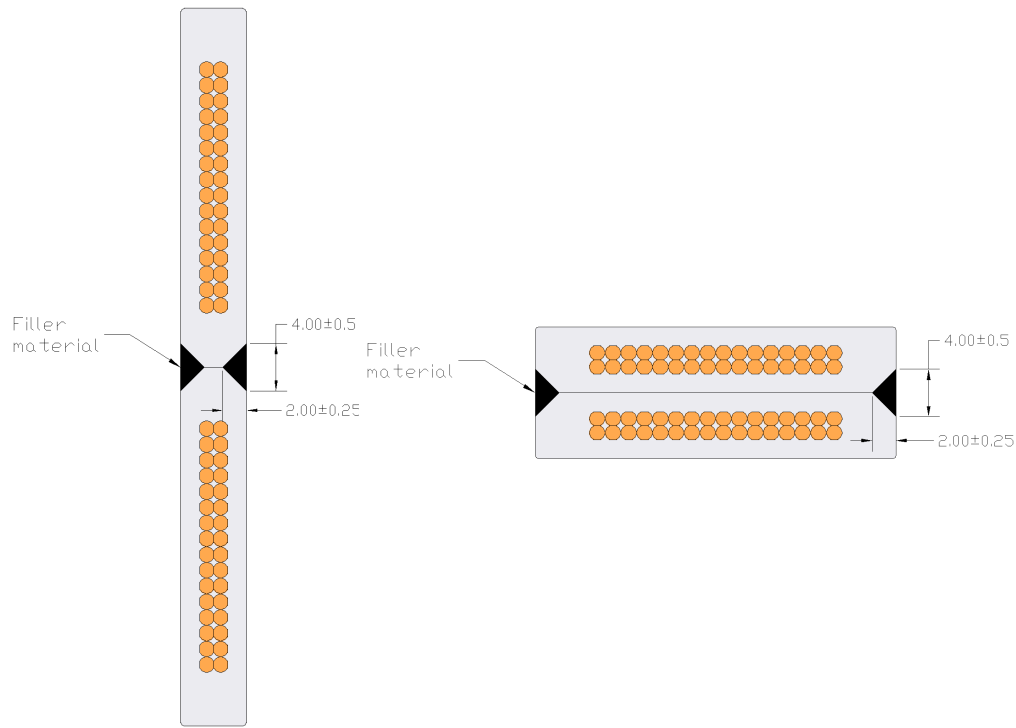

Figure 6.11. Splice joints: layer-to-layer (left) and coil-to-coil (right).

### 6.3.1.3   Cryostat

The PS cryostat consists of the following components and systems:

- Structural supports for the magnetic coils and vacuum vessel to ground
- A 4.7 K cooling circuit
- An 80 K thermal shield
- An 80 K cooling circuit
- A vacuum vessel with a warm bore
- Multi-layer insulation (MLI)
- Interface to the TS cryostat
- Interface to the cryogenic transfer line
- Interface to the HRS in the warm bore
- Interface to the vacuum enclosure at the non-TS end

The dimensions and materials for the cryostat components are described in [10]. The interfaces are discussed separately in Section 6.3.1.5.

***Thermosiphon Cooling System***

The PS magnet will be conduction cooled using a thermosiphon cooling scheme. The thermosiphon consists of two manifolds, one at the top and the other at the bottom of the cold mass, connected by 17 siphon tubes that provide parallel liquid helium flow paths to cool the cold mass. Sizing and attachment details of the cooling tubes are critical for proper performance of the system. A detailed thermal analysis of the coil, as will be described later in Section 6.3.1.6, is required to ensure the coil temperature is within the specified thermal margin.

*Heat Loads*

The PS cold mass assembly will be cooled with saturated helium at 4.7 K flowing in a thermosiphon circuit. The heat load contributions at 4.7 K originate from the following sources:

- 80 K radiation from the inner and outer thermal shields
- Conductive heat loads from the axial and radial supports
- Heat generation or dynamic heat load from particle irradiation

Table 6.6 summarizes the different heat loads at 4.7 K.

Design of the Thermosiphon System

The thermosiphon cooling works on the basis of a flow circulation induced by the density difference between the heated liquid/vapor and the unheated coolant. Conceptually, saturated helium is filled from the bottom supply manifold, through the siphon tubes to the top return manifold. The liquid helium in the siphon tubes conductively absorbs the heat from the cold mass and as the helium temperature increases, the density of the helium decreases, pushing the heated liquid up while forcing the colder liquid to the bottom in a natural circulation loop.

As the thermosiphon system is essentially driven by gravity, the siphon tubes must be oriented vertically. The siphon tubes are semi-circular segments that are welded to the cold mass itself and attached at the bottom and top to the supply and return manifolds, respectively. The thermosiphon cooling concept has been successfully demonstrated and used on the ALEPH solenoid [25] and CMS detector solenoid [26].

The advantage of the thermosiphon system is reliability, as it does not include any moving parts such as cold pumps. It is also efficient because the temperature is uniform due to the fact that the cooling helium flow spontaneously adapts to the heat load distribution.

Table 6.6. Summary of heat loads at 4.7 K.

**80 K radiation heat load to 4.7 K surfaces**

| Surface | Surface area ($m^2$) | Number of MLI layers | Heat flux ($W/m^2$) | Heat load (W) |
|---|---|---|---|---|
| Cold mass outer surface | 28.4 | 30 | 0.2 | 5.68 |
| Cold mass inner surface | 21.4 | 30 | 0.2 | 4.28 |
| Total heat load from 80 K radiation | | | | 10 |

**Conductive heat load from supports to 4.7 K components**

| Component | Heat load per support (W) | Quantity | Total heat load (W) |
|---|---|---|---|
| Axial supports (TS end/short) | 0.143 | 6 | 0.86 |
| Axial supports (non-TS end/long) | 0.2183 | 6 | 1.31 |
| Radial supports | 0.282 | 16 | 4.51 |
| Total conductive heat load at 4.7 K | | | 6.7 |

**Dynamic heat load on the cold mass due to particle irradiation**

| | Total heat load on cold mass (W) |
|---|---|
| MARS V15 particle simulation | 28 |
| Total heat load to 4.7 K components (static + dynamic) | 45 |

*Siphon Tube Locations*

An important design note is that the locations of the siphon tubes are important to the cooling of the PS magnet. The reference locations are described in [10]. One of the criteria for locating the siphon tubes is that the first tube is placed at a convenient distance from the Non-TS end of the coil housing so as to allow for bolt insertion and ease of welding. The rest of the tubes shall be placed at equal intervals. The positional tolerance for the tube locations will be ±2 mm.

*Design of the Siphon Tubes*

As described in Section 6.3.1.2, the major contributors to the cooling of the PS coils are the high-purity aluminum strips known as thermal bridges, which thermally connect the magnet coils to the cooling tubes. In order to best utilize the design arrangement, the siphon tubes have been designed with a special geometry as shown in Figure 6.12.

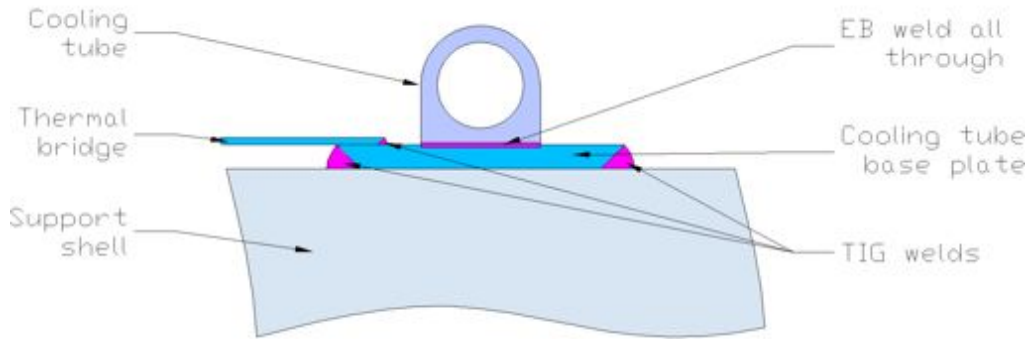

Figure 6.12. Siphon tube connections.

Because almost the entire heat load from the magnet is transmitted to the cooling tubes through the thermal bridges, any appreciable ΔT must be recovered. For this reason, the siphon tubes are extruded with the special "D" shape to have a flat face at the bottom, and the base plates of 99.999% pure (5N) aluminum are Electron Beam Welded (EBW) to the flat faces of the siphon tubes. The EBW assures the full weld penetration through the tube-plate interface, and the presence of the highly conductive aluminum plates minimizes the ΔT between the thermal bridges and the liquid helium.

The base plates are TIG welded all along to the support shells as shown in Figure 6.12 using Al 1199-O as the filler prior to the final machining of the shell inner surfaces and the coil insertion. The thermal bridges are TIG welded to the base plates as shown in Figure 6.12 after the coil and shell assembly but prior to the final cold mass assembly.

### Thermal Shield
The PS magnet cold mass is shielded from the room temperature (300 K) thermal radiation by means of 6 mm thick 6061-T6 aluminum cylinders both at the inner and outer surfaces. The thermal margins for the shields are designed to be around 85 K. The thermal shields themselves will be cooled with Liquid Nitrogen (LN2).

Pressurized two phase nitrogen will be supplied from a liquid nitrogen dewar at about 95 psia (65 kPa absolute) to a distribution box that is located above the feedbox level for each magnet of the Mu2e experiment. At the distribution box, the two phase nitrogen will be routed through a phase separator where it will be sub-cooled to about 90 K. Following this, the sub-cooled LN2 will be throttled to about 82 K with about 9% vapor content at the inlet to the solenoid magnet. Table 6.7 summarizes the heat loads in the PS at LN2 temperatures.

Table 6.7. Summary of heat loads at 80 K.

**300 K radiation heat load to 80 K surfaces**

| Surface | Surface area (m$^2$) | Number of MLI layers | Heat flux (W/m$^2$) | Heat load (W) |
|---|---|---|---|---|
| Outer thermal shield outer surface | 32.1 | 60 | 1.5 | 48.15 |
| Inner thermal shield inner surface | 20.85 | 60 | 1.5 | 31.3 |
| Total heat load from 300 K radiation | | | | 79.5 |

**Conductive heat load from supports to 80 K components**

| Component | Heat load per support (W) | Quantity | Total heat load (W) |
|---|---|---|---|
| Cold mass axial supports (TS end/short) | 0.75 | 6 | 4.50 |
| Cold mass axial supports (non-TS end/long) | 1.145 | 6 | 6.87 |
| Cold mass radial supports | 1.4825 | 16 | 23.7 |
| Thermal shield radial supports | 0.75 | 16 | 11.9 |
| Total conductive heat load to 80 K | | | 47.0 |

**Dynamic heat load on the thermal shield due to particle irradiation**

| | Total heat load on thermal shield (W) |
|---|---|
| MARS V15 particle simulation | 2 |
| Total heat load to 80 K components (static + dynamic) | 128.5 |

Both the inner and the outer thermal shields are cooled in series by a .5" ID extruded tube that is skip welded to the cold mass side of the shields. The tube has been sized to occupy the least space while providing for a low pressure drop along its entire run length and maintaining a positive pressure at the end of its run as it vents to the atmosphere.

***Multilayer Insulation***

Multilayer insulation or MLI, as it is commonly known, will be used to intercept the thermal radiation from the room temperature (300 K) level to the thermal shields' (80 K) level and from the 80 K level to the liquid helium or 4.7 K level. Optimal numbers of layers of MLI have been proven to be very effective [27] in intercepting thermal radiation, and their use is necessary for superconducting applications such as solenoid magnets and RF cryomodules. The recommended numbers of layers for the PS magnet are shown in Table 6.8.

Table 6.8. Recommended number of MLI layers.

| Location | Number of layers | Layer density |
|---|---|---|
| Between Cryostat walls and thermal shield, i.e. between 300 K level and 80 K level | 60 | 3 layers/mm |
| Between thermal shield and cold mass, i.e. between 80 K level and 4.7 K level | 30 | 3 layers/mm |

### *Cold Mass Suspension System*

The suspension system serves as the structural attachment of all cryostat systems to the vacuum vessel, which in turn anchors them to the experimental area floor. At the same time, it needs to impose as little thermal load on the cryogenic system as possible. These two requirements, structural integrity and low thermal load, are generally at odds with one another.

The PS cryostat cold mass suspension consists of axial and radial support systems for both the coil and thermal shield assemblies designed to resist axial magnetic and off-center forces as well as gravity loads. The maximum axial force on the cold mass is estimated in Section 6.3.1.6 to be 1427 kN toward the TS. The maximum force away from the TS is 115 kN and the weight of the cold mass plus the force due to possible misalignment is 158 kN. The complete cryostat assembly and the location of the coil and thermal shield are shown in Figure 6.13.

### *Coil Assembly Axial Anchor*

In order to minimize the length of any penetration out of the end of the vacuum vessel, the PS axial anchors run the entire length of the cryostat and anchor the cold mass near the center of the vacuum vessel. Twelve axial rods, six running from each end of the vacuum vessel and attaching to the cold mass about one-third of the overall length from the TS end constitute the anchor system. The axial anchor coil attachment point allows for radial shrinkage of the cold mass during cool-down.

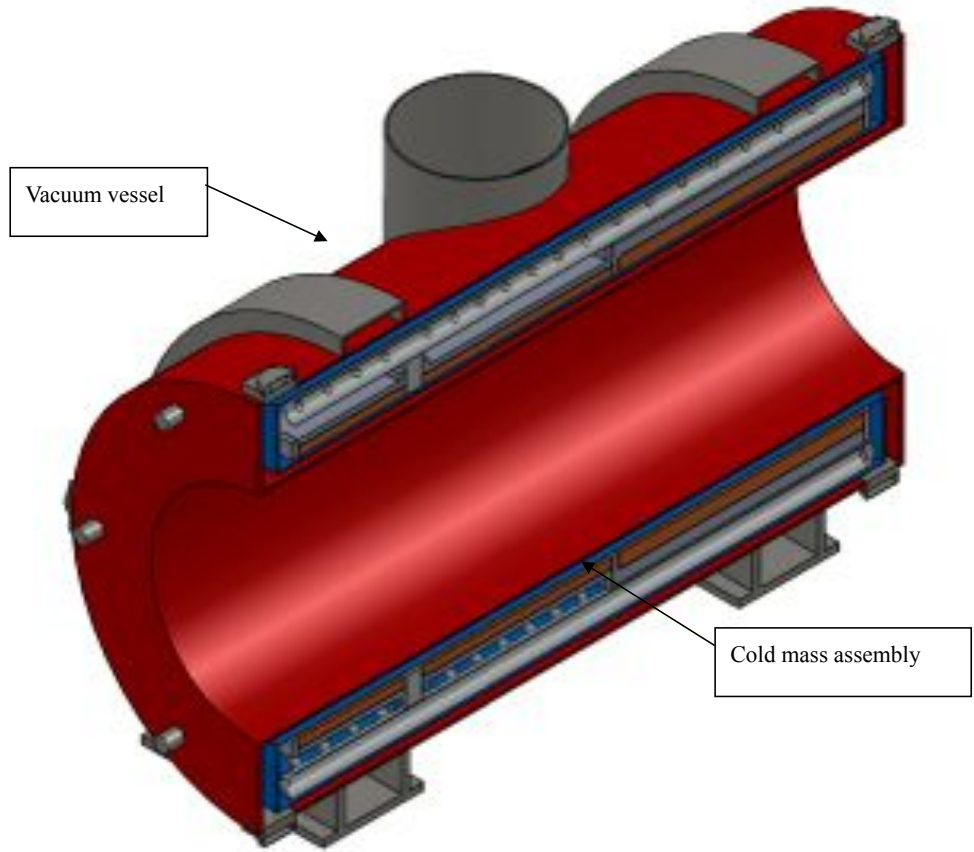

Figure 6.13. PS cryostat assembly cross-section.

A series of Belleville springs at the vacuum vessel end of each rod accommodates a maximum of 3 mm of axial contraction of the rods themselves during cool-down. In this configuration, each set of six rods resists axial loads in each of two directions. In both cases, the set not in use is allowed to slide, unloading the Belleville springs rather than allowing them to be compressed. The details of the axial anchor center and end attachments are described in [10].

Several candidate materials were studied for the anchor rods themselves. As mentioned earlier, they must be structurally strong and present a relatively low heat load to the cryogenic system. Early in the design process, we elected to consider an all-metal anchor system for reasons of ease of attachment and overall reliability. Starting with materials commonly used in superconducting magnets and using a figure of merit defined as the ratio of allowable stress divided by thermal conductivity to 4.7 K, the best choices were determined to be titanium and Inconel 718 with the latter being chosen primarily due to availability and ease of fabrication, especially welding. The allowable stress for Inconel 718 in the annealed and precipitation hardened condition is 530 MPa.

To resist the 1427 kN maximum axial load acting toward the TS with an adequate safety factor requires 6 28.58 mm diameter rods. A much smaller 115 kN axial force acts away from the TS during some fault conditions. This force is resisted by 6 12.7 mm diameter rods.

*Coil Assembly Radial Support*

The radial supports oppose the weight of the magnet and any off-center load that exists due to misalignment of the coil inside the vacuum vessel. The assumed maximum radial load is 2 g's or 20 tonnes. As with the axial supports, Inconel 718 was chosen for a series of 16 tension rods arranged in pairs at each end of the cold mass. The coil end is attached by a spherical clevis. A spherical washer pair attaches the vacuum vessel end. Belleville springs at the warm end accommodate thermal contraction during cool-down. The details of the radial support attachment to the cold mass and the vacuum vessel are described in [10].

Figure 6.14 shows the complete coil assembly and the axial and radial supports.

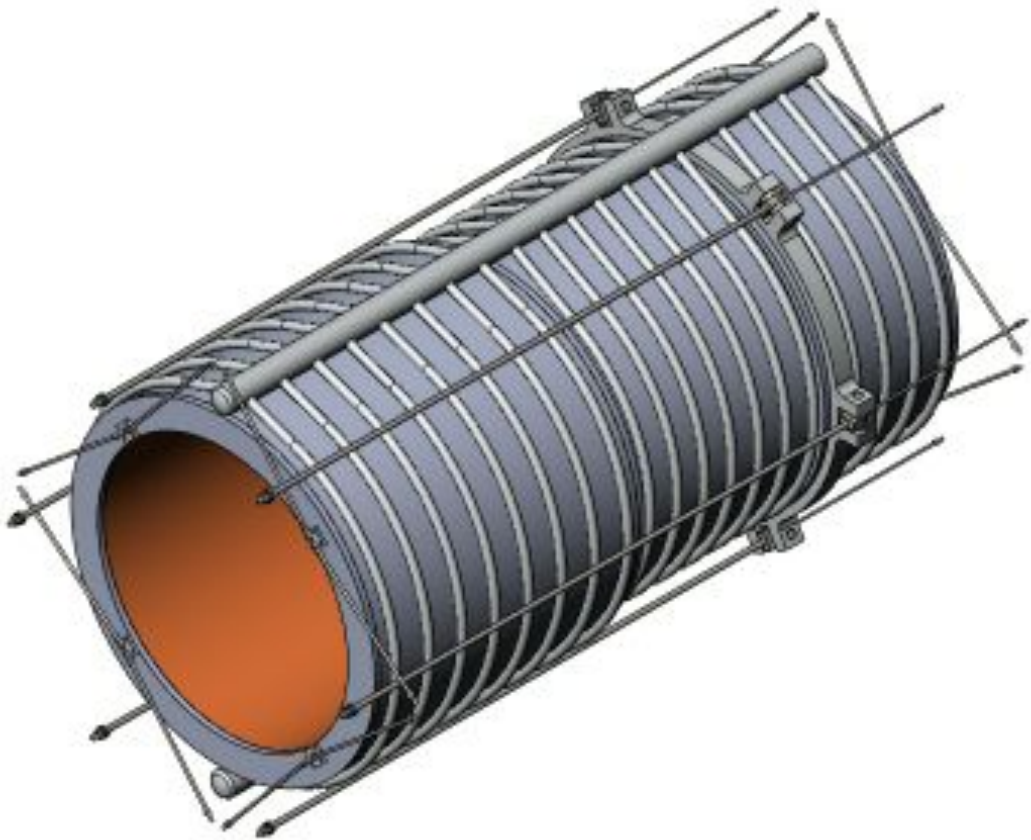

Figure 6.14. Figure PS cold mass suspension system.

*Thermal Shield Axial and Radial Supports*

The thermal shield supports resist gravity and shipping and handling loads. The weight of the complete thermal shield assembly is approximately 1 tonne. The thermal shield is attached to the cold mass axial supports at the 80 K intercept on the 28.58 mm support rods, providing a natural axial support point for the thermal shield assembly. The radial supports are lightly loaded and are identical to those on the cold mass shown above. They share the design with the cold mass supports for commonality of design. Since they only have connections to the vacuum vessel and thermal shield, they don't contribute to the 4.7 K heat load.

*Summary*

The coil and thermal shield suspension systems have been designed to support these structures when they are subjected to magnetic forces, misalignment inside the vacuum vessel, magnet quenches, gravity, shipping and handling, and loads from the radiation absorber. Each set of components has been sized to meet the unique requirements of these loads.

### 6.3.1.4  Instrumentation

The production solenoid requires a large number of instrumentation sensors and associated wiring to be installed during fabrication. This instrumentation falls into four groups:

- Quench protection
- Cryogenic monitoring and controls
- Mechanical characterization
- Conductor resistance monitoring

***Voltage Taps***

Voltage taps are critical elements of the magnet protection system. Voltage tap connections and cabling must be robust, vacuum compatible, radiation hard and able to withstand voltages up to 1.7 kV (during the Hipot test). Voltage tap cables must be bundled and routed to minimize pickup and cross-talk.

*Tap Locations*

A schematic of the coil and leads voltage tap connections and cabling is described in [10]. Voltage taps are the physical connections to the bus in which the instrumentation wires for quench protection and resistance measurements are terminated. A tap can be made by drilling a hole into the aluminum stabilizer of the solenoid coils or leads so non-magnetic screws can be used to secure the wires. Taps are made on either side of each splice between the layers of each coil, on either side of each splice between coils, and at the ends of each lead including the trim lead. Each tap will have two wires connected to them.

### Superconducting Wire Sensors

In addition to the voltage taps across the solenoid leads superconducting NbTi twisted pairs bonded to the leads will be used to detect quenches. The wires must be bonded to the leads in a way that provides electrical isolation while maintaining good thermal contact. During normal operation, the wires will be superconducting. In the event the lead quenches, the wire should also heat above the transition temperatures, so quenches in the leads may be detected by monitoring the wire resistance.

### Temperature Monitoring

The temperature of the magnet coils, heat shield, and support rods must be monitored. Resistive Temperature Devices (RTDs) will monitor the temperatures of the following solenoid components:

- Cold-mass during cool-down and warm-up
- Cold-mass near support rod mounts
- Magnet Warm spots
- Supply manifold
- Return manifold
- 80 K Heat Shield
- Radial Support Rods
- Axial Support Rods

Cernox RTDs will be used to monitor components operating at temperatures close to that of liquid helium, such as the cold mass. Platinum sensors will be used for components operating at higher temperatures, such as the 80 K thermal shield. A description of the temperature sensors is given in [10].

### Strain Gauges

Temperature compensated strain gauge full bridges installed on each radial and axial support post will monitor the forces on the cold mass. A description of the strain gauges is given in [10].

- Strain gauges can be connected using a 32 AWG phosphor bronze triple twisted pair.
- Strain gauge cables must be thermally anchored to at least one point.

- The recommended gauge, Vishay Micromeasurement Model EK-06-250TB-1000 dual strain gauge rosette, provides a flexible polyimide backing to allow the gauges to be mounted easily on the curved surfaces of the rod in combination with a nickel-chromium alloy sensitive to temperatures as low as -195 °C. The 0.25 inch length/1000 Ohm resistance gauge is the longest, highest resistivity gauge available with these characteristics.

**Position Sensors**

The position of the coil with respect to the vacuum vessel in all three dimensions shall be monitored at both ends of the magnet using a total of 12 non-contact, non-magnetic sensors, for example Philtec DMS-RC290 optical sensors or equivalent. The position sensors must have an operating range of at least 40 mm and an integral linearity and a long term stability of 100 μm or better. Two position sensors at each end of the cold-mass in the horizontal symmetry plane shall monitor the X position of the coil and two sensors in the vertical symmetry plane at each end will monitor the Y position of the coil. The Z position will be monitored by 4 sensors mounted longitudinally in the horizontal symmetry plane.

**RRR Monitors**

High radiation levels in the PS during operation are expected to degrade the residual resistance ratio (RRR) of the copper and aluminum stabilizers. RRR monitors (RRRM), described in [10], consisting of aluminum and copper wires wound around circuit boards shall be attached to the cold-mass to allow for measurement of radiation-induced resistance changes.

### 6.3.1.5  Interfaces

*Cryostat Inner Shell*

The PS cryostat inner shell is part of the PS Cryostat detailed in Section 6.3.1.3 of this document. The inner shell is a primary component of the PS insulating vacuum system (Section 6.3.1.3) on its OD as well as a primary component of the Muon Beamline Vacuum System on its ID; therefore, it must be designed and manufactured with vacuum performance in mind.

All surfaces must be thoroughly cleaned and degreased before welding. Shell longitudinal and circumferential welds must be designed to minimize the chance of virtual leaks in the vacuum systems. All inner bore welds must be ground smooth.

The PS cryostat inner shell provides the interface for a Heat and Radiation Shield (HRS) that will be supplied, installed, and connected by Fermilab. The HRS is a large cylindrical weldment that will be installed by sliding it into the upstream end of the PS bore.

The PS inner shell must support the HRS gravity loads, as well as the HRS installation and operating forces as described in [28].  The HRS is undergoing design simulations, so the specification of the PS interface features described below will be finalized once the HRS design has been finalized.

*HRS Downstream Connection*
After the HRS is inserted into the PS bore and alignment measurements are taken, the downstream end of the HRS will be shimmed and welded to the PS cryostat inner shell. A feature required in the PS is a weld ring in the cryostat bore. The HRS weld ring must be manufactured from 316L stainless steel and welded into the PS cryostat bore 597.0 mm from the downstream end plate as shown in Figure 6.15. The HRS weld ring thickness must be 0.76 mm, and the final inner diameter of the ring after welding must be 1295 mm.

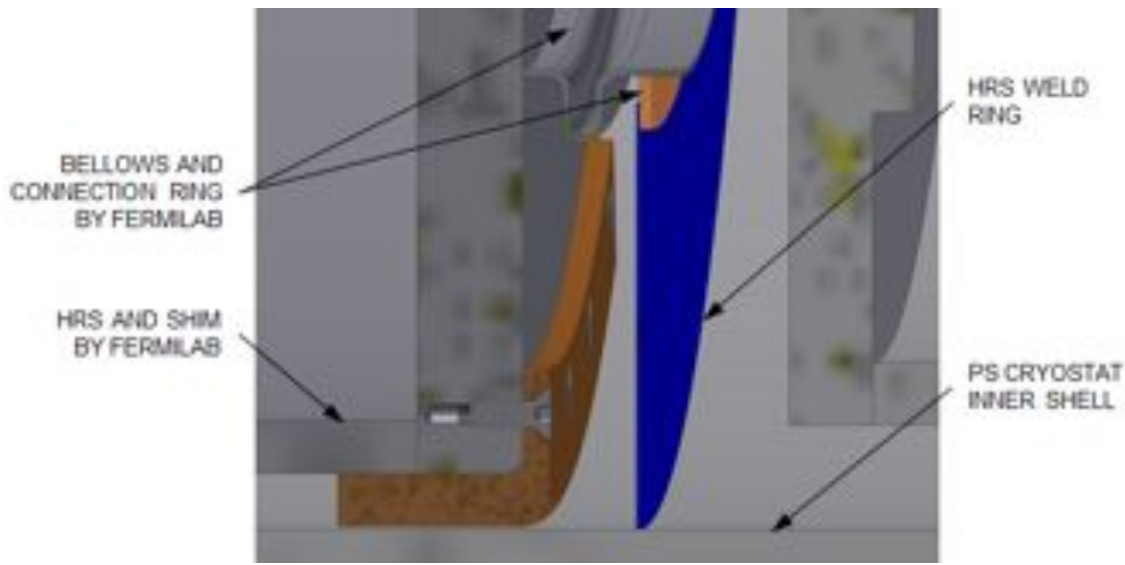

Figure 6.15. HRS downstream connection to PS.

The HRS weld ring is also a component of the muon beamline vacuum system and must be designed and manufactured with vacuum performance in mind.  All surfaces must be thoroughly cleaned and degreased before welding.  The weld must be continuous on the downstream face of the ring; skip welds are allowed on the upstream face.

*HRS Upstream Connection*

The upstream end of the HRS will also be welded to the PS after HRS installation and alignment. The feature required in the PS is a stub extension of the PS cryostat inner shell that allows a heavy structural and vacuum weld to be made far enough away from the PS to protect the cold-mass insulation from weld heat. Figure 6.16 and Figure 6.17 show a possible stub connection to the PS, and the HRS flange.

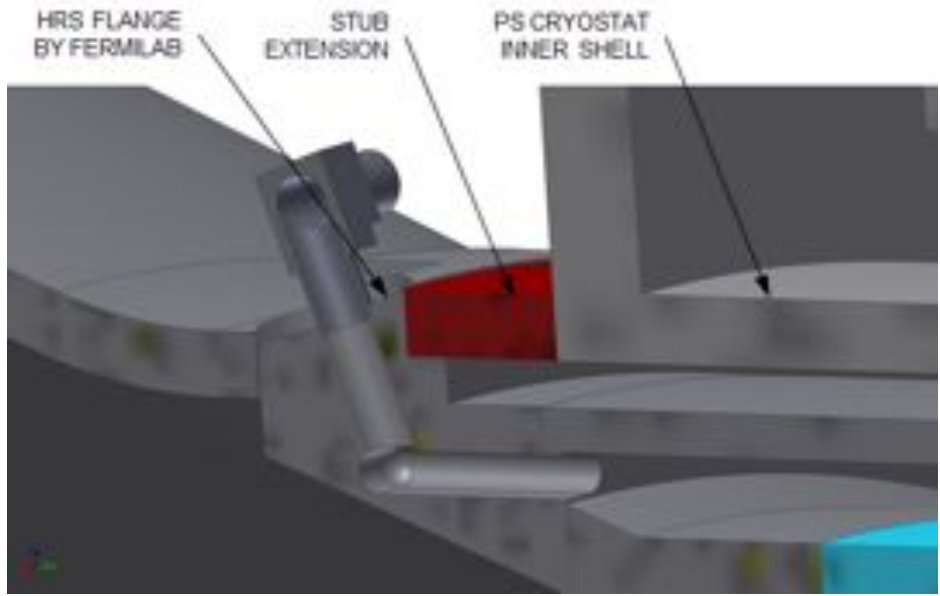

Figure 6.16. Figure 7.16 HRS upstream connection to PS.

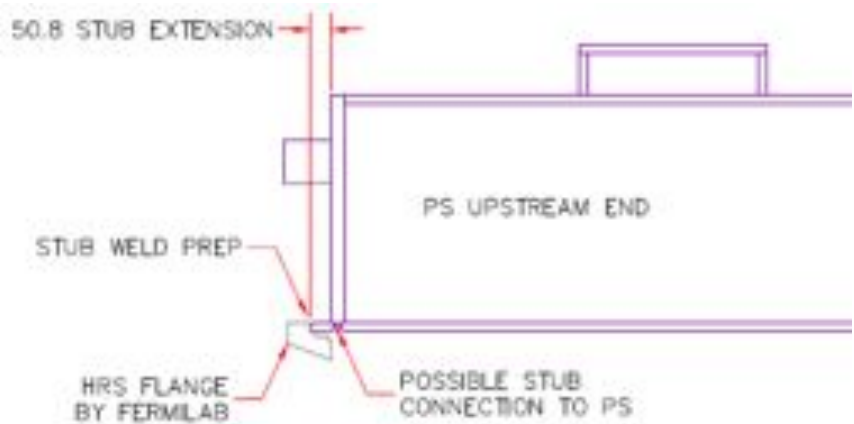

Figure 6.17. Stub extension geometry and connections.

The PS inner shell stub extension must be manufactured from 316L stainless steel and welded to the PS cryostat. The stub must extend 50.8 mm from the upstream cryostat end plate and the stub shell thickness must be 20.0 mm. The upstream stub end must have an outer diameter weld prep for attaching to the HRS flange.

The final inner diameter of the stub after welding must be 1500.0 mm and the final outer diameter of the stub after welding must be 1540.0 mm. The inner diameter of the stub and the PS cryostat inner shell must provide a contiguous bore diameter to allow the HRS to slide into place as described in Section 6.3.1.5.

### Connection to TS

The PS Solenoid must connect to the TSu Solenoid to provide a common muon beamline vacuum system volume. The connection will be made with a flexible bellows to allow any combination of axial and lateral movement produced by cool-down and magnetic forces between solenoids. The bellows will be procured and welded in place to both the PS and the TSu by Fermilab.

A feature required in the PS is a weldable face on the downstream cryostat end plate. The downstream face of the cryostat end plate in the area between 1800.0 mm diameter and 1850.0 mm diameter must be free from steps, discontinuities, or voids.

Any welds on the downstream face of the PS cryostat end plate must be ground smooth in this area. Fermilab will align TSu and PS and make the light vacuum weld to the PS face so that the bellows is in a neutral, un-flexed position as shown in Figure 6.18.

### Connection to Transfer Line

The PS Solenoid must connect to a transfer line that supplies the magnet with power and cryogenic fluids. The transfer line also provides a pump-out duct for the PS insulating vacuum and acts as a conduit for instrumentation signals entering and leaving the magnet.

A section of the transfer line that will connect to this extension port is shown in Figure 6.19.

The feature required in the PS is a transfer line extension port protruding from the side of the PS cryostat chimney. The PS transfer line extension port must be manufactured from 316L stainless steel tube with a 254.0 mm (10.0 inch) OD and a 2.8 mm (0.109 inch) wall.

The PS transfer line extension port must be positioned 1005.0 mm in the negative X direction from the PS vertical centerline and 1294.0 mm above the horizontal centerline of the PS bore as shown in Figure 6.20. The port must extend 550.0 mm in positive Z from the center of the PS chimney.

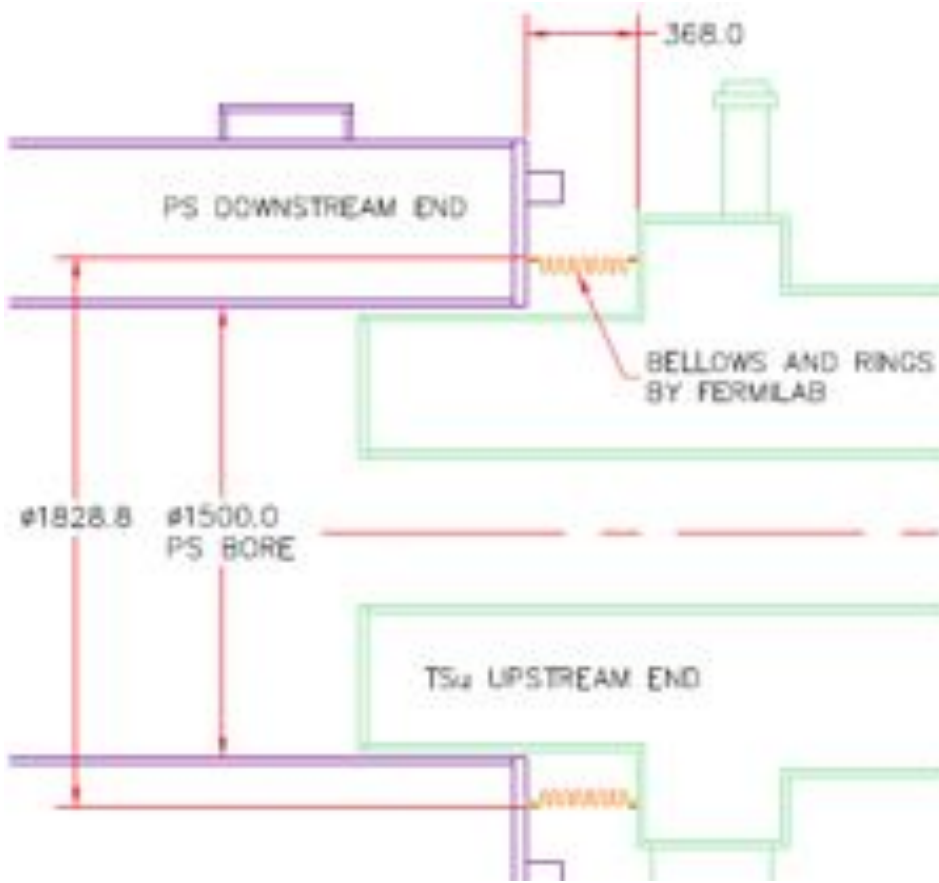

Figure 6.18. Connection between PS and TSu.

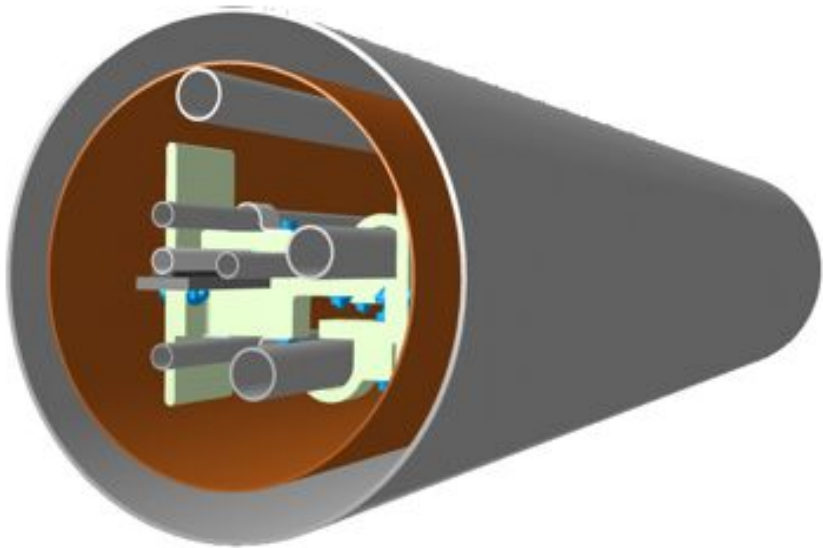

Figure 6.19. Section of transfer line.

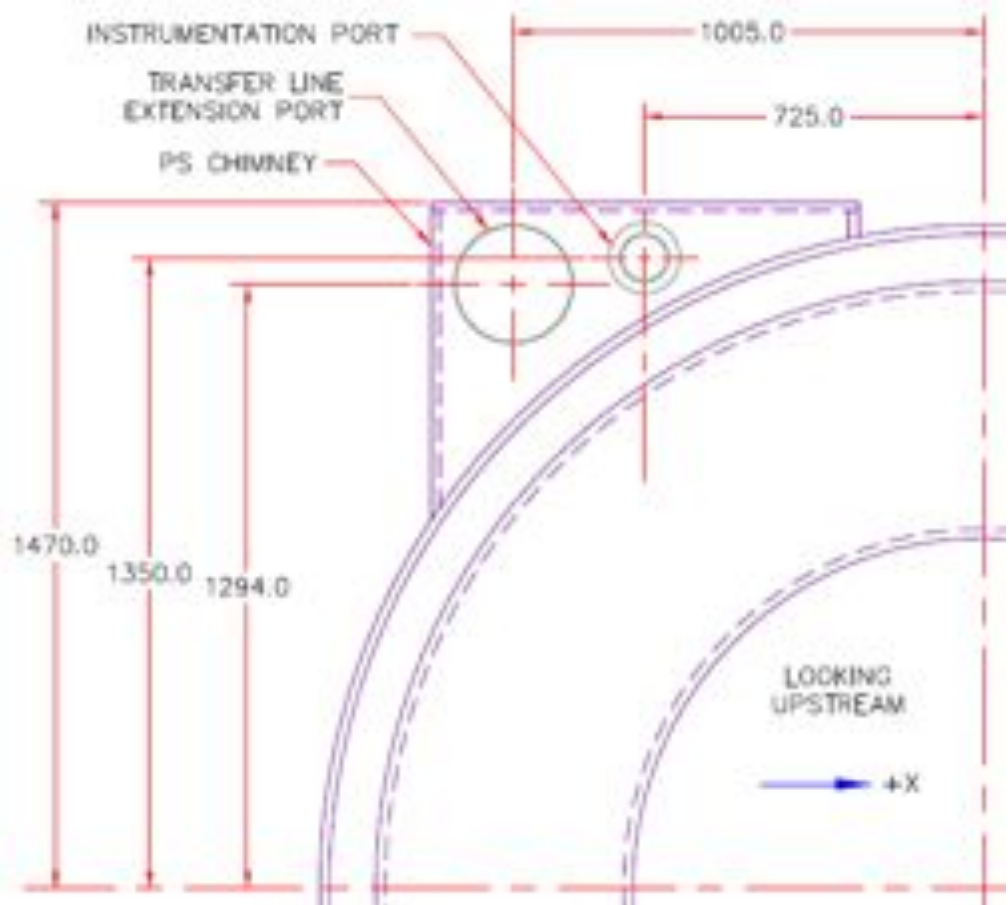

Figure 6.20. PS chimney ports end view.

The PS transfer line extension port is also a component of the PS insulating vacuum system and must be designed and manufactured with vacuum performance in mind. All surfaces must be thoroughly cleaned and degreased before welding. The weld must be continuous outside the PS chimney; skip welds are allowed inside the chimney.

*Superconductor and Piping*
The PS superconductor and all the cryogenic pipes required for magnet operation will enter the PS through the transfer line. The PS features necessary to make these connections are the extension of the PS conductor and piping out the transfer line extension port described above. See also Section 6.3.1.2 and Section 6.3.1.3 of this document.

The PS must be delivered with 2.0 ± 0.1 m of each superconducting cable, and 1.0 ± 0.1 m of each cryogenic pipe extending out beyond the end of the transfer line extension port. The pipes must be sealed and superconductor supported and protected from contamination and damage in shipping and handling.

*Thermal Shields and Insulation*

The thermal shields in the PS must be thermally connected to the thermal shields in the transfer line extension port. This may be accomplished by connecting the thermal shield sections together through the thermal shield in the PS chimney. The PS must be delivered with 0.50±0.05 m of thermal shield extending out beyond the end of the transfer line extension port. The thermal shield must be protected from contamination and damage during shipping and handling. See also Section 6.3.1.3 of this document.

Likewise, the 4 K and 80 K MLI blankets in the PS must be connected and continuous through the transfer line extension port. This also may be accomplished by connecting the blanket sections together through the MLI blankets in the PS chimney. The PS must be delivered with 0.50±0.05 m of MLI extending out beyond the end of the transfer line extension port. The MLI must be protected from contamination and damage during shipping and handling.

*Instrumentation*

All instrumentation cables coming into or out of the PS must be routed through the transfer line extension port. This may be accomplished by routing all instrumentation through the PS chimney in a way that protects the cables from mechanical and thermal damage. Section 6.3.1.4 of this document specifies that some cables must be routed along the conductor leads. All cables must be strain relieved and secured inside the solenoid.

The PS must be delivered with 2.0±0.1 m of all instrumentation cables extending out beyond the end of the transfer line extension port. All instrumentation cables and labeling must be protected from contamination and damage during shipping and handling.

**Instrumentation Port**

It may not be feasible to route all solenoid instrumentation through the transfer line as described above. For this reason, an additional instrumentation port is required coming out the side of the PS cryostat chimney. The PS instrumentation port must be manufactured from 316L stainless steel tube, 101.6 mm (4.0 inch) OD x 2.11 mm (0.083 inch) wall.

The PS instrumentation port must be positioned 725.0 mm in the negative X direction from the PS vertical centerline and 1350.0 mm above the horizontal centerline of the PS bore as shown in Figure 6.20. The port must extend 700.0 mm in the positive Z direction from the center of the PS chimney.

***Insulating Vacuum System***

The PS cold mass resides in an insulating vacuum space formed between the inner and outer cryostat shells and the upstream and downstream cryostat end plates. The inner bore of the PS cryostat is also part of the Muon Beamline Vacuum space. Therefore, the PS cryostat must be designed to operate under any combination of vacuum and atmospheric pressures including: vacuum in the insulating vacuum space and atmospheric pressure in the PS bore; atmospheric pressure in the insulating vacuum space and vacuum in the PS bore; and vacuum in both the bore and insulating vacuum spaces.

The specified operating pressure of the PS insulating vacuum system is $1 \cdot 10^{-5}$ Torr at maximum. The specified operating pressure of the Muon Beamline Vacuum system is $1 \cdot 10^{-4}$ Torr at maximum. The insulating vacuum features required on the PS are pump-out ports. The transfer line extension port above is one pump-out port, and an additional series of radial ports on the outer cryostat shell must be provided. Possible locations of these ports are shown in Figure 6.21.

The PS radial pump-out ports must be manufactured from 316L stainless steel tube, 203.2 mm (8.0 inch) OD.

***Shipping Restraints***

The PS axial and radial supports described in Section 6.3.1.3 are not intended to restrain the maximum forces that might be induced from shipping and handling. Therefore, additional shipping restraints must be added to the PS cryostat to keep the cold mass positioned within the cryostat and protect the axial and radial supports.

A possible implementation of these shipping restraints is shown in Figure 6.21. The ports on the outer cryostat shell for the insulating vacuum may be used for connection of radial shipping restraints. Axial threaded holes may be provided in each cryostat end plate.

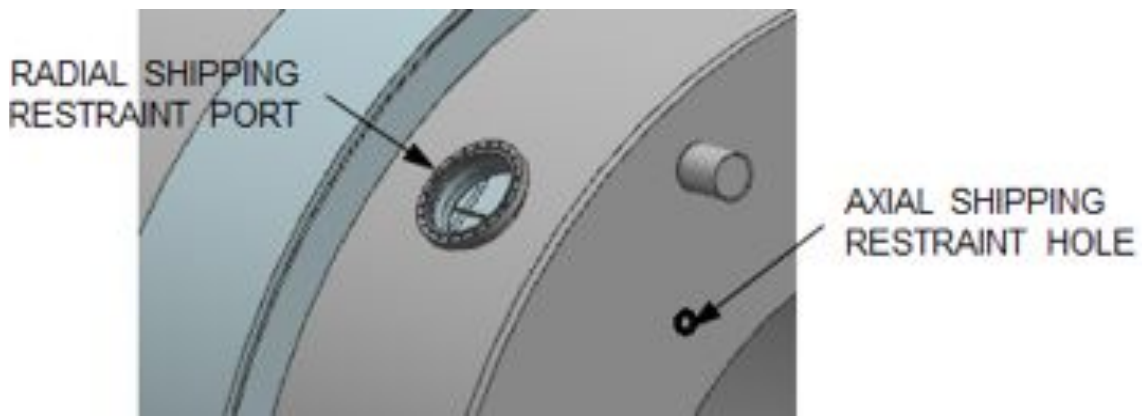

Figure 6.21. PS axial and radial shipping restraints.

***Magnet Support***

The PS Solenoid gravity and magnetic loads are ultimately transferred through the PS support saddles shown in Section 6.3.1.3 of this document to the floor. The PS support feet will be bolted to a support frame by Fermilab that in turn will be bolted to pads provided in the floor of the Mu2e building. A side view of the PS and support frame is shown in Figure 6.22. The interface features required as part of the PS are the two support feet that weld to the reinforced saddles of the cryostat. The support feet must be manufactured from 316L stainless steel with a length of 2528 mm, a width of 700.0 mm, and a thickness of 50.0 mm.

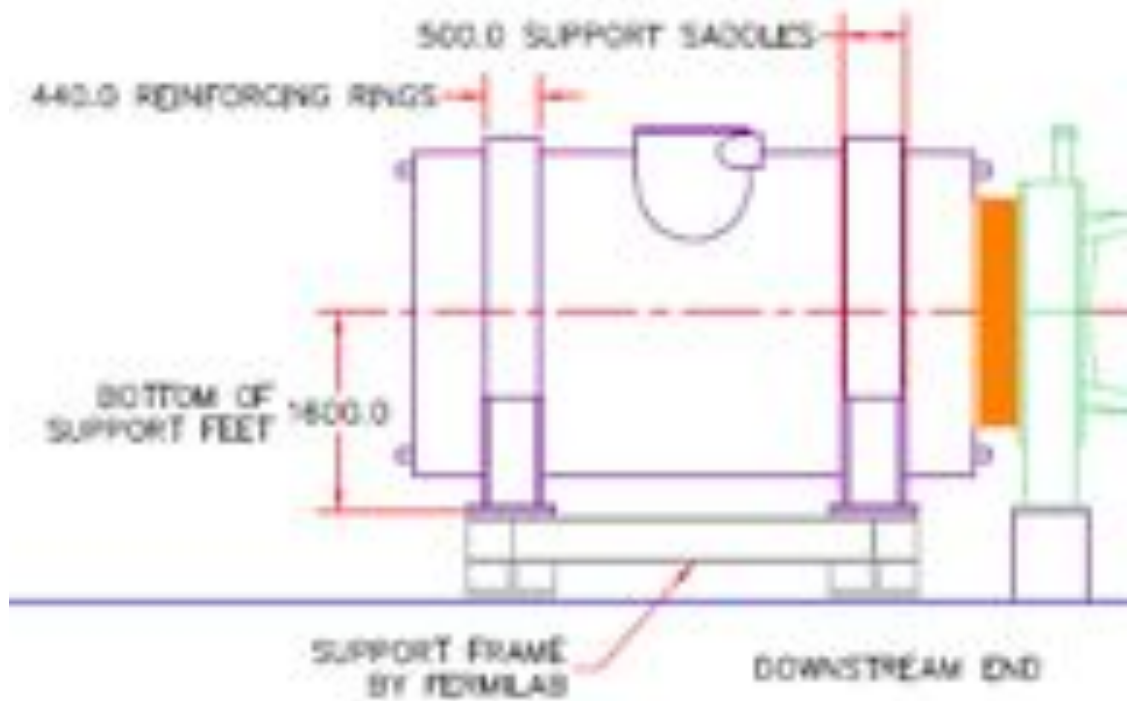

Figure 6.22. PS mounted on support frame.

### *6.3.1.6   Analyses*

***Magnetic Analysis***

According to the requirements document [2], the PS magnet should have magnetic field gradient along the magnet axis with the peak field on the axis of 4.6 T and the field at the interface with the Transport Solenoid (TS) of 2.5-2.7 T. It implies that the current per unit of length must vary by a factor of ~2 along the magnet.

In order to satisfy this requirement, the magnet consists of three sections with 3, 2 and 2 layers of the cable described in Figure 6.6 and Table 6.3, wound in the hard way around the aperture.

Figure 6.23 shows the COMSOL Multiphysics FEM magnet model with the flux density diagram.

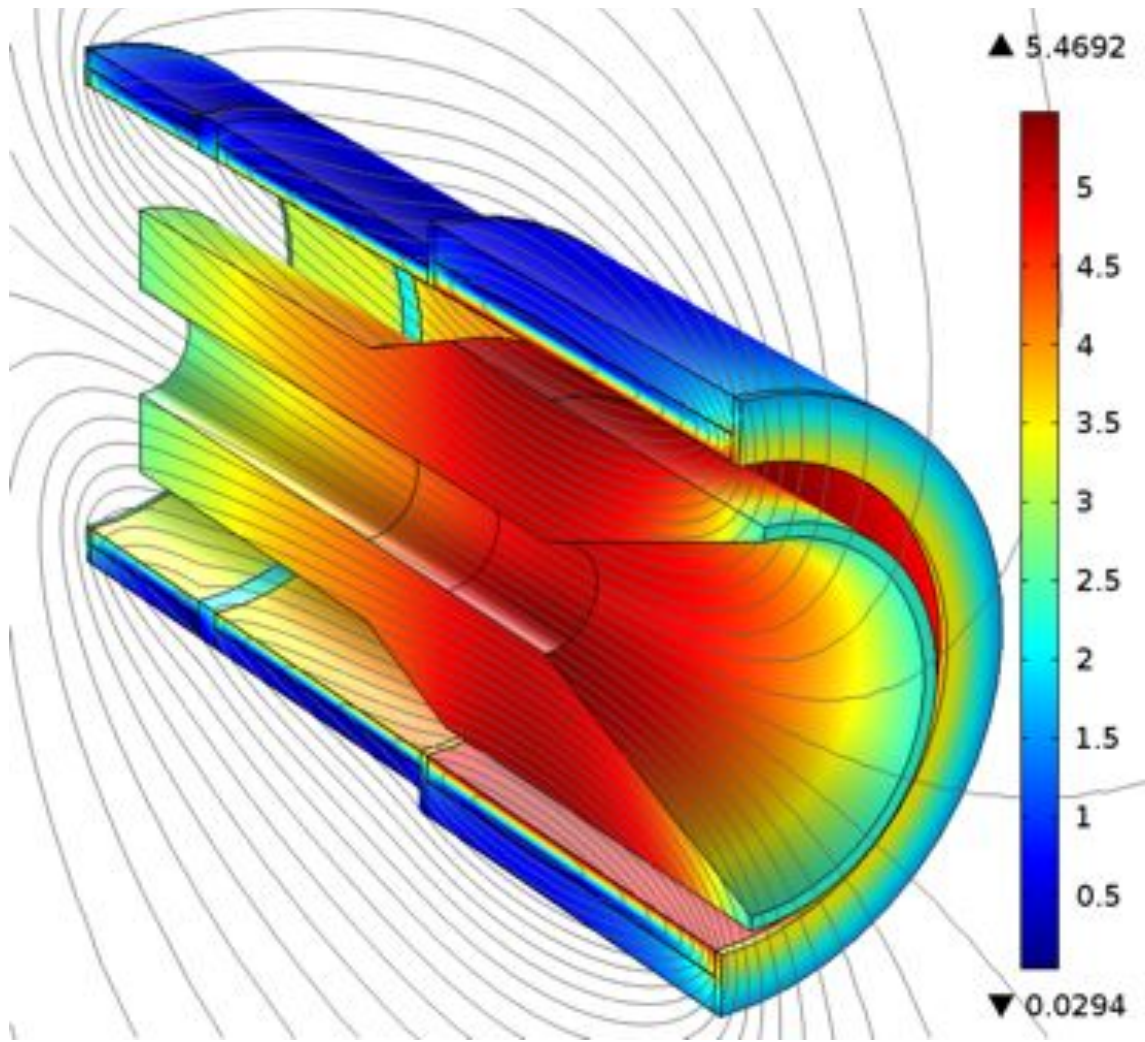

Figure 6.23. PS magnet model with the flux density in Tesla at the maximum current. HRS and the coil support shells are also shown.

Note that the pictures also show the straight section of the TS isincluded in the model for the field matching purposes. Also included is the Heat and Radiation Shield (HRS) [30] made from C63200 bronze with a nominal relative magnetic permeability of 1.04. Since it is not an electro-technical material with well controlled magnetic properties, the magnetic permeability was increased to 1.1 during modeling to account for possible property variations.

In order to guarantee meeting the peak field requirement of 4.6 T, the magnet is designed to operate with 5.0 T on the axis while meeting all other requirements specified in [2]. Since the PS should provide the same field at the interface with the TS regardless of the peak field on the axis, adjustment of the peak field (and consequently the axial field gradient) is accomplished by a trim power in TS. The field in the central part of the magnet is shown in Figure 6.24 together with the specification.

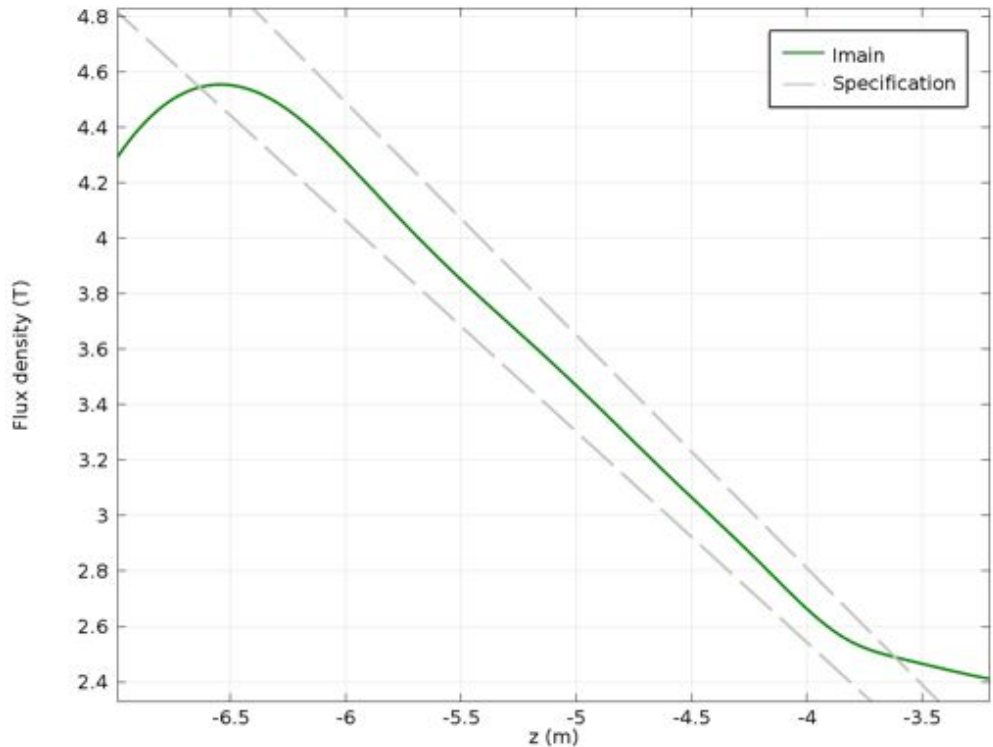

Figure 6.24, Flux density and the field specification in the central part of the magnet in the global coordinate system at the nominal operating current.

The magnet is designed to operate at 63 % of the short sample limit (SSL) along the load line at the nominal field and the temperature of 4.7 K, as shown in Figure 6.25. At higher temperatures, the SSL shifts closer to the operating point and passes through it at a temperature of 6.60 K, which is the current-sharing temperature at the nominal operating current. In order to maintain the temperature margin requirement of 1.5 K at that current, the coil temperature cannot exceed 5.10 K. In that case, the magnet operates at 68 % with respect to the SSL at that temperature. The major PS parameters are summarized in Table 6.11 for the nominal and maximum operating conditions.

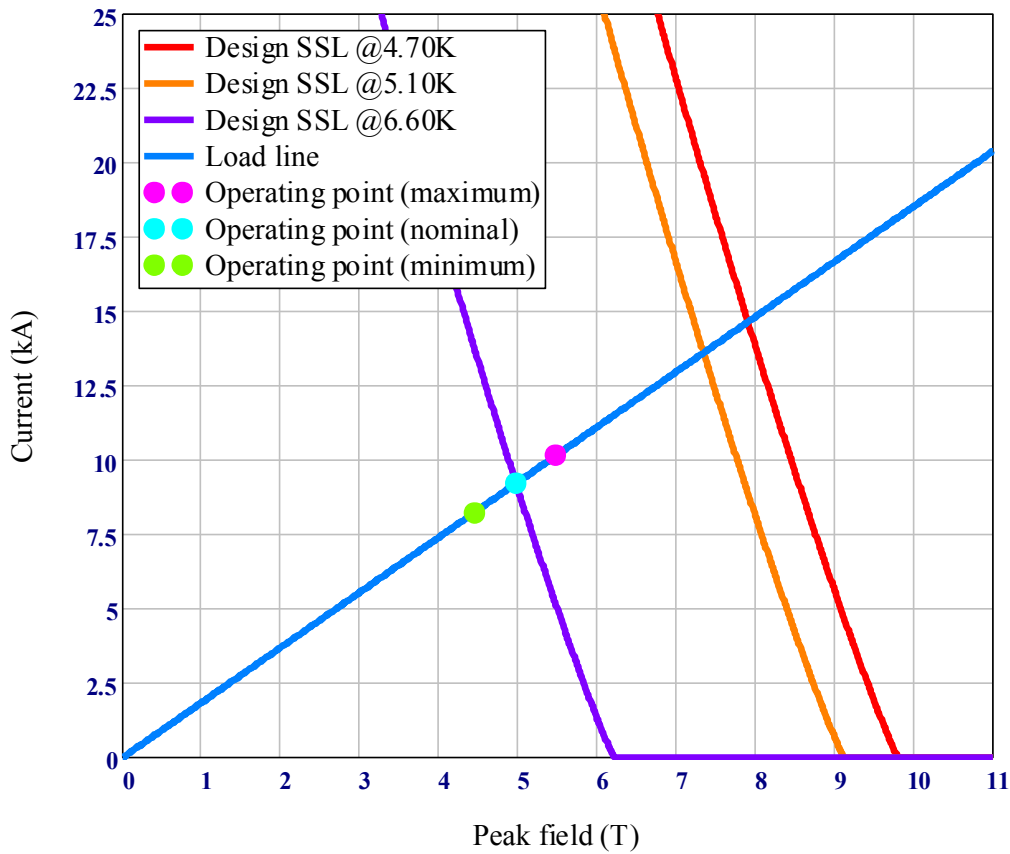

Figure 6.25. Critical currents and the magnet load line.

During normal operation, the axial force on the PS coil acts toward the TS coil. The axial and transverse force sensitivity due to possible PS to TS and PS to HRS misalignment was studied using the COMSOL and OPERA 3D models. The forces were calculated at the nominal coil position as well as during the coil displacement. Table 6.9 and Table 6.10 list the results of this analysis. The forces scale linearly with the displacements much smaller than the transverse coil size.

Table 6.9. Forces on the PS coil due to interaction with the TS coil and the force sensitivity due to coil misalignment when PS operates at 5.0 T on the axis and the TS is at the nominal field. HRS is not present.

| Force on PS coil | At the nominal position, kN | Sensitivity to the coil displacement, kN/cm |
|---|---|---|
| $F_x$ | -15.7 | -9.0 |
| $F_y$ | 0.0 | |
| $F_z$ | 1361.9 | 18.0 |

Table 6.10. Forces on the PS coil due to interaction with the HRS and the force sensitivity due to coil misalignment when the PS operates at 5.0 T on the axis and the HRS has a magnetic permeability of 1.1. TS is off.

| Force on PS coil | At the nominal position, kN | Sensitivity to the coil displacement, kN/cm |
|---|---|---|
| $F_x$ | 0 | |
| $F_y$ | 0 | 19.5 |
| $F_z$ | 29.0 | -18.6 |

Table 6.11. Magnet parameters.

| Parameter | Unit | Operating current | |
| | | Nominal | Maximum |
|---|---|---|---|
| Liquid helium temperature ($T_{LHe}$) | K | | 4.7 |
| Operating current ($I_{op}$) | A | 9200 | 10150 |
| Peak axial field at $I_{op}$ | T | 4.56 | 5.01 |
| Peak coil field at $I_{op}$ | T | 4.96 | 5.48 |
| Quench current at $T_{LHe}$ | A | | 14649 |
| Current-sharing temperature at $I_{op}$ | K | 6.60 | 6.30 |
| Minimum temperature margin | K | | 1.50 |
| Maximum allowable temperature ($T_{maxall}$) | K | 5.10 | 4.80 |
| Fraction of SSL at $T_{LHe}$ | | 0.628 | 0.693 |
| Fraction of SSL at $T_{maxall}$ | | 0.676 | 0.705 |
| Stored energy | MJ | 66.83 | 79.74 |
| Self field inductance | H | | 1.58 |
| Fast dump resistance | mW | | 59.11 |
| Peak dump voltage | V | 544 | 600 |
| Initial time constant of fast discharge | s | | 26.7 |
| Cable length | km | | 8.67 |
| Cryostat inner diameter | m | | 1.50 |
| Cryostat length | m | | 4.50 |
| Cold mass | tonnes | | 13.1 |
| Cryostat mass (excl. HRS) | tonnes | | 10.7 |

The force sensitivities to the PS coil displacement relative to those of the TS and HRS have opposite signs and would partially cancel each other if the HRS to the TS position were fixed. It is, however, likely that the HRS will be displaced with the PS coil since it is supported by the PS cryostat wall and the cancellation will not happen. Moreover, the HRS may already be displaced in relation to the PS coil in opposition to PS-TS displacement.; thus, there is a possibility of a worst-case scenario in which the displacement forces are additive, regardless of the sensitivities' signs.

### Radiation Analysis

Simulations were performed with the MARS15 code [31] on the model shown in Figure 6.26. The neutrons were propagated down to 0.001 eV. The full set of critical radiation quantities was calculated. The superconducting coil material was described using a homogeneous approach (the material was represented as a mix of all elements with appropriate weight factors). The origin of the MARS15 coordinate system in the global coordinates is at $X_0 = 3.904$ m, $Y_0 = 0$ m, $Z_0 = -9.068$ m.

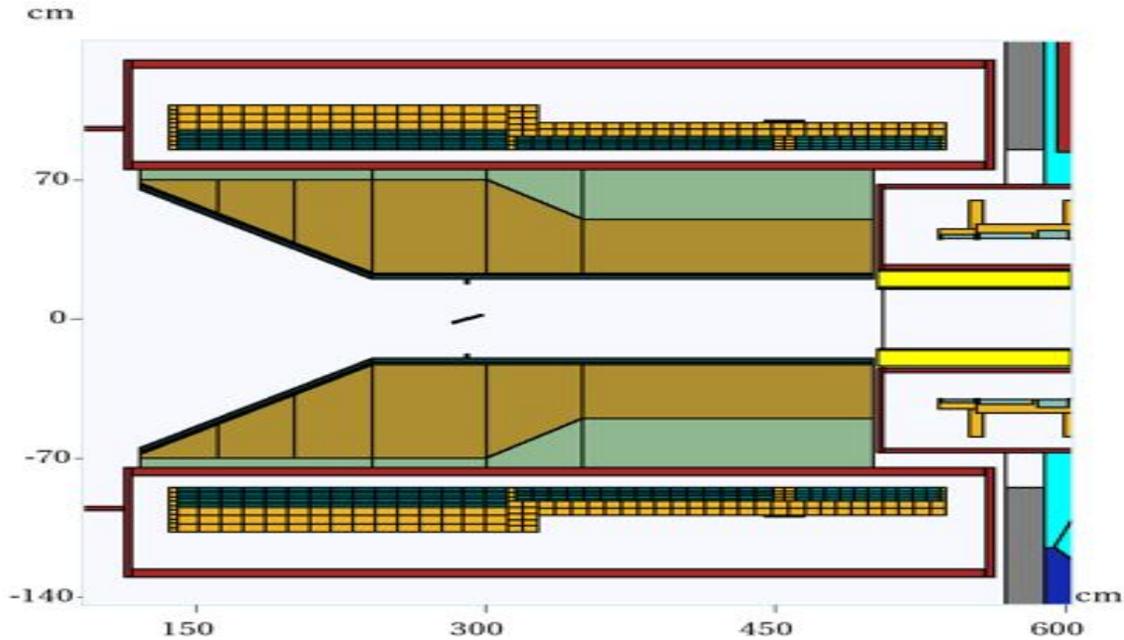

Figure 6.26. MARS15 model of the PS magnet and the HRS.

The peak calculated Displacement per Atom (DPA), which characterizes radiation damage in metals, was ~$2.5 \cdot 10^{-5}$ yr$^{-1}$ and the peak power density was 12.6 µW/g [32]. The corresponding peak absorbed dose for the coil insulation is 240 kGy/yr; total dynamic heat load in the cold mass is 28 W.

While these quantities are a factor of 2-3 less than the limits set in the requirements [9], the power density distribution in the cold mass is strongly non-uniform as shown in Figure 6.27, which creates a localized hot spot in the middle of the first coil. Because of this and the degradation of material properties under irradiation, a careful thermal analysis (presented later in this section) is required to determine if the design meets the peak temperature requirements.

### Degradation of RRR

Residual Resistivity Ratio (RRR) is an important parameter for superconducting magnet design that affects the magnet performance during operation in superconducting mode and irreversible transition to the normal state (quench). RRR of aluminum and copper stabilizers used in the coil is considerably degraded under neutron irradiation at cryogenic temperatures. A number of experiments have been performed in order to estimate the effect of neutron radiation on metal resistivity [33][34], however, the different (reactor) neutron energy spectrum in these experiments requires the numerical analysis to interpret the data for the Mu2e energy spectrum, which carries a large (factor of two) uncertainty.

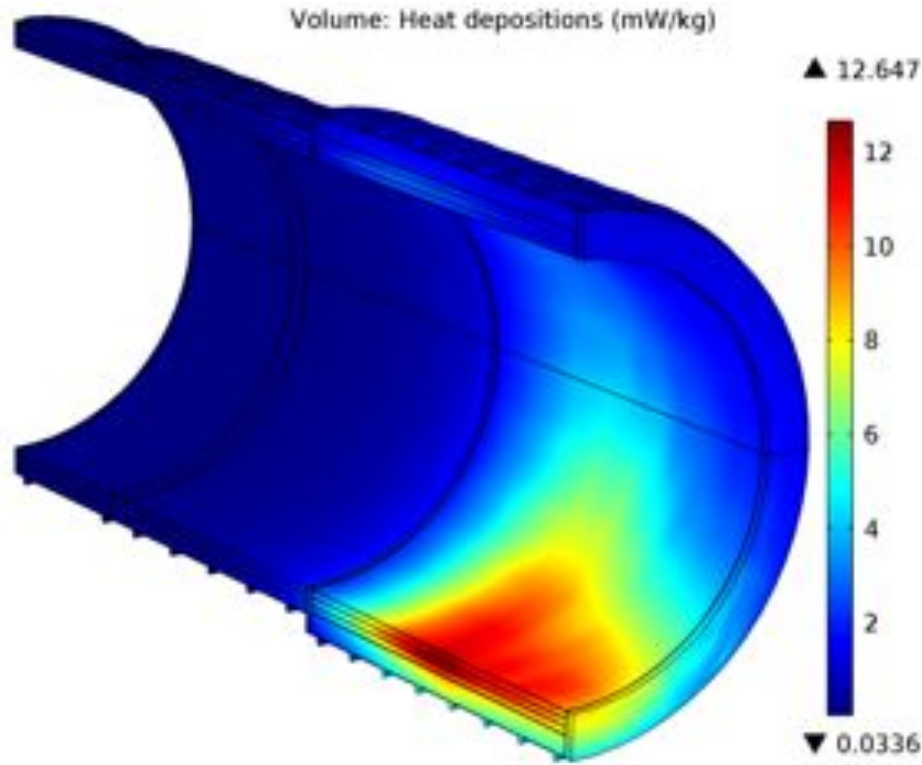

Figure 6.27. Figure Radiation power distribution in the cold mass.

The RRR degradation of aluminum and copper calculated using the above results is shown in Figure 6.28. Taking conservative safety factors into account, the calculated DPA translates into the degradation of RRR in Al and Cu from the initial values of 600 and 80 to 100 and 50 respectively in about one year of the experiment's operation. These numbers are regarded as the minimum allowable values for RRR.

The magnet will be equipped with RRR gauges that monitor the material property changes during operation. Once the critical resistivity degradation is detected, the magnet will be thermo-cycled to room temperature, restoring the original resistivity in Al and ~87% of that in Cu [33][34].

***Quench Protection Analysis***

The purpose of the quench protection system is to limit the peak coil temperature to 130 K and the peak coil to ground voltage to 600 V during any normally protected quench. This is achieved by detecting the resistive voltage rise associated with quench development and extracting the stored energy to an external dump resistor.

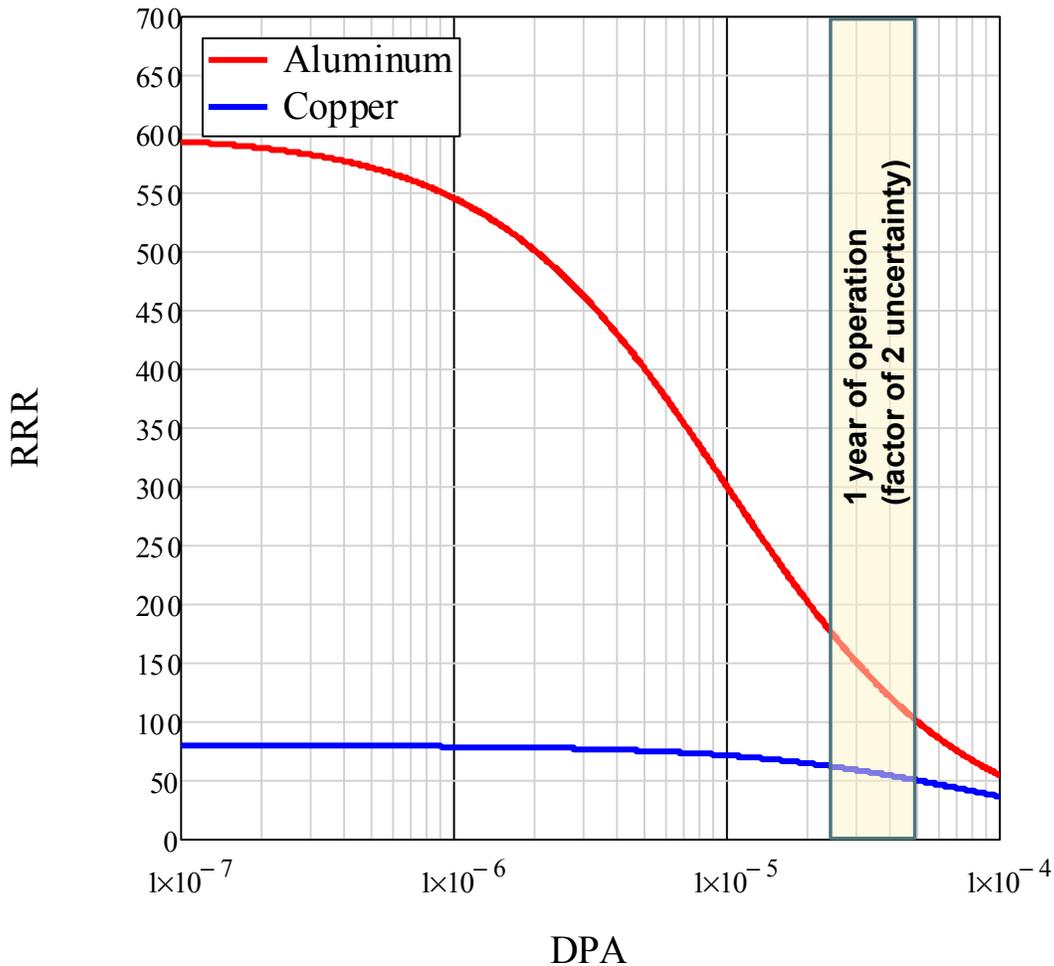

Figure 6.28. Degradation of aluminum and copper RRR in PS coil as a function of DPA.

***Quench Protection System (QPS)***

A schematic of the magnet electrical circuit is shown in Figure 6.29. The magnet is powered by a two-quadrant thyristor-based power converter with a maximum voltage of 20 V. The power converter has an internal DC current transformer (DCCT) for current regulation. In the case of a quench, the energy is extracted to the fast dump resistor permanently connected between the magnet leads with a resistance of 59 mΩ, chosen to limit the voltage across the magnet leads to 600 V at the maximum operating current. The power converter reverses the voltage when the slow ramp-down is requested.

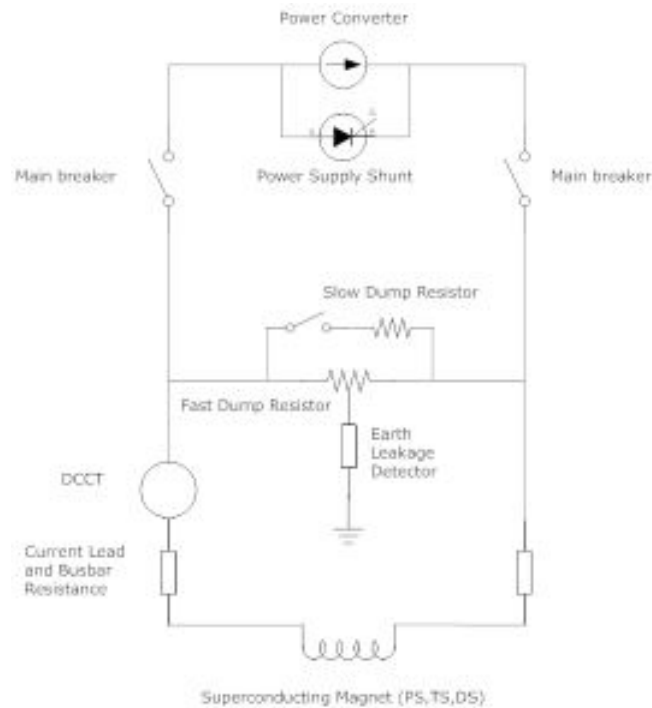

Figure 6.29.  Schematic of the magnet electrical circuit.

Each leg of two buses leading to the power converter has an independent, electronically controlled solid-state breaker. The bus-bars are made of copper. Their total resistance is significantly less than the resistance of the circuit in the case of a slow ramp-down, but sufficient for converter regulation. The grounding circuit symmetrically divides the voltage across the magnet terminals through a 1 kΩ resistor. The leakage current to ground is continuously monitored and triggers the magnet discharge if the preset limit is reached.

Each coil layer is equipped with redundant voltage taps connected to the QPS as described in Section 6.3.1.4. The QPS continuously monitors the magnet voltages during operation. If the resistive voltage exceeding the detection threshold of 0.5 V is detected, the QPS waits for 1 s to eliminate false signals and to give the magnet a chance to recover if the perturbation energy was less than the minimum quench energy (MQE) at a given location. If the resistive voltage remains above the detection threshold, the QPS activates the main breaker to disconnect the power converter from the magnet and the current discharges through the fast dump resistor.

*Quench Analysis Strategy*
The goal of the quench analysis was to determine the peak temperatures and voltages in the magnet at different operating conditions. The primary tool for modeling the quench development in the PS magnet was COMSOL Multiphysics code that was benchmarked against QLASA code [35] at an early stage of the analysis.

It is safe to assume that the quench may occur at any location of the magnet and at any stage of the operation. Because of the potentially time consuming nature of the simulations, several representative cases have been carefully selected. Two strategic quench locations have been chosen for analysis: the high-field region in the inner layer of the 3-layer coil and the low-field region at the magnet end adjacent to the TS as shown in Figure 6.30. All simulations were performed at the maximum operating current specified in Figure 6.11.

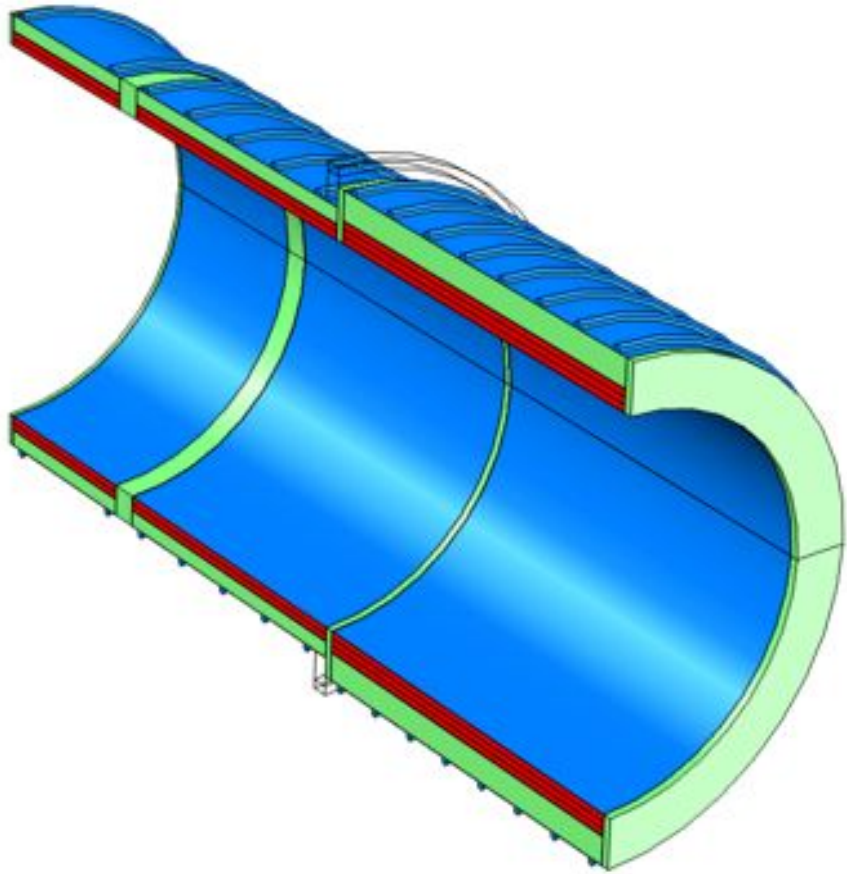

Figure 6.30. FEM quench analysis model showing two characteristic quench locations.

The PS to TS magnetic coupling was not included in the analysis, which essentially means the TS current remained constant during the PS discharge (it could be either on or off); however, it is possible that the PS quench will induce sufficiently large eddy currents in the TS structure to trigger the TS quench. Interaction of different magnets and the corresponding impact on the QPS will be considered in the next stage of the analysis.

*Magnet Commissioning Stage*

During the commissioning stage, the magnet will be tested in the stand-alone mode before and after the HRS is installed, and then as part of the Mu2e magnet system. During this stage, the RRR will remain at its maximum value. Figure 6.31 shows the resistive zone for the quench start at the high field and low field locations when the quenches are induced at multiple locations due to heating of the coil support structure by the eddy currents, also known as quench-back. The peak coil temperatures and voltages are shown in Figure 6.32.

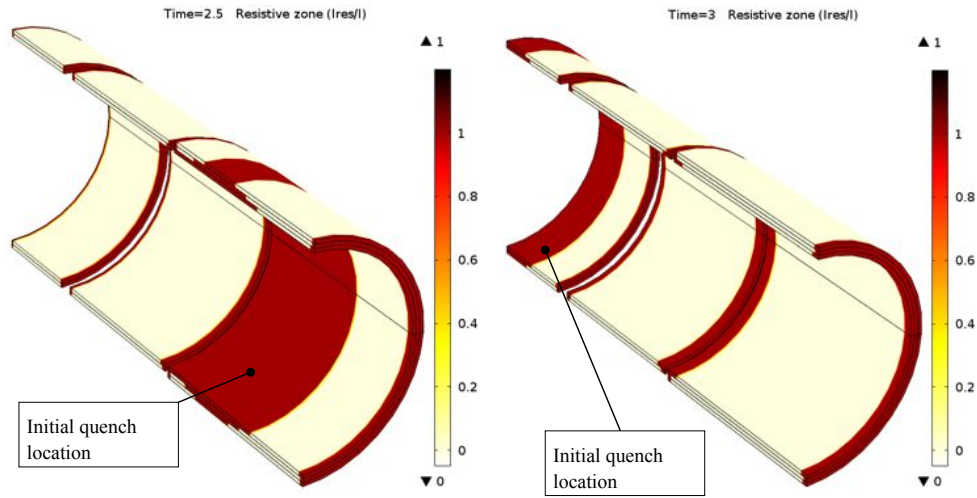

Figure 6.31. Resistive zones (dark regions) from the quenches originating at the HF (left) and LF (right) locations at the beginning of the quench-back from the structure.

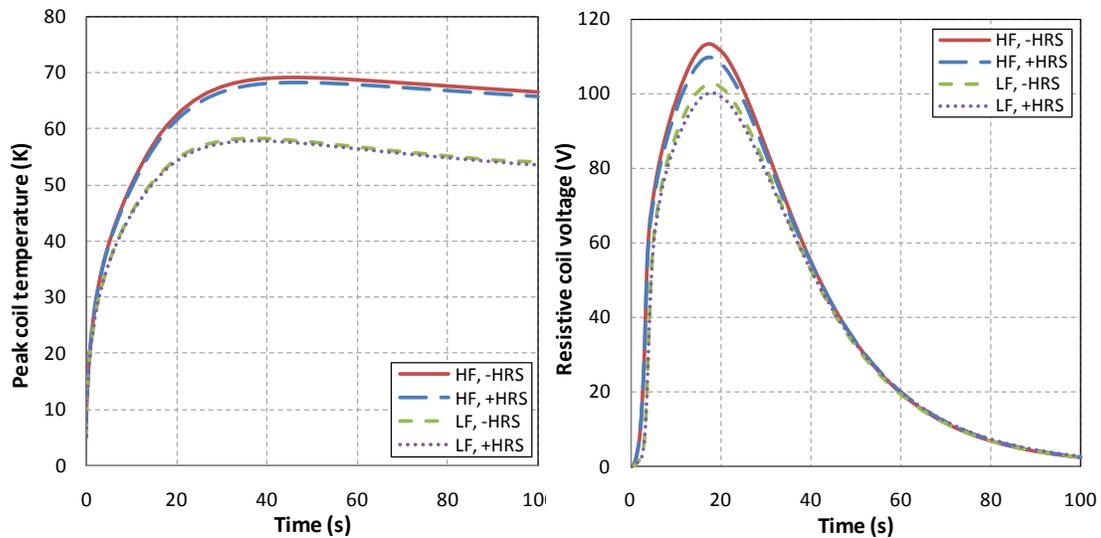

Figure 6.32. Peak coil temperatures (left) and voltages (right) during the quench at the maximum RRR of the cable stabilizer.

The maximum peak temperature of 69 K occurs during the quench at the high field location without the HRS. Adding the HRS reduces the peak quench temperature by 0.9 K. Although a part of the stored energy dissipates in the HRS volume, the effect on the peak coil temperature is small because this energy dissipates outside of the cryostat and does not contribute to coil heating and accelerating the quench propagation. In case of quenches at the low field location, the peak quench temperature is further reduced to 58 K, and diminishes further by 0.5 K when the HRS is installed. The peak resistive voltage of 100-112 V, generated in the coil depending on the location, is offset by a factor of six higher inductive voltage. The dump voltage, which is the sum of the resistive and inductive coil voltages, is always less than 600 V.

*Normal Operation Stage*
The HRS will always be present during normal operation stage, and the cable RRR will gradually decrease from the maximum to the minimum allowed values in about one year. The RRR degradation will not be uniform because of the non-uniform distribution of the radiation flux in the cold mass; however, for the purpose of this study, it was assumed that the RRR of the cable stabilizer will uniformly degrade to the minimum allowed values specified above in the whole coil volume.

The case of the maximum RRR has already been considered in the previous section. Figure 6.23 shows the peak coil temperatures and voltages when the RRR is degraded to the minimum allowed value. The peak coil temperatures are 82 K and 78 K for the quenches at the high field and the low field locations, respectively. While the absolute temperatures increased by 14-20 K with respect to the maximum RRR case, the temperature difference between the high field and the low field quenches reduced by more than a factor of two. The peak resistive voltages increased by almost a factor of three, but are still a factor of two less than the peak voltage across the dump.

Despite the practically negligible effect on the peak quench temperature, the presence of the HRS during a quench creates an additional force on the cold mass due to interaction of the induced eddy currents with the magnetic field. The force evolution on the cold mass due to interaction with the HRS is shown in Figure 6.34 (left). Because the considered HRS material is a general purpose bronze with no strict control over the magnetic and electrical properties, its nominal magnetic permeability was increased to 1.1 and the electrical resistivity reduced to $1 \times 10^{-7}$ $\Omega \cdot m$ to account for the possible property variations. The maximum dynamic force of 115 kN in the direction away from the TS is reached at ~3.5 s after the quench start. The 2-way cold mass axial support system described in Section 6.3.1.3 is designed to counteract this additional force.

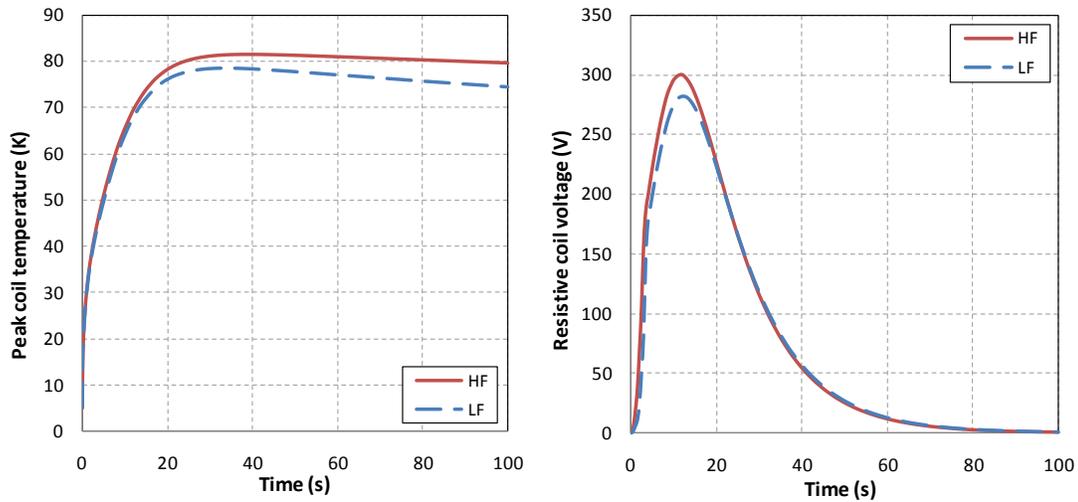

Figure 6.33. Peak coil temperatures (left) and voltages during quench (right) at the minimum RRR.

There is a strong dependence of the initial quench propagation speed and the coil voltages on the quench location and material properties. Figure 6.34 (right) presents the resistive coil voltages before the current discharge for different quench scenarios. The slowest voltage growth is for the quench at the low field location and the maximum RRR, where it takes ~1 s to reach the quench detection threshold. The fastest voltage growth is for the quench at the high field location and the minimum RRR, where it takes ~0.2 s to reach the same threshold.

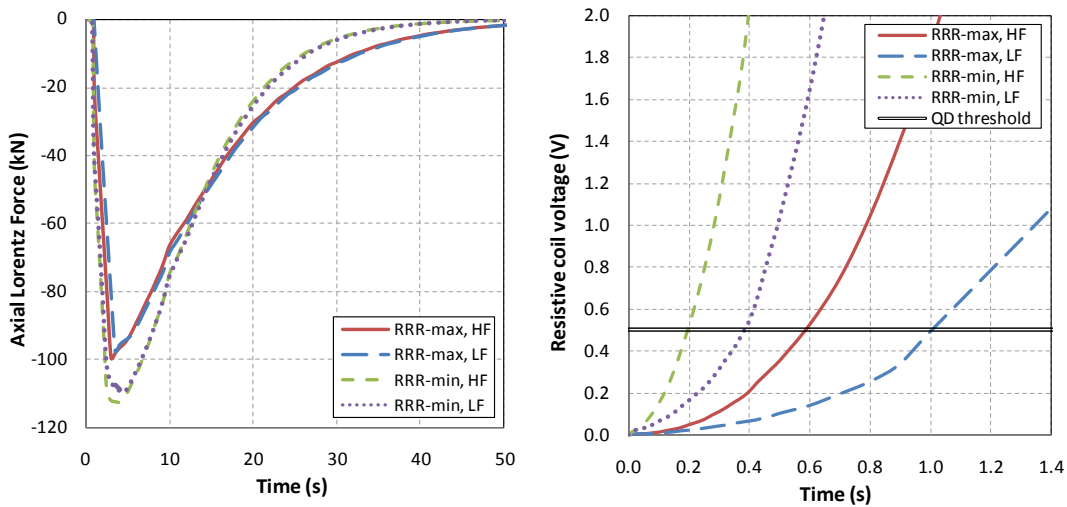

Figure 6.34. Axial Lorentz force on the cold mass (left) and the resistive voltages before the current discharge (right).

*QPS Failure*

It has been shown in the previous section that proper operation of the QPS limits the peak coil temperatures and voltages well below the acceptable values at all stages of the operation. A catastrophic failure of the QPS was considered as a part of the risk assessment. It was assumed that the energy extraction circuit would fail completely and all of the stored energy dissipates in the cold mass. In reality, it represents a simultaneous failure of both dump switches, or an electrical short between the bus-bars and the subsequent arc with nearly zero resistance. The only mechanism of the energy dissipation is the normal zone propagation and the quench-back from the structure when the field change becomes sufficiently high.

The peak coil temperatures and voltages are shown in Figure 6.35. In contrast to the protected quench behavior, the maximum temperature of 240 K is for the maximum RRR quench at the low field location because of the slowest propagation speed. The minimum temperature of 130 K is for the minimum RRR quench at the high field location, which has the fastest propagation speed. The low-field quenches have a low probability of occurring due to a high MQE.

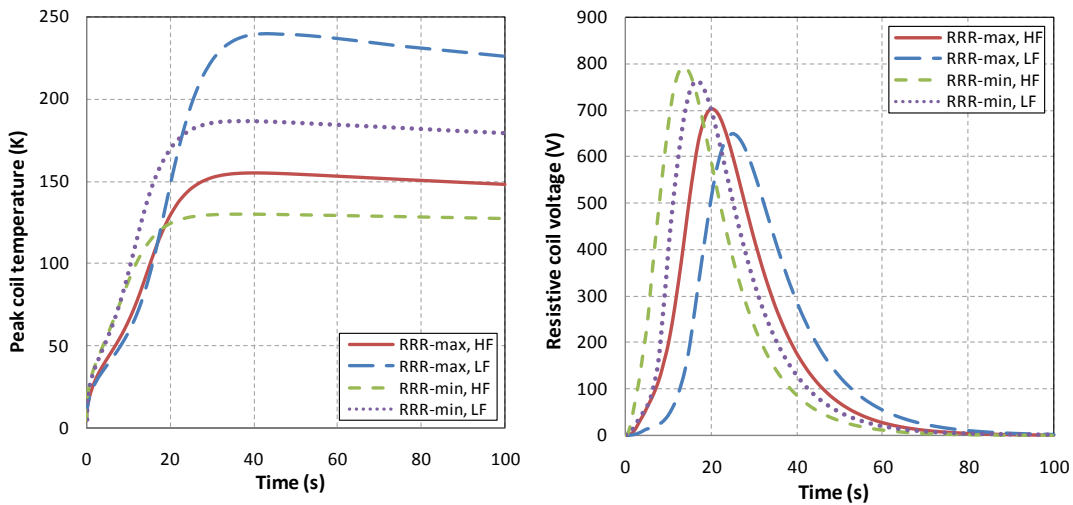

Figure 6.35. Peak coil temperatures (top) and voltages (bottom) during an unprotected quench.

The resistive coil voltage increases to a maximum of 800 V during the minimum RRR quench at the high field location; however, the voltage across the dump resistor is zero, meaning the inductive voltage is the mirror image of the resistive voltage. These voltages mostly cancel each other inside of the coil.

*Quench Energy Balance*

The energy dissipations in different magnet components are presented in Table 6.12 for various quench conditions discussed earlier. During the protected quenches, most of the energy is extracted to the dump resistor with 12-18% of the total stored energy dissipating in the coils in the case of maximum RRR and 32-35% of that dissipating in the case of minimum RRR.

When the energy extraction fails, all the stored energy dissipates in the cold mass, because contribution from the HRS is practically negligible as a result of the slow current decay and consequently low dB/dt, with ~91% of the total stored energy dissipating in the coils.

Table 6.12. Quench energy balance.

| Quench condition | | | | | Dump activation delay (s) | Total energy dissipated, MJ | | | |
| RRR | | Location | | | | | | | |
| max | min | High Field | Low Field | HRS | | Coils | Shells | Dump | HRS |
|------|------|------|------|------|------|------|------|------|------|
| **Normally protected** | | | | | | | | | |
| x | | x | | | 1.6 | 14.33 | 5.03 | 60.45 | 0 |
| x | | x | | x | 1.6 | 14.00 | 4.96 | 59.97 | 0.87 |
| x | | | x | | 2.0 | 10.74 | 4.95 | 64.11 | 0 |
| x | | | x | x | 2.0 | 10.23 | 4.89 | 63.82 | 0.86 |
| | x | x | | x | 1.2 | 27.97 | 6.73 | 43.92 | 1.19 |
| | x | | x | x | 1.4 | 25.70 | 6.54 | 46.41 | 1.16 |
| **Unprotected** | | | | | | | | | |
| x | | x | | | ∞ | 73.14 | 6.66 | 0 | 0 |
| x | | | x | | ∞ | 73.65 | 6.15 | 0 | 0 |
| | x | x | | | ∞ | 72.20 | 7.60 | 0 | 0 |
| | x | | x | | ∞ | 72.57 | 7.24 | 0 | 0 |

### Forces on the Cold Mass

The total forces on the PS cold mass that will be intercepted by the cryostat suspension system and transferred to the experiment's floor. Forces on the cold mass include static Lorentz forces between the PS and TS coils, the static and dynamic electromagnetic forces between the PS and the HRS, the forces due to misalignment between the above components, the cold mass weight and the dynamic forces during transportation.

Assuming that the PS cold mass position can be kept accurate to 1 cm in any direction relative to the TS and the HRS, and taking the worst-case scenario from Table 6.9, Table 6.10 and Figure 6.34, the peak axial force on the cold mass is 1427 kN toward the TS and 115 kN away from the TS.

The peak static transverse force on the cold mass comes from the cold mass weight of 129 kN and the force due to worst-case PS coil misalignment of 29 kN; however, it is reasonable to assume 2g acceleration for transportation purposes, which results in a peak dynamic transverse force of 258 kN.

Because of the different number of layers and shell thicknesses, the weight distribution is not uniform along the cold mass. The cold mass center of gravity is 1.75 m from the non-TS end flange (or 2.27 m from the TS end flange), which is exactly at the end of the 3-layer coil. It results in a weight distribution of 0.435:0.565 between the TS and non-TS end flanges. The peak dynamic transverse force is therefore applied to the TS end flange is 112 kN and that applied to the non-TS end flange is 146 kN.

### Thermal Analysis

#### Thermal Interfaces
Secondary particles generated from the 8.3 kW proton beam interacting with the production target will deposit energy in the Production Solenoid coils. A heat shield lining the warm bore of the PS made of bronze will intercept most of the energy [36]. However, an expected 28 W of continuous beam power will still be deposited in the cold mass. The radiation heat is extracted from the coil through a system of thermal bridges. Because of the irradiation-induced degradation, one cannot take advantage of thin layers of high-purity Al as is often done in conduction-cooled magnets; the thermal bridges need to be sufficiently thick to conduct the heat even after irradiation.

The thermal bridges, made of 1.5-mm Al sheets of the same composition as the cable stabilizer, are installed on the inner coil surface and extend throughout each coil length as described in Section 6.3.1.2. The ends of the thermal bridges are bent around the coil and the outer shell assembly and connected to the base plates of the cooling tubes by means of weld joints as described in Section 6.3.1.2.

The coil insulation details are described in Section 6.3.1.2. The composite ground insulation with a total thickness of 500 μm separates the insulated cables from the metal parts on all sides as shown in Figure 6.7. There are two 25 μm layers of Kapton embedded between the layers of fiberglass in the ground insulation. The 250 μm thick dry E-glass sheets placed between the coil layers are to be filled with epoxy during coil impregnation. The outer coil surface is wrapped with a sufficient thickness of fiberglass prior to impregnation that will be machined down to an average of 2 mm after the coils are impregnated.

The ends of the thermal bridges are stress-relieved by providing a clearance at the corners of the support shells to accommodate the differential contraction between the coils and shells due to cooling down and Lorentz forces. Layers of mica paper are introduced between the thermal bridges, flanges and support shells to avoid accumulation of shear stresses at these interfaces.

*3D Thermal Model*
A 3D FEM model created within NX 7.5/ANSYS 14.5 interface, shown in Figure 6.36, was finalized to the level of lumped cable turns to form each coil layer and included all the cooling/insulation features described earlier. The simulation was performed for the worst expected case, when the RRR of all Al elements (excluding the support structure made of Al 5083-O) was degraded to the minimum allowable value of 100. The rule of mixtures was used to define the equivalent thermal conductivities of the insulated cables, the interlayer insulation, and the ground insulation. All other elements had the actual thermal properties of the corresponding materials.

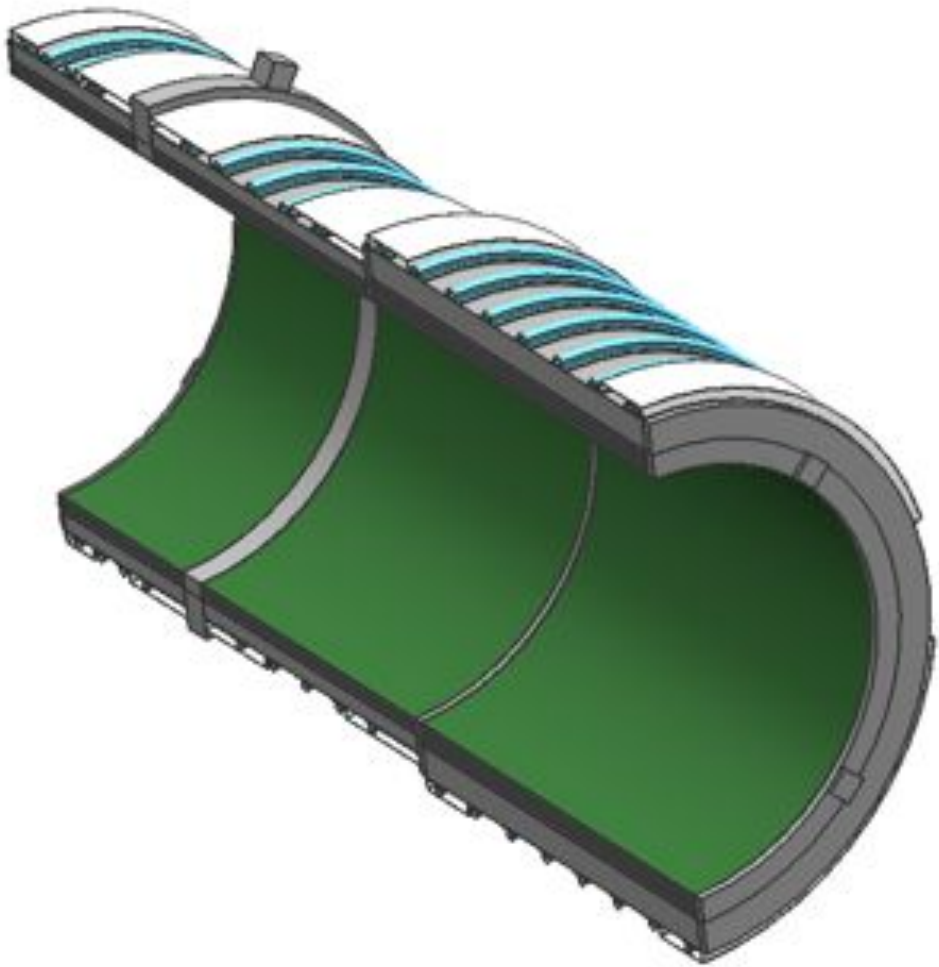

Figure 6.36. 3D thermal model of the PS cold mass.

The 3D thermal analysis was performed for the radiation heat load in case of the baseline absorber design described above. In addition to that, the static heat loads at 4.7 K listed in Table 6.6 were applied to all external surfaces to model the thermal radiation/gas conduction; to the middle support ring to model the heat load through the axial supports; and to the end flanges to model the heat load through the transverse supports. It was assumed that the cooling tubes were kept at the constant temperature $T_0$ by the cryogenic system. One-half of the cold mass was modeled since the particle production target lies in the horizontal plane and the heat deposition map is statistically symmetric with respect to that plane. The distribution of the radiation power is strongly non-uniform in the azimuthal and axial directions as shown in Figure 6.27 and Figure 6.37.

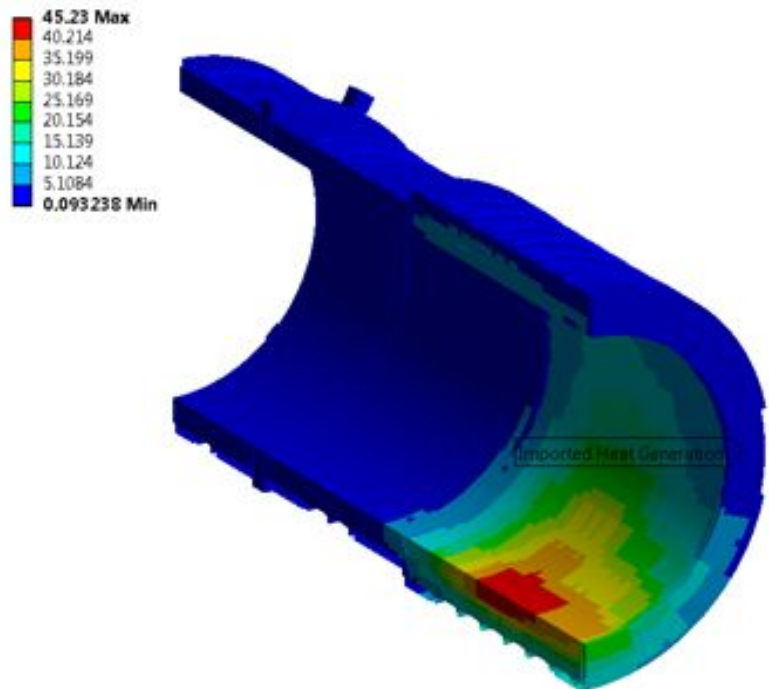

Figure 6.37. Dynamic heat load (W/m$^3$) in the cold mass as generated by MARS code. The model is rotated by 90 degrees around the magnetic axis for clarity (i.e. the vertical model plane is the horizontal plane of the experiment).

The resulting cold mass temperature is shown in Figure 6.38 for $T_0 = 4.7$ K. The maximum temperature is in the middle of the inner surface of the thickest (3-layer) coil; that location coincides with the peak field location, and, therefore, directly affects the thermal margin.

Under the static heat load, the peak temperature difference between the coil and the cooling tubes is ~125 mK, indicating that most of the static heat load is intercepted by the thermal bridges and plates before it enters the coil. Under the nominal dynamic heat load, the temperature increment increases up to ~369 mK. Nevertheless, the peak coil temperature is below the maximum allowable temperature (including 1.5 K thermal margin) by ~31 mK at the maximum current corresponding to the nominal 4.6 T field on the axis. It provides an additional thermal margin to offset the uncertainty in calculating the power depositions, fabrication tolerances and material properties.

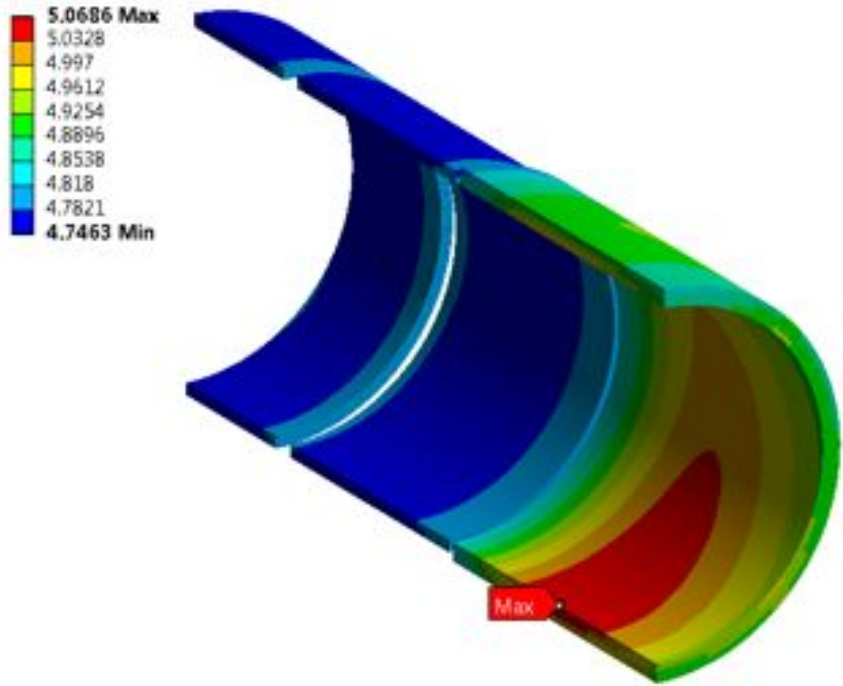

Figure 6.38.  Coil temperature distribution due to static and dynamic heat load at $T_0 = 4.7$ K.

### Cold Mass Structural Analysis
The cold mass structural elements are designed according to the structural criteria specified in Section 6.3.1.2. Because the cold mass includes the insulation and the cable consisting of different materials, the structural properties of these components have to be determined from a separate analysis. The structural properties of constituent insulating and metallic materials of the cable and the calculation of average orthotropic properties of the insulated cable stack using FEM models are described in [10].

#### Support Shell Properties
The material of choice of the coil outer support shells is Al 5083-O.  That particular grade of aluminum was selected because the ASME BPV code allows taking advantage of the material strength enhancement at cryogenic temperatures, and because it does not exhibit reduction of strength after welding. Table 6.13 lists the maximum allowable stresses derived from [37].

Table 6.13. Maximum allowable stress of Al 5083-O at cryogenic temperatures (≤77K).

| Thickness range, mm | Maximum allowable stress, MPa |
|---|---|
| 1.30 - 38.10 | 107.6 |
| 38.13 - 76.20 | 101.3 |
| 76.23 - 127.00 | 95.8 |

*Elastic Cold Mass Analysis*

The coils are supported against Lorentz forces by the external shells made of Al 5083-O. Each coil with its support shell forms a separate assembly. The coil assemblies are bolted together through the end flanges that constitute the cold mass assembly.

A preliminary elastic structural analysis [38] was performed using the material properties defined in [10]. The coils were modeled as objects with uniform orthotropic material properties. The buffer layer between the coils and the shells described in Section 6.3.1.2 was modeled separately with the properties of G10. It was assumed that any gaps between the coils and the support shell are filled with epoxy. Consequently, the FEM model was created with zero coil-shell gaps and zero pre-stress at room temperature. The coil sections are to be assembled with layers of mica paper between the coils and the end flanges to allow free sliding at these interfaces. The model included frictionless contact elements at these locations.

Figure 6.39 shows deformation and equivalent stress in the cold mass assembly due to cooling down and energizing with the maximum operating current defined in Table 6.11. The equivalent stress distribution in the coil and the support structure is shown in Figure 6.40 The peak equivalent stresses are 21 MPa in the coil and 51 MPa in the support structure after the cool-down. Energizing the PS with the maximum operating current brings the stresses to 73 MPa and 96 MPa for the coil and the support structure, respectively. Thus the shell stresses are within the maximum allowable values listed in Table 6.13 at all conditions. The shell thicknesses determined from this analysis were 83 mm, 50 mm and 50 mm for the 3-layer coil and two 2-layer coils respectively.

*Plastic Structural Properties of Aluminium Stabilizer*

Because of the difference in the thermal contraction coefficients between the NbTi/Cu strands and Al stabilizer, the Rutherford cable will be under compression and the Al stabilizer will be under tension after the cool-down. The large difference in the thermal contractions results in the plastic deformation of Al stabilizer already after the cool-down. The Lorentz forces during the magnet excitation increase the tension in the stabilizer and hence the amount of plastic deformation; hence, the plastic deformation plays an important role in the structural properties of the cold mass and must be carefully studied.

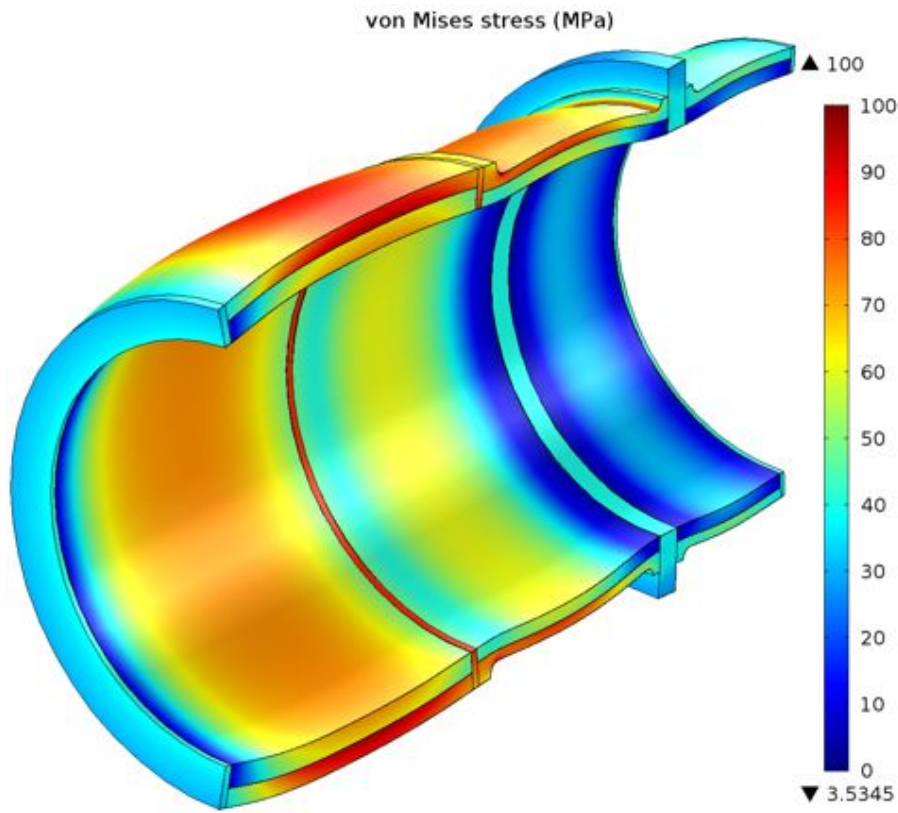

Figure 6.39. Deformation and equivalent stress in the cold mass due to cooling down from room temperature to 4.7 K and energizing with the maximum operating current.

Because the degree of plasticity in Al stabilizer depends on the amount of cold work applied during cable fabrication, the Ramberg–Osgood equation in the following form was fitted to the available stress-strain measurements from the ATLAS R&D to describe the stress-strain relationship in the Al stabilizer [24]:

$$\varepsilon(\sigma) = \frac{\sigma}{E} + \varepsilon_0 \left(\frac{\sigma}{\sigma_0}\right)^n$$

where $\varepsilon$ is strain, $\sigma$ is stress, $\varepsilon_0$ is reference strain (normally 0.002), $\sigma_0$ is reference stress at $\varepsilon_0$ strain (normally the offset yield stress), $E$ is elasticity modulus and $n$ is the fitting parameter. Once the fitting is established, it is possible produce stress-strain curves for other yield stresses by changing the reference stress as shown in Figure 6.41.

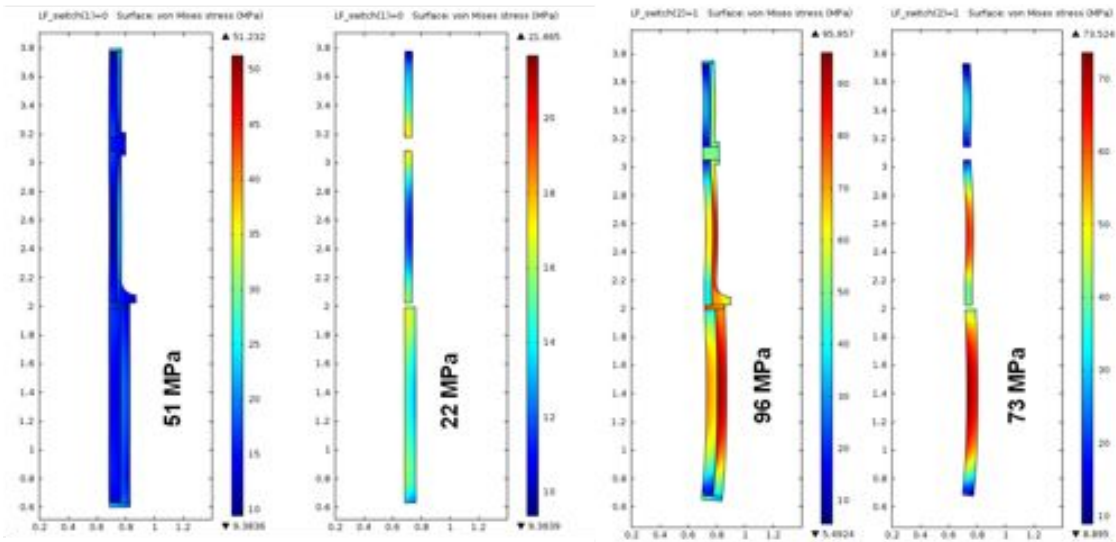

Figure 6.40. Equivalent stresses in the coil and the support structure due to cooling down from room temperature to 4.7 K (two left plots) and subsequently energizing with the maximum operating current (two right plots).

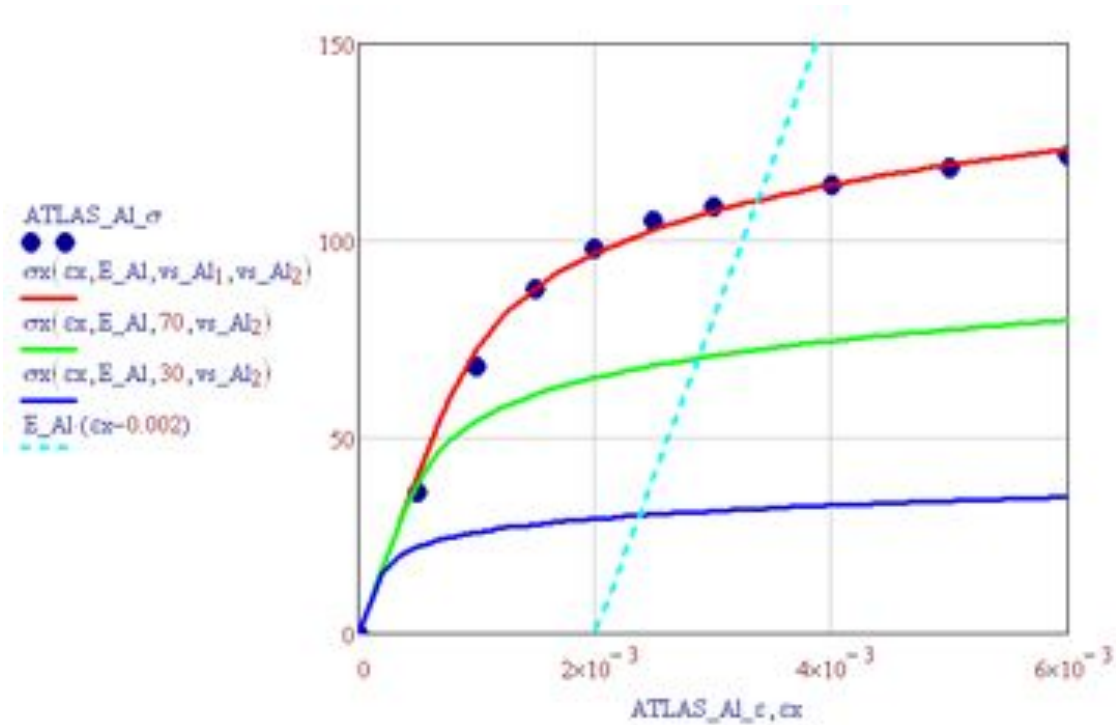

Figure 6.41. Fitting the Ramberg–Osgood equation to the measured stress-strain curve of the Al stabilizer and scaling the curve to other offset yield stresses.

*Elastoplastic Analysis of the Complete Coil*

Since it was found that plasticity of Al stabilizer plays an important role in the structural properties of the cable stack, it was necessary to include it in analysis of the complete coil in order to determine the impact on the thicknesses of the support shells. Since every cable in the coil, in general, experiences different Lorentz forces, the amount of plastic strain is also different. Therefore, the plastic properties of the coil cannot be simply described with some equivalent properties as was done in the case of the elastic analysis.

Instead, a complete FEM model has been created using COMSOL Multiphysics code. Every cable in the coil was explicitly modelled and contained the Rutherford cable, stabilizer and insulation. The Al stabilizer was assigned the plastic properties described above.

The equivalent stresses in the support shells with the thicknesses defined above after cool-down were within the limits, but the peak stress when energized was considerably higher than the maximum allowable value listed in Table 6.13 due to the impact of the stabilizer's plasticity that reduces the effective coil modulus and therefore transfers larger forces to the support shells. To alleviate this problem, shell thicknesses were increased to 125 mm, 70 mm and 70 mm for the 3-layer coil and two 2-layer coils respectively. The resulting equivalent stresses shown in Figure 6.42 are within the maximum allowable values. These shell thicknesses are the nominal design values.

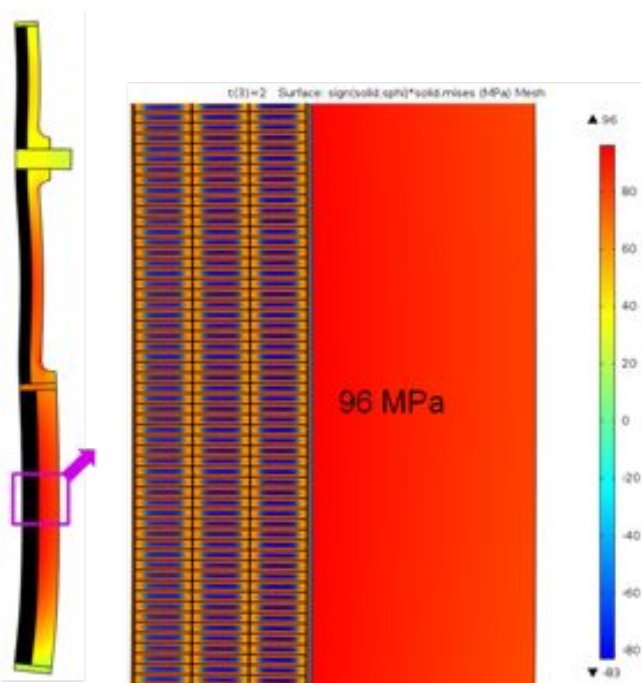

Figure 6.42. Equivalent stress in the cold mass after cool-down to 4.7 K and energizing with the maximum operating current for the increased shell thicknesses.

*Displacements*

Displacements of the cold mass key locations determined in elastoplastic analysis after cool-down and energizing with the maximum operating current are described in [10]. Information from this analysis was used in defining the warm and cold dimensions of the cold mass components.

## 6.3.2   Transport Solenoid

### 6.3.2.1   Introduction and General Requirements

The Transport Solenoid consists of a series of wide aperture superconducting solenoid rings contained in two cryostats. Each cryostat has a chimney for superconducting leads, helium supply and return lines and instrument ports. Internal mechanical supports in each cryostat transmit forces to external mechanical supports that connect to the experiment enclosure structure. The Transport Solenoid does not have an iron return yoke.

As shown in Figure 6.43, the Transport Solenoid is segmented into the following set of components:

- TS1 - Straight section that interfaces with the Production Solenoid.
- TS2 - Toroid section downstream of TS1.
- TS3 - Straight section downstream of TS2 TS3u coils are in the TSu cryostat and TS3d coils are in the TSd cryostat.
- TS4 - Toroid section downstream of TS3.
- TS5 - Straight section downstream of TS4 that interfaces with the Detector Solenoid.

The Transport Solenoid performs the following functions:

- Pions and muons are created in the production target in the Production Solenoid. The Transport Solenoid maximizes the muon yield by efficiently focusing these secondary pions and subsequent secondary muons towards the stopping target located in the Detector Solenoid. High energy negatively charged particles, positively charged particles and line-of-sight neutral particles will be almost completely eliminated by the two 90° bends combined with a series of absorbers and collimators.
- The TS1 field must be merged with the field of the Production Solenoid at the interface for optimum beam transmission.
- There must be a negative axial gradient at all locations in the straight sections (TS1, TS3 and TS5) for radii smaller than 0.15 m to prevent particles from becoming trapped or losing longitudinal momentum.

- Through the first toroid section (TS2) the beam will disperse vertically, allowing a collimator in TS3 to filter the particles based on sign and momentum.
- The second toroid section (TS4) will nearly undo the vertical dispersion, placing the muon beam in the center of the TS5 axis.
- The TS5 field must be merged with the field of the Detector Solenoid at the interface for optimum transmission of the muon beam to the stopping target.
- The Transport Solenoid acts as a beamline interface for the antiproton window and various collimators, including the rotating collimator in TS3.

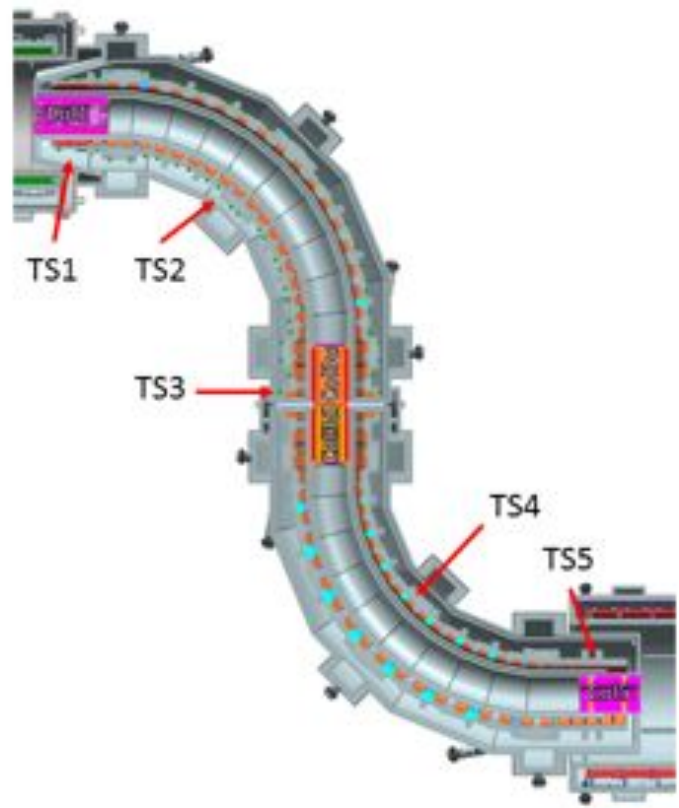

Figure 6.43. The Transport Solenoid with the significant components identified.

The Transport Solenoid consists of two independent cryostats and power units. The TS1, TS2 and TS3u coils are assigned to the TSu cryostat. The TS3d, TS4 and TS5 coils share the TSd cryostat. Each cryostat will have its own superconducting link, feed box, power converter and extraction circuit.

All TS coils use the same conductor design and similar cooling schemes. The TSu unit and the TSd unit are nearly identical, so only the preliminary design of TSu will be presented.

### 6.3.2.2   TSu Design

The TSu is shown in Figure 6.44 and includes the following design features:

- A single cryostat is employed to avoid gaps between coils and reduce complexity and cost.
- The coils are powered in series to minimize the number of leads and the complexity of the power and quench protection systems.
- The quench protection strategy is based on extracting most of the energy and delivering it to external dump resistors.
- Coils are preassembled and tested inside modules, with two coils per module in most cases, in order to reduce complexity during cold mass assembly.
- The mechanical support system consists of 17 supports including: four supports along the toroid main radius, four axial supports close to each end, and three gravity supports.

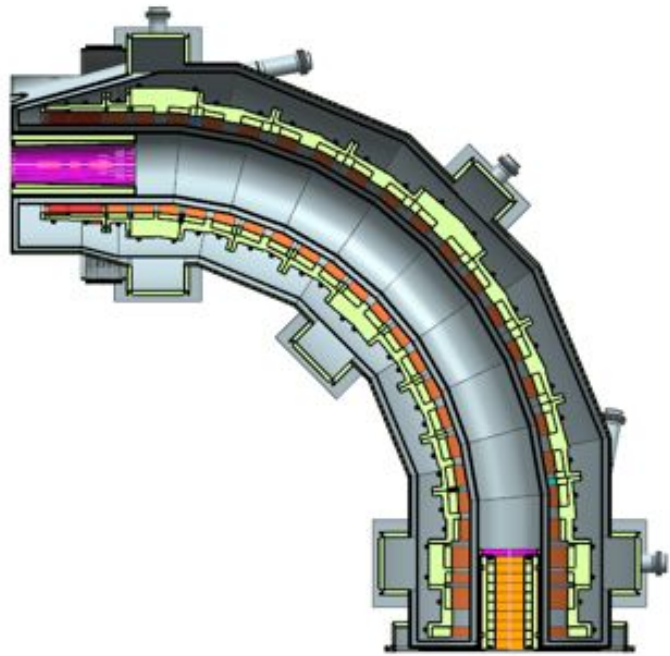

Figure 6.44. TSu cross section along the horizontal plane.

There is a gap between the TS3u and the TS3d coils as a result of the cryostat interfaces and the mechanical hardware necessary to actuate the rotating collimator and to insert the anti-proton window. To allow for a 220 mm gap, the inner radii of the TS3 coils have been increased to 465 mm, compared to inner radii of 405 mm for the remaining TS coils. Further details about the TSu design can be found in reference [39].

### 6.3.2.3   Conductor Design and Performance

The conductor used for the TS coils is an aluminum stabilized NbTi Rutherford cable. This kind of conductor is typically used for detector systems in particle accelerators and colliders. The TS cable design is based on the conductor used for the BELLE detector solenoid at KEK [40]. The TS conductor was designed to have sufficient temperature and enthalpy margins to accommodate possible energy releases in the coils during energization. The same cable design will be used for all the TSu and TSd coils. The main cable design parameters are summarized in Table 6.14. The TS cable conforms to the specifications described in [55] - [57]. A cross section of the cable is shown in Figure 6.45. The insulation is made of braided fiberglass or tape with some overlap with a thickness of 0.25 mm per side. Epoxy impregnation completes the insulation process.

Table 6.14. Summary of TS conductor main design parameters.

| Conductor Parameter | Unit | Design Value | Measured Value |
|---|---|---|---|
| Cable critical current at 5T, 4.2K | A | 5900 | 5950-6300 |
| Number of strands | | 14 | |
| Strand diameter | mm | 0.67 | within tolerances |
| Strand copper/SC ratio | | $1 \pm 0.05$ | 0.97-1.02 |
| Initial RRR of Cu matrix | | 150 | 100-104 |
| Filament size | μm | < 30 | 25.5-25.7 |
| Strand twist pitch | mm | $15 \pm 2$ | 15.8-15.9 |
| Rutherford cable width | mm | $4.79 \pm 0.01$ | within tolerances |
| Rutherford cable thickness | mm | $1.15 \pm 0.006$ | within tolerances |
| Al-stabilized cable width (bare) at room temperature | mm | $9.85 \pm 0.05$ | within tolerances |
| Al-stabilized cable thickness (bare) at room temperature | mm | $3.11 \pm 0.03$ | within tolerances |
| Initial RRR of Aluminum stabilizer | | > 800 | 925-1160 |
| Aluminum 0.2% yield strength at 300 K | MPa | > 30 | 45-56 |
| Aluminum 0.2% yield strength at 4.2 K | MPa | > 40 | 74-84 |
| Shear strength between Aluminum and NbTi strands | MPa | > 20 | 35-46 |

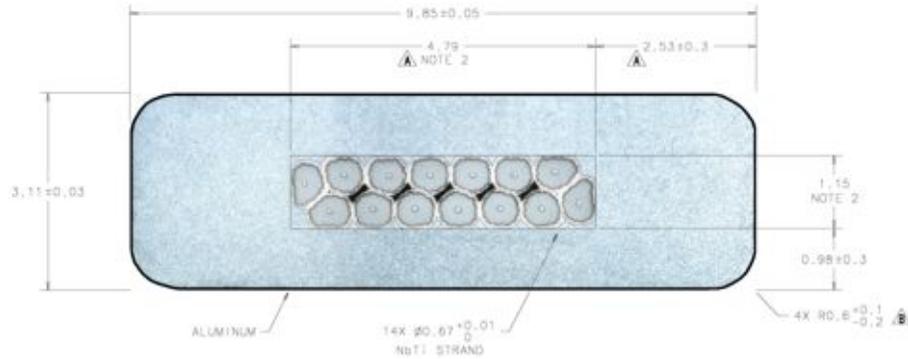

Figure 6.45. Cross-section of TS Al-stabilized cable.

The TSu is powered by a dedicated power supply with an operating current of 1730 A. With this configuration of conductor and insulation, the critical engineering current density is 47 A/mm$^2$ and has a peak field of 3.4 T. The operating current fraction on the load line at 5.1 K is 58%. The temperature margin at 5.1 K and 3.4 T is 1.82 K.

A prototype length (3.2 km) of the TS conductor was fabricated and measurements of this cable are summarized in Table 6.14. Figure 6.46 presents the critical current test results performed on full Al-stabilized conductors. The results show that the design parameters can be reached in a consistent manner.

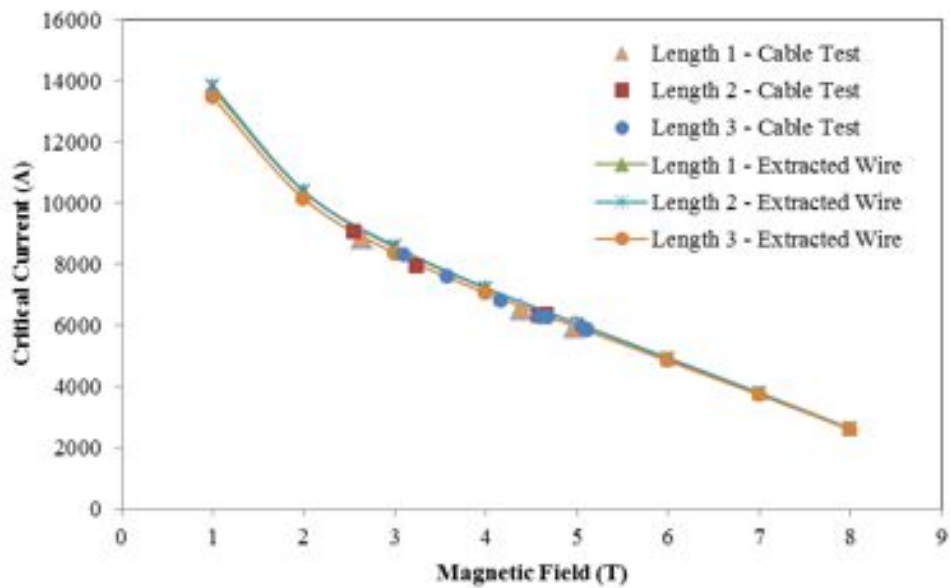

Figure 6.46. Summary of critical current test results performed on full Al-stabilized cables and strand samples extracted from TS cables after chemical etching of the stabilizer.

### 6.3.2.4   TS Magnetic Design

As shown in the requirements (Section 6.2) the straight sections and the toroid sections have unique field requirements. The straight sections require a field gradient that is negative everywhere inside a radius of 0.15 m. In the toroid sections the field ripple must be small and the radial gradient must satisfy $|dBs / dr| > 0.275$ T/m.

A computer model of all Mu2e Solenoids was generated using the design coil geometries. There is no iron yoke around any Mu2e magnets and the fringe field of the PS and DS magnets has a significant impact on the trajectory of the particles in the Transport Solenoid, which causes some horizontal drift. This drift is corrected by a small rotation around the vertical axis of the TS2 and TS4 coils. The results of the magnetic analysis are shown in Figure 6.47 through Figure 6.50. In each straight section (TS1, TS3 and TS5) the generated fields and gradients are shown and compared to requirements (black lines). For the toroid sections (TS2 and TS4), the field ripples are compared to requirements. In all cases, the field specifications, represented by the black lines, are met.

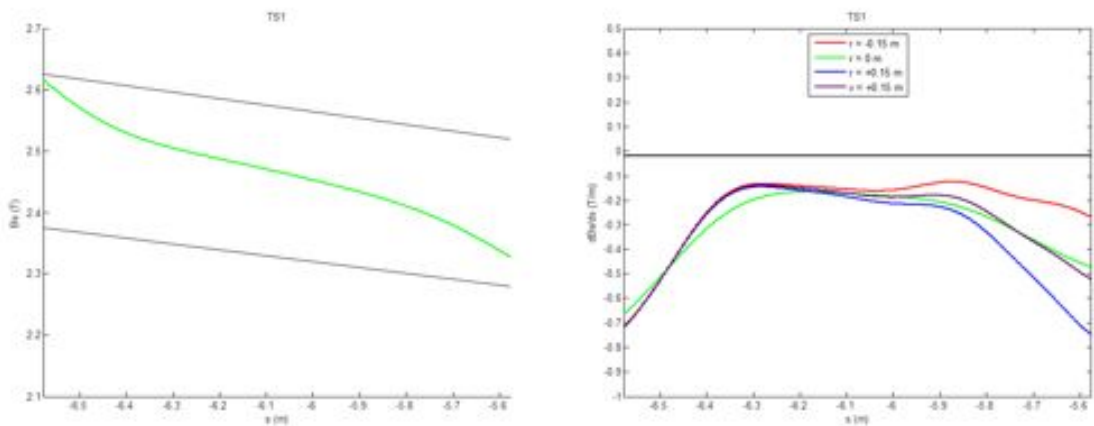

Figure 6.47. Axial field distribution at the center of TS1 (left). Axial gradient along TS1 (right).

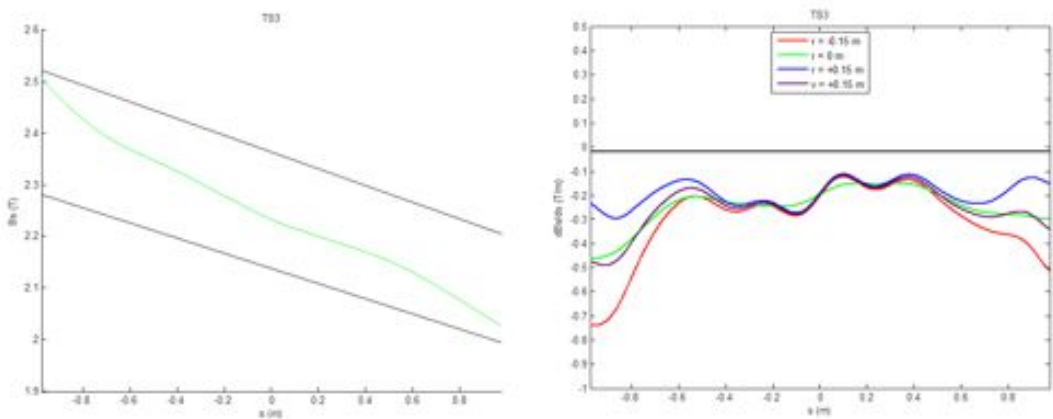

Figure 6.48. Axial field distribution at the center of TS3 (left). Axial gradient along TS3 (right).

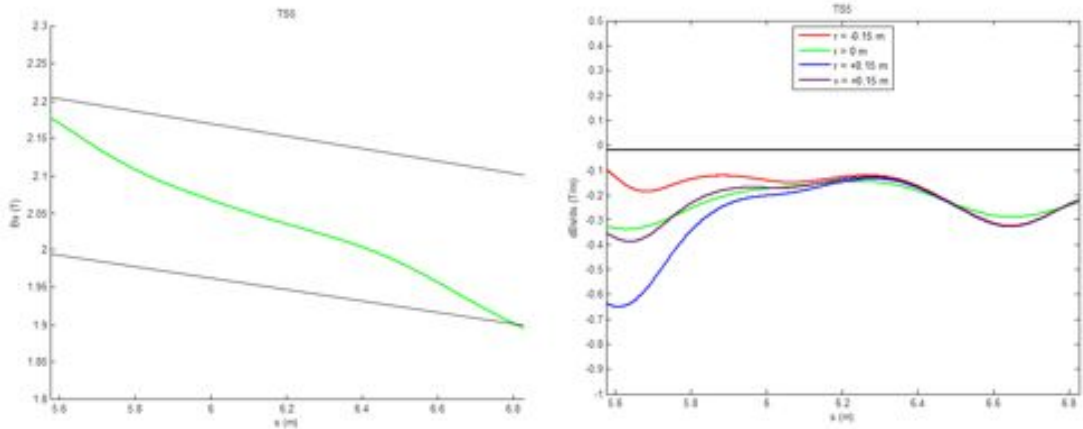

Figure 6.49. Axial field distribution at the center of TS5 (left). Axial gradient along TS5 (right).

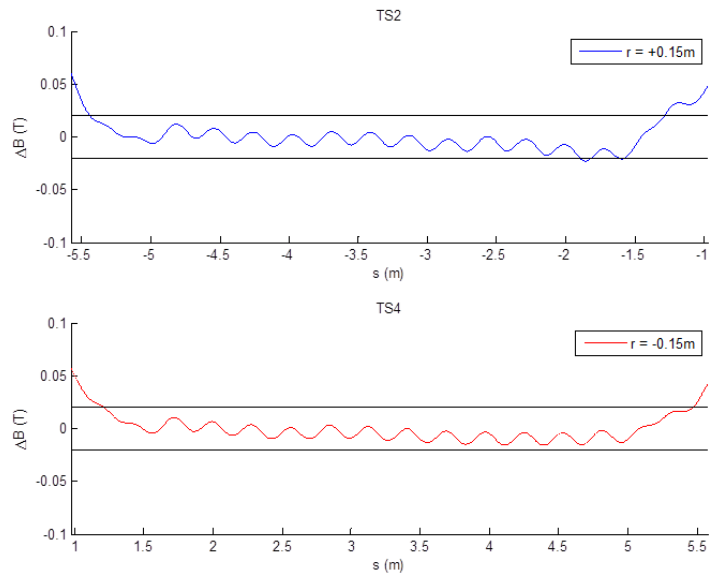

Figure 6.50. Ripple in the TS2 and TS4 curved sections.

### 6.3.2.5 TSu Coil Design

The TS coils will be wound on collapsible mandrels and then inserted into aluminum shells (modules). The modules are assembled into a single cold mass and power unit. TS1 is a straight solenoid made of 3 coils with different outer diameters and separated by flanges. TS2 is a quarter of a toroid made of 18 coils. TS3u is a straight solenoid made of four coils. Figure 6.51 shows the distribution of these coils and Table 6.15 lists the main coil parameters.

Table 6.15. TSu Coil parameters.

| Coil # | Section | Coil Inner Radius [mm] | Coil Outer Radius [mm] | Coil Length [mm] | Number of layers | Number of turns per layer |
|--------|---------|------------------------|------------------------|------------------|------------------|---------------------------|
| 1  | TS1  | 405.00 | 423.00 | 172.60 | 5  | 17 |
| 2  |      | 405.00 | 430.30 | 284.20 | 7  | 28 |
| 3  |      | 405.00 | 444.90 | 162.40 | 11 | 16 |
| 4  | TS2  | 405.00 | 448.60 | 172.60 | 12 | 17 |
| 5  |      | 405.00 | 448.60 | 172.60 | 12 | 17 |
| 6  |      | 405.00 | 448.60 | 172.60 | 12 | 17 |
| 7  |      | 405.00 | 463.20 | 172.60 | 16 | 17 |
| 8  |      | 405.00 | 463.20 | 172.60 | 16 | 17 |
| 9  |      | 405.00 | 463.20 | 172.60 | 16 | 17 |
| 10 |      | 405.00 | 463.20 | 172.60 | 16 | 17 |
| 11 |      | 405.00 | 466.80 | 172.60 | 17 | 17 |
| 12 |      | 405.00 | 466.80 | 172.60 | 17 | 17 |
| 13 |      | 405.00 | 466.80 | 172.60 | 17 | 17 |
| 14 |      | 405.00 | 466.80 | 172.60 | 17 | 17 |
| 15 |      | 405.00 | 470.50 | 172.60 | 18 | 17 |
| 16 |      | 405.00 | 470.50 | 172.60 | 18 | 17 |
| 17 |      | 405.00 | 470.50 | 172.60 | 18 | 17 |
| 18 |      | 405.00 | 470.50 | 172.60 | 18 | 17 |
| 19 |      | 405.00 | 470.50 | 172.60 | 18 | 17 |
| 20 |      | 405.00 | 477.80 | 172.60 | 20 | 17 |
| 21 |      | 405.00 | 448.60 | 172.60 | 12 | 17 |
| 22 | TS3u | 465.00 | 523.20 | 172.60 | 16 | 17 |
| 23 |      | 465.00 | 512.20 | 81.20  | 13 | 8  |
| 24 |      | 465.00 | 519.50 | 172.60 | 15 | 17 |
| 25 |      | 465.00 | 621.70 | 81.20  | 43 | 8  |

The TSu cold mass is comprised of thirteen coil modules. Each module consists of bobbins made of 5083-0 Al. Most coil modules contain two wound coils inside the inner diameter. There are twenty-five coils total. Modules are bolted together at flanges to create a rigid structural unit. That unit is mounted in the cryostat using seventeen Inconel support rods, each of which have a spherical bearing at each end. The TS cold mass and cryostat components are shown in Figure 6.52.

Each module can house up to two coils, which are inserted into each end. A typical module can be seen in Figure 6.53. Each coil is wound in a collapsible mandrel over an aluminum strip, which is used to provide cooling for the coil and ground insulation (around the whole coil). Each module will be warmed up, allowing sufficient clearance (typically 1 mm) for coil insertion followed by a shrink fit. The modules can be fabricated by using a 5-axis industrial milling machine and a CNC lathe. The cross section of all TSu coils and modules is shown in Figure 6.54.

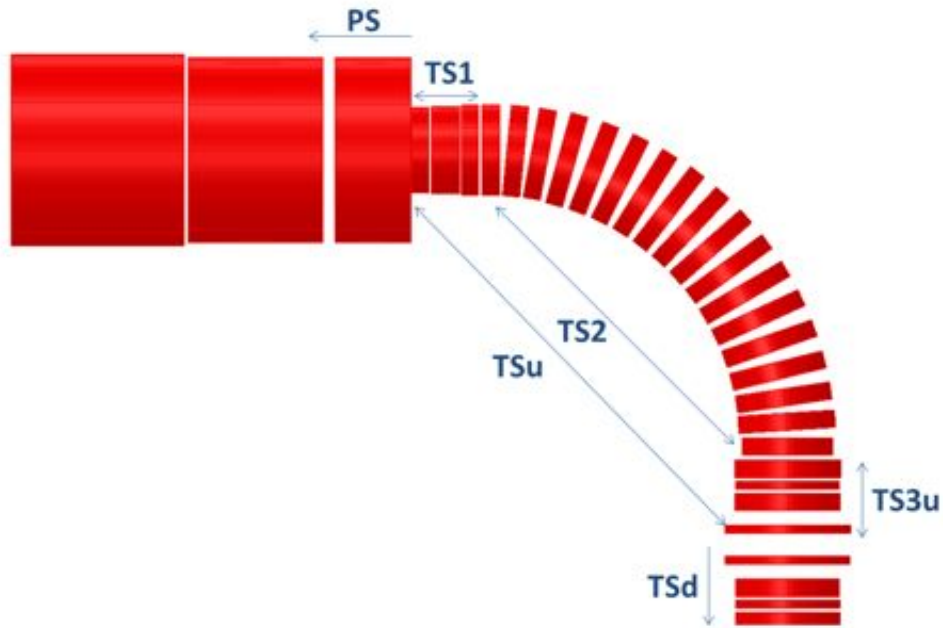

Figure 6.51. TSu coil locations with respect to the adjacent magnets

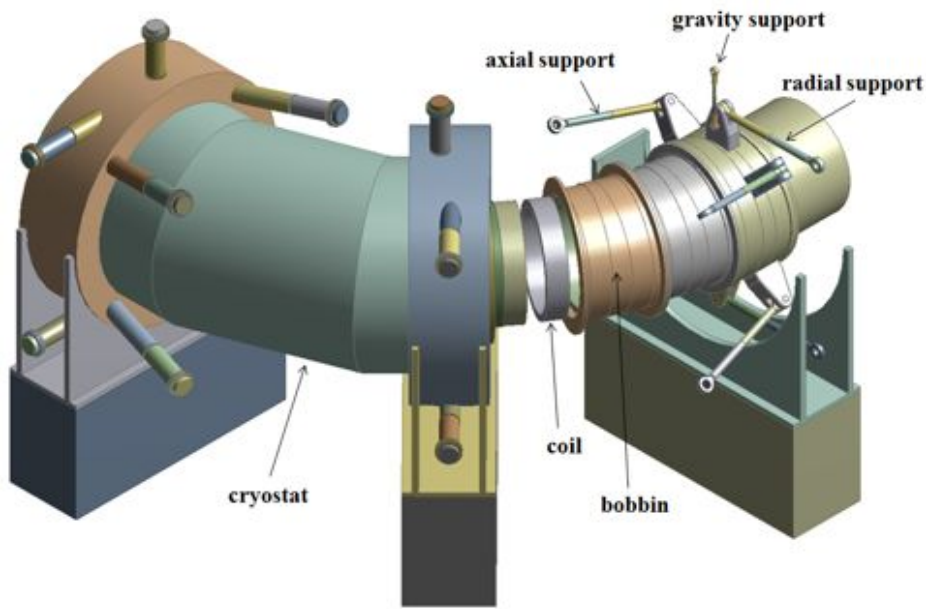

Figure 6.52. TSu components (partial cutaway).

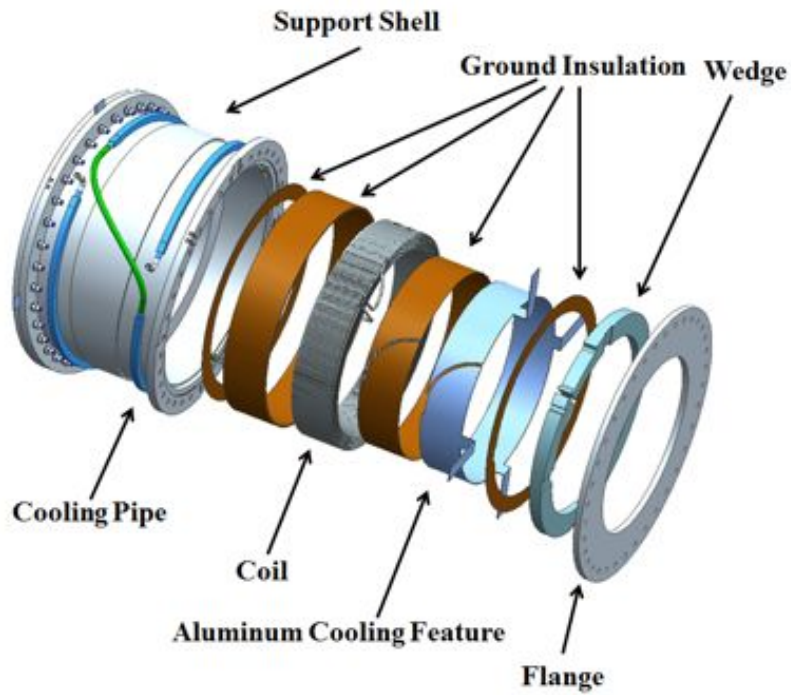

Figure 6.53. A typical TS coil module.

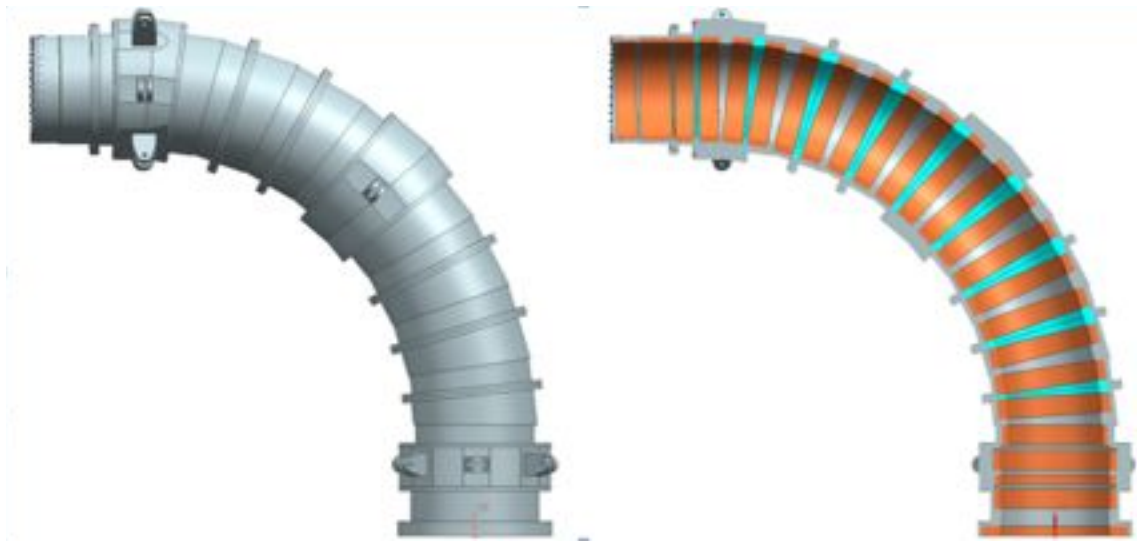

Figure 6.54. TSu modules assembled together (left) and cross section showing modules and their coils (right).

TS1 is a straight section with a length of 704 mm and a free end flange that interfaces with the Production Solenoid (Figure 6.55). TS2 is a toroid with a global centerline bend radius of 2.929 m. TS3u is a 750 mm long straight solenoid with a free end flange that interfaces with the TS3d (Figure 6.57).

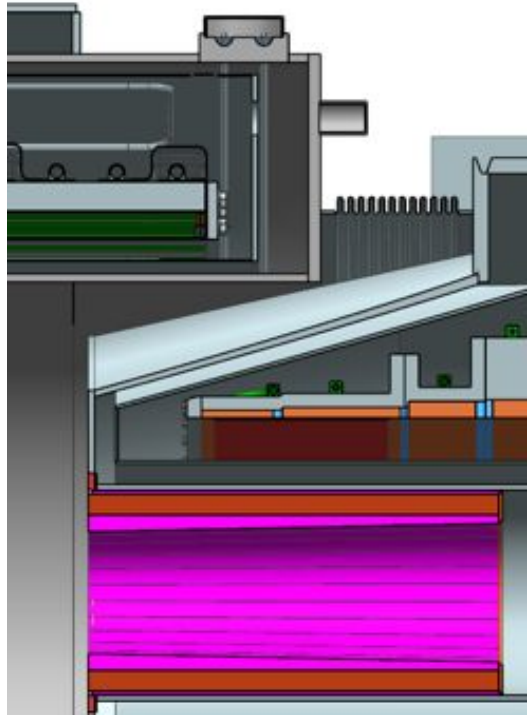

Figure 6.55. TSu-PS cryostat interface.

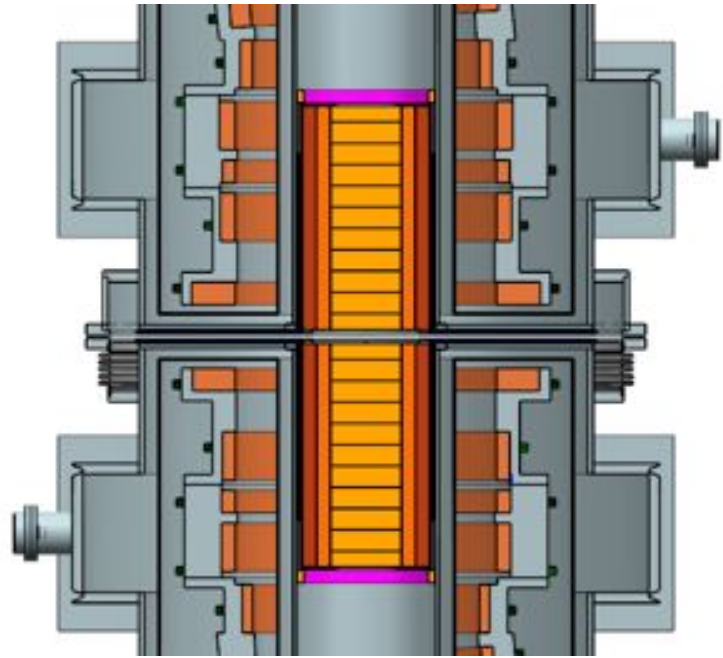

Figure 6.56. TSu interface with TSd.

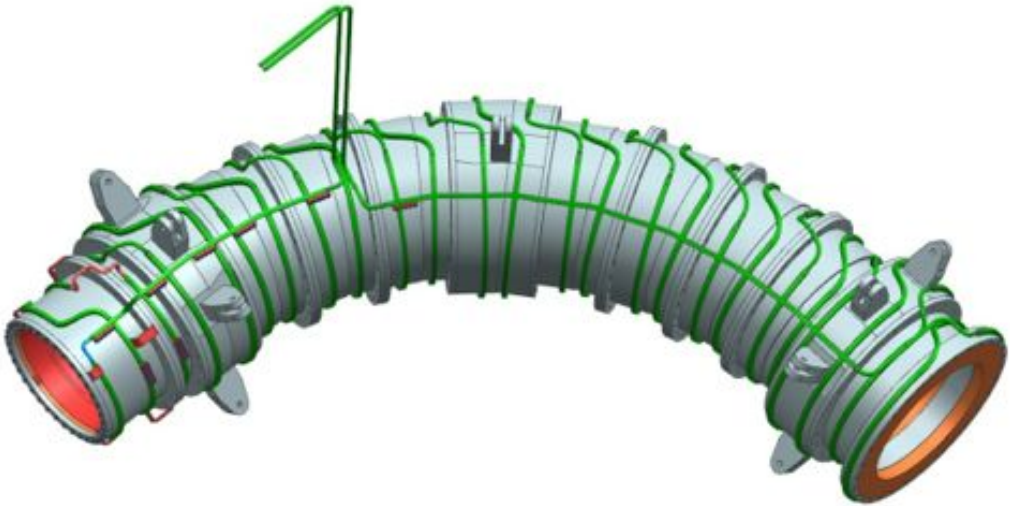

Figure 6.57. TSu Cooling scheme sketch.

The TS coils will be cooled down to cryogenic temperatures through an indirect cooling method. Each module will have an aluminum tube welded to it (Figure 6.54) where the liquid helium flows providing cooling for the aluminum shell that houses the coil. Pure aluminum strips coming from the inner bore of the coils are attached to this tube providing cooling to the coil. The tube can also provide cooling for the splice boxes.

During the final assembly, modules will be bolted together and cryogenic connections will be made in between the modules, as well as the splices. Figure 6.57 shows a sketch of the TSu cooling scheme

### 6.3.2.6   *TSu Mechanical Design*
There are several methods currently under investigation for the splices:

- Rutherford-to-Rutherford splicing (shown in Figure 6.58) can be done effectively with good electrical resistance and mechanical strength; however, the removal of the aluminum from the cable can be complicated because of the difficulty involved in manipulating an object as large as the module.
- Soldering the two cables together (Al-to-Al) is possible using Indium or a Tin Silver alloy. This procedure can also be done relatively fast with minimal preparation of the surface of the conductor, as no aluminum removal is necessary. This type of joint usually provides good electrical resistance but a very poor mechanical bond. This problem can be overcome through the use of the design splice cooling box.

- Weld of the two conductors is also possible. This process yields good mechanical bonds and, typically, good electrical resistance; however welds have the potential to degrade the superconductor given conductor's overall size.

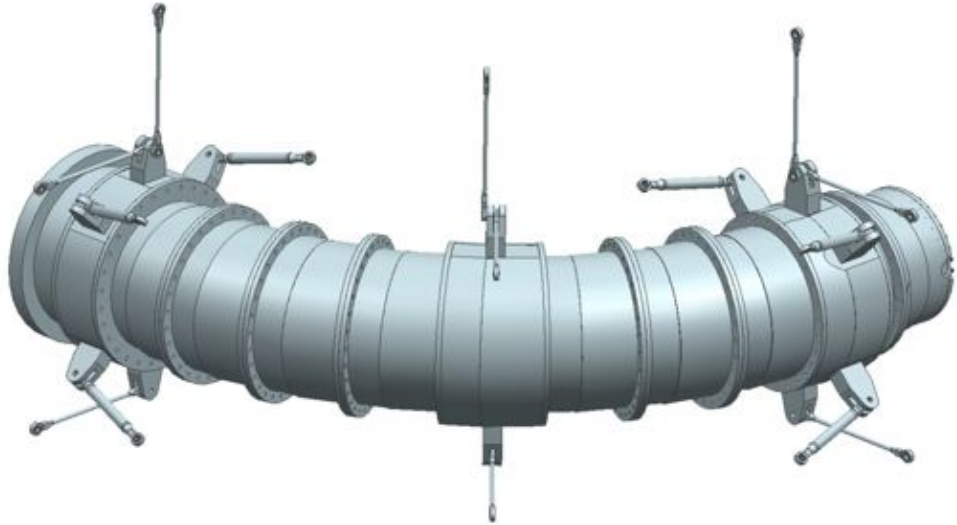

Figure 6.58. Example of a Rutherford-to-Rutherford splice.

The mechanical support system for TSu consists of 4 radial supports (in the direction of the toroid main radius), 8 axial supports and 3 gravity supports, as shown in Figure 6.59.

The radial supports react only against tensional loads. The axial supports operate under both tension and compression. The orientation and configuration of the axial supports are designed to allow a clear path for the proton beam tube that passes nearby as it intersects the PS. The supports are made of Inconel 718.

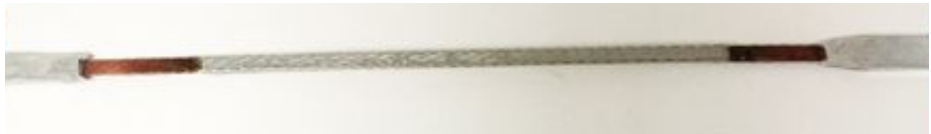

Figure 6.59.  TSu support structure (cooling pipes not shown).

The structural analyses were performed in order to reveal the maximum loads the supports will see during normal operation and for a range of fault and misalignment scenarios. Because they are simple two-force members, the support capacities can be determined by well-understood closed-form calculations. The magnetic and gravity supports are shown in Figure 6.60 and Figure 6.61, respectively.

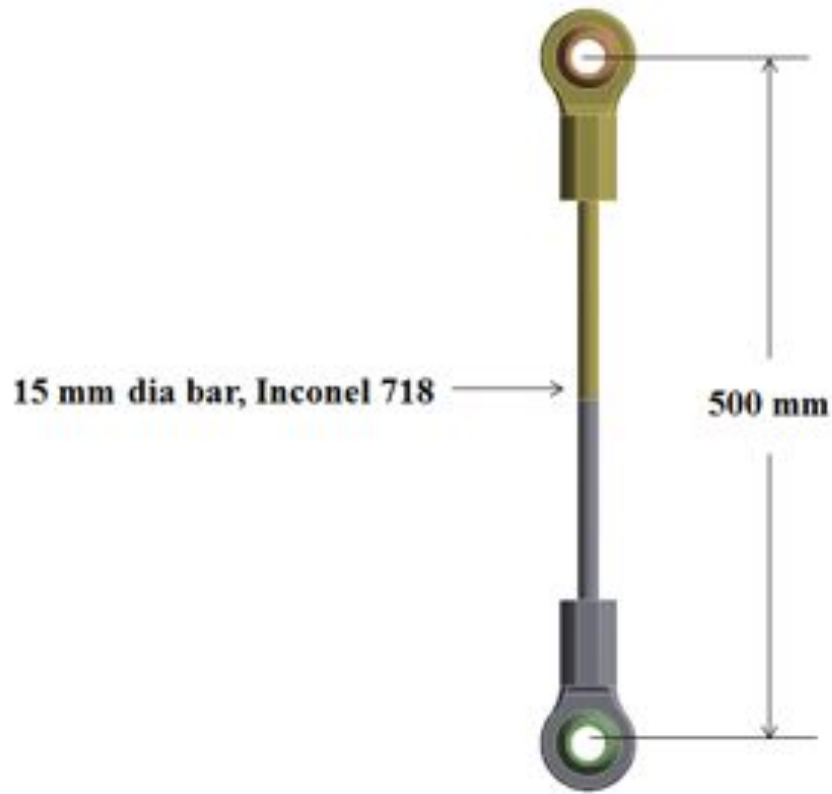

Figure 6.60. Magnetic force support.

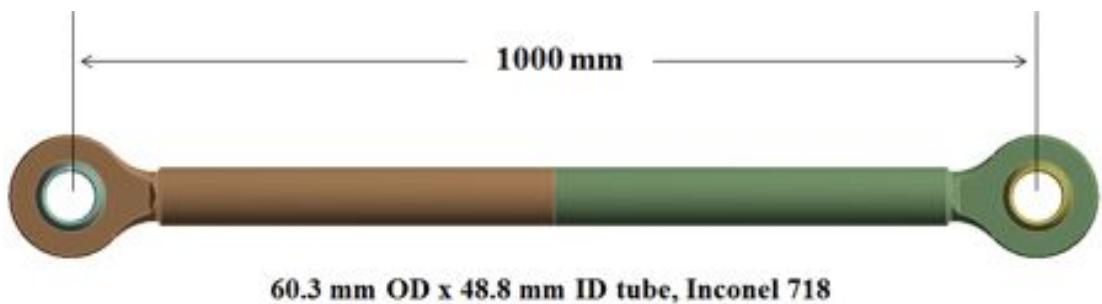

Figure 6.61. Gravity support.

The seventeen Inconel support rods are configured in such a way that three supports bear only the weight of the cold mass (approximately 70 kN), while the remaining supports, which are considerably more robust, oppose the magnetic forces. Under normal operation, these forces tend to act toward the center of the toroid's axis of revolution in the horizontal plane, and total 1915 kN.

Of the fourteen magnetic force supports, only the four axial supports at the downstream end are designed to resist both tension and compression. The remaining supports can resist only tension.

The magnetic supports attach at their cold ends to three reinforced bobbins on the cold mass and at their warm ends to three reinforced support rings on the cryostat. The warm support ends attach to the cryostat at nozzles. The radial supports lie in the horizontal plane; the axial supports are tipped outward 15 degrees from the toroid magnetic axis to prevent interference with the cryostat shell.

The cryostat sits on three rectangular boxes that attach to the floor in the experimental hall. These boxes are ultimately responsible for safely transferring the total magnetic force into the structure of the floor.

Current thoughts on the alignment and cool down strategy are as follows:

- The cold mass is suspended on the gravity supports and adjusted so that the center of the toroid lies slightly below the centers of the PS solenoid and the TSd magnet. This vertical offset compensates for the small upward motion caused by contraction of the gravity supports on cool down. All other supports, including the downstream axial supports, are loose during this operation.
- When vertical alignment is achieved, the downstream axial supports (which can resist both tension and compression) are locked in place.
- The magnet is cooled down.
- After cool down, the tension-only supports are tightened just enough to remove any play.

Several load cases were simulated in order to study the TS coil displacements, the stresses in the coils, the structure and support rods during normal operation and various failure scenarios (Table 6.16). Large variation of the forces could be generated when the magnetic systems adjacent to TSu are powered off.  These conditions will be avoided during normal operation; nonetheless, they will occur during test of TSu as stand-alone magnet and may occur in the case of a quench or the failure of adjacent systems. The TSu structure and supports have been designed for normal operation with sufficient margin to withstand these special and failure modes.

After cooldown (Figure 6.62) and during excitation (Figure 6.63) the von Mises stress in the coils is less than 25 MPa when stress concentration points due to the mesh are excluded Table 6.16 shows the stress in the supports during normal operation (all magnets at operating current) and in case of failure scenarios (PS off or TSd off, with all other magnets on). The radial supports see the highest tensile load during normal operation. The axial supports could see the highest tensile stresses or some compression if adjacent magnets fail. The supports are made of Inconel 718, which has an allowable stress of 531 MPa when cold, and are designed to withstand the full load of these failure modes In all conditions the stresses are below the allowable stress for Al-5083 (107 MPa at 4 K).

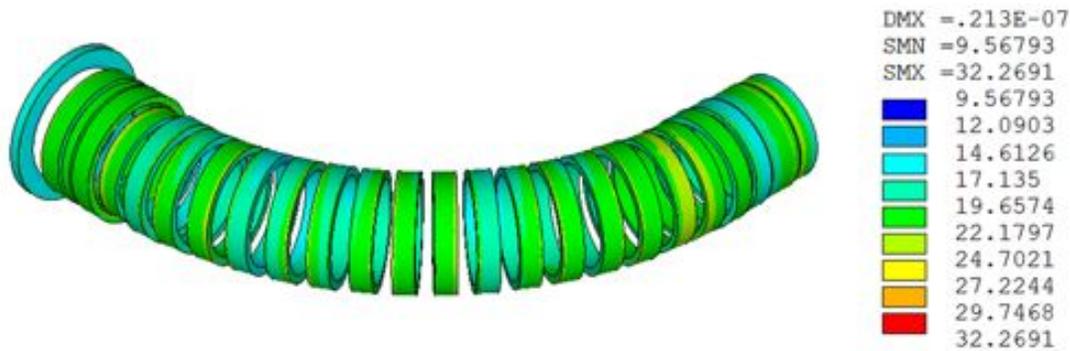

Figure 6.62. Stresses (in MPa) in the TSu coils after cool down.

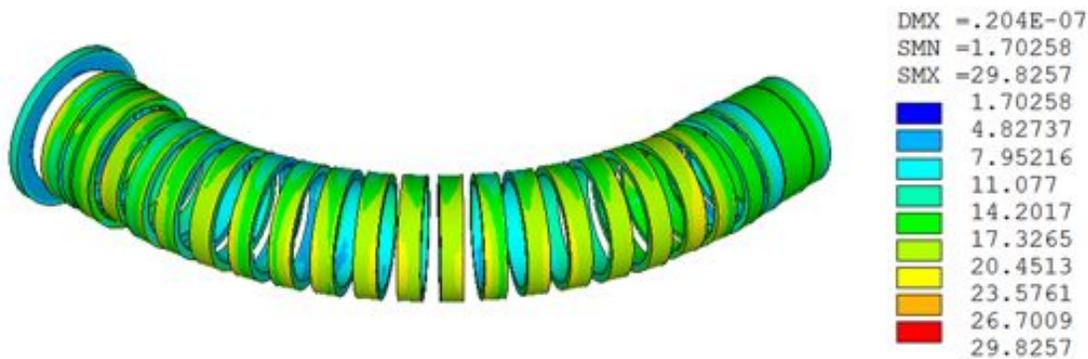

Figure 6.63. Stresses (in MPa) in the TSu coils during excitation.

The global displacements (vector sum) of the cold mass for the cool down, energized and warm up load cases are shown in Figure 6.64 and Figure 6.65, respectively. The cool down and warm up displacements are given relative to the original warm position. The contribution from energizing is small compared to the thermal contractions, and so is shown with the cool down load step subtracted out; in other words, the energizing deformations are relative to the cool down position. The cool down deformations are consistent with the lengths and thermal contractions of the cold mass. Because the downstream axial supports are fixed in axial translation, the motion of the downstream end of the cold mass is primarily that of the support contractions and the contraction of the last bobbin. This displacement amounts to about 4-5 mm. At the upstream end, total displacements of up to 20-21 mm occur.

Table 6.16. Stresses in the support rods (in MPa) under different power conditions.

| Support | Load Case | | |
| --- | --- | --- | --- |
| | **Normal Operation** | **PS Failed** | **TSd Failed** |
| Upstream Axial 1 | 158 | 0 | 250 |
| Upstream Axial 2 | 169 | 0 | 262 |
| Upstream Axial 3 | 151 | 0 | 208 |
| Upstream Axial 4 | 164 | 0 | 220 |
| Downstream Axial 1 | 175 | 371 | -121 |
| Downstream Axial 2 | 176 | 374 | -119 |
| Downstream Axial 3 | 192 | 294 | -62 |
| Downstream Axial 4 | 204 | 309 | -53 |
| Upstream Radial 1 | 159 | 51 | 70 |
| Upstream Radial 2 | 154 | 18 | 64 |
| Downstream Radial 1 | 140 | 0 | 94 |
| Downstream Radial 2 | 134 | 3 | 88 |
| Center Radial 1 | 283 | 14 | 165 |
| Center Radial 2 | 263 | 18 | 145 |
| Gravity 1 | 24 | 24 | 24 |
| Gravity 2 | 28 | 27 | 28 |
| Gravity 3 | 20 | 19 | 20 |

The deformations due to energizing, seen in Figure 6.65, show that displacements of about 1.8 mm will occur at approximately the midpoint of the toroidal arc. The axial and radial supports provide good stiffness against the attractive forces from the PS and TSd, with displacements of less than a millimeter at each end. Later in this report the individual coil motions from the cryostat simulation will be given.

The TSu cryostat is shown in Figure 6.66 and Figure 6.67. Table 6.17 lists the dimensions and materials for the various cryostat components. The TSu cryostat consists of the components and systems listed below:

- Structural supports for the magnetic coils and the vacuum vessel.
- A 4.5 K cooling circuit.
- An 80 K thermal shield.
- A vacuum vessel with a warm bore.
- Interface to the PS cryostat.
- Interface to the proton beam line.
- Interface to the TSd cryostat and to the antiproton window.
- Interface to a cryogenic transfer line.
- Support for a Collimator in the warm bore at the PS interface.
- Support for a Rotatable Collimator in the warm bore at the TSd interface, and interface with rotating mechanism.

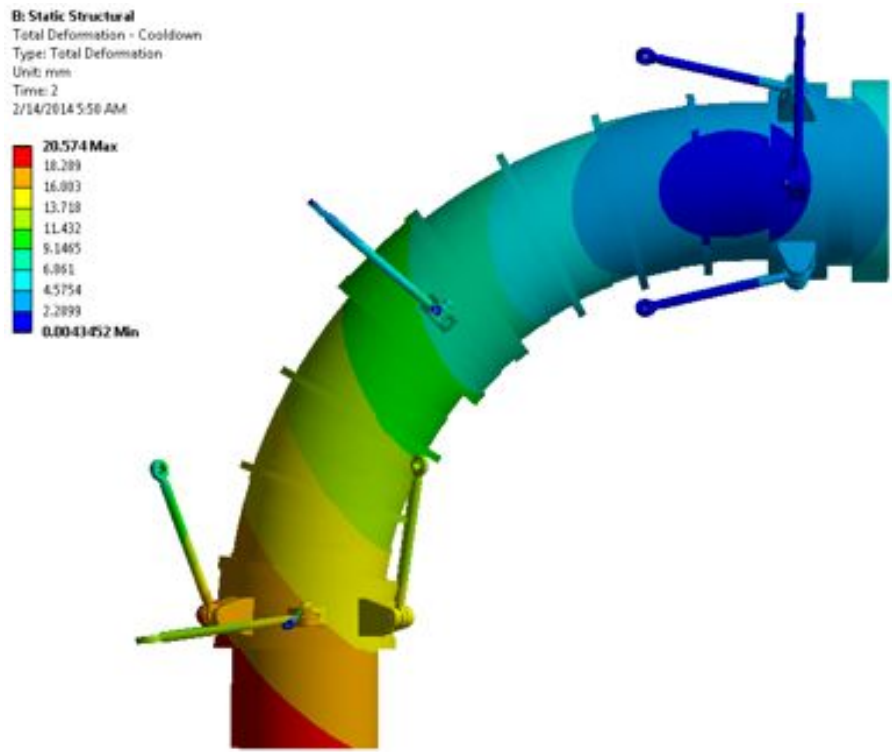

Figure 6.64. Cold mass displacements after cool down.

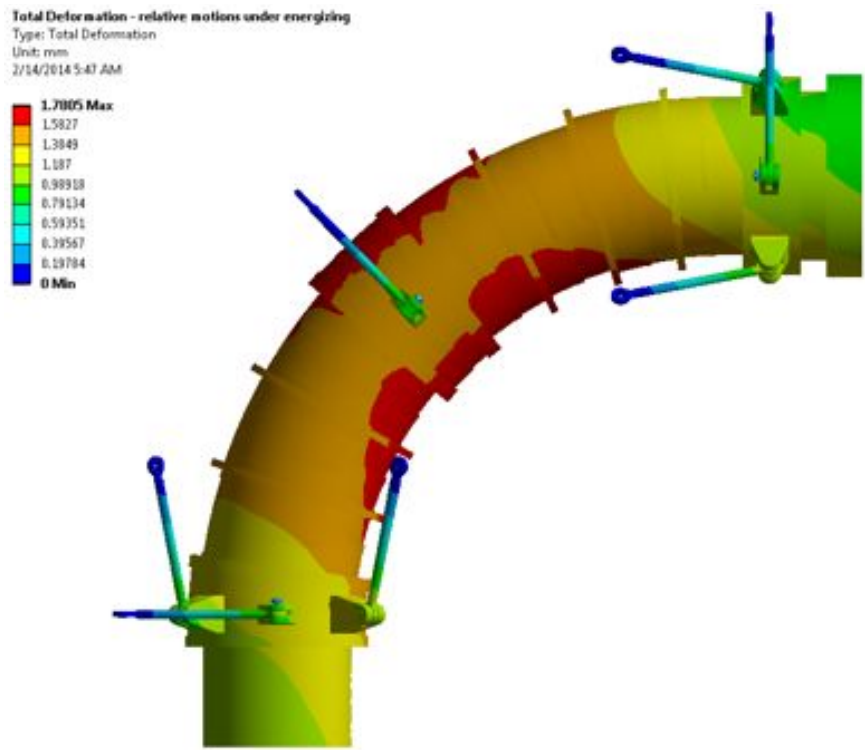

Figure 6.65. Cold mass displacements during the energization relative to the cool down.

Table 6.17. TSu Cryostat Parameters.

| Cryostat Component | Dimension (mm) | Material |
|---|---|---|
| Vacuum Vessel Outer Shell (OD / wall thickness) | 1350 / 25 | Stainless Steel |
| Vacuum Vessel Inner Shell (ID / wall thickness) | 500 / 12.7 | Stainless Steel |
| Vacuum Vessel End Wall (upstream / downstream thickness) | 30 / 15 | Stainless Steel |
| Thermal Shield Outer Shell (OD / wall thickness) | 1100 / ~2 | Aluminum |
| Thermal Shield Inner Shell (ID / wall thickness) | 650 / ~2 | Aluminum |
| Thermal Shield End Wall (thickness) | 3 | Aluminum |

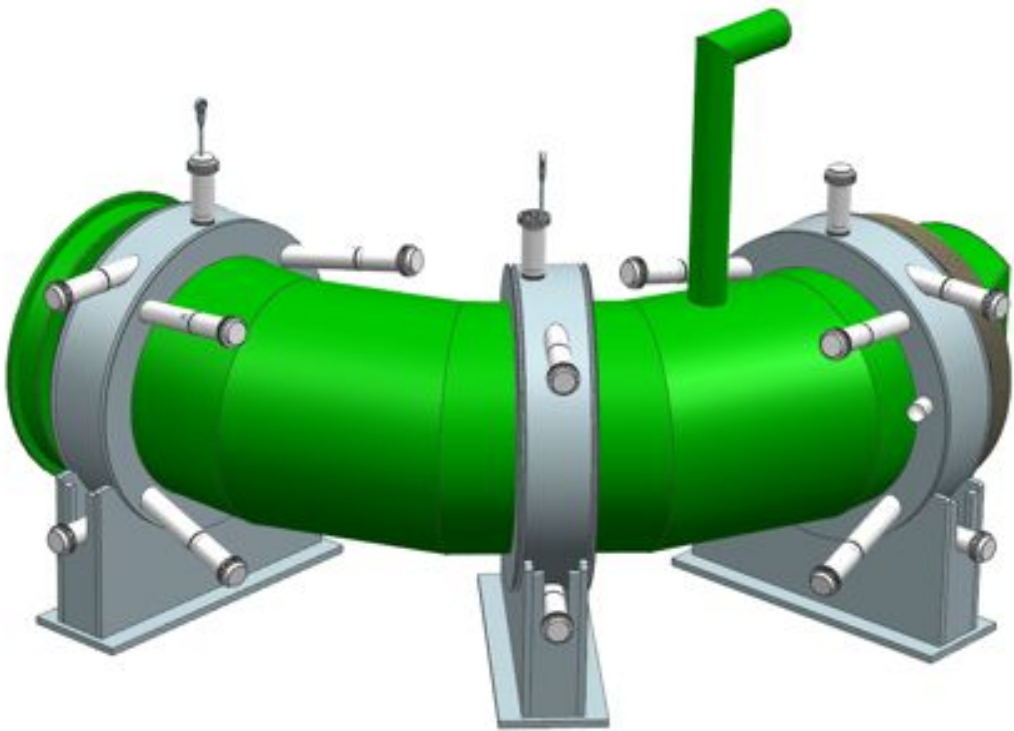

Figure 6.66. View of TSu cryostat. The proton beam line can be seen on the right.

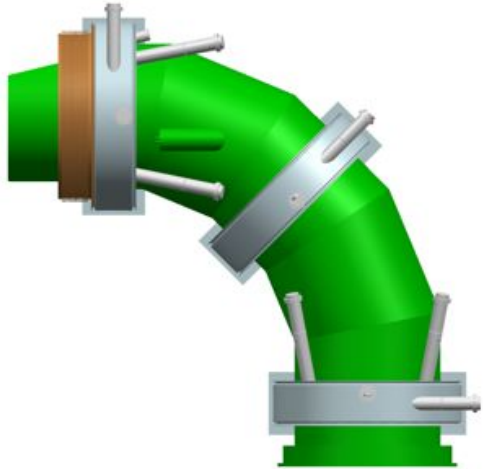

Figure 6.67. Top view of the TSu cryostat.  The proton beam tube can be seen near the top of the figure.

The interface between TS3u and the PS cryostats will consist of flanged connections with bellows. The incoming proton beam line passes through the TSu cryostat where a large section of the cryostat houses the axial supports at the ends of the PS. The cryostat around the TS1 coils has a cutout to prevent interference with the proton beam. This can be seen in Figure 6.55 and Figure 6.67.

The interface between TS3u and TS3d cryostats will consist of flanged connections with bellows housing the frame of the antiproton window, which is also used to separate upstream and downstream vacuum, between mating flanges. The bellows will allow for up to 20 mm of axial offset. Details of this interface can be seen in Figure 6.56.

### 6.3.2.7  TSd Design Concept
The TSd coil and cryostat designs are very similar to those of TSu. TS3d coils are connected to a toroidal section (TS4) and to a straight section (TS5) in a single cryostat and powered in series. Part of TS5 is housed inside the DS, and this can be seen in Figure 6.68.

### 6.3.2.8  TSu Quench Protection and Analysis
The quench protection strategy in the Transport Solenoid is based on extracting most of the energy to external dump resistors rather than relying on the quench to propagate through a series of small coils, which is inefficient and could damage the magnets. If the resistive voltage component exceeds the quench detection threshold of 0.5 V for more than 1 second, the current will be shunted through a 0.34 Ohm dump resistor and the power supply will be switched off.  Symmetric grounding will be used to keep the maximum coil-to-ground voltage well below the requirement with 600 V at the leads when the dump resistance is activated in the circuit. Symmetric grounding may be disconnected in case of a single-point coil-to-ground failure. The ground insulation must therefore be 2 mm thick and be made of several layers (minimum 6 layers) of fiberglass cloth or G10. The layer-to-layer insulation will be 0.25 mm thick and made from 0.125 mm thick fiberglass tape with a 45% overlap.

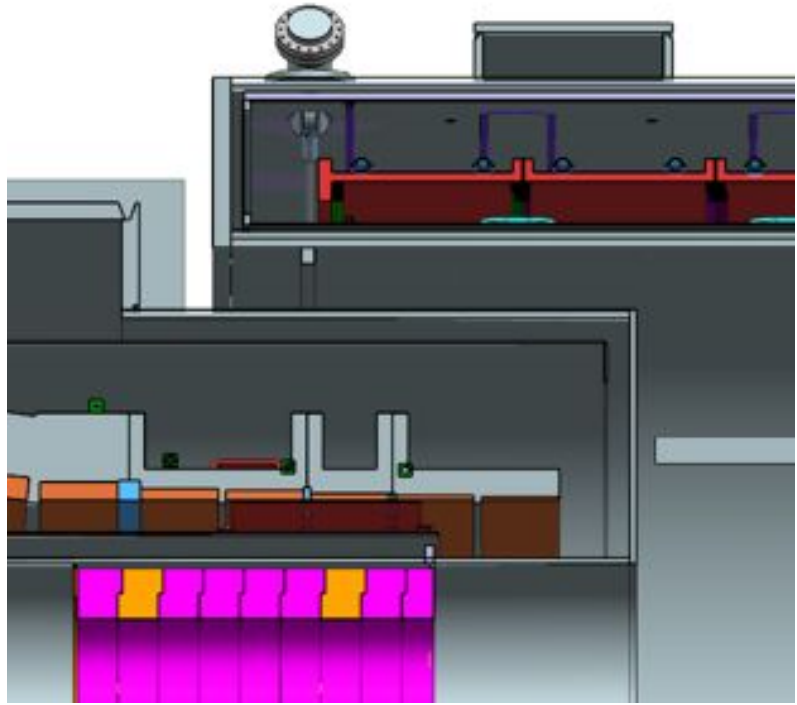

Figure 6.68.  TSd-DS interface.

The main parameters of the TSu quench protection system are shown in Table 6.18. The TSu energy stored at nominal operating current (1730 A) is 7.1 MJ with the adjacent magnets powered off. In this analysis, 10.4 MJ of stored energy was used to account for coupling with the adjacent magnets and to provide some margin. The inductance was scaled accordingly.

The hot spot temperature was computed using the numerical code QLASA [1] in the adiabatic condition, assuming that the quench propagates only in the quenching coil. The "High Field" column shows the results when the quench starts in the peak field area of the coil with the highest field (3.4 T). The "Low Field" column shows the results for the case when the quench starts in a low field (1.0 T) area of a coil with a low peak field (2.2 T). The hot spot temperature is well below the maximum acceptable temperature (120 K) in both cases, even with these conservative assumptions.

Table 6.18. TSu quench protection parameters. The *High Field* column shows the results for a quench that starts in the peak field area of the coil with the highest field. The *Low Field* column shows the results for a quench that starts in a low field (1.0T) area of a coil with a low peak field.

| | Units | High Field | Low Field |
|---|---|---|---|
| Copper RRR | | 100 | |
| Aluminum RRR | | 200 | |
| Operating current | A | 1730 | |
| B max in quenching coil | T | 3.4 | 2.2 |
| B where quench starts | T | 3.4 | 1.0 |
| Energy (with coupling & margin) | MJ | 10.4 | |
| Inductance (with coupling & margin) | H | 6.95 | |
| Dump resistance | Ohm | 0.34 | |
| L/R | s | 20.4 | |
| Quench detection threshold | V | 0.5 | |
| Time above threshold before QPS activation | s | 1 | |
| Voltage at leads at dump insertion | V | 590 | |
| Hot spot temperature (QLASA) | K | 70 | 69 |
| $J_{eng}$ | A/mm$^2$ | 47 | |
| $I_{op}$ / $I_c$ (5.1 K) on load line | | 58% | |
| Temperature of heat generation (3.4 T, $I_{op}$) | K | 6.92 | |

### 6.3.2.9   TS Module Testing

Upon receipt from the vendor, an incoming inspection will be made to certify that each coil module meets specification. This inspection will include measurements of the overall module geometry, inductance, resistance, hipot to ground, and room-temperature magnetic field angle using a Single Stretched Wire (SSW) system and survey, in addition to instrumentation checks. All of these measurements must be performed on each coil. The magnetic center axis of each coil with respect to the module alignment surfaces must be measured at room temperature with a precision of 1 mm and 1 mrad to check that module machining and coil insertion meet tolerances.

After SSW measurement of the individual modules, preparation for each cold test will follow. Generally this involves making internal electrical splices, adding voltage taps, bolting modules together, completing the electrical joint and cryogenic connection between modules, adding cryogenic fittings for test stand connections, preparing and securing the current leads and final leak, pressure and electrical testing. The test subject will then be delivered to the Fermilab solenoid test facility and mounted to the top plate assembly using four support rods that will adapt to a module bolted flange. Current lead connections to the test stand superconducting bus and cryogenic supply and return connections will then be made and tested. Finally, internal instrumentation and quench protection connections are made and checked before the entire assembly is installed into the STF vacuum vessel. This test preparation plan assumes that a permanent full coverage thermal shield with multi-layer superinsulation will exist in the vessel before production testing begins; this shield will obviate the need for custom, labor-intensive and time consuming shielding and insulation of each test subject. Once the magnet is installed and the vessel evacuated, final electrical, leak and pressure tests are followed by controlled helium cool down to 4.5 K.

The cryogenic quench performance test will be performed on all coils, tested in two (or three) module assemblies. Because of the TS magnet complexity and the large number of electrical joints between coils, the baseline plan is to perform cryogenic power testing of all modules. The modules must reach a current 20% above the 1730 A operating current in order to be accepted from the vendor. Hall probes will monitor the peak field on each coil during each test. Internal coil-to-coil electrical joint resistances will be measured during each cold test to ensure that they do not exceed the required maximum value. Coil and conduction-cooling component temperatures will be monitored to certify performance of the cryogenic elements of each module assembly.

A total of 14 tests will be performed in the STF at the Fermilab Central Helium Liquifier, of either single coil module or pair of modules, and these tests must occur at an estimated rate of one test per month. The STF cryostat with two modules assembled can be seen in Figure 6.69. Figure 6.70 identifies the module combinations for each test, numbered in the order they will be tested.

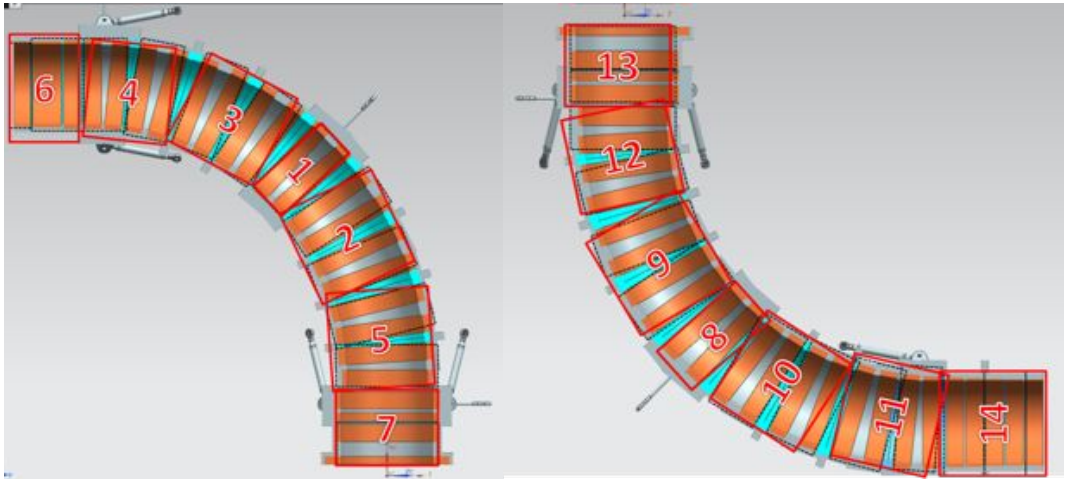

Figure 6.69. Solenoid test facility with two TS modules in it (left) and a detailed view of the supports for the modules (right).

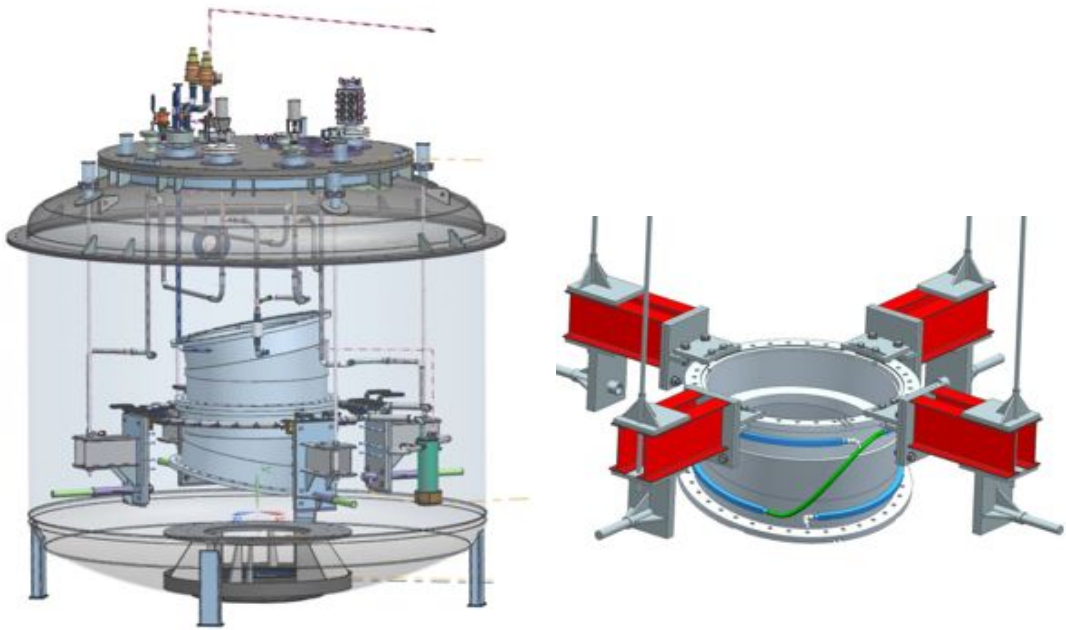

Figure 6.70.  TSu (left) and TSd (right) module pairing and ordering for cold tests.

During the power testing, internal splice resistances will be measured, and conductor RRR will be measured for each coil during the warm up to room temperature. After removing the test subject from the vacuum vessel, disassembly follows the assembly procedure in reverse

### 6.3.3  Detector Solenoid

#### 6.3.3.1  Overview and General Requirements

The main functions of the Detector Solenoid (DS) are to provide a graded field in the region of the stopping target and to provide a precision magnetic field in a volume large enough to house the tracker downstream of the stopping target. The inner diameter of the magnet cryostat is 1.9 m and its length is 10.9 m.  The inner cryostat wall supports the stopping target, tracker, calorimeter and other equipment installed in the Detector Solenoid. This warm bore volume is under vacuum during operation. It is sealed on one side by the VSP and instrumentation feed through bulkhead, while it is open on the other side where it interfaces with the Transport Solenoid. The last section of the Transport Solenoid protrudes into the DS cryostat.

The Detector Solenoid is designed to satisfy the field and operational requirements defined in the DS requirements document [3]. The overall structure of the solenoid is shown in Figure 6.71.  It consists of two sections: a "gradient section", which is about 4 m long, and a "spectrometer section" of about 6 m. The magnetic field at the entrance of the gradient section is 2 T and it decreases linearly to 1 T at the entry to the spectrometer section, where it is then uniform over 5 m.

The Detector Solenoid coil design is based on a high purity aluminum sheath surrounding a NbTi Rutherford cable. This type of conductor has been used successfully in many similar superconducting detector solenoids. Aluminum has very small resistivity and a large thermal conductivity at low temperatures providing excellent stability. Furthermore, aluminum stabilized conductors can be extruded in lengths of several kilometers. Precise rectangular conductor shapes can be obtained, allowing for high accuracy in the coil winding.

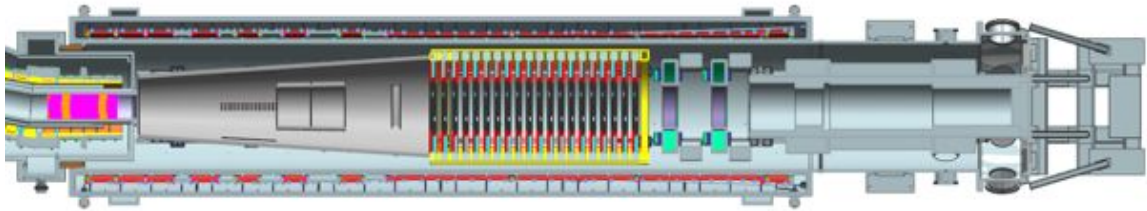

Figure 6.71. Overall structure of the Detector Solenoid coils and cryostat.

Two types of conductor are required, both 20 mm in height. The "narrow" (5.25 mm in width) DS1-type conductor will be used in the DS gradient section, while the "wide" (7 mm in width) DS2-type conductor is used in the spectrometer section. The dimensions are optimized to give the required field when identical current is transported in both conductors. The conductors contain Rutherford-type NbTi cables with 12 and 8 strands, respectively. The strands have a diameter of 1.3 mm, a SC/Cu ratio of 1, and a critical current of 2750 A/mm$^2$ (4.2 K, 5 T). As a result, the conductors have significant stability and safety margins in case of a quench.

In the baseline design, the gradient section is wound in two layers using the "narrow" DS1-type conductor (20 mm × 5.25 mm), which is necessary to obtain a field of 2 T. The field gradient is obtained by introducing several sets of spacers between coil turns. The field uniformity in the spectrometer section is achieved with a "wide" DS2-type conductor (20 mm × 7 mm), wound in a single layer coil.

It is envisaged that the DS coil will be wound in standardized modules on accurately machined collapsible mandrels. After curing, the winding mandrels are extracted and the outer aluminum support cylinders are placed over each module and the assembly epoxy bonded. The preassembled modules are then electrically connected and bolted together with spacers in a single cold mass before installation in the cryostat. The Detector Solenoid (cold mass and cryostat) weights about 42 tonnes. Other parameters of the magnet are summarized in Table 6.19.

### 6.3.3.2   Magnetic Field Design

Figure 6.72 shows the coil structure of the Mu2e solenoid system. DS is composed of 11 coils. DS overlaps the three last coils of TS. Figure 6.73 shows the details of DS coils and the overlap with TS.

Table 6.19. Summary of the Detector Solenoid parameters

| Parameter | Units | Value |
|---|---|---|
| **Coil** | | |
| Inner radius | mm | 1050 |
| Thickness (two layer coil) | mm | 43 |
| Length | mm | 10,150 |
| Mass (cold mass) | Kg | 10,000 |
| **Cryostat** | | |
| Inner diameter | mm | 1900 |
| Outer diameter | mm | 2656 |
| Length | mm | 10,750 |
| Mass | Kg | 32,000 |

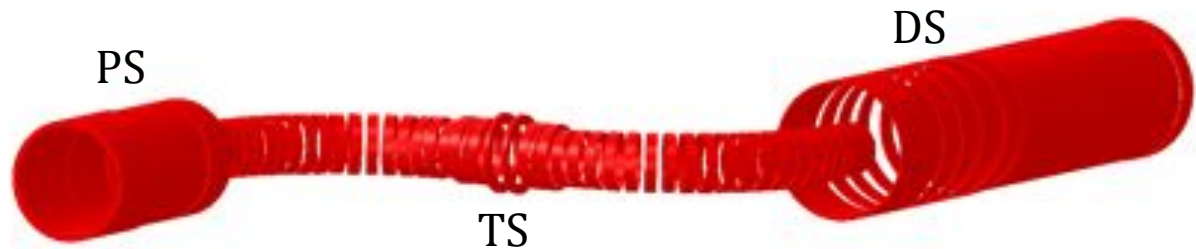

Figure 6.72. Coil layout for the Production Solenoid (PS), Transport Solenoid (TS) and Detector Solenoid (DS).

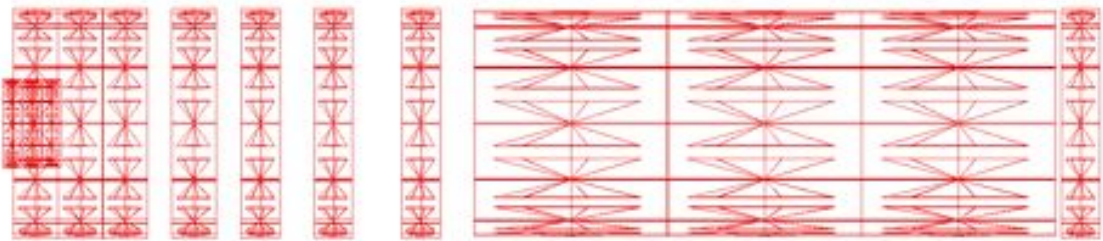

Figure 6.73. Detail of the DS coils and the three TS coils overlapped by DS.

### Magnetic Requirements

The magnetic field in the DS is divided into three main sections: the gradient region (DS1), the transition region (DS2) and a nearly-constant field region (DS3-4). The overall magnetic field distribution along the DS can be seen in Figure 6.74. The field at the upstream end of DS is around 2 T. This field is only achievable when the TS coils are in operational condition. The field linearly decreases to approximately 1 T. The nearly-constant field region is subdivided in two different regions: the Spectrometer region (DS3) and the Calorimeter region (DS4), the last has slightly relaxed tolerances. Table 6.20 summarizes the magnetic requirement in all of these sections.

In general, local minima need to be avoided as much as possible; local minima could trap particles that could be a source of backgrounds.

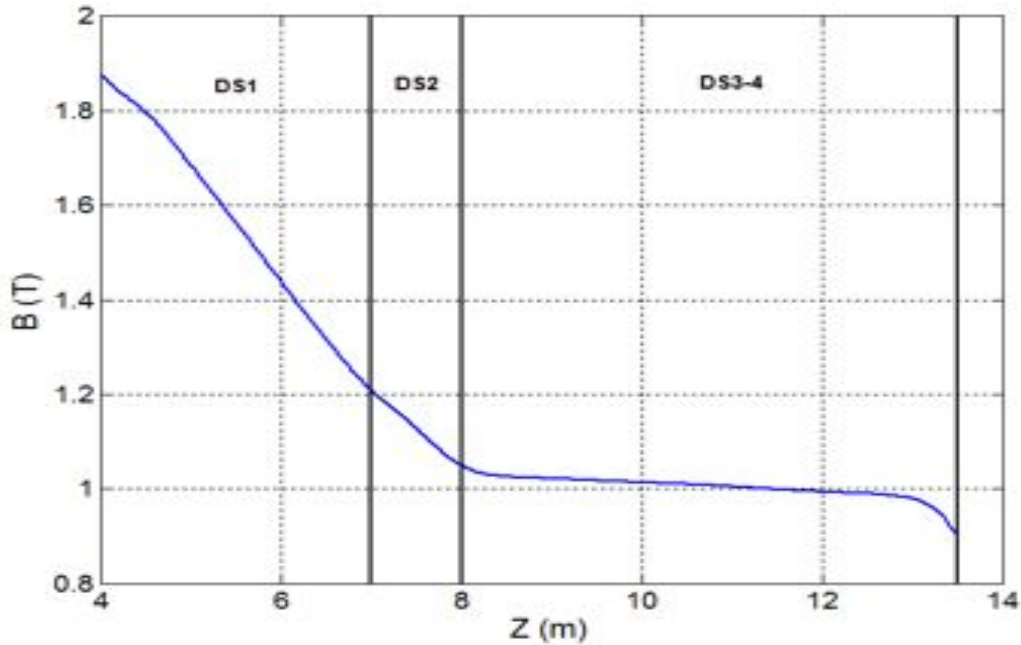

Figure 6.74. DS magnetic field on-axis.

Table 6.20. Detector Solenoid Magnetic Requirements.

| Region | L (m) | z (min) | z (max) | $R_{max}$ (m) | dBs/ds (T/m) | dB/B max | dB/B where |
|---|---|---|---|---|---|---|---|
| DS1 Gradient | 3.0 | 3.93 | 6.93 | 0.3-0.7cone | -0.25±0.05 | 0.05 | r <$R_{max}$ |
| DS2 Transition | 1.2 | 6.93 | 8.13 | na | gradient magnitude decreasing | na | na |
| DS3 Uniform | 3.6 | 8.13 | 11.73 | 0.7 | negative gradient | 0.01 | r<$R_{max}$ |
| DS4 Uniform | 1.5 | 11.73 | 13.23 | 0.7 | negative gradient | 0.05 | r<$R_{max}$ |

*Magnetic Design and Performance*
Figure 6.75 shows the distribution of coils that form the detector solenoid. DS is composed of 11 coils. The TS coils were suppressed from the picture for clarity. However the field contribution from the TS is significant and it has to be taken into account in the design. Table 6.21 summarizes the geometry of the DS coils at the nominal operating temperature and coils are energized.

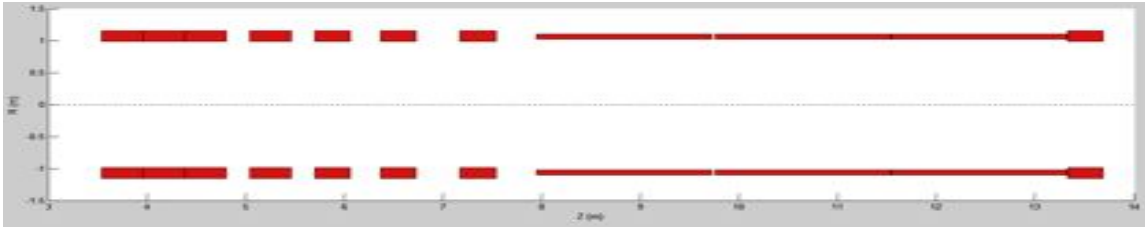

Figure 6.75. DS coils.

Table 6.21. Coils geometry summary at 4.2K and when the coils are powered.

| Coil # | Coil IR [m] | Coil OR [m] | Coil Length [m] | #ZC [m] | # Layers | #Turns Per Layer | Total #Turns |
|--------|-------------|-------------|-----------------|---------|----------|------------------|--------------|
| 1 | 1.0500 | 1.0915 | 0.42075 | 3.7489 | 2 | 73 | 146 |
| 2 | 1.0500 | 1.0915 | 0.42075 | 4.1739 | 2 | 73 | 146 |
| 3 | 1.0500 | 1.0915 | 0.42075 | 4.5989 | 2 | 73 | 146 |
| 4 | 1.0500 | 1.0915 | 0.42075 | 5.2519 | 2 | 73 | 146 |
| 5 | 1.0500 | 1.0915 | 0.36325 | 5.8801 | 2 | 63 | 126 |
| 6 | 1.0500 | 1.0915 | 0.36325 | 6.5701 | 2 | 63 | 126 |
| 7 | 1.0500 | 1.0915 | 0.36325 | 7.3971 | 2 | 63 | 126 |
| 8 | 1.0500 | 1.0705 | 1.8310 | 8.8178 | 1 | 244 | 244 |
| 9 | 1.0500 | 1.0705 | 1.8310 | 10.6528 | 1 | 244 | 244 |
| 10 | 1.0500 | 1.0705 | 1.8310 | 12.4883 | 1 | 244 | 244 |
| 11 | 1.0500 | 1.0915 | 0.36325 | 13.6425 | 2 | 63 | 126 |

There are single and double layer coils. The double layer coils can be found in two different lengths. The coils are wound from two conductor sizes. Double-layer coils use DS1-type conductor and single-layer coils use DS2-type. The single-layer coils are the longest coils, and they are located in the spectrometer/calorimeter region. The design takes into consideration a 0.250 mm thick insulation around the conductor.

DS is powered using a single power supply. The operational current is 6114 A.

### 6.3.3.3  Conductor Design

The design of the Detector Solenoid coils is based on two conductors (DS1-type and DS2-type) composed of a high purity aluminum case surrounding a Nb-Ti Rutherford cable. Aluminum has very small resistivity and a large thermal conductivity at low temperatures providing excellent stability. Furthermore, aluminum stabilized conductors can be extruded in lengths of several kilometers. Precise rectangular conductor shapes can be obtained, allowing high accuracy of coil winding. The dimensions of the DS1-type and DS2-type conductors are optimized to produce the required field distributions when identical current is transported in both conductors.

The manufacturing of the aluminum stabilized conductor is a multi-step process that can be briefly summarized as follows:

- Fabrication of NbTi strands
- Cabling of NbTi strands into rectangular Rutherford Cables of prescribed size
- Fabrication of aluminum billets with prescribed high purity
- Cladding of Rutherford cables with aluminum
- Cold working of aluminum clad cable to reach required size and mechanical properties
- Testing of final cable and sub components

The DS cable conforms to specifications described in [58] - [60].

### DS Strands

DS strand features and technical specifications are summarized in Table 6.22.

### Cable Design

#### Technical Specifications for the DS Rutherford Cables

The Rutherford-type NbTi cables for the Detector Solenoid contain 12 and 8 strands, respectively. As detailed in the following, to meet the specifications with the required margins, a maximum allowable cabling degradation of 5% is factored into the design, reducing the single strand $J_c$ to 2600 A/mm$^2$. In Table 6.23 and Table 6.24 the technical specifications for DS1-type and DS2-type Rutherford cable are summarized. The number of strands for DS1 and DS2 is optimized to satisfy the magnetic design and the margins required by the experiment, while having the same level of current transported by the two cables.

Table 6.22. Strand mechanical and electrical specifications (for reference).

| Parameter | Unit | Value | Tolerance |
|---|---|---|---|
| Diameter | mm | 1.303 | ± 0.005 |
| Cu /SC ratio | | 1 : 1 | ± 0.1 |
| Filament Diameter | μm | 40 | - |
| Surface coating | | None | |
| Minimum critical current (at 4.22 K, 5 T) | A | 1850 | - |
| Minimum RRR | | 80 | |
| Twist direction | | Left | |
| Twist pitch | mm | 30 | ± 4 |

Table 6.23. Cable 1 (DS1) mechanical and electrical specifications.

| Parameter | Unit | Value | Tolerance |
|---|---|---|---|
| Number of strands | | 12 | - |
| Cable width | mm | 7.88 | ± 0.01 |
| Cable thickness at 5 kPsi | mm | 2.34 | ± 0.01 |
| Transposition angle | degree | 15 | ± 0.5 |
| Lay direction | | Right | - |
| Minimum critical current (at 4.22 K, 5T) | A | 20900 | - |
| RRR | | ≥ 60 | |
| Cable residual twist | degree | < 45 | |
| Minimum bending radius | mm | 20 | |

*Technical Specifications for DS Al-Stabilized Cables*

To meet the specifications with the required margins on conductor performance, as explained in the following, a maximum allowable extrusion plus cold work degradation of 10% is factored in the design. The requirements of the field uniformity in the gradient and spectrometer sections impose similar dimensional tolerances of the conductors, as specified in Figure 6.76 and Figure 6.77. Cable parameters with their tolerances are summarized in Table 6.25 and Table 6.26. A final cold work step in the manufacturing process is envisioned to insure final cable dimensions as well as to meet the required properties of pure Aluminum. To meet the magnet design specifications in terms of mechanical requirements and quench protection, minimum values for Aluminum and Copper RRR, Aluminum yield strength and bonding strength between the Aluminum case and the Rutherford cable are explicitly specified for each of the cable designs.

Table 6.24. Cable 2 (DS2) mechanical and electrical specifications.

| Parameter | Unit | Value | Tolerance |
|---|---|---|---|
| Number of strands | | 8 | - |
| Cable width | mm | 5.25 | ± 0.01 |
| Cable thickness at 5 kPsi | mm | 2.34 | ± 0.01 |
| Transposition angle | mm | 15 | ± 0.5 |
| Lay direction | | Right | - |
| Minimum critical current (at 4.22 K, 5T) | A | 13900 | - |
| RRR | | ≥ 60 | |
| Cable residual twist | degree | < 45 | |
| Minimum bending radius | mm | 20 | |

Table 6.25. Mechanical and electrical specifications of the DS1 aluminum stabilized cables in final state (i.e. after cold work).

| Parameter | Unit | Value | Tolerance |
|---|---|---|---|
| Cable critical current (at 5 T, 4.22 K) | A | > 18800 A | |
| Cable width (after cold work) at 293K | mm | 20.1 | ± 0.1 |
| Cable thickness (after cold work) at 293K | mm | 5.27 | ± 0.03 |
| Copper RRR | | > 80 | |
| Aluminum RRR | | > 800 | |
| Aluminum 0.2% yield strength at 293 K | MPa | > 30 | |
| Aluminum 0.2% yield strength at 4.2 K | MPa | > 40 | |
| Shear strength between aluminum and strands | MPa | > 20 | |

DS1 and DS2 conductors have been designed keeping in mind that the magnet is required to operate at least at 45% of conductor quench current, allowing a temperature margin of 2.5K with respect with the conductor temperature of 5K.

*Production Lengths*

The detector solenoid coil modules will feature a combined total length of 17 km of Al-stabilized NbTi cable, including one spare unit per cable design. Number of units and unit lengths are summarized in Table 6.27.

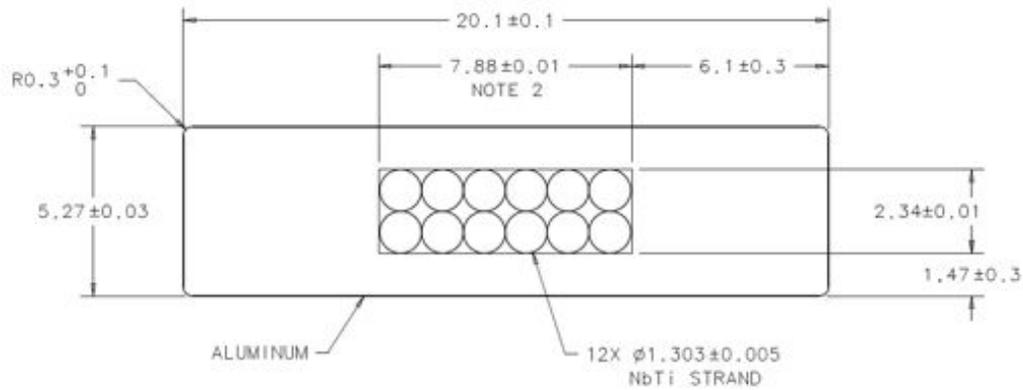

Figure 6.76. Cross-section of DS1 Al-Stabilized Rutherford Cable at 293K.

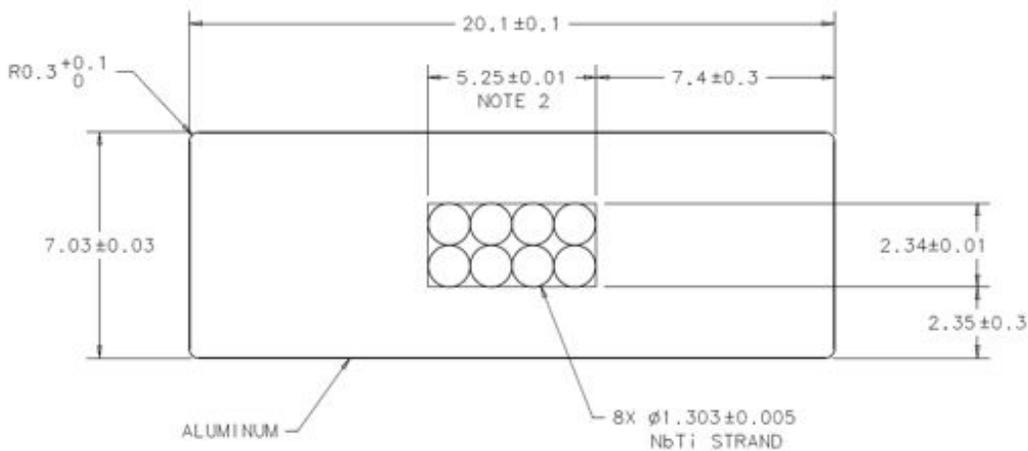

Figure 6.77. Cross-section of DS2 Al-Stabilized Rutherford Cable at 293K.

Table 6.26. Mechanical and electrical specifications of the DS2 aluminum stabilized cables in final state (i.e. after cold work).

| Parameter | Unit | Value | Tolerance |
|---|---|---|---|
| Cable critical current (at 5 T, 4.22 K) | A | > 12500 A | |
| Cable width (after cold work) | mm | 20.1 | ± 0.1 |
| Cable thickness (after cold work) | mm | 7.03 | ± 0.03 |
| Copper RRR | | > 80 | |
| Aluminum RRR | | > 800 | |
| Aluminum 0.2% yield strength at 293 K | MPa | > 30 | |
| Aluminum 0.2% yield strength at 4.2 K | MPa | > 40 | |
| Shear strength between aluminum and strands | MPa | > 20 | |

Table 6.27. Summary of Production Lengths for DS Al-stabilized Cable.

| Cable | Number of Units | Unit Length |
|-------|-----------------|-------------|
| DS1 | 9 | 1100 meters |
| DS2 | 4 | 1750 meters |

### 6.3.3.4  Coil Design

**Coil Design Overview**

Figure 6.78 shows the layout of all 11 DS coils. Table 6.28 lists coil parameters at room temperature.

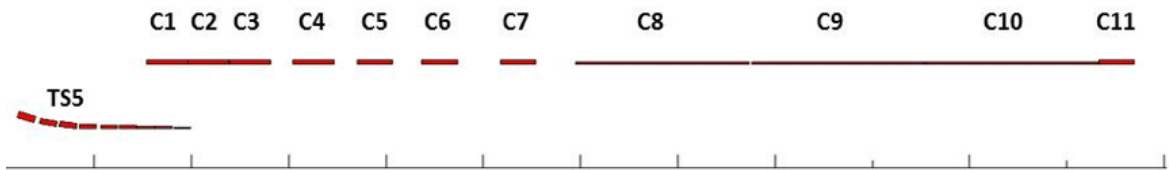

Figure 6.78. Layout of the DS Coils.

**Insulation Scheme**

The DS employs a composite cable insulation made of polyimide and pre-preg glass tapes as shown in Figure 6.79. Additional ground and coil insulation is shown in Figure 6.80.

Table 6.28. Parameters of the Detector Solenoid coil segments at 300 K. Both DS1 and DS2 cables included the 0.25 mm composite cable and 0.5 mm ground insulation.

| Coil Number | Center Z (mm) | Length (mm) | Length Tolerance (mm) | Inner Radius (mm) | Radius Tolerance (mm) | Turns |
|-------------|---------------|-------------|-----------------------|-------------------|-----------------------|-------|
| 1 | 241 | 422.5 | 2 | 1053.5 | 1 | 2x73 DS1 |
| 2 | 668 | 422.5 | 2 | 1053.5 | 1 | 2x73 DS1 |
| 3 | 1095 | 422.5 | 2 | 1053.5 | 1 | 2x73 DS1 |
| 4 | 1751 | 422.5 | 2 | 1053.5 | 1 | 2x73 DS1 |
| 5 | 2382 | 364.5 | 2 | 1053.5 | 1 | 2x63 DS1 |
| 6 | 3075 | 364.5 | 2 | 1053.5 | 1 | 2x63 DS1 |
| 7 | 3905 | 364.5 | 2 | 1053.5 | 1 | 2x63 DS1 |
| 8 | 5332 | 1838.5 | 7 | 1053.5 | 1 | 1x244 DS2 |
| 9 | 7175 | 1838.5 | 7 | 1053.5 | 1 | 1x244 DS2 |
| 10 | 9018 | 1838.5 | 7 | 1053.5 | 1 | 1x244 DS2 |
| 11 | 10177 | 364.5 | 2 | 1053.5 | 1 | 2x63 DS1 |

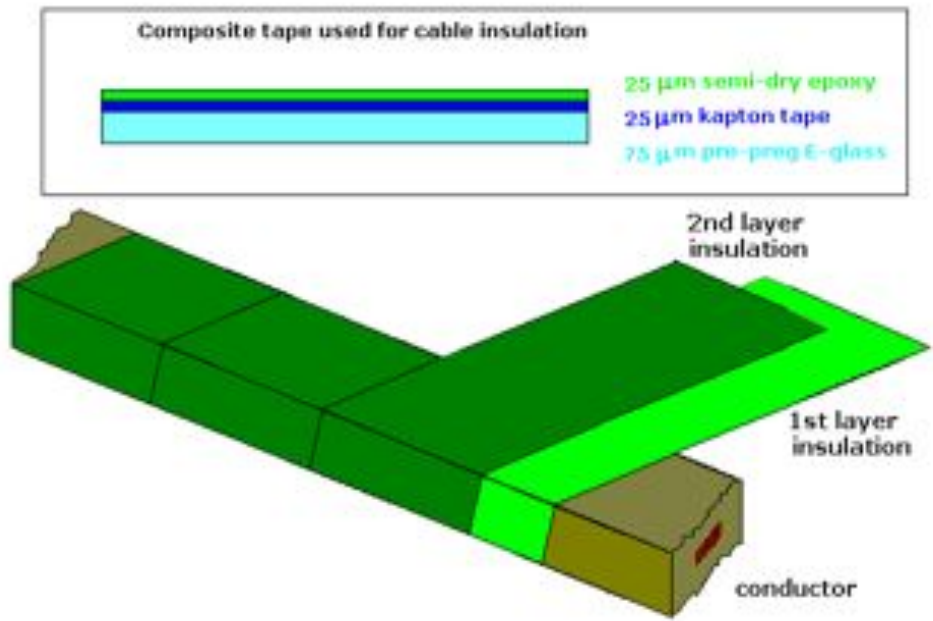

Figure 6.79.  Composite tape for cable insulation.

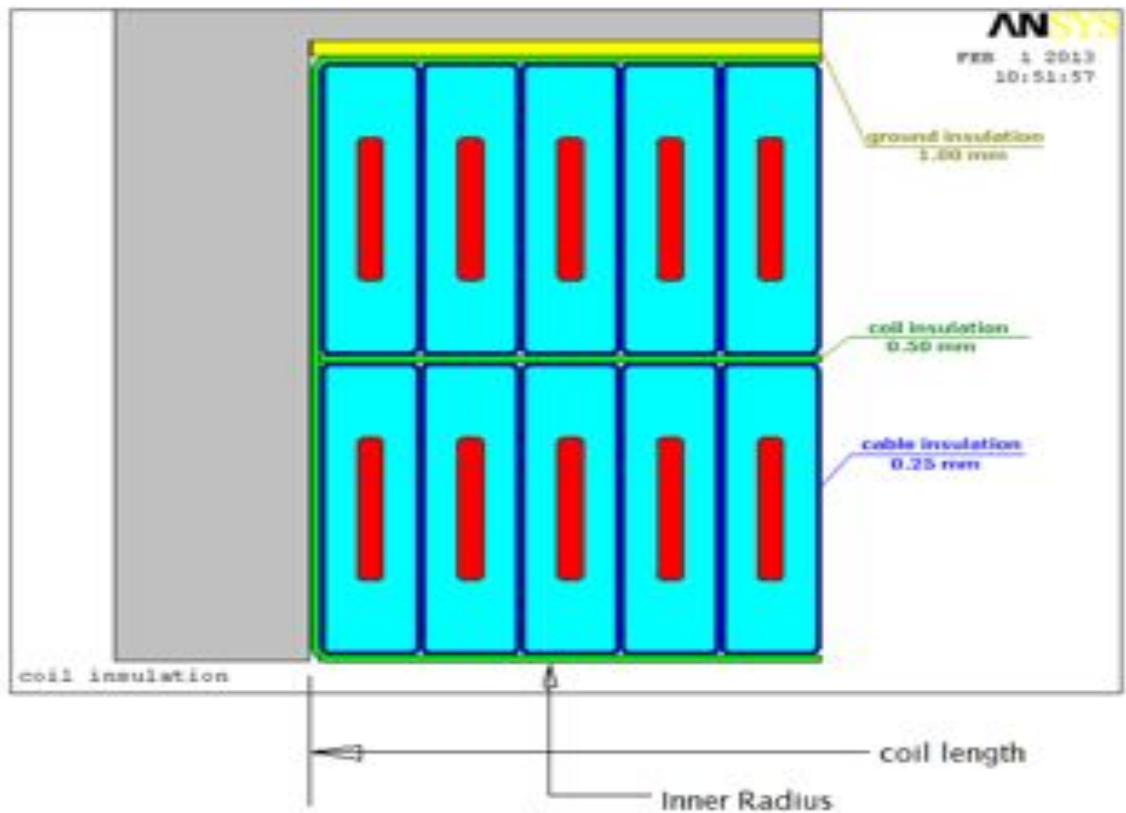

Figure 6.80. Insulation Scheme.

*Coil Winding*

The most productive way to produce compact and rigid coils is by applying the outer coil winding technique, utilizing a collapsible inner mandrel as shown in Figure 6.81.

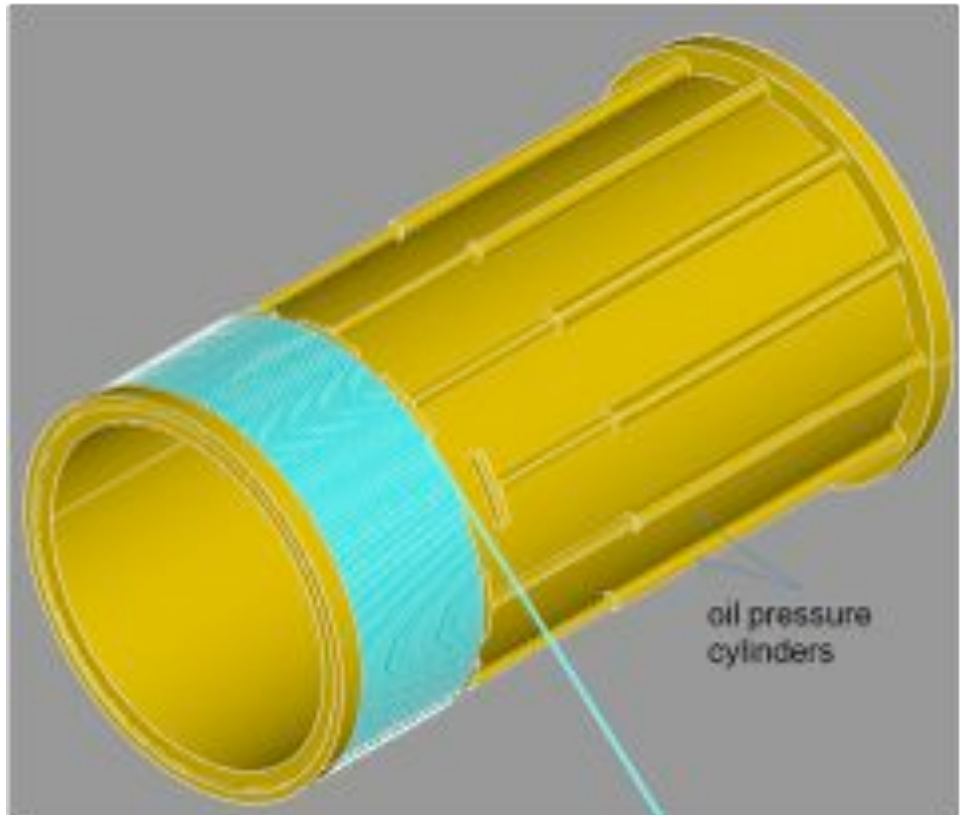

Figure 6.81. Winding around a collapsible mandrel and under axial pressure.

Radial and axial positions of the coil during winding need to be controlled to obtain the specified tolerances. The radial position can be maintained by a stiff mandrel with correct radial dimensions and by applying appropriate cable tension during the winding process. For axial position control, compression tooling is required to apply pressure as shown in the figure. This will ensure that the winding is under axial compression at all times, resulting in the appropriate packing factor.

The winding mandrel needs to be collapsible, so it can be removed from the coil bore after the coil is installed into the support cylinder. The mandrel should be stiff enough that it will not become significantly deformed during coil winding. The winding surface of the mandrel should have precise radial dimensions and is required to be smooth.

During winding, axial pressure is always applied, and the tension in the cable is maintained. The radial and axial positions of each turn must be controlled within the specified tolerance. The cable insulation must not be damaged.

### Coil Curing

The coils need to be cured and vacuum-impregnated. The curing is necessary to solidify the epoxy used in the composite insulation, and the vacuum impregnation is necessary to fill the voids between the layers and to form a buffer on the outer coil surface for further machining. These two processes can be performed simultaneously or in series. The mandrel will remain in the coil during these manufacturing processes.

#### 6.3.3.5   Splice Design

The DS coil cable (lead) connections/splices are welded joints. The aluminum matrix of the coil conductor on both sides of the narrow edges is welded together. The splice length is 700 mm (see Figure 6.82). Weld penetration depth needs to be 3 mm. No additional solder material will be used. Welding will be performed in an argon gas atmosphere. This welding procedure will need special tooling to keep the conductors in the right position and prevent heating of the NbTi superconductor above 350 C for more than 15 minutes.

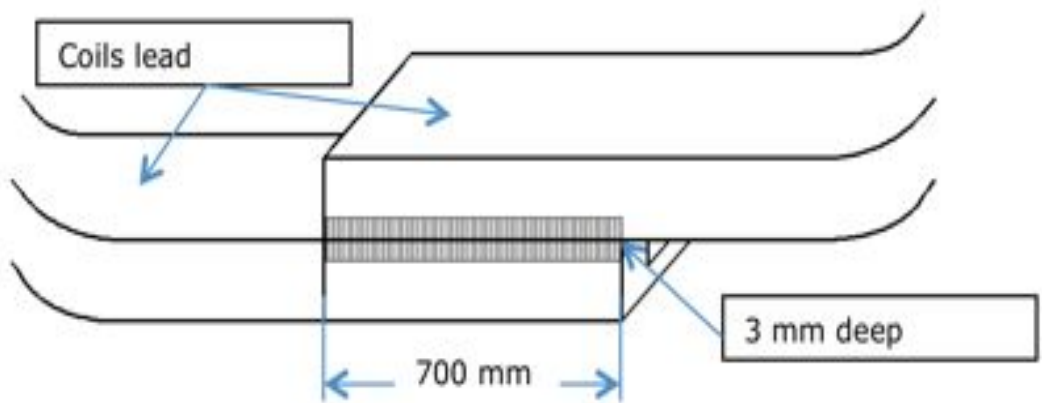

Figure 6.82. Splice schematic.

Splice boxes are used to support the splices. The splice box bases will be welded to the helium cooling pipes (Figure 6.83).

#### 6.3.3.6   Cold Mass Design

Coils, their support cylinders and the spacers form the cold mass.

### Support Cylinders and Spool Pieces

Support cylinders are used on the outside of the coil to take the magnetic load and maintain the coil position. These support cylinders are made of aluminium 5083. When coils are inserted into the support cylinder they form a spool piece. This operation requires precise machining and shrink fitting activity. The coils and the support cylinders need to be shrink fitted with interference so there is always radial pressure between the two.

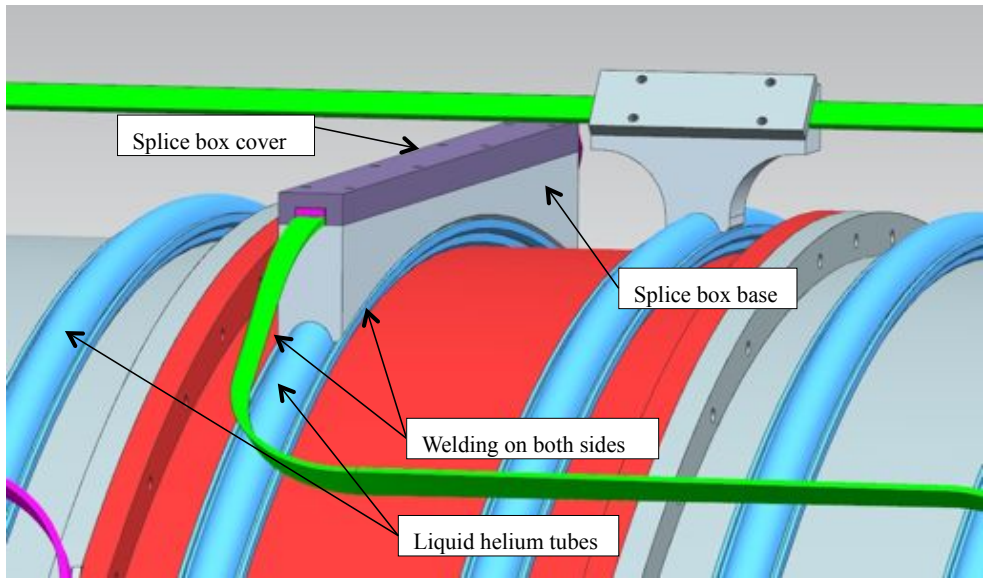

Figure 6.83. Splice box design.

***Spacers***

Figure 6.84 shows the Spool Pieces and spacers; the spool pieces are not labeled separately, only the coils are labeled with C1 through C11.

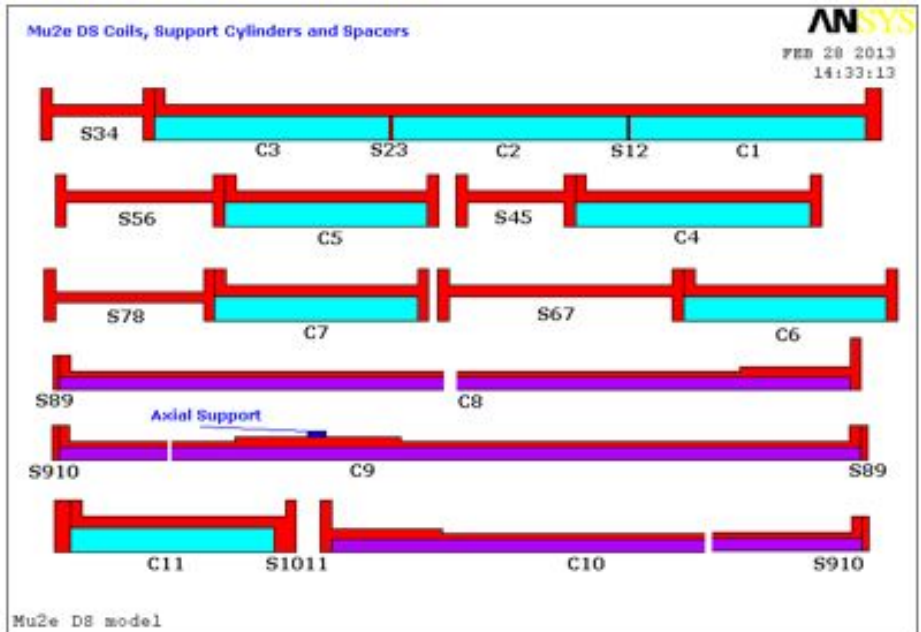

Figure 6.84. Coils, support cylinders and spacers.

Spacers need to be placed between the coils (spool pieces) to obtain the correct magnetic profile. Spacers have nominal dimensions, but the actual length of each spacer will be determined by the actual length of each coil.

### Cold Mass Assembly

Aluminum alloy bolts will be used to connect coil modules, taking into account the weight of the cold mass, magnetic forces, and dynamic forces during assembly and shipping. After coil modules are connected together, cooling tubes will be welded to the cold mass.

### Cold Mass Support System

Inconel 718, a 52% nickel alloy, is used for the cold mass radial and axial supports. The DS cold mass support system (see Figure 6.85) uses 16 tangentially arranged metallic rods, eight on each end, to support against dead weight and lateral de-centering forces. Eight metallic rods at the downstream end only provide support against the axial magnetic forces.

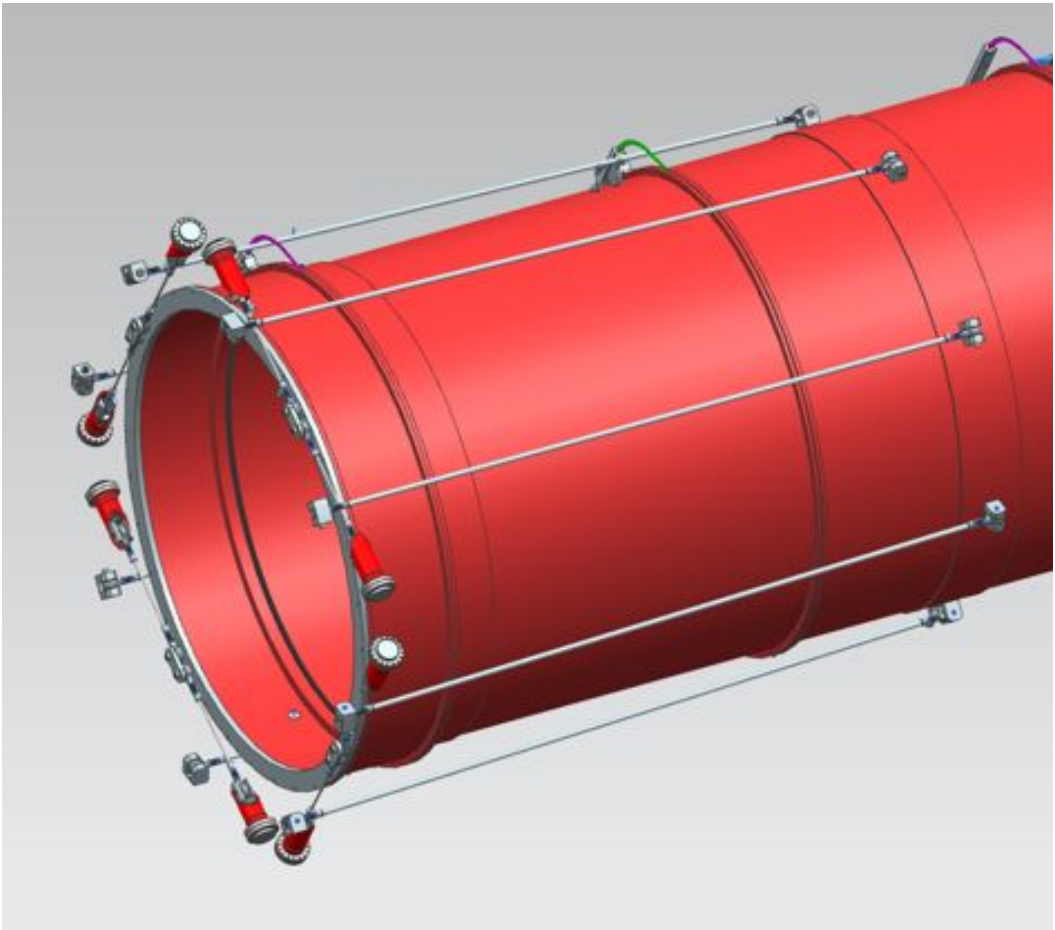

Figure 6.85. Down-stream End of DS (cryostat end-ring removed for clarity).

The radial support rods are 630 mm long and 12.7 mm in diameter. The axial support rods are 3295 mm long, 25.4 mm in diameter. Both support types are made from Inconel 718, and all supports are connected to their 5083-0 Al brackets at the cold mass by one inch diameter Inconel 718 pins. Both support systems use spherical rod ends to adjust for motion during cool-down, including the axial thermal contraction of 36 mm at the upstream end.  The design loads are summarized in Table 6.29.

Table 6.29. Design Load Summary.

| Load (kN) | Axial Support | Upstream Radial Support |
|---|---|---|
| Nominal | 964 | 3.1 |
| Decentering | 17 | 7.6 |
| Weight | | 60.5 |
| Maximum | 981 | 68.1 |
| Design load | 1000 | 70 |

### 6.3.3.7   Thermal Design

#### Introduction

The cold mass assembly and the thermal shields of the DS are indirectly conductively cooled. The cold mass is cooled by saturated helium at 4.7 K flowing in a Thermosiphon scheme and the thermal shields are cooled by pressurized 2 phase nitrogen containing approximately 90% liquid nitrogen at around 82 K. The thermal margin allocated for the magnet requires that the superconducting coils of the magnet to be maintained at temperatures less than 5.1 K and the thermal shields at temperatures less than 85 K.

#### Heat Loads

Table 6.30 through Table 6.32 summarize the heat loads at 4.7K.

Table 6.30. Radiation Heat Load at 4.7 K.

| | Surface Area (m^2) | Number of MLI layers | Heat Flux (W/m^2) | Heat Load (Watt) |
|---|---|---|---|---|
| Cold mass outer surface | 84.7 | 20 | 0.2 [41] | 16.94 |
| Cold mass inner surface | 66.5 | 20 | 0.2 [41] | 13.3 |
| Total Radiative Load on cold mass | | | | 30.24 |

Table 6.31. Conductive Heat Load from Suspension System at 4.7 K.

| Component | Cross Section (m^2) | Length (80 K to 4.7 K) (m) | Heat Load per support (W) | Quantity | Total heat load (W) |
|---|---|---|---|---|---|
| Axial | 0.507 | 2.5 | 0.07 | 8 | 0.56 |
| Radial | 0.127 | 0.5 | 0.09 | 16 | 1.44 |
| Total conductive load at 4.7 K | | | | | 2 |

Table 6.32. Summary of Heat Load at Helium Temperatures (4.7 K).

| | |
|---|---|
| Radiation Heat Load | 30.24 W |
| Suspension System | 2 W |
| Transfer line heat load (from feedbox to magnet) | 4 W |
| Feedbox Heat Load | 10 W |

***Thermosiphon Cooling Scheme for Cold Mass Assembly***

Figure 6.86 shows a schematic representation of the thermosiphon concept.

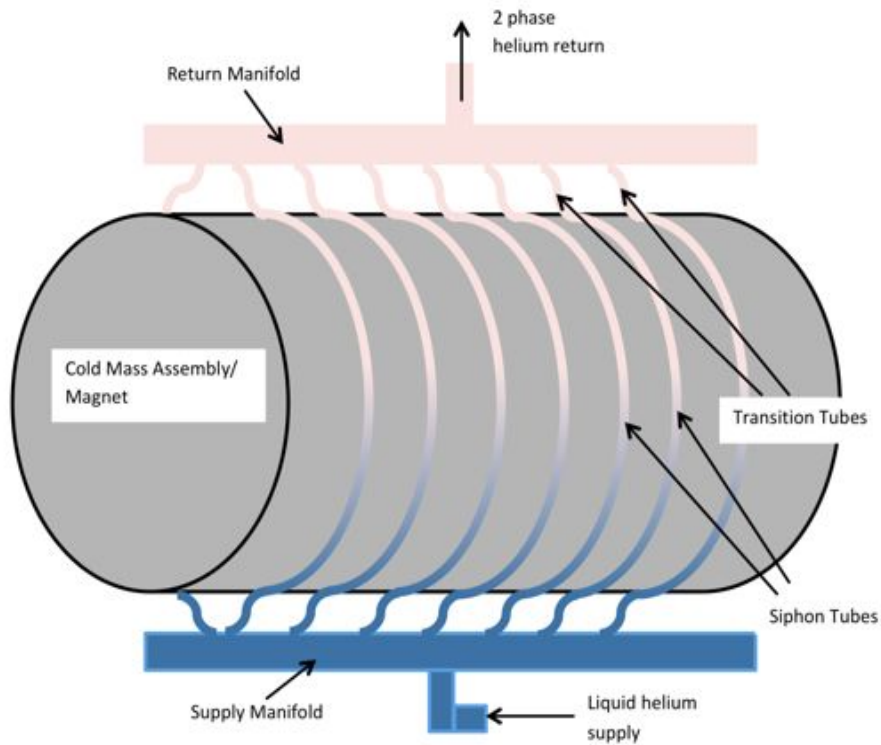

Figure 6.86. Thermosiphon Cooling Principle.

Saturated helium is filled from the bottom supply manifold and then travels through the siphon tubes to the top return manifold. The liquid helium in the siphon tubes conductively absorbs the heat from the cold mass and as the helium temperature increases, the density of the helium decreases, pushing the heated liquid up while forcing the colder liquid to the bottom in a natural circulation loop.

As the thermosiphon system is essentially driven by gravity, the siphon tubes must be oriented vertically. The siphon tubes are semi-circular segments that are welded to the cold mass itself and attached at the bottom and top to the supply and return manifolds.

Table 6.33 through Table 6.35 summarize the heat loads at 80 K. Table 6.36 lists piping parameters for the thermosiphon scheme.

Table 6.33. Radiation Heat load to the Thermal Shields at 80 K.

|  | Surface Area (m^2) | Number of MLI layers | Heat Flux (W/m^2) | Heat Load (Watt) |
|---|---|---|---|---|
| Outer thermal shield outer surface | 87.6 | 60 | 1.5 [42] | 131.4 |
| Inner thermal shield inner surface | 66.34 | 60 | 1.5 [42] | 99.51 |
| Total Radiative Load on thermal shields |  |  |  | 231 |

Table 6.34. Conductive Heat Load from Suspension Systems at 80 K.

|  | Component | Heat Load per support (W) | Quantity | Total heat load (W) |
|---|---|---|---|---|
| Cold mass Suspension System Intercept | Axial Support | 4.25 | 8 | 34 |
|  | Radial Support | 1.125 | 16 | 18 |
| Thermal shield suspension system | Center Connector | 2.85 | 48 | 137 |
|  | Radial Connector | 0.31 | 384 | 119 |
| Total conductive heat load |  |  |  | 308 |

Table 6.35. Summary of Heat Loads at Nitrogen Temperatures (80 K).

| | |
|---|---|
| Radiative Heat Load | 231 W |
| Conductive heat load | 308 W |
| Transfer line heat load | 80 W |
| HTS current lead heat load | 60 W |

Table 6.36. Thermosiphon Piping Parameters.

| Component | Outer Diameter (inch) | Wall Thickness (inch ) | Material |
|---|---|---|---|
| Bottom Supply Header | 3.5 | .25 | Aluminum 6061-T6 |
| Top Return Header | 3.5 | .25 | Aluminum 6061-T6 |
| Siphon Tubes | 1.25 | .125 | Aluminum 6061-T6 |
| Transition Tubes | 1.0 | .065 | Aluminum 6061-T6 |
| Inlet Line | 1.25 | .065 | Aluminum 6061-T6 |
| Return Line | 1.5 | .065 | Aluminum 6061-T6 |

### *Two Phase Forced Flow Cooling for the Thermal Shields*

Both thermal shields will be cooled with a LN2 circuit supplying two phase nitrogen at around 1.7 bar of pressure. Table 6.37 summarizes piping parameters for the thermal shields.

Table 6.37. Piping parameters for Thermal Shields.

| Component | Outer Diameter (inch) | Wall Thickness (inch) | Material |
|---|---|---|---|
| Outer thermal shield LN2 piping | .625 | .0625 | Aluminum 6061-T6 |
| Inner thermal shield LN2 piping | .625 | .0625 | Aluminum 6061-T6 |
| Inlet LN2 Line | .625 | .0625 | Aluminum 6061-T6 |
| Return LN2 Line | .625 | .0625 | Aluminum 6061-T6 |

### 6.3.3.8   *Cryostat Design*

The length of the cryostat is 10900 mm. The outer diameter is 2656 mm, and the inner diameter is 1900 mm. The material chosen is stainless steel 316L, which has yield strength of 172 MPa, and allowable stress for the pressure vessel design is 115 MPa. The cryostat consists of concentric 2 cm thick cylindrical shells connected by annular end rings (which are 4 cm thick). The shells are sized according to ASME Section VIII, Div. 1 rules for cylindrical shells under external pressure. Given that the bore of the solenoid may be evacuated while the magnet is warm (at room temperature and pressure). The outer shell is designed for 1 atmosphere external pressure; the inner shell is designed for both internal and external pressure of 1 atmosphere. The cryostat sits on two saddles, positioned very close to the ends of the vessel. The configuration is shown in Figure 6.87.

The bore of the cryostat must accommodate approximately 10 tons of detectors, shielding and other equipment. This load rests on rails attached to the inner shell of the vacuum vessel. The cryostat provides the load path for cold mass reactions (weight and magnetic force) through its support system. The axial supports bear directly on the cryostat outer shell and transmit the forces to the saddle support. This arrangement produces essentially no stress on the cryostat. The warm ends of the radial supports attach to the cryostat through towers, which transmit the load through the outer shell to the saddles.

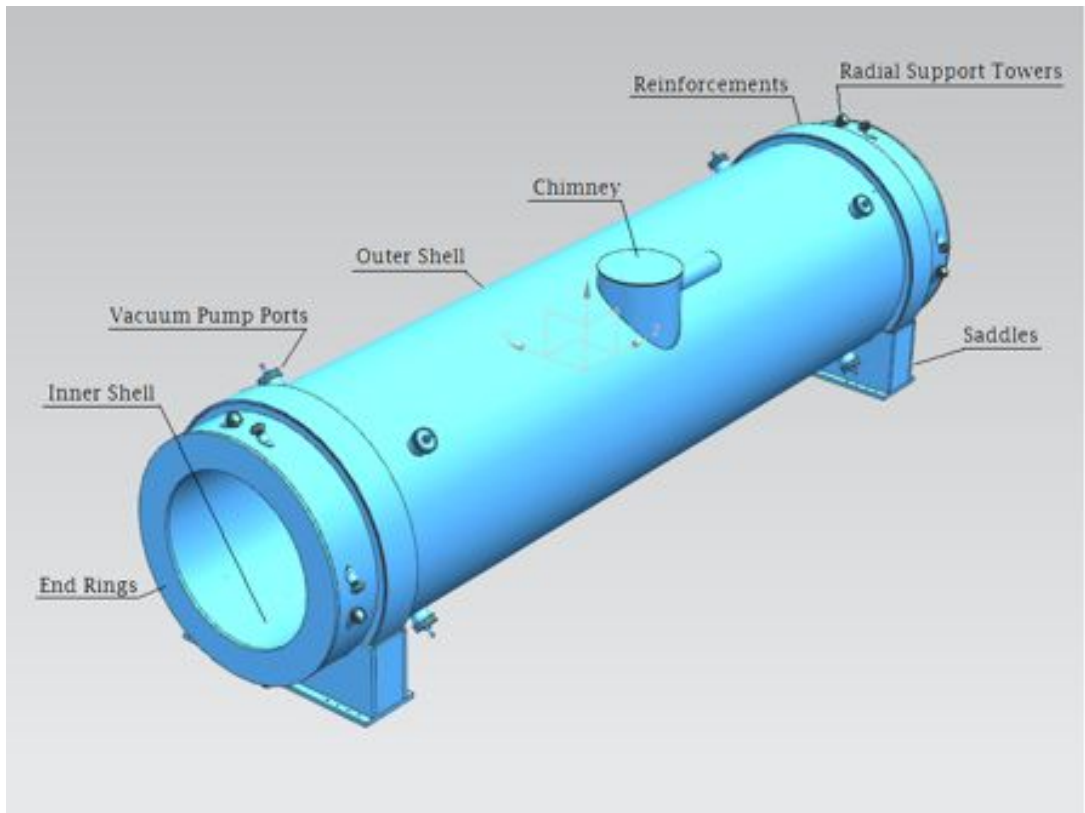

Figure 6.87. DS Cryostat features.

### *Openings on Outer Shell*

There are several openings in the outer shell of the cryostat. The largest one is the chimney, with a diameter of 36 inches (915 mm). Electrical conductors and cooling pipes as well as sensor wires, will all go through this chimney. Openings are also needed for the vacuum pump ports: there are 8 of them, four on each end. The diameter of the pump ports is 200 mm. The smallest ones are radial support towers with diameter of 100 mm. There are 16 of them, eight on each end. Prior to fabrication, all of them need to be evaluated to determine whether reinforcement is needed to comply with the ASME pressure vessel code.

### *Loads on the Inner Shell*

The Mu2e detectors will sit inside the cryostat on two rails. The DS cryostat will provide a support platform for the rails. These platforms are attached to the inside of the inner vacuum shell as shown in Figure 6.88.

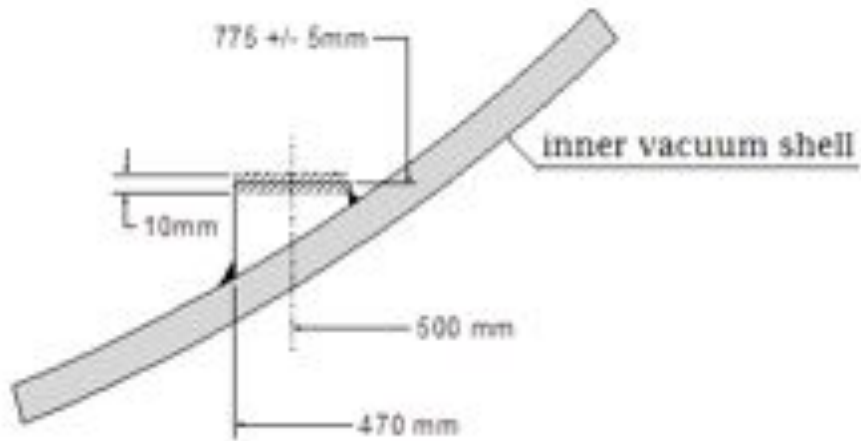

Figure 6.88. Support platform inside the DS bor**e**

The inner vacuum shell of the cryostat will support the tracker, calorimeter, muon stopping target, the upstream end of the muon beamline and internal detector shielding loads. These loads, which total approximately 70 kN, are shown in Figure 6.89.

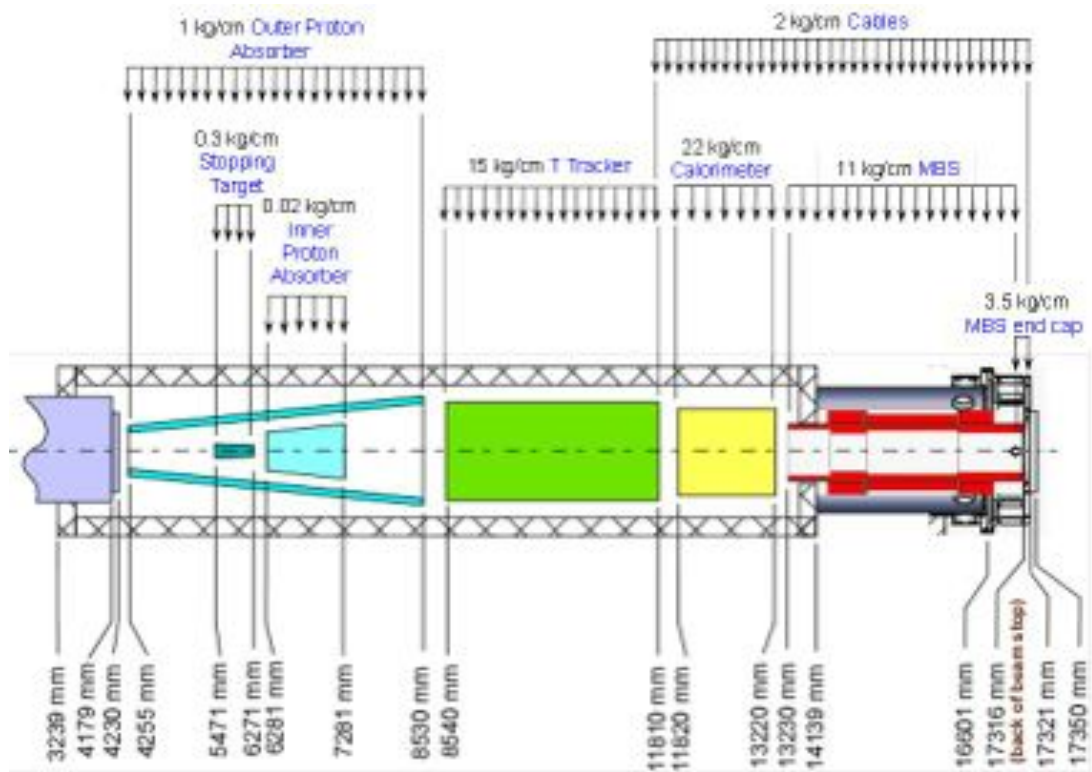

Figure 6.89. Loads on the Inner Cryostat Shell.

***Thermal Shield and Insulation***

In order to reduce the radiation heat load to the solenoid cold mass and to provide thermal interception for cold mass supports, there will be a thermal shield between the vacuum vessel and the cold mass as shown in Figure 6.90.

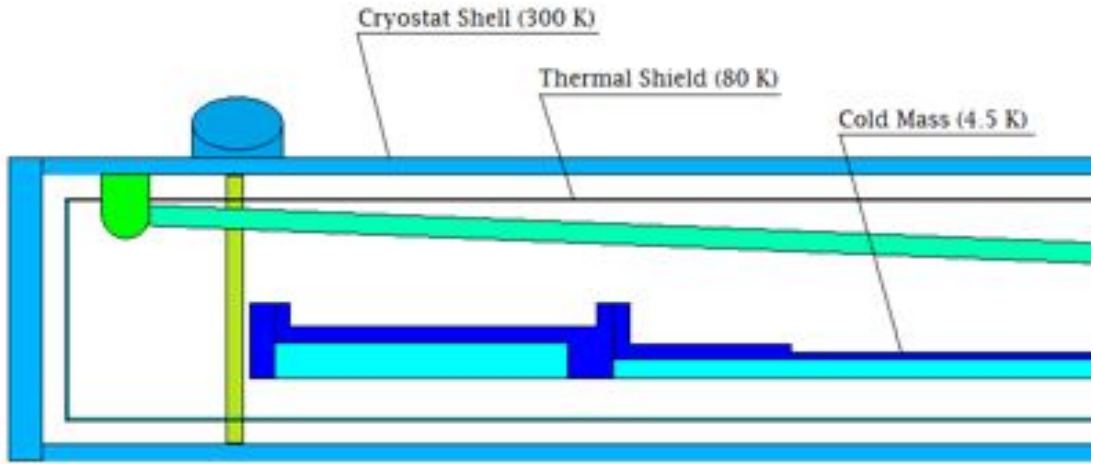

Figure 6.90. Schematic of 80K thermal shield.

The thermal shield is made of 3 mm thick aluminum sheets. Both the inner and outer shields are divided into 6 segments axially and 4 segments in the circular direction. The pieces are electrically separated in order to reduce eddy currents and accommodate thermal shrinkage; however, they form a complete thermal radiation cover for the 4.5 K cold mass. They are fastened to the vacuum vessel walls by G10 connectors. For each aluminum sheet, the center connector fixes the position and the radial connectors at the corners allow thermal contractions.

Cooling tubes with 80 K helium are welded to the thermal shield. The design of the zigzag cooling tube allows some flexibility at connections between thermal shield pieces to allow thermal contraction.

Thermal connections are made between the cold mass supports and the thermal shield so that most of the heat from the 300 K along the supports will be intercepted.

Multi-layer super-insulation will be put between the cryostat and the thermal shielding to further reduce radiative heating from the 300 K vacuum vessel. The thickness of the super-insulation will be about 15 mm, and its apparent conductivity will be no more than 0.062 mW/m-K. This results in a heat load to the thermal shielding of no more than 1 W/m$^2$. The other heat loads are all listed in Table 6.38. There will also be some multi-layer insulation between thermal shield and cold mass to reduce the heat load to the cold mass.

Table 6.38. Heat loads to the thermal shield.

|  |  | **Intensity** | **Quantity** | **Heat (W)** |
|---|---|---|---|---|
| Radiation | surface | 1 W/m$^2$ | 180 m$^2$ | 180 |
| Conduction | Center connector | 2.85 W/each | 48 | 137 |
|  | Radial connector | 0.31 W/each | 384 | 119 |
|  | Axial support | 4.25 W/each | 8 | 34 |
|  | Radial support | 1.125 W/each | 16 | 18 |
|  |  |  | Total | 488 |

### *Reinforcement of the Vacuum Vessel Outer Shell*

A reinforcement ring is added to the outer shell of the vacuum vessel on top of the supporting saddle. This will add stiffness to the vacuum vessel, especially during the installation of the cold mass before the vessel is closed by the end flanges. The reinforcement ring is made of 2 cm thick steel plate. This is the same thickness as the outer shell. The reinforcement ring can be welded to the outer shell.

### 6.3.3.9   Instrumentation

Instrumentation sensors are installed to provide quench protection, cryogenic monitoring and control, and mechanical characterization. Details are provided in reference [11].

### *Quench Protection Instrumentation*

The quench protection system monitors the coils and superconducting current leads for resistive growth and provides a trigger to shut down the power supply in order to limit the rise in conductor temperature due to Joule heating. The primary sensors for quench detection are voltage taps. Redundant voltage taps must be installed across each power lead and across each splice. The splice taps will also allow for monitoring of the splice resistances.  The superconducting leads may be monitored by superconducting wire sensors in addition to voltage taps. Superconducting NbTi twisted pairs bonded to the leads will be used to detect quenches. The wires must be bonded to the leads in a way that ensures electrical isolation while maintaining good thermal contact. During normal operation, the wires will be superconducting. In the event that the lead quenches the wire should also heat above the transition temperatures so that quenches in the leads may be detected by monitoring the wire resistance.

### *Cryogenic Monitoring and Controls Instrumentation*

Instrumentation will be implemented for cryogenic monitoring and controls. Temperature sensors are required for monitoring and controlling the solenoid's cool down, warm up, operational steady state, and changes in response to a heat load such as a magnet quench. Temperature sensors will be installed for monitoring the temperatures of the magnet's warm spots and delta-Ts during cool-down and warm-up, the thermal shields, the coil near the support posts, and the supply and return manifolds.

### Mechanical Characterization Instrumentation
The mechanical state of the solenoid will be monitored using strain gauges and position sensors. Several strain gauges will be installed on the upstream, downstream, and longitudinal supports. Position sensors will be mounted between the coil and cryostat to monitor coil displacement.

### 6.3.3.10 Interfaces

### Cryostat Inner Shell
The inner shell is a primary component of the DS insulating vacuum system on its OD, as well as a primary component of the Muon Beamline Vacuum System on its ID.

### Connection to Vacuum Pump Spool Piece
The Vacuum Pump Spool Piece (VPSP) is an element of the Muon Beamline Vacuum System that attaches to the downstream end of the DS cryostat. The VPSP is shown in Figure 6.91.

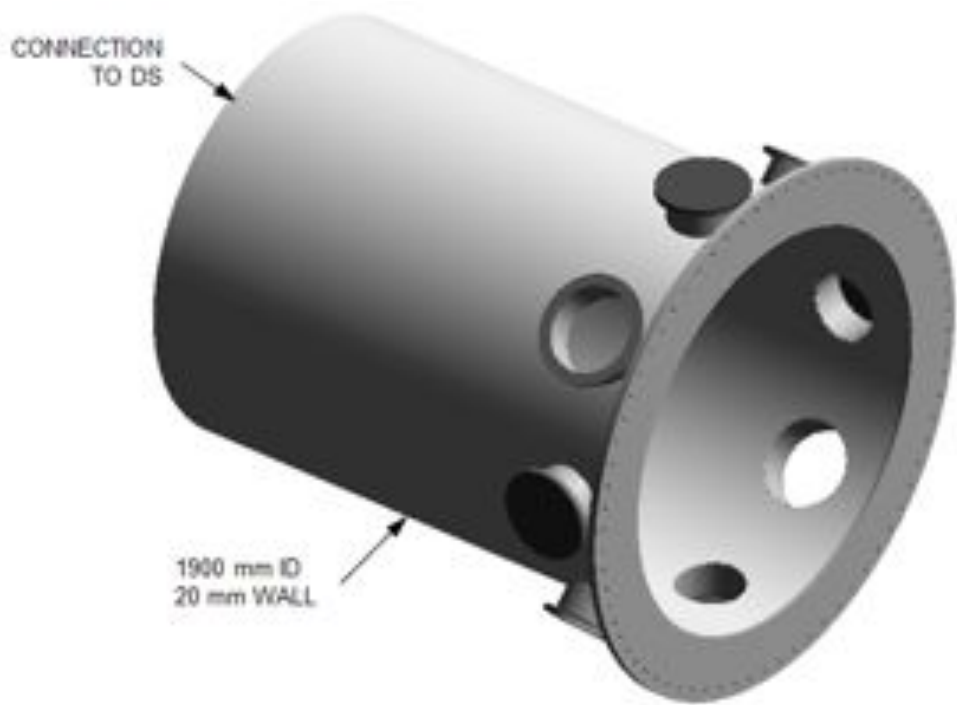

Figure 6.91. Features of the Vacuum Pump Spool Piece.

The feature required in the DS is a stub extension of the DS cryostat inner shell that allows a heavy structural and vacuum weld between the VPSP and the DS to be made far enough away from the DS to protect the cold-mass insulation from weld heat.

***Connection to TS***

The DS Solenoid must connect to the TS Solenoid to provide a common Muon Beamline Vacuum System volume. The connection will be made with a flexible bellows (Figure 6.92) to allow any combination of axial and lateral movement produced by cool-down and magnetic forces between solenoids.

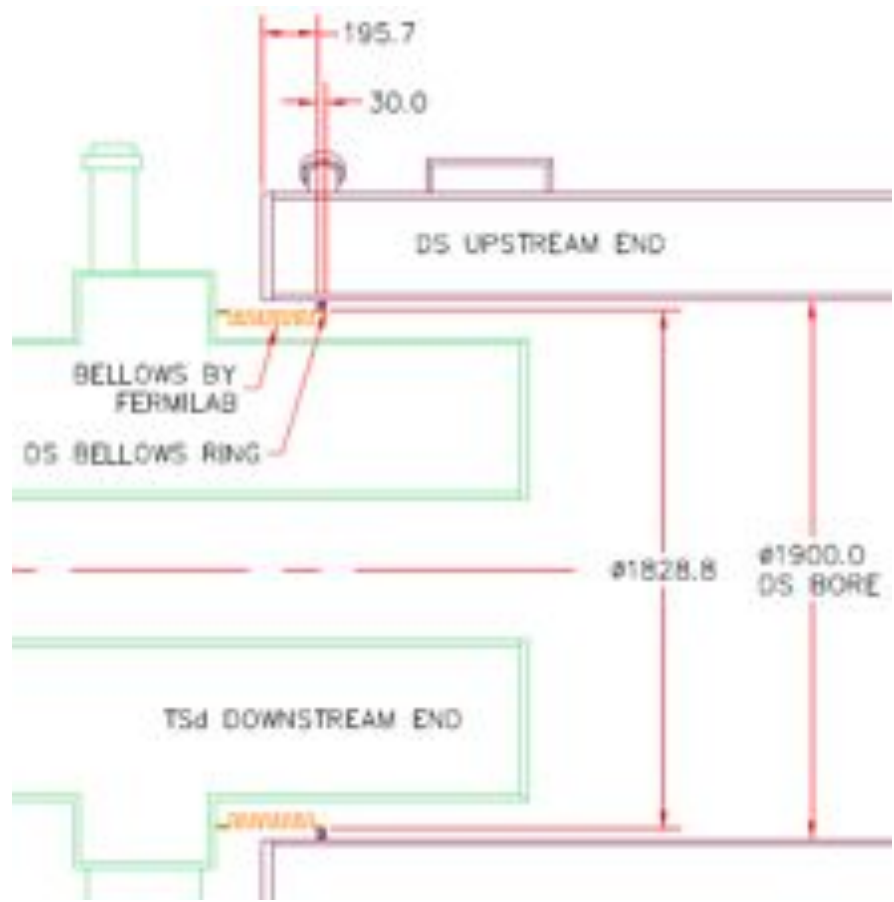

Figure 6.92. Connection between DS and TSd.

***Connection to Transfer Line***

The DS Solenoid must connect to a transfer line that supplies the magnet with power and cryogenic fluids (Figure 6.93). The transfer line also provides a pump-out duct for the DS insulating vacuum, and acts as a conduit for instrumentation signals entering and leaving the magnet.

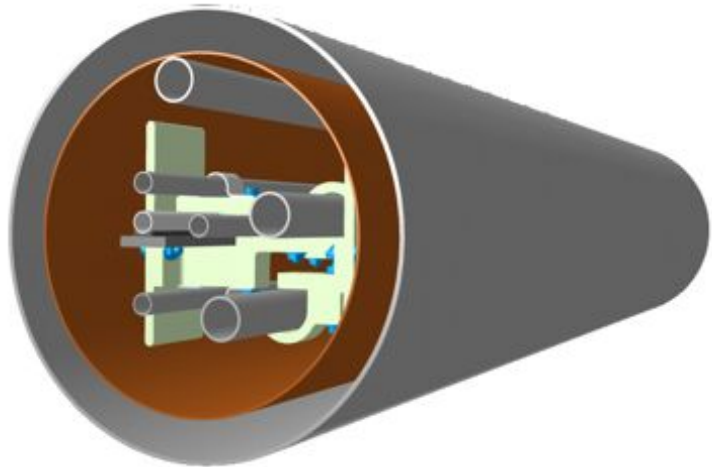

Figure 6.93.  Section of Transfer Line.

***Shipping Restraints***

Additional shipping restraints will be added to the DS cryostat to keep the cold mass positioned within the cryostat and protect the axial and radial supports (Figure 6.94, Figure 6.95).

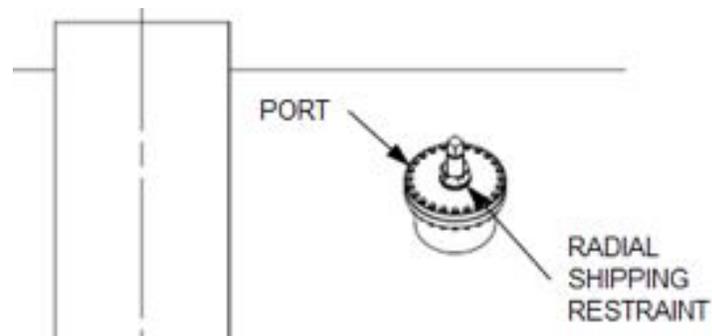

Figure 6.94. DS Radial Shipping Restraints.

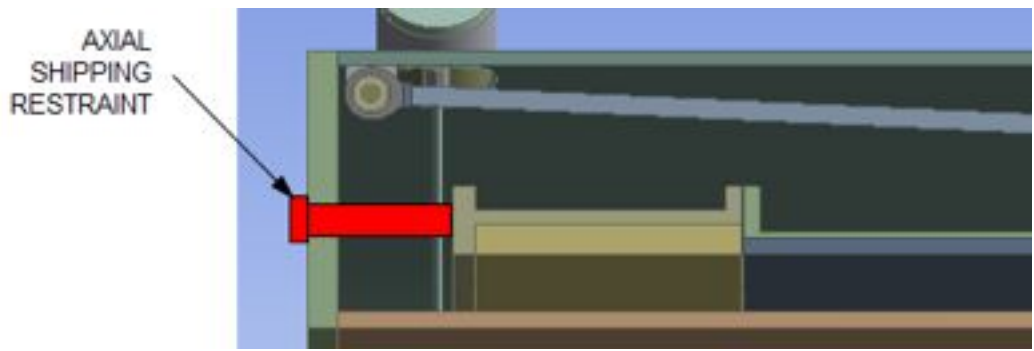

Figure 6.95. DS Axial Shipping Restraints.

***Insulating Vacuum System***

The DS coldmass resides in an insulating vacuum space formed between the inner and outer cryostat shells and the upstream and downstream cryostat end plates. An additional DS feature required to support the insulating vacuum system is pump-out ports. The transfer line extension port is one pump-out port, and the radial shipping restraint ports can provide additional ports when the restraint rods are removed and replaced with ducts or pumps.

***Magnet Support***

The DS Solenoid gravity and magnetic loads are ultimately transferred through the DS support saddles to the floor. The DS support feet will be bolted to a support frame, which in turn will be bolted to pads provided in the floor of the Mu2e building. A side view of the DS and support frame is shown in Figure 6.96.

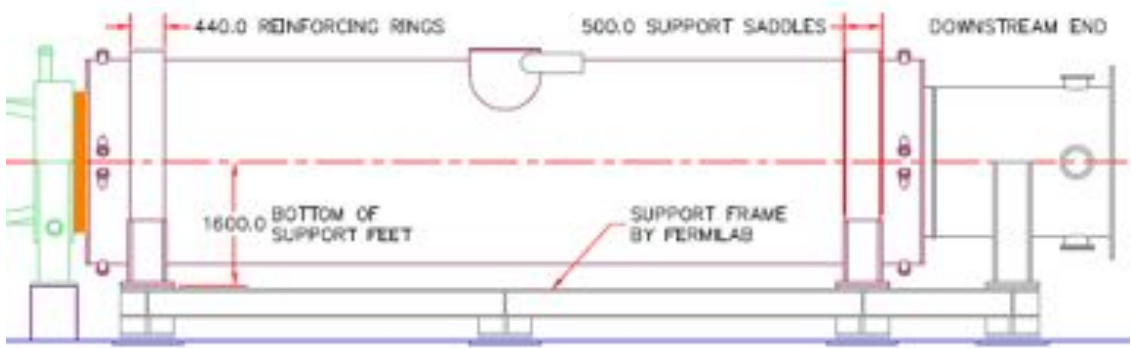

Figure 6.96. DS Mounted on Support Frame.

## 6.3.4   Cryogenic Distribution

### 6.3.4.1   Introduction

The superconducting solenoids require a cryogenic distribution system and supporting cryoplant for liquid helium and liquid nitrogen. The requirements for this system are described in Section 6.2 and reference [5]. The scheme is to divide the solenoids into 4 semi-autonomous cryostats. Cryostats can be cooled down or warmed up independent of the state of the other cryostats. Each cryostat will require 4.5 - 4.7 K liquid helium as well as 80 - 90 K liquid nitrogen for the cryostat thermal shields. This system is shown in block diagram form in Figure 6.97 and described in detail in the sections below.

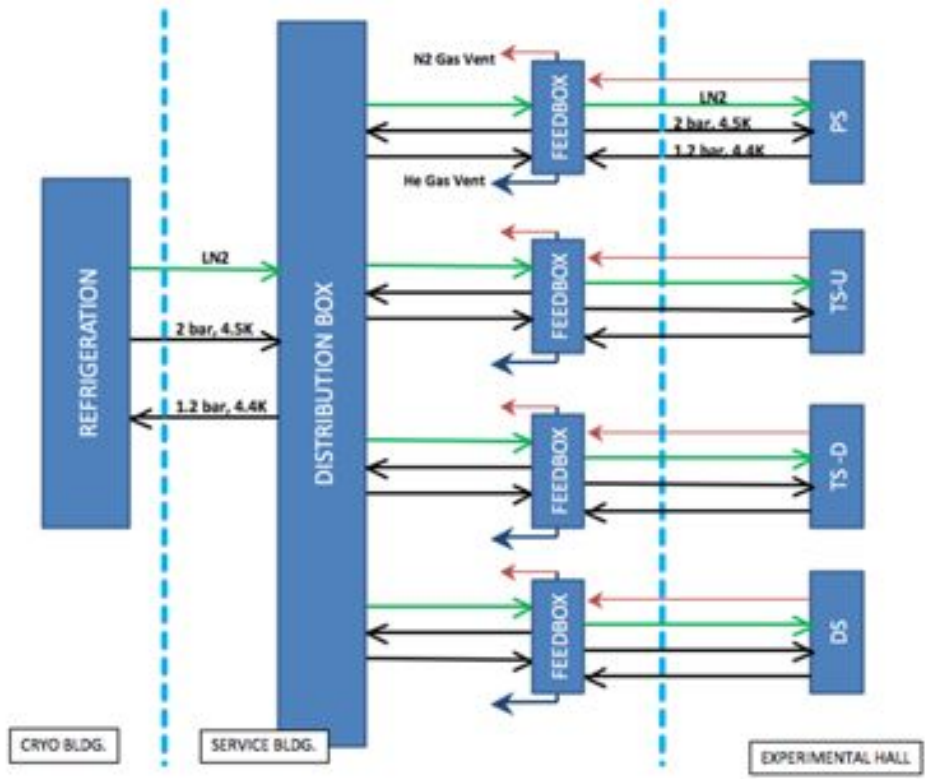

Figure 6.97. Block Diagram of the Mu2e Cryogenics System. "Cryo Bldg." refers to the Muon Campus Cryo Building. "Service Bldg." refers to an above-ground portion of the Mu2e experimental site used for cryogenics and power infrastructure.

### 6.3.4.2   *Cryoplant Description*

While the cryoplant is beyond the scope of this project, it is an important interface to the cryogenic distribution system. Thus, a brief description of the system is provided here. Refrigerators and refrigerator components from the Tevatron will be recycled and refurbished to provide liquid helium for the solenoids. As shown below, one satellite refrigerator appears to be sufficient for steady state operation with a second satellite for redundancy (hot spare) as well as added capacity for cool-down and quench recovery. The Refrigeration vs. Liquefaction curve for a satellite refrigerator is shown in Figure 6.98. As shown, a steady state operation of 350 Watts with 0.8 g/s liquefaction can be comfortably achieved with one refrigerator.

A separate building will be dedicated to the refrigerator and support equipment for Mu2e and other Muon Campus experiments. Four satellite refrigerators will be installed to support operations of the experiments. Modified Tevatron valve boxes will connect the refrigerators to a new distribution system.

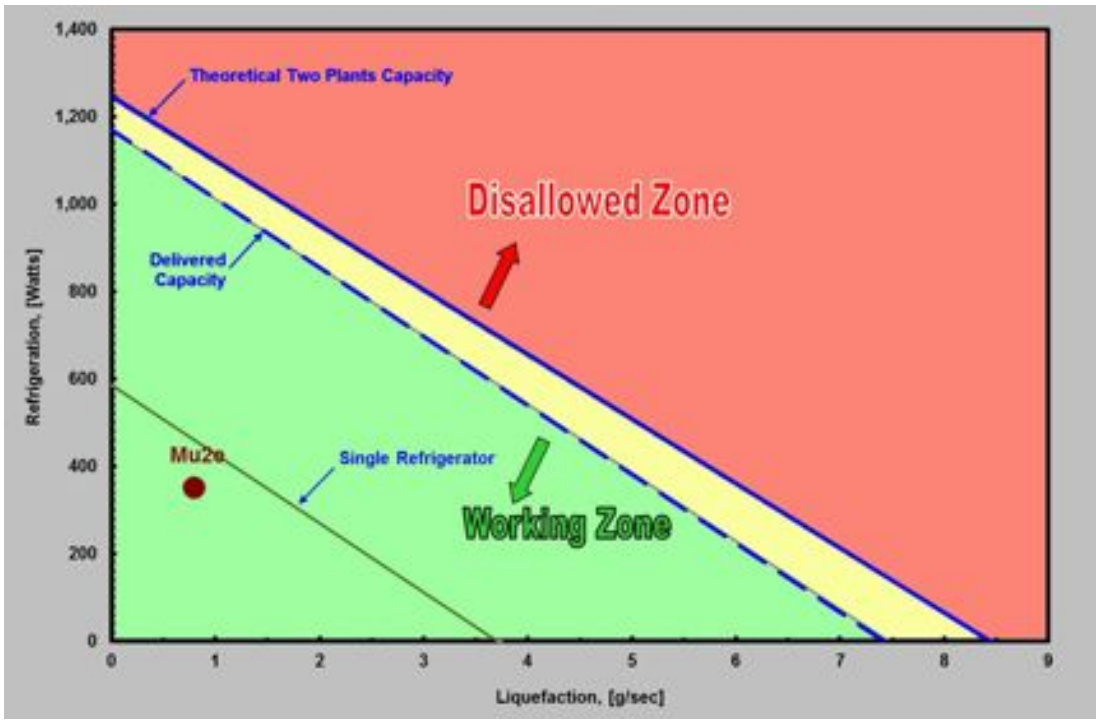

Figure 6.98. Satellite Refrigerators Stand Alone Capacity.

### 6.3.4.3  Cryogenic Distribution System

The cryogenic distribution box will be located between the feedboxes and the refrigerator building, shown schematically in Figure 6.97. The distribution box will contain cryovalves for controlling the flow of liquid helium and nitrogen to the individual cryostats. In this way, magnets can be individually warmed up or cooled down, as required. The distribution box will require liquid nitrogen to cool the 80 K thermal shields. Figure 6.99 is a model showing the cryogenic distribution system. The feedboxes will be located in a room in the above grade detector building. From the feedboxes, cryogenic distribution lines will run to each cryostat. A chase will be designed for this above-grade to below-grade transition to minimize the line of sight for radiation from the detector.  The horizontal runs will be located near the ceiling of the below-grade detector hall. The locations of the cryostat penetrations will depend on the individual cryostat, but will be chosen to avoid interference with mechanical supports or required radiation shielding.

### 6.3.4.4  Feedbox Distribution

The cryo distribution feedbox is modeled after previous designs of similar systems. Aside from the local cryogenic distribution, it serves as the cryo-to-room-temperature interface for magnet power supplies and instrumentation for thermal and quench systems. A feedbox schematic is shown in Figure 6.100.  Note that the feedboxes for the PS and DS magnet systems are configured for thermal siphoning conduction cooling. The TSu and TSd magnet systems are configured for forced flow application.

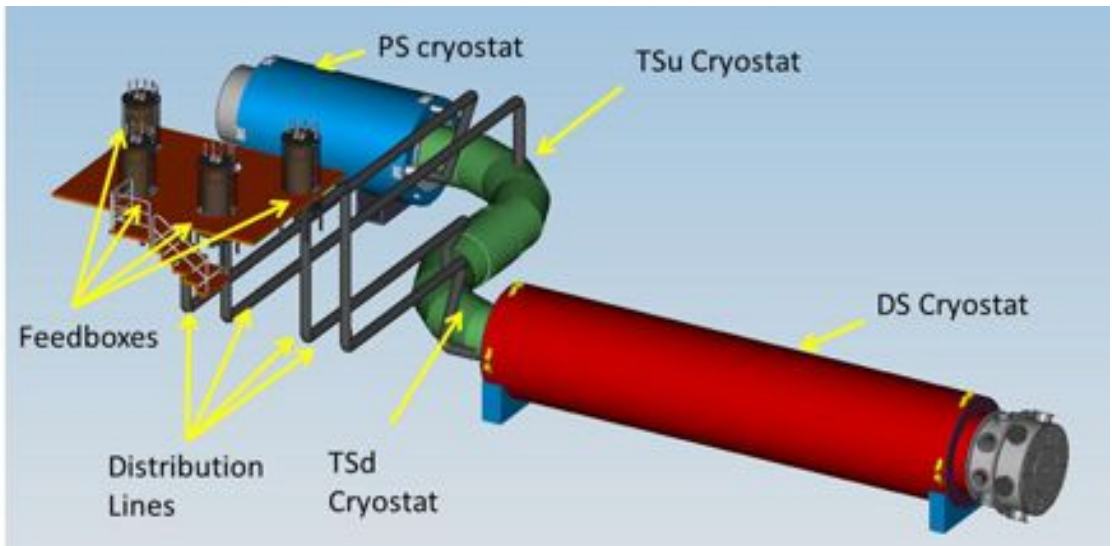

Figure 6.99. Layout of Cryogenic Distribution System.

Recycled high temperature superconducting (HTS) leads from the Tevatron will be used for the high current Detector and Production Solenoids. It has been demonstrated that these leads are capable of 10 kA DC operation.  Liquid nitrogen will be required to cool the HTS section of the leads.

### 6.3.4.5  Cryogenic distribution lines

The cross section for the cryogenic distribution line is shown in Figure 6.101.  The line contains two 0.625" lines for liquid helium.  These lines, while supplying helium to the cryostat, will conductively cool the magnet supply and return electrical bus.  Additional lines are shown for the 4 K return and 80 K $LN_2$ for the cryostat shield.

The distribution lines will be vacuum jacketed to minimize the heat load. Each line will be approximately 20 meters in length.

### 6.3.4.6  Cryogenic Analysis of Magnets

Studies have been performed on each of the magnet subsystems (PS, TS and DS) to evaluate the basic cooling schemes, more specifically thermal siphoning as opposed to forces flow, as well as the geometry, sizing of pipes, etc.  Estimates of the static and dynamic heat loads and cool-down times were obtained from these studies. Details of these are summarized here.

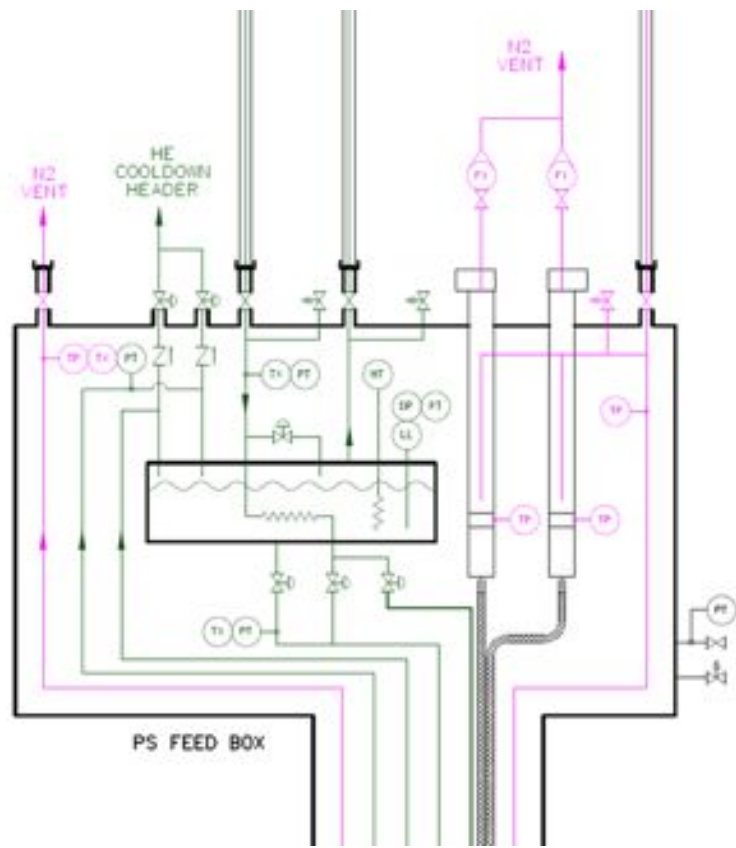

Figure 6.100.  Schematic of Feedbox with thermal siphoning.

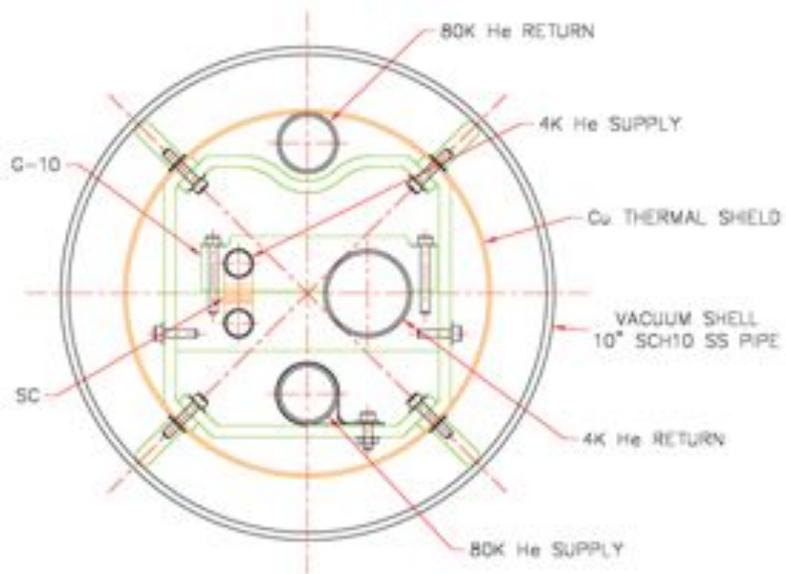

Figure 6.101. Cross Section of cryogenic distribution line.

Figure 6.102 shows a schematic of a thermal siphoning piping scheme for the Production and Detector Solenoids. The piping on the upper right is a continuation of the piping shown on the bottom right of Figure 6.100. Note that the liquid supply control valves shown are not physically located on the cryostat. They must be placed in a low radiation area so they can be serviced in the event of a failure.

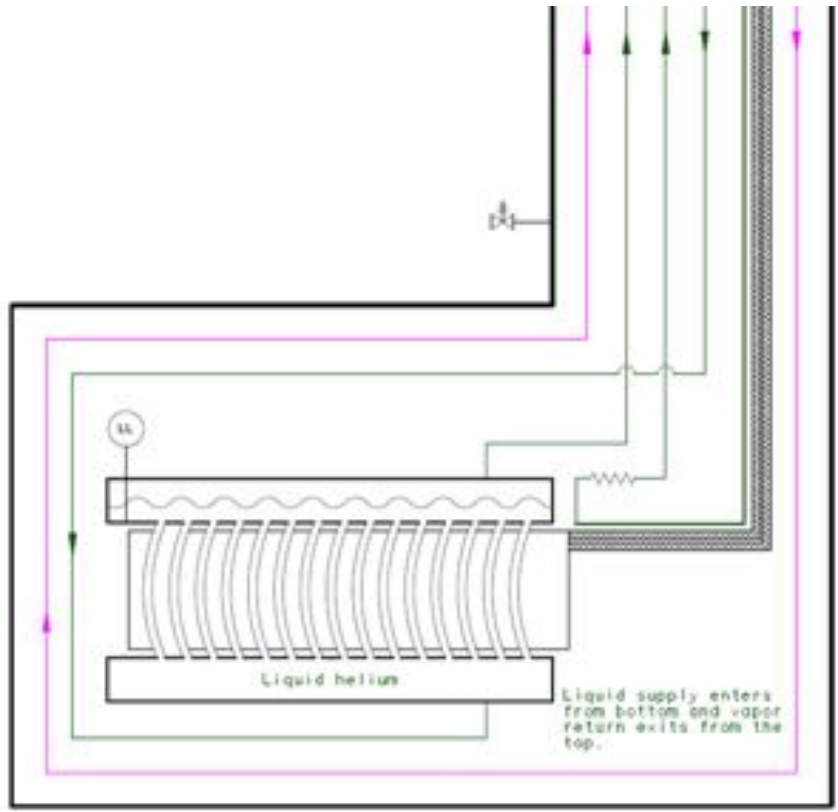

Figure 6.102. Thermalsiphon magnet cooling scheme for the Production Solenoid.

Table 6.39 and Table 6.40 include estimates of the heat loads and liquid helium requirements for the solenoid system. In this study, the PS and DS will utilize thermal siphoning while the TS will use a forced flow system. As shown, the total heat load to 4.7 K is 321 W. As is shown in Figure 6.98, this load is well within the capacity of one satellite refrigerator running in steady state. The estimate assumes a heat load of 67 W in the PS, which is dominated by dynamic (beam) heating. This value is strongly tied to the heat shield design and is subject to uncertainties in the particle production model used in simulations of the proton beam interacting in the production target. The heat load to 80 K is dominated by the cryostat design and cryostat surface area. As is shown, the heat load is 1.73 kW, which translates into a required liquid nitrogen flow of ~950 liters/day.

Table 6.39. Heat load estimate at 80 K.

| Best Estimates (no contingency) Nominal Temperature | Production Solenoid 80 K | TSu | TSd | Detector Solenoid | Total |
|---|---|---|---|---|---|
| 80 K Magnet Heat (W) | 128.5 | 252.0 | 252.0 | 539.0 | 1171.5 |
| 80 K Feedbox and Transfer Line[*] Heat (W) | 140.0 | 140.0 | 140.0 | 140.0 | 560.0 |
| Total 80 K Heat (W) | 268.5 | 392.0 | 392.0 | 679.0 | 1731.5 |
| Nitrogen usage for Magnet (liquid liters/day) | 147.42 | 215.22 | 215.22 | 372.80 | 950.67 |
| Number of 10kA HTS Leads | 2 | 0 | 0 | 2 | 4 |
| Number of 2kA HTS Leads | 0 | 2 | 2 | 0 | 4 |
| N2 10kA lead flow per magnet (g/s) | 2.2 | 0 | 0 | 2.2 | 4.4 |
| N2 usage for 10kA leads (liquid liters/day) | 235.54 | 0.00 | 0.00 | 235.54 | 471.08 |
| He vapor 2kA lead flow per magnet (g/s) | 0 | 0.16 | 0.16 | 0 | 0.32 |
| He vapor usage for 2kA leads (liquid liters/day) | 0 | 110.592 | 110.592 | 0 | 221.18 |

[*]Transfer Line length only from feedbox to magnet considered

Table 6.40. Heat load at 4.7K.

| Best Estimates (no contingency) | PS | TSu | TSd | DS | Total |
|---|---|---|---|---|---|
| Nominal Temperature | 4.7 K | | | | |
| 4.7 K Magnet Heat (W) | 66.7 | 44.0 | 42.0 | 32.2 | 184.9 |
| 4.7 K Feedbox and Transfer Line[**] Heat (W) | 14.0 | 14.0 | 14.0 | 14.0 | 56.0 |
| Thermosiphon | | | | | |
| Total heat load (W) | 80.7 | 0 | 0 | 46.2 | 126.9 |
| Total helium flow (g/s) | 4.78 | 0.00 | 0.00 | 2.74 | |
| 3.0 bar to 2.7 bar forced flow | | | | | |
| Helium inlet temperature (K) | | 4.7 | 4.7 | | |
| Total heat added (W) | | 58.0 | 56.0 | | |
| Selected flow rate (g/s) | | 50.0 | 50.0 | | |
| Exit temperature (K) | | 4.82 | 4.81 | | |
| Circulating pump real work (W) | | 25.0 | 25.0 | | |
| Circulating pump system static heat (W) | | 15.0 | 15.0 | | |
| Total load for forced flow (W) | 0 | 98.0 | 96.0 | 0 | 194.0 |
| Total refrigerator cooling load at 4.7 K (W) | | | | | 320.9 |

[**]Transfer Line length only from feedbox to magnet considered

## 6.3.5   Magnet Power Supply Systems (PSS)

### 6.3.5.1   Introduction

The Mu2e experiment's solenoid power supply system has been designed to provide DC current to four major solenoids and smaller trim supplies that will be connected to sections of the major solenoid coils. The power supply system for the high current superconducting elements will have the latest Fermilab Accelerator Division E/E Support voltage and current regulation system in order to provide the best control and regulation used in the FNAL complex. The operation of the power supplies will be managed by the Magnet Control System (MCS), which will control the turn on, ramping and coordination of all power supplies.   Each solenoid will have an energy extraction system (dump system) installed that will be interlocked with the Quench Protection System (QPS) for solenoid protection.   AD E/E Support's latest versions of the electronic dump switches and controller will be used for the energy extraction switch with a reused TeV and new dump resistors.

### 6.3.5.2   Requirements

The power supply system will power four different solenoids that are in close proximity to each other. Table 6.41 lists the power supplies/magnet loops that will be used in this system.   In addition to the main coil power supply system, trim power supplies will be used to add or subtract current for part of two of the Solenoids to allow for smaller field corrections as needed.

Table 6.41. Power Supply Summary.

| Power Supply Name | Solenoid Name | Voltage | Current |
|---|---|---|---|
| E:PS | Production (using 2 TeV Low Beta 375kW) | 50/25 | 15,000 |
| E:TSu | Transport Upstream (using TEL 1) | 20 | 2,500 |
| E:TSuT | Transport Upstream Trim New Supply | 10 | 250 |
| E:TSd | Transport Down Stream  (using TEL 2) | 20 | 2,500 |
| E:TSdT | Transport Down Stream New Supply | 10 | 250 |
| E:DS | Detector Solenoid (1 Low Beta 375kW) | 50/25 | 7,500 |

### 6.3.5.3   Technical Design

In an effort to manage cost the plan is to reuse as much existing equipment from the TeV as possible.  This will include three of the TeV Low Beta Quad 375kW power supplies, 2 for PS and 1 for DS, with simple modification for two quadrant operation.   The TeV electron lens, TEL-1, 2 power supplies and dump switches are a direct match for the TS solenoids and can be used as is but will require a higher power Dump Resistor.   New supplies and isolation switches will be needed for all the trim power supplies.

***Reference Design***

A typical power system with energy extraction is shown in Figure 6.103. The high current power supplies will consist of multiple full wave bridges in parallel to get the current required for each load. The PS power supply will consist of two paralleled low beta quad power supplies that have four internal parallel SCR bridges. The DS magnet will use one low beta quad power supply. Fermilab has 8 operational supplies of this type with one spare unit.

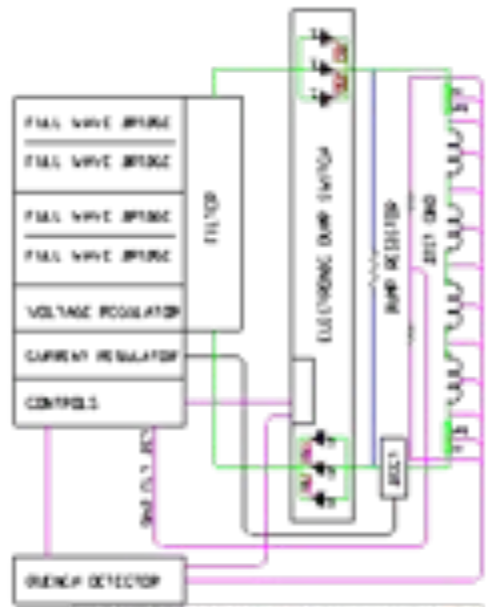

Figure 6.103. Power supply block diagram.

Current sharing between bridges will be performed using a series choke internal to the supply at the summing point of each bridge; this will provide impedance, which is used to filter the output voltage and reduce the voltage ripple applied to the load. The paralleled power supplies of the PS magnet will also use the bus work to resist force sharing between the supplies. The configuration of the power supplies has the added benefit of reducing the cross section of the bus from each supply, making the bus simpler to install. It is expected that the bus will be routed to the power feed can and the final parallel connection will be at the power leads.

The filter is used on superconducting loads to improve the voltage monitoring for the Quench Detection allowing for improved and faster cell resistive voltage detection by allowing for lower cell voltage trip limits. The impedance of the chokes will also allow us to add common mode filtering to the supply to reduce capacitive currents flowing into the magnets from the common mode voltage of each bridge. The filter capacitor banks will have been in storage for a few years before being installed in the experiment so all the capacitors will be replaced before the banks are used.

The TEL power supplies and dump switches are a direct match for the TS solenoids. They can be moved as is but we will make some improvements to the internal control logic in the power supplies to make them more compatible with new style of current regulator. One of the power supply systems has been upgraded to the latest dump switch system; the other has not been, but it will be improved before installation.

During the last few years of TeV operation, we upgraded some of the dump switches to modern electronic switches and removed the mechanical DC breakers due to their high maintenance cost. We will move those electronic switches to the experiment and construct additional units as required for higher current. The dump switch controllers are new and will be reused, and we have enough on hand for the experiment. The dump switches in the two TEL supplies are not identical, but we will modify the TEL-1 system to be the same before installation in the experiment. We will also modify the dump switches so that there is one on each leg of the power supply instead of both on one, as is currently the case.

***Criteria for Constraints on Power Supply Layout***
The power supply will need to provide enough voltage (RAMPING) to overcome the dump switch series and high current bus voltage drops and provide enough L(di/dt) voltage to ramp up in a reasonable time. A minimum of two dump switches is used in these systems to provide both redundancy and load voltage to ground balance during a quench. The power supplies are required to maintain a low current (STAND-BY) level to allow for system check out before ramping up to operation levels. There will be both dv/dt and di/dt limiters in the regulation electronics to minimize coupling during turn on and limit fast voltage pulses that could have an effect on the quench detector. The tracking is expected to be less than $+/-10^{-3}$ with steady state regulation better than $+/-10^{-4}$ of full-scale current. All of the solenoids are expected to track during normal ramping; therefore, the solenoid with the slowest Lid/dt voltage will define the speed for the entire system.

The power supplies that are planned to be used from the TeV are 50 volt @7500 amps units. The requirements document, Mu2e-docDB-1237 Rev2, calls for only 12 volts to be applied to the magnets. We plan to modify the supplies by either placing the primaries of the internal transformers in series or adding a line voltage lowering transformer in front of the supplies to reduce the available voltage. This will reduce the risk of applying too much voltage to the magnets in the event of loss of control. This will lower the avaiable voltage and therefore the maximum di/dt in the magnets.

### Normal Discharge

The power supplies will operate in two quadrant mode, which indicates +/- voltage, positive current as shown in Figure 6.104.  Figure 6.104 shows the maximum ramping speed of the PS solenoids using the power supply at half voltage; the dump systems will have a 5 volt drop and a 0.0006 Ohm Bus resistance.  This shows the maximum speed of the power systems, but in this operating mode, the current is fully regulated and therefore can operate at any speed slower than what is shown.  It is important to limit the ramping current to less than the conductor di/dt limit, and this will be controlled by the ramp generators.  The system will have the ability to ramp down faster (NORMAL DISCHARGE) than up using the bus, semiconductor voltage drops plus -90% of the power supplies convert voltage.  This will be useful if there is a quench in another coil that increases the heat load on the coils and a possible cascade of quenches without using the dumps.  This mode of operation can also be used at any time the solenoid currents need to be ramped down and will allow for tracking of the systems.  It is expected that this will be used to reset the fields in the coils or change operating levels as needed.

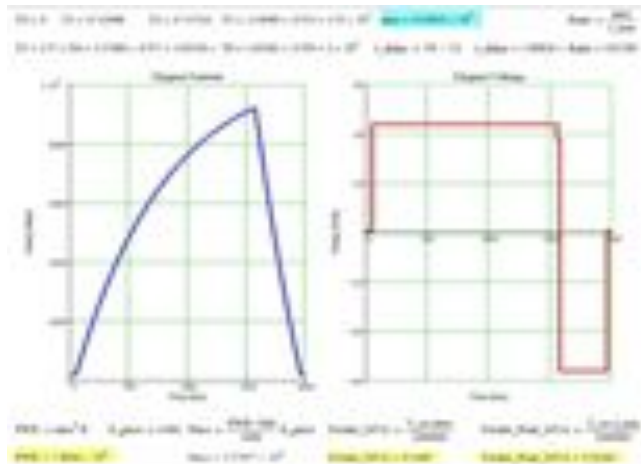

Figure 6.104. Normal discharge (ramp down).

### Slow Discharge (Bypass)

During slow discharge the bus work, bypass SCR and dump switch semiconductor voltage drops will use the back EMF from the magnet to discharge the magnet.   This is an unregulated decay in the current.   It is completely controlled by the bus work resistance and semiconductor forward voltage drops.  This type of ramp down is initiated by turning off the bridge SCRs and turning on the bypass SCRs in the power supply.  This will be used if a ground fault is detected, turning off the dump switches and resulting in a much higher voltage being applied to the magnets and increasing the risk of further damage, particularly if the fault is in the coil and not the power supply system.

***Emergency discharge (fast discharge)***

The dump switches will provide fast energy extraction during full quench detection and will use a much higher voltage and therefore higher di/dt than the power supply. The dump resistor is connected close to the magnet power leads and will generate an L/R time constant to discharge the solenoids. The semiconductor voltage drops are removed from the circuit by the dump switches, so this is a purely passive decay. The dump resistors are located outside of the building to avoid adding an excessive heat load to the building during a dump. The time constants for each magnet will be: 31 seconds for PS, 14.5 seconds for TSu, 11.5 seconds for TSd and 14 seconds for DS.

### 6.3.5.4   Cooling Requirements

The large power supplies and dump switches are water cooled equipment that will require LCW. The building layout for the large supplies should be such that the bridge, filter and dump switch cabinets are all in close proximity to each other and as close to the feed can as is reasonable. All of the Control and Regulation Electronics should be in an air-conditioned room.

### 6.3.5.5   Uninterruptible Power Supply

The quench detection and all of the control electronics will need an uninterruptible power supply, or UPS, installed because of the long decay time. The power supplies will need enough power to get the bypass SCRs, which are stored in internal cap banks in the power supplies, turned on. The dump switches are redundant devices and will need a UPS system for each switch, or 2 total for all of the dumps. The present design calls for four small UPS systems to be installed: one for the power supplies, two for the dump switches and one for the QPS. The QPS and PC-104 based current regulator computers have transient recorders built in, so we will need to maintain power to them long enough to allow them to back up the data using a non-volatile medium. The UPS system will be backed up by a generator so the holding time for the systems will need to be long enough for the generator to start and switch over, which will take 2-5 minutes.

### 6.3.5.6   Solenoid Magnet Energy Details

A summary of the stored energy for each magnet is shown in Table 6.42.

Table 6.42. Magnet Summary

| Solenoid | Maximum Current | Inductance | Stored Energy | Peak Dump Voltage | Dump Resistance |
|---|---|---|---|---|---|
| Production PS | 9,200 Amps | 1.58H | 66.7MJ | 600 Volts | 0.05 Ohms |
| Transport US | 1,730 Amps | 4.77H | 7.1MJ | 600 Volts | 0.33 Ohms |
| Transport DS | 1,730 Amps | 3.79H | 5.7MJ | 600 Volts | 0.33 Ohms |
| Detector | 6114 Amps | 1.40H | 26.1 MJ | 600 Volts | 0.1 Ohms |

### 6.3.5.7   Current Regulation Electronics Details

There are two things we need to be able to trust in the current regulation system; one is the current measuring device and the other is the stability of the current reference. Each of the main power supplies will have a total current monitoring DCCT (Direct Current Current Transformer) installed that will be used for both regulation and the QPS system. In addition to the main DCCT, there will be current measuring devices internal to each supply that will be used to back up the main device and will be monitored independently. A high quality DCCT will also be needed in each trim circuit for both regulation and QPS signals. The current regulation system will be a version of the E/E Support latest design constructed for the last MI upgrade. This system is a Digital/Analog combined regulator built using a PC-104 embedded processor system that provides both the feedforward and feedback voltage reference to the power supply. Each PC-104 regulator supports four power supplies in a single package and provides all of the voltage reference drive for the main current including any correction needed. The DCCTs used for the feedback will be commercial devices procured at a level sufficient to meet the long term stability requirements of the experiment. Typical regulator systems of this type operate at the +/4ppm [43] of FS level as designed but can be improved so that they operate at the +/-0.25ppm level by procuring a high performance DCCT. This system is not intended for fast ramping power supplies and has a di/dt limiter used during startup.

 Due to the coupling nature of the magnets we need to have the currents track each other with a certainty of $10^{-3}$. This tracking will be primarily managed by the current reference provided by the MCS. Fermilab will add an alarm circuit to the PC-104 controller to ensure that the currents are tracking.

Fermilab will need to develop a functionality table for this added control that includes the di/dt limit, which is controlled by the slowest magnet loop, response table, which represents actions to take if one falls behind or trips, and whether or not we allow for recovery if one of the loops quenches, among other factors.

### Regulation System Grouping

The PS, TSu and TSuT power supplies will be controlled using one PC-104 regulator so that we can add a coupling feedforward signal between supplies as needed to maintain regulation. The DS, TSd and TSdT will be controlled using a second regulator system. The PC-104 regulator is a digital current regulation system that can pass information internal to itself in order to provide coupling correction; however, if we need correction between the TSu and TSd regulation systems we expect to pass this information between PC-104 regulators using analog signals. These regulators are operated through a network connection, but do not require the connection to maintain regulation. The analog correction signals between regulators would allow the coupling regulation to be maintained without the network.

### Theory of Operation

The current regulator has one PC-104 processor that sends an 18 bit digital reference to four Sigma Delta DACs in temperature regulated modules, one for each magnet loop. Each DAC module receives an analog current from a current output DCCT and converts the current to a voltage using burden resistors in the temperature regulated module. The reference and current signals are subtracted to generate an error signal, which is then amplified 100 times. This amplified error is then sent to the PC-104 module that adjusts the drive to the power supply to minimize the error signal. The reason for the amplification is to reduce the sensitivity of the PC-104's AD converter; looking at small signals with only 16 bits is limiting, and the larger the signal, the more bits are used. The magnet parameters are loaded into the PC-104 along with the current loop bandwidth and maximum gain limit. The PC-104 then uses this information to provide the correct voltage reference to each power supply. It is expected that the MCS/ramp generator will be a digital system; therefore, sending the ramp reference to the PC-104 using digital communications will provide the most stable system. We have not installed magnet coupling corrections to the present design but are developing a model that will be used to make the first design attempt and implement it into the PC-104 regulator.

### Current Reference

In order to reduce the amount of analog noise in the system, the reference to each power supply will be supplied from a ramp generator over an Ethernet connection to the PC-104 regulator. If the ramp generator cannot provide the digital signal, we will use an analog input to the temperature regulated DAC module instead of the internal DAC. This will require that we locate the ramp generator as close to the PC-104 regulator as possible to reduce the noise in the reference signal. The current reference is one of the signals that we need to trust in order for the stability of the reference to be the limiting factor in the current regulation loop. All ramping speeds and changes are provided by an independent ramp generator controlled by or provided by the Magnet Control System. This ramp generator will need to take into account the magnet coupling and dynamic range of the power supply voltage in construction of the reference.

### Transient Recorder

An additional feature of this system is that the PC-104 has a transient recorder built in that will record trip events and provide data for analysis and is very useful during single event trips that happen very infrequently. A second option built into this regulator is a window detector that can be set up to monitor current, DAC settings and the current error signal to ensure that they are in a set range. These limits are set up using an independent path into the processor. This setup is used in the MI to prevent beam transfers if the signals are not in the correct operational range, but can also be used as an indicator that one of the circuits is out of operational range for the experiment. There are other features built into this regulator but they will not be useful on an air core solenoid.

### 6.3.5.8  *Energy Extraction Electronics Detail*

The main dump switches will be used to disconnect the power supplies from the magnets when a quench is detected. The dump resistor will always be connected to the magnet and the dump switch will disconnect both leads of the power supply from the circuit, allowing the magnet energy to decay into the resistor. New dump resistors will be procured for all the magnet systems; the PS requires 0.05 ohms at 90 MJ and the DS requires 0.1 ohms at 45 MJ so we will procure three 0.1 ohm resistors. We will parallel two for the PS and the third will be installed on the DS system. The TS magnet loops will each have a 0.33 ohm, 14 MJ resistor installed. The maximum temperature of the dump resistor will be 200 degrees C, so they will be located outside of the service building. Locating the resistors outside of the building will result in long cables between the feed can and the resistor. In order to limit the voltage spike that would result from the fast switching and current diversion into the cable inductance a dv/dt cap bank will be installed at the output of the dump switches.

Each dump switch will have separate electronics for control with parallel paths to the QPS for improved redundancy. This redundancy will extend all the way to the power provided by the two independent UPS units. Monitoring electronics in the power supply system PLC should also monitor the status of the dumps and be able to turn the dumps off if needed. Even though the power supplies will not have enough voltage to damage the resistors, this monitor will provide added error checking for correct operation. There are several types of potential dump failures that can occur. The first type is having two independent failures in the different dumps and not detecting them, which would place the magnets at risk. Both dump switches will always turn off during any dump command; however, to ensure that both switches are operational, we will use a delay of 30 milliseconds between the two switches so that we can detect a voltage rise and verify operation of each switch on every other dump. A second type of dump failure would be a failure of the dump to turn on, which would cause the power supply to be powering an open circuit and reaching full voltage. If this happens, we will bypass the power supply until the dump issue is corrected. Turning on the dump switches with full voltage from the power supply will cause a quench to be detected but will not put the system at risk.

A System PLC will monitor the auxiliary status on the power supplies, dump switches, water pressure and UPS for correct operation. All other dump switch control will be under the direct control of the quench detection system. The voltage on the dump switches will also be monitored in the dump controllers to look for voltage when the dumps are told to be on. If any of the monitored statuses of the dumps raises a red flag, the system should not turn on or the QPS will turn the entire system off in a controlled fashion.

### 6.3.5.9  *Power Supply Topology Details*

The power supply topology uses a phase to neutral bridge configuration so that we have one SCR per on phase in the bridges to limit the internal DC loss. The power supply will use a filter choke and will provide the impedance needed for both the filter and the current sharing between the bridge sections. For the PS magnet loop, we will use two supplies in parallel to provide the high current and the choke and bus work resistance to assist in the sharing between supplies. (Figure 6.105) The output of the PS power supplies will be paralleled at the feed can close to the input power leads. The use of parallel bus work allows for installation with reasonable manpower effort as well as reduced cost for copper. The power supply for the DS loop will use one of these supplies and one bus routing. The TS power supplies are expected to use cable to deliver power to the load.

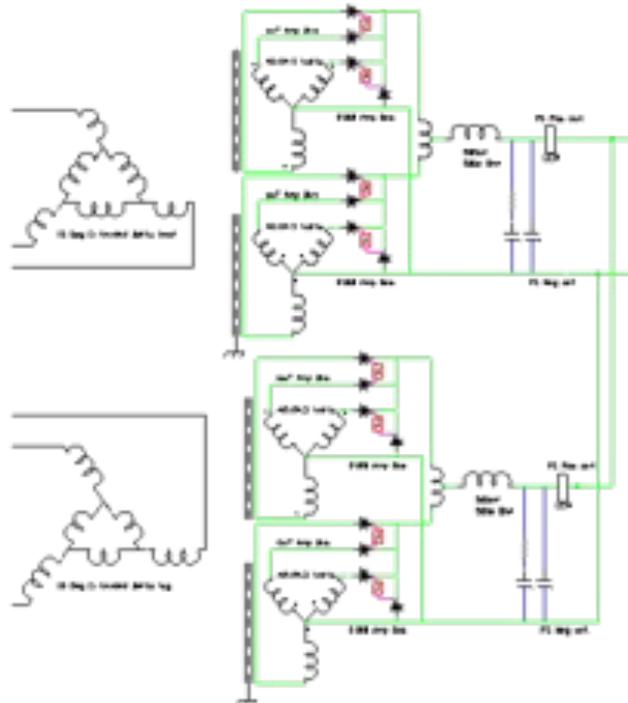

Figure 6.105. PS Power Supply Topology

This design of the power supply uses a pair of delta primary transformers with dual inverted Wye connect secondaries connected in parallel through interphase transformers. Each power supply has internal monitoring for temperature and door interlocks. Because of the long time delay and potential high voltage from the Dump Resistor the doors for the power supply, dump switches and filter cabinets will need to be locked with a controlled key to delay access.

All of the power supplies in this system will need to be of a 12 pulse design to avoid sub harmonic loading on the main power transformer. Also, the main transformer should not be shared with house power electronics due to the chopping noise on the line and the fact that the average power level is 463kVA. A typical house power transformer used in the MI and Muon Campus is 750kVA, so the entire power margin would be used just for the power supplies and cooling system. With the reuse of the TeV low beta quad power supply system, we plan to take the main power transformer and power panel to provide the power needed for these supplies as well as the power needed for the Muon campus beam line supplies that are located in the Mu2e service building.

We plan to reuse as much bus work from the low beta systems as is reasonable. This is 2x2square water cooled bus that can be used for straight runs between power supply cabinets and the dump switch systems. As is the case with the low beta system, we expect to be able to support the bus work with the building steel structure. All of the bus work will need to be covered due to the 600 volts that would be generated during a dump. The power supplies, dump switches, main transformer, power panel and bus are being held in reserve under an agreement in the Accelerator Division. This is all large equipment that will require coordination with the other equipment in the building during installation.

### 6.3.5.10 Voltage Regulation Electronics Details

Each power supply will be a voltage regulated source, which will correct for line voltage changes. Over the years E/E Support has procured many different power supplies from many different manufacturers. This has resulted in many different voltage regulation designs from the manufactures, and because of these unique regulators, the load on the engineering staff to keep them operational has been heavy. When we constructed the Main Injector, we chose to start installing the FNAL voltage regulator in all SCR style power supplies; therefore, when we procure new power supplies, we include the requirement that the supplier use our regulator in our specification. This requirement indicates that the power supply must have a firing circuit that is hard fire and compatible with the drive from our regulator. Using this regulator in all 120 or more systems has greatly reduced the load on the engineering staff. These regulators are very reliable, and because there are many copies on hand, we have technician experts that can troubleshoot these devices. The low beta quad power supplies were upgraded to this style of voltage regulator and will be part the new system. The only change to the present installation is that the PS needs two supplies in parallel, so we will modify one of the supplies to be a slave to the main voltage regulator to ensure firing angle balance between the supplies with the bus work creating equal loads.

### 6.3.5.11 Power Supply Control

All of the control and status read backs are provided through the PC-104 current regulator using a single E-net connection. This system uses a programmable logic controller to collect all the trip data from the power supply control PLC and provide it to the PC-104 using an E-net connection. The information is collected in the PC-104 and converted to ANCET format. The PC-104 also provides the on, off and reset functions to the power supplies through the PLC. In addition, a System PLC will provide the necessary management of the paralleled supplies to ensure they act as one power supply. If either of the supplies trips, the other will also need to turn off, so the PLC will be coded to provide this function. Each power supply will have current monitors internal to the power supplies to verify current balance; these current monitors will be installed after the filter connected to the internal power supply PLC/control card and are used to check for over currents. These monitors will provide an independent over current trip beyond the PC-104 and QPS monitoring of the main DCCT. The power supply's 7,500 amp DCCT was used for regulation, and we will also bring these units with each power supply. The main DCCT used will be a total current unit that will be a new procurement.

### 6.3.5.12 Power Feeder Loading

There will be a power transformer with a range of 13.8kV to 480V installed to provide power to these supplies. At a DC current of 11,000 amps, the line current will be 315 amps, yielding a transformer and feeder load of 298kVA for the PS, 153kVA for the DS and 62kVA for the two TS loops. The internal DC power consumption is expected to be 87kW for the PS, 43kW for the DS and 19kW for the two TS loops, which is a total of 149kW needed for cooling. This analysis assumes that the total high current bus work is less than 200 feet in length and that we use 4sq-in of copper, which has a resistance of 2.3uOhms/foot. The TS coil loops will use cable for the connection to the feed can but 6-500mcm cables each way will be required for each magnet. Making the connection to the power lead will be problematic with this number of cables so we may choose to install the "Main Injector Quad Bus" to save space. The trim power supplies will all be connected using cables.

The 13.8kVAC feeder for the man power transformer dedicated to the power supplies will come from the Muon Campus feeder 24 line. Lock Out Tag Out points will need to be provided for enclosure access and maintenance because providing a disconnect for each power supply will be costly and time consuming for operations. In other experiments we have provided a central LOTO point at the 13.8kVAC level. To make this simple, we used a motor drive switch with remote operation, verification and lockout.

The final inspection for the LOTO is a viewing window in the switch. A LOTO controller will need to be constructed to ensure that the power supplies are ramped down and turned off in a safe manner before the LOTO key is issued. The Muon campus beam line power supplies will also have a local LOTO disconnect for all power supplies powered from the main transformer that will be located near the beam line supplies. This LOTO disconnect will allow for the lock out of each beam line independent of the experiment depending on the type of access.

We plan to recover a 1,500kVA transformer and power panel from the TeV to use for this application but it will require that oil containment be constructed. The present plan for the service building is to use a 750kVA oil filled transformer for house power so the containment will be in place. The present plan for the building calls for two transformers to be installed; keeping the chopped line power of the large supplies on a separate circuit will help to keep the building house power line "clean". The separate transformer will also allow us to add a power factor correction and line filtering if needed to manage the harmonic distortion to the AC power.

### 6.3.5.13 LCW Cooling System

The LCW cooling water will be routed from the Muon Campus. We will need 149k Watts of cooling with a minimum pressure range of 60-100 psi and flow of 48 GPM. Most of the power dissipation in the system will result from the electronic dump switch voltage (5 volts) and the power supply internal SCR (1.5 volts). The high flow rate is needed to supply the parallel cooling paths for the SCRs in power supplies and IGBTs in the dumps switches. To ensure stability, the copper bus should be water cooled with minimal flow; however, as the bus gets close to the power leads, we will need to ensure that the water will not freeze if the circulating pump is turned off. This means that we will need a larger cross section of bus close to the leads to keep the heat load on the lead in a safe operating area; we should not source bus heat into the leads from the water. In the past we have used the water system as a heat source to keep ice off the leads, but the problem with this setup was the difficulty of not being able to turn off the water system when the magnets were cold (as the non-flowing water would freeze along with the leads). As part of the interface between the power leads and power supply connections, we will need to address this issue.

### 6.3.5.14 Transport Trim Power Supplies

The trim power supplies will add current to one section of the main coil as needed to correct the field. These power supplies will need to be protected from the voltage of the main power supply and dump resistor. This will be accomplished in two ways: The power supply voltage will be greater than the ½ of the main power supply voltage, and the power supply will use diodes/switches to isolate the trim supply. This can be a problem during turn on and quenches because multiple regulators on the same bus can cross couple through the magnet current all trying to make corrections at the same time. This supply will also need to be isolated to greater than the dump voltage to ground to avoid an imbalance during quenches. By installing a small version of the main electronic dump, we can provide the isolation from the coil and dump voltage. This small dump will also need to be controlled by the QPS so that when the main dump switches are turned off, the trims are also isolated from the magnets. The QPS will also need a DCCT to measure the current in the trim supply and make the correct calculation for the resistive voltage. Startup testing will need to be performed to check the leakage to ground. Care will be taken to ensure that trim is also tested during these checks; the switches will be turned off before the main supply is turned on. A large snubber circuit will need to be installed across each switch to manage the energy stored in the stray inductance of the trim loop and will need to be rated greater than the hipot voltage just in case there is a ground fault in the trim. The snubber circuit will cause voltage transients on the coil that will confuse the QPS if the isolation switch is turned off when the current is not zero. It is expected that the trim current references will be set to zero before the dump switches are turned off during a slow discharge. The difficulty of procuring commercial power supplies with ground referenced control and output voltage isolation greater than their operating voltage makes these power supplies custom devices. We should be able to use a commercial supply by powering it using an isolation transformer and isolated controls, but the cost of this may be higher than a custom power supply. As the final design is completed, we will find the best solution for these systems.

Connecting two sources of high power together will require care in managing access to the equipment for personnel and will require written LOTO procedures. This has been done in the past on many different systems, including SC loads. The long time constants of decay for SC loads increase the risk that someone can open an access door and make contact with the bus while voltage is back feeding from the magnet and dump resistor. A lockout key system will be needed to lock the doors on equipment until the correct equipment is turned off. This will be required for all systems rather than just the trim. The bridge, dump switch and filter cabinets will be physically separate elements and the correct power supply needs to be off and verified before access can begin.

### 6.3.5.15 Controls System Interlock Details

Hardware interlocks will be necessary for the individual power supplies and, at the system level, are things that are common to all the supplies. Due to the long decay time constant, personnel will need to be protected from any back feed voltage from the load. Control of access to the power supplies and dump switches will need to be locked using a key system that will prevent opening of the door before the decay time is expired. Thermal protection for the transformers, internal and bypass SCR heat sinks, load cable and bus and dump switch SCRs will be part of the power supply internal system and be monitored by the PLCs. Water cooling will be used in the systems and differential pressure switches will be installed and used to prevent the power supplies from being turned on if the water pressure is low. In the past, the pressure switches have been heavily filtered to prevent unnecessary trips, and the power supplies should not be tripped off after the current is established. The thermal switches will be the backup for pressure switches.

During a quench, the power supplies will be bypassed using two sets of SCRs, one internal and one external. A bypass failure detector will be installed in each power supply that will turn off the main contactor in the supply if current is still in the transformer 30 milliseconds after being set to bypass. This will result in a bridge bypass using the bridge SCRs if a true bypass failure were to occur. Ground faults will need to be handled differently than other trips. If a ground fault is detected, the supplies should be bypassed, but the dump switch should not be turned off unless a quench is detected. The higher voltage of the dump to ground can cause damage to the conductor if the fault is in the conductor; if the fault is in the warm bus, this can be investigated after the system has decayed. An offset ground fault detector will be used in these systems; this type of detector allows ground fault detection before the main power supply is turned on. The offset detector applies a positive voltage of 5V to the system and will allow for detection of a hard fault before turn on.

A response table needs to be constructed for every type of trip for each magnet loop due to the coupling, and trips can be managed by the QPS. Internal trips to the power supplies will trip the main contactor directly and inform the QPS system only after the trip has occurred. The trip matrix will need to define the effect of the QPS on the other magnet systems during any of these types of trips that are not directed by the QPS.

### 6.3.5.16 Procurement Strategy

In order to manage the number of procurements we will use our recent history of procurement of power supplies. The Muon Campus is being upgraded to supply beam to the experiment and as part of the construction of the M4-M5 beam lines. Some new supplies and all new current regulation and control electronics are being procured. We will take advantage of procurements of a similar size for this project. New PC-104 current regulators will need to be procured and constructed and the experiment equipment will be grouped with the larger number of campus units. This will consist of many small procurements and an assembly contract.

Specifications will be written for each power supply design based on recent design procurements in 2010 for the ANU upgrade for the trim power supplies. During the ANU upgrade, we updated all the regulation electronics to be compatible with components still being manufactured and this required some new circuit board updates. We will use the upgraded electronics and be able to procure identical equipment for the project with little or no rework necessary. The assembly of the regulation electronics will be contracted to a local company with final testing and calibration done by FNAL electrical technicians. To maintain quality control of the components used in the electronics, FNAL technicians will procure all components and inspect them before shipping them to the assembler. Inspections and testing will be witnessed at the vendor/s for each design type. As part of the specification a minimal set of tests is defined and will be witnessed.

### 6.3.5.17 Contingency for Reuse of Existing Equipment

The power supplies being reused have been in the TeV for a long time. All of the TeV equipment was maintained at a very high level, so this equipment is a good match for the experiment. The cost estimate includes improvements to the filter and new firing circuits internal to the power supplies; however, a risk factor that can be used for rebuilds or replacement should be added to cover any unforeseen failures that may occur. In other experiments we have used a risk equal to 30% of the cost of procuring new equipment. The only reused equipment will be the power supplies themselves; all other equipment is new. This would result in a risk factor of $110k, but this would not be needed until late in the project when we are ready to incorporate the power supplies. This would be late in the project.

### 6.3.5.18 Quality Assurance

We will be reusing existing equipment from the TeV including the power supplies and filters. The equipment being reused was used in a very similar manner during TeV collider mode of operation, making this installation a good match for the experiment. New equipment will be constructed for the dump switches and will use a proven new design from the Tevatron. There will be no new designs for the power systems; all equipment will be copies of existing proven designs presently installed in the accelerator complex.

### 6.3.5.19 Installation and Commissioning

The power supply system consists of large equipment that will be connected to other large equipment in the building. This will require coordination of the equipment installation with other large equipment in the Mu2e service building. This equipment includes the control racks, cable tray and infrastructure in the building. The infrastructure will be installed first in the building and will include the relay racks and all power distribution including cable tray and wire ways. This will be followed by the feed can and its support hardware including vacuum pumps and piping. The power system will consist of four large components: the power supply, the filter, a dump switch, dump resistors and DC buswork. The control cable and monitoring cable for the power supply system will be installed last and will allow commissioning to begin. The power supply control cable is expected to be constructed in the AD E/E support shop by AD techs or a local contractor. This cable will be inspected and tested before installation in the service building to decrease the time required for system testing.

## 6.3.6   Quench Protection and Monitoring

### 6.3.6.1   Introduction

The Mu2e superconducting magnets and leads need to be protected against heat-induced damage during quenches. The Mu2e quench detection system is a critical component of the quench protection system for the Mu2e solenoids. The quench detection system monitors signals from sensors installed in the Mu2e magnet, leads, and superconducting bus, and generates an interlock signal if signals from the sensors are consistent with those expected during a quench. The interlock signal will trigger the shutdown of the power supplies and switch dump resistors into the magnet power circuits to extract the stored energy from the magnets.

The primary goal of the Mu2e quench detection system is to reliably detect true quenches as rapidly as possible. The secondary goal is to minimize the number of false quenches detected, because rapidly extracting the power in the absence of a true quench will produce unnecessary stress on the magnets and delays in the experiment.

A variety of methods, both optical and electrical, have been used to detect quenches in superconducting leads and magnets. The most common approach monitors voltage taps in the magnets and shuts off the power if the voltage across the taps exceeds a predetermined threshold. Multiple taps are often employed and the voltages from the taps can be weighted and summed to improve detection thresholds and suppress noise and pickup.

Detection of quenches in superconducting leads using voltage taps alone can be difficult because of the small signals generated during a quench. Thin superconducting wires in thermal contact but electrically isolated have been successfully used to detected quenches in superconducting leads.

Quench detection in the Mu2e system is complicated by the mutual coupling between the multiple magnets. As individual magnets are ramped or quenched, the changing magnetic field can induce large voltages in the other coils. If the quench detection system is not able to account for the coupling between magnets, true quenches may take longer to detect or the rate of false detections may increase.

### 6.3.6.2 Requirements

The Mu2e quench detection system is shown schematically in Figure 6.106. The Mu2e quench detection system includes the hardware, firmware, and software required to detect quenches using the sensors installed in the Mu2e Solenoids and to generate the interlock signals to ramp down the current to the magnets and trigger energy extraction.

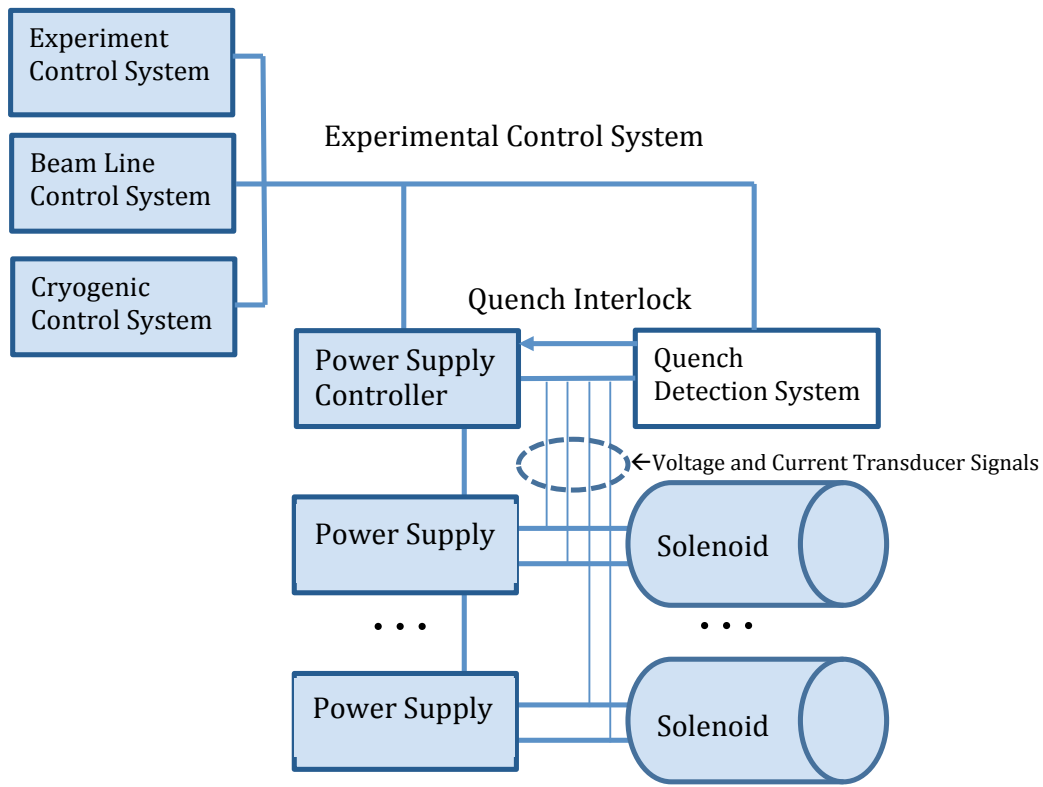

Figure 6.106. Mu2e Quench Detection System

### Power Distribution and Solenoid Systems

The Mu2e solenoid system consists of the four magnets forming the Mu2e solenoid system. Each magnet consists of a number of coils connected in series. For the purpose of this document, a magnet consists of the set of all coils powered by a single power supply. Each magnet is connected to its respective power supply by a system of normal conducting copper leads and high temperature superconducting leads, followed by a low temperature superconducting link.

The copper leads, HTS leads, LTS links, trim power leads, and the magnets themselves will be instrumented with redundant voltage taps for quench detection. Superconducting twisted pair sensors in good thermal contact with but electrically isolated from the LTS links over their entire length will provide additional information to the quench detection system.

When a quench is detected, the system must turn off the magnet power supplies and send a trigger signal to the energy dump extraction circuit. The quench detection system must be able to detect quenches in any or all of the HTS leads, the LTS links, the trim leads, and the magnets themselves. Failure to detect a quench within the required response time could lead to irreversible damage to the magnets, the links or the leads.

Components of the solenoid system relevant to quench detection are described in more detail below and their important characteristics are summarized.

### *Expected Solenoid Quench Characteristics*

The characteristic signals expected from quenches in each of the Mu2e solenoids were modeled using QLASA quench detection simulation program [35].

The peak temperatures and resistive voltages that may be expected to develop in each magnet during a quench were simulated for a range of detection thresholds. The resulting curves were used to determine the detection sensitivity and response time required to prevent the voltage developed across the coils from exceeding 600V and to prevent the peak coil temperatures following a quench from exceeding 130K.

Table 6.43 lists the quench velocities along with the heat generation and critical temperatures calculated by QLASA. Precise predictions of peak temperatures using QLASA are difficult because the program does not simulate thermal coupling between the coils and because of the assumptions it makes about the distribution of materials within the conductor when calculating transverse propagation speeds.

Table 6.43. Quench Velocities

| Solenoid | $V_{Longitudinal}$ m/s | $V_{Radial}$ m/s | $V_{Axial}$ m/s | $T_o$ K | $T_a$ K | $T_c$ K |
|---|---|---|---|---|---|---|
| PS | 4.987 | 0.1036 | 0.1036 | 4.5 | 6.77 | 7.17 |
| TS | 4.491 | 0.0959 | 0.0959 | 5.0 | 7.03 | 7.86 |
| DS | 4.395 | 0.2104 | 0.2104 | 4.5 | 7.25 | 8.37 |

The time evolution of the voltage tap signals is determined primarily by the axial velocity of the quench, which is sensitive to the distribution of materials. Peak temperatures are sensitive to assumptions about the thermal coupling between coils. Furthermore, the development of resistive voltages during a quench may be expected to vary by up to an order of magnitude depending on the precise location and magnetic field at the initial quench location.

The thresholds and response times listed here were calculated for quenches initiated at the innermost radius of the coil in the highest field region. Thermal conductivities and quench velocities calculated by QLASA were compared to an independent calculation that made more realistic assumptions about the distribution of materials and were found to agree within a factor of 2 or better. The initial time development of the resistive voltages was compared to semi-analytic approximations for the maximum and minimum that might be expected and found to be of comparable magnitude.

*Threshold Voltage*
Peak coil temperatures for the PS, TS, and DS solenoids as a function of the quench detection threshold for the resistive voltage are plotted in Figure 6.107.

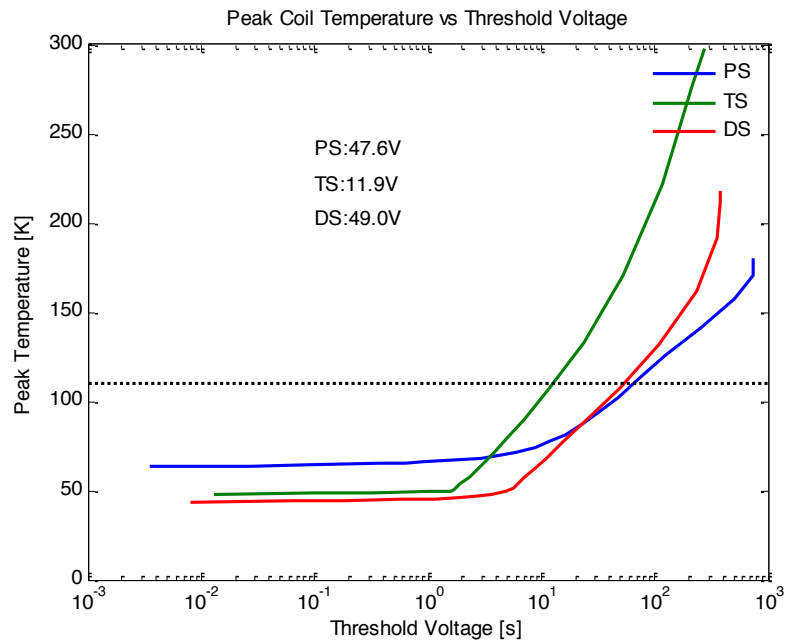

Figure 6.107. Peak Temperature vs. Quench Detection Threshold.

*Response Time*
The response time and quench detection thresholds vary between the magnets, the LTS links and the HTS leads. Peak coil temperatures for the PS, TS, and DS solenoids as a function of the quench detection response time for the resistive voltage are plotted in Figure 6.108.

The response time and expected resistive voltages for each of the magnets are summarized in Table 6.44. For the reasons discussed above, these numbers should be interpreted as a guide to the order of magnitude to be expected. Calculations of peak temperatures are extremely sensitive to assumptions about heat sharing, the distribution of materials within the conductor, the precise location of the quench, and the magnetic field at the initial quench location.

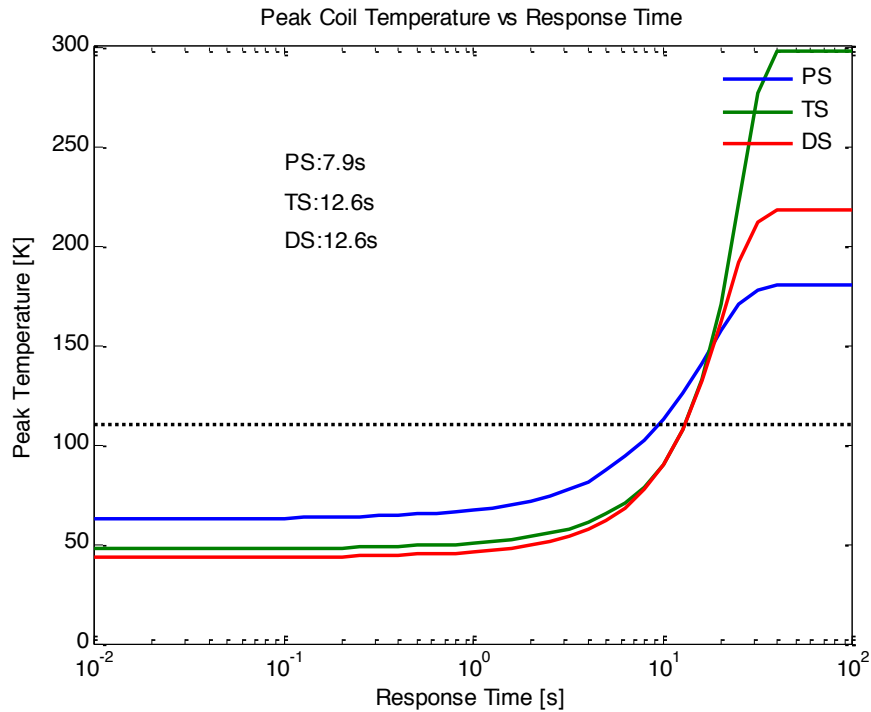

Figure 6.108. Peak Coil Temperature vs. Response Time.

Table 6.44. Expected Quench Detection Thresholds and Response Times for the Mu2e Solenoids

| Magnet | Threshold V | Response Time s |
|--------|-------------|-----------------|
| PS | 47.6 | 7.9 |
| TSU | 11.9 | 12.6 |
| TSD | 11.9 | 12.6 |
| DS | 49.0 | 12.6 |

### 6.3.6.3   Technical Design

#### Quench Protection and Monitoring System

All superconducting magnets can transition from the superconducting state to the normal conducting state if their operating parameters exceed the critical surface. If all of the energy stored in the any of the magnets is allowed to dissipate in the coils, the resulting temperature rise could lead to thermal stresses that may damage that solenoid; therefore, the quench protection system must continuously monitor the solenoids, the leads, and the superconducting links for any sign of non-zero resistance. If a quench is detected, the system must initiate a rapid controlled extraction of the energy from the magnet by the power supply system. The power supply circuit is shown schematically in Figure 6.109. When a controlled extraction is initiated the supplies are switched out, forcing the current to flow through the dump resistors. The energy must be extracted rapidly enough that the peak coil temperature in any of the solenoids will not exceed 130K but not so rapidly that voltage induced on any coil exceeds the insulation standoff limit of 600V.

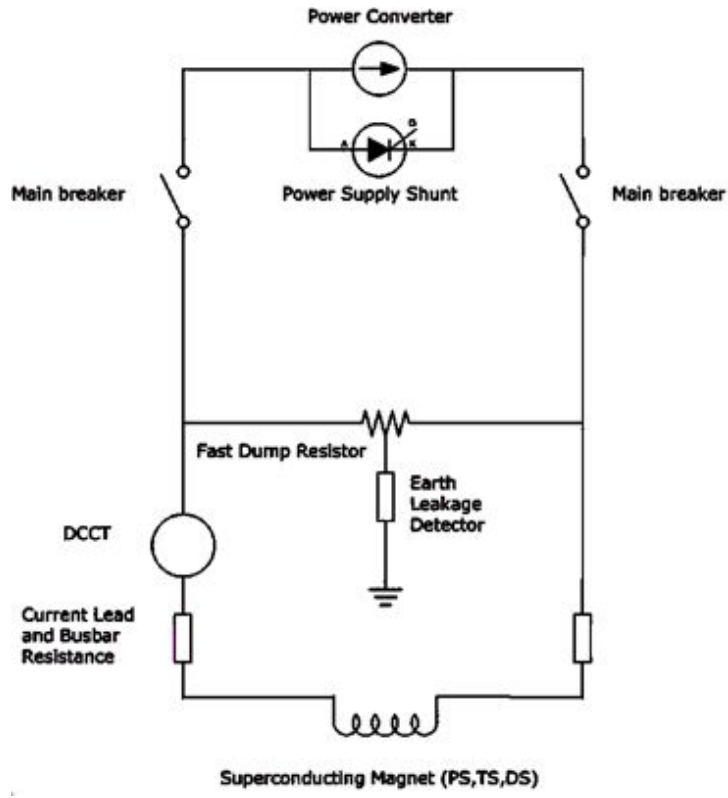

Figure 6.109. Magnet Power Supply and Protection Circuit.

During a controlled extraction both the voltage and the temperature remain within safe ranges. Since even controlled shutdowns can stress the magnets, the system must also reliably discriminate between quenches and other events that may generate similar signals. Quench detection is complicated in the Mu2e solenoid system by the inductive coupling between the magnets. Changes in the current in one solenoid can induce voltages on the other solenoids that may mask signals induced by a quench. To limit false quenches, the current ramp rate can be limited to minimize coupled voltages well below the quench voltage thresholds or the detection system can be designed to reject the coupled inductive voltages while remaining sensitive to small resistive voltages.

*Quench Development and Detection*

Solenoid and power supply system parameters relevant to quench detection are summarized in Table 6.45. The currents, inductance, and stored energy all vary by almost an order of magnitude from solenoid to solenoid.

Table 6.45. Mu2e Solenoid Parameters Relevant to Quench Detection

| Parameter | Unit | PS | TSu | TSd | DS |
|---|---|---|---|---|---|
| Operating temperature | K | <5.1 | 5.0 | 5.0 | 4.7 |
| Maximum design current | A | 10150 | 1730 | 1730 | 6114 |
| Peak field in coil | T | 5.48 | 3.4 | 3.4 | 2.15 |
| Current fraction along load line at 4.5 K | % | 63 | 58 | 50 | 45 |
| Inductance | H | 1.58 | 7 | 7 | 1.4 |
| Stored energy | MJ | 79.7 | 10.4 | 9 | 26.1 |
| Cold mass | tons | 10.9 | 13 | 13 | 10 |
| E/m | kJ/kg | 7.31 | 0.8 | 0.7 | 3.6 |
| Dump Resistance |  | 0.05 | 0.33 | 0.33 | 0.05 |

Quench development in each of the Mu2e solenoids has been simulated using both quench specific computer codes such QLASA and more general simulation packages such as COMSOL Multi-physics. Table 6.46 lists expected quench characteristics for each of the solenoids.

Table 6.46. Expected Quench Characterstics for the Mu2e Solenoids

|  | Units | PS | TS | DS |
|---|---|---|---|---|
| Longitudinal Quench Velocity | m/s | 5.0 | 4.5 | 4.4 |
| Radial Quench Velocity | m/s | 0.1 | 0.1 | 0.2 |
| Axial Quench Velocity | m/s | 0.1 | 0.1 | 0.2 |
| Response Time | s | 0.3 | 0.8 | 1 |
| Threshold | V | 0.5 | 1 | 1 |
| Peak Resistive Voltage | V | 330 | 16 | 170 |
| Peak Temperature | K | 90 | 47 | 55 |

Physical taps in the coils serve as the primary quench detection sensors. In the superconducting state, the voltage between successive taps should be zero. If a quench develops between successive taps, the voltage will grow as the size of the resistive zone grows. When the resistive voltage across any pair of successive taps exceeds a predefined threshold, the quench detection system will initiate a controlled extraction. Voltage tap signals are protected against shorts and over-currents by in-series limiting resistors. Magnet leads and superconducting links will be instrumented with both voltage taps and redundant superconducting wire sensors. The small resistive voltages that develop across the leads and links during a quench may be difficult to detect. Small diameter NbTi twisted pair cables connected thermally to but electrically isolated from the leads and links serve as independent redundant quench detection sensors. In the event the conductor goes normal and the temperature rises, the wires will also heat and go normal. If the wires are excited with a small current and routed to minimize inductive voltages it is possible to detect quenches in the leads and links long before the resistive voltages would reach detectable levels.

Tap voltage signals will be combined with the precision current measurements to track the internal state of each solenoid in real time. If these signals indicate that any segment of the solenoid has become resistive, a controlled extraction will be initiated. As will be discussed below, knowledge of the internal system states and of the couplings can be used to distinguish between the resistive voltages that characterize a quench and the voltages induced by changes in the currents of other coupled coils.

### Quench Detection System Architecture

To ensure reliable detection of true quenches while remaining immune to false quenches, the Mu2e quench detection system is organized into three heterogeneous redundant tiers (Figure 6.110). Redundant independent voltage taps and superconducting wire sensors are monitored by redundant independent heterogeneous processors. Redundancy ensures that quenches will be detected even in the event of single point failures and that any single point failures can be detected and corrected when they occur. The use of heterogeneous components minimizes the likelihood that undetected design errors will be duplicated in the redundant channels. Table 6.47 lists the technologies that will be used to implement each tier.

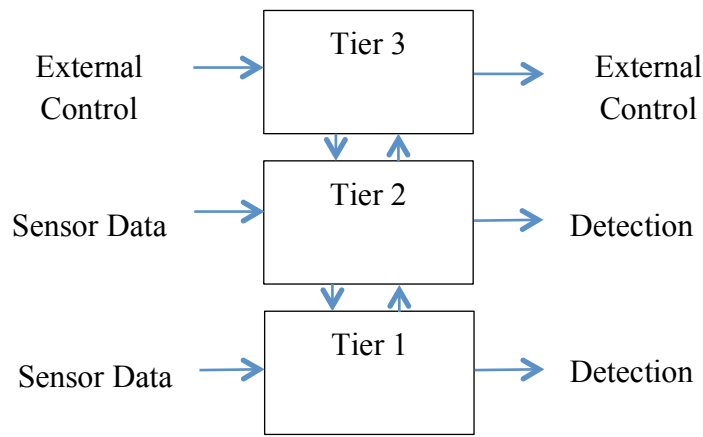

Figure 6.110. Quench Detection System Architecture

Table 6.47. Quench Detection Functionality and Implementation

| Tier | Function | Sensors | Hardware | Firmware | Software |
|------|----------|---------|----------|----------|----------|
| 1 | Primary Quench Detection | Primary Voltage Taps Primary SCWS DCCT | FPGA | VHDL Verilog | None |
| 2 | Secondary Quench Detection | Secondary Taps Secondary SCWS DCCT | DSP and/or Analog Threshold Detection Circuits | None | C |
| 3 | Long Term Monitoring and Control | None | PC | None | C++ Java Labview Matlab Python |

### Analog Processing

The voltage taps' analog signal processing chain is shown in Figure 6.111. Twisted pair cables convey signals from voltage taps in the coils to racks in the control room. Each tap is protected against over currents by limiting resistors. The signals from each twisted pair cable are connected to a low drift amplifier through an anti-aliasing filter and network of protection diodes and powered by a battery backed-up isolated supply. The output of each amplifier will be digitized by a high resolution ADC at the kHz level sampling rate. Each ADC can be controlled and read out via an isolated digital interface.

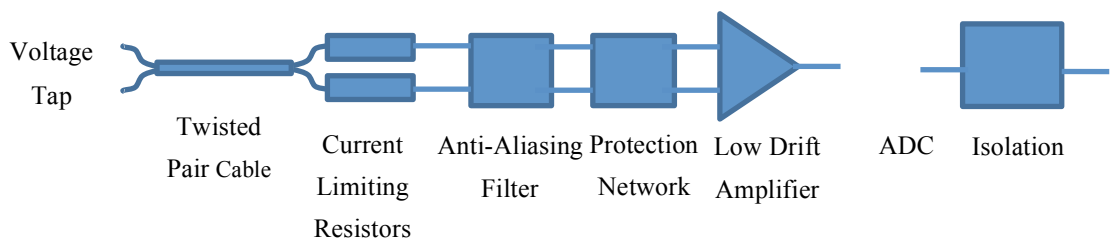

Figure 6.111. Analog Signal Processing Chain

### Tier-1

Tier-1 processing will be implemented in an FPGA to provide rapid real-time response. The Tier-1 processing will:

- Monitor digitized sensor signals via SPI from the

      a. Primary voltage taps
      b. Primary SCWS
      c. DCCT current sensors
      d. Primary copper bus current monitors
- Generate the primary extraction signal for the power supply system
- Relay system state and status information to Tier 2 and Tier 3 processors via the SPI bus

The FPGA firmware will implement a digital model of coupled solenoids including inductive coupling between coils, allowing tap voltages and currents to be compared to the digital model to detect the development of resistive zones while discriminating against other events.

Tier-1 hardware and software will be designed to provide continuous quench protection even in the event that Tier-2 and/or Tier-3 processors fail or are power cycled.

### *Tier-2*

Tier-2 processing will be implemented in analog quench detection hardware circuitry and DSP to provide flexibility while still providing sufficiently rapid real-time response to protect the magnets.

The Tier-2 processing will:

- Monitor analog threshold detection circuitry (analog quench detection) from
      a. Secondary voltage taps
      b. Secondary SCWS
      c. DCCT current sensors
      d. Secondary copper bus current monitors
- Monitor digitized sensor signals via the SPI from the
      e. Secondary voltage taps
      f. Secondary SCWS
      g. DCCT current sensors
      h. Secondary copper bus current monitors
- Implement a digital model of coupled solenoids including inductive coupling between coils, allowing tap voltages and currents to be compared to the digital model to detect the development of resistive zones while discriminating against other events
- Implement additional more sophisticated quench detection algorithms if required
- Generate the redundant extraction signal for the power supply system

- Acquire Tier 1 state and status information via the SPI bus and compare it to Tier 2 state and status information in order to identify and report failures at the earliest possible opportunity
- Relay system state and status information to Tier 2 processors via SPI bus

Tier-2 hardware and software will be designed to provide continuous quench protection even in the event that Tier 1 and/or Tier 3 processors fail or are power cycled.

### Tier-3

Tier-3 processing will be implemented in a standard PC to provide maximum processing power and programming flexibility.

Tier 3 will

- Not provide real-time capability
- Retrieve state and status information from Tier 1 and Tier 2 processors
- Provide long term monitoring of the state and health of the quench detection system
- Provide a user interface for local control of the quench detection system
- Archive state and health data for offline retrieval and processing
- Interface to other Mu2e subsystems

### Magnet System Monitoring

Temperature sensors splice voltage monitors and other sensors not used for quench detection will be monitored by Programmable Logic Controllers (PLCs).

### External Interfaces

The quench detection system will interface to the following three major subsystems in the Mu2e experiment:

- The power supply system
- The beam-line control system
- The experimental control system

### Power Supply System Interface

The quench detection system will receive digitized data via the SPI from the power supply system for:

- Power supply voltages
- Current Transducers

- Copper bus voltage drops
- Power supply status information

The quench detection system will generate:

- Redundant failsafe interlock signals for the power supply system
- Relay status information to the power supply system.

***Beam Line Control System Interface***

The quench detection system will relay state and status information to the ACNET based beam-line control system. Details of this interface are yet to be determined.

***Experimental Control System Interface***

The quench detection system will relay state and status information to the EPICS based experimental control system. Details of this interface are yet to be determined.

***System Identification***

Discriminating between induced voltages and quench related resistive voltages requires accurate measurements of the impedances of each solenoid and the couplings between them. Impedances and coupling will be measured in situ by modulating the current in each magnet in turn with slow (<1Hz) sine waves and recording the voltage induced in the magnet itself and on the other magnets in the system. Simulations indicate that inductances and coupling can be measured with less than one percent error using this technique.

***System Validation***

The quench detection system will be validated prior to commissioning using a real-time simulator implemented in an FPGA. The simulator will provide simulated signals for all of the quench protection sensors over the SPI bus and will allow simulated quenches to be initiated in any segment of the solenoids.

## 6.3.7   Magnetic Field Mapping

### *6.3.7.1   Requirements*

The field mapping system's functional, technical, operational, and ES&H requirements are detailed in reference [8]. There are several main elements of the field mapping system whose technical design is discussed here:

- A system to map the field within the production solenoid heat & radiation shield warm bore and into the TS1 collimator

- A system to map the field within the detector solenoid warm bore and into the TS5 collimator
- A system of in-situ magnetic field sensor arrays to monitor the axial field strength and gradients through each of the TS collimators (1, 3u, 3d, 5) during all powered operation
- A system of in-situ magnetic field sensors to monitor the field strength in the detector solenoid spectrometer region during DS powered operation
- A system to test the transport of low energy electrons from the production target through the collimators to the stopping target region.

### 6.3.7.2 Technical Design

#### PS/TS1 Field Mapping System Design

A mechanical design concept for the PS field mapper is shown in Figure 6.112. It consists of a drive system mounted on a stable base, which can translate a cantilevered beam into the HRS bore, at the end of which is a piezo motor-driven rotational positioning stage on which the Hall probe array is mounted. The piezo stage will be re-used for source positioning during the electron source test. All components within the bore will be non-magnetic to avoid perturbing the field and minimize use of metal parts to reduce forces on the moving system caused by eddy currents. In order to directly measure field strength and gradient and determine the magnetic center versus Z, the Hall probe array will include a pair of 2D probes (radial and axial) separated by a precisely known distance in Z (100 mm). Both probes will be located at the largest radius capable of passing through the limiting aperture of the bore and collimator.

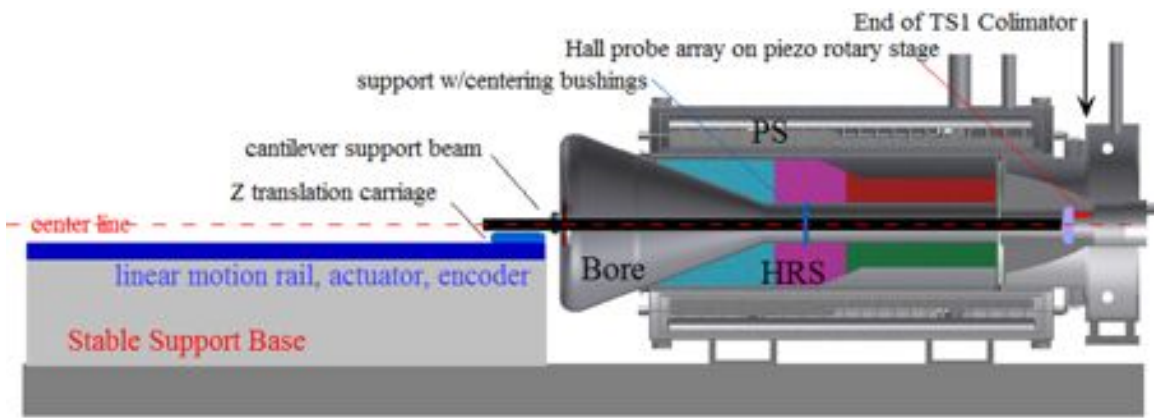

Figure 6.112. Conceptual mechanical design for PS Field Mapper.

The positioning system will be situated at the non-TS end of the PS and must be capable of moving the probes approximately 5 m in Z, from the flange of the HRS bore to the upstream end of the TS1 collimator (1 meter into the TS). The beam cantilever length may be less than this, and it can be made it more stable through use of a support bearing against the HRS bore tube just upstream of the target support ring. The Z translation mechanism will have a step size of 25 mm or less, and will encode the Z position to an accuracy and resolution better than 0.1 mm over the full Z range. Stage rotation will be continuous over 360 degrees in both directions; the angle encoder will have a resolution and accuracy of 0.1 milliradian or less. The mechanism will be adjustable to align and translate the probe array along the HRS bore geometric center. This is a challenge, and it is likely that survey of the mechanism system and tilt sensor measurements will be used to correct data for positioning errors.

Due to the large stray magnetic field downstream of PS (2 T to 0.1 T), the Z and angular motions will most likely be actuated by air motors; however, it may be an option to use electric motors, shielded or positioned in a low field region, using a non-magnetic cable to couple the motor to gears on the drive system.

The Hall array will use 2D probes to measure the axial field and gradient and radial field in relation to rotation angle for magnetic axis determination. These probes will have a full scale range of at least 5 T with 0.1% linearity and must be insensitive to the planar Hall effect. The SENIS 2D magnetic sensors meet these requirements and are the current best candidate probes for this purpose. Further evaluation of probe options will be made, as commercial Hall probe technology is expanding and quickly improving. Probes for PS mapping will be calibrated to 5 T using a Tevatron dipole in the Fermilab magnet test facility.

The probe motion control and data acquisition system will be based upon the architecture of the Fermilab Technical Division/Test & Instrumentation Department magnetic measurement system currently under development. A script-driven system running LabView, it will utilize compact Rio components for real time motion control, and PXI-controller instruments for digitization and acquisition of probe voltage data. Configuration and measurement data will be archived to a content management system database. The system will have an interface to the Mu2e power system and digitize the real time magnet current signals synchronously with Hall sensors and position encoders. The location of the control/daq hardware during measurements is intended to be as close as possible to the PS mapping transporter so that noise levels on probe signals and cable costs can be kept low. The actual location is not yet determined, although the Remote Handling Room or the proton beam line tunnel are candidate locations. The electronics may need to be shielded from the high stray field levels as shown in Figure 6.113. It may be that signal cabling must be routed all the way to a control room/electronics area; in any case, control cabling and some (NMR) signals will need to run there. Electronics rack space, routing and lengths of the cabling must therefore be determined; a preliminary estimate is 50 m long cables are needed.

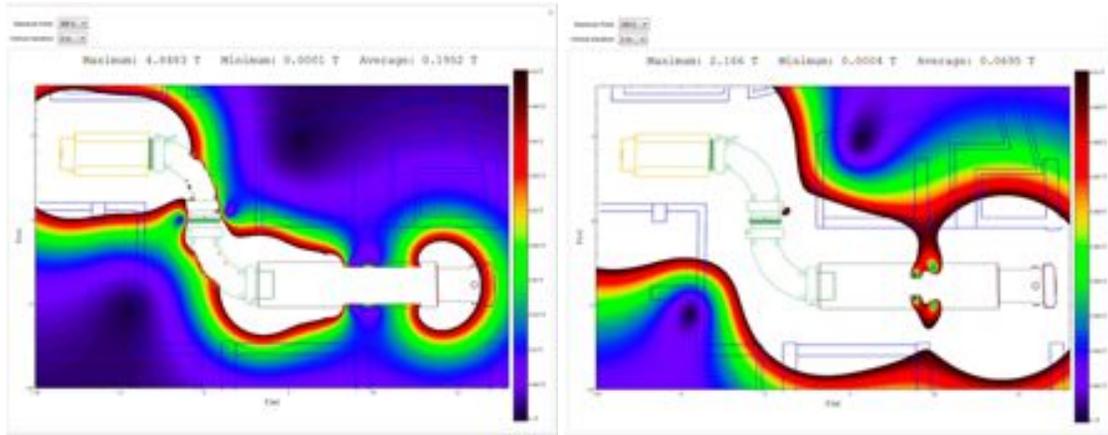

Figure 6.113. Map of stray magnetic field strength in Mu2e Hall [61] on the magnet vertical center below 500 G (left) at 1 m off center below 100G (right).

The PS field mapping will occur in two steps. First, a coarse-grained map will be made during the powered commissioning of the PS magnet system alone, to measure and compare with the expected field profile, determine the PS magnetic axis, and watch for any HRS magnetic anomalies. The PS field mapping system itself will be scrutinized for any problems under real, high field operating conditions. Subsequently, during operation in the "final magnetic configuration" with TS and DS on, fine-grained field maps of the interface region will be made with nominal operating currents and again with the TS/PS trim power supply operated at its high and low values. Following any adjustments to the TS magnets, which might occur after an electron source test, additional field mapping may be required, unless the in-situ Hall probe array measurements are deemed satisfactory.

### DS/TS5 Field Mapping System Design

A mechanical design concept for the DS field mapper is shown in Figure 6.114. It consists of a transport carriage supported by bearing blocks that will ride along the DS rails, which will also support the Mu2e detector train and therefore will be leveled and made parallel to a high degree of precision. The transport carriage will position Hall and NMR probes within the DS bore volume. Linear motion of the carriage along Z will be actuated by an air motor and gears that engage a belt anchored at the upstream and downstream ends of the DS warm bore. Another air motor and gearing system will actuate angular motion of a shaft to control the probe angular positions. All components within the bore will be non-magnetic to avoid perturbing the field and minimize use of metal parts to reduce forces on the moving system caused by eddy currents. Three 3D

Hall probes will be mounted along each of two "propeller" plates on opposite ends of the carriage, which rotate with a shaft centered on the DS bore. The probes will be positioned at radii of 0.3, 0.5, and 0.7 m on the upstream propeller, and 0.2, 0.4, and 0.6 m on the downstream propeller.  An extension of the rotating shaft will support a set of two 3D probes positioned 180 degrees apart at a radius of 0.1 m, capable of passing through the limiting aperture of the TS5 collimator (12.8 cm). 4 NMR probes will be mounted to the carriage frame, at a radius of 0.1 m, with active elements near the plane of each propeller; two will be range 4 probes for absolute field measurement from 0.34 to 1.05 T, and the others range 5 probes for field measurement from 0.7 to 2.1 T. All probes will be oriented to allow measurement of the axial + radial total field, and will utilize gradient coils to compensate for the large axial gradient in the DS1-3 regions, in order to get an NMR signal lock there.

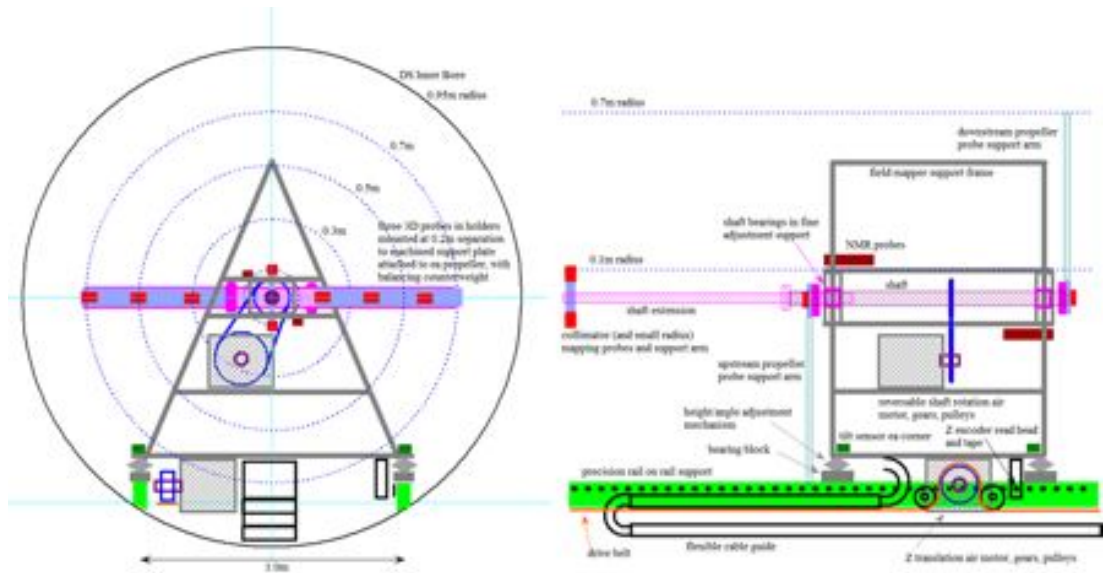

Figure 6.114. Schematic of the DS field mapper.

The carriage must be capable of moving the probes approximately 10 m in Z, between the downstream and the upstream ends of the DS, sending the shaft extension 1 meter into the TS5 collimator. The Z translation mechanism will have a step size of 25 mm or less, and will encode the Z position to better than 0.1 mm accuracy and resolution over the full Z range. Probes on the two propellers should be separated by an integral number of Z steps, with a goal being within 1.0 mm. Shaft rotation will be 360 degrees in both directions, and will have a step size of 5 degrees or less over that range; continuous motion would be preferred, if possible. The angle encoder will have a resolution and accuracy of at least 0.1 milliradian. The mechanism will be adjustable so that the probe array can be aligned and translated along the DS bore geometric center. The propeller rotation over 360 degrees should be planar, with a goal of maintaining planarity to within 0.1 milliradian; counterweights will be used to balance and eliminate torques on the shaft that may cause lateral positioning errors. Survey of the system will be used to adjust the mechanism and to make measurements that can be used to correct data for positioning errors; thus, probes must be visible from the downstream end of DS. Space will be limited for installation of the mapper at the downstream end of the DS to about 1 meter on the external rails. The actual carriage length is therefore constrained to be less than one meter; the shaft extension will be removable so that it can be installed after the carriage is in place on the rails.

Commercial Hall probe technology continues to improve, and costs may decline with time; a final choice of DS mapper probes can be made a couple of years in the future. The preliminary choice among 3D probes currently available is the SENIS F3A series probe, which has 0.1 degree orthogonality of Hall elements, compensation for planar Hall effect, relatively low noise, drift, and temperature coefficients, good linearity, and a small active element with a well-determined position. Probes for DS mapping will have field angles and response calibrated up to 2 T in the Fermilab magnet test facility.

The same motion control and data acquisition system will be used for both the PS and DS, so it will be possible to map only one of these magnets at a time (this is a value engineering choice). The interface between sensors and the control/daq system will be designed to accommodate both PS and DS sensor configurations. The DS will require the additional NMR readout instrument, which will be the Metrolab PT-2025 system and multiplexer. Probes will be of the type with built-in amplifier (model 1062) that are capable of driving a signal up to 100 m.

The DS field mapping will to occur in two steps. First, a "coarse" map (with reduced granularity of measurement points) will be made when only the DS is first powered for some length of time at nominal current; this will ensure that the DS field follows the expected profile and has no obvious flaws and will provide information on the magnetic axis. It also will be the opportunity to complete commissioning of the field mapping system itself, exercise the field map analysis tools, and check that all is ready and working for mapping the final magnetic configuration. The second step will be to map the fields in the operating configuration, with PS and TS powered, with fine granularity. Additionally, coarse-grained maps of the DS will be made at the physics calibration field settings. The scheduling of these final maps is an integration issue.

### TS In-Situ Field Monitor System Design

To monitor the axial field strength and to ensure negative gradient meets requirements in the TS straight sections, the in-situ field monitor system must be able to measure the field strength with sufficient precision at discrete positions with well-determined separation.

The requirement is specified at a radius of 15 cm, which results in the need to embed sensors into the collimators; measurements outside the collimators are not useful because proximity to the coils results in large ripple of the field there. In the TS1 and TS5 collimators, three linear arrays of Hall sensors are needed: one along the bottom at the vertical midplane, and one on each side of the horizontal midplane. In the TS3 collimators, due to the collimator design, only two arrays will be installed, one on each side of the horizontal midplane; as this collimator is rotatable by 180 degrees, the field profile at top and bottom can be measured by 90 degree rotation.

The design is to embed Hall sensors into the collimators during their construction. Each array will consist of to 10 "transverse" Hall probes spaced along the collimator length and oriented to measure axial field. An RTD to monitor and correct for temperature variations will be embedded with each array. The probes for each collimator will be internally ganged (e.g., on a printed circuit connection board at an end of the collimator) into two current chains to provide some redundancy in case of an open circuit failure during operation and to reduce the number of required feedthrough pins. The signal wires (2 per probe) and excitation current wires (2 per chain) will be brought through vacuum feedthrough connections on the TS inner bore adjacent to each collimator, then routed through the cryostat insulating vacuum space, along with other instrumentation wires there, through the chimney to the instrumentation trees at the feedboxes. For practical reasons of performing electrical quality checks and final collimator installation, intervening cable connectors will be needed between the feedthrough and the collimator arrays. To avoid cable interference with the collimator operation, a self-retracting reel could be utilized; custom design of non-magnetic, non-outgassing materials, such as brass or aluminum, will be needed. All signal and current wire pairs will be twisted and shielded to minimize electrical noise. The number of needed current sources, operating at typically 100 mA, will depend upon the total resistance of the chains, including the resistance of current-carrying wires and input resistance of probes, which is typically about 2 Ohms each. The readout plan is that Hall probe, RTD, and current chain precision shunt resistor voltage signals will be multiplexed and digitized by a high resolution voltmeter, as part of the solenoid instrumentation monitoring electronic system.

There are many potential vendors of single-axis Hall probes suitable for the purpose of in-situ field monitoring. These will be evaluated and selected at an appropriate time to allow them to be tested and ready for installation into collimators. Probes for use in TS will have their response calibrated up to 2 T in the Fermilab magnet test facility.

The collimators in each region have somewhat different geometries, materials and construction details, so the details of the Hall probe arrays are also slightly different. The TS1 collimator is fabricated from copper wedges in an epoxy impregnated structure. For this collimator, 8 slots for the Hall probes will be machined into three of the wedges (left, right, top) at 100 mm intervals, with a machined groove to route the wires toward the downstream end. These slots and grooves will be filled with RTV during the epoxy impregnation (at 200 C), then cleaned out for installation of the Hall probes (which would be damaged by the curing temperature). These Hall probes will therefore be positioned at about 17.5 cm radius. Figure 6.115 gives an illustration of the TS1 collimator and Hall probe array locations. Three arrays of 8 probes on 2 current chains will require 2x2+3x8x2 = 52 pins on feed-throughs and wires routed to electronics. Three RTDs will require additional 2+2x3=8 pins, for a total of 60 pins. Standard 19-pin connectors are preferred, so four connectors and feed-throughs are needed. Additional RTDs or Hall probes could be instrumented, or more current chains introduced, with the available extra pins.

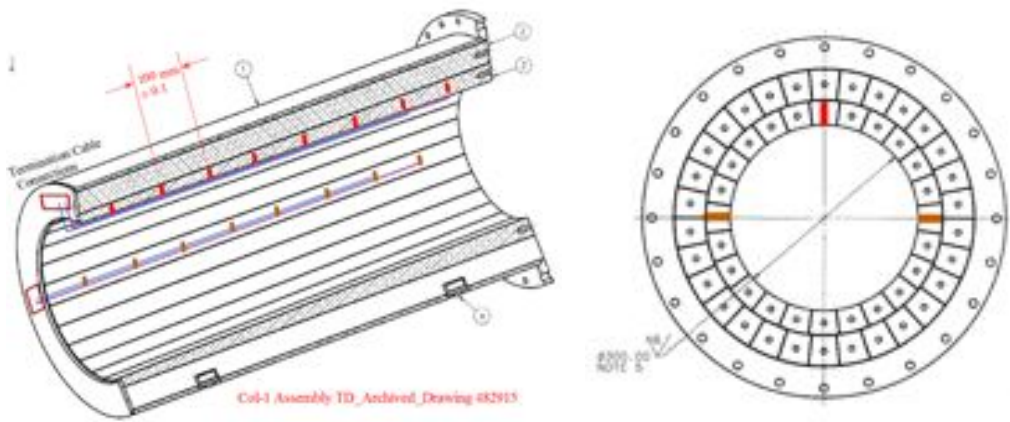

Figure 6.115. TS1 collimator assembly and face showing Hall probe slot locations and wire routing to termination point for ganging of current chains and making cable connections. (Array shown at top will actually be positioned at bottom, to avoid any risk of wires falling out into the beam region).

Next, TS3 upstream and downstream collimators, which are identical, will each have their own arrays of Hall probes installed, using the same design and a similar approach to that of TS1. Figure 6.116 shows the TS3 collimator assembly and Figure 6.117 shows the TS 3 collimator with Hall probe locations indicated. Because the copper segments are stacked, it would not be possible to install the probes after epoxy impregnation into interior elements; therefore, the only possible locations accessible and at or very close to the proper 15 cm radius are those shown. Here slots machined into the core slice faces would be blocked to prevent epoxy filling them during impregnation, then cleaned and the Hall probes installed afterward. Special consideration of the cable management is needed for the TS3 collimators, which are designed to rotate 180 degrees: A swiveling pulley or cable reel could work in this case, as shown in this figure. It is critical that this be designed to prevent any failure that could result in material such as cable or connector becoming an obstruction to the beam.

Two arrays of 10 probes on 2 current chains will require 2 x (2i) + 2 x 10 x (2v) = 44 pins on feedthroughs and wires routed to electronics for both of the TS3 collimators. Two RTDs will require an additional 2+2x2=6 pins, for a total of 50 pins. Standard 19-pin connectors are preferred, so three connectors and feed-throughs are needed for each TS3 collimator.

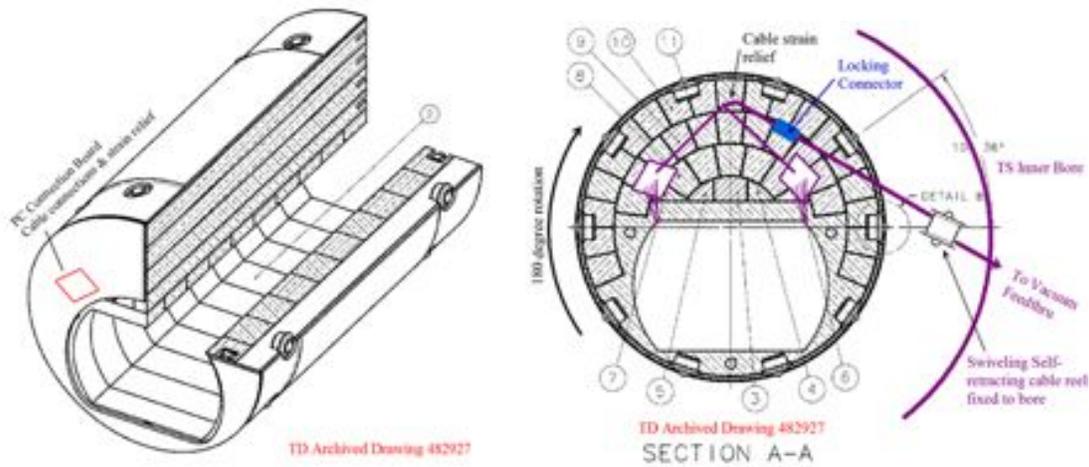

Figure 6.116. TS3 collimator assembly in rotating support shell, and face section showing construction from multiple elements. Wires will be routed to connection boards mounted on end of the collimator shell where cable connections and strain relief are made.

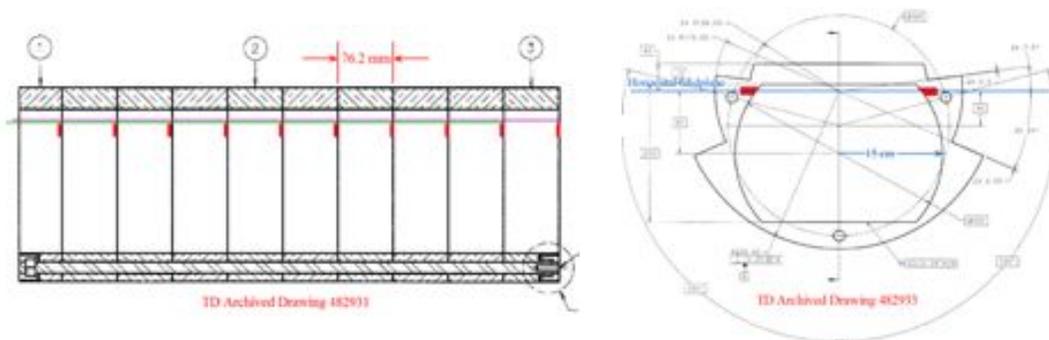

Figure 6.117. TS3 collimator core assembly showing location of 10 Hall probes in two parallel arrays at the horizontal midplane.

Last, the TS5 collimator has a very different construction, made up of polyethylene disks. As there is no epoxy impregnation of this device, the Hall probes will be installed during the collimator assembly. Figure 6.118 shows the TS5 collimator assembly with Hall probe locations indicated. Here again the design will utilize three linear arrays, along the left, right, and bottom as in TS1. Slots will be machined on the disk faces in which to position the probes, and along the outer radius to route the wires to connection blocks on the upstream end of the collimator housing; there, current chains and cable connections will be made up. In this case, with a collimator aperture of 128 mm radius, the probes can be positioned at the desired 150 mm radius. In order to keep the probes aligned, an alignment key must be added to the collimator disk design, and perhaps a feature to orient it within the housing to be discussed with Muon Beamline. Three arrays of 8 probes on 2 current chains will require 2x2+3x8x2 = 52 pins on feedthroughs and wires routed to electronics. Three RTDs will require additional 2+2x3=8 pins, for a total of 60 pins. Standard 19-pin connectors are preferred, so four connectors and feed-throughs are needed; additional RTDs or Hall probes could be instrumented, or more current chains introduced, with the available extra pins.

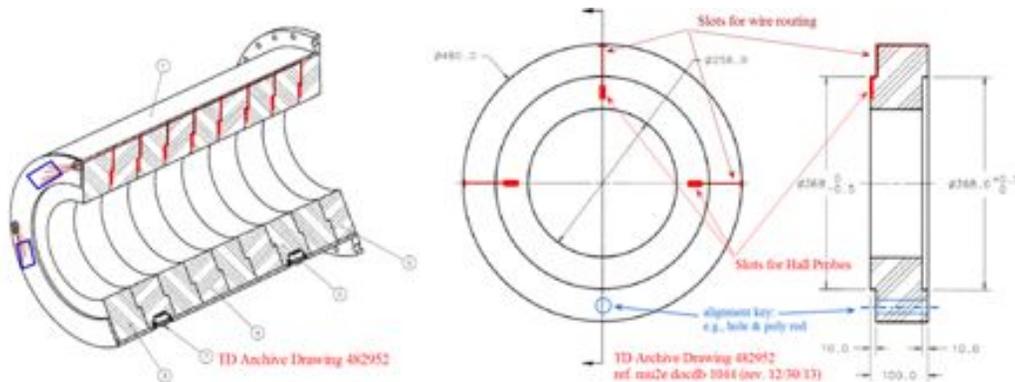

Figure 6.118. Concept for the TS5 collimator In-Situ Hall probe arrays. (Array shown at top will actually be positioned at bottom, to avoid any risk of wires falling out into the beam region).

### 6.3.7.3  DS In-Situ Field Monitor System Design

Two sets of NMR probes will be used to monitor the magnetic field strength in the DS uniform field region. One set will be positioned permanently in two locations on the inner bore of the DS, with range 4 and range 5 probes at each location to monitor the field under both normal operating and calibration modes. The locations will be at the upstream and downstream ends of the tracker uniform field region, probably near the top, or along one side of the bore. These probes must be installed and used during the DS field mapping. The RF and signal cables will be terminated with connectors that either couple to the field mapping NMR readout system, or to the IFB feedthroughs. The NMR readout system will continue to be used for the In-Situ probes during Mu2e experiment operations and calibrations.

A second set of four NMR probes (range 4 and 5) will be mounted to the upstream and downstream ends, near the outer radius, of the tracker frame. The axial and radial fields vary most rapidly with position in these locations; variation in the measured field can result from either motion of the tracker, or from actual DS field variations, so comparison with the probes fixed to the bore will identify the cause of a change in strength. Signal and RF cables will be routed to the IFB with the tracker cables and detector train; these probes cannot be monitored during the field mapping.

All probes will be of the same type used for the DS field mapper, oriented to allow measurement of the axial+radial total field; gradient compensation coils will not be needed here as field gradients are small enough in the uniform field region to acquire an NMR signal lock.

***Electron Source Test System Design***

The final step of the field mapping effort will take place after solenoid system commissioning is completed, but before installation of shielding around the solenoids takes place. This will be a test of low energy electron transport from the PS target, through the TS1, 3, and 5 collimators, to the DS stopping target region. It is planned to occur in two stages. First, a detector placed at the TSu/TSd will detect the transverse position where electrons traverse the TS3u collimator, as a function of the starting transverse position of a source at the production target location. Second, this detector will be removed, and a second detector at the DS stopping target location will detect the transverse position where electrons arrive, as a function of their starting position. For this test to occur, the pbar windows must not be installed.

The electron source could be a beta or electron conversion isotope with electron or positron energies up to a few MeV; further work is needed to identify the best source options. The source will be mounted to a motion stage that can move the source transversely: a combined rotary stage to be used for the PS field mapping with linear radial stage on top. Clearly the stages must operate in a ~5 T magnetic field, which will require they be made of non-magnetic materials with ceramic bearings and piezo motors. Such stages are commercially available, but further study is needed to define the requirements before specifying the desired solution, such as estimating desired range of transverse motion and source weight. To install the source, we envision two options. In order of preference: 1) The PS field map positioning system will be designed with a provision to mount a vacuum sealing flange to the cantilevered beam, so that the motion stage and source are moved to the target location, the seal is made and control signals are connected via a vacuum feedthrough connector on the flange. 2) The production target installation/replacement robot can be used to position, and the target support ring used to support, or anchor, electron source motion stage; a vacuum sealing flange with feedthrough connector for control signals is also needed.

The distance from the production target to the TSu/TSd interface is ~2.7m+$\pi$R$_{ts}$/2+1m, where R$_{ts}$=2.93m, or 8.3 m. From the TSu/TSd interface to DS stopping target is ~1m+ $\pi$R$_{ts}$/2+1.9m, or 7.5m; thus, electrons must travel 8.3m (15.8m) from the source to the TSu/TSd interface or stopping target. Figure 6.119 shows the range of electrons in air and helium at 1 atm. pressure 760 Torr. The 1 MeV range in helium is 0.5 g/cm$^2$, which, with a density 0.1786 g/cm$^3$, corresponds to a distance of 2.8 m. The range in air is only 0.42m. Therefore, the beam line space must be evacuated: 20 Torr of air gives a range of 15.8 m, so we should be well below that, say ~1 Torr or less. To evacuate the beam line and be able to provide control and readout signals will require vacuum blank-off flanges with feedthroughs at the PS and DS, and the ability to install a detector in the 0.5 cm gap at the TSu/TSd interface with a vacuum seal and signal feedthroughs also at that location.

For low energy electron detectors at both TS3 and DS locations, we envision using a plane of scintillation tiles with fiber light guides to SiPMs to detect scintillator hits. This technology has been developed at NIU/NICADD, where sufficient light can be generated with a 2 mm thick scintillator and SiPMs have been shown to work in high magnetic fields [49] - [52]. Details of the scintillator tile pixel size and number of channels is under study, in association with the source positioning requirements.

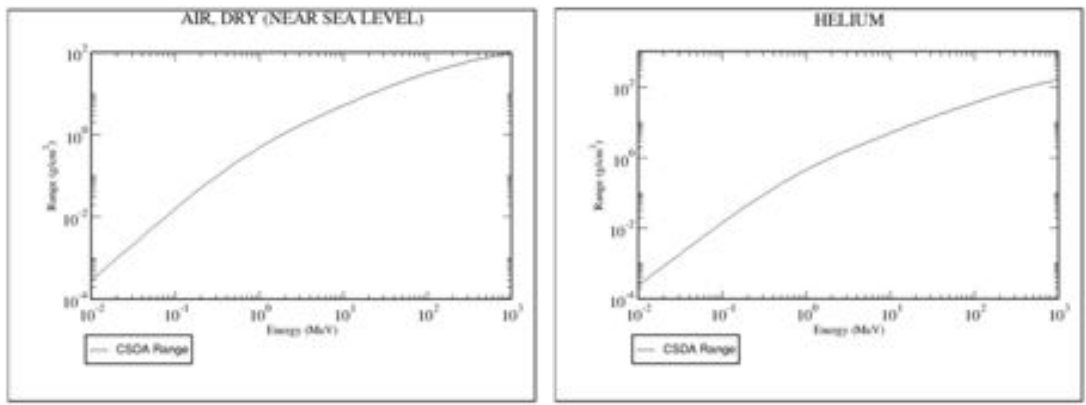

Figure 6.119. Range of electrons in air (left) and helium at 1 atm, versus energy, in the continuous slowing down approximation (CSDA).

# 6.4  Risks

There are several potential risks associated with the acquisition of a large, superconducting magnet system. These risks are understood and have documented and proven mitigation strategies. The primary identifiable risks are listed below with an estimate of their severity. A more detailed analysis for the solenoids along with the rest of the project is covered in the Mu2e Risk Analysis Documentation [53].

**Schedule delay (high)**

Unanticipated technical difficulties or overly aggressive schedule estimates can lead to schedule delays.  This is the highest risk to the solenoid system. The solenoids are on the project critical path, so any delay to the solenoid schedule will likely result in a delay in the entire project. This can be mitigated through careful analysis of vendor's capabilities prior to awarding contracts and monitoring vendor progress throughout the project. Built into the project plan is the ability to install and commission individual magnet components in the order that they arrive from the vendor.

**Interface problems (moderate)**

There is a moderate risk that solenoids built by different vendors might not fit together mechanically or magnetically. Detailed and carefully reviewed interface specifications, QA and QC requirements built into the vendor contracts are important elements in mitigating this risk.

**Insufficient testing of DS and/or PS at Vendor (moderate)**

The DS and PS are superconducting magnets with large stored magnetic energy.  A full test of the magnets requires test infrastructure comparable to the Fermilab Magnet Test facility.   Magnet fabrication vendors will very likely not have the capability to perform these tests at the fabrication site. This risk can be largely mitigated by designing the magnets with large operating margins in the superconductor and by monitoring the conductor vendor and fabrication vendor QA and QC to ensure that the magnet meets all of the design specifications. In the case of the PS and DS, the operating current/critical current is always less than 70 percent along the magnet load line.

**Production Solenoid must be installed through PS hatch using a large rented crane (high)**

The installation plan calls for the PS to the installed through the crane covered interior building hatch.  The PS and the heat and radiation shield must therefore be installed prior to TS installation.  If there is a schedule delay in PS installation, the TS may have to be installed first.  The alternative installation plan would be to install the PS through the PS hatch, using a large rented crane, at an additional cost to the project.  This risk can be mitigated through careful monitoring and coordination of the PS/TS fabrication schedule.

**Tevatron HTS leads cannot be used (moderate)**

The baseline plan is to use the HTS designed and fabricated for the Tevatron. One set of leads was used successfully in powering Tevatron circuits with currents on the order of 5 kA. The other leads have been in storage for the past 10 years. For Mu2e, the currents will go as high as 10 kA. Bench tests of two pairs of leads demonstrate that these leads can operate at this current. Others have only been tested to the nominal Tevatron operating current. This risk is mitigated by re-testing the Tevatron leads early in the project under the expected Mu2e conditions.

**PS conductor "first article" does not meet specifications (high)**

The superconducting cable for PS is acknowledged to be the most difficult conductor to fabricate. Approximately 200 meters of this cable has been made during a prototype manufacturing run. However, the vendor has not yet had the opportunity to fabricate a full production unit length. Because of the procurement of long lead items, specifically, the special Ni doped aluminum and the NbTi strand, it is expected to take 18 months to complete this first unit length. Considering the fabrication schedule for the PS solenoid, any significant delays beyond this 18 month period could delay the completion of the PS magnet and eventually cause this activity to be on the Mu2e project critical path. This can be mitigated partially by carefully monitoring the vendor's progress, providing timely analysis and test results for PS conductor QC particularly for the conductor final acceptance, and working with the vendor to prepare ahead of time the paperwork for the shipment of the conductor to the PS magnet fabrication vendor.

# 6.5 Quality Assurance

Quality Assurance (QA) plays a very important role in all phases of the Mu2e solenoid program: design, procurement, production and inspection/testing. Several of the deliverables will be "one-of-a-kind" objects critical to the experimental program and will be extremely difficult to service once the experiment is in operation; therefore, it is imperative that an effective QA program be implemented to guarantee that the solenoids meet their detailed specifications and operate reliably for the duration of the experimental program.

The Mu2e solenoid project, for both in-house and vendor-provided components, will be executed in a manner that is consistent with the Fermilab Integrated Quality Assurance (IQA) Program as well as the Fermilab Technical Division TD-2010 Quality Management Program. Details of the program will be described in the Mu2e QA plan [54]. Highlights of the QA implementation for the Mu2e solenoids are outlined below.

### 6.5.1 Document Control and Records Management

Control documents will be created and maintained at a level commensurate with the level of work performed. At the highest level, documents must go through version control, and change approval by project managers. These documents include engineering specifications, engineering drawings, travelers and operating procedures. The Fermilab Engineering Data Management System (EDMS) and the Mu2e document database will be used as the repository for these documents. Records of less formal documentation, such as presentations from in-house design meetings, vendor contacts and in-house reviews will also be stored in the document database.

It is important to keep records to provide evidence of the work performed. As much as possible, the records should be kept in an electronic format so all interested parties can access them. Records include documentation from all phases of the project: design, procurement, production and inspection/testing.

### 6.5.2 Functional and Interface Specifications

Engineering requirements will be conveyed to designers, procurement and production staff through functional and interface specification documents. Documents will specify the component performance requirements, mechanical tolerances and in some cases the required material properties. These documents will be formally controlled documents, subject to review and approval by project management.

### 6.5.3 Testing

Because of the high degree of reliability required of the solenoid components, a program of testing will be implemented throughout the production process. Specific to the solenoid fabrication, the following tests will be performed:

- Electrical measurements will be performed to assure that the coils have the proper electrical insulation to withstand the anticipated high voltage from coil to ground as well as between coil turns. Tests will be performed both at room temperature and in a cryogenic helium environment.
- Vendors will perform magnetic measurements after the coils have been fabricated and installed in cryostats but prior to shipment to Fermilab. Measurements will be compared to calculated values for acceptable field quality as defined by the functional specifications.
- Mechanical dimensions will be measured during fabrication to ensure that the magnet coils have the proper dimensions. This is critical due to the tight mechanical tolerances between components. Additional material tests will be performed to verify structural integrity at cryogenic temperatures.

### 6.5.4 Calibration and Monitoring

Instrumentation will be developed and/or commercially acquired for evaluating the electrical, magnetic and mechanical properties of the solenoid components. This instrumentation will be periodically calibrated. Calibration records will be kept along with the instrumentation data in an approved database. A subset of the instrumentation will be used post-production, as part of final vendor tests prior to shipment, as part of acceptance tests upon arrival at Fermilab, and during installation and commissioning.

### 6.5.5 Review Process

Progress on design, procurement and production will be periodically evaluated through a series of reviews. Reviews serve to verify that the project is ready to proceed to the next phase. Reviews will range from formal Mu2e project reviews to informal in-house reviews. Review presentations, review evaluations and review responses will be reported to the Mu2e project management. They will also be recorded in the Mu2e document database.

### 6.5.6 Vendor Responsibility

Key provisions of the Mu2e QA plan will be written into contracts with vendors. The level of QA implementation will be sized according to the project risk associated with the procurement. For procurements greater than $250k (the PS, DS, TS, power supplies, cryo feed boxes, etc.) vendors will be required to submit their own written QA plan.

## 6.6 Installation and Commissioning

### 6.6.1 Mu2e Building Layout

A detailed and fully integrated installation procedure will be developed for the Mu2e solenoids and supporting equipment as part of the final design process. The solenoids will be constructed elsewhere and installed in the Mu2e detector enclosure at the appropriate times. The hi-bay area above the detector enclosure will provide two independent 30 ton cranes for handling the components. These two cranes will have the capability to be coupled and controlled as one 60-ton crane for handling large components such as the solenoids and heat and radiation shield (HRS). The layout of the underground detector enclosure is shown in Figure 6.120.

### 6.6.2 Solenoid and Supporting System Installation

The installation plan is broken down into magnets and all other solenoid systems. The magnet installation sequence is constrained while the infrastructure can generally be worked on in parallel. The HRS is not part of any solenoid system but is mentioned in this section because it is an important constraint on the solenoid installation sequence. The DS solenoid will be installed through the main hatch, but the HRS, PS, TSu and TSd solenoids will be installed through the TS hatch shown in Figure 6.120.  Installing these heavy components through the TS hatch eliminates the need for a large outdoor crane rental and the potential for weather-related delays, but dictates the order of installation for those components.  All HRS and solenoid installation steps below include installation of the support frame required under each component.

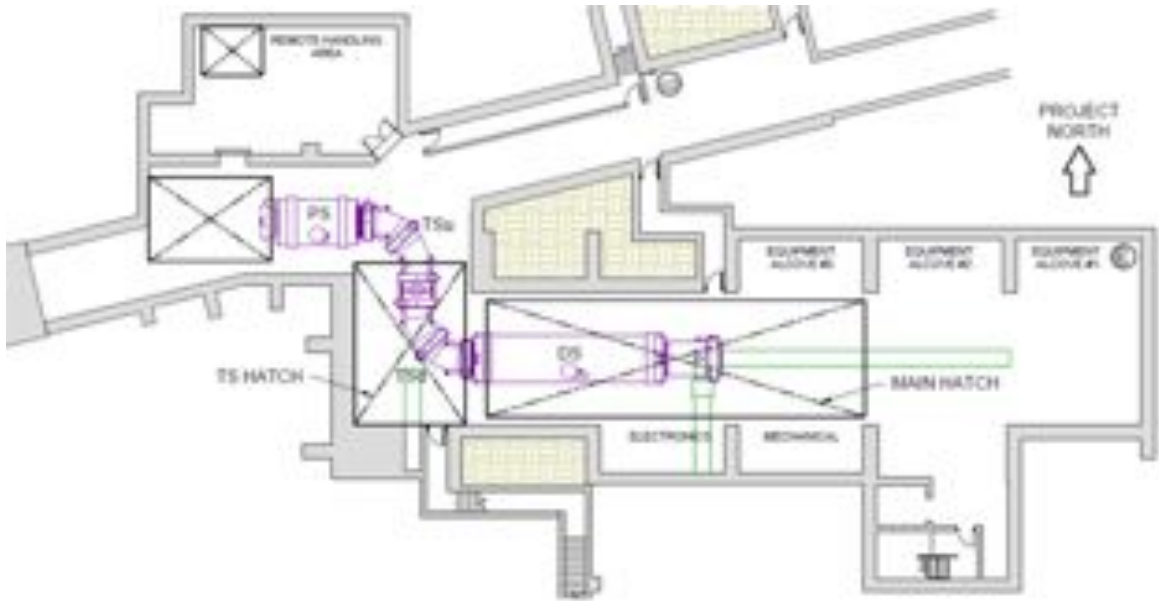

Figure 6.120. Layout of the Mu2e detector enclosure.

Figure 6.121 shows the floor pads the solenoids will mount to and the floor tracks required for transporting components to their final positions.  The HRS, PS, and TSu solenoids will be lowered onto the shaded north-bound tracks and hydraulically transported on rollers to the intersection of the shaded west-bound tracks.  The HRS will be transported to the west end of those tracks to allow the PS to be transported along the same installation path.  The HRS must then be installed into PS before TSu installation. TSu will follow the same installation path while TSd will require only an east-bound transport to engage the DS solenoid.  All transports will use the same hardware and equipment.  The initial installation plan is as follows.

- Install DS magnet (order flexible, install prior to TSd).
- Install HRS assembly.
- Install PS magnet.
- Install HRS into PS bore.
- Install TSu magnet.
- Install TSd magnet.

- Connect magnets to power, cryogenic, and vacuum systems

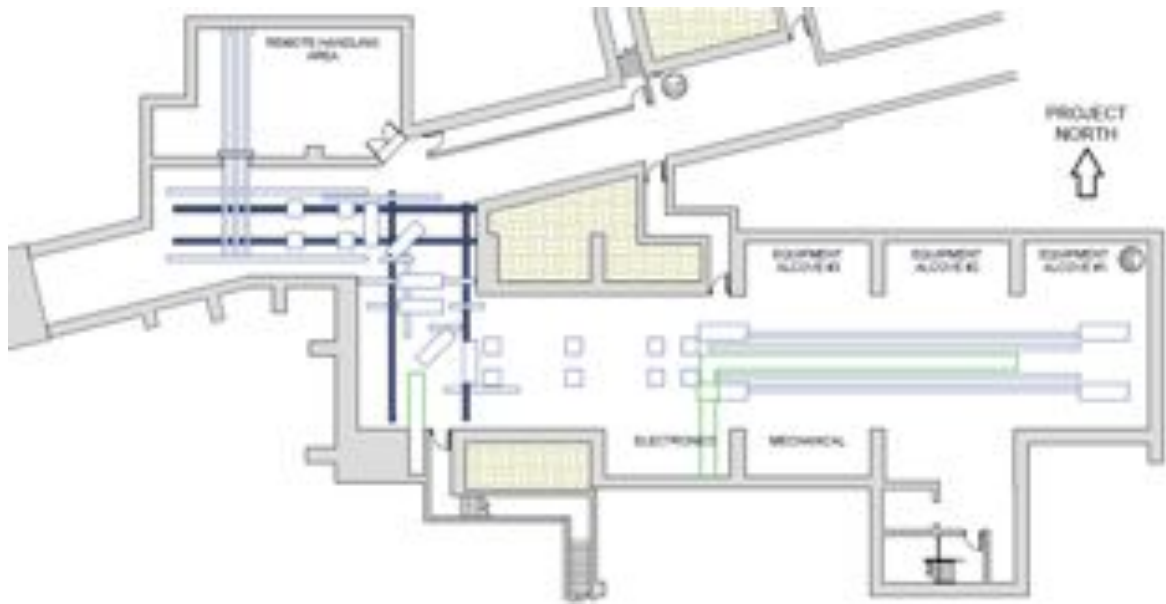

Figure 6.121. Layout of the Mu2e floor track system.

In order to connect the solenoids to the power, cryogenic, and vacuum systems, these other system installations, which have proceeded in parallel with solenoid installation, must also have been completed:

- Feed box installation
- Transfer line installation
- Magnet power system installation
- Cryogenic instrumentation and controls
- Insulating vacuum system installation

### 6.6.3   Solenoid System Commissioning

At this point in the installation sequence, commissioning of the solenoid power, cryogenic, vacuum, and control systems must occur.  This commissioning is required to ensure that the solenoids can be safely and reliably operated, that all ESH&Q and systems controls are in place, and that all equipment and personnel are protected during subsequent solenoid operation.  These steps complete the on-project phase of solenoid installation and commissioning:

- Commissioning of solenoid power, cryogenic, insulating vacuum, and control systems
- Preliminary field mapping
- Adjust magnet alignment

After the magnets have been aligned, the final steps in solenoid installation and commissioning are as follows:

- Make interconnects between TSu and PS, TSu and TSd, and TSd and DS
- Install PS Enclosure and DS blank flange to establish Muon Beamline vacuum.
- Perform electron source tests.
- Perform final TS coil adjustments.

# 7    Muon Beamline

## 7.1    Introduction

The fundamental goals of the muon beamline are to deliver a stopped muon rate of approximately $10^{10}$ per second to the muon stopping target, located in the Detector Solenoid (DS), and to reduce the background in the tracker, calorimeter and cosmic ray veto detectors to a level sufficient to achieve the desired experimental sensitivity.

It is important to identify all possible background sources and equip the muon beamline vacuum space with elements that can produce a muon beam with the requisite cleanliness while guiding negatively charged muons to the stopping target. The stopping target and the surrounding absorbers must be designed to maximize the capture of muons and transmission of conversion electrons to the detector while minimizing particles that can result in background.

The muon beamline is essentially a vacuum space, which serves as a free path for negatively charged muons within the desired momentum range. The muons spiral to the detector area along the streamline of the B-field (created by superconducting solenoid magnets). The S-shaped Transport Solenoid muon channel is surrounded with coils that form a toroidal B-field in the two curved sections and solenoidal fields in the three straight sections. The upstream toroidal field, with strategically placed collimators and absorber, filters the particle flux producing a momentum- and charge-selected muon beam, with good reduction in contamination from charged particles: $e^{\pm}$, $\mu^{+}$, $\pi^{\pm}$, p and neutral particles. A thin window resides in the central straight section of the muon beamline. The window stops antiprotons from reaching the muon stopping target and creating background. It also serves to separate the upstream and downstream muon beamline vacuum volumes to prevent radioactive ions or atoms from the production target from contaminating the detector solenoid volume. The muon stopping target consists of thin aluminum discs placed in the graded magnetic field region of the Detector Solenoid warm bore. The muons have a high probability of stopping in the muon stopping target. Muons not stopped in the target mostly bypass the detectors and are transported to the muon beam stop. Protons and neutrons originating from the muon capture process in the stopping target, final collimator and muon beam stop are attenuated by absorbers to minimize detector background rates. Stray particles produced in the vicinity of the primary production target and muon beamline are suppressed from impacting the cosmic ray veto detectors by substantial shielding surrounding the muon beamline.

The muon beamline Level 2 system is responsible for a variety of items essential to the effective transmission of the muon beam and performance of the Mu2e detectors. Muon beamline (WBS 5) is subdivided into eleven Level 3 sub-projects:

- WBS 5.1 Muon Beamline Project Management
- WBS 5.2 Muon Beamline Vacuum System
  - The end enclosures and the infrastructure to evacuate the upstream and downstream muon beamline
- WBS 5.3 Muon Beamline Collimator
  - The collimators and antiproton absorbers that reside inside the bore of the Transport Solenoids
- WBS 5.4 Upstream External Shielding
  - The muon beamline shielding external to the Production Solenoid (PS) and the Upstream Transport Solenoid (TSu)
- WBS 5.5 Muon Stopping Target
  - The muon stopping target and its support frame
- WBS 5.6 Muon Stopping Target Monitor
  - The muon stopping target monitor and associated infrastructure
- WBS 5.7 Detector Solenoid Internal Shielding
  - Shielding internal to the Detector Solenoid (DS) primarily intended to reduce background rates incident upon the tracker and calorimeter.
- WBS 5.8 Muon Beam Stop
  - Muon beam stop and supports
- WBS 5.9 Downstream External Shielding
  - The muon beamline shielding external to the Downstream Transport Solenoid (TSd) and the Detector Solenoid (DS).
- WBS 5.10 Detector Support and Installation System
  - The system designed to support the assembled detector train, including the muon stopping target, the DS Internal Shielding, tracker, calorimeter and muon beam stop. This system will facilitate relative alignment of the detectors, insertion into the DS bore, and extraction from the DS bore for detector servicing when required.
- WBS 5.11 Muon Beamline System Integration

The muon beamline WBS dictionary defines the technical objectives, scope, deliverables, relationships and assumptions for these various efforts [1]. The following sections provide a summary of the requirements and technical design for the muon beamline elements. Interfaces with the other Level 2 subsystems are documented in [2].

## 7.2    Vacuum System

The muon beamline vacuum system consists of two distinct vacuum volumes that are physically separated by a sealed vacuum window located between sections of the Transport Solenoids as indicated in Figure 7.1.

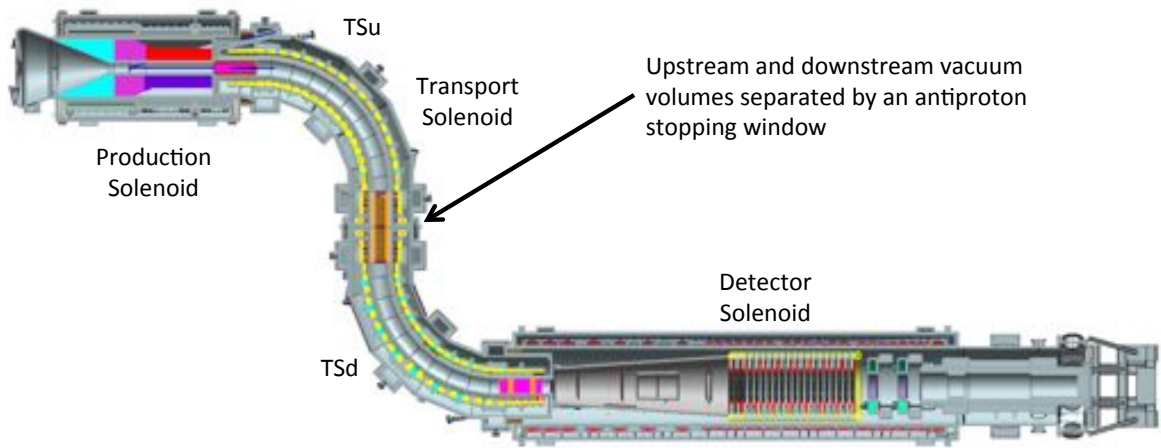

Figure 7.1. Overview of the Mu2e solenoids highlighting the location of the antiproton stopping window assembly located at the interface between the TSu and TSd.

This vacuum window also serves as an antiproton stopping window [3].  (Additional details can be found in Section 7.3.)  The upstream vacuum volume is defined by the stainless steel inner cryostat wall of the Production Solenoid, the Heat and Radiation Shield (HRS), plus the stainless steel inner cryostat wall of the upstream Transport Solenoid (TSu).  For convenience we denote this space as PS+TSu.  The downstream vacuum volume is defined by the stainless steel inner cryostat walls of the downstream portion of the TS (TSd), and the Detector Solenoid (DS).  For convenience we denote this space as TSd+DS.

Under normal operating conditions, the operating pressure shall be maintained at or below the required vacuum levels listed below.  The pressure in the PS+TSu is set by target lifetime criteria.  The requirement on the pressure in the TSd+DS is intended to minimize secondary interaction in the detector volume.  The time required to pump down from one atmosphere to the required pressure in both the upstream and downstream vacuum volumes is expected to be on the order of one week.

Both vacuum volumes will need to be prepared and cleaned using standard high vacuum cleaning practices in order to achieve the required vacuum levels.  However, neither the PS+TSu nor the TSd+DS have stringent operational cleanliness requirements. That means that minor oil contamination from the vacuum pumps will be acceptable.  These volumes do not require accelerator quality vacuums or integrated silicon chip manufacturing cleanliness standards in operation.

A performance analysis of the muon beamline vacuum system is provided in Section 7.2.2. High vacuum pumps used for the PS + TSu volume will be located in magnetic fields on the order of 500 Gauss. There is little concern that the magnetic field in the PS region will be adversely affected by magnetic material in the vacuum pumps located in the remote handling room. Therefore, these pumps and associated valves and instrumentation do not need to be non-magnetic as long as the equipment can function in those fringe fields and magnetically induced mechanical forces on those objects can be addressed. Because the pumps used for the PS + TSu volume will be located several meters from the PS vessel, the pumping speed (and therefore the ultimate system pressure) is greatly affected the line diameter and length. This line will need to be at least 0.6 meter in diameter.

Gas load contributions due to material outgassing, permeation across the vacuum boundary, leaks from the multiple vacuum windows, leaks from the many feed-throughs in the vacuum enclosures, and leaks from the detector straw tubes have been estimated and used to size the high vacuum pumps. Bake-out of the assembled vacuum system components is impractical. Therefore, outgassing rates used in this assessment are based on minimally cleaned, non-baked surfaces.

### 7.2.1 Requirements

The requirements for the muon beamline vacuum system are provided in [4] and can be summarized as follows:

- Required vacuum level:
  - PS + TSu; $\leq 1 \times 10^{-5}$ torr
  - DS + TSd; $\leq 1 \times 10^{-4}$ torr
- Required vacuum pump down time:
  - PS + TSu; approximately 100 hours
  - DS + TSd; approximately 100 hours
- Magnetic field anticipated in vicinity of high vacuum pumps:
  - PS + TSu; approximately 500 Gauss
  - DS + TSd; approximately 600 Gauss
- Magnetic field anticipated in vicinity of backing and roughing vacuum pumps:
  - PS + TSu; approximately 100 Gauss
  - DS + TSd; approximately 100 Gauss
- Required pre-operational cleanliness for vessels:
  - PS + TSu; standard high vacuum cleaning and degreasing
  - DS + TSd; standard high vacuum cleaning and degreasing
- Required operational cleanliness:

- o  PS + TSu; minimize, but not eliminate vacuum pump oil back-streaming
- o  DS + TSd; minimize, but not eliminate vacuum pump oil back-streaming
- • Appropriate windows and access ports must be provided as part of the enclosure for the Production Solenoid
- • Appropriate window, ports and feedthroughs must be provided as part of the enclosure for the Detector Solenoid
- • Vacuum lines (and other services) should be located to minimize penetrations in the detector shielding
  - o  Penetrations to the transport and detector solenoids should if at all possible come through the bottom, and if they cannot be through the bottom, then they should be away from the target region. In no cases should they penetrate the top [5].

Few, if any components of the vacuum system will be ASME code stamped since the ASME code applies to pressure vessels designed with an internal pressure of at least 15 psig. The Mu2e vacuum vessels will not be designed for such a high internal pressure. However, the external pressure criteria in the ASME Boiler and Pressure Vessel Code Section VIII, Division 1 will be applied to the vacuum vessels to ensure the vacuum vessels will not collapse. Applying the external pressure portions of the ASME pressure vessel code to the vacuum vessels is in accordance with the standards of the Fermilab Engineering Manual [6] and Fermilab ES&H Manual (FESHM) [7], particularly the Vacuum Vessel Safety chapter 5033 [8]. Some welding details critical to good vacuum practice (avoiding trapped volumes) do not meet the ASME code requirements for pressure vessels and therefore preclude code stamping of the vacuum vessels. Vacuum windows (thin membrane used for allowing the beam to enter the vessels, not optical windows) will meet FESHM chapter 5033.1 – Vacuum Window Safety [9].

An appropriately sized relief valve will be required on each vacuum volume to prevent the volume from being pressurized.

### 7.2.2   Technical Design

The vacuum system must be sized and located to address the above cited requirements. The anticipated gas loads are a key factor is this assessment. Outgassing data rates use published literature values for minimally prepared samples. Where values in the CDR version of the requirements document 1481 are more optimistic than published values, the published values will be adopted. Where insufficient published data exists (borated polyethylene) outgassing measurements will be made.

- • Gas Loads after one hour for the two vacuum volumes calculated by applying the material outgassing rate per unit area times the estimated area in each volume:
  - o  PS + TSu; 0.9468 torr-l/sec.
  - o  DS + TSd; 32.06 torr-l/sec.

| Material | $q_1$ Value in Mu2e doc 1481-v4[10] | $q_1$ Value in O'Hanlon [11] or Dayton[12] | Published $q_1$ Value converted |
|---|---|---|---|
| | Torr-liters /second-cm$^2$ | W/m$^2$ | Torr-liters /second-cm$^2$ |
| Stainless steel | 8.28e-9 | 2333e-7 | 1.75e-7 |
| Tungsten | 5.19e-9 | 5.33e-4 (Dayton) | 4.0 e-7 (Dayton) |
| Copper | 4.01e-8 | 5.33e-5 | 4.0 e-8 |
| Aluminum | 6.22e-9 | 84 e-7 | 6.3e-9 |
| Lead | 1e-7 | No value given | n/a |
| Kapton | 1.9e-7 | No value given | n/a |
| Polyethylene | 2.0e-7 | | 2.30 e-7 (Elsey) [13] |

***PS+TSu Vacuum***

Realistic Ultimate pressure Calculations of the PS+TSu considering Conductance Values:

Calculate the net pumping speed for the high vacuum mounted with a 10 meter long, 0.6 meter diameter tube using air as the gas at 22 C temperatures:

From O'Hanlon equation (3.21) [11], the conductance of an aperture is:

$C_{aperture}$ (l/s) = 11.6 * Area (in cm$^2$)
For a 0.6 meter diameter aperture, A = $\pi$/4 * (60$^2$ cm$^2$) = 2827 cm$^2$
$C_{aperture}$ (l/s) = 11.6 * 2827 = 32,798 l/s

From O'Hanlon equation (3.26) [11], the conductance of a long tube is:

$C_{long\ tube}$ (l/s) = 12.1 * D$^3$ (cm$^3$) / Length (in cm)
For a 0.5906 meter diameter tube, 10 meters long,
$C_{long\ tube}$ (l/s) = 12.1 * 59.06$^3$ (cm$^3$) / 10*100 (in cm) = 2492 l/s

Net pumps speed at the chamber for a 10,000 l/s pump (combined speed of a 20,000 l/s pump and a cold trap) and a 4949 l/s conductance is:

$1 / S_{net} = 1/ S_{pump} + 1/ C_{total}$

$1/S_{net} = 1/10000 + 1/2492$

$S_{net} = 1880.5 \ l/s$

For the PS+TSu at 1 hour:

$P = Q / S = $ pressure (torr) $= 5 \times 10^{-4}$ torr

Where,

$Q = $ gas load torr-liters per second $= 1$ torr-liters per second

$S = $ pump speed (net at the boundary of the chamber $= 1880 \ l/s$)

For the PS+TSu at 10 hours:

$P = Q / S = $ pressure (torr) $= 5 \times 10^{-5}$ torr

Where,

$Q = $ gas load torr-liters per second $= 0.1$ torr-liters per second

$S = $ pump speed (net at the boundary of the chamber $= 1880 \ l/s$)

For the PS+TSu at after many, many hours when outgassing is negligible:

$P = Q / S = $ pressure (torr) $= 5 \times 10^{-6}$ torr

Where,

$Q = $ gas load torr-liters per second $= 0.01$ torr-liters per second

$S = $ pump speed (net at the boundary of the chamber $= 1995 \ l/s$)

***DS+TSd vacuum***

Realistic calculations of the DS+TSd considering conductance values:

Calculate the net pumping speed for the high vacuum mounted with a 0.4 meter long short tube using air as the gas at 22 C temperatures:

From O'Hanlon equation (3.21) [11], the Conductance of an aperture is:

$C_{aperture} \ (l/s) = 11.6 * $ Area (in $cm^2$)

For a 0.4 meter diameter aperture, $A = \pi/4 * (40^2 \ cm^2) = 1256 \ cm^2$

$C_{aperture} \ (l/s) = 11.6 * 1256 = 14,576 \ l/s$

Note that the conductance of an aperture is the maximum conductance. As noted on O'Hanlon page 35 [11], any structure longer than an aperture will have a lower conductance.

From O'Hanlon equation (3.26) [11], the conductance of a short tube is:

$C_{short \ tube} \ (l/s) = 11.6 * a * $ Area (in $cm^2$)

For a 0.4 meter diameter aperture, $A = \pi/4 * (40^{\wedge 2}\ cm^2) = 1256\ cm^2$

The value "a" comes from a transmission probability (see table 3.1 in O'Hanlon [11]). For l/d = 1.0, "a" = 0.514

$C_{short\ tube}$ (l/s) = 11.6 * 0.514 * 1256 = 7,492  l/s

Combined conductance for a short tube and an aperture (see O'Hanlon eqn. 3.24) [11]:

$1 / C_{total} = 1/ C_{tube} + 1/ C_{aperture}$

$C_{total}$ = 4948.8 l/s

Net pumps speed at the chamber for a 6000 l/s pump and a 4949 l/s conductance is:

$1 / S_{net} = 1/ S_{pump} + 1/ C_{total}$

$1/S_{net}$ = 1/6000 + 1/4949

$S_{net}$ = 2712 l/s

Use two pumps in parallel, so the combined speed is 2 * 2712 l/s = 5424 l/s

For the DS+TSd at 1 hour:

$P = Q / S$ = pressure (torr) = $5.9 \times 10^{-3}$ torr

Where,

Q = gas load torr-liters per second = 32 torr-liters per second

S = pump speed (net at the boundary of the chamber = 5424 l/s)

For the DS+TSd at 10 hours:

$P = Q / S$ = pressure (torr) = $5.9 \times 10^{-4}$ torr

Where,

Q = gas load torr-liters per second = 3.2 torr-liters per second

S = pump speed (net at the boundary of the chamber = 5424 l/s)

For the DS+TSd after many, many hours when outgassing is negligible:

$P = Q / S$ = pressure (torr) = $5.9 \times 10^{-5}$ torr

Where,

Q = gas load torr-liters per second = 0.32 torr-liters per second

S = pump speed (net at the boundary of the chamber = 5424 l/s)

***Piping and Instrumentation Diagram***

A piping and instrumentation diagram for the muon beamline vacuum system is available in Figure 7.2. The high vacuum pumps are expected to be diffusion pumps. A view of the proposed implementation of the PS+TSu high vacuum system is shown in Figure 7.3.

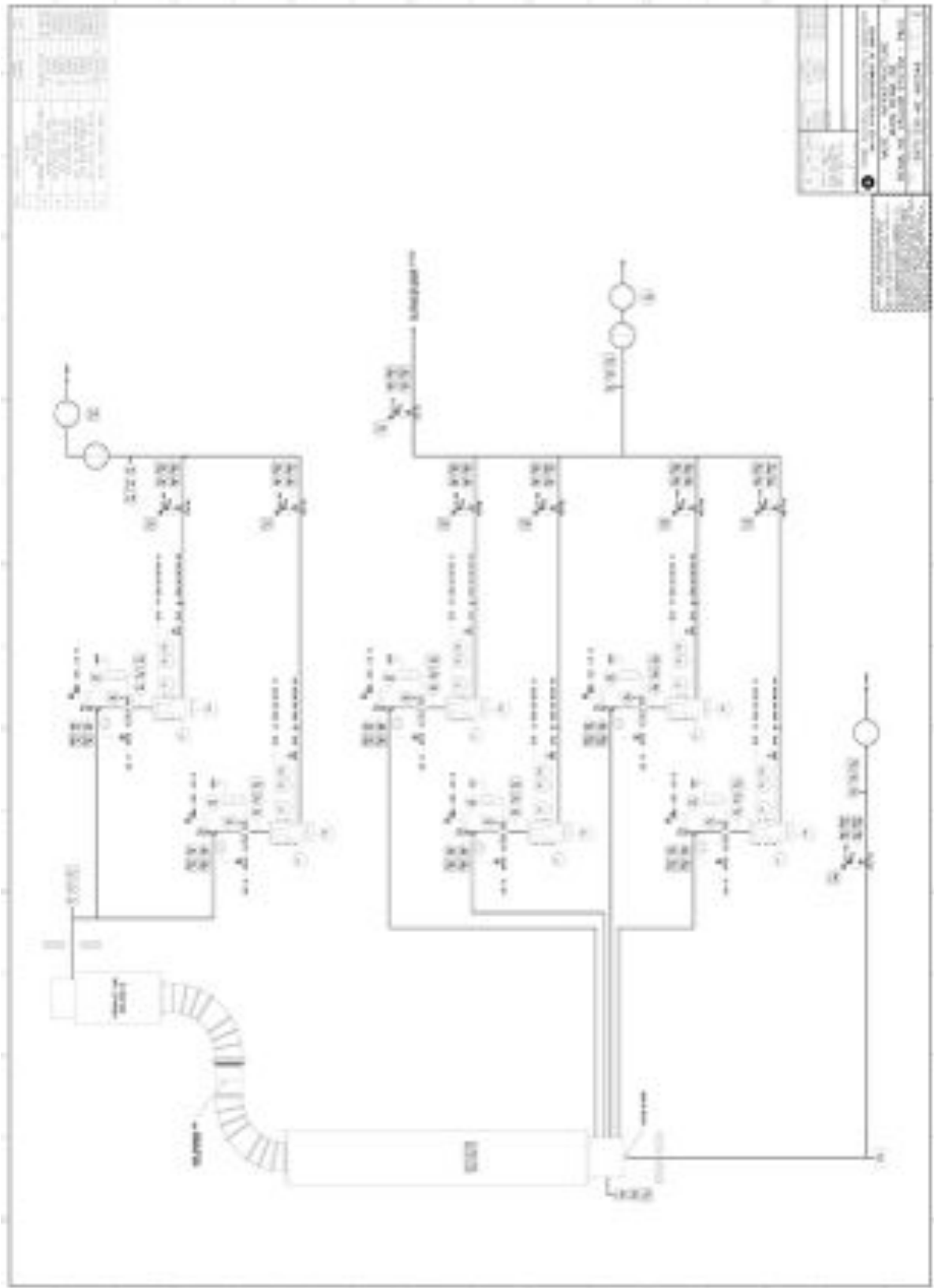

Figure 7.2. Vacuum System Piping and Instrumentation Diagram [14].

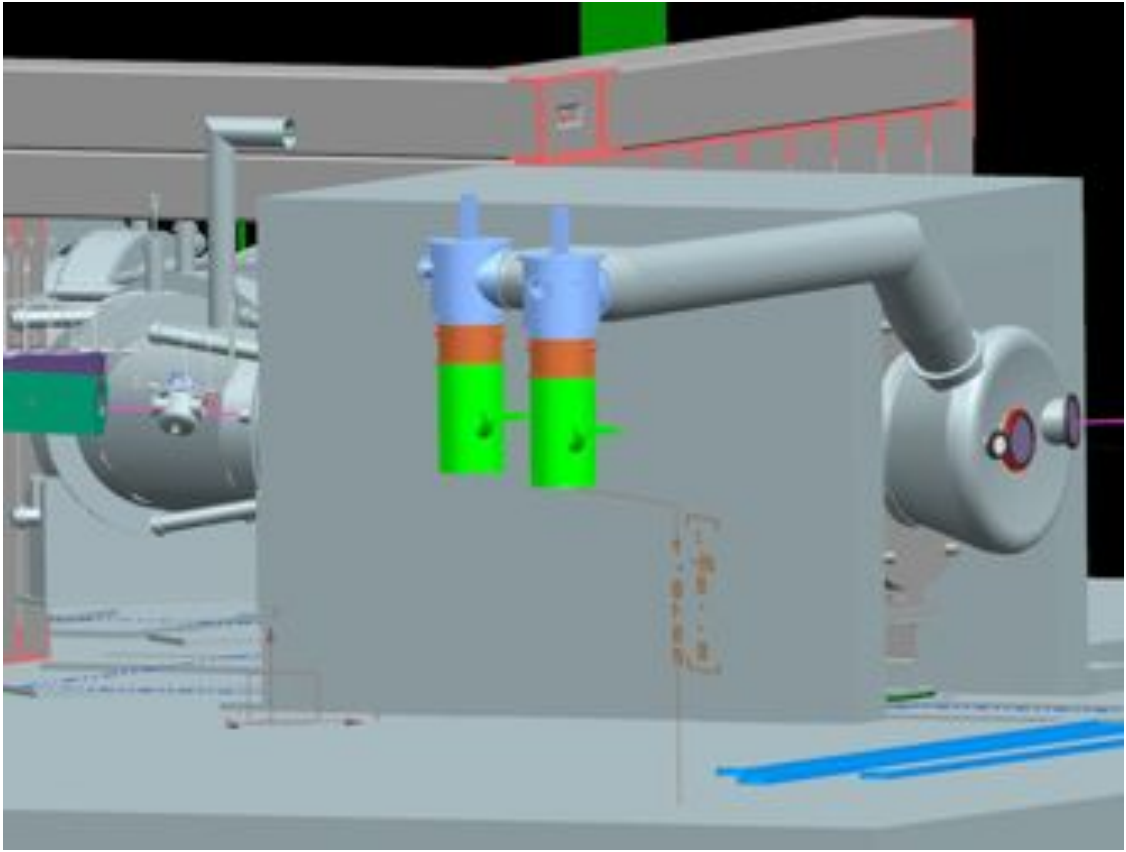

Figure 7.3. View of the production solenoid region (from the north west) illustrating the location of the PS+TSu high vacuum pumps and associated ductwork. The PS is surrounded by shielding. TSu is visible on the left hand side of this figure.

### *Vacuum Vessel Enclosures*

Two vacuum vessel assemblies are included in this system: the Production Solenoid enclosure, PSE and the Detector Solenoid enclosure. The DS enclosure is composed of two elements: the VPSP (vacuum pump spool piece) and IFB (Instrument Feedthrough Bulkhead).

Both vessel assemblies are stainless steel shells. Both require design and analysis to show they will restrain the vacuum loads in accordance with the safety standards (FESHM 5033) and applicable portions of the ASME boiler and pressure vessel code. The DS Enclosure will require a seal between the VPSP and the IFB. The radiation environment will need to be considered in the selection of the seals. Flange bolting will depend on the seal requirements.

### *Pump-Down Sequence*

In order to achieve the operating pressures a series of stages must be taken:

- Stage 1: Operate the roughing pumps to reduce the pressure to less than the allowable pressure across the antiproton stopping window assembly at the TSu/TSd interface. Use a control system to open and close valves on the PS and or DS side in order to avoid a differential pressure across the anti-proton stopping window.  Rough the diffusion pumps at the same time.

- Stage 2: After achieving a pressure less than the maximum allowable differential across the antiproton stopping window assembly at the TSu/TSd interface, operate the roughing pumps until the pressure stops decreasing.  Continue to evacuate the diffusion pumps.

- Stage 3: Energize the heaters on the diffusion pumps with the inlet valve closed. Cool down the cold traps. Once the diffusion pumps are up to temperature and the cold traps are ready, close the roughing line to the vessel and open the diffusion pump inlet valves.

***Vacuum System Controls***

- Vacuum equipment is nearly always on-off type control.  This applies to pump and valves.

- Since vacuum equipment uses on-off control, proportional, integral, and derivative control (P-I-D) algorithms are not required.

- Prior to selecting hardware for implementing the control logic, a written interlock description must be prepared to correctly identify the I/O and logic complexity.

- Controls for the vacuum system should be implemented in a manner similar to the implementation of other controls used on Mu2e.

- Programmable Logic Controllers should be selected to be compatible with PLC's that the Fermilab group assigned responsibility for long term operation of the experiment is already using and prepared to maintain for the duration of the experiment.

- Because the logic used for vacuum equipment is very simple, implementation on a PLC may not be essential, however, PLC usage allows a human-machine interface (HMI) that provides graphical displays similar to those on other systems.

## *7.2.3* **Risks**

The greatest risk related to the vacuum system is that the gas loads are found to exceed the estimates used to size the high vacuum pumping system.  The VPSP will be designed with additional hillside ports to accommodate additional pumping capacity if required.

The failure risk for windows designed to conform to FESHM 5033.1 should be low to moderate.  After completion of the final assembly and leak checking (which will occur after the magnetic field mapping [15]), the risk of additional leaks is expected to be low.

## *7.2.4* **Quality Assurance**

Quality Assurance will follow the Fermilab Quality Assurance Manual (QAM) [16] and the Requirements of the Fermilab Engineering Manual (FEM) [6].  The following process will be followed:

- o  Designs will be developed and documented in engineering notes and drawings by the originators.

o Engineering notes will be reviewed and checked by a second, competent individual.
o Drawings will follow the checking, approval and release process.
o Material requisitions will be prepared based on the sizing and requirements developed in the engineering notes and drawings.
o Purchasing will generate purchase orders based on the requisitions.
o Received and delivered material will be inspected, usually by the design engineer, for conformance to the purchase orders, requisitions, drawings and engineering note calculations.
o When formal quantitative inspections are required due to application of the "Graded Approach" chapter in the QAM, the need will be described in the applicable engineering note and the design engineer will arrange for qualified individuals to perform these inspections. Helium Mass Spectrometry leak testing of vacuum vessel shells is an example of a quantitative inspection necessary to confirm the maximum leak rates assumed in the design calculation are achieved and the actual gas load due to leaks is below that used in the system sizing.

### 7.2.5   Installation and Commissioning

The installation of most muon beamline vacuum system components will occur as part of the project. Final installation of the end cap enclosures as well as commissioning and initial operation will occur after CD-4 due to magnetic field mapping schedule constraints [15].

All vacuum pumps, electrical power and controls and cooling utilities will be installed to the greatest possible extent during the construction portion of the project. Systems will be tested, leak checked, and instrumentation calibrated. As much of the vacuum ductwork as possible will be installed and leak tested.

Spatial constraints in the Mu2e building will require the scheduling of the muon beamline vacuum components with other work within the building. The muon beamline vacuum components installation will be scheduled around work for systems that set the critical path.

## 7.3   Muon Beamline Collimators

The muon beamline collimators, in conjunction with the Transport Solenoids, will maximize the transport of negative muons that stop in the stopping target in the Detector Solenoid (DS) and strongly suppress all other particles delivered to the DS. As illustrated in Figure 7.4, there are four collimators located within the warm bore of the muon beamline transport solenoids.

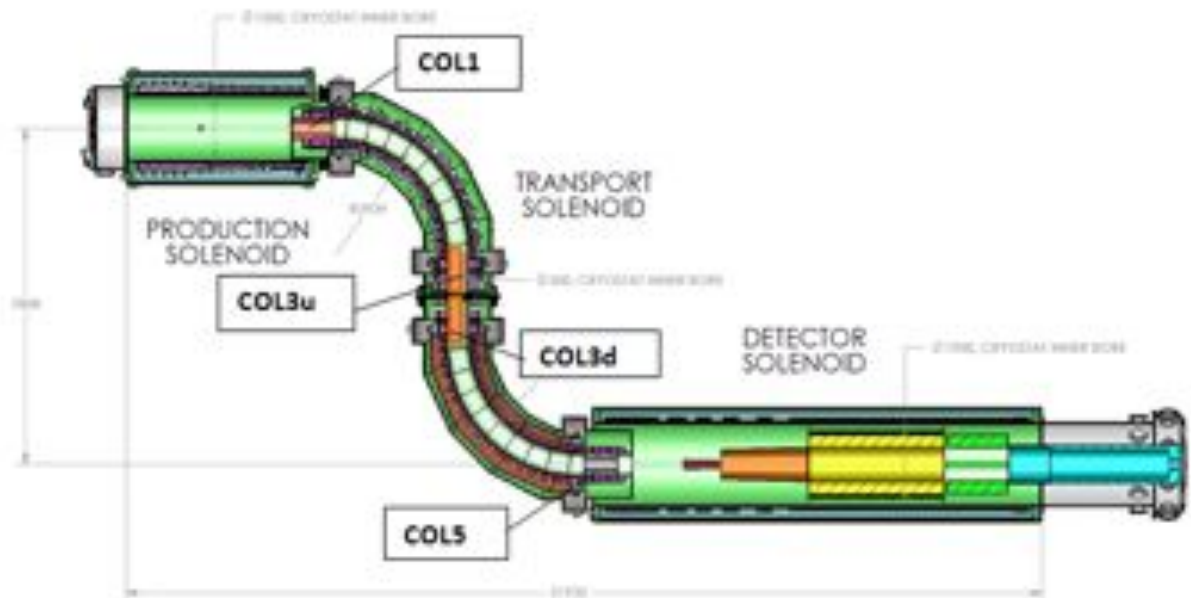

Figure 7.4. Overview of the muon beamline highlighting the locations of the collimators positioned within the warm bore of the Mu2e Transport Solenoid cryostats.

The muon beamline collimators filter the beam as it passes through the Transport Solenoids (TSu and TSd), selecting muons of the desired charge and momentum range to optimize the probability of their capture in the muon stopping target. This selection is achieved by a vertical offset of the apertures in the two collimators (COL3u and COL3d) with respect to the horizontal center plane to take advantage of "curvature drift" in the two transport solenoid bends formed by the two toroidal sections (see Figure 7.5). In this curved B-field negatively charged particles are deflected upward in the first curved section, and pass through the offset apertures in COL3u and COL3d, and are deflected back onto the nominal beam line in the second curved section [17][18]. Since the drift direction depends on the charge, positively charged particles are driven downward in the first curved section and stopped in COL3u and COL3d. The extent of a particle's vertical displacement at the TSu/TSd interface is momentum dependent, thus the size and location of the vertical aperture will also select the beam by momentum. COL3u and COL3d are designed so that they can be rotated on demand to efficiently transport positive muons in that rotated configuration to facilitate the collection of detector calibration data.

In addition to the role of beam filter the upstream collimator, COL1, protects the first several TS coils from radiation originating in the production target.

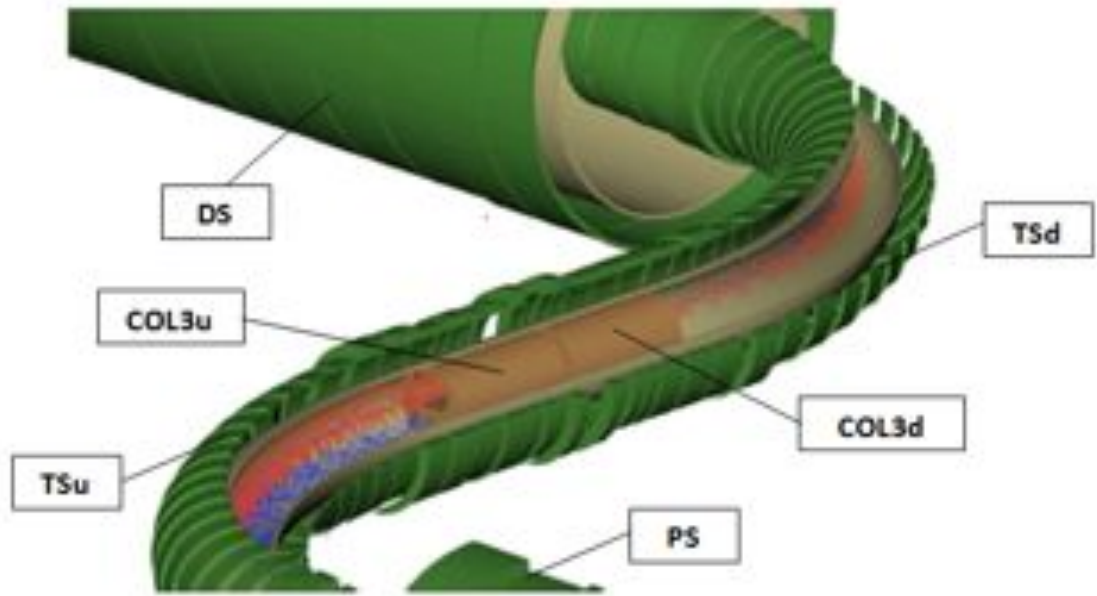

Figure 7.5. Cutaway view of the Transport Solenoids and the collimators COL3u and COL3d showing the offset apertures in those collimators. The upper spiraling negative muons (red) pass through the aperture while the positive muons (blue) are stopped by these collimators.

The detector area is required to be shielded from neutron background. This is the primary purpose for introducing COL5 in the Mu2e muon beamline and as seen in Figure 7.4, COL5 is located near the upstream end of the DS.

A thin window assembly is installed between COL3u and COL3d to absorb antiprotons in the beam and to separate the muon beamline vacuum space into independent upstream and downstream volume. More recent studies have indicated that an additional thin absorber will also be required to sufficiently suppress backgrounds from high energy antiprotons [19]. This window is proposed to be mounted near the upstream end of COL1, and is referred to here as the COL1 antiproton window.

### 7.3.1   Requirements

The Mu2e system of collimators, in conjunction with the TS magnet system, will maximize the transport of negative muons that stop in the stopping target in the Detector Solenoid (DS) and strongly suppress all other particles delivered to the DS. As described in [20], the collimators and the antiproton stopping windows are required to:

- Charge and momentum select particles, preferentially muons from the beam
- Absorb antiprotons at the stopping windows
- Inhibit migration of radioactive molecules downstream into the detector solenoid
- Protect the detectors from neutron background and low momentum muons which might not hit the stopping target

- Reduce particle debris incident upon TSu originated from the production target

The primary purpose of the collimators is to select charge and momentum by exploiting the drift, perpendicular to the plane of the S-shaped TS magnet, in opposite directions by positive and negative charges. In the upstream curved solenoid portion, as shown in Figure 7.5, the spiraling positive (blue) and negative (red) muons are deflected downwards and upwards respectively, by amounts that depend on their momentum. The vertical displacement midway through the S-shape TS magnet is:

$$D[m] = -\frac{Q}{e}\frac{\pi}{0.6B[T]}\frac{P_L^2 + 0.5P_T^2}{P_L[GeV/c]}$$

where: $e$ is the magnitude of the charge of the electron, and $P_L$ ($P_T$) is the component of the momentum along (transverse) the magnetic field. The particles also execute fast gyrations with radius:

$$a[m] = \frac{P_T[GeV/c]}{0.3B[T]}$$

Using the above equations the collimators can be designed to filter the beam favoring low-momentum particles. The collimators need to be optimized to remove or heavily suppress electrons above 100 MeV. By using rotatable collimators (COL3u and COL3d) with offset apertures the selection cited above can be achieved. These collimators will allow passage of the low momentum negative particles, including the desired low momentum negative (or positive – if the collimators are rotated 180 degrees) muons and strongly suppress positives (or negatives). The particle trajectories are re-centered on the solenoid axis by the second curved section of the TS.

Antiprotons (p-bar), if allowed to continue on to the muon stopping target, would produce a serious physics background [21]. There are several ways to decrease p-bar induced backgrounds. The Mu2e approach is to introduce thin windows in the transport solenoid region. The windows need to be thick enough to reduce the probability for p-bar passage through the windows but also thin enough to not substantially decrease the muon yield. Secondary particles emerging from the antiproton windows need to be filtered out. These requirements will drive the locations and sizes of the antiproton windows. The antiproton windows should not be placed further downstream than the TSu/TSd interface. Requiring an antiproton window near the upstream end of COL1 and at TSu/TSd interface sandwiched between COL3u and COL3d will satisfy the above requirements [19].

Radioactive molecules will be generated in the production solenoid area. These molecules can be long-lived and with time they can migrate downstream to the detector area producing undesirable physics background. To prevent this migration the upstream PS+TSu vacuum volume will be isolated from the downstream TSd+DS vacuum volume by the antiproton stopping window assembly at the TSu/TSd interface, which will be a vacuum tight separator as well. This vacuum window assembly needs to suppress molecule diffusion through the window material.

### 7.3.2   Technical Design

The collimators will be supported by the inner walls of the TS cryostats and the flanges of those cryostats. The collimators will each be assembled into individual housings that will have rollers to facilitate insertion inside the Transport Solenoid cryostat bores and flanges to connect the collimator housings to the TS cryostats thus providing support for the collimators. The insertion clearance between the outside diameter of each collimator housing and the inner TS cryostat wall should be 1-2 mm radially.

There are two types of collimator assembly configurations, the stationary collimators (COL1 and COL5), and the rotatable collimators (COL3u and COL3d). Several parameters of these collimators are summarized in

Table 7.1.

#### *Stationary Collimators---COL1 and COL5*
COL1 and COL5 are illustrated in Figure 7.6 and Figure 7.7. The specified collimator material will be assembled into a stainless steel container that serve as the housing for the collimator.

Table 7.1.  Main parameters of the muon beamline collimators.

| Collimator Name | Inner Bore Radius (cm) | Outer Radius (cm) | Length (cm) | Material | Mass (kg) |
|---|---|---|---|---|---|
| COL1 | 15.0 | 24.0 | 100.0 | Cu , C | 1020 |
| COL3u | 15.0/10.0 | 24.0 | 80.0 | Cu | 1012 |
| COL3d | 15.0/10.0 | 24.0 | 80.0 | Cu | 1012 |
| COL5 | 12.8 | 24.0 | 100.0 | Polyethylene | 248 |

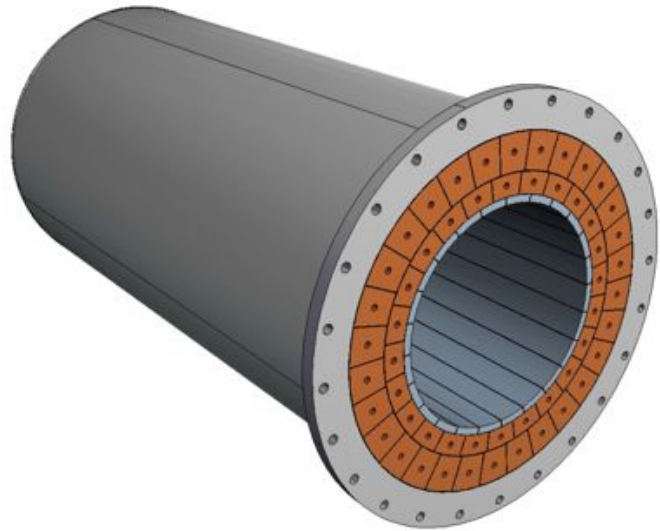

Figure 7.6. View of the COL1 as seen from the upstream end, illustrating the individual copper pieces inserted into the stainless steel housing.  Note the segmented innermost graphite layer.

The preliminary design of COL1 uses trapezoid shape copper bars to form the collimator body and the segmented graphite shells for the innermost layer of the collimator.  Each copper bar and graphite shell will be individually wrapped in glass insulating tape and vacuum impregnated after insertion inside the stainless steel housing.  This procedure will mechanically and thermally connect all parts together and eliminate voids inside the collimator body, which is important for performance in the muon beamline vacuum.

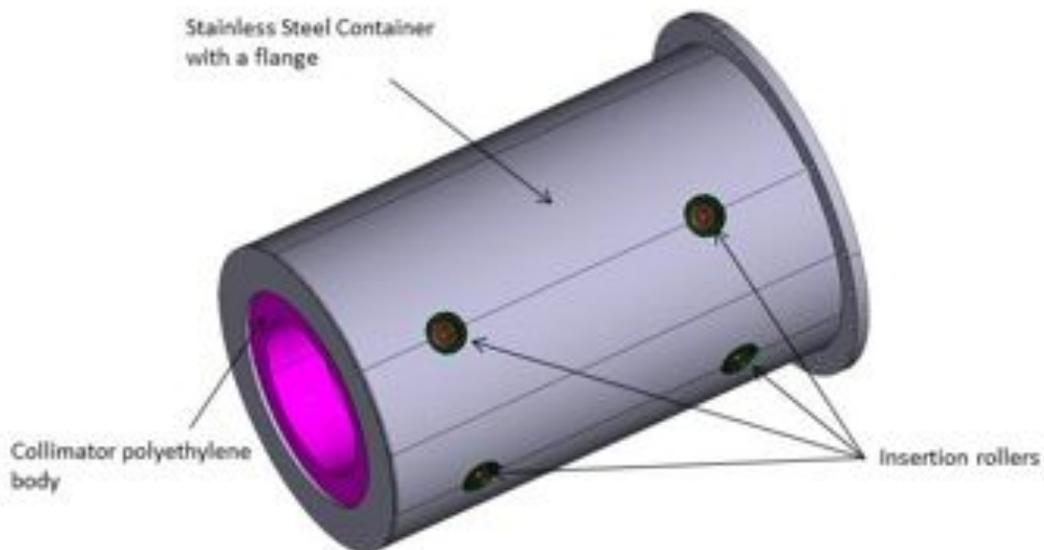

Figure 7.7. View of COL5 as seen from below, highlighting the insertion rollers embedded in the stainless steel collimator housing.

COL1 serves dual functions. Besides filtering high momentum particles it also reduces the particle flux into the TSu coils. The unwanted energy from this particle debris that is deposited in the TSu coils could ultimately lower the quench limit of the superconducting solenoid magnet. COL1 must be optimized together with any additional shielding material that may introduced into the bore of the TSu to reduce the peak power density deposited into the TS coils to less than 5 μW/g at the highest proton beam intensity. Current estimates of the anticipated power density at the TS1 coil are about 0.5 μW/g [22].

Although the heat loads in COL1 are quite modest in comparison to those in the Production Solenoid [22], the COL1 design needs to accommodate them. Heat generated in the inner collimator parts due to incident radiation will be distributed through the body of the collimator because of the epoxy impregnation of all parts that will improve thermal connections between part surfaces in vacuum conditions. The intent is to rely upon only passive cooling for COL1.

The electrical isolation of the individual trapezoid shape copper bars will reduce the eddy-currents during a quench of the solenoid and reduce the axial magnetic forces as well. This collimator will be located inside the warm bore of the TSu and the flange of the stainless steel housing will be tightly bolted to the flange of the upstream end of the TSu cryostat to transfer axial forces.

Recent efforts to further suppress antiproton transmission [19] incorporated an additional 20cm long arc of graphite at the downstream end of COL1 as illustrated in Figure 7.8. This additional material extends inward to a radius of 12cm, from Z=-325.4cm to Z=-305.4cm in the Standard Mu2e Coordinate System [23], covering the lowest 120 degrees in phi.

COL5 will have a polyethylene body, which will serve as an effective shield for the neutron backgrounds. Polyethylene rings will be machined and installed inside the stainless steel housing. Four insertion rollers are integrated into the bottom side of the collimator housing to facilitate insertion of the collimator into the TSd solenoid warm bore. The view of the collimator in Figure 7.7 is rotated by ~90 degrees around the beamline axis to show these insertion rollers.

The materials employed in the fabrication of the collimators must withstand the anticipated radiation dose. The absorbed dose in the COL1 is estimated to be ~1.1MGy/yr. [24]. The absorbed dose in the vicinity of COL5 is estimated to be ~0.2MGy/yr. [25]. At this level of radiation dose, the polyethylene in COL5 is expected to perform acceptably [26].

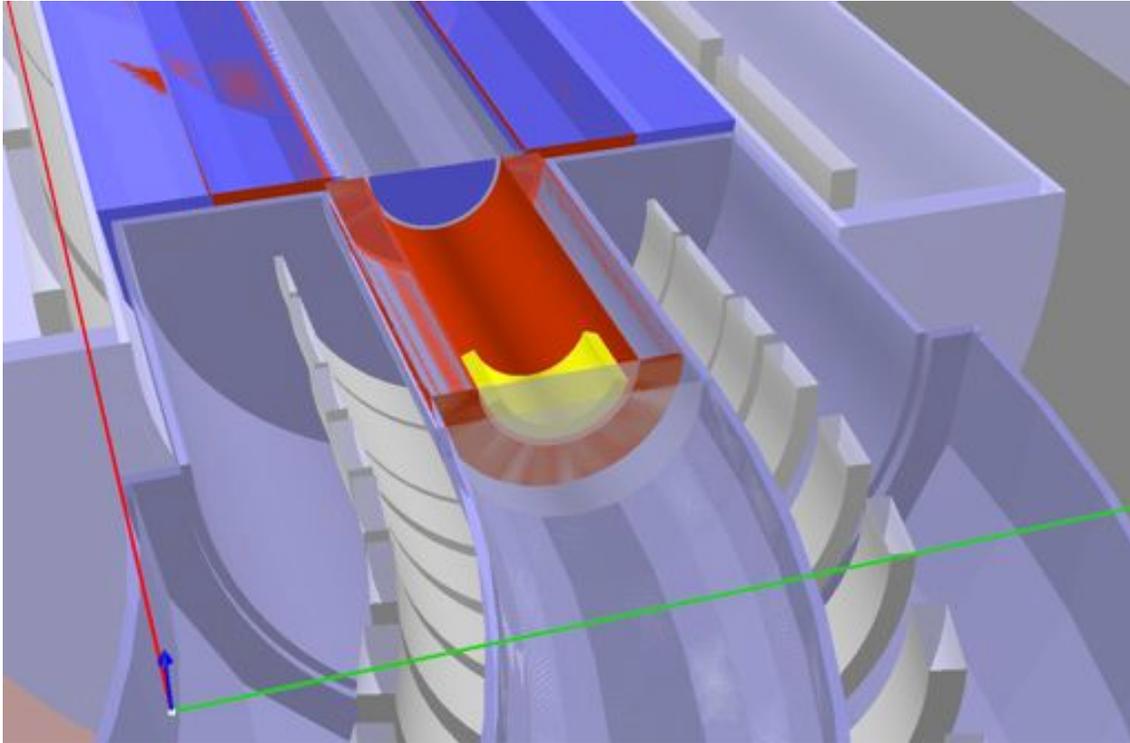

Figure 7.8. Cutaway view of the PS and the upstream end of TSu, showing COL1 (in red and yellow) with the additional graphite insert at the downstream end of COL1 highlighted in yellow.

***Rotatable Collimators---COL3u and COL3d***

The copper body of the COL3u and COL3d (see Figure 7.9 and Figure 7.10) will be fabricated in a manner very similar to COL1, and the bodies will be assembled and epoxy impregnated inside inner housings. The inner housing will be inserted inside an outer stainless steel container and supported by two bearings. The inner housing and collimator body will be capable of rotating with respect to the outer container.

A large gear will be attached to the inner housing of each of these collimators to provide azimuthal rotation of the inner housing containing the collimator with respect to the outer container that will be attached to the TS cryostat by the container flange. This rotation will be driven by electrical servomotors (see Figure 7.11) and will be electronically controlled for proper positioning.

There is 54 mm space between cryostat flanges of the TSu and TSd magnets and part of this space is devoted to a drive mechanism for a small spur gear that will engage the large gear attached to the collimator inner housing. A servomotor will drive the small gear via a drive shaft that passes through a vacuum feedthrough in the TS cryostat.

A cut-away view of COL3u, antiproton stopping window assembly and COL3d inserted into the Transport Solenoids is shown in Figure 7.11.

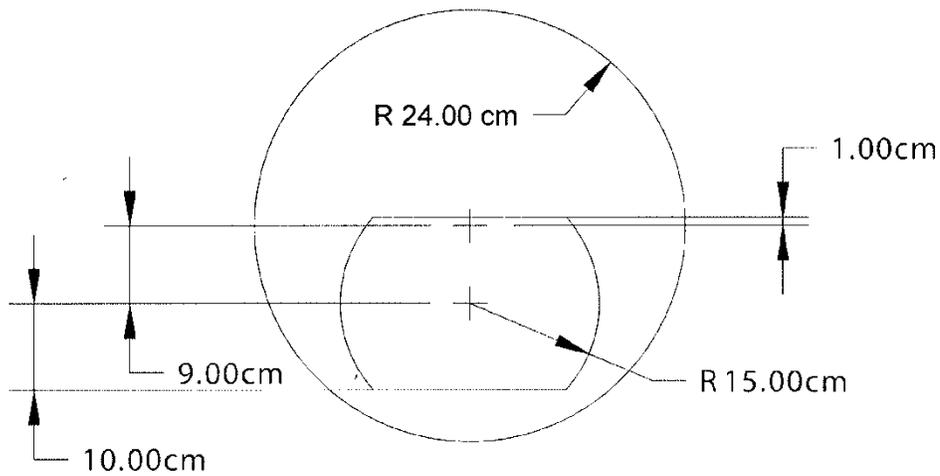

Figure 7.9. Cross-sectional view of the central collimators COL3u and COL3d showing the dimensions of the aperture (with the aperture oriented in the position to transmit positive muons).

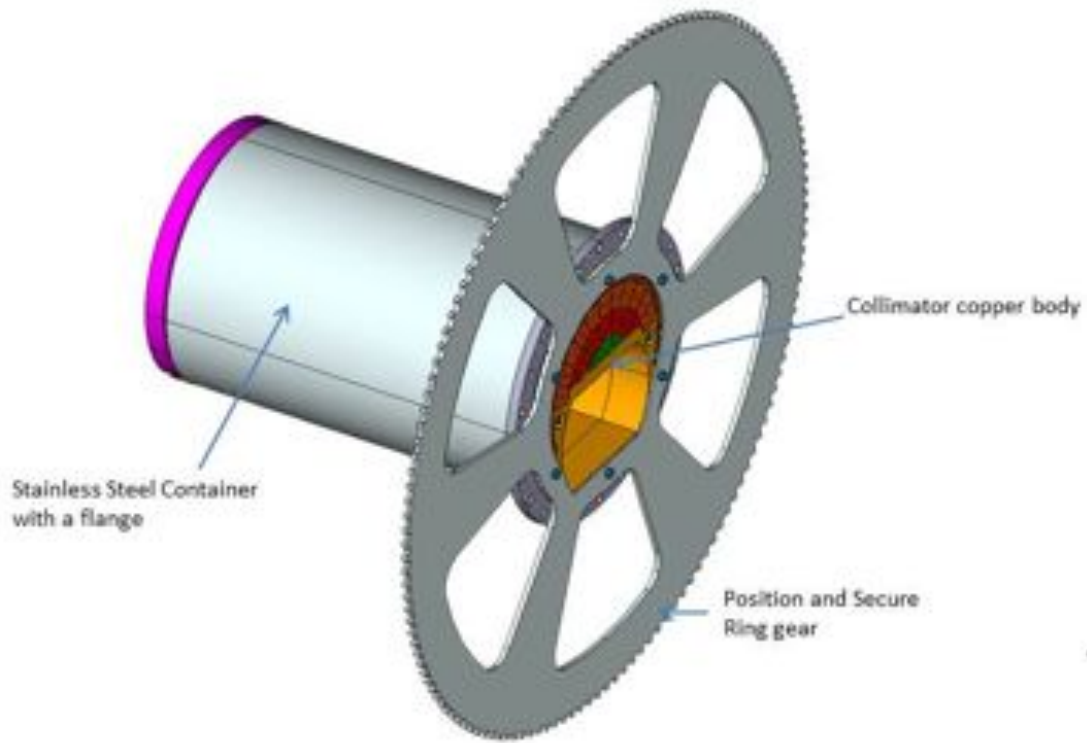

Figure 7.10. Preliminary design of the collimators COL3u/COL3d.

The materials employed in the fabrication of the collimators must withstand the anticipated radiation dose. The absorbed dose in the vicinity of COL3u and COL3d is estimated to be ~0.5MGy/yr. [25].

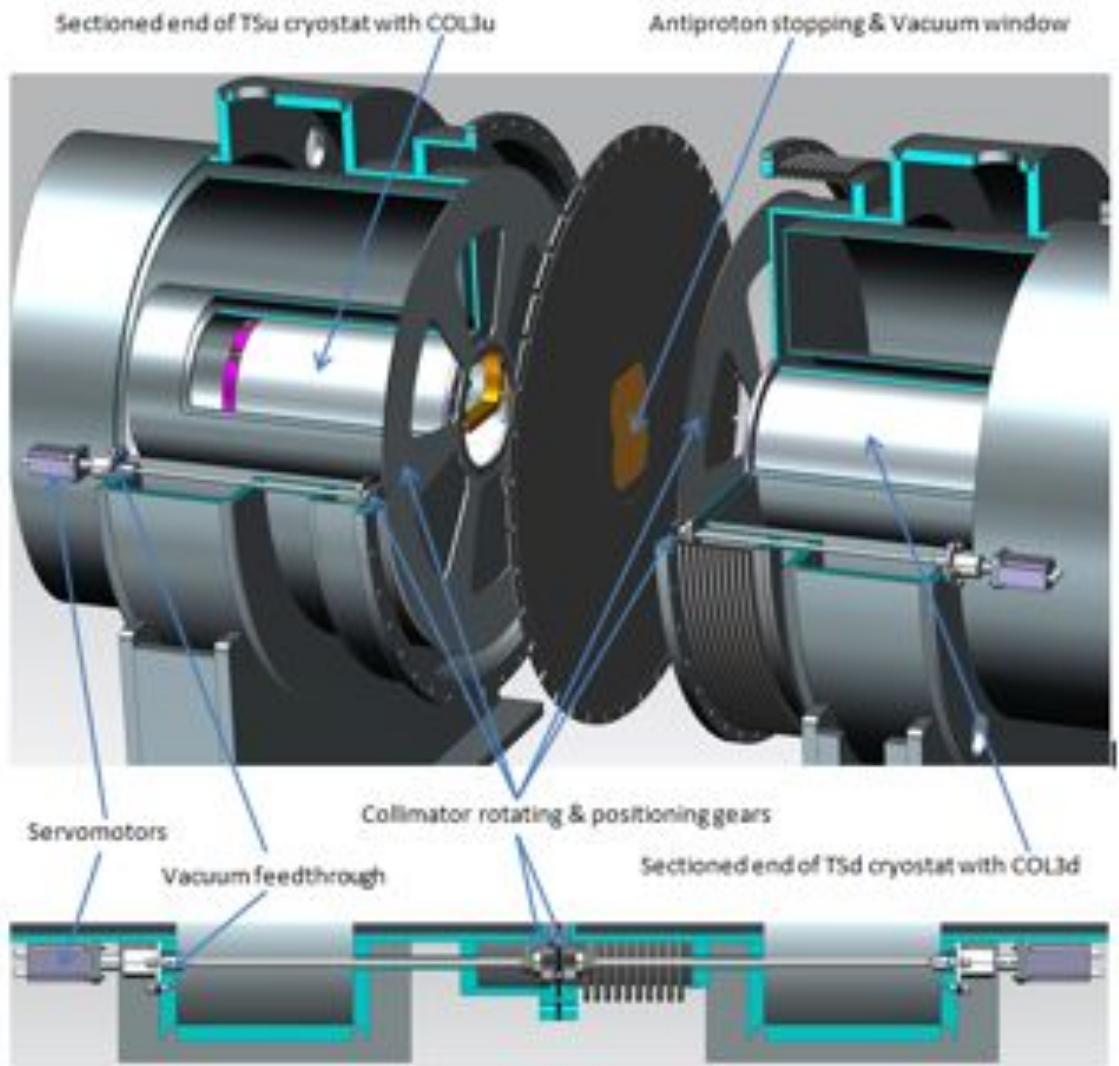

Figure 7.11. Preliminary design of the collimators COL3u and COL3d assembly inside TS cryostat shown in an exploded view above and in an assembled cross section view. Also note the antiproton stopping window assembly.

### *Antiproton windows*

The anti-proton stopping window assembly positioned between COL3u and COL3d is shown in Figure 7.11. The profile of that window was optimized in a study reported in [3]. As cited above, an additional thin absorber to be mounted near the upstream end of COL1 will also be required to sufficiently suppress backgrounds from high energy antiprotons [19].

The thin window at the TSu/TSd interface is currently specified as a beryllium plate whose thickness is 200μm below the nominal aperture (with the collimators oriented to transport negative beam). The thickness of the window increases to a maximum of 1301μm above the top of the collimator aperture [19]. This beryllium window will be installed in the center of the stainless steel support disc (see Figure 7.11). Vacuum tight seals of the window are provided by a flange. The operating pressure inside the bore upstream of the antiproton stopping window assembly at the TSu/TSd interface is specified to be $10^{-5}$ torr [4]. Downstream of that antiproton stopping window assembly the operating pressure is specified to be $10^{-4}$ torr [4].

The version of the COL1 antiproton absorber, as represented in the most recent simulations [19], is a 350 μm thick kapton window. This window is centered on the muon beamline and positioned just 1mm upstream of COL1. The diameter of this window is 30cm. Due to the radiation levels expected in this region, a more robust material should be adopted in the final configuration. Aluminum and beryllium are candidates for the window material. The 1mm gap is also likely to be modified since it serves to baffle the pump down of the TSu region.

### 7.3.3   Risks

The most significant risk in the collimator system are likely to be operational risks associated with the rotation of the collimators (problems during rotation or loss of orientation), the insertion of the antiproton stopping window assembly at the TSu/TSd interface, or the difficulties with the COL1 antiproton absorber. The collimator rotation mechanism will be tested in advance to ensure that it is fully functional. The collimators will be designed to ensure that the orientation can be re-established, and the possibility of ports to directly inspect the orientation will also be investigated.

### 7.3.4   Quality Assurance

Quality Assurance will follow the Fermilab Quality Assurance Manual (QAM) [16] and conform to the requirements of the Fermilab Engineering Manual (FEM) [6], similar to the procedures outlined in Section 7.2.4.

### 7.3.5   Installation and Commissioning

#### Collimators

The collimators will be installed inside the TSu and TSd cryostat bores prior to delivery of the TS assemblies to the Mu2e detector hall. Magnetic field mapping instrumentation [15] will need to be installed in the collimators prior to insertion of the collimators into the Transport Solenoid.

Radiation exposure and activation of collimator components will need to be monitored before servicing or removal of components. Studies of anticipated dose [25] and dose rate [27] are underway.

Access to the collimators after the muon beamline is established would be a long and complicated process. As a consequence, the collimators will be designed so that access for service is not required during the lifetime of the experiment.

***Antiproton Windows***
Due to plans for magnetic field mapping, the antiproton windows are not expected to be installed until the field mappings are complete [15].

Installation and removal of the antiproton stopping window assembly at the TSu/TSd interface requires opening the TSu/TSd bellows connection by about 5mm and extracting the window assembly from between solenoids. The window support disc will have seals on each side located close to external diameter for the vacuum sealing to the TS cryostat flanges. This antiproton stopping window assembly (which also serves as a vacuum window) must be serviceable and/or replaceable to address the remote possibility of window failure. It is expect that disassembly and removal of the antiproton stopping window assembly might be accomplished within three working days. However, access to this window will require removal of upstream Cosmic Ray Veto and a substantial amount of shielding, which will take additional time and effort, and these items will need to be re-installed to resume operation. Consequently, the intent is to minimize the need to access the antiproton stopping window assembly at the TSu/TSd interface.

## 7.4    Upstream External Shielding

The upstream external shielding of the Muon Beamline surrounds the Production Solenoid and isolates the primary proton beamline from the Detector Solenoid hall while shielding the Cosmic Ray Veto (CRV) from radiation generated at the production target located in the PS and resulting from the secondary beam transported by the upstream Transport Solenoid (TSu). The primary purpose of the Upstream External Shielding is to shield the Mu2e detectors, reducing the rate of neutron and gamma background impinging on the CRV to an acceptable level. This shielding also serves to isolate the primary proton beamline from the downstream muon beamline, dividing the lower level of the Mu2e Experiment Hall into two independent zones.

The current configuration of this shielding is composed of 138 tons of reinforced normal density concrete and 243 tons of reinforced higher density (barite) concrete. This concrete is supported by the lower level floor of the Mu2e Experiment Hall. The shielding as implemented in the current simulation is illustrated in Figure 7.12.

Additional information on recent Fermilab experience with barite and barite concrete is summarized in [28].

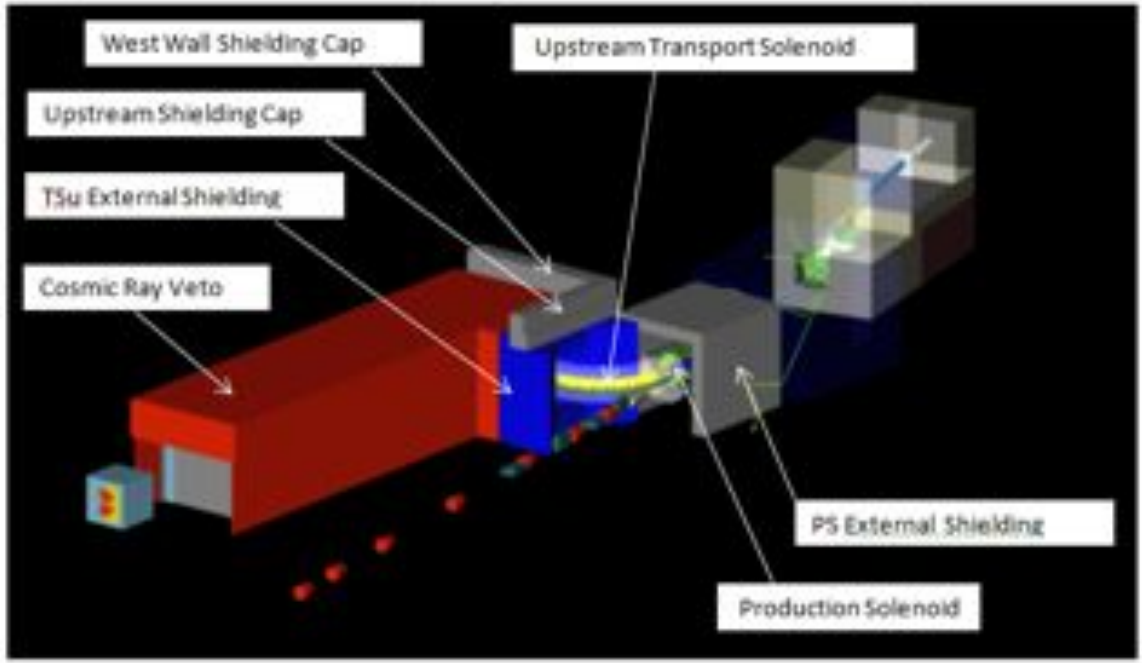

Figure 7.12. Overview of Mu2e Muon Beamline illustrating the Upstream External Shielding elements designed to limit the radiation rates from the primary proton target incident on the Cosmic Ray Veto as implemented in the current simulation. Normal density concrete is shown in grey, while high density (barite) concrete is shown in blue.

### *7.4.1* **Requirements**

The Mu2e experiment will require a negative muon beam, stopping muons in an aluminum target and search for evidence of the neutrinoless conversion of stopped muons into electrons in the field of a nucleus [29]. During muon production, besides the required negatively charged muons, many other particles are created. The primary purpose of the TS magnets and collimator system is to perform charge and momentum selection on the particle flux exiting the PS. The S-shaped beamline is quite an effective filter of neutral particles however, to even further reduce the flux of particles headed downstream, additional shielding is required around the solenoid system. The amount and type of shielding required for the desired detector performance is dependent upon the configuration of other materials in the path of the emerging particles (such as the HRS [30]), as well as on the sensitivity of the particle detectors being shielded.

The primary purpose of the upstream external shielding around the PS and TSu is to reduce the rate of particles incident on the Mu2e detectors, and the requirements for the upstream external shielding are described in [31]. The rate incident upon the Cosmic Ray Veto must be suppressed to a low enough level to support efficient operation of the CRV, so that potential sources of background due to incident cosmic rays can be identified and suppressed by the CRV. The simulations have gone through numerous generations in an effort to arrive at a configuration that provides sufficient shielding capability implemented in (relatively) affordable materials. A recent version of results from the shielding simulation including CRV rates is available in [32]. The current understanding of the CRV rate requirements and results of implementing CRV detector response can be found in Chapter 10 and in [33], which also relies on parameters specified in [34]. An overview of the methods used to evaluate the CRV response is provided in [35]. An update on the most recent design of the shielding resulting from the simulations studies intended to match to the details of the civil construction plans is available in [36]. Simulations have illustrated the sensitivity of the CRV to significant gaps in coverage [37]. Efforts must be made to minimize open penetrations and direct line-of-sight cracks wherever feasible.

The upstream external shielding also isolates the primary proton beamline from the DS hall, providing a natural radiation zone break and serves as an element of the airflow control system for the Mu2e Experiment Hall.

The upstream external shielding is entirely external to the muon beamline. The temperature expected in this area is between 20 and 30 degrees Celsius (between 68 and 85 degrees Fahrenheit) [38]. Material must be dimensionally stable and maintain structural integrity over the operating life at these temperatures.

The reinforced concrete assemblies must be demonstrated to be stable under anticipated operating conditions, including the forces resulting from the solenoid fields. The design must also take into consideration appropriate seismic related constraints.

### *7.4.2 Technical Design*

The upstream external shielding surrounds the upstream portion of the Mu2e muon beamline, isolating the primary beamline from the DS detector hall, reducing the particle flux on the Mu2e detectors. The upstream external shielding is composed of several elements as illustrated in Figure 7.12.

The PS External Shielding (shown in Figure 7.13) is currently composed of 22 inch thick reinforced concrete surrounding the sides and the top of the PS. The top of this shielding is 174 inches above floor level (Y), and the shielding is 207 inches long (Z), for a total mass of 90 tons. The cryo services to the PS penetrate this shielding in the south east corner just under the top of the shielding. An additional relief in this shielding may be required to accommodate the upstream muon beamline vacuum line.

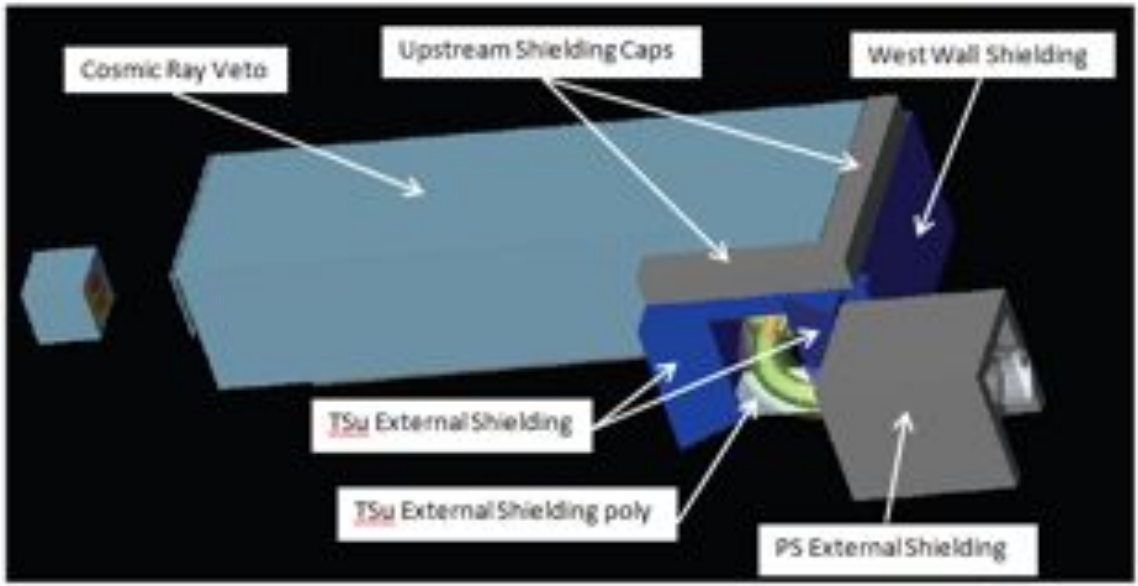

Figure 7.13. Overview of the Mu2e muon beamline highlighting components of the upstream external shielding as implemented in the current simulation (in grey for normal density concrete and dark blue for high density concrete) and the Cosmic Ray Veto (in light blue covering the downstream external shielding).

The TSu External Shielding is primarily high density (barite) concrete surrounding the downstream end of the TSu. The shielding is anticipated to be composed of 152 inch tall (Y) upright columns (shown triangular in profile in Figure 7.13 as per the current simulation), and 152 inch tall (Y) rectangular columns on each side of the TSu (for a total of 111 tons) and a high density concrete beam (39.5 inches wide (X), 23 inches tall (Y) and 264 inches long (Z), 15 tons) placed across the rectangular columns. In addition, there is currently a 24 inch wide (X), 132 inch tall (Y), and 15 inch long (Z) 3 ton barite concrete column in the north west corner between the PS External Shielding and the west TSu External Shielding column. This column is shorter than its neighbors to accommodate passage of services into the west wall relief. Note that the TSu services are routed near the ceiling, and pass above the PS external shielding.

The space below TSu between the TSu support stands is not yet fully defined and will likely be difficult to access. It is currently expected to be filled with perhaps 2 tons of poly beads (either bagged or retained via an aluminum or stainless steel frame), estimated at about half the density of poly and located approximately as indicated in Figure 7.13.

The West Wall Shielding completes the coverage of the relief in the west wall through which the cryo and vacuum services will be routed from the alcove below the Solenoid and Power Supply Room to the PS and TSu, and provides additional shielding of the Mu2e detectors from the radiation emerging from the primary proton target located in the PS. The West Wall Shielding is currently composed entirely of high density (barite) concrete, but based upon available simulation results, it is anticipated that perhaps some of this wall may eventually be shielded using normal density concrete. The west wall shielding is currently 264 inches long (X), 175 inches tall (Y), and 38.5 inches thick (Z), for a total of 113 tons. The precise dimensions of this West Wall Shielding must be revisited once the dimensions of the CRV and its support scheme have been finalized.

The Upstream Shielding Cap completes the isolation of the primary proton beam from the DS hall, and is composed of normal density concrete blocks placed on top of the TSu External Shielding (39.5 inches long (X), 56 inches tall (Y), and 262 inches thick (Z), for a total of 24 tons) and on top of the West Wall Shielding (262 inches long (X), 56 inches tall (Y), and 38.5 inches thick (Z), for a total of 24 tons).

The upstream external shielding is made primarily of concrete reinforced with steel bars. The Fermilab stock of blocks is unlikely to have sufficient quantity to fill the needs of the project. Furthermore, the stock blocks are unlikely to meet the challenging space constraints, so new blocks will need to be designed and procured. While the TSu external shielding columns implemented in the simulation (and shown in Figure 7.13) are composed of substantial blocks including unconventional (triangular) shapes, the actual implementation of the TSu external shielding is more likely to be realized utilizing an assembly of smaller blocks which are better matched to the lifting capacity available in the building. Figure 7.14 illustrates the details of the proposed implementation.

It is also anticipated that the alcove below the Solenoid and Power Supply Room may require shielding (to reduce backshine from the west wall relief).

### 7.4.3   Risks

Since the shielding materials will be in the stray field of the solenoids, the materials must not distort the field inside the solenoid bores, and must be selected to minimize the impact on the solenoid coil supports. The baseline material for the reinforcing bars, lifting points and brackets is carbon steel, but stainless steel is an option if demonstrated to be necessary. The shielding must also withstand the structural loads resulting from the magnetic fields. Sensitivity has been investigated by simulations, and it is not anticipated that stainless steel rebar or fixtures will be required [39].

Unfortunately, the shielding is much more massive than the Mu2e detector, so the installation of the shielding may influence the alignment of the muon beamline. The anticipated sensitivity of the muon beamline to solenoid alignment is under evaluation, but the Mu2e detector hall floor is designed as a single 3.5 foot thick slab to minimize differential alignment complications among various elements of the muon beamline.

Since the shielding installation should not be completed before the solenoid magnetic field mapping is complete, this risk is unlikely to be realized or retired prior to that time.

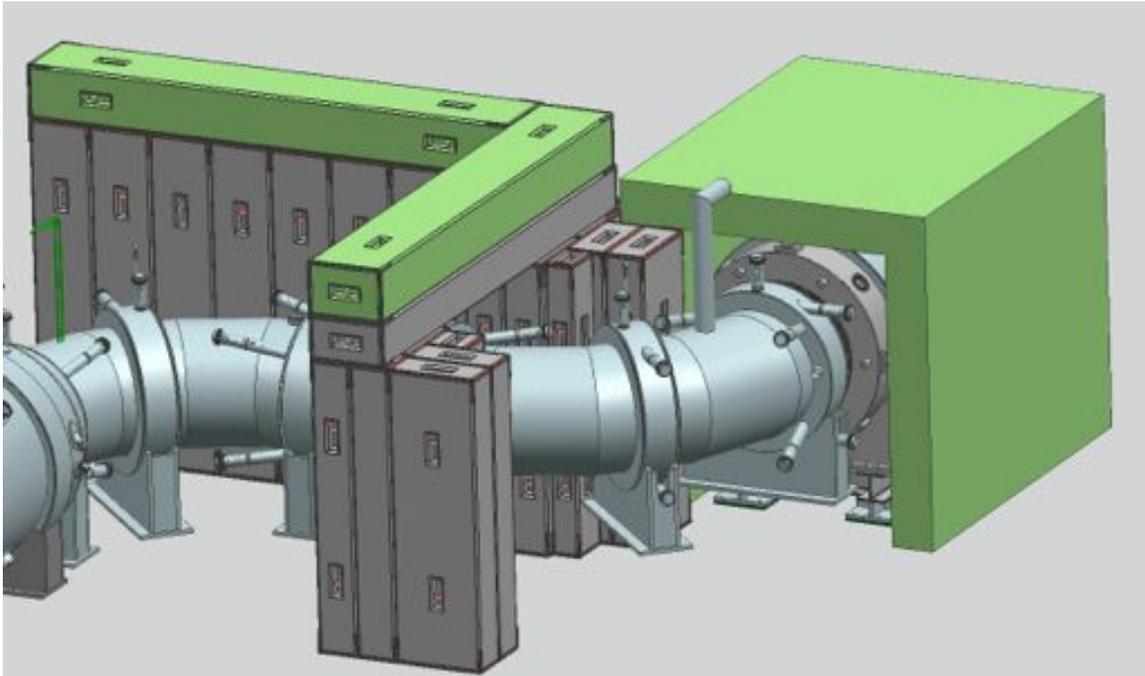

Figure 7.14. The upstream external shielding as viewed from the north east (with the PS on the right hand side and the DS on the left hand side of the figure). The grey blocks are high density barite concrete blocks, while the green elements are composed of normal density concrete.

Staging of these blocks will likely require outside storage and handling capacity for these large blocks. Unfortunately, barite concrete has been reported to have poor frost resistance, losing 45 to 60% of its strength after 25 frost cycles [40], so special considerations for the storage of this high density concrete may be necessary if this behavior is confirmed. Efforts will be made to investigate this issue.

### 7.4.4 Quality Assurance

Structural and magnetic analysis will be completed to ensure all components will meet the requirements outlined in [31].

Close collaboration between Fermilab personnel and the vendor who makes the barite blocks will take place to ensure that the barite mix is appropriate with respect to density and manufacturability.

All blocks, both concrete and barite, will undergo inspection by the vendor before shipment to Fermilab. Test pieces may be required to ensure that manufacturing processes will be viable. All components will be inspected by FNAL personnel upon arrival at Fermilab, and discrepancies will be documented.

### 7.4.5   Installation and Commissioning

The installation will be complicated since much of this shielding is beyond building crane coverage, and the shielding needs to be positioned near delicate equipment and fill significant space, and is intended to minimize line of sight cracks. Note that the PS External Shielding is not under crane coverage. The PS External Shielding must either be cast in the floor space under the PS hatch or cast elsewhere and lowered down the PS hatch and then translated to the east to surround the PS. The PS External Shielding will have to cross rails installed for the primary target remote handling system.  The current plan is to cast the PS External Shielding under the PS hatch, and then roll that shielding into place on rollers riding on steel floor plates embedded in the topping slab of the building floor and bridging across the tracks for the remote handling system.

Parts of the TSu External Shielding are beyond the building crane coverage, so additional installation equipment may be required.  The TSu external shielding will be assembled (primarily) from individual re-enforced barite concrete blocks, which will be sized to allow the possibility of using a spreader and counterbalancing blocks to position the blocks that are beyond direct crane coverage.

The installation sequence will be dictated by external constraints. It is anticipated that the solenoids and solenoid services will all be in place prior to final positioning of this shielding, and the initial round of solenoid alignment and field mapping must be completed prior to final positioning of the shielding to facilitate access to the various solenoid support and transport solenoid coil supports.

The anticipated installation sequence is to install the blocks making the west TS shield column, move the PS external shielding to its final position, install the blocks making the east TS shield column, then install the beam that spans these two columns and the north part of the Upstream Shielding Cap above the TS beam.  These shielding blocks should be installed after preliminary solenoid field mapping and completion of the electron source test.  Before installing the downstream external shielding, the air flow isolation of the primary beamline from the DS hall should be completed. This divider might be a thin aluminum plate attached to the blocks, or even a thin plastic sheet. After the downstream external shielding (WBS 5.9) and the CRV-U support installation is complete, then the West Wall shielding and the part of the Upstream Shielding Cap above the West Wall can be installed.

Note that the part of the Upstream Shielding Cap above the West Wall and the West Wall Shielding may need to be removed to provide access to the rotating collimators and the antiproton stopping window area if/when service is required. Staging of these blocks will likely require outside storage and handling capacity for these large blocks should that situation arise. Arrangements should be made to clearly label the blocks to distinguish the high density blocks from the standard blocks, as well as to indicate the location of the various blocks in the final assembly.

The PS and TS cryostats include external features that must be taken into account in the design and installation of the shielding, so the stay clears must be scrupulously respected [41][42]. The concrete block assembly is sensitive to variations in floor elevation, and must not interfere with the solenoids. Substantial variations in the floor level could result in significant complications during the assembly of the shielding. So, the flatness of the floor provided by conventional facilities will need to be taken into account.

## 7.5 Muon Stopping Target

The muon stopping target is a central component of the Mu2e experiment. Interactions in the stopping target cause energy loss and the capture of the beamline muons after they enter the DS. The stopped muons form muonic atoms in the stopping target, where they can potentially undergo neutrino-less conversion of muons to electrons. The stopping target design goal is to maximize the number of stopped muons while minimizing the amount of material traversed by conversion electrons that enter the acceptance of the downstream detector.

### 7.5.1 Requirements

The target material selection is based upon several criteria:

- For low Z nuclei, the sensitivity to electron conversion is roughly proportional to Z, therefore Z should be maximized.
- In order to avoid backgrounds (mainly from radiative pion capture), the measurement window when the search for the conversion electrons commences in the detectors should start on the order of 700 ns after the proton pulse. For low Z nuclei, the capture rate scales roughly with $Z^4$, therefore Z should not be too large.
- A potential background can come from radiative muon capture, via the reaction $\mu^- + {}^A_Z X_N \rightarrow {}^A_{Z-1} Y^*_{N+1} + \nu_\mu + \gamma$. The photon can have a maximum energy (for example in the case where the neutrino energy is zero) in the vicinity of the conversion electron energy. The photon can pair produce a fake conversion electron in the target or nearby materials. With the proper choice of the target, the mass of the daughter nucleus exceeds that of the parent nucleus, $m(Y^*) > m(X)$, pushing the photon energy below the conversion electron energy. An energy difference $m(Y^*) - m(X)$ greater than 2 MeV is desirable.
- The target material should be readily purified and chemically stable to avoid muon stops in foreign nuclei. In particular, contaminants that produce higher conversion electron energies can pose a background threat from electrons produced by muon decay in orbit and should be avoided.

* Ideally the target material is self-supporting.

Additional details are available in [43].

### 7.5.2   Technical Design

The location of the muon stopping target relative to other elements in the DS is shown in Figure 7.15.

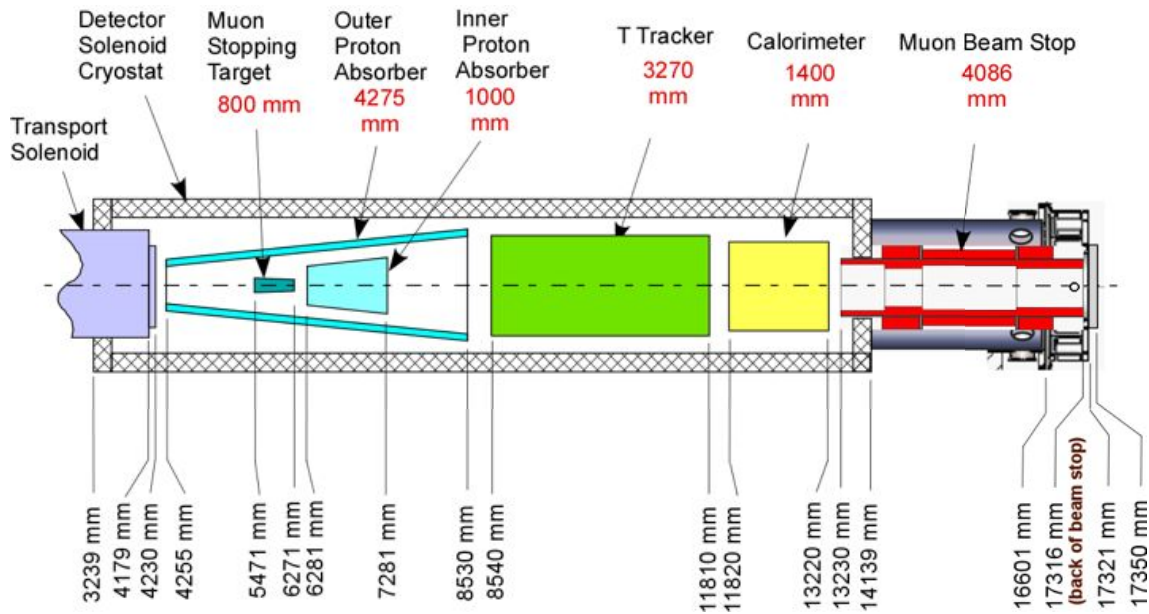

Figure 7.15. Cross section view showing locations of components within the Detector Solenoid. The muon stopping target is located between Z=5471mm and 6271mm. The TSdA is the small disk at Z coordinate ~4230 mm [23]. The proton absorbers are conical frusta illustrated in light blue between Z= 4255 and Z = 8530mm.  Note that the muon stopping target and the IPA are surrounded by the OPA.

### *Baseline Muon Stopping Target*

The target for the first run of the Mu2e experiment will be pure aluminum, which satisfies all of the above criteria.  The electron conversion energy is 104.97 MeV. The lifetime of muonic aluminum, 864 ns, is relatively long, enabling the use of a delay of 700 ns for the muon beam flash to die out before taking data [44][45][46].

The baseline design of the muon stopping target consists of 17 aluminum foils of thickness 200 μm with radii decreasing uniformly from 83 mm to 65 mm in the downstream direction, as illustrated in Figure 7.16.

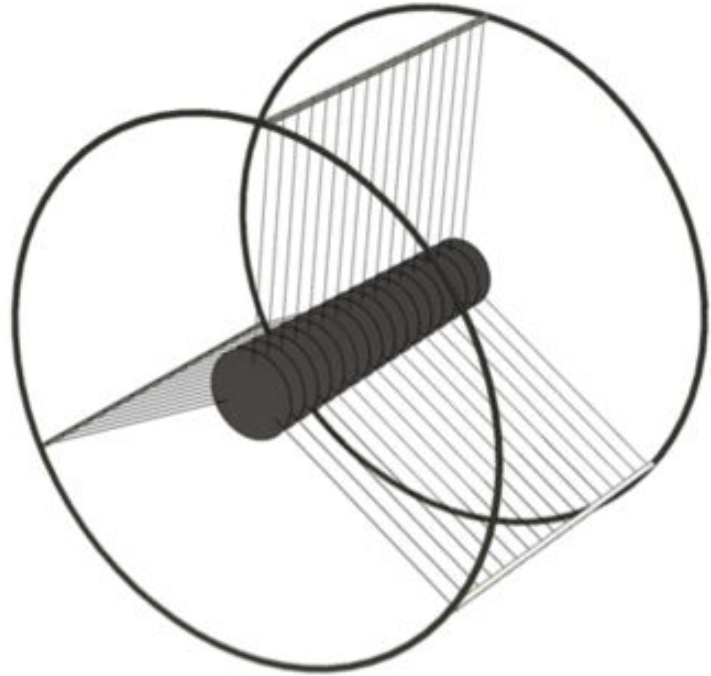

Figure 7.16. Illustration of the preliminary design of the muon stopping target composed of 17 aluminum foils and its mechanical support frame.

The foils are placed with an equidistant spacing of 50 mm along the axis of the detector solenoid in the center of the muon beam, with the center of the target assembly at a Z position of 5871 mm that is 4329 mm in front of the center of the tracker. Because of the diffuse nature of the muon beam a significant number of muons can strike the structure supporting the stopping target, producing DIO electrons at large radius where the acceptance for reconstruction in the detector is high. Because the endpoint energy of the DIO spectrum decreases for higher Z materials, and because the lifetime of the muons in muonic atoms decreases for higher Z materials, the supports must be constructed from high Z materials. Each foil of the stopping target is mechanically fixed by three tungsten wires that are attached to a support structure located near the inner surface of the detector solenoid. The baseline design employs 3 mil diameter wire to support the target foils. Simulations show that the tungsten has negligible effect on the energy straggling of conversion electrons [47].

*Optimization Studies*

In the interplay with the muon beam, the magnetic field and the tracker, the physical configuration and geometry of the muon stopping target directly affect the achievable sensitivity of the Mu2e experiment. Alternative geometries to the baseline design are currently being studied in order to improve the performance of the muon stopping target and to enhance the overall sensitivity of the experiment. Several competing effects must be considered in the optimization studies. First, modifications of the total amount of the traversed material or its distribution may increase the number of stopped muons per proton pulse, thereby increasing the sensitivity of the experiment. However, increasing the target mass, for example, results in more multiple Coulomb scattering and more energy loss of the outgoing conversion electrons. These processes may alter the position and resolution of the conversion electrons before they enter the tracker. In addition the mass and the matter distribution of the stopping target determine the magnitude and the experimental acceptance for background contributions originating from decay-in-orbit (DIO) electrons and other background sources. Moreover the technical feasibility of a mechanically stable configuration puts tight constraints on the target design. The ongoing optimization studies of the muon stopping target concentrate on improvements of the foil target design, and explore the potential of other geometric topologies and physical configurations. A selection of results from Monte Carlo simulations for different foil target configurations are shown in Figure 7.17. The examples shown already demonstrate the potential to improve the achievable sensitivity of the Mu2e experiment for foil target configurations with a larger number of thinner foils than the baseline design.

As alternative configuration, a target composed of low density aluminum wool, shows promise of further improvement in experimental sensitivity by providing additional flexibility in tailoring the mass and density distribution of the aluminum target material.

### 7.5.3   Risks

The target supports are delicate and could be damaged during installation or during installation of the surrounding outer proton absorber. Design of the surrounding outer proton absorber is being optimized in an effort to minimize this risk. Ongoing prototyping studies should reduce risk of failure of the target supports.

### 7.5.4   Quality Assurance

Materials certifications will be required for the key components. Components of the muon stopping target and support that are received from vendors will be inspected at Fermilab by quality control technicians. They will check to verify that the components have been built to specifications.

### 7.5.5   Installation and Commissioning

The target assembly will be installed on the external rails at the appropriate time. This installation must be performed with the appropriate care due to the delicate nature of the target suspension system.

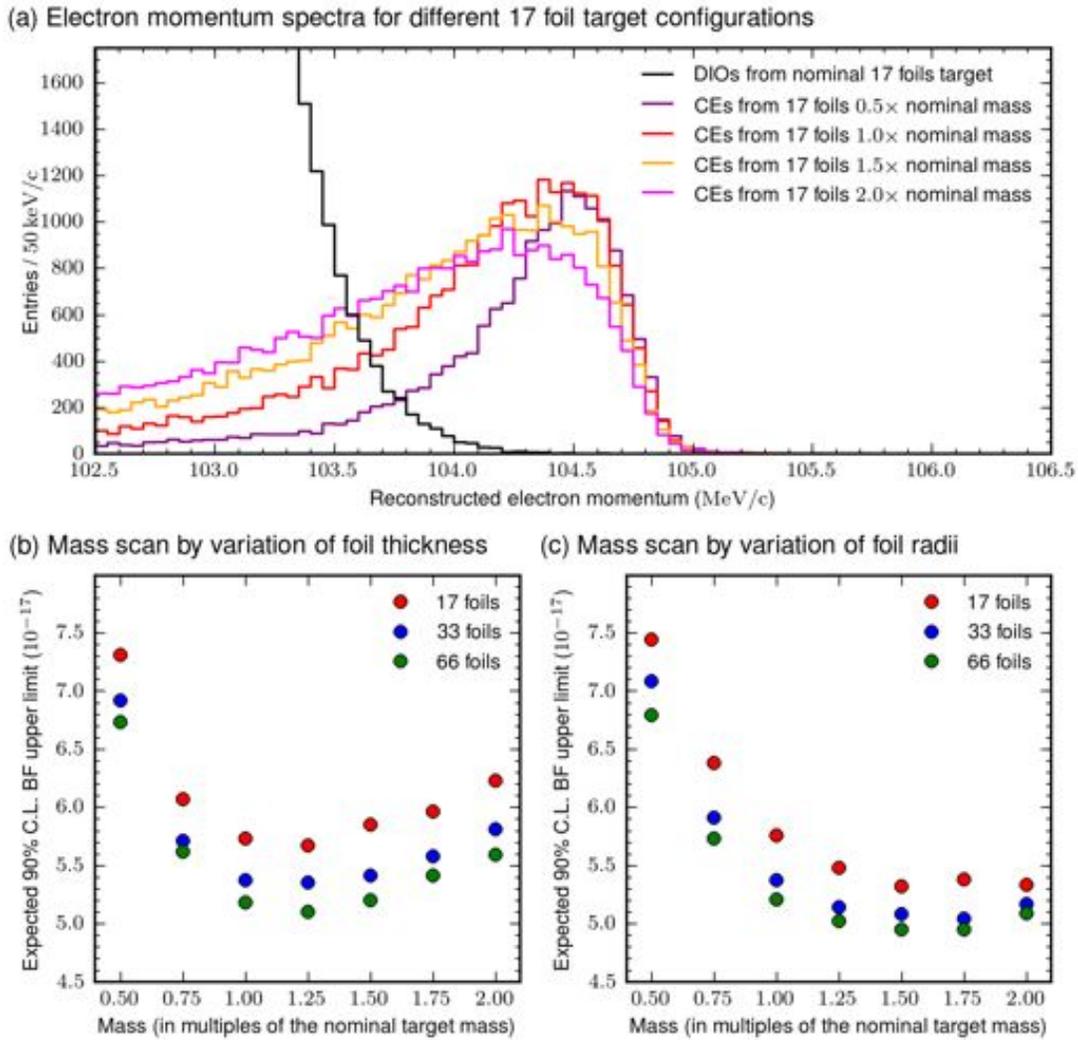

Figure 7.17. Examples of the results for the optimization studies of the muon stopping target. The results are obtained by full MC simulations including the simulation of hits in the tracker, pattern recognition and track reconstruction. Top: Momentum spectra for conversion electrons (CEs) for target configurations composed of 17 foils at different target masses, and for decay-in-orbit electrons (DIOs) from the 17 foils nominal target. The CE spectra are normalized to the same number of incident muons and have been scaled up by a common arbitrary factor to increase the visibility. Bottom: Expected Feldman-Cousins 90% C.L. upper limit on the muon to electron conversion branching fraction for configurations composed of 17, 33 and 66 foils at different target masses [48]. In the left (right) figure the mass variations are realized by modifying the thickness (radii) of the foils.

## 7.6  Muon Stopping Target Monitor

Given the complexity of generating and collecting low energy negative muons in the production solenoid, and transporting them via the transport solenoids to the target in the detector solenoid, it is evident that some means of confirming the rate and integral number of negative muons that stop in the muon-stopping target is necessary. It is equally evident that such a device should prove to be useful in the initial process of tuning conditions for the proton beam and the solenoids.

### *7.6.1* **Requirements**

The goal is to determine the number of stopped muons to an accuracy of 10% (at 1 standard deviation) over the course of the experiment [49].

It is possible to imagine a "dead-reckoning" calculation based on a simulation of the beamline, modeling of the stopping target, etc. However, it seems prudent to directly monitor some process that signals the capture of a muon on an aluminum nucleus.

One could monitor the X-rays emitted during the formation of a muonic atom and the associated cascade to the 1s state. The highest yield X-ray is the 2p →1s radiative transition that confirms the arrival of a muon in the ground state. Other observable X-rays having substantial yields and signaling arrival in the 1s state, are the 3p→1s, and the 4p→1s. Typically the 3d→2p transition that populates the 2p state also appears in the energy spectrum. In addition, muons that stop in impurities in the target, such as oxygen, generate X-rays with their own characteristic energies, potentially allowing us to determine the fraction of muons that stop in impurity nuclei.

A solid-state detector such as Ge is typically used for detecting such X-rays. However, these detectors have a relatively slow time response and are not normally used in an environment as intense as expected at Mu2e. The Mu2e beam has an intense electron component at early times (beam flash); the same aluminum that stops muons will then produce bremsstrahlung photons (from those electrons) at a rate we calculate to be 51 MHz/cm$^2$ with a mean energy of 1.4 MeV. The interesting muons arrive about 100 nsec after the flash and produce their X-rays within picoseconds of their arrival. Commercial off-the-shelf detectors and their electronics cannot manage these rates (MeV/sec limits) and their associated electronics are not fast enough [50]. This solution is still under study as of this writing but is not the baseline. The detector would also experience radiation damage from the beam flash and the neutron flux is problematic, requiring annealing once per day or less.

To eliminate the rate and radiation issues created by the flash, the baseline approach is to detect delayed gamma rays from the decays of radioactive nuclei produced in the muon capture process, rather than muonic X-rays. These delayed gammas, like the X-rays, are unique to the target material, and information on their energies and intensities is available in the literature. For $^{27}$Al, a $^{27}$Mg nucleus is produced in 13% of muon captures. The excited $^{27}$Mg beta-decays back to an excited state of $^{27}$Al with a half-life of 9.5 minutes. This 9.5 minute half-life gives us an opportunity to eliminate the problems of the beam flash. The proton beam structure of Mu2e will be a stream of proton pulses for 0.5 s followed by 0.8 s idle. The gamma spectrum would be acquired during the relatively background-free idle period and the beam would be blocked using a thick collimator during the time the beam is on, eliminating the problems created by the beam flash [51]. After the beta emission, the excited $^{27}$Al returns rapidly to the ground state, producing an 844 keV gamma ray 72% of the time.

Three requirements determine the best location for the Ge detector to view the muon stopping target:

- The detector should only view the target, if possible. Hence the first requirement is for good collimation ahead of the detector.
- Because of the extraordinarily high X-ray and gamma rates the detector must be far from the source, along a low attenuation path for photons.
- The detector must lie beyond the Detector Solenoid (DS) magnetic field where it can be serviced periodically and annealed to repair radiation damage.

The Ge detector performance must be such that annealing to recover from radiation damage is not required more often than once per 2 months of calendar time.

### 7.6.2   Technical Design

A preliminary spectrum of delayed gammas from muon capture in an aluminum target measured with a germanium detector is shown in Figure 7.18. The gamma ray can be detected with a high resolution photon detector. The gamma ray is unique to the target, and no other material in view of the detector will consist of aluminum. Measurements by the AlCap experiment [52] will establish the normalization between the number of stopped muons and the rate of 844 keV gammas. Good energy resolution is desirable in order to deliver good signal to noise and to resolve the 844 keV peak from gamma rays with nearby energies. Commercially available intrinsic germanium detectors will be used because they deliver excellent resolution (~2 keV@1.33 MeV) together with high efficiency.

Figure 7.19 shows a preliminary plot of the singles spectrum (self-triggered spectrum) from the AlCap experiment. Muons are stopped at the estimated rates of 3 kHz in Aluminum and 4 kHz in lead shielding and other materials. Despite the large number of stops in materials other than aluminum and the absence of any cuts on the gamma time relative to the muon stopping time, the 347 keV gamma line is clearly visible above the background. In the case of Mu2e, the live window will be delayed in time from the stopped muons, therefore we expect the signal to noise to improve. This will be quantified in the future in the AlCap data by vetoing any Ge data within a few muon lifetimes after the muon's arrival. Care will be taken in Mu2e to avoid viewing materials that are activated or stopping locations other than the target, therefore we expect that the background situation will be improved relative to that at AlCap.

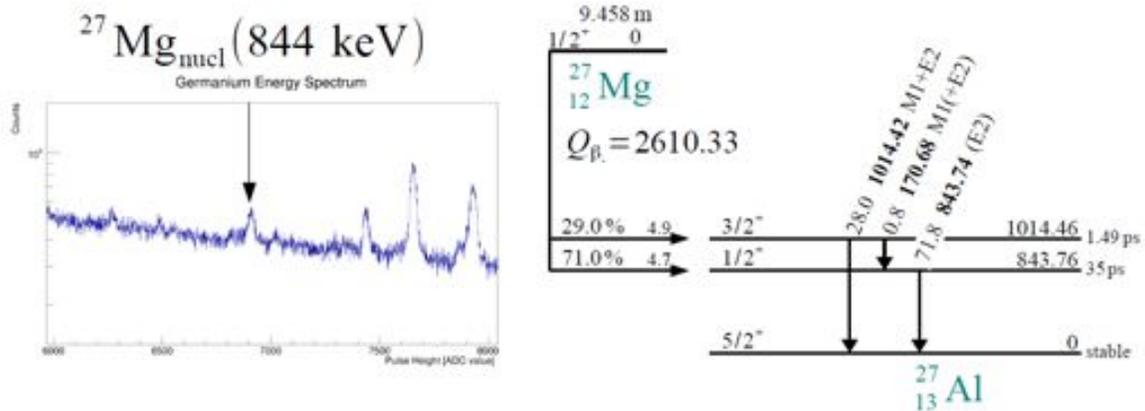

Figure 7.18. Preliminary singles germanium spectrum from the AlCap experiment at PSI. When muons stop in aluminum, they capture on the nucleus 60% of the time. A fraction of the captures produce $^{27}$Mg in the ground state, which has a half-life of 9.5 minutes. In the decay, an 844 keV gamma is produced 72% of the time.

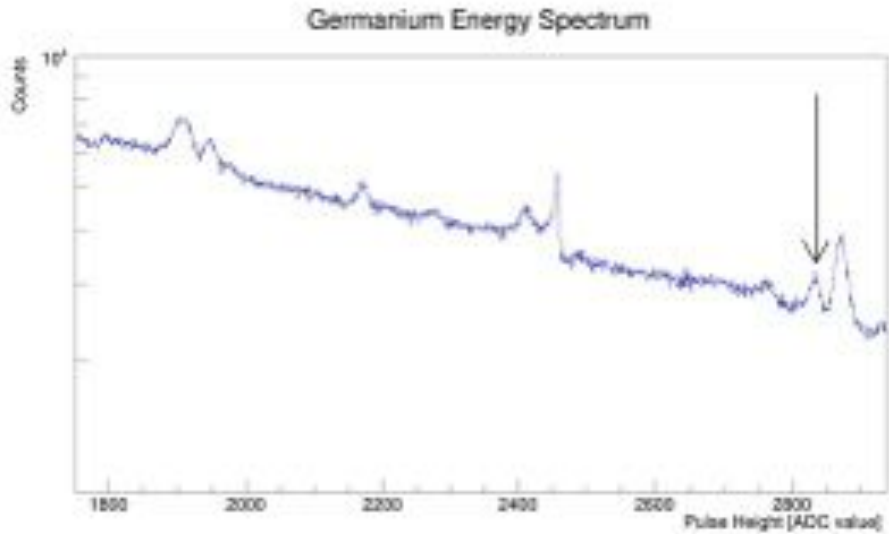

Figure 7.19. Preliminary singles germanium spectrum from the AlCap experiment at PSI. Muonic aluminum x-ray at 347 keV is clearly visible.

Germanium is moderately susceptible to radiation damage from neutrons and ionizing radiation (photons, charged particles). Most of the damage is readily annealed by heating the Ge for about a day. To provide continuous monitoring, multiple detectors would have to available to be rotated into the beam and annealed very often, which is undesirable. In order to circumvent this, we propose to take advantage of the beam cycle and delayed nature of the gamma rays to block most of the harmful radiation from impinging upon the Ge detector by employing a blocking shutter. The primary proton beam arrives in a stream of 260 ns wide pulses separated by 1695 ns, for a period of 0.5 s, followed by 0.8 seconds during which there is no beam. A shutter will block the beam for the 0.5 s when the proton pulses arrive, but will allow transmission during the 0.8 second beam-off period. For a 5 cm diameter active area in the detector, positioned 15 m downstream from the target, the fractional geometric acceptance is $1\times10^{-6}$. With a muon stopping rate of $1\times10^{10}$ Hz, the incident rate of 844 keV gammas is 480 Hz.

In the Mu2e experiment, the Ge detector will be located near the axis of the Detector Solenoid, downstream of the bore vacuum, muon beam stop and endcap shielding. The detector is enclosed in a small concrete house to provide shielding (see Figure 7.20). It will view the stopping target through a 10 cm diameter, 3 m long stainless steel pipe connected to the center of the downstream end of the Detector Solenoid. Collimators inside the pipe will allow full view of the stopping target while limiting the view of other materials such as the last collimator in the TS solenoid (TS5). A dipole magnet will sweep away electrons in the beam. The beamline will include a shutter that will be triggered to move into the beamline during the beam, and move out to provide a clear aperture between beam spills. A vacuum window will separate the Detector Solenoid from the collimation pipe end window.

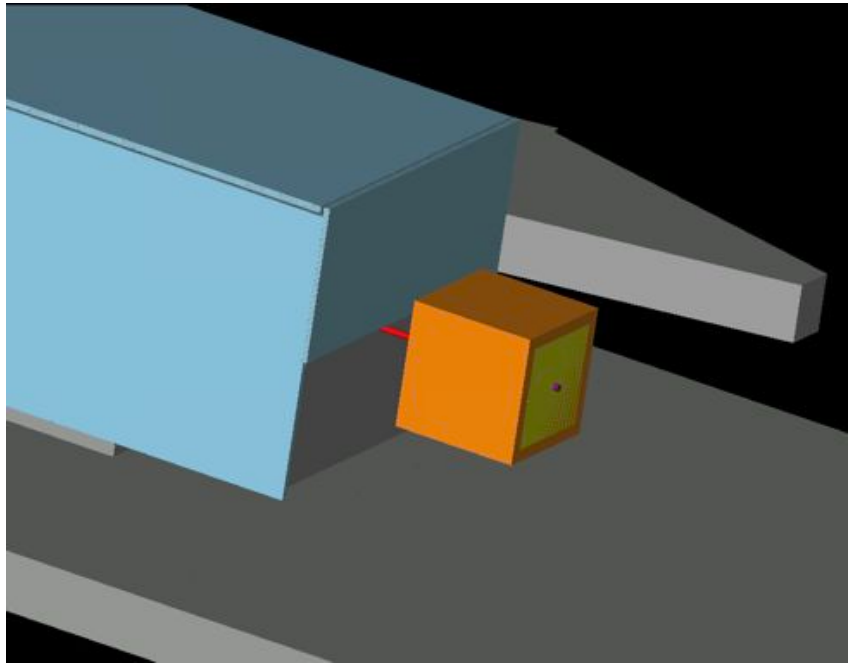

Figure 7.20. The Ge detector will be housed in a concrete shield, placed downstream of the outer concrete shield of the end of the Detector Solenoid. The germanium views the stopping target through the collimated tube in red.

Radioactive sources, such as Europium-152, are required for calibration purposes. The calibration and absolute efficiency of the Ge detector will be determined following ANSI N42.14-1999 "American National Standard for Calibration and Use of Germanium Spectrometers for the Measurement of Gamma-Ray Emission Rates of Radionuclides". We expect a few percent measurement based on standard practice, well within the 10% requirement.

### 7.6.3   Risks

If the simulations indicate that a substantial mass is required for the shutter, reliably cycling that shutter every beam spill may represent a mechanical challenge. The Germanium detector will require liquid nitrogen cooling. The nitrogen supply will need to be appropriately routed to minimize the ODH risk.

### 7.6.4   Quality Assurance

Components of the muon stopping target monitor system that are received from vendors will be inspected at Fermilab by quality control technicians.  They will check to verify that the components have been built to specifications.

### 7.6.5   Installation and Commissioning

Since the muon stopping target monitor system is located downstream of the DS bore, it will not be possible to install this system until the detector train has been inserted into the DS bore. The detector train will not be inserted into the DS bore before the solenoid magnetic field mapping is completed [15].  The muon stopping target monitor system must be designed to be readily and reliably extracted and re-inserted since the entire system must be removed to allow access to the detector train inside the DS bore for servicing when required (see Section 7.10).

## 7.7   Detector Solenoid Internal Shielding

A collection of absorbing materials is placed inside the DS warm bore to suppress the rates of protons and neutrons.  Excessive amounts of these particles will result in undesirable backgrounds in the tracker. The DS internal shielding consists of three objects:

- TSdA, Transport Solenoid downstream absorber
- IPA, Inner Proton Absorber
- OPA, Outer Proton Absorber

### 7.7.1   Requirements

The detailed requirements are described in [53], but can be summarized as follows:

- Sufficiently reduce the background rates at the tracker so that electron tracks may be reliably reconstructed.
- Minimize the energy loss and multiple scattering of electrons that travel within the acceptance of the tracker, including those that pass through the inner proton absorber and its supports.
- Minimize muons stopping on the proton absorbers.
- Minimize contributions to the background rates in the calorimeter.

### *7.7.2* **Technical Design**

Figure 7.15 shows the placement of the TDR baseline design DS internal shielding components within the DS bore. The TSdA is mounted on the downstream end of the TSd cryostat vacuum jacket. The IPA is located just downstream of the muon stopping target. The OPA surrounds both the stopping target and the IPA. It extends from just downstream of the TSd to just upstream of the tracker. Table 7.2, Table 7.3, and Table 7.4 list the parameters for the IPA, OPA, and TSdA, respectively, as implemented in the current Mu2e simulation.

***Transport Solenoid Downstream Absorber (TSdA):***

This is an absorber in the shape of a disk. A hole in the center allows the muon beam to pass through. Simulations indicate that this disk reduces the hit rate in the tracker by ~30% while having only a small impact (~0.2%) on the acceptance for conversion electrons [56]. The TSdA has a mass of ~22 kg, assuming it is made of borated polyethylene. It will be mounted on the end of the TSd cryostat vacuum jacket using retaining clips. There do not appear to be any significant technical issues associated with its manufacture or installation.

Table 7.2. Parameters of the IPA based on the current Mu2e simulation [55]. The IPA is aligned with the axis of the DS bore. Z-start and z-end are in the Mu2e coordinate system [23]. The mass shown in this table does not include the supports.

| | |
|---|---|
| Z-start | 6281 mm |
| Z-end | 7281 mm |
| Length | 1000 mm |
| Inner radius at Z-start: | 334.9 mm |
| Inner radius at Z-end: | 352.8 mm |
| Thickness | 0.5 mm |
| Material | HDPE |
| Density | 0.95 g/cm$^3$ |
| Mass (calculated): | 1.0 kg |
| Volume | 0.0011 m$^3$ |
| Surface area | 4.3 m$^2$ |

Table 7.3. Parameters of the OPA based upon the current Mu2e simulation. The OPA is centered on the axis of the DS bore. Z-start and Z-end are in the Mu2e coordinate system [23]. The mass shown in this table does not include the supports.

| | |
|---|---|
| Z-start | 4255 mm |
| Z-end | 8530 mm |
| Length | 4275 mm |
| Inner radius at Z-start: | 452.4 mm |
| Inner radius at Z-end: | 728.4 mm |
| Inner radial tolerance | ±5.0 mm |
| Thickness | 20 mm |
| Material | 5% borated polyethylene |
| Density | 0.95 to 1.08 g/cm$^3$ |
| Mass (calculated): | 306 to 348 kg |
| Volume | 0.323 m$^3$ |
| Surface area | 32.5 m$^2$ |

Table 7.4. Parameters of the TSd absorber based upon the current Mu2e simulation. The TSdA is centered on the axis of the DS bore and mounted on the downstream face of the TSd cryostat vacuum jacket. Z-start and Z-end are in the Mu2e coordinate system [23].

| | |
|---|---|
| Z-start | 4180 mm |
| Z-end | 4230 mm |
| Length | 50 mm |
| Inner radius: | 250 mm |
| Outer radius: | 450 mm |
| Material | 5% borated polyethylene |
| Density | 0.95 to 1.08 g/cm$^3$ |
| Mass (calculated) | 20.9 kg to 23.8 kg |
| Volume | 0.022 m$^2$ |
| Surface area | 1.1 m$^2$ |

***Inner Proton Absorber (IPA):***

The purpose of the IPA is to reduce the rate of protons intercepting the tracker while minimizing energy loss and straggling of conversion electrons. The IPA should not intercept the muon beam. The IPA is a thin (0.5 mm thickness) conical frustum. It is required to be made of low-Z materials. Identifying a thin material that will hold the required shape is non-trivial, but it is expected that the requirements can be met using carbon fiber. Less expensive thermoplastic material such as HDPE is less likely to have sufficient mechanical rigidity at this thickness but the higher hydrogen content is preferred. The IPA will be supported from the OPA by tungsten wires. Three wires would support the IPA near its upstream end and another three wires would support it near its downstream end.

Based upon the most recent simulations [54], the preference is for an IPA with a uniform thickness. This can be achieved most effectively in designs in which the IPA is self-supporting or requires only a minimal support structure, e.g. support wires. Carbon fiber can be made very stiff so it will satisfy the rigidity requirements. One issue associated with this option is that a mandrel is required on which to wind the carbon fibers. The cost of the mandrel would be a substantial fraction of the IPA manufacturing cost.

Assuming the IPA is realized in carbon fiber, it could be made at an internal facility at Fermilab. On the other hand, if the IPA is made of thermoplastics for considerations of cost or other reasons, it will be manufactured by an outside vendor.

### Outer Proton Absorber (OPA):

This is also a conical frustum, but much thicker than the IPA. Its thickness is required to be at least 20 mm. It will have to be made in sections to accommodate support of and access to the muon stopping target (see Section 7.5). Slots will be made in the OPA to allow for the stopping target support wires to pass through and attach to the stopping target support frame.

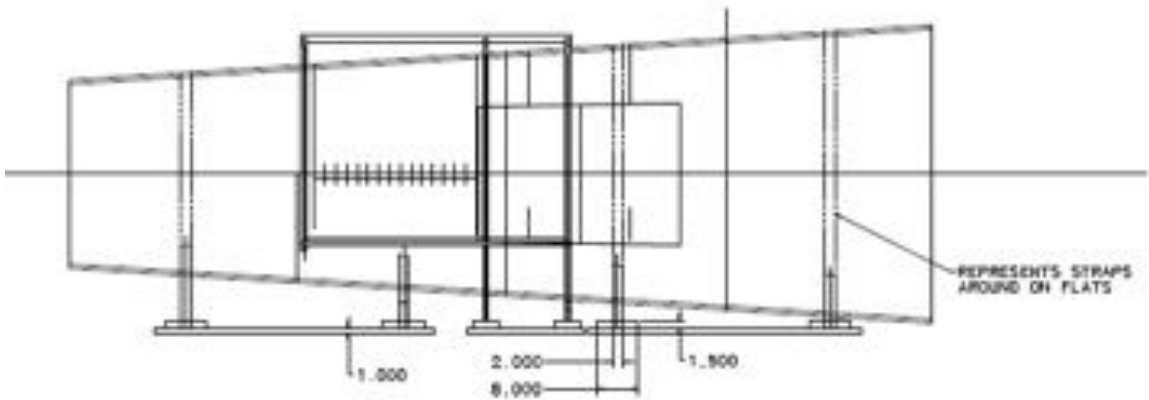

Figure 7.21. Elevation view of the OPA and its support structure. Also shown are the stopping targets and their support frame. The IPA is visible just downstream of the stopping targets.

The OPA will be made of two sections joined at about the same longitudinal location as the downstream end of the muon stopping target and the upstream end of the IPA. Each section will be supported by two sets of support feet, or cradles. The cradles will most likely be made of stainless steel. The feet will be mounted to bearing blocks that slide on the DS internal rails (see Section 7.10). This arrangement is shown in Figure 7.21.

Borated polyethylene is readily available in 4 ft. x 8 ft. sheets. We would require 8 sheets to make the complete OPA. Among poly forming vendors that have been contacted to date, there appears to be little or no experience with forming borated polyethylene sheets into special shapes. These sheets are normally used as flat sheets in shielding applications. We anticipate, however, that the required bending, welding, and/or joining of the sheets can be accomplished and plan to verify this via prototyping.

Slots will need to be cut into the OPA at the location of the stopping target supports. This will allow tungsten support wires to be strung from the stopping targets to the support frame. The support frame is located outside the volume enclosed by the OPA. Three slots separated by 120° are required. The width of the slots is to be determined, but should be as narrow as possible without risk of damaging or breaking wires during assembly. Figure 7.22 shows a view of the support structure for the OPA.

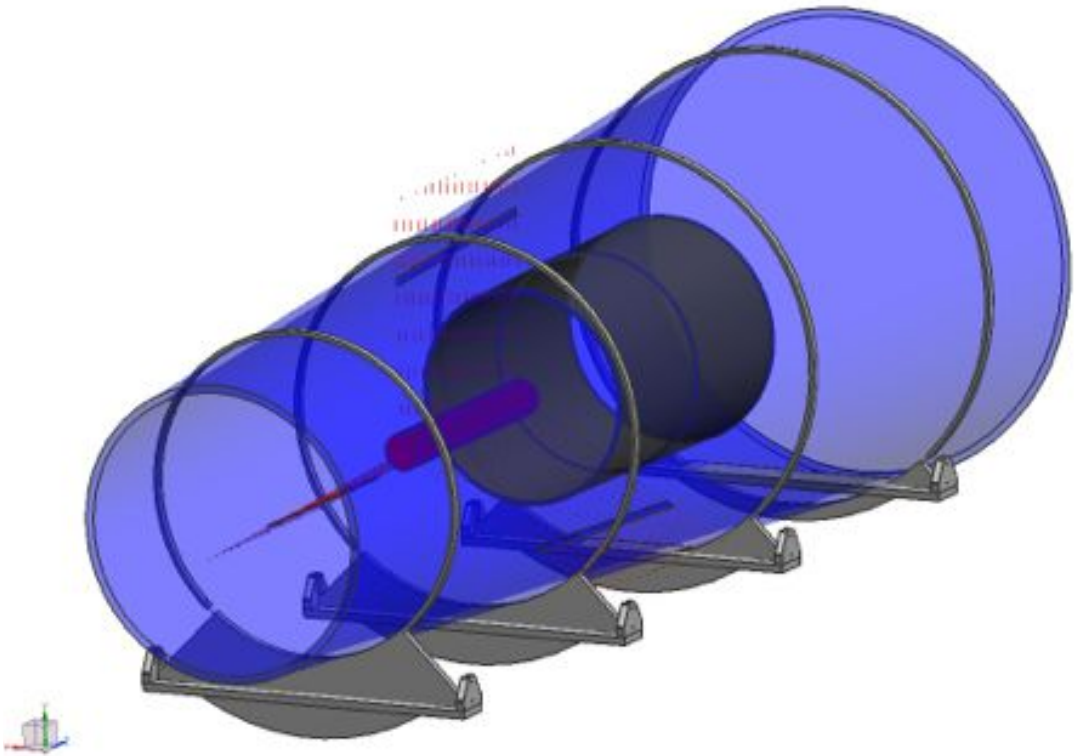

Figure 7.22. View of the OPA surrounding the stopping target and the IPA. Note the OPA supports.

### 7.7.3  Risks

The major risk for the IPA is the possibility that the selected material will not hold its required shape. We will investigate whether options such as support ribs are acceptable; this can be checked via the Mu2e simulation.

Another risk for the OPA is that it may be difficult to form borated polyethylene into the required shape. A fallback would be to assemble the OPA from flat segments into the form of a barrel. An important consideration here is whether the increased surface area would result in significant outgassing. Studies of outgassing properties of borated polyethylene are planned.

### 7.7.4   Quality Assurance

Components of the DS internal shielding that are received from vendors will be inspected at Fermilab by quality control technicians. They will check to verify that the components have been built to specifications.

### 7.7.5   Installation and Commissioning

The proton absorber system will be installed on its supports, and the assemblies will be mounted on the external rail system (consult Section 7.10) at the appropriate time, and subsequently rolled into the DS bore. The bearing blocks that support the OPA will include adjustment capability to facilitate alignment of the proton absorber assembly. The OPA will surround the muon stopping target. Therefore, close communication with the Muon Stopping Target (WBS 5.5) and the Detector Support and Installation System (WBS 5.10) is essential. The installation of the proton absorbers will be incorporated into the assembly procedure for the other DS components.

## 7.8   Muon Beam Stop

The muon beam stop (MBS) will be located within the bore of the DS, and is designed to absorb beam particles that reach the downstream end of the solenoid while minimizing the background to the upstream detectors resulting from muon decays and captures in the beam stop. An illustration of the muon beam stop in the installed position with the other DS internal components is shown in Figure 7.23.

### 7.8.1   Requirements

As described in [57], the primary requirement for the MBS is to limit the induced rates in the Tracker, the Calorimeter and the Cosmic Ray Veto due to backsplash and secondary interactions. These requirements can be summarized as follows:

- Backsplash particles from the MBS should not produce delayed signals from the tracker that could be mistaken for conversion electrons.
  - o   Based on GEANT simulation, products of muon capture in the beam stop will produce negligible background rates relative to other sources.
- The MBS should not produce secondary particles with a larger radiation impact on the calorimeter than those that arise from the muon stopping target.
- Shielding should ensure that the rate seen by the CRV from particles originating in the MBS should not be larger than the rate from the stopping target.

- o   Simulations of cosmic ray background sources define the CRV requirements on the combined performance of the MBS and the surrounding downstream external shielding.
- •   The muon beam stop should also provide a clear line-of-sight from the muon stopping target to the muon stopping target monitor, which is located well downstream of the MBS.

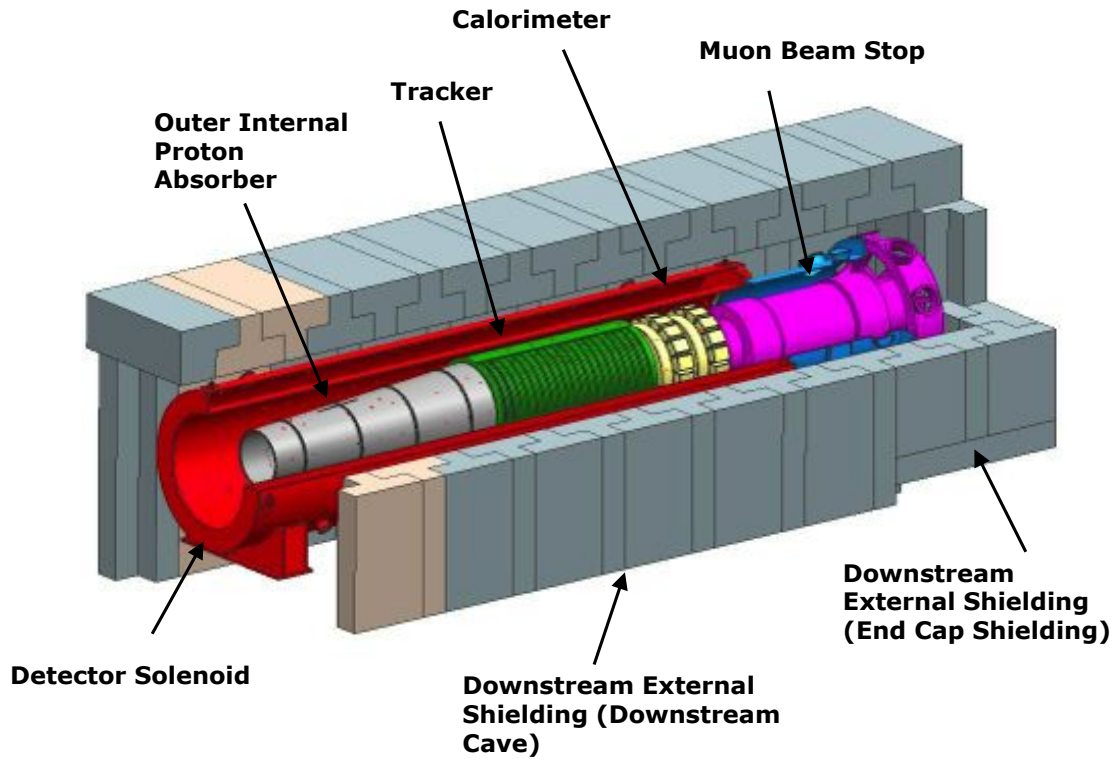

Figure 7.23. Cutaway view of the Detector Solenoid and surrounding downstream external shielding highlighting the relative locations of various elements inside the DS bore. The muon beam stop is located at the downstream end of the DS bore.

The following requirements are incorporated into the optimization of the muon beam stop:

- •   The surface that intercepts the beam directly should be far from the detectors.
- o   This insures that the solid angle from the beam stop to the detectors is small thereby minimizing the flux of backscattered particles that could interact in the detector volume.
- •   Beam muons that arrive at the end of the beamline should be stopped quickly, but the resulting neutron flux in the calorimeter and tracker from muon capture must be kept to a reasonable value (for example it should not exceed the neutron flux from the stopping target).
- o   A layer of low Z material such as polyethylene should be used to thermalize and absorb some of the neutrons produced in muon capture.

- The inner radius of the beam stop walls closest to the calorimeter should not be much smaller than the inner radius of the calorimeter.
- Aluminum and titanium should be avoided as candidates for MBS materials, since these are primary stopping target candidates; muonic x-rays from these materials should not reach the stopping target monitor located downstream.

### 7.8.2 Technical Design

The preliminary design of the MBS consists of stainless steel and high density polyethylene. The backbone of the MBS is a stainless steel tube of 10mm wall thickness, outfitted with two integral reinforcing rings, each 40mm thick. Inside the tube is polyethylene, and the outside of the tube is also partially covered by polyethylene pieces.

The relative positions and descriptions of the components of the muon beam stop are shown in Figure 7.24 (suppressing the details of the supports).

As illustrated in Figure 7.24, the upstream polyethylene element (labeled b-i) covers the entire 360 degrees in azimuth. As shown in the small inset, the downstream elements of the external polyethylene absorber (labeled b-ii and b-iii) do not cover the bottom 45 degrees of the MBS. There is no Cosmic Ray Veto coverage at the bottom of the detector, so this part of the exterior absorber has been omitted to reduce the total mass (and cost) of the MBS, as well as to facilitate the sideways extraction of the external rail stands during insertion of the detector train (see Section 7.10).

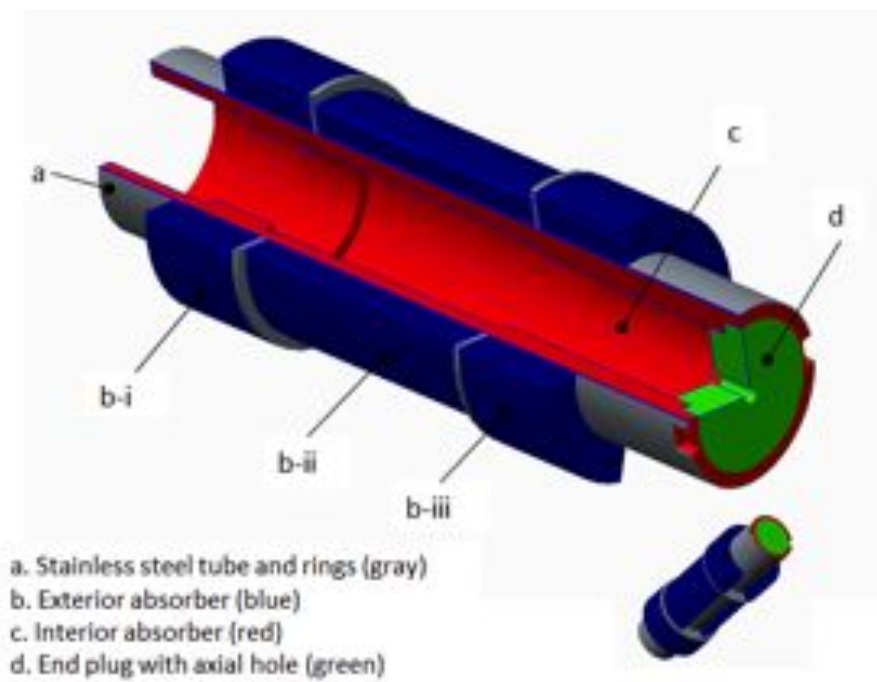

a. Stainless steel tube and rings (gray)
b. Exterior absorber (blue)
c. Interior absorber (red)
d. End plug with axial hole (green)

Figure 7.24. Cutaway view illustrating the Muon Beam Stop components.

The end plug (labeled d in Figure 7.24) contains a 10 cm diameter hole through the center to provide a line-of-sight for the muon stopping target monitor located on the Detector Solenoid axis downstream of the MBS. The precise sizes, volumes and masses of the muon beam stop elements are described in [57], and resulted from simulation studies and respecting the mechanical constraints [58].

### *Manufacturing and Assembly*

The stainless steel outer shell will consist of a three-piece weldment, with an inner tube that is rolled and welded, and two steel flanges. The polyethylene pieces which cover the inside diameter of the tube will be made of a series of rings, separately machined and bonded together with no line-of-sight cracks, except for the end plug, which will be made from a single sheet. The pieces that cover the outside surface will be made in the same manner and fitted from the ends on the outside of the tube. The center-outside rings will be made in more than one piece and bolted or attached together to fit on the outside surface. The thickness of the rings will be limited to a maximum of 100 mm, the manufacturing limits on the raw material.

The MBS components will be manufactured at outside facilities, assembled at Fermilab, and then transported as an assembly to the experiment hall. Since there is sufficient time available, the stainless steel tube assembly will be manufactured and shipped to Fermilab for inspection prior to procurement of the HDPE. Measurements of the inside diameter of the stainless steel tube will be used to determine the precise outside diameter of the HDPE parts that will fit inside it. The HDPE parts will then be manufactured, shipped to Fermilab and inspected. Finally, the stainless steel structure and the HDPE parts will be assembled.

### *Support and Alignment*

In the final configuration, the MBS will be supported and aligned by a combination of two support mechanisms, one on the upstream end and one on the downstream end. The upstream "Gimbal" support shown in Figure 7.25 will rest on a linear rail system and bearings, manufactured to fit inside the Detector Solenoid warm bore [57]. The downstream end of the MBS includes trunnion sockets and that end of the MBS will rest on trunnions (shown in Figure 7.26 and Figure 7.27), which are integrated into the Instrumentation Feedthrough Bulkhead (IFB), which will be supported by the IFB support structure [59].

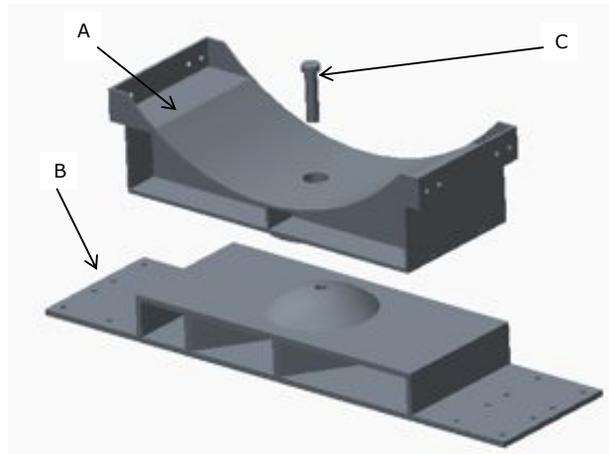

Figure 7.25. Gimbal Support including the saddle (A), the bridge (B) and the retaining pin (C). The saddle is fitted with a bronze bearing surface that will mate with the spherical section mounted on the bridge. The retaining pin is designed to ensure that the MBS body will not separate from the upstream support. The bridge will be seated on bearings that mount on the rail system [59].

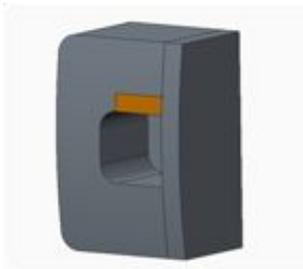

Figure 7.26. Details of a trunnion socket to be integrated into the back end of the MBS. Note the bronze insert and the removable cover plate.

### 7.8.3   Risks

There are no moderate or high risks that involve the Muon Beam Stop.

There is a low risk that the fit of the polyethylene parts into and over the stainless tube will complicate the assembly of the MBS. This risk will be mitigated by procuring the stainless steel tube first, and using the measured values to specify the sizes of the HDPE parts (as described above). If time does not allow this to be done, careful tolerances and inspection of the parts before arrival will mitigate this risk.

There is a low risk that the "gimbal and trunnion" assembly, which is designed to accommodate relative motion of the IFB on Hilman rollers will not work as planned. Relative motion is anticipated since the upstream end of the MBS will be supported on the rail system while the downstream end will be supported via the IFB on the floor plates in the detector hall.  This risk will be mitigated by a test of the system at an existing rail system mockup.

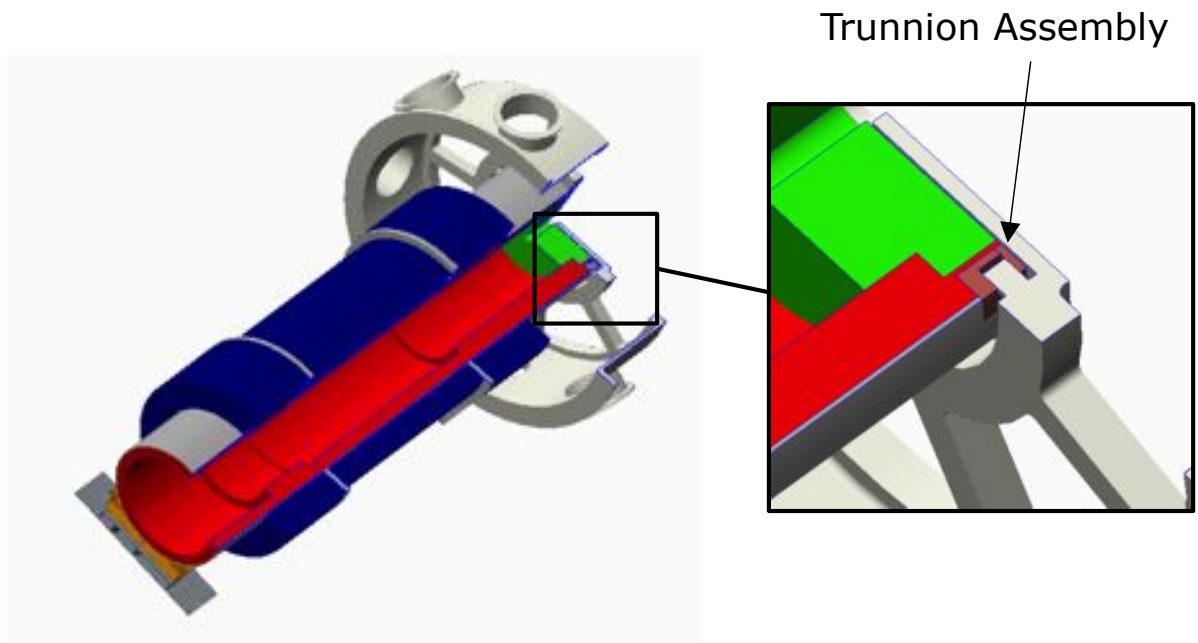

Figure 7.27. Muon Beam Stop connected to the Gimbal support on the upstream end and the IFB through the Trunnion support on the downstream end.

There also is a low risk that the weld between the stainless steel structural tube and the reinforcing rings will have a magnetic permeability higher than the requirements specify. This risk will be mitigated by performing tests on weld samples with the materials before manufacturing, and testing the MBS welds for permeability as part of the incoming inspection process.

### 7.8.4   Quality Assurance

Thorough structural, thermal and magnetic analysis will be completed to ensure all components will meet the requirements described in [57].

Test pieces will be manufactured to ensure that manufacturing processes will be viable. Tests of weld samples will be made to ensure that the magnetic properties of the weld between stainless steel segments will have acceptably low magnetic permeability.

After manufacturing, all components will be inspected by FNAL personnel upon arrival at Fermilab, and discrepancies will be documented.

A full size rail system mockup has been assembled.  Tests of the support mechanism, including the trunnion and gimbal, will be completed and documented.

### 7.8.5   Installation and Commissioning

Prior to installation of the MBS, the external rail system will be measured with respect to the DS geometric bore.  During the initial installation, the MBS will be lowered onto the external rail system.  The upstream end will be placed onto the "gimbal" support and the downstream end will be placed onto a temporary support that rests on the external rail system.  The MBS will then be rolled into the DS bore and placed into its approximate final axial position. The components will be measured in X and Y with respect to the geometric bore of the detector solenoid.  Adjustment is not expected to be necessary, because the tolerance of the MBS and support system is adequate to achieve the alignment criteria [57][59].

After the various detector components have been (similarly) aligned, the IFB will be moved into position, and the trunnions on the IFB will engage the trunnion socket on the MBS (see Figure 7.26 and Figure 7.27).  The temporary downstream MBS support will then be extracted. The MBS will then be coupled to the Calorimeter, Tracker, Proton Absorbers and Stopping Target. The axial attachment of all components will complete the "detector train".  The services for the tracker and calorimeter (cables and tubes) will be routed and connected to the IFB. Then the entire "detector train," consisting of all components, can then be inserted into the DS bore after the detectors are commissioned.

## 7.9   Downstream External Shielding

Shielding needs to be placed around the downstream Transport Solenoid (TSd) and the DS to limit the number of neutrons and gammas reaching the Cosmic Ray Veto Counter. This shielding will be supported on the lower level floor of the Mu2e Experiment Hall. The primary purpose of this shielding is to reduce the rates of neutrons and gammas incident on the CRV to an acceptable level.

### 7.9.1   Requirements

The physics requirements for the downstream external shielding are aimed at facilitating detection of the experimental signature events by reducing the experimental background rates.  In the Mu2e experimental setup the downstream cave is strategically placed to surround the TSd and DS cryostats.  The end cap shielding encloses the downstream end of the muon beamline vacuum enclosure.  The requirements for the downstream external shielding [60] are as follows:

- Reduce the neutron and gamma background incident upon the Cosmic Ray Veto (CRV) Counters.
- Allow a line of sight to the muon stopping target monitor and reduce the neutron and gamma background incident upon the muon stopping target monitor.

In addition to the above physics requirements, the downstream external shielding must satisfy the following mechanical and structural requirements:

- Provide a base for support of the CRV modules (as well as facilitate temporary storage of additional CRV modules removed from the upstream end during access to the antiproton window assembly at the TSu/TSd interface). The materials used in the downstream external shielding must therefore be able to withstand the loads imposed on them by other parts within the assembly as well as the loads of the CRV modules and associated mounting fixtures [61].
- Facilitate access to the IFB and the detector train inside the DS.
- Facilitate access to the antiproton stopping window assembly and the collimator rotation mechanism located at the TSu/TSd interface.
- Accommodate passage of power, cryo and vacuum services to the DS while reducing rates of particles escaping through this penetration.
- Reduce rates of particles escaping through the TSd trench as appropriate.
- Reduce rates of particles escaping through the DS trenches as appropriate.
- Stay clear of the solenoids during installation, and satisfy the constraints imposed by the building geometry.
- Provide a relatively economical structure that is mechanically stable and serviceable while allowing for adequate exit pathways.

Material requirements depend upon a number of factors beyond the mass of the components, including forces induced by effects of the solenoid magnetic fields on the steel reinforcement bars [39], and capacity to temporarily store CRV modules on top of the shielding while accessing the antiproton window assembly at the TSu/TSd interface.

### 7.9.2   Technical Design

The Downstream External Shielding is constructed of concrete and high density (barite) blocks, and will be completely external to the TSd and DS cryostats.

The current configuration of this shielding is primarily composed of reinforced concrete T blocks assembled into two independent (but mated) sections. The "Downstream Cave", which surrounds the body of the TSd and DS, and the "End Cap Shielding", which serves to enclose the downstream end of the muon beamline vacuum space. The fully assembled downstream external shielding is illustrated in Figure 7.28. The upstream open end of this shield will be placed against the downstream end of the TSu External Shielding (WBS 5.4, see Section 7.4) [31].

As described above and shown in Figure 7.29 below, the downstream cave is placed around the TSd and DS, and the end cap shielding surrounds the downstream end of the muon beamline vacuum enclosure. All downstream external shielding is supported independently of the DS.

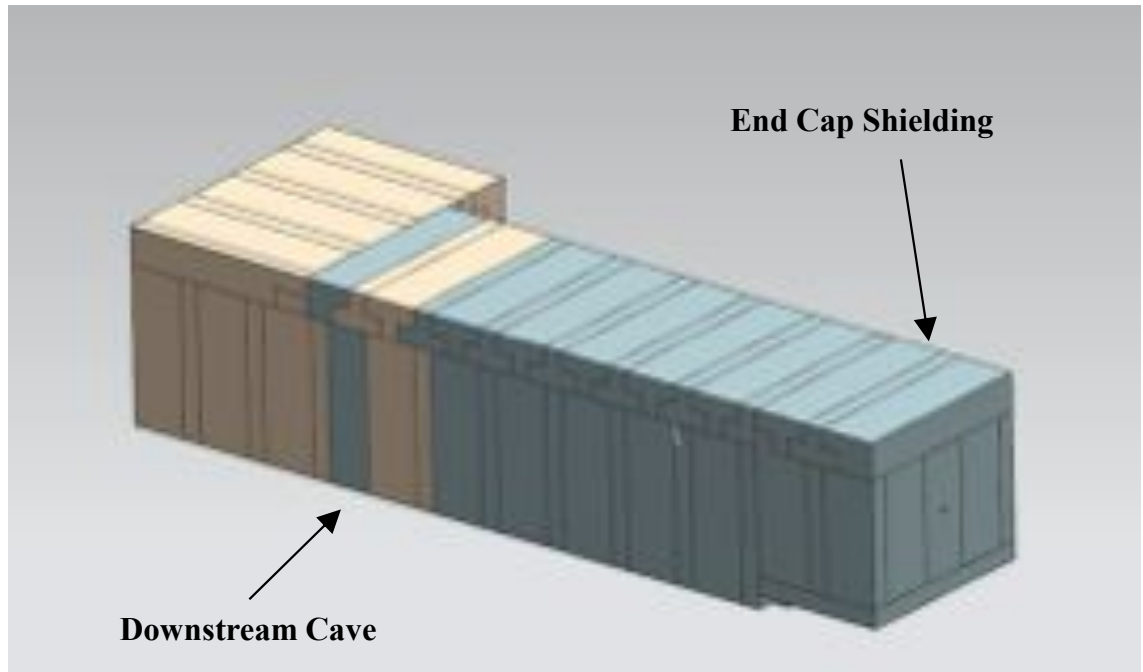

Figure 7.28. An overview of the Downstream External Shielding, including the downstream cave and end cap shielding. Note that the Hilman rollers under the end cap shielding to facilitate movements are not represented in this view.  The TDR baseline design employs high density concrete at the upstream end and in the vicinity of the muon stopping target, indicated by the brown blocks, and normal density concrete elsewhere, indicated by the blue blocks.

The current design of the shielding requires 36-inch thick concrete to sufficiently suppress the rates at the CRV [35].  The upstream end of the downstream cave, which surrounds the TSd is composed of high density (barite) concrete [28]. The concrete blocks surrounding the muon stopping target region of the DS (~ 2m) are also to be composed of high density (barite) concrete, while the remainder of the concrete blocks are normal density concrete.

The downstream external shielding is constructed primarily of concrete blocks.  The blocks include steel reinforcement bars and brackets.  An illustration of a "typical" concrete block is shown in Figure 7.30

The concrete blocks include steel reinforcing bars and steel angles to protect the block's edges (not shown in Figure 7.30). The default plan is to employ normal carbon steel reinforcement bars. Stainless steel reinforcement bars will only be used if demonstrated to be necessary due to magnetic field constraints.  In an attempt to minimize line of sight cracks, the current plan is to assemble the downstream external shielding primarily from T blocks.  Detailed drawings of similar T blocks are available at [63].

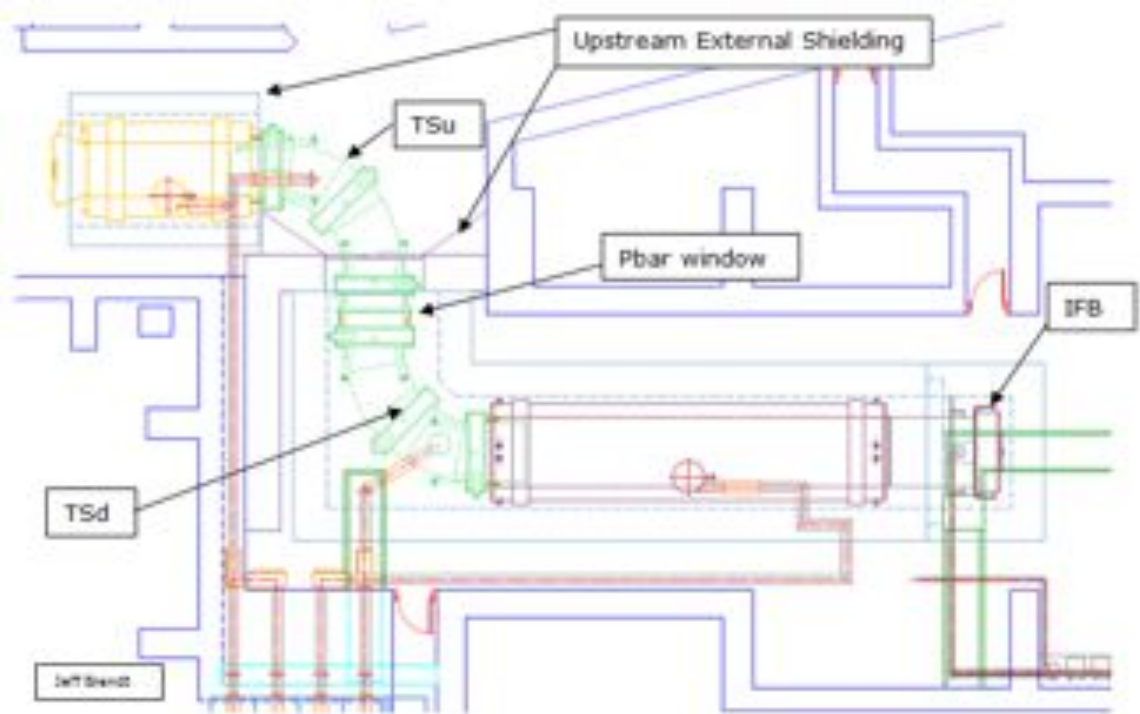

Figure 7.29. Plan view of the Upstream External Shielding (in light blue surrounding the PS and purple isolating the primary proton beamline from the DS hall) and the Downstream External Shielding (in blue surrounding the TSd and DS).

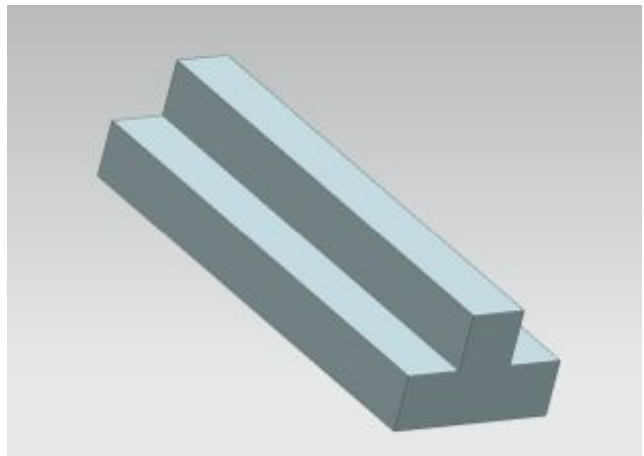

Figure 7.30. Illustration of a typical concrete shielding block.

### Downstream Cave

The downstream cave is composed of 409 tons of high density concrete and 312 tons of normal density concrete.

The nominal dimensions for the downstream cave are shown in Figure 7.31, viewed from above and in Figure 7.32 viewed from the south side.  Figure 7.31 and Figure 7.32  also show the location of the downstream cave with respect to the Mu2e coordinate system
[23]

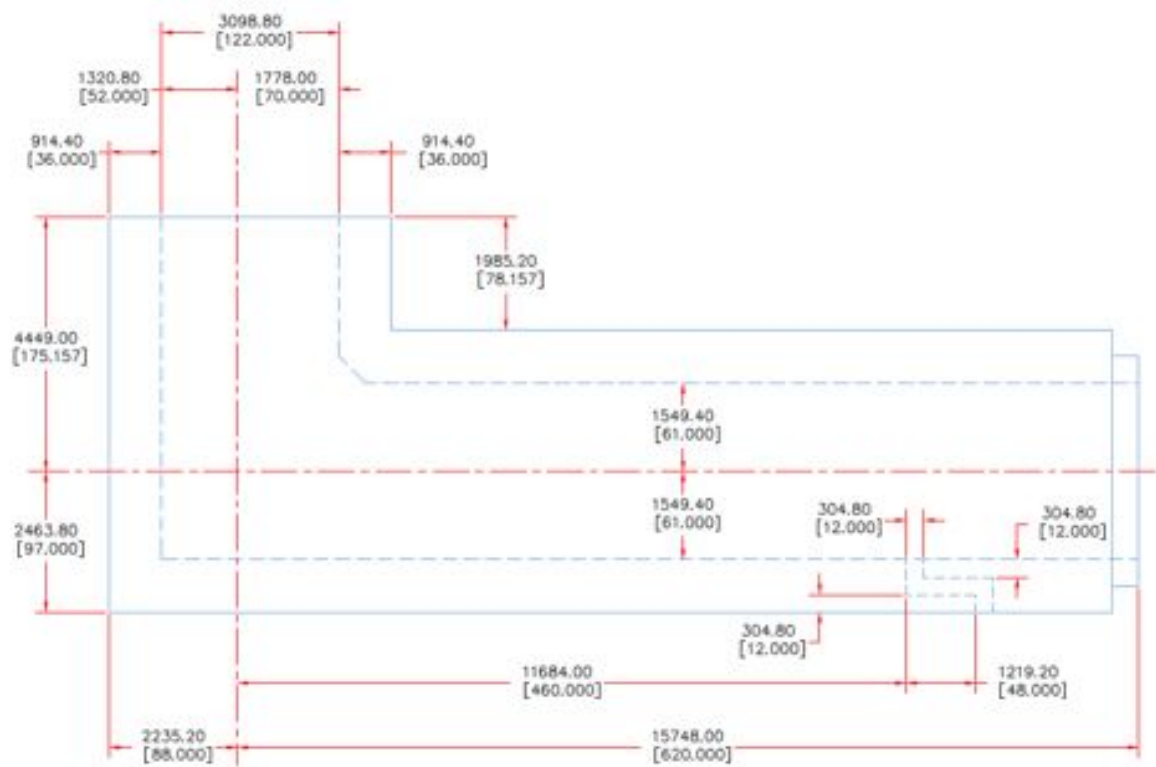

Figure 7.31. Plan view of the downstream cave. Note the penetration labyrinth for the services
for the DS. Red dashed lines represent centerlines of the TS and DS. Dimensions in mm [and in
inches].

### End Cap Shielding

The end cap shielding is composed of 117 tons of normal density concrete.

The end cap shielding encloses the IFB. The IFB serves as the interface for services,
cables and pipes from the Tracker and Calorimeter in the downstream muon beamline
vacuum space. A view of the end cap shielding is shown in Figure 7.33. A side view
(looking north) of the end cap shielding (without supporting rollers) is shown in Figure
7.34, and an end view (looking west) is shown in Figure 7.35.

### Alignment and Construction

The side walls of the downstream cave must be installed after the solenoids are already in
place, so they will all need to be installed from the outside of the solenoid envelope, and
will be positioned close to the solenoids without contacting them. To facilitate
installation, the plan is to fix standoff angles on the floor just inside the nominal location so that the side wall blocks can be lowered within a few inches of the floor and then moved up against the stop for reliable and efficient alignment.

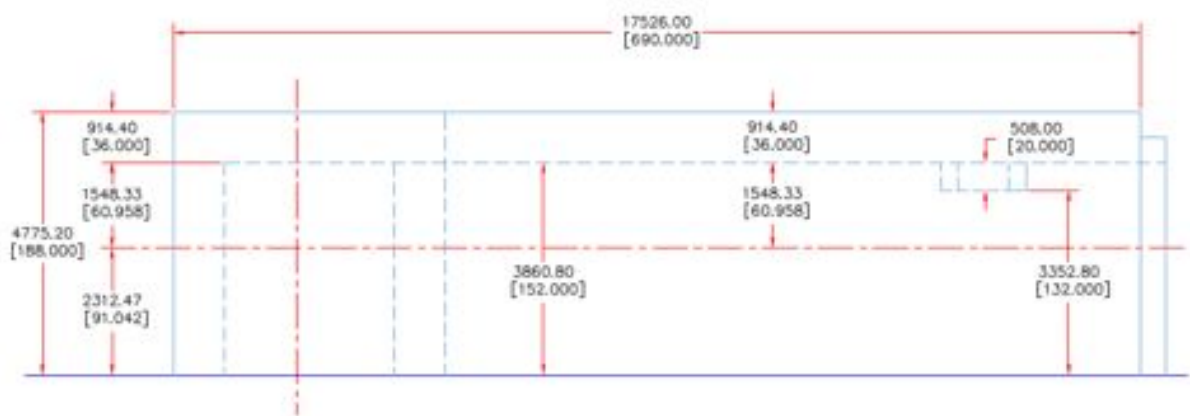

Figure 7.32. Elevation view of the downstream cave (as seen from the south side). Red dashed lines represent centerlines of the TS and DS. Dimensions in mm [and in inches].

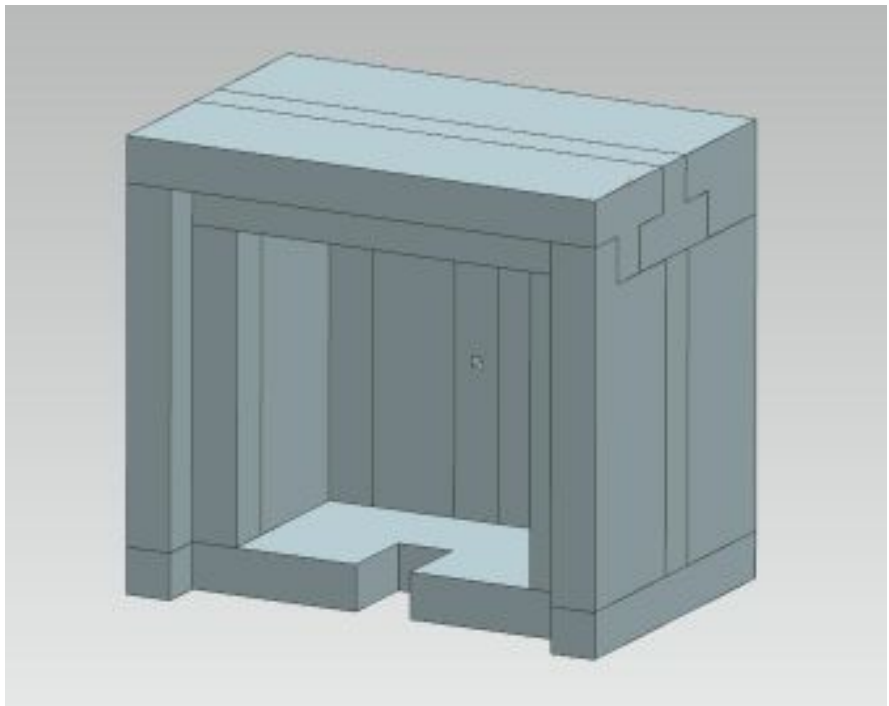

Figure 7.33. End cap shielding as seen from the upstream end.

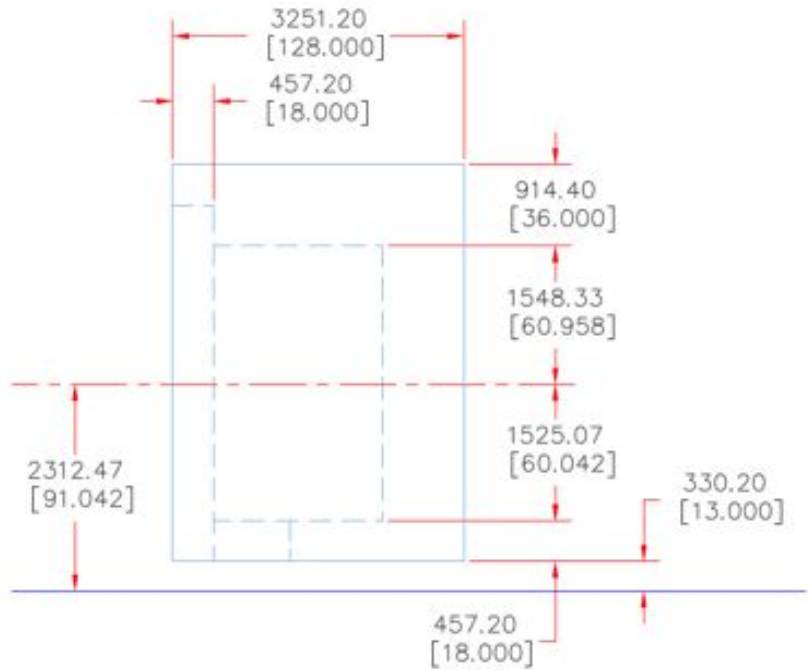

Figure 7.34. Elevation view of the end cap shielding (rollers supporting the end cap shielding in the 13 inch gap above the floor not shown). The red dashed line represents the centerline elevation of the DS. Dimensions in mm [and in inches].

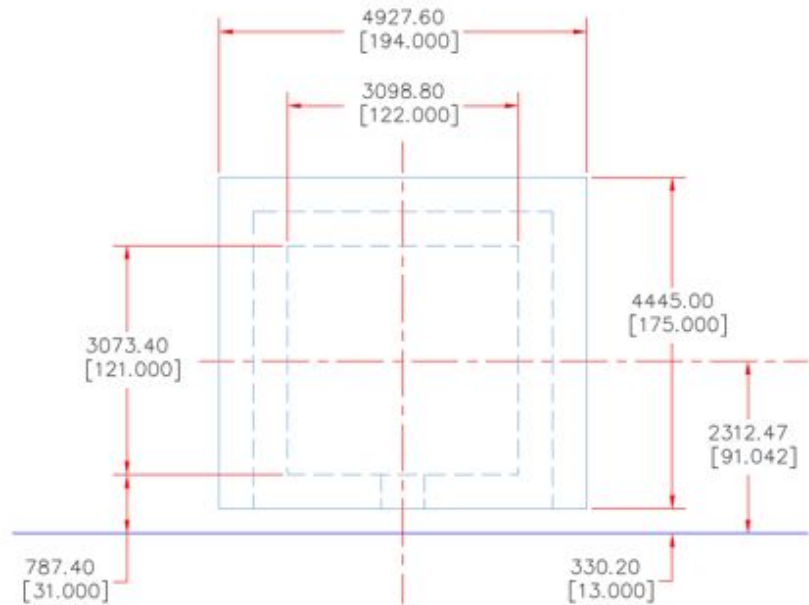

Figure 7.35. End view of the end cap shielding. Red dashed lines identify the centerline of the DS. Rollers supporting the end cap shielding are not shown. Dimensions in mm [and in inches].

Since the location of the downstream cave is constrained on the upstream end by the upstream external shielding and on the north, west and south sides by the solenoids and on east side by the north south DS trench, and the intent is to minimize line of sight cracks, the tolerances on the blocks and the spacing between blocks must be carefully controlled to accommodate these various constraints. The plan is to build in an intentional gap between neighboring block surfaces to account for the tolerances. Since the blocks are T blocks, this will result in local thinner regions in the shielding, but will not generate line of sight cracks (with the exception of the line of sight cracks between the side walls and the top blocks).

The assembly is also sensitive to variations in floor elevation. Substantial variations in the floor level will result in significant complications during the assembly of the downstream cave. If it turns out the floor is not level within the anticipated tolerances, the blocks dimensions may need to be adjusted by shimming and grouting as required [60].

Since the CRV modules will be supported from the downstream external shielding, and that shielding will be a concrete block assembly, a stay clear needs to be incorporated in the planning for the CRV modules. The CRV team should plan on mounting the CRV no closer that one inch from the nominal surface locations illustrated in Figure 7.31, Figure 7.32, Figure 7.34 and Figure 7.35.

Concrete block tolerances should accommodate the above cited assembly constraints.

The shielding is to be assembled from many pieces, and the components are designed such that line-of-sight cracks are minimized. Nevertheless, where cracks are unavoidable, such as penetrations for services and/or lifting fixture openings, it may be necessary to pack such spaces with filler material (such as poly beads or sand or borax) as needed.

The downstream external shielding is made primarily of concrete blocks reinforced with steel bars. It is very unlikely that a sufficient quantity of blocks in appropriate sizes will be available from the Fermilab stock to address these shielding needs and conform to the specialized space and size constraints. So, due to these space constraints in the building, as well as the specialized sizes of the blocks involved, new blocks will need to be procured. If simulations show that the steel reinforcement bars impact the magnetic fields to an unacceptable level, blocks with stainless steel reinforcement bars may need to be procured.

Individual blocks will be sized so that they can be manoeuvred using a 30 ton building crane whenever possible.

### 7.9.3   Risks

There is a risk that the shielding may be inadequate, allowing a flux of neutrons and gammas that exceeds specifications.  The muon stopping target is a significant source of neutrons that could impact the detector performance.  If background rates in the TSd and DS regions are higher than anticipated, the particles incident on the cosmic ray veto counters will exceed the expected levels. To mitigate this risk, detailed simulations using G4beamline and MARS have been performed, and comparisons provide some indication of the level of uncertainty. Shielding has been significantly enhanced as a consequence of these studies and the CRV has been redesigned to rely upon four layers so that accidental coincidences can be suppressed.

Also, rates of neutrons and photons incident on the CRV in the vicinity of the MBS may generate a need for additional shielding in the vicinity of the MBS. CRV livetime is dependent upon the incident particle rates, and will suffer if those rates become too high. Rates can be reduced by increasing the steel in the MBS or enhancing the external shielding in the vicinity of the MBS [64].

Since the shielding is much more massive than the Mu2e detector, the installation of the shielding may influence the alignment of the muon beamline. Civil construction plans attempt to take into account the anticipated loads. No substantial degradation in performance is anticipated assuming the muon beamline remains in a plane since the plan is to adjust the incident proton beam to accommodate shifts in the primary target elevation. Substantial local variations might require a re-alignment of the incident primary proton beamline, or else incur an iterative process of assessing performance and moving enough shielding to facilitate shimming and then putting the shielding back into place.  Hydrostatic levels will be installed on the solenoids to provide monitoring of the relative local changes in elevation.

A substantial mass and volume of shielding will need to be installed in tight spaces and in some places beyond crane coverage after the solenoids are in place.  This creates a threat of damage to surrounding elements. The elements most likely at risk are solenoids (and CRVs during re-installations after access to the TSu/TSd interface region). Mitigation includes constructing brackets to facilitate placement of shielding components and carefully planning the installation sequence and methods.

This shielding is not expected to be installed until after the magnetic field mapping is complete, so none of these risks have the potential to be realized prior to that time.

### 7.9.4   Quality Assurance

Thorough structural, thermal and magnetic analysis will be completed to ensure all components will meet the requirements described in [60].

Close collaboration between Fermilab personnel and the vendor who makes the barite blocks will take place to ensure that the barite mix is appropriate with respect to density and manufacturability.

All blocks, both concrete and barite, will undergo inspection by the vendor before shipment to Fermilab. Test pieces may be required to ensure that manufacturing processes will be viable. All components will be inspected by FNAL personnel upon arrival at Fermilab, and discrepancies will be documented.

A study will be completed and documented with regard to the resistance of high density concrete to freezing cycles, and the storage methods of the barite blocks will be designed to accommodate this issue.

Tests of shimming methods, space available for components and for servicing after block installation, will be carefully considered. Assembly procedures will be verified to the fullest extent possible.

### 7.9.5   Installation and Commissioning

The installation will be complicated because the shielding is heavy and large and needs to be positioned accurately and near delicate equipment. Furthermore, some of this shielding is not directly under building crane coverage due to the ceiling beam between the TS hatch and the DS hatch.

The installation sequence will be dictated by external constraints. It is anticipated that the solenoids and solenoid services will all be in place prior to final positioning of the downstream external shielding, and the initial round of solenoid alignment and field mapping must be completed prior to final positioning of at least some of this shielding to facilitate access to the solenoid supports and the transport solenoid coil supports.

A description of the installation/assembly sequence for the downstream cave is as follows:

- Install the floor level block alignment feature, (the standoff angle mentioned above).
- Begin the sequence of installation of side wall blocks starting from the upstream end and moving towards the downstream end in repeated steps as follows:

- o Install first vertical T block (call this block α for the moment) with long side nearest solenoid
- o Temporarily position appropriate spacer next to this block
- o Install next T block (call this block γ) with long side nearest solenoid
- o Extract temporary spacer from between the two blocks α and γ just installed
- o Install T block (call this block β) with long side away from solenoid between T blocks α and γ
- o And repeat this sequence until the side walls are complete.

There are at least two special cases associated with the installation of the side wall blocks. Both are addressed in [60].

1. Due to the hatch beam that supports the TS hatch blocks, direct crane access to the region near the upstream end of the DS is occluded, complicating the lifting process for blocks in this area.
2. The blocks which are to be located under the DS cryogenic line may require a special lifting fixture or tools to move into final position, and hand stacked blocks may be required to generate part of the labyrinth around the service line.

Once the side walls of the downstream cave are installed, the ceiling blocks can be installed, starting with the TSd ceiling support cross beam near the upstream end of the DS. Pairs of T blocks with the long sides down must be installed before the mating block can be installed in between the pair. The ceiling blocks under the hatch beam may be installed using a coordinated lift and spreader.

Part of the downstream cave will need to be removed to provide access to the COL3u and COL3d collimator/antiproton stopping window area if/when service is required. Staging of these blocks will likely require outside storage and handling capacity for these large blocks should that situation arise. Unfortunately, barite concrete has been reported to have poor frost resistance, losing 45 to 60% of its strength after 25 frost cycles, so special considerations for the storage of this high density concrete may be appropriate [40].

If necessary, the end cap shielding assembly/installation could be completed after the detector train has been inserted into the DS bore so that the space is available during initial installation and commissioning.

Since it is anticipated that at least some of this downstream external shielding may have to be moved after initial installation, the blocks will be clearly labeled to distinguish the high density blocks from the standard blocks, as well as to designate the location of the individual blocks in the final assembly.

# 7.10  Detector Support and Installation System

## *7.10.1*  Requirements

The detector support and installation system is required to transport and align components within the Detector Solenoid warm bore.  The muon stopping target, proton absorbers, tracker, calorimeter, and muon beam stop must be transported accurately and safely into position and aligned with respect to the standard Mu2e coordinate system [23]. The components vary significantly in mass (from less than 3 kg to nearly 5000 kg) as well as in their alignment accuracy requirements. These components will be supported by the inside wall of the Detector Solenoid cryostat.

Physics requirements dictate the overall size, location and placement accuracy of the individual components within the DS bore. In addition, the support structure must not impede particle trajectories or lead to an enhancement of detector rates or physics background as a result of interacting particles.  A detailed description of the requirements and specifications for the detector support and installation system can be found in [59].

## *7.10.2*  Technical Design

An overall view of the detector train and the detector support and installation system is shown in Figure 7.23 and Figure 7.36.  The components are supported by two rails and transported on linear ball bearings. Two separate rail systems will be implemented, the "internal" and "external" systems.  Once installed, the alignment of all components will be maintained by the internal rail system.  Figure 7.23 shows the components inside the Detector Solenoid bore supported by the internal rail system, while Figure 7.36 shows the components on the external system (or staging area) before insertion into the DS.

### *Internal Rail System*

Each component is enclosed within an individual support structure.  Descriptions of the individual support structures are included in the preliminary design sections for each component (see for example Section 7.7).

The support structure for each component will be mounted onto the rail system and aligned to the Detector Solenoid cryostat inner wall independently in the vertical (Y) and lateral (X) directions.  Vertical and lateral adjustment of components will be done using adjustment mechanisms illustrated in cross section in Figure 7.37. The rails will be shimmed/aligned via the 2$^{nd}$ tier bars and attached to stainless steel support platforms that are welded onto the inside wall of the DS cryostat.  The rails and the cryostat wall will support the weight of each component, allowing all alignment criteria to be achieved.  The rails and bearings are made exclusively of non-magnetic components described in [59]. The alignment criteria for each component are specified in [59]. A detailed breakdown of issues that affect the alignment is given in [66].

Structural analysis will verify that the loads of the objects imposed on the rail system will not exceed the load capacity of the rail system or the DS bore tube, respectively.

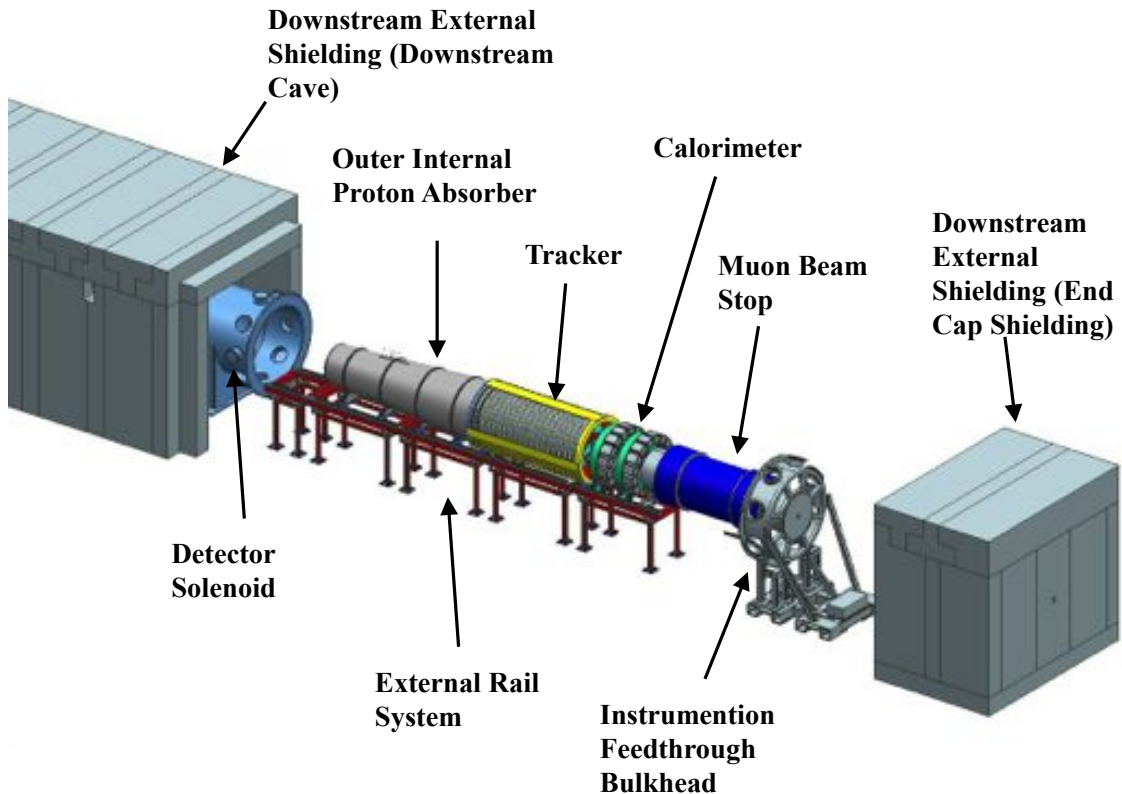

Figure 7.36. Detector components positioned on the external rail system.

*External Rail System*

The external rail system is positioned outside of the Detector Solenoid as the detector train is extracted and is used to support the detector train during servicing, and to move components into position inside the DS warm bore for operation. The external rail system consists of a series of 6 stands, each of which can be installed or removed as needed. The external stands are made of aluminum, each with sections of rails mounted to the top surface that can be connected and disconnected accurately. The rails will be identical in cross section to those used for the internal system, and the last stand, closest to the cryostat, will be attached to the internal system during installation. Figure 7.38 shows a single external stand with the rail attached. Figure 7.39 shows a connection between rails on two stands, using a rail "link", and shows a plan view of the external rail system with the components in position to begin insertion of the detector train into the DS.

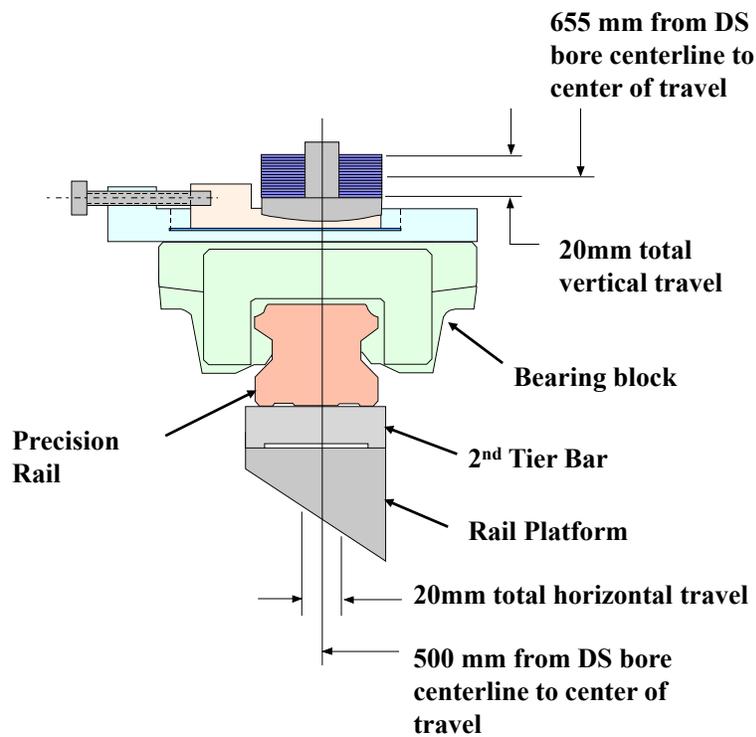

Figure 7.37. Cross section of an internal rail mounted on the rail platform, and supporting a bearing block outfitted with a custom adjustment mechanism.

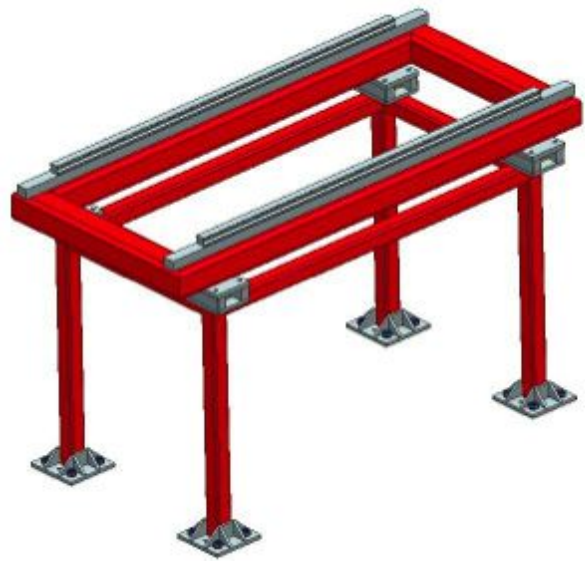

Figure 7.38. Individual external rail system stand.

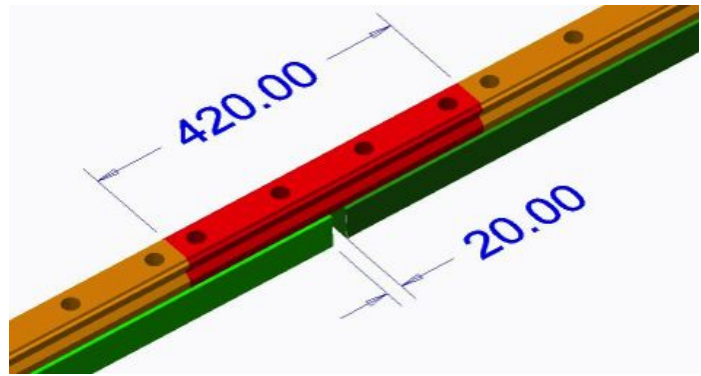

Figure 7.39. External rail system "link" attachment, which spans between two external rail stands (in green).

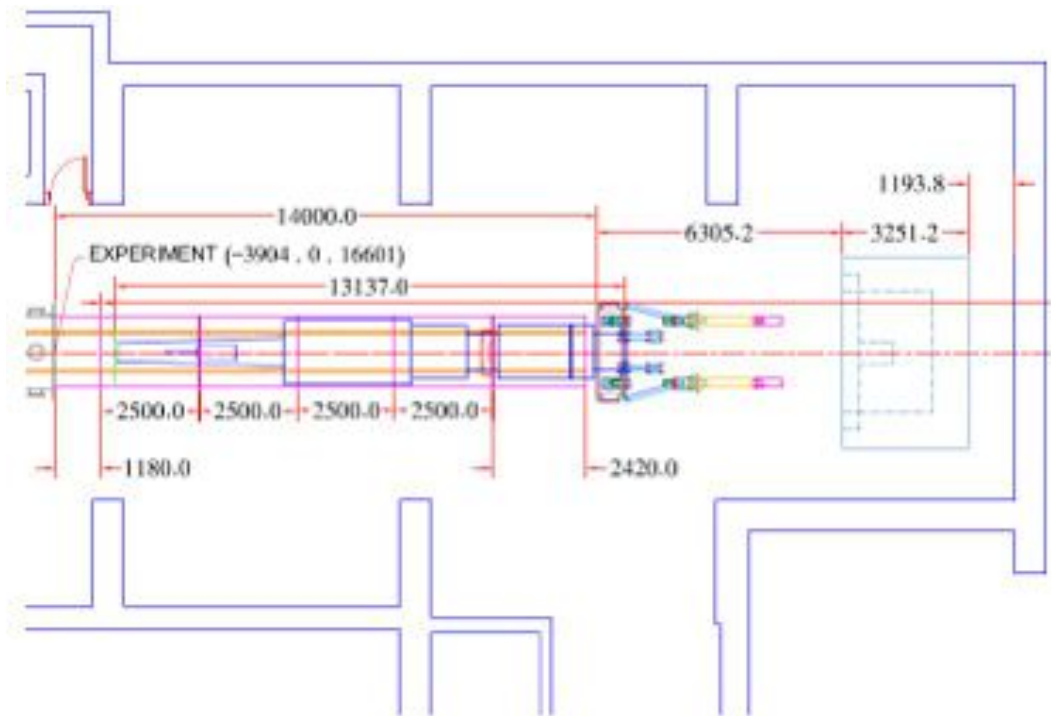

Figure 7.40. Plan view of the complete external rail system, showing the detector train in the fully extracted position.

Cables, fibers, cooling tubes and source tubes from the tracker and calorimeter will be installed and attached to the Instrumentation Feed-Through Bulkhead (IFB) before moving the entire assembly, including the detector components, into position [59][4]. The cables and tubes will pass over the muon beam stop before being terminated in the IFB as shown in Figure 7.41. As a result, the tracker, the calorimeter, the muon beam stop and the IFB must be rigidly attached to one another in the axial direction and moved into place as a single unit. In addition, to facilitate axial alignment and transportation, the proton absorbers and the muon stopping target will also be axially attached to the other components. After being individually aligned in X and Y, all components will be connected axially by coupling their respective bearing blocks, as shown in Figure 7.42. Each component will be measured axially with respect to the VPSP/IFB flange position and adjusted before the axial position is locked into place.

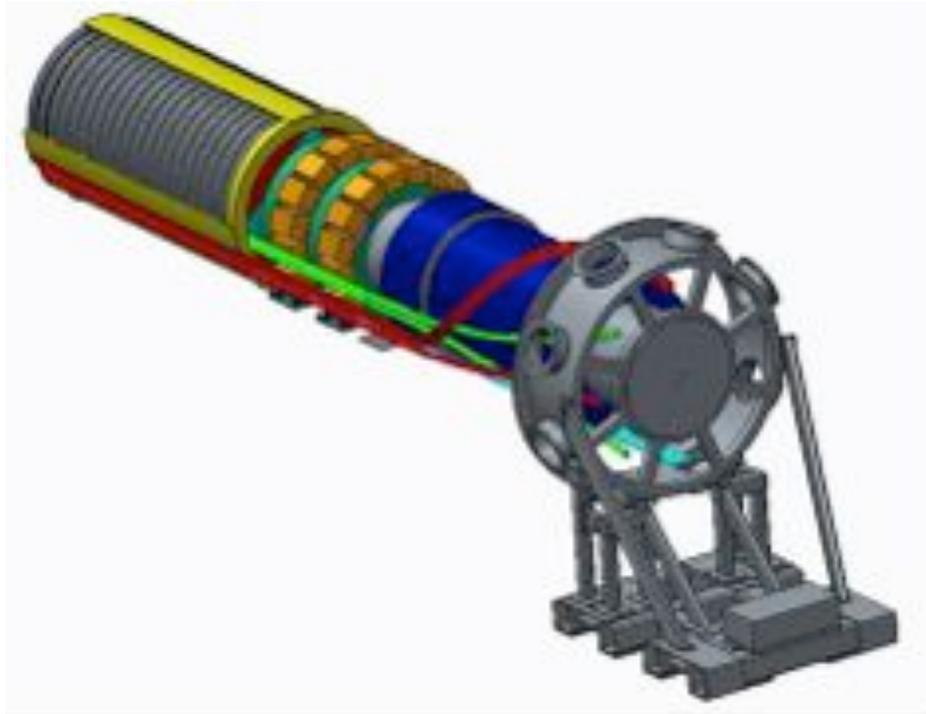

Figure 7.41. Cables and cooling tubes over tracker, calorimeter and muon beam stop, terminated at the IFB.

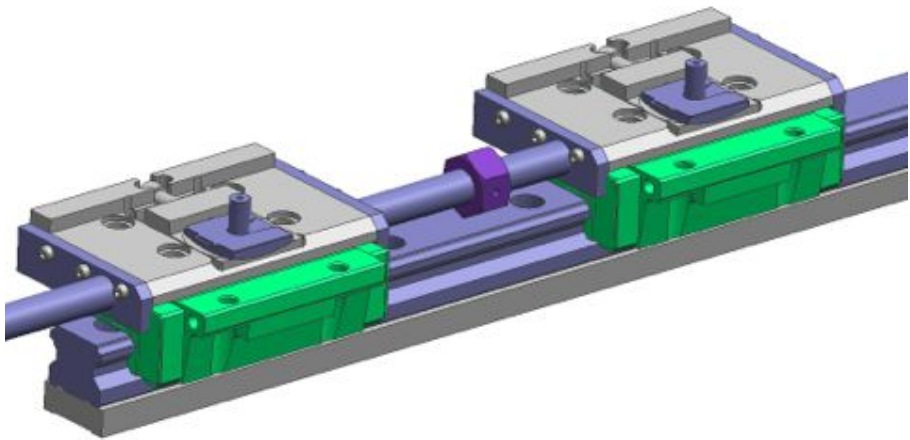

Figure 7.42. Two bearing blocks mounted on a rail segment connected via the axial coupling mechanism.

*Manufacturing and Assembly*

Several manufacturers have been identified who make rail systems that will potentially fit the requirements of the project with respect to accuracy and load, but achieving both of these criteria while simultaneously satisfying the strict magnetic criteria of the Mu2e project has proven to be challenge, although vendors have been identified.

The supports for the internal rail system will be welded to the inside wall of the Detector Solenoid by the cryostat manufacturer before the DS is delivered to FNAL. The rail system and bearing blocks will be manufactured outside Fermilab by the rail system vendor and shipped to Fermilab. The parts will then be mounted by Fermilab personnel to the existing supports, aligned with respect to the center of the DS geometric bore and installed.

### *7.10.3* Risks

Due to the complexity and scope of the Mu2e installation activities, and given the numerous constraints imposed by the parallel installation of other components, the installation activities could take longer than expected, resulting in delays. To mitigate this risk, the installation schedule will be worked through in detail, including access limitations and potential incompatibilities between activities. Any measurements, shimming and installation of rails within the solenoid bore that can be done before installation in the assembly hall will be considered. If schedule slippage occurs, a second shift for installation activities can be added.

There is also a moderate risk that the tracker and/or calorimeter could move due to forces generated by the solenoid magnetic fields, or other changes in the operating environment. Movement would cause misalignment of either detector, introducing particle momentum measurement error. The mitigation plan for this risk is to complete FEA calculations of all forces and stresses expected within the bore area during operation, including measurements of prototype weights on a test mockup. Also, the tracker will be instrumented to monitor orientation including magnetic field measurement capability.

### *7.10.4* Quality Assurance

Thorough structural, thermal and magnetic analysis will be completed to ensure all components will meet the requirements described in [59].

All components will be inspected by FNAL personnel upon arrival at Fermilab, and discrepancies will be documented. Rails and bearings will also undergo documented inspections and testing by the manufacturer before being delivered to Fermilab.

A full size rail system mockup (8 meters long) has been assembled. Tests of shimming methods, rail and bearing deflections, coefficients of friction, space available for components and for servicing, binding issues, movement for thermal contraction, adjustment of components and alignment of components are being completed and documented. Assembly procedures will be verified to the fullest extent possible.

### *7.10.5* **Installation and Commissioning**

A brief summary of the installation process follows:

The Detector Solenoid, including the rail platforms provided by the DS vendor, will initially be measured with respect to fiducials placed within the detector hall. The internal rails will be installed accurately with respect to the DS bore by use of specially machined 2$^{nd}$ tier bars bolted to the rail platforms as shown in Figure 7.37 and described in [59]. All components will then be placed on the rails and aligned.

During the initial installation, each component will be lowered onto the external rail system, rolled into the DS bore and placed into its approximate final axial position. The components will be measured in X and Y with respect to the geometric bore of the detector solenoid, comparing targets placed on the component structures to the initial bore measurements.

Several iterations may be required to achieve the alignment criteria. A final measurement will then be performed [59].

After the initial installation has been completed, the detector train can be rolled out of and back into that position on the set of rails provided by the detector support and installation system. The rail system is designed to maintain the appropriate level of reproducibility every time the components are extracted and re-inserted into the DS bore.

This page intentionally left blank

# 8    Tracker

## 8.1    Introduction

The Mu2e tracker provides the primary momentum measurement for conversion electrons. The tracker must accurately and efficiently identify and measure 105 MeV/c electrons while rejecting backgrounds and it must provide this functionality in a relatively unique environment. The tracker resides in the warm bore of a superconducting solenoid providing a uniform magnetic field of 1 Tesla[i]; the bore is evacuated to $10^{-4}$ Torr. A key feature of Mu2e is the use of a pulsed beam that allows for elimination of prompt backgrounds by looking only at tracks that arrive several hundred nanoseconds after the proton pulse (see Figure 8.1). The tracker must survive a large flux of particles during the early burst of "beam flash" particles that result from the proton pulse striking the production target, but it does not need to take data during this time. The Mu2e signal window is defined as $700 < t < 1695$ (Figure 8.1), where $t = 0$ is the arrival of the peak of the beam pulse at the stopping target. However, in order to study backgrounds such as radiative pion capture, the tracker must be fully efficient during the interval $500 < t < 1700$ nsec; we take this as the tracker's live window. To calibrate using positrons from $\pi^+ \rightarrow \nu e^+$ decays the tracker must also be able to collect data, during special runs with reduced beam intensity, for $300 < t < 1700$ nsec. An overview of the key tracker parameters appears in Table 8.1.

Table 8.1. Overview of key tracker parameters

| | |
|---|---|
| Number of Straws | 23,040 |
| Straw Diameter | 5 mm |
| Straw Length | 430 – 1200 mm, 910 mm average |
| Straw Wall | 15 µm Mylar (2×6.25µm plus adhesive) |
| Straw Metallization | 500Å aluminum, inner and outer surface 200Å gold overlaid on inner surface |
| Gas Volume (straws only) | $4 \cdot 10^8$ mm$^3$ (0.4 m$^3$) |
| Sense wire | 25 µm gold-plated tungsten |
| Drift Gas | Ar:CO$_2$, 80:20 |
| Gas gain | $3\text{-}5 \cdot 10^4$ (exact value to be set later) |
| Detector Length | 3196 mm (3051 mm active) |
| Detector Diameter | 1620 mm (1400 mm active) |

---

[i] The "uniform" field actually includes a controlled gradient to avoid trapping soft particles in accidental inhomogeneities. The gradient is small and well controlled; it is accounted for in track fitting but does not otherwise affect the tracker performance.

The dominant interactions for stopped muons are radiative muon capture, $\mu^- N \rightarrow \gamma \nu N'$, and decay in orbit: $\mu^- \rightarrow e^- \nu_e \nu_\mu$. The former frequently leads to ejected protons, neutrons, and photons from nuclear breakup. The protons are a source of background hits which, due to high ionization rates, are a concern for crosstalk but are also easily distinguished from hits expected for a signal electron. Photons, directly from the decay or delayed from ejected neutrons, create low energy electrons from scattering and conversions. These electrons do not traverse a large number of straws but generate hits which, unlike proton hits, cannot be rejected before pattern recognition.

Decay in orbit, or DIO, produces electrons that are distinguishable from the signal only by their momentum. The differential energy spectrum of DIO electrons, shown in Figure 8.2, falls rapidly near the endpoint, approximately[ii] proportional to $(E_{endpoint} - E_e)^5$. The most prominent feature is the well-known Michel Peak. Nuclear recoil slightly distorts the Michel peak and gives rise to a small recoil tail that extends out to the conversion energy. The Mu2e tracker is optimized to distinguish conversion electrons from DIO electrons. The key to this is momentum resolution, with particular emphasis on DIO electrons that smear up in energy and appear to be signal electrons. A precision, low mass tracking detector in a magnetic field is the most practical way of achieving the required precision. A tracking system has been designed to meet this requirement while being blind to most of the rate from DIO electrons.

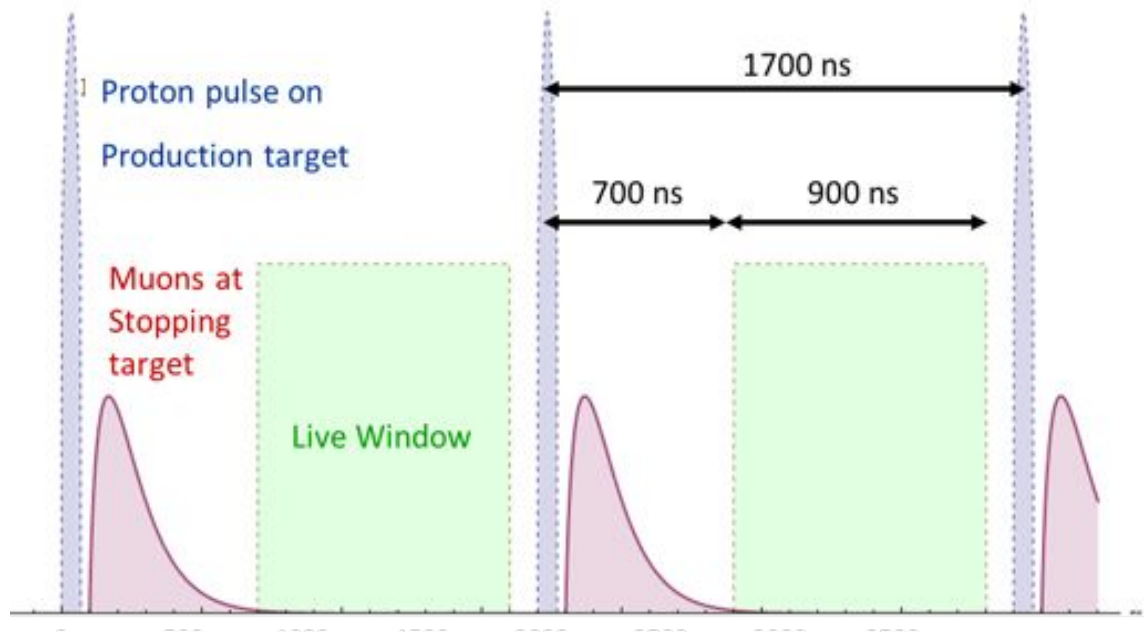

Figure 8.1. The structure of the beam sent to the Mu2e detector.

---

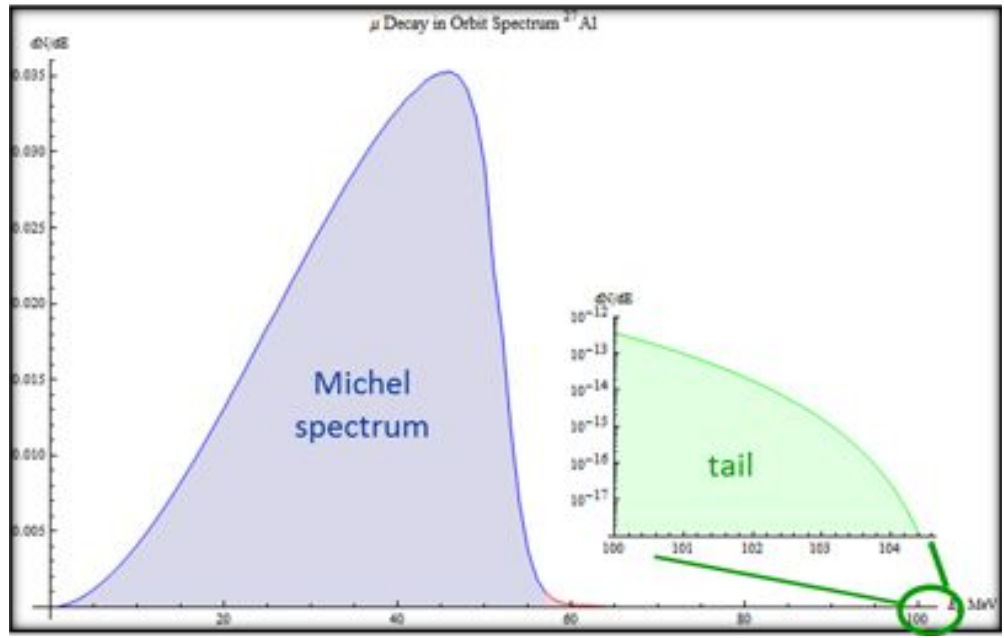

Figure 8.2. Electron energy spectrum from muon decay in orbit. Recoil against the nucleus results in a small recoil tail that extends out to the conversion energy (inset).

## 8.2 Requirements

Requirements for the tracker have been documented elsewhere [1] and are only summarized here.

The Detector Solenoid [2] provides a uniform 1 Tesla field in the region occupied by the tracker. To have good acceptance for signal electrons without being overwhelmed by DIO electrons (including electrons scattered into the active region), the active area of the tracker extends from about 40 < r < 70 cm (where radius r is measured from center of the muon beam)[iii]. Mechanical support, readout electronics, etc. are to be placed at r >70 cm, out of the way of both signal and DIO electrons. These dimensions depend on the size and geometry of the muon stopping target, the size of the muon beam, and the magnetic field properties; they have been optimized to maximize the acceptance to conversion electrons while minimizing the number of low energy electrons that intersect the tracker.

The momentum resolution requirement is based on background rejection: the signal is sharply peaked, whereas backgrounds are broad (cosmic rays, radiative pion capture) or steeply falling (DIO electrons). For a Gaussian error distribution the requirement is that $\sigma$ < 180 keV/c. This is simply a convenient reference point; the actual resolution is not

---

[iii] 105 MeV/c tracks can reach 71 cm, but those exceeding 70 cm have a helix that is too tight to be effectively fit.

Gaussian and may be asymmetric. Furthermore, scattering and straggling in material upstream of the tracker are significant contributors to the final resolution.

The detector is surrounded by vacuum. This improves resolution by minimizing scattering through the measurement region and d$E$/d$x$ straggling throughout the entire path for signal electrons. It reduces scattering of DIO electrons into the tracker. And it avoids stopped muons from interaction between the residual gas and remnant muon beam.

Because of the need to break and later re-establish vacuum, access to the detector is expected to require several days of downtime. Therefore the tracker will be designed for an overall mean time to failure (MTTF) of >1 year[iv]. A few dead channels do not constitute a tracker failure if they can be kept isolated. The ability to isolate local failures is an important part of the design. In particular, the capability to remotely disconnect high voltage to each straw will be implemented so that a few broken or otherwise defective wires will not necessitate an access.

The device must tolerate (but need not take useful data during) a "beam flash" prior to the live window, and be efficient at the peak rates the beginning of the live window. Actual rates depend on detector geometry and are presented later.

During the live window, of order half the hits in the detector are from slow protons ($\lesssim$100 MeV/c momentum or $\sim$5 MeV kinetic energy) ejected from the stopping target. The tracker must have d$E/$d$x$ capability to distinguish such protons from electrons.

## 8.3   Mechanical Construction

The selected design for the Mu2e tracker is a low mass array of straw drift tubes aligned transverse to the axis of the Detector Solenoid, referred to as the T-tracker. The basic detector element is a 25 μm sense wire inside a 5 mm diameter tube made of 15 μm thick metalized Mylar®, referred to as a straw. This choice is based on several points.

- The straw can go from zero to 1 atmosphere pressure differential (for operating in a vacuum) without significant change in performance.
- Unlike other types of drift chambers, each sense wire is mechanically contained within a straw. Thus, failures remain isolated, improving reliability.

---

[iv]   An MTTF of 1 year leaves a significant probability of needing an unscheduled repair. However, the resulting data loss is expected to be <1% [1].

- The transverse design naturally places mechanical support, readout electronics, cooling, and gas distribution at large radii.

The detector has ~23,000 straws distributed into 20 measurement stations across a ~3 m length. Each station provides a ~200 μm measurement[v] of track position.

Each straw is instrumented on both sides with preamps and TDCs. Each straw has one ADC for d$E$/d$x$ capability. To minimize penetrations into the vacuum, digitization is done at the detector with readout via optical fibers. Electronics at the detector will *not* require an external trigger: all data will be transferred out of the vacuum to the DAQ system, and a trigger may be implemented as part of the DAQ.

### 8.3.1   Straws

The T-tracker is made from 5 mm diameter straws, the assembly of which is shown in and Figure 8.4. Each straw is made of two layers of ~6 μm (25 gauge) Mylar®, spiral wound, with a ~3 μm layer of adhesive between layers, for a total wall thickness of 15 μm. The inner surface has 500 Å aluminum overlaid with 200 Å gold as the cathode layer. The outer surface has 500 Å of aluminum to act as additional electrostatic shielding and reduce the leak rate.

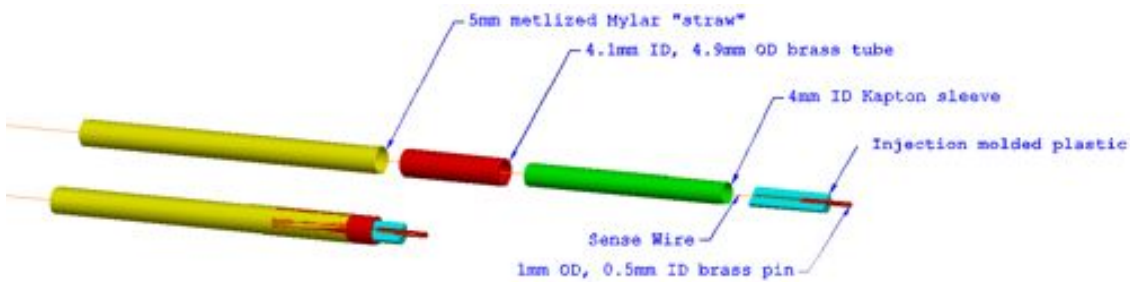

Figure 8.3. Straw termination, shown exploded and assembled. The brass tube connects to the straw with silver epoxy. The green insulator slips inside a brass tube (red) to prevent breakdown near the tube end. The sense wire is soldered into the brass pin, and epoxied to the injection molded plastic. After assembly the brass tube allows connection to the cathode while the brass pin allows connection to the anode.

A 4.95 mm outer diameter brass tube is mechanically and electrically connected to each straw end using silver epoxy. Inside the brass tube is an extruded Kapton® tube to protect against breakdown at the edge of the brass tube. Inside the Kapton® tube is an injection molded plastic insert. Attached to a groove in the insert is a small, U-shaped brass pin. A 25 μm gold plated tungsten wire is soldered to the pin as well as epoxied to the plastic insert. Both brass parts are gold-plated to ensure good solder and epoxy joints.

---

[v] Resolution measured along the drift direction, which is a line from the sense wire to the point of closest approach to the sense wire.

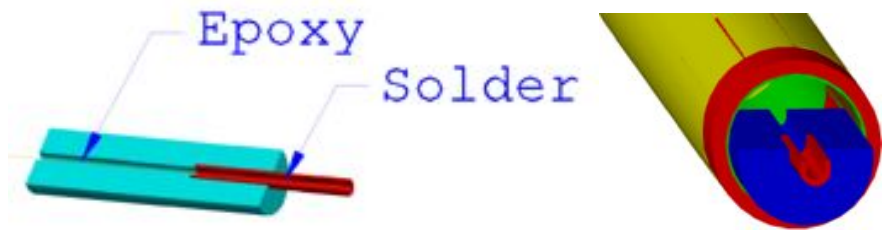

Figure 8.4. Straw termination details. On the left, the plastic insert and brass pin used to hold the sense wire. On the right, a close-up of the completed straw. Note the wire is externally aligned, then soldered and epoxied to lock the position.

Sense wire location is set by external fixturing during assembly, not the straw termination. The wire position is then locked in by solder and epoxy. The brass pin allows up to 250 μm mispositioning of the straw body without impacting the sense wire location.

Straws vary in length from 334 mm to 1174 mm active length. The straws are supported only at the ends and kept straight with an initial tension of 700 g. The primary mechanical component of the straw is Mylar[®]. As with any plastic, Mylar[®] under stress creeps (gradually stretches), and is sensitive to humidity. However, both creep and hygroscopic expansion are less with Mylar[®] than the more widely used Kapton[®] straws. Due to stress relaxation, the straw tension will relax to ~400 g over the lifetime of the experiment, as seen in .

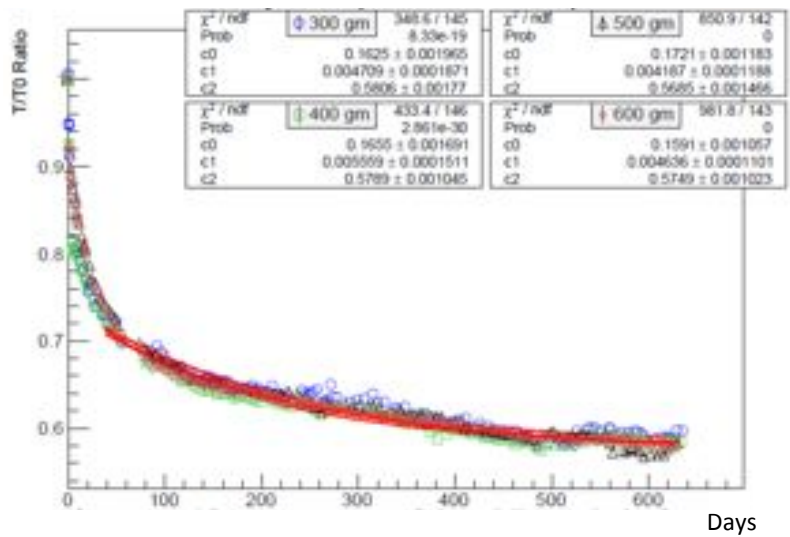

Figure 8.5. Tension versus time, fit to the form $T(t)/T(0) = c0e^{-c1t} + c2$. Independent of initial tension, the tension for $t \rightarrow \infty$ is ~58% of the initial tension.

### 8.3.2   Straw Assemblies

Groups of 96 straws are assembled into ***panels*** as shown in Figure 8.6. Each panel covers a 120° arc with two layers of straws, as shown in Figure 8.7. The double layer improves efficiency and helps determine on which side of the sense wire a track passes (the classic "left-right" ambiguity). A 1.25 mm gap is maintained between straws to allow for manufacturing tolerance and expansion due to gas pressure. This necessitates that individual straws be self-supporting across their span.

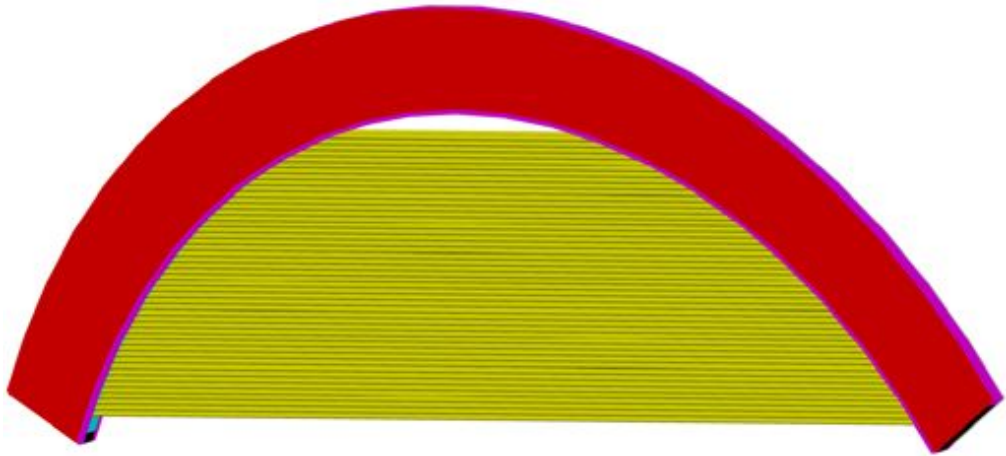

Figure 8.6. Completed panel, with covers shown in red. Screws to attach covers not shown.

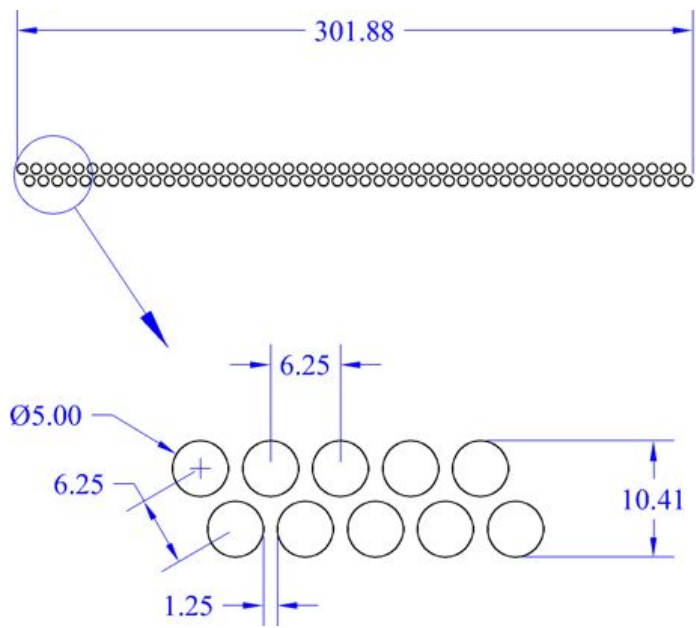

Figure 8.7. Edge view of a panel showing the arrangement of straws within a panel. Dimensions are in millimeters.

The structure of a panel includes the following parts, as seen in Figure 8.8. Materials are selected to minimize cost without unacceptable degradation in rigidity.

- Inner ring.
    - 1/8″ upper and 3/16″ thick lower stainless steel rings
    - Two plastic inserts, created on a 3D printer, with 96 holes (plus alignment features) each
    - 316 Stainless steel filler in the gap between plastic inserts
- Base plate. 3/16″ aluminum
- Outer ring. 5/8″ aluminum
- Covers. 3/16″ aluminum (seen only in Figure 8.6).
- O-ring and screws (not shown) to complete the gas seal

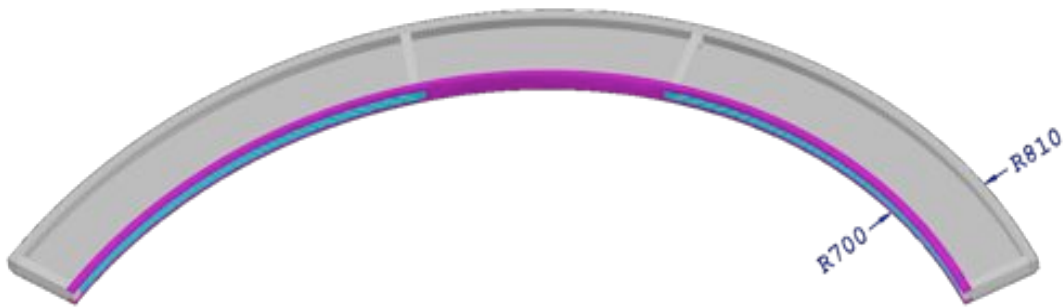

Figure 8.8. Panel without covers or straws. Magenta regions are stainless steel alloy 316; gray are aluminum; and cyan is plastic. Dimensions are in millimeters.

Deformation in the x and y directions are shown in Figure 8.9. The panel is assumed to be held flat in z. In the final assembly, a similar constraint in z is achieved by assembling panels into planes, and attaching covers, while holding the panels flat. Wires run in the x direction; movement in x affects tension, but not wire position. The ≲300 μm change in length from x distortion will need to be compensated for by slightly higher wire tension during assembly. The more critical dimension for alignment is y. The worst-case straw movement is ~120μm. FEA results will be used to correct for this movement, and the correction will be checked by the X-ray survey of wire positions as described below.

As noted in Section 8.3.1, straws and sense wires are aligned separately to avoid accumulating errors. This is done with the Panel Assembly and Alignment System (PAAS). The PAAS, as used for straws, is shown in Figure 8.10; as used for wires, in Figure 8.11.

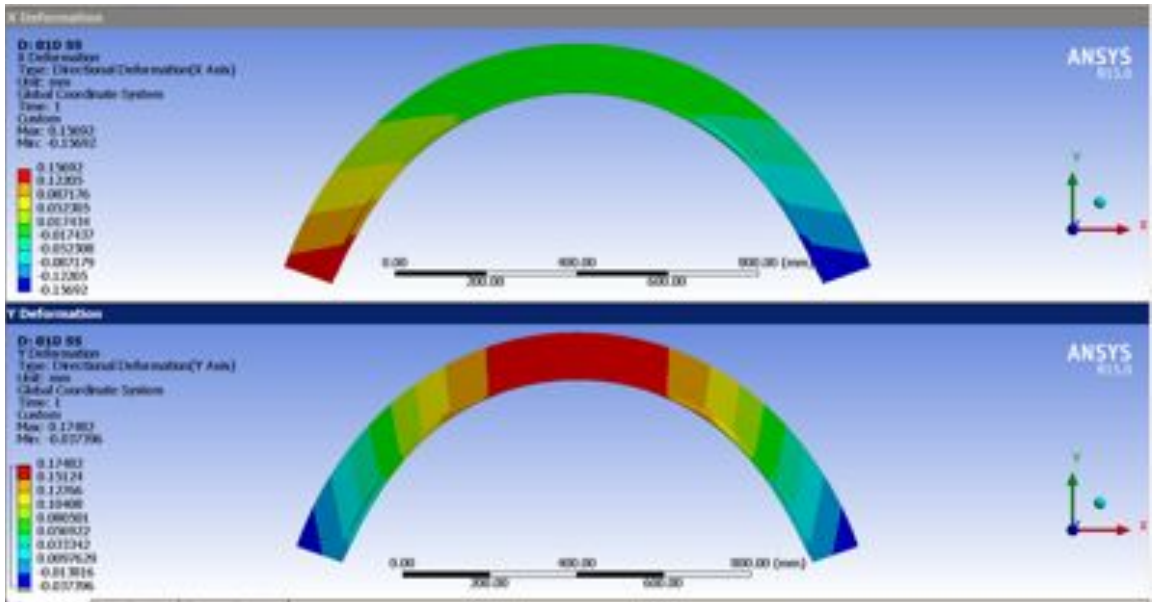

Figure 8.9. Distortion of a panel in x and y while constrained to remain flat in z.

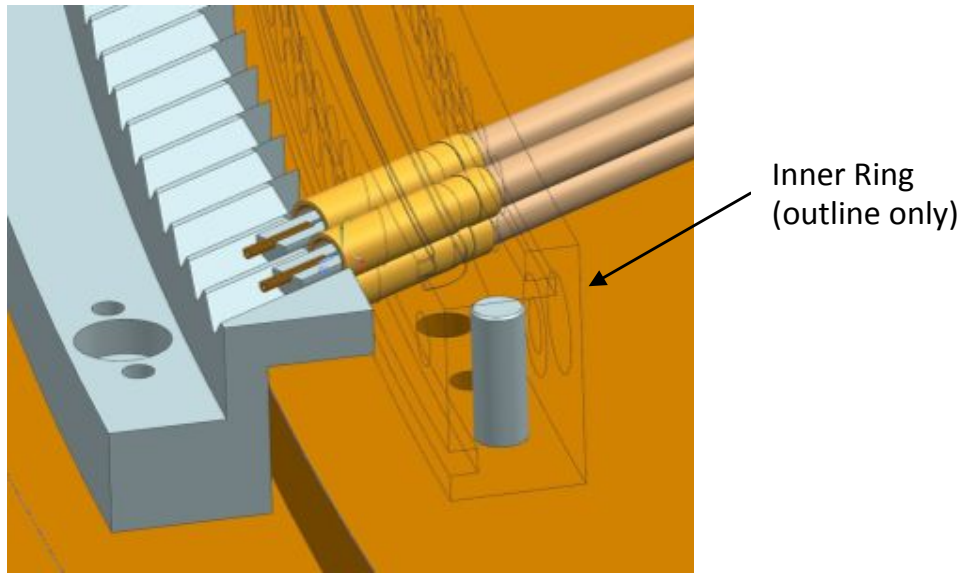

Figure 8.10. Straw alignment on the PAAS. 3D printing is used to make V-grooves to align the straw end. For simplicity, the inner ring is shown as an outline only.

Each panel requires the following utilities to enter from the exterior vacuum side to the interior gas side.

- Gas. 2×¼" lines (supply and return). Stainless steel tube epoxied into a hole in the outer ring.
- Power. 2×8A (supply and return). Standard vacuum feedthroughs.

- High voltage. Single line, standard vacuum feedthrough.
- Copper signal lines. Number not yet determined. For high-speed lines (calibration), vacuum rated SMA connectors. For low speed lines (CANBUS, monitoring), vacuum rated DB25 or DB9 connectors.
- Optical signal lines. 4 individual lines. Made by stripping the jacket from a standard optical cable, then epoxying the bare fibers into a hole in the panel. With triple clad fiber, light loss is estimated to be comparable to a standard feedthrough. However, robustness of this connection is yet to be tested.

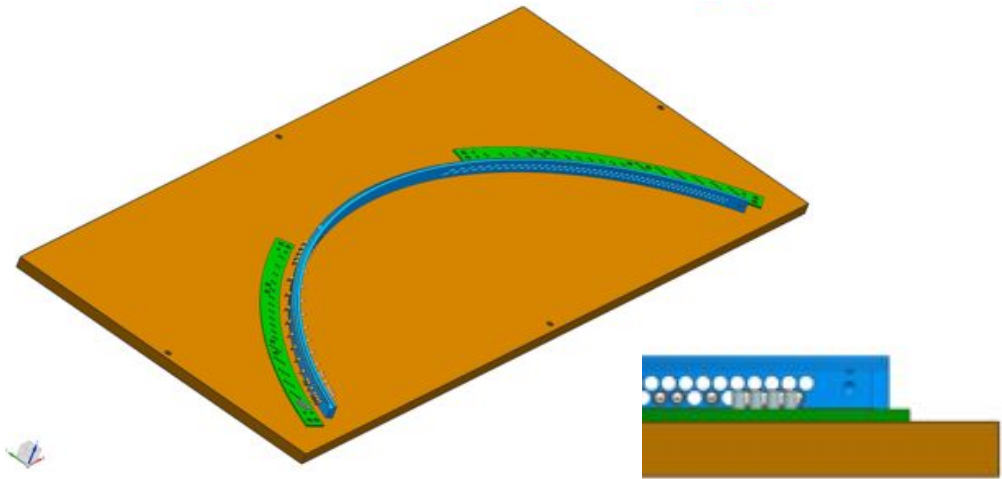

Figure 8.11. Wire alignment on the PAAS structure. Crossed dowel pins (vertical and horizontal) inserted in a wire alignment jig, are used to position the wire. For simplicity, straws are not shown, and pins for only four wires are included.

After panels are completed, wire positions are measured relative to survey points on the panel. This is done using an X-ray machine developed at Duke for ATLAS TRT straws [3]. Straw and wire tension will be re-measured shortly before using panels in the next step of assembly.

After panels are completed, wire positions are measured relative to survey points on the panel. This is done using an X-ray machine, Figure 8.12, developed at Duke for ATLAS TRT straws [3]. Wire (and potentially straw) positions can be measured to a precision of 25 μm (Figure 8.13). Straw and wire tension will be re-measured shortly before using panels in the next step of assembly.

Six panels are assembled into a **_plane_** as shown in Figure 8.14. Three 120° panels complete the ring of one face; another three panels, rotated by 30°, complete another ring on the opposing face. After the plane is assembly, a cooling ring is attached around the outer diameter. This arrangement has been found to give the best stereo performance [4] and is chosen despite the mechanical complications compared with 60° rotation.

A pair of planes forms a ***station***. The two planes are identical. During assembly the 2nd plane is rotated 180° around a vertical axis. The completed station is shown in Figure 8.14.

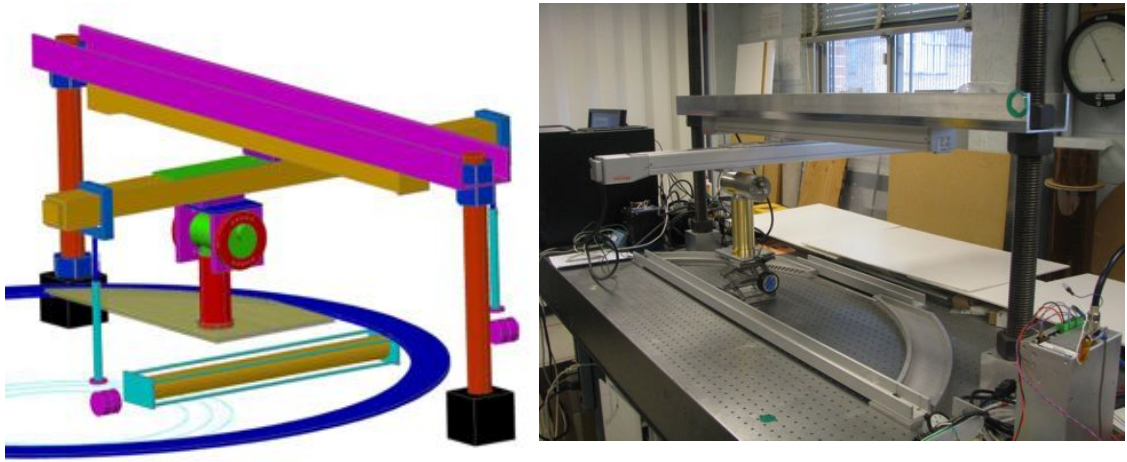

Figure 8.12. X-ray machine for measuring wire positions. The X-ray tube and collimator move on linear slides for scanning the panel.

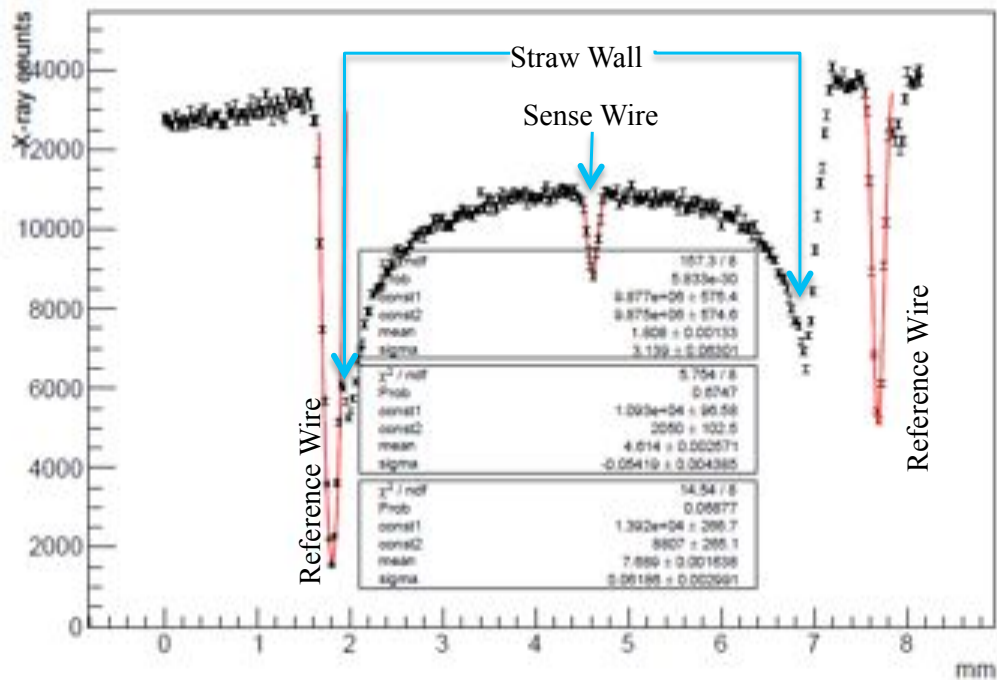

Figure 8.13. Measurements from X-ray machine scanning a sense wire inside a straw. Note the straw (cathode) location can also be measured, if desired.

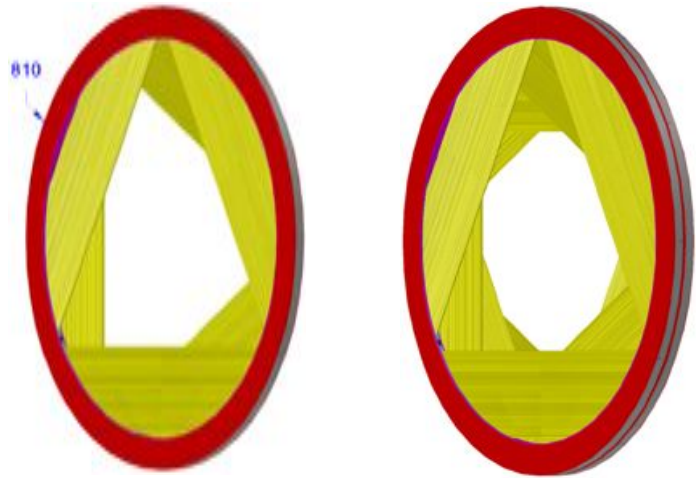

Figure 8.14. Isometric view of a tracker plane (left) with three panels each on the front and back face; and a station (right) consisting of two planes. Dimensions are in millimeters.

Panel orientations within a station seen in Figure 8.14 are tabulated in Table 8.2. φ is measured from the origin (detector center) to the middle of the panel, i.e. a panel with vertical wires will have φ = 0 or 180°. The local z coordinate, $z_L$, is measured from the station center to the center of the panel (midway between layers).

Table 8.2. Orientation of straws within a station. ZL refers to the local z coordinate, i.e. relative to the center of the station.

| | | Φ (degrees) | | | | | | | | | | | |
|---|---|---|---|---|---|---|---|---|---|---|---|---|---|
| | | 15 | 45 | 75 | 105 | 135 | 165 | 195 | 225 | 255 | 285 | 315 | 345 |
| | **-34.5** | | ✓ | | | | ✓ | | | | ✓ | | |
| **$z_L$** | **-12.5** | | | ✓ | | | | ✓ | | | | ✓ | |
| **(mm)** | **12.5** | | | | ✓ | | | | ✓ | | | | ✓ |
| | **34.5** | ✓ | | | | ✓ | | | | ✓ | | | |

### 8.3.3  Tracker Frame

Twenty stations are assembled into the completed tracker, shown in Figure 8.15. A breakdown of the number of components that make up the tracker (stations, planes, panels, etc.) is shown in Table 8.3. Horizontal beams maintain longitudinal alignment of the rings. The thicker ring seen at the upstream end, and the two thinner rings placed between stations at the downstream end, stiffen the structure. The frame distortion is shown in Figure 8.16. Stiffening rings and beams are stainless steel, pending further analysis and value engineering.

The entire tracker rests on four bearing blocks (not shown) placed near the horizontal beams and attached to the stiffening rings. The connection between the tracker and the bearing blocks are kinetic to avoid over-constraining and distorting the tracker frame, except all four points are constrained in the vertical direction (the frame is not rigid enough to support the weight without this). Vertical adjustment screws are included to level and center the frame.

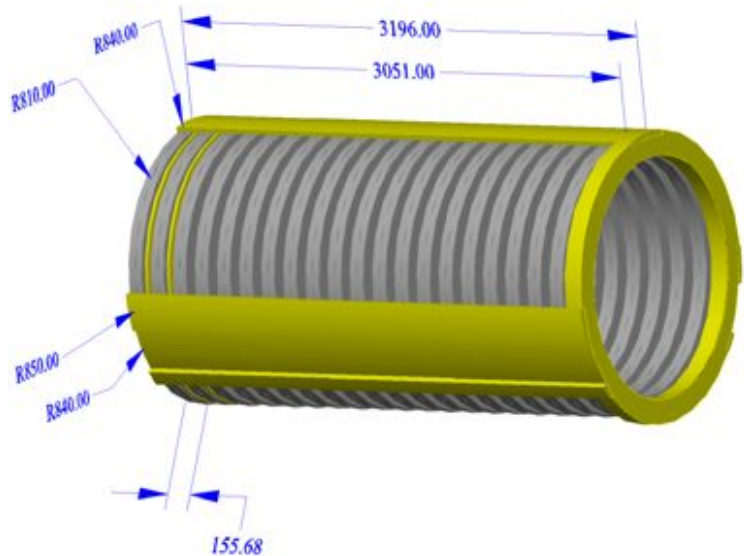

Figure 8.15. The assembled tracker, with 20 stations. Stations are shown in grey and support structure in yellow. Dimensions are in millimeters.

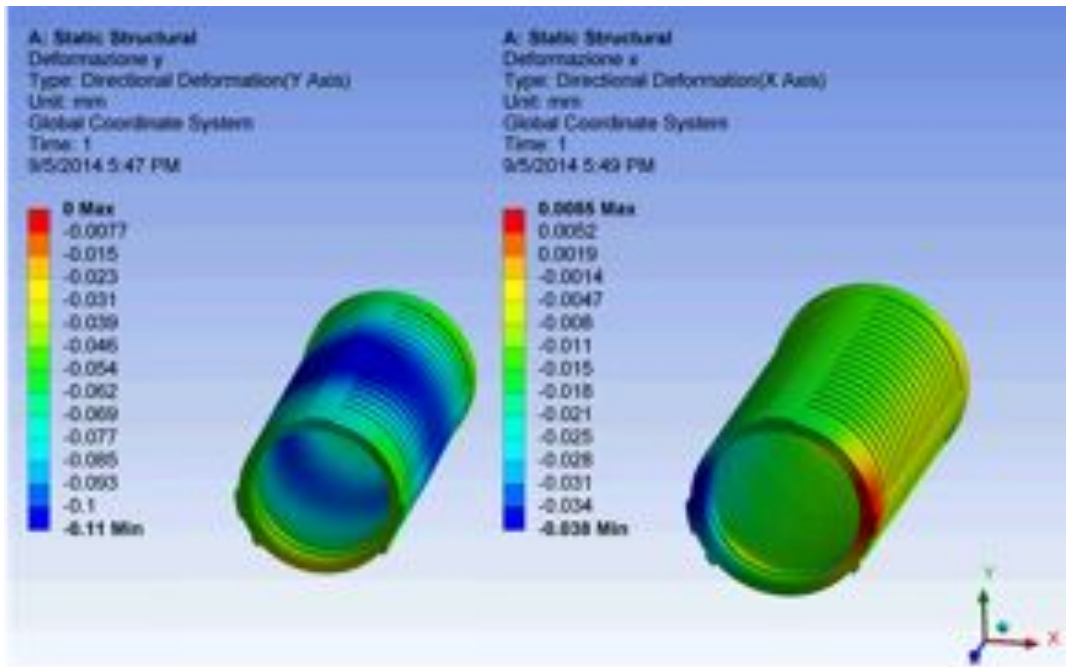

Figure 8.16. Frame distortion in the vertical (left) and horizontal (right). Maximum distortion is 100 μm in the vertical, 38 μm in the horizontal.

Table 8.3. Breakdown of the number of components in the Tracker. The straw total of 23,040 at the bottom of the second column is the product of the numbers above.

| Stations | 20 |
|---|---|
| Planes per station | ×2 |
| Panels per plane | ×6 |
| Layers per panel | ×2 |
| Straws per layer | ×48 |
| **Total Straws** | **23,040** |

The completed tracker will be surveyed before being moved to the mu2e building. Each panel will include monuments for this purpose, which are also used during the X-ray measurement of wire positions. Once installed, but outside the DS, a subset of the markers will be surveyed. Survey while inside the DS is of limited value as the original monuments cannot be sighted. Precision levels on the tracker frame will be used instead.

## 8.4 Front End Electronics

To minimize penetrations through the cryostat, digitizers and zero-suppression logic are located on the detector. Requirements for the electronics [5] and DAQ [6] system are documented separately. Here we summarize the implementation of on-chamber (front end) electronics. The flow of signals through the readout chain is shown in Figure 8.17.

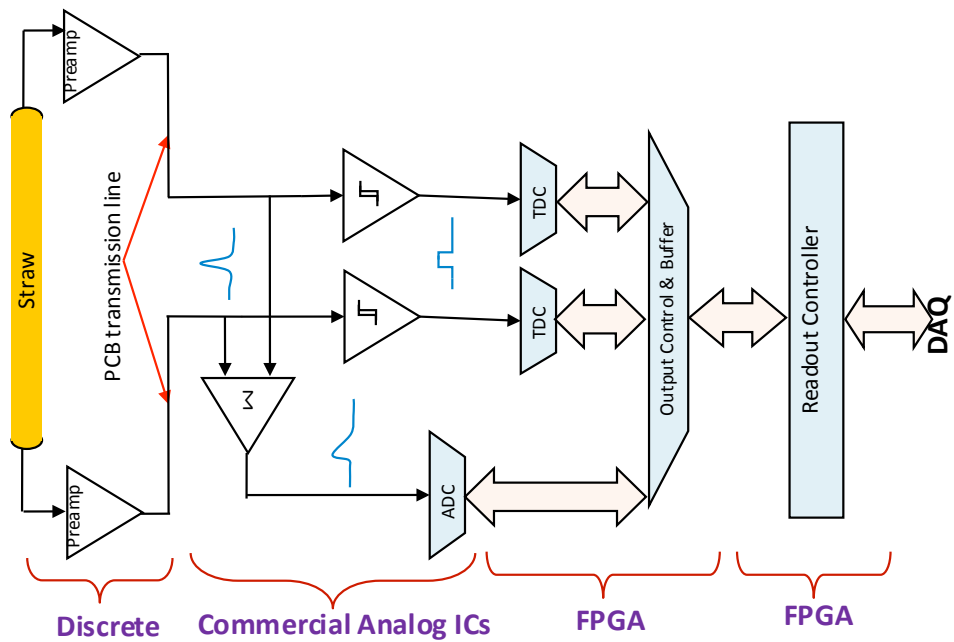

Figure 8.17. Signal flow through front end electronics.

The tracker uses "time division": pulse timing is measured at each end of the straw in order to measure the position along the wire. As seen in Figure 8.18 [7], resolution is ~3cm; this is used for pattern recognition before using stereo for getting a more accurate position. Pulse height is measured to provide d$E$/d$x$ for particle identification. Therefore each straw has:

- 2 preamp channels, 1 for each end.
- 2 TDC channels, 1 for each end.
- 1 ADC channel, measuring sum of both ends.
- 1 High voltage feed, with disconnect.

There are a total of 46,080 preamp and TDC channels, and 23,040 ADC channels.

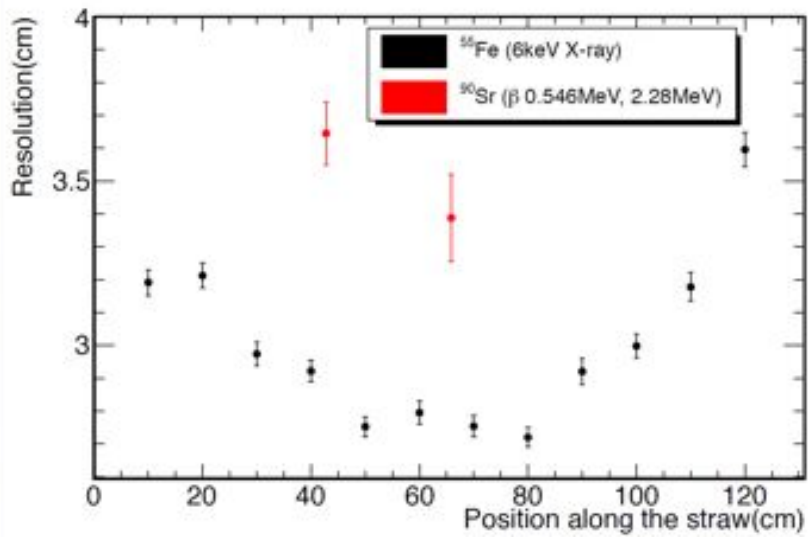

Figure 8.18. Time division resolution for $^{55}$Fe (X-ray) and 90SR (electrons) using a 500Msps commercial digitizer.

### 8.4.1   Preamp

As shown in Figure 8.19, preamps are located at each straw end; this is required to get proper (~300Ω) termination of the straw. There is one channel per preamp board. Half the boards provide high voltage, the other half provide calibration pulsing. A schematic for the preamp (excluding high voltage disconnect and calibration) is given in Figure 8.20

High voltage passes through a "fuse" that can be blown remotely to isolate broken wires without the need for an access. The fuse consists of a miniature beryllium-copper spring with one side soldered to the PC board using a low-temperature solder. To blow the fuse, that joint is heated by a resistor till the solder melts. A layer of Teflon over the connection prevents splattering of molten solder without preventing the spring from pulling back. Current to each heating resistor is controlled by an addressable switch with a unique address. To avoid accidentally blowing fuses, power for this circuit is separate from the preamp power and is normally locked out.

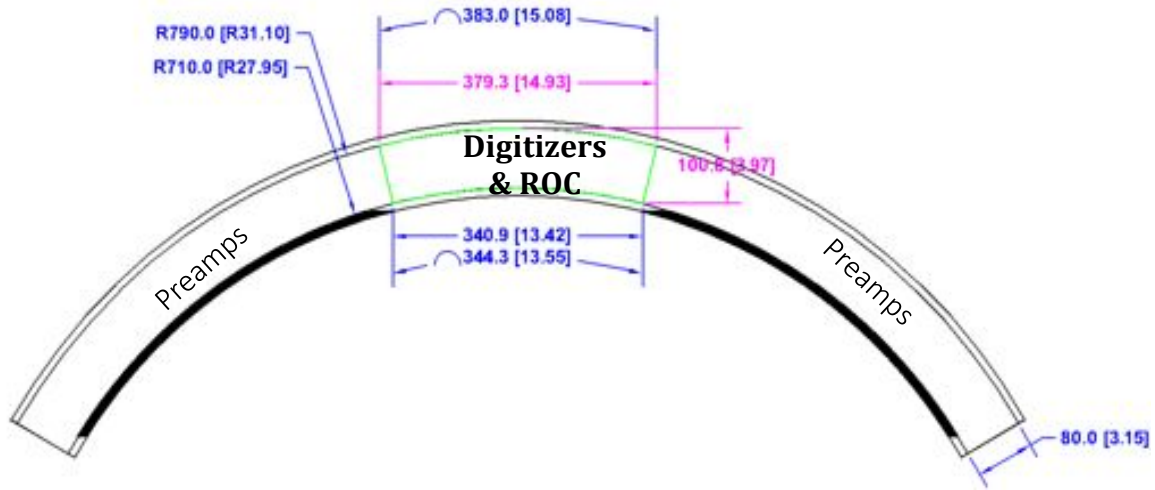

Figure 8.19. One panel in the mu2e tracker. The complete tracker contains 240 panels.

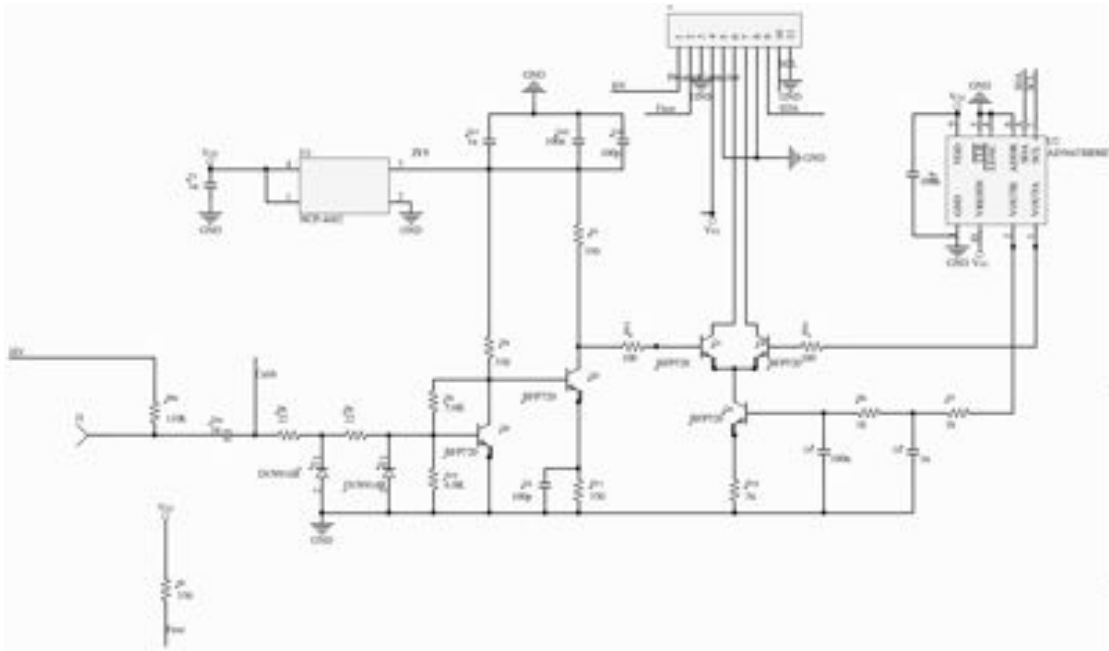

Figure 8.20. Schematic of preamp. A nanoDAC[®] on each preamp board allows offset and gain to be remotely controlled. Calibration and HV disconnect circuitry not shown.

### 8.4.2   Digitizer

After amplification and shaping, the analog signal is sent on micro-strip transmission line to the digitizers. Bringing signals from the two sides to a single digitizer board reduces concerns of clock synchronization at the sub-nanosecond level needed for time division.

A panel requires six digitizer boards, each servicing 16 straws. Each digitizer board contains: one FPGA with 32 TDC channels; two ADC chips, 8 channels each; and associated analog circuitry. Rate leveling across digitizer boards is done at the preamp to digitizer transmission line, mapping straws to digitizer boards as shown in Table 8.4 [8]:

Table 8.4. Mapping of straws to digitizer channels.

| Digitizer | Straws (0 ↔ inner most) |
|:---:|:---:|
| 0 | 0  6 12 … 90 |
| 1 | 1  7 13 … 91 |
| … | |
| 5 | 5 11 17 … 95 |

The analog input stage of the digitizer is shown in Figure 8.21. There are two comparators to send signals for timing measurement. These work with the preamp pulses at full bandwidth. In addition, there is a summation and integration amplifier to feed a reduced-bandwidth signal to the ADC.

A completed board is shown in Figure 8.22. The ADCs are a commercial 8 channel, 50MSPS, 12 bit devices with high speed serial output. *In-situ* measurements of the completed board show 11.4 Effective Number of Bits [9]. However, since we do not require 12 bits of precision, only 10 bits are retained. An FPGA is used to read the ADC and to function as a TDC [10].

Each straw has one ADC, fed by the sum of the two sides. The intrinsic d$E$/d$x$ fluctuations are too large for pulse height information to be beneficial to fitting. The main purpose of the ADC is to reject hits from protons, a significant source of "noise" hits. A secondary purpose is supplementing the calorimeter in distinguishing muon from electron tracks [11].

The d$E$/d$x$ for protons from the muon stopping target is higher than for an electron by as much as ×50. However, it is not necessary to measure protons without saturation. "Too large," i.e. saturated, is sufficient to flag a pulse as being from a proton. The dynamic range of electrons is large due to intrinsic d$E$/d$x$ fluctuations and the variable path length of a track through a straw, but it is also not necessary to precisely measure charge at the low end. "Too small", i.e. no ADC response, is sufficient to flag a pulse as *not* from a proton. The planned system is therefore modest: 50 MHz, 12-bit ADC.

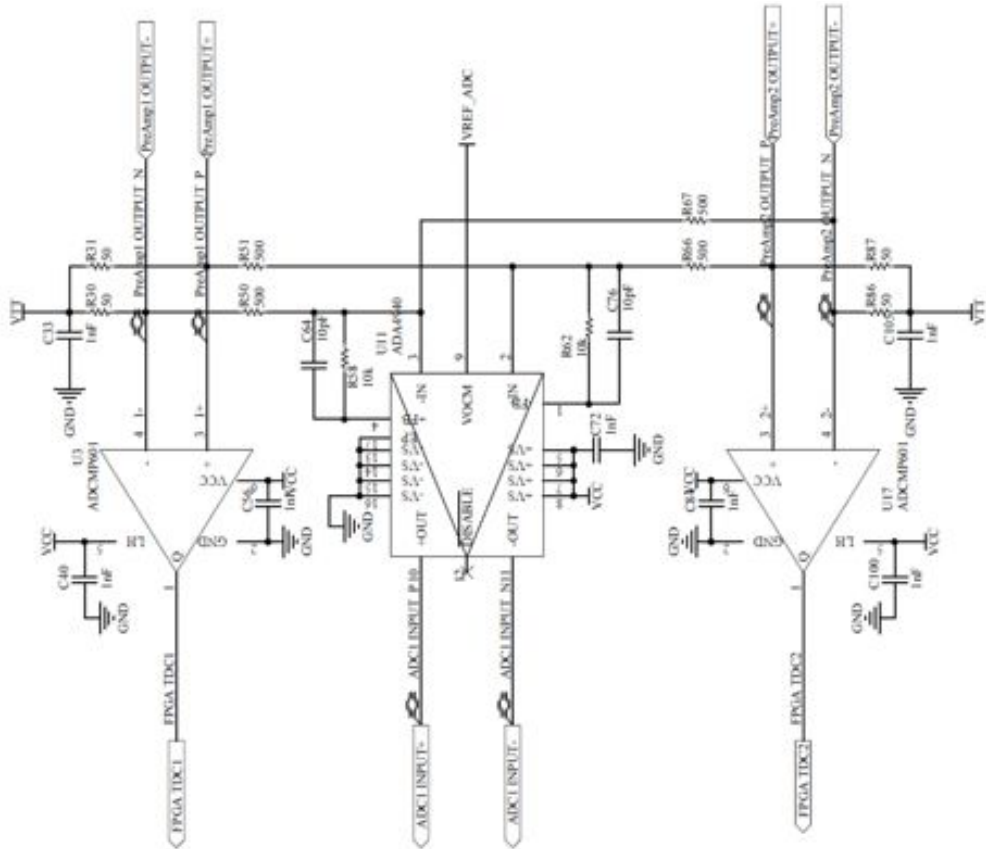

Figure 8.21. Digitizer input stage.

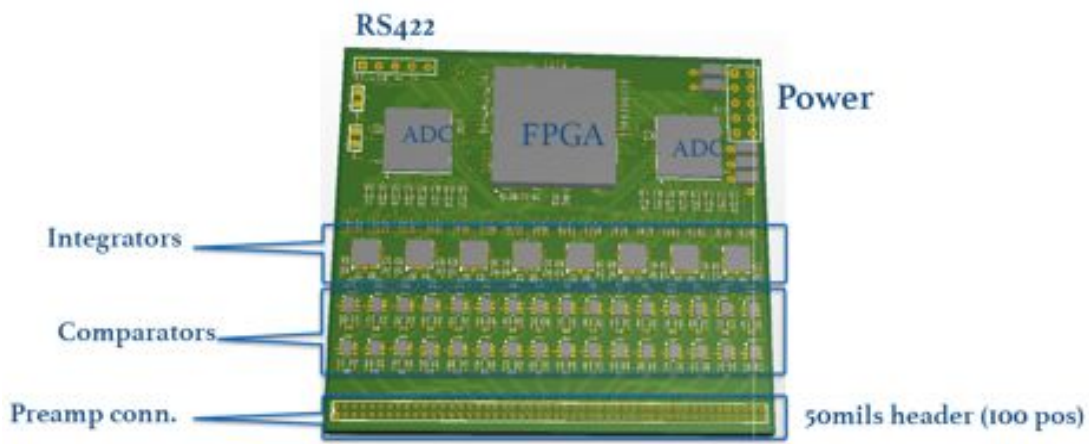

Figure 8.22. 16-straw digitizer board. The final board will not include RS422 and other test/debug capability, and will be ~10% smaller.

Each ADC digitizes continuously and sends data, without zero-suppression, to the digitizer FPGA. This FPGA also functions as the TDC for the corresponding straw. The ADC clock comes from the digitizer FPGA to ensure the TDC and ADC data remains synchronized. The FPGA combines internal TDC data with external ADC data. ADC zero suppression is achieved by sending only ADC data associated with a TDC hit.

The TDC is implemented in the FPGA as a combination of a delay chain for fine timing, and a counter running at 62.5 MHz for coarse timing. Per-stage delay in an FPGA varies with temperature and voltage. An automatic calibration procedure is built into the FPGA to measure and correct for this. The TDC intrinsic resolution is ~25 psec. Folding in comparator jitter, noise, and other external effects the final resolution is ~40 psec. For comparison, the resolution from time division is >80 psec (Figure 8.18).

Communication between the digitizer and ROC is via LVDS signals. For the current development we have chosen to use four lines per 8 straws: clock, frame, and two data lines. However, depending on final FPGA selection, and board layout issues, we are exploring options such as using 8b/10b or similar SERDES (self-synchronizing) data transfer.

Since both the digitizer and ROC are FPGA-based, data format is flexible. We have tentatively chosen to transfer data with the following fixed-length per hit format [8]:

- 16 bit header - The header contains information to uniquely specify this as a packet header, a channel identifier to specify the channel so the ROC can assign the hit to a wire number, and a packet checksum
- 16 bit -TDC left straw end
- 16 bit -TDC right straw end
- 8×10 bit ADC

Hit data is followed by a trailing "end of file" packet with status and error information. Including rate-leveling, and averaging over a microbunch, the highest rate for any 4-straw group (corresponding to one digitizer data line to the ROC) is 240 kHz [12] or 30 Mbps (at 128 bits/hit). The maximum rate allowed by the LVDS lines is 200 Mbps per data line. There is plenty of head room for adjusting data format, or unexpectedly high rates.

### 8.4.3   Readout Controller

The ROC's primary function is to receive data from the digitizer boards, buffer the data, and then transmit it to the DAQ system. Buffering is needed to continue transferring data during the beam inter-spill time (836 msec out of each 1333 msec). (We expect to take cosmic ray data during the inter-spill time – the front end must remain live – but rates are very low.)

Since all elements of the chain – digitizer, ROC, DAQ – are programmable, communication is flexible. Tentatively we have settled on the same fixed-length format, 128 bits per hit, as used for digitizer to ROC data transfer, with an additional packet (content and length under discussion) for status and error information.

The connection from ROC to DAQ is via 2.5 Gbps full-duplex fiber optic links arranged in rings with multiple ROCs per ring, as seen in Figure 8.23. This allows lossless data taking with a defective fiber link while maintaining, on average, one full-duplex link per ROC [13]. Rate leveling within each ring is done by pairing high rate (upstream) with low rate (downstream) planes.

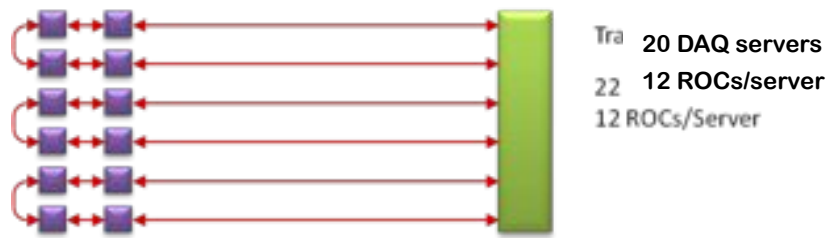

Figure 8.23. ROC connection to DAQ. One (of 20) servers shown.

The ROC includes external DRAM for buffering; this allows data transmission to continue over a full 1.333 sec Main Injector cycle. Since the Main Injector supplies us beam only 32% of the time, the rates presented in Section 8.4.2 translate to an expected output from the ROC of

$$30 \text{ Mbps/4 straws} \times 96 \text{ straws} \times 32\% = 230 \text{ Mbps}$$

compared to a single optical link's 2.6 Gbps capability: A single optical fiber readily handles several ROCs, motivating the ring architecture shown in Figure 8.23. With 240 controllers, the total rate from the T-Tracker is 55 Gbps.

The ROC also links the experiment's Slow Controls system to the digitizers and preamps [14]. DACs, ADCs, and sensors are distributed through each panel and connect to the ROC via SPI and $I^2C$. To reduce the number Chip Enable lines (SPI normally uses one per chip), SPI port expanders will be used. Some critical functions of Slow Controls:

- Set the preamp DACs to control threshold and gain.
- Set switches to control HV fuses (see preamp section),

- Monitor voltages, temperature, pressure.

## 8.5    Infrastructure

The tracker must bring power, signals, gas, and cooling through vacuum penetrations and past the beam stop and calorimeter; see Figure 8.24. Except for cooling, utilities use the horizontal support beams as "cable trays". A tentative layout is shown in Figure 8.25.

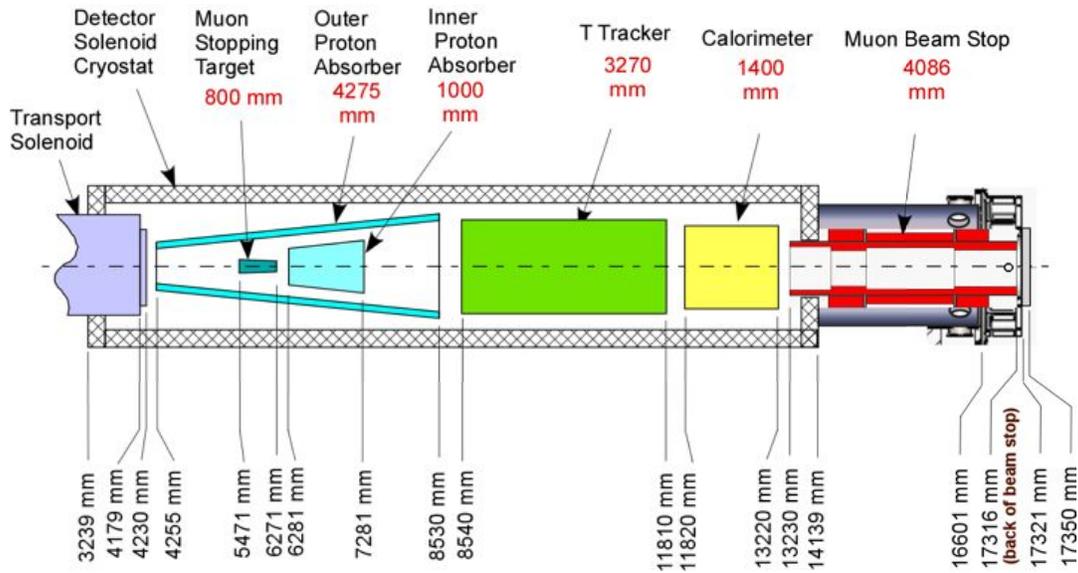

Figure 8.24. Longitudinal positioning of components in the DS bore.

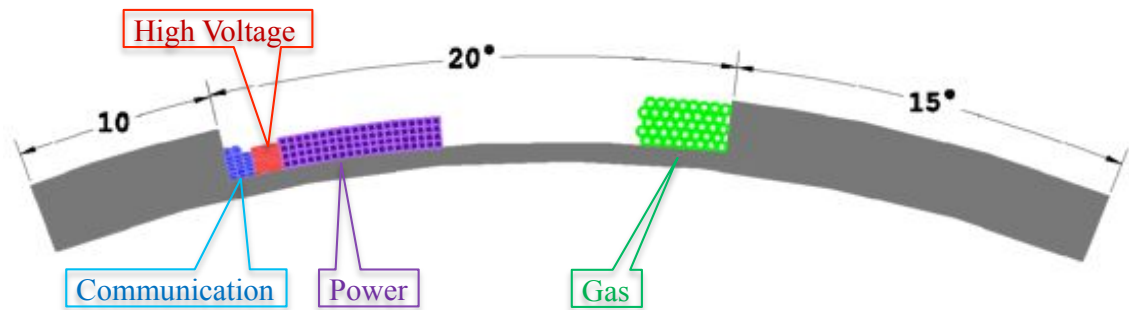

Figure 8.25. Layout of utilities within the horizontal support beams. There are two such beams as seen in Figure 9.13.

### 8.5.1    Gas

A separate gas line is brought in for each plane for a total of 20 pair (20 each supply and exhaust). Gas flows over electronics to assist with cooling, as seen in Figure 8.26. We rely on the high flow impedance presented by the straw termination [15] to maintain uniform flow through all straws.

To avoid the risk of gas contamination, a Partlene coating will be applied to electronics on the gas inlet side. Digitizers are on the exhaust side of the gas flow and do not need to be coated.

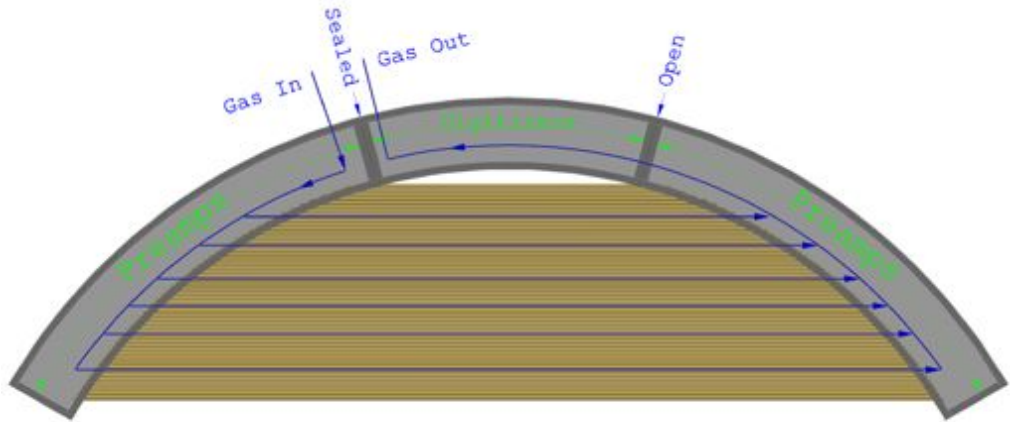

Figure 8.26. Gas flow through a panel. Digitizers are on the exhaust side to minimize concerns about outgassing from these boards.

## 8.5.2  Low Voltage

The power budget per component is given in Table 8.5 [5]. Power is brought in at 48 VDC and stepped down first by buck convertor [16], then by charge pump regulators to the various other required voltages. Assuming 90% conversion efficiency at each step, the total power including step-down losses is 9.5 kW. Total current is ~200A.

Each of the two beams will carry 80 × 3.5 mm square solid copper wire with 0.25 mm Nomex[®] insulation, as shown in Figure 8.27. This gives a total of 80 pair (supply and return) of power cables to carry the required 200 A, or 2.5 A per wire. Resistance per pair is 27 mΩ for a 10 meter run (sufficient to exit the vacuum) for a voltage drop of ~67 mV. Power loss on the cables will adds ≲15 W heat load.

Table 8.5. Power consumption of tracker components.

|  | mW per Straw | W per Panel | Total Power (W) |
|---|---|---|---|
| Preamp | 2×20 | 3.84 | 922 |
| Digitizer | 150 | 14.4 | 3,456 |
| Controller |  | 14 | 3,360 |
| Total |  |  | 7,738 |

In addition to voltage drop and power loss, a concern in the low voltage power distribution is perturbations to the magnetic field. This can be minimized by balancing supply and return currents in a "quadrupole" array as shown in Figure 8.27 [17].

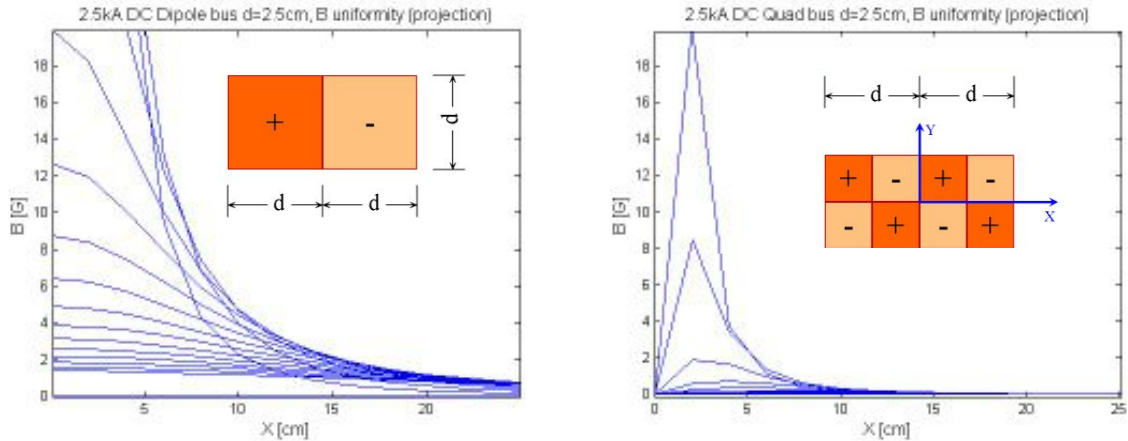

Figure 8.27. Magnetic field due to power buses in dipole and quadrupole configuration with 2.5kA/line, >100× the tracker's nominal. Curves are at 2 cm slices in Y. The tracker active region is at X ≳ 10cm.

### 8.5.3   High Voltage

High voltage is distributed using 0.05″ pitch silicone ribbon cable rated for 5kV compared to our operating point of ≲1.5kV. A separate line (conductor) will be sent to each panel for a total of 240 conductors, allowing each panel to be monitored and, if necessary, isolated without breaking the DS vacuum.

### 8.5.4   Cooling

Since the detector is in a vacuum, heat must be removed with an active cooling system. To maintain the most uniform temperature with the smallest volume in plumbing and smallest vacuum penetrations we use a Suva® system. Entering the vacuum is one pair of lines (supply and return) as shown in Figure 8.28 [18]. Each plane has a cooling loop at the outer periphery, as seen in Figure 8.14, which taps into the cooling lines. The expected temperature gradient is shown in Figure 8.29.

## 8.6   ES&H

The Mu2e tracker is similar to other gas-based detectors that are commonly used at Fermilab. Potential hazards include power systems and compressed gas. These hazards have been identified and documented in the Mu2e Hazard Analysis Report [19].

The detector requires both low voltage, high current (~48 V @ 200A total) and high voltage, low current (~1.5 kV @ 500 mA) power systems. During normal operation the tracker will be inaccessible, inside the enclosed and evacuated Detector Solenoid. Power will be distributed to the tracker through shielded cables and connectors that comply with Fermilab policies. Fermilab will review the installation prior to operation.

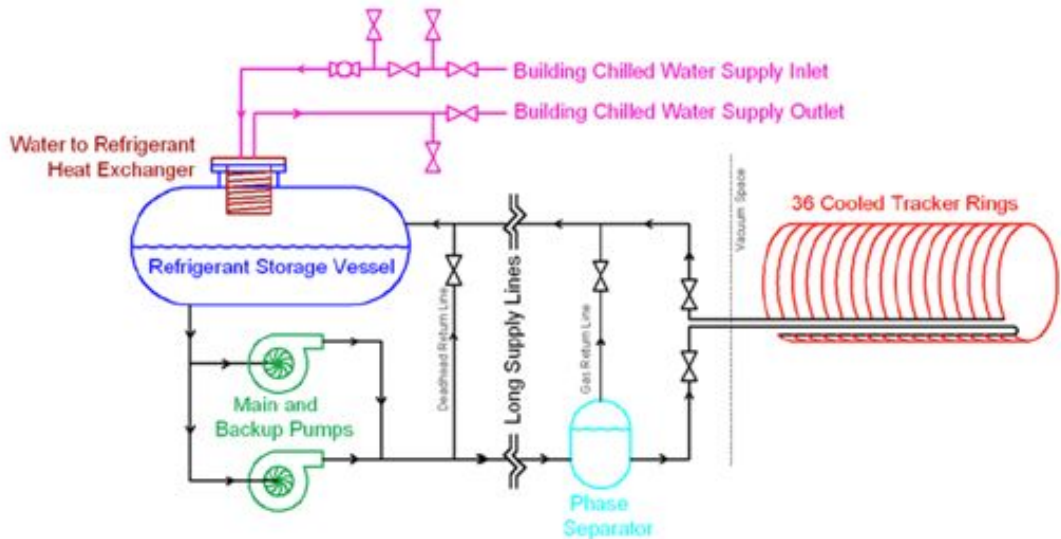

Figure 8.28. Conceptual design of the tracker cooling system.

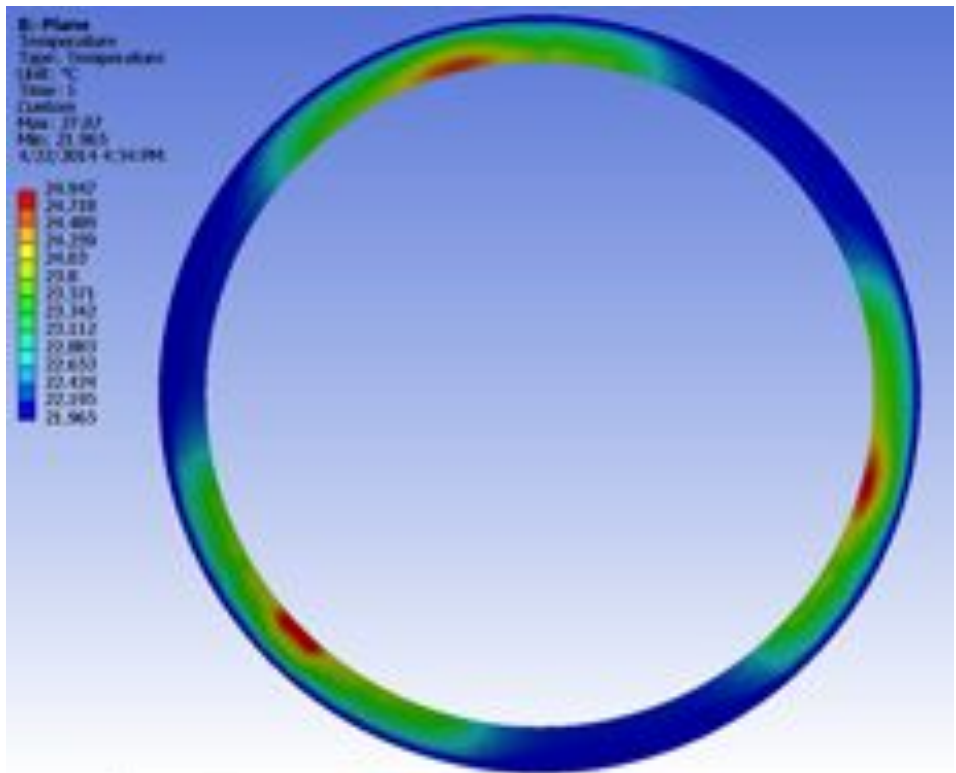

Figure 8.29. Temperature gradient across support ring for 20°C coolant and heat distributed according to Table 8.5. The hot spots are from the ROC.

The tracker will require a supply of chamber gas to be supplied from compressed Ar:$CO_2$, 80:20. To minimize variation over time, a single trailer will be used for the entire run. The trailer will be parked in a dedicated location appropriate to the type of gas being used. The installation, including all associated piping and valves, will be documented and reviewed by the Fermilab Mechanical Safety Subcommittee.

The cooling system requires SUVA® or similar refrigerant. Plumbing will be designed to tolerate a wide range of pressures, allowing for a cost effective and environmentally friendly solution based on product availability and EPA guidelines at the time of operation.

The T-Tracker itself does not have any radioactive sources; however, there will be sources used in monitoring chambers. Usage of radioactive sources will be reviewed to ensure adherence to Fermilab safety policies.

## 8.7 Risks

A complete list of risks, opportunities, and mitigation strategies are maintained in the Mu2e Risk Registry [20].

### 8.7.1 Performance Risk.

Extensive simulations of the tracker have been performed, providing confidence that it will deliver the required efficiency and resolution. However, work on pattern recognition continues. There is a risk that it will prove difficult to identify tracks with the required efficiency. Progress on simulations and pattern recognition could lead to design modifications, if needed. This translates to a cost and schedule risk.

The tracker performance is sensitive to beam properties as well as upstream components like the proton absorber and stopping target. Work on beam simulation, and optimization of the target and proton absorber continue. Although these risks are not strictly part of the tracker, changes elsewhere (for instance, eliminating the proton absorber) could result in higher rates in the tracker. This risk is mitigated by working closely with other groups to ensure the combined system meets requirements.

Inability to hold alignment in the detector would lead to degraded resolution and unaccounted for background. The problem can be solved using in-situ, track-based alignment. This is a difficult and slow process for Mu2e, resulting in delays in publishing results. This risk is mitigated by precision electronic levels and Hall probes on the tracker frame to monitor twist and skew.

### 8.7.2   Technical and Operational Risk

The tracker technology chosen for Mu2e is well established and has been implemented in other high energy and nuclear physics experiments [21] [22]. The risk of catastrophic problems inherent to straws is ruled out. However, to optimize resolution at low momentum, we are using thinner wall straws than in the past. To tolerate 15 psid with thin walls, we use Mylar[®] rather than the more common substrate, Kapton[®]. Risks associated with wall thickness and material are mitigated by extensive long-term tests [23] [24] as well as choosing to operate at the same hoop stress level as NA62 [22].

Chamber aging is a poorly understood phenomenon and requires testing. We have done this up to the expected total dose, including beam flash, of 0.9 Coulomb/cm on the upstream inner straws. The final materials intended for the straw construction were included in the test. We observed no measurable degradation [25]. Note, however, changes in the proton absorber may require repeating or extending this test to higher total dose.

Contaminated gas is a serious risk for any drift chamber. This risk is mitigated in several ways. First, $Ar:CO_2$ is one of the least prone to harmful contaminants. Second, by purchasing a single batch of gas and performing detailed analysis. Never the less, there is risk of contamination of the gas system. Monitoring chambers will be included in the system, illuminated with radioactive sources, to give early warning of problems.
Wire chambers always suffer the risk of a broken wire.  The loss of a single wire is a minor problem, but the wire shorting other electrodes can leave a large region of the detector inoperable. The selected straw tracker mechanically contains each wire; however the resulting short to the cathode surface would upset the high voltage distribution.  The ability to disconnect individual wires from high voltage without the need to access the detector has been included in the design to mitigate this risk.

## 8.8   Quality Assurance

Proper quality assurance is essential to construct a tracking detector that meets Mu2e's requirements for performance and reliable operation. Quality Assurance will be integrated into all phases of tracker work including design, procurement, fabrication and installation.

All straws[vi] will be leak tested before assembly using a fast $CO_2$ sensing technique [26]. A subset will be cross-checked by measuring leak rate in vacuum [27]. After assembly, each panel will be similarly leak-tested: 100% with $CO_2$, a subset under vacuum [28]. All straws will also be visually inspected, and end-to-end resistance measured. From the leak

---

[vi] If the failure rate is low, we may switch to testing a subset of straws. Panels will still be 100% tested.

test onward, straws are kept in individual storage tubes with bar codes to allow tracking each straw until it is made part of a panel. Panels will also have bar codes. Panel assembly will include recording the straw bar code and where in which panel it is placed. The appropriate tension must be applied and maintained for both the straw tube and the sense wire. Both wire and straw tension can be measured after assembly via vibration resonance [29]. The straw body (cathode) is electrically isolated from the tracker structure, allowing this test (as well as cathode resistance check) to be performed after assembly. Tension values over time will be recorded.

All panels will go through an X-ray machine developed by Duke for use on ATLAS' TRT straws [3]. Positions of all wires, relative to survey monuments on the panel, will be recorded.

All electronics components will be tested prior to installation on the tracker, including a suitable burn-in period. High voltage boards will be tested for leakage current. The threshold characteristics of each channel will be tested with a threshold scan. A noise scan will be performed for various threshold settings to identify channels with large noise fractions.

## 8.9    Value Management

Use of off-the-shelf electronics opens opportunities for cost saving as new chips become available, or existing chips come down in price. The tradeoff between improved chips versus board redesign must be considered before final production.

3D printing is used in the manufacture of panels, both as part of the final panel and in the assembly fixtures. This is a new and rapidly advancing technology. Improved materials and precision may become available which expand the range of applications.

Full production as budgeted assumes only already designed tooling and fixtures, with only minor revisions. The tradeoff between designing and building more elaborate apparatus, versus the saving in labor, must be reviewed at each step.

Our plan is to have straws and straw assembly work done at universities. Although beneficial in many ways, distributed processing will require greater care in recording all QA/QC data. With guidance from Fermilab's Computing Division, we will set up a database to allow multiple institutions to both access and enter this information.

### 8.9.1    Number of Stations

This chapter, and the sensitivity and background estimates described in chapter 3, assume the nominal 20 station tracker. Because of their modular design, the number of tracker stations is not strongly constrained by the other components of Mu2e. Geometric constraints, power limits, and heat removal capacity limit the maximum possible number of stations that can fit inside the cryostat to roughly 36. In general, we expect more stations will improve the tracker performance by increasing the number of measurements on each track. However, more stations also means more material in the conversion electron (CE) path, which will degrade the momentum resolution due to increased scattering and reduce the reconstruction efficiency due to the increased probabilty of bremsstrahlung. The optimum performance will be obtained when these effects balance. The number of stations is also constrained by the Mu2e cost cap. As the cost of the tracker is roughly linear with the number of stations, any predicted performance improvement must be weighed against its cost.

To inform a final decision on the number of stations, we repeated the CE yield and DIO background estimates for different numbers of tracker stations. These studies were done using the same physics inputs, simulation, and reconstruction sequence as for the nominal case (Section 3.5). Tracker hit backgrounds were computed using a G4 simulation that assumed a 22-station tracker. CE and DIO were simulated using the exact number of stations being tested. Background hits in 'missing' stations were ignored in the reconstruction. Tracks were selected using the same quality cuts for all cases, with the largest differential effect coming from requiring at least 25 hits/track. The momentum selection window lower edge was adjusted for each case to yield approximately the same number of DIO backgrounds as the nominal case. The results for 18, 20, and 22 stations are shown in Figure 8.30. For a constant DIO rate, we see that the relative CE yield increases 3.5% for each additional station. The momentum resolution over this range of stations was found to be unchanged, within errors, as shown in Figure 8.31. Additional details of this study are provided in [30].

To understand the aymptotic behavior, we also estimated the CE yield dependence on the number of stations without background hit overlays, as shown in Figure 8.32. The broad optimum between 21 and 24 stations results from the approximate cancelation of increased selection efficiency against increased brehmsstrahlung. Additional details of this study are provided in [31].

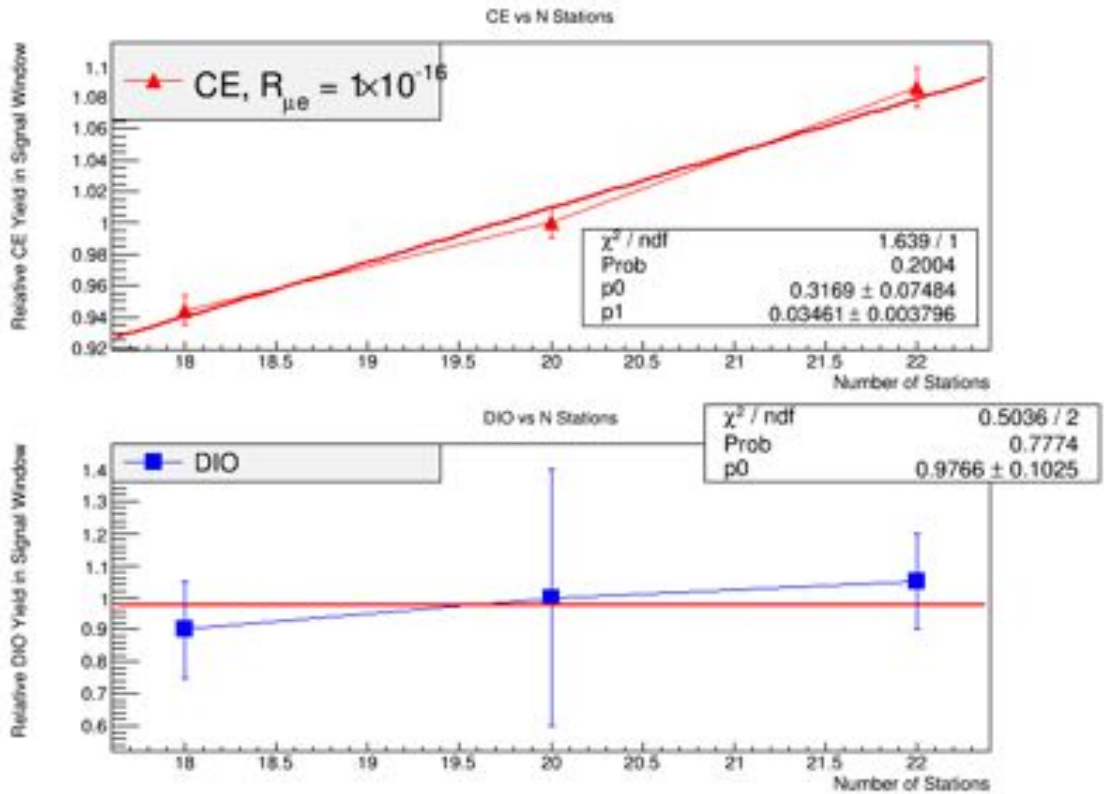

Figure 8.30. The Conversion Electron (CE) yield for Rμe=10⁻¹⁶ (top) and the relative DIO background rate (bottom) as a function of the number of stations from a full simulation of the Mu2e tracker, including hit backgrounds. The same track quality selection cuts are made for each point, but the lower edge of the momentum selection window was adjusted to give a constant DIO rate.

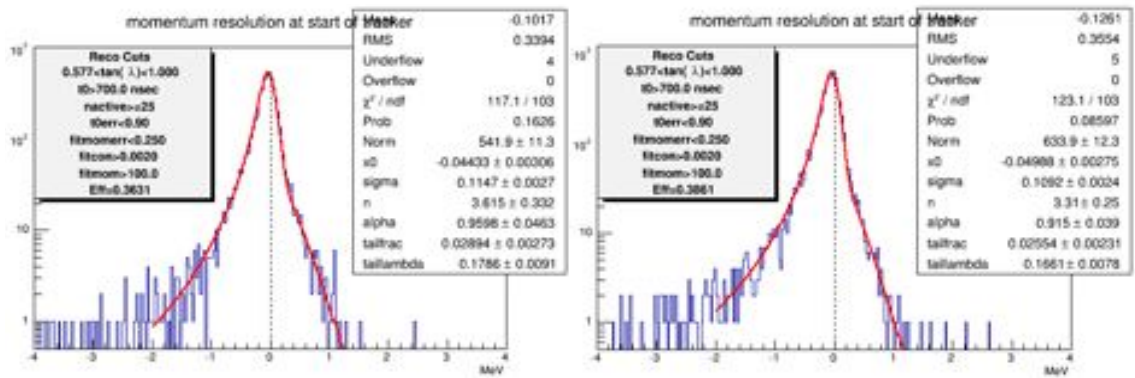

Figure 8.31. Simulated CE momentum resolution with full hit background overlay for 18 stations (left) and 22 stations (right).

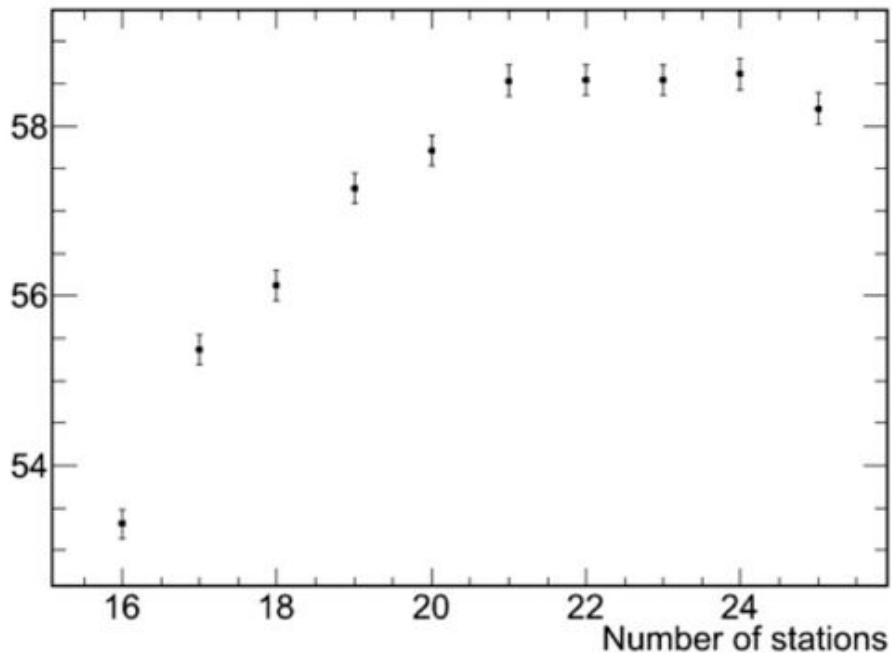

Figure 8.32. Relative Conversion Electron yield as a function of the number of stations, not including hit background effects. The same track quality selection cuts and the same momentum selection lower edge are made for each point. The Y axis is an arbitrary scale.

This page intentionally left blank

# 9      Calorimeter

## 9.1     Overview

The design of the Mu2e detector is driven by the need to reject backgrounds to a level consistent with a single event sensitivity for $\mu \rightarrow e$ conversion of the order of $3 \times 10^{-17}$. The calorimeter system is a vital link in the chain of background defenses. A background of particular concern is false tracks arising from pattern recognition errors that result from high rates of hits in the tracker. The accidental hits could combine with, or obscure, hits from lower energy particles, to create a trajectory consistent with a higher energy conversion electron. Thus a primary purpose of the Mu2e calorimeter is to provide a second set of measurements that complement the information from the tracker and enable us to reject background due to reconstruction errors. Another source of background is cosmic ray muons, not vetoed by the CRV system, that stop in the calorimeter and produce a backward-going electron track within the 105 MeV acceptance window. A calorimeter with excellent time resolution can reject such tracks.

The energy resolution of a crystal calorimeter complements, but is not competitive with, that of a tracking detector. Even a coarse confirmation of track energy by the calorimeter will, however, help reject backgrounds from spurious combinations of hits from lower energy particles. The Mu2e simulation is not yet at the stage where this can be explicitly demonstrated, but 5% energy resolution has been achieved by other experiments operating in a similar energy regime [1].

For real tracks, activity in the tracker and in the calorimeter will be correlated in time. The time resolution of the calorimeter should be comparable to the time resolution of extrapolated tracks from the tracker, estimated to be of ~1 nanosecond. A calorimeter timing resolution of < 1 ns is consistent with the tracker and can be easily achieved.

## 9.2     Design concept

In the 100 MeV energy regime, a total absorption calorimeter employing a homogeneous continuous medium is required to meet the resolution requirement. This could be either a liquid such as xenon, or a scintillating crystal; we have chosen the latter. Two types of crystals have been considered in detail for the Mu2e calorimeter: lutetium-yttrium oxyorthosilicate (LYSO) and barium fluoride ($BaF_2$). The baseline design selected for the Mu2e calorimeter uses an array of $BaF_2$ crystals arranged in two annular disks. Electrons following helical trajectories spiral into the front faces of the crystals, as shown in Figure 9.1. Photo-detectors, electronics and services are mounted on the rear face of the disks. The crystals are of hexagonal shape, 33 mm across flats and are 200 mm long; there are a total of 1860 crystals. Each crystal is read out by two large-area APDs; solid- state photo-detectors are required because the calorimeter resides in a 1 T magnetic field. Front-end electronics is mounted on the rear of each disk, while voltage distribution, slow controls and digitizer electronics are mounted behind each disk. A laser flasher system provides light to each crystal for relative calibration and monitoring purposes. A circulating liquid radioactive source system provides absolute calibration and an energy scale. The crystals are supported by a lightweight carbon fiber support structure. Each of these components is discussed in the sections that follow.

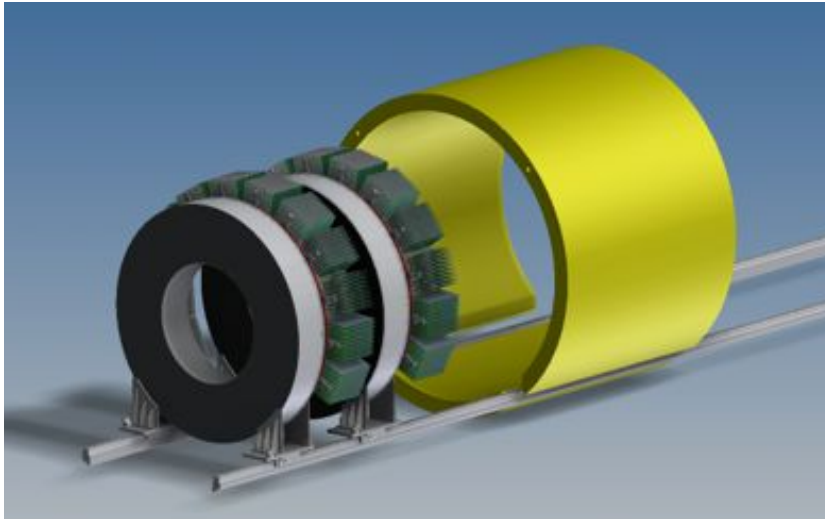

Figure 9.1. The Mu2e calorimeter consisting of an array of $BaF_2$ crystals arranged in two annular disks. Electrons spiral into the upstream faces.

## 9.3    Requirements

The requirements for the calorimeter have been documented by the Mu2e collaboration [2]. The primary functions are to provide energy, position and timing information to confirm that events reconstructed by the tracker are well measured and are not the result of a spurious combination of hits. Moreover, the calorimeter should also provide the experiment's trigger. This leads to the following requirements:

- An energy resolution of 5% at 100 MeV is desirable to confirm the electron momentum measurement from the tracker, which is much more precise.
- A timing resolution better than ~ 0.5 ns is required to ensure that energy deposits in the calorimeter are in time with events reconstructed in the tracker.
- A position resolution better than 1 cm is necessary to allow comparison of the position of the energy deposit to the extrapolated trajectory of a reconstructed track.
- The calorimeter should provide additional information that can be combined with information from the tracker to distinguish muons from electrons.

- The calorimeter must provide a trigger, either in hardware, software, or firmware that can be used to identify events with significant energy deposits.
- The calorimeter must operate in the unique, high-rate Mu2e environment and must maintain its functionality for radiation exposures up to 20 Gy/crystal/year and for a neutron flux equivalent to $10^{11}$ n_1MeV eq /cm$^2$.

The energy resolution of a crystal calorimeter complements, but is not competitive with, that of the tracking detector. Even a coarse confirmation of track energy by the calorimeter will, however, help reject backgrounds from spurious combinations of hits from lower energy particles. The Mu2e calorimeter group [2], as well as other experiments operating in a similar energy regime [1], has achieved an energy resolution of 5% at 100 MeV.

## 9.4    Performance – meeting the requirements

To provide a guide for this discussion, a large sample of DIO events ($25 \times 10^6$) has been simulated in the momentum range of 100 - 105 MeV/$c$. The DIO sample has been produced with the expected energy spectrum and normalized to the expected rate for a 3 year run. $10^5$ conversion electrons (CE) were also produced and normalized to the number of events expected for a $\mu^-N \rightarrow e^-N$ conversion rate of $10^{-16}$. For each simulated event, tracks and clusters have been reconstructed with the official Mu2e framework that also estimates their momenta $p$ and deposited energy $E$. For the purpose of this test, perfect reconstruction is assumed, so pileup of background hits in track and cluster reconstruction have not been included. Figure 9.2 shows the distribution of energy and momentum for DIO and CE events together with the integral of CE events (for a BR of $10^{-16}$).

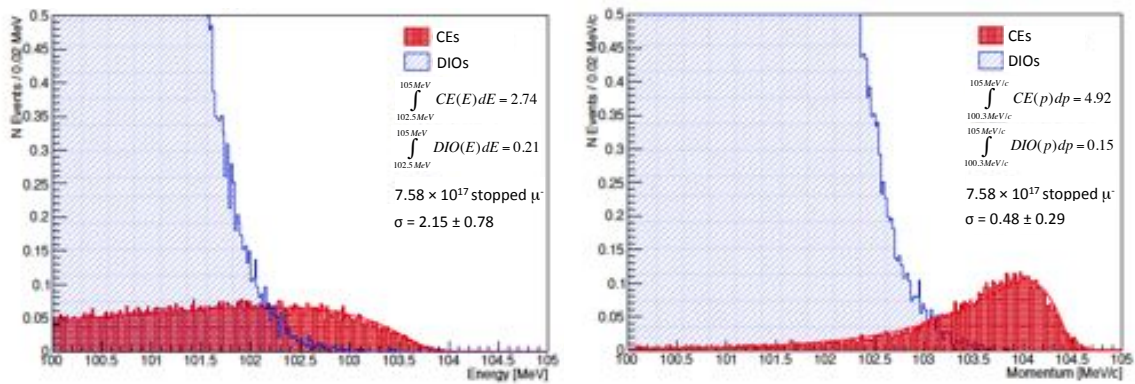

Figure 9.2. Energy (left) and momentum (right) distribution for DIO and CE events. DIO spectrum is in blue while CE spectrum is in red. Perfect reconstruction and no additional hits due to overlap from backgrounds are assumed.

To use this information to discriminate against the DIO background, a pseudo-$\chi^2$ variable, $\xi^2 = \text{signed}((p-\mu_p)/\sigma_p)^2 + \text{signed}((E-\mu_E)/\sigma_E)$, has been developed where $\mu$ is the most probable value and σ represents the FWHM/2.35, respectively. The sign +(-) is assigned to events above (below) the most probable values. In Figure 9.3 (left), the distribution of ξ is shown for the DIO and CE events. Cutting at $\xi^2 < 3.5$, as indicated in the picture, results in $N_{DIO}$=0.23 and $N_{CE}$=4.11, respectively. These results have to be compared with $N_{DIO}$=0.15 and $N_{CE}$=4.9, estimated by using tracker-only information. In Figure 9.3 (right), the scatter plot of $p_{trk}$ *vs.* $E$ is shown after the application of this cut. It is evident that the requirement of energy information does not improve the S/N while reducing the efficiency by ~20%. The situation becomes slightly worse as energy resolution deteriorates. A summary of these results is reported in Table 9.1, where additional Gaussian smearing has been added to the simulated calorimeter resolution.

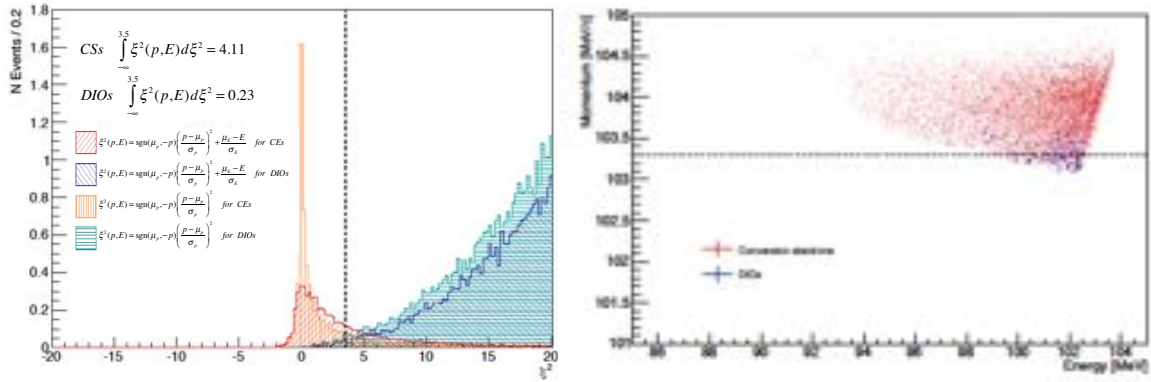

Figure 9.3. Distribution of the ξ variable described in the text (left), scatter plot of $p$ vs $E$ for events with ξ < 3.5 (right).

In Figure 9.4, the momentum distributions for CE+DIO events selected by the tracker are compared with the ones obtained with the combined information; the left (right) plot displays the case for an energy resolution of 2.1%, (3.1%) respectively. We conclude that the combined information does not improve the signal-over-noise ratio and slightly reduces the tracker-based reconstruction efficiency. However, it adds a confirmation to the CE candidate in case of a wrong track reconstruction.

The tracker and calorimeter hits produced by the same particle should be close in time. The calorimeter timing resolution should be comparable to the time resolution of extrapolated tracks from the tracker, estimated to be ~0.5 ns. A calorimeter timing resolution of about 0.5 ns is consistent with the tracker and will not spoil the joint calorimeter track performance. The requirement on the calorimeter's position resolution is based on the error associated with extrapolating a track from the tracker to the calorimeter, shown in Figure 9.5. There is no need for the calorimeter position resolution to be better than the extrapolation error, which is driven by multiple scattering in the tracker. Based on this study, a position resolution of 0.5 cm is sufficient.

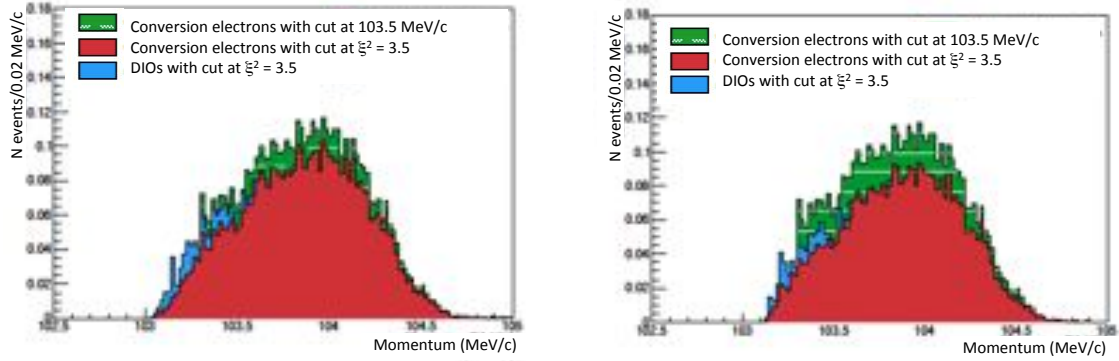

Figure 9.4. Momentum distributions for the CE candidate and DIO background for selections with the ξ variable (red histogram), additional track only candidate (green stacked histogram). DIO events are displayed in blue: the left plot is for a calorimeter resolution of 2.1%, the right one for 3.1%.

Table 9.1. Number of CE and DIO candidates after the application of a ξ cut at 3.5 as a function of the calorimeter energy resolution.

| EMC $\sigma$ $(E_{dep} - E_{in})$ [MeV] | Energy $\sigma$ [MeV] | Smearing | DIOs | CEs | $\frac{S}{N}$ |
|---|---|---|---|---|---|
| 0.939 | 2.15 | 0 | 0.227 | 4.099 | 18.057 |
| 1.429 | 2.40 | 0.5 % | 0.231 | 4.103 | 17.762 |
| 1.761 | 2.50 | 0.6 % | 0.239 | 4.140 | 17.322 |
| 2.027 | 2.59 | 0.7 % | 0.243 | 4.186 | 17.226 |
| 2.083 | 2.69 | 0.8 % | 0.246 | 4.224 | 17.175 |
| 2.364 | 2.90 | 1.0 % | 0.253 | 4.302 | 17.004 |
| 2.604 | 3.08 | 1.2 % | 0.263 | 4.363 | 16.589 |
| 2.964 | 3.27 | 1.5 % | 0.277 | 4.399 | 15.881 |

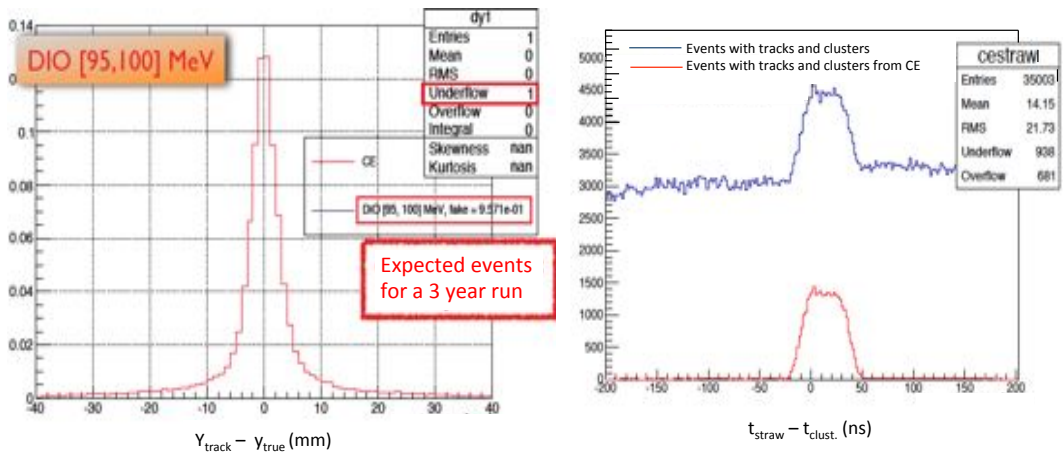

Figure 9.5. Distribution of the difference between tracks extrapolated from the tracker to the calorimeter vs. the true calorimeter position (left). The tracks were fitted with a Kalman filter and extrapolated to the calorimeter using the parameters of the reconstructed track. Shown on the left is the $\Delta t$(straw-cluster) distribution for all hits (blue) and the CE related hits (red).

Mu2e Technical Design Report

The calorimeter timing information can be used by the cluster reconstruction algorithm in several ways. For the cluster reconstruction itself, good time resolution helps in the connection/rejection of cells to the cluster and in the cluster merging. This, however, depends strongly on the geometry and granularity choice, and will be discussed further after a presentation of the baseline detector layout. Timing information can also be used to improve the pattern recognition in the tracker Figure 9.6 and add discriminating power to the identification of μ with respect to the electrons (PID).

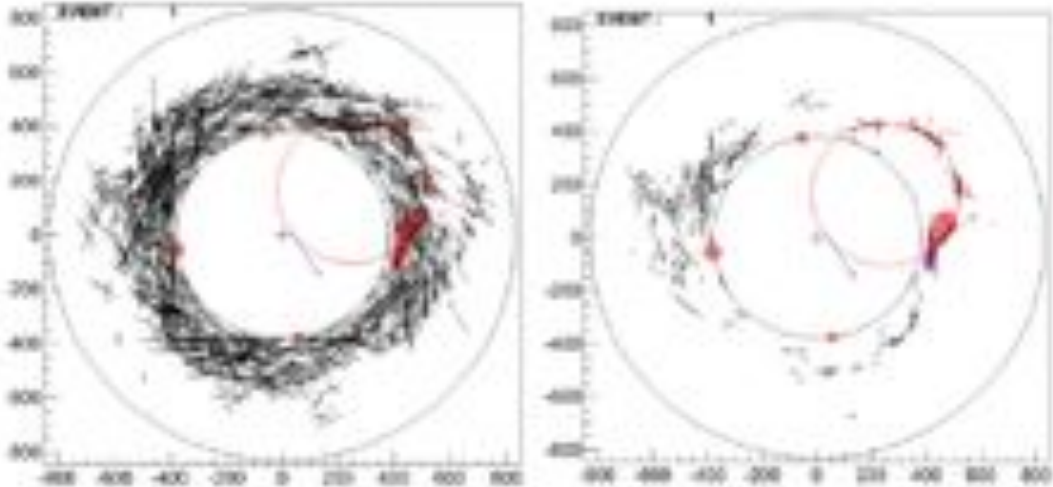

Figure 9.6. Distribution of the hits in the tracker before (left) and after (right) the application of a timing window based on timing information in the calorimeter. The situation for the pattern recognition is dramatically improved.

## 9.4.1    Particle Identification and Muon Rejection

Cosmic rays generate two distinct categories of background events: muons trapped in the magnetic field of the Detector Solenoid and electrons produced in a cosmic muon interaction with detector material. According to the most recent studies of the cosmic background [4], after 3 years of data taking one could expect about 2.2 events in which negative cosmic muons with $103.5 < P < 105$ MeV/c enter the detector bypassing the CRV counters and surviving all analysis cuts. To keep the total background from cosmics at a level below 0.1 events, a muon rejection of 200 is required (Section 10.2). Timing and $dE/dx$ information from the Mu2e tracker allows for limited PID capabilities [6]. However for a muon rejection factor of 200, the efficiency of the electron identification based on the tracker-only information could be 50% or even below. The energy and timing measurements from the Mu2e calorimeter (see Figure 9.7) provide information critical for efficient separation of electrons and muons in the detector. The calorimeter acceptance has been optimized such that (99.4+/-0.1)% of conversion electron (CE) events with tracks passing "Set C" quality cuts have a calorimeter cluster with E > 10 MeV produced by the conversion electron. A reconstructed CE candidate event is therefore required to have a calorimeter cluster, pointed to by the track. A track-cluster matching $\chi^2_{\text{match}} = (\Delta U/\sigma_U)^2 + (\Delta V/\sigma_V)^2 + (\Delta T/\sigma_T)^2$ is defined, where $\Delta U$ and $\Delta V$ are the track-to-cluster coordinate residuals in directions parallel and orthogonal to the track, and $\Delta T$ is the difference between the track time extrapolated to the calorimeter and the reconstructed cluster time. The estimated resolutions are $\sigma_U = 1.5$ cm, $\sigma_T = 0.8$ cm, and $\sigma_T = 0.5$ ns. For the background occupancy level exepected in the Mu2e detector during the data taking, a requirement $\chi^2_{\text{match}}$ <100 is 98% efficient for the expected CE signal. Events are also required to be consistent with the electron hypothesis such that they have $|\Delta T| < 3$ ns and E(cluster)/P(track)<1.15. After the cleanup cuts, the log likelihoods of the electron and muon hypotheses are defined: ln $L_{\text{e},\mu}$ = ln $P_{\text{e},\mu}(\Delta t)$+ln $P_{\text{e},\mu}(E/P)$, where $P_{\text{e},\mu}(\Delta t)$ and $P_{\text{e},\mu}(E/P)$ are $\Delta t$ and E/P probability density distributions for electrons and muons correspondingly. These distributions are shown in . A ratio of the likelihoods of the two hypotheses ln $(L_e/L_\mu)$ = ln $L_e$ - ln $L_\mu$ determines the most likely particle mass assignment. Figure 9.8 (left) shows the muon rejection factor plotted vs the CE identification efficiency for different background levels: CE only, CE plus nominal expected background, CE plus two times the expected background. For the nominal background expectation and muon rejection factor of 200, the electron identification efficiency is (96.5 +/- 0.1)%. This number includes the geometrical acceptance and efficiency of all cuts and demonstrates a high efficiency of the PID procedure. Figure 9.8 (right) shows dependence of the electron identification efficiency for different values of the calorimeter energy and time resolution in the range $0.02 < \sigma_E/E < 0.2$ and $0.05 < \sigma_T < 1$ ns. The value of the muon rejection factor is fixed at 200. One can see that in the expected operational range, $\sigma_E/E < 0.1$ and $\sigma_T < 0.5$ ns, the PID is robust with respect to the calorimeter resolution, with the electron identification efficiency variations below 2% in this region of parameter space.

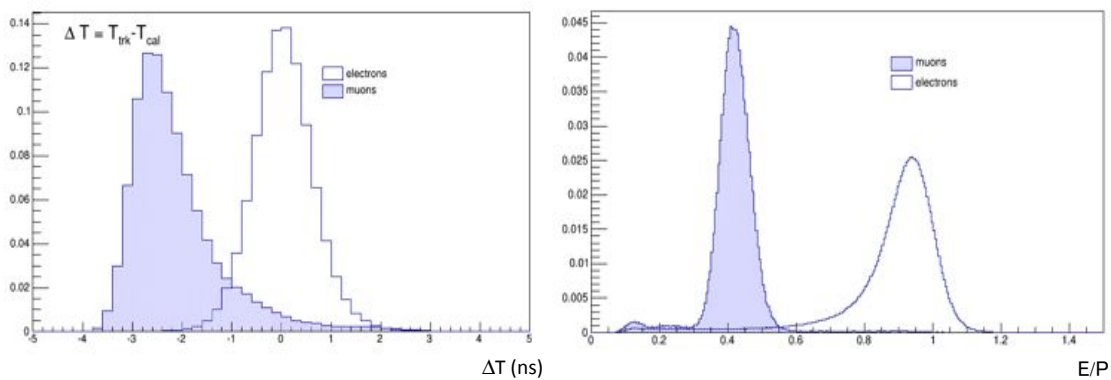

Figure 9.7. Distributions of $\Delta t$ (left) and E/P(right) for 105MeV/c electrons and muons used to build the PID likelihood.

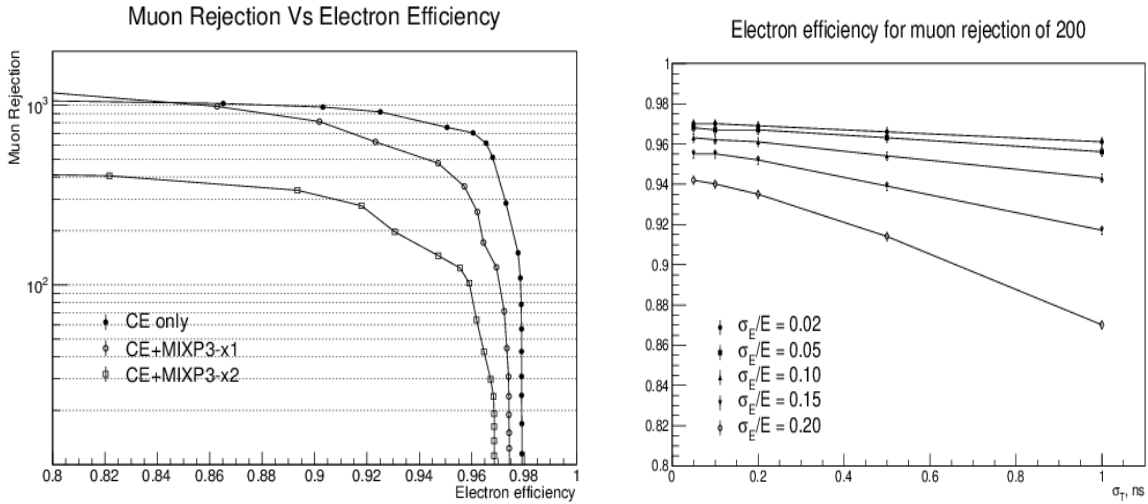

Figure 9.8. (left) PID efficiency for CE vs. muon rejection for different background levels: no background, expected background, twice the expected background; (right) PID efficiency for CE for muon rejection factor of 200 and different assumptions about the calorimeter energy and timing resolution.

### 9.4.2    Calorimeter Trigger

The calorimeter system can also generate a fast, efficient trigger for the experiment that is independent of the tracker. This trigger will take the form of an offline HLT/L3-like filter that can be used after streaming the events to the online computing farm, but before storing data on disk. The DAQ will read events from the tracking and calorimeter digitizers at a maximum throughput of 20 GByte/sec (Table 12.4) and the online farm will be able to fully reconstruct nearly all the streamed data. The calorimeter filter should be able to help/improve the processing in the online farm while restricting the data stored to disk to a maximum of $O(10)$ PB/year, *i.e.,* to 2 kHz.

The most important aspect of this filter is that it is fully independent of the tracker, with completely different systematics due to environmental backgrounds. The latter point is particularly important for smooth start-up of the experiment when running conditions will not be perfectly known. Indeed, while the overlapping hits in the tracker make pattern recognition difficult, a calorimeter-based filter that depends only on the applied energy threshold will see the additional hits only as increased energy. This will translate to higher throughput for the background without substantially affecting the trigger efficiency. The offline application of the $\Delta t_{\text{HITS}}$ cut will/can also be used to speed up the tracker reconstruction.

In [5], the study of the DIO rejection and signal efficiency for a simple calorimeter cluster-based trigger has been revised. In Figure 9.9, the DIO survival rate as a function of the trigger efficiency is shown for different values of energy resolution. It is clear that the requirement to bring down the data storage rate to 2 kHz while keeping a filter efficiency of > 90% implies building a calorimeter with an energy resolution better < 7%.

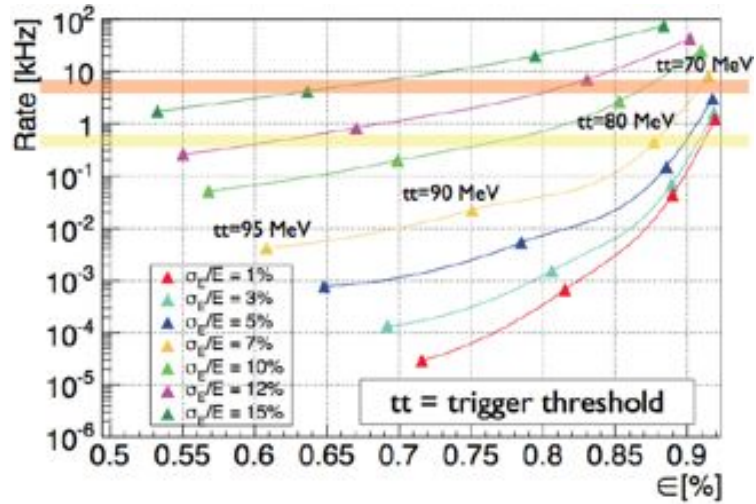

Figure 9.9. DIO rejection versus calorimeter trigger efficiency for different calorimeter energy resolution. The two horizontal bands correspond to storage on disk at 600 Hz or 4 kHz.

### 9.4.3    Summary of Calorimeter System Parameters

Table 9.2 collects the parameters of a calorimeter that meets the Mu2e requirements.

Table 9.2. Summary of calorimeter parameters.

| | |
|---|---|
| Number of Disks | 2 |
| Disk Inner and Outer Radius | 351 mm, 660 mm |
| Crystal Type, density, $X_0$, $R_M$ X0, Rm | BaF2, 4.9 g/cm$^3$, 2.0 cm, 3.0 cm |
| Crystal Shape | Hexagonal |
| Crystal Length | 180 (200) mm |
| Crystal Transversal Area | 33 mm between parallel faces |
| Total number of crystals Disk 1+2 | 1860 |
| Crystal weight | 1 Kg |
| Total scintillation mass | 2000 kG |
| Number of APD/crystal | 2 |
| APD transversal dimension | 10x10 mm$^2$ |
| Total number of APDs | 3720 |
| Total number of LV/HV boards | 240 |
| Total Number of Digitizers | 240 |
| Total number of preamplifiers | 3720 |
| Power Dissipation Preamp | 260 x 3720 mW = 967 W |
| Power Dissipation LV/HV | 1 W x 240 = 240 W |
| Power Dissipation Digitizer | 10 W x 240 = 2400 W |
| Distance between disks | 700 mm |

The calorimeter system should also be radiation hard and have a cell occupancy at a level of 10-20%, in order not to impact the data acquisition. For the radiation hardness it is estimated [7] that the hottest region of the disks will get a dose of ~ 20 krad/year. The neutron fluence will be a few x $10^{11}$ $n$/cm$^2$/year [8].

In summary, an energy resolution of ~5% is a reasonable goal for the calorimeter. A time resolution of better than 500 ps is required to be useful for PID, and a 0.5 cm position resolution is desirable for track matching.

## 9.5    Crystals and Photosensors

### Crystal baseline design

At the start of the Mu2e project, the crystal considered for the calorimeter was lead tungstate (PbWO$_4$). The low light output required running the calorimeter at -25$^o$C with very tight tolerances on the temperature stability, and the radiation dose dependence of the light output made for a difficult calibration problem. At the time of the Mu2e CDR, PbWO$_4$ had been replaced with lutetium-yttrium oxyorthosilicate (LYSO) crystals. LYSO is an excellent match to the problem at hand: it has a very high light output, a small Molière radius, a fast scintillation decay time, excellent radiation hardness, and a scintillation spectrum that is well-matched to readout by large-area avalanche photodiodes (APDs) of the type employed in the CMS and PANDA experiments. LYSO is also the preferred option for the KLOE-2 upgrade. LYSO crystals are commercially available from Saint-Gobain, SICCAS (Shanghai Institute of Ceramics), SIPAT (Sichuan Institute of Piezoelectric and Acousto-optic Technology) and other producers. Despite an active R&D program at Caltech, in cooperation with SICCAS and SIPAT, aiming to reduce the commercial price of LYSO crystals, the large increase in Lu$_2$O$_3$ salt price over the past two years has made the cost of a LYSO calorimeter unaffordable. The only downside of LYSO is the cost, which is driven by the cost of the Lu$_2$O$_3$ salt. Manipulation of the price of rare earths by the Chinese government has over the past several years resulted in an increase of the price Lu$_2$O$_3$ of by a factor of more than three (see Figure 9.10). At current prices we have concluded that an LYSO calorimeter in unaffordable. We have therefore chosen barium fluoride crystals for the calorimeter baseline.

Table 9.3 shows a comparison of the properties of BaF$_2$, LYSO, pure CsI and PbWO$_4$. Several points are worth discussing. It is clear that LYSO is the superior alternative, based on radiation length, Molière radius and light output. BaF$_2$ has much less light than LYSO, but much more than PbWO$_4$, and does not suffer from rate-dependent light output. It also has a substantially larger Molière radius and radiation length, which are disadvantages. The emission spectrum of LYSO peaks at 402 nm, a wavelength compatible with APD readout. LYSO has a scintillation decay time of $\tau = 40$ ns; the optimal integration time of 200 ns is compatible with the expected signal and background rate. The greatest advantage of LYSO crystals is the relatively high light yield, which provides excellent energy resolution at room temperature. This greatly simplifies the calorimeter design. Temperature stability requirements are also substantially less stringent.

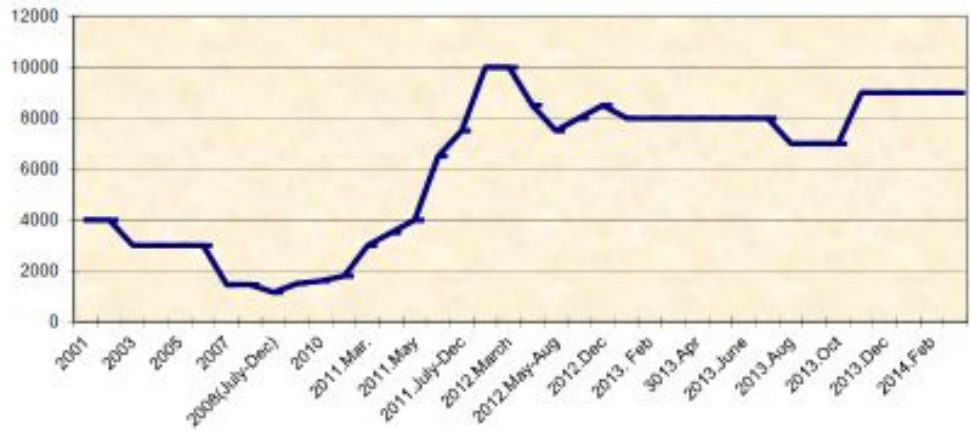

Figure 9.10. Market price of $Lu_2O_3$ as a function of time.

Table 9.3. Comparison of crystal properties for $BaF_2$, LYSO, CsI and $PbWO_4$.

| Crystal | $BaF_2$ | LYSO | CsI | $PbWO_4$ |
|---|---|---|---|---|
| Density (g/cm$^3$) | 4.89 | 7.28 | 4.51 | 8.28 |
| Radiation length (cm) $X_0$ | 2.03 | 1.14 | 1.86 | 0.9 |
| Molière radius (cm) Rm | 3.10 | 2.07 | 3.57 | 2.0 |
| Interaction length (cm) | 30.7 | 20.9 | 39.3 | 20.7 |
| $dE/dx$ (MeV/cm) | 6.5 | 10.0 | 5.56 | 13.0 |
| Refractive Index at $\lambda_{max}$ | 1.50 | 1.82 | 1.95 | 2.20 |
| Peak luminescence (nm) | 220, 300 | 402 | 310 | 420 |
| Decay time $\tau$ (ns) | 0.9, 650 | 40 | 26 | 30, 10 |
| Light yield (compared to NaI(Tl)) (%) | 4.1, 36 | 85 | 3.6 | 0.3, 0.1 |
| Light yield variation with temperature (% / °C) | 0.1, -1.9 | -0.2 | -1.4 | -2.5 |
| Hygroscopicity | None | None | Slight | None |

The much larger LYSO signals provide flexibility in the choice of photosensors and front-end electronics (FEE). There are several possible alternatives: (a) the use of a simple voltage amplifier in place of a charge sensitive amplifier and shaper. Our tests show that even a single APD/crystal can provide an ENC of 100 keV; (b) using the FEE together with a large APD can reduce the ENC due primarily to the APD leakage noise to

~30-40 keV; and (c) retaining the FEE developed for PWO-2 and reducing the APD size from $10 \times 10$ mm$^2$ to $5 \times 5$ mm$^2$ can keep the ENC at the level of ~150 keV. For any of these options, the overall noise for a group of 25 crystals will be below 1 MeV, allowing the energy resolution to be pushed as close as possible to the intrinsic photoelectron statistics limit of 1%.

A third advantage of LYSO is the excellent radiation hardness, which has been measured for both $\gamma$'s and neutrons. Negligible deterioration of signals (10% loss in light yield) is observed with $\gamma$ exposures of 10,000 Gy (*i.e.* 15 years of Mu2e running). Therefore, with LYSO there is no need any stimulated recovery mechanism and there will be no reduction of running time.

Figure 9.11 shows the response of a LYSO crystal read out by a conventional PMT to a $^{22}$Na source. The energy resolution is excellent. The same technique is used to measure the LRU (Longitudinal Response Uniformity) by scanning the crystal along its axis [2]. Control of the cerium concentration in the growing process has brought the longitudinal response uniformity in current production LYSO crystals to better than 2-3%.

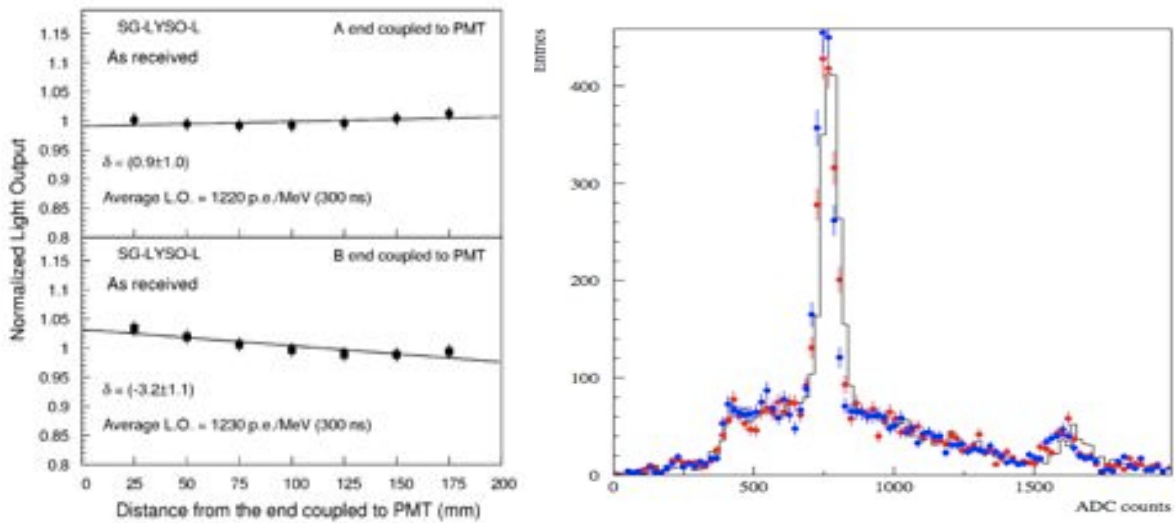

Figure 9.11. Longitudinal response uniformity measurement for a Saint-Gobain crystal (left). Charge response to a $^{22}$Na source for a LYSO crystal readout by a PMT (right).

However, given the extraordinary expense of LYSO, it has been necessary to seek alternatives. Barium fluoride (BaF$_2$) has been selected as the baseline scintillating crystal. As shown in Table 9.3, the light yield is much greater than PbWO$_4$, although smaller than that of LYSO. The presence of a very fast scintillation decay time component at 220 nm (<1 ns) is very useful in background rejection, providing compensation for the larger shower size. If rates are not too high, it may be possible to use the larger slow component (650 ns) at 300 nm as well.

The emission spectrum of $BaF_2$ is shown in Figure 9.12. The short wavelength of both the fast and slow scintillation components presents a difficult readout problem. Photomultiplier tubes with quartz windows and perhaps solar-blind photocathodes are well-matched to the $BaF_2$ spectrum, but will not work in the field of the Detector Solenoid. Channel plate PMTs are at present far too expensive, although spinoffs from the LAPPD project are still being pursued. Our main thrust, however, is to use solid-state photosensors, either APDs, SiPMs or MPPCs, with extended UV response.

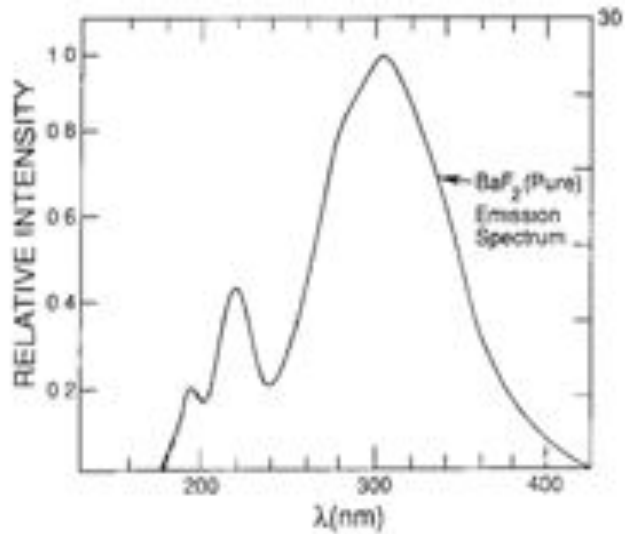

Figure 9.12. $BaF_2$ emission spectrum. The fast component (900 ps) peaks at 220 nm. The slow component (650 ns) peaks at ~300 nm.

## 9.6    Photosensors

There are adequate photosensor candidates for $BaF_2$ readout: typical large APDs have poor quantum efficiency in the $BaF_2$ spectral region (see  (left)). However, APDs and MPPCs from Hamamatsu and RMD are made without the normal protective epoxy coating, and are therefore somewhat fragile, can have quantum efficiencies in the 200 nm region of ~17% (see  Figure 9.13 (right)), but do not discriminate between the 220 nm fast component and 300 nm slow component of $BaF_2$. The presence of the slow component limits the rate capability of the calorimeter and can therefore be an issue in high rate conditions.

There is a straightforward approach to both improving the photosensor quantum efficiency and the slow component discrimination, but one that requires some R&D. A Caltech/JPL/RMD consortium has been formed to develop a modified RMD large-area APD (Figure 9.14 (left)) into a delta-doped superlattice APD [9]. This device will also incorporate an atomic layer deposition antireflection filter [10] that will provide 60% quantum efficiency at 220 nm and ~0.1% efficiency at 300 nm, thereby enabling us to not only have a larger number of photoelectrons/MeV (×3), but also to take full advantage of the fast decay time component of BaF$_2$. The greatly reduced undepleted region of this device will also result in substantially improved rise time.

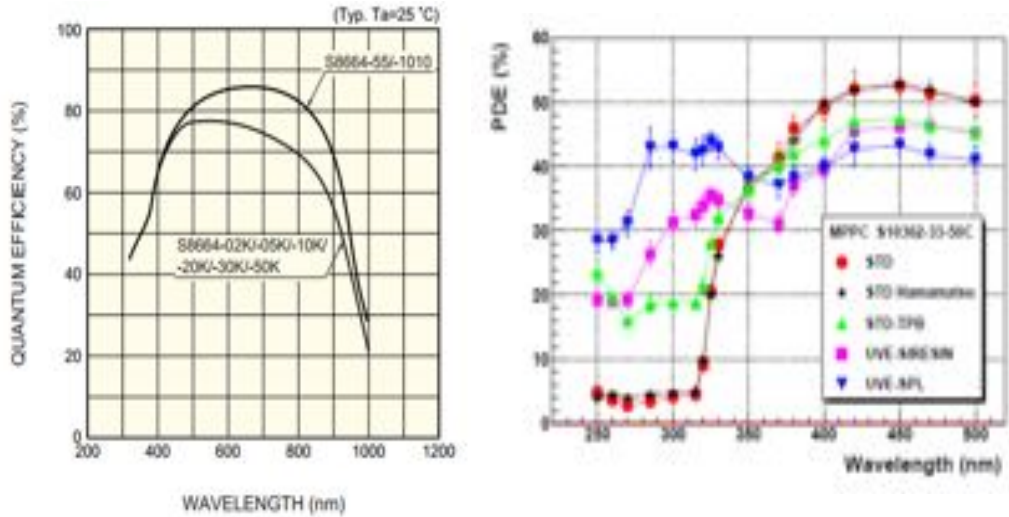

Figure 9.13. Spectral response of a conventional Hamamatsu APD (left); Photon detection efficiency (PDE) as a function of wavelength for four SiPM prototypes (right). The typical PDE values of the standard MPPC S10362-33-50C from Hamamatsu are shown for comparison. These measurements were performed at 25°C and include effects of cross-talk and after-pulses.

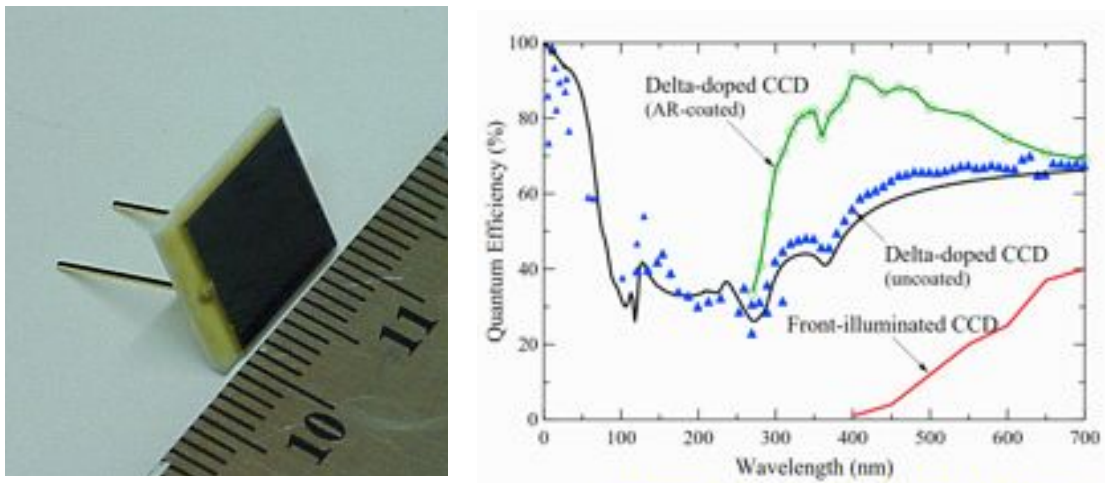

Figure 9.14. (left) A standard RMD 9x9 mm APD (left); QE versus wavelength for a CCD before and after delta-doping (right) [12]. The black line is the QE theoretical (1-R) limit.

Figure 9.14 (right) shows the response of a CCD imaging device whose response has been modified at JPL by delta-doping and the application of an antireflection (AR) filter. The improvement in the UV response over the basic front-illuminated CCD is evident. The same procedure will be applied to a large-area RMD APD, for which the Caltetch/JPL/RMD consortium has received an SBIR grant. Conventional RMD APDs will be thinned to remove the surface and undepleted region before the avalanche layer, and the superlattice structure and optimized antireflection coating will then be deposited at the JPL Microdevices Lab. Figure 9.15 shows the calculated QE response of the resulting APD as a function of wavelength. For a five-layer AR coating, the QE at the fast component of BaF$_2$ is nearly 70%, and the extinction at the slow component wavelength is nearly complete. The greatly reduced undepleted region also improves the time response of the device, as shown in Figure 9.16.

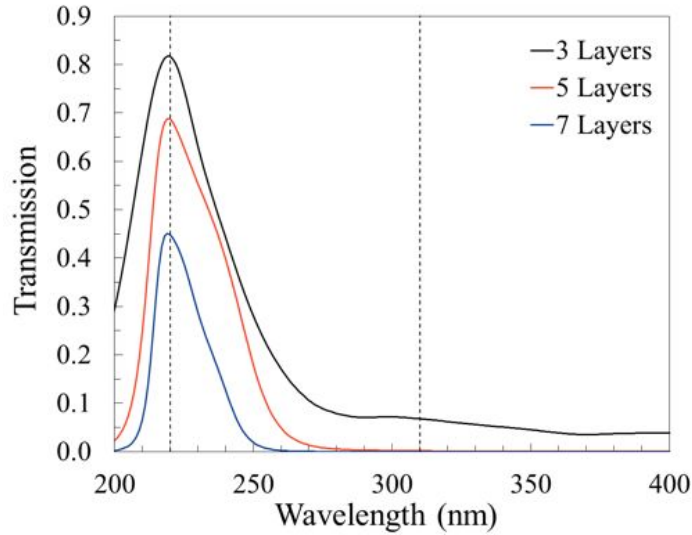

Figure 9.15. Calculated response of 3, 5 and 7 layer combination Al$_2$O$_3$/Al interference filters on a Si substrate. The blocking ratios for 220 vs. 310 nm are 12:1, 400:1 and 15,000:1.

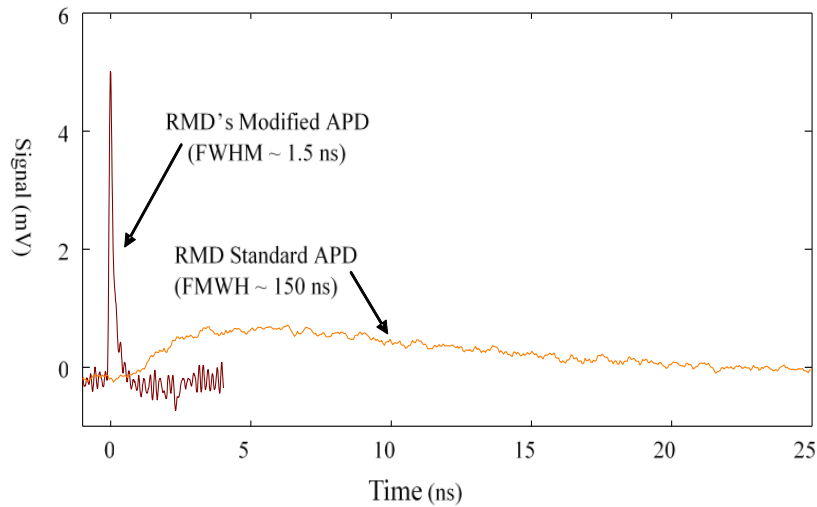

Figure 9.16. The rise times from two APDs, directly measured on a digital oscilloscope, while illuminated with a pulsed 405 nm laser. The red trace is the thinned APD (FWHM ~ 1.5 ns) while the orange trace is a standard APD (FWHM ~ 150 ns).

# 9.7    Radiation Hardness

Four scintillating crystals, cerium doped lutetium oxyorthosilicate ($Lu_2SiO_5$:Ce, LSO) [11] or lutetium yttrium oxyorthosilicate ($Lu2_{(1-x)}Y_{2x}SiO_5$:Ce, LYSO) [12], barium fluoride ($BaF_2$) and pure CsI [13] are under consideration by the Mu2e experiment to construct a crystal electromagnetic calorimeter. Mass production capability exists in industry for all four crystals.

All known crystals suffer from radiation damage. There are three possible damage effects in crystal scintillators: (1) damage to the scintillation-mechanism, (2) radiation-induced phosphorescence and (3) radiation-induced absorption [14]. A damaged scintillation mechanism would reduce scintillation light yield and light output, and may also change the light-response uniformity along the crystal if the dose profile is not uniform along the crystal. Radiation-induced phosphorescence (commonly called afterglow) causes an increased dark current in photo-detectors, and thus an increased readout noise. Radiation-induced absorption reduces light attenuation length and thus light output. It may also change light-response uniformity if light attenuation length is shorter than twice the crystal length [14]. No scintillation-mechanism damage was observed in crystals listed in Table 9.3. The main radiation damage in these crystals is radiation-induced absorption, or color-center formation. Radiation-induced color centers may recover at the application temperature through color-center annihilation, leading to a dose-rate dependent damage [14]. If so, a precision light monitoring system is mandatory to follow variations of crystal transparency *in situ*. Radiation-induced absorption in all of the crystals listed in Table 9.3 does not recover at room temperature, and is thus not dose-rate dependent. Radiation-induced damage in these crystals was measured for samples grown recently with a length long enough for the Mu2e calorimeter application. Both transmittance and light output were measured. Radiation damage in these crystals manifests itself as a loss of transmittance and light output as a function of the integrated γ-ray dose.

## 9.7.1    Radiation dose in crystals

The expected dose deposited in each crystal is estimated using the full Mu2e simulation, which includes contributions from particles produced by the beam flash, electrons from muons decaying in orbit, neutrons, protons, and photons. The dose per year is shown in Figure 9.17 for the front and back disks. The average dose is around 3 kRad/year (0.5 kRad/year) in the front (back) disk, increasing to 15 kRad/year for the innermost crystals of the front disk.

Figure 9.18 shows the dose from neutrons originating from the beam flash and from the stopping target incident on the calorimeter disks together with the total dose as a function of radial position on the disk.

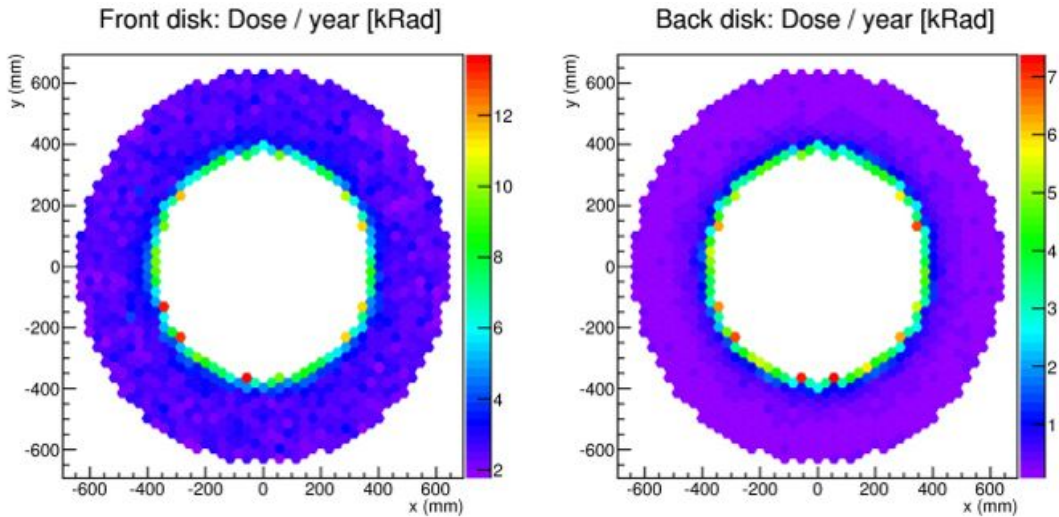

Figure 9.17. Expected dose in each crystal of the front (left) and back (right) disks. The dose is given in kRad/year.

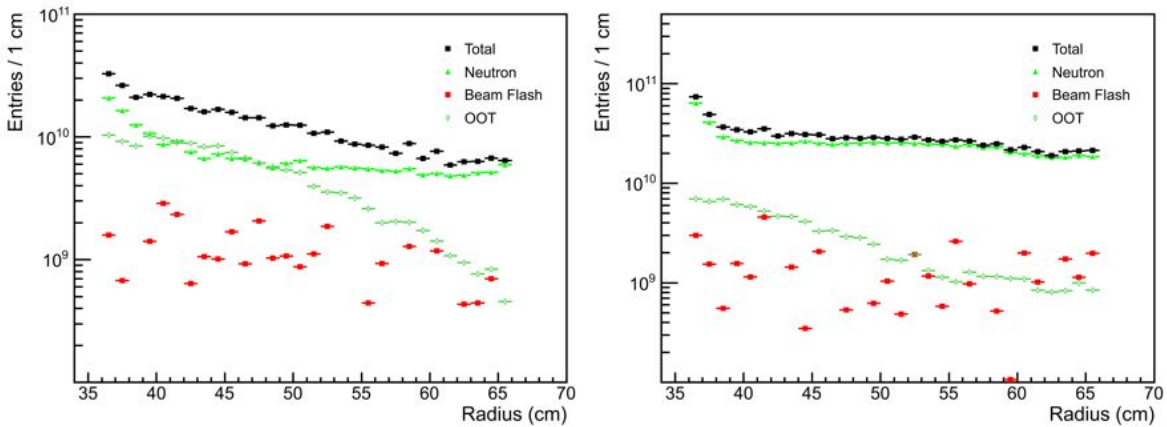

Figure 9.18. The dose from neutrons, from the beam flash and from the out of the target (OOT) particles incident on the first (left) and second (right) calorimeter disks, together with the total dose, as a function of radial position on the disk. The units are 1 MeV/year/cm$^2$.

## 9.7.2    Radiation damage in LSO/LYSO

Because of their high stopping power ($X_0 = 1.14$ cm), bright (4 times BGO) and fast ($\tau = 40$ ns) scintillation light LSO/LYSO crystals have attracted broad interest in the HEP community. Their radiation hardness against γ-rays [15], neutrons [16] and charged hadrons [17] has been thoroughly studied. As a result of these studies, LSO/LYSO crystals have been considered by the Mu2e and SuperB experiments for use in their total absorption calorimeters. A LSO/LYSO crystal Shashlik sampling calorimeter is also proposed for future HEP experiments in severe radiation environments, such as the CMS endcap calorimeter upgrade for the proposed HL-LHC [18]. As stated above, radiation damage in LYSO does not recover at room temperature and is thus not dose rate-dependent, but can be repaired by thermal annealing [15].

Figure 9.19 shows the longitudinal transmittance spectra (left) and the normalized light output (right) measured by a XP2254 PMT as a function of the integrated dose up to 1 Mrad for five 20-cm long LSO/LYSO samples from CTI, CPI, SG, SIC and SIPAT. Also shown in the left plot are the photo-luminescence spectra without internal absorption (blue dashes) as well as the values of the emission-weighted longitudinal transmittance (EWLT). All five tested samples have consistent radiation resistance with a loss of EWLT and light output at a level of about 12% for an integrated γ-ray dose up to 1 Mrad. This excellent radiation hardness is the best among all known inorganic crystal scintillators.

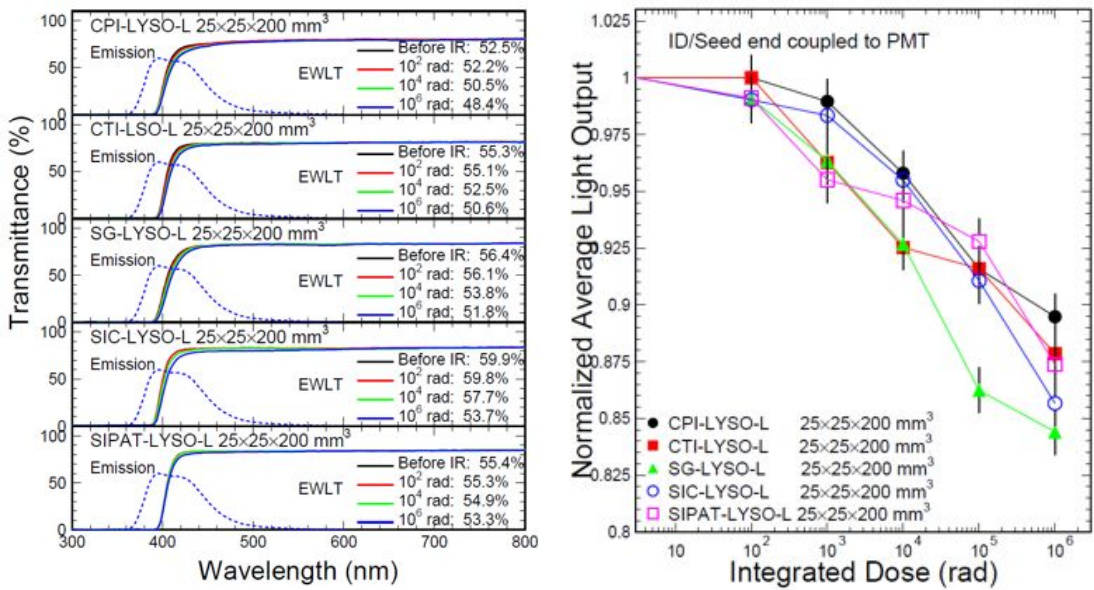

Figure 9.19. The longitudinal transmittance spectra (left) and the normalized light output (right) are shown as a function of the integrated dose up to 1 Mrad for five LSO/LYSO samples.

### 9.7.3 Radiation damage in BaF$_2$

Table 9.3 shows that BaF$_2$ is a unique crystal scintillator, with a fast scintillation light component with sub-nanosecond decay time and a brightness of about 5% of LYSO. Its radiation hardness was thoroughly investigated twenty years ago when this material was proposed for the GEM experiment for the proposed SSC. In that study, radiation damage in BaF$_2$ was found to be dose-rate independent [19], and that it could be thermally annealed or optically bleached for recovery [20]. This feature reduces the cost for radiation damage investigation, and provides a possibility to cure radiation damage *in situ* by *e.g.* optical bleaching. Importantly, it was also found that radiation damage in 25-cm long BaF$_2$ crystals saturated after about 10 krad [19].

Figure 9.20 shows the longitudinal transmittance spectra (left) and the light output as a function of integration time (right) for a 25-cm $BaF_2$ sample grown at SIC in 2012 as a function of the integrated dose up to 1 Mrad. The light output was measured using a R2059 PMT with a bi-alkali cathode and a quartz window that measures both the fast ($A_0$) and slow ($A_1$) components. Also shown in the left plot is the x-luminescence spectrum (blue dashes) and the corresponding EWLT values for the fast (220 nm) and slow (300 nm) scintillation component. Radiation damage at a level of 33% and 40% is observed in, respectively, the EWLT and light output for the fast scintillation component after an integrated dose of 10 krad. No further damage was observed beyond 10 krad, indicating that the defects in this $BaF_2$ crystal are fully activated to form color centers at this radiation level. This saturation effect is consistent with the result observed twenty years ago [19].

Since radiation damage in halide crystals is caused by oxygen contamination [14] it is expected that an R&D program aiming at reducing oxygen contamination will further improve crystal quality and reduce the level of radiation damage in $BaF_2$.

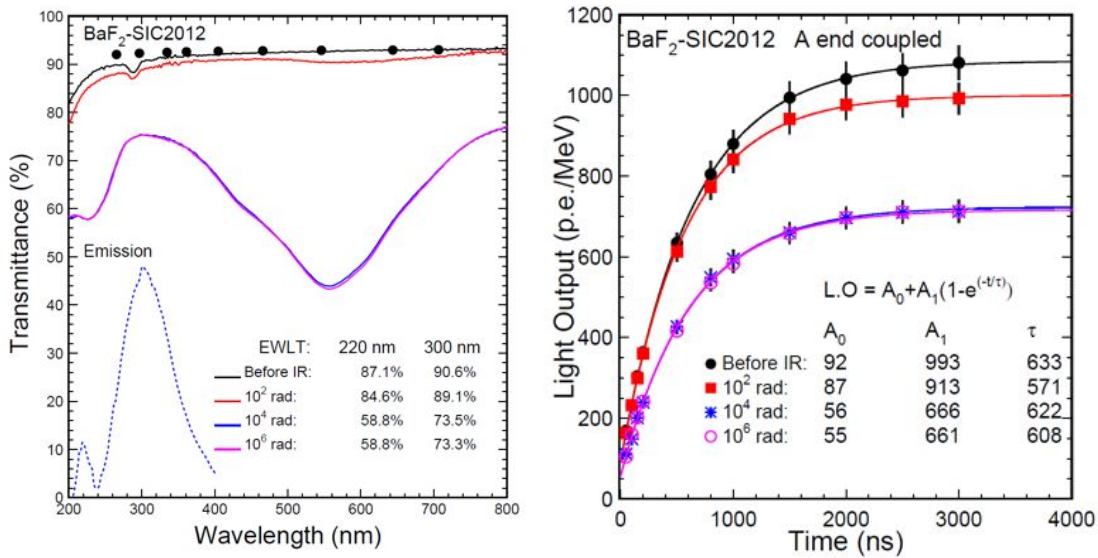

Figure 9.20. The longitudinal transmittance spectra (left) and the light output as a function of integration time (right) are shown as a function of the integrated dose up to 1 Mrad for a $BaF_2$ sample of 2.5 x 2.5 x 25 $cm^3$.

### 9.7.4    Radiation damage in pure CsI

Because of its low melting point and raw material cost, pure CsI is a low-cost crystal scintillator. Table 9.3 shows that it has a fast scintillation light peaked at 310 nm, with a decay time of about 26 ns and a brightness that is similar to the fast component of $BaF_2$. Its radiation damage has been found to be dose rate-independent [21]. Unlike $BaF_2$, thermal annealing and optical bleaching were not found to be effective for CsI [16].

Radiation damage study for CsI is thus a costly exercise, since crystal samples after testing are unusable. It was also found that radiation damage in 20-cm long pure CsI crystals showed no saturation, with light output loss of 70 - 80% after 1 Mrad [20].

Figure 9.21 shows the longitudinal transmittance (left) and light response uniformity (right) for a pure CsI sample of 5 x 5 x 30 $cm^3$ grown at SIC in 2013 and irradiated up to 1 Mrad. Its light output was measured by using a R2059 PMT with a bi-alkali cathode and a quartz window. Also shown in the left plot is the photo-luminescence spectrum (blue dashes) and the corresponding EWLT values. Radiation damage at a level of about 60% and 80% was observed respectively in EWLT and light output after an integrated dose of 1 Mrad. The damage, however, shows no saturation up to 1 Mrad, indicating a high density of defects in this crystal. The result of this measurement is consistent with the data obtained twenty years ago for 20-cm long pure CsI crystal samples from Kharkov [21].

Since radiation damage in CsI crystals is caused by oxygen contamination [16], it is expected that an R&D program aiming at reducing oxygen contamination could improve crystal quality and reduce the level of radiation damage in CsI.

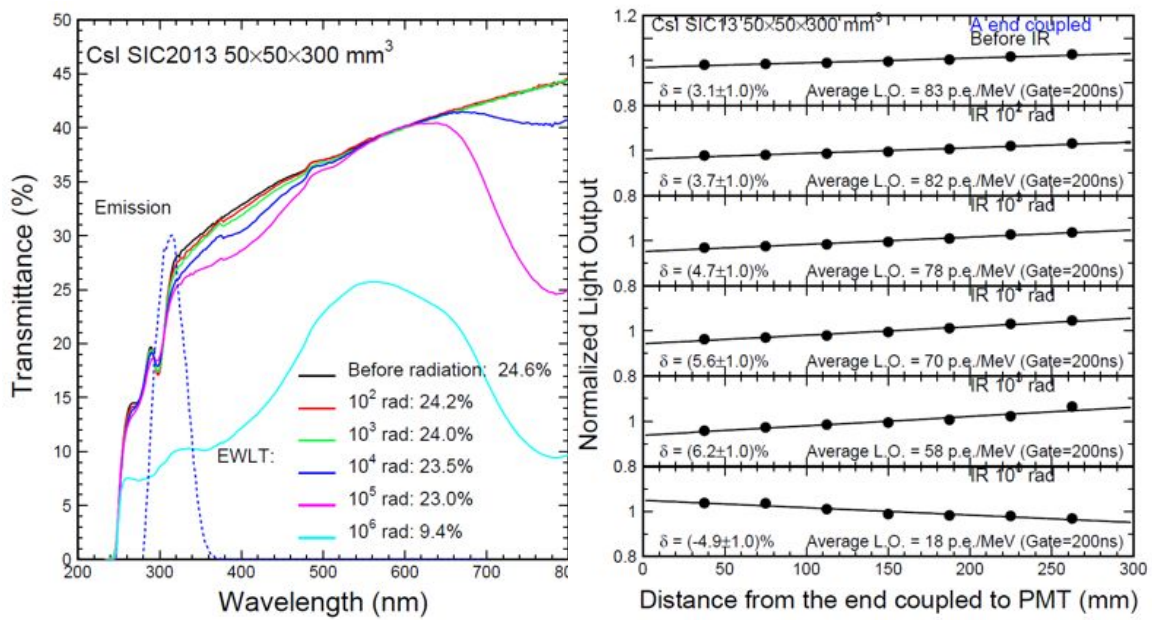

Figure 9.21. The longitudinal transmittance (left) and the light output as a function of integration time (right) are shown as a function of the integrated dose up to 1 Mrad for a pure CsI sample of 5 x 5 x 30 $cm^3$.

### 9.7.5    Summary

All three inorganic crystal scintillators discussed in this section can be used to construct a total-absorption electromagnetic calorimeter for the Mu2e experiment. All materials suffer from radiation damage in the form of radiation-induced absorption or color-center formation. The radiation damage in all three crystals does not recover, so is not dose rate-dependent in the manner of $PbWO_4$. While radiation damage in LYSO and $BaF_2$ can be thermally annealed, this is ineffective for pure CsI, indicating a high R&D cost to pursue pure CsI. Results obtained with $BaF_2$ and pure CsI samples grown recently at SIC are consistent with data published twenty years ago.

Figure 9.22 shows a comparison of the radiation hardness for these three crystal scintillators up to a 1 Mrad dose. The losses in EWLT (top left), light output (bottom left) and radiation-induced absorption coefficient (RIAC) at the peak of their radio-luminescence spectra are shown. LSO/LYSO crystals are clearly the best in both light output and radiation hardness. The high cost of $Lu_2O_3$ raw material, however, makes its price prohibitive. Both $BaF_2$ and pure CsI have comparably fast light and much lower cost. These two materials are, however, significantly more susceptible to radiation damage than LSO/LYSO. Because of low defect density, radiation damage in $BaF_2$ saturates after about 10 krad, promising a stable detector for high integrated doses. Radiation damage in pure CsI is small at low doses, but shows no saturation at high doses, indicating continuous degradation under irradiation. One additional advantage of $BaF_2$ is that it is possible to cure radiation damage in $BaF_2$ *in situ* through optical bleaching.

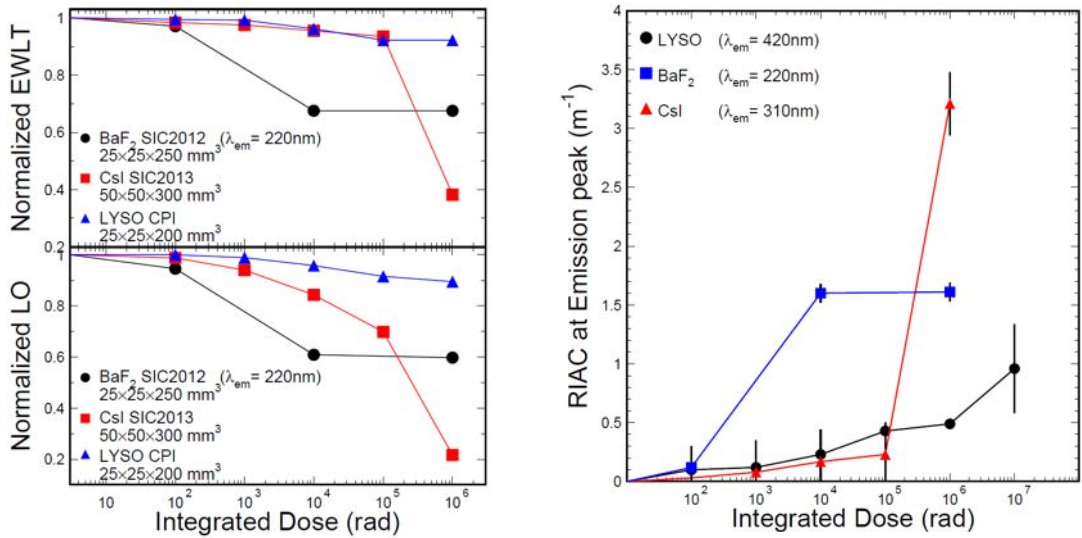

Figure 9.22. The normalized EWLT and light output (left) and the RIAC at the emission peak (right) are shown as a function of the integrated dose up to 1 Mrad for LYSO, BaF2 and pure CsI crystals

The quality of both BaF$_2$ and pure CsI can be improved through systematic R&D programs aimed at reducing oxygen contamination during crystal growth. A close collaboration with crystal growers will be crucial for this effort.

## 9.8     Simulation

### 9.8.1     Calorimeter optimization

The baseline calorimeter design for Mu2e consists of two annular disks [22] separated by approximately a half-wavelength of the typical conversion electron helical trajectory. This configuration minimizes the number of low-energy particles that intersect the calorimeter from the Transport Solenoid, the muon stopping target or the muon beam stop, while maintaining excellent signal efficiency. Hexagonal-faced crystals have been selected to tessellate the annular disk, as these provide a more natural tiling and offer better coverage than square crystals. Hexagonal crystals also offer superior light collection efficiency and more closely approximate the shape of electromagnetic showers. In optimizing the disk design the inner and outer radii of the disks, their placement and relative separation, and the dimensions of the crystals have been considered.

The dimensions of the disks were the first issue to be addressed.

Figure 9.23 (left) shows the calorimeter efficiency for conversion electrons that have been reconstructed in the tracker as a function of the disk inner and outer radii. Only clusters with a deposited energy above 60 MeV were considered. The separation between the disks is set to 70 cm, corresponding approximately to a half-wavelength. The crystal size is taken to be 33 mm across flats. The efficiency reaches a maximum at an outer radius of about 67 cm for inner radii of both 35 and 36 cm. A similar conclusion holds considering disks with varying outer radii. The design was further refined by minimizing the empty space between the crystals and the disk boundaries as a function of the crystal size and the disk radii. Figure 9.23 (right) shows empty space as a function of the disk inner radius for a crystal dimension of 33 mm across flats. A crystal size of 33 mm across flats with disk inner and outer radii of 351 mm and 660 mm, respectively was selected. This choice ensures sufficient space to mount the readout at the back of the crystals while maintaining efficiency and limiting the number of readout channels. The crystal layout is shown in Figure 9.24.

Finally, the separation between the disks was optimized. As shown in Figure 9.25, a separation of 70 cm is optimal, being independent of the energy deposited in the calorimeter. The position of the disk with respect to the tracker has a negligible impact on the efficiency, as expected from translational invariance.

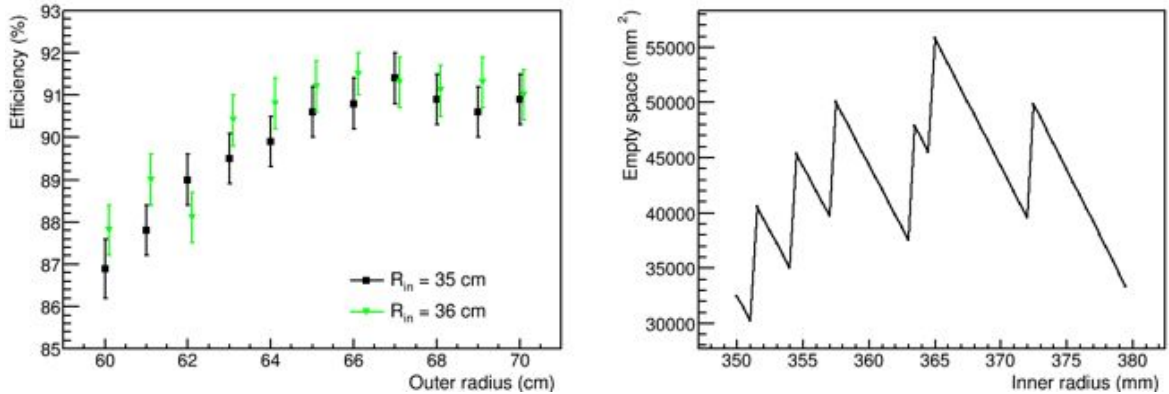

Figure 9.23. Calorimeter efficiency for detection of good signal tracks initially found in the tracker as a function of the disk outer radius for different values of inner radius (left); Empty space between the crystals and the disk inner bore (right).

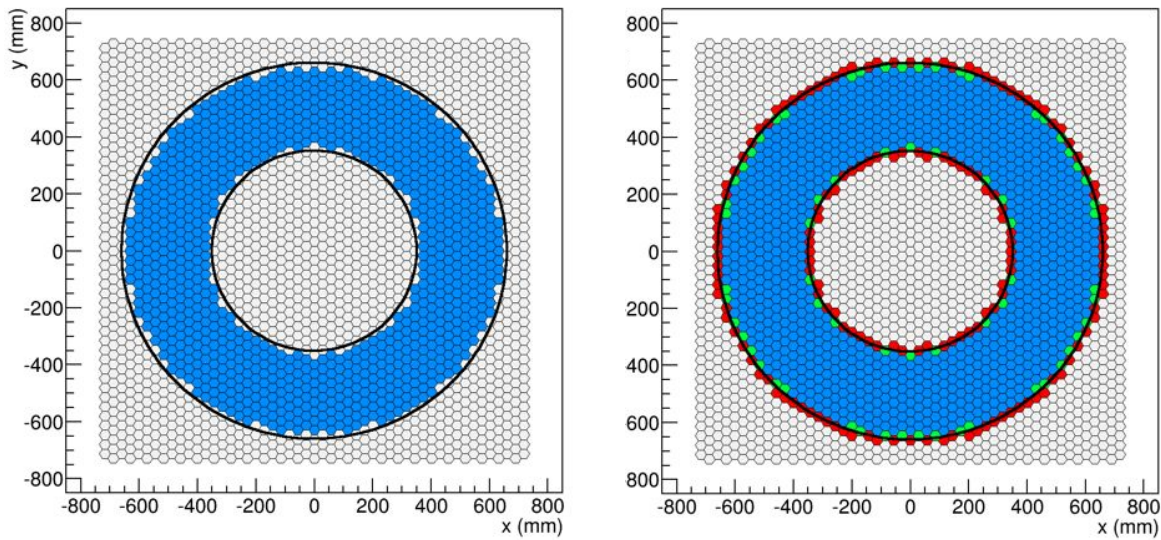

Figure 9.24. Crystal layout for a crystal size of 33 mm across flats with disk radii of 351 mm and 660 mm (left). The crystals in the disk are colored in blue. Similar layout, together with crystals intersecting the disk boundaries colored in green (blue) if their center lies inside (outside) the disk boundaries.

### 9.8.2    Calorimeter Resolution

Tracks do not enter the calorimeter normal to its face, but at an angle close to 45 degrees. As the interaction depth is not known, the distance $\Delta y = y_{track}\text{-}y_{cluster}$ depends on the track direction, as well as on the shower depth. To remove this dependence, track-to-cluster residuals are calculated in the direction orthogonal to the track; the corresponding distribution is shown in Figure 9.26 (left). A coordinate resolution of about 1 cm can be achieved with BaF$_2$ crystals.

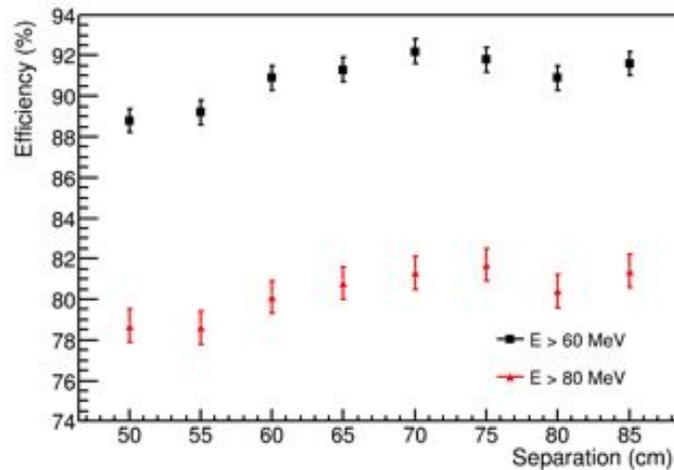

Figure 9.25. Calorimeter efficiency for detection of good signal tracks first found in the tracker, as a function of the separation between the disks, for two thresholds of the energy deposited in the calorimeter.

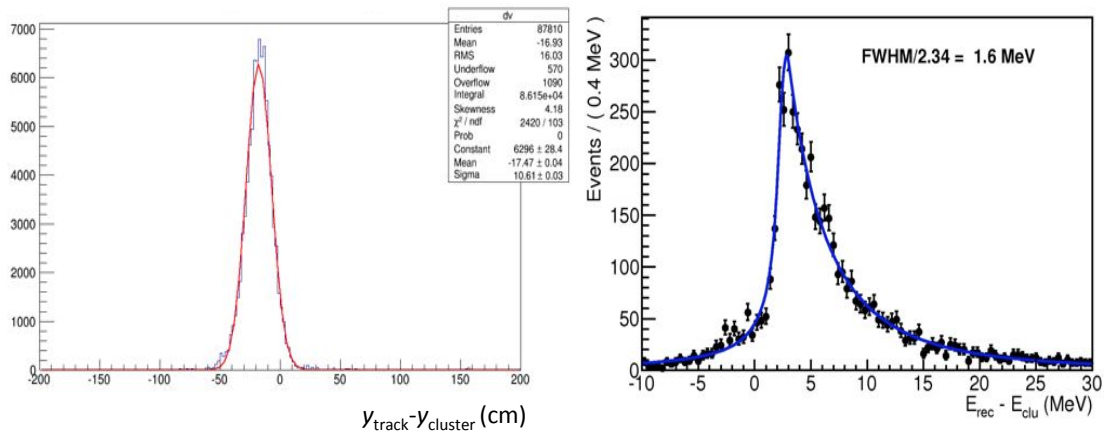

Figure 9.26. Distribution of residuals between the reconstructed track and the calorimeter cluster in cm (left). Residuals are calculated in the direction orthogonal to the track. Difference between the input energy and the reconstructed cluster energy for a LYSO-based calorimeter (right). Positive tail in this coordinate is due to longitudinal leakage and albedo. Negative tail is due to the pileup of environmental background. For the LYSO, a simulation of ~1000 p.e./MeV and an electronic noise of 30 keV has been used.

Event reconstruction in the calorimeter proceeds in several stages. The interaction of the incident particle with the crystals is first simulated by GEANT4, recording the energy, position and time of each step. Each energy deposit is converted into photons, taking into account corrections from non-linearities in the light production and non-uniformities in the longitudinal response. The response of each APD is then simulated, including the related electronic noise. A final version of signal digitization and pile-up identification remains to be implemented. To simulate these effects, hits within a time window of 100 ns are grouped together to form crystal hits.

Finally, the crystal hits are finally used to form calorimeter clusters. The clustering [22] algorithm starts by taking the crystal hit with the largest energy as a seed, and adds all simply connected hits within a time window of ± 10 ns and a threshold in energy of 3 times the electronic noise. Hits are defined as connected if they can be reached through a series of adjacent hits. The procedure is repeated until all crystals hits are assigned to clusters. Additional low-energy deposits that are disconnected from the main cluster are recovered by dedicated algorithms. These fragments are usually produced by the shower, or by low-energy photons emitted by incident particles. Recovering these split-off deposits significantly improves the energy resolution.

The energy resolution is estimated by simulating conversion electrons distributed at random in the stopping target foils, together with the expected neutron, photon and DIO backgrounds. The distribution of the difference between the true signal electron energy obtained by simulation and the reconstructed cluster energy is plotted in Figure 9.26 (right) for a "LYSO-based" calorimeter. This variable accounts for the energy lost by the electron before hitting the calorimeter. The low-side tail is due to background pile-up with the cluster. The distribution is fit with a Crystal Ball function to extract the resolution. A full width at half-maximum of 1.6 MeV is observed. The fraction of pile-up background for cluster energies between -10 and 0 MeV is found to be 9%. This has not been optimized; many improvements can be applied at the hit-reconstruction and cluster level to reduce this contamination. The first level of rejection is in the applied threshold and on the cluster formation technique; a higher threshold level and a more refined technique for joining cells (based on a time resolution parameterization instead of a single 10 ns cut) would improve this substantially.

Since the choice of crystal in the baseline calorimeter design has changed, the algorithm has not yet been re-optimized for the $BaF_2$-based calorimeter. With the Hamamatsu devices having a gain of ~50, the electronic noise would be larger than in the LYSO case (~400 keV w.r.t. 30 keV). The modified APD being developed, based on an RMD APD with a gain of 500, reduces the noise so that the threshold level is effectively similar to that for a LYSO calorimeter. The spurious hit contribution coming from the environmental background will be much smaller in $BaF_2$ compared to LYSO, since the pulse width will be smaller (50 ns compared to 200 ns). Indeed, the average energy of the environmental noise is at a level of a few hundreds of keV for $BaF_2$.

To guide the discussion, the energy resolution expected with the $BaF_2$ calorimeter has been studied as a function of two variables: (a) the light yield and (b) the electronic noise of the preamplifier. The distribution of the energy resolution is shown in Figure 9.27 (left), for the case of a light yield of 30 p.e./MeV and for an electronic noise of 60 keV.

An energy resolution of 3.6% is found. Figure 9.27 (right) shows a compilation of the resolution expected for different values of the electronic noise, with and without nominal background.

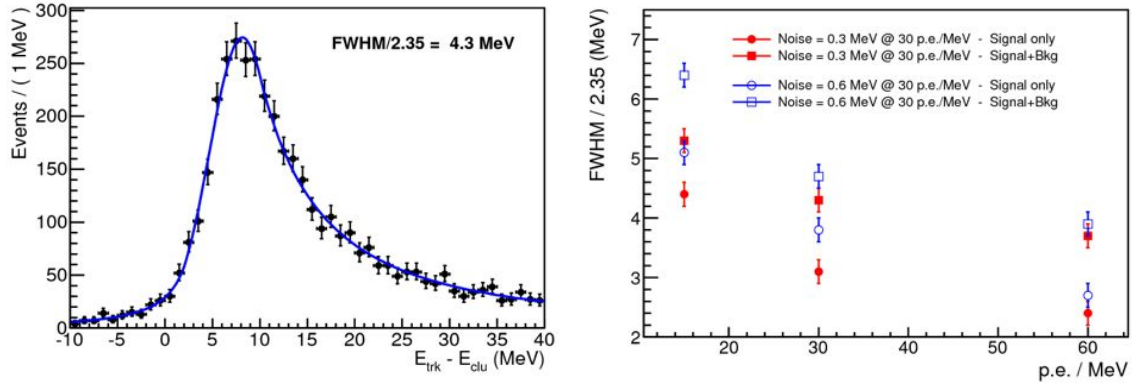

Figure 9.27. Difference between the input energy and the reconstructed cluster energy for a BaF$_2$ based calorimeter (left). The case presented is for a light yield of 30 p.e./MeV and 60 keV electronic noise. Dependence of the energy resolution for a BaF$_2$ based calorimeter as a function of the light yield for different values of electronic noise, and with and without nominal background (right).

### 9.8.3    Calorimeter-driven Track Finding

In addition to improved background rejection, the calorimeter provides a robust approach to track reconstruction. Mu2e doesn't have an "event time"; all straw hits reconstructed within a micro-bunch therefore have to be considered by the track-finding algorithm and the track time is reconstructed as a track fit parameter. The standalone Mu2e track reconstruction attempts to find the 100 ns time slice within the microbunch with the maximum number of hits in it, and uses those hits to find a track. In the presence of the correlated in-time background produced by δ-electrons, such an approach relies strongly on the δ-electron hits being identified and excluded before execution of the track reconstruction, which at present uses a neural network-based procedure. A cluster produced by a track and reconstructed in the calorimeter can be used as a seed for the track finding. Figure 9.28 shows the momentum distributions for tracks found by the standalone track finding algorithm and for tracks that are missed by the standalone algorithm but reconstructed after incorporating the calorimeter clusters. The calorimeter-driven track finding improves the overall track finding efficiency by 18%. More details can be found in [23].

## 9.9    Electronics

The overall scheme for the calorimeter readout electronics is shown in Figure 9.29. The front-end electronics (FEE) consists of two discrete and independent chips (Amp-HV) for each crystal that are directly connected to the back of the photosensor pins. These provide both the amplification stage and a local linear regulation for the photosensor bias voltage. Each disk is subdivided into twelve similar azimuthal sectors of 78 crystals. Groups of 16 Amp-HV chips are controlled by a dedicated ARM controller that distributes the LV and the HV reference values, while setting and reading back the locally regulated voltages. Groups of 16 amplified signals are sent to a digitizer module where they are sampled and processed before being optically transferred to the DAQ system.

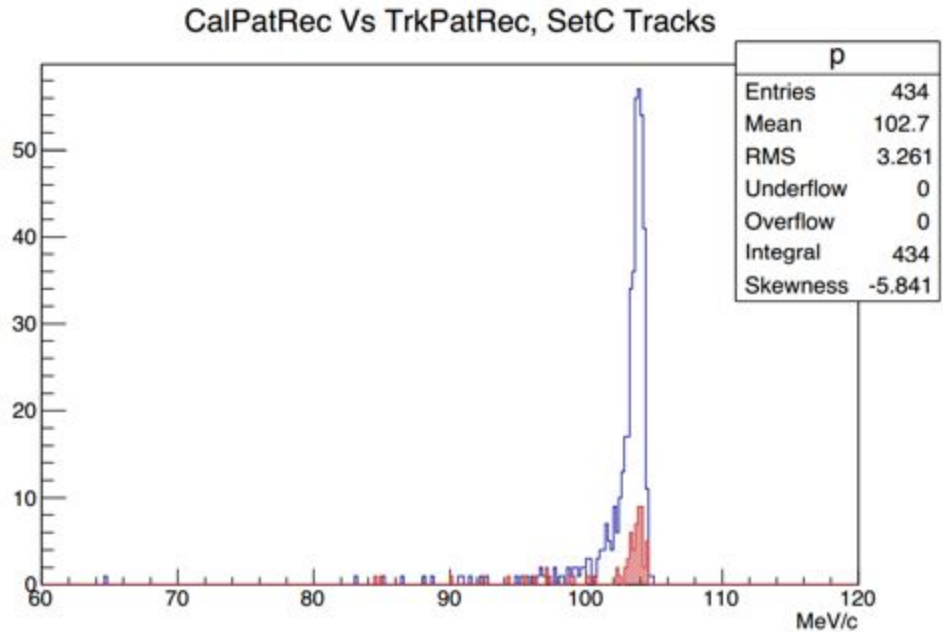

Figure 9.28. Distribution of the reconstructed momentum for CE candidates provided by the standalone track-based pattern recognition (blue) and the track candidates provided by the combined calorimeter and track pattern recognition that are missed by the standalone track-based algorithm (red).

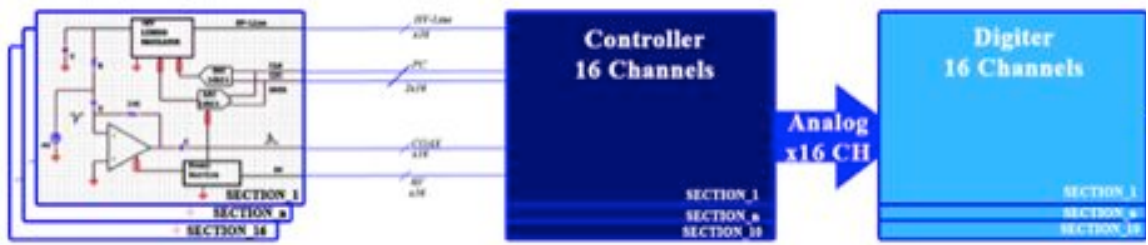

Figure 9.29. Overall schematic of the calorimeter electronics. The drawing represents the distribution of the electronics and the connection for one dodecagonal sector of one disk. 10 ARM controllers and 5 WFD digitizer boards go to a single crate. The WFD are optically connected in a ring for providing data to the DAQ readout controller.

The required characteristics for the preamplifier are (i) high amplification with low noise, (ii) fast signal rise and fall times for good time resolution and pileup rejection, (iii) a low detection threshold at the MeV level, (iv) operation in a rate environment of 200 kHz/channel, and (v) low power consumption.

The average input current depends the background hit distribution. The calorimeter hits are due to two sources: (i) the flash of particles, called the "beam flash", produced within 200 ns of the interaction of the proton beam in the production target, and by the cascade of decays and interactions with the surrounding material, and (ii) the background events generated by the muon beam interacting with the collimators, muon stopping target and the beam dump. From the simulation, the sum of the *prompt* beam flash hits per channel corresponds to an equivalent energy deposition of ~ 5 MeV for each micro-bunch. For the second background source, the arrival time of the related beam background follows an exponential decay curve with a time constant similar to that of the signal.

The largest source of background in the calorimeter is from neutrons generated by muon capture, which produces an occupancy of more than 1 hit/channel. In Figure 9.30 (left), the calorimeter cell occupancy as a function of disk radius is shown. In Figure 9.30 (right), the occupancy is reduced to ~10% by applying a 1 MeV threshold. This environmental background corresponds to an average current input to the photosensors, which is dominated by the beam flash and equivalent to $I = N_{pe} \cdot MBR \cdot e$, where $N_{pe}$ is the average number of photoelectrons, MBR is the micro-bunch rate and e is the elementary electrical charge. In the LYSO case, using $N_{pe}$= 5 MeV x (2000 p.e./MeV) = 10000 and MBR = 200 x$10^3$ Hz, we obtain I=200 pA;  this translates to 10 pA for $BaF_2$. The typical APD dark current ranges from 10 to 100 nA when working at the operating point of gain = 50. For the LYSO case, the beam flash-related current is comparable to the leakage current; it is negligible for $BaF_2$. To protect the system against a higher beam flash dose in the future, a bias voltage regulation scheme will be implemented, tied to microbunch gating, to lower the bias in the first 200 ns, ramping it back to full voltage over the next 100 ns.

## 9.9.1    The Amp-HV chip

The Amp-HV is a multi-layer double-sided discrete component chip that carries out the two tasks of amplifying the signal and providing a locally regulated bias voltage, thus significantly reducing the noise loop-area. The two functions are each independently executed in a single chip layer, named the Amp and HV sides, respectively.

The development of the Amp-HV chip has been done by the Laboratori Nazionali di Frascati (LNF) Electrical Design Department. A detailed description of the system can be found in [24]. Forty prototypes were built during 2013 and have been used for testing a LYSO matrix prototype. A picture of an Amp-HV prototype is shown in Figure 9.31.

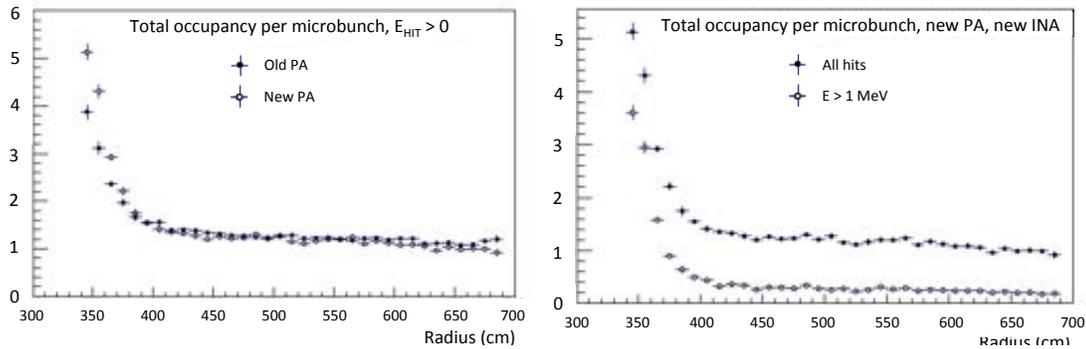

Figure 9.30. Occupancy (hit crystals/microbunch) as a function of the calorimeter radius (left) and the dependence of the occupancy on the applied threshold (right). Data points are shown for different Proton Absorber (PA) and Inner Neutron Absorber (INA) configurations.

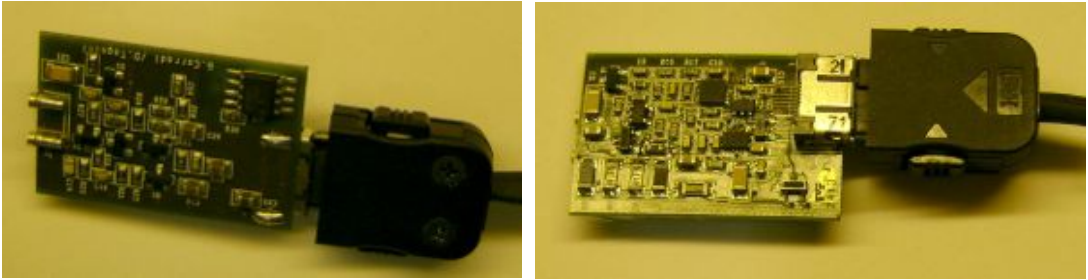

Figure 9.31. An Amp-HV prototype.

### The Amplification layer

The specifications for the amplification layer have been developed and tuned to work with a Hamamatsu S8664 APD connected to a LYSO crystal. Minor adjustments to the gain and power dissipation parameters will be implemented in the next production run. The electronic scheme is that of a double stage transimpedance preamplifier, with a final trans-impedance gain of 14 kΩ (voltage equivalent, $V_{out}/V_{in}$ of 300) while maintaining an equivalent noise charge (ENC) level of about 1000 electrons with no input capacitor source. The basic characteristics are described in Table 9.4; the preamplifier circuit schematic is shown in Figure 9.32.

### The linear regulator layer

The linear regulator is required to provide extremely precise 16-bit voltage regulation and long-term stability of better than 100 ppm. The current limit of the APD is conservatively set to about 300 μA; this value will be optimized in a latter design stage. The list of measurement characteristics of the prototype are summarized in Table 9.4. The basic schematic of the linear regulator is shown in Figure 9.33. The high voltage required for the APD is produced by a primary generator residing on the controller board that generates a voltage of 530 V and a current of 5.5 mA using low-noise switching technology, sufficient to power 16 channels in parallel. The ON and OFF states of the channel are controlled by an ARM processor, as described below. The input voltage of 530 V is followed by a constant current generator, programmable through appropriate adjustment resistors, which provides the current to the next stage in parallel. This provides a stabilized voltage with local feedback to the APD detector. The output voltage is regulated by a DAC and is then read out again via an ADC, with 16-bit accuracy.

Table 9.4. Characteristics of the Amp-HV chip: (left) for the amplification side and (right) for the linear regulator side.

| | | | |
|---|---|---|---|
| • Dynamic | 2.5 V | • Adjustment range Vout | 250V to 500V |
| • Bandwidth | 70 Mhz | • Accuracy, reading and writing, Vout | 16 bit |
| • Rise Time | 6 ns | • Currency limiter can be adjusted | tpv. 300 μA |
| • Polarity | Reversed | • Noise total | 2 mVpp |
| • Output impedance | 50 Ω | • Long-term stability | 100 ppm |
| • Stability with source capacity - max | 300 pf | • Settling time | < 500 μs, ρ<1 |
| • Coupling output end source | AC | • Typical power dissipation | 135 mW |
| • Noise, with source capacity of 1 pf | 1000 enc | • Double filter high voltage, attenuation | 56 db |
| • Power dissipation | 14 mW | | |
| • Power supply | 6 V | | |
| • Input protector over-voltage | 10 mJ | | |

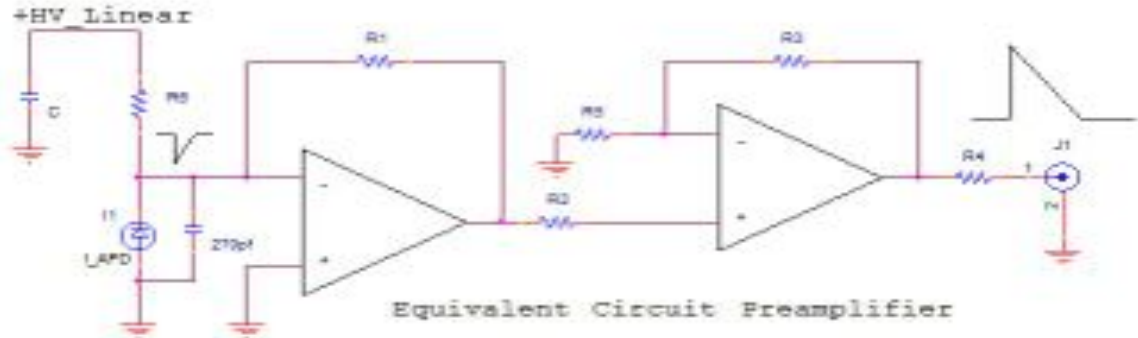

Figure 9.32. Simplified schematic of the Amp-HV amplifier.

### Amp-HV cooling

Integrating the cooling and the mechanics has not been yet fully addressed for the Amp-HV chip, but a viable solution is under design. Since the average power to be dissipated is ~150 mW per channel, the ground cannot be connected directly to the shielding surface or to the cold fingers. Therefore the use of a bulk bridge resistor, with a 1 pf capacitance, capable of transferring heat from the Amp-HV chip to the nearby mechanical structure, is foreseen.

A first CAD drawing of how this solution can be implemented is shown in Figure 9.34. In the actual scheme, we foresee connecting and cooling at least 16 channels together. Details are being worked out together with the mechanical engineering integration.

Figure 9.33. Simplified schematic of the linear regulator.

Figure 9.34. CAD drawing showing an Amp-HV with bridge resistors inside a shielding box.

### 9.9.2    The ARM controller

The design of the CPU system architecture, consisting of a series of Cortex M3-ARM processors, is shown in Figure 9.35. The ARM processor controls each of the 16 connected Amp-HV cards. The HV card is used for setting voltages, with independent voltage drops. In this way, the bias can be adjusted from 250 to 500 V, with a 16-bit regulation range. Similarly, the channel settings can be directly read out, using a 16-bit ADC.

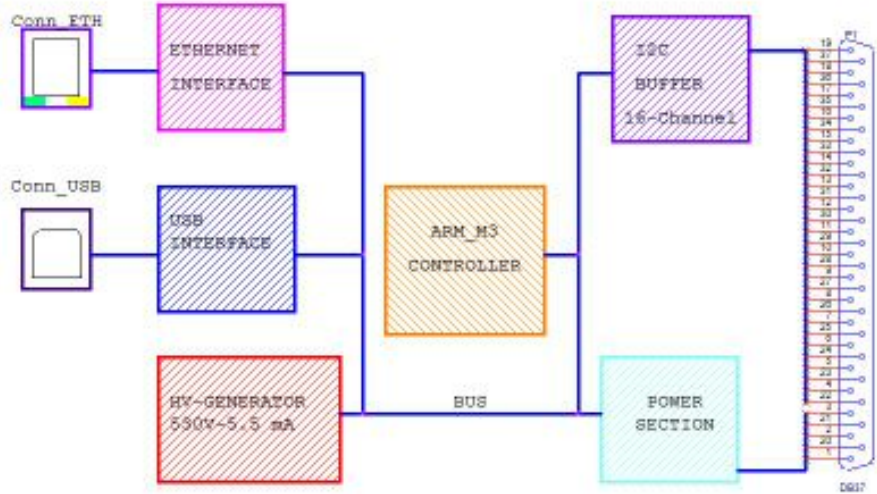

Figure 9.35. Design of the M3-ARM controller.

The primary HV generator is switched ON under CPU control. It is also possible to monitor the primary power supply, which feeds the 6 V for the preamplifiers and 12 V for the primary high voltage generator. These operations are done through firmware in the CPU via a standard Ethernet connection. Multiple systems are connected via a hub. In the near future, this will be implemented as an optical link. The final version of the controller will also measure the temperature of each APD detector and its Amp-HV card, as well as the CPU temperature. Each Amp-HV chip is connected to the ARM controller through a 50-200 cm cable. The cables are required for the transport of feeds for low and high voltage, as well as for the signals of the I2V control.

The power dissipation of the controller board is about 5W, and requires two voltages: +8V/100 mA to power the Amp-HV cards and 12V/300 mA for the high-voltage primary generator. Figure 9.36 shows a prototype of the AMP-HV card with its connections to the ARM controller.

***HV Generator Circuit Description***

The architecture of the switching power supply is a bust flyback design. This configuration provides good regulation with low noise. The circuit diagram of the HV-generator is shown in Figure 9.37. The switching frequency is about 250 kHz, with a long-term stability of about 0.5%, and a peak noise of less than 50 mV-pp at maximum load. The efficiency is better than 85%. This circuit was used to test prototypes, but will not work in a magnetic field. Two options are therefore being studied: (i) implementation of the same design without using a transformer or (ii) relocating the HV generator outside of the Detector Solenoid and running the HV signal to the electronics on cables through the end-plate feedthroughs. In the latter case, each HV cable would serve 4 boards, *i.e.*, 64 channels.

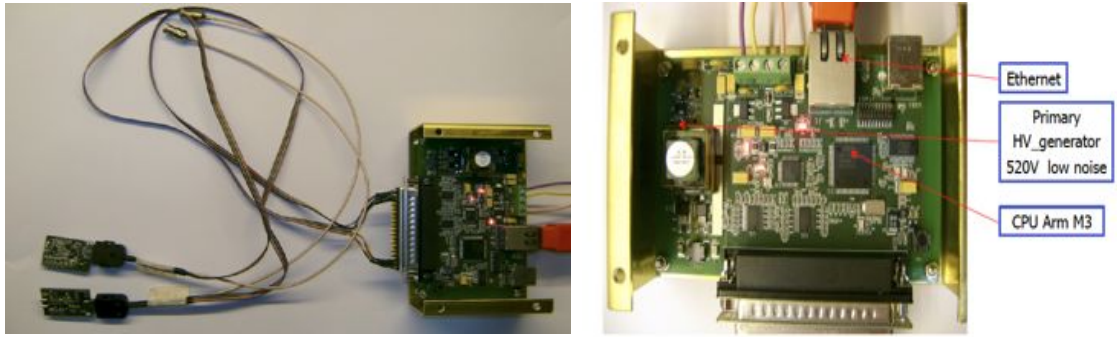

Figure 9.36. Connection between an ARM controller and an Amp-HV card (left) and a picture of the ARM (right).

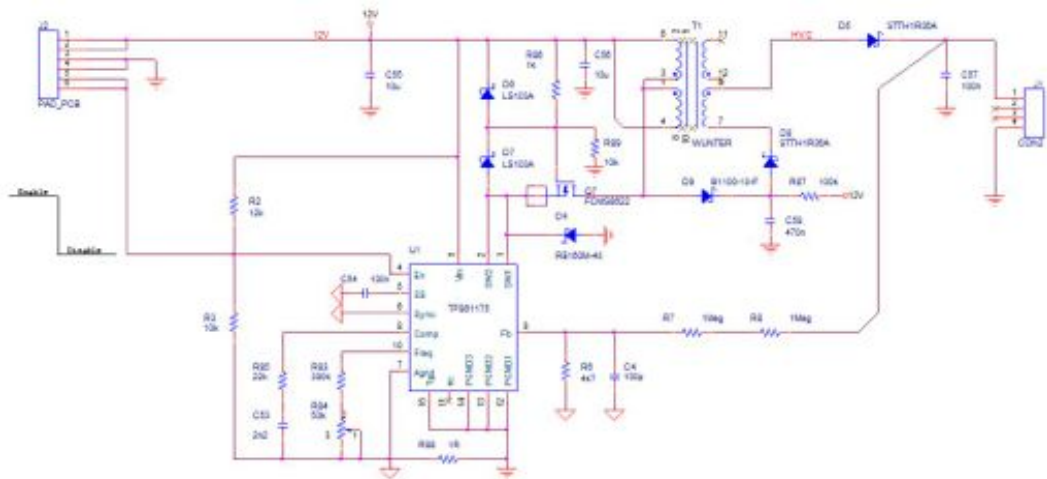

Figure 9.37. Schematic of the HV generator.

### 9.9.3    Measurement of signal characteristics

The pre-amplifier has a bandwidth of 70 MHz, corresponding to a signal rise time of 14 ns. To check this with the complete electronic chain, the photo-sensitive area of an APD was illuminated with a green laser with a narrow pulse width of 2 ns. The preamplifier output is shown in Figure 9.38 (left). A 16 ns rise time is measured, in agreement with expectations.

The photosensor, a Hamamatsu APD S8664-1010, has an area of $10 \times 10$ mm$^2$, resulting in a detector capacitance of 270 pF. This is the highest source of noise in the apparatus, requiring the use of a low-noise amplification stage. To achieve a high detection efficiency at the MeV level, the noise performance of the preamplifier is crucial and has to be accurately determined. ENC measurements have been carried out in several different configurations: with all voltages off, with low voltage on and with/without the high voltage on. For each configuration, the ENC is measured by estimating the *rms* of the output distribution. The ENC was determined to be ~1000 $e^-$ with negligible input capacitance, growing nearly linearly up to 270 pF. Table 9.5 shows the ENC measurement for different configurations. The measurement related to the amplifier itself has the LV on and the HV off, which corresponds to ~11,500 $e^-$ for a 500 ns integration window. To confirm this measurement, the energy dependence has been extracted from the noise term found using a $^{22}$Na source with a LYSO crystal read out by an S8664-1010 APD, followed by the Amp-HV chip. The APD gain was set to 150 and a light yield of 2400 photoelectrons/MeV was measured. Figure 9.38 (right) shows the noise distribution in counts. The *rms* is 2.6 counts, which corresponds to 36 keV after correcting for the Na$^{22}$ energy peak. To extract the ENC($e^-$), the energy dependence has to be included for the light yield and APD gain. An ENC of 13,000 $e^-$ was measured.

Table 9.5. Measurement of the amplifier ENC for different conditions.

| Conditions | Gate width (μsec) | *rms* Noise (Counts) |
|---|---|---|
| LV off | 0.5, 1 or > 1 | 1 |
| LV on | 0.5 | 2.2 |
| LV on | 1 | 2.8 (HV off), 3.5 (HV on) |
| LV on | 1.5 | 3.6 (HV off), 3.9 (HV on) |

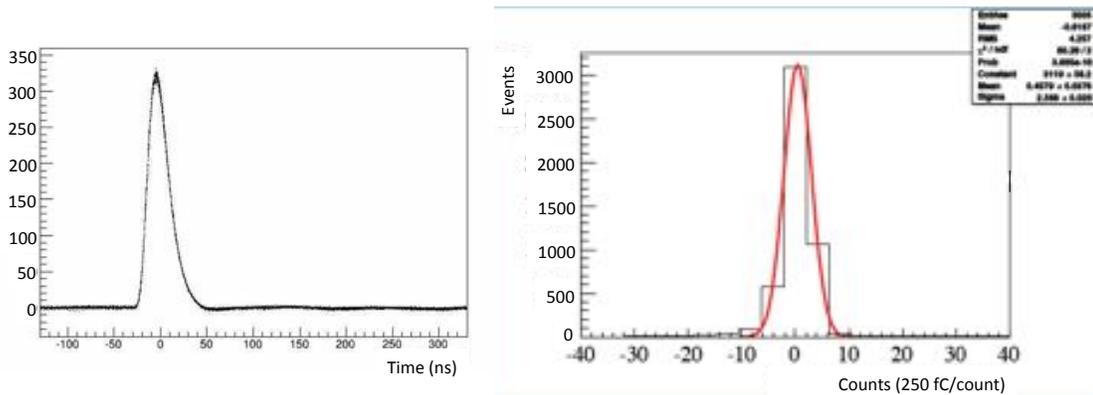

Figure 9.38. Pulse shape for an APD + amplification system fired with a green laser (left) and the noise distribution for a LYSO crystal read out by an APD + amplifier (right).

For the BaF$_2$ calorimeter, the photosensor will be the 9 x 9 mm$^2$ RMD/JPL device that has two improved characteristics with respect to the S8664: (i) a capacitance of ~60 pF, and (ii) an operational gain of 500. The ENC($e^-$) will be ~5000 e-. Assuming a light yield of 30 p.e./MeV, the expected noise level is ENE=ENC($e^-$)/(G∗LY) =5000/(30 x 500) = 0.33 MeV. After receiving a dose of >10 krad, the LY will be reduced by 30-40%, so that the noise will increase to 0.55 MeV. The signal peak in mV will be equivalent to 1/9 that of LYSO, corresponding to a voltage output of 4 mV/MeV.

### 9.9.4     The Digitizer modules

The calorimeter is composed of 1860 crystals, each equipped with 2 APD photosensors, for a total of 3720 fast analog signals that must be digitized after being amplified and shaped by the FEE. The exact shape of the signal is a function of the crystal material, the APD and the FEE parameters. For the $BaF_2$ case, pulses of 50 ns width and a maximum pulse height of 200 mV is expected. This requires very fast digitization and good resolution. 200 Msps and 11 bits of resolution are a good compromise between power dissipation and cost. The Mu2e Calorimeter Waveform Digitizer subsystem (Cal_WFD) is an electronic printed circuit board that digitizes analog data, serializes it and sends it upstream to the DAQ via a fiber optic transceiver. The Cal_WFD must also perform some digital signal processing (DSP) operations, removing data below threshold as well as providing the mean charge and time for each channel by means of running averages. A prototype board has been designed and is currently being tested. The experience gained with the prototype will be the baseline for the final Cal_WFD design. The requirements described are applied both to the prototype and the final board design.

From the occupancy plot of Figure 9.30, the expected data throughput is derived when for a zero-suppression threshold of 1 MeV, corresponding to ~20 % channel occupancy, *i.e.*, an average of 40 kHz/channel of random hits. A 50 ns signal width provides 10 samples after zero-suppression. In our estimate, we add a factor of two safety margin for the determination and monitoring of the signal baseline. In this case, the data throughput is 3720 (chan.) x 11 (bits) x 20 (samples) x 40,000 (hits/s), corresponding to 32.2 Gbits/sec (4 GBytes/sec), which matches the capabilities of the DAQ system. For a total of 120 WFD boards reading out the full calorimeter, 260 Mbits/sec (32 MBytes/sec) per board is expected. Since the rings are limited by a 2.5 Gb/sec throughput, the calorimeter is organized into twelve sectors with 5 boards/ring per sector. The calorimeter consists of a total of 24 rings.

***Waveform Digitizer Prototype***
The Calorimeter Waveform Digitizer Prototype (Cal_WFD_Proto) converts analog signals to digital, performs zero suppression, adds metadata, and combines individual channels into a single block of data. The Readout Controller serializes and translates the data into the correct protocol, and sends the data out on optical transceivers to the DAQ. While the production version must operate in a difficult environment (high radiation/high magnetic field), the knowledge gained from using this prototype will be essential to understanding the needs of the production version.

*Requirements*

The purpose of the Calorimeter Digitizer Prototype is to create a hardware and software development platform that incorporates the Texas Instruments ADS58C48 ADC, a WFD, and a Readout Controller into a single PCB. The Cal_WFD_Proto will aid the development of VHDL and Slow Controls coding, and allow us to understand the needs of a functional production WFD and Readout Controller.

*Hardware configuration*

The Cal_WFD_Proto is an electronic PCB that measures 25.4 cm wide x 25.4 cm high (10 in wide x 10 in high). A block diagram is shown in Figure 9.39. A full description of the board with electronics scheme can be found in [25].

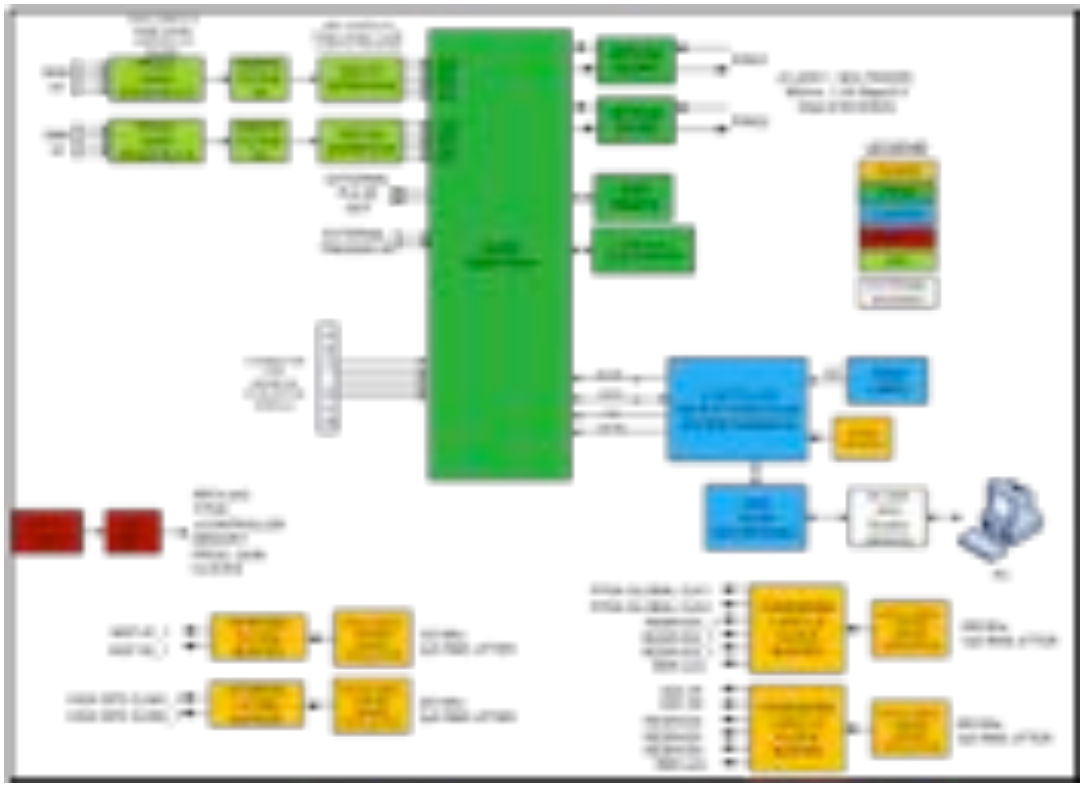

Figure 9.39. Block Diagram of the Cal_WFD_Proto Board.

*Analog Inputs*

The Cal_WFD_Proto has eight differential analog channels that are digitized. The analog channel design was based on the Texas Instruments ADS58C48EVM Evaluation Module.

*ADC EVM Connector*

A Samtec high-speed ground plane socket is mounted on the bottom side of the board to allow the use of ADS58C48EVM evaluation module. The ADS58C48 has an additional four channels that go to the same Xilinx Spartan-6 FPA as the analog inputs. The ADS58C48EVM allows for an evaluation of the analog input signal chain.

### External Trigger Input

The digitization of the analog signals is driven by the rising edge of a differential external trigger clock.

### Optical Transceivers

Optical transceivers convert the serialized data from the FPGA to an optical signal to be delivered to the DAQ system. The use of two optical transceivers provides redundancy. If one of the transceivers fails, the other can continue sending data without having to repair or swap out the board. A second role for the optical transceivers is to provide a path for slow controls communication that allow the board to be configured remotely and to be interrogated to monitor system voltages and other environmental parameters. The selected optical transceivers are class 1, multimode, 850 nm devices. The data rate for the transceivers varies from 1.25 Gbps to 2.5 Gbps, using a gigabit Ethernet protocol with 8b/10b encoding.

### Mobile Low Power Dual Data Rate (LPDDR) SDRAM

A Micron MT46H64M16LFBF 1Gb mobile, lpddr, sdram is connected to the FPGA. It can buffer up to one second of digitized data before transfer to the DAQ.

### CAN Transceiver Module

The CAN Transceiver module is a redundant communications path for slow controls in the event that the fiber optics fails or the FPGA programming is corrupted. The CAN module is connected to a 16-bit flash-based microcontroller that implements the CAN protocol. An external PC or controller communicates with the microcontroller through the CAN Interface.

### Temperature Sensor

A Texas Instruments LM82 Local Digital Temperature Sensor is attached to the microcontroller via a two-wire serial interface to measure the board temperature.

### Microcontroller

A general-purpose, 16-bit, flash-based microcontroller is connected to the FPGA to provide environmental data (temperature) as well as configuration data (slow controls) through the CAN Interface. The microcontroller uses ANSI C as the programming language. It uses a 10 MHz discreet crystal as the clock for operation.

### Field Programmable Gate Array (FPGA)

The Xilinx Spartan-6 Field Programmable Gate Array (FPGA) has several functions:

- Translation, zero-suppression and serialization of data to the optical transceivers.
- Configuration of the Cal_WFD_Proto board.

- Communication with the microcontroller for environmental data (temperature) and as a secondary path for slow controls.

The FPGA uses the VHDL programming language to implement the Waveform Digitizer and the Readout Controller. Separate clocks drive various aspects of the FPGA. One clock is used for the gigabit transceivers on the FPGA. Another clock is used for the main system clock. A third clock is provided for redundancy and to allow other parts of the FPGA to be driven from a different clock domain.

Clocks are generated on the Cal_WFD_Proto board. All clocks are differentially and a/c coupled to each device. Four sets of clocks are generated on the board.

### 9.9.5    Preliminary design of final Digitizer modules

The prototype uses a flash ADC with a maximum sample rate of 200 Msps with 11 bits of resolution. Each Waveform Digitizer (WD) will be a custom board residing inside the evacuated warm bore of the Detector Solenoid. The boards are 15 x 15 cm$^2$ and support 16 analog channels. 240 boards will be needed to read out the entire calorimeter.

Each WD board will include an FPGA that performs several functions:

- Provide control signals to the ADCs.
- Data sparsification.
- Timestamp to the data.
- Write the data to memory. Data will be set to DAQ in the period between spills.
- Read data from memory and serialize.
- Serialized data, 8b/10b encoded, are sent to the DAQ through an electrical-optical converter.
- To limit the number of fibers reaching the DAQ, the WD will be daisy-chained in rings. Each FPGA will receive data from the previous WD, add its own data and pass data onto the next board.
- Slow control commands will be received and acted on by the FPGA through the same fibers used for data transmission.

***Technical design***

The WD follows the same block diagram as the prototype (Figure 9.39), but it will be required to function inside the Detector Solenoid vacuum vessel, a harsh environment that requires special design considerations. In particular, the boards will be operated in vacuum ($10^{-4}$ Torr) and a magnetic field (1 Tesla) and access will be difficult. This implies design rules quite similar to those used for space flight and other difficult-to-access applications:

- Highly reliable design (high MTBF). Some components will be of military grade and others will be COTS, but chosen from lists of components known to have been operated in vacuum.
- Heat will be removed from the vacuum by conduction, so the board mechanics (and the mechanics of the supporting brackets and crates) will have to be designed to take this into account. The design will require extremely low power dissipation. The same SUVA based cooling system used for the Tracker (see Section 8.5.4) will be used to remove heat from the ADC and the FPGA. A thermal simulation will be performed as part of the final design.

The expected radiation dose to the WD modules will be 15-20 krad/year and $\sim10^9$ $n_{1MeVeq}$/year. It will be necessary to choose components that are rad-hard, resistant to Single Event Upsets (SEU) and latch-up free. It will be necessary to perform one or more irradiation tests before finalizing the design. We assume the following:

- ADC: the model used in the prototype appears to be adequate for the production board, both in terms of dynamics, sample rate and low power requirements. The cost is around $100 in quantity and supports 4 channels, bringing the cost to a reasonable $25 per channel. No radiation hardness data exists. The same is true for the programmable amplifier.
- FPGA: a high speed, low power, radiation resistant FPGA is required with an internal serializer and a DDR controller. Currently, the best choice seems to be the Microsemi SmartFusion2 M2S150T-1, which is SEU free, flash-based, and includes some mechanism for SRAM protection. A hard ARM processor is included, which will be useful for decoding slow control commands. In 2015, Microsemi will release a new family of SmartFusion2 devices that will be resistant to radiation exposures up to 100 Krad. This FPGA will be evaluated when it becomes available.

## 9.10  Calorimeter Mechanics

The arrangement of the two calorimeter disks inside the detector solenoid is shown in Figure 9.40. Each disk has an inner radius of 351, an outer radius of 660 mm, and consists of 930 trapezoidal $BaF_2$ crystals. The crystals are 200 mm long with hexagonal base whose apothem is 16.2 mm. Each crystal will be wrapped with a single 65 μm thick layer of 3M ESR reflective film.

Each disk will be supported by two coaxial cylinders (see Figure 9.41, top). The inner cylinder must be as thin as possible in order to minimize the passive material in the region where spiraling background electrons are concentrated. The outer cylinder can be as robust as required to support the load of the crystals. Each disk has two cover plates. The plate facing the beam will be made of low radiation length material to minimize the degradation of the electron energy deposition, while the back plate can be very robust. The back plate will also support the photosensors, the front-end electronics, HV/LV supply and digitizers.

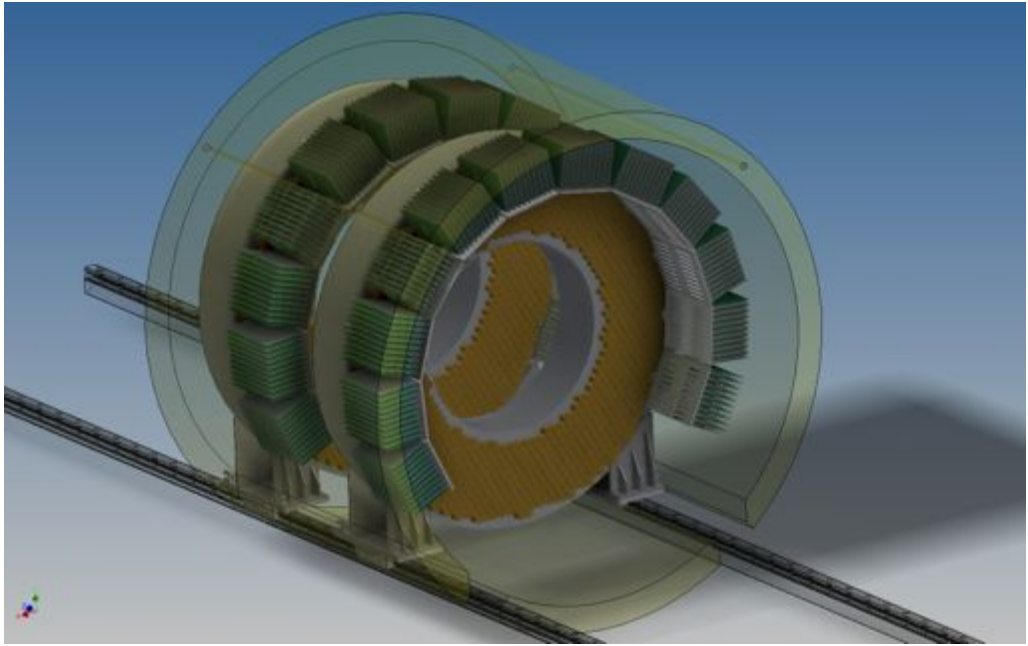

Figure 9.40. Placement of the two calorimeter disks on the rails in the Detector Solenoid.

The crystal arrangement will be self-supporting, with the load carried primarily by the outer ring. A catenary structure resembling a Roman arch will be constructed to reduce the overall load on the inner cylinder. The mechanical properties of the crystals are critical for this type of configuration. These include the Young's modulus, tensile modulus, Poisson ratio (or torsional modulus of elasticity), yield strength and ultimate strength. A Finite Element Model, using the crystal properties as input, will be constructed to optimize the design. The boundary conditions of this layout will be fixed and the structural analysis will be used to verify displacements and deformations of the various components.

The back plane will most likely be built of stainless steel or aluminum. It provides support for the whole mechanical system, but also provides access to the back of each individual crystal. A readout unit is composed of a crystal, two APDs and two AMP-HV chips. The back plate will provide access to each crystal and will support the APDs and electronics. An example of the concept is shown for a small prototype in Figure 9.42.

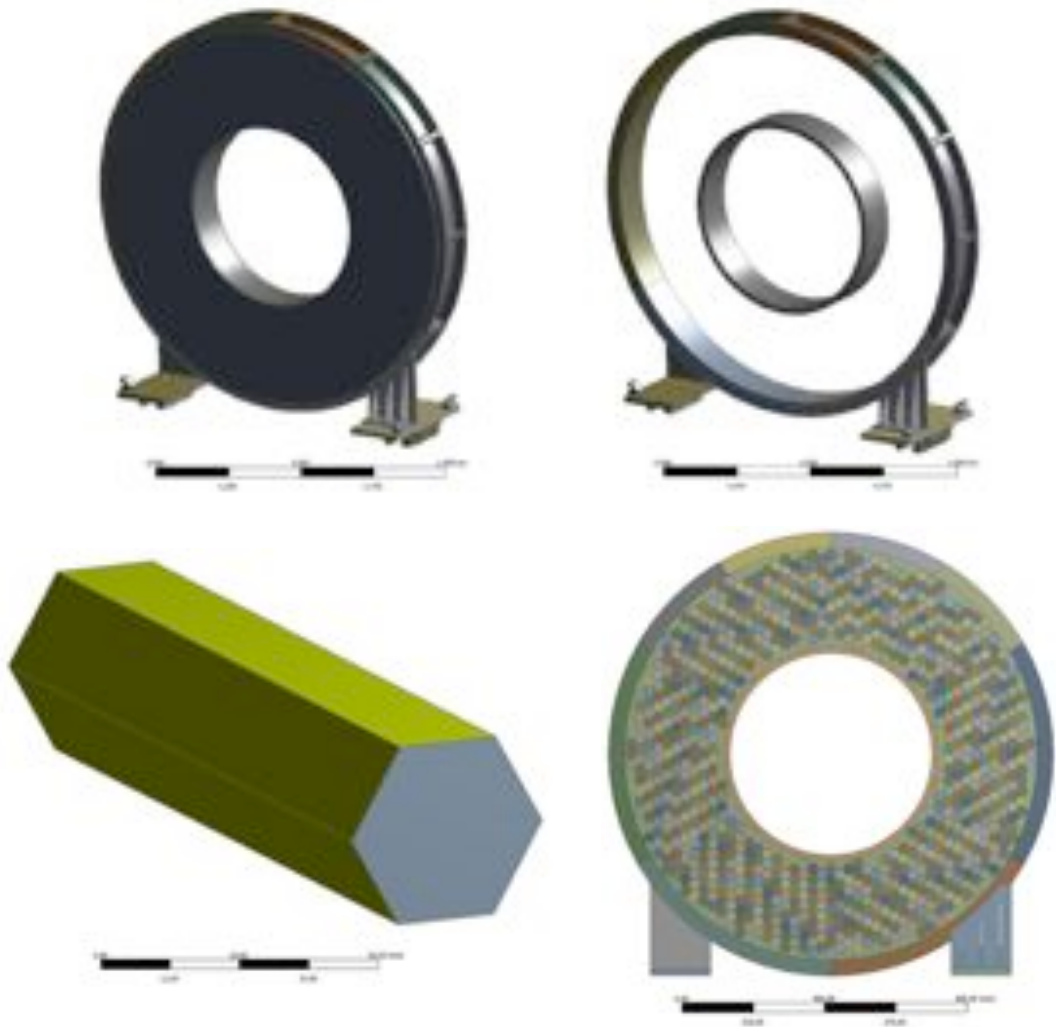

Figure 9.41. CAD layout of the calorimeter mechanical support structure (top left), details of the inner and outer cylindrical shells (top right), hexagonal crystal view (bottom left) and placement of the crystals inside the disk (bottom right).

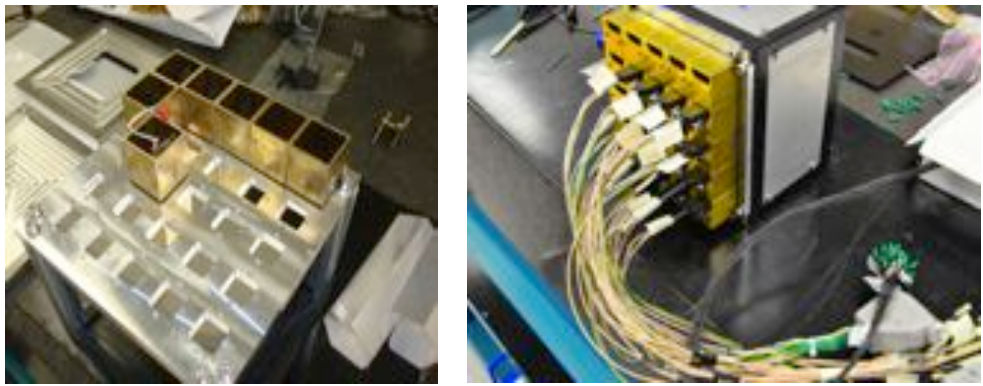

Figure 9.42. A prototype crystal array illustrates the details of the mounting structure for the APDs and Front End Electronics.

Mu2e Technical Design Report

The FEE boards are located at the back of each disk. Figure 9.43 shows how the boards will be installed on the disk. With a granularity of 16 electronic channels per board and a total of 930 crystals per disk, each disk can be subdivided into 12 sectors with ~78 crystals each. This allows for the electronics to be distributed into 12 crates per disk, where each crate houses 8 sets of AMP-HV and Waveform Digitizer boards

In order to gain as much room as possible between the disks when servicing of the electronics or APDs is required, the crates are placed at the outermost region of each disk. The crates are mounted on a pneumatic cartridge that allows them to be extended radially, thereby completely exposing the area behind the crystals when required. The crates are designed to provide heat dissipation for the APDs and electronics boards; they will have metal fingers in contact with a cooling pipe routed circularly below the bottom of the rack's connection mechanism to the disk. The cooling system will be connected to the same cooling circuit used by the tracking system.

Each disk will also have six two-inch diameter integrating optical spheres. Each sphere will distribute homogeneous laser light via two/three bundles of quartz fibers connected to each crystal through the APD holder. A more dedicated description of the mechanics and a collection of drawings can be found in [26].

## 9.11   Test beam and experimental measurements

To validate the inputs to the calorimeter simulation, several different sets of measurements were carried out. A summary of previous measurements, described in the Mu2e Conceptual Design Report [27] for the LYSO crystals, is reproduced here. The most recent measurements performed with both a new LYSO matrix prototype and single *alternative* crystals are reported in the following sub-sections. A perspective on the next round of R&D planning is also presented.

A LYSO array was exposed to a tagged photon beam at MAMI (the Maintz Microtron) in March 2011. For this test, 9 SICCAS LYSO crystals, $20{\times}20{\times}150$ mm$^3$, were assembled into a matrix and read out using a single S8664-1010 APD per crystal. The LYSO array was surrounded by a leakage recovery matrix of PbWO$_4$ crystals, read out by bialkali photocathode Hamamatsu PMTs. The total matrix coverage was ~2.5 R$_M$. Each channel was calibrated to approximately 2% using cosmic rays. The APDs were operated at an average gain of ~150. The crystals were exposed to a tagged photon beam with energies ranging from 20 up to 400 MeV. Data were taken at twelve different energies over a period of 2 days. Figure 9.44 shows the dependence of the energy resolution as a function of beam energy for test beam data (black) and for the simulation (red).

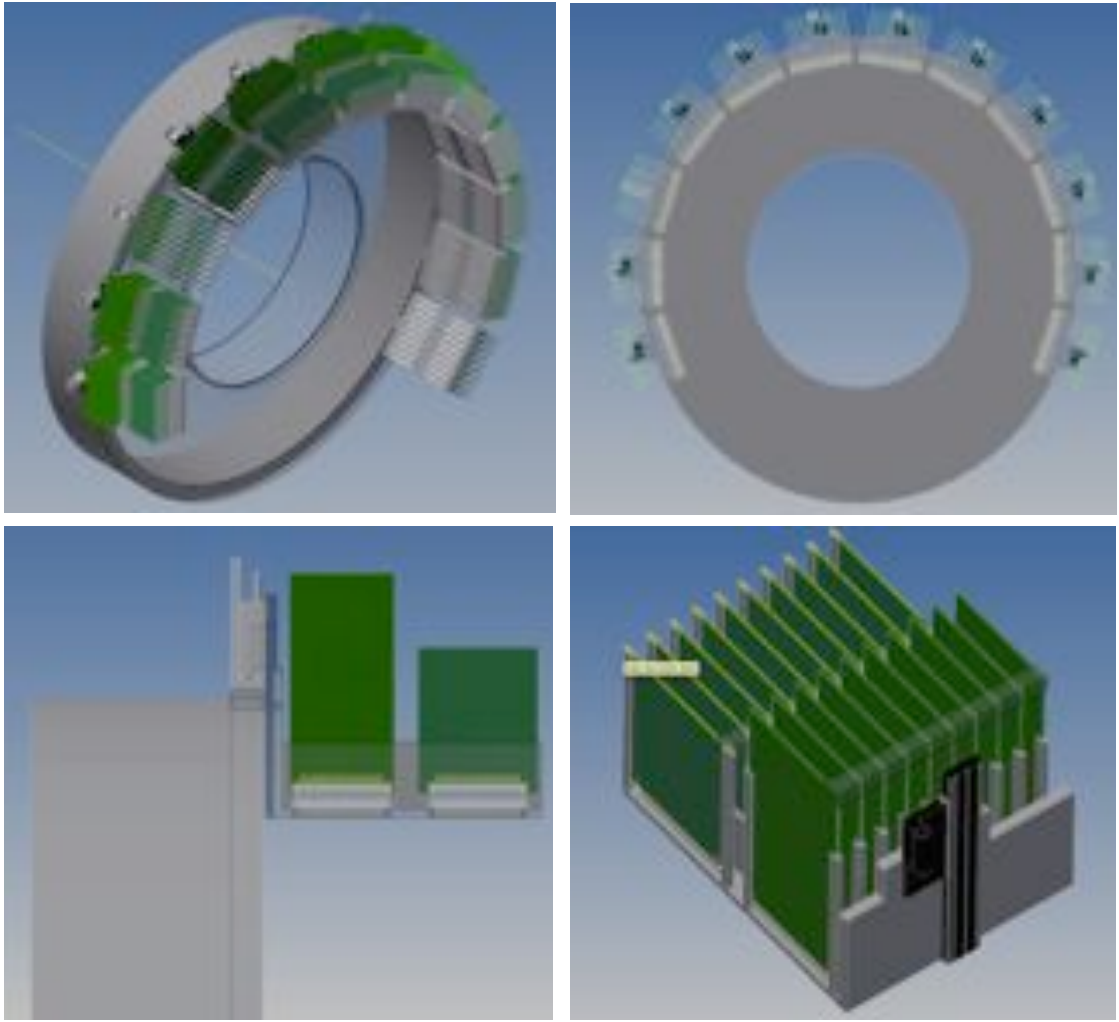

Figure 9.43. Arrangement of the AMP-HV and Waveform Digitizer boards in custom-made crates (top), and details of the support system for the boards to allow opening of the boards that allows access the back side of the disk in case repairs are required.

The resolution dependence was fit with the following parameterization:

$$\frac{\sigma}{E} = \frac{a}{\sqrt[1/4]{E(GeV)}} \oplus \frac{b}{E(GeV)} \oplus c$$

where $a / \sqrt[1/4]{E(GeV)}$ is the usual parameterization of energy dependence replacing the expected stochastic term, $b$ is the coefficient of the noise term and $c$ is the constant term. The experimental points are well represented by an $a$ value of 2.4% for the $E^{1/4}$ dependence, a negligible electronic noise term and a constant 3.2% term due to shower leakage. To obtain reasonable agreement with the data, the energy response of each crystal was additionally smeared by 4% in the simulation using a Gaussian distribution. Details of the measurement can be found in [2]. The smearing was introduced as the simplest way to simulate the expected crystal non-uniformity. All in all, this test resulted in a 5.3% energy resolution that was still improvable, due to the small dimension of the matrix and to the quality of the crystals. Two other sets of measurements were carried out with the prototype crystal matrix. A position resolution measurement was performed using the MAMI test beam that resulted in an observed resolution of ~3 mm for photons at normal incidence. The measurement was limited by the size of the beam. A timing resolution measurement was also carried out with electron beam at the BTF facility of Laboratori Nazionali di Frascati, LNF, Italy. As reported in [28], the LYSO timing resolution was measured to be of ~200 ps in the energy range 100 - 500 MeV.

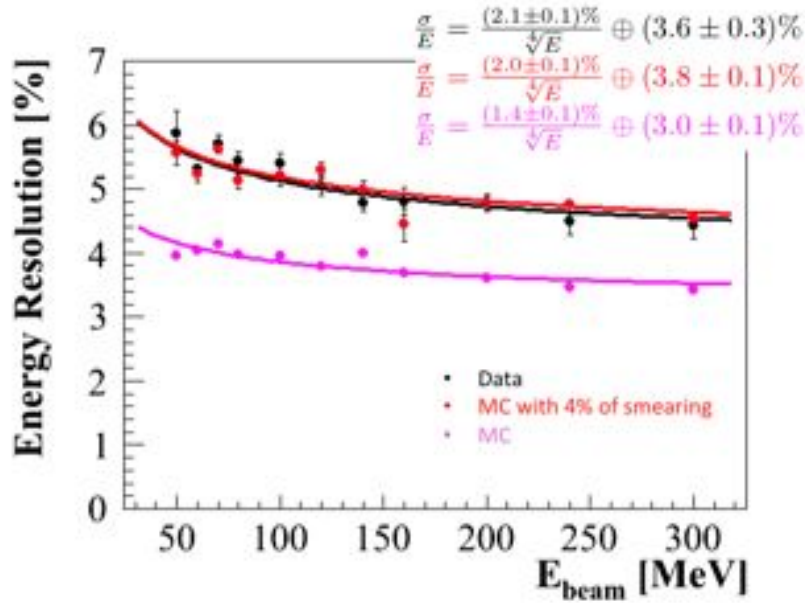

Figure 9.44. Test Beam results from MAMI. The measured energy resolution of the overall LYSO crystal matrix (black points) is compared to simulation (red). To obtain reasonable agreement with the data, the energy response of each crystal was additionally smeared by 4% in the simulation (blue).

### 9.11.1 Measurements with a new LYSO array

In order to complete the LYSO R&D program, a larger and more uniform crystal matrix has been built [29]. A CAD drawing and a picture of the assembled prototype are shown in Figure 9.45. A few differences can be noted with respect to the old prototype. The 25 square crystals have dimensions that are much closer to those required for the final detector, 30x30x130 mm$^3$. The ratio, $R_{APD}$, between the photosensor active area and the area of the transverse face of the crystals is consistent with the hexagonal shape of the baseline design. The crystals are wrapped with the improved ESR-3M reflector, which provides 30% more light yield than Tyvek. The light yield and longitudinal response uniformity (LRU) was measured for each crystal. Uniformity and transmittance were found to be excellent. The longitudinal length, in $X_0$, is slightly smaller than in the final detector when considering the shower-length correction due to the average angle of incidence for conversion electrons of 50°. The FEE, shown in Figure 9.46, is composed of 16 prototypes of the Amp-HV chip, with the HV set and monitored by two readout ARM controllers, as envisioned for the final detector. To screen the FEE from external noise sources, the APD and FEE were inserted in a brass box Faraday cage. All chips are connected to a common external ground. The DAQ readout uses an array of four CAEN-1720 digitizers (Waveform digitizers) with 12-bit resolution and a sampling rate of 250 msps. The Waveform digitizer characteristics are very similar to those of the prototype boards under development at the University of Illinois and Pisa University.

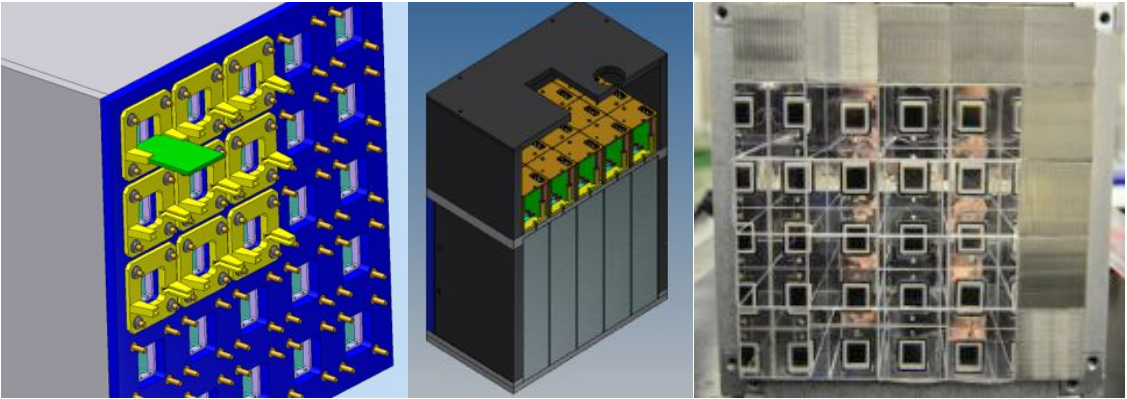

Figure 9.45. CAD drawing of the readout side of the 25-crystal prototype matrix (left), a CAD view of the assembled matrix with crystals, APD and FEE boxes in a light-tight box (center), and a picture of the actual matrix during assembly.

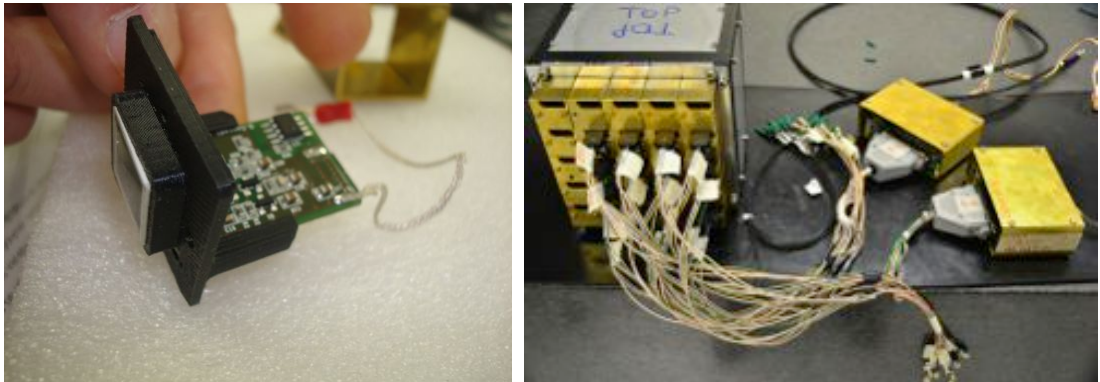

Figure 9.46. Picture of a FEE prototype board connected to an APD and inserted into a plastic APD support produced with a 3D printer (left). Connection of the amp-HV chips to the ARM readout controllers is shown in the picture on the right. The black bundle, with green caps, is composed of Fused Silica fibers running from the diffusing sphere of the Laser prototype system.

A first version of the laser calibration system has also been implemented as part of the prototype matric. The laser system consists of a 0.5 μJ/pulse green laser (at 530 nm) synchronized with an external trigger, followed by a Thor-Lab 2" diffusing sphere and a bundle of fused silica fibers. The fibers are inserted by means of a dedicated connector into the APD holders, thus illuminating the crystal. By reflection and diffusion, the transmitted light allows calibration and monitoring of the APD gains. Figure 9.47 shows the test system, with optical fibers entering from the back side of the crystals. A simplified cooling system has also been implemented during the test by flowing cold air in the FEE region, aiming only to keep the calorimeter at constant temperature. This system was not intended to be a prototype of the cooling system that must work in vacuum and is therefore based on conduction. A temperature monitor, based on a PT-100 sensor, was inserted close to the central crystal and photosensor box.

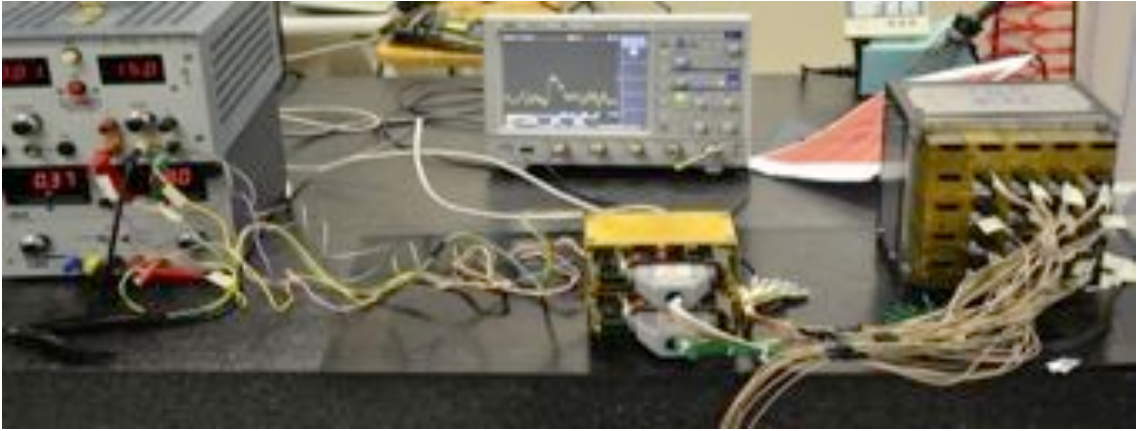

Figure 9.47. The FEE controller connected to the 16 FEE chips, optical fibers (green caps) entering the back of the APD supports and a scope for signal monitoring.

A week of data-taking was planned for the second week of February 2014 at the Frascati BTF facility with e⁻ beams between 100 and 300 MeV. For this test, only 16 crystals were available. However, due to a large leak of the main water supply in Frascati, the cooling systems and most of the Laboratory infrastructure were shut down for three weeks, and the beam test had to be re-scheduled. Another test with a complete matrix will be carried out at BTF at a later date; one week of data-taking is planned at MAMI in September 2014.

In the following sections we summarize the results obtained with cosmic ray and laser testing. Using the WFD, both the pedestal values and the charge were obtained by integrating the pulse shape for 400 ns in out-of-time and in-time windows with respect to the signal maximum amplitude. The results are expressed in pC. Two techniques were used to determine the timing associated to a pulse: a simple algorithm based on the centroid technique, and one based on a fitting of the signal shape.

### Determination of noise and coherent noise

The noise of each single readout channel has been evaluated with an out-of-time gate. Typical distributions are shown in Figure 9.48.

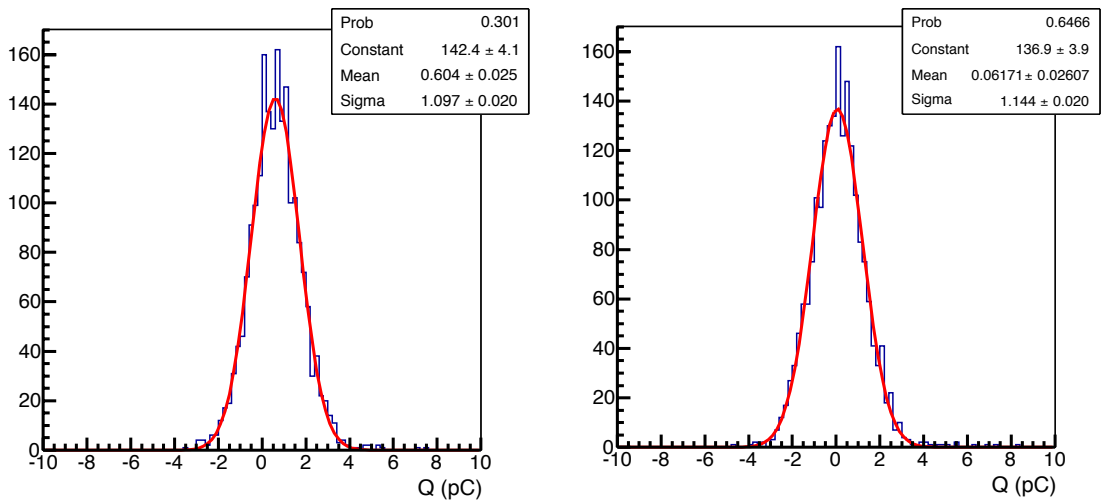

Figure 9.48. Pedestal distribution for two calorimeter channels with an out-of-time gate to determine noise levels.

***Calibration with the Laser system***

Using the laser prototype, a light pulse was sent to each APD to calibrate and monitor the calorimeter response. In Figure 9.49, the distribution of the laser charge is shown for the first calorimeter channel; the resolution of ~3% is dominated by the residual temperature variation of the APD. All channels were fit with a Gaussian, and the distribution of the mean and sigma are shown as a function of channel number in Figure 9.50. The spread of the averages is consistent with our knowledge of the APD settings and the equalization of the fibers in the bundle. Gain equalization is in progress. The pedestal distributions are well represented by a Gaussian fit with a mean close to 0 and σ of ~1.2 pC. The distribution of the sum of the charge for 1 board (*i.e.*, 8 channels) still has a Gaussian shape and a sigma of ~4 pC, very close to σ × √N_{ch}, thus demonstrating that coherent noise is practically negligible.

***Calibration with Cosmic Rays***

A cosmic ray test stand constructed from two NE-110 plastic slab scintillators (50x50x200 mm$^3$) positioned above and below the crystal matrix prototype (Figure 9.51) is used to test and calibrate the crystals by triggering on the coincidence of the two counters. Several days of cosmic ray data were taken using the test stand. At the same time, the laser calibration system was operated at a rate of about 1 Hz to continuously monitor the APD gains. The average event rate for cosmic rays is 0.3 Hz over all calorimeter cells, which is reduced by a factor of 10 when applying a tight calorimeter column selection cut. In three days of running ~80,000 cosmic triggers were recorded, resulting in 2400 well-selected minimum ionizing events/column. In Figure 9.52 (left), the distributions of the minimum ionizing peaks (MIP) for 4 out of the 16 channels are shown, while the energy sum for the four columns is presented in Figure 9.52 (right). The MIP peaks are determined with a Gaussian fit restricted to the peak region, resulting in an average value of 220 pC, a resolution of ~12 % and an error on the average of better than 0.5%. Since the noise contribution also has a Gaussian shape with 1.2 pC *r.m.s* and the average energy deposition for a MIP corresponds to ~30 MeV, resulting in an equivalent noise of ~130 keV/channel.

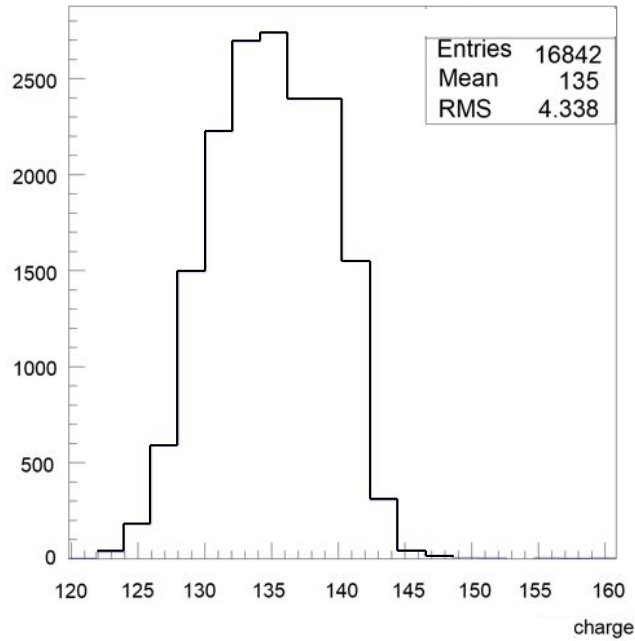

Figure 9.49. Distribution of the laser charge for one calorimeter channel.

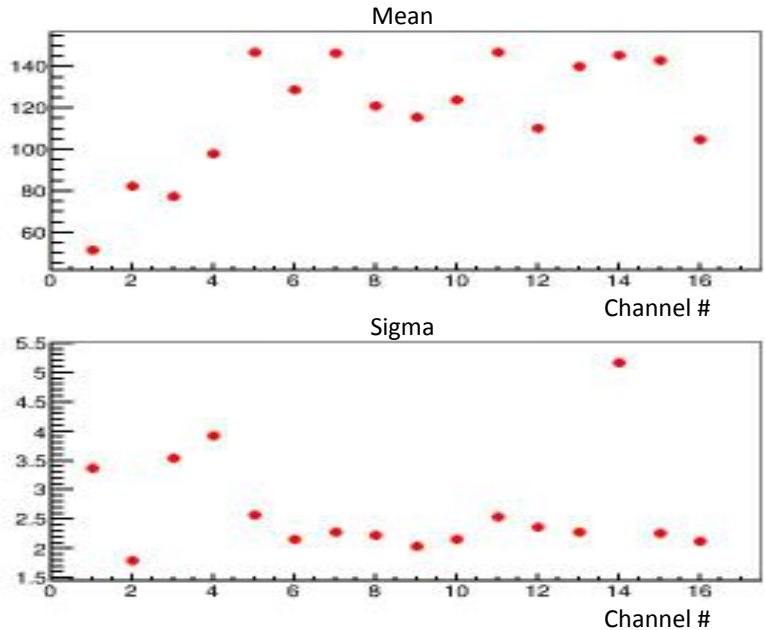

Figure 9.50. Measurement of the mean (top) and sigma (bottom) of the laser charge distribution for each of 16 calorimeter channels.

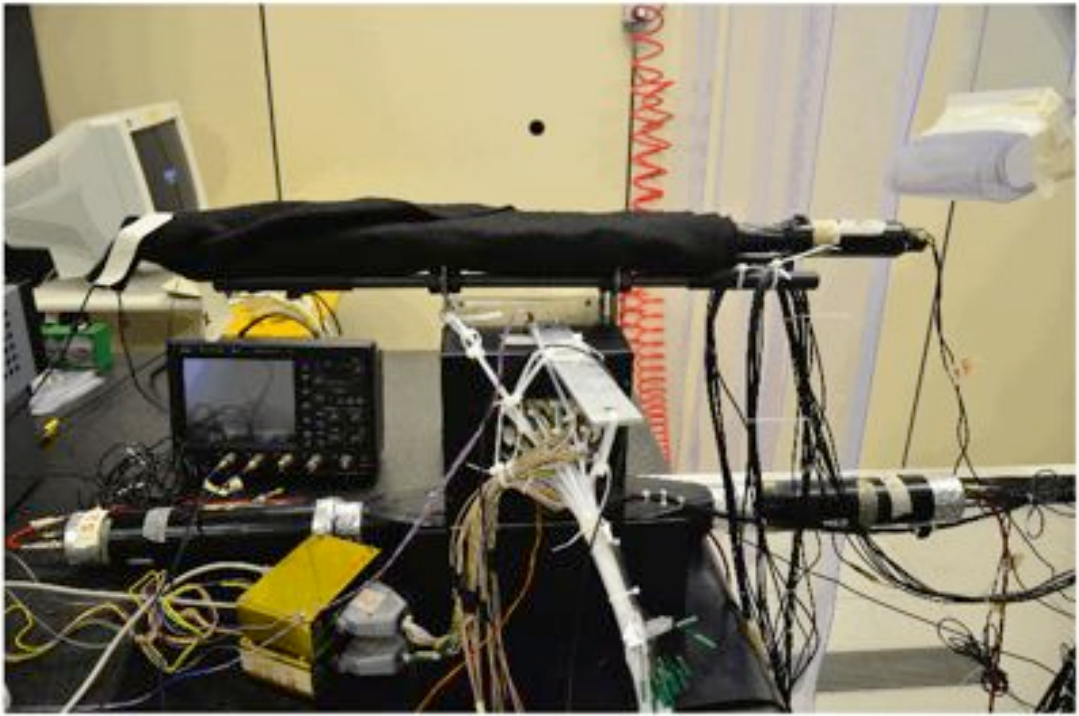

Figure 9.51. Cosmic ray test stand used to calibrate the calorimeter prototype.

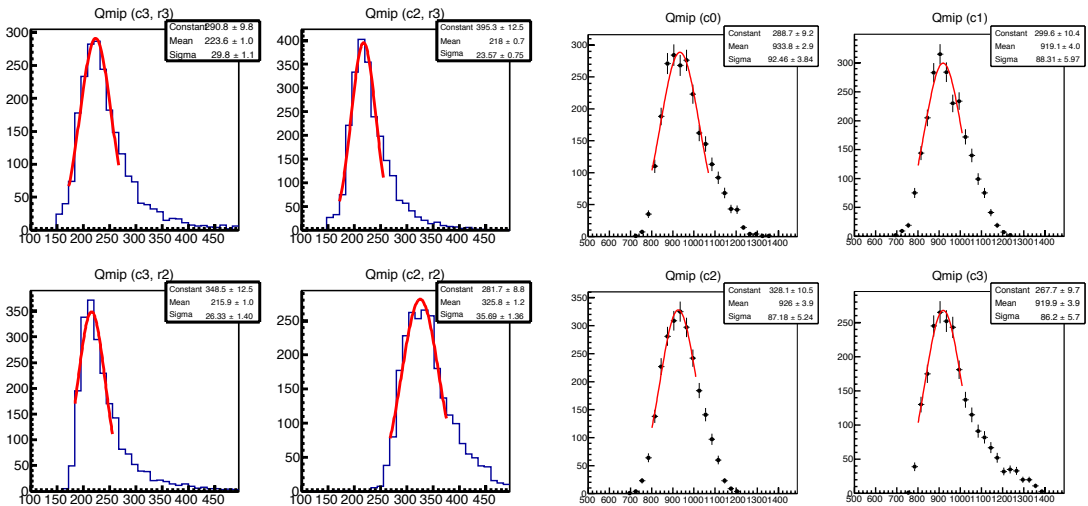

Figure 9.52. Distributions of events selected with the cosmic ray trigger and a tight column selection. The charge distribution for four calorimeter channels is shown on the left. The energy sum for the four calorimeter columns is shown on the right.

### *Measurement of timing resolution*

Two different methods are employed to determine the timing resolution achievable with the calorimeter prototype. Laser pulses were first used to tune the algorithm and check the timing response of the FEE and digitization systems. A selected MIP sample was later used to determine the calorimeter timing. In the laser case, the time resolution was determined for a single channel and for the average of up to 8 channels. In Figure 9.53 (left), the signal shape for a laser pulse is shown with a Log-N fit superimposed. The $t_0$ parameter represents the time when the maximum amplitude occurs and it is used as the best time estimate since no further corrections on pulse height are needed. In Figure 9.53 (right), a similar fit is performed on the shape of the selected MIP events. The fit has a slightly worse $\chi^2$/dof. A better parameterization of signal shape is in progress.

In Figure 9.54 the distribution of the time resolution for the laser run (left) and the time distribution for the MIP (right) sample are shown. The laser pulse provides an estimate of the time resolution limit of 70 ps due to the electronic setup. The MIP data provides a time resolution of 600 ps for a corresponding energy deposition of ~30 MeV, in reasonable agreement with a previous determination at 100 MeV with e⁻ data [24].

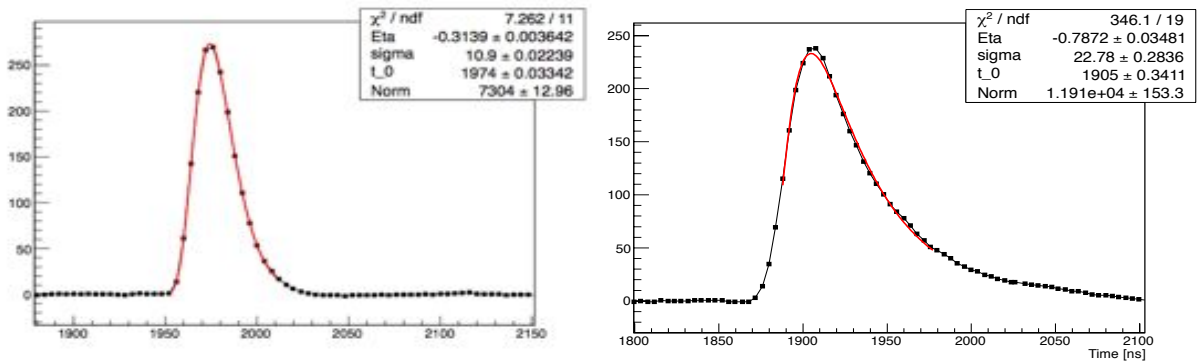

Figure 9.53. Fit to the signal shapes for laser pulses (left) and cosmic ray events (right).

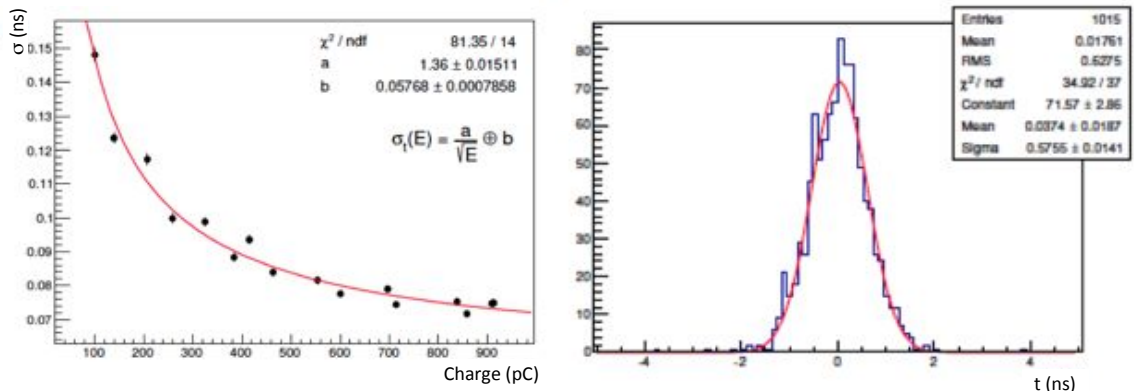

Figure 9.54. (left) Distribution of the time resolution for the laser run when summing many calorimeter channels; (right) difference between the timing of a single channel and the sum of two other channels for a cosmic ray run after applying a column selection.

## 9.11.2 Measurement of alternative crystals

The cosmic ray and laser tests of the LYSO prototypes have shown that this kind of crystal, together with a large-area APD and the developed FEE, is well matched to the calorimeter requirements. Completion of dedicated test beams running will be carried out before the end of 2014 to complete this prototyping phase. In the meantime, using the experience gained, an R&D program has begun for BaF$_2$. Part of this program will be to compare its performance with the backup alternative of pure CsI [30]. These studies will be based both on source and cosmic rays tests. The BaF$_2$ R&D program includes development and evaluation of the UV-extended, solar-blind APDs from RMD/JPL. For the CsI, currently available MPPCs and APDs can be used. Single crystals as well as a small matrix will be exposed to an electron beam in the winter of 2014. A final technology choice is foreseen for the first quarter of 2015.

***Measurement with a radioactive source and UV extended PMT***

BaF$_2$ and pure CsI crystals have been characterized using a test stand at LNF. The crystals were 3x3x20 cm$^3$ and were produced by SICCAS. The measurements were performed with a $^{22}$Na source. The crystal signal shape and the corresponding light output were determined for each crystal. By applying different integration gates, the contributions of the fast and slow components was determined. Due to the slight hygroscopicity of CsI, the tests were performed in a clean room with a relative humidity of 33%. A vacuum bag was used to encapsulate the crystal when not under measurement. All crystals were wrapped with reflective/diffusive material such as Teflon and 3M-ESR. In the CsI case, the effect of a standard optical grease (Bicron BC-630 and silicon Paste-7 from Rodhorsil) has been tested with good efficiency down to 300 nm. For the BaF$_2$ case, all reported measurements have been performed with an air gap. Future tests will include UV transmitting optical grease (DC 200 [31], viscosity > 600). In Figure 9.55, an example of the pulse shape is shown for the BaF$_2$ and CsI samples. Figure 9.56 shows the measurement of the fast and slow components for a BaF2 crystal. Figure 9.57 shows the measurement of the fast and slow components of pure CsI for different wrapping materials and optical couplings.

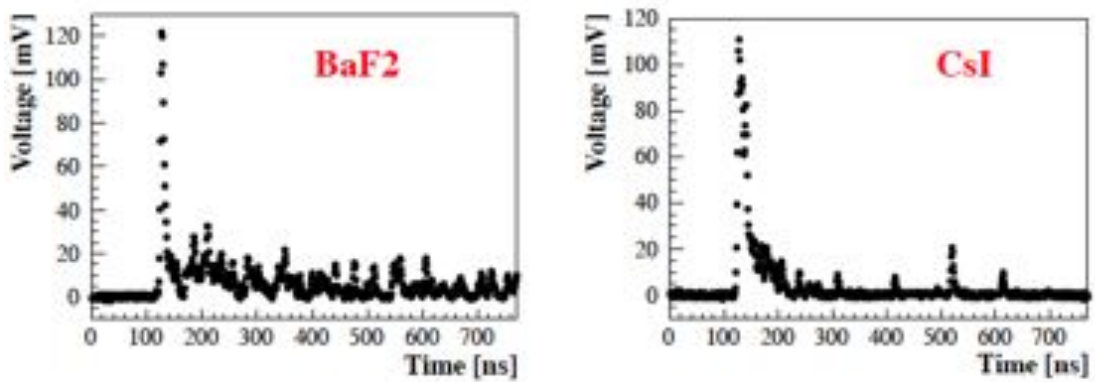

Figure 9.55. Signal shapes for BaF$_2$ and pure CsI crystals.

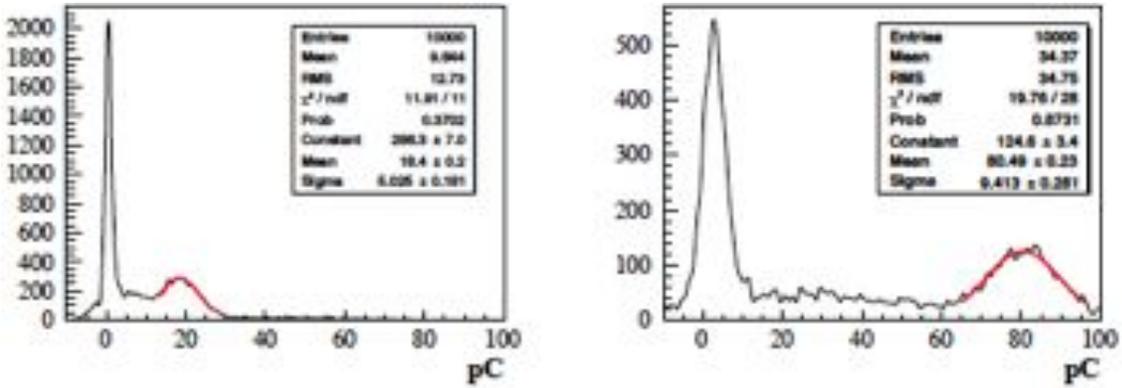

Figure 9.56. Response of a BaF$_2$ crystal to 511 keV photons from a $^{22}$NA source showing the fast (left) and slow (right) components.

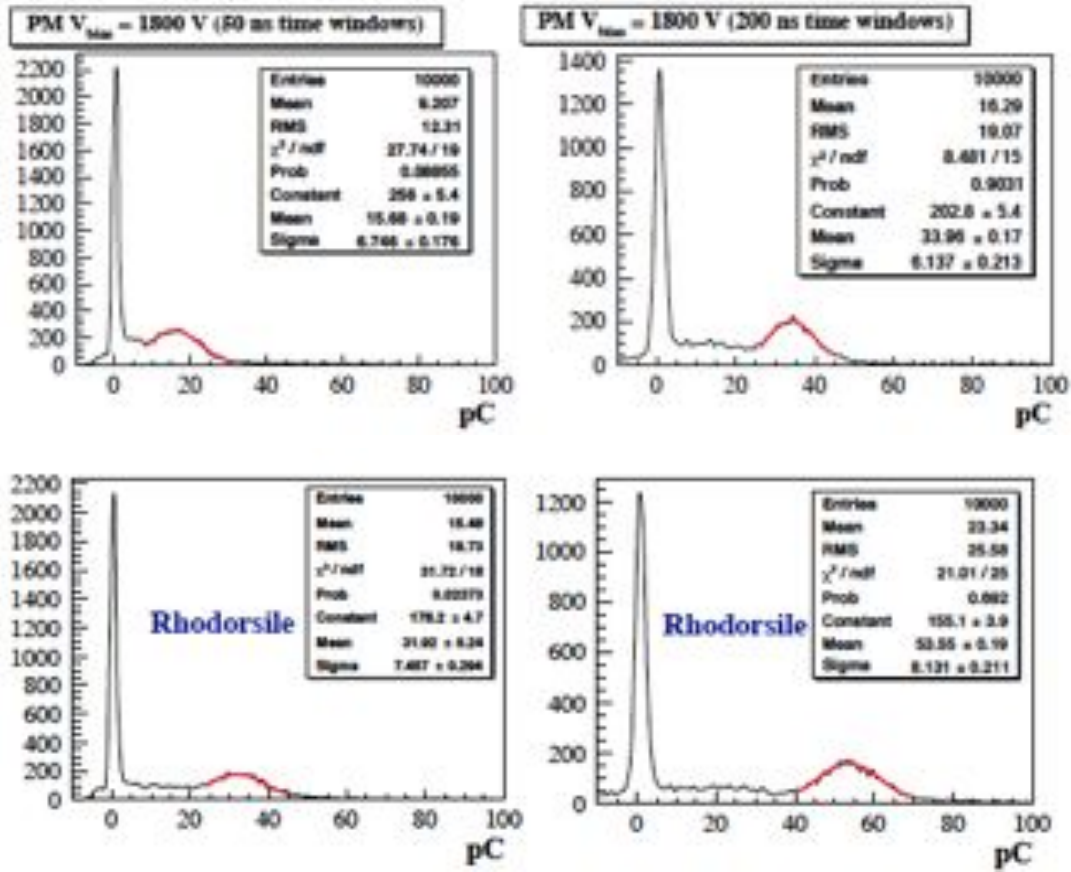

Figure 9.57. Response of pure CsI to 511 keV photon: (top) teflon wrapping and no grease, (bottom) teflon wrapping and Rhodorsil paste. Right (left) plots are for the fast (slow) emission component.

The light yield has been determined for the two crystals from the measured peak charge (in pC) by correcting for the gain of the photomultiplier, determined by comparison with a calibrated APD. From the digitized signal shape it is confirmed that the $BaF_2$ has a fast component (< 1ns) followed by a ~650 ns slow tail. The pure CsI spectrum ends 160 ns after the start. The fast components have been measured using a 50 ns gate, while the slow component was measured with a 650 (200 ns) gate for the $BaF_2$ (CsI) crystal. In Table 9.6, the results of the measurements are summarized. For the best wrapping case, 35 p.e./MeV is observed for the fast $BaF_2$ component, which corresponds to 30-35 p.e./MeV for readout with a UV-extended APD, assuming a factor of two improvement due to the optical grease. For the pure CsI ~20 p.e./MeV is estimated for the MPPC readout using the best configuration of wrapping and grease.

Table 9.6. Summary of measurements for pure CsI and $BaF_2$ crystals.

| Case | Slow Peak | Slow ($\sigma$/P) | Fast Peak | Fast ($\sigma$/P) |
|------|-----------|-------------------|-----------|-------------------|
| $BaF_2$ (air) | 18 pC | 30% | 81 pC | 12% |
| CsI (air) | 16 pC | 42% | 34 pC | 19% |
| CsI (Rodh.) | 31 pC | 22% | 54 pC | 15% |

The measurements performed with a source and a PMT show that both the $BaF_2$ and the pure CsI crystals have a reasonable light output with the expected ratio of fast and slow components. Given the small size (3x3 $mm^2$) of UV extended SiPMs available, the best performing crystal and photo-sensor combination was pure CsI readout by an MPPC with 50 $\mu$m pixels and an active area of 12x12 $mm^2$.

The pure CsI crystal was wrapped with the 3M-ESR reflective film and a 16-anode MPPC was used as the photosensor. The anode signals were input to a discrete chip that provided the analog sum and a x2 amplification stage. A single-pole RC shaper was also used. This electronics contributed a total signal width of 100 ns (Figure 9.58, right), determined by illuminating the MPPC with a blue laser with a 50 ps pulse. The photosensor was optically connected to the crystals by means of BC-630 optical grease. Similar to the PMT case, the signal has a rise-time of 15 ns and a decay time of 26 ns (Figure 9.58, left).

### Measurement with cosmic rays and a large-area MPPC

The pulse height distribution obtained with an integration gate of 200 ns is shown in Figure 9.59 (right) for cosmic ray events selected by a coincidence between two finger scintillators positioned above and below the crystal. The pulse height exceeds 600 mV when running the SiPM at 73 V, while the collected charge corresponds to ~300 pC. The average energy deposition is equivalent to 15 MeV and the energy resolution for a MIP is 14% with an equivalent noise of 130 keV. In Figure 9.59 (left), the time difference between the two finger scintillators and the difference between a finger scintillator and the CsI timing are shown. A time resolution of ~800 ps is measured for a MIP signal, with an associated trigger time-jitter of ~300 ps. Scaling this result to 100 MeV, a resolution of 300 (200) ps is expected when using one (two) MPPCs per channel.

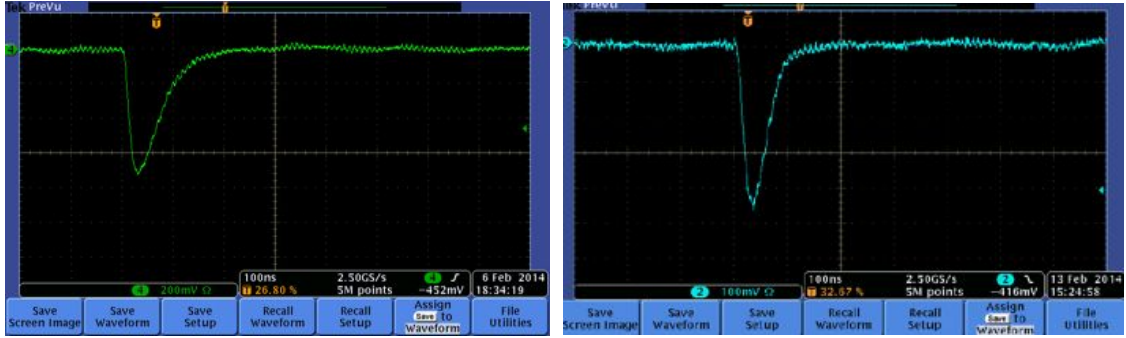

Figure 9.58. Signal shape for a single MIP event in a pure CsI crystal readout by a 12x12 mm$^2$ Hamamatsu MPPC array (left) and the response of the Hamamatsu MPPC to a 50 ps blue laser pulse (right).

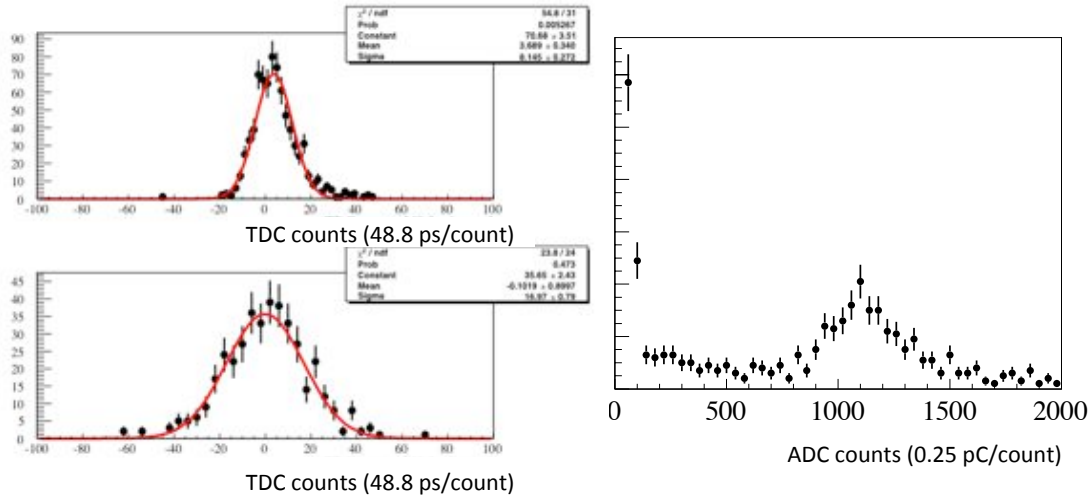

Figure 9.59. Distribution of cosmic ray data obtained with the pure CsI crystal readout by a 16 anode MPPC. The time difference between (top) the two finger scintillators, arranged above and below the crystal, appears at the top left. The time difference between the CsI crystal and one of the counters appears at the bottom left. The pulse height response is shown on the right.

## 9.12   Laser Monitoring System

In order to continuously monitor variations of the crystal transmittance and the APD gains, a laser system has been designed similar to the one used for the CMS calorimeter [32]. The use of solar- blind photosensors from RMD requires a laser with a wavelength where the sensor has reasonable quantum efficiency. The laser light is transmitted by a distribution system and optical fibers on the readout side of the detector. The fiber end has a ferrule connector positioned between the two photosensors and held in place by a small screw in a reproducible way. The light will be transmitted through the crystal and then reflected and diffused by the crystal and the wrapping material before it illuminates the active area of the photosensor. As shown in 13 (left)**,** an upper limit for the detectable light is at ~270 nm. Deterioration of the crystal transmittance due to the irradiation is usually concentrated at the lowest wavelengths and can be controlled by the source response assuming a tight control of the photosensor gain. We are evaluating different options for a laser emitting DUV light between ~220 and 260 nm.

A schematic of the overall system is shown in Figure 9.60. A high-precision, high-power, pulsed laser sends light through standard collimation optics to an optical splitting system, done with mirrors, to subdivide the beam into 8 equal parts. By means of eight 1-mm diameter, 20 m long quartz fibers, the light is brought to the Detector Solenoid bulkhead and through a vacuum feed-through, to the back face of the calorimeter disks. On each disk, there are four 2-inch diameter integrating spheres (see Figure 9.61) with one input for the incoming fiber and three outputs. Running from two of the outputs is a bundle of 150 200-$\mu$m diameter fused silica fibers, for a total of 1200 fibers/disk. Of the 1200 fibers/disk, 930 are used for gain calibration, 8 for monitoring; the remaining 264 are replacements in case fibers are broken during handling or installation

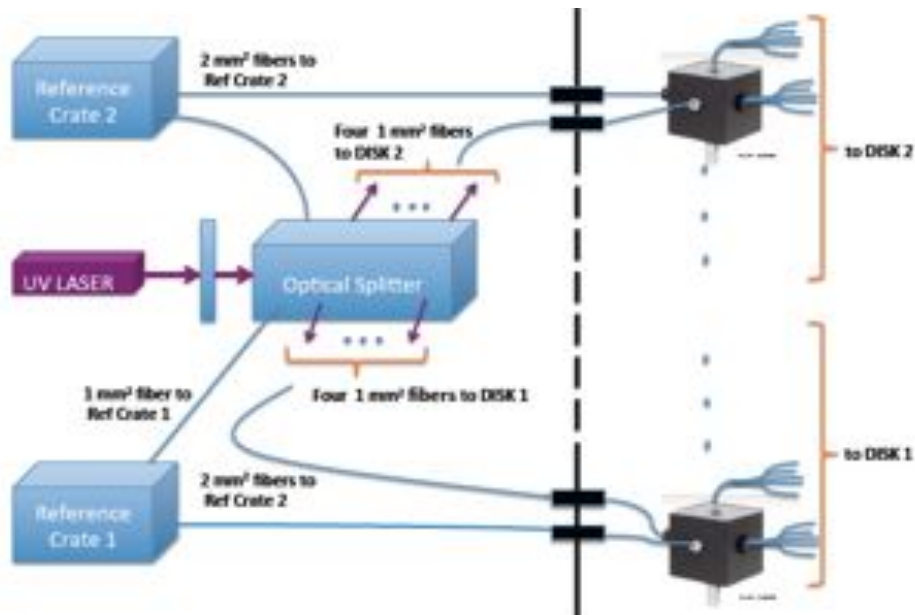

Figure 9.60. Schematic of the laser monitor system.

The light output from the laser system is monitored with pin-diodes that measure the output light from the laser and the returning light from the integration spheres. A total of 50 pin diodes are needed. The pin diode monitors are required to track pulse amplitude variations larger than 1%. The laser and the monitor boxes will be temperature controlled to reduce the variation of the laser to a few percent and to minimize the pin-diode temperature correction. In order to monitor calorimeter response linearity, a neutral filter wheel with gradually changing absorption values is inserted between the primary beam and the light distribution system.

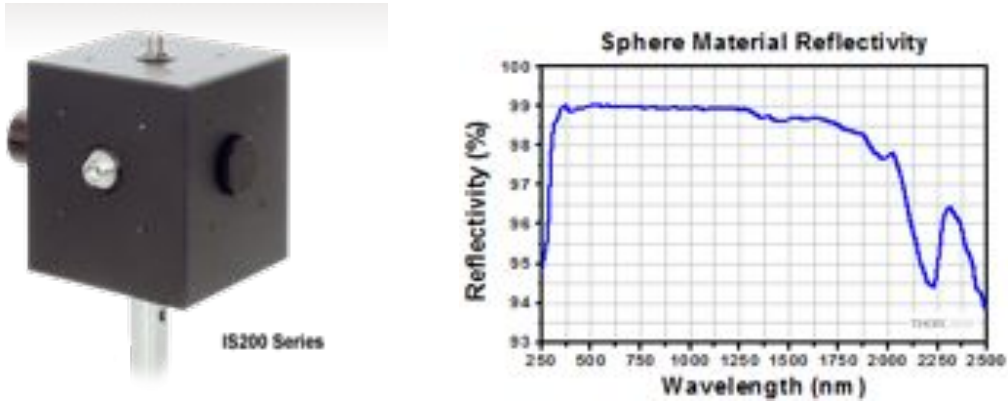

Figure 9.61. Picture of the ThorLab IS-200 integrating sphere (left); and the sphere's reflectivity dependence on wavelength.

There is not a stringent requirement on the laser pulse width, since the APD readout electronics has a rise time between 6 to 8 ns, thus setting an upper limit on the width of 10 ns. Similarly, the pulse frequency is not strongly constrained since, as shown in the prototype test, running at 1 Hz provides better than per-mil statistical precision in one hour of data-taking. It is instead mandatory to synchronize the laser pulse with an external trigger to allow the light to reach the detector at the correct time relative to the proton beam pulse so that laser data can be taken during the time when the calorimeter is acquiring physics data as well as during the gaps between beam when the calorimeter is quiet. The laser pulse energy is strongly attenuated by the distribution system. However, the laser signal is required to simulate a 100 MeV energy deposition. For $BaF_2$ this corresponds to ~10,000 p.e. in each photosensor. This roughly translates to a 10-20 nJ energy source. A safety factor of 20 is designed into the system to account for the eventual degradation of the signal transmission with time, resulting in an energy pulse requirement of ~ *0.5* μJ.

There is a stringent requirement on the fibers. They should have high transmission at 200-260 nm, a small attenuation coefficient and they must be radiation hard up to O(100 krad). The best choice is fused silica fibers, both for their transmission properties (see Figure 9.62, left), a nearly flat wavelength dependence down to 150 nm, a long attenuation length and high radiation tolerance.

### 9.12.1   Laser monitor prototype for the LYSO crystals

The setup used for the transmission test and for the calibration of the LYSO calorimeter prototype is shown in Figure 9.62 (right). The light source was an STA-01 solid-state pulsed laser emitting at 532 nm with a pulse energy of 0.5 μJ, a pulse width < 1 ns, good pulse-to-pulse stability (3%), and synchronization to an external trigger for frequencies up to 100 kHz. Table 9.7 summarizes the performance of equivalent STA-01 lasers emitting in the UV region that are being evaluated for the final implementation. The prototype distribution system uses a 2" integrating sphere, the ThorLab-IS200, with one input port and 3 output ports. Each of the output ports has a 0.5" diameter. Pictures of the sphere and of its reflectivity diagrams are shown in Figure 9.61**.** A Hamamatsu Pin-Diode S1722-02 is mounted in one sphere port to monitor the laser pulse variation, while a bundle of fifty 2 m long, *Leon*i fused silica fibers of 200 (400) μm diameter core (core plus cladding) is inserted with an SMA connector to another port.

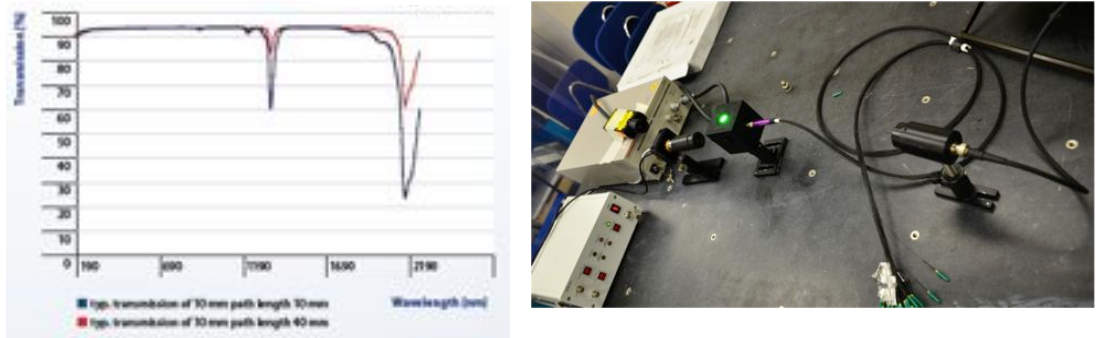

Figure 9.62. Transmission as a function of wavelength for fused silica fibers (left) and a picture of the light distribution system prototype (right).

Table 9.7. Main properties of STANDA Lasers operating in the UV region

| Models | STA-01-TH | STA-01-FH |
|---|---|---|
| Wavelength, nm | 354 | 266 |
| Average output power (max), mW | 15 | 20 |
| Pulse energy, μJ | > 1.5 | 20 |
| Pulse duration, ns | < 0.5 | < 0.5 |
| Repetition rate (max), Hz | 10,000 | 0.1 - 1000 |
| Beam profile | | $M^2 < 1.2$ |
| Pulse spectral structure | | Single longitudinal mode |
| Polarization ratio | | > 100:1 |
| Beam waist diameter inside the laser head $1/e^2$, μm | | 25 - 200 |
| | | < 5 (near transform limited) |
| Pulse spectrum FWHM, pm | | < 0.6 |
| Pulse-to-pulse energy stability rms | | < ± 1.5% |
| Power stability over 6 hours | | 100 - 240 |
| External power supply voltage, VAC | | 15 - 40 |
| Operating temperature °C | | USB, External trigger (TTL rising edge) 1 Hz max |
| Interfaces | | repetition rate |

The number of photoelectrons, Npe, observed at the end of the transmission line has been determined by a direct measurement of the APD charge seen in the calorimeter. The input laser source was first reduced by a factor $T_{filter}$ = 200 by means of a neutral density filter, in order to avoid signal saturation. The average APD charge, with the APD gain set to 50, was around 120 pC, with a channel-by-channel spread of ± 10%. This corresponds to Npe = 33,600, a factor of 3 more than required in the $BaF_2$ case. However, this determination does not take into account the reduction factor of 14 that results from the initial optical splitting system and for the factor of 2 in the energy ratio between UV and green light. The measured Npe is consistent with the pulse energy and distribution losses. One photon at 520 nm corresponds to $4 \times 10^{-19}$ J, so that in a single laser pulse $\sim 10^{12}$ photons are produced. Using the measured $T_{fiber}$ and $T_{filter}$, the light transmitted at the end of the chain is estimated to be $N_{photon} = 10^{12} \times (7 \times 10^{-5}) \times 0.005 = 3.5 \times 10^5$. Correcting this estimate for the APD quantum efficiency of 70% and for the APD/crystal area ratio of 1/9, 27,000 detected photoelectrons are expected, in reasonable agreement with the measurement.

The prototype calibration system has been tested by measuring the transmission at one of the output ports, $T_{port}$, and by measuring the transmission at the end of the fiber bundle, $T_{fiber}$. The transmission in one port can be written as $T_{port} = (S_{port}/S_{sphere}) \bullet M$, where S represents the surfaces and M=R/(1-R• (1-f)) is the sphere multiplication factor. R is the sphere reflectivity and f is the ratio between the ports and the sphere surfaces. At $\lambda > 400$ nm, R is 98%, f is ~5% and M is ~16, so that the transmission factor is ~0.012×16=0.192. The first measurement was performed by calculating the ratio between the light emitted by the laser and the light exiting from the sphere port. A calibrated photocell of 13 mm diameter has been used, positioned at zero distance from the hole. The transmission measured is ~0.12, in reasonable agreement with our simplified model. Similarly, the transmission at the end of the fiber bundle has been measured, resulting in an average factor of $T_{fiber}$=7x10$^{-5}$ that, as expected, is much better than the product of the simple geometrical ratio, $10^{-5}$, and the fiber numerical aperture. The spread of transmission values for the best 43 fibers in the bundle has a σ = 8%. The seven remaining fibers were accidentally cracked before the test, showing deviations worse than a factor of two.

Finally, in Figure 9.63 (left), the variation of the observed laser pulse as a function of the running time is shown. The average laser fluctuation observed in 12 hours of running has a σ ~5% and is mainly due to the variation of the APD response. This is shown by comparison with the reference pin-diode (green circles), which is much flatter than the calorimeter response. The residual fluctuation of the calorimeter to pin-diode ratio is 3.5%. This is much worse than the ratio between two calorimeter channels (red points) that is at a level of 0.4% and of the PIN-diode, which is 1.6%. To confirm this, the dependence of the calorimeter response on the temperature has been studied by measuring the temperature in the APD region with a PT-100 probe (Figure 9.63, right).

The APD gain dependence on temperature is consistent with the observed residual calorimeter/pin fluctuation as shown by the anti-correlation between the APD temperature and the calorimeter response in Figure 9.63 (right). The gain variation of the APD corresponds to ~ -4 %/C°.

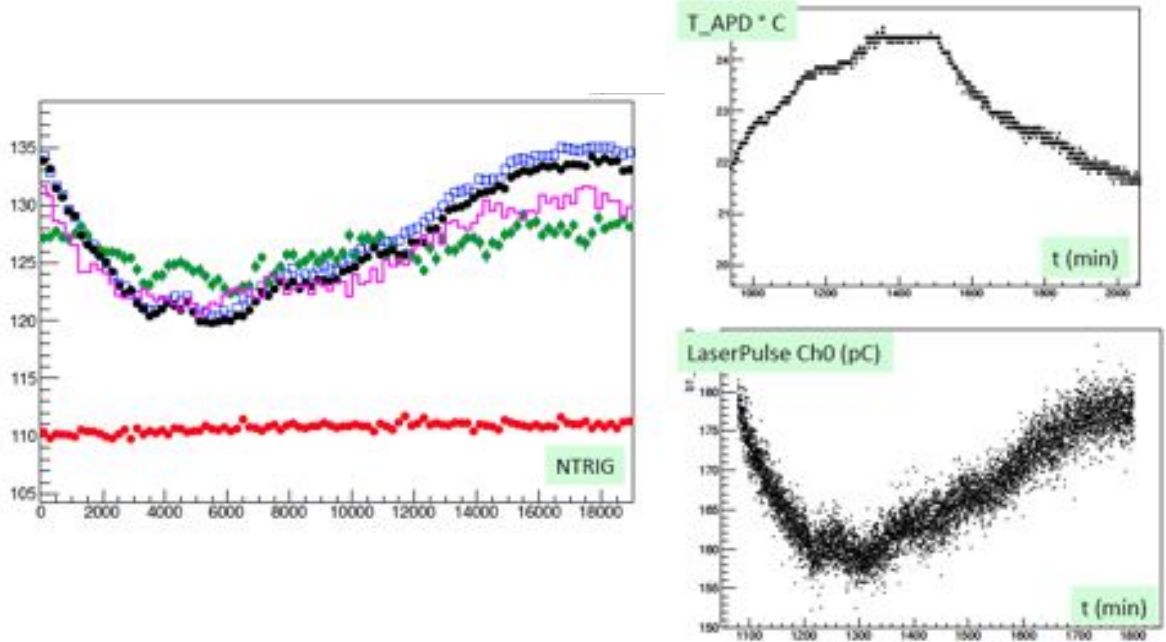

Figure 9.63. The left plot shows the distribution of the average laser pulse energy as seen by two calorimeter channels as a function of the trigger number (black and blue points), of the average pin diode response (green points) and of the ratio between the calorimeter channels (red points). The purple histogram shows the ratio between calorimeter channels and the PIN-diode, isolating the residual fluctuations of the APD gain. Also shown is the distribution of the temperature as a function of running time in minutes (top right), and the variation of the laser pulse for channel 1 during the same period (bottom right).

## 9.13   Source Calibration System

Calibration and monitoring while physics data is being accumulated is an important ingredient if the best possible performance of the calorimeter is to be realized. A suitable system must provide precise, independent crystal-by-crystal calibration. The use of radioactive sources is a proven technique for accomplishing such a calibration. However, most long-lived sources are limited to an energy around 1 MeV, which makes it difficult to secure a signal that is significantly above electronic noise, and sources that must be deployed individually are not practical with a system of ~2000 crystals. Mu2e has adopted an approach formerly devised for the *BABAR* electromagnetic calorimeter [33]. In this system, a 6.13 MeV photon line is obtained from a short-lived $^{16}$O transition that can be switched on and off as desired. This system was successfully used for routine weekly calibrations of the *BABAR* calorimeter. It is an ideal match to the Mu2e requirements, and we have started the process of salvaging the *BABAR* system in order to refurbish it for use in Mu2e.

The decay chain producing the calibration photon line is:

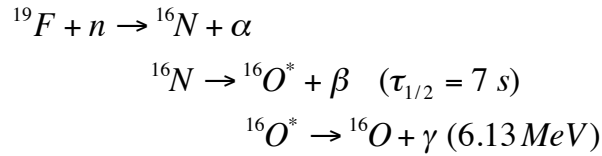

The fluorine, a component of Fluorinert™ coolant liquid, is activated with a fast neutron source, producing the $^{16}$N isotope. This isotope then $\beta$-decays with a half-life of seven seconds to an excited state $^{16}$O$^*$, which in turn emits a 6.13 MeV photon as it cascades to its ground state. A source spectrum collected with a *BABAR* CsI(Tl) crystal with PIN diode readouts is shown in Figure 9.64. There are three principal contributions to the overall energy distribution: one peak at 6.13 MeV, another at 5.62 MeV and a third at 5.11 MeV, the latter two representing $e^+e^-$ annihilation photon escape peaks. Since all three peaks have well-defined energies, they simultaneously provide both an absolute calibration and a measure of the linearity of response at the low end of the calorimeter energy scale.

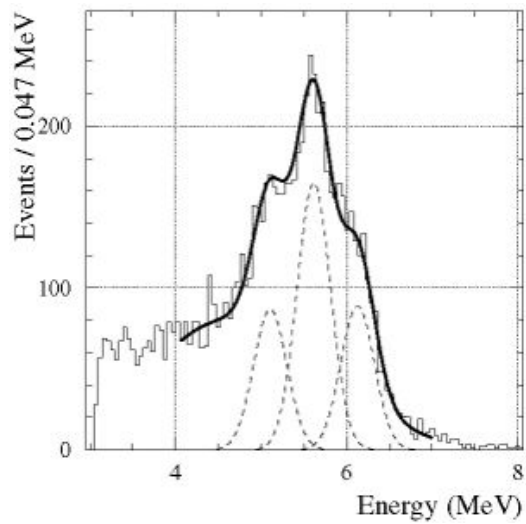

Figure 9.64. Typical source calibration spectrum from a *BABAR* CsI(Tl) crystal showing the 6.13 MeV peak, along with two escape peaks.

The fluorine is activated using neutrons provided by a commercial deuterium-tritium (DT) generator producing 14.2 MeV neutrons, at typical rates of several times $10^8$ neutrons/second, by accelerating deuterons onto a tritium target. The DT generator is surrounded with a bath of the fluorine-containing liquid Fluorinert™, which is then circulated through a system of manifolds and pipes to the calorimeter crystals. Many suitable fluorine-containing liquids are commercially available; Fluorinert™ "FC-77" was used in *BABAR* and stored in a reservoir near the D-T generator. When a calibration run is started, the generator and a circulating pump are turned on. Fluid is pumped from the reservoir through the DT activation bath and then to the calorimeter. The system is closed, with fluid returning from the calorimeter to the reservoir. A schematic of the Mu2e system, based on the *BABAR* system, is shown in Figure 9.65.

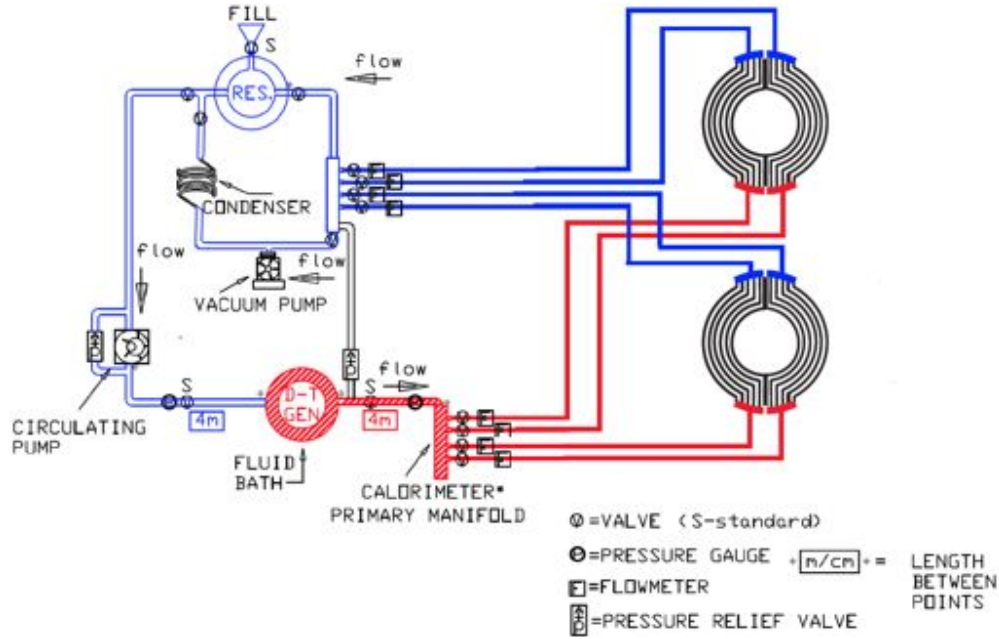

Figure 9.65. Schematic Layout of the EMC calorimeter source.

The DT neutron generator is a small accelerator. Radiation safety protocols factor into the design of the calibration system, and operation of the source will be done remotely in a no-access condition. The half-life of the activated liquid is 7 seconds; residual radioactivity is thus not a substantial concern when the DT generator is not operating. The DT generator will be shielded according to FNAL safety regulations. The shielding will be interlocked such that the DT generator cannot be operated if the shielding is not in place. The fluid reservoir is capable of holding the entire volume of FluorinertTM fluid required for operation of the system. In the event of a fluid leak, the maximum exposure for the *BABAR* system was calculated to result in a maximum integrated dose of less than 1 mrem. For Mu2e, a detailed hazard analysis will be performed in collaboration with Fermilab radiation safety experts. Operation of the system is anticipated to be approximately weekly during Mu2e running.

In *BABAR*, the fluid was pumped at 3.5 liter/second, producing a counting rate of ~40 Hz in each of the ~6500 crystals, which were an average distance of about 12 m from the DT generator. This produced a calibration with a statistical uncertainty of ~0.35% on peak positions in a single crystal in a 10-15 minute calibration run. The fluid transport manifold consisted of thin-wall (0.5 mm) aluminum tubing (3/8-inch diameter); 1 mm of Al represents 1.2% of a radiation length. The tubes were placed in front of the *BABAR* crystals, with an additional 2 mm of Al in the structural support for the tube assemblies. A similar system of thin-wall aluminum tubing mounted on a supporting structure will be implemented for each of the two Mu2e calorimeter disks.

### 9.13.1   Salvage of System Components from SLAC

Many of the main components of the source calibration system used at *BABAR* have been preserved in good condition during the *BABAR* detector decommissioning process and have been requested from SLAC. These items include:

- the *BABAR* DT generator, model ING-07, manufactured by the All-Russia Institute of Automatics (shown partially disassembled prior to installation at *BABAR* in Figure 9.66) , including HV power supply, PC-interface controller card and cabling;
- elements of the fluid distribution system, including the primary outgoing and incoming manifolds, valves and pressure gauges, and the main distribution panel on which many of these items are mounted;
- pumps for the fill and fluid activation loops; and
- any remaining stocks of Fluorinert™ FC-77.

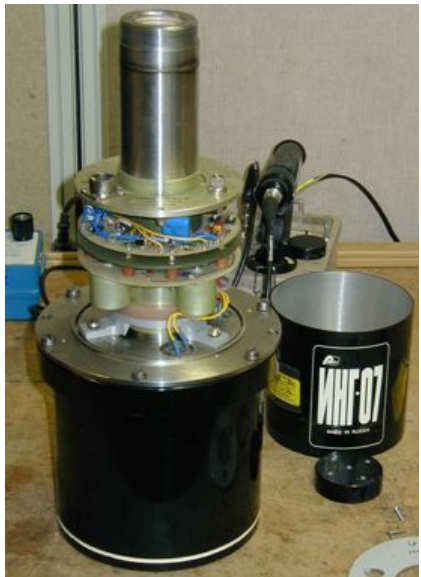

Figure 9.66. The model ING-07 DT generator.

Prior to installation at Mu2e, these salvage items will be tested, refurbished as required, and assembled into a prototype system at the California Institute of Technology. This will allow assurance of performance of the salvaged items prior to their incorporation into the final calorimeter source calibration system, which will be developed and built at Caltech, and ultimately transported to FNAL for installation at Mu2e

## 9.13.2   Implementation for Mu2e

The source calibration system for Mu2e is designed to provide a weekly calibration of the entire calorimeter [34] in about 10 minutes of data acquisition. The design precision is better than 0.1 MeV at the 6.13 MeV line, or better than 1.4%. This is a negligible contribution to the overall resolution of the calorimeter.

The number density of fluorine in Fluorinert™ FC-77 is approximately $4 \times 10^{28}$ $m^{-3}$, essentially all in the desired $^{19}F$ isotope.  There is some uncertainty in this number density as the proprietary formulation is not precisely known; we work with a worst-case assumption. The viscosity, at 0.8 centiStokes, is similar to that of water. The radiation length of FC-77 is approximately 20 cm.

The relevant $^{19}F(n,alpha)^{16}N$ cross section is about 24 mb [35]. The total inelastic cross section is around 80 mb, dominated by $^{19}F(n,2n)^{18}F$. The elastic cross section is much larger, at about a barn.

The bath irradiated by the DT generator has a volume of about 20 liters, with the fluid pumped at a rate of 3.5 l/s; for a neutron rate of $10^9$ n/s, the density of $^{16}N$ at the bath exit is thus about $1.5 \times 10^9$ $m^{-3}$. With decays, this is attenuated by a factor of 0.7 by the time the fluid reaches the furthest crystals in the calorimeter.

The conceptual layout of the source calibration components is shown in Figure 9.67. The basic plumbing design consists of 4.1-cm ID transport pipes of about 15 m length to the calorimeter disks, where 3-cm manifolds are located. Each disk has two such manifolds, one for supply and one for return. Connecting the manifolds are the thin-wall tubes that carry the irradiated fluid over the face of the calorimeter disk. There are 12 of these for each disk, arranged in a concentric pattern and ranging in length from 1.5 to 1.7 m. The tubes are 0.5-mm wall-thickness round aluminum tubing with an inside diameter of 3/8 inch, similar to the tubing used at *BABAR*.

 shows the layout of Al pipes at the front face of a calorimeter disk, along with the manifolds at bottom (red) and top (blue) that lead to and from the fluid activation bath. Studies using as a figure-of-merit the number of photons passing through a surface element at the front face of a crystal have been performed to optimize the spacing between pipes and the perpendicular distance from the pipes to the front surface of the crystals. Figure 9.69 illustrates the ±10% variation in illumination as a function of increasing radial distance for an inter-pipe spacing of 60 mm and 30 mm between pipes and the front surface of the crystals.

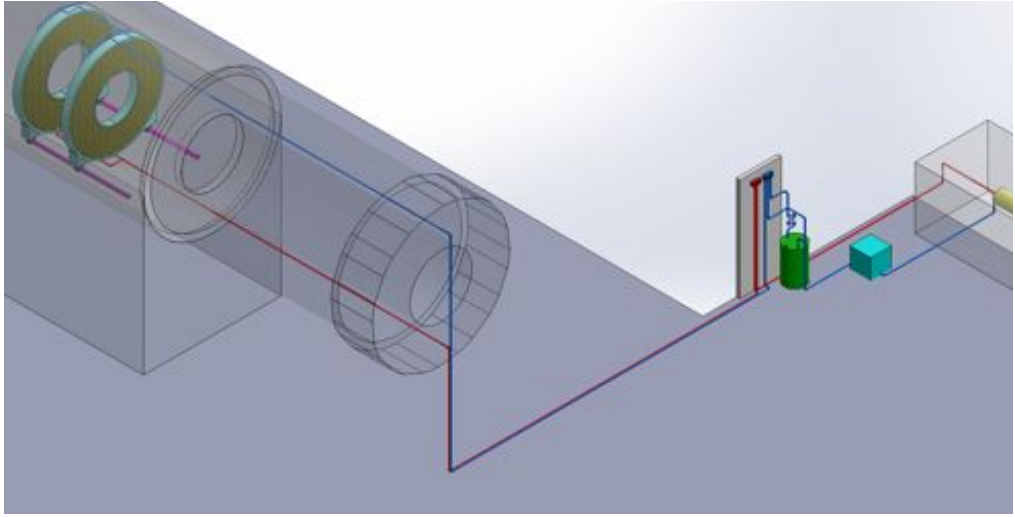

Figure 9.67. Physical layout of the calibration source components in the Mu2e experimental hall.

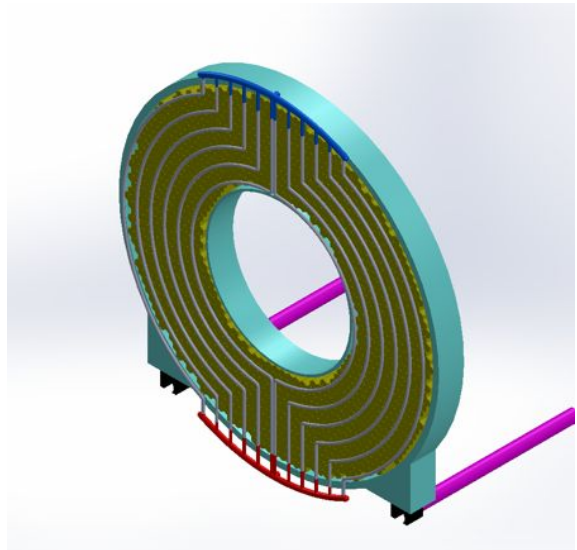

Figure 9.68. Calorimeter disk with aluminum source calibration pipes.

Figure 9.70 shows the relative intensity for different values of the distance between pipes and the crystal front surface as a function of radial distance. Based on these studies, we have chosen to set the distance between pipes to 60 mm, which essentially allows one pipe to pass between every other crystal. The distance from the pipes to the crystal front surface should be minimized but, as can be seen from the figure, any distance between approximately 10-30 mm is reasonable. The final value for the perpendicular distance will be determined within this window taking into account any engineering constraints.

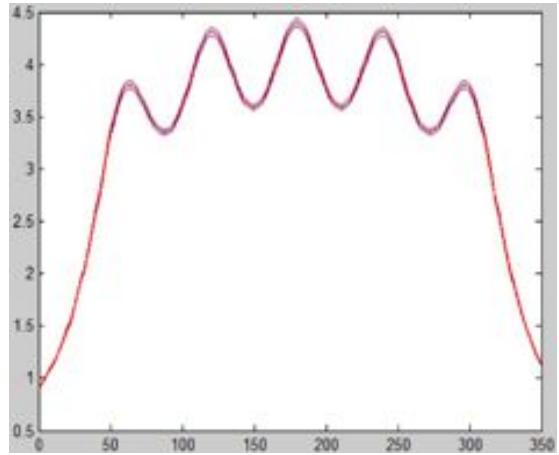

Figure 9.69. Radial variation in illumination with an inter-pipe spacing of 60 mm with 30 mm between the pipes and the crystals.

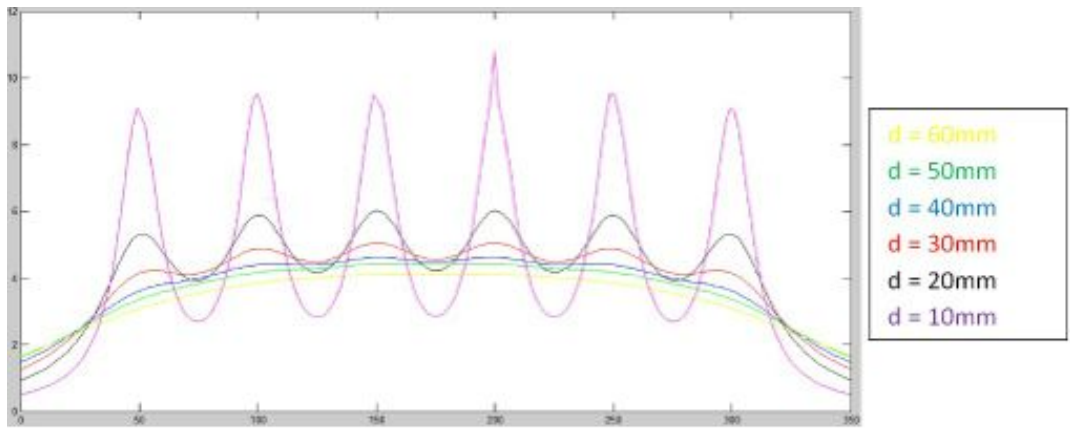

Figure 9.70. Simulation of the relative intensity of illumination as a function of radial distance for several values of the gap between the pipes and the front surface of the crystals.

## 9.14 Risk Management

There are several risks that could jeopardize the success of the calorimeter subproject. The risks as well as potential mitigation strategies are described below.

There is a risk that we cannot develop the UV extended solid-state photo-detectors that are blind to solar wavelengths on the schedule required by the project. $BaF_2$ has a long component at a wavelength of 300 nm. Without the development of these new photosensors, the rate capability of the calorimeter might be compromised. To mitigate this risk, work on solar blind photodetectors is currently underway at Caltech, JPL and RMD. In the meantime, there is a method to directly reduce the production of the $BaF_2$ long component by doping the crystal with 1% La. Interactions with the vendors are currently underway. If these mitigations do not come together before the final calorimeter technical review and the start of the production, a cheaper alternative, though one with poorer performance, will be implemented. As reported, a parallel R&D program has begun to study the feasibility of using pure CsI crystals with large-area SiPM read-out. In this case, a complete demonstration of timing and radiation hardness capability has to be delivered in a timely fashion.

The large neutron flux in Mu2e poses several risks to the performance of the calorimeter. One such risk is associated with the radiation hardness of the photosensors. Neutrons incident on either APDs or SiPMs could increase the dark current and deteriorate the calorimeter's performance. The use of the disk geometry greatly reduced this problem with respect to the original vane geometry. Indeed, the highest neutron flux estimated by the simulation in the disk readout-area is below $2 \times 10^9$ $n_{1MeVeq}/cm^2$, a factor 3-4 better than in the vane case. Moreover, the photosensors will be connected through *bridge-resistors* to their external shielding so that metallic cooling fingers attached to the main cooling system can be used to cool them down ($\sim 0°C$) to increase their radiation hardness. Neutrons interacting in the crystals can also degrade the energy resolution of reconstructed clusters. For large neutron fluxes the pileup in and around the reconstructed cluster could become very important, depending upon the timing characteristics of the selected crystals. Pulse shape analysis can mitigate this risk to a large degree. If the situation becomes intolerable, an additional mitigation will be to enhance the neutron shielding inside the Detector Solenoid.

Finally, there is a risk that INFN might not be able to commit to the calorimeter construction by the time of CD-2. There is the associated risk that the INFN commitment might not be as large as originally assumed. The number of INFN physicists participating in Mu2e can also limit the funds that INFN is willing to commit. Calorimeters are, by their nature, expensive devices, thus challenging the standard Euro/FTE formula used at INFN. From the practical point of view, some delay on the decision can be tolerated, due to the existence of a parallel path of approval that is well underway. In the worst-case scenario of INFN dropping from the calorimeter construction, the mitigation will be to reduce the construction to one disk only or to fill a reduced area of the detector, thereby losing up to 35% in relative acceptance. In this way, the risk, corresponding to O($1M), can be minimized.

## 9.15   Value Management

Even though a performance baseline has now been established for the calorimeter, value management will continue through final design and construction. In particular, a careful examination and validation of detector requirements coupled with evaluation of alternative engineering and design choices will continue, with special attention to cost.

The option of using large-area SiPMs is more attractive due to the low light yield of the crystals under consideration (relative to LYSO) and to the fact that most producers are currently developing blue or UV extended devices for application in other fields, such as astro-particle physics. The first results obtained with the pure CsI crystal and a SiPM matching the emission spectra, are very encouraging. The inherent high gain and lower noise of SiPMs might allow for a simpler design of the front-end electronics, reducing the HV needs and simplifying the amplifier design. The basic layout of the FEE chain will be kept unchanged but there would be no need to have a DC-DC converter working in the magnetic field and the amplifier gain requirement would be much easier to meet.

## 9.16   Quality Assurance

For a calorimeter of this complexity, Quality Assurance is a fundamental component of the procurement, fabrication and assembly phases. Quality Assurance will be applied to all components and subsystems, building on the relevant experience from the *BABAR*, CMS and PANDA calorimeters and from the Mu2e group itself. Indeed, within the Mu2e collaboration, the expertise that already exists in the construction of the KLOE-2 calorimeter upgrade, as well as the *BABAR* and Super*B* calorimeters have been useful for carrying out the calorimeter R&D program where the QA procedures have been developed.

### 9.16.1   QA for Crystals

In order to construct a high performance calorimeter that satisfies the Mu2e physics requirements, strict requirements are imposed on various crystal parameters that must be controlled both at the production sites, and upon receipt by Mu2e. The calorimeter is composed of ~2000 $BaF_2$ or pure CsI crystals of hexagonal shape. Each crystal has to satisfy three different QA tests in order to be accepted by Mu2e. These include

1.   an optical and dimensional inspection,
2.   a measurement of the emission spectra and transmission quality, and
3.   a test of light yield and longitudinal uniformity of response, LRU.

To ensure timely feedback on crystal production (~100 /month), the use of automated stations is required. In the following, the organization of this effort is described in some detail. Many of these techniques were developed during our studies of LYSO crystals. These acceptance criteria have now been applied to twenty $BaF_2$ crystals.

***General inspection and validation of dimensions***
Each crystal is required to satisfy the following:

1.   To be free of cracks, chips and fingerprints. They shall be inclusion-free, bubble-free and homogeneous.

2. To deviate from a perfect 3-dimensional hexagonal prism by less than 50 μm.

3. Mechanical tolerance of ±50 μm per side with a 0.3 mm chamfer on all edges.

A generic visual inspection will be done upon receipt of the crystals and the packaging will be opened in a dedicated clean room. Crystals will only be handled by experienced technicians wearing gloves. Each crystal will then have its dimensions checked using a Coordinate Measuring Machine (CMM). This facility allows, through optical scan and precise dimension determination, controlling the hexagonal shape at better than 10 μm.

### *Transmission and light emission tests*
We require each LYSO crystal to:

1. Have a longitudinal transmission above 75% at 420 nm and 80% at 440 nm.

2. Have a transverse transmission above 75% at 420 nm.

3. Pass the overall longitudinal uniformity test that consists of scanning the crystals at 9 points (3 rows of 3 points) along the transverse face;

4. Pass the overall transverse uniformity test by scanning the crystals at 27 points (3 rows of 9 points each) along the longitudinal axis;

5. Pass a test of the emission spectra performed with a UV LED.

These tests have already been developed and executed using measurement stations at both Caltech and LNF. The station at Caltech is a commercial device from Hitachi able to measure in the range 200-900 nm, which can measure the transmittance for $BaF_2$ or pure CsI crystals. The transmission station at LNF, LATTER (Longitudinal and Transversal Transmission Emission Response), was designed and assembled during 2013 and is tuned to operate in the range 350-900 nm. The CAD drawings and an illustration of the basic principle are shown in Figure 9.71. A light source uniformly illuminates the back of the crystal while a spectrophotometer from Ocean-Optics, with special focusing optics, is able to read light coming from a narrow ellipse of 1.5, 2 mm radii. The crystal can be positioned in front of the spectrophotometer and translated longitudinally, adjusted vertically or rotated around its axis by precise step-motors. The measurement of transmittance takes five seconds per point; when multiplied by the number of testing positions, it corresponds to an elapsed time of 10 minutes per crystal. The program is written in LabView and controlled by means of a laptop. At the end of the transmission test, the crystal emission spectrum is also measured with a UV LED (350 nm). A more detailed description of the stations can be found elsewhere [35][36]. All 25 LYSO crystals used for construction of the medium-size prototype have been tested and qualified for longitudinal transmission by using these stations. 15 crystals were tested at LNF and 10 crystals were tested at Caltech. Examples of acceptable and unacceptable spectra are shown in Figure 9.72. Similar tests have been carried out with the Caltech station. The transmittance of six LYSO crystals is shown in Figure 9.73. Out of 25 LYSO crystals, only one was found to be unacceptable from the transmission point of view. All 20 BaF2 crystals satisfied the acceptance criteria.

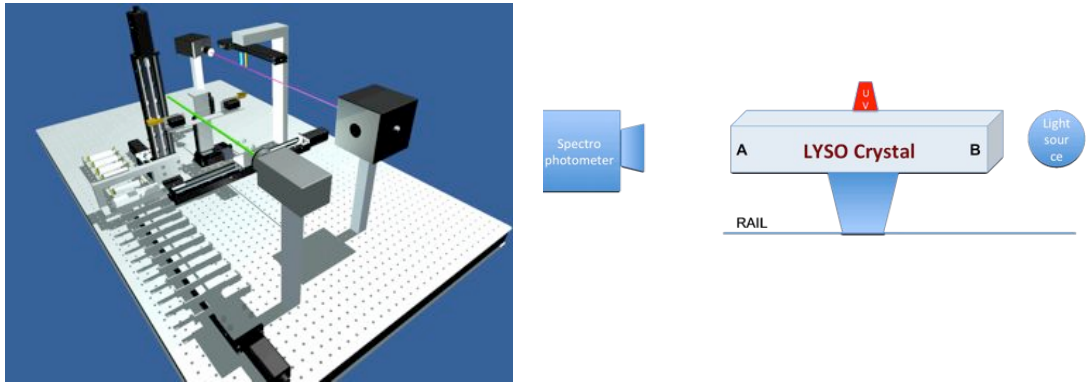

Figure 9.71. CAD Drawing of the LATTER test station at Frascati (left) and a schematic of the light transmission test setup (right).

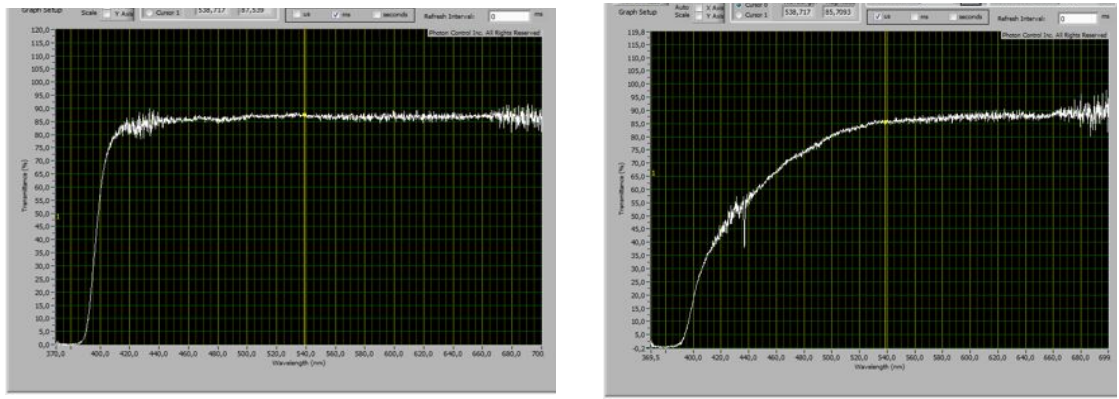

Figure 9.72. Longitudinal transmittance (%) as a function of wavelength in nm for an acceptable crystal (left) and an unacceptable crystal (right).

An example of an emission spectrum is shown in Figure 9.74 (left). Interestingly enough, we have also learned how to extract the Moyal law, which describes the emission spectra of the crystals or a generic plastic scintillator [37], by fitting it with the equation

F = n1×M1(λ1, σ1) + n2×M2(λ2, σ2)

where n is the amplitude, λ is the average wavelength, σ is the spread in nm and M is the following law:

$$a \cdot \exp\left(-\frac{1}{2}\left(\frac{(\lambda - \mu)}{\sigma} + e^{-(\lambda - \mu)/\sigma}\right)\right).$$

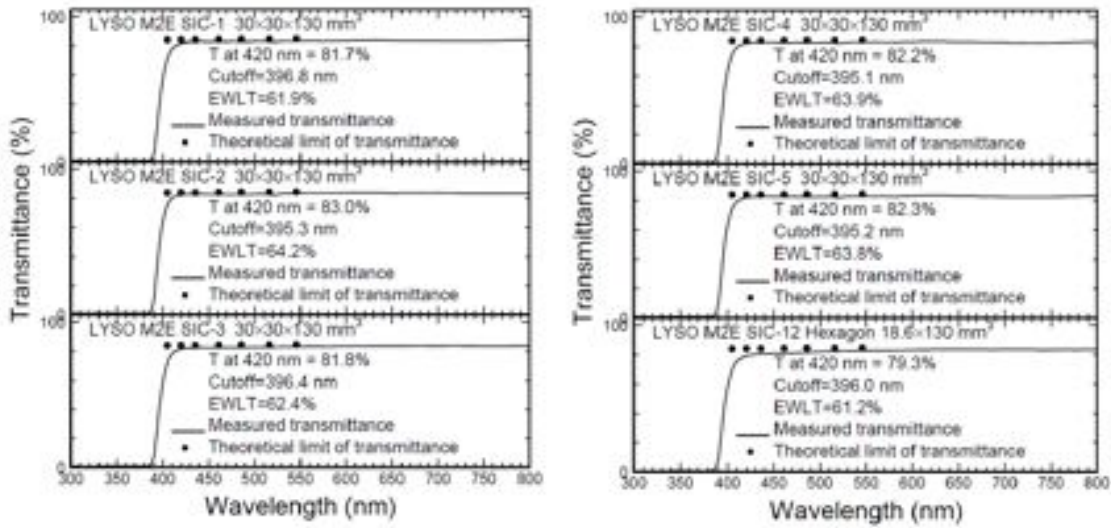

Figure 9.73. Longitudinal transmission parameters for six crystals measured at Caltech. The solid curve is the measured transmittance, the black points are the theoretical limit of transmittance.

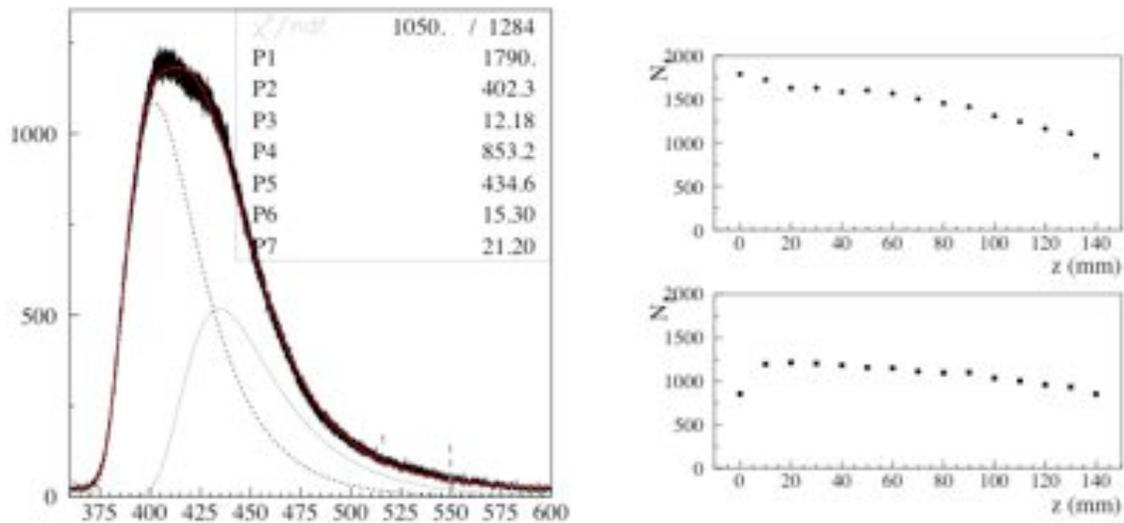

Figure 9.74. Emission spectra for one of the SICCAS LYSO crystals (left). The most important component is at 400 nm. A relevant component also exists at 435 nm. Dependence of the self-absorption along the z-axis (right). Top plot is for the amplitude of the 400 nm component and the bottom plot is for the 435 nm component. While the absorption in 1.5 m is almost a factor of two for the 400 nm component, the absorption at 435 nm is smaller than 15% and therefore practically negligible.

***Measurement of Light Yields and LRU***

The light yield and Longitudinal Response Uniformity (LRU) tests were set up using LYSO crystals. Each crystal was fully wrapped with an ESR-3M reflective sheet and was required to:

- Have a light yield above 2000 p.e./MeV when readout with a 2" PMT;
- Have a correct signal shape, 99% of signal below 200 ns;
- Have an energy resolution (FWHM) at 511 keV < 13 % with a 2" PMT;
- Have a reasonable LRU, defined as the RMS of the uniformity measurement in 7 or more points along the axis, RMS_LRU< 4%.

All measurements are carried out in a temperature and humidity controlled environment. Both the LATTER (LNF) and Caltech stations use a collimated $^{22}$Na source that illuminates the crystals over a region of a few mm$^2$. Each station is equipped to detect the two 511 keV annihilation γ's produced by this source; one γ is tagged by means of a small monitor system consisting of a LYSO crystal (3x3x10 mm$^3$) readout by a 3x3 mm$^2$ MPPC, while the second γ is used for calibrating the crystals. Crystals are wrapped in the same way as they will be wrapped for the final detector and are read out by means of a 2" PMT or by a 1 cm$^2$ APD. In the case of the PMT readout, no additional amplifier is used, while for the APD the signal is amplified by a commercial CAEN charge-amplifier. The tag and test signals are both acquired by means of a CAEN digitizer system running at 1 Gsps. A large effort has been made to make these QA stations user-friendly. In the LNF case, all data-taking is done by changing the position of the crystals while running a DAQ program at 500 Hz that allows collection of 10,000 events in 20 sec. A ROOT macro reconstructs, analyzes and fits the data in less than 30 sec. The Caltech unit, shown in Figure 9.75, is aimed at large-scale testing, displaying data to the user as it is being acquired, along with Go/no-Go acceptance criteria.

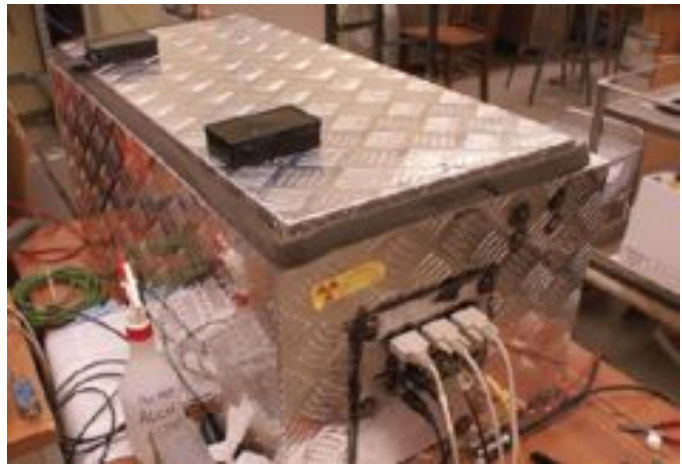

Figure 9.75. The Caltech QA station for measuring crystal light response.

All analysis and fitting are done in an automated manner. Typical reconstructed 511 keV photon peaks for PMT and APD readout are shown in
Figure 9.76. A plot of the LRU measured at Caltech and LNF for a few crystals is summarized in Figure 9.77. Adjustment of the photosensors is in progress to optimize these stations for measurements of BaF2 and pure CsI. A 2" UV-extended PMT such as the ET-9813 QB will be used.

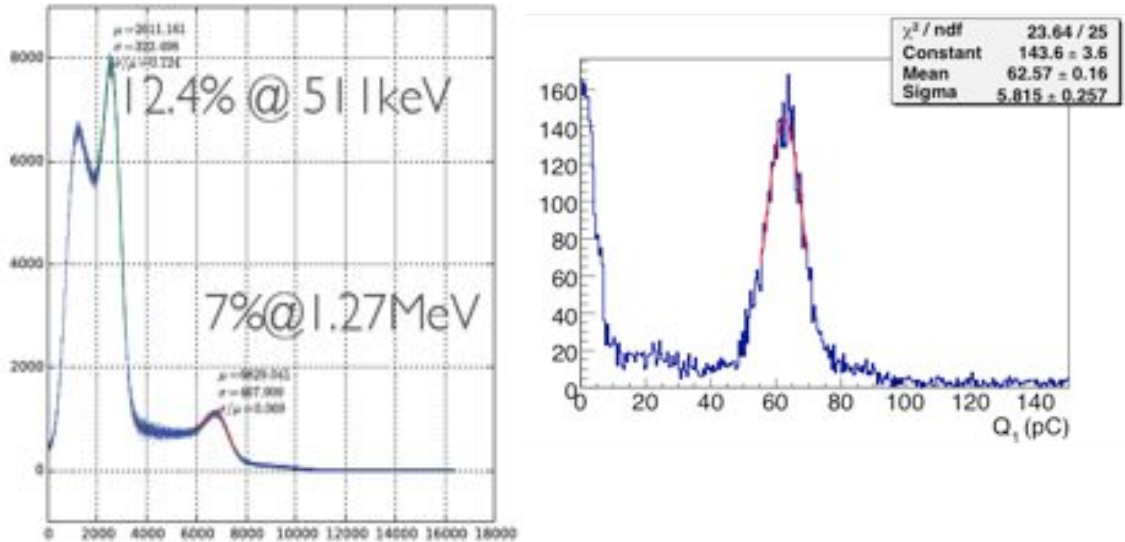

Figure 9.76. Distribution of the response of a LYSO crystal to a 511 keV gamma from a Na22 source for the Caltech QA station with APD readout and a charge amplifier (left), and for the LNF station with PMT readout (right).

In order to also test the linearity of response as a function of the deposited energy for low values (< 2 MeV) of energy deposition, a linearity-station has also been setup at JINR in Dubna. This station uses many different radioactive sources that produce particles at different energies ($^{241}$Am, $^{55}$Fe, $^{22}$Na, $^{137}$Cs, $^{60}$Co) to illuminate the crystals. We have compared the response of a LYSO crystal from SICCAS in the energy range 522-2500 keV has been compared to a LYSO crystal from Zecotek LFS in the energy range 356-2500 keV. A LYSO crystal from St. Gobain was taken as reference. The observed non-linearity is below 0.5 % from 50 to 2000 keV [38]. Similar measurements will be carried out for the BaF$_2$ and CsI crystals.

During the production phase, the crystal characterization will be done at FNAL in a clean room, and one or both setups will be used for the QA procedure.

### Acceptance Procedure
Having described the technique, the acceptance procedures are summarized below.

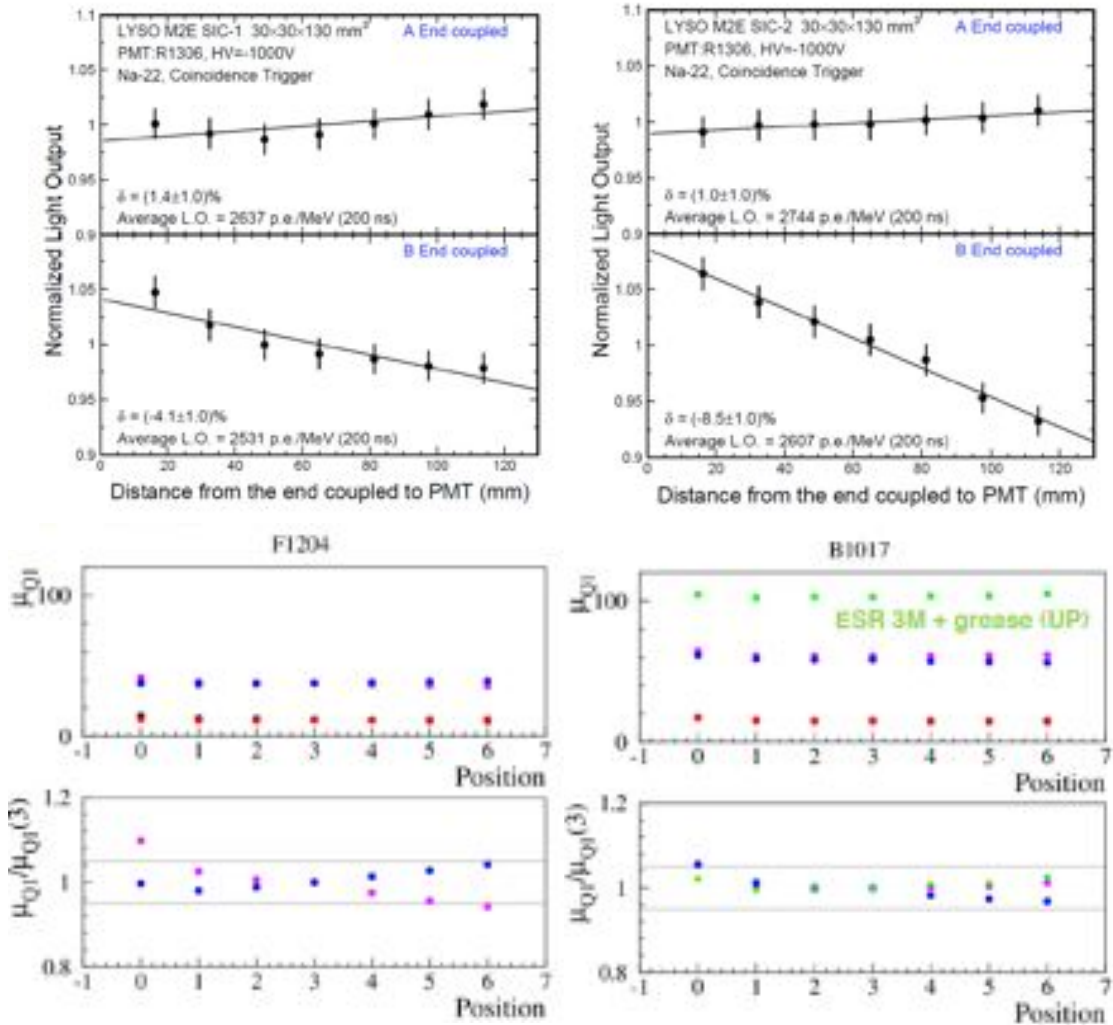

Figure 9.77. The Longitudinal Uniformity of Response (LRU) of two SICCAS LYSO crystals, as measured by the Caltech station (top), and the LNF station (bottom). The crystal response is more uniform when connecting the photosensor at one of the two ends. This is due to the Ce concentration gradient (CeC) produced during crystal growth; the light yield is related to the CE concentration while the uniformity is only related to the reflection and collection of the scintillation light.

## 1.  QA at the site of production:

Crystals will be tested at the vendor site before they are shipped to Mu2e. A test station will be provided to the vendor along with the set of specifications that each crystal must pass before it can be shipped. The test station will consist of a light tight box and a stepping motor assembly for moving a radioactive source ($^{137}$Cs) along the crystal axis to measure light yield and LRU.

**2. QA upon receipt by Mu2e:**

Mu2e will repeat the QA tests performed by the vendor for each delivered crystal. Additional tests will also be performed, including a measurement of the crystal dimension and its transmission properties as described above. ***Final crystal acceptance will be based on the tests made by Mu2e.***

**3. Radiation Hardness testing**:

A random sample of ~1% of the production crystals will be tested for radiation hardness. Gamma irradiation will be performed using a high-intensity [137]Cs source at Caltech where a motorized source mount is already available. Neutron irradiation can be performed at JINR, Dubna.

### 9.16.2   QA for photosensors

The QA for the silicon photosensors will be based on measurements of leakage current and gain, as well as on their temperature dependence and bias voltage. The APDs will be illuminated by means of a continuous blue, green or UV Laser for testing purposes. A schematic of an existing the measurement station in Udine for testing single photosensors is shown in Figure 9.78. If SiPMs are selected as the final read-out choice, it will also be necessary to measure their singles rate as a function of threshold scan (see Figure 9.78, bottom-right).

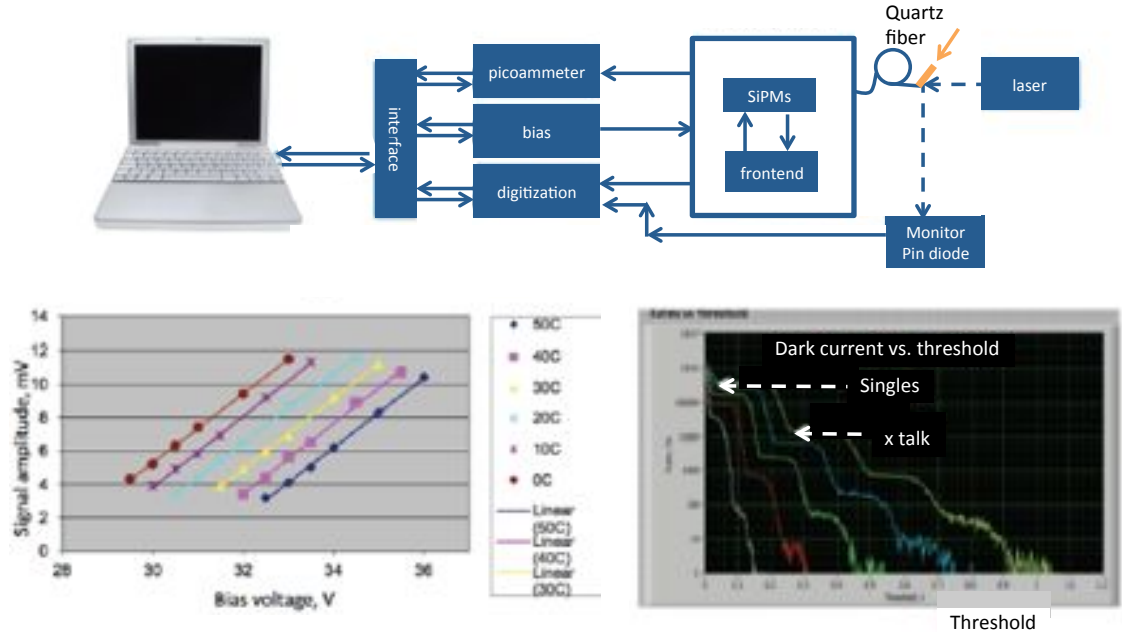

Figure 9.78. Schematic layout of the Udine QA station for testing single photosensors (top). To illustrate the capabilities of the test station, a measurement of the dependence of the gain of an APD on the bias voltage is shown on the bottom left and a measurement of the singles rate in a SiPM as a function of threshold is shown on the bottom right.

The conceptual design of a QA station that can measure multiple photosensors at one time and characterize the uniformity of response over their active area is shown in Figure 9.79. The setup is composed of a box housing the photosensors, a Keithley 6487 picoammeter, a voltage source, a light distribution system and a PIN diode. The housing box has a zero insertion force (ZIF) socket supported by a mechanical structure, which allows two-dimensional movements via two encoders and two step-motors. The Keithley 6487 supplies both the bias voltage to the photosensors and the measurement of the current. A relay is used to switch the Keithley between photosensors.

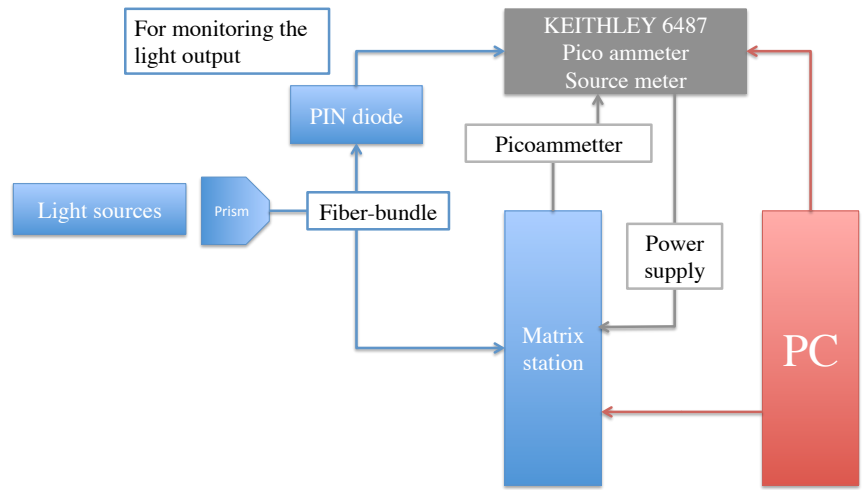

Figure 9.79. Generic schematic of the QA station for photosensors.

The light system depends on the choice of photo-sensor; for APDs, it is based on a sodium lamp that provides a broad spectrum from 200 to 800 nm in several nanometer bins. For SiPMs, three pulsed lasers can be used: one UV, one blue and one green. The optical system adopted for the light distribution consists of one fiber bundle with 250 μm plastic fibers for bringing the light to the photosensor socket, and finally one PIN diode used for monitoring the intensity of the light generated. For the SiPM option, a small plastic diffusor ($2\text{x}2\text{x}5$ mm$^3$) is placed between the fiber and the SiPM to avoid saturation of the photosensor.

### Gain measurement

The gain of each photosensor is calculated by measuring the current and the dark current as a function of the bias voltage:

$$G(V) = \frac{I(V) - I_{dark}(V)}{I_0 - I_{0,dark}}$$

Here $G(V)$ is the resulting gain at the bias voltage $V$, $I(V)$ is the current measured at a bias voltage $V$, $I_{dark}(V)$ is the dark current at bias $V$, and $I_0$ and $I_{0dark}$ are respectively the current and the dark current at the starting voltage. The measurement of both the station temperature and the supplied light intensity by means of the PIN diode allows for redundant control of the overall system. As shown previously, both the gain of APDs and SiPMs, have a non-negligible temperature dependence.

***Quantum efficiency***
The setup to measure Quantum efficiency is very similar to the setup for measuring gain, except for the light source. For APDs, a sodium lamp is well-suited for QE measurement, while for SiPMs additional pulsed lasers are needed in order to provide a wider light frequency range. A reference photosensor will be used to obtain a normalization for the QE determination.

***Excess noise factor***
For measuring the excess noise factor, the light source is replaced with a [90]Sr source and an amplification system.

***Acceptance criteria***
As in the case with the crystals, interaction with the photosensor vendor will be important in order to obtain devices that satisfy the Mu2e requirements. Vendors will be expected to provide test data but final acceptance will depend on tests performed by Mu2e. There will be one measurement station in Italy and one measurement station at Caltech to split the characterization of the sensors. A random sample of ~5% of the sensors will be irradiated with neutrons to determine radiation hardness.

### 9.16.3   Other QA activities

The preamplifiers and HV boards will be validated using standard bench test measurements of amplification and noise. A burn-in test of the HV board will also be performed.

A final test will be performed on crystals that have been fitted with APDs and readout electronics before they are inserted into a disk. The response of each individual assembly will be tested with a radioactive source. A system test will be performed on the assembled calorimeter using cosmic rays prior to installation in the Detector Solenoid.

## 9.17   Installation and commissioning

After testing of the individual components, the two disks will be assembled in a clean room with a controlled temperature and humidity environment. The assembly will be done with the disks in a vertical position with the supports mounted on the mechanical structure. Special tooling will keep them safely in stable equilibrium.

The insertion of crystals into a disk will start from the bottom; as soon as a *dodecagonal-like* sector is completed, the photosensors and the amplifiers will be mounted on the rear. A light-tight box will cover the disk in order to allow a test of the completed sector by means of a laser pulser and cosmic rays. A dedicated set of 160 preamplifier cables will be connected in a temporary manner to the Amp-HV chips to facilitate a three-day cosmic ray test using a standalone DAQ system. Similarly, a bundle of 80 optical fibers will bring a calibration signal to the crystals, allowing us to check both the crystal transmittance and the photosensor gain. This integration test will be synchronized with the delivery schedule for crystals and photosensors. A production rate of 160 crystals/month will roughly correspond to the completion of two dodecagonal sector/month. The whole operation would conservatively require 1.5 years, assuming a half-year long learning curve.

One year from the start of assembly the first disk will be completed by closing its ends with light-tight covers on which are mounted a set of radfets and temperature monitors, and by connecting the distribution spheres for the laser calibration system and the mechanical support for the final FEE and digitizers. After this operation, the routing of FEE cables, optical fibers and cooling fingers will take place in the back side of the disk, thus allowing a first complete debugging of the overall system. The signal cables from the FEE to digitizers and the local service cables from the ARM controller to the amplifier will be laid down in a configuration as close as possible to the final one. After this operation, the calorimeter is basically ready, needing only to receive the main service cables, the input quartz fibers for the laser calibration system and the fiber optics for the DAQ readout. Once the two disks are ready, the transfer from the assembly area to the Mu2e building will take place. The insertion on the rails will be carried out one disk at a time, with the FEE mechanics included. Once the two disks are over the rails, the support structure holding each single disk in a vertical position will be taken apart and the two disks joined. The final connection of services and debugging will take place as the final step before a system test with cosmic rays. A 3-4 month cosmic ray run will be carried out with all detector systems to exercise the final DAQ system and confront any system issues.

This page intentionally left blank

# 10   Cosmic Ray Veto

## 10.1   Introduction

Cosmic-ray muons are a known source of potential background for experiments such as Mu2e. A number of processes initiated by cosmic-ray muons can produce 105 MeV particles that appear to emanate from the stopping target. These muons can produce 105 MeV electrons and positrons through secondary and delta-ray production in the material within the solenoids, as well as from muon decay-in-flight. The muons themselves can, in certain cases, be misidentified as electrons. Such background events, which will occur at a rate of about one per day, must be suppressed in order to achieve the sensitivity required by Mu2e. Backgrounds induced by cosmic rays are defeated by both passive shielding, including the overburden above and to the sides of the detector hall, as well as the shielding concrete surrounding the Detector Solenoid, by particle identification criteria using the tracker and calorimeter, and by an active veto detector whose purpose is to detect penetrating cosmic-ray muons.

The cosmic ray veto consists of four layers of long extruded scintillator strips, with aluminum absorbers between each layer (see Ref. [1] for a detailed description of the parameters of the cosmic ray veto). The nomenclature associated with the Cosmic Ray Veto is described in Table 10.1. The scintillator surrounds the top and sides of the Detector Solenoid (DS) and the downstream end of the Transport Solenoid (TSd), as shown in Figure 10.1 and Figure 10.2. The strips are 2.0 cm thick, providing ample light to allow a threshold to be set sufficiently high to suppress most of the backgrounds. Aluminum absorbers between the layers are designed to suppress punch through from electrons. The scintillation light is captured by embedded wavelength-shifting fibers, whose light is detected by silicon photomultipliers (SiPMs) at each end (except those counters closest to the TSd). A track stub consisting of at least three adjacent hit strips in different layers within a 5 ns time window signals the presence of a cosmic-ray muon. A conversion-like electron candidate within 125 ns of such a track stub is assumed to be produced by a cosmic-ray muon and is vetoed in the offline analysis.

The cosmic ray veto must operate with an excellent efficiency, about 0.9999, in an intense radiation environment that consists of neutrons produced at the production target, stopping target, and muon beam stop, as well as gammas produced largely from neutron capture.

Unlike the other backgrounds to the Mu2e conversion signal, the cosmic-ray background scales as the detector live time. A direct measurement of the cosmic-ray background can and will be done when the beam is not being delivered. During normal beam operations, 16 times more cosmic-ray-muon induced conversion-like background events will be collected out-of-spill than will be collected in-spill.

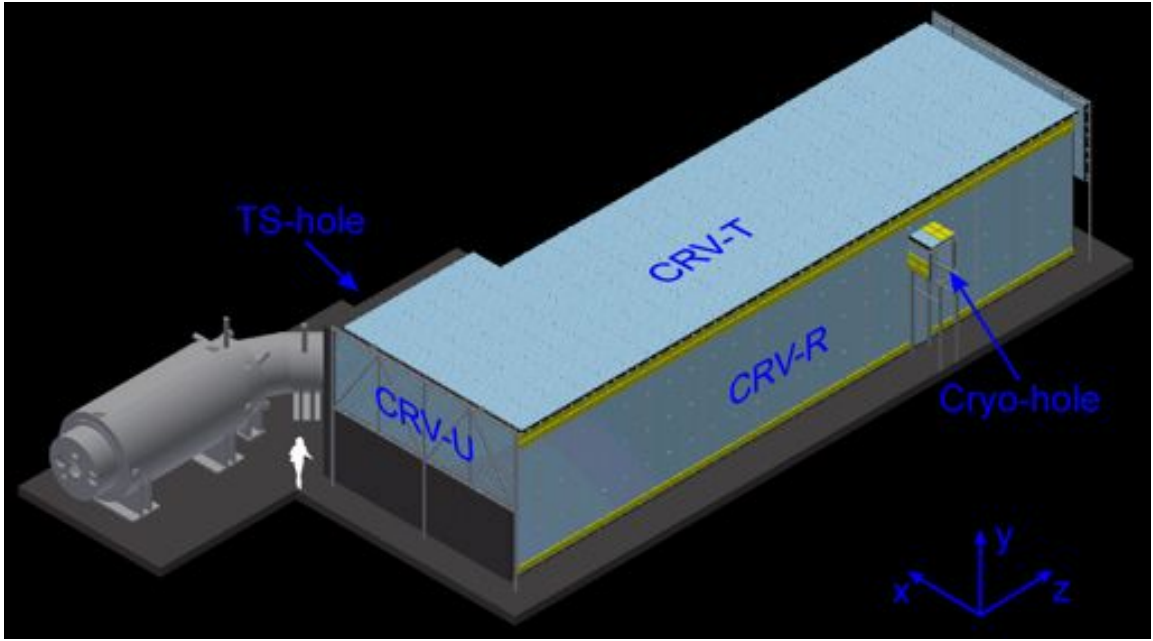

Figure 10.1. The cosmic ray veto covering the Detector Solenoid looking upstream, showing the downstream (CRV-D), left (CRV-L), top (CRV-T) sectors, as well as the hole where the transport solenoid enters the enclosure.

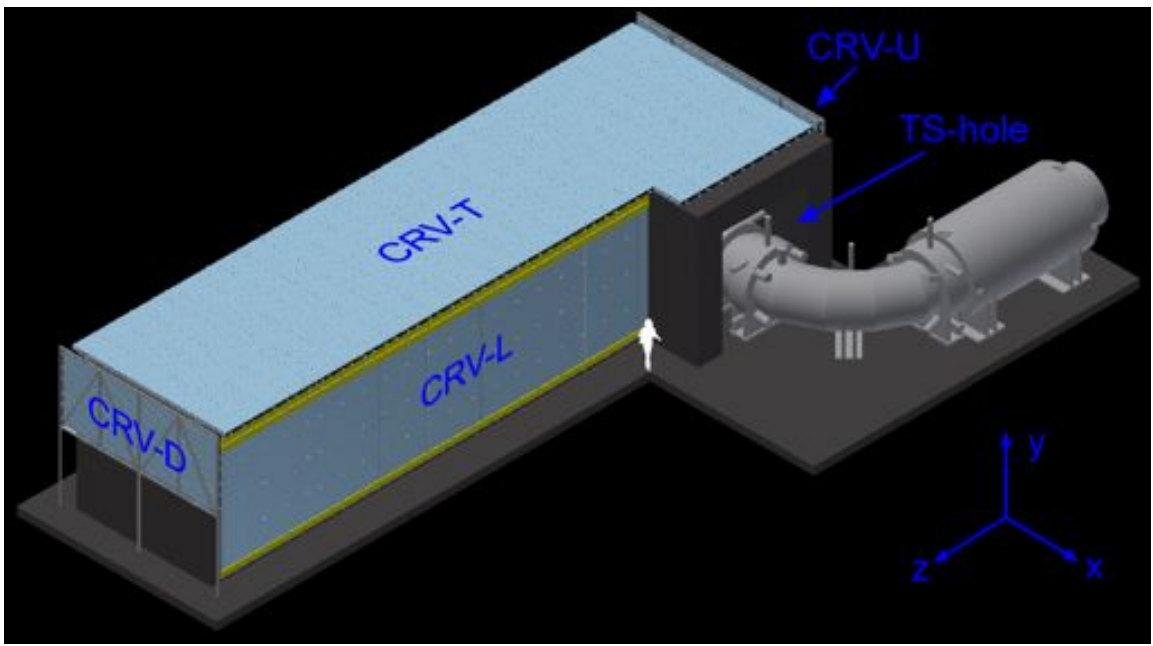

Figure 10.2. The cosmic ray veto covering the Detector Solenoid looking downstream, showing the upstream (CRV-U), right (CRV-R), and top (CRV-T) sectors.

Table 10.1. Cosmic ray veto nomenclature.

| Cosmic Ray Veto Nomenclature | | |
|---|---|---|
| Extrusion | | The raw scintillator strips with channels for fibers used to fabricate counters. |
| Counter | | The fundamental unit of the CRV: a scintillator extrusion outfitted with two fibers and photodetectors. |
| Di-counter | | Two extrusions glued together and fitted with a manifold on each end. |
| Module | | The fundamental mechanical element of the CRV, consisting of either 32 or 64 counters. |
| Silicon Photomultiplier | SiPM | The photodetector, also called a Geiger-mode APD (G-APD) or multi-pixel photon counter (MPPC). |
| Fiber Guide Bar | FGB | A plastic block glued to the end of an extrusion in which the fibers are potted and their ends polished. |
| Counter Motherboard | CMB | The electronics board on which the SiPMs, flasher LEDs, and temperature sensor are mounted. |
| SiPM Mounting Block | SMB | The plastic block inserted over the fiber guide bar onto which the counter motherboard is mounted. |
| Counter Readout Manifold | CRM | The assembly mounted on the readout end of the di-counters. |
| Counter Mirror Manifold | CMM | The assembly with a mirror on the non-readout end of selected di-counters read out on only one end. |
| Front-end Board | FEB | The electronics that reads out the SiPMs, temperature sensor, drives the LEDs, and sets the SiPM bias. |
| Readout Controller | ROC | The electronics that receives data from the FEBs, sends it to the DAQ, and provides power to the FEBs. |

## 10.2  Requirements

The Mu2e collaboration has developed a complete set of requirements for the cosmic ray veto [2]. The outstanding technology-independent performance requirements of the cosmic ray veto are that it (1) limit the conversion-electron background to less than 0.1 events over the duration of the run; (2) produce cosmic-ray muon trigger primitives for detector calibration purposes; (3) produce less than 10% dead time; and (4) not use more than 20% of the DAQ bandwidth.

### 10.2.1   Overall Cosmic Ray Veto Efficiency

A detailed simulation of conversion-like electron background from cosmic-ray muons has been done and is continually updated with improvements in the apparatus model, as well as with increased statistics. It is described in Chapter 3 of this TDR and in detail in Ref.

[3]. The incident muon flux is modeled with the Daya Bay code [4], which uses a modified Gaisser spectrum [5]. The muon flux is about 60% positive. Note that the rate of upward-going muons is negligible at $\approx 2 \times 10^{-13}$ cm$^{-2}$s$^{-1}$sr$^{-1}$ [6], and can be ignored.

A comprehensive GEANT4 model includes the detector hall, a description of the solenoids and their magnetic fields, the collimators, the shielding surrounding the detector solenoid, the tracker, the calorimeter, proton and neutron absorbers, stopping targets, and muon beam stop. The simulations include all relevant muon decay and secondary particle production mechanisms. Charged particles that produce a minimum number of hits in the tracker are fed to a Kalman-filter based track-finding package [7]. Standard track quality cuts are applied [8]. A total of 28 billion cosmic-ray muon events have been generated in a large area over the apparatus. This corresponds to roughly 2% of the total live running period. Targeted simulations of the cosmic-ray flux over the entire running period have also been done in order to explore the hole in the cosmic ray veto enclosure needed for the muon beamline where there is no coverage, as well as the CRV-U and CRV-D regions.

Using the standard track-finding cuts, with no calorimeter requirement, the following results were obtained. In a momentum window between 100 – 110 MeV, a total of 201 events passed cuts; of these 140 were electrons, 7 positrons, 21 negative muons, and 33 positive muons. All of these events produce hits in the cosmic ray veto and hence would be vetoed. The dominant process is delta-electron production by muons, which accounts for about half of the background events. Primary muons that have not yet decayed, account for another quarter of the background events. The largest sources of background events are the proton and neutron absorbers.

Studies ([3] [9]) show that the tracker and calorimeter can reject negative and positive muons and positrons with a good efficiency. Hence, assuming that all 61 muons and positrons can be rejected by the tracker and calorimeter, a total of 140 events remain that must be vetoed by the CRV. Note that the targeted simulation of cosmic-ray muons entering the TS-hole shows that 0.8 negatively charged muons are not detected by any CRV sector during the course of the entire run and pass the standard track-finding cuts. Hence the tracker and calorimeter must be used to eliminate such events. To determine the required veto efficiency, we assume the following: (1) a signal window of 103.75 – 105.0 MeV; (2) a total veto live time for a three-year run of $1.48 \times 10^7$ s (which includes a 125 ns pre-spill veto period beyond the nominal 995 ns live spill gate (see Figure 10.3); (3) a 95.5% detector live time; (4) a desired upper limit on the background of 0.10 events; and (5) a 90% confidence level result of 150 background events passing all cuts. The 90% CL conversion-electron-like background is then

$$93.5 \times 10^3 \,\mu/s \cdot 1.48 \times 10^7 \,s \cdot 0.955 \cdot \frac{1.25\,MeV}{10\,MeV} \cdot \frac{150\,evts}{27.9 \times 10^9 \,\mu} = 888 \ evts,$$

where $93.5 \times 10^3$ muons/s is the cosmic ray muon rate at the surface over the area of Monte Carlo muon generation. To reduce this number of events to the desired background of 0.10 requires an overall veto inefficiency of

$$\bar{\varepsilon} = \frac{0.10}{888} = 1.1 \times 10^{-4}.$$

Note that if the 61 muons and positrons rejected by the tracker and the calorimeter are not subtracted from the above tally, then the overall veto inefficiency requirement is (at 90% CL) $8 \times 10^{-5}$.

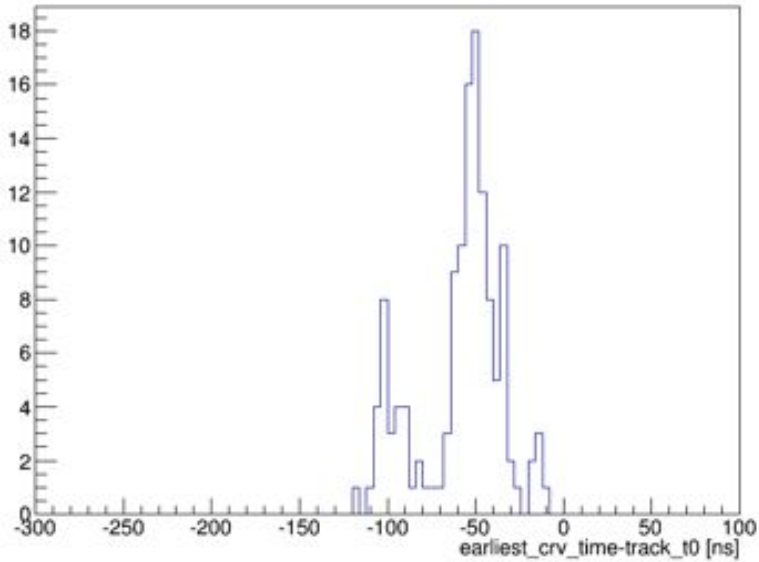

Figure 10.3. Difference between the time measured by the tracker and the impact time of the muon at the CRV for the cosmic-ray muon induced background events.

### 10.2.2   The required coverage of the cosmic ray veto

The required coverage is shown in Figure 10.4 through Figure 10.8. Two types of simulations have been performed to determine the coverage: the general simulation described above (Section 10.2.1), and simulations targeting smaller areas where there are either unavoidable holes in the coverage, or where the extents of the coverage need to be known. To date, the general simulation has only covered 2% of the anticipated cosmic-ray flux over the entire running period, whereas the targeted simulations each have covered the entire running period. In the general simulation, no cosmic-ray muons producing conversion-like background events have been found that would not be vetoed by the CRV. In the targeted simulations, 0.8 cosmic-ray muons have been found that would not be vetoed by the CRV. They are all vetoed by the calorimeter and tracker particle ID cuts. These simulations are ongoing with the goal of simulating the entire running period many times over.

The most important cosmic ray veto sector is CRV-T. Figure 10.4 shows the points of impact at the plane of the CRV-T for those cosmic-ray muons that produce conversion-like background events. Green markers indicate muons that only impact the CRV-T plane and none other. The two clusters of green markers are over the stopping target, proton and neutron absorber regions and the calorimeter, the largest sources of background electrons.

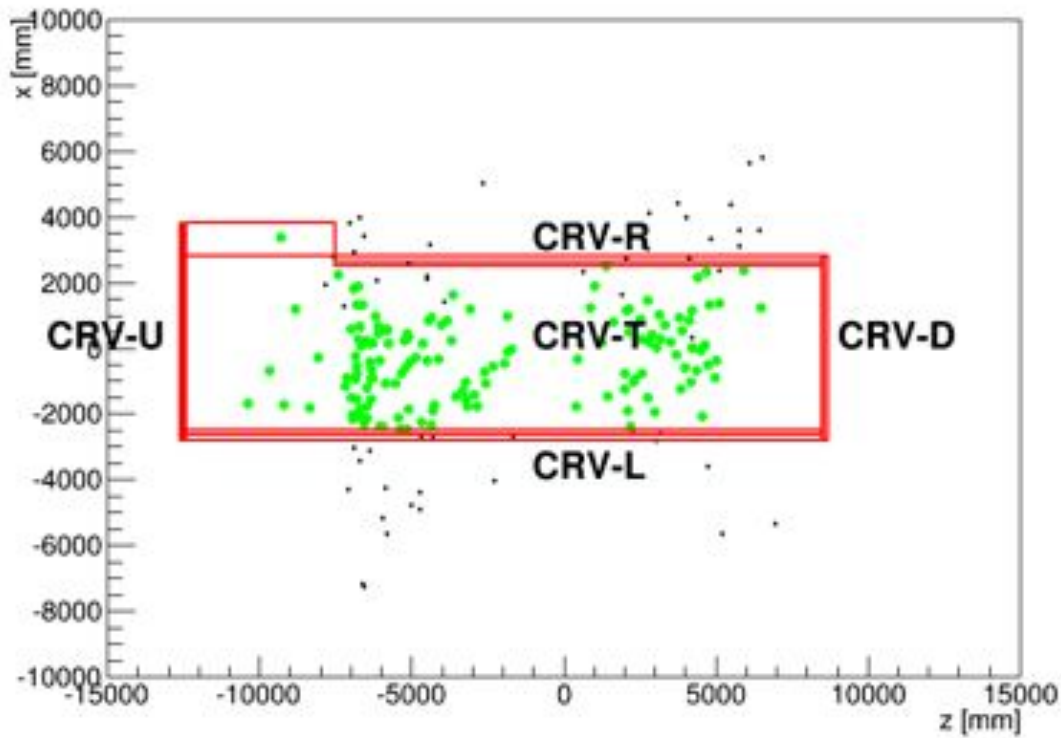

Figure 10.4. The *xz* points of impact at the plane of the CRV-T sector for those cosmic-ray muons that produce conversion-like background events. Green markers indicate muons that only impact the CRV-T plane and none other.

Figure 10.5 and Figure 10.6 show respectively the points of impact at the planes of the CRV-L and CRV-R for those cosmic-ray muons that produce conversion-like background events. Green markers indicate muons that only impact the CRV-L or CRV-R planes and none other. It appears that there are areas where either no coverage is needed or shorter counters could be used. However, simulations using relaxed track-finding cuts indicate otherwise [3], and we caution that only a small fraction of the full running period has been simulated in the general simulation. Note too that longer side counters give another chance of vetoing muons that are exiting the DS.

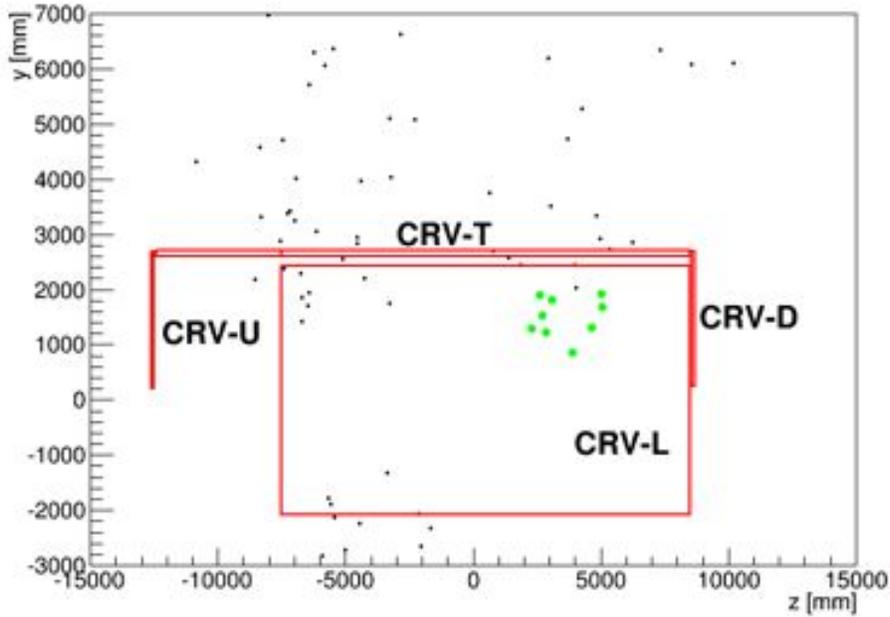

Figure 10.5. The *yz* points of impact at the plane of the CRV-L sector for those cosmic-ray muons that produce conversion-like background events. Green markers indicate muons that only impact the CRV-L plane and none other.

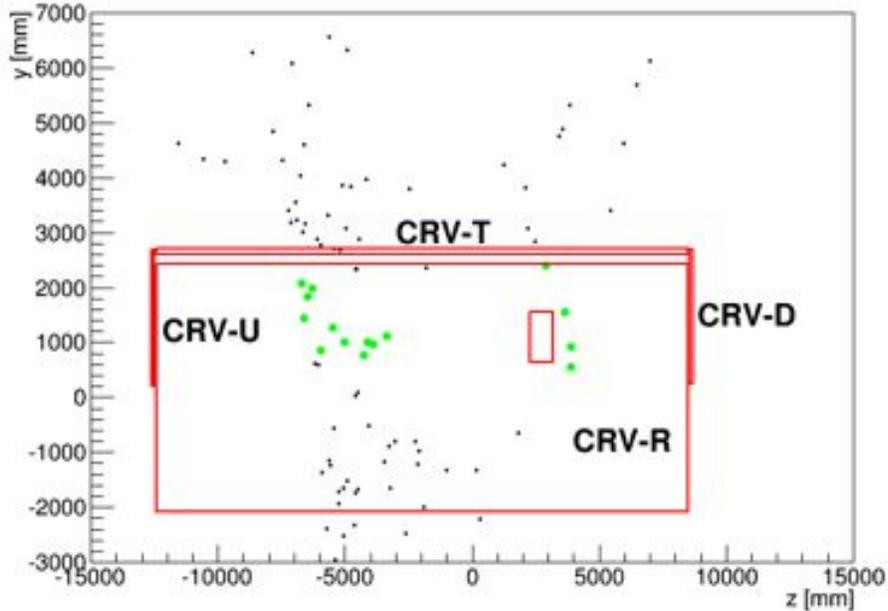

Figure 10.6. The yz points of impact at the plane of the CRV-R sector for those cosmic-ray muons that produce conversion-like background events. Green markers indicate muons that only impact the CRV-R plane and none other.

Figure 10.7 (Figure 10.8) show the points of impact at the plane of the CRV-U (CRV-D) sector for those cosmic-ray muons that produce conversion-like background events.

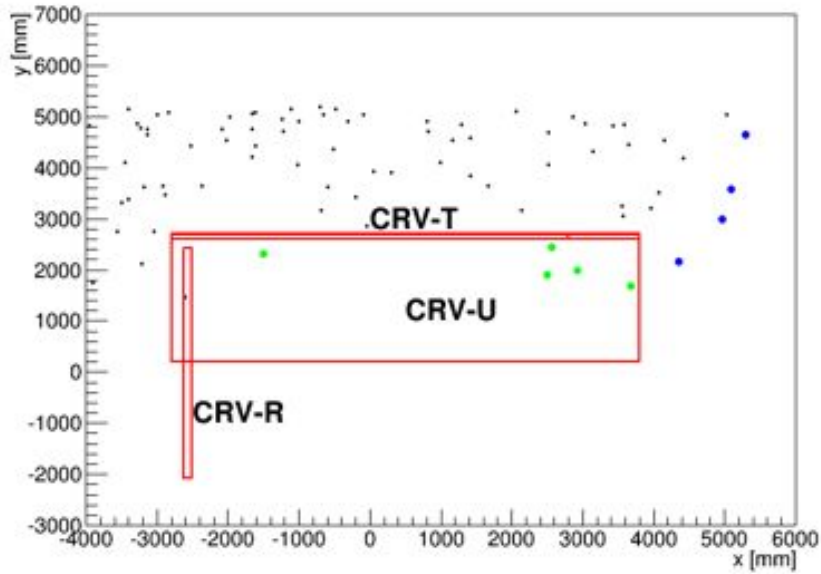

Figure 10.7. The *xy* points of impact at the plane of the CRV-U sector for those cosmic-ray muons that produce conversion-like background events. Green markers indicate muons that only impact the CRV-U plane and none other, while blue markers indicate muons that are not vetoed by any part of the CRV (muons that pass track-finding cuts but are vetoed by the calorimeter)

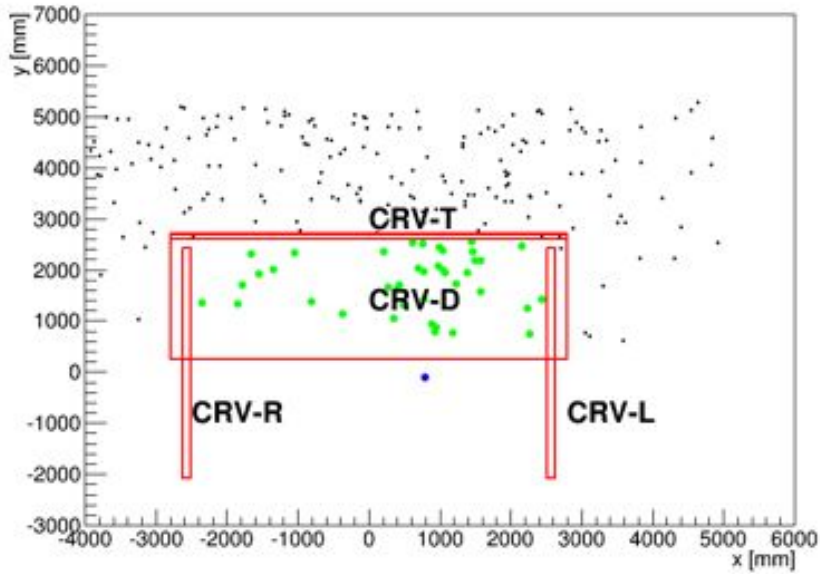

Figure 10.8. The *xy* points of impact at the plane of the CRV-D sector for those cosmic-ray muons that produce conversion-like background events. Green markers indicate muons that only impact the CRV-D plane and none other, while blue markers indicate muons that are not vetoed by any part of the CRV (muons that pass track-finding cuts but are vetoed by the calorimeter).

Green markers indicate muons that only impact the CRV-U (CRV-D) plane and none other, while blue markers indicate muons that are not vetoed by any part of the CRV. Note that the muons not vetoed by any sector of the CRV do not produce electrons through decay or other processes, but are themselves mis-reconstructed by the track-finding code as conversion electrons. However, they are all vetoed by the calorimeter/tracker particle ID cuts.

As described above, there is an unavoidable hole in the CRV due to the penetration needed for the muon beamline. To determine the effect of this hole on the conversion-like background a special targeted simulation was made of the flux expected for the duration of the run. Six events that pass track-finding cuts and do not impact any cosmic ray veto counters were found in a 10 MeV momentum window, corresponding to 0.8 events in the 1.25 MeV wide signal window. All are muons that can be eliminated by tracker and calorimeter particle identification requirements.

### 10.2.3   Meeting the Efficiency Requirement

A cosmic-ray muon is defined to be a coincidence of hits in three locally adjacent counters in the four layers of the cosmic ray veto. Naively, to produce an overall inefficiency of $1 \times 10^{-4}$ requires a single-plane inefficiency of 0.4% for a four-layer cosmic ray veto, as shown in Figure 10.9. Unavoidable gaps between counters would produce a larger than 0.4% single-plane inefficiency for normally incident cosmic-ray muons. However, the spread in the cosmic-ray muon angular distribution is such that this requirement can be met. To understand how the efficiency requirement can be attained, it is useful to translate the single-plane efficiency requirement into a counter photoelectron (PE) yield requirement as explained below.

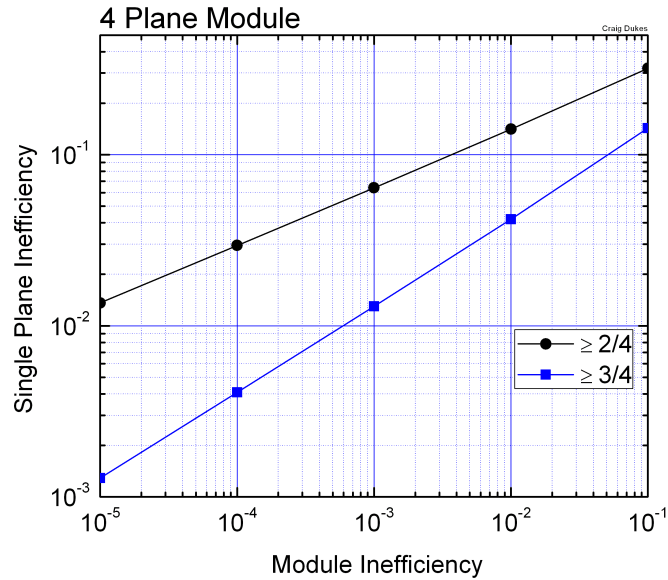

Figure 10.9. Single plane efficiency vs. required coincidence inefficiency for 2/3, and 3/4 required hits/planes. The CRV requirement is 3/4.

A Monte Carlo simulation was performed in which cosmic-ray muons were generated with the same angular distribution as those muons producing conversion-like backgrounds [10]. The observed photoelectron yield in each counter was generated following a Poisson distribution using a user-defined mean photoelectron yield per centimeter of scintillator. The counter response was assumed to be uniform over its transverse profile, which test-beam studies have borne out to a good approximation [11]. The simulation was used to determine the required photoelectron yield per cm of muon track length needed to achieve the required efficiency given above. Since that efficiency depends critically on the CRV geometry (see Figure 10.10), the simulation first optimized the offset distance between layers, and then determined the maximum gaps (internal and external) that can be tolerated between counters.

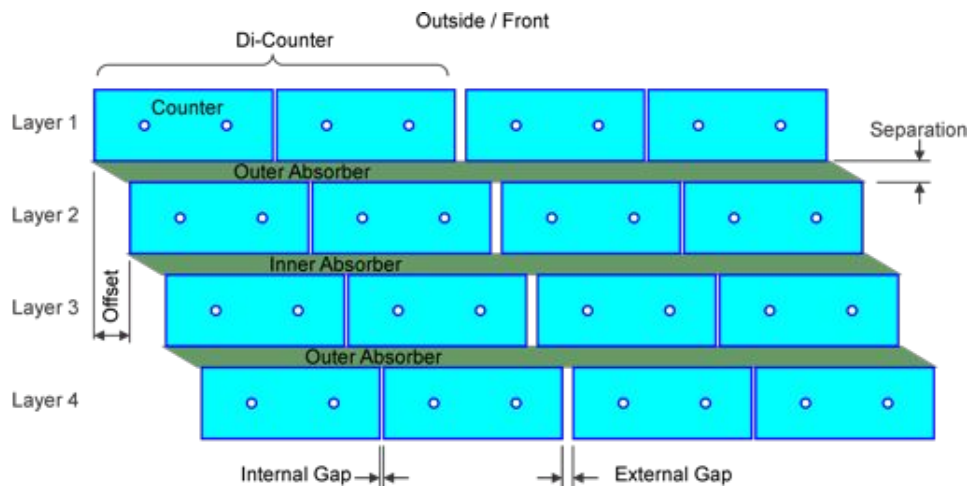

Figure 10.10. The CRV module geometry and nomenclature. Internal gaps are those between the counters in a di-counter.

Using an internal (external) gap between counters of 1 mm (3 mm), we find that an offset of 10 mm is required [10]. Using this result, the overall track-stub efficiency for different photoelectron yields (PE) can be found. The results of the simulation are shown in Figure 10.11 for light yields ranging from 8 to 18 PE/cm. The overall CRV efficiency requirement can be met for all PE/cm light yields shown in the figure, if the minimum PE threshold is low enough.

There are three constraints that set the required PE threshold: (1) the intrinsic noise rate of the SiPMs, (2) the maximum threshold allowed to achieve the desired CRV efficiency, and (3) the minimum equivalent energy needed to defeat the background rate from neutron and gamma-induced processes. Constraints (1) and (3) put a floor on the PE threshold, whereas (2) cannot be exceeded. Constraint (1) also applies individually to each SiPM, whereas (2) and (3) apply to the summed signal of the two SiPMs at the counter ends. Note that this summed signal is made in the offline analysis.

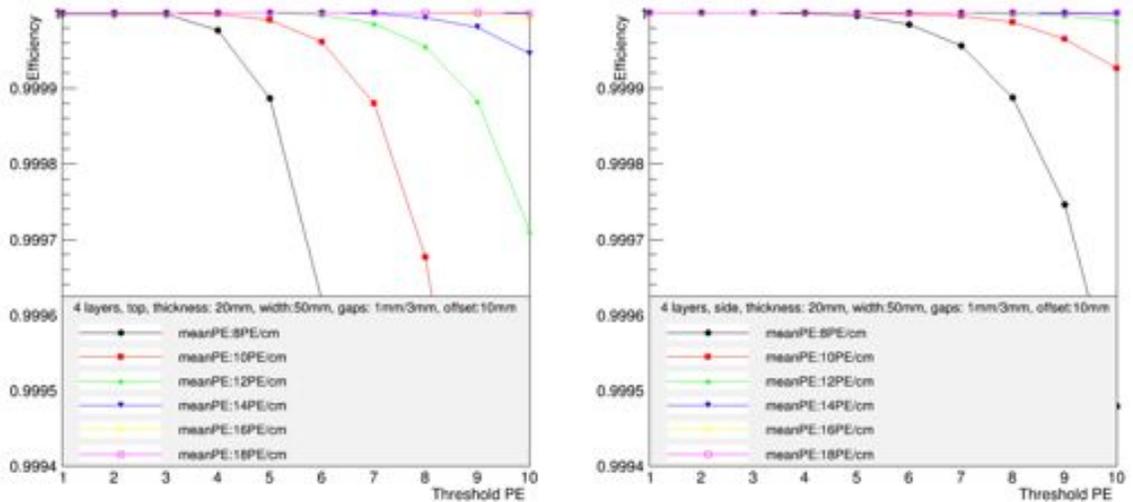

Figure 10.11. Efficiency vs. photoelectron threshold for photoelectron yields at the photodetector of 8 to 18 per cm. An internal (external) gap of 1 mm (3 mm) has been assumed. The left plot is for top counters and the right plot is for side counters. The optimal layer offset of 10 mm is used.

The intrinsic single photoelectron noise rate for most SiPMs is about 1 MHz, and declines by roughly an order of magnitude for each additional photoelectron (PE). Hence, to reduce the intrinsic noise rate to an acceptable level we require a threshold of PE ≥ 3. Again, this threshold applies separately for each of the two SiPMs on the end of the counters.

As described in Section 10.10.2, we require an effective energy threshold of between 0.5 MeV to 1 MeV in order to reduce neutron and gamma-induced rates in the counters to a level that does not produce an inordinate amount of deadtime. Since a muon deposits about 2 MeV/cm in a counter, a 1 MeV energy threshold corresponds to the energy deposited in half a cm of scintillator.

Figure 10.11 shows that a light yield of 14 PE/cm, with the corresponding 1-MeV-equivalent threshold of 7 PE gives the required track-stub efficiency with a safety factor of roughly 25% in decreased light yield needed to account for manufacturing variations, fiber attenuation differences, and other such effects. This requirement comes from the top counters, whose muons deposit less light due to their more normal incident angles. Higher thresholds can be used for the side counters. Note that these thresholds will be applied in offline analysis: lower thresholds will be applied in real time at the front-end electronics.

We require that the light yield at the far end of the longest counters without mirror reflectors, 5.6-m long, meet this PE requirement. The light yield from the far end of a counter is reduced by two factors: (1) the fiber attenuation length, and (2) fiber ageing. The attenuation length of prototype fiber obtained from Kuraray (double-clad, Y11, 175 ppm, non-S type, 1 mm diameter) has been measured using a jig set up and operated by Michigan State University for the NOvA experiment [12]. The result is shown in Figure 10.12. The attenuation factor from the far end is 0.397. MINOS has found that their light yield, using similar counters, has decreased by 3% per year due to scintillator ageing effects [13]. Based on this, a 10-year time from extrusion fabrication until the end of the run will produce an additional decrease in light yield of 0.737. Putting the two attenuation factors together gives a total decrease in light from the near end, at the time of fabrication, to the far end, after a ten-year period, of 0.293 for a 5.5-m-long counter.

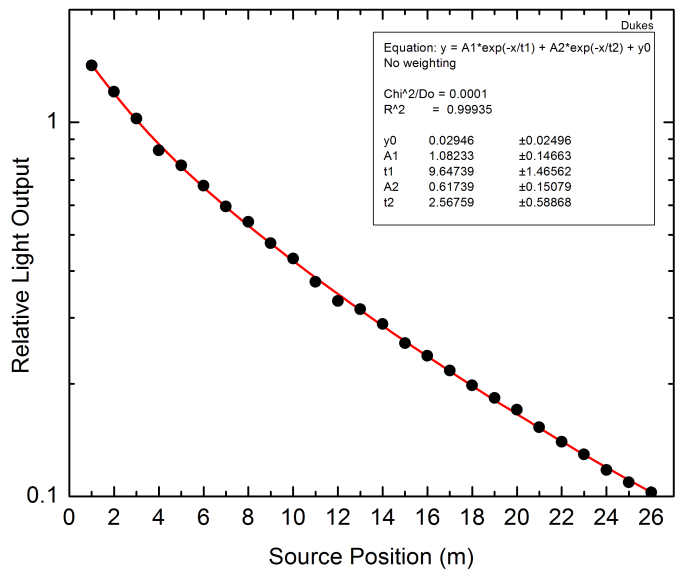

Figure 10.12. Fit to the light yield measurements of prototype wavelength-shifting fiber.

In order to determine the light yield a test beam run was made in the fall of 2013 at the Fermilab Meson Test Beam Facility (MTBF) with two prototype di-counters equipped with several different types of SiPMs and a three-layer, 1-m long module. As 5 cm wide extrusions were not yet available, $4 \times 2$ cm$^2$ extrusions from an existing die were used, each with two fiber channels embedded along the length of the extrusion. The 1-mm-diameter SiPMs were read out using the Preamplifier and Digitizer with Ethernet (PADE) system, a 32-channel board designed by P. Rubinov and T. Fitzpatrick that provides SiPM bias control, signal pre-amplification, analog to digital conversion, and high-speed serial links.

Figure 10.13 shows the response of a counter to normally incident protons at the center of the counter between the two fibers, which were not glued into the channels [11]. The dashed curves give the response of the individual SiPMs; the solid curve gives the sum of the two. Events clustered about zero photoelectrons correspond to instrumental problems. The mean light yield was found to be 31 PE/cm (Strictly speaking, SiPMs measure hit pixels, not photoelectrons. For the purposes of this discussion, photoelectrons will be used. For low PE yields, the two are equivalent). A $1 \times 1$ mm$^2$ SiPM with 100 µm-pitch pixels (Hamamatsu Part no. S10362-11-100) was used in this test. We plan to use SiPMs with pixels with no larger than 50 µm. Recent Hamamatsu devices of 25 µm and 50 µm pitch have effective quantum efficiencies (PDEs) that are comparable to the 100 µm device used in this test.

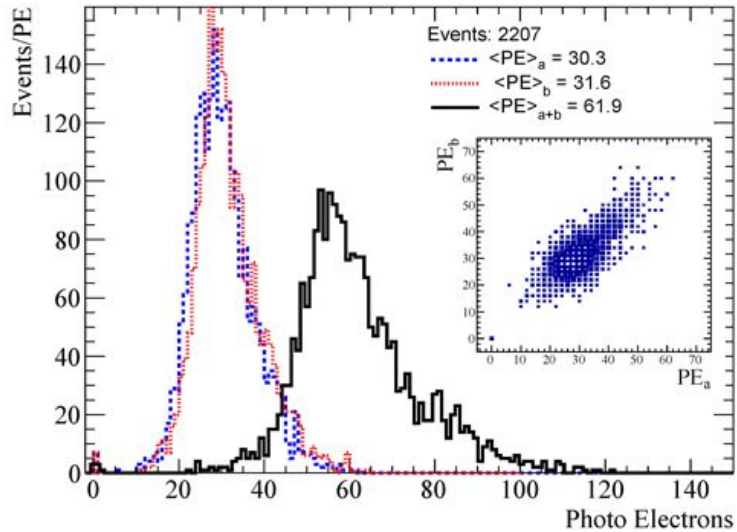

Figure 10.13. Response in photoelectrons from two Hamamatsu $1 \times 1$ mm$^2$ (S10362-11-100P) SiPMs with a $100 \times 100$ µm$^2$ pixel size to protons incident normal to a $2 \times 4$ cm$^2$ counter outfitted with unglued 1.0-mm diameter fibers.

The measured light yield corresponds to 9.4 PE/cm at the far end of a 5.6-m-long counter. This is below the 14 PE/cm that is our minimum requirement. Hence, an increase in light yield of 1.5 is needed to produce 14 PE/cm. With this we are able to meet the efficiency requirement.

In order to boost the light output to meet requirements, the fiber diameter and the corresponding area of the SiPM must be increased. The MINOS collaboration made studies of the light yield vs fiber diameter; their results are plotted in Figure 10.14 [14]. To get an increase in light yield of 1.5 a fiber diameter of 1.4 mm is needed.

Several important points should be emphasized. First, the two knobs used to tune the counter light yield in order to achieve the efficiency requirement of 0.9999 are the fiber diameter and the corresponding photodetector size, both of which can be obtained from manufacturers in a range of custom sizes at modest cost. Second, the 14 PE/cm light yield requirement at the far end of a 5.6-m-long counter corresponds to a 35 PE/cm light yield at the near end.

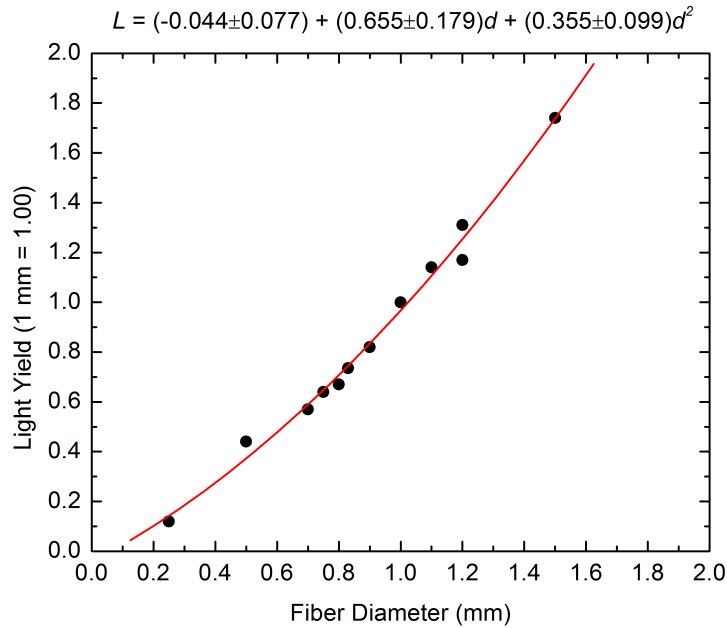

Figure 10.14. Measurements made by the MINOS collaboration of light output versus fiber diameter.

## 10.3    Technical Design

### 10.3.1    Overview

The baseline design of the cosmic ray veto is described below. The layout of the cosmic ray veto is shown in Figure 10.15 through Figure 10.18 and the system parameters are listed in Table 10.2. A complete list of parameters is given in Ref. [1].

The cosmic ray veto is made of four layers of inexpensive extruded scintillation counters with embedded wavelength-shifting fibers, each read out by a single silicon photomultiplier (SiPMs), most on both ends. The design is simple, minimizes gaps, allows access to the electronics, is robust, and should operate with a high efficiency. The design can be tuned to give the required veto efficiency by adjusting the photoelectron yield by increasing or decreasing the wavelength-shifting fiber diameter, in conjunction with the associated SiPM size. There is considerable experience with this technology in recent high-energy physics experiments, in particular, at Fermilab. Similar detectors have been fabricated for the MINOS [15] and MINERvA [16] experiments and quite a bit of tooling and infrastructure exists at Fermilab for the fabrication of the detector components, in particular the scintillator extrusions. The quantities of components needed for the cosmic ray veto are well within the demonstrated capabilities of Fermilab: 5152 scintillator counters of 1248 m$^2$ area and 50 km of fiber are needed compared to 100,000 counters over 28,000 m$^2$ and 730 km of fiber for MINOS.

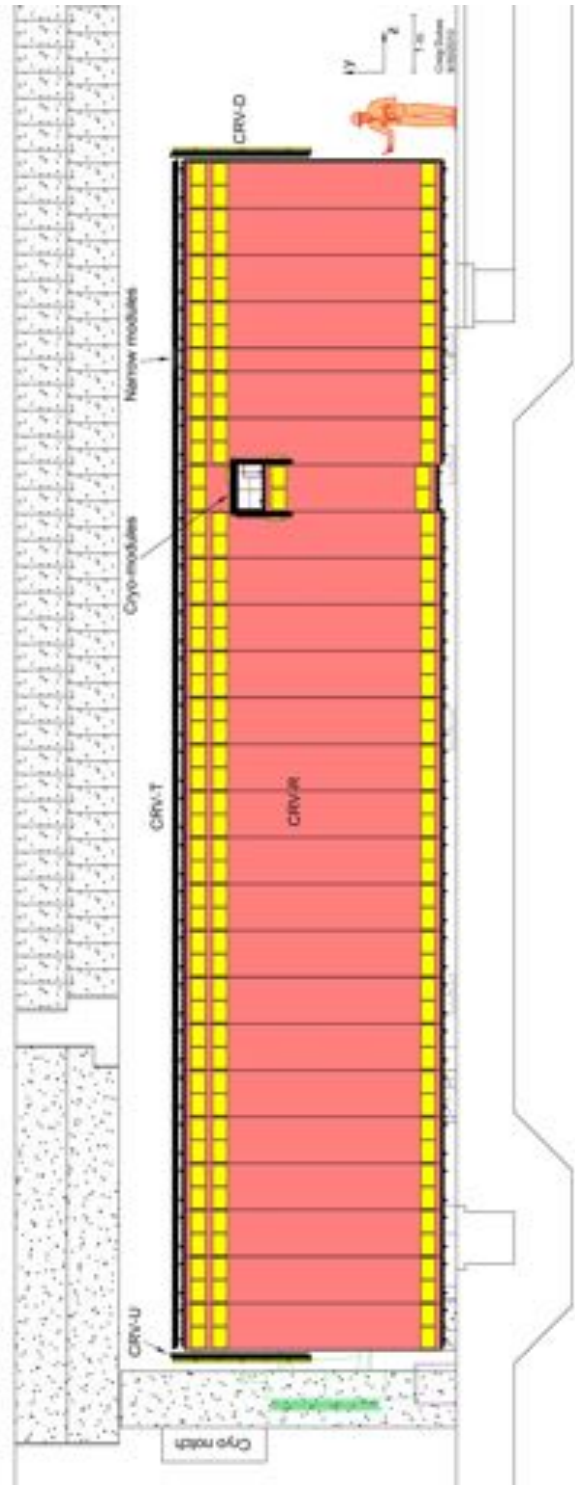

Figure 10.15. Elevation view of the right side of the CRV. Module supports are not shown. Front-end board enclosures are shown in yellow. The extra set of front-end boards at the top of the side counters are used to read out the top modules.

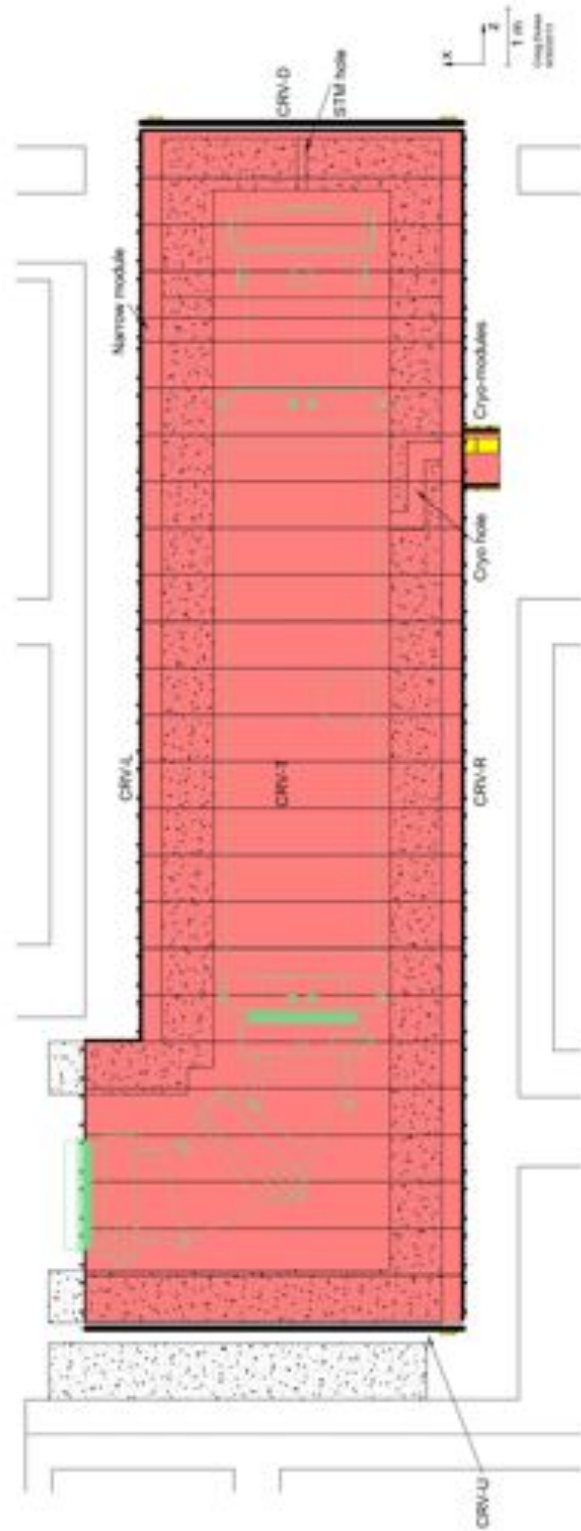

Figure 10.16. Plan view of the CRV showing the outlines of the shielding and TS and DS. The module supports are not shown. Front-end board enclosures are shown in yellow.

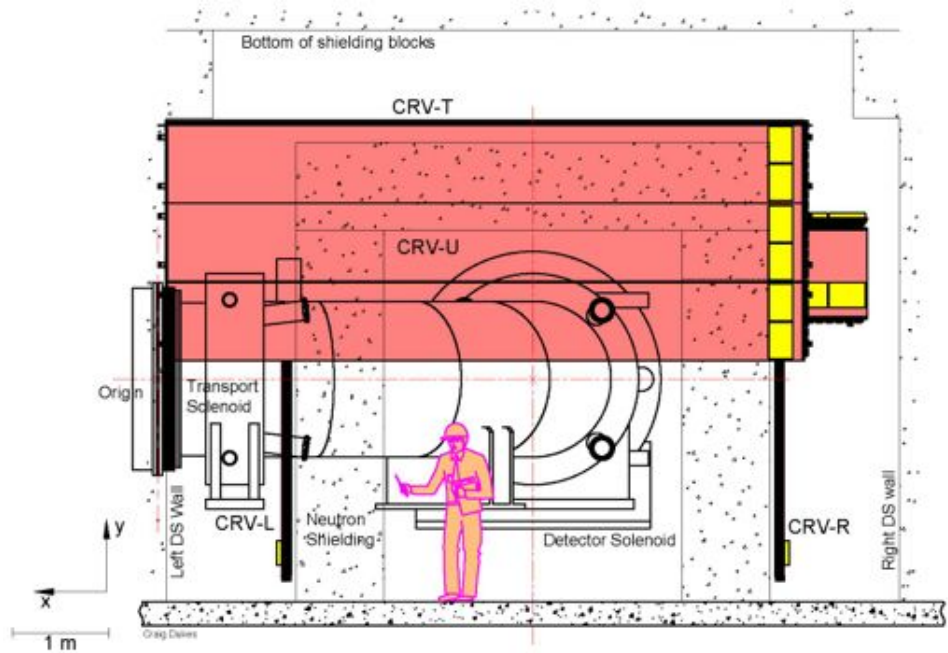

Figure 10.17. Elevation view of the upstream end of the CRV showing the outlines of the top and side shielding and TS and DS. Module supports are not shown. Front-end board enclosures are shown in yellow.

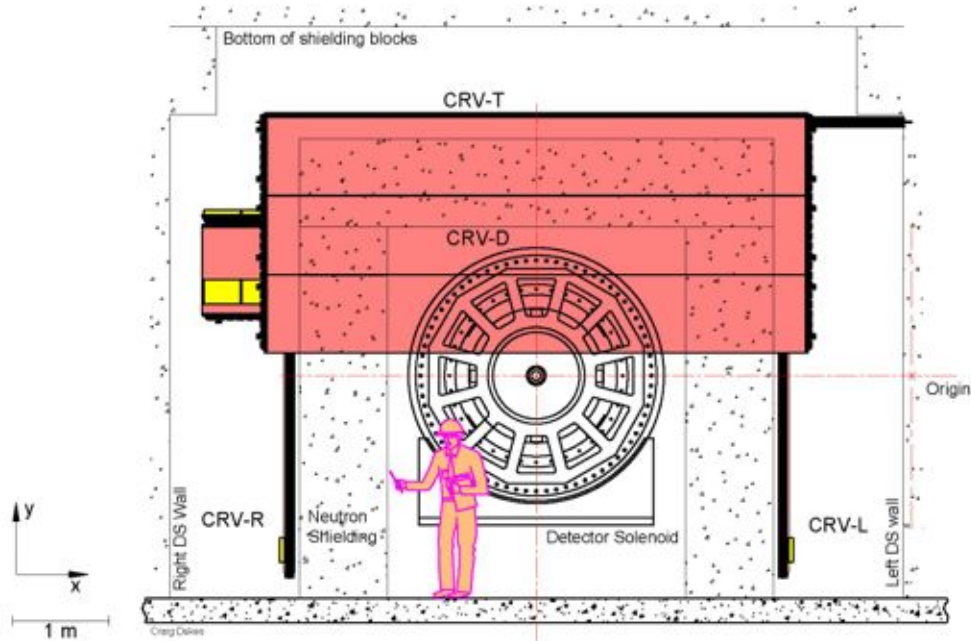

Figure 10.18. Elevation view of the downstream end of the CRV showing the outlines of the top and side shielding and DS. Module supports are not shown. Front-end board enclosures are shown in yellow.

Table 10.2.  Cosmic ray veto parameters by sector.

| | Sector | | | | | |
|---|---|---|---|---|---|---|
| | CRV-U | CRV-T | CRV-L | CRV-R | CRV-D | Total |
| Extra-Long | 3 | 6 | 0 | 0 | 0 | 9 |
| Long | 0 | 19 | 0 | 0 | 3 | 22 |
| Long-Narrow | 0 | 1 | 0 | 0 | 0 | 1 |
| Medium | 0 | 0 | 19 | 24 | 0 | 43 |
| Medium-Narrow | 0 | 0 | 1 | 1 | 0 | 2 |
| Short | 0 | 0 | 0 | 1 | 0 | 1 |
| Cryo | 0 | 0 | 0 | 4 | 0 | 4 |
| Total | 3 | 26 | 20 | 30 | 3 | 82 |
| Sector active area ($m^2$) | 16 | 123 | 73 | 97 | 14 | 323 |
| Scintillator active area ($m^2$) | 63 | 476 | 281 | 374 | 54 | 1,248 |
| Counters | 192 | 1,632 | 1,248 | 1,888 | 192 | 5,152 |
| Di-counters | 96 | 816 | 624 | 944 | 96 | 2,576 |
| Extrusions per layer | 48 | 408 | 312 | 472 | 48 | 1,288 |
| Extrusion length (m) | 1,267 | 9,523 | 5,616 | 7,478 | 1,075 | 24,960 |
| Scintillator mass (kg) | 1,343 | 10,095 | 5,953 | 7,927 | 1,140 | 26,458 |
| Total sector mass (kg) | 2,818 | 21,176 | 12,488 | 16,202 | 2,391 | 55,075 |
| Fibers | 384 | 3,264 | 2,496 | 3,776 | 384 | 10,304 |
| Fiber length (m) | 2,544 | 19,129 | 11,295 | 15,053 | 2,160 | 50,182 |
| SiPMs | 384 | 5,760 | 4,992 | 7,040 | 768 | 18,944 |
| Fiber guide bars | 192 | 1,632 | 1,248 | 1,888 | 192 | 5,152 |
| Counter reflector manifolds | 96 | 192 | 0 | 128 | 0 | 416 |
| Counter mother boards | 96 | 1,440 | 1,248 | 1,760 | 192 | 4,736 |
| SiPM mounting blocks | 96 | 1,440 | 1,248 | 1,760 | 192 | 4,736 |
| Front-end boards (FEB) | 6 | 90 | 78 | 110 | 12 | 296 |
| Readout channels | 384 | 5,760 | 4,992 | 7,040 | 768 | 18,944 |
| FEBs per readout controller | | | 24 | | | |
| Readout controllers | 1 | 4 | 4 | 5 | 1 | 15 |

The cosmic ray veto is mounted just outside of the concrete neutron and gamma shield surrounding the detector and transport solenoids and extends up almost to the midpoint of the transport solenoid.

The scintillation counters are grouped into modules, with each module having four layers of counters. There are 82 modules, of seven different sizes: their parameters are given in Table 10.3.  Full-size modules contain 64 total scintillation counters, 16 per layer, and are 0.859 m wide, whereas narrow modules have half that number of counters, and are 0.443 m wide. Besides these two different widths, the only difference between modules are their lengths, which range from 0.9 m to 6.6 m. Module masses range from 128 kg to 939 kg. The modules are deployed into six geographical sectors: the upstream (CRV-U) and downstream (CRV-D) regions, the right (CRV-R) and left (CRV-L) sides, and the top sector (CRV-T). Table 10.2 lists the sector parameters.

The nine extra-long modules are 6.6 m in length and are used to cover the CRV regions that cover the downstream portion of the transport solenoid. The high radiation levels in this area and lack of accessibility dictate single-ended readout. The counters ends most upstream are outfitted with fiber mirrors. These, and the cyro-modules described below, are the only ones with counters that are not read out on both ends. Note that by length we always mean the active scintillator length: the overall module is slightly longer.

Table 10.3. Cosmic ray veto module parameters.

| | Extra-Long | Long | Long-Narrow | Medium | Medium-Narrow | Short | Cryo |
|---|---|---|---|---|---|---|---|
| Layers | | | | 4 | | | |
| Counter length (m) | 6.600 | 5.600 | 5.600 | 4.500 | 4.500 | 3.000 | 0.900 |
| Counter width (m) | | | | 0.050 | | | |
| Counter thickness (m) | | | | 0.020 | | | |
| Overall modules width (m) | 0.859 | 0.859 | 0.443 | 0.859 | 0.443 | 0.859 | 0.859 |
| Module external thickness (m) | | | | 0.120 | | | |
| Surface area layer (m$^2$) | 5.471 | 4.642 | 2.313 | 3.731 | 1.859 | 2.487 | 0.746 |
| Inner gap (mm) | | | | 1.0 | | | |
| Outer gap (mm) | | | | 3.0 | | | |
| Layer offset (mm) | | | | 10.0 | | | |
| Fibers/counter | | | | 2 | | | |
| Fiber diameter (mm) | | | | 1.40 | | | |
| Counters/layer | 16 | 16 | 8 | 16 | 8 | 16 | 16 |
| Counters total | 64 | 64 | 32 | 64 | 32 | 64 | 64 |
| Di-counters total | 32 | 32 | 16 | 32 | 16 | 32 | 32 |
| Outer absorber thickness (mm) | | | | 9.525 | | | |
| Inner absorber thickness (mm) | | | | 12.700 | | | |
| Cover thickness (mm) | | | | 0.762 | | | |
| Total module mass (kg) | 939 | 797 | 398 | 640 | 320 | 427 | 128 |
| Number of fibers | 128 | 128 | 64 | 128 | 64 | 128 | 128 |
| Light yield (far/near) | 0.273 | 0.311 | 0.311 | 0.362 | 0.362 | 0.454 | 0.655 |
| Light transit time (ns) | 38 | 32 | 32 | 26 | 26 | 17 | 5 |
| Fibers/SiPM | | | | 1 | | | |
| Fiber ends read out | 1 | 2 | 2 | 2 | 2 | 2 | 1 |
| SiPMs/module | 128 | 256 | 128 | 256 | 128 | 256 | 128 |
| SiPMs per counter mother board | 4 | 4 | 4 | 4 | 4 | 4 | 4 |
| Counter mother boards | 32 | 64 | 32 | 64 | 32 | 64 | 32 |
| SiPM mounting blocks | 32 | 64 | 32 | 64 | 32 | 64 | 32 |
| Counter reflectors | 32 | 0 | 0 | 0 | 0 | 0 | 32 |
| Channels per front-end board | | | | 64 | | | |
| Front-end boards | 2 | 4 | 2 | 4 | 2 | 4 | 2 |

The top CRV modules are the long type, 5.6 m in length, with the exception of the aforementioned extra-long modules in the upstream reagion of the CRV-T that covers the TSd. The sides of the CRV employ the medium-length modules, 4.5 m long. On the right side of the CRV are several special-length modules to cover the cryo feed for the detector solenoid, and the associated penetration in the shielding. Four of these modules, called cryo-modules, are 0.9-m long, and read out only on one end. An additional one-off short module of 3.0-m length is situated below the cryo penetration.

Although the greater light yield from the short cryo-modules would allow smaller diameter fibers and SiPMs to be used, in order to ease fabrication we employ the same size in every module.

Three narrow modules are placed just before the moveable endcap. The two end regions, CRV-U and CRV-D, each have three modules with extra-long counters oriented horizontally. The upstream (downstream) endcap employs extra-long (long) modules.

### *10.3.2*   **Counter Design**

The fundamental elements of the cosmic ray veto are the 5152 scintillation counters, which range in length from 6600 mm to 900 mm, are 50 mm wide, and 20 mm thick. The counter width is a compromise between the need for high efficiency, which favors a wide counter in order to reduce the effect of cracks, and the difficulties of the extrusion process, which favor low-aspect ratio narrow counters. Each counter contains 2 channels of 2.6 mm diameter into which the wavelength-shifting fibers are inserted (Figure 10.19). (Note that the dimensions given in the figure are nominal:  the actual dimensions will be somewhat different.) Fibers will not be glued into the extrusion channels. Details on the extrusions and their production are found in Section 10.4.

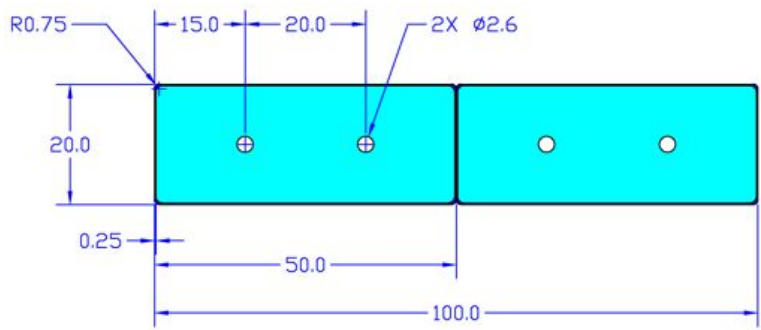

Figure 10.19. Di-counter showing the scintillator extrusion profile and nominal dimensions (mm).

The extrusions have 1.4-mm diameter wave-shifting fibers embedded in each channel. Two extrusions are glued side-by-side to make a di-counter. A fiber guide bar is glued at each di-counter end (Figure 10.20) with four fiber funnels which on the extrusion end are larger than the fiber channel, and then narrow down to a diameter slightly larger than that of the fibers themselves. In the fiber guide bar there are two light tunnels for flasher

LEDs, and two screw holes, which also serve to register the counter motherboard (CMB), and hence the SiPMs to the fiber guide bar.

Each fiber is read out by surface-mount SiPM, four of which are mounted on a single counter motherboard. The counter motherboard, described in more detail in Section 10.7, besides holding the SiPMs, also has two flasher LEDs, and a temperature sensor. A manifold cover, shown in Figure 10.21, with an O-ring seal provides modest pressure to keep the SiPMs flush to the fibers, as well as providing a light-tight seal. A micro-HDMI connector serves as the link to the front-end boards. The manifold is designed to be easily removable as it is expected that SiPM failures will require occasional counter motherboard replacement.

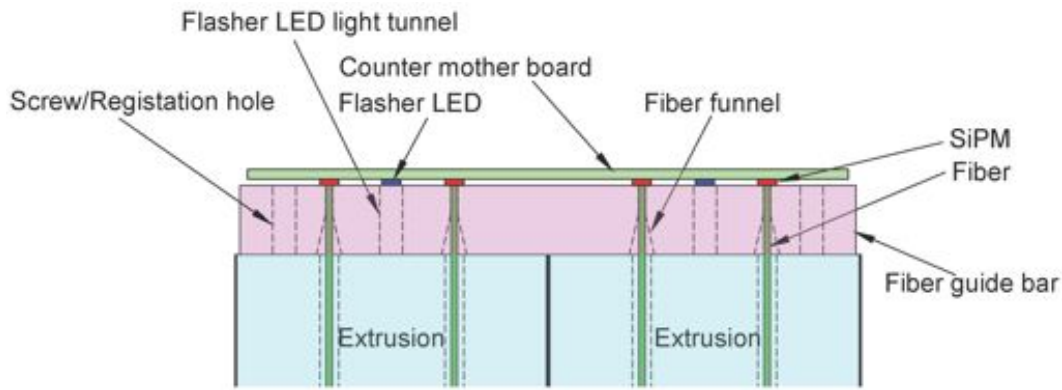

Figure 10.20. Counter readout manifold showing the fiber guide bar and counter motherboard. The manifold cover and micro-HDMI connector are not shown.

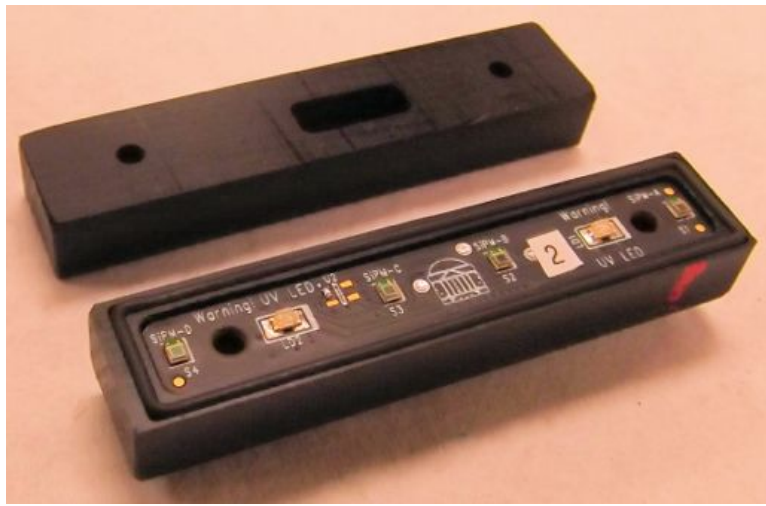

Figure 10.21. The counter manifold cover top (above) and bottom (below). The O-ring and counter motherboard with its SiPMs and flasher LEDs can be seen.

In all but the extra-long modules, the two di-counter ends are identical. For the extra-long modules one end of the di-counter has a counter mirror manifold affixed to it. This manifold has an aluminum-coated kapton film mounted flush to the fiber guide bar that serves as a light reflector. Preliminary studies show that reflectivities of 50% can be obtained with this simple design, which is sufficient to allow the photoelectron yield from the far end to meet requirements [17].

### 10.3.3   Module Design

Four layers of counters are grouped together to form a module (Figure 10.22), which come in two widths and five lengths (Table 10.3). Normal modules are 16 counters wide and contain a total of 64 counters. Narrow modules have half that number of counters, but are otherwise identical. The di-counters in each layer are glued to aluminum absorber plates, which serve two purposes: to provide mechanical support for the modules and to inhibit electrons from gamma interactions from traversing multiple counters, producing spurious coincidences. The inner absorber is 12.7 mm (1/2″) thick, whereas the two outer absorbers are 9.525 mm (3/8″) thick. The counter layers within a module are offset by 10 mm in order to minimize inefficiencies due to projective cracks. The absorber sides are beveled and protrude out by 1 mm in order to protect the counters on the sides from being damaged when the modules are installed. Thin hard rubber strips are glued to the exposed side counters to protect them from damage and to make them light tight. The front and back layers of counters are covered by a thin (0.8128 mm) Al. sheet onto which the enclosures for the front-end boards are attached.

The modules are designed to mate together with minimal gaps between di-counters of different modules. On the top and bottom of the front sides of each module adjustable draw latches bring adjacent modules together during installation. Black electrical tape is used to make them light tight.

The inner aluminum absorber that is used to mount the modules has two tabs that protrude from the top and bottom (see Figure 10.23). They are used to hang the modules mounted on the sides (CRV-L and CRV-R) and to mount the endcap modules (CRV-U and CRV-D) onto their strongbacks.

### 10.3.4   Module Support Structure

The module support structure is designed to minimize gaps between modules, allow the modules to be installed and removed without undue difficulty, and to allow access to the electronics without need to move or remove the modules. It is anticipated that modules will rarely, if ever, need replacing, whereas defective electronics will have to be replaced periodically. The shielding blocks onto which the cosmic ray veto is mounted do not present even surfaces, moreover, those in the TS and upstream DS region will need to be removed and reinstalled to allow access to the rotating collimator and antiproton windows at the TSu/TSd interface. Their position after reinstallation will not be within the precision required by the cosmic ray veto.

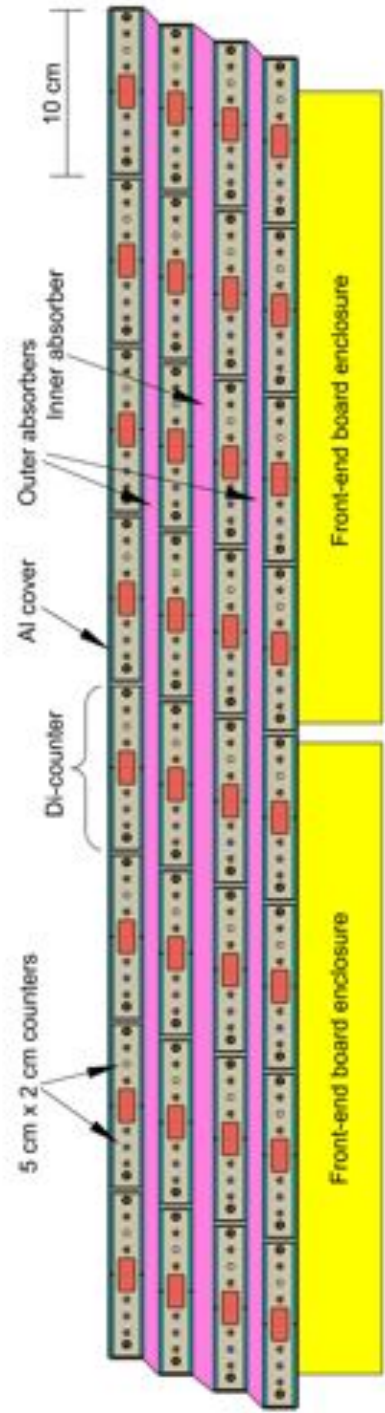

Figure 10.22. Top view of a module. Cables and support structure are not shown. Narrow modules are identical except they have half the number of counters.

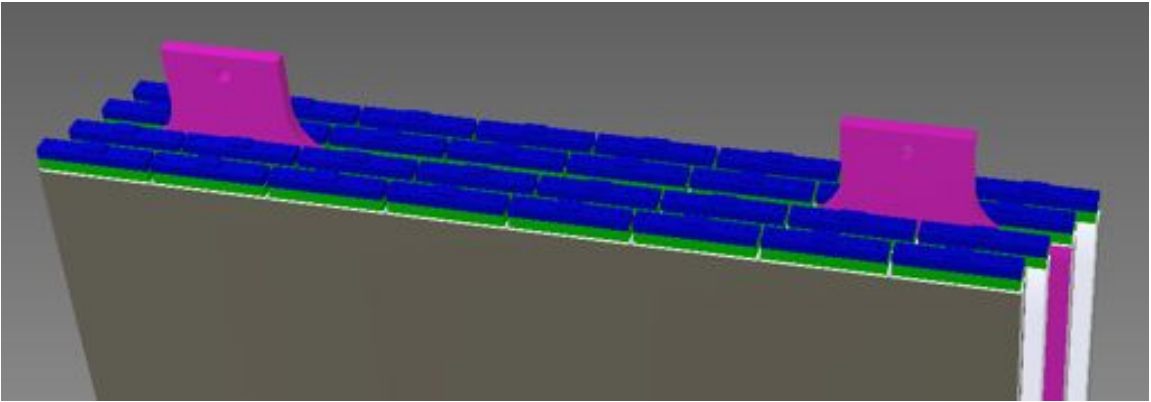

Figure 10.23. Isometric view of the top of a module showing the support tabs.

Details of the top and side support structures are shown in Figure 10.24 and Figure 10.25. C-channels will be secured with Hilti concrete bolts to the top of the shielding. On each C-channel are a series of ball rollers. Shims will be placed between the bottom of the C-channels and the concrete to level all of the ball rollers.

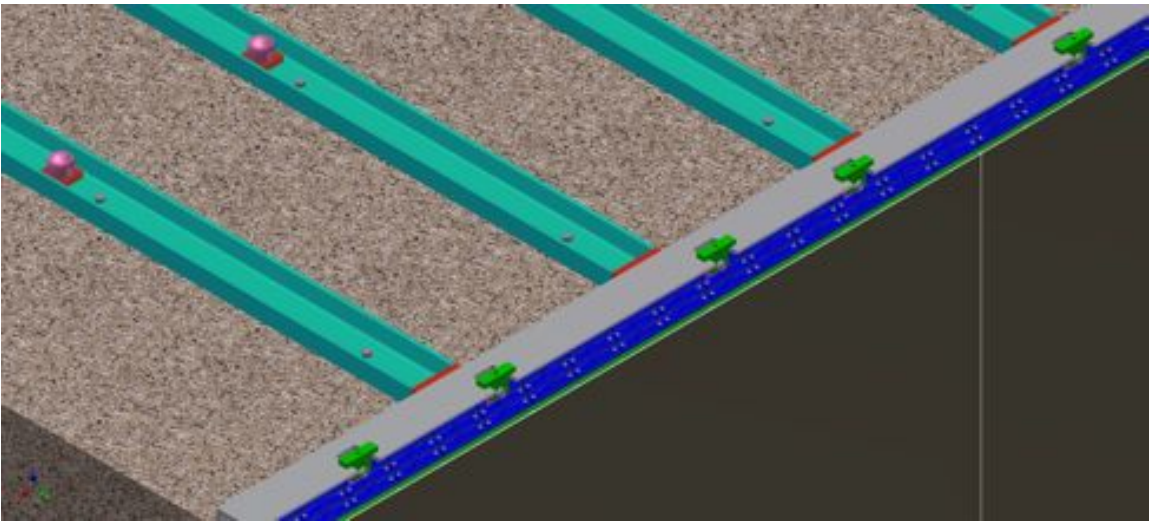

Figure 10.24. Detail of the module support structure showing the C-channels bolted to the concrete shields used to support the top modules, and the side module support structure.

Modules will be stacked and laid flat during shipping. A vacuum lifting fixture that supports the modules along their length will be used to remove the modules from the shipping crates and place them onto the rollers. Top modules can then be easily pushed along the top of shielding into their proper place.

The modules on the sides of the concrete shielding will be hung from hangers that bolt onto the tabs that extend from the ½" thick center absorber of the module (Figure 10.26).

A stainless steel angle will be bolted into plates that are welded into the ends of the C-channels that support the top modules. Slots will be cut into the angle where the lifting tabs can be placed

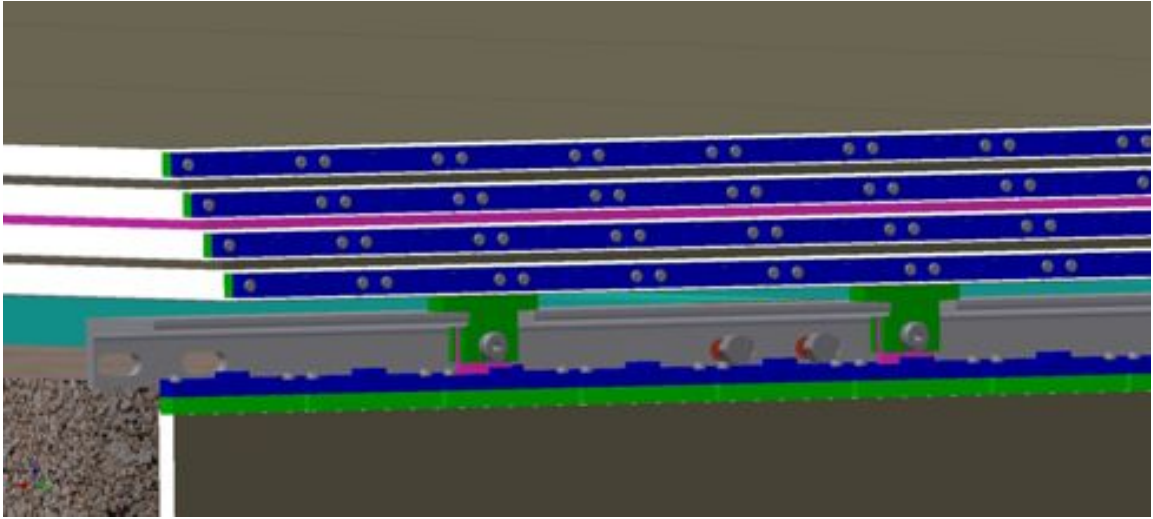

Figure 10.25. Detail of the module support structure showing top and side modules.

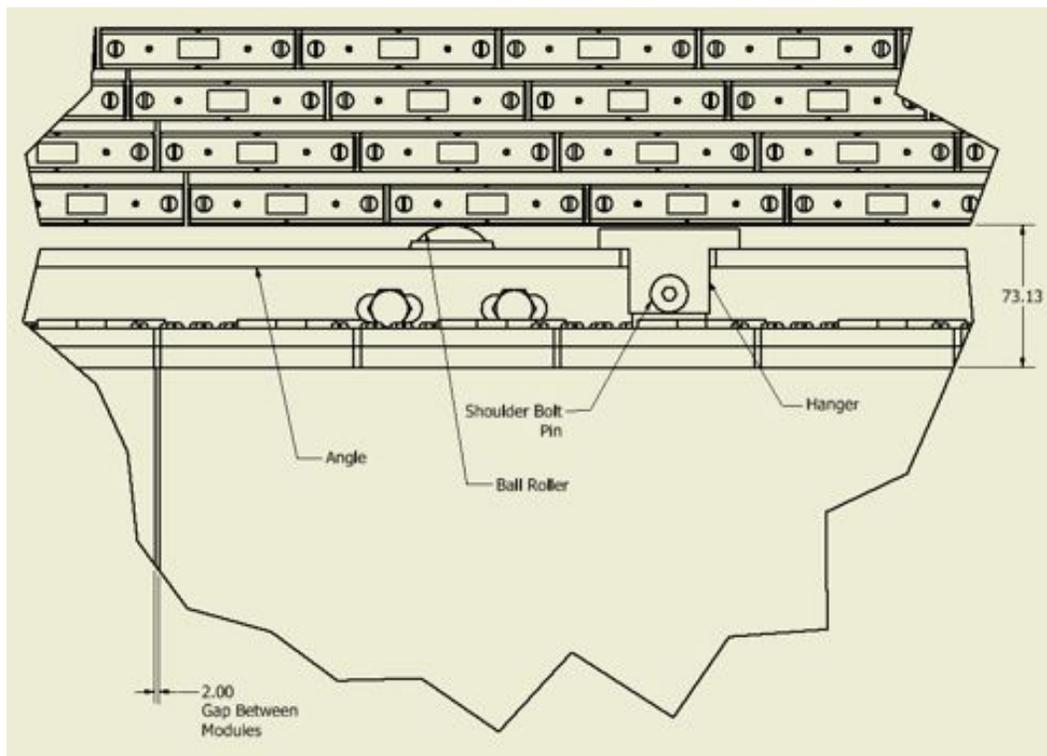

Figure 10.26. Detail of side and top module support structure.

The three extra-long modules at upstream and downstream ends, CRV-U and CRV-D, are mounted onto a stainless steel strongback, which is lowered into position, and secured by posts on either end (Figure 10.27).

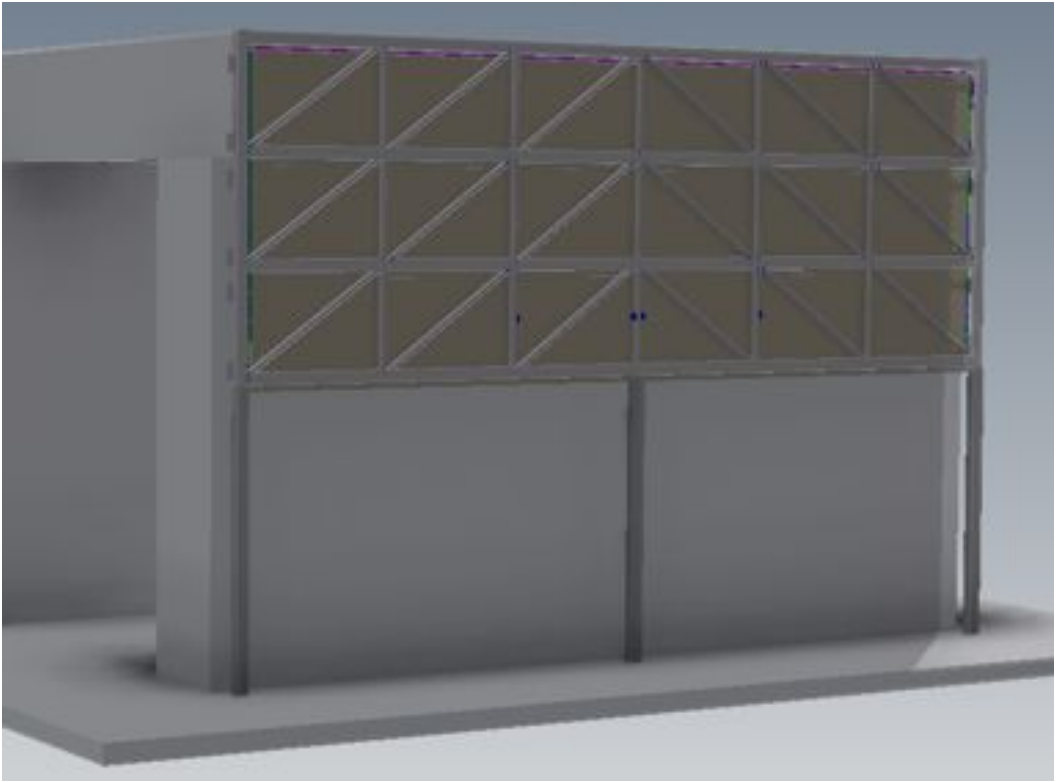

Figure 10.27. The strongback supporting the three extra-long modules in the CRV-U sector.

### *10.3.5* **Structural Studies**

The modules and their supports have been examined as a composite structure, with the details of this analysis given in Ref. [18]. The analysis calculated the shear forces due to bending, the stresses at the holes in the tabs used to hang the side modules, the shear stresses due to the hanging weight, and the stresses due to thermal expansion. The analysis also examined the maximum spacing of support points along the length of the module that would be allowed when the modules are lying down, as they would be on the top or when being lifted from a shipping crate using a vacuum lifter. The analysis showed that the shear stresses due to bending in each layer are very small. The normal stresses in the outside cover, however, increase rapidly and the concern would be buckling of the bottom cover that is in compression. If the spacing between supports is kept less than 5 ft. then the stresses are very small, less than 2 psi shear stress in the adhesive, and the deformation between supports is approximately 0.005". This analysis assumes 100% coverage of the adhesive. The stresses increase as the adhesive coverage is decreased.

The internal stresses that develop when the module is subjected to a 30C temperature change were also calculated. The scintillator has a greater expansion coefficient than aluminum and is therefore constrained by the aluminum. The normal stress in the aluminum is 136 psi and the stress in the spacers is 3800 psi for a 30 C temperature change.

A simple finite element analysis (FEA) model was made of the hanging assembly as shown in Figure 10.28. A section of the angle and top C-channels were modeled. The center ½" thick aluminum absorber from which the module weight is transferred to the tabs was also modeled with the weight of the entire module applied to the bottom of the spacer. The average von Mises stresses in the assembly were small. The stresses in the hangars that support the center aluminum spacer are consistent with hand calculations. The deflections in the vertical direction are shown in Figure 10.29 and are less than 0.015".

In the layout shown, the C-channels are distributed on 18" centers. Not shown in the figure are the horizontal mounting rails to which the C-channels attach, making the design independent of the size of the T-blocks.

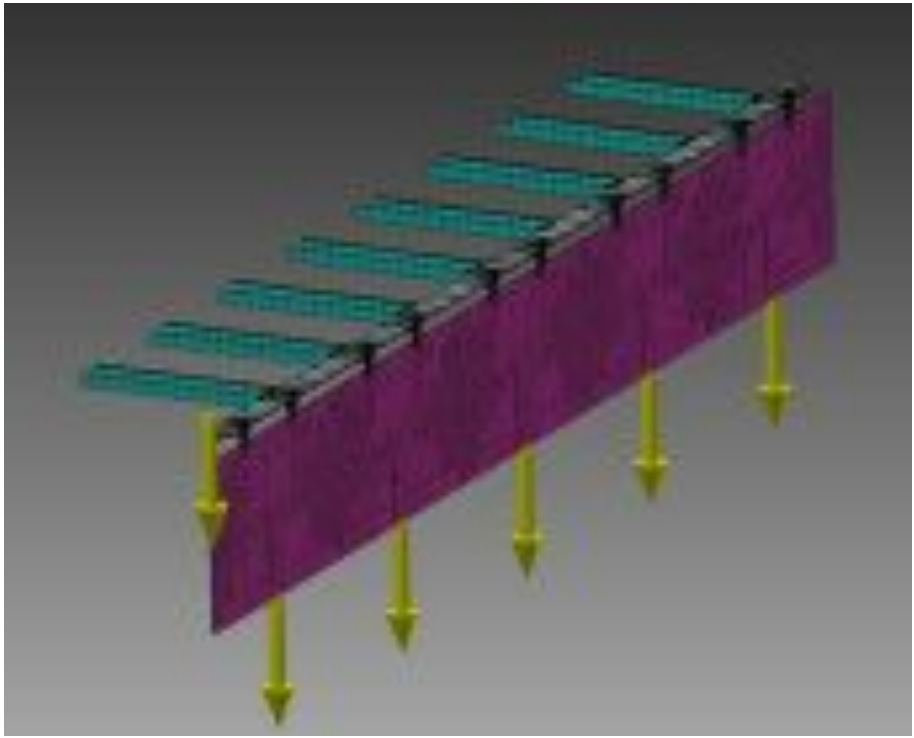

Figure 10.28. The finite element analysis model of the module hanging assembly.

Devcon HP-250 epoxy [19] will be used to glue the di-counters to the Al absorbers. Hence the counters on either side of the inner absorber must bear the entire load of the outer absorbers, as well as the outer layers of di-counters. Glue tests have shown that the epoxy strength is several orders of magnitude beyond what is needed [20]. Further tests will be done before fabrication proceeds, and during fabrication glue samples will be taken and tested at every stage of fabrication (see Section 10.8.7). Epoxies are known to exhibit structural weakness at $10^{18}$ n(>0.1 MeV)/cm$^2$, or $10^7$ Gy [21], many orders of magnitude higher than the highest doses expected in the hottest regions of the cosmic ray veto.

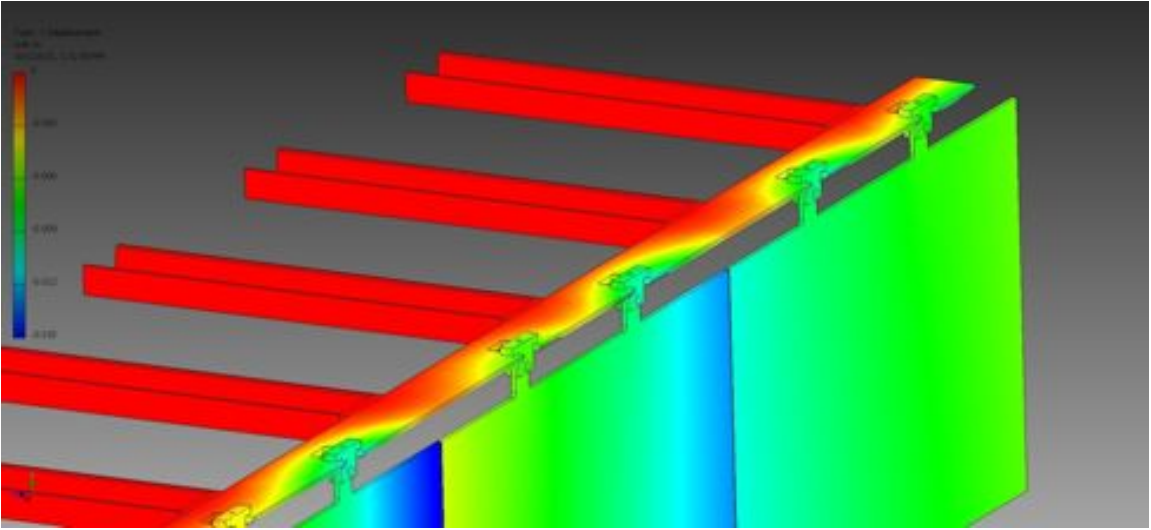

Figure 10.29. Vertical deflections of the side module support structures.

## 10.4   Scintillator Extrusions

Particle detection using extruded plastic scintillator and optical fibers is a mature technology. MINOS [15] and MINERvA [16] at Fermilab as well as experiments elsewhere have shown that co-extruded solid scintillator with embedded wavelength-shifting (WLS) fibers and photomultiplier readout produces adequate light for minimum-ionizing particle tracking and that it can be manufactured with excellent quality control and uniformity in an industrial setting. Mu2e intends to use this same extruded scintillator technology for the active elements of its cosmic ray veto.

The extruded scintillator elements will be produced at Fermilab using the FNAL-NICADD Extrusion Line Facility [22]. The extrusion line was purchased by NICADD (Northern Illinois Center for Accelerator and Detector Development) in 2003. The co-extruder line was purchased in 2005 [23]. The goal of the facility is to ensure that the high energy physics community has access to high-quality, low-cost extruded scintillator. The NICADD facility has produced extrusions for several experiments since it was commissioned. The quantity needed for Mu2e is well within the demonstrated capabilities of Fermilab: 5152 scintillation counters comprising 26,458 kg are needed for Mu2e, compared to 25,000 kg that were extruded for MINERvA and 28,000 kg for Double Chooz. Note that MINOS has 100,000 counters [15].

The scintillator extrusion has a rectangular profile of 5×2 cm$^2$, with two 2.6-mm-diameter holes for the wavelength-shifting fibers. The scintillator extrusions consist of a polystyrene core (Dow Styron 665 W) doped with PPO (1% by weight) and POPOP (0.03% by weight) and a white, co-extruded, 0.25-mm thick titanium dioxide (TiO$_2$) reflective coating, which is introduced in a single step as part of the co-extrusion process. Figure 10.30 shows the front of the assembled extrusion die and a detail of the die section where the co-extruded material is delivered over the core material. The composition of this capstocking is 15% TiO$_2$ (rutile) in polystyrene. In addition to its reflectivity properties, the outer layer facilitates the assembly of the scintillator strips into modules as the ruggedness of this coating enables direct gluing of the strips to each other and to the module absorbers.

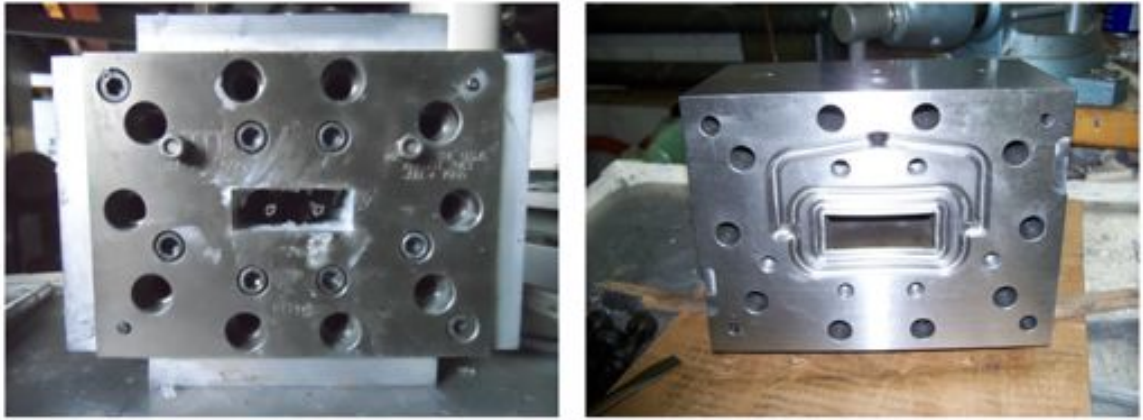

Figure 10.30. Assembled extrusion die for a $5 \times 2$ cm$^2$ profile scintillator counter. The right-hand photo shows the channels for the co-extruded reflective titanium dioxide coating.

The extrusion production is characterized by a continuous "in-line" extrusion process as opposed to a batch process. The polystyrene pellets are dried in a nitrogen atmosphere and automatically conveyed to a gravimetric feeder. The dopant mixture is added periodically to a different gravimetric feeder that is surrogate to the pellet feeder. These feeders have the necessary precision and reliability to ensure a constant delivery ratio. The pellet feeder is controlled by computer to match the output of the twin-screw extruder to ensure the correct composition and processing. The extruder is responsible for melting and mixing the polystyrene pellets and the dopants. A twin-screw extruder provides the highest degree of mixing to achieve a very homogeneous concentration. The outer reflective coating is added through material injected from a second extrusion machine (co-extruder) that mixes the polystyrene and TiO$_2$ pellets. The co-extruder is manually operated to start-up and to vary the thickness of the reflective coating. As the plastic emerges from the die, it goes directly into the cooling tank. There it is formed into the final shape using the sizing tooling and vacuum. It continues to be cooled with water and air until it can be handled.

Approximately 33,000 kg of extruded scintillator is needed for the CRV detector including spares and scrap at the extrusion and assembly factories. A total of 5152 counters will be extruded with lengths ranging from 0.9 m to 6.6 m. Note that the shorter counters will be cut at the module fabrication factory from longer extrusions. The extrusion rate for the prototype runs is 75 kg/h. Although the die has been tested at 100 kg/h, it becomes harder to cool down the extruded part as the extrusion rate increases, which results in deformation of the fiber channel shape. Hence the schedule was developed using the slower 75-kg/h rate. About 19 weeks of solid extrusion are needed to produce all of the counters. Figure 10.31 shows an extruded sample representative of this R&D stage. Currently, the die is being tested and machined to produce counters with the size and tolerances required by the module fabrication group.

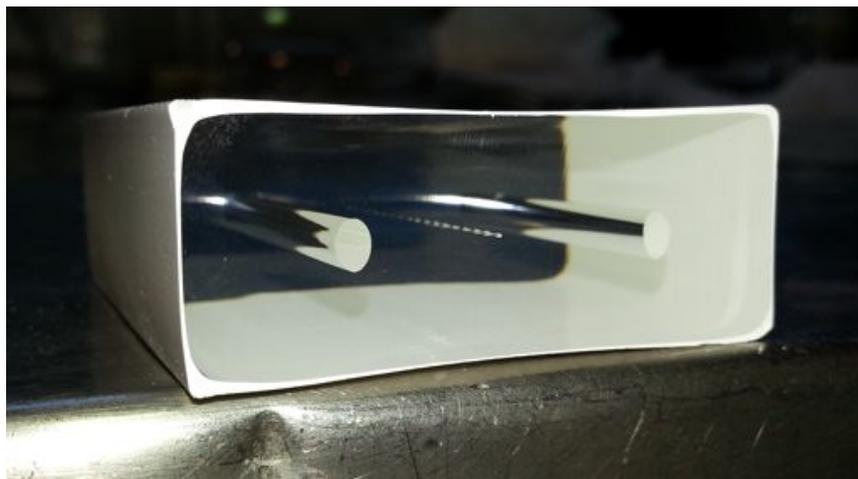

Figure 10.31. A prototype CRV scintillator extrusion (5×2 cm$^2$) with reflective titanium dioxide coating and two channels for the wavelength-shifting fibers.

Quality assurance and quality control (QA/QC) procedures to ensure the final product meets requirements are in place and have been extensively used in previous extrusion projects. Two parameters will be checked during production at the extrusion facility: dimensions and light yield. Dimensions will be checked every 45 minutes using a caliper. The data will be entered in a computer and the file submitted weekly to the Mu2e document database. The light yield will be tested using a radioactive source. A reference sample will be measured to monitor the stability of the equipment as well as to provide a minimum acceptable value (Figure 10.32). The results will also be uploaded weekly into the Mu2e document database. These measurements will be carried out on a QC sample (15 cm long). This short QC sample will be cut once every 5 to 10 full length scintillator strips. These two tests will be the basis of the quality control program. Additional testing will be performed if a problem is noticed.

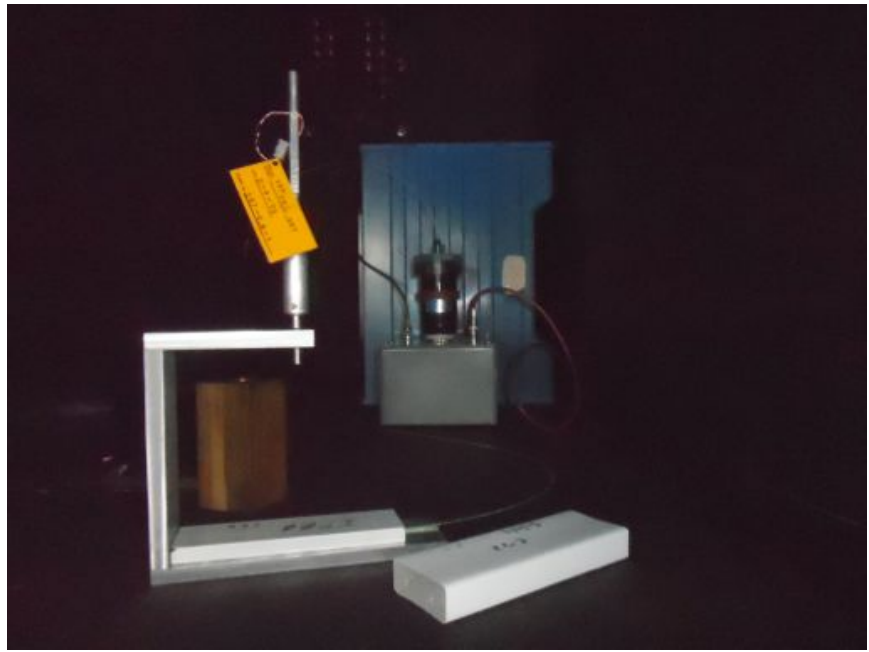

Figure 10.32. Setup inside a dark box for the quality control tests of the scintillation light yield.

## 10.5   Wavelength Shifting Fibers

The wavelength-shifting (WLS) fibers provide an efficient method for collecting the blue light (~425 nm) generated in the scintillation counters by charged particles. A fluorescent compound in the WLS fiber absorbs the scintillating light and re-emits it in the blue-green (500-600 nm) spectral region (Figure 10.33), trapping a fraction of the light by total internal reflection. As the light propagates down the fiber, the shorter (<520 nm) wavelengths of the emission spectrum are significantly attenuated, while the longer wavelengths are less affected (Figure 10.34).

Suitable WLS fiber is available from Kuraray [24], which has produced high-quality fibers for MINOS, MINERvA and NOvA and many other experiments. The WLS fiber is double-clad, 1.4-mm diameter with a polystyrene core covered with an inner cladding, approximately 3% of the fiber diameter, of polymethylmethacrylate (PMMA or acrylic) and an outer cladding of a fluorinated-polymer, approximately 1-3% of the fiber diameter. The relative indices of refraction are 1.59, 1.49, and 1.42 for the core, inner, and outer cladding respectively.

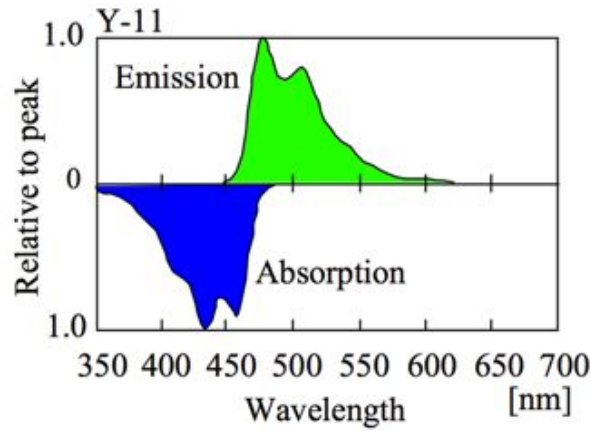

Figure 10.33. Absorption and reemission light spectra of Y11 WLS fiber.

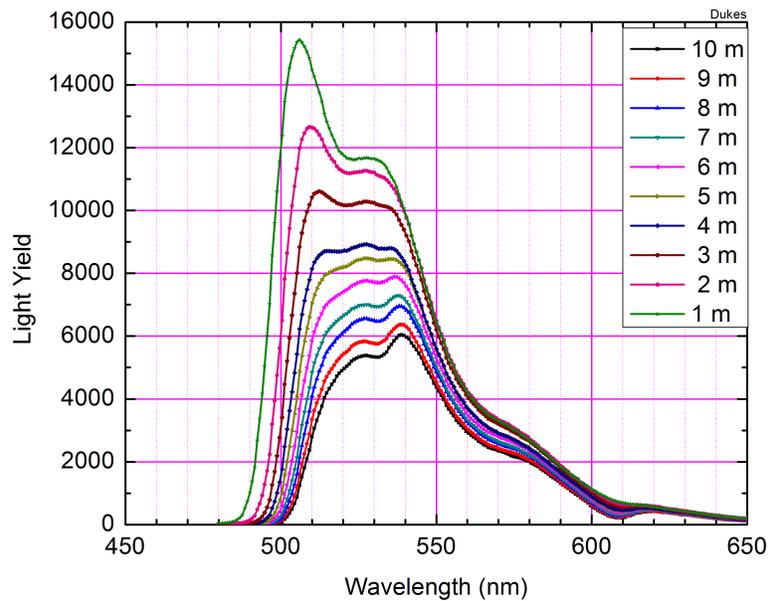

Figure 10.34. The waveshifting fiber light yield as a function of wavelength. The measurement was performed with the NOvA fiber tester, using 1.0 mm diameter double clad Y11 fiber purchased for CRV prototypes [12]. The curves correspond to measurements from 1 m to 10 m from the readout end.

The polystyrene is doped with the fluorescent dye, K27. A higher concentration of K27 increases the light-capture fraction but worsens the attenuation length. In the concentration range of 150-300 ppm, as studied by NOvA and MINERvA, there is only a small (<5%) impact on the light yield and attenuation properties. A concentration of 175 ppm is selected, which was found optimal by MINERvA. S-type fiber has the polystyrene chains oriented longitudinally making the fiber more flexible, but at a cost of 10% lower attenuation length. The module assembly plan and quality assurance procedures do not require the additional flexibility given by the S-type fibers, therefore non-S-type fiber has been selected for the better light attenuation properties.

Kuraray will perform quality control (QC) measurements of the fiber diameter, eccentricity, light yield and attenuation length. Kuraray guarantees stringent specifications on the fiber diameter of $3\sigma/D < 2.5\%$. Occasional bumps of 2% and 4% of the fiber diameter have been found to occur at the rate of 50 and 15 bumps per km, respectively. The position of the bumps will be recorded and provided by Kuraray. The eccentricity is expected to be less than 1.5%.

The fiber will be delivered on large spools. Upon delivery, measurements of the fiber diameter, light yield and attenuation length will be performed as part of the Mu2e quality assurance (QA) program. To measure the light yield and attenuation at different wavelengths (see Figure 10.35), the apparatus shown in Figure 10.36 will be used. Note that the NOvA fiber tester cannot be used because the fiber diameter is too large; this new design will work with essentially any fiber diameter. The test jig consists of the factory supply spool, a large diameter transfer pulley, and a take-up spool carrying the optical readout hardware. The transfer pulley is driven by a stepping motor and controlled for gentle acceleration and deceleration. A blue LED light source illuminates the WLS fiber and is read out by a USB-enabled spectrophotometer and a photodiode. The apparatus can increment at any predetermined distance, stopping for measurements of the light spectra and intensity. The measurements will be performed on the first 25 m of every spool, each of which contains 800 m of the fiber. The tested fiber will be rewound onto the spool for the later use in the module fabrication factory. After the fibers are installed in the counters at the module fabrication factory, each fiber will be tested by a custom-made apparatus to ensure that it was not damaged, either during the insertion procedure or during the fly-cutting process.

## 10.6   Photodetectors

### *10.6.1*   Introduction

Readout of the cosmic ray veto is simplified by the fact that it resides on the outside of the Detector and Transport Solenoids and is easily accessible, the channel count is small, and the single channel hit rates are low relative to many HEP experiments. The baseline photo-detector is the so-called silicon photomultiplier (SiPM) [25]. SiPMs are pixilated avalanche photo-diodes that operate in Geiger mode. Each SiPM pixel is separated from the bias voltage by a current limiting quenching resistor. A photon of proper wavelength incident on a pixel can excite an avalanche that causes the pixel to discharge. The resulting SiPM output signal is quantized and is proportional to the pixel capacitance and the number of discharging pixels.

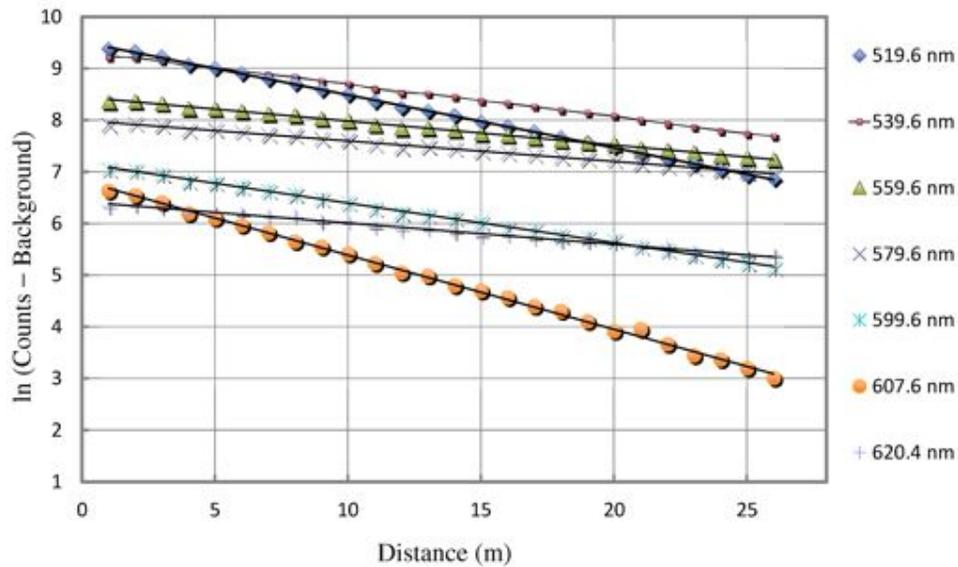

Figure 10.35. Light yield measurements made by the NOvA fiber test jig of 1mm diameter double-clad Y11 fiber for various wavelength values as the function of distance from the readout end.

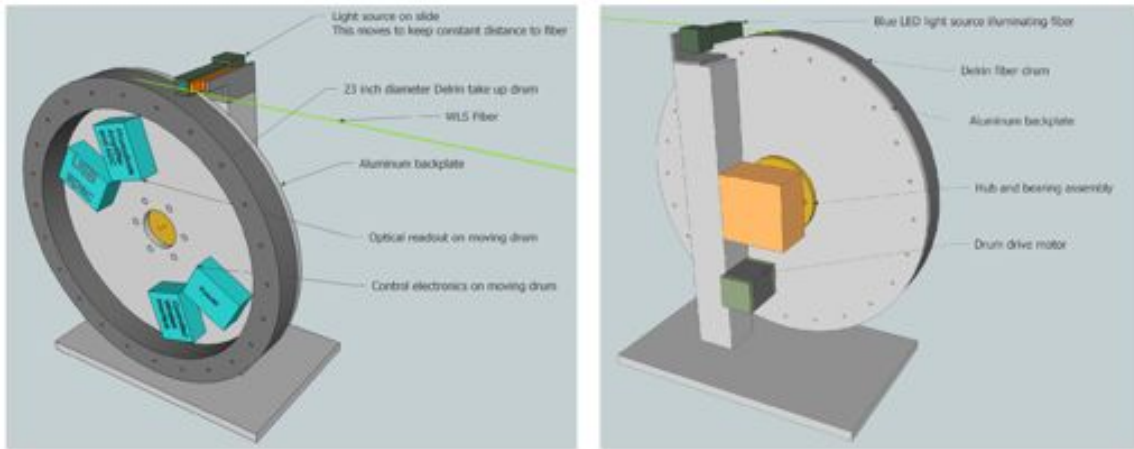

Figure 10.36. Front (left) and back (right) schematics of the fiber QA test jig. A blue LED illuminates the fiber, which is read out by a photodiode and spectrophotometer. The fiber spool is not shown.

SiPMs offer excellent reliability and ruggedness, immunity to the magnetic field, high photo-detection efficiency, high gain, and simple self-calibration. They are also extremely compact in size, having photosensitive areas that typically range from $1 \times 1$ mm$^2$ to $3 \times 3$ mm$^2$. Their spectral response is 320 - 900 nm, making them ideal for coupling directly to individual wavelength-shifting fibers. SiPMs also can be customized in both size and packaging at low cost. They typically have higher effective quantum efficiency than most photomultipliers in the red end of the spectrum. Their gain ($>10^5$) is sufficient to make the readout relatively simple. The excellent signal dispersion (the ability to distinguish one avalanche from two, three and so on) makes the calibration and monitoring of the photodetector straightforward. In addition, since it is a fully solid state device, the operating bias, which is a few volts above breakdown, is relatively low (30 V-100 V). This eliminates the need for expensive high voltage power supplies and cables that are required for hybrid-photodiodes and standard photomultiplier tubes.

The SiPM dark-noise rate due to Geiger discharge from thermally generated carriers in the conduction band is typically on the order of a hundred kHz per square millimeter. While the dark-noise rate is very high when compared to, for example, photomultipliers, it will be reduced to a manageable level by applying a threshold of ≥3 PE: typically the noise rate falls by one order of magnitude for each additional PE. Other features of SiPMs that produce false signals are crosstalk and after-pulsing. Crosstalk occurs when optical photons from an avalanche in one pixel causes a signal in a neighboring pixel. After-pulsing is caused by trapped carriers that have a delayed release. SiPM manufacturers have reduced crosstalk by introducing optical trenching, which is a physical barrier or break between pixels. Vendors have decreased their after-pulsing rates by reducing impurities and introducing blocking layers. These modifications also improve the radiation tolerance of the devices.

SiPMs are relatively new devices that are primarily used in medical instrumentation, in particular PET and CT scanners. Hence, there are many manufacturers actively improving their devices; the result is a large number of vendors with a broad array of product lines. While the use of SiPMs is relatively new in particle physics, it is an attractive technology and there has been much recent R&D work in collaboration with commercial manufacturers to develop devices that will meet specific detector requirements, e.g. SiPMs with reduced after-pulsing, tuned spectral responses toward the ultra-violet, higher radiation tolerance, and larger dynamic range (high pixel count and fast recovery time). Several experiments and projects, including the near detectors for T2K (INGRID and ND280) [58][62][63][64], the Calorimeter for the Linear Collider Experiment (CALICE) [59], the CMS Hadron Calorimeter (HCAL) upgrade [60][61], and the Jefferson Lab Hall D Barrel Calorimeter [30] have done extensive R&D on SiPMs and Mu2e is benefiting from their experiences. Vendors with candidate devices that are being evaluated include Hamamatsu [26], KETEK [27], CPTA [28], and AdvanSiD [29].

### 10.6.2   SiPMs for Mu2e

The cosmic ray veto requirements for the photodetector do not push the SiPM performance beyond what is currently available commercially from multiple vendors. Important factors that impact the SiPM choice are the need to maintain a high light yield and good time resolution over the lifetime of the experiment and the need to have a stable, well understood PE threshold during operation. SiPMs with high effective quantum efficiencies are commercially available: a PDE of 30% or larger should be sufficient for Mu2e, where the PDE is defined as the product of quantum efficiency, geometric fill factor, and avalanche probability.

The key performance requirements for the SiPM are the following:

- The device must operate in a 0.10 T magnetic field without degradation to performance or reliability.

- The nominal photosensitive area of the device must have a diameter of at least 1.65 mm. This size accommodates the specified fiber diameter of 1.4 mm with an additional 0.25 mm to allow for misalignment between the fiber and the device, although not so large as to unnecessarily increase the dark count rate. Many vendors produce $2 \times 2$ mm$^2$ devices, which will be used should they meet the dark count and radiation hardness requirements. Custom sizes can be obtained at modest increased cost.

- The device must have a PDE that is at least as high as the 100 μm-pitch reference device tested in the studies described in Section 10.2.3. The greatest PDE variation for SiPMs comes from the geometrical fill factor, which decreases with pixel size. Recent improvements by vendors in device processes have yielded remarkably high PDEs for small pixel sizes. Currently, multiple vendors have 25 μm and 50 μm-pitch devices that have PDEs equal to or higher than our reference test-beam device. The availability of devices with higher PDE values will allow smaller fiber diameters to be used, as well as smaller diameter SiPMs.

- The device should have a gain $\geq 10^5$.

- The number of pixels should be sufficient to cover the entire dynamic range (up to 200 PE) without significant non-linearity in the response.

- The device must operate during exposure to $10^{10}$ neutrons/cm$^2$ @1 MeV neq (see Section 10.10.3). In particular, the single photoelectron and secondary PE peaks must be distinguishable in order to maintain the in-situ calibration of the devices.

- Any degradation to the device response after 4 years of operation in the Mu2e detector environment must not compromise the efficiency requirement.

- The after-pulsing characteristics must be small enough so as not to cause excess detector deadtime. Note that after-pulses primarily have 1 and 2 PE amplitudes and hence lie below the ≥3 PE SiPM threshold.

- The device must be packaged in a small form factor/low profile package that allows tight direct coupling of the SiPM to the fiber.

Each SiPM will be coupled directly to the ends of a single wavelength-shifting fiber. This requires the photodetector to operate in a magnetic field of 0.1 T and in a radiation environment of both neutrons and gammas. Ionizing radiation from gamma interactions is not a problem, as studies have shown no change in SiPM performance at doses up to 20 Gy [30], which is higher than the expected dose (see Section 10.10.3). Neutrons can cause overall bulk damage to the silicon lattice, increasing the dark count and degrading the signal-to-noise of the device. The highest dose the devices will receive is just less than $10^{10}$ neutrons/cm$^2$ (1 MeV neq) in a few hot regions of the cosmic ray veto (see Section 10.10.3). Figure 10.37 shows the degradation of the photoelectron peaks after neutron irradiation at a fast neutron source reactor of $2\times10^{10}$ neutrons/cm$^2$ (1 MeV neq) and the recovery of the device after annealing [31]. (For more results on SiPM annealing, see Ref. [30]).

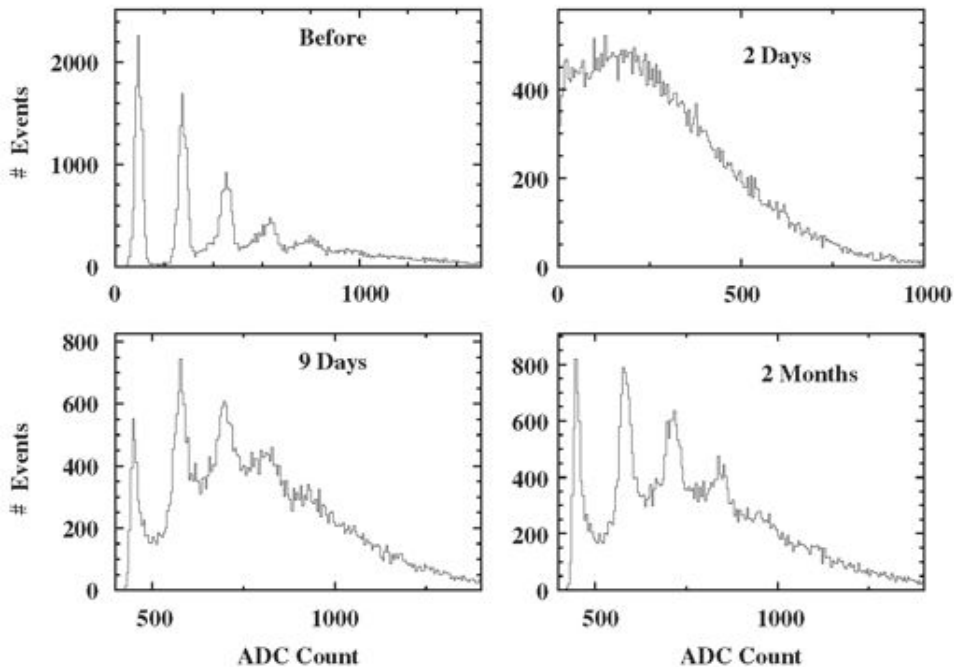

Figure 10.37. Response of a Hamamatsu SiPM before $2\times10^{10}$ n/cm$^2$ (1 MeV neq) irradiation and after two days, nine days, and two months showing recovery from annealing [31]. Recovery from the dose is not complete.

Neutrons can also cause anomalous signals in the SiPM when slow neutrons become captured in the boron-10 of the silicon and when neutrons interact in the hydrogen-rich epoxy encapsulation of the device. High energy physics experiments have studied these effects extensively [31][32][33][34], providing feedback to vendors who have modified their processes by reducing impurities and adding blocking layers to improve the first effect. Minimizing the thickness of the epoxy encapsulation of the package helps reduce the latter effect, and many manufacturers are working toward a packaging option that eliminates the wire bond and reduces the need for a thick protective epoxy layer.

Surface mount and canned (TO-18) Hamamatsu (MPPC S10362-11-100U) and AdvanSiD (ASD-SiPM1C-M-40) SiPMs were evaluated in a test beam at Fermilab in the fall of 2013. In addition to gaining experience in the operation and in-situ calibration of the SiPMs, the test revealed several mechanical issues related to details of the design of the coupling of the SiPM to the fiber. These have led to the choice of a surface mount package and the current SiPM mount scheme described in this TDR.

Hamamatsu (S12892-PA-50(X)) and KETEK (PM2250-B63T75S-P4) both have improved $2 \times 2$ mm$^2$ devices that will be tested. One improvement is a reduction in the thickness of the clear epoxy layer on the SiPM face used to protect the wire bond, which is typically ~300 µm thick. This epoxy layer can degrade after irradiation, reducing the light transmittance. Newer packaging options, such as through silicon via (TSV), with thinner coatings (~100 µm), are becoming available and will be evaluated. Figure 10.38 shows a typical surface mount packaged SiPM. These devices should meet all basic requirements. Radiation tolerance to neutrons will be evaluated in a reactor for total dose effects, and with Cf-252 or Am-Be sources for performance degradation at lower dose rates.

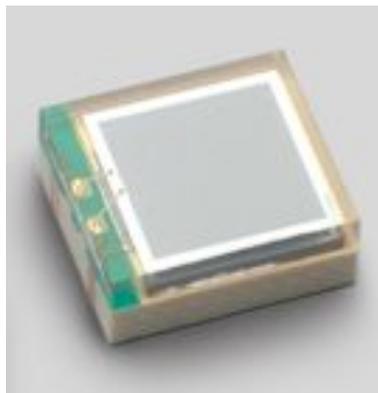

Figure 10.38. Typical surface mount package. The SiPM surface is coated with a thin epoxy layer to protect the photosensor and the wirebonds (seen on the left).

The small size, high PDE, and low cost of the SiPM makes double-ended readout of a single fiber over the majority of the cosmic ray veto modules possible without unacceptably large dead regions in the coverage. This adds redundancy and makes the overall detector more reliable as each counter end will have a completely different readout chain. Hence in the majority of the modules, single device failures will not preclude attaining the desired detector efficiency requirement. The exceptions to the double-ended readout are the extra-long modules in the CRV-U and CRV-T sectors, as well as the cryo-modules. For the extra-long modules, the upstream end of the counters will see radiation doses up to $10^{11}$ n/cm$^2$ (1 MeV neq) as shown in Figure 10.71. At those radiation levels, the single PE peak that is required to calibrate the SiPMs is expected to significantly degrade. In addition, the shielding in this region makes access difficult and time consuming. Therefore there are no SiPMs in this region. Rather the SiPMs will be outfitted with counter manifolds with reflective mirrors. Single-ended readout sacrifices redundancy, hence, SiPM and front-end board failure in these modules will require immediate replacement during operation. The CRV-modules have a sufficiently high light yield that each SiPM on the readout end separately meets the photoelectron yield requirement.

The SiPMs needed for the experiment by sector are listed in Table 10.2. The total number of devices needed, not including spares and wastage, is 18,944.

### *10.6.3*  **Calibration**

The SiPM dark count and after-pulsing rates are very sensitive to the operating bias; less so are the PDE and gain, where the operating point is typically several volts over the breakdown voltage. Because of this sensitivity, the operating bias must be maintained to about 10 mV. The operating bias, however, is not fixed but is dependent on environmental parameters; in particular, the temperature and radiation dose. Controlling the SiPM temperature would be difficult, if not impossible, given the need to keep the counter readout manifolds as small as possible. And although cooling SiPMs reduces the dark count rate by about an order of magnitude for each 25°C reduction, at a threshold of PE ≥3 we expect that the rates will be dominated by signals from neutrons and gammas produced in the beamline. Finally, controlling the temperature would not eliminate drifts in the operating point given the increasing radiation dose. For these reasons it is mandatory that in-situ calibrations be periodically made to determine the operating point. Maintaining the correct operating point will be achieved by tracking the gain of the single PE peak and measuring the relative dark count rates of the single and double PE peaks. This will be done automatically during interspill periods. Bias voltages for the SiPMs will be adjusted so that the single PE pulse heights are uniform for all channels. In T2K, the operating voltage is adjusted once a month in normal operation [35].

To track the sensitivity of the operating point on temperature (typically dV/dT ~60 mV/$^\circ$C), the temperature will be measured with sensors mounted on each counter motherboard. This information will be recorded in a database. A Mu2e hall temperature excursion of ±2$^\circ$C is expected during Mu2e operations based on experience with similar buildings with similar HVAC requirements [37]. As changes are expected to be slow, the SiPM operating voltage will require infrequent adjustments based on temperature.

### 10.6.4 Quality Control and Quality Assurance

The large number of SiPMs in the cosmic ray veto demands a clear workable plan for photodetector quality assurance. In order to avoid board rework and to reduce the need for intervention in the experiment after installation, the quality of the devices will need to be verified before they are assembled onto the counter motherboards. In addition, verifying that the devices will survive the commissioning period and the three years of physics operation will require longevity and radiation testing of a subsample of devices before the batches of SiPMs are accepted from the vendor.

The plan for device testing is largely based on experience from other experiments. The T2K experiment has had ~60,000 devices in operation for several years [35]. Because they were one of the first HEP experiments to use SiPMs, they extensively tested these devices, measuring parameters such as gain, breakdown voltage, dark count rate, after-pulsing and PDE at three different temperatures [36]. In the end, they found that the key parameters that needed to be measured and tracked in order to operate the devices in the detector were the breakdown voltage at one specific temperature and the temperature dependence of the gain. The temperature dependence of the gain for a device is common to the fabrication process, so measuring a small percentage of devices provides the needed temperature correction function for all devices [65]. T2K found that all of their SiPMs had identical temperature dependencies [36]. Figure 10.39 shows the T2K measurements of the gain and breakdown voltage for three different temperatures.

The vendor will assure the SiPMs will perform within the parameters specified in the device data sheet, and will provide basic device parameters, including the operating voltage for a specific gain and dark count rate at a set temperature. The Mu2e SiPM testing plan is based on the experience from the T2K testing at Kyoto University, which found only 0.05% failures out of ~18,000 devices [36] after rigorous characterization of each device. Based on this low failure rate and the finding that parameters measured at Kyoto matched those from the manufacturer, rigorous testing of all devices has been deemed unnecessary and vendor provided parameters are sufficient. The plan for Mu2e is to test only a ~10% subsample of the devices to validate the individual batches, on the order of 650-1000 devices per batch, before acceptance from the manufacturer. The device I-V curve and dark-count rate will be measured at a fixed temperature with the Mu2e custom SiPM tester and cross-checked with the operational information provided by the vendor before batch acceptance. Since the temperature dependence is process specific, a small number of devices (1%) will be used to determine that function before mounting on the counter motherboards.

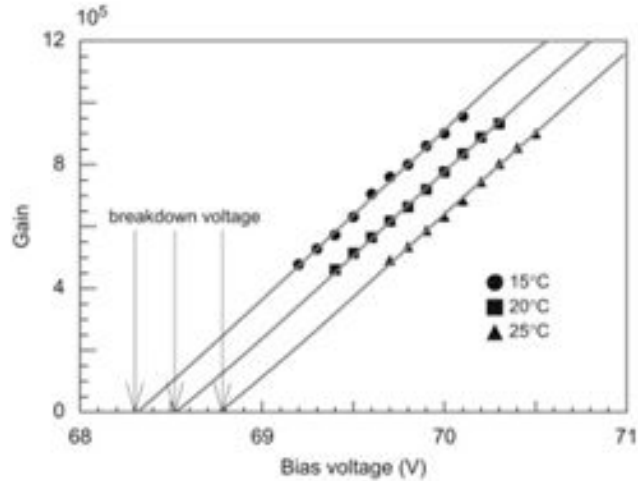

Figure 10.39. T2K measurements of the breakdown voltage from gain vs. voltage extrapolation for 3 different temperatures [36].

Mu2e will use surface mount or TSV SiPM packages. Because those small form factor packages do not have space for serial numbers, it will be difficult to track the operational parameters for individual devices once the devices are sent to the assembler for mounting on the counter motherboard. This motivates the need to be able to measure the I-V curves in-situ with CRV readout electronics after the devices are mounted on the counter motherboard.

The SiPM tester (see Figure 10.40 and Ref. [38]) that is being developed for Mu2e is based on experience gained from testing several thousand SiPMs for the NIU Proton Tomography Project [39]. The SiPMs will be placed into a 32-device custom waffle pack that provides electrical contacts on the pads of the SiPM. The waffle pack will be placed in a thermally stabilized dark box that is connected to the tester box via cable feed-thru patch panel. The tester applies bias voltages to 32 SiPMs and simultaneously measures the dark current, taking roughly a hundred points to determine the full I-V curves for the devices. For quality assurance, the tester will be checked on a regular basis with a set of 32 SiPMs that will be used as a "standard" for calibrating the testing device.

In addition, a subset of the production devices (0.5%) will undergo additional radiation and lifetime tests to validate the vendor process. These are assumed to be destructive tests. Consequently, these devices will not be installed in the detector. The devices will be characterized before and after these tests. For radiation testing, the devices will be dosed to several exposures ($1\times10^9$ n/cm$^2$, $5\times10^9$ n/cm$^2$, and $1\times10^{10}$ n/cm$^2$) to study the performance degradation and to determine the rate of annealing/recovery. Of particular interest is the increase in dark count, the effect on the I-V curve, and the ability to discern the single PE peak, since those are the parameters that will be used to monitor and calibrate the devices.

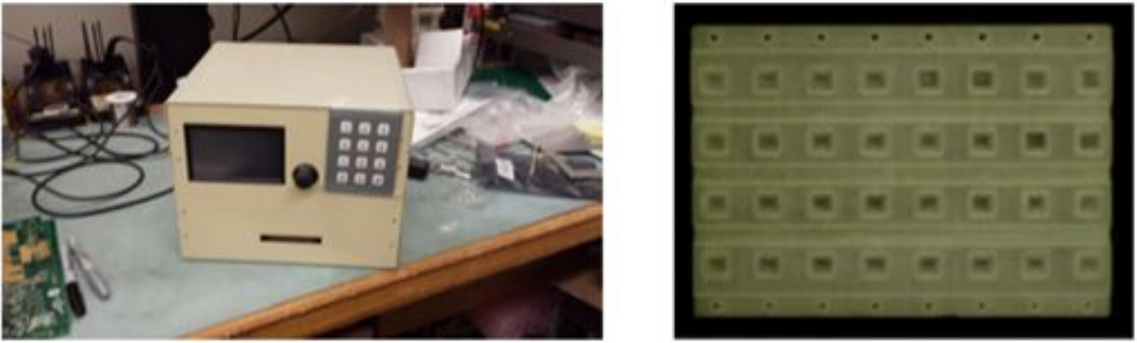

Figure 10.40. SiPM tester (left) and a 32-SiPM waffle pack holder (right).

Accelerated aging studies also will be performed. Electronics device time-to-failure is well-modeled by the Arrhenius equation, which describes the relationship between failure rate and temperature in accelerated aging studies. By elevating the device operating temperature, a shortened lifetime study can be performed. SiPMs that survive at an elevated temperature of $70^\circ$C over 64.5 days are sufficient to validate device longevity for the CRV, which will be operating with nominal temperature of $25^\circ$C and over 4 years (1 year of commissioning and 3 years of physics operations) in the Mu2e experiment.

## 10.7 Electronics

The electronics (Figure 10.41 and Table 10.4) consists of: (1) a counter motherboard mounted directly on the ends of the di-counters and onto which the SiPMs, flasher LEDs, bias gate circuit, and temperature sensors are connected; (2) a front-end board (FEB), which reads out and digitizes signals from the SiPM, both in time and charge, controls the flasher LED, runs calibrations, and provides bias voltage to the SiPMs; and (3) a readout controller, which takes the data from the front-end boards and sends it to the data acquisition system and provides a means of communication with the front-end boards. The system has been designed to be simple, redundant, and inexpensive. It can operate in either a streamed mode, for diagnostic and calibration purposes, with sufficient storage to stream the CRV data over an entire 1.333-s spill cycle, or in zero-suppressed mode. The electronics must have the ability to see single photoelectrons, a dynamic range of 2000, a time resolution of 1 ns and the ability to handle hit rates of up to 1 MHz. In addition, the CRV electronics in the detector hall exposed to the harshest magnetic and radiation fields will need to operate in magnetic fields up to 0.1 T, and suffer a radiation dose of up to $10^{10}$ neutrons/cm$^2$ (1 MeV neq) with no untoward effects.

The Mu2e DAQ is designed to operate with software-only triggers run on a farm of online processors, with a latency of about one second, or about one spill cycle length. A trigger accept, which during normal data taking comes from track-finding criteria, is sent to the front-end boards which sends the entire micro-pulse of data to the readout controllers and then to the DAQ.

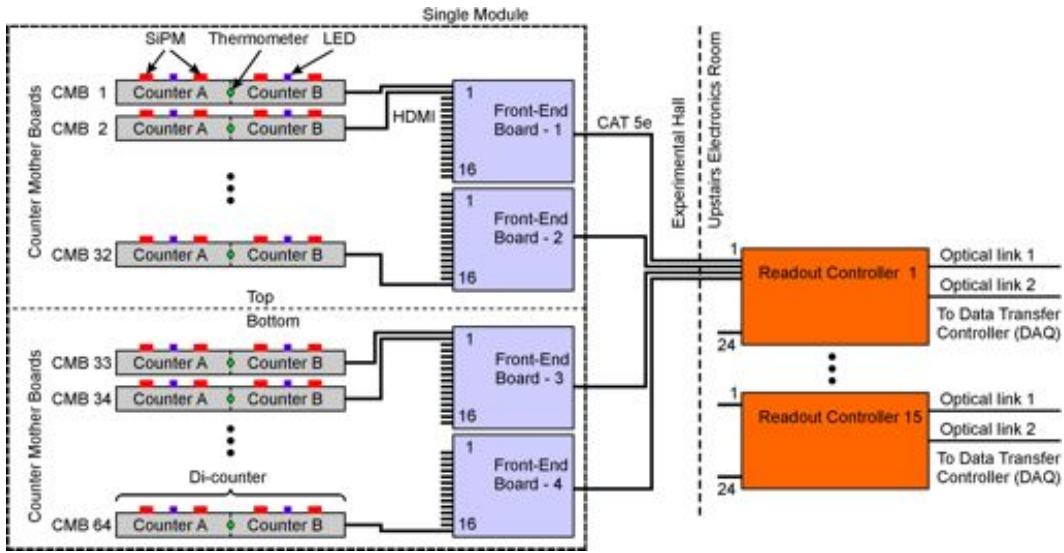

Figure 10.41. Layout of the readout electronics for a normal-width module read out on both ends. A typical module has 64 counters grouped in 32 di-counters. Each di-counter has 4 SiPMs for a total of 128 SiPMs at each end of the module. Each front-end board reads out 64 SiPMs, or one end of 16 di-counters: two front-end boards are on each end of the module, for a total of four. Power-over-ethernet cables send the data from the front-end boards to the readout controllers, which then send their data to the DAQ. Each readout controller services up to 24 front-end boards.

### 10.7.1    Counter Motherboard

The counter motherboard, which is mounted directly on the end of a di-counter and provides direct electrical connection to the photodetectors, has the following components: (1) four SiPMs, (2) two flasher LEDs, and (3) a temperature sensor. A circuit diagram of the counter motherboard is shown in Figure 10.42; a photograph of a prototype board used for a beam test was shown earlier (Figure 10.21). The bias voltages for the SiPMs, the Low Voltage Differential Signaling (LVDS) trigger signals for the LEDs, and the power for the temperature sensor are supplied from the front-end board to the counter motherboard over a micro-HDMI cable. A low voltage digital-to-analog (DAC) on the front-end board is used to trim each SiPM bias to the desired value. A total of 4,736 counter motherboards are needed.

Table 10.4. Electronics parameters.

| | |
|---|---|
| Channels (SiPMs) per counter motherboard | 4 |
| Total number of counter mother boards | 4736 |
| Channels (SiPMs) per front-end board (FEB) | 64 |
| Counter motherboards per FEB | 16 |
| ADC digitization clock (2/3 RF frequency) | 79.65 MHz |
| ADC samples/hit | 4 |
| Octal analog front end/ADC | TI AFE5807 |
| ADC size | 12 bits (4096) |
| Event size/hit | 12 bytes |
| Power per FEB | 17 W |
| Total number of FEBs | 296 |
| Maximum FEBs per readout controller (ROC) | 24 |
| Average FEBs per ROC | 19.7 |
| Maximum data rate: FEB to ROC | 12.5 MB/s |
| Maximum data rate: ROC to DTC | 312.5 MB/s |
| Total number of ROCs | 15 |
| Maximum data rate:  DTC to DAQ | 1 GB/s |
| Total number of DTCs | 2 |
| Maximum trigger latency (buffer size) | 1.0 s |

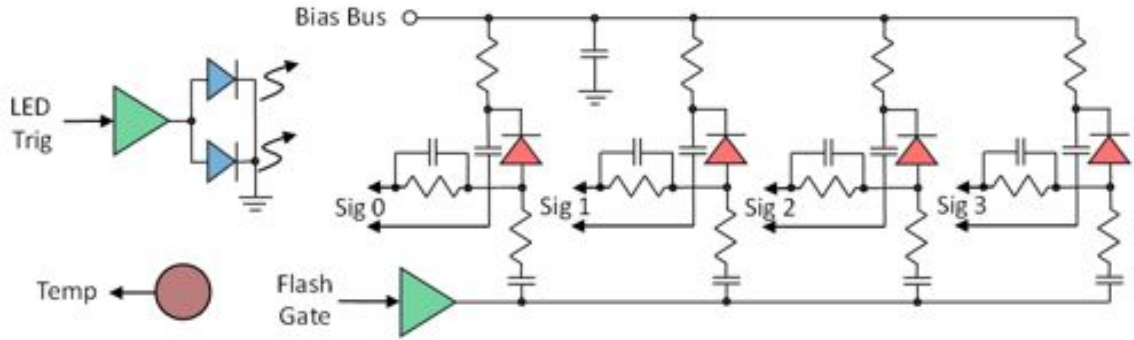

Figure 10.42. Diagram of the counter motherboard.

As explained above, we expect to employ a SiPM that is considerably larger ($2 \times 2$ mm$^2$) than the wavelength-shifting fiber diameter (1.4 mm). In that case, the positioning of the surface-mount SiPM during soldering will be straightforward, with sufficient self-alignment being achieved during reflow. However, should a SiPM whose size more closely matches the fiber diameter be used then a precise coupling of the SiPMs on the counter motherboard to the wavelength-shifting fibers will be important to maintain a sufficient PE yield. This requires good lateral alignment and consistent height for the four SiPMs on the counter motherboard. In order to maintain the lateral alignment of the SiPMs during reflow soldering, stainless steel precision templates that are registered to the counter motherboard registration holes will be used. The height uniformity will be maintained by uniform application of solder. In either the self-alignment or template soldering case, the SiPMs will be soldered onto the boards after the other components using tin-bismuth (Sn-Bi) eutectic low temperature (138°C) solder. This assembly process is based on previous experience from the CMS HCAL Outer Barrel SiPM project, which recently had ~3000 SiPMs assembled onto PCBs. With this procedure, a SiPM placement of ±50 microns was achieved [40].

### *10.7.2*   **Front End Board**

The front-end readout board (FEB) supplies the SiPM bias voltages, receives and digitizes the SiPM signals, controls the LED flasher and bias reduction gate, reads the temperature sensor, and performs SiPM calibrations. It is directly connected to up to 16 counter motherboards via micro-HDMI cables. One of the salient features of the design is the use of commercial off-the-shelf parts; no custom integrated circuits are employed. A block diagram of the front-end board is shown in Figure 10.43.

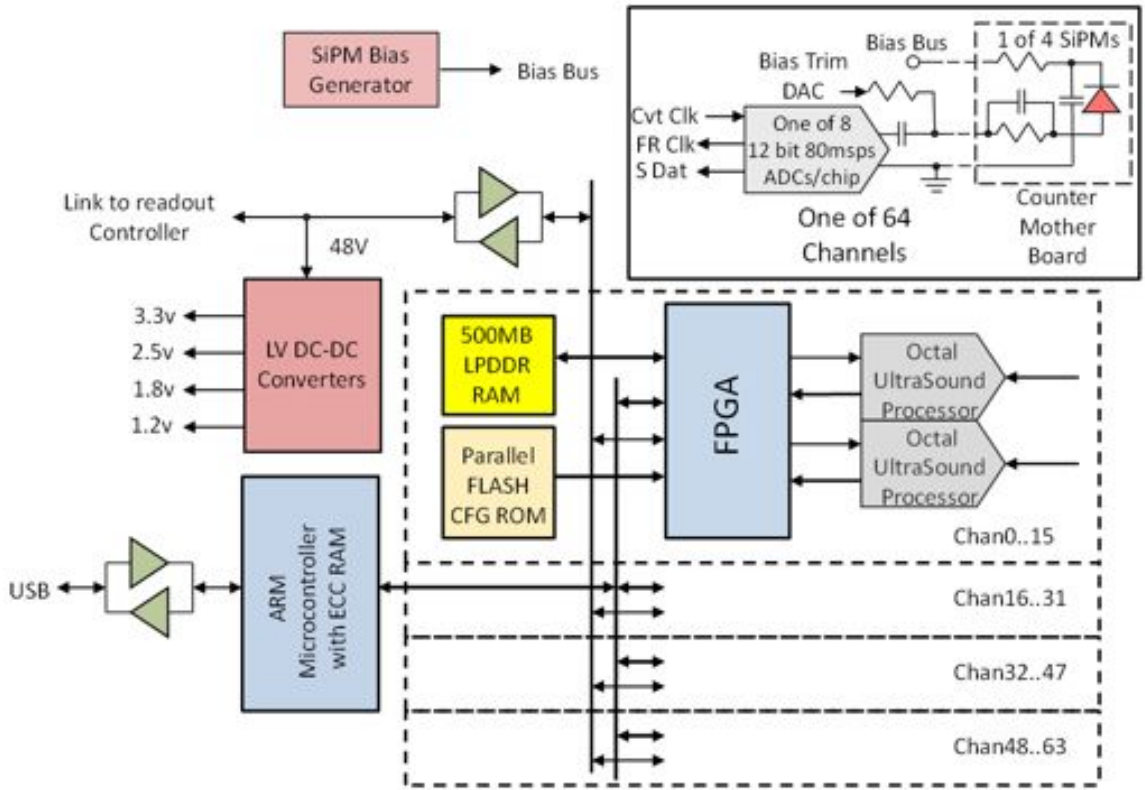

Figure 10.43. Block diagram of the front-end board.

The SiPM signals are fed to a commercial ultrasound processing chip on the FEBs, each of which has eight sets of low-noise preamplifiers, programmable gain stages, programmable anti-alias filters and a 80 Msps, 12-bit ADC [41]. A block diagram of the chip is shown in Figure 10.44.

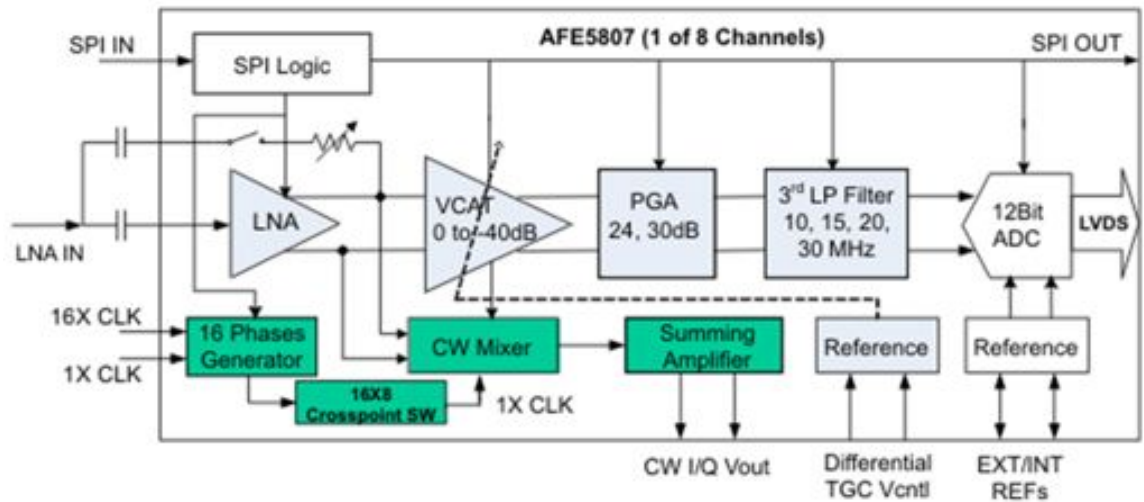

Figure 10.44. Block diagram of the ultrasound chip.

The SiPM signal is AC coupled at the ADC input on the front-end board. An AC-coupled input allows for a stable pedestal even if the SiPM suffers induced increased leakage current from radiation damage. With AC coupling, a SiPM produces a bipolar pulse that causes a voltage undershoot and requires time to recover. At high counting rates a substantial number of pulses would ride on the voltage undershoot from previous pulses. A pole-zero circuit will be implemented to reduce the undershoot recovery time and improve the double-pulse resolution. Figure 10.45 shows a simulation of a 10 photoelectron signal (green) from a SiPM similar type to what will likely be employed by Mu2e. The arrival of each photoelectron spread over the time constant of the wavelength shifter is visible. The blue trace shows the same signal after the pole-zero and the low-pass filter. The red trace shows the output of the sample-and-hold portion of the ADC. With four samples of the waveform and an 80 Msps sampling rate, 1 ns timing resolution will be achievable locally.

During Mu2e operations, the beam flash will illuminate a majority of the SiPM pixels, which would result in a large number of after-pulses during the live-gate period of the pulse. In addition, this large signal would cause a significant voltage undershoot that requires time to recover and could cause sagging of the SiPM bias voltage. In order to avoid this during Mu2e operations, a signal (flash gate) that reduces the SiPM bias to near, or below, breakdown during the flash will be implemented, as shown schematically in Figure 10.46. This scheme was tested; the results are shown in Figure 10.47. A SiPM was illuminated by a ~20 PE light pulse while the SiPM bias was reduced to below breakdown. The induced noise from the application of the bias reduction gate was found to be small, as was the pulse height variation immediately after full bias restoration. The pulse heights varied by roughly 5% over a wide range of times.

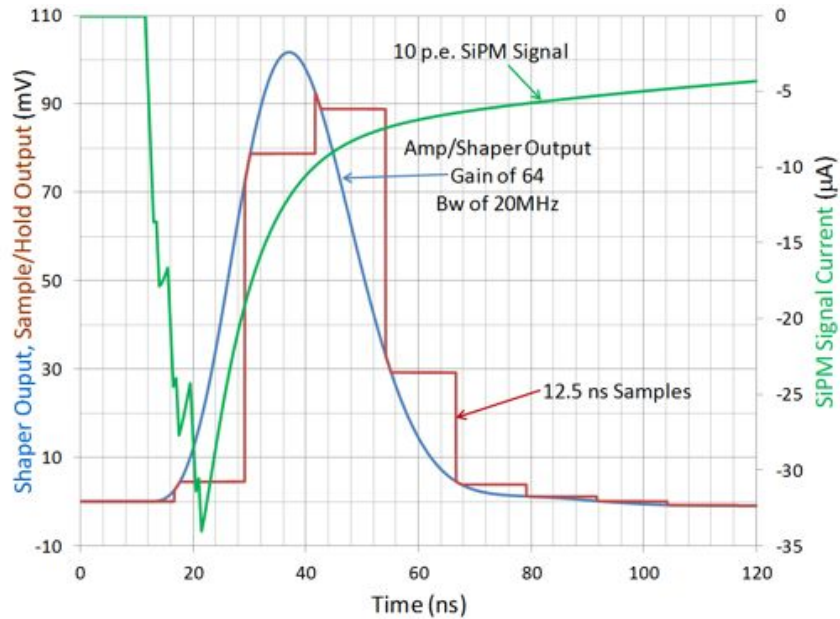

Figure 10.45. Simulation of the ultrasound chip response to a 10 PE SiPM pulse.

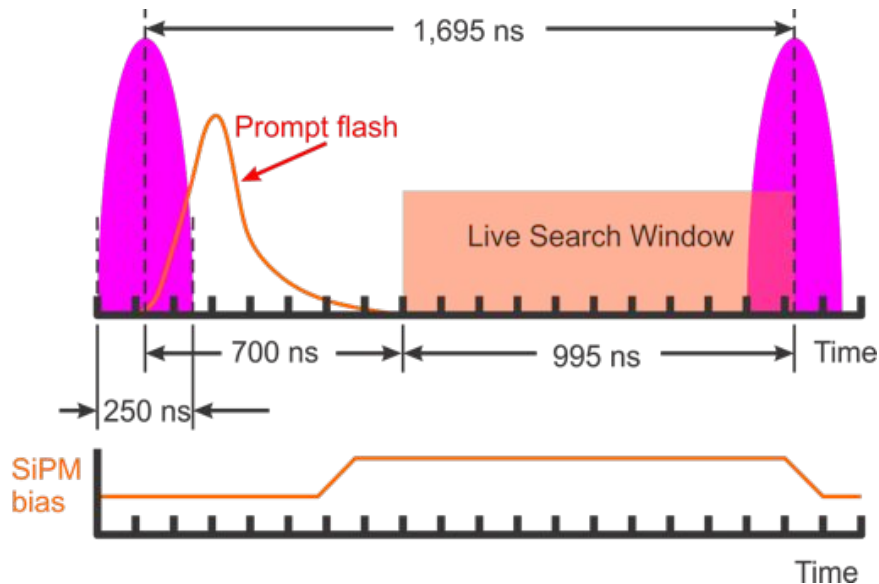

Figure 10.46. Schematic of the micropulse structure, showing the timing of the SiPM bias.

The digital portion of the front-end board uses FPGAs to convert the bit-serial ADC data to parallel, to apply thresholds to the digitized data for zero suppression, and to control the buffer memory. Low-power double-data-rate (LPDDR) synchronous dynamic random-access memory (SDRAM) is used for buffering data and has the capacity to buffer un-zero-suppressed data for a time equal to the trigger latency of the DAQ system. This gives us the option of deferring the zero suppression to offline event processing where more sophisticated techniques could be applied, and allows it to operate in "oscilloscope" mode for diagnostic purposes. The data from the FPGA are serialized and sent to the controller card using 100 mbit Ethernet Physical Layer Device (PHY) chips. A microcontroller is used for status and slow control.

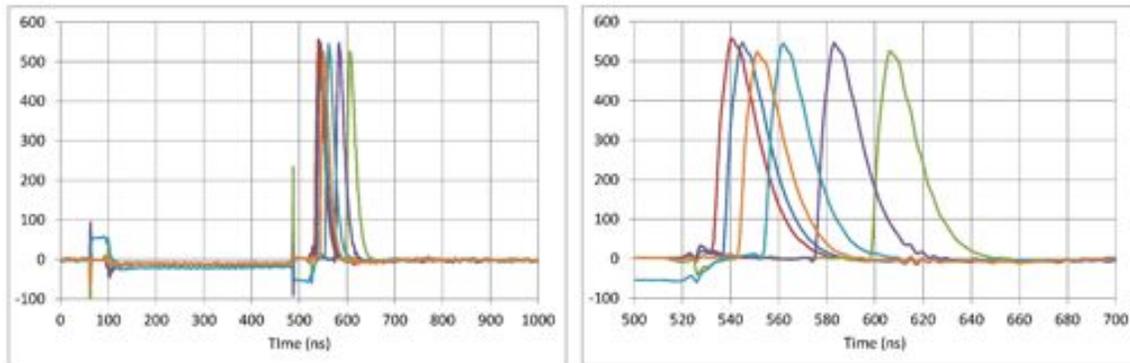

Figure 10.47. Effect of a fast bias reduction on the response of a SiPM. The SiPM is continually flashed with light pulses of constant amplitude. From 70 ns to 480 ns the bias was reduced below breakdown and restored at 480ns. In the left plot the SiPM shows no response to the light during the reduced bias period, and little induced noise from the rapid change in bias. The right plot shows the response of the SiPM just after the full bias has been restored.

Power is supplied over the same Category 5 cables used for the data link to the readout controllers, simplifying the cable plant. Commercial Power over Ethernet (PoE) protocol will be used. The stated power consumption of the ADCs is 117 mW per channel. A precise estimate of the FPGA power consumption is difficult without a detailed design, but a reasonable supposition is that the consumption is equal to the ADC's. Further, if we assume 1 W for the microcontroller and 1 W for the DRAM, then board consumption is then 17 W. If the DC-DC converters on board have an aggregate efficiency of 75%, the delivered power to the FEB is ~23 W, which is under the PoE Category 5 limit of 36 W [42]. The total system of 296 boards will require 6.8 kW.

The front-end boards must operate in significant radiation levels and ambient magnetic fields over much of the cosmic ray veto. Fringe fields will be up to 400 Gauss, which will saturate any magnetic materials used in the front-end board. To abandon the use of magnetic materials in the power supply section would increase the power consumption by at least three-fold, and require separate power cables rather than PoE. Instead, we will fit a magnetic shield around the susceptible components. To deal with the expected radiation levels we will use a microcontroller with error corrected memory and FPGAs that flag corruption of their configuration bits. For specific critical portions of the FPGA logic, such as control registers, voting logic will be used. Parallel FLASH ROMs that can configure the FPGAs in about 30 ms, will be used for storing configuration data. The quoted configuration RAM single-event upset (SEU) rate for an FPGA of the size we will use is 760 failures-in-time (FIT), where a FIT is one failure per billion hours per $1.4\times10^{10}$ n/cm$^2$. This is about 39 SEUs per hour for the 1200 FPGAs in the cosmic ray veto front-end boards, assuming that all FPGAs are subject to the maximum radiation dose of $1.0\times10^{10}$ n/cm$^2$ (1 MeV neq). For a typical configuration, one SEU in five will alter the behavior of the FPGA. The parameter storage registers in the ultrasound chips will be monitored and reloaded if corrupted. Radiation hard gallium nitride FETs will be used for the power switches in the power supplies. In order to radiation qualify the electronics design, we will need to validate the total ionizing dose tolerance and measure the single event upset rate of all components used on the boards.

Although the cosmic ray veto is an inherently digital device – it is only necessary to know that a counter has fired – for calibration purposes it is vital that the number of SiPM pixels that have signals be known in order to monitor the photoelectron yield.

### 10.7.3    Readout Controller

Digital data and power (24W × 25 channels) is transmitted from the front-end boards to readout controllers in the relay racks in the electronics room via Cat 5 cables and from the readout controllers to the data transfer controllers of the DAQ via optical fibers. The readout controllers serve as the link between the front-end boards and the slow and fast DAQ systems and provide power to the front-end boards. Figure 10.48 shows a block diagram of the readout controller. Each readout controller has 24 ports for front-end boards and two fiber-optic links [43]. Fifteen readout controllers serve the entire cosmic ray veto. The topology of the fiber-optic links depends on the data rates and desired redundancy.

 As a first step toward a complete readout controller, a prototype daughter card for use with a commercial FPGA evaluation board has been built for a proof-of-concept test. A block diagram of the setup is shown in Figure 10.49 and a photo of the daughter card is shown in Figure 10.50. Visible at the lower left of the photo is a four-port front-end card link. At the lower right are the two fiber optic connections for the DAQ link. At the upper right is the FPGA mezzanine card (FMC) connector for attachment to the FPGA evaluation board.

### 10.7.4    Timing

The high background rate in the CRV demands a small coincidence time window between counters to reduce the false veto rate. Hence, we need 1 ns timing over the entire system. The timing distribution consists of a 1.695 μs signal sent to the readout controller from the DAQ system. Messages defining the stages of the accelerator super cycle will be sent on the down-going DAQ link in advance of the 1.695 μs microspill period of the Delivery Ring. The stated jitter on this signal is 200 ps. There are 90 1/53.102 MHz periods in the Delivery Ring. A phase lock loop (PLL) on the readout controller using a voltage controlled crystal oscillator (VXO) will multiply the reference to a convenient frequency, and encode accelerator status information onto this signal using a punched clock scheme in a manner similar to what was used on the DØ detector. At the front-end board, a clock and data recovery PLL running at 3/2 of the 53.102 MHz clock (79.65 MHz) will supply the ADC clocks and recover the accelerator status information. The PLLs on the readout controller and FEB will have sub-nanosecond jitter. The propagation down the clock tree will vary primarily due to temperature variations, which affect the propagation delay of the line drivers and receivers and change the length of the interconnecting cables. In order to monitor synchronization and time in the detector components, cosmic rays can be used to measure the time of arrival of particles with respect to the local clock. A subset of these muons will traverse other detectors and will allow a cross-correlation of the timing. A significant portion of the CRV will see muons that do not go through other detector subsystems. These modules can be cross-correlated to CRV sections that have been linked to other detectors. Beam particles are another source of timing calibrations, but again there are CRV sections that will not be exposed to beam particles. Both schemes will be needed to ensure time synchronization across the entire system.

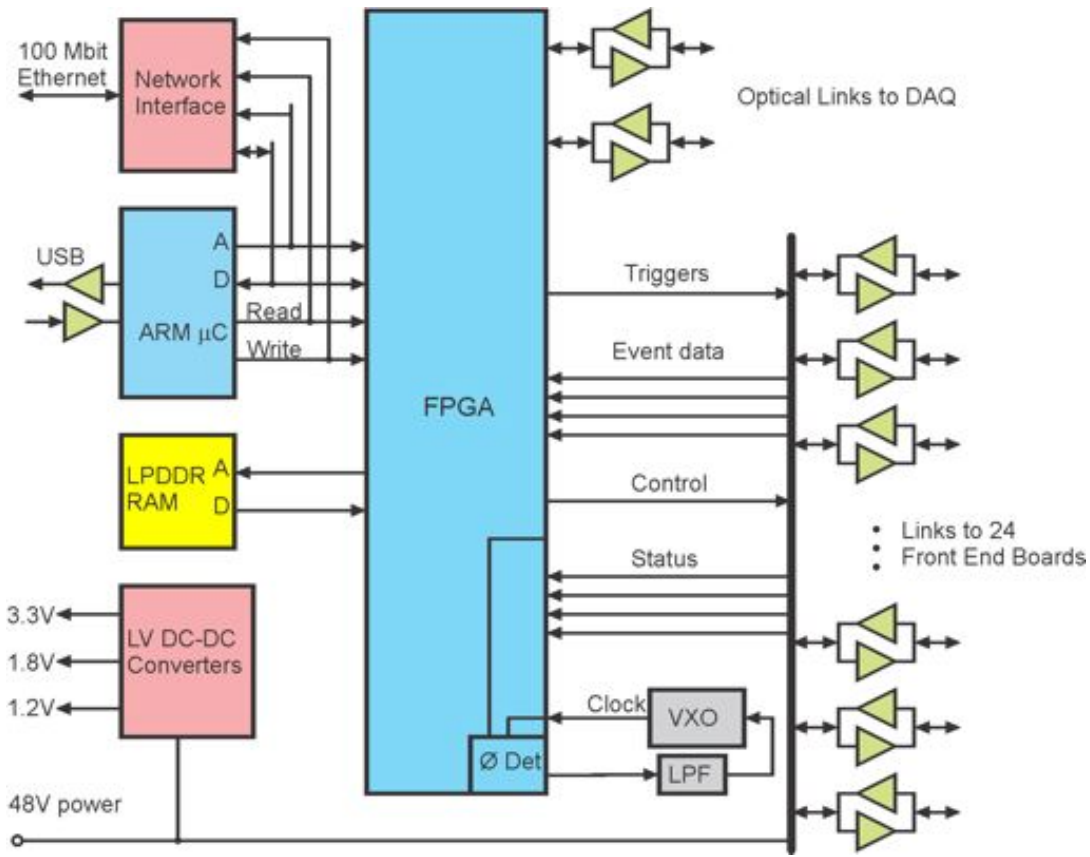

Figure 10.48.  Block diagram of the readout controller.

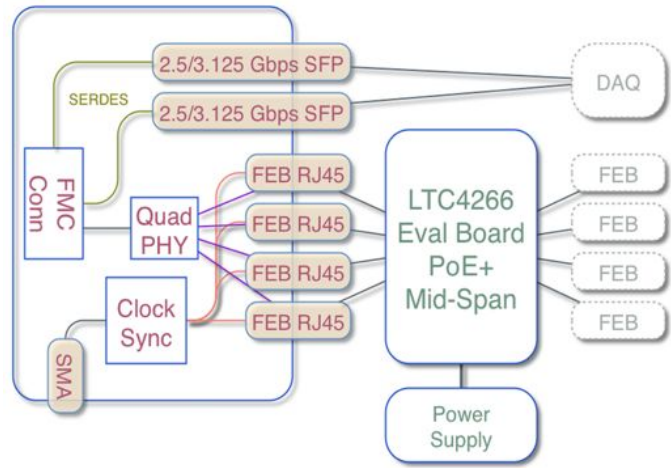

Figure 10.49.  Block diagram of the proof-of-concept readout controller mezzanine card.

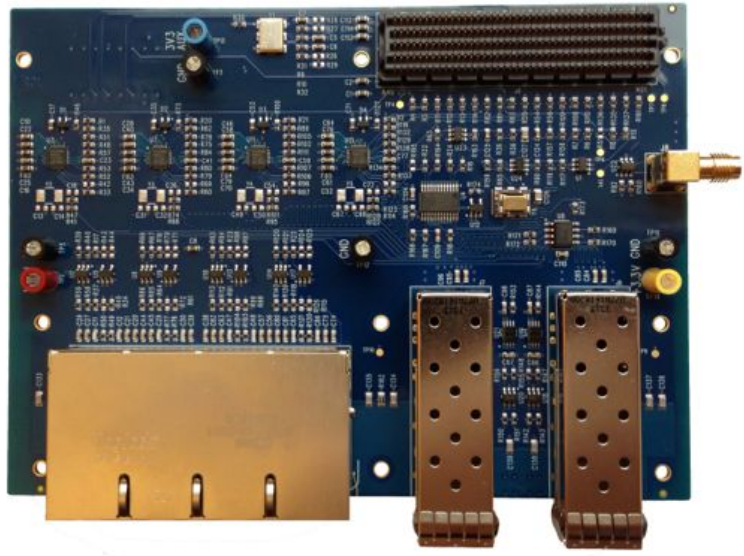

Figure 10.50.  Photo of the proof-of-concept readout controller daughter card.

### *10.7.5* **Status and Slow Control**

Status and control information is divided into two types, fast and slow. Slow status information such as power supply voltages, temperature, and run configuration will be pooled in the FEBs in a single memory block. This block will be sent to the readout controller at a rate of order 1 Hz. The readout controller in turn will collect the blocks from the FEBs and form an aggregate data pool accessible by the DAQ. This circumvents the need to send requests from the DAQ and acknowledges back from the queried device for each requested item. Writes to this same slow data pool in the readout controller by the DAQ will be periodically sent to the FEBs. Fast status information such as loss of event synchronization will be written into event headers. The times at which this class of error occurs will be latched and appear in the slow data pool.

### 10.7.6   Gain Monitoring

The SiPM bias voltage operating point is temperature dependent. Rather than controlling the SiPM temperature, which would be extremely difficult given the physical constraints of the system, we have opted to change the bias in response to changes in temperature. Thermometers on each counter motherboard will allow the temperature to be monitored. Note that other parameters, such as radiation damage, can also affect the operating point and hence we cannot solely rely on known temperature dependencies in adjusting the SiPM bias. The best means to track these changes is to monitor the single PE peak, as well as the relative rates of the single and double PE peaks. This will be done in the front-end board FPGAs by occasionally histogramming signals from the SiPMs, SiPM dark noise counts, during the inter-spill periods at zero-threshold and recording the low PE spectrum. Cosmic-ray muons will also provide a redundant means of calibration. There are sufficient resources in the FPGAs for this task.

### 10.7.7   System Performance

The expected data rates in the cosmic ray veto are given in Table 10.5. The beam spill has a duration 0.497 s and contains several hundred thousand pulses of 1.695-μs in duration. The overall duty factor during the spill is 0.574. After the spill is a 0.836-s-long period without beam. This cycle repeats every 1.333 s. During the period between spills when no beam is being delivered, normal data taking will continue. This will allow a direct measurement of the cosmic-ray induced background rate.

During the spill the average (maximum) rate in a counter, and hence the SiPMs, is expected to be about 127 kHz (685 kHz), primarily from background neutrons and gammas, as discussed in Section 10.10.2. The cosmic-ray muon rate itself over the entire CRV is 14.5 kHz [10]. The system is designed to handle a maximum rate of 1000 kHz per SiPM. The interspill rates are dominated by SiPM dark currents, which we expect to average 10 kHz at a threshold of 3 PE. Assuming a 12-byte hit size (see Table 10.6), we find the average (maximum) data rate per FEB to be 26 MB/s (213 MB/s). This is greater than the maximum data rate of 12.5 MB/s out of a FEB, and hence a trigger is needed. The average (maximum) total data rate produced by the CRV is 7.6 GB/s (62.9 GB/s). Simulations show that a rejection of 100 is expected from the software track-finding trigger. Hence, the triggered rates into the readout controllers and the DAQ are 76.1 MB/s (629 MB/s) average (maximum). The total amount of CRV data written out for the entire three-year run is expected to be 1 PB.

Table 10.5. Date rates in the cosmic ray veto. Average and design values are given. The total includes the data taken in the spill on and spill off periods. The spill on rates are from neutron and gamma interactions in the CRV; spill off rates are from SiPM noise. The cosmic-ray muon rate over the entire CRV is 14.5 kHz.

| | Item | Average | Design |
|---|---|---|---|
| **Spill-On Period** | Instantaneous hit rate/counter (channel) | 127.0 kHz | 1000.0 kHz |
| | Hit event size | 12 bytes | |
| | Instantaneous data rate/channel | 1.5 MB/s | 12.0 MB/s |
| | Spill on Time (s) | 0.497 | |
| | Spill duty factor | 0.574 | |
| | Average data rate/FEB | 56.0 MB/s | 441.1 MB/s |
| | Total data per FEB per live spill | 27.8 MB | 219.2 MB |
| **Spill-Off Period** | Instantaneous hit rate/counter (channel) | 10.0 kHz | 100.0 kHz |
| | Instantaneous data rate/channel | 0.1 MB/s | 1.2 MB/s |
| | Spill off Time (s) | 0.836 | |
| | Spill off duty factor | 1.000 | |
| | Average data rate/FEB | 7.7 MB/s | 76.8 MB/s |
| | Total data per FEB per interspill | 6.4 MB | 64.2 MB |
| **Total** | Average data rate/FEB | 25.7 MB/s | 212.6 MB/s |
| | Average data rate/CRV | 7.6 GB/s | 62.9 GB/s |
| | Total data per FEB per cycle | 34.3 MB | 283.5 MB |
| | Total data CRV per cycle | 10.1 GB | 83.9 GB |
| **FEB to DAQ** | Trigger rejection | 100 | |
| | Data rate out per FEB | 0.3 MB/s | 2.1 MB/s |
| | Average data rate to a ROC | 6.2 MB/s | 51.0 MB/s |
| | Total CRV data rate to DAQ | 76.1 MB/s | 629.3 MB/s |
| | Total CRV data for run | 1.1 PB | 9.3 PB |

Writing non-zero-suppressed data during the beam spill, and zero-suppressed during the spill off period, would increase the average triggered data rate per FEB to 16.1 MB/s; this would result in a total rate to the DAQ of 4.8 GB/s and produce 70 PB of data for the entire run. At present, a total data sample of that size makes this option unfavorable. It would also require five, rather than two data transfer controllers (DTCs). When data taking starts in several years, the price of storage may be such that this would be a viable option.

We have assumed a 12-bit ADC word, which is awkward from a data storage point of view. Compression to eight bits is desirable. The full ADC resolution is only important for 10 PE signals or less. An eight bit piece-wise linear scheme (e.g. halving the ADC data every 30 counts) can be used to maintain a precision that varies from 3% to 6% over the range of the ADC, with the exception of the first 30 counts. Other schemes are possible. The one requirement is that the implementation be feasible within the FPGAs.

Table 10.6. Hit format, assuming four samples per hit.

| Word | Byte | Data | Comments |
|------|------|------|----------|
| 1 | 1 | 53.1 MHz Clock | 18.8E-9 Clock Period |
|  | 2 |  | 1.23E-3 Total Period |
| 2 | 1 | FEB # | 9 bits 512 max |
|  | 2 |  |  |
| 3 | 1 | Channel # | 6 bits 64 max |
|  | 2 | ADC Count | 4 bits 16 max |
| 4 | 1 | ADC 1 | 12 bits 4096 max |
|  | 2 |  |  |
| 5 | 1 | ADC 2 | 12 bits 4096 max |
|  | 2 | ADC 3 | 12 bits 4096 max |
| 6 | 1 |  |  |
|  | 2 | ADC 4 | 12 bits 4096 max |

### *10.7.8* **Trigger**

The cosmic ray veto will be used to produce cosmic-ray muon triggers for calibration purposes. A simple time coincidence between pairs of adjacent counters on different layers will be used to indicate the presence of a cosmic-ray muon. The spatial positions of the hit counters will allow a crude track stub to be formed. A timing coincidence gate of 5 ns on individual counter hits will greatly reduce the false coincidence rate from noise and backgrounds. A window of ~125 ns will be formed around the coincidence time. Although such a trigger could be implemented in the FEB, it will be easier to do it with the analysis framework in the trigger buffer nodes, as is done with the tracker trigger.

### *10.7.9* **Safety**

There are two aspects of electrical safety for the CRV: (1) human safety and (2) detector equipment safety. Although the SiPM voltages are relatively low (< 100 V) compared to voltages required by PMTs, any device with an applied voltage under 50 V with significant capacitance could present a thermal hazard, and those with voltages greater than 50 V could present both an electrical and/or thermal arc hazard. The CRV electronics system will comply with recommended design and procedures as stated in the Fermilab safety procedures FESHM 5040 document [44]. Detector equipment safety will be achieved by monitoring procedures as described in Section 10.7.5, appropriate fusing/current limit settings for low voltage power supplies, and appropriate relay rack protection systems. In addition, devices such as the SiPMs have series quenching resistance that limits the overall current. These devices are self-limiting and cannot be damaged even with exposure to ambient light.

## 10.8   Module Fabrication

### 10.8.1   Introduction

The cosmic ray veto (CRV) module fabrication takes the scintillator extrusions, waveshifting fibers, and photodetectors, and assembles them into working modules. A total of 82 modules of seven different sizes will be fabricated (see Table 10.3), as well as 9 spare modules. Normal width modules consist of 64 scintillator counters, narrow modules have 32 counters, the counters arranged in 4 identical layers. The scintillator lengths range from 0.900 m for the cyro-modules to 6.600 m for the extra-long modules. Details of the module assembly factory, including drawings, can be found in Ref. [45].

The module assembly factory will be located at the University of Virginia (UVA). At the factory, individual polystyrene scintillator extrusions will first be glued into di-counter pairs. Then four fibers will be inserted into each di-counter and the fiber guide bar will be glued to each end to precisely position the fibers. The fibers will not be glued into the scintillator channels. A fly cutter polishes the fiber ends. The quality of the light transmission through each fiber will be measured with a special jig. Then the SiPM and counter motherboards are installed at each end of the di-counter and, using the LED flasher on the counter motherboard, the response of each SiPM will be measured to ensure it meets requirements before they are assembled into modules. The four layers of di-counters will be glued onto three aluminum plates, each layer staggered by 10 mm with respect to the nearest one in order to minimize inefficiencies due to projective cracks, as shown in Figure 10.22.

The photoelectron yield at each end of each module will be measured to ensure it meets requirements. This will be done with cosmic-ray muons using a test stand designed by Brookhaven National Laboratory and being fabricated at Fermilab. The modules will then be crated and shipped to Fermilab. The module assembly factory will produce an average of six modules per month during the production phase.

A short prototype module was fabricated at UVA in the summer of 2013 to be used in the test-beam studies described in Section 10.2.3 (See Figure 10.51). Experience from building this prototype provided important information used in the design of the module assembly factory and the fabrication steps detailed in this section.

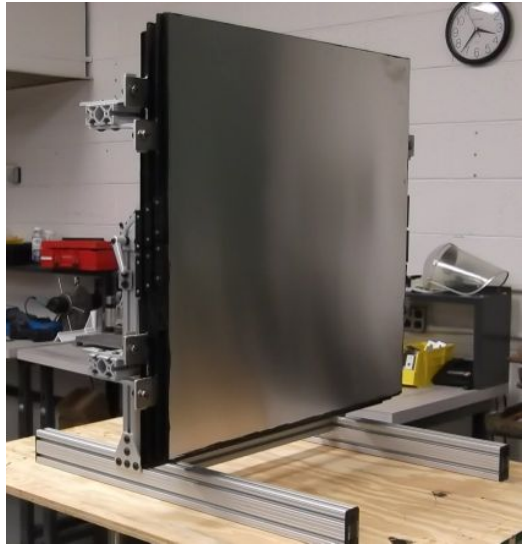

Figure 10.51. Photo of the short prototype module fabricated at UVA for use in test-beam measurements at Fermilab.

The module assembly factory will consist of 5000 ft$^2$ of rented commercial space. Table 10.7 itemizes the area required to fabricate and test the modules: we estimate that 5000 ft$^2$ will be adequate. The layout for the module assembly factory is shown in Figure 10.52.

The factory has five main areas: receiving and storage; counter assembly/QA, module assembly, module QA, and module crating/shipping. These are described in detail below.

### 10.8.2   Storage/receiving Area

A receiving area is required where trucks will be loaded and off-loaded. Depending on the facility resources two options are possible: a loading dock, which will require pallet jacks for loading and unloading, or no loading dock, which will require a forklift. Facility access for a large truck and room on the loading dock or an exterior room to maneuver a forklift or pallet jacks is required. After the delivery truck is off-loaded, materials will be moved to the storage area.

In addition to the module components that have already been described (scintillator extrusions, aluminum plates, and fiber) other module parts will be obtained and carefully inventoried. These include fiber guide bars, SiPM mounting blocks, counter motherboards, reflector manifolds, module end-caps, draw latches, front-end board enclosures, edge strips, etc. Three distinct aluminum pieces are required for each module type: the covers, inner and outer absorbers, and middle absorber. For the 7 module types 21 different aluminum parts are needed.

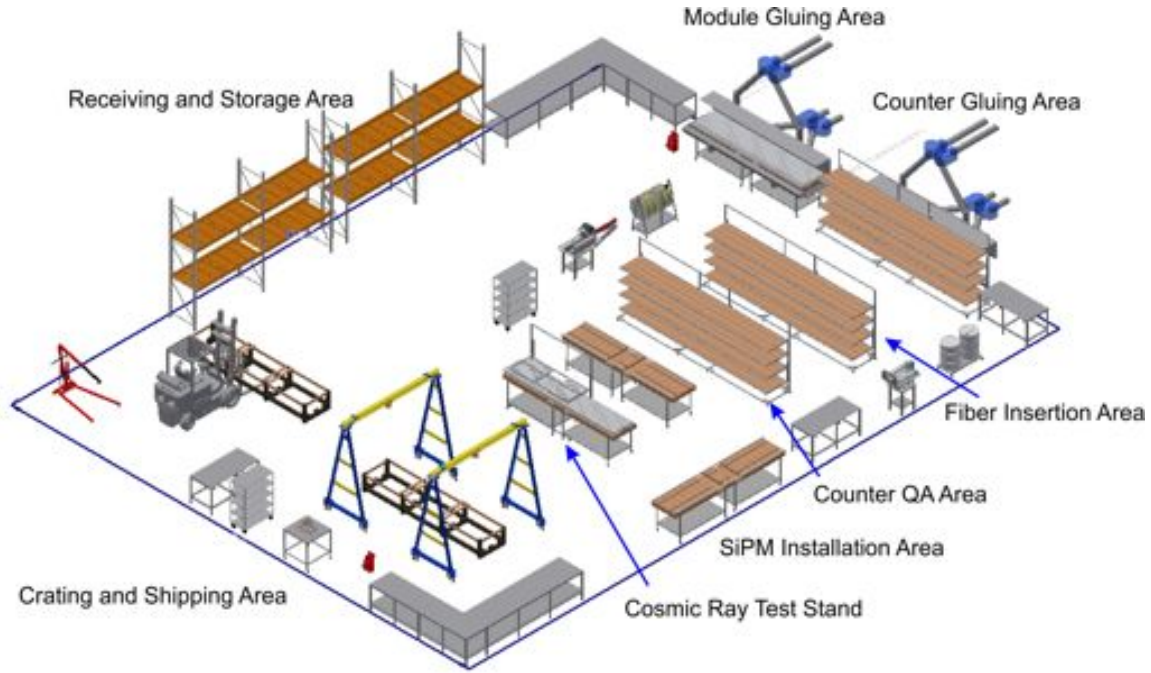

Figure 10.52. The layout of the module assembly factory showing storage areas, three counter assembly tables, two module assembly tables, and a module testing and crating area.

Table 10.7. Required space to house the module assembly factory.

| Required space | |
| --- | --- |
| Storage/receiving | 1500 ft$^2$ |
| Counter assembly/QA | 900 ft$^2$ |
| Module assembly | 400 ft$^2$ |
| Module QA test stand | 400 ft$^2$ |
| Crating/shipping | 1000 ft$^2$ |
| Total production space | 4200 ft$^2$ |
| Target space | 5000 ft$^2$ |

The storage area is large enough to store raw the raw materials, but not of the assembled modules. Pallets of aluminum will be approximately 4 ft. by 20 ft. long and weigh up to three tons. About 25 tons of aluminum is needed, and will be shipped in 9 pallets. The stacked aluminum pallets will require an area of 26 ft. by 26 ft. for egress, or about 700 ft$^2$. The Fermilab extrusion facility has enough space to store all the extrusions, which will be shipped to UVA in five batches of 8 to 10 crates each. One crate will hold 130 extrusions, enough for two modules. Approximately 45 crates of extrusions will be required for all the modules. Each crate weighs approximately 1400 lbs. and measures 3 ft. by 20 ft. by 2 ft. high. An area of 300 ft$^2$ is sufficient to store nine crates, which will be stacked three high. Storage for tools, fiber spools, machined parts and electronics is estimated to be 200 ft$^2$.

### 10.8.3    Counter Assembly/QA Area

The counter assembly area is where di-counters are made and tested. It consists of three counter-assembly tables, a chop-saw, a fly-cutter, a dust vacuum, two fume hoods, di-counter glue guides, the wavelength-shifting fiber spool, and fiber guide bar glue jig. The counter assembly table, shown in Figure 10.53, consists of four vertically moving shelves of work space approximately 3 feet wide by 20 feet long. The four counter shelves are mounted to 80/20 [46] rails that allow the shelves to be moved vertically, raising and lowering each shelf to the level of the saw and fly-cutter. The shelves are moved via a winch and pulley system to raise and lower each shelf to a 48" work/machine height. The counter assembly tables are mounted on locking casters and are moved under the fume hood during application of the adhesive.

The work done on each shelf will be performed on eight di-counter pairs at a time (one layer of a full-width module). A prototype counter assembly table has been constructed at UVA and its functionality will be tuned through the fabrication of a series of prototype modules.

A chop-saw and fly-cutter are mounted on a heavy duty machine cart, with locking casters that allow it to be rolled from one end of counter assembly table to the other and locked in place before energizing machines (see the photos in Figure 10.54). This eliminates the need to purchase multiple saws or to rotate the di-counters by 180 degrees.

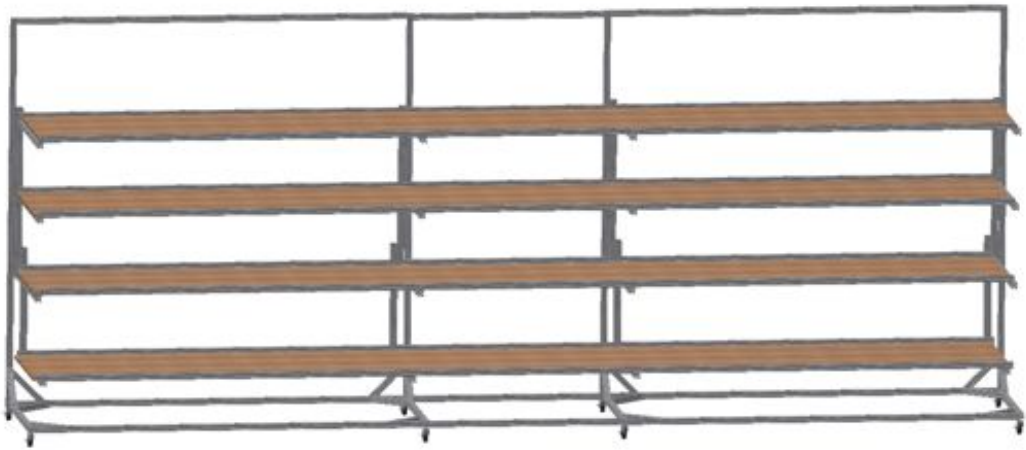

Figure 10.53. A sketch of the counter assembly table.

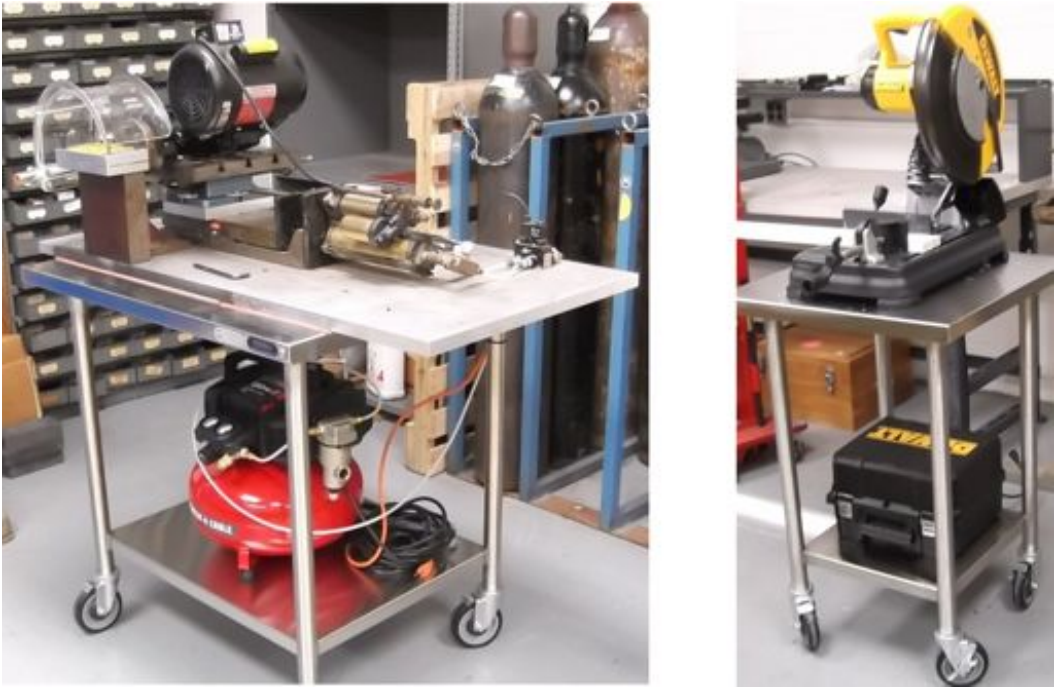

Figure 10.54. Photos of the fly-cutter saw (left) and chop-saw (right). Both are mounted on tables with castors so that they can be easily moved around the module assembly factory.

The chop-saw (DeWalt Multi-Cutter Saw, model DW872) with a 14-inch, 1300 RPM, 90 tpi carbide blade is used to cut extrusions to size. The prototype chop-saw machine table will be modified to perform the requisite 36-inch cross-cut for a layer of module di-counters. The chop-saw cross slide will be actuated manually by machine operator and an aluminum stage and clamping system will be fabricated in house.

A fly-cutter (used for the MINOS experiment) has been obtained from Argonne National Lab. It is used to polish the wavelength-shifting fibers. The fly-cutter needs to be modified to meet our needs. First the fly-cutter cross-slide stage needs to be extended to facilitate a 36-inch travel. This modification involves custom fabrication of a new cross-slide mount and counter clamping system out of aluminum, purchasing a new 36-inch precision cross slide, purchasing a new hydraulic system, and installing system components. A 80× digital microscope will be mounted on the new 36-inch fly-cutter cross slide to image the polished fiber ends in real time, eliminating the need to dismount the di-counters for the fiber optical QA step. The modified fly-cutter will retain the electric motor and spindle from the original Argonne machine and will be able to cut and polish one entire end of one full-width module layer of 32 inches. At 1.5 inches per minute it will take about 40 minutes to make the two required cuts. Including time for mounting, it will take one hour to polish each end of a module layer, and hence eight hours of fly-cutting effort per module. Assuming one out of eight module layer ends will need to be re-cut due to failed QA it will take nine hours of effort on each module.

Without the modifications to fly cutter listed above the total time would be 27 hours per module for fly cutting and fiber QA.

The following steps are performed to the scintillator extrusions to make di-counters.

1. Loose extrusions (quantity 16-20) will be placed onto each of the four shelves of the counter assembly table.

2. Extrusions will be prepped for gluing with scotch-brite and isopropanol.

3. Devcon HP-250 [19] epoxy will be applied to the extrusion edges.

4. Pairs of extrusions will be placed into glue guides and clamped into place until the epoxy sets (overnight).

5. The counter assembly table shelf position will be raised or lowered to the chop-saw machine height and the di-counters will be cut to length.

6. Glue test specimens (counters) will be cut and glued (see the QA section).

Once the extrusions are glued into pairs, each pair (32 in total per assembly table; 8 di-counters per shelf) will have their fibers and fiber guide bars installed as explained below.

1. The extrusion ends will be prepped for fiber guide bar mounting.

2. A waste fiber will be inserted into each channel to remove obstructions and to ensure a good fit.

3. The wavelength-shifting fiber spool will be positioned next to the counter assembly table, and four good fibers will be inserted into each di-counter.

4. The fiber guide bars will be de-burred, inspected and prepped for gluing with scotch-brite and isopropanol.

5. Devcon HP-250 epoxy will be applied to the di-counter ends, the fiber guide bars, and into the fiber channels of the fiber guide bars.

6. A glue jig clamping device will hold the fiber guide bars in place until the glue sets (3 days).

7. After at least three days of curing time, the fiber ends will be manually trimmed flush to the fiber guide bar with a straight X-Acto razor blade.

After the di-counters have had their fibers and fiber guide bars installed, the following steps are used to cut and polish the fibers and install the SiPMs and counter motherboards.

1. The counter assembly table shelf position will be raised or lowered to the fly-cutter machine height and the di-counters will be aligned and squared with it.

2. The di-counters will be fly cut: one 0.015" trim cut and one 0.004" polish cut.

3. During the fly-cutting process, the fiber ends will be imaged via the 80× microscope, and the image stored in a database.

4. Di-counters with fibers that fail the optical QA will be re-fly cut until they pass the optical transmission QA.

5. Di-counters will be tested with the fiber light transmission tester.

6. Di-counters that pass the optical transmission QA will have the manifold with SiPM and counter motherboards installed and will be LED flash tested.

7. The di-counter ends at the fiber guide bar-counter interface will be painted using black latex-enamel paint.

8. The di-counters will be moved to the module assembly area.

The counter assembly stage has time bottlenecks due to the curing time for the counter pairs (one day), the curing time for gluing the fiber guide bars (three days), and the fiber QA time (one day). Di-counter pairs for each module will require no less than five days at the counter assembly table stage. Therefore, it is highly desirable to have two or three counter assembly areas for the production facility. This allows the flow of assembly to continue during the three-day curing period for the fiber guide bars.

### *10.8.4*   **Module Assembly Area**

The module assembly area consists of two module assembly tables, a crane or portable lift (gantry hoist) access, and enough egress to safely move materials onto and off the tables. The tables have casters that allow them to be moved under a fume hood upon application of the adhesive. The module assembly area egress allows for open-air transit of materials from the counter assembly table, the storage area, and transit to the crating/shipping area. The area needed for each table is about 200 ft$^2$. The area for the material transport region is another 400-600 ft$^2$.

The module assembly table (as shown in Figure 10.55) consists of the following items.

• Two heavy-duty 3 ft. by 20 ft. machine tables with locking casters. Tables are rated for 3000 lbs. capacity.

• An 80/20 [46] frame attaching the two tables.

• A 2"x6" lumber frame with reliefs for fork lift access points.

• A 3/4" plywood top.

• The table has guides to locate the counter pairs onto the aluminum absorber plates and guides to offset counter layers 10 mm from each other during gluing.

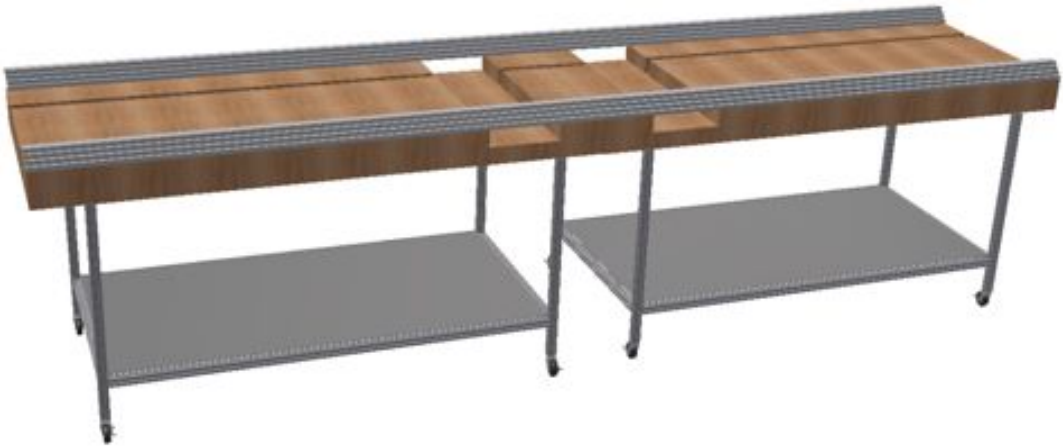

Figure 10.55. The module assembly table.

The steps to assemble each module are as follows.

1. A 0.032"-thick aluminum cover is placed onto module assembly table.

2. Devcon HP-250 epoxy is applied to the cover plate and spread to a uniform 0.005" thickness with a squeegee.

3. Eight di-counters are placed onto the wet aluminum cover plate and positioned using the module glue jig. The di-counters are shimmed so that they are uniformly spaced keeping the outside counters aligned with the edge of the aluminum cover plate.

4. Devcon HP-250 epoxy is applied to the surfaces of the di-counters and spread to a uniform 0.005" thickness with a squeegee.

5. A 0.250"-thick aluminum outer absorber plate is placed onto the wet di-counter layer and positioned using the module glue jig

6. Devcon HP-250 epoxy is applied to the absorber plate and spread to a uniform 0.005" thickness with a squeegee.

7. Eight di-counters are placed onto the wet aluminum plate and positioned using the module glue jig.

8. Steps 4-7 are repeated two more times to complete the four layers of di-counters.

9. Devcon HP-250 epoxy is applied to the surfaces of the di-counters and spread to uniform 0.005" thickness with a squeegee.

10. A 0.032"-thick aluminum cover is placed onto the wet di-counter layer and positioned using the module glue jig.

11. A plywood weight is placed on top to provide light contact pressure while curing.

12. Three days are allotted for curing before removing from module assembly table. Note that four glue-test specimens will be epoxied using the same batch as used for each module.

## 10.8.5   Module Quality Assurance Area

After curing, the module is lifted from the module assembly table to a special module transport table using a vacuum lifter. The table allows the module to be rotated if access to the other side is needed. A visual inspection is then made and the module is outfitted with the on-counter electronics. It is then moved to the module test stand (cosmic-ray test stand) for light-leak tests and to measure the photoelectron yields. Light-leak tests will be done by measuring the low-end photoelectron spectrum. Photoelectron yield measurements are described below.

The module QA area consists of a cosmic-ray test stand with two multi-layer cathode strip chambers, described in more detail below. An area of approximately 400 ft$^2$ (16 feet by 25 feet) will accommodate this function. The cosmic-ray tests will take two nights. After the module passes the QA tests, it will be moved to the crating area. Modules that do not pass QA tests in general cannot be repaired if the problems lie with the counters or fibers. If the failure is egregious the module will be tossed out. If the module just fails to meet the performance requirements it may be delegated as a spare, or used in an area of the CRV where the requirements are relaxed.

## 10.8.6   Crating and Shipping Area

The crating and shipping area consists of an open space large enough to load finished modules into shipping crates for transport to Fermilab. Before crating, module aluminum end-caps will be installed to protect the delicate counter manifolds. Each shipping crate is approximately 3 ft wide by 20 ft long by 3 ft tall and weighs up to 4,000 lbs when loaded with two modules. They can be stacked three to five high. Approximately 40 shipping crates are needed. Although the baseline plan is that the shipping crates will be fully enclosed plywood boxes, an alternative heavy-duty pallet system is being considered.

The crates will be delivered as needed. An area of 600 ft$^2$ is needed to store crates and crating materials. An additional 300 ft$^2$ is needed for equipment access so the total crating and shipping area should be at least 900 ft$^2$.

## 10.8.7   Quality Assurance at the Module Assembly Factory

At the module assembly factory several procedures will ensure the quality of the CRV modules. These are summarized below and described in detail in the quality assurance program document for the module assembly factory (Ref. [47]). Note that the active module components – fiber, scintillator, photodetectors, and electronics – will have all passed quality control requirements before being shipped to the module assembly factory.

### *Counter Quality Assurance*

Fibers and counters are glued into the modules and cannot be replaced after production, so quality assurance on the counters before module fabrication is critical. For this reason, several QA tests will be performed on the di-counters before they are glued into modules. After polishing, the fiber ends will be imaged by a microscope and digital camera with a setup similar to the prototype setup shown in Figure 10.56. This is used to assure a smooth polished end for each fiber. After this visual inspection, one end of the fiber will be flashed with a calibrated light source and the yield measured on the other end. This test ensures that light can pass through the fiber without significant loss of intensity. A schematic for the fiber transmission tester to be used for this test is shown in Figure 10.57. It is designed by Dean Shooltz (of Shooltz Solutions LLC [48]), who built a similar device for Michigan State University for testing NOvA fiber after it was potted and fly-cut [49].

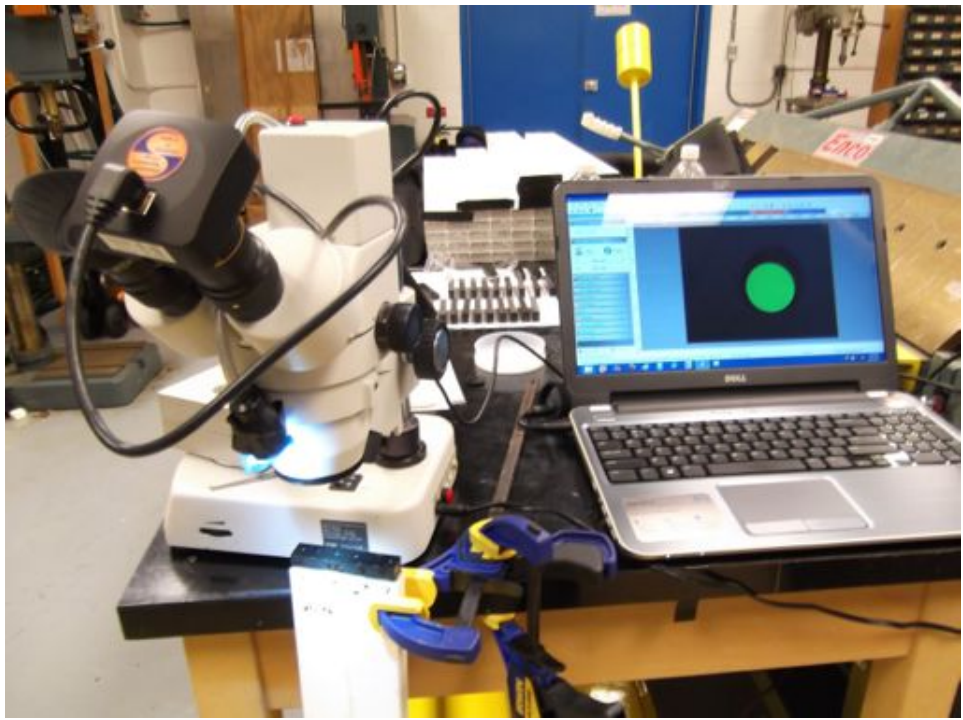

Figure 10.56. Photo of a fiber imaging setup at UVA. The microscope and digital camera image the fiber and the resulting image can be seen on the laptop.

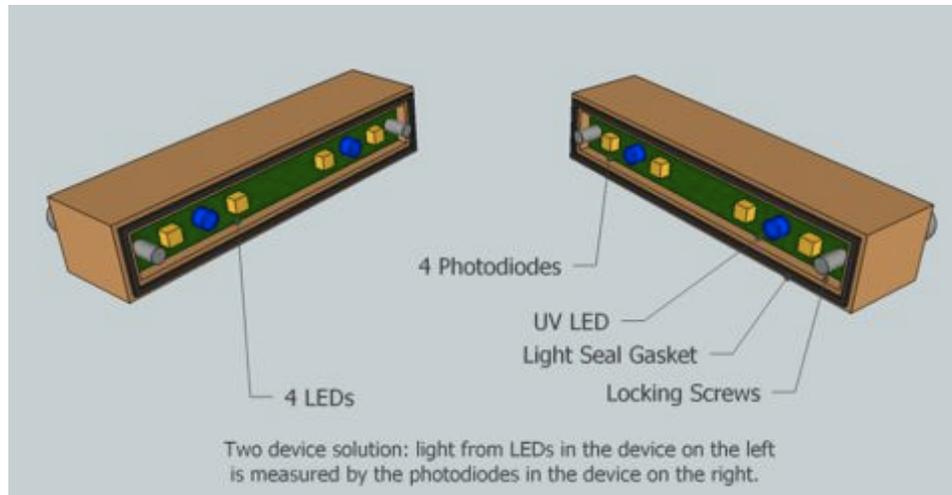

Figure 10.57. Diagram of the emitter-receiver device pair that will be used to test light transmission through each fiber.

***Module Quality Assurance***

The assembly of the CRV module critically relies on the quality of the glue bond between counters and the aluminum absorbers. Procedures outlined in Ref. [50] will be followed. Four samples of short extrusions glued to aluminum plates will be taken from each glue batch. Two will be subjected to shear and peel strength tests and two will be kept for further tests if needed.

The assembled modules will undergo performance tests using cosmic-ray muons before they are declared good and suitable for shipment to Fermilab. The tests will be done with a cosmic-ray test stand employing cathode-strip chambers (CSCs) designed by Brookhaven National Laboratory (Figure 10.58). The CSCs are being fabricated at Fermilab using jigs employed for the fabrication of similar CMS chambers, and will be shipped to UVA where details of the test procedure will be optimized using module prototypes. The test stand consists of two multi-layer CSCs, each with a 1×1 m$^2$ active area, a gas system, large scintillation counter trigger paddles, and readout electronics. The CSC width is greater than the module width: the module being tested will move along the cosmic-ray test stand. The tests will be done using the on-module counter motherboards and prototype front-end boards and readout controllers. The key performance parameter is the photoelectron yield, which will be measured at both ends of each module for each counter. Two runs of 10-hour duration each will be taken at each module end, allowing several hundred thousand vertical cosmic-ray muons to be accumulated.

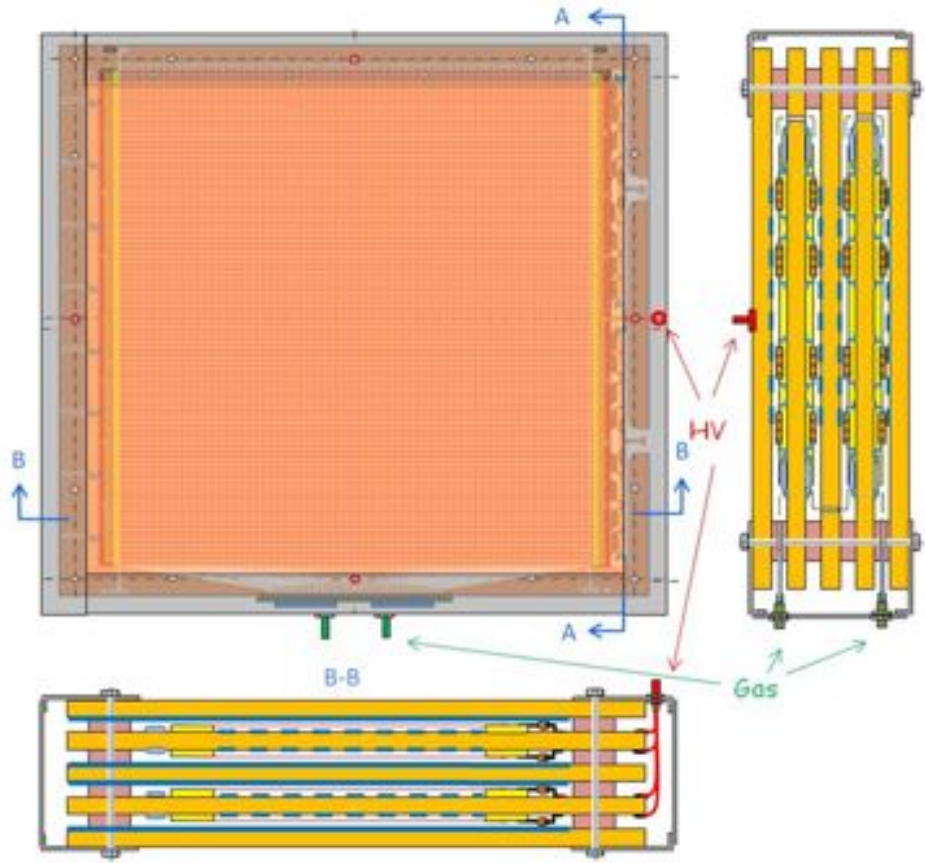

Figure 10.58. Cathode strip chamber design to be used in the cosmic-ray test stand.
In addition to the quality assurance tests described above, the following QA steps will be performed at the module assembly factory on materials used in the module production.

- Scintillator extrusions - dimensional tolerances, visual inspection for defects.

- Devcon HP-250 Epoxy - Glue test protocol, peel, shear, durability, hardness.

- Fiber guide bars (FGB) - dimensional tolerances, visual inspection for defects.

- Counter motherboards (CMB) - dimensional tolerances, electronics tests.

- SiPM mounting blocks (SMB) - dimensional tolerances, visual inspection for defects.

- Front end boards (FEB) - dimensional tolerances, electronics tests.

- Aluminum absorbers - dimensional tolerances, visual inspection for defects.

- Light tight paint and bumpers - visual inspection for defects.

- Module end caps - dimensional tolerances, visual inspection for defects.

All quality assurance tools will be calibrated and certified to the relevant national standards.

### 10.8.8    Personnel at the Module Assembly Factory

The module assembly procedure has been designed so that it is possible to carry out almost any step with a single person. The fabrication of prototypes and prototype assembly stations is being done at UVA by a single technician leader with occasional help from students and scientific staff. For the main module production effort the technician leader will be aided by another mechanical technician, and the two will do the majority of the work. Student labor will be employed for some repetitive tasks such as applying glue, de-burring fiber guide bars, etc. Graduate students, postdocs, and faculty will commission the QA tools, establish metrics, and analyze data from QA measurements. The technician leader will be in charge of the module assembly factory operation, the factory quality assurance program, and factory employee management. He will report daily on progress and fabrication issues to the faculty member charged with managing the fabrication task.

### 10.8.9    Environmental Safety and Health at the Module Assembly Factory

There are several materials and tasks at the module assembly factory that require training and special procedures to ensure the safety of factory employees. All employees and students in the UVA Mu2e group must study and sign the UVA standard operating procedures (SOP) document before working in the HEP laboratory or module assembly factory. Each employee will be trained in all current OSHA, Fermilab and UVA ES&H safety guidelines for identifying, mitigating and managing all relevant hazards at the job site. Operation of the chop-saw or fly-cutter or access to the machine shop will require UVA machine shop training. The cosmic-ray test stand uses compressed gas bottles and all personnel will be trained in their safe handling. The handling of the modules and heavy components using cranes, vacuum lifters, and fork lifts will only be done by the technician leader.

A significant amount of epoxy will be used to assemble each module. All epoxy contains toxic chemicals that require engineering controls to mitigate occupational exposure levels. Devcon HP-250 is less harmful than many other epoxies due to the absence of the known carcinogen MDA, which is used as a hardener. Nevertheless, all epoxies still contain a suspected carcinogen, MBCHA [51] that also fully vaporizes during curing. The National Institute for Occupational Safety and Health requires using ventilation for curing epoxy. The Department of Labor OSHA standard 1926.57(b) describes ventilation requirements [52]. The ventilation must remove airborne contaminants to or below acceptable levels. The ventilation system shown in Figure 10.59 will be verified via air sampling tests during prototype assembly activities. A similar design will be adapted for the counter assembly tables.

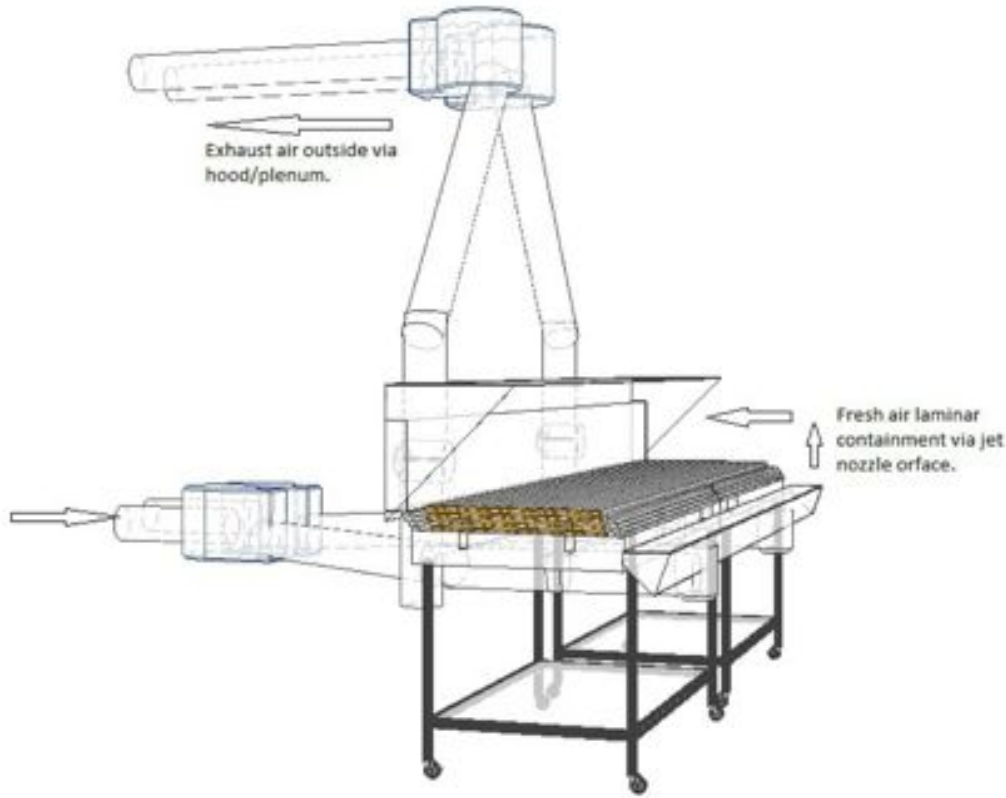

Figure 10.59. Drawing of the ventilation system for the module assembly table.

### 10.8.10  Summary of Module Factory Procedure and Rate

Assembly factory jigs are being constructed at the HEP laboratory at UVA, including a counter assembly table and a module assembly table. They will be used to fabricate the following prototypes.

- A 4.7-m-long module, not outfitted with electronics, to study fabrication procedures and make time and motion studies.

- Two cryo-length modules, outfitted with electronics, to study mounting techniques and to serve as test platforms for the CRV electronics. One will be shipped to Fermilab.

- Two medium-length modules, outfitted with electronics and the production scintillator and fiber. They will be shipped to Fermilab to test installation and mounting procedures for the side modules.

Module production will begin with a pre-production phase in the HEP laboratory at UVA using the assembly stations that will be fabricated to produce prototypes. During this phase we will produce modules at a reduced rate. Due to limited space in the HEP building we will start with shorter modules (two short and five cryo) modules. Other specialty modules (three medium narrow and two long narrow) will also be produced in the pre-production phase.

The space in the HEP laboratory at UVA is insufficient for the production module assembly factory; hence a larger commercial space will be rented and outfitted as described above. During the production phase, the module assembly factory will produce six modules per month. This will make it possible to produce all of the remaining modules (10 extra-long, 24 long, and 45 medium) in less than 15 months. Time and motion studies during fabrication of prototype modules will improve this estimate.

## 10.9    Detector Installation and Commissioning

### 10.9.1    Introduction

Modules are received from the University of Virginia module assembly factory and staged in a storage facility at Fermilab. On arrival they will be tested using front-end boards at the Fermilab facility to check for damage during shipment. Modules will then be re-crated and stored until installation. After the shielding around the TS and DS has been installed, the module support system will be installed on the shielding. The modules will be transported to the Mu2e detector hall via Fermilab truck for installation.

The Mu2e facility has dual overhead crane coverage that will be used to install the modules. A combination of a vacuum lifter and transport jigs will be used in the installation. Final testing and commissioning will commence when installation is complete.

### 10.9.2    Module Receiving, Staging, & Storage

Before module fabrication begins, two prototype side modules will be received from the module assembly factory used to test the installation support structure and installation procedures. A mockup of a small part of the shielding will be constructed using existing shielding blocks. Module supports will be installed and a module lifting jig, shown in Figure 10.60, will be used to rotate the modules to the vertical position. Module sizes range from $6.60 \times 0.86 \times 0.12$ m$^3$ to $0.90 \times 0.86 \times 0.12$ m$^3$ and module weights range from 2071 lb. to 282 lb.

A total of 91 modules (82 + 9 spares) will be shipped to Fermilab in wooden crates; two to four modules per crate depending on the module size. The crates will be designed so that three to five can be stacked on top of each other. The staging area has not yet been identified: one possible location is the Wide Band Storage area where dual overhead cranes of sufficient capacity are available. An area of 4000 ft$^2$ to 5000 ft$^2$ is required and will be needed for several years. For the 9 spare modules an area of about 500 ft$^2$ is needed for indefinite storage. After arrival at the staging area, individual modules will be taken from the crates using a vacuum lifter and placed on two tables. Front-end boards will be installed and cabled and a set of simple electronic heartbeat tests using the readout controllers will be done to ensure no damage has occurred.

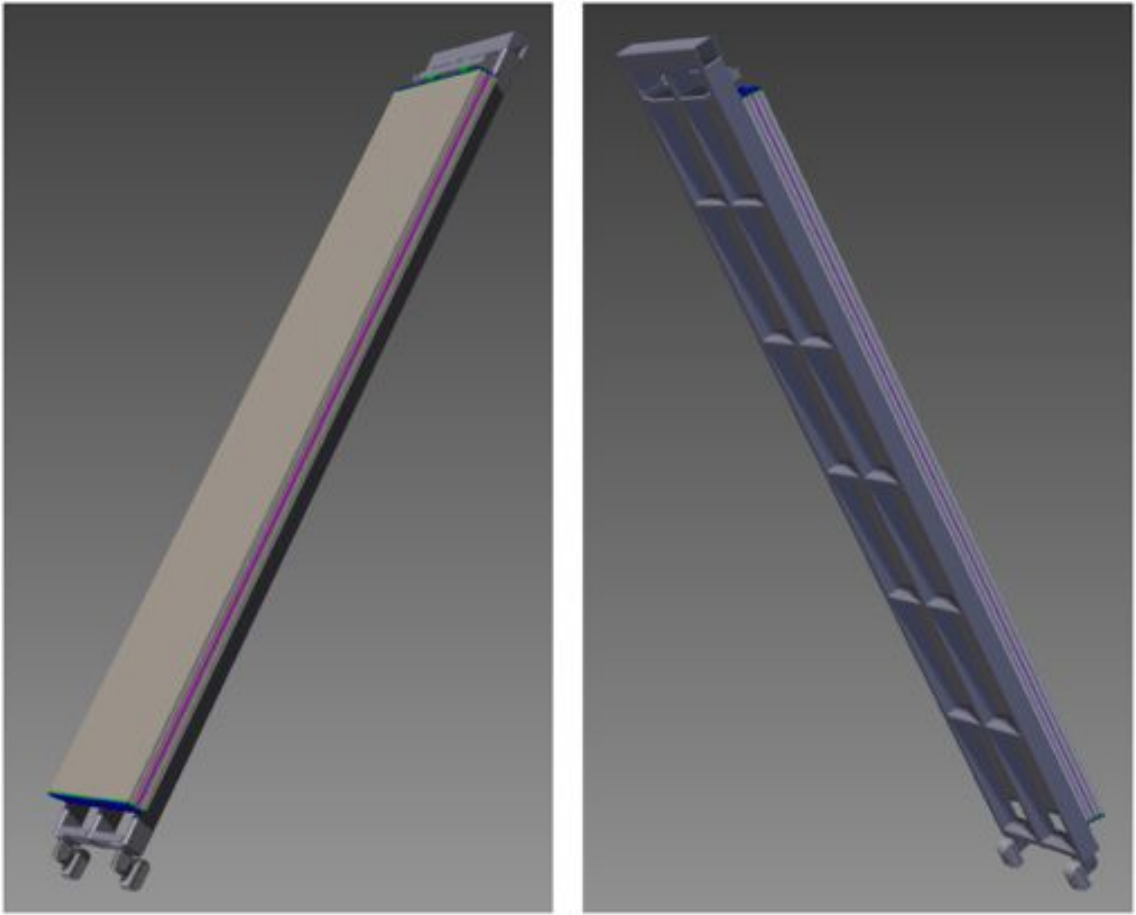

Figure 10.60. Front and back views of the module lifting jig.

### 10.9.3    Module Support Structure

The module support structure is designed to minimize gaps between modules, allow the modules to be installed and removed without undue difficulty, and to allow easy access to the electronics. It is anticipated that the need to replace modules will seldom if ever occur, although replacing defective electronics will be needed. The module support structure for the top and side modules consists of the top support bars and side hanger bars as shown in Figure 10.24. The top supports are leveled with adjustable feet and attached to the top of the shielding with concrete anchors. The hanger bars bolt to the top supports. Top modules rest on rollers mounted to the top supports. Side modules are bolted to the hanger bars and may be shimmed to adjust for variations in the shield block dimensions.

Endcap modules will be mounted to a strongback that will be positioned and mounted to the shield blocks, as shown in Figure 10.27. Cryo-modules will be mounted using a commercial framing system, such as Unistrut, that will be positioned around the cryo hole and attached to the shielding blocks.

### *10.9.4* **Installation**

Modules will be transported from storage to the loading dock of the surface building on Fermilab trucks in their shipping crates. The crates will be off loaded and modules will be removed from the crates with the vacuum lifter. The side modules will be installed first. They will be transferred to the module lifting jig in the horizontal position and then rotated to the vertical position using a crane, after which they will be lowered, moved into position, and hung from the side supports.

The two endcaps, CRV-U and CRV-D, will then be installed. The endcap modules first will be mounted onto their strongbacks in the horizontal position at the staging area, and then sent to the surface building. The assembly will then be rotated to the vertical and lowered into the detector hall as a unit by one of the building cranes.
The top modules will be installed after the side and endcap modules. They will be positioned with the vacuum lifter.

Finally, the modules covering the cryo feed at the DS will be installed onto their support structure.

### *10.9.5* **Testing & Commissioning**

After installation the modules will be cabled up and tested. The front-end electronics will have already been integrated into the DAQ and tested. A one-month commissioning period is anticipated using cosmic rays.

## 10.10 Effects of Neutron and Gamma Radiation on the CRV

The cosmic ray veto must operate in a high-rate background environment composed of neutrons and gammas, where the gammas are almost entirely produced by neutron capture. This radiation can have several deleterious effects on the CRV: it can produce spurious hits in the counters leading to false coincidences and excessive deadtime, and it can cause radiation damage to the electronics, scintillator and wavelength-shifting fibers.

Sources of these backgrounds are the following:

- The primary 8 GeV proton beam striking the production target and directly producing gammas and neutrons;

- The pions, muons, and other components of the muon beam that interact with the collimators in the transport solenoid to produce neutrons;

- Muons captured in the stopping target and the muon beam stop (MBS).

Stopped negatively charged muons when captured on nuclei produce about one neutron per capture for light nuclei such as aluminum, with energies up to 100 MeV. About 10 billion neutrons per spill second are produced from the stopping target alone.

Several physics processes contribute to energy deposition in the CRV counters from this background radiation. Fast neutrons can produce recoil protons from collisions with hydrogenous material that deposit visible energy through ionization. Slow neutrons can be captured, producing gammas with energies of up to 10 MeV. Gammas are also produced from bremsstrahlung and muon decay-in-orbit in the MBS and the last collimator. The gammas reaching the CRV can then deposit visible energy through Compton scattering, pair production, or the photoelectric effect.

### 10.10.1  Shielding Simulation Methodology

Extensive simulation studies to optimize the radiation shielding were made by the Mu2e Neutron Working Group. They are summarized in Ref. [53]. In order to facilitate rapid changes to the shielding model the G4beamline simulation package was employed [54]. The QGSP_BERT_HP physics list was used in order to correctly model low-energy neutron processes. The simulations were carefully cross-checked and found to agree reasonably well with MARS, as well as the GEANT4 package. The geometry module included detailed descriptions of the solenoids, the detector components, the shielding, and the detector hall.

The simulations were done in three steps. First the flux of particles produced in the production target was determined by simulating the proton beam interactions in the target. The results were checked whenever possible with existing measurements. The particles were tracked to the plane of Collimator 1 and the particle type, momentum, and position were stored. Using the stored information from the first step, a second simulation was done by tracing the particles and any produced secondaries until they either died or reached the counters of the CRV, where the time of arrival was recorded. Finally, using the fluxes at the CRV, the hit rate in the counters was determined using neutron and gamma efficiencies derived from an independent simulation. Other particles that made it to the CRV, such as electrons, arrive at low enough rates that they were ignored.

The shielding design, shown in Figure 10.61 and described in Section 7.9, is a result of a compromise between cost exigencies and meeting the radiation requirements of the CRV, tracker, and calorimeter. Care has been taken to reduce gaps in the shielding: those that are unavoidable, such as the labyrinth that houses the cryo feed for the detector solenoid, have been shown to produce no untoward effects. In the CRV region the main components of the shielding are 36-inch-thick (907 mm) concrete blocks that surround the transport and detector solenoids and onto which the CRV is mounted. Shielding near high radiation sources, such as the middle collimator and the stopping target employs barite-loaded concrete blocks to provide additional reduction. The design of the muon beam stop was optimized to minimize radiation to the CRV, as well as to the calorimeter and tracker.

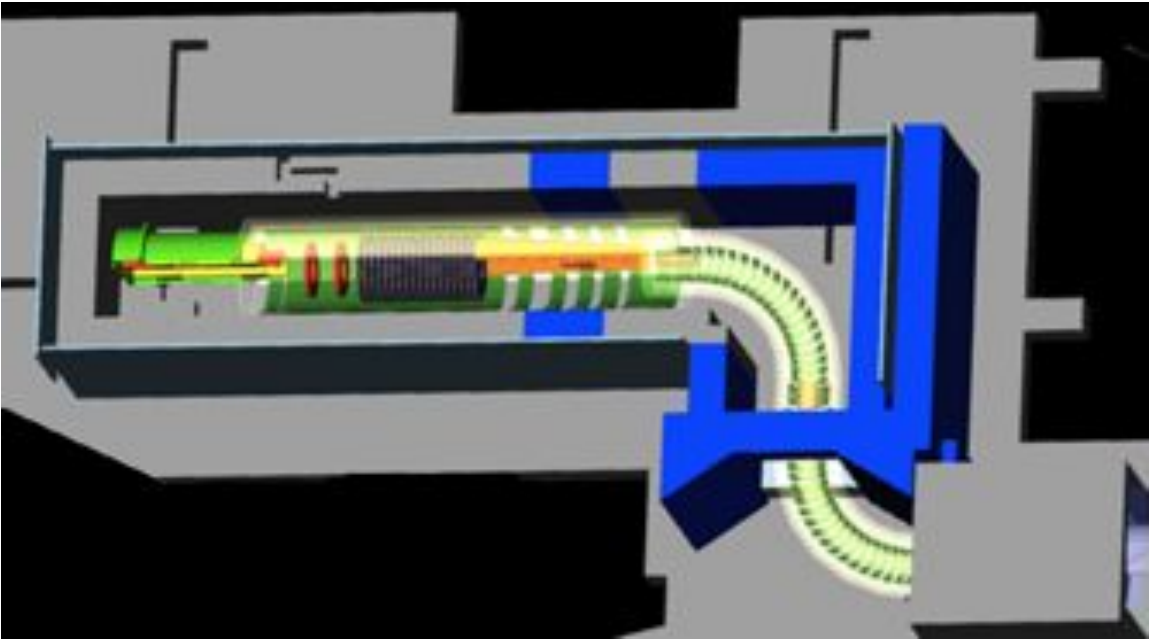

Figure 10.61. The shielding surrounding the TS and DS. Blue areas are barite-loaded shielding blocks. The top shielding blocks and CRV-T have been removed.

### 10.10.2  Rate and Deadtime Estimation

We only consider the backgrounds arriving at the CRV during the live gate, as the prompt background components have dissipated. However, particles are tracked through future bunches to include the effects from particle interactions in later signal windows. Note that different sources of background radiation have different time distributions:

- Fast neutrons from the proton solenoid (PS) region are prompt and have a steeply falling time distribution;

- Neutrons from the stopping target have a distribution given by the lifetime of the aluminum muonic atom;

- Gammas that originate from neutron capture have a delayed component due to the neutron thermalization time in the detector hall.

A muon stub in the CRV is defined to be the coincidence of hits in at least three of the four counter layers within a 5 ns time window. Background radiation can produce a false stub as shown in Figure 10.62 by either: (a) producing a coincidence of hits in three different counters from three separate source particles, (b) a coincidence of a hit in two counters from a single source particle with a hit in another counter from a second source particle, or (c) three hits in three counters from a single source particle. We denote these respectively as accidental, semi-correlated, and correlated backgrounds.

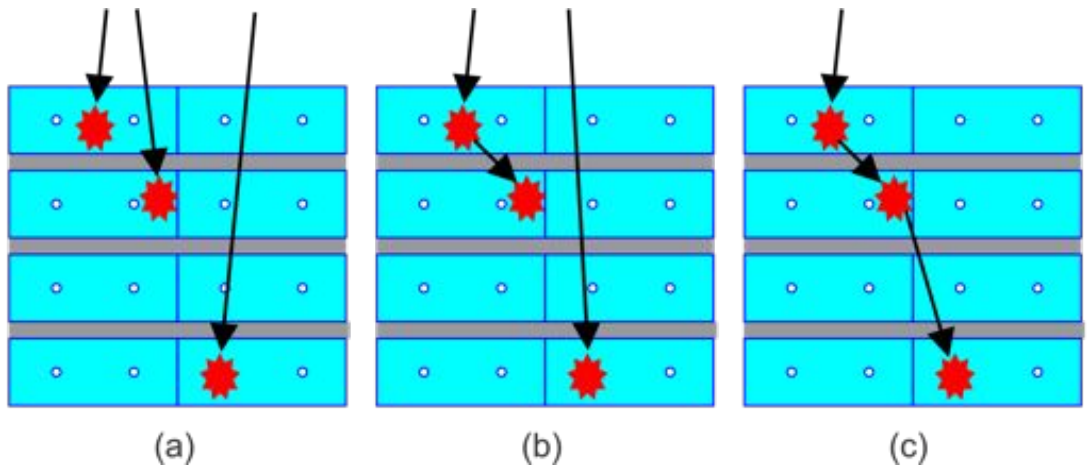

Figure 10.62. Sources of background track stubs in the cosmic ray veto: (a) accidental coincidences; (b) semi-correlated coincidences; and (c) fully correlated stubs.

To estimate the response of the CRV to neutrons and gammas a separate simulation was done, again using G4beamline. Neutrons or gammas are sent into a model of a CRV module at a normal angle of incidence. The visible energy deposition was scored in each layer from which efficiencies were determined. Figure 10.63 shows the energy spectrum from 10 MeV neutrons and gammas. The spectra were made for incident neutron energies of 0.025 eV (thermal) to 1.0 GeV; and gamma energies from 0.050 MeV to 50 MeV.

The efficiencies for various layer coincidence requirements are shown in Figure 10.64 for neutrons and gammas. At neutron energies below about 1 MeV the capture process dominates, producing gammas whose energy is independent of the parent neutron energy. Hence the efficiency is relatively constant. At energies above 1 MeV proton recoil begins to dominate, although the efficiency is suppressed by Birk's rule.

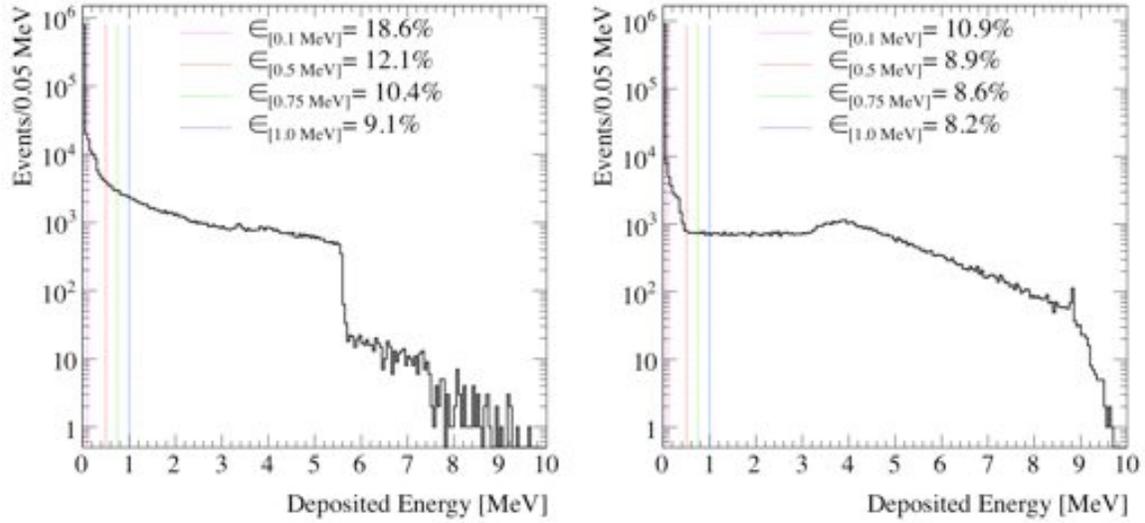

Figure 10.63. Energy deposition of 10 MeV neutrons (left) and gammas (right) normally incident on a CRV module. The spectrum at left falls off at 5.5 MeV due to Birk's suppression of the recoil protons, and the energy deposition above 5.5 MeV occurs from additional neutron capture predominantly on hydrogen or aluminum. The spectrum at right is dominated by Compton scattering and pair production processes. The percentages of events depositing more than 0.1, 0.5, 0.75, and 1.0 MeV are given.

There are two sources of deadtime from the CRV: that caused by real cosmic-ray muons and that caused by backgrounds. The former rate is 14.5 kHz, which corresponds to a deadtime of 0.2%. The latter is calculated as follows. The fractional dead time, $f_v$, for accidental, semi-correlated and correlated hits is given respectively by the following equations,

$$f_v = \sum_{i=1}^{N_{ct}} 3n_{1i}^3 \Delta t_v N_1 \Delta t_c^2 \qquad = 3\Delta t_v N_1 \Delta t_c^2 \sum_{i=1}^{N_{ct}} n_{1i}^3 \qquad (10.1)$$

$$f_v = \sum_{i=1}^{N_{ct}} 2n_{12i} n_{2i} \Delta t_v N_2 \Delta t_c \qquad = 2\Delta t_v N_2 \Delta t_c \sum_{i=1}^{N_{ct}} n_{12i} n_{2i} \qquad (10.2)$$

$$f_v = \sum_{i=1}^{N_{ct}} n_{3i} \Delta t_v \qquad = \Delta t_v \sum_{i=1}^{N_{ct}} n_{3i} \qquad (10.3)$$

where (1) $\Delta t_v$ is the veto time window of 125 ns; (2) $\Delta t_c$ is the coincidence time window of 5 ns; (3) $N_1$ is the total number of combinations to form a 3/4 coincidence from accidental hits per outer counter (108); (4) $N_2$ is the total number of combinations to form a 3/4 coincidence from semi-correlated hits per outer counter (14); (5) $n_1$ is the singles rate in any given counter (derived from [1-fold]/4); (6) $n_{12}$ is the singles OR doubles rate in any given counter (derived from the combined efficiency [1-fold]/4 + [2-fold]/2); (7) $n_2$ is the doubles rate ([2-fold]); and $n_3$ is the triples rate in any given counter including quadruples ([3=fold]+[4-fold]), and (8) $N_{cl}$ is the total number of counters per layer (1288).

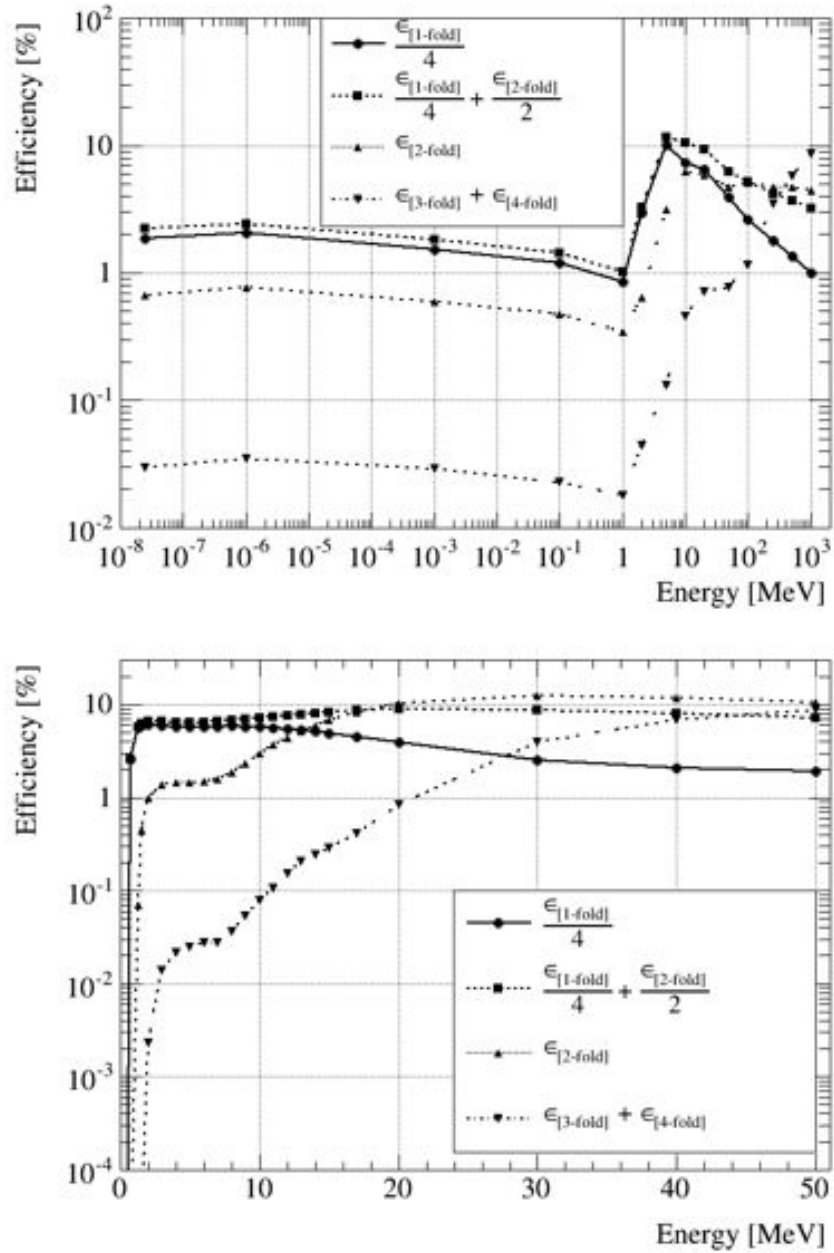

Figure 10.64. Detection efficiencies for neutrons (top) and gammas (bottom) producing an n-fold coincidence in a CRV module, using a visible energy requirement of at least 0.500 MeV for each required hit layer.  The [1-fold]/4 curve gives the efficiency for an exclusive single hit in a specific layer; [2-fold]/2 is the efficiency to get a hit from an exclusive doublet event in a specific layer;  [2-fold] is the efficiency to get a hit from an exclusive doublet event in any two layers; [3-fold] + [4-fold] is the combined efficiency for hits in any three layers or all four layers.

The hit rates per CRV counter, $n_1, n_{12}, n_2, n_3$ are estimated using background radiation fluxes folded in with the factorized n-fold efficiencies from Figure 10.64. For convenience, we make the assumption that hit rates are uniformly distributed across the CRV counters: the sums over the number of counters in equations 10.1 through 10.3 become simply $N_{cl}$. We can then rewrite the equations to obtain the maximum rates at the CRV for a given dead-time fraction:

$$n_1 = \sqrt[3]{\frac{f_v}{3\Delta t_v N_1 \Delta t_c N_{cl}}} \tag{10.4}$$

$$\sqrt{n_{12}n_2} = \sqrt{\frac{f_v}{2\Delta t_v N_2 \Delta t_c N_{cl}}} \tag{10.5}$$

$$n_3 = \frac{f_v}{\Delta t_v N_{cl}} \tag{10.6}$$

From these equations the maximum instantaneous and integrated rates for accidental, semi-correlated, and correlated hits for a given deadtime fraction can be determined. They are listed in Table 10.8 and shown in Figure 10.65. The total beam live time over the three years of data taking is $1.5 \times 10^7$ s, and the average dimension of the CRV counters that have been used to find the rates is $490 \times 5$ cm$^2$.

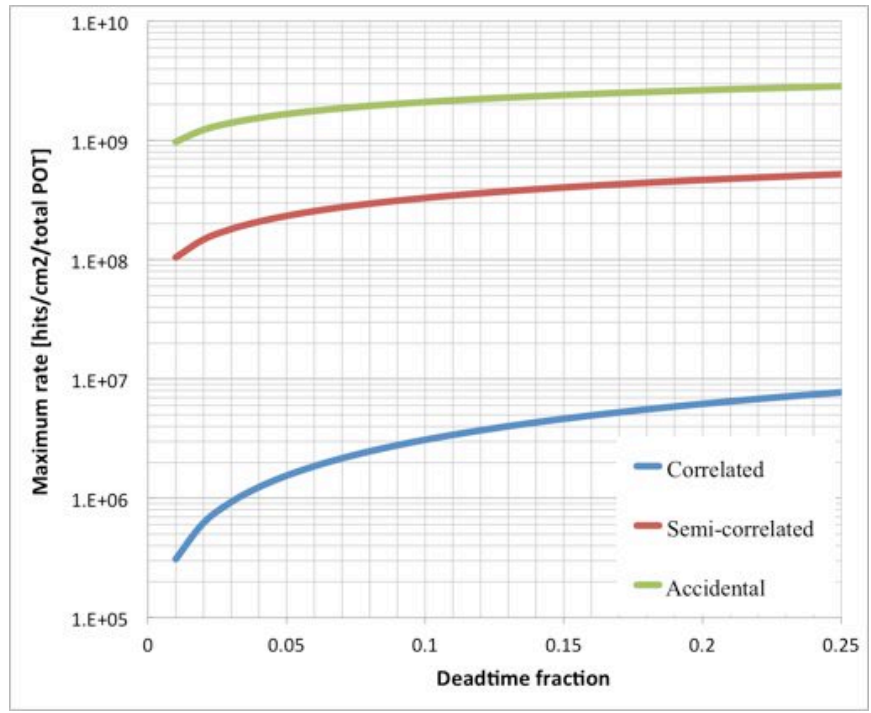

Figure 10.65. Maximum integrated rates as a function of dead-time fraction.

Table 10.8. Maximum instantaneous and integrated rates for various dead time fraction scenarios.

| Deadtime fraction | Maximum instantaneous rate [kHz/counter] | | | Maximum integrated rate [hits/cm2/total POT] | | |
|---|---|---|---|---|---|---|
| | Accidental | Semi-Correlated | Correlated | Accidental | Semi-Correlated | Correlated |
| 0.01 | 198 | 21 | 0.063 | $10\times10^8$ | $1\times10^8$ | $0.003\times10^8$ |
| 0.05 | 339 | 47 | 0.315 | $17\times10^8$ | $2\times10^8$ | $0.015\times10^8$ |
| 0.10 | 427 | 67 | 0.629 | $21\times10^8$ | $3\times10^8$ | $0.031\times10^8$ |

The integrated hit rates determined from the G4beamline simulation, and broken down by process, are shown in Figure 10.66, assuming a 0.500 MeV energy threshold. The average energy deposition from a minimum-ionizing particle in the CRV counters at normal incidence is 4 MeV. Simulations show that this proposed energy threshold will provide sufficient efficiency while test beam studies prove that sufficient light yield is achievable (see Section 10.2.3 on CRV requirements).

To estimate the rate at the front-end boards the same simulation was done with a 0.100 MeV threshold, which corresponds to about 3 PE at the near end of the extra-long counters and 1 PE at the far end. The results are shown in Figure 10.67.

The hit rates in the DS region have the following properties.

- The high rates at z = −2 m are due to PS sources. The impact on the correlated hit rate is smaller since the dominant component of the particle flux is the gammas below 10 MeV that originate from neutron capture.

- The bump at z = 4 m corresponds to the region in between the last collimator and the stopping target that is shielded by regular concrete.

- The dip at z = 6 m corresponds to the location of the stopping target which is shielded by barite-loaded concrete. The correlated hit rates show smaller impact from the barite region. Neutrons with energies up to 100 MeV originating from the stopping target are not effectively attenuated by the barite. However, the gamma component of the total flux is significantly reduced.

- The muon beam stop (MBS) is located at z = 16 m. The correlated rates in the MBS region are dominated by the high-energy gammas from muon decay in the polyethylene of the MBS.

The sources of the gammas and neutrons interacting in the cosmic ray veto are shown in Figure 10.68. The effect of the barite-loaded concrete around the stopping target region shielding gammas is clearly visible. The outline of an alcove near the upstream portion of CRV-R can be seen. Neutrons produced in the PS region traverse a notch in the shielding used for cryo-feeds to the PS. They stop in the alcove walls producing capture gammas that make it to the cosmic ray veto counters.

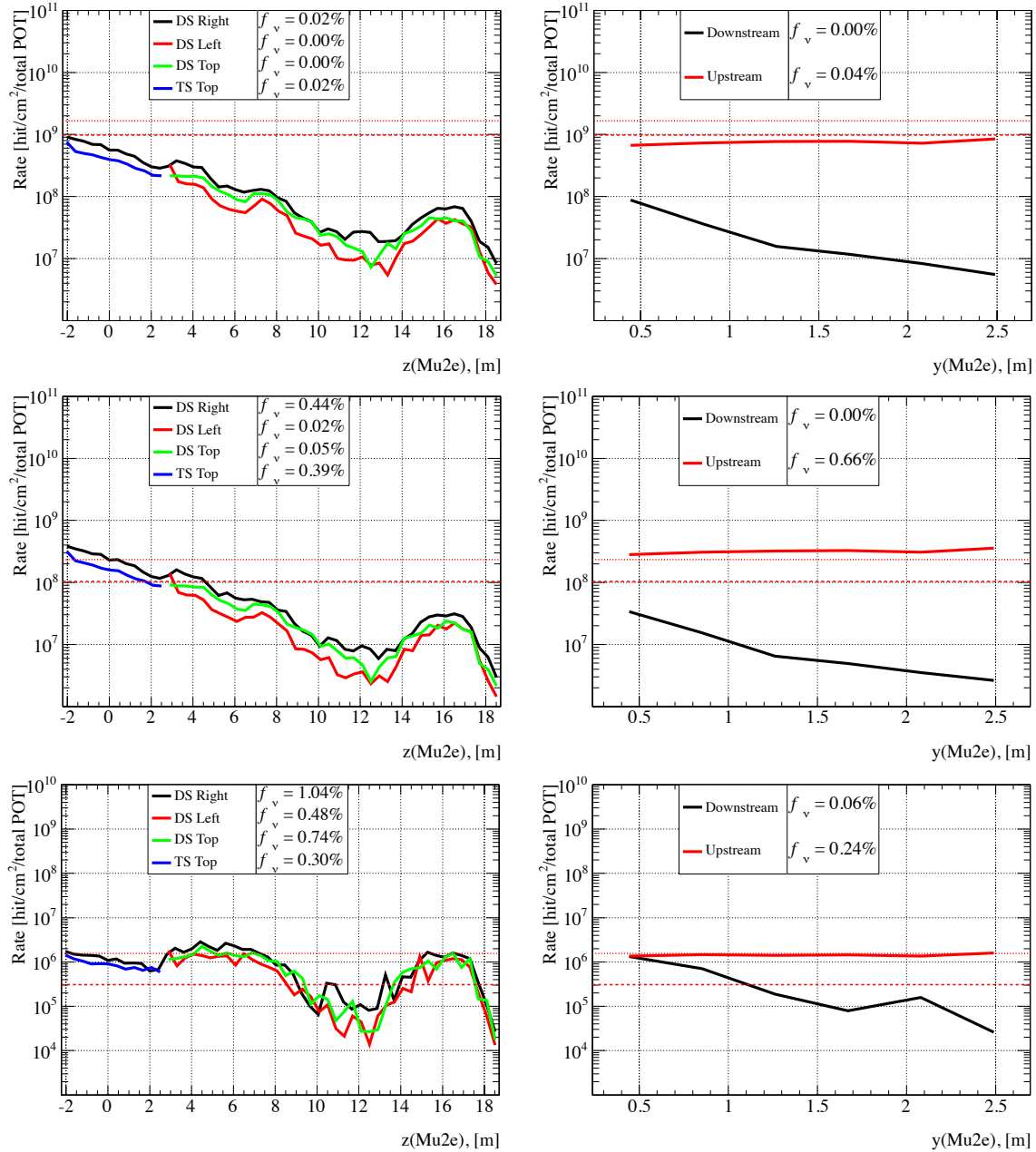

Figure 10.66. Accidental (top), semi-correlated (middle) and correlated (bottom) hit rates per unit area over the entire running period. Dashed and dotted red lines correspond to 1% and 5% fractional dead time assuming uniform flux distribution. In these plots, the Mu2e coordinate system is used, with its origin at the center of the Transport Solenoid.

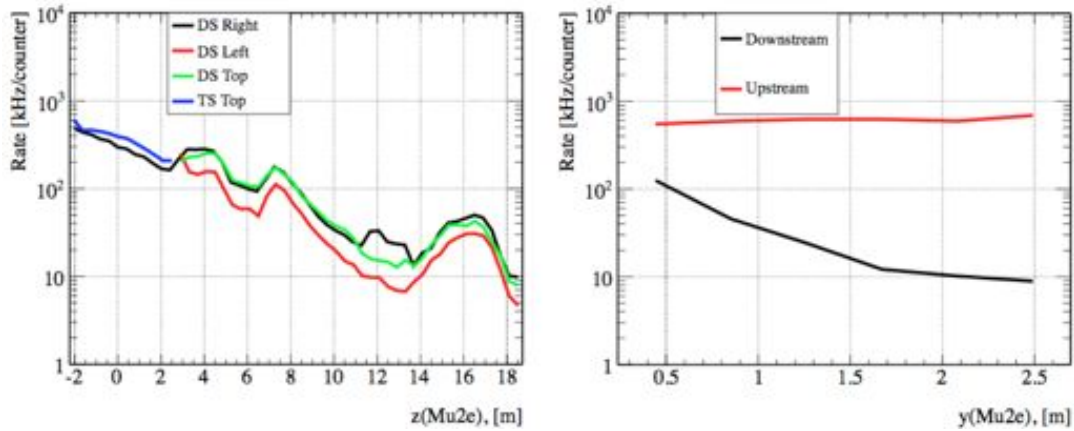

Figure 10.67. Expected instanteneous rates per counter for 100 KeV energy threshold.

The deadtime has been calculated for three different energy thresholds, and is given in Table 10.9. The cosmic ray veto has been designed to meet requirements with an energy threshold of 1.0 MeV. At this threshold the deadtime is 1.1%; a threshold of 0.500 MeV corresponds to a deadtime of 4.5%. These thresholds are applied offline.

Table 10.9. Fractional dead time for various energy thresholds

| Energy threshold [MeV] | Dead time [%] |
|---|---|
| 0.10 | 100 |
| 0.50 | 4.5 |
| 1.00 | 1.1 |

A significant fraction of the dead time is produced by the correlated hit rates from a single high-energy particle producing hits in multiple layers forming a 3/4 coincidence. Figure 10.69 shows the source breakdown for the correlated hit rates. The largest contributions to the dead time are driven by high-energy neutrons and primary or secondary gammas originating from the stopping target and the muon beam stop. An additional important source of the dead time originates from gammas produced from neutrons escaping the proton solenoid (PS) area directly through the shielding or through the cryogenics penetration notch in the upstream region. Finally, the collimators in the transport solenoid also produce a non-negligible source of dead time.

Table 10.10 gives the counter rates for three different thresholds. As mentioned above, a threshold of 0.10 MeV corresponds to between 1 and 3 PE in an extra-long counter, depending where the hit is along the counter. The 0.10 MeV numbers are used in determining the front-end board rates.

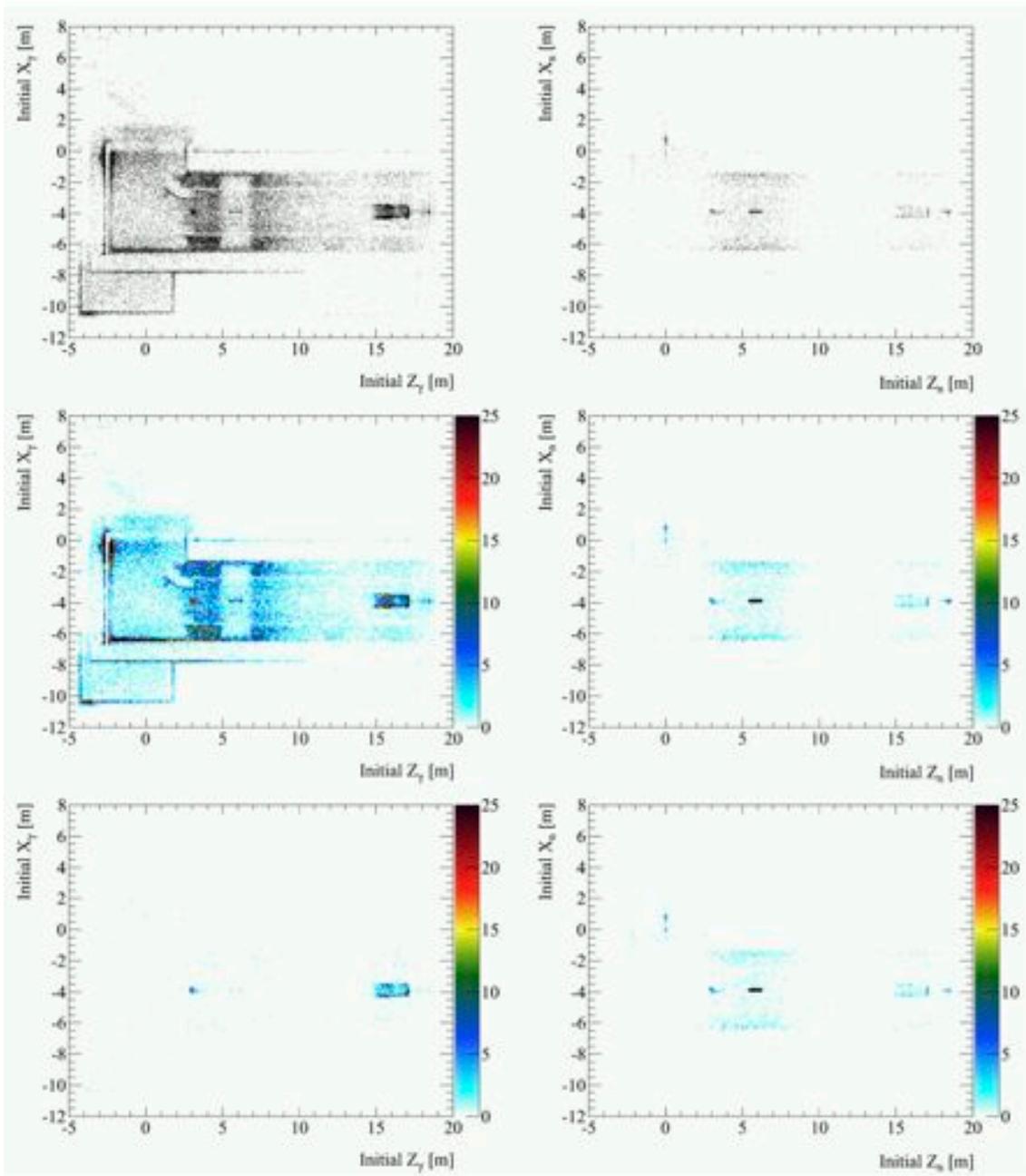

Figure 10.68.. Plan views of the origins of gammas (left) and neutrons (right) depositing energy in the CRV. The top set of plots are unbinned and show all points of origin, while the bottom two sets of plots are binned with events per bin color coded. Events in the time period before the live gate are removed from the plots. A kinetic energy threshold of 5 MeV is applied to each point on the top and middle plots. Bottom distributions are produced using a 10 MeV energy threshold that removes the gammas originating from neutron capture. The total number of simulated events corresponds to $5 \times 10^9$ POT.

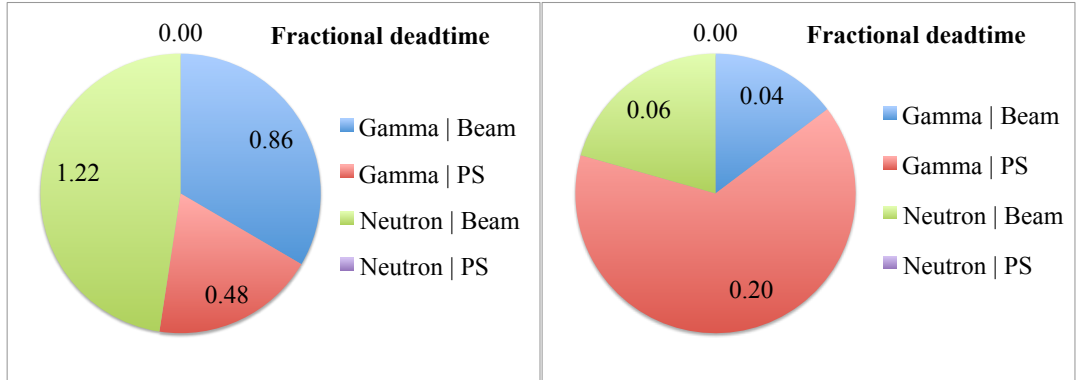

Figure 10.69. Source breakdown of the fractional dead time for correlated hit rates in the DS (left) and up/downstream (right) regions.

Table 10.10. Maximum and average expected rates per counter.

| Threshold [MeV] | Max Rate [kHz] | Average Rate [kHz] |
|---|---|---|
| 0.10 | 685 | 127 |
| 0.50 | 260 | 48 |
| 1.00 | 160 | 30 |

### *10.10.3* **Radiation Damage**

The damage effects by energetic particles in the bulk of any material can be described as being proportional to the displacement damage cross section. This quantity is equivalent to the non-ionizing energy loss (NIEL) and hence the proportionality between the NIEL value and the resulting damage effects is referred to as the NIEL-scaling hypothesis. As recommended for the LHC silicon detector study [56], the equivalent neutron radiation damage to silicon detector normalized to 1 MeV neutron is shown on Figure 10.70.

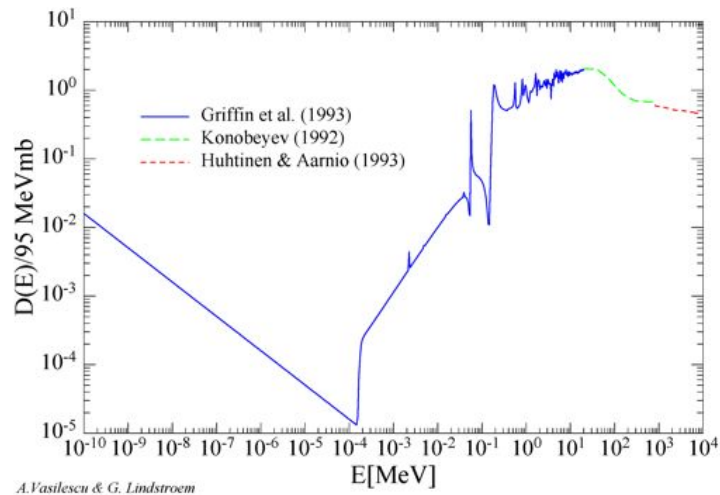

Figure 10.70. Neutron induced displacement damage in silicon [56].

In order to determine the radiation damage the neutron and ionizing radiation doses were simulated using the entire beam spill fluxes, rather than just the live spill window flux, as was done in determining the counter rates. The energy spectrum of the neutrons and gammas reaching the CRV are shown in Figure 10.73. The radiation damage to the electronics readout components and to the SiPMs is estimated from the total expected neutron flux normalized to the 1 MeV neutron induced displacement equivalent (see Figure 10.70). The damage rates expected at various regions in the CRV are shown in Figure 10.71. At the ends of the counters where the readout electronics would be mounted, if the dose is higher than $1\times10^{10}$ n/cm$^2$, passive reflector manifolds will be used.

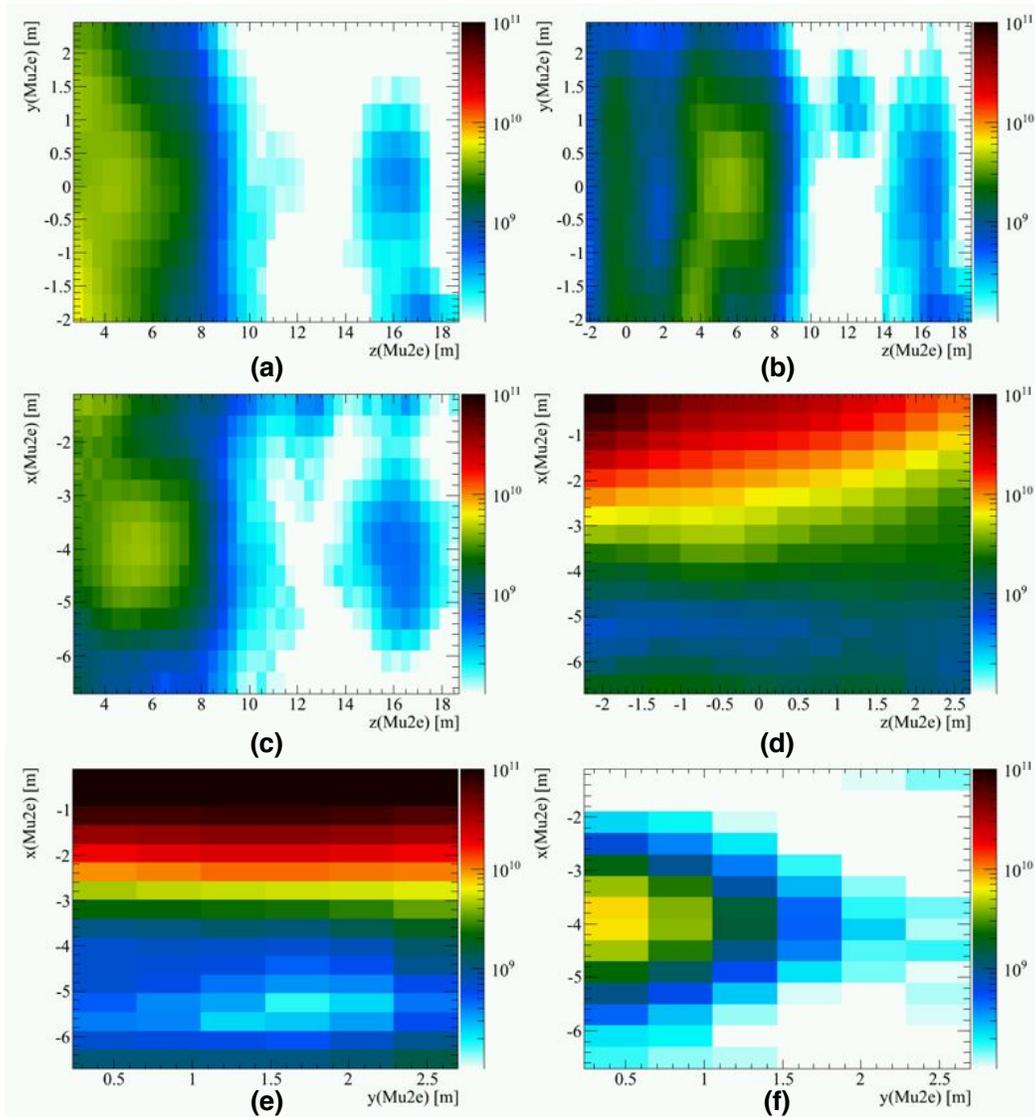

Figure 10.71. Expected non-ionizing radiation doses in: (a) CRV-L, (b) CRV-R, (c) CRV-T, (d) CRV-T extra-long modules, (e) CRV-U, and (f) CRV-D regions. The color scale units are expressed in n/cm$^2$ (1 MeV neq) over the total Mu2e running period.

The radiation damage to the polystyrene scintillator comes primarily from ionizing radiation from gammas [55]. Figure 10.72 shows the expected ionizing dose to the polystyrene integrated over the lifetime of the Mu2e experiment. The total ionizing dose level is expected to be less than 10 Gray (1 kRad), which is less than the dose level expected to produce significant damage to the scintillator or WLS fiber.

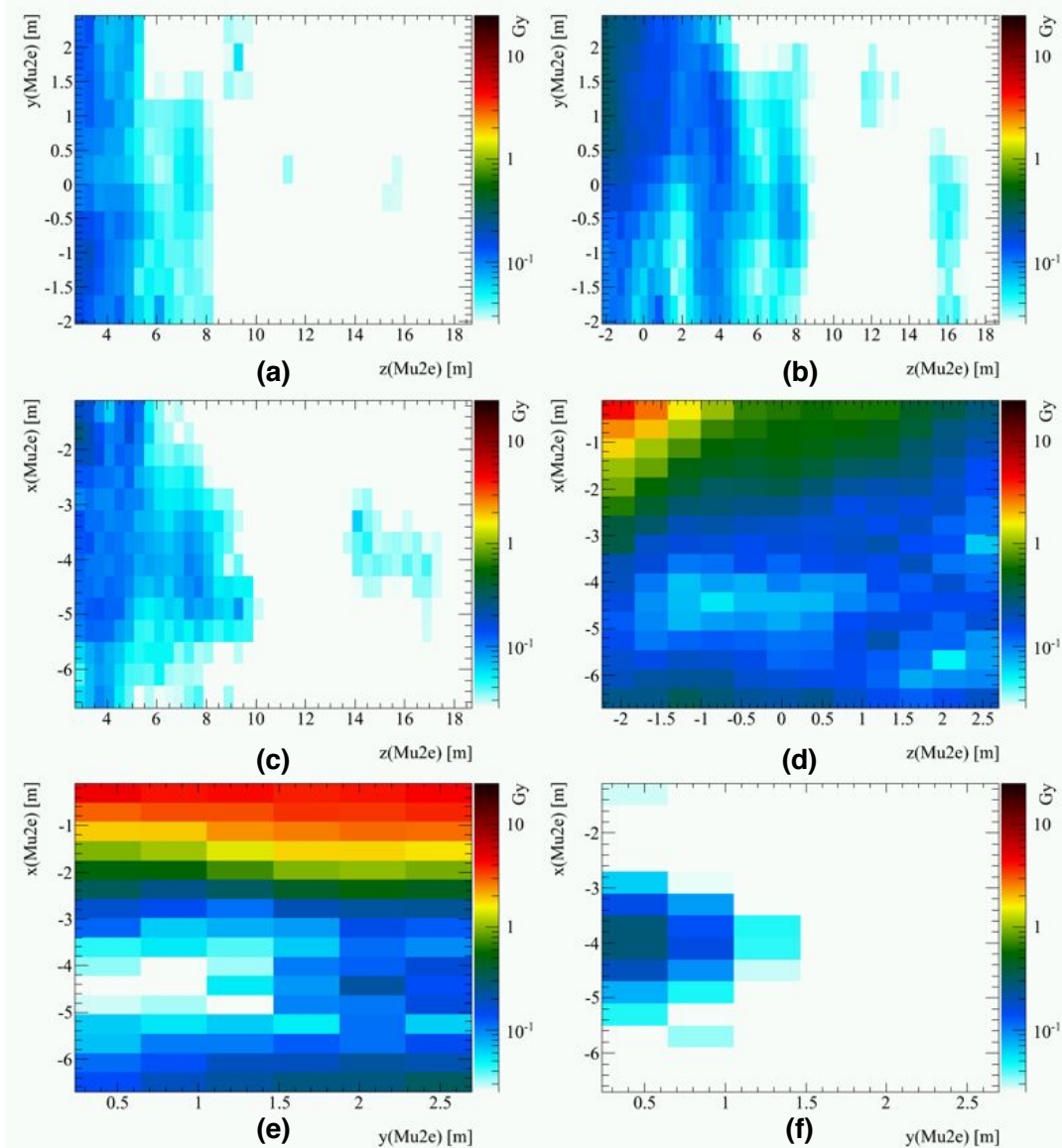

Figure 10.72. Ionizing radiation doses to the polystyrene in (a) CRV-L, (b) CRV-R, (c) CRV-T, (d) CRV-T extra-long modules, (e) CRV-U, and (f) CRV-D regions. The expected doses are below the level required to produce the visible damage to scintillating counters or WLS fibers.

## 10.11  Calibration

The spatial, temporal, and energy response of the CRV needs to be known and tracked with time. The counters are up to 6.6 meters long and 5 cm wide, providing crude spatial resolution. This puts rather loose requirements on the survey values needed after installation, and there is no need to track the survey values, except after modules have been removed and reinstalled, for example, to access the neutron and gamma shielding.

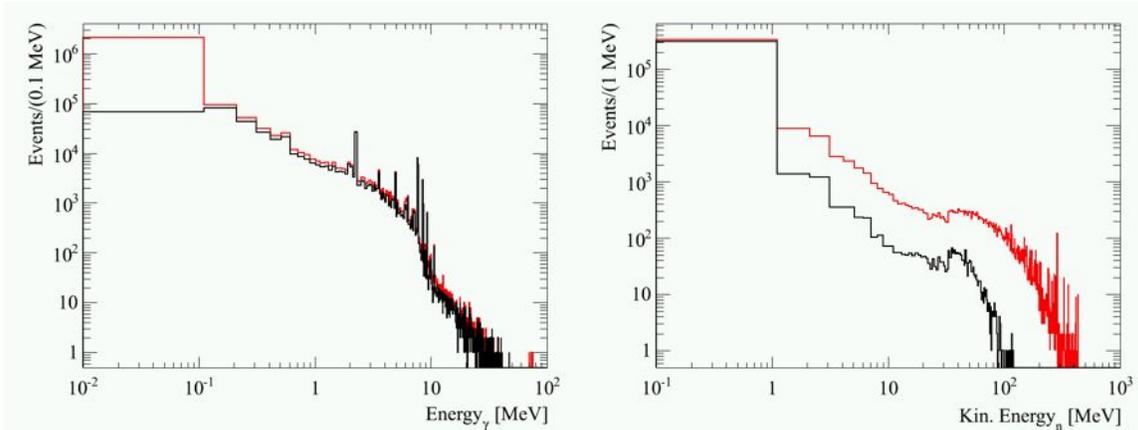

Figure 10.73. Energy spectrum of gammas (left) and neutrons (right) reaching the CRV. Black curves correspond to the radiation sources expected in the live spill, excluding first 670 ns after POT. (In the Mu2e analysis, the first 700 ns after protons strike the target will be excluded). Red curves correspond to all sources.

The SiPM efficiencies and time resolution within a counter can be measured and monitored by comparing the response of the two SiPMs on the same end. The copious rate of cosmic-ray muons traversing the CRV will allow local time variations to be measured and taken out. This can be bootstrapped over the entire detector. We note that although local time variations between SiPMs need to be small, global time variations do not. All cosmic-ray muons producing hits in the tracker and calorimeter traverse the CRV, and hence the timing of the various detectors can be well-determined relative to each other.

Although the CRV is essentially used as digital device, that is, all we need to know is whether a cosmic-ray muon traversed a counter, we do need to maintain a stable threshold of between 0.5 to 1.0 MeV in order to reduce the copious number of background hits in the counters and preserve a high experimental live time. This is done by first measuring the photoelectron yield of each counter at the module factory to normally incident cosmic-ray muons, which deposit 2 MeV/cm in the scintillator. The current plan is to do this at each end of each counter. The ability of SiPMs to resolve single photoelectrons makes this feasible. The LED flashers on each end of the counters facilitate monitoring changes in the SiPM response over time.

About 14,500 muons traverse the CRV every second. Their path length, related to their energy deposit, is shown in Figure 10.74 for both top and side modules, where all cosmic-ray muons have been included. Clear peaks are seen which will be used to monitor the calibration throughout the running period. The peaks can be sharpened by make simple cuts on the track-stub combinations.

The most important performance parameter of the CRV is the rejection of cosmic-ray muons producing conversion-like electrons. This can be determined by measuring the number of electrons in the signal window when the beam is off and monitoring the fraction that are rejected by the CRV.

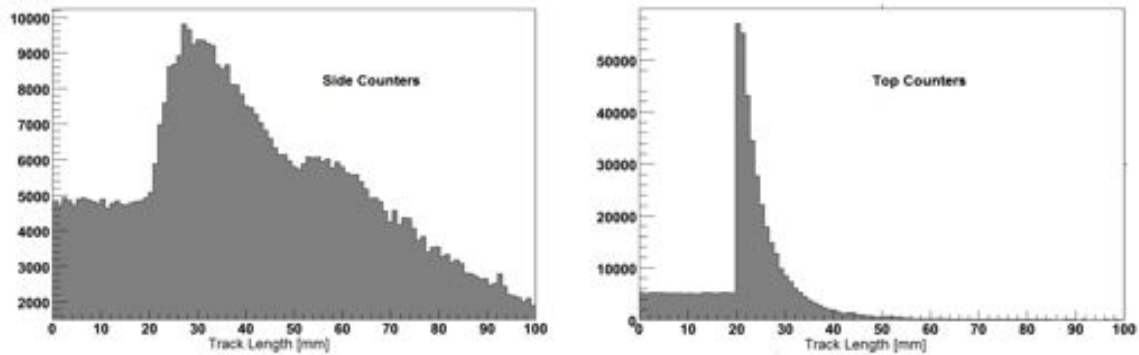

Figure 10.74. The path length for cosmic-ray muons through side (left) and top (counters).

## 10.12  ES&H

Polystyrene, the scintillator base and fiber core material, is classified according to DIN4102 as a "B3" product, meaning highly flammable or easily ignited. It burns and produces a dense black smoke. At temperatures above 300° C it releases combustible gases. This will be taken into account during the cosmic ray veto production, assembly, storage, and operation phases. Polystyrene scintillator and fibers are commonly used in experiments at Fermilab and the required safety precautions are well documented and understood.

Small quantities of adhesive will be used in fabricating counters and modules. Ventilation appropriate for these quantities will be installed in the module assembly factory and personnel working with adhesives will wear the appropriate personal protective equipment. For more details see Section 10.8.7.

The Fermilab-NICADD Extrusion Facility has a documented set of ES&H procedures that will be followed.

The size and weight of the modules require special precautions during handling. Explicit procedures for safely handling modules will be developed as part of a series of time-and-motion studies.

The photodetectors and electronics systems do not present any special safety issues. There will be no exposed low or high voltages and lockout/tag-out procedures will be used to ensure that systems are de-energized when they are being worked on. Everyone involved in the work on these systems will receive the required electrical safety training.

## 10.13 Risks and Opportunities

The risks associated with the performance of the baseline design for the cosmic ray veto are few as the technology is mature and has been used successfully by several Fermilab experiments. Potential risks are listed below. They all appear in the Mu2e Risk Registry [57].

### 10.13.1 Risks

1. More neutron and gamma shielding is needed, because either more refined simulations show that the neutron and gamma rate in the cosmic ray veto is larger than anticipated, or SiPM radiation damage tests show them more susceptible to radiation damage than present studies indicate. The calculations of the background rates and radiation doses described in Section 10.10 are quite mature and have been vetted using other software tools by the Neutron Working Group. The production mechanism for neutrons from captured muons is well measured and the simulations are straightforward. Nevertheless, the simulations are being improved with better models of the cosmic ray veto, shielding, and apparatus, and there is a chance that the rates will be found to be larger than present estimates. This risk that the SiPMs will not be able to operate at the required dose rates is small as vendors are aggressively competing to improve all aspects of SiPM performance. A higher-than-anticipated neutron rate would be mitigated by placing additional shielding in hot regions, or by replacing the present shielding with more effective materials, such as additional barite-loading shielding blocks.

2. More coverage will be needed with CRV-D. A recent simulation targeting cosmic-ray muons entering in the downstream sector show that three events over the course of the entire running period enter just below the module coverage [3]. All are vetoed by the calorimeter. Simulating with more statistics may show that some events are not vetoed by the calorimeter or CRV, and hence more coverage is needed.

3. A custom SiPM size will be needed, most likely due to the need to keep the dark-noise rate, which scales are SiPM active area, within specifications.

4. The photoelectron yield is less than expected. This would require a larger diameter fiber (and larger associated SiPM only in the case that a custom SiPM size is required).

5. There are a limited number of vendors of wavelength-shifting fiber, and only Kuraray, which has been the vendor of choice for every major experiment for several decades, produces fiber of sufficient quality. Their facility represents only a small part of a large enterprise; it could be shut down due to financial or other reasons. Inferior fibers of larger diameter would have to be employed to produce the required light yield, and the SiPM size would also have to be increased.

### *10.13.2* **Opportunities**

1. Collaborators at Brookhaven National Laboratory (BNL) have developed an ASIC to be used in the upgraded ATLAS muon chambers, called the VMM2 chip. This sophisticated ASIC has several very desirable features. It is designed to be radiation hard and to operate in a high magnetic field. Firmware development would be greatly reduced as it is designed to produce a peak amplitude and time stamp. The power consumption is less and the hit size is smaller. The cost is low as the development costs have all been amortized by BNL. The present design would require several modest changes to be suitable, which BNL has indicated they are willing to do. Samples of the ASIC will be obtained and tested.

2. If space can be found to fabricate the modules at on-campus facilities at UVA, this eliminates the need to rent space for the module fabrication factory. The search for suitable sites is being actively pursued by UVA collaborators.

3. The photoelectron yield is greater than expected due to the availability of SiPMs with higher PDEs. This would allow a smaller diameter fiber and perhaps a smaller SiPM to be used.

4. The gaps between counters are too large to allow the efficiency requirement to be met. The easiest way to mitigate this would be to fabricate wider counters, 6 cm to 7 cm, which allow wider gaps without adversely affecting the efficiency requirement. This would require minor redesigns to various components of the CRV, including items such as counter motherboards, modules, and the support structure. The increased cost would be offset by the smaller number of fibers, SiPMs, and front-end boards.

This page intentionally left blank

# 11    Trigger & DAQ

The Mu2e Trigger and Data Acquisition (DAQ) subsystem provides necessary components for the collection of digitized data from the Tracker, Calorimeter, Cosmic Ray Veto and Beam Monitoring systems, and delivery of that data to online and offline processing for analysis and storage. It is also responsible for detector synchronization, control, monitoring, and operator interfaces.

## 11.1  Requirements

The Mu2e Collaboration has developed a set of requirements for the Trigger and Data Acquisition System [1]. The DAQ must monitor, select, and validate physics and calibration data from the Mu2e detector for final stewardship by the offline computing systems. The DAQ must combine information from ~500 detector data sources and apply filters to reduce the average data volume by a factor of at least 100 before it can be transferred to offline storage.

The DAQ must also provide a timing and control network for precise synchronization and control of the data sources and readout, along with a detector control system (DCS) for operational control and monitoring of all Mu2e subsystems. DAQ requirements are based on the attributes listed below.

- Beam Structure
  The beam timing is shown in Figure 11.1.  Beam is delivered to the detector during the first 492 msec of each Supercycle.  During this period there are eight 54 msec spills, and each spill contains approximately 32,000 "micro-bunches", for a total of 256,000 micro-bunches in a 1.33 second Supercycle. A micro-bunch period is 1695 ns.  Readout Controllers store data from the digitizers during the "live gate". The live gate width is programmable, but is nominally the last 1000 ns of each micro-bunch period.

- Data Rate
  The detector will generate an estimated 120 KBytes of zero-suppressed data per micro bunch, for an average data rate of ~70 GBytes/sec when beam is present. To reduce DAQ bandwidth requirements, this data is buffered in Readout Controller (ROC) memory during the spill period, and transmitted to the DAQ over the full Supercycle.

- Detectors
  The DAQ system receives data from the subdetectors listed below.

*Calorimeter* – 1860 crystals in 2 disks. There are 240 Readout Controllers located inside the cryostat. Each crystal is connected to two avalanche photodiodes (APDs). The readout produces approximately 25 ADC values (12 bits each) per hit.

*Cosmic Ray Veto system* – 10,304 scintillating fibers connected to 18,944 Silicon Photomultipliers (SiPMs). There are 296 front-end boards (64 channels each), and 15 Readout Controllers. The readout generates approximately 12 bytes for each hit. CRV data is used in the offline reconstruction, so readout is only necessary for timestamps that have passed the tracker and calorimeter filters. The average rate depends on threshold settings.

*Extinction and Target Monitors* – monitors will be implemented as standalone systems with local processing. Summary information will be forwarded to the DAQ for inclusion in the run conditions database and optionally in the event stream.

*Tracker* – 23,040 straw tubes, with 96 tubes per "panel", 12 panels per "station" and 20 stations total. There are 240 Readout Controllers (one for each panel) located inside the cryostat. Straw tubes are read from both ends to determine hit location along the wire. The readout produces two TDC values (16 bits each) and typically six ADC values (10 bits each) per hit. The ADC values are the analog sum from both ends of the straw.

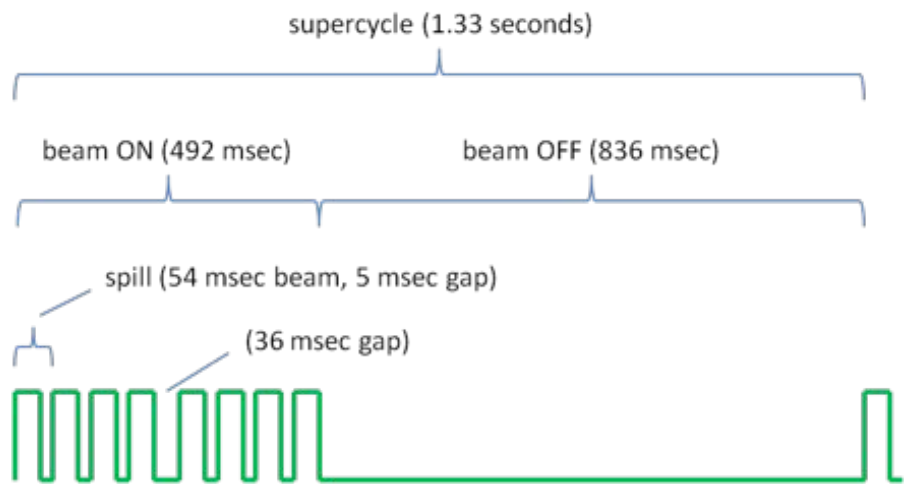

Figure 11.1. Mu2e Beam Structure.

- Processing
  The DAQ system provides online processing to perform calorimeter and tracker filters. The goal of these filters is to reduce the data rate by a factor of at least 100, limiting the offline data storage to less than 7 PetaByte/year. Based on preliminary estimates, the online processing requirement is approximately 30 TeraFLOPS.

- Environment
  The DAQ system will be located in the surface level electronics room in the Mu2e Detector Hall and connected to the detector by optical fiber. There are no radiation or temperature issues. The DAQ will however, be exposed to a magnetic fringe field from the detector solenoid at a level of ~20-30 Gauss.

## 11.2  Technical Design

The Mu2e DAQ is based on a "streaming" readout. This means that all detector data is digitized, zero-suppressed in front-end electronics, and then transmitted off the detector to the DAQ system. While this approach results in a higher off-detector data rate, it also provides greater flexibility in data analysis and filtering, as well as a simplified architecture.

The Mu2e DAQ architecture is further simplified by the integration of all off-detector components in a "DAQ Server" that functions as a centralized controller, data collector and data processor.  A single DAQ Server can be used as a complete standalone data acquisition/processing system or multiple DAQ Servers can be connected together to form a highly scalable system.

To reduce development costs, the system design is based almost entirely on commercial hardware and, wherever possible, software from previous DAQ development and open source efforts.

### 11.2.1   Interfaces

The DAQ subproject interfaces to all other subprojects (Table 11.1).

### 11.2.2   Scope

The DAQ subproject includes five main elements, listed in Table 11.2.

### 11.2.3   Schedule

Since much of the DAQ development effort involves software and firmware, the schedule is organized in three phases, each with a series of releases. Software components are developed incrementally and in parallel, so for most releases there is not a unique deliverable (e.g., a certain number of lines of code or a specific set of software functions). The first release is the general framework for all of the components and each subsequent release adds features and improvements. The three development phases are listed in Table 11.3.

Table 11.1.  DAQ Subproject Interfaces.

| Subproject | Interfaces |
|---|---|
| Tracker, Calorimeter, CRV | The DAQ connects to detector readout controllers via optical links that carry fast control, slow control and data. The Timing system supplies micro bunch frequency clocks to each detector subsystem. |
| Solenoids, Beamline | The DAQ provides the infrastructure for slow control and monitoring, and readout of the target monitor. |
| Accelerator | The DAQ receives beam timing and status information from the accelerator for timing system synchronization. The DAQ also provides the infrastructure for slow control and readout of the extinction monitor. |
| Civil | The civil construction subproject provides the surface level electronics room, power, and air conditioning for the DAQ. It also supplies cable chases for connecting the detector hall electronics to the electronics room. |

Table 11.2.  DAQ Subproject Elements.

| WBS | Element | Scope |
|---|---|---|
| 475.09.01 | Management | General subproject management, cost, scheduling, reporting, project controls. |
| 475.09.02 | System Design | Overall system requirements, architecture, design reports. |
| 475.09.03 | Data Acquisition | Hardware, firmware, and software for transport of data from the detector readout controllers to the online processing. Timing system and event building network. |
| 475.09.04 | Data Processing | Servers and software for online processing, data filters, and local data storage. |
| 475.09.05 | Controls and Networking | Software and hardware for general-purpose networking infrastructure, detector control system (slow control), and control room. |

## *11.2.4*  **DAQ Architecture**

Figure 11.2 shows the basic Mu2e DAQ system architecture.  Readout Controllers digitize and zero-suppress data at the detector.  The data is then transmitted over optical links to DAQ Servers in the surface level electronics room.  Control information is sent from the DAQ Servers to the Readout Controllers over the same bidirectional optical links. Data is exchanged between DAQ Servers (via the Event Building Network) to form complete events. The DAQ Servers filter these events and forward a small subset of them to offline storage.

Table 11.3.  DAQ Subproject Development Phases.

| Phase | Scope |
|---|---|
| Prototype | Purchase of prototype system components. Implementation and test of interfaces for high-speed data transfer (firmware, software). Benchmarks of online processing filters. Initial port of *artdaq* software to Mu2e.  Timing system characterization. Interface tests for prototype detector readout controllers. |
| Pilot | Purchase of pilot system components.  Development of a small test system (6 servers) with complete end-to-end readout, event building, and processing. Initial Detector Control System implementation. Basic networking infrastructure. Start of formal Mu2e *artdaq* software and system firmware releases. Online filter development and performance testing.   Readout controller interface and bandwidth tests. |
| Production | Purchase of production system components.  Installation and test of full DAQ system. Full timing system interface to accelerator and detector. Completion of networking and control room infrastructure and local data storage.  Completion, debug and optimization of *artdaq* and slow control software.   Completion of online filter software. |

The software architecture (Figure 11.3) is based on *artdaq* [2]. This software runs on DAQ servers and on dedicated control and monitoring computers.  *Artdaq* is a toolkit of C++ 2011 libraries and programs for use in the construction of DAQ systems.  It provides functionality that includes the following:

- Management of the readout and configuration of the DAQ hardware.  This makes use of experiment-supplied software components.
- Routing of data between threads within a process, between different processes, and between different machines, and for assembling complete events from these data.
- Encapsulation of the data being routed, and support for experiment-specific raw data formats to provide type-safe data access.
- Event analysis and filtering using the *art* event-processing framework.
- Basic control and monitoring applications.
- Infrastructure for distributing configuration data to DAQ processes.

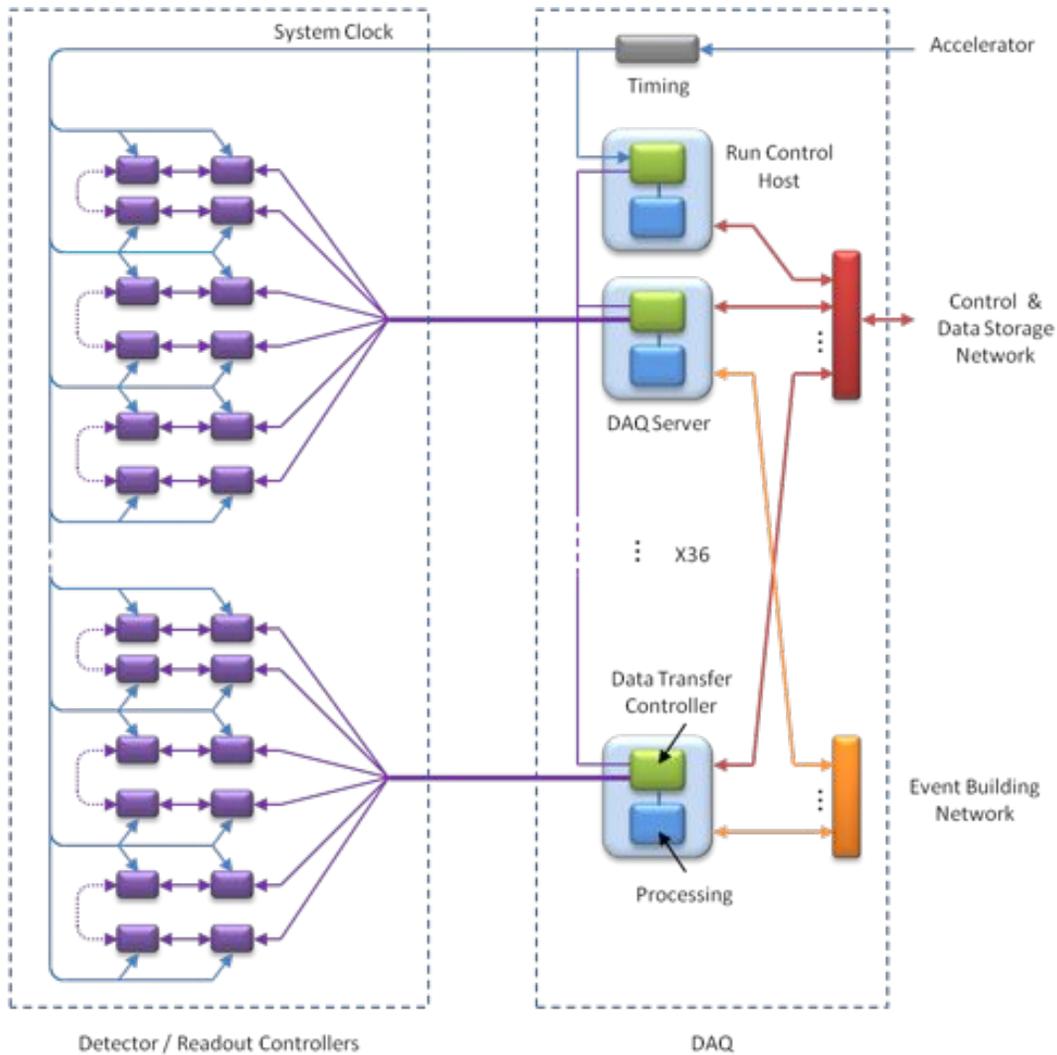

Figure 11.2. Mu2e DAQ Architecture.

### *11.2.5*  **Readout Controllers**

The number of Readout Controllers and the estimated data rate for each subdetector are listed in Table 11.4.

Readout Controllers (ROCs) are not part of the DAQ system, but rather are included separately in each detector subsystem. The ROC specifications listed here are intended to provide a common interface to the DAQ.

Readout Controllers have an FPGA for data collection, buffer management and processing. This FPGA also provides the high-speed serial transceivers (SERDES) for the optical links. Each ROC also has a microcontroller, which handles Detector Control System (DCS) "slow control" operations and is responsible for initializing the FPGA. The microcontroller is integrated into the FPGA.

Table 11.4.  Readout Controllers and Optical Links by Subdetector.

| Detector | Number of ROCs | Average Rate per ROC | Total Data Rate | Number of Optical Links | Number of DAQ Servers |
|---|---|---|---|---|---|
| Tracker | 240 | 50 MBytes/sec | 12 GBytes/sec | 120 | 20 |
| Calorimeter | 240 | 30 MBytes/sec | 8 GBytes/sec | 72 | 12 |
| Cosmic Ray Veto | 15 | 20 MBytes/sec | 1 GByte/sec | 15 | 3 |
| Extinction and Target Monitors | 2 | low | < 1 GByte/sec | 4 | 1 |

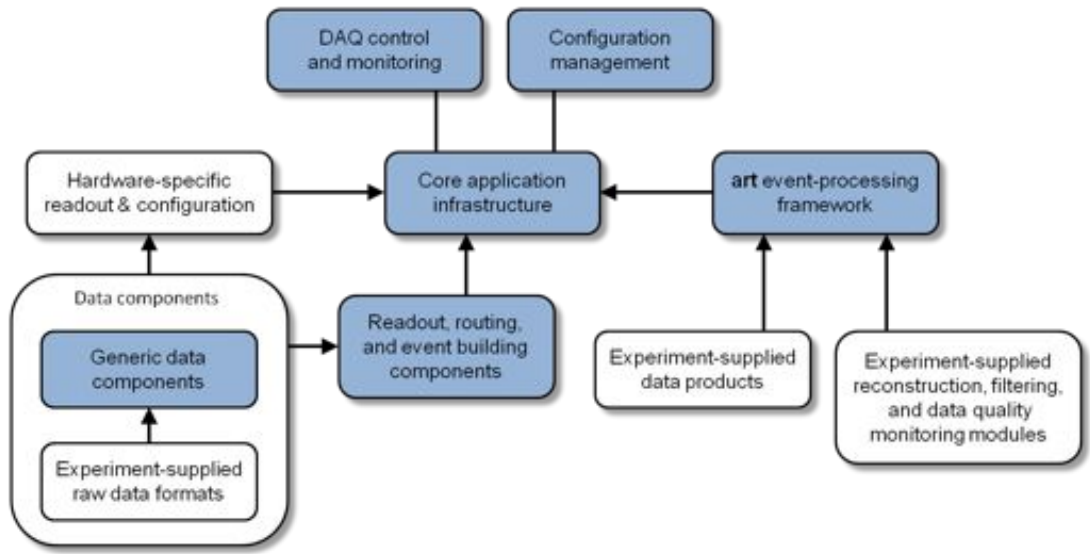

Figure 11.3. *artdaq* Architecture. Core components are shown with a blue background; experiment-supplied components are shown with a white background.

Readout Request and Data packets are handled by the FPGA. DCS packets are routed through the FPGA to the microcontroller. Because all communication is normally routed to or through the FPGA, there must be a failsafe way to reload the FPGA in the event of firmware corruption. A watchdog timer will restart the microcontroller on loss of system clock, or if any of several FPGA and microcontroller check signals are outside nominal timing windows. This will automatically reload a "golden" version of the FPGA and microcontroller firmware from dedicated SPI memory, providing a known DCS connection. DCS commands can then be used to remotely load new software/firmware into the application program memory. A DCS "run" command must be sent to the microcontroller to cause it to switch from golden mode to application program memory.

Readout Controllers in or near the detectors will be exposed to a high neutron flux [3]. SRAM based FPGAs are sensitive to radiation-induced single-event upset (SEU) in the configuration and application memory. Mu2e Readout Controllers in higher radiation areas will use Microsemi SmartFusion2 series FPGAs [4] that provide on-chip microcontroller and SERDES and a number of features to mitigate SEU, including flash based configuration and ECC protected memory and registers [5]. Commercial integrated circuits can typically tolerate a total dose of at least 100 Gy without significant degradation. In the region where the Tracker and Calorimeter ROCs are located, total dose is estimated at 5 Gy/yr. [6].

A block diagram for a digitizer/ROC is shown in Figure 11.4. It receives (and phase aligns as necessary) a system clock at the micro-bunch frequency of 590 KHz. A clock generator multiplies the system clock to drive the digitizer sample clocks (typically 50-100 MHz). The leading edge of the system clock is the time-zero reference for a micro-bunch. A local timestamp counter (driven by the sample clock) measures the time offset within the micro-bunch. The microcontroller and FPGA interface logic operates from a local oscillator, independent of the System Clock.

The ROC receives a Readout Request packet for each micro-bunch. This packet contains micro-bunch readout control information, along with the System Timestamp.

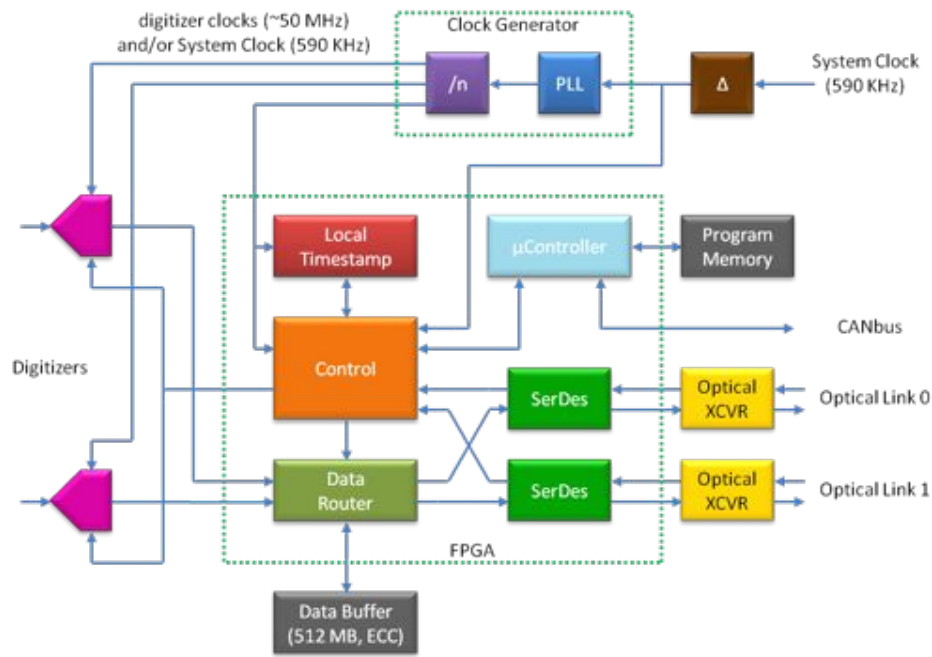

Figure 11.4. Basic Digitizer/Readout Controller Architecture

Data from the digitizers is zero-suppressed, formatted and written to the ROC Data Buffer during the beam spill. Data packets are read from the Data Buffer and transmitted on the optical link during the full accelerator Supercycle. The buffer is large enough to hold at least 1 second of ROC output data, and uses ECC memory for SEU mitigation.

### 11.2.6    DAQ Server

The central component of the Mu2e DAQ system is a commercial 3U server, which manages data collection from the Readout Controllers, Event Building, and Online Processing (Figure 11.5). There are a total of 36 DAQ servers, occupying four racks in the electronics room.

The servers used for prototype system development are Supermicro X9DRE-TF+ with dual E5-2630 processors. Servers used in the final system will be based on the most cost-effective processing available at the time of production orders. Production orders will be as late as possible to take advantage of traditional performance improvements.

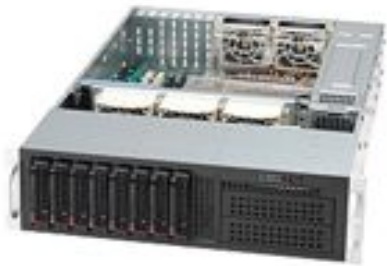

Figure 11.5. DAQ Server.

Servers must meet the following minimum requirements listed below.

- Two x8 double-wide PCIe slots for the Data Transfer Controller(s).
- ECC memory support (for 36 servers, each with 16 GBytes of memory, the system error rate would be unacceptable without error correction [7]).
- 10G Ethernet or Infiniband port for software event building.
- 1G compatible Ethernet port for connection to the general-purpose network.
- 100M Ethernet IPMI port (must be a dedicated port, shared IPMI ports are unreliable).

### 11.2.7    Data Transfer Controller

The Data Transfer Controller (DTC) collects data from multiple detector Readout Controllers. It may optionally perform event building and data pre-processing. For Mu2e, the DTC (Figure 11.6) is implemented using a commercial PCIe card (for example, [8]) located in the DAQ Server.

High-speed serial ports are provided by an adapter module (for example, [9]) that plugs into the FMC connector on the PCIe card. This adapter has eight bidirectional SFP+ ports, and can be used with optical or copper cabling. Six of the ports are used to connect to Readout Controller optical links. One port can be used to connect to the Event Building

Network to exchange data between DTCs. The last port is used to communicate with the Run Control Host computer.

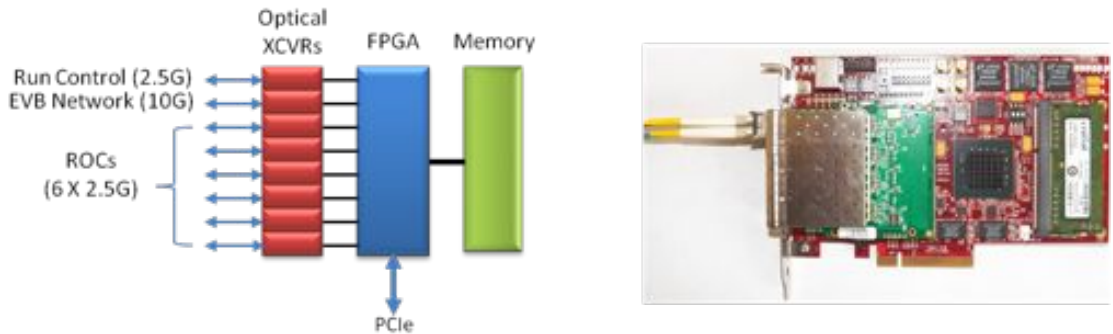

Figure 11.6. Data Transfer Controller (PCIe FPGA card and 8 port SFP+ FMC adapter).

The DTC receives Readout Request packets from the Run Control Host. These packets are forwarded on each attached ROC link. Data packets from the Readout Controllers are returned on the same links. The DTC multiplexes data from six links into one timeslice that is then transferred to the Server over PCIe, or to other DTCs via the Event Building Network.

Data packets from the ROCs are written to DTC memory. If software event building is used, the Data packets are read from DTC memory by the server processor via the PCIe interface, and then exchanged between servers to form complete events in server memory. If hardware event building is used, the Data packets are first exchanged between DTCs to form complete events in DTC memory, and then read from DTC memory by the server processor via PCIe.

DCS packets are transferred between the DTC and DAQ Server over the PCIe interface, and then over the general-purpose network to the DCS Host.

### *11.2.8*  **DTC/ROC Interface**

The detector Readout Controllers connect to the DAQ Servers via redundant optical links. Six bidirectional links are bundled in one 12-fiber MTP cable. For Readout Controllers inside the detector vacuum, the fiber is brought out through a sealed feedthrough at the end of the DS cryostat.

The boundary between the detector and DAQ is defined as the optical connector outside the detector. MTP-LC breakout cables run upstairs to the electronics room where the DAQ servers are located (Figure 11.7). Each cable contains six bidirectional data links. For the full Mu2e system, there are 36 of these cables (a total of 216 links). Each link can support a data rate of 300 MBytes/sec, for an aggregate bandwidth of approximately 60 GBytes/sec.

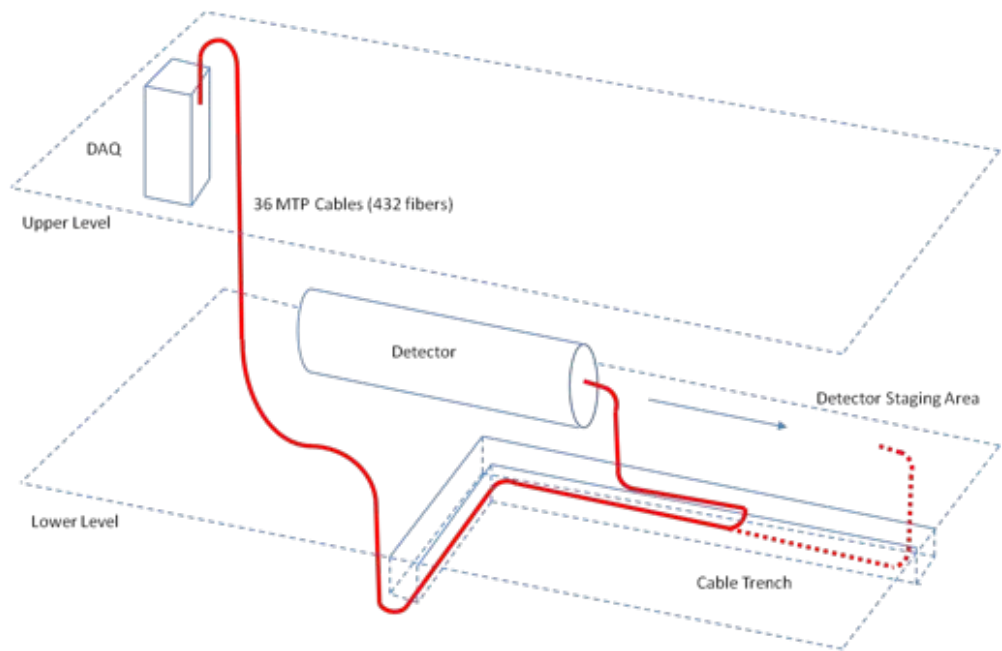

Figure 11.7. Optical Cabling from Detector to DAQ.

Twenty cables (120 links) are used for the Tracker, with four Tracker Readout Controllers sharing each pair of redundant links (Figure 11.8). Twelve cables (72 links) are used for the Calorimeter, with six or eight Calorimeter Readout Controllers sharing each pair of links. Three cables (15 links) are used for the CRV, with one CRV Readout Controllers per link. The Extinction and Target Monitors use one cable (4 links).

Optical links are used for several reasons;

- Optical fiber minimizes the vacuum feedthrough cross-section.
- The cable distance from the detector (in the staging area) to the DAQ is approximately 60 meters.
- Optical fiber provides noise immunity and electrical isolation.

The fiber is 50/125 micron multimode, OM3. Optical transceivers are class 1, multimode, and 850 nm. Coding is 8B/10B. Links operate at 2.5 Gbps (3.125 Gbps encoded). This data rate was chosen as the most cost-effective based on current FPGA and optical transceiver technology. The attenuation of optical fiber increases with radiation exposure [10]. For Mu2e, the attenuation is expected to be less than 0.01dB/m.

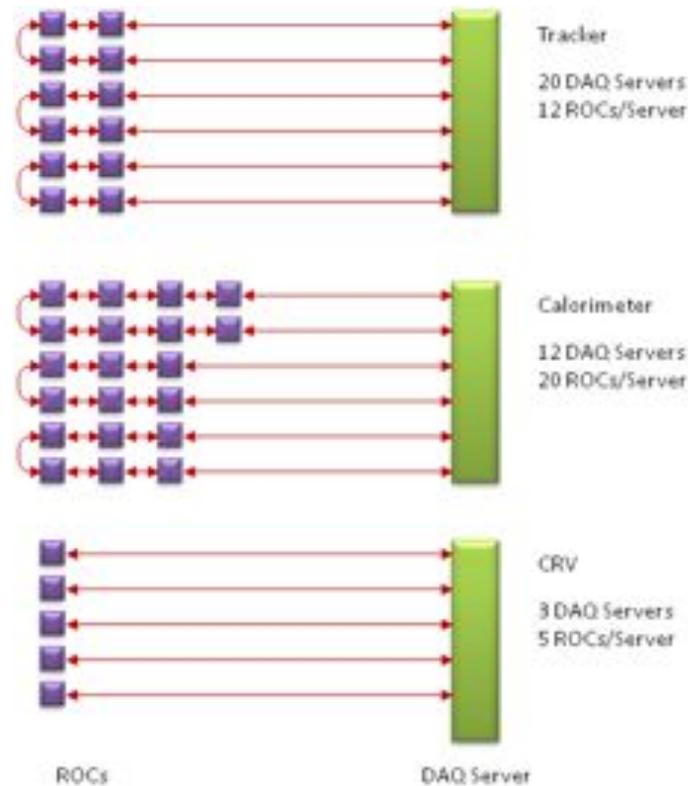

Figure 11.8. ROC Connections to DAQ Servers (DTCs).

Redundant links are used to increase system reliability (Figure 11.9), since repair would require removal of the entire detector. Each ROC has two ports and either port can be used for control/readout.  In normal operation, half of the ROCs in a redundant loop will operate on one side of the loop and half on the other. The data rate will average 100 MBytes/sec per link with two Tracker or four Calorimeter Readout Controllers. If a link fails, the ROCs attached to that link can be read from the other port.

Optical links carry both control and data packets.  The link interface is implemented in FPGA firmware. This means that the ROC FPGA must be operational in order to download new microcontroller software or FPGA firmware via the optical link DCS channel.  To prevent loss of the DCS connection, the ROC microcontroller boot program is located in protected memory. This program contains the DCS software and basic FPGA firmware to operate the link interface.  An independent watchdog circuit will restart the ROC in failsafe mode if the external system clock is interrupted, or if any of several internal watchdog signals are outside nominal timing windows.

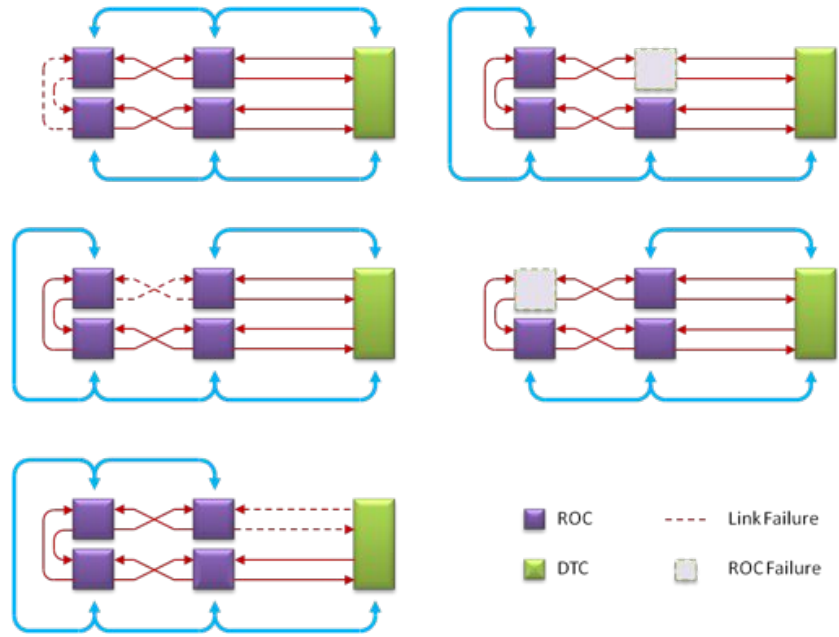

Figure 11.9. Redundant Link Configurations.

### 11.2.9 DTC/ROC Link Packet Format

All packets are 8 words X 2 Bytes in length and are delimited by 8B/10B K characters. When no packet data is available to send, the transmitter generates pairs of comma characters (K28.1/K28.5) to maintain link synchronization and bit/word alignment. There may be any number of comma words between or within packets. The receiver discards them. Each packet is preceded by a control word containing two K characters to designate the packet type (Table 11.5). These control words are used by the receiving logic to route the packet to the correct input buffer, but are not stored with the packet. A detailed description of each Packet Type from Table 11.5 is given below.

Table 11.5.  Packet Types

| Packet Type | Direction | K character (high) | K character (low) |
|---|---|---|---|
| DCS Request | DTC → ROC | K28.0 | K28.2 |
| Readout Request | DTC → ROC | K28.0 | K28.3 |
| Data Request | DTC → ROC | K28.0 | K28.4 |
| (reserved) | DTC → ROC | K28.0 | K28.6 |
| DCS Reply | ROC → DTC | K28.0 | K23.7 |
| Data Header | ROC → DTC | K28.0 | K27.7 |
| Data | ROC → DTC | K28.0 | K29.7 |
| (reserved) | ROC → DTC | K28.0 | K30.7 |

- DCS Request - the DCS Request packet is used to read or write DCS information in the ROC. For a DCS read request, the ROC internal address is provided. For a DCS write request, the ROC internal address and a data word (8 bytes) are provided.

- Readout Request - Readout Request packets are broadcast. They are processed by all ROCs on a link. Readout Request packets are sent immediately after the rising edge of the system clock. They are generated by the Run Control Host (Control Fanout module) and must reach all ROCs before the leading edge of the next system clock.  The Run Control Host sends one Readout Request packet for every system clock (1695 nsec period).

  o The Readout Request packet contains the System Timestamp, along with partition information needed to configure the ROCs for the NEXT system clock period (μBunch).  This information is used to control settings that must be synchronized to a specific micro-bunch (synchronous resets, digitization enable/disable, zero-suppression enable/disable, calibration signal injection, live gate expansion, etc.). Readout Request packets can also be used to synchronize commands to precise times within a micro-bunch, by specifying the ROC internal timestamp (time offset from the start of the μBunch). This would be used, for example, to request calibration signal injection at a particular data sample time. In this case, the partition number in the Readout Request packet indicates that it is a calibration event and another field in the packet dynamically selects the internal timestamp.

- Data Request - to transfer data from the ROCs, the DTC sends a Data Request packet to each ROC. The ROC replies with a Data Header packet and additional Data packets as needed. The Data Request packet contains the System Timestamp.

- DCS Reply - the DCS Reply packet is used to return DCS information from the ROC.  It contains the ROC internal address of the requested data and a data word (8 bytes).

- Data Header - the Data Header packet contains the System Timestamp, the partition number, a Data Packet Count (number of Data packets for this timestamp/μBunch, which can be zero) and optional status information.  If the ROC receives a Data Request packet, it must return a Data Header packet even if it has no data for that timestamp.

- Data - a Data Header packet can be followed by up to 255 Data packets. Data packets are used to transmit detector data. The last Data packet in a group includes a cumulative CRC and is padded to 16 bytes if necessary.

For multiple ROCs on a link, ROC addressing is accomplished using a 4-bit hop count (HC) field in the packet. If a packet with a HC value of 1-15 is received on either ROC port, the HC is decremented by one and the packet is retransmitted on the other port. If it is a broadcast packet type (e.g., Readout Request), the packet is processed, otherwise it is ignored. If a packet with an HC value of 0 is received on either port, the packet is processed and is NOT retransmitted on the other port. If a request packet is processed by the ROC, any corresponding reply packets are transmitted on the same port as the received request.

Reply packets (DCS Reply, Data Header, Data) do not contain a hop count, since the destination is always the DTC. If a Reply packet is received on either ROC port, it is retransmitted on the other port.

Figure 11.10 illustrates the packet flow for Data Readout and Data Transfer operations. These are independent and asynchronous. Readout activity occurs primarily during the first $3^{rd}$ of the each Supercycle while data transfer takes place over the full Supercycle. For the Tracker and Calorimeter ROCs the number of Data Requests and Readout Requests is the same since all data is transferred to the DAQ. For the CRV, Data Requests are only sent for those timestamps that pass the online filters, or are needed for calibration.

### *11.2.10* **Run Control Host**

The Run Control Host receives beam status and timing information from the Accelerator Controls network and operator commands from the remote control room.

The Control Fanout (CFO) module in the Run Control Host is responsible for generating and synchronizing Readout Requests. It sends a Readout Request control packet for each system clock. The CFO contains a set of standard Readout Request packet templates (normal readout, calibration, no operation, etc.), and a default list mapping these packets to each of the ~785,000 system clock periods in a 1.33 second Supercycle. The CFO also maintains the System Timestamp that it sends with each Readout Request packet. The Run Control Host can instruct the CFO to override the default packet on any clock or series of clocks.

The system can be partitioned at the level of a single DAQ server. Each server is given a list of other servers in the same partition for event building. In normal operation all servers are assigned to the primary DAQ partition. The Run Control Host provides global synchronization of the DAQ servers via run control state commands (Figure 11.11) for each partition.

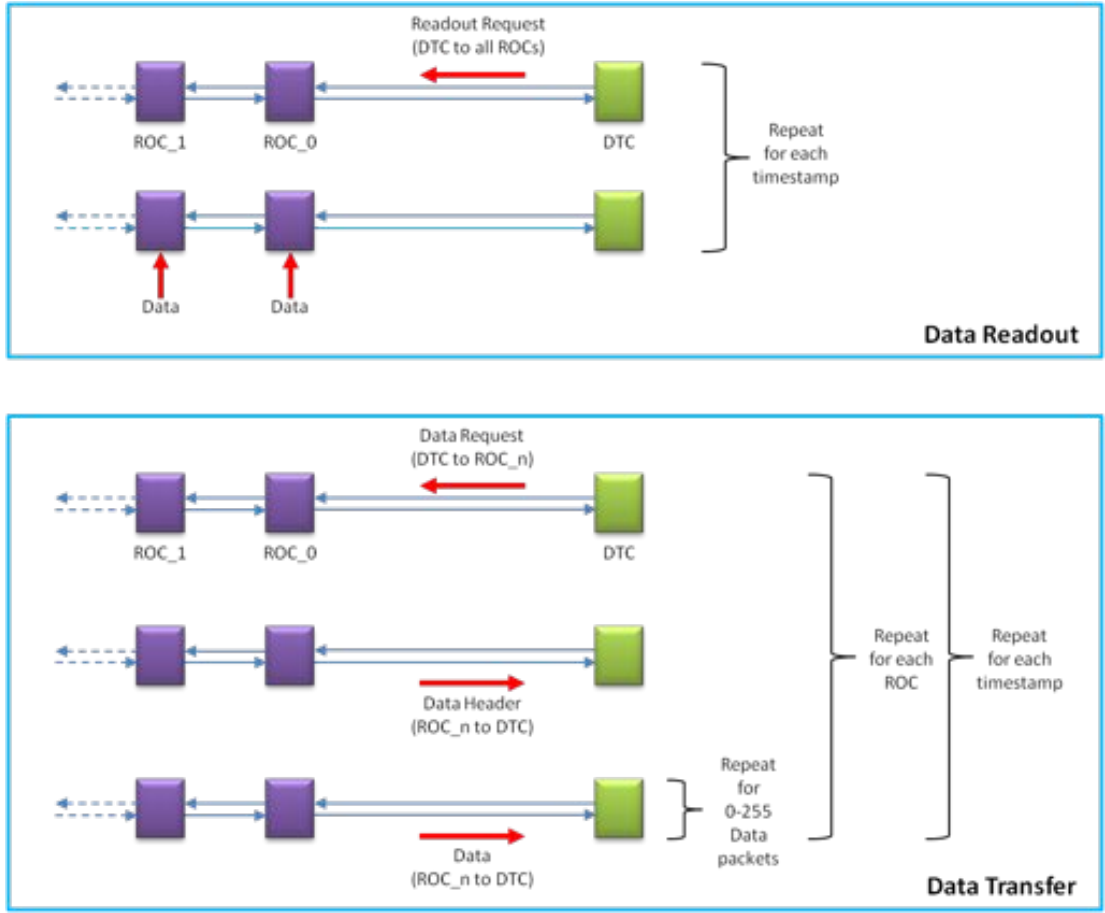

Figure 11.10. Data Readout and Data Transfer Operations.

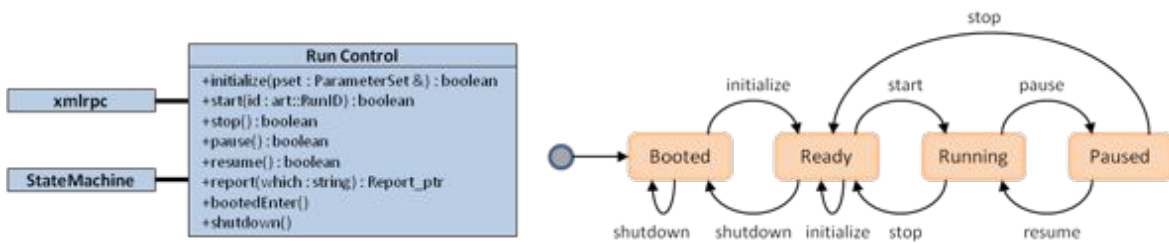

Figure 11.11.  Run Control States.

### *11.2.11*  **Run Control Host/DTC Interface**

The DTCs receive control and timing information from the Run Control Host on the DTC Control rings (Figure 11.12). These rings operate at 2.5 Gbps. The Control Fanout (CFO) card/optical link adapter in the Run Control Host is physically identical to that used for the DTCs.

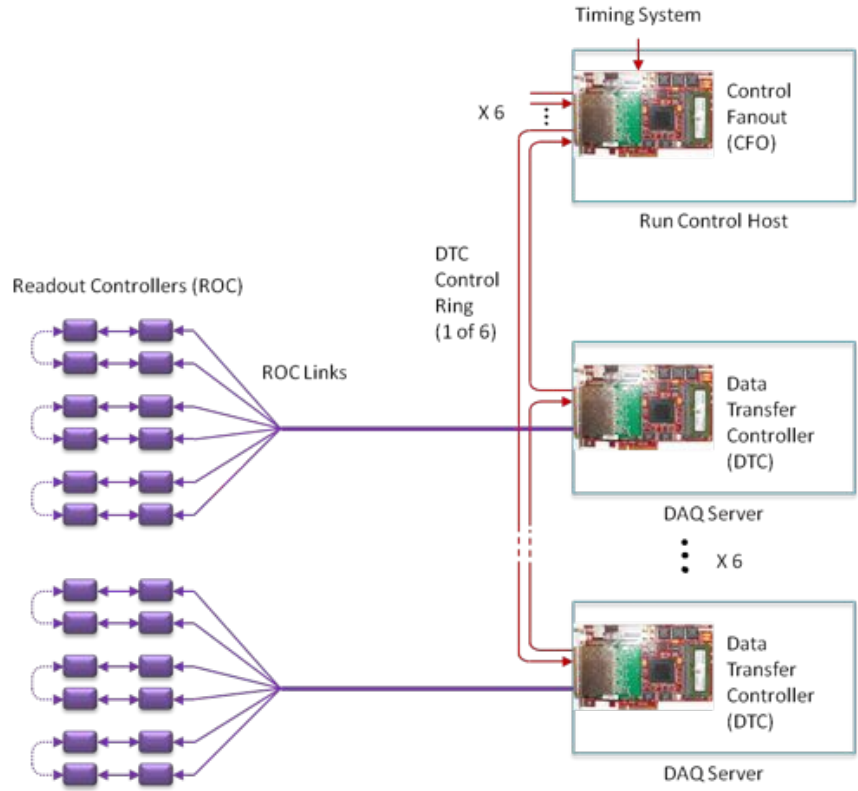

Figure 11.12. Control hierarchy.

### 11.2.12  Timing System

The DAQ supplies a continuous low-jitter system clock to the Readout Controllers and the Run Control Host (CFO module). This clock is synchronized to the accelerator Delivery Ring frequency (590,018 Hz), and the leading edge of the clock is aligned with the start of each micro-bunch.

The DAQ provides a limited number of phase aligned clock signals that are distributed inside the detector. Phase alignment at individual ROCs is accomplished by an adjustable delay at the ROC clock input. Additional alignment is provided by software calibration. Each ROC generates its own internal high-speed digitization clocks, phase locked to the system clock. Each ROC also generates an internal timestamp for data within the micro-bunch timing window. For ROCs outside the detector solenoid, the Timing system supplies a separate clock to each ROC.

Figure 11.13 shows the Timing system clock distribution. A Clock Generator on the ROC is programmed via the DCS connection to drive the digitizers at any N/M multiple of the system clock. Clock outputs can be phase aligned in 25 psec increments. The CFO uses Supercycle Start and/or beam pickup signals to determine the micro-bunch schedule.

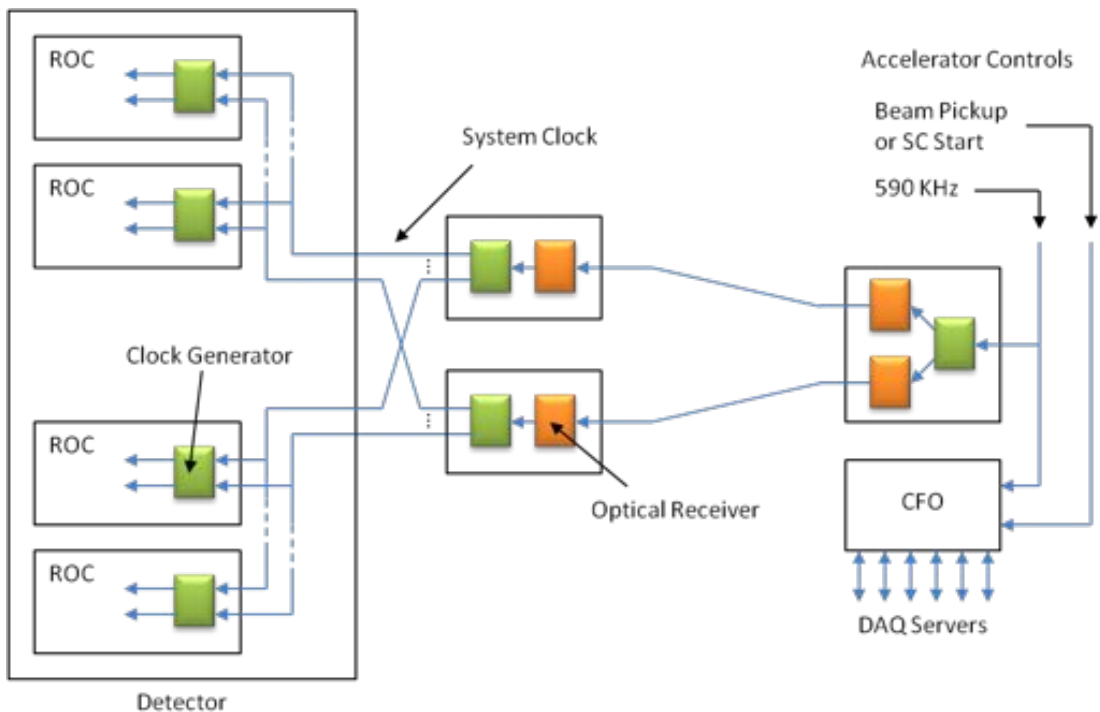

Figure 11.13. Timing System.

### *11.2.13* **Timestamps**

Each Readout Controller generates its own internal timestamp for data within a micro-bunch. This internal timestamp counter is driven by the digitization clock and is reset at the rising edge of the system clock. It may be 1 or 2 bytes, depending on the resolution of the detector. The digitization clock frequency is determined by the ROC and can be different across different detectors.

In addition to the internal timestamp generated by each ROC, there is a System Timestamp generated by the Control Fanout Module in the Run Control Host. This is a six-byte value that increments at the 590 KHz system clock rate. It has a range of 15 years. It can be stopped and restarted at any value as long as the new start value is higher than the previous stop value. The System Timestamp can be correlated with actual calendar time, or the high bytes can represent a Run Number, Supercycle, etc.

The System Timestamp is sent by the CFO to the DTCs at each system clock. The DTCs broadcast the timestamp to all attached ROCs in a Readout Request packet. The DTCs also send a System Timestamp as part of Data Request packets, and the ROCs return the System Timestamp in the Data Header packet. Sending the System Timestamp directly to the ROCs for each system clock (instead of relying on a timestamp generated from the clock itself) avoids loss of ROC event synchronization with the rest of the system as a result of missing or extra clocks.

The system clock runs continuously, whether or not there is beam. This allows readout (e.g., acquisition of calibration or pedestal data) at any time in the accelerator Supercycle. The System Timestamp is the only value used for event identification in the DAQ system. Events are NOT renumbered following various stages of filtering.

### *11.2.14* **Event Building**

Each Readout Controller collects data from a small subset of the detector. The Event Building (EVB) function combines these subsets to form a complete detector data set for analysis by an online processor. Event building is typically done in a switching network.

Figure 11.14 shows two options for placement of the EVB switching network. In the first case, the switch is located between the Readout Controllers and the Servers. The EVB input queues are managed by the Readout Controllers and the Servers manage the output queues. In the second case, the Readout Controllers send data directly to the Servers, which then manage both EVB input and output queues. The second option is used for Mu2e.

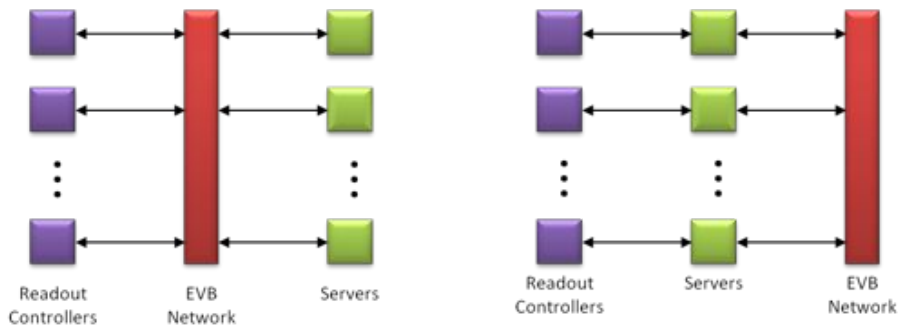

Figure 11.14. Placement of Event Building (EVB) Network.

The event building process is relatively simple. For Mu2e there are 36 servers. Thirty-six input buffers and thirty-six output buffers are allocated in the memory of each server (Figure 11.15). Variable length timeslices of data from the detector are copied to the input buffers, with timeslice 0 going to input buffer 0, timeslice 1 going to input buffer 1, etc. Packets are then transferred between servers via the EVB network. Packets from server X, input buffer Y go to server Y, output buffer X. Thirty-six packets are transferred in parallel.

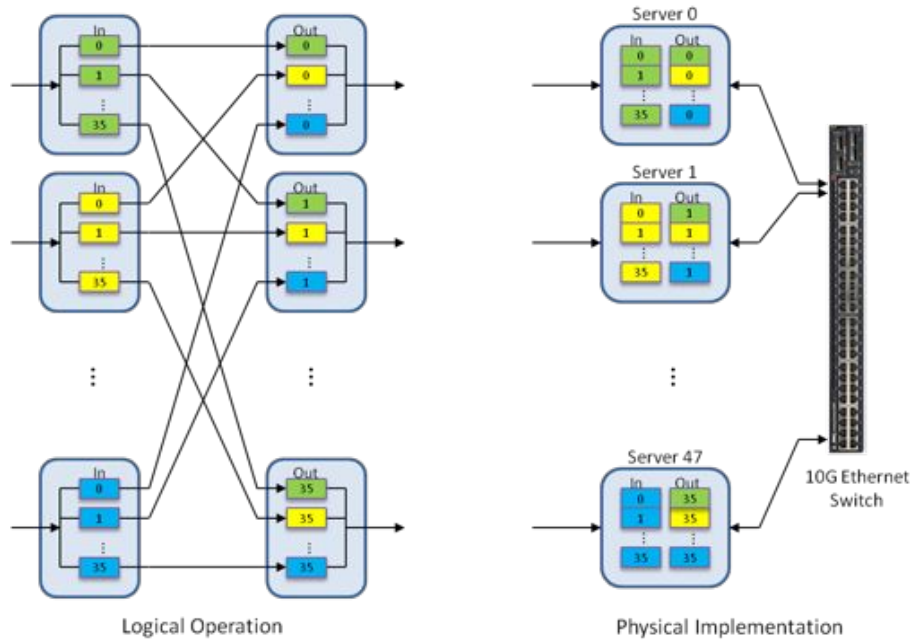

Figure 11.15. Event Building.

The event building function can be performed by software in the Server or by firmware in the DTC FPGA:

- Server option ("software" event building) - the event building network is a commercial 10Gbase-T Ethernet or Infiniband switch (Ethernet is more widely used, but Infiniband switches are less expensive with less software overhead). Event fragments are copied from the DTC to the Server over PCIe. EVB input and output buffers exist in Server memory, and the processor handles all buffer management.

  o The advantage of the Server option is that the event building software is part of the existing *artdaq* framework and is therefore easier to maintain. The disadvantage is the additional processing load for event building and management of the network interface. Standard Linux Ethernet or Infiniband drivers can be used, or the driver can be replaced with a low-level frame buffer interface to eliminate much of the overhead. Internal buffering in the switch (with standard network flow control) is used to automatically order packets.

  o The software event building option will be implemented first. If performance and scaling are satisfactory, then this is the preferred method. Tests of *artdaq* on a small, four-server system with Infiniband networking have demonstrated a throughput of approximately 900 Mbytes/sec per server.

    o In the *artdaq* EventBuilder process, the *fragment receiver* layer receives data from the Data Transfer Controller, and is responsible for sending the data to the correct *event builder* process, using standard Message Passing Interface (MPI) protocols. The *event-building* layer receives data from the fragment receivers, collating them into complete events. Complete events are then sent to another thread in the same process for *event processing*. The *event processing* layer runs the *art* event-processing framework, which performs the data filtering.

- DTC option ("hardware" event building) - the event building network is a commercial 10G SFP+ Ethernet switch. A port on each DTC card is connected to the switch via a direct-attach SFP+ copper cable. Input and output buffers are in DTC memory, and FPGA firmware handles the buffer management. Complete events are then copied from the DTC to the server over PCIe. No IP stack is necessary. The switch is programmed with a static MAC address table (one entry for each DAQ server).

    o The EVB network operates synchronously. Each server loops through its input buffers sending up to 2 KBytes per buffer per system clock. The servers are programmed to start this rotation based on their switch position (server 0 starts with input buffer 0, server 1 starts with input buffer 1, etc.). Synchronous operation provides several advantages; 1) the synchronous network is inherently non-blocking so no flow control is necessary, 2) the amount of buffering needed in the switch is minimized, and 3) it makes diagnostics easier since all transfers are deterministic.

    o The advantages of the DTC option are that it offloads the processor in the DAQ Server and allows use of the DTC FPGA for pre-processing or triggering on fully assembled events if needed. The disadvantage is that the FPGA firmware for event building is more difficult to develop and maintain than the *artdaq* event building software.

### *11.2.15*  Online Processing

All data filtering and triggering in the MU2e DAQ architecture is done in software. The production DAQ will use 36 dual-CPU servers. The online processing system must handle a total rate of 192,000 micro-Bunches per second, an average of 5400 events per second per server. Options considered for online processing include;

- Conventional (multi-core) CPU - A server based on current generation dual Intel XEON E5-2687 CPUs is rated at ~0.4 TFLOPS. We can expect at least a 2-3X improvement in performance before production orders are placed in 2017.

- XEON PHI - Peak floating point performance of the current XEON PHI co-processor is 1 TFLOPS, but overall throughput is limited by the PCIe I/O and scalar core. The next version will remove some of these limitations and has an estimated peak performance of 2-3 TFLOPS [11].

- GPU - GPU cards with peak floating point specifications of 2 TFLOPS are available today in the target cost range. They are somewhat more difficult to program. A 10X improvement on small tracking codes using GPUs vs. CPUs is possible [12]. Next generation GPUs have estimated peak performance of 4-5 TFLOPS.

- µServer CPUs based on ARM, Atom (Avoton) or Tilera cores - These are not yet under active consideration due to limited floating point performance.

Data for each micro-bunch is expected to average 120 KBytes. This should fit into cache on a multi-core CPU, but may be problematic for many-core processors and GPUs.

The offline Tracking filter has been optimized for online use, and verified to produce equivalent results [13]. The following measurements (Table 11.6) were made at various stages of optimization, for the main filter routines (99% of the execution time) running on a single core under the mu2e-art software framework. The data set consists of 1000 events with an average of 3240 hits/event. The E5-2687v2 processing times were determined using an E5-2630 processor and scaling by 0.63X to reflect the higher clock speed of the E5-2687v2.

These results show that the processing performance is already very close to the 192,000 events/sec production system requirement. Another 2X performance increase is expected in early 2015 as both processor platforms move from 22nm to 14nm technology. Additional improvements from code optimization and vectorization are also expected.

A Calorimeter filter has been proposed that provides up to 50X rejection [14]. This will be used in conjunction with the Tracking filter to generate an average trigger accept rate of less than 1%.

The option also exists to do data pre-processing in the DTC FPGAs. Each FPGA has ~800 Digital Signal Processing (DSP) cores, with several hundred GFLOPS of processing capability. A simple Calorimeter filter could be implemented in the DTCs using OpenCL. If needed, the FPGAs can perform fix-to-float conversion, sorting, unpacking, array construction and other overhead operations to reduce burden to the processors.

Table 11.6.  Tracking Filter Optimization.

|  | XEON E5-2687v2 | XEON PHI 5510P |
|---|---|---|
| **Stereo Hits** | | |
| 0) Reference code (gcc compiler) | 83.6 msec | - |
| 1) Algorithmic improvements (gcc compiler) | 4.3 msec | - |
| 2) Intel compiler, loop vectorization | **1.4 msec** | **4.8 msec** |
| | | |
| **Background Hits** | | |
| 0) Reference code (gcc compiler) | 9.0 msec | - |
| 1) Intel compiler | 5.1 msec | 123.0 msec |
| 2) Refactoring | 3.4 msec | 38.1 msec |
| 3) Double $\rightarrow$ single precision | **2.1 msec** | **23.9 msec** |
| | | |
| **Overhead** | | |
| 0) Reference code (gcc compiler) | 0.9 msec | - |
| 1) Intel compiler (estimated) | **0.3 msec** | **2.0 msec** |
| | | |
| Total processing time | 3.8 msec | 30.7 msec |
| Events/sec (single core) | 260 | 32 |
| Number of cores (36 servers) | 720 | 4,320 |
| Events/sec  (36 servers) | **187,000** | **138,000** |

## *11.2.16*  DAQ Software

As mentioned previously, the *artdaq* data acquisition toolkit will be used to build the Mu2e DAQ software system. *artdaq* provides software applications for managing the data flow as well as libraries and applications for encapsulating the data, analyzing the data, and performing other basic data acquisition functions. The core data-flow applications in *artdaq* consist of the following:

- BoardReaders that configure and read out hardware modules, and send data fragments to EventBuilders,
- EventBuilders that assemble full events and pass the events to instances of the art analysis framework for reconstruction and filtering, and

- Aggregators that organize events in time order, write them to disk, and analyze them to monitor the quality of the data.

These applications are shown in Figure 11.16 along with additional components that are part of *artdaq*. The additional components include infrastructure for sending and receiving control messages, managing the state of individual processes and the full system, logging messages to central loggers and viewers, and the sending and parsing of configuration parameters.

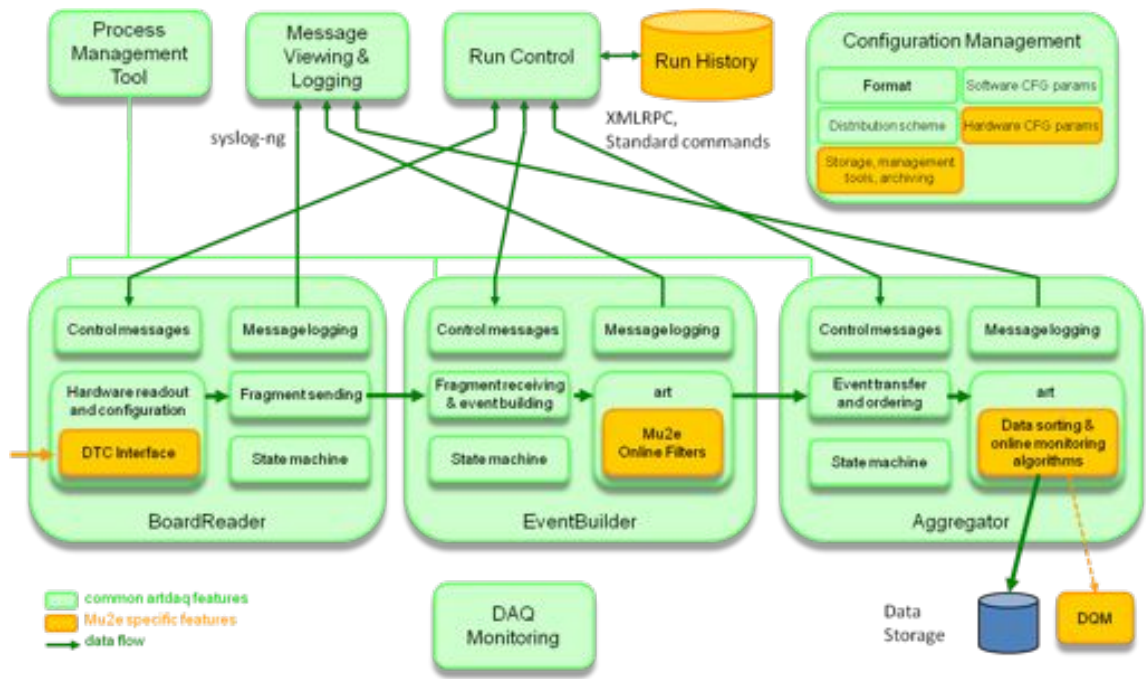

Figure 11.16. *artdaq* components. Applications and infrastructure components that are shown in green are part of the core *artdaq* toolkit. Components shown in orange are modules that experiments provide to read out their specific hardware and perform their specific analyses and monitoring.

The toolkit is designed to provide core functionality while allowing experiments to customize the hardware readout and event analysis as needed. Figure 11.16 illustrates this separation of responsibilities by showing the core applications and functions in green and experiment-provided components in orange.

In a typical *artdaq* system, one BoardReader process reads out one hardware module, so the number and location of BoardReader processes are determined by the connections from the electronics on the detector to the computing cluster. In the Mu2e experiment, this corresponds to one BoardReader per DTC. The number of EventBuilder processes will be determined by the computing needs of the reconstruction and filtering algorithms and the number of processor cores that are available on each of the computers in the reconstruction farm. The number of Aggregator processes is somewhat fixed: one for writing the accepted events to disk and one for running the online monitoring. The location and number of each of these types of processes is configurable, and a sample *artdaq* deployment for Mu2e is shown in Figure 11.17 and Figure 11.18.

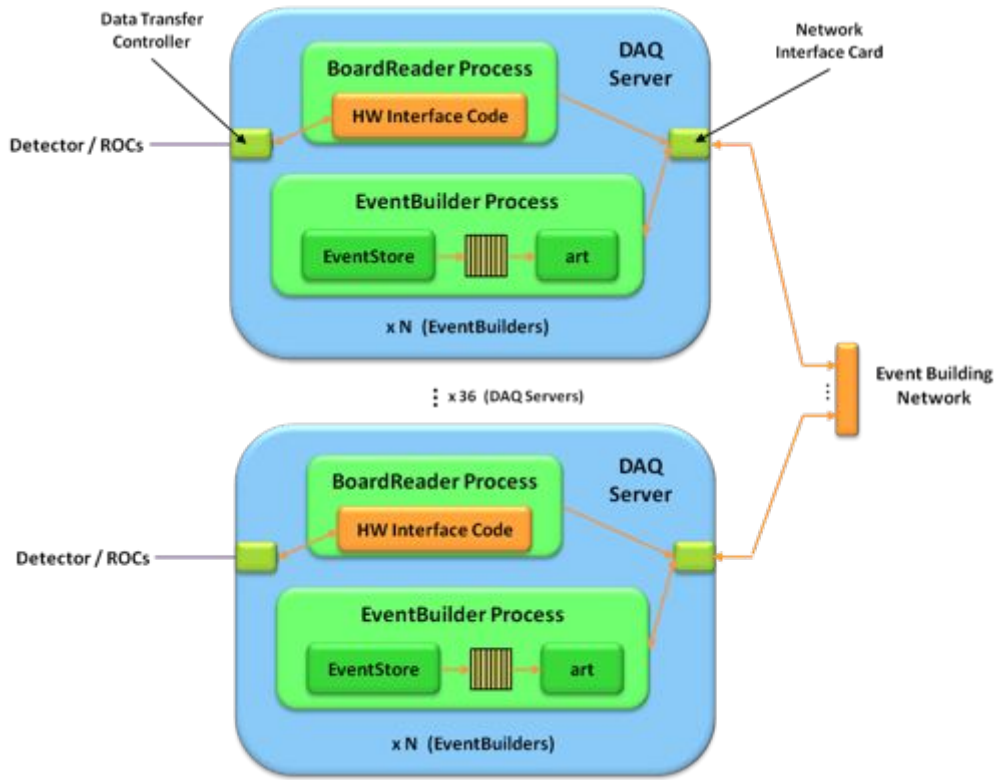

Figure 11.17. Sample deployment of *artdaq* BoardReader and EventBuilder processes on DAQ Servers. Data fragments are sent from the BoardReaders to the EventBuilders where they are assembled into full events and analyzed.

Figure 11.17 focuses on the readout and analysis of the data. It illustrates that one BoardReader and a configurable number of EventBuilder processes will be run on each of the DAQ Servers. Data fragments that are read out from each hardware module will be tagged by their time window and sent to the EventBuilders in a round-robin pattern. Fragments from a specific time window will all be sent to a single EventBuilder so that a full event can be built.

Figure 11.18 focuses on the transfer of the accepted reconstructed events to the Aggregator(s) and the subsequent writing of the data to disk and running of online monitoring. The Aggregator processes will run on a dedicated server.

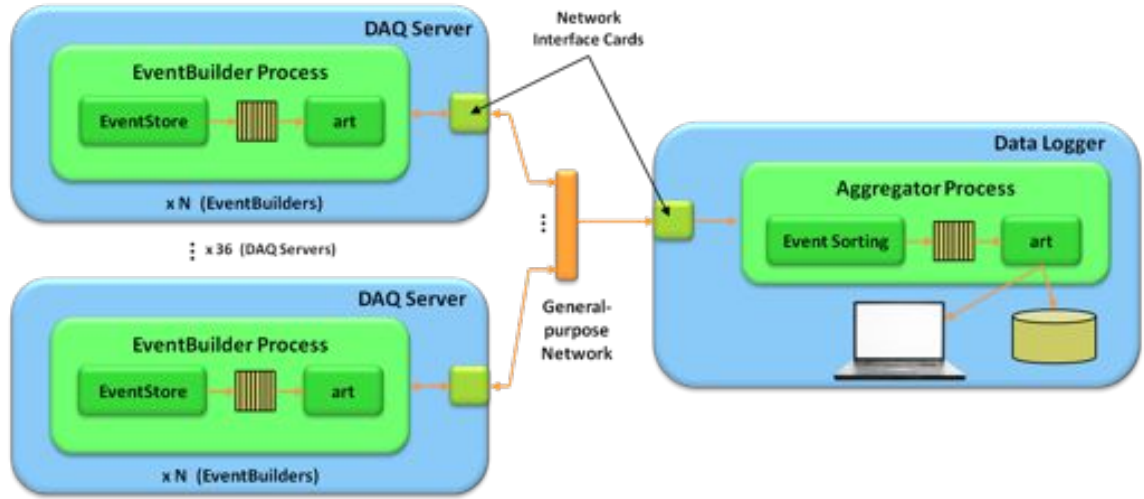

Figure 11.18. Sample deployment of *artdaq* EventBuilder and Aggregator processes. Accepted events are sent from the EventBuilders to the Aggregator(s) for storage and monitoring.

The *art* analysis framework will be used as the environment in which the online processing algorithms that are described in Section 11.2.15 are executed. *art* is a generic C++-based modular analysis framework for event data analysis. It provides the infrastructure for running software modules that are provided by experimenters and managing the data that is analyzed and produced by the analysis modules. It has been developed at Fermilab for use in current and future intensity frontier and cosmic frontier experiments, and it is currently used in the offline environments of the Mu2e, NOvA, LBNE, and other experiments. It is also currently used in the DAQ system of the DarkSide-50 experiment, which is also *artdaq*-based. The use of the same analysis framework online and offline has substantial advantages, most notably the ability for physicists to develop algorithms independently of the full DAQ system and move them to the online environment when they are ready. Within the DAQ system, EventBuilder processes handle the starting of *art* threads and the transfer of full events to *art* for analysis. It also handles the configuration of the *art* framework and the analysis modules using the configuration parameters that it receives from Run Control. The same configuration language is used to configure *artdaq* processes as is used to configure *art*.

As part of the software interface to the DTC, a Linux device driver for communicating over the server PCIe bus is being developed. The driver will be responsible for managing the buffers into which the data is written when it is received from the ROC, responding to the interrupts when DMA transfers complete, notifying the user code that data is available, and delivering the data to the user code.

### 11.2.17  Detector Control System (DCS)

The Mu2e DCS will be similar to that used in NOvA. The DAQ subproject will supply the controllers, networking and an assortment of generic I/O modules (A/D, D/A, digital). Specialized sensors and controllers (temperature, pressure, high voltage, etc.) are supplied by the other subprojects. The DCS does not provide any safety related control or monitoring.

The primary DCS connection to the Readout Controllers is a virtual channel on the optical links. DCS commands can be sent over either of the two links. A backup CANbus interface is provided in case the optical link interfaces malfunction.

Detector hall DCS electronics and power supplies containing transformers, cooling fans and Ethernet or DC-DC converter magnetics may be affected by the solenoid fringe fields. Ethernet switches and controllers should be located as far as possible from the solenoids, preferably in fields of 200 Gauss or less.  Figure 11.19 shows the calculated magnetic field throughout the Detector Enclosure area. The white area of the plot (essentially all of the detector hall) is above 100 G [15].

The higher-level DCS controllers will be commercial devices connected via Ethernet. All DCS electronics located in the magnetic field should operate without cooling fans and be remotely powered. Commercial endpoints and endpoint controllers may require some modification or shielding.

The DCS software is based on a combination of EPICS [16] and Control System Studio [17] (Figure 11.20).

### 11.2.18  Data Logger

The Data Logger collects and buffers processed data from the DAQ servers. It has enough disk capacity for several days of continuous operation without transferring data to offline storage. The disk storage may be centralized in the Data Logger, or distributed in the DAQ servers and managed by the Data Logger.

Offline storage is part of Mu2e operations.

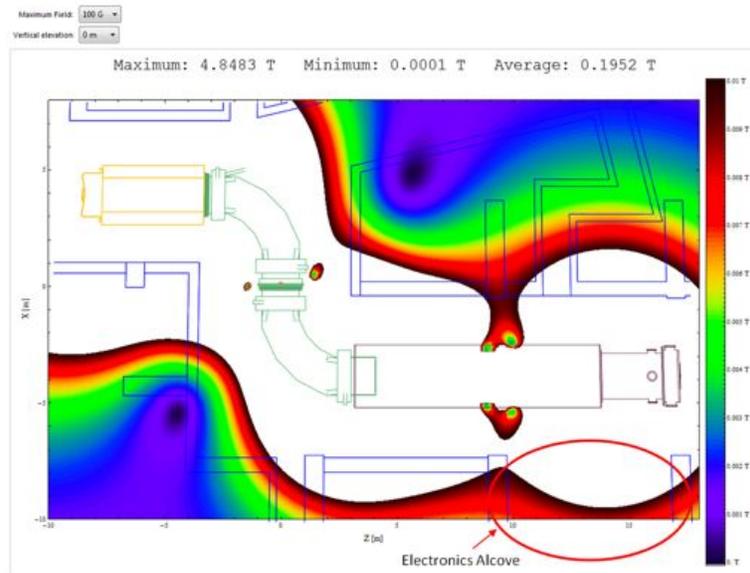

Figure 11.19. Magnetic fringe field throughout the Detector Enclosure.

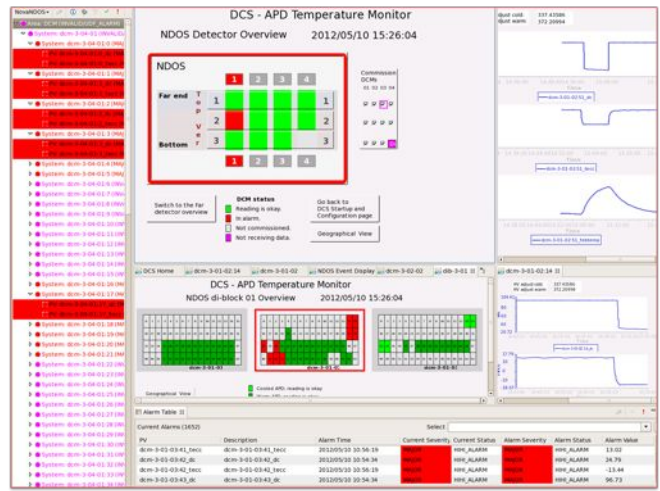

Figure 11.20. Control System Studio (example from NOvA DCS).

### *11.2.19*  **General-Purpose Networking**

The general-purpose networking (Figure 11.21) includes separate local networks for Data/Run Control and DCS/Management. The required bandwidth for output to offline storage is 3 Gbps. The management network provides connections to server IPMI ports, EVB and Data switch management ports, and PDUs.

The server IPMI ports support KVM and virtual media over LAN for remote access. Networking equipment will be monitored using Nagios and SNMP. Network configuration and security will be handled by Fermilab Network and Communication Services.

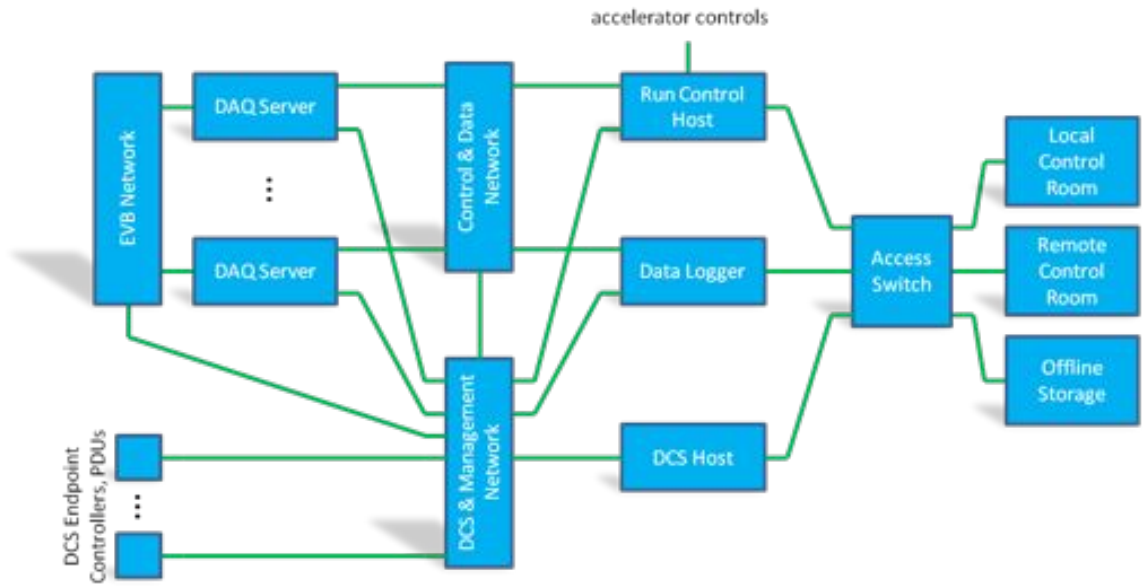

Figure 11.21. General-purpose Networking.

### *11.2.20*  **Local Electronics Room**

The DAQ system is located in the surface level electronics room, and occupies five racks (Figure 11.22). All power and data cabling is overhead.

There are two 208VAC, 20A, three-phase power feeds from the service panel to each DAQ rack. Two PDUs in each rack convert the power to six single-phase 120VAC circuits. The PDUs have network connections for remote monitoring and control. The total DAQ system power requirement is estimated at 30KW.

The DAQ subproject is responsible for electrical wiring from the panel to the DAQ racks, and for electronics room cable trays carrying data and power cables.

The fringe magnetic field in the electronics room is estimated at 20-30 Gauss. Existing steel in the racks and chassis should provide adequate magnetic shielding.  Fan speed and temperature will be monitored via IPMI/SNMP.

### *11.2.21*  **Remote Control Room**

The Mu2e remote control room will be located in the Experimental Operations Center (Figure 11.23) on the 1st floor of Wilson Hall. It will be shared with other Intensity Frontier experiments. There is room for up to six experiments, with two operator consoles each.

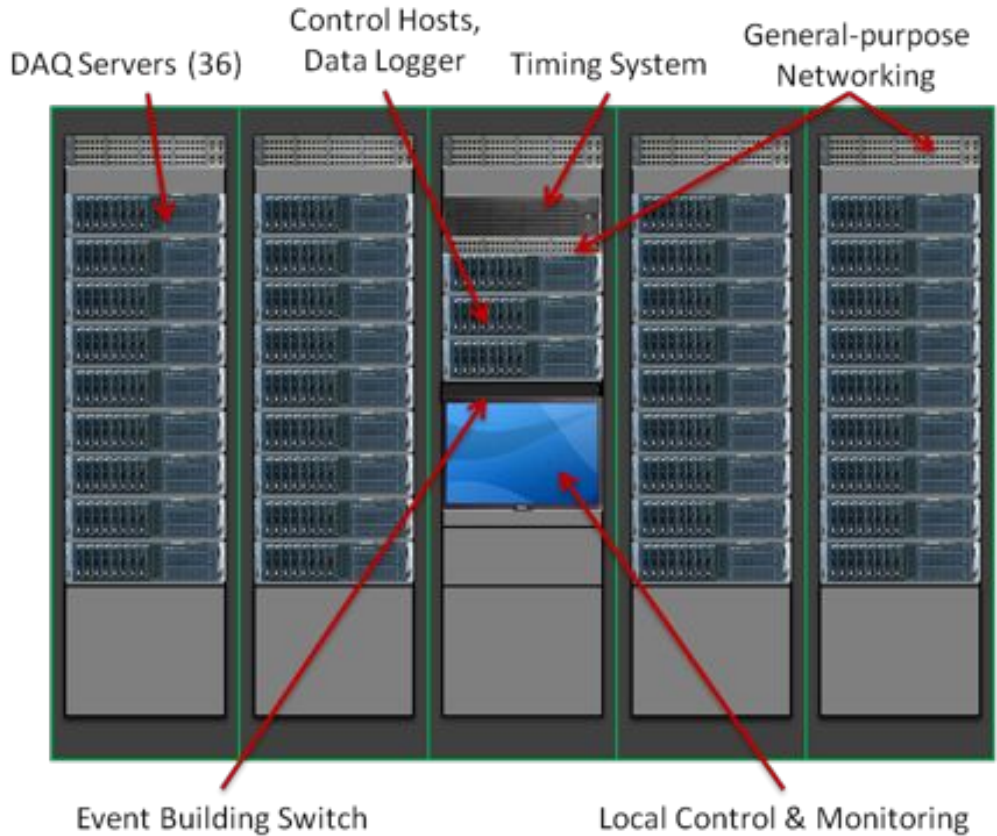

Figure 11.22. DAQ Electronics racks.

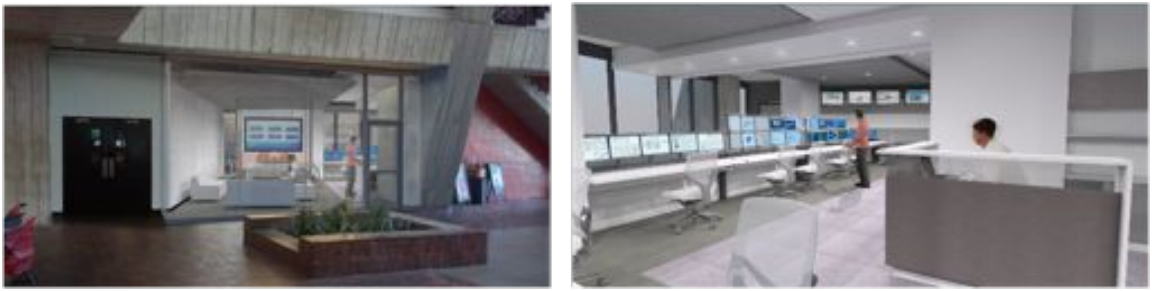

Figure 11.23. Experimental Operations Center (XOC).

Some level of remote control and monitoring is also possible via office desktop computers and VPN connections. Capability and access will be determined by the experiment. As proposed, all remote control room infrastructure other than the experiment-specific computers, will be provided by Fermilab.

A 96-fiber cable runs from the Muon Campus to Wilson Hall. The DAQ system will use two single-mode pairs. Existing fiber between Wilson Hall and GCC will be used for connection to the offline data storage system.

The Experimental Operations Center (XOC) is also the central key checkout location for access to the Mu2e detector building.

### 11.2.22 Development and Test

A goal of the DAQ system design is to reduce or eliminate the need for dedicated test stands. Most development work can take place on the user's desktop. A single DAQ server can provide a complete, self-contained 1 GByte/sec DAQ system. This should facilitate early integration with detector electronics.

There will be an emphasis on in-situ testing. All optical links will include power monitoring and bit-error rate test capability using pseudo-random pattern generators and checkers. Microcontrollers and FPGAs will include self-test and memory diagnostic routines.

For the production system, a series of FMEA (failure mode effects analysis) tests will be performed. Basic faults will be introduced in electronics, power and data connections, and a log/database kept of observed symptoms and error messages for help in diagnosing operational problems.

After the full production system (detectors, front-end electronics and DAQ) is installed in the detector hall, an extended system test will be performed with cosmic ray data. Tests using beam will occur during the initial operational phase of the experiment.

### 11.2.23 System Management

Copies of release software and firmware developed for the Mu2e DAQ system will be maintained in a Teamcenter [18] repository. Releases will be scheduled at approximately two months intervals during development. Full documentation for all custom and commercial hardware used in the system will also be maintained in Teamcenter.

Online PostgreSQL databases are used for luminosity, calibration, filter and run conditions. Offline PostgreSQL databases are used for hardware and configuration management. All online and offline databases, software and firmware repositories will include automatic backup.

### 11.2.24 DAQ System Parameters

The DAQ system is highly scalable. If necessary, it can be expanded by increasing the number of DAQ Servers. Parameters for the initial 36-server configuration are listed in Table 11.7. The average event size depends on CRV threshold settings.

The output event size includes CRV data appended to events that pass the online filter. Each DAQ Server adds approximately 1 GByte/sec of bandwidth and 1 TFLOPS of online processing capability. The incremental cost of each DAQ Server is $8K.

Table 11.7.  DAQ System Parameters.

| Parameter | Value |
|---|---|
| DAQ Servers | 36 |
| Detector Optical Links | 216 |
| System Bandwidth | 40 GBytes/sec |
| Online Processing | 40 TFLOPS |
| Input Event Size (average) | 120 KBytes |
| Input Event Rate | 192 KHz |
| Input Data Rate | $\leq$ 22 GBytes/sec |
| Rejection Factor | $\geq$ 100 |
| Output Event Size (average) | 130 KBytes |
| Output Event Rate | $\leq$ 2000 Hz |
| Output Data Rate | $\leq$ 260 MBytes/sec |
| Offline Storage | ~7 PByte/year |

# 11.3  Performance

Each DAQ server is essentially an independent 1 GByte/sec DAQ system. Thirty-six servers are tied together through the event building network to form the complete Mu2e DAQ system.  The system performance scales linearly.

All DAQ data interfaces have been tested and verified to operate at the required bandwidth (Figure 11.24). Timing system characterization is underway (Figure 11.25). Optical transceivers have been tested in loopback mode and in a high magnetic field. CANbus communication for the backup DCS connection has been tested.

Firmware is currently under development to link components together for end-to-end prototype system tests (Figure 11.26).  The LVDS interface between the Tracker digitizer and Readout Controller has been tested. The Readout Controller emulator and digitizer emulator will soon be replaced by actual prototype modules.

Each front-end system has sufficient internal buffer for approximately 1 second of digitized data.  For test and calibration, this provides a convenient method of exercising the system at full bandwidth.   A large sample data set (~100,000 events) can be downloaded to the front-end buffers using the DCS, then continuously replayed and verified.

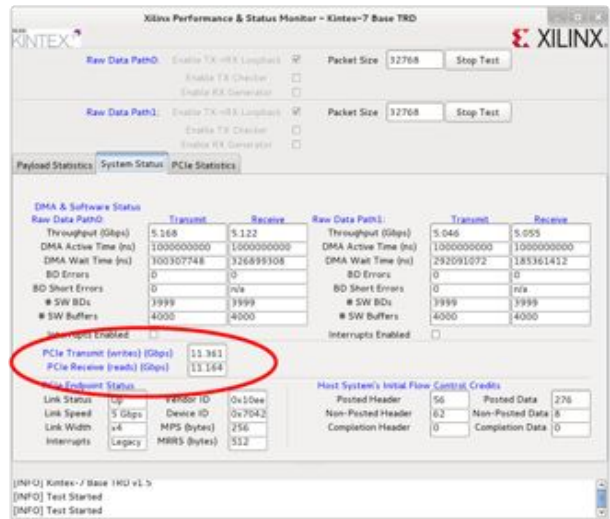

Figure 11.24. PCIe (DTC) Bandwidth Tests.

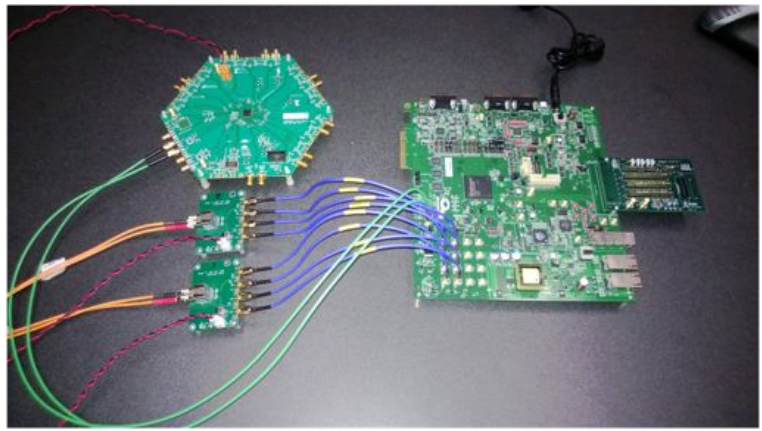

Figure 11.25. Prototype ROC & Timing System Tests.

# 11.4  Risks

A moderate risk associated with the DAQ is the difficulty in estimating resource requirements for software and firmware development. This could result in increased labor costs, but should not impact schedule. Additional scientific labor would be employed as a first step in mitigation.

There are lower-level risks associated with estimating system bandwidth and processing requirements. The system is easily expanded, so the worst-case impact would be the cost of additional DAQ Servers or reduction in initial duty-cycle.  There are numerous options for mitigation, including adjustments to data thresholds, additional software optimization, FPGA pre-processing, and triggering.

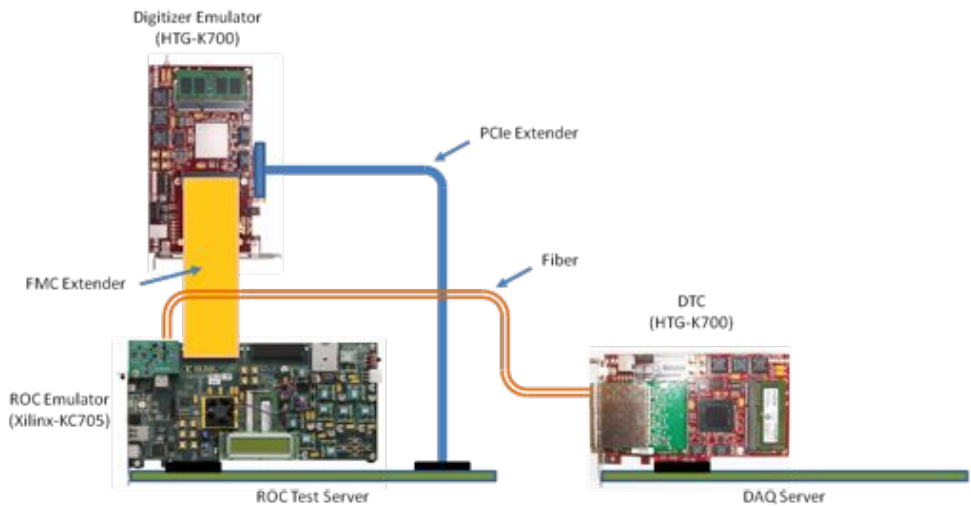

Figure 11.26. Prototype "End-to-End" Data Transfer Test.

## 11.5 Quality Assurance

The DAQ system incorporates a number of features and procedures to ensure correct operation.

- Prior to delivery, DAQ Servers undergo initial hardware test at the vendor.
- Following delivery, DTC cards are installed and the DAQ Servers undergo 100 hours of burn-in and loopback test.
- All control and data packets are time-stamped and include CRCs.
- Data links include built-in bit-error rate test logic and power monitors.
- FPGAs include self-test logic.
- Embedded software/firmware is monitored by watchdog timers.
- DAQ Servers are monitored using IPMI.
- Networking equipment is monitored using SNMP.
- PDUs include network connections for power monitoring.

## 11.6 Installation and Commissioning

Prototype and pilot DAQ system development and testing will take place on the third floor of the Feynman Computing Center. Since there is not sufficient electrical power at this location to test the full production DAQ, development will move to the Mu2e detector building as soon as beneficial occupancy is available and production system orders are placed.

This page intentionally left blank

# 12 Satisfying the Physics Requirements

The Mu2e experiment is aiming for sensitivity to the $\mu^- N \rightarrow e^- N$ conversion process that is several orders of magnitude better than anything that has come before it. This ambitious goal leads to a set of technical requirements, described in a number of Requirements Documents and a list of physics requirements, summarized in Section 3.8. The preliminary designs for the accelerator, the Mu2e solenoid systems, and the Mu2e detector sub-systems have been developed to satisfy these requirements and have been discussed in detail in Chapters 4 - 11. A description of how each Physics Requirement is satisfied by the current design appears below.

*To suppress prompt backgrounds from beam electrons, muon decay-in-flight, pion decay-in-flight and radiative pion capture requires a pulsed beam where the ratio of beam between pulses to the beam contained in a pulse is less than $10^{-10}$. This ratio is defined as the beam extinction. The spacing between beam pulses should be about twice the lifetime of muonic aluminum (>864 ns) and the beam pulse should not be wider than 250 ns.*

- The accelerator scheme discussed in Chapter 4 will deliver protons to the Mu2e production target in pulses that are approximately 230 ns wide and spaced 1695 ns apart (center-to-center). The technique for generating the required bunch structure in the Recycler Ring will naturally suppress protons between bunches. The number of out-of-time protons is further suppressed by employing a fast AC dipole to sweep clean the inter-pulse beam. The AC dipole will reside in the beam line that brings the protons from the Debuncher to the Mu2e production target (cf. Section 4.9). Simulations estimate that this design achieves an extinction of better than $10^{-11}$ while transmitting 99.7% of the in-time beam.

*In order to suppress backgrounds from decays of muons in atomic orbit in the stopping target, the reconstructed width of the conversion electron energy peak, including energy loss and resolution effects, should be on the order of 1 MeV FWHM or better with no significant high energy tails.*

- The Tracker design discussed in Chapter 8 consists of 40 planes of straw tubes. Each plane consists of multiple straw layers in order to improve efficiency and to resolve left-right ambiguities in the drift distance. Every effort is made to minimize the amount of material in the fiducial volume of the Tracker. The straws will be kept thin (e.g. the total thickness of the straw walls is 15 μm) and the gas manifolds, front-end electronics, and associated cooling are placed at large radii, outside the active fiducial volume. The intrinsic high-side resolution (ignoring the effect of the target and all other material upstream of the tracker), which is the most important for distinguishing conversion electrons from backgrounds, has a core component sigma of 118 KeV/c, and a 2% high-side exponential tail. The net resolution is significantly less than the estimated resolution due to energy loss in the upstream material.

- Using a full tracker simulation that includes occupancy effects from background events, energy loss, straggling, detector resolution and a prototype pattern recognition and track fitting algorithm an overall resolution function of about 1 MeV FWHM is achieved (Figure 3.17).

*To suppress backgrounds from beam electrons, the field in the upstream section of the Detector Solenoid must be graded so that the field decreases toward the downstream end. This graded field also serves to increase the acceptance for conversion electrons.*

- As discussed in Chapter 6, the field in the Detector Solenoid starts at 2.0 Tesla just after the transition from the Transport Solenoid. The field then decreases linearly down to 1.0 Tesla over roughly five meters. This graded field largely eliminates background from beam electrons that scatter in the last collimator or stopping target by pitching them forward, thus reducing their transverse momentum so that they fall out of the geometric acceptance of the Tracker. The stopping target is approximately centered in this graded region, which nearly doubles the geometric acceptance for electrons from the $\mu^- N \rightarrow e^- N$ conversion process.

*Suppression of backgrounds from cosmic rays requires a veto surrounding the detector. The cosmic ray veto should be nearly hermetic on the top and sides in the region of the collimator at the entrance to the Detector Solenoid, the muon stopping target, tracker, and calorimeter. The overall efficiency of the cosmic ray veto should be 0.9999 or better.*

- Simulations show that cosmic ray induced background events are initiated along the tops and sides of the Detector Solenoid and also along the tops and sides of the Transport Solenoid straight section that interfaces with the Detector Solenoid. The veto system discussed in Chapter 10 provides coverage along the entire length of the Detector Solenoid and the adjacent half of the Transport Solenoid for the tops and sides. The design consists of four layers of scintillator and will employ a 3-out-of-4 veto logic. Full hit-level simulations of this design with accidental overlays result in less that 0.1 cosmic ray induced background events (Table 3.4).

*Suppression of long transit time backgrounds places requirements on the magnetic field in the straight sections of the Transport Solenoid. The field gradient in the three straight sections of the Transport Solenoid must be continuously negative and the gradient must be relatively uniform.*

- The conceptual design for the Transport Solenoid focused on satisfying these complex field requirements. A 3D model of the Transport Solenoid field was implemented using OPERA. The field in the straight sections is always negative and the gradient is within the specifications (cf. Sec 6.3.2 and Figs. 6.47-6.50).

*The ability to separate muons and pions from electrons with high reliability and high efficiency is required to eliminate backgrounds from ~105 MeV/c muons and pions.*

- To keep the total background from cosmic rays at a level below 0.1 events, a muon rejection of 200 is required (Section 10.2). A combination of timing and dE/dx information from the tracker combined with energy and timing information from the calorimeter results in the required rejection power, with the calorimeter providing the bulk of the discrimination (Figure 9.7).

*To mitigate backgrounds induced from antiproton annihilation, thin windows in appropriate places along the muon beam line are required to absorb antiprotons.*

- To reduce backgrounds from antiprotons, two thin absorbers are installed in the Transport Solenoid. The absorbers are kept thin as possible to maximize transmission of muons to the stopping target. A full hit level simulation indicates that this design maintains the antiproton-induced background below 0.05 events (see Section 3.6.4).

*The capacity to take data outside of the search window time interval must exist.*

- The DAQ has been designed to handle the higher rates that are present at earlier times in the spill cycle. This capacity is required to determine backgrounds using data, in particular, electrons from radiative pion capture.

*The capacity to collect calibration electrons from $\pi^+ \rightarrow e^+ \nu$ is required.*

- The Muon Beamline, discussed in Chapter 7, is optimized to efficiently transport negatively charged low momentum muons from the production target to the stopping target. It can be made to efficiently transport positively charged particles by rotating the middle collimator. The design discussed in 7.3.2 has a rotating central collimator for the purpose of selecting and transporting positively charged pions to the stopping target. Stopped $\pi^+$ can decay to $\pi^+ \to e^+\nu$ providing a mono-energetic source of $e^+$ that can be used to calibrate the absolute momentum scale of the spectrometer. The Tracker is designed in a charge symmetric way (cf. Chapter 8) and the reconstruction efficiency and resulting momentum resolution for $e^-$ and $e^+$ are expected to be comparable. The Trigger and DAQ system can be configured to trigger on high energy $e^-$ or $e^+$ or both simultaneously (cf. Chapter 11).

*The capacity to measure the beam extinction to a level of $10^{-10}$ with a precision of about 10% over about a one hour time span must exist.*

- Extinction at the level of $10^{-10}$ is critical to the success of Mu2e. It is important to monitor the extinction level to know that this requirement is being achieved. Pulse-to-pulse measurements of extinction at this level are neither feasible nor required. A statistical measurement is adequate so long as the time required to accumulate adequate statistics is relatively short to avoid long periods of operation with poor beam extinction. The design for the Mu2e Extinction Monitor satisfies these requirements (cf. Section 4.10.2).

*The capacity to determine the number of ordinary muon captures with a precision of order 10% must exist.*

- The number of muon captures is the normalization for the muon-to-electron conversion measurement. The stopping target monitor for Mu2e will measure the spectrum of delayed gammas from muon capture in an aluminum target using a germanium detector (cf. Section 7.6).

*The muon beam line is required to have high efficiency for the transport of low energy muons (~0.002 stopped negative muons per 8 GeV proton on target). To mitigate backgrounds from muon and pion decay-in-flight, it must suppress transport of high-energy muons and pions. It must also greatly suppress the transport of high energy electrons.*

- Simulations of the Mu2e apparatus using the QGSP-BERT model of particle production in a GEANT4 simulation results in ~0.0019 stopped muons per incident proton (Section 3.4).

- The requirement to reduce the transmission of high-energy pions, muons and electrons is driven by the need to reduce backgrounds from pion decay-in-flight, muon decay-in-flight and beam electrons. The TS collimators and the graded field in the upstream section of the PS are designed to accomplish this. The estimated backgrounds from pion decay-in-flight, muon decay-in-flight and beam electrons are shown in Table 3.4. Combined they contribute less than 2% of the total estimated background for the experiment.

*The muon beam line should avoid a direct line-of-sight path of neutral particles (mainly photons and neutrons) from the production target to the muon stopping target.*

- One of the roles of the S-shaped Transport Solenoid is to eliminate line-of-sight neutral particles from reaching the muon stopping target.

*The detectors must be able to perform in a high-rate, high-radiation environment.*

- The Tracker, calorimeter and Cosmic Ray Veto (CRV) have all been designed to operate in the Mu2e environment. Prototype straw tubes have been tested for ageing at the expected nominal dose rate without effect. Calorimeter crystals have been selected based on their known radiation hardness characteristics. The shielding between the solenoids and the Cosmic Ray Veto has been optimized to limit the rates in the CRV. All electronics components will be tested for radiation hardness before a final selection is made. This includes photosensors, digitizers and FPGAs. Much of this radiation testing remains to be done, but the anticipated radiation is at level where other rad-hard components have successfully operated.

*The muon beam line should be evacuated.*

- The warm bore of the solenoids, from the PS through the DS, will be evacuated by the Muon Beamline Vacuum System. The PS and TSu will be evacuated to approximately $10^{-5}$ Torr and the TSd and DS will be evacuated to $10^{-4}$ Torr (cf. Section 7.2).

The Mu2e preliminary design described in this Technical Design Report satisfies the full set of physics requirements and currently yields a single event sensitivity of $2.87 \times 10^{-17}$ using our current set of algorithms and a full hit-level simulation that includes accidental activity in the detectors from all known background processes. This represents an improvement of about 4 orders of magnitude over the current world's best limit on the $\mu^- \, N \rightarrow e^- \, N$ conversion process and provides discovery sensitivity over a broad range of new physics models including Supersymmetry, Little Higgs, Extra Dimensions, generic 2HDM, and Fourth Generation models.